\documentclass[dissertation,physics,nochapterblankpages]{puthesis}

\usepackage{multicol}
\usepackage{float}

\title{Two- and Three-Particle Jet-Like Correlations}

\author{Jason Glyndwr Ulery}{Ulery, Jason Glyndwr}

\pudegree{Doctor of Philosophy}{Ph.D.}{December}{2007}

\majorprof{Fuqiang Wang}

\campus{West Lafayette}

\newcommand{\be}{\begin{equation}}
\newcommand{\ee}{\end{equation}}

\newcount{\myi}
\newcommand{\Repeat}[2]{%
    \myi=0
    \loop
        \ifnum\myi<#2
        #1
        \advance\myi by 1
    \repeat
}

\begin{document}

\volume

%
%
%
%
%




\begin{acknowledgments}
I would first and foremost like to thank my advisor Fuqiang Wang.  He has provided me with encouragement and support throughout my research and this thesis would not have been possible in its present form without his help.  I would like to thank Olga Barannikova, Alex Cardenas, and Levente Molnar for their help with computing and programing.  I would like to thank the other members of the heavy-ion group at Purdue: Andrew Hirsch, Norbert Porile, Rolf Scharenberg, Brijish Srivastava, and Blair Stringfellow, for the advice they have given me.  I would like to thank the members of my committee: Fuqiang Wang, Andrew Hirsch, Denes Molnar, and Matthew Jones.  I would also like to thank all of the graduate students I shared the office with: Alex Cardenas, Timothy Herston, Levente Molnar, Terence Tarnowsky, Michael Skoby, Aoqi Feng, Quan Wang, and Joshua Konzer for their friendship.  I would like to thank my family for their support and my grandparents for paying for my undergraduate education.

\end{acknowledgments}





\tableofcontents

\listoftables

\listoffigures





\begin{abstract}
  We present results of 2-particle jet-like correlations, with high $p_T$ $h^{\pm}$ triggers and identified $\pi^{\pm}$, $p$, and $\bar{p}$ triggers in d+Au and Au+Au collisions and 3-particle jet-like azimuthal correlations in $pp$, d+Au, and Au+Au collisions at $\sqrt{s_{NN}}$=200 GeV.  These results use data from the STAR TPC during RHIC runs II, III, and IV.

Modifications in 2-particle correlations are observed in Au+Au collisions.  These modifications are not seen in $pp$ or d+Au collisions.  This demonstrates that the modifications are due to final state nuclear effects.

High $p_T$ protons, anti-protons and charged pions are identified by the relativistic rise of dE/dx in the STAR TPC. Correlations of charged hadrons with high $p_T$ $p$, $\bar{p}$, and $\pi^{\pm}$ show no discernible difference.  The results post challenges to recombination and coalescence models which are otherwise very successful in explaining the large baryon/meson ratio and the splitting of the elliptic flow at intermediate $p_T$.  

In central Au+Au collisions, the away-side 2-particle correlation is significantly broadened and even double humped in selective kinematic ranges.  Three-particle correlations were employed to identify the underlying physics mechanism(s).  Results in $pp$, d+Au and peripheral Au+Au collisions show dijet structure with away-side $k_T$ broadening.  Results in mid-central and central Au+Au collisions are consistent with a near-side jet and on the away-side a combination of conical emission and large angle gluon radiation and deflected jets.  The associated $p_T$ independent emission angle suggests Mach-cone shock waves being the underlying physics mechanism for the conical emission.  The emission angle is measured to be $1.39\pm0.01$ (stat.) $\pm0.04$ (sys.) in ZDC triggered 0-12\% Au+Au data.
\end{abstract}

\chapter{INTRODUCTION}

The search for the basic building blocks of matter has been around for a long time\cite{atom}.  The Greek philosopher Leucippus (490-? B.C.) insisted that if one were to divide matter one would eventually reach pieces so small they could not be divide further.  One of his students, Democritus (460-370 B.C.), called such fragments atomos which means unbreakable.  Democritus considered all matter to exist of atomos and if there was space in between the atomos it contained nothing.  The first recorded experiments to test this were done by Robert Boyle in 1662\cite{boyle}.  He studied the relationship between pressure and volume.  In 1913, Jean Perrin calculated the size of atom from Einstein's equations for Brownian motion\cite{perrin}.  By 1955, atoms could actually be seen\cite{muller}.  In 1898, Thomson suggested atoms may have a net positive charge with electrons embedded into it.  In 1904, Nagaoka suggested that the atom may have a small positive center with electrons orbiting it.  In 1911 Rutherford suggested an even smaller and more massive center to the atom with the electrons orbiting.  Rutherford called this center a nucleus and was able to measure its size.  In 1932, Chadwick discovered a neutral particle that was massive like a proton was required to explain radiation observations.  The particle was named neutron after the neutral particle that was being looked for to explain the charge and spin of the nitrogen-14 nucleus.  Later results from particle accelerators suggested there was more structure.  In 1964, Gell-Mann and Zweig published a papers theorizing that there would be three fundamental particles, which Gell-Mann called quarks, along with three anti-particles that make up all of the know hadrons and anti-hadrons\cite{gellmann,zweig1,zweig2}.  Gell-Mann called these quarks up, down, and strange.  Starting in 1964 several people, including Greenberg, Nambu, and Han, assigned a new distinction to explain how three of the same type of quark could make up a hadron without violating the Pauli exclusion principal.  They called this distinction color.  A new theory was worked out to explain the interactions of the then six quarks and six antiquarks with three colors called quantum chromodynamics (QCD).  This theory called for an exchange particle for the strong interaction between quarks, Gell-Mann called this particle a gluon because it is the glue that hold the quarks together.  It has been theorized that at high enough temperatures the quarks and gluons will no longer be confined in hadrons.  This deconfined phase of quark matter has been given the name quark-gluon plasma\cite{wong}.  To search for the quark-gluon plasma continues with particle accelerators creating high energy and high density conditions in the laboratory.  

\section{Quantum Chromodynamics and Quark Gluon Plasma}

Quantum chromodynamics or QCD is a theoretical model introduced to explain the known hadrons in terms of elementary constituents, quarks\cite{eisb}.  In 1954 Yang and Mills developed a gauge theory for the strong interaction.  This was an SU(2) theory that tried to make a local symmetry out of global isospin invariance.  The Yang-Mills theory is non-Abelian and requires a charged force carrier.  Since charged force carriers had not been detected this theory was not well accepted at the time.  In 1963, it was suggested by Gell-Man that there were fundamental particles call quarks.  Mesons were expected to be bound states of quark-antiquark pairs while baryons were expected to consist of three quarks.  All of the charges and quantum numbers of all of the known hadrons at that time could be explained by three quarks which they called up, down and strange.  There were still problems with the QCD.  One problem was that free particles with fractional electric charge could not be found.  Another problem was that the certain baryons, such as the $\Delta^{++}$ which is composed of three up quarks in the same spin state, required the wavefunction of three quarks to be symmetric under interchange of quark flavor and spin; however the quarks must have spin 1/2 and obey Fermi-Dirac statistics, which requires the total wavefunction to be antisymmetric.  To resolve this Han, Nambu, Greenburg and Gell-Mann proposed that quarks have an additional quantum number which they called color.  The problematic baryon wave functions were then made to be antisymmetric in the color quantum numbers.  Currently QCD includes six quarks.  The charm, bottom, and top quarks were introduced to explain additional hadrons later discovered.  The six quarks are listed with their charge and mass in table~\ref{tab:qmass}.  The quarks make up an SU(3) group of flavor.  The theory for the strong interaction in QCD is a Yang-Mills theory based on the SU(3) of color.  The symmetry of this field is a color symmetry if every red quark becomes yellow, every yellow quark blue and every blue quark red then all hadrons are still colorless.  The quanta of this SU(3) gauge field are called gluons.

\begin{table}[hbtp]
\centering
\caption{Charge and current mass of the quarks\cite{pdg}.}
\begin{tabular}{|l|l|l|l|} 
\hline
Quark&Symbol&Charge (e)&Current Mass ($MeV/c^2$)\\
\hline
Up&{\it u}&$+\frac{2}{3}$&$1.5-3.0$\\
Down&{\it d}&$-\frac{1}{3}$&$3-7$\\
Charm&{\it c}&$+\frac{2}{3}$&$1,250\pm90$\\
Strange&{\it s}&$-\frac{1}{3}$&$95\pm25$\\
Top&{\it t}&$+\frac{2}{3}$&$174,200\pm3,300$\\
Bottom&{\it b}&$-\frac{1}{3}$&$4,200\pm70 (\overline{MS})$\\
 & & &$4,700\pm70 (1S)$\\
\hline
\end{tabular}

\label{tab:qmass}
\end{table}

The QCD Lagrangian has the form\cite{wong},
\begin{eqnarray}
\mathcal{L}_{QCD} &=& \sum_{f} i\bar{\psi}_{f,k}\gamma^{\mu}D_{\mu}\psi_{f}-m_{f}\bar{\psi}_{f,k}\psi_{f,k}-\frac{1}{4}F_{\mu\nu}^{i}F^{\mu\nu,i}\\
D_{\mu}\psi &=& (\partial_{\mu}-igA_{\mu}^{\alpha}T_{\alpha})\psi \\
F_{\mu\nu} &=& \partial_{\mu}A_{\mu}-\partial_{\nu}A_{\nu}-ig[A_{\mu},A_{\nu}].
\end{eqnarray}
where $f$ is the flavor index (labels up, down, strange, charm, top, and bottom), $k$ is the color index, $D_{\mu}$ is the covariant derivative, $F_{\mu\nu}$ is the field strength tensor, $A_{\nu}$ and $A_{\mu}$ are the gauge field operators, $m_{f}$ is the quark mass, and $g$ is the coupling constant. The theory is similar to quantum electrodynamics (QED) but with additional complications from the additional quantum number of color.  We have the 6 quarks and their 6 respective anti-quarks as our fermions, each with one color.  The boson force carriers are a color octet of gluons each with a pair of colors, 
\begin{equation}
r\bar{b}, r\bar{g}, b\bar{r}, b\bar{g}, g\bar{r}, g\bar{b}, \frac{1}{\sqrt{2}}(r\bar{r}-b\bar{b}), \frac{1}{\sqrt{6}}(r\bar{r}+b\bar{b}-2g\bar{g}).
\end{equation}
There is another possible combination (since we have 3 colors and 3 anticolors there are 9 combinations).  This combination, $\frac{1}{\sqrt{3}}(r\bar{r}+b\bar{b}+g\bar{g})$, is color neutral and excluded by construction.  The hadrons are color neutral; therefore, they must contain combinations of quarks that are color neutral.  The two lowest energy states of color neutral combination of quarks are {\it qqq} where each quark is a different color (baryons) and $q\bar{q}$ where the antiquark is the anticolor of the quark (mesons)\cite{perkins}.   QCD theory is much more complicated to calculate than QED because the force carriers in QCD, gluons, are charged (carrying color charge) while the QED force carriers, photons, are not.  This results in interactions between the QCD force carries that have no analog in QED.

\subsection{Hadron Bag Model}

There have been no experimental observations of a free quark.  This leads to the concept that at large distance scale quarks are confined in hadrons.  A bag model provides a phenomenological description of quarks inside hadrons~\cite{bag}.  In the MIT bag model, quarks are massless when inside a bag of finite dimensions and are infinitely massive when outside the bag.  If the quarks are confined in the bag than the gluons also are confined.  The total color charge inside the bag must be zero due to Gauss's Law therefore our baryon ($qqq$) and meson ($q\bar{q}$) states are the lowest order states allowed.  The energy of a system of $N$ quarks in a bag of radius $R$ and pressure $B$ is given by\cite{wong},
\begin{equation}
E=\frac{2.04N}{R}+\frac{4\pi}{3}R^{3}B.
\end{equation}
The equilibrium radius of the bag can be found using the condition $dE/dR=0$ to be,
\begin{equation}
R=\left(\frac{2.04N}{4\pi}\right)^{\frac{1}{4}}B^{-\frac{1}{4}}.
\end{equation}

Our bag can be interpreted to give rise to an internal bag pressure, $B$ (for a baryon of radius 0.8 fm $B^{1/4}=204$ MeV).  The wavefunctions of the quarks inside the bag balance this internal pressure.  If the pressure of the quarks is increased to the point it is greater than the internal pressure of the bag then the quark cannot be confined.  A new phase of matter is then possible.  This form of matter has deconfined quarks and gluons and is called quark gluon plasma, (QGP).  A large quark pressure in our bag can be created by increasing the temperature and by increasing the baryon number density\cite{wong}.

\subsection{Quark Gluon Plasma}

\begin{figure}[htbp]
\centering
\includegraphics[width=0.6\textwidth]{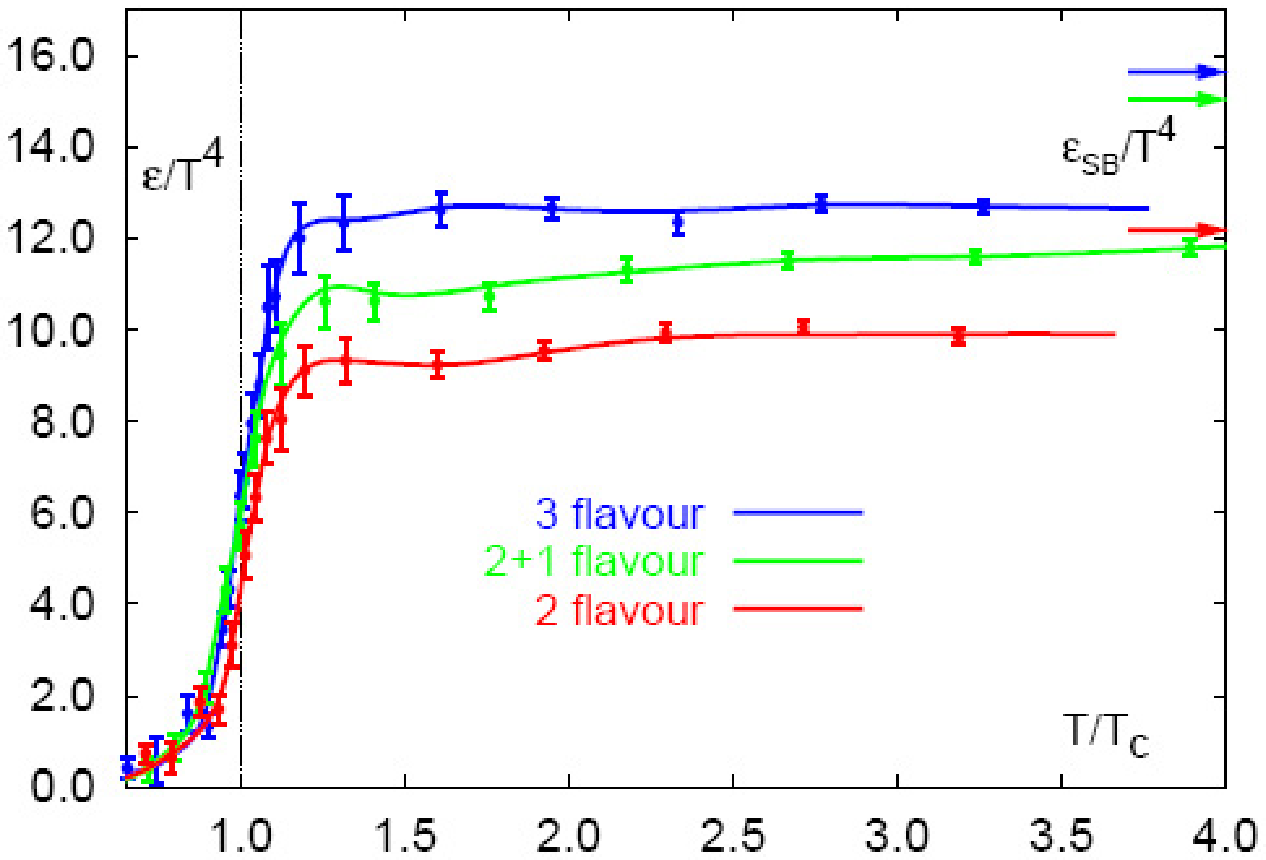}
\caption{Lattice QCD results for the energy density over temperature to the fourth power as a function of temperature over the critical temperature.  The y-axis $\epsilon/T^{4}$ is proportional to the effective number of degrees of freedom.  The sharp rise in the number of degrees of freedom signals a phase transition.  Plot is from \cite{lattice}.}
\label{fig:lattice}
\end{figure}

Using the hadron bag model we found that at high enough temperature and density quarks and gluons can be deconfined to create a quark gluon plasma.  QGP is a property of QCD at small momentum scales, at which the calculation is notorious. One method of caluclation is lattice QCD, which can give numerical nonperturbative results for QCD and the QGP.  This is done though by discretising space-time coordinates on a lattice.  Figure~\ref{fig:lattice} shows the energy density over temperature to the fourth power plotted as a function of the temperature.  The energy density over temperature to the fourth power is proportional to the effective number of degrees of freedom.  At the critical temperature, $T_c$, there is a sharp rise in the effective number of degrees of freedom indicative of a phase transition.  This phase transition is a transition from hadronic matter to QGP.  The different curves represent different numbers of flavors of quarks used in the calculation.  Two flavor uses two flavors of light quarks, 2+1 flavor uses two flavors of light quarks and one flavor of heavy quark (with a mass closer to the strange quark mass), and 3 flavor uses three flavors of light quarks.  The temperature of the phase transition from these calculations is $173\pm8$ MeV for two quark flavors and $154\pm8$ MeV for three quark flavors.  The 2+1 flavor temperature is close to the 2 flavor\cite{lattice}.  

\begin{figure}[htb]
\centering
\includegraphics[width=0.6\textwidth]{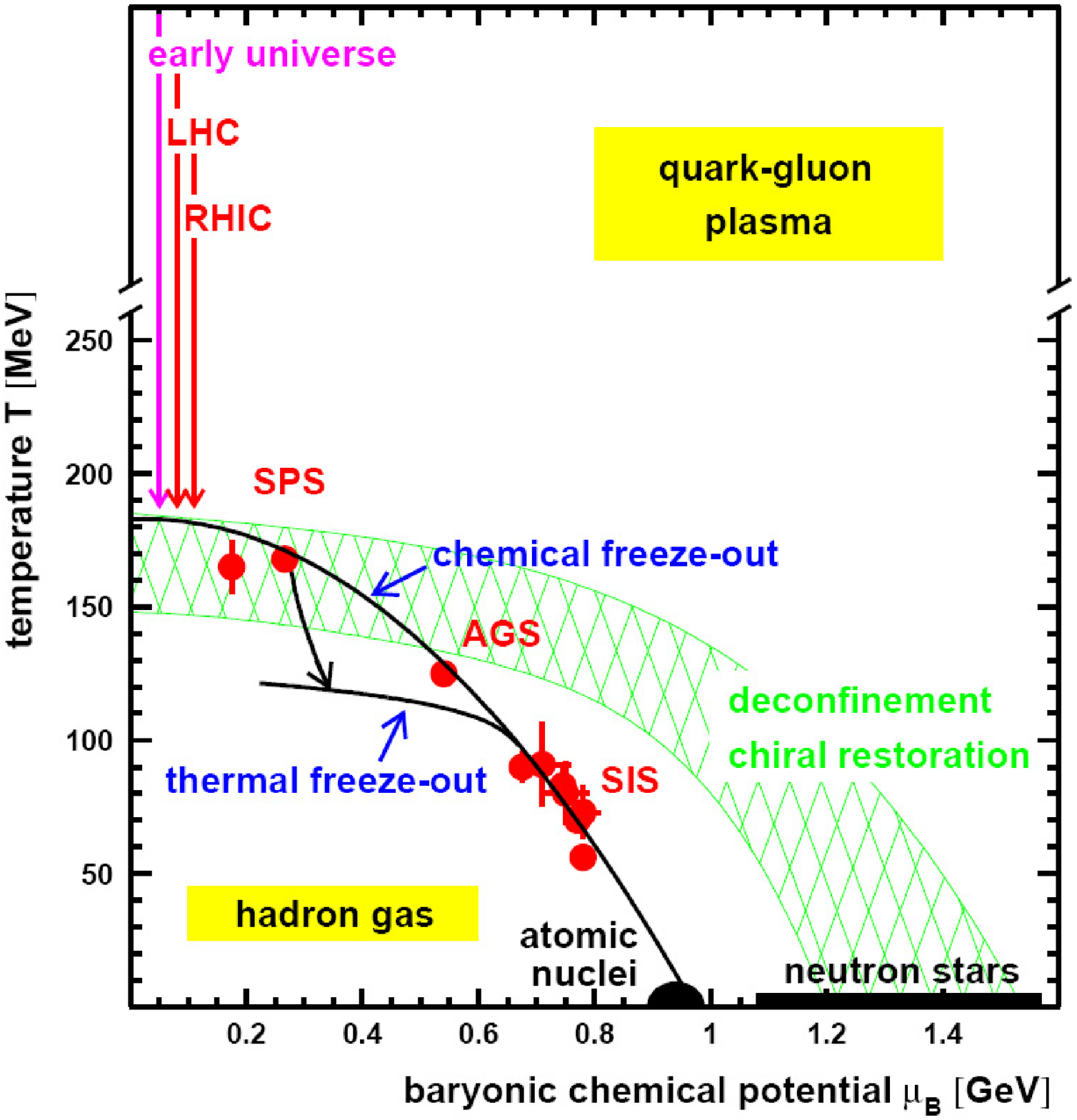}
\caption{Expected phase diagram of matter based on experimental data and the thermal model.  The hashed region represent lattice QCD predictions for the phase transition between hadronic and partonic matter.  Plot is from \cite{phase}.}
\label{fig:phase}
\end{figure}

The phase diagram for hadronic matter and partonic matter (matter of quarks and gluons) is shown in figure~\ref{fig:phase}.  The green hashed area shows the lattice QCD predictions for the phase transition.  At high baryon chemical potential (the change in internal energy with the change in the number of baryons) and low temperature a deconfined state of quarks and gluons is predicted to exist in neutron stars.  The region currently used experimentally to find a deconfined state of matter is in the low chemical potential and high temperature region as shown by markers on the plot.

\subsection{Hard Scattering and Jets}

Jets are cones of hadrons produced from a quark or gluon in a relativistic collision.  Jets are the result of 2-body parton-parton hard scattering~\cite{peskin}.  A parton is either a quark or a gluon.  Two-body interactions in QCD can be with any combination of quarks and gluons leading to the interactions $qq \rightarrow qq$, $qq' \rightarrow qq'$, $q\bar{q} \rightarrow q\bar{q}$, $q\bar{q} \rightarrow q'\bar{q}'$, $q\bar{q} \rightarrow gg$, $gg \rightarrow q\bar{q}$, $qg \rightarrow qg$, and $gg \rightarrow gg$ where $q$ is any quark or antiquark, $\bar{q}$ is its corresponding antiquark if it is a quark or its corresponding quark is it is an antiquark, $q'$ is a different quark or antiquark from $q$, and $g$ is a gluon.  The cross sections for these processes are~\cite{peskin},
\begin{eqnarray}
\frac{d\sigma}{d\hat{t}}(qq \rightarrow qq) &=&
\frac{4\pi\alpha_{s}^{2}}{9\hat{s}^2}\left[\frac{\hat{u}^{2}+\hat{s}^{2}}{\hat{t}^{2}}+\frac{\hat{t}^{2}+\hat{s}^{2}}{\hat{u}^{2}}-\frac{2\hat{s}^{2}}{3\hat{u}\hat{t}}\right] \nonumber \\
\frac{d\sigma}{d\hat{t}}(qq' \rightarrow qq') &=&
\frac{4\pi\alpha_{s}^{2}}{9\hat{s}^{2}}\left[\frac{\hat{s}^{2}+\hat{u}^{2}}{\hat{t}^{2}}\right]\nonumber \\
\frac{d\sigma}{d\hat{t}}(q\bar{q} \rightarrow q\bar{q}) &=&
\frac{4\pi\alpha_{s}^{2}}{9\hat{s}^{2}}\left[\frac{\hat{s}^{2}+\hat{u}^{2}}{\hat{t}^{2}}+\frac{\hat{t}^{2}+\hat{u}^{2}}{\hat{s}^{2}}-\frac{2\hat{u}^{2}}{3\hat{s}\hat{t}}\right] \nonumber \\
\frac{d\sigma}{d\hat{t}}(q\bar{q} \rightarrow q'\bar{q}') &=&
\frac{4\pi\alpha_{s}^{2}}{9\hat{s}^{s}}\left[\frac{\hat{t}^{2}+\hat{u}^{2}}{\hat{s}^{2}}\right]\nonumber \\
\frac{d\sigma}{d\hat{t}}(q\bar{q} \rightarrow gg) &=& \frac{32\pi\alpha_{s}^{2}}{27\hat{s}^{2}}\left[\frac{\hat{u}}{\hat{t}}+\frac{\hat{t}}{\hat{u}}-\frac{9}{4}\left(\frac{\hat{t}^{2}+\hat{u}^{2}}{\hat{s}^{2}}\right)\right]\nonumber \\
\frac{d\sigma}{d\hat{t}}(gg \rightarrow q\bar{q}) &=& \frac{\pi\alpha_{s}^{2}}{6\hat{s}^{2}}\left[\frac{\hat{u}}{\hat{t}}+\frac{\hat{t}}{\hat{u}}-\frac{9}{4}\left(\frac{\hat{t}^{2}+\hat{u}^{2}}{\hat{s}^{2}}\right)\right]\nonumber \\
\frac{d\sigma}{d\hat{t}}(qg \rightarrow qg) &=& \frac{4\pi\alpha_{s}^{2}}{9\hat{s}^{2}}\left[-\frac{\hat{u}}{\hat{s}}-\frac{\hat{s}}{\hat{u}}+\frac{9}{4}\left(\frac{\hat{s}^{2}+\hat{u}^{2}}{\hat{t}^{2}}\right)\right]\nonumber \\
\frac{d\sigma}{d\hat{t}}(gg \rightarrow gg) &=&
\frac{9\pi\alpha_{s}^{2}}{2\hat{s}^{2}}\left[3-\frac{\hat{t}\hat{u}}{\hat{s}^{2}}-\frac{\hat{s}\hat{u}}{\hat{t}^{2}}-\frac{\hat{s}\hat{t}}{\hat{u}^{2}}\right]
\end{eqnarray}
where $\hat{s}=x_{1}x_{2}E^{2}_{cm}$, $\hat{t}=-Q^{2}$, and $\hat{u}=\sum{m_{i}^{2}}-\hat{s}-\hat{t}$ are the partonic Mandelstam variables and $\alpha_{s}$ is the QCD coupling constant.  Here $x_i$ is the fraction of the energy carried by the incoming parton, $i$, $E_{cm}$ is the center of mass energy of the collision, $-Q^{2}$ is the square of the momentum transfer (between and incoming and outgoing parton), the $m_{i}'s$ are the parton masses (which are generally approximated as 0 for light quarks) and the sum is over the incoming and outgoing parton masses.  Experimentally one cannot distinguish between different scattering processes.  For this reason the sum of all processes is used.  We can then use the parton distribution functions, $G(x,Q^{2})$, to obtain the jets cross section,
\begin{equation}
\sigma_{A+B \rightarrow c+X} = \sum_{abd}\int{}dx_{a}dx_{b}d\hat{t}G_{A}(x_{a},Q^{2})G_{B}(x_{b},Q^{2})\frac{d\sigma_{a+b \rightarrow c+d}}{d\hat{t}}.
\end{equation}
where A and B are the incoming hadrons.  This is the cross section for partonic jets.  The parton distribution functions give the distributions of a given parton as a function of $x$ and $Q^2$.  In experiment we can only measure hadrons.  Fragmentation functions are used as a model of how the parton fragments into hadrons.
\begin{equation}  
\sigma_{A+B \rightarrow C+Y}=\sum_{abcd}\int{}dx_{a}dx_{b} dz d\hat{t} D_{c \rightarrow C}(z) G_{A}(x_{a},Q^{2})G_{B}(x_{b},Q^{2})\frac{d\sigma_{a+b \rightarrow c+d}}{d\hat{t}}.
\end{equation}
where $D_{c->C}(z)$ is the fragmentation function, {\it z} is the fraction of the parton {\it c}'s energy that is carried by the hadron {\it C}.  Since the calculation is done here only to leading order a factor {\it k} is multiplied by our cross section to roughly account for higher order terms.  The factor {\it k} can be obtained through comparison with data.

\subsection{Strongly Coupled QGP}
The earliest predictions of QGP were of a plasma with weak coupling.  There has been experimental evidence at RHIC for a strongly interacting QGP (sQGP).  This includes evidence for strong collectivity at low $p_{T}$ through agreement with hydrodynamic models and large energy loss of jets at high transverse momentum (jet quenching).

\subsection{Hydrodynamics and Statistical Models}

\begin{figure}[htb]
\hfill
\begin{minipage}{.4\textwidth}
\centering
\includegraphics[width=0.9\textwidth]{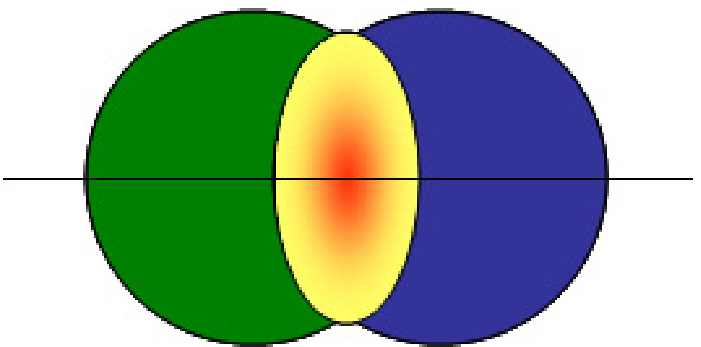}
\end{minipage}
\hfill
\begin{minipage}{.58\textwidth}
\centering
\includegraphics[width=1.0\textwidth]{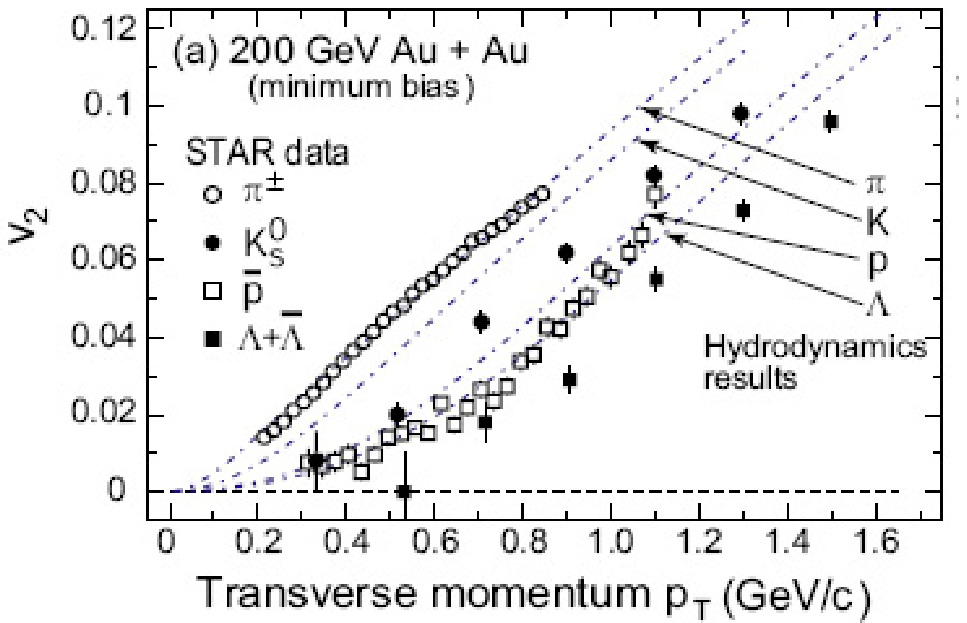}
\end{minipage}
\caption{Left:  Illustration of the collision of two nuclei.  The central gradient shaded area represents the overlap region.  The direction of the beams is into and out of the page.  The line represents the reaction plane (the plane goes into and out of the page).    Right:  STAR experiment (see Chapt. 2.2) results for the elliptic flow of identified hadrons from \cite{inflow}.  Curves are hydrodynamic calculations.}
\label{fig:introflow}
\end{figure}

Hydrodynamic behaviors are an important signature of QGP.  This is because agreement with hydrodynamics implies strong collectivity and local thermal equilibrium.  One important observable in heavy ion collisions is the elliptic flow, $v_{2}$.  The flow is a measure of the azimuthal anisotropy, relative to the reaction plane, of the collision.  Figure~\ref{fig:introflow}, left, shows a cartoon of two colliding nuclei.  The elliptical area in the center is our collision overlap region.  The reaction plane is the plane defined by the line connecting the center of the two nuclei (black line) and the beam direction (out of the page). The anisotropic flow is typically expanded with a Fourier expansion with the second order term, elliptic flow being the dominant term at RHIC (Relativistic Heavy Ion Collider, see Chapt. 2.1). As the system expands, the initial spacial asymmetry is reduced by the positive elliptic flow (momentum asymmetry) which is itself generated by the spacial asymmetry.  This self-quenching effect makes elliptic flow sensitive to the early dynamics of the system~\cite{sorge}.  Figure~\ref{fig:introflow} shows $v_{2}$ measurements compared with hydrodynamics calculations.  The hydrodynamic calculations agree reasonably well with the data.  The best agreement appears to be with an early thermalization ($\tau<1$ fm/c)\cite{white}.  The elliptic flow is discussed in more detail in chapter 3.1.2.

\begin{figure}[htb]
\begin{minipage}{.59\textwidth}
\centering
\includegraphics[width=1.0\textwidth]{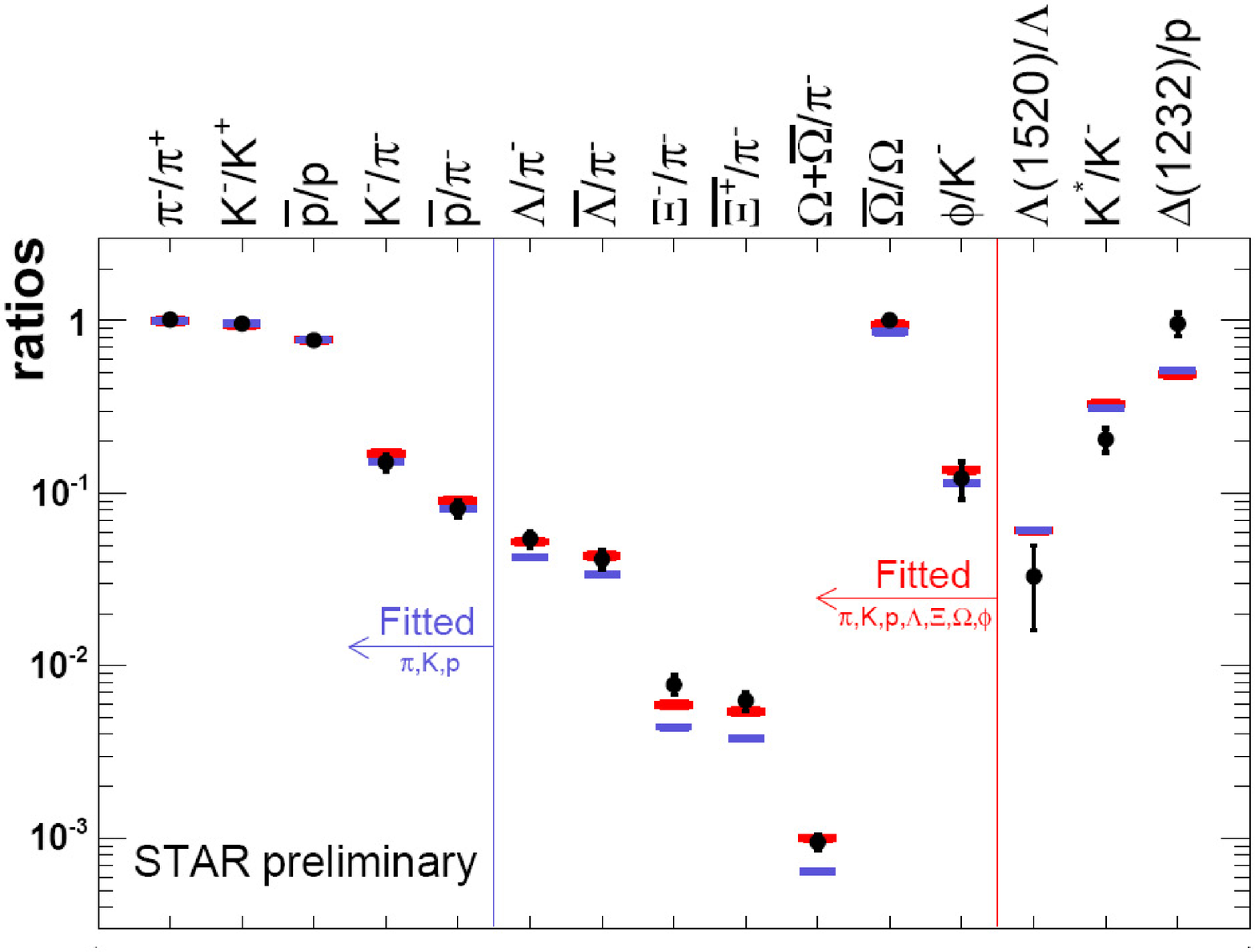}
\end{minipage}
\hfill
\begin{minipage}{0.39\textwidth}
\includegraphics[width=1.0\textwidth]{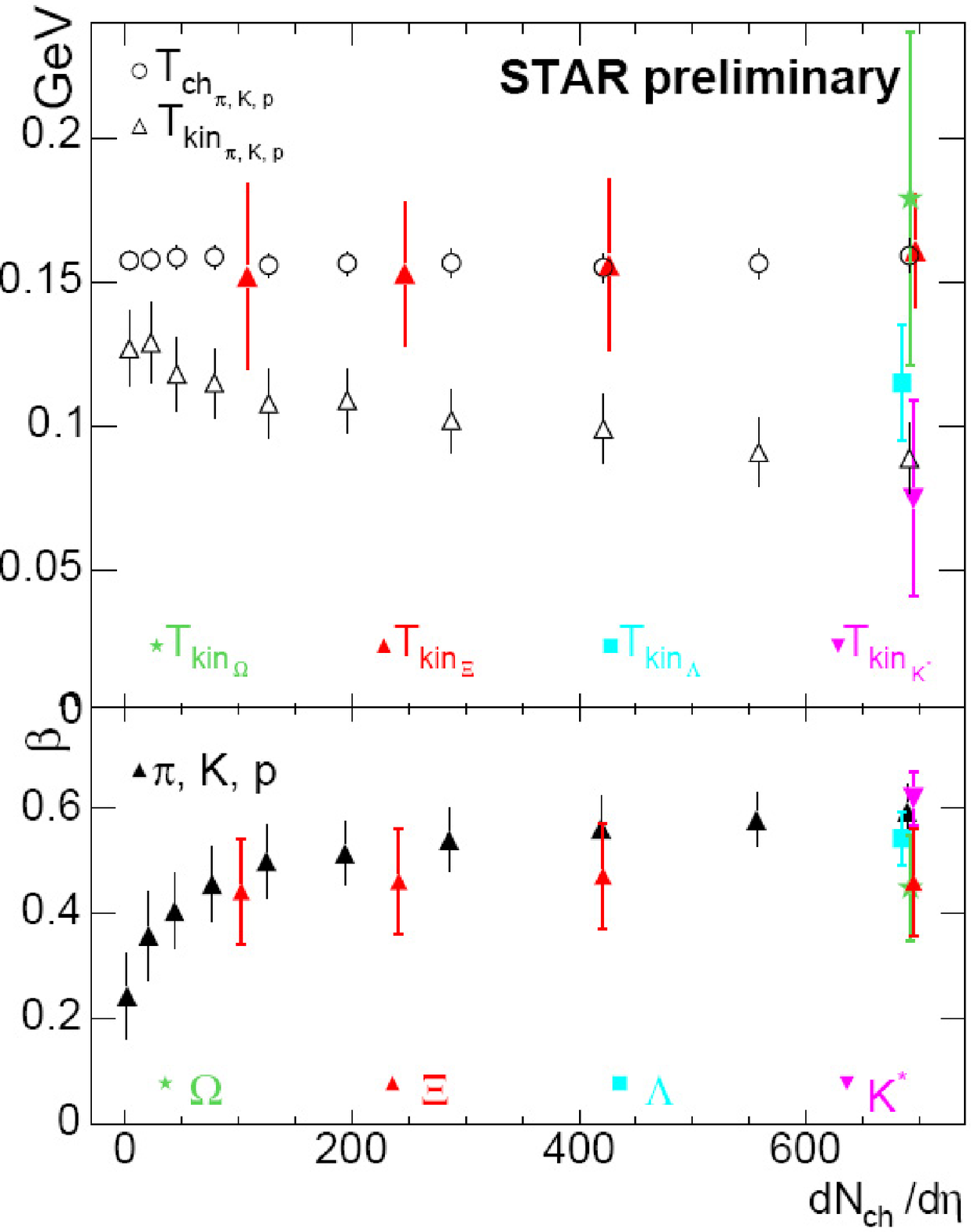}
\end{minipage}
\caption{Left:  Measured particle ratios in symbols.  Lines represent statistical model fits to $\pi$, {\it K}, and {\it p} measurements (blue) and $\pi$, {\it K}, {\it p}, $\Lambda$, $\Xi$, $\Omega$, and $\phi$ measurements (red).  Upper Right:  Kinetic and chemical freezeout temperatures from fits.  Lower Right:  Radial flow velocity, $\beta$.  Black shows the $\pi$, {\it k}, {\it p} result and the $\Lambda$ (blue), $\Xi$ (red), $\Omega$ (green) and $K^{*}$ (purple) results are shown separately.  $dN_{ch}/d\eta$ is the number of charged particles per unit rapidity (see Chapt. 1.2.1).  Results are from Au+Au collisions at $\sqrt{s_{NN}}$ = 200 GeV.  Plots are from \cite{spectra,olga}}.
\label{fig:ratio}
\end{figure}

For a system in equilibrium we can use statistical models to extract freezeout temperatures and radial flow velocities.  These are done with fits to the particle spectra and ratios.  Figure~\ref{fig:ratio} (left) shows statistical model fits to a large variety of measured particle ratios.  The fits give a chemical freezeout temperature of about $160\pm6$ MeV\cite{spectra,olga} which is near the predicted value of the phase transition temperature from lattice QCD, $T_c$.  The kinetic freezeout temperature and the radial flow velocity are shown as a function of centrality in figure~\ref{fig:ratio} (right).

\subsection{Partonic Energy Loss and Jet Quenching}

The partons that travel through our medium may experience elastic collisions and medium induced radiative energy loss from gluon radiation.  It was first suggested by Bjorken\cite{bjorken} in 1982 that partons traveling through the medium may experience significant energy loss due to elastic collisions with the medium and that this might be an observable effect.  Later calculations have shown elastic energy loss to be quite small for light quarks, but have shown radiative energy loss should be large enough in heavy ion collisions to have an observable effect\cite{quench2,quench}. This leads to what is known as ``jet quenching''. 

\begin{figure}[htb]
\centering
\includegraphics[width=0.55\textwidth]{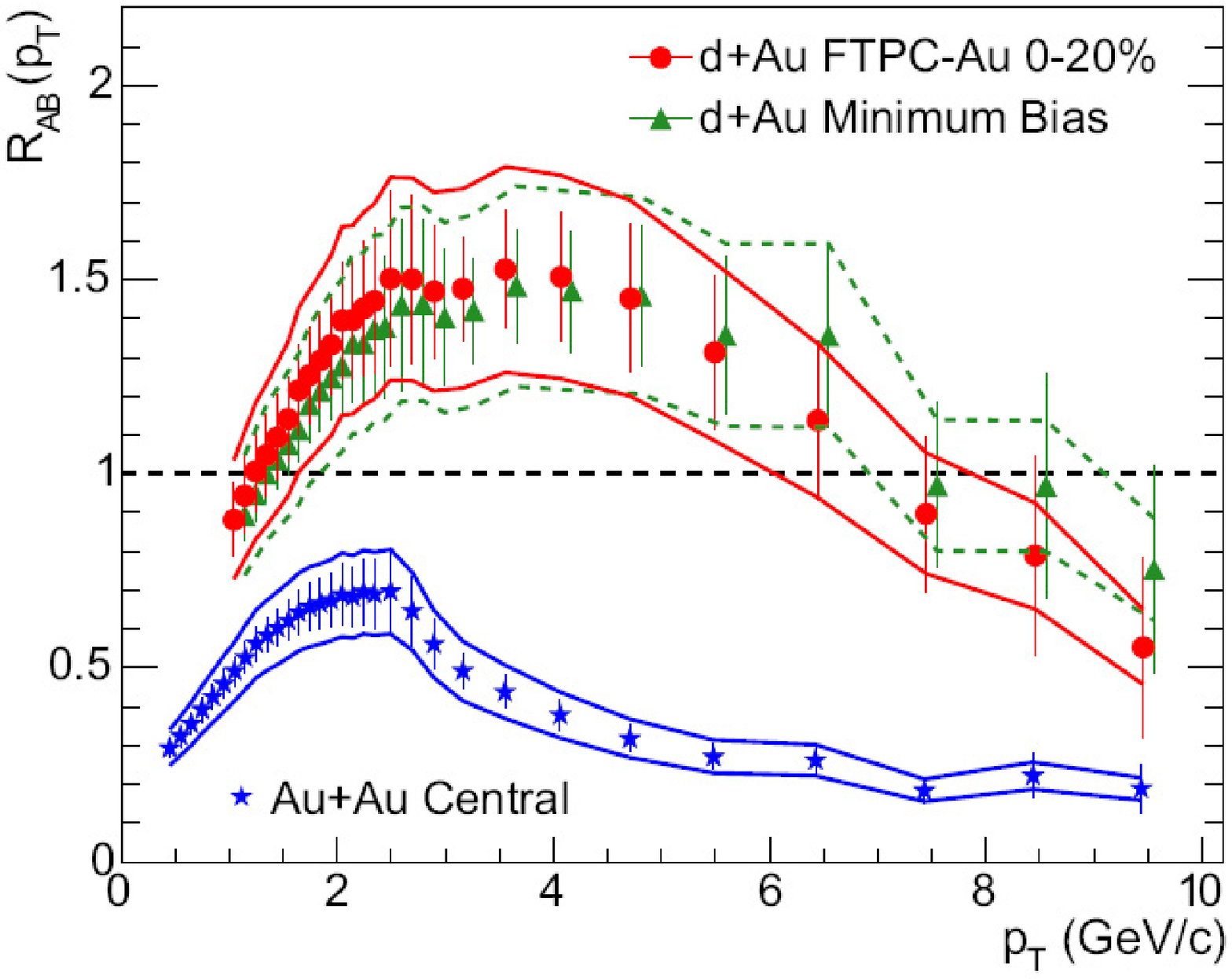}
\caption{Nuclear modification factor in central Au+Au collisions and d+Au collisions at $\sqrt{s_{NN}}$=200 GeV from the STAR experiment.  The two d+Au curves are for minimum bias and central d+Au.  Plot is from \cite{rab}.}
\label{fig:raa}
\end{figure}

One effect of jet quenching is the high-$p_{T}$ hadron cross sections in heavy ion collisions are suppressed from those found in elementary collisions.  A useful observable to quantify jet quenching is the nuclear modification factor given by,
\begin{equation}
R_{AB}(p_{T})=\frac{\frac{dN_{AB}}{d\eta d^{2}p_{T}}}{T_{AB}\frac{d\sigma_{NN}}{d\eta d^{2}p_{T}}}
\end{equation}
where A and B are our two colliding nuclei and $T_{AB}=\langle N_{bin} \rangle /\sigma^{pp}_{inelastic}$, where and $<N_{bin}>$ is the number of binary nucleon-nucleon collisions in A+B.  Figure~\ref{fig:raa} shows the nuclear modification factor in central Au+Au collisions and d+Au collisions.  The nuclear modification factor is less than one at high $p_{T}$ in central Au+Au collisions where hard processes dominate.  This suppression of the high $p_{T}$ particles relative to the binary scaled {\it pp} collisions is a signature of jet quenching.  At low $p_T$ $R_{AB}$ is less than one because the low $p_T$ region is dominated by large cross-section soft physics where particle production does not scale with the number of binary collisions but rather approximately with the number of participants. In addition, gluon saturation may come into play. 

\begin{figure}[htb]
\hfill
\begin{minipage}{.4\textwidth}
\centering
\includegraphics[width=1.0\textwidth]{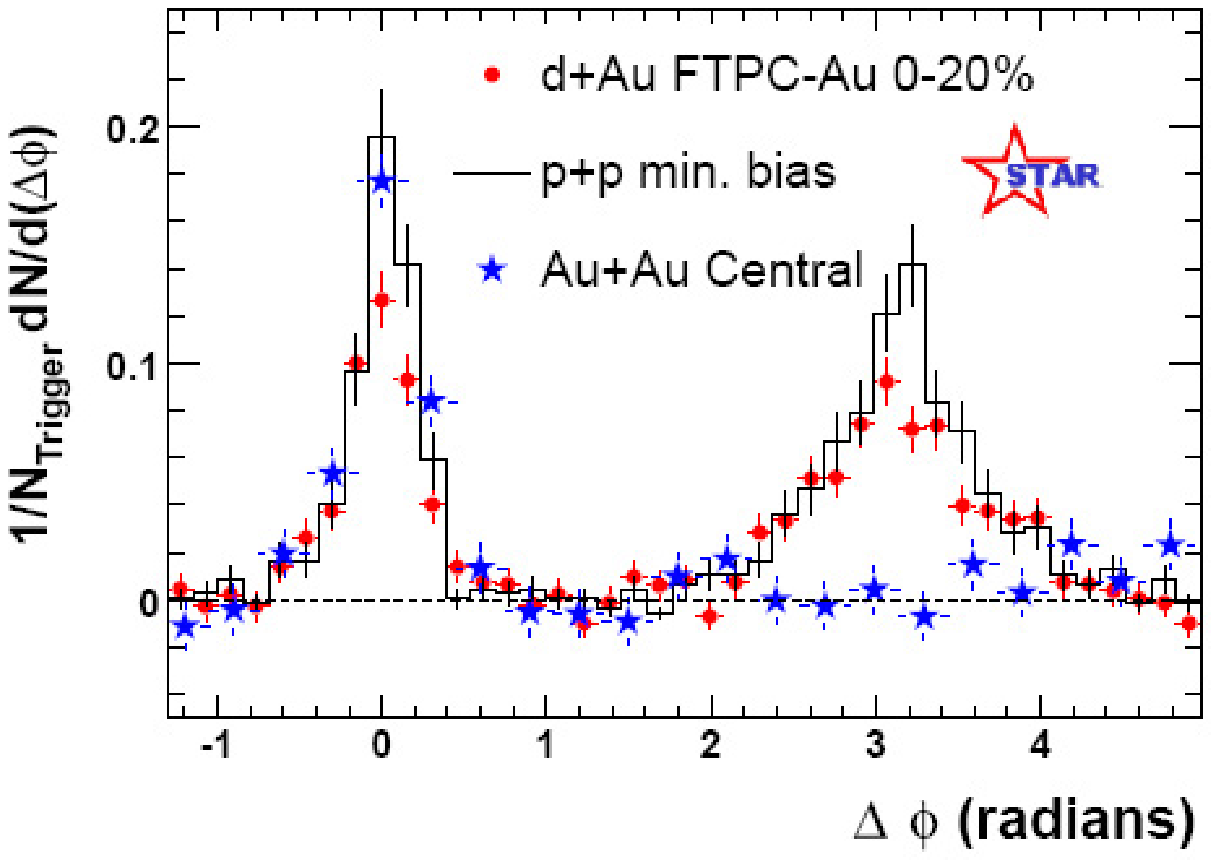}
\end{minipage}
\hfill
\begin{minipage}{0.4\textwidth}
\centering
\includegraphics[width=1.0\textwidth]{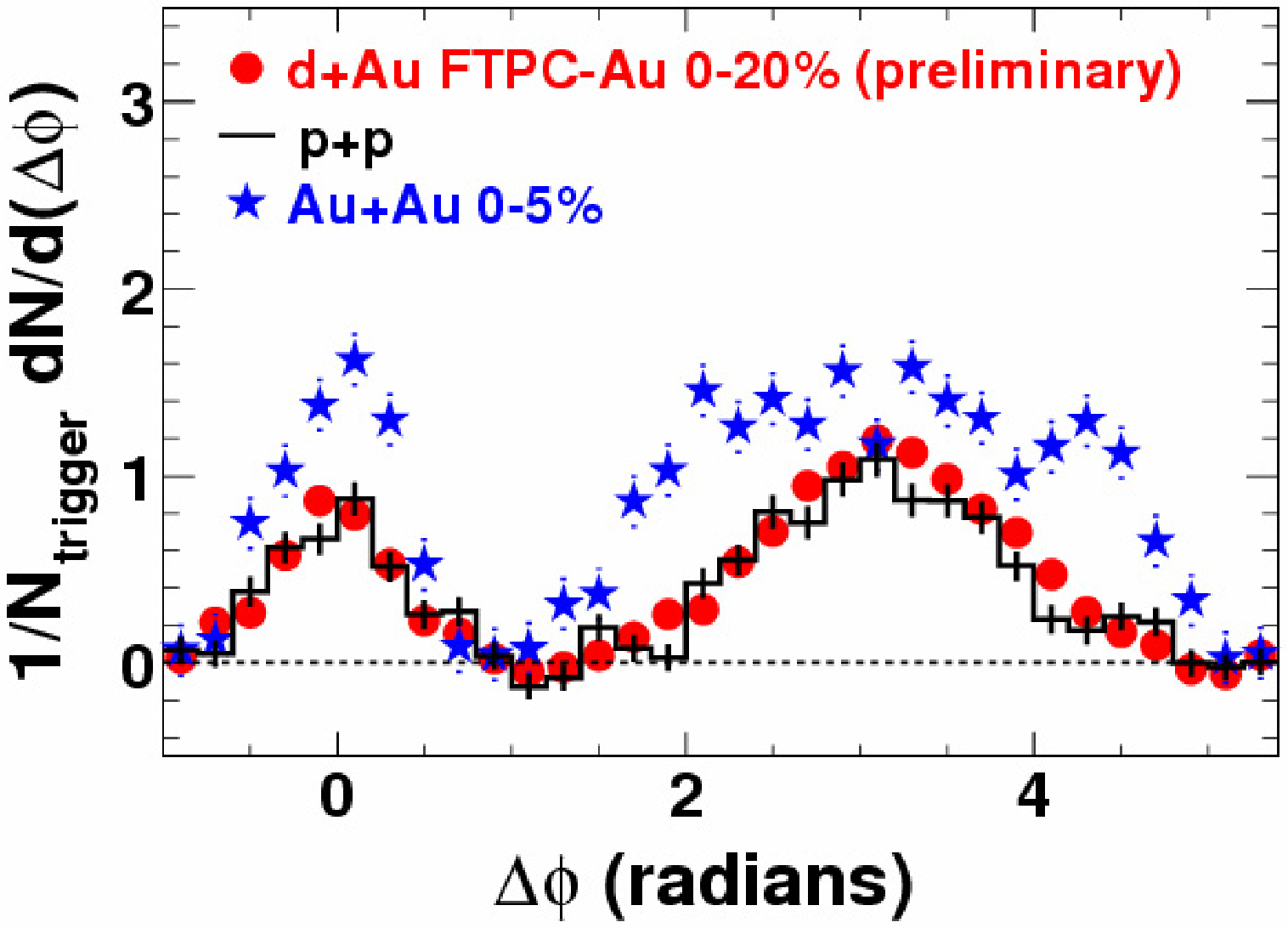}
\end{minipage}
\begin{minipage}{0.18\textwidth}
\centering
\includegraphics[width=1.0\textwidth]{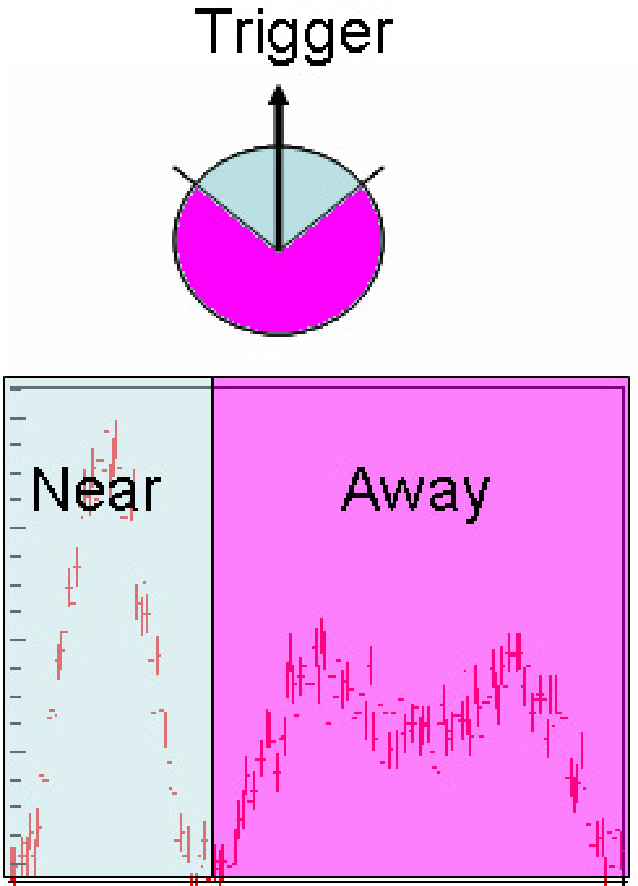}
\end{minipage}

\caption{Left:  Background subtracted 2-particle correlations for trigger particles of $4<p_{T}^{trig}<6$ GeV/c and associated particle of $2<p_{T}<p_{T}^{trig}$ GeV/c from \cite{rab}.  Middle:  Background subtracted 2-particle correlations for trigger particles of $4<p_{T}^{trig}<6$ GeV/c and associated particles of $0.15<p_{T}<4$ GeV/c in {\it pp} and Au+Au collisions from \cite{fuqiang} and d+Au collisions discussed in Chapter 3.2.  Right:  Illustration of the locations of the near side and away side.}
\label{fig:b2bfuq}
\end{figure}

Another effect of jet-quenching is that the jet structure in heavy ion collisions is modified from {\it pp} collisions.  We can therefore study the jets themselves to examine jet quenching.  This is done statistically because there is too much background in a heavy ion environment to reconstruct jets event by event.  Figure~\ref{fig:b2bfuq} shows 2-particle azimuthal correlations with a high $p_{T}$ trigger ($4<p_{T}^{trig}<6$ GeV/c) with intermediate $p_{T}$ associated particles ($2<p_{T}<p_{T}^{trig}$ GeV/c), left, and with associated particles going down to low $p_{T}$ ($0.15<p_{T}<4$ GeV/c), right.  In the left panel, we see that for the higher $p_{T}$ associated particles the away-side (particles about $\pi$ radians from the trigger particle) is greatly suppressed in central Au+Au collision when relative to {\it pp} and d+Au collisions.  This suppression shows that the away-side jet is quenched when transversing the medium.  The near-side jet is not suppressed so it must have originated from near the surface and traveled through relatively little medium.  The middle panel is dominated by the lower associated $p_{T}$ particles.  Here the away-side distribution in central Au+Au collisions is enhanced relative to {\it pp} and d+Au collisions implying that the energy lost in the suppression of the higher $p_{T}$ particles is transfered to lower $p_{T}$.  The away-side peak is also significantly broadened in the central Au+Au collisions relative to {\it pp} and d+Au collisions and may even be double peaked.  Different physics mechanisms have been suggested to explain this and are discussed in chapter 4.

\section{Relativistic Heavy Ion Collisions}

\begin{figure}[htb]
\centering
\includegraphics[width=0.5\textwidth]{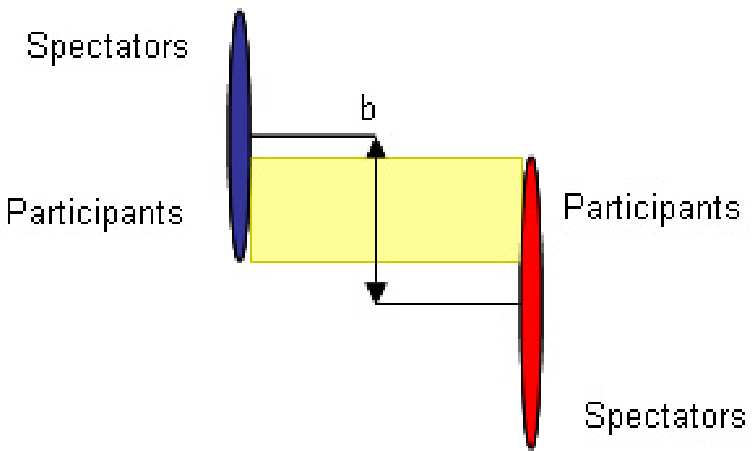}
\caption{Cartoon diagram of two colliding nuclei.  The colliding nucleons are known as participants while the nucleons that do not collide are spectators.  The impact parameter, {\it b}, is the distance between the centers of the two nuclei, at the moment of the closest distance.}
\label{fig:collision}
\end{figure}

An effective way to deposit large amounts of energy in a small volume in an attempt to produce a QGP is through relativistic heavy ion collisions.   By colliding many nucleons together in a small volume we can increase the energy deposited in contrast to collisions between only two elementary particles.  Figure~\ref{fig:collision} shows a cartoon of two colliding nuclei.  The nuclei are Lorentz contracted, in the center of mass frame, into thin disks.  The contraction reduces the volume in which the energy is deposited, helping to increase the energy density.  The nucleons involved in the impact are termed participants.  The nucleons that do not participate in the collisions are termed spectators.

\begin{figure}[htb]
\centering
\includegraphics[width=0.6\textwidth]{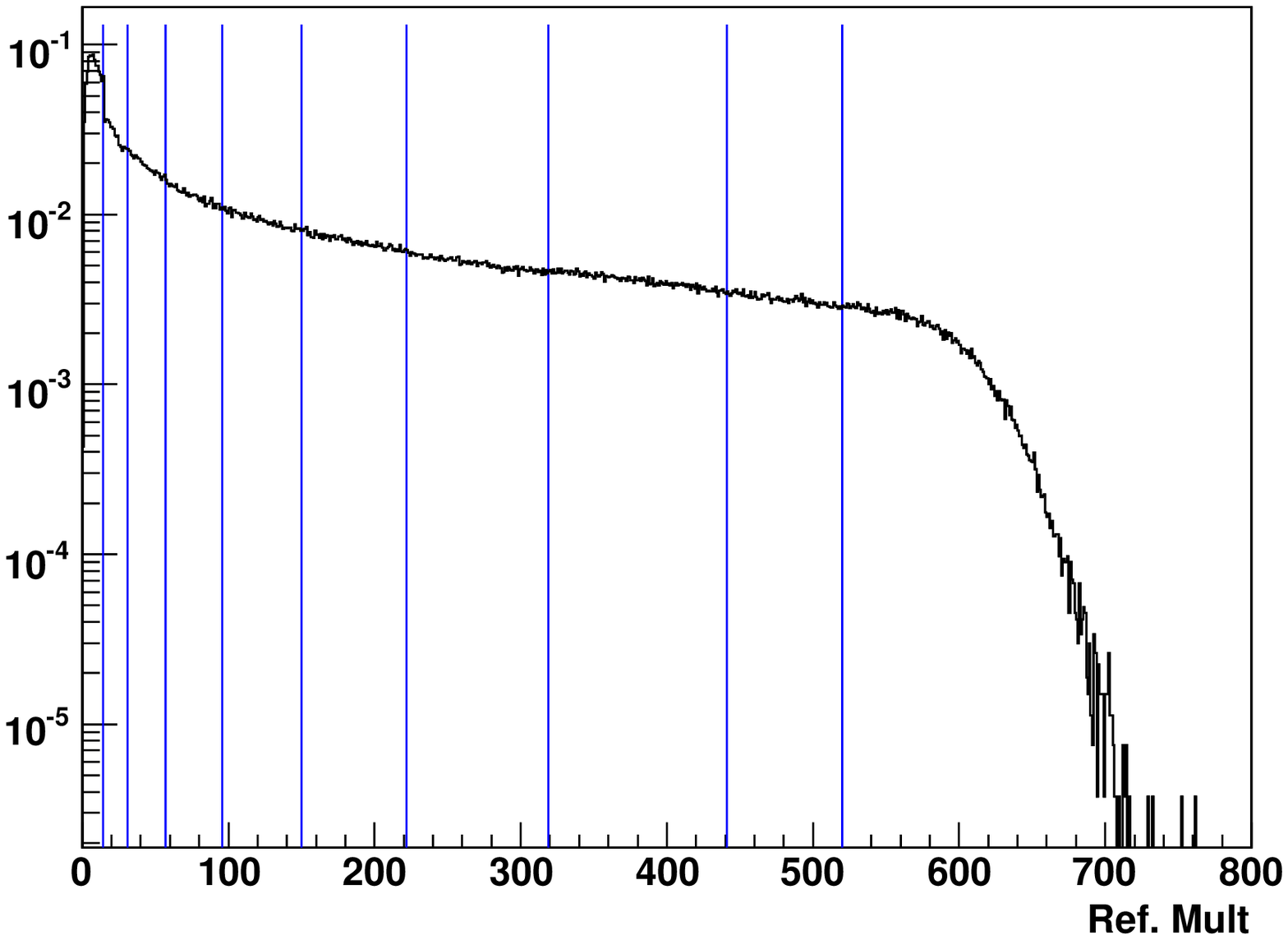}
\caption{Reference multiplicity distribution for Au+Au collisions.  Blue lines show the centrality bin divisions.  The bins are 80-100\%, 70-80\%, 60-70\%, 50-60\%, 40-50\%, 30-40\%, 20-30\%, 10-20\%, 5-10\%, and 0-5\% most central collisions from left to right.  The y-axis is in arbitray units and the x-axis is uncorrected number of charged particles per unit rapidity taken in one unit of rapidity.  For Au+Au collsions at $\sqrt{s_{NN}}=200$ GeV.}
\label{fig:refmult}
\end{figure}

In experiment, the impact parameter is unknown so we use the number of particles to determine centrality.  In the STAR experiment (see Chapt 2.2), the uncorrected number of particles in the center of the Time Projection Chamber, TPC, (see Chapt. 2.2.3) ($|\eta|<0.5$) is used.  A large number of particles corresponds to a small impact parameter (head on collisions) and peripheral collisions correspond to a small number of particles.  Figure~\ref{fig:refmult} shows the reference multiplicity distribution from Au+Au collisions.  The blue lines show the division into multiplicity bins that are used to study different ranges of impact parameter.  The correspondence between multiplicity and impact parameter can be studied in a model dependent fashion using Glauber models~\cite{glauber}.      

\subsection{Relevant Variables in Relativistic Heavy Ion Collisions}

\begin{figure}[htb]
\centering
\includegraphics[width=1.0\textwidth]{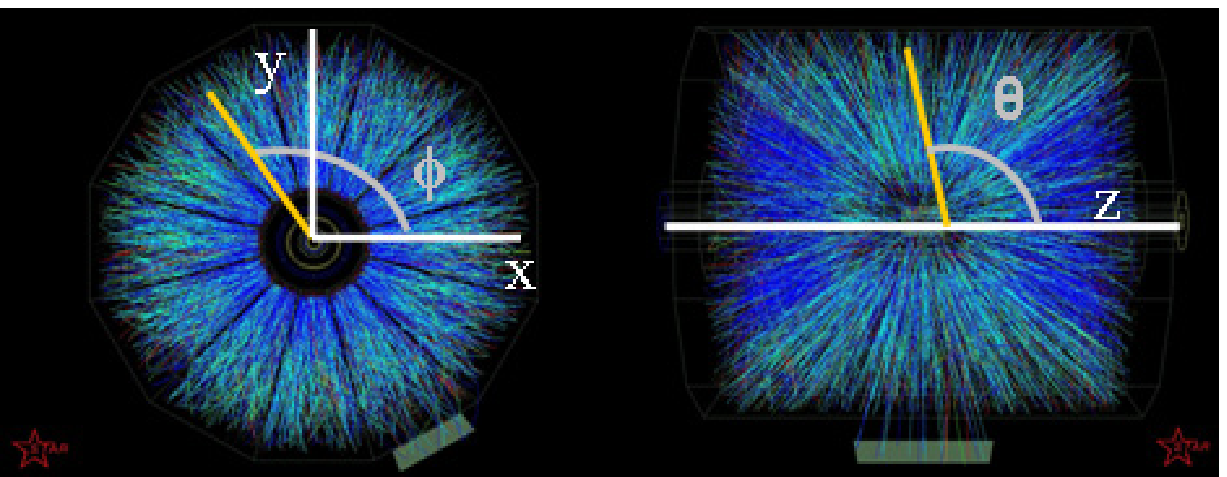}
\caption{Axes superimposed on reconstructed tracks in the STAR detector.  Yellow is the particle direction.  Left:  Beam direction is into and out of the page.  Right:  Beam direction is left and right.}
\label{fig:axis}
\end{figure}

It is useful to define some variables that are commonly used in heavy ion physics.  The reference frame is generally defined such that the z-axis is the beam line.  The azimuthal angle $\phi$ goes around the beam pipe.  The transverse direction is radially out from the beam.  This frame is illustrated in Figure~\ref{fig:axis}.  The transverse direction and the azimuthal angle will be used extensively in this thesis.  The momentum in the transverse direction will be denoted $p_{T}=\sqrt{p_{x}^{2}+p_{y}^{2}}$.  Another commonly used variable in heavy ion collisions is the rapidity,
\begin{equation}
y=\frac{1}{2}ln\left(\frac{E+p_{z}}{E-p_{z}}\right).
\end{equation}
The rapidity is useful in that one can switch between reference frames along the z-axis and the only change in rapidity is an additive constant.  However, since the energy of the particle is not generally readily available but the momentum is, the pseudorapidity is generally used.  The pseudorapidity is defined as,
\begin{equation}
\eta=\frac{1}{2}ln\left(\frac{|\vec{p}|+p_{z}}{|\vec{p}|-p_{z}}\right)=-ln\left[tan\left(\frac{\theta}{2}\right)\right].
\end{equation}
When the momentum is large the pseudorapidity is approximately the rapidity.
\chapter{EXPERIMENT}

\section{Relativistic Heavy Ion Collider}
The data used in this thesis is from collisions that were carried out at the Relativistic Heavy Ion Collider (RHIC) at Brookhaven National Laboratory (BNL).  This collider can in principle collide any nuclei as long as a suitable ion source is available.  So far it has collided protons (A=1), deutrons (A=2), copper nuclei (A=63) and gold nuclei (A=197).  The top energies are $\sqrt{s_{NN}}=200$ GeV for Au+Au and $\sqrt{s_{NN}}=500$ GeV for {\it pp} collisions where $\sqrt{s_{NN}}$ is the center of mass energy per nucleon pair.  Current experimental data has been taken at $\sqrt{s_{NN}}=200$ GeV for Au+Au, Cu+Cu, d+Au, {\it pp}, and polarized {\it pp}.  Additional lower energy runs have been carried out at $\sqrt{s_{NN}}=130$ GeV, for Au+Au collisions, at $\sqrt{s_{NN}}=62.4$ GeV in Au+Au, Cu+Cu and {\it pp} collisions, and $\sqrt{s_{NN}}=22$ GeV for Au+Au and Cu+Cu collisions.  There has also been higher energy {\it pp} collisions at $\sqrt{s_{NN}}=400$ GeV.  There are future plans to run {\it pp} collisions at the full energy and to do low energy scans ($\sqrt{s_{NN}} \approx 5-50$ GeV) with Au+Au collisions.

\begin{figure}[htb]
\centering
\includegraphics[width=.99\textwidth]{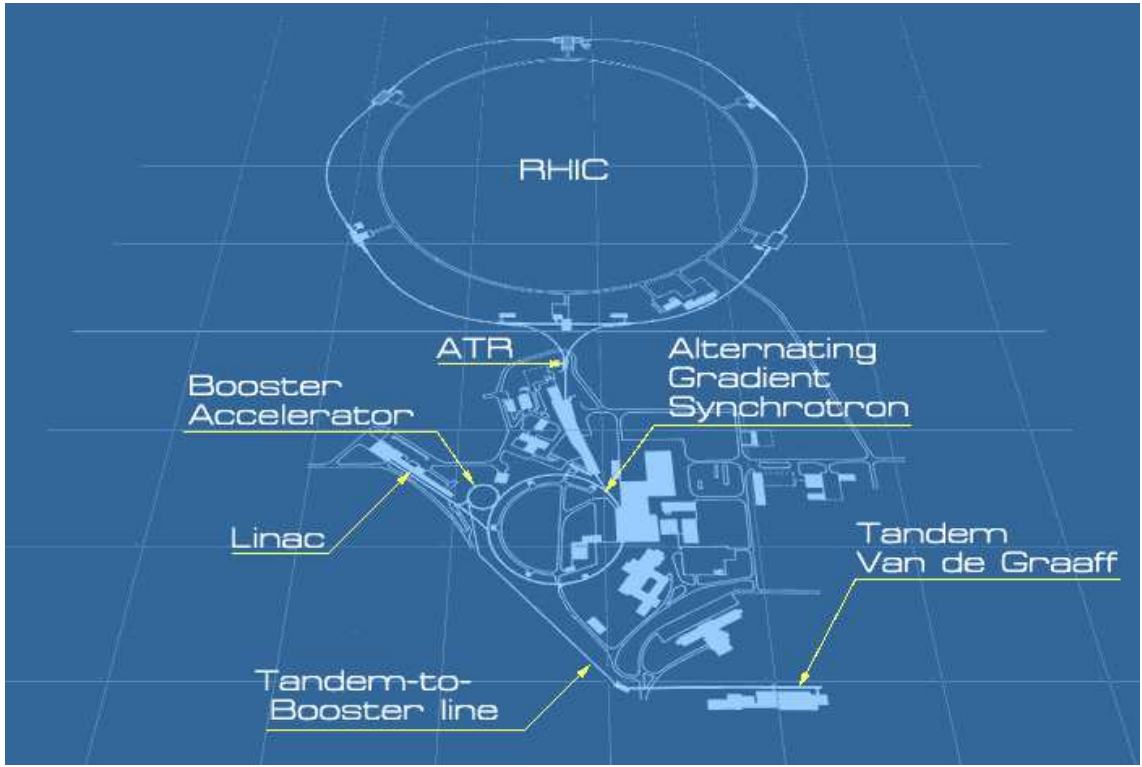}
\caption{Diagram of the RHIC collider and the accelerators that feed the beam into RHIC.  Diagram from \cite{rhicweb}}
\label{fig:rhic}
\end{figure}

Figure~\ref{fig:rhic} shows the chain of accelerators at RHIC.  The RHIC beam for ions begins with a negative ion source at the Tandem Van de Graaff.  The Tandem Van de Graaff accelerates ions to 1 MeV/nucleon.  The ions are partially stripped of electrons in the Tandem Van de Graaff and again on exiting.  A bending magnet is then used to make a charge selection on the ions.  The ions of a particular charge (+32 for Au) are then further accelerated by the Booster Synchrotron to 95 MeV/nucleon.   After exiting the Booster Synchrotron, the ions are further stripped of charge (+77 for Au) and then injected into the Alternating Gradient Synchrotron (AGS).  In the AGS, the ions are accelerated to 10.8 GeV/nucleon (for Au) and then fully stripped on exit.  Polarized protons (for use in spin studies) are injected from the Proton Linac at 200 MeV directly into the Booster Synchrotron.  Siberian Snakes\cite{snakes} are used in the AGS and RHIC to preserve the polarization.  After leaving the AGS, the beam is transfered to RHIC.  Particles are injected into RHIC with a common magnetic rigidity value B$\rho$=81.1141 Tm, where B is the strenght of the magnetic field and $\rho$ is the radius of a charged particle in circular motion in the field.  B$\rho$ is also equal to the momentum perpenduclar to the field divided by the charge.  Ions can then be accelerated to a maximum magnetic rigidity of 839.5 Tm which corresponds to a magnetic field of 3.458 T.  This gives the maximum kinetic energy of 100 GeV/nucleon for Au ions.

There are two beam lines throughout the entire acceleration procedure, including two Tandem Van de Graaffs.  This allows for two different types of ions, one in each beam line.  With two different types of ions we can get asymmetric collisions such as deuteron and gold which has been run at RHIC.  The RHIC ring has a 3.834 km circumference and has a total of 1700 superconducting magnets cooled to $<4.6$ K.  There are six interaction points, four of which have dedicated heavy-ion detectors.  The detectors are STAR, PHENIX, PHOBOS and BRAHMS\cite{exprhic}.

\section{STAR Experiment}
\begin{figure}[htbp]
\centering
\includegraphics[width=1.0\textwidth]{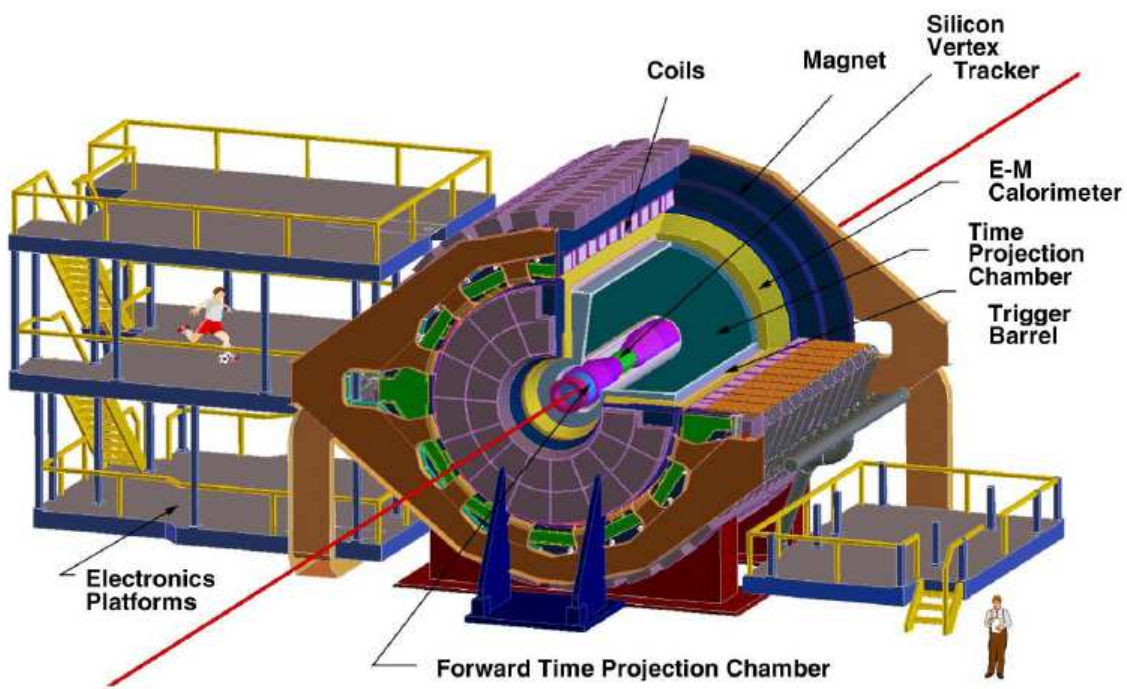}
\includegraphics[width=1.0\textwidth]{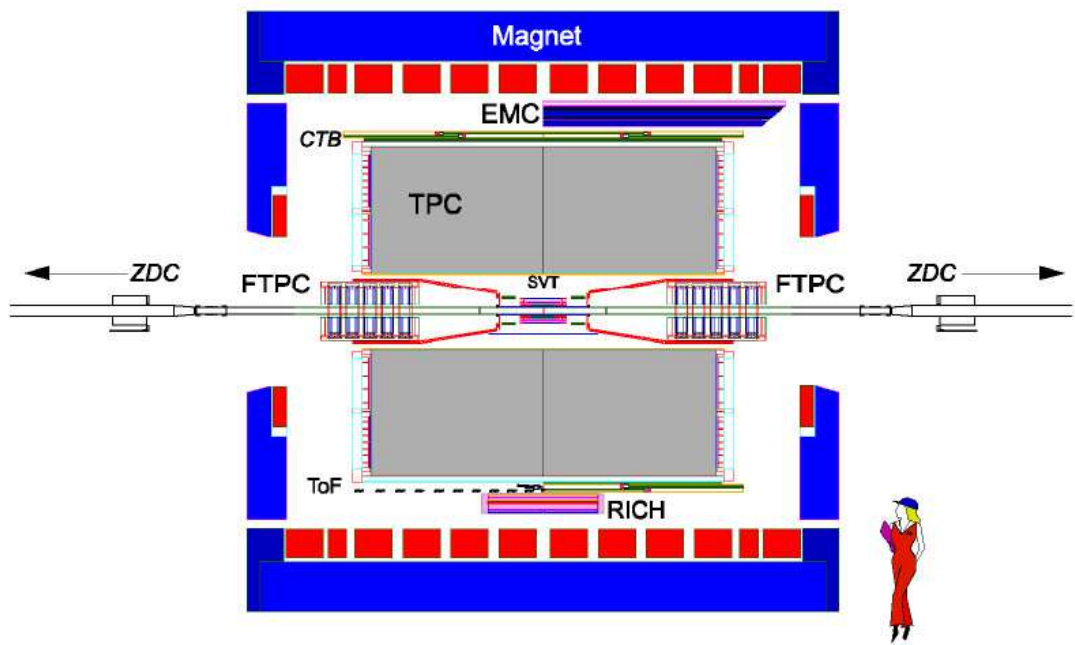}
\caption{Two cutaway views of the STAR detector.  The bottom figure shows the configuration as of 2001.  Figures are from \cite{expstar}.}
\label{fig:STAR}
\end{figure}

One of the two large experiments at RHIC is the Solenoidal Tracker at RHIC (STAR)\cite{expstar}.  The STAR experiment was designed to have a large acceptance (including full, 2$\pi$, azimuthal acceptance), high precision tracking, momentum determination, and particle identification at midrapidity ($|\eta|\sim1.5$). Figure~\ref{fig:STAR} shows two different cutaway views of the STAR experiments.  

The entire detector is surrounded by a 0.5 T solenodial magnet\cite{expmagnet}.  The magnet allows for momentum determination of charged particles.  Near the beam pipe resides a three layer Silicon Vertex Tracker (SVT)\cite{expssd} surrounded by an additional layer of Silicon Strip Detector (SSD)\cite{expssd}.  These detectors cover full azimuth and $|\eta|<1$ in pseudo-rapidity.  The SVT and SSD were to enhance the measurement of hadrons with a short lifetime by providing inner tracking.  The primary subdetector of STAR is a large volume Time Projection Chamber (TPC)\cite{expTPC} which provides tracking and particle identification though ionization energy loss.  The TPC has full azimuthal coverage and covers $|\eta|<1.8$ in pseudo-rapidity.  For particle tracking in the forward direction (large $|\eta|$) there are two Forward Time Projection Chambers (FTPC)\cite{expFTPC}.  These are radial drift TPCs covering full azimuth and $2.5<|\eta|<4$ in pseudo-rapidity.  There is also a Time of Flight (TOF) detector in STAR that provides extended particle identification.  Currently it only covers a small region but is in the process of being upgraded to a full barrel TOF at mid-rapidity.  There are two ElectroMagnetic Calorimeters (EMC) at STAR.  These allow for the measurement of transverse energy and for triggering on high $p_{T}$ photons, electrons, and electromagnetically decaying hadrons.  The two EMCs are the Barrel EMC (BEMC)\cite{expbemc} and the Endcap EMC (EEMC)\cite{expeemc}.  The BEMC surrounds the TPC while the EEMC is in front of the TPC.  These two EMCs provide full azimuthal coverage for the combined pseudo-rapidity coverage of $-1<\eta<2$. There are also additional detectors that are used for trigger input.  The Central Trigger Barrel (CTB)\cite{expTrig} triggers on the flux of charged-particles in the midrapidity region.  The Zero-Degree Calorimeters (ZDCs)\cite{expZDC} are used for energy determination of neutral particles in the forward direction.  The Beam Beam Counters (BBC)\cite{expBBC} are used to determine luminosity in {\it pp} collisions.

\subsection{STAR Magnet}

\begin{figure}[htbp]
\hfill
\begin{minipage}[t]{0.49\textwidth}
\centering
\includegraphics[width=1.0\textwidth]{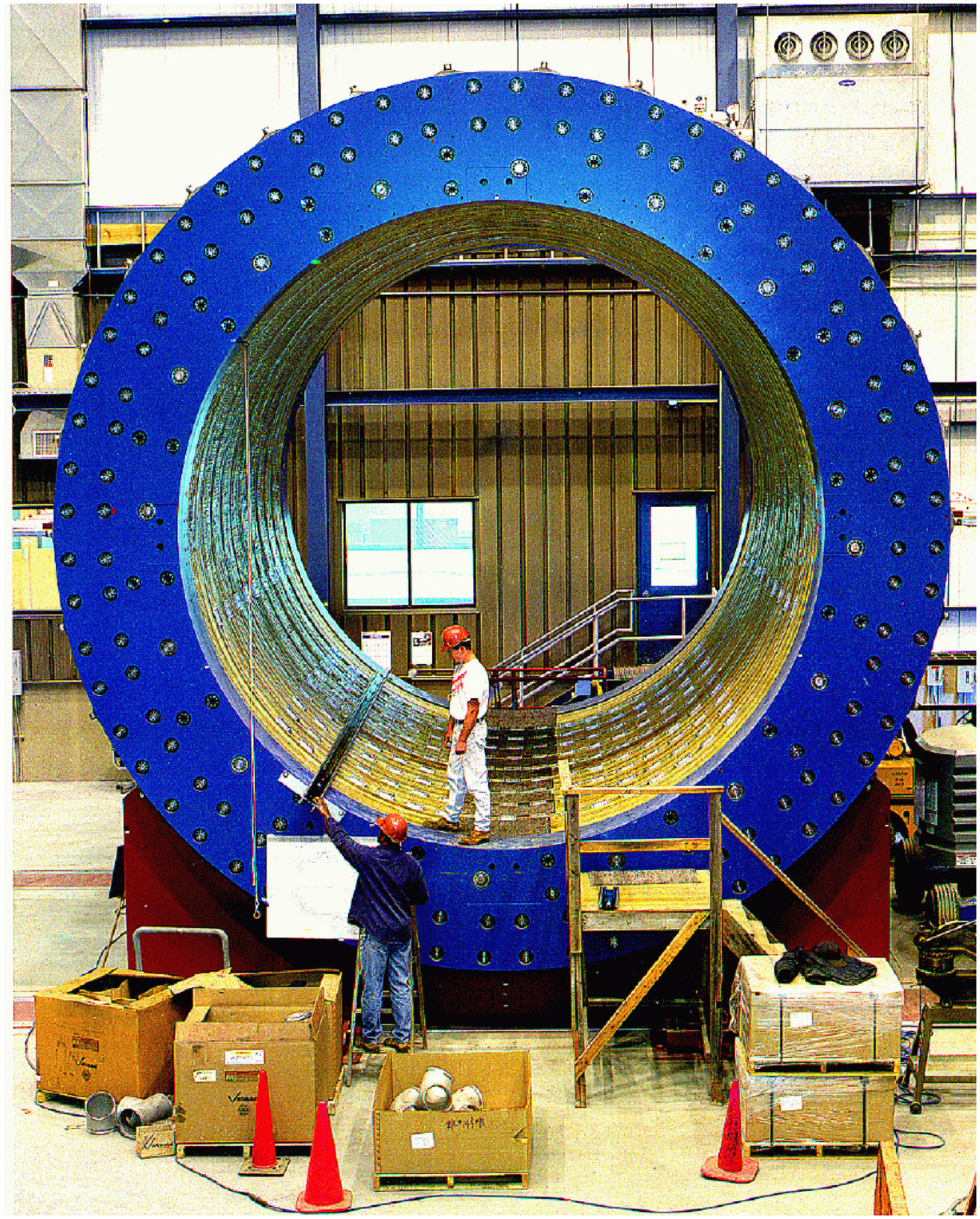}
\end{minipage}
\hfill
\begin{minipage}[t]{0.49\textwidth}
\centering
\includegraphics[width=1.0\textwidth]{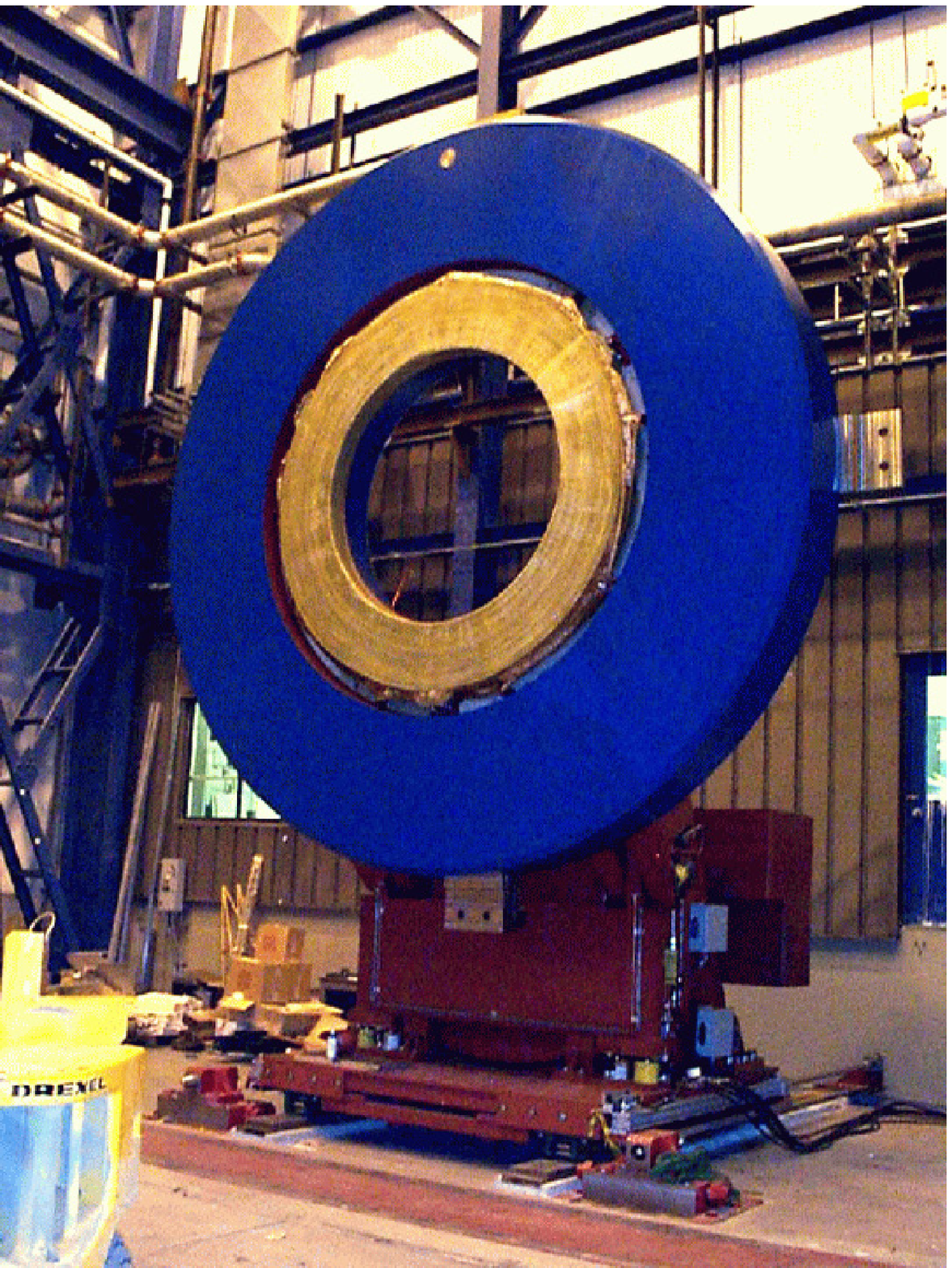}
\end{minipage}
\hfill
\includegraphics[width=1.0\textwidth]{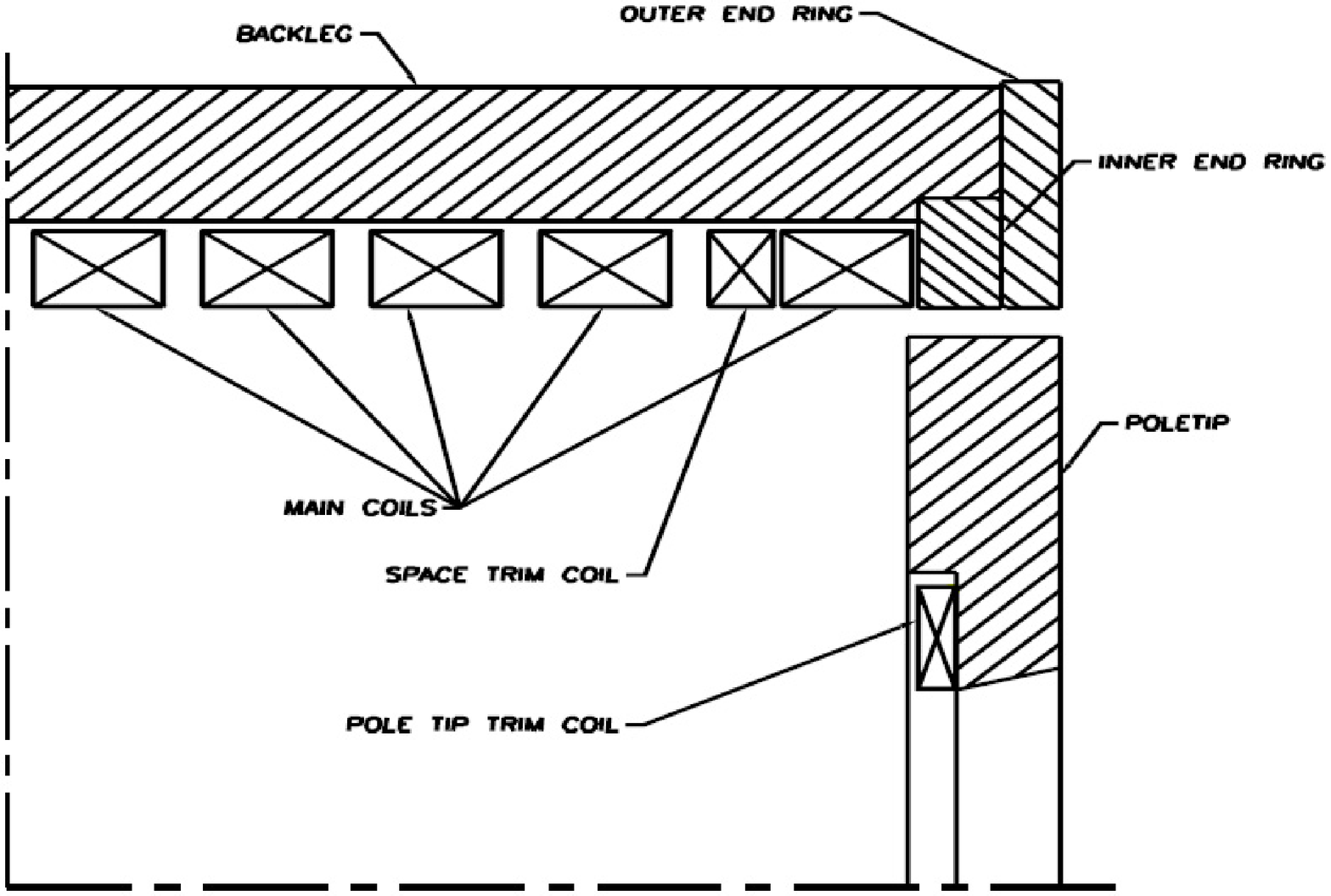}
\caption{Top Left: Main and Space Trim Magnet coils.  Top Right:  Poletip Trim Magnet coils.  Bottom:  Diagram of magnet coils.  Photos and diagram are from \cite{expmagnet}.}
\label{fig:magnet}
\end{figure}

A large mostly solenodial magnet surrounds the STAR experiment and is used for particle momentum reconstruction.  The specifications required for the STAR magnet come from the physics goals.  The field has to be large enough to measure the momentum of high-energy electron tracks.  However the larger the field the higher the low momentum cut off for tracks that can be measured in the STAR TPC.  The field must also be very homogeneous to reduce distortions on the drift electrons in the TPC.  The magnet can provide a near uniform magnetic field over the range $0.25<|B_{z}|<0.5$ Tesla.  The magnet consists of three types of coils:  Main, Space Trim and Poletip Trim.  Most of the field of the magnet is produced from the Main coils.  The Trim coils are used to reduce distortions.  The coils are cooled by a liquid cooling system.  The magnet and positions of the coils are shown in Figure~\ref{fig:magnet}

Magnetic field mapping is done for all three field components $B_{r}$, $B_{\phi}$, and $B_{z}$.  This was done in 36 azimuthal points, 57 axial locations.  This was measured with a steerable array of Hall probes\cite{tipler} from CERN (European Organization for Nuclear Research) and supplemented by NMR measurements.  The reproducibility of the absolute field is better than $\pm$ 0.5 Gauss.  For full magnetic field (0.5 T) the maximum radial field value is $\pm$ 50 Gauss and the maximum azimuthal component is less than 3 Gauss.

Momentum determination from the magnet is done with the Lorentz Force Law,
\begin{equation}
\vec{F}=q\vec{v} \times \vec{B},
\end{equation}
where $F$ is the force on a particle with charge $q$ in magnetic field $B$.  Our magnetic field is, to a high precision, entirely along the z-axis so our equation becomes,
\begin{equation}
F=q v_{T} B
\end{equation}
where $v_T$ is the velocity perpendicual to the magnetic field.  Since our magnetic field is along the z-axis the velocity is in the transverse direction.  This force provides the centripetal force for circular motion,
\begin{equation}
F=\frac{p_{T}}{r}v_{T}.
\end{equation}
The two forces must be equal so,
\begin{eqnarray}
qv_{T}B&=&\frac{p_{T}v_{T}}{r}\nonumber \\
p_{T}&=&qBr
\end{eqnarray}
This gives the transverse component of the particle momentum.  The axial component and total momentum can be found using the measured angle of the particle with respect to the beam (z-axis), $\theta$.
\begin{eqnarray}
p&=&\frac{p_{T}}{sin(\theta)}\nonumber \\
p_{z}&=&\frac{p_{T}}{tan(\theta)}
\end{eqnarray}

\subsection{Trigger Detectors}

The purpose of the trigger detectors is to enable event selection criteria to be applied at a rate greater than that at which the slow detectors operate.  This is done because RHIC has a crossing rate of about 10 MHz and the slow detectors can only operate at rates of about 100 Hz.  The fast detectors must be used to provide a rate reduction of 5 orders of magnitude and intelligently select desired events.  Three triggering detectors were used for the data contained in this thesis.  They are the Zero Degree Calorimeters (ZDCs), the Central Trigger Barrel (CTB), and the Beam-Beam Counters (BBCs).  In Au+Au minimum bias collisions, cuts are made on the signals in both ZDCs (east and west) and the CTB.  There is also a cut on the primary event vertex obtained from the ZDCs.  The ZDC cuts required a coincidence between the two ZDCs of summed signal greater than about 40\% of a single neutron signal.  The CTB cut is used to reject nonhadronic events (which removes events with a very low number of charged particles in the CTB).  This cut rejects some of the desired events that are very perpherial and is why we do not use centralities below the 80\% most perpherial.  In central Au+Au collisions, there are much higher cuts on the ZDCs and the CTBs.   There is a cut on the primary vertex obtained from the BBC.  The cuts are tuned such that the events taken are about 10\% of the total cross section and such that the multiplicity distribution matches the minimum bias distribution for the top 5\% most central collisions.  In d+Au minimum bias collsions, the trigger cut was on the east (Au side) ZDC only.  For {\it pp} collisions there is a cut on the BBC signal.

\subsubsection{Zero Degree Calorimeters}

The Zero Degree Calorimeters detect neutrons and measure their total energy in a cone about the beam line to both sides of the collisions.  The total energy is used to calculate the neutron multiplicity.  The neutron multiplicity is correlated with the event geometry and can be used for centrality determination.

\begin{figure}[htb]
\centering
\includegraphics[width=.9\textwidth]{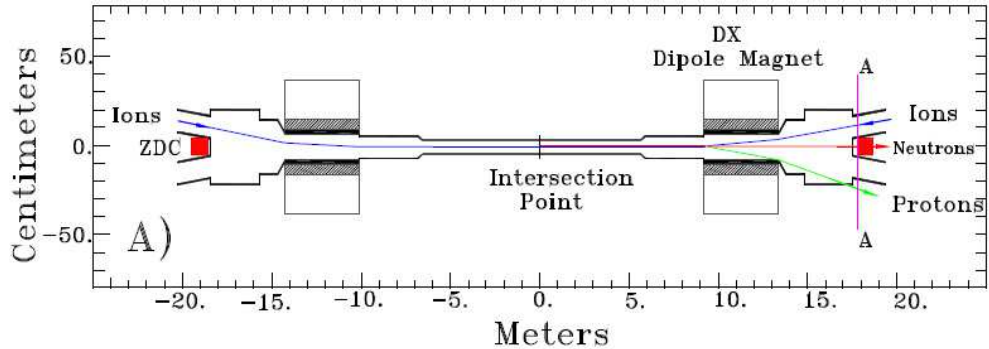}
\caption{Position of the ZDC and DX magnets in the beam pipe.  The path of the ions, protons and neutrons are shown.  Figure is from \cite{expZDC}.}
\label{fig:ZDC}
\end{figure}

The ZDCs are hadron calorimeters.  Each one consists of three modules.  The modules have a series of tungsten plates with layers of wavelength shifting fibers that route \v{C}erenkov light\cite{jackson} to a photomultiplier tube.  The \v{C}erenkov light is produced by the shower particle moving though the optical fibers.  The DX dipole magnets bend charged particles (protons and ions) away from the ZDCs; leaving only neutral particles (neutrons) in the ZDCs.  The ZDCs make up a 2 mrad cone about the beam direction and are about 18 meters from the interaction point.  The setup of the ZDCs is shown in Fig.~\ref{fig:ZDC}.

\subsubsection{Central Trigger Barrel}

\begin{figure}[htb]
\centering
\includegraphics[width=.99\textwidth]{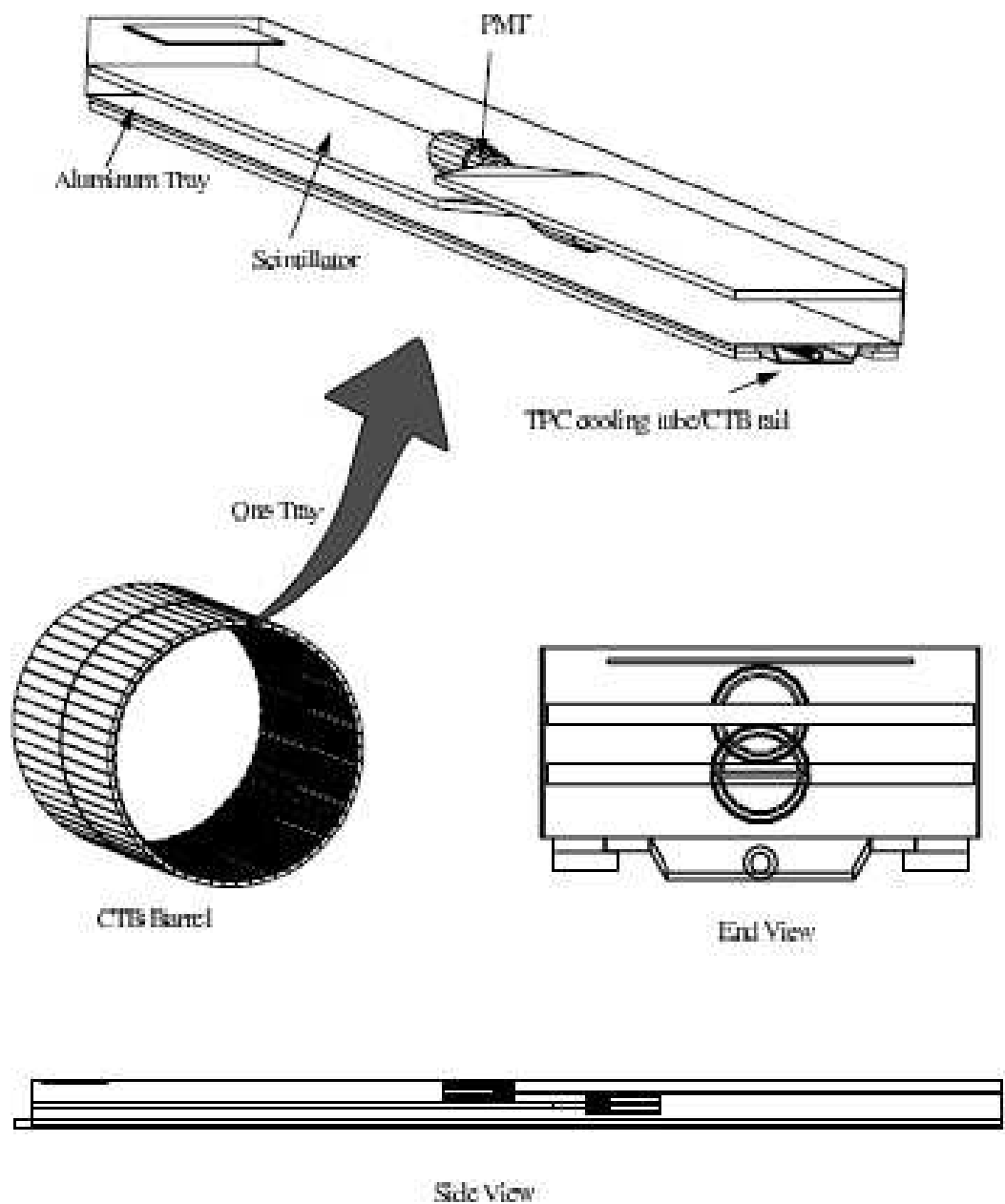}
\caption{Diagram of the Central Trigger Barrel.  The CTB cylinder is shown along with detailed views of the slats.  Diagram is from \cite{expTrig}.}
\label{fig:CTB}
\end{figure}

The CTB measures the charged particle multiplicity at midrapidity ($|\eta|<1$).  The CTB consists of four cylindrical bands with 60 scintillator slats each.  Each band covers the full azimuth and one-half unit in pseudo-rapidity.  The CTB is positioned just outside the TPC at about 4 meters from the beam pipe.  Each one of the slats consists of a scintillator, a light guide and a photomultiplier tube.  Two slats are arranged end to end in an aluminum tray for mounting and handling.  Figure~\ref{fig:CTB} shows the setup of the CTB.

\subsubsection{Beam-Beam Counters}

\begin{figure}[htb]
\centering
\includegraphics[width=.4\textwidth]{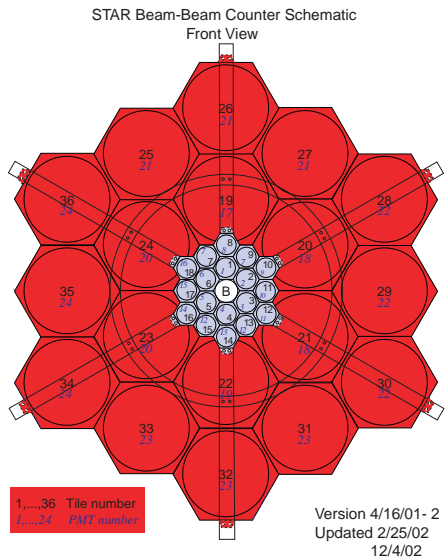}
\caption{Diagram of the arrangement of the BBC scintillator tiles.  Diagram is from \cite{expBBC}.}
\label{fig:BBC}
\end{figure}

The STAR BBCs are scintillator tiles mounted outside of the pole tip magnets.  There are 16 small tiles near the beam pipe surround by 16 larger tiles on each side.  The tiles are hexagon in shape.  The smaller tiles can be inscribed in a circle with 9.64 cm diameter.  The larger tiles can be inscribed in a circle with 4 times the diameter of the smaller tiles.  The scintillators are connected to photomultiplier tubes.  The arrangement of the tiles for each of the BBCs is shown in Figure~\ref{fig:BBC}.  When the {\it pp} data used in this thesis was taken a total of 18 of the smaller scintillator tiles were installed and used.  These were connected to 8 photomultiplier tubes.  The multiplicities determined from the BBCs are used for {\it pp} luminosity measurements and triggering.  

\subsection{Time Projection Chamber}

The TPC is the primary tracking device in the STAR detector.  It provides tracking, momentum measurement, and particle identifications.  The momentum measurements come from the curvature of the tracks in the magnetic field as previously discussed.  The particle identifications is accomplished though measurements of the ionization energy loss (dE/dx).  The detector covers full azimuth and $|\eta|<1.8$.

\begin{figure}[htb]
\centering
\includegraphics[width=1.0\textwidth]{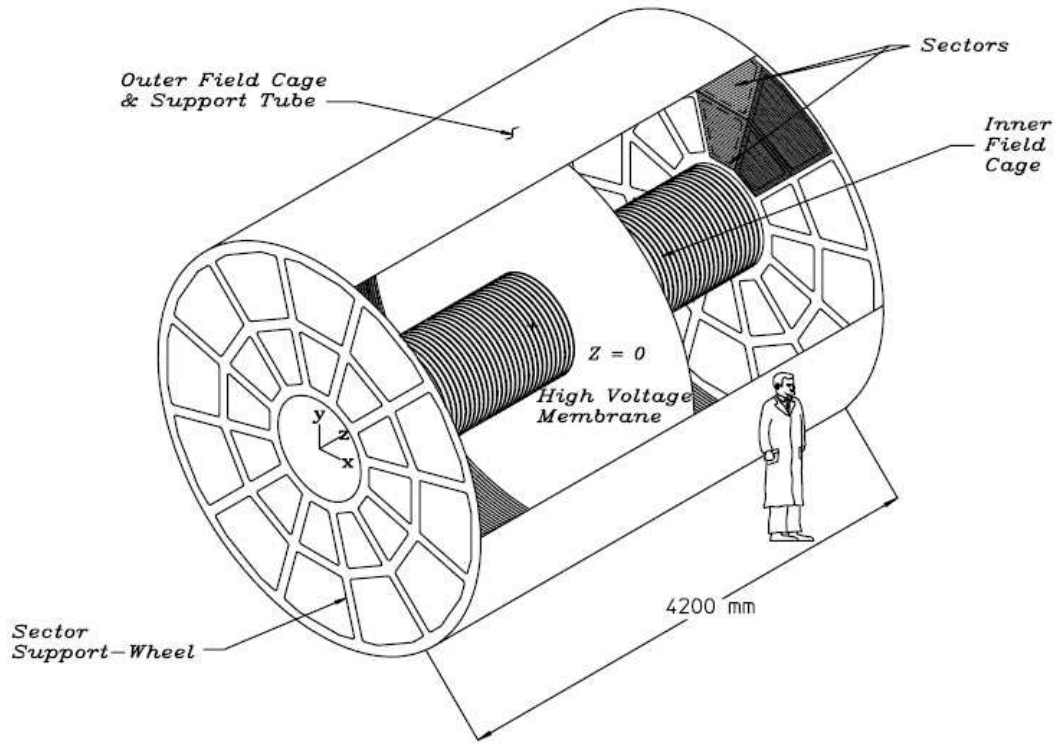}
\caption{Diagram of the STAR TPC from \cite{expTPC}.}
\label{fig:TPC1}
\end{figure}

Figure~\ref{fig:TPC1} shows a diagram of the TPC.  The TPC sits in the middle of the STAR magnet.  It is 4.2 m long and has inner and outer diameters of 1 and 4 m, respectively.  The TPC volume is filled with P10 gas (90\% argon and 10\% methane).  The gas is maintained to a high purity by the TPC gas system.  This is required because water and oxygen in the gas will absorb electrons.  The oxygen is kept below 100 parts per million and the water less then 10 parts per million.  At this level of purity the absorption of electons is only a few percent.  The transverse diffusion ($\sigma=\sqrt{2Dt}$ where $D$ is the diffusion coefficient and $t$ is the drift time) of the gas is 230 $\mu m / \sqrt{cm}$ at full field (0.5 T).  For an electron drifting 2.1 m this give a transverse drift of 3.3 mm.  The drift velocity of the gas is 5.45 $cm/\mu s$.  There is a longitudional drift spread of about 230 ns (FWHM).  The diffusion sets the scale for the readout.  

\begin{figure}[htb]
\centering
\includegraphics[width=0.8\textwidth]{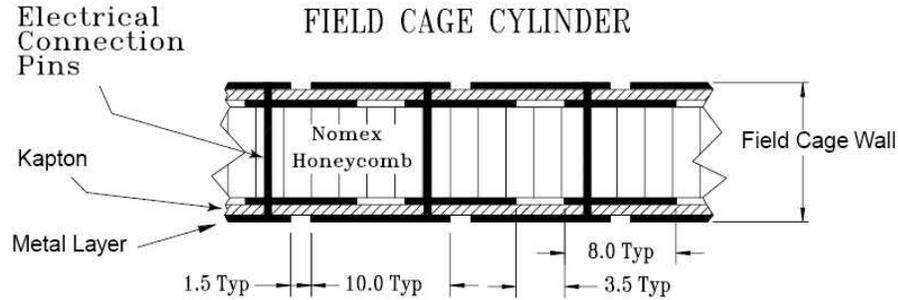}
\caption{Cutaway view of the field cage cylinder from \cite{expTPC}.  Distances (Typ) are in mm.}
\label{fig:cage}
\end{figure}

A uniform electric field is applied to the TPC which is what causes the electrons to drift to the ends.  The electric field is about 135 V/cm.  The electric field is provided by a thin high voltage conductive Central Membrane at the center of the TPC, the inner and outer concentric field-cages cylinders and readout end caps on the ends of the TPC.  Uniformity of the electric field is crucial to achieve submillimeter track reconstruction precision for electron drift paths of up to 2.1 m.  The Central Membrane (CM) is a disk with a central hole.  The CM is a cathode and kept at a potential of 28 kV.  It is made of 70 $\mu m$ of carbon-load Kapton film.  It is mounted inside the outer cage cylinder and the inner cage cylinder runs though the hole in the center.  Thirty six narrow stripes of aluminum have been attached to each side to provide targets for lasers for calibration.  The field cages keep the electric field uniform and provide containment for the gas.  As shown in Figure~\ref{fig:cage} the field cage cylinders consist of two layers of metal coated Kapton.  The layers are separated by a NOMEX honeycomb.  The metal is etched into 10 mm strips with a 1.5 mm separation so that the required voltage difference between rings can be maintained.  This design was optimized for reduced mass, minimization of track distortions from multiple Coulomb scattering and reduction of background from secondaries.    

\begin{figure}[htb]
\centering
\includegraphics[width=.6\textwidth]{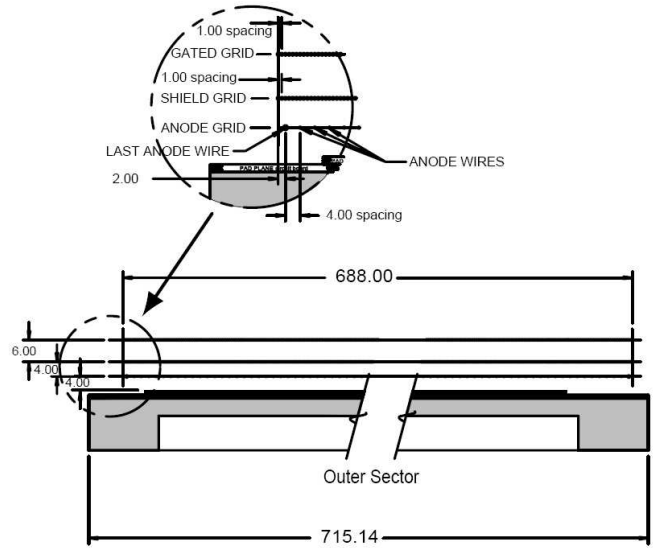}
\caption{Cutaway view of the outer subsector pad and wire planes.  All dimensions are in mm.  Diagram is from \cite{expTPC}.}
\label{fig:wire}
\end{figure}

\begin{figure}[htb]
\centering
\includegraphics[width=.9\textwidth]{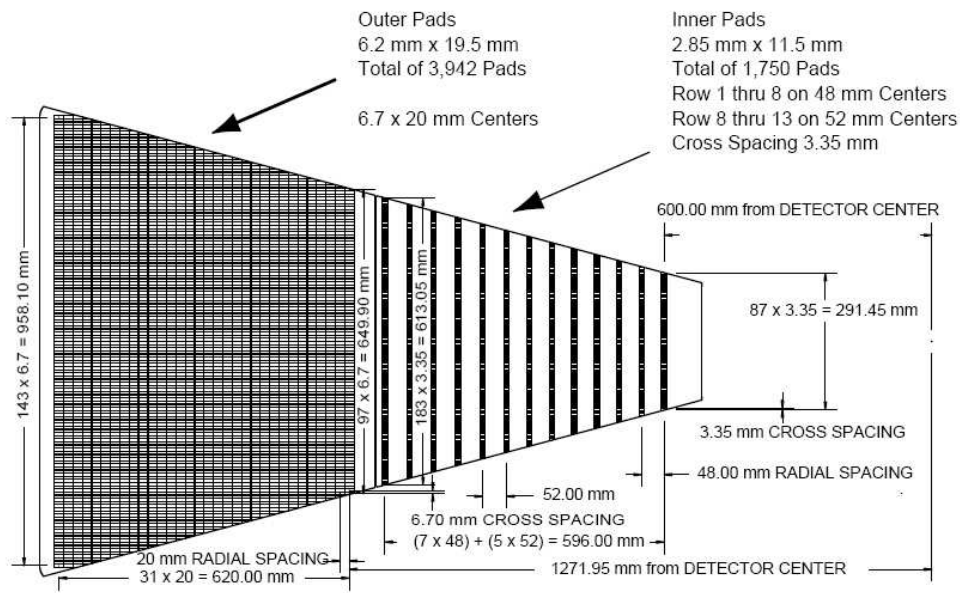}
\label{fig:pads}
\caption{The pad plane of one sector.  The inner subsector is to the right and the outer subsector is to the left. Diagram is from \cite{expTPC}.}
\end{figure}

The TPC end-caps contain anodes and pad planes.  The readout planes are multi-wire proportional chambers (MWPCs) with pad readout.  The chambers consist of three wire planes and a pad plane.  Figure~\ref{fig:wire} shows a cutaway view of the positions of the wires and the pads.  The amplification and readout are done by the anode wire planes which consist of 20 $\mu m$ wires.  The wire direction is chosen to best determine the momentum of very high transverse momentum particles whose tracks do not curve much in the magnetic field.  This places the anode wires roughly perpendicular to the radial direction.  In the other direction the resolution is limited by the wire spacing (4 mm).  The pad dimensions are also optimized to the best position resolution perpendicular to radial tracks.  The width of the pad is chosen such that the induced charge from an avalanche point on the anode wires shares most of its signal with three pads.  The outer radius pad subsectors have continuous pad coverage to achieve the best possible energy loss (dE/dx) resolution.  The outer radius subsectors are arranged on a rectangular grid with a pitch of 6.7 mm along the wires and 20.0 mm perpendicular to the wires and a 0.5 mm gap between pads.  There is a 4 mm separation between the pad plane and the anode wires.  The inner subsector pads are optimized for good two-hit resolution due to the high track density in that region.  The pads in the inner subsector are 3.35 mm along the wire and 12 mm perpendicular to the wires.  The inner and outer pads for a sector is shown in Figure~\ref{fig:pads}.  The distance to the anode wires is reduced to 2 mm to put most of the signal on three pads as it is in the outer sector.  The smaller pads give better tracking of the low momentum particles.  The smaller pads in the inner sector however require the use of separate pad row instead of continuous coverage due to constraints on the front end electronics.  The anode wires voltage is set independently for the two sectors to maintain a 20:1 signal to noise ratio for tracks from the center of the TPC.  A ground grid plane of 75 $\mu m$ wires is used to terminate the field in the avalanche region and provide additional shielding for the pads.  The anodes, pads, and grounding grid make up the MWPC.

The outermost wire plane is the gating grid and is located 6 mm from the ground grid.  The gating grid is used to control the entry of electrons from the TPC drift volume into the MWPC.  It also serves to keep positive ions from the MWPC from entering the TPC drift volume.  When the grid is open it is transparent to the drifting electrons.  The grid opens when the event is being recorded by putting all wires at the same voltage.  The grid goes to alternating positive and negative voltages ($\pm$75 V) when the grid is closed.  Positive ions from the MWPC are too slow to go though during the open period and are captured during the closed period.  The combined delays of the trigger and the opening time of the gating grid reduce the active length of the TPC from 210 cm to 198 cm due to electrons lost before the grid is opened.  

The MWPCs and gating grids are on 12 sectors on each end of the TPC.  These sectors are arranged around as circle.  There is 3 mm of dead space between the sectors.  Because of the dead space there is an azimuthal dependence to the particle detection efficiency.  

The x and y coordinates of a cluster are reconstructed assuming a Gaussian distribution for most tracks.  Tracks with a large crossing angle are reconstructed using a weighted mean algorithm because ionization is deposited on many pads.  The z coordinate of a cluster is determined by measuring the drift time of the secondary electrons from their origin to the anodes and dividing by the average drift velocity.  The arrival time of the cluster is calculated by measuring the arrival time of electron in discrete time intervals of about 100 ns each.  The charge weighted average time interval is used for the arrival time.  The drift velocity must be known to high precession for accurate z position reconstruction.  The drift velocity changes with pressure and small changes to the gas composition.  This is minimized by setting the cathode voltage so the electric field in the TPC is at the peak in the velocity vs. electric field / pressure curve.  The peak is broad and flat so small pressure changes will have little change on the drift velocity.  The drift velocity is independently measured every few hours using tracks created from laser beams.  The collision time can be offset by trigger delay, drift time from the gating grid to the anode wires, and the shaping of the signal in the front end electronics.  The timing offset can be adjusted by reconstructing the interaction vertex using data from each side of the TPC separately and matching them.  

The positions of secondary electrons are distorted by non-uniformities in the electric and magnetic fields and from global misalignments.  The typical size of the distortions is $\leq$1 mm.  This size of distortion can have an effect on the transverse momentum determination of high $p_{T}$ particles.  To correct for the distortions the magnetic field was measured using Hall probes and NMR probes.  The electric field was calculated from the geometry.  The hit position distortions are calculated and corrected for.  After correction the point to track fit error is about 50 $\mu m$ and the absolute error on a point is about 500 $\mu m$. 

\begin{figure}[htb]
\centering
\includegraphics[width=0.99\textwidth]{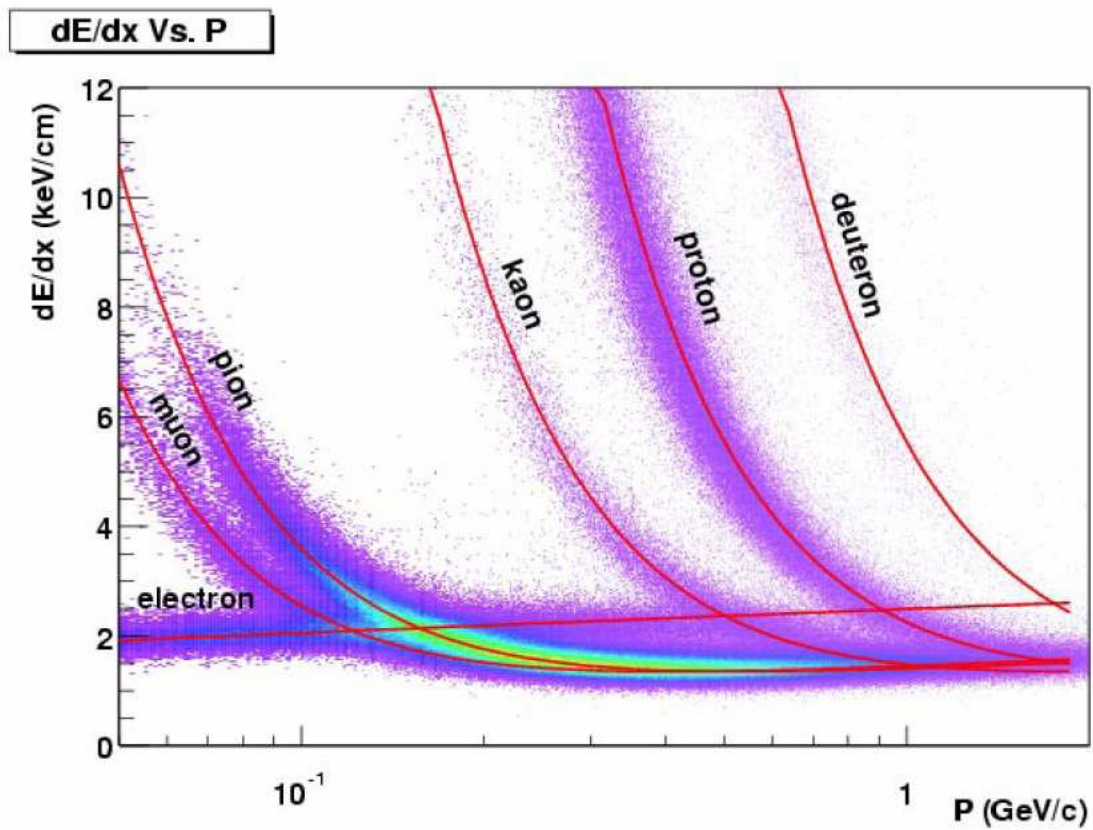}
\caption{Energy loss distribution of charged particles in the STAR TPC as a function of $p_T$ for 0.5 T magnetic field.  Plot is from \cite{expTPC}.}
\label{fig:introdedx}
\end{figure} 

The energy loss in the TPC gas can be used for identifying particles at low $p_{T}$ (below about 1.2 GeV/c) and at high $p_{T}$ (above about 3 GeV/c).  For the region in between the dE/dx bands cross so identification from energy loss is not possible.  The energy loss as a function of $p_T$ is shown in Fig.~\ref{fig:introdedx}.  A resolution of 7\% in relative dE/dx is required to distinguish protons and pions up to 1.2 GeV/c.  A resolution of 8\% in the relative dE/dx have been achived in the data.  The resolution depends on the gas gain which is pressure dependent.  The pressure is kept at 2 mbar above atmospheric pressure so it is time dependent.  A wire chamber with a $^{55}Fe$ source measurses the gas gain.  Local variations are calibrated by averaging at the pad-row level.  The energy loss is measured on up to 45 padrows for each track.  This has too few points to average out ionization fluctations.  Beacuse of this a most probable energy loss instead of an average energy loss is used for particle identification.
\chapter{TWO-PARTICLE JET-LIKE CORRELATIONS}

Jets make a good probe because their properties in elementary collisions can be calculated reliably by pQCD so they are well calibrated and in heavy ion collisions they interact with the created medium.  Since they interact with the medium we can study the effect of the medium on jets and the effect of jets on the medium.  In heavy-ion collisions it is not possible to reconstruct jets event by event due to large backgrounds.  To study jets in heavy-ion collisions we reconstruct jet-like correlations statistically through angular correlations.  In two-particle correlations this is done by triggering on an intermediate or high $p_T$ particle and studying the angular distributions of the other particles in the event with respect to the trigger particle.  In this thesis we will concentrate on azimuthal correlations.  

\section{Analysis Procedure}
We select our trigger particle such that it has transverse momentum, $p_T$, much greater than the average $p_T$ of the produced particles.  These high $p_T$ particles may predominantly come from jets of hard-scattering partons.  The selection of high $p_T$ particles, thus, preferentially triggers on jets.  The azimuthal distributions of lower $p_T$ particles, associated particles, in the event are investigated with respect to the trigger particles.  We shall denote the azimuthal angle of the associated particles and trigger particles as $\phi$ and $\phi_{Trig}$, respectively.  We shall denote the distribution of the lower $p_T$ particles with respect to the trigger particle as $J_2(\Delta \phi)$ where $\Delta \phi$ = $\phi-\phi_{Trig}$.  

For each trigger particle a $\Delta \phi$ distribution is constructed for all particles within a given $p_T$ window in the same event.  All trigger particles within a given $p_T$ range and all events within a given centrality are accumulated.  The correlation is normalized per trigger particle, not per event, because we are interested in quantities on a per-jet basis. The background $\Delta\phi$ distribution is constructed through event mixing.  The associated particles are taken from a different event as the trigger particle to construct the $\Delta \phi$ distribution.  The mixed events must be of the same centrality as the trigger particle but otherwise no conditions are place on them and will be referred to as inclusive events.  Our mixed events will take care of $\phi$ dependent effects due to detector efficiency to first order.  However, in a given event all particles are correlated with the reaction plane.  This is the flow correlation.  The flow correlation between the trigger particle and the background particles is lost in event mixing, because there is no correlation between the reaction plane of the triggered event and the reaction plane of the inclusive event.  This correlation is put in by hand using the measured elliptic flow ($v_2$) values.  In principal, one can mix events with the same (or similar) event planes; however, the measured event plane is not equal to the reaction plane due to event plane resolution.  The constructed mixed event background would need to be corrected for the event plane resolution.  This is more complicated than the current procedure.

The level of the background created from event mixing is not exact.  There are two reasons the level is incorrect  The first is that choosing events with a trigger particle gives us a bias in our event selection towards higher multiplicity events which have larger background multiplicity.  The background created through event mixing is too low to represent the background in the triggered events.  This bias is especially significant in {\it pp}, d+Au, and peripheral Au+Au collisions.  Another reason is that the total multiplicity used to determine centrality in the triggered events is of the underlying event plus the jet.  Thus the true background is lower than that from mixed events constructed from inclusive events of the same centrality window.  This effect is usually smaller than the other but can become the dominate effect in central Au+Au collisions.  To correct for these biases we introduce a normalization factor $a$.  An assumption must be made to the level of the background normalization to determine $a$.  The assumption used in our analysis is that the correlation signal is zero at $\Delta\phi=1$ (Zero Yield At 1 or ZYA1) which is the minimum of the correlation signal.  This assumption is known as an upper limit on the background since the signal is positive definite.  A assumption must be made because the true level of background is unknown a priori.

\subsection{Flow Correction}

In non-central Au+Au collisions, the overlap geometry is not isotropic.  This anisotropy results in a non-uniform pressure gradient in the initial stage of the collision, which in turn results in a momentum anisotropy in the final state.  This anisotropy is characterized by Fourier expansion,
\begin{equation}
	\frac{dN}{d\phi}=\frac{N}{2\pi}[1+2\sum_{n=1}^{\infty}v_{n}\cos n(\phi-\Psi)]
\label{fourier}
\end{equation}
where $\phi$ and $\Psi$ are the azimuthal angles of the particle and the reaction plane, respectively.  At mid-rapidity, the measured directed flow ($v_{1}$) is consistent with zero as expected due to symmetry.  For symmetric collisions, odd orders of $v_{n}$ are expected to be zero at midrapidity.  Elliptic flow ($v_{2}$), however, can be large especially at mid-rapidity.  For 2-particle correlations the elliptic flow is the only term that significantly contributes to our background; all higher orders of $v_n$ are negligible.  Keeping terms up to order $v_{2}$ we get the azimuthal distributions of the associated particles and the trigger particles respectively:

\begin{equation}
\frac{dN}{d\phi}=\frac{N}{2\pi}[1+2v_{2}\cos 2(\phi-\Psi)],
\label{v2}
\end{equation}
\begin{equation}
\frac{dN_{Trig}}{d\phi_{Trig}}=\frac{N^{Trig}}{2\pi}[1+2v_{2}^{Trig}\cos 2(\phi_{Trig}-\Psi)].
\label{v2t}
\end{equation}

\begin{figure}[htb]
\hfill
\begin{minipage}[t]{.49\textwidth}
\centering
	\includegraphics[width=1\textwidth]{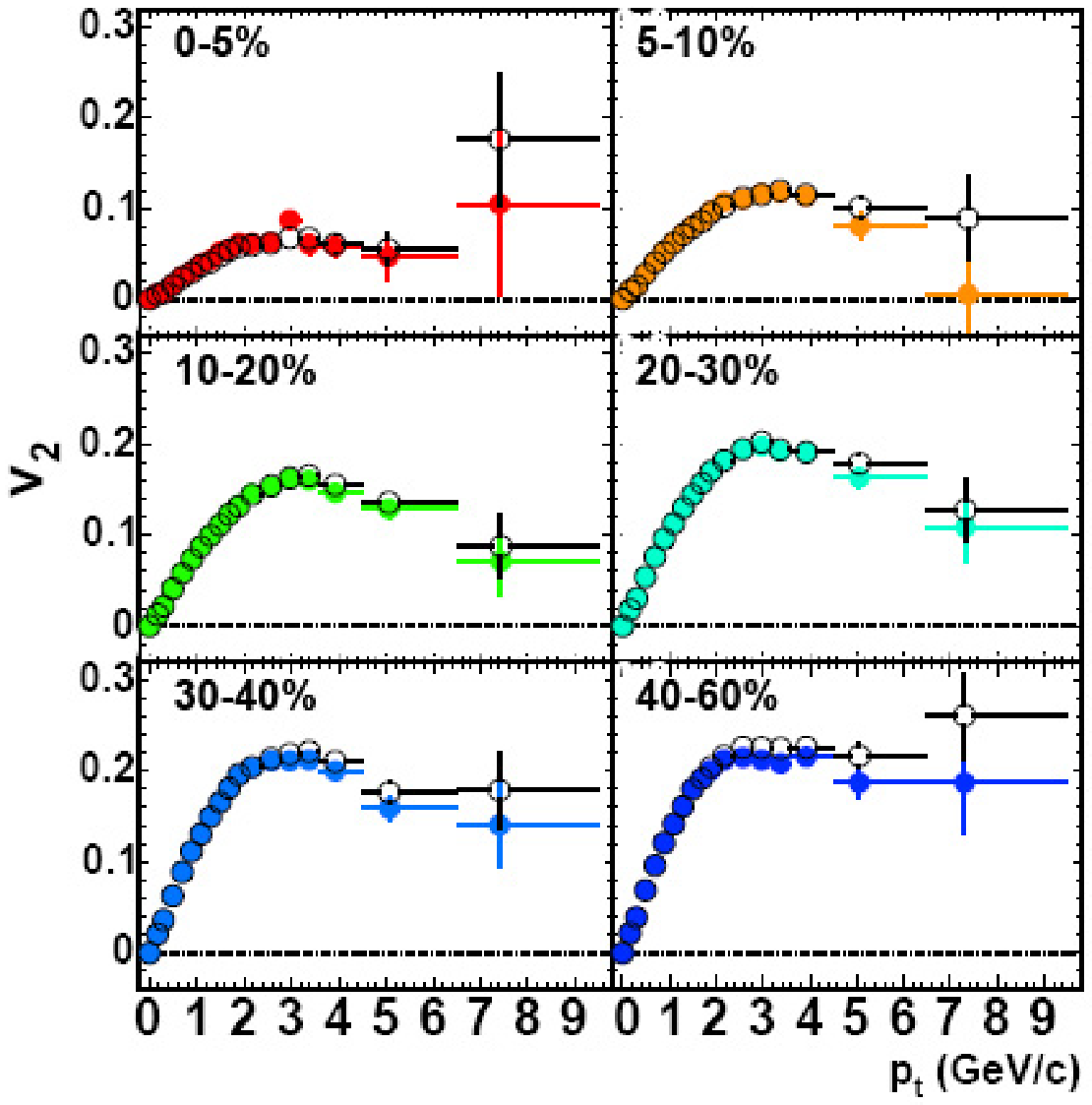}
\end{minipage}
\hfill
\begin{minipage}[t]{.49\textwidth}
\centering
\includegraphics[width=1\textwidth]{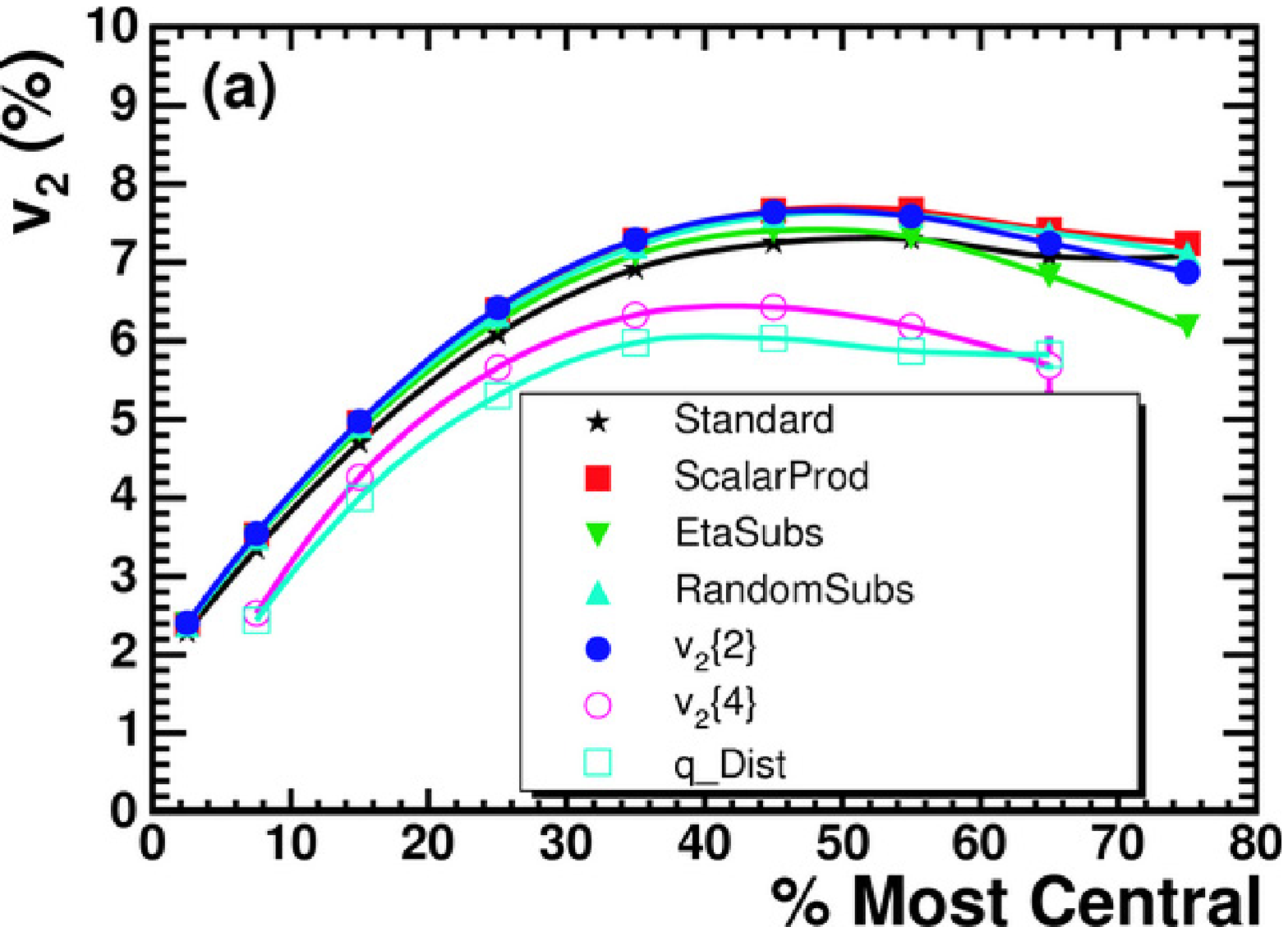}
\end{minipage}
\hfill
\caption{Left: Reaction plane and modified reaction plane $v_{2}$ results as a function of $p_T$ in different centrality windows.  Right:  Comparison of $v_{2}$ values using different flow measurement techniques as a function of centrality with `standard' signifying the reaction plane results and `$v_{2}$\{4\}' signifying the 4-particle cumulant results.  Flow measurements are from Au+Au collisions at $\sqrt{s_{NN}}=200 GeV$.  Plots are from reference\cite{flow}}
\label{fig:flowfigure}
\end{figure}

\begin{figure}[htb]
\centering
\includegraphics[width=0.6\textwidth]{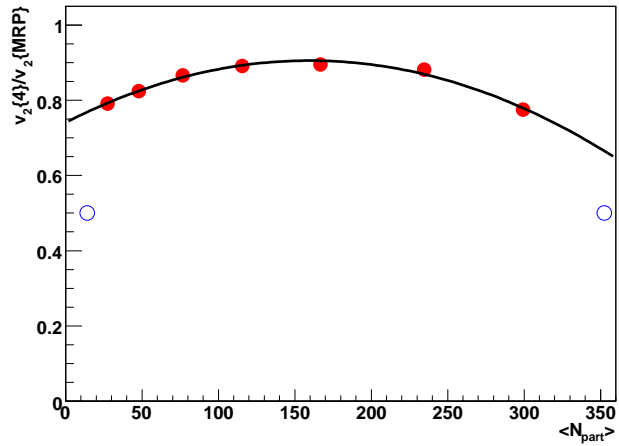}
\caption{Ratio of 4-particle cumulant and reaction plane $v_{2}$.  Solid points are measured.  The line is a fit to a second order polynomial.  The open points are a conservative estimate of the ratio for the centrality bins without 4-particle cumulant $v_{2}$ measurement.  Elliptic flow values are from \cite{flow}.}
\label{fig:flowratio}
\end{figure}

The correlation functions we are interested in are expressed in terms of $\Delta\phi=\phi-\phi_{Trig}$.  To obtain the flow correction in $\Delta\phi$, we integrated over $\Psi$, $\phi_{Trig}$, and $\phi$ as,
\begin{eqnarray}
\frac{dN}{d\Delta\phi} &=& \int_{0}^{2\pi}\int_{0}^{2\pi}\int_{0}^{2\pi}\frac{N N_{Trig}}{4\pi^{2}}[1+2v_{2} \cos 2(\phi-\Psi)]\nonumber \\  &&[1+2v_{2}^{Trig} \cos 2(\phi_{Trig}-\Psi)]\delta(\Delta\phi-(\phi-\phi_{Trig}))d\Psi d\phi d\phi_{Trig}\nonumber \\
&=& \frac{N N_{Trig}}{2\pi}[1+2v_2 v_2^{Trig}\cos(2\Delta\phi)]
\label{v22}
\end{eqnarray}
In this thesis, we use the average of the measured $v_{2}$ from the modified reaction plane, $v_{2}\{MRP\}$, and 4-particle cumulant, $v_{2}\{4\}$, methods from \cite{flow}.  The reaction plane method\cite{RP} over predicts the $v_{2}$ due to contributions from non-flow effects (such as jets).  The modified reaction plane method excludes particles with $|\Delta\eta|<0.5$ to reduce the non-flow contribution from jets.  The 4-particle cumulant method\cite{cumulant} is able to suppress the additional contributions from non-flow but under predicts the $v_{2}$ signal in the presence of $v_{2}$ fluctuations.  Figure~\ref{fig:flowfigure}, left, shows the $v_{2}$ values from the reaction plane and modified reaction plane methods as a function of $p_T$ for different centrality windows.   The right panel compares the $v_2$ values for the reaction plane and the 4-particle cumulant methods as a function of centrality (along with the values from some other methods).  Since there are no $v_{2}$ measurements for the top 5\% and 70-80\% for the 4-particle cumulant method we assume $v_{2}\{4\}=v_{2}\{MRP\}/2$ for these two centralities.  This is a very conservative estimate to account for any systematic uncertianties associated with extrapoliation.  The ratio $v_{2}\{4\}/v_{2}\{MRP\}$ is plotted in figure~\ref{fig:flowratio} for the measured and estimated values.  For our systematic uncertianty on the flow correction we will vary the $v_{2}$ value used between these two measurements.  The $v_{2}$ values for the two measurements are taken from published STAR data\cite{flow}.  The background normalization and flow subtraction proceedue has been previously used in Au+Au and {\it pp} collisions\cite{fuqiang}.

\begin{table}[hbt]
\centering
\caption{Elliptic flow parameterization.  For particles below 4 GeV/c $v_{2}=P_1 p_{T}^{P_2} e^{-(p_{T}/P_3)^{P_4}}$.  Above 4 GeV/c the high $p_{T}$ $v_{2}$ is used.  The systematic uncertainty on the $v_{2}$ is given in the last column.}
\begin{tabular}{|l|l|l|l|l|l|l|} 
\hline
Centrality&$P_1$&$P_2$&$P_3$&$P_4$&High $P_T$ $V_{2}$&Sys. Uncert.\\
\hline
70-80\%&0.43383&1.587&0.514&0.568&0.19265&$\pm$33\%\\
60-70\%&0.26470&1.410&1.376&0.809&0.23667&$\pm$12\%\\
50-60\%&0.19412&1.258&2.241&0.981&0.19747&$\pm$9\%\\
40-50\%&0.20927&1.303&2.002&0.981&0.19267&$\pm$7\%\\
30-40\%&0.16241&1.208&2.750&1.098&0.18034&$\pm$6\%\\
20-30\%&0.14261&1.213&2.846&1.027&0.17471&$\pm$6\%\\
10-20\%&0.12437&1.308&2.326&0.883&0.13340&$\pm$6\%\\
 5-10\%&0.08216&1.341&2.357&0.924&0.09434&$\pm$12\%\\
 0-5\% &0.03727&1.273&3.133&1.352&0.04797&$\pm$33\%\\
\hline
\end{tabular}

\label{tab:v2}
\end{table}

\subsection{Other Corrections}
\begin{figure}[htb]
\centering
\includegraphics[width=0.6\textwidth]{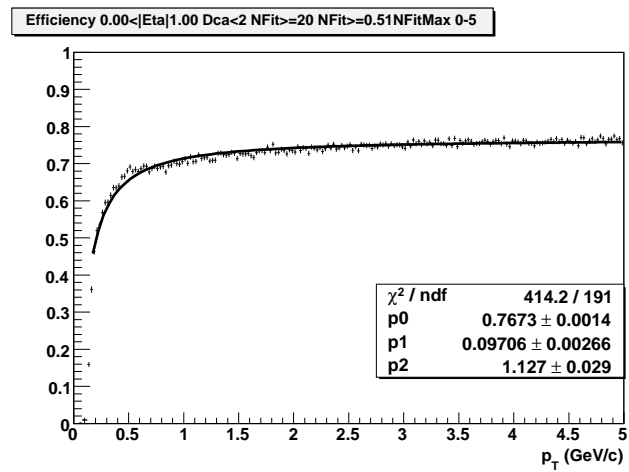}
\caption{Detector efficiency for 0-5\% most central Au+Au collisions at $\sqrt{s_{NN}}=200$ GeV as a function of $p_T$.  Errors are statistical.}
\label{fig:Eff}	
\end{figure}

The overall detector efficiency is obtained using a track embedding technique.  Monte Carlo tracks are placed within real events and then propagated through simulations of the detector responses\cite{levente}.  The probability the embedded track can be reconstructed determines the efficiency.  The tracks are run through a simulation of the detector geometry and then through the tracking software to determine the overall efficiency of the detector and the reconstruction algorithm.  The efficiency is determined as a function of $p_T$ using the same track quality and pseudorapidity cuts as is used in real data.  The charged pion efficiencies are used for all tracks because the majority of tracks are pions.  At low $p_T$, the pions dominate; while at high $p_T$, the proton contribution is no longer small, but the efficiencies for protons and pions are similar\cite{levente}. Efficiency corrections are only performed on the associated particle tracks only since the correlation functions are normalized per trigger and thus trigger particle efficiency cancels.  Figure~\ref{fig:Eff} shows the parameterization to the detector efficiency for 0-5\% central Au+Au collisions.  Effiencies for the other centralities of Au+Au collisions and d+Au collsions may be found in the appendix (Fig.~\ref{fig:Eff2}).  The  fit function is

\begin{equation}
	{\rm p}_0 e^{-(\frac{p_1}{p_T})^{p_2}}.
	\label{effeqn}
\end{equation}In the analysis the resultant fit function is then evaluated on a particle by particle basis and each particle is assigned a weight corresponding to the inverse of the efficiency. 

\section{d+Au Collisions}

Two-particle jet-like correlations have been previously studied in Au+Au and {\it pp} collisions\cite{fuqiang}.  These results have shown a broadened and enhanced away-side peak at low associated $p_T$ and a suppressed away-side peak at high associated $p_T$ in central Au+Au collisions with respect to {\it pp} collisions.  Two-particle azimuthal correlations in d+Au collisions at high $p_T$ have been studied before\cite{rab}. In this thesis, we analyze d+Au collisions with associated particles going down to low $p_T$.  These collisions are interesting because they will allow us to disentangle initial state nuclear effects from other effects seen in Au+Au collisions when compared to $pp$ collisions. 

\subsection{Results and Discussions}

\begin{figure}[htbp]
\hfill
\begin{minipage}[t]{0.4\textwidth}
	\centering
		\includegraphics[width=1\textwidth]{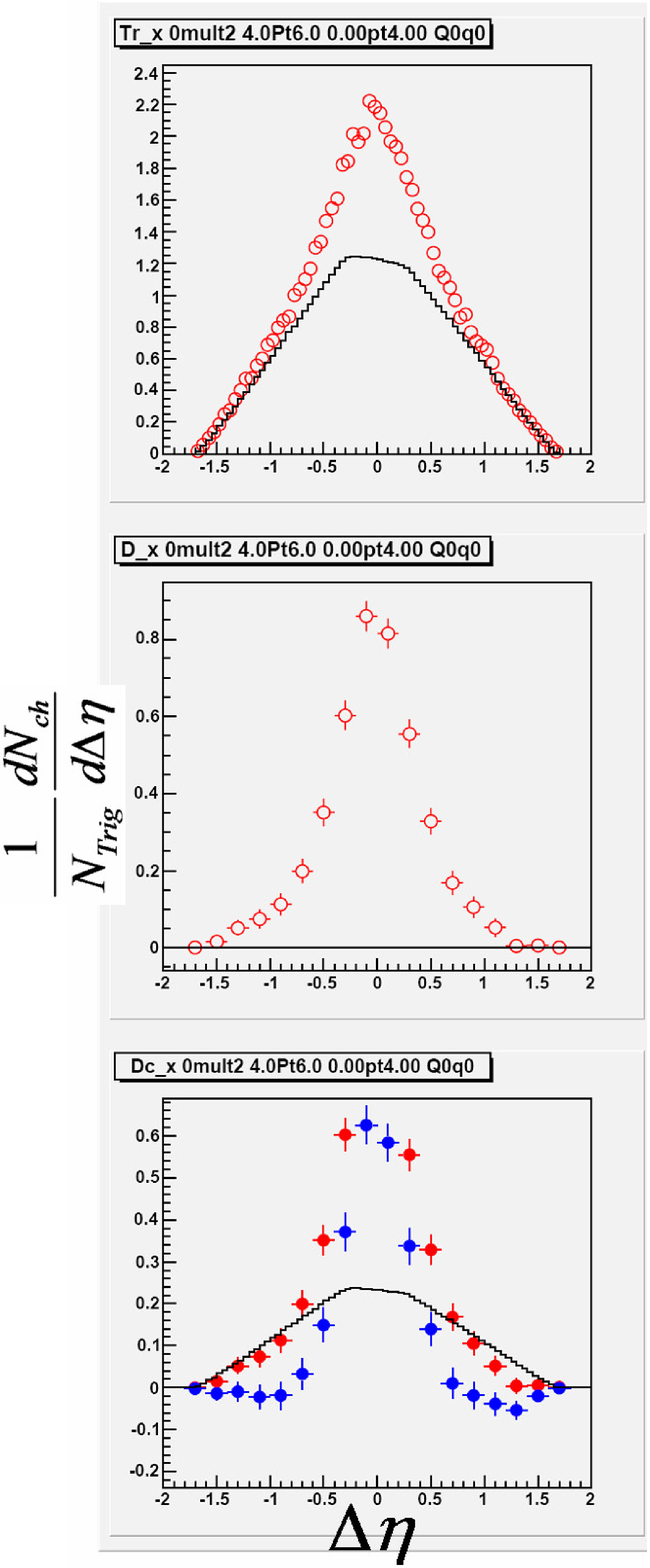}
\end{minipage}
\hfill
\begin{minipage}[t]{0.57\textwidth}
\centering
\includegraphics[width=1\textwidth]{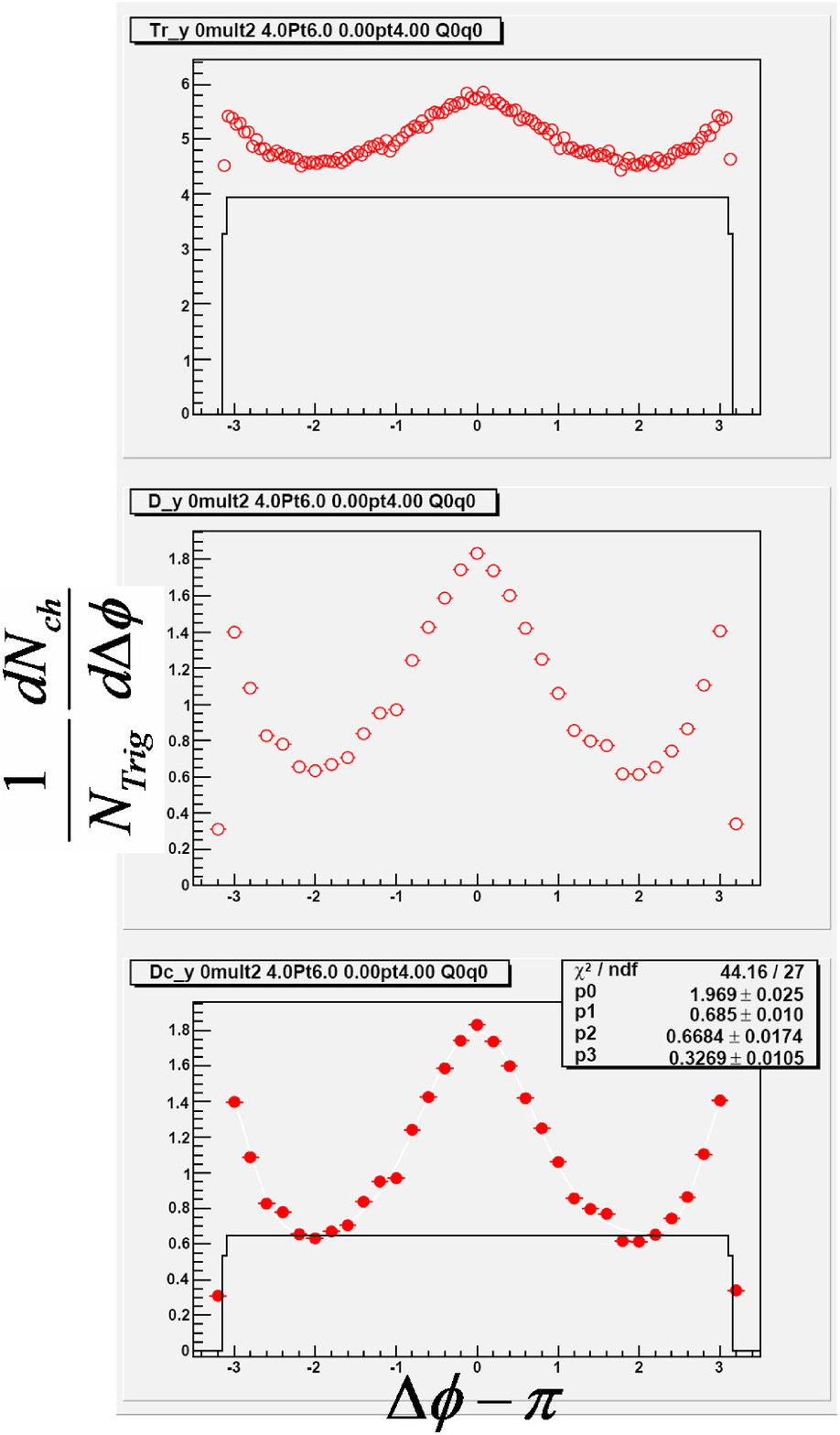}
\end{minipage}
	\caption{Analysis plots for minimum bias d+Au at $\sqrt{s_{NN}}=200$ GeV/c.  Top Left: raw signal in $\Delta\eta$ (open red points) and background from mixed events (black line).  Bottom Left:  background subtracted signal in $\Delta\eta$ (red points), background subtracted signal with additional normalized background subtracted (blue points), and residual background obtained from ZYA1 (black line).  Top Right: raw signal in $\Delta\phi$ (open red points) and background from mixed events (black line).  Bottom Right:  background subtracted signal in $\Delta\phi$ (red points) and residual background obtained from ZYA1.  The $p_T$ ranges of the trigger particles is $4<p_T<6$ GeV/c and that of the associated particles is $0.15<p_T<4.0$ GeV/c.  Errors are statistical.}
	\label{fig:dAuSig}
\end{figure}

We study these correlations as a function of $\Delta\eta$ and $\Delta\phi$, the $\Delta\phi$ correlations are studied in greater detail.  Figure~\ref{fig:dAuSig} shows the raw correlation functions in $\Delta\eta$ on the left and $\Delta\phi$ on the right in the top panels with the open red signals.  The $\Delta\eta$ correlations are shown only for particles on the near side (particles with $|\Delta\phi|<1$).  The black lines in the top panels show the backgrounds from event mixing.  Since this is d+Au no $v_{2}$ correlation is present so our $\Delta\phi$ background is just a flat line.  The effect of the TPC sector boundaries is too small to be visible on this plot.  The triangle-like shape to the $\Delta\eta$ background is a result of the detector 2-particle acceptance.  The triangle-like shape is not symmetric about $\Delta\eta=0$ because d+Au is an asymmetric system.  The trigger particle is constrained in $|\Delta\eta|<0.7$ while the associated particle are in $|\Delta\eta|<1.0$.  The associated particles are given the larger $\Delta\eta$ range so that most of the near-side jet-cone will be within our acceptance.
The bottom left and right panels show the background subtracted signal for $\Delta\eta$ correlations and $\Delta\phi$ correlations in red. The background we subtracted does not quite match the magnitude of the true background, as seen in the upper right and lower left panels.  There are 2 competing reasons for this mismatch.  The first (and the dominate one in d+Au collisions) is when we trigger on a high $p_T$ particle we bias our events towards higher multiplicity.  This leads to the background being too low.  The second arises from the centrality selection by cutting on the total reference multiplicity.  Since the total reference multiplicity includes jet-like correlated particles, the true background level is smaller than the mixed-event multiplicity.  We scale our mixed event background such that after subtraction the jet-like correlation signal is zero at $\Delta\phi = 1$.  This is done over a fixed range $|\Delta\phi\pm1|<0.2$.  The same scaling factor has been used for $\Delta\phi$ and $\Delta\eta$ correlations.  The final correlation functions are then obtained by subtracting the normalized background.  The blue points in the $\Delta\eta$ correlation have this additional mixed event background subtracted. 

\begin{figure}[htb]
\hfill
\begin{minipage}[t]{.49\textwidth}
	\centering
		\includegraphics[width=1\textwidth]{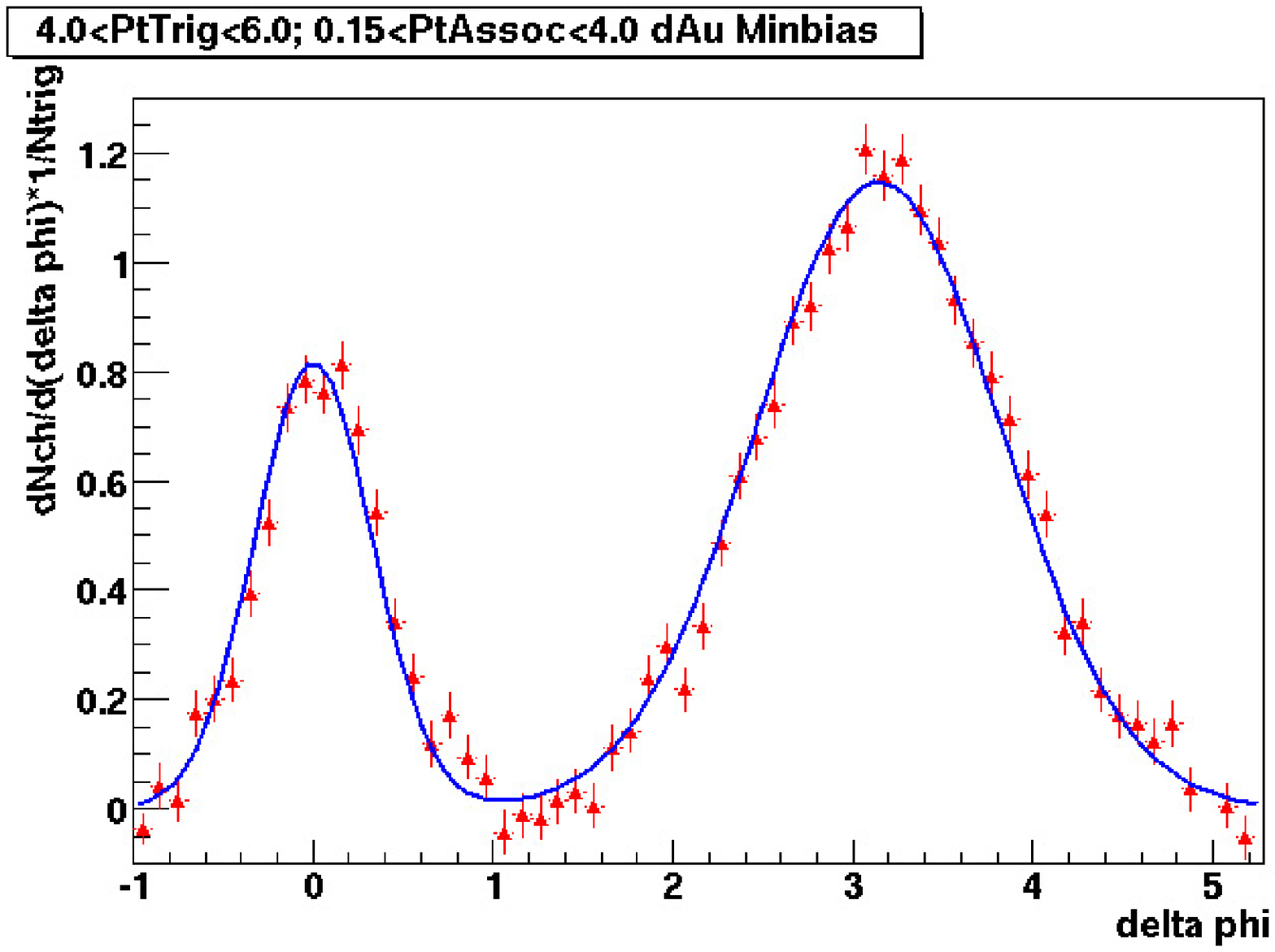}
			\end{minipage}
	\hfill
\begin{minipage}[t]{.49\textwidth}
	\includegraphics[width=1\textwidth]{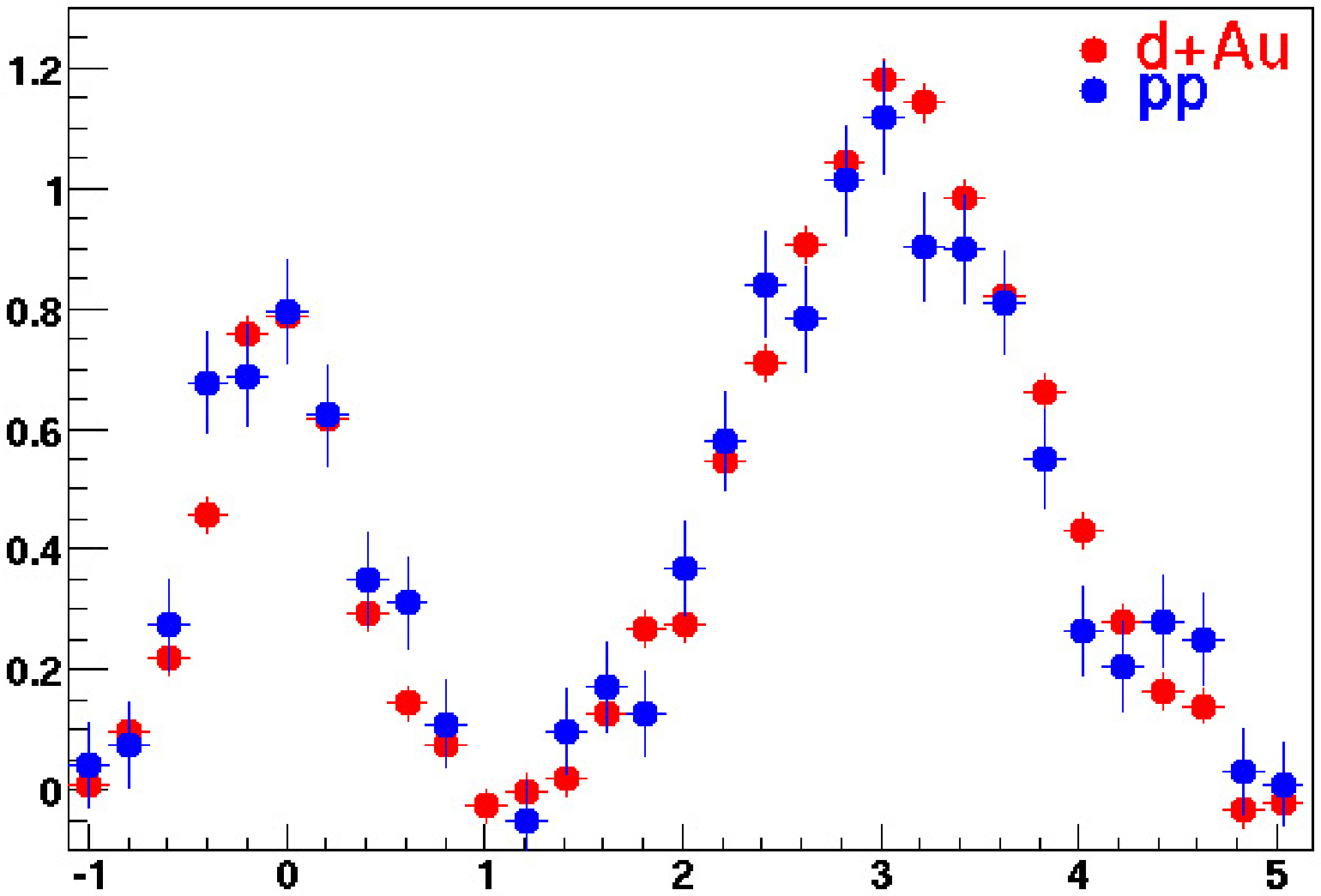}
			\end{minipage}
	\hfill
	\caption{Background subtracted $\Delta\phi$ distributions for d+Au collisions at $\sqrt{s_{NN}}=200$ GeV/c (red) for $4<p_{T}^{Trig}<6$ and $0.15<p_{T}^{Assoc}<4$ GeV/c.  Left:  Fit with two Gaussians with centroids fixed at 0 and $\pi$.  Right:  Compared with $pp$\cite{fuqiang} (blue).  Errors are statistical.}
	\label{fig:dAuPhi}	
\end{figure}

The $\Delta\phi$ signal with full background subtraction for d+Au collisions is shown if Figure~\ref{fig:dAuPhi} in the red symbols for $4<p_{T}^{Trig}<6$ GeV/c and $0.15<p_{T}^{Assoc}<4$ GeV/c.  The left panel has been fit to Gaussians with centroids fixed at 0 and $\pi$ for comparison.  The right panel shows a comparison to the  published $pp$ correlation (shown in blue points). The d+Au data has been rebinned in this plot.  The d+Au and $pp$ correlations are consistent which indicates that there is not significant initial nuclear effect in d+Au collisions.  It implies that the modification of the away-side jet-like correlation shape observed in central Au+Au collisions relative to $pp$ collisions is not an initial nuclear effect.

\begin{figure}[htb]
\hfill
\begin{minipage}[t]{.49\textwidth}
	\centering
		\includegraphics[width=1\textwidth]{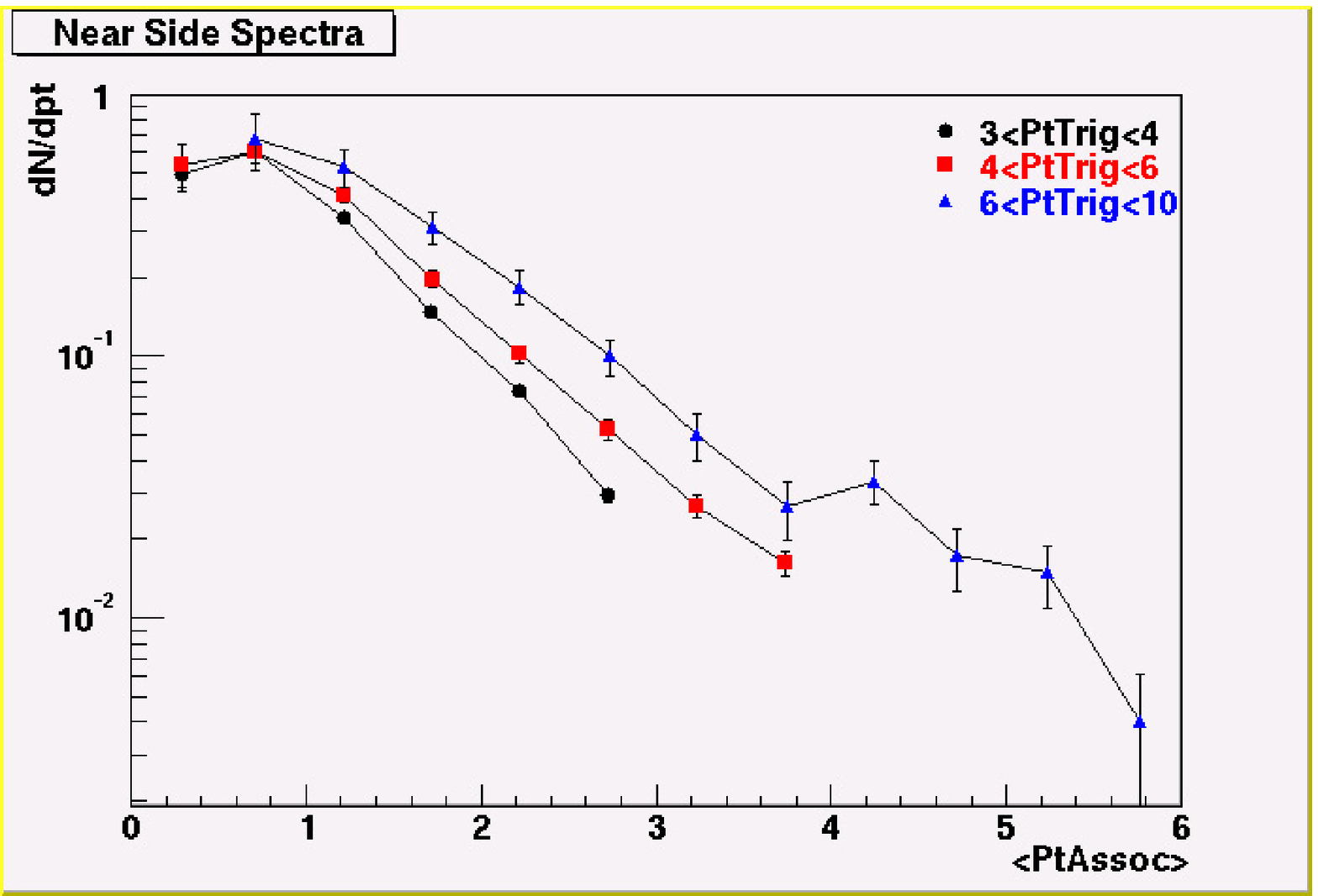}
		\includegraphics[width=1\textwidth]{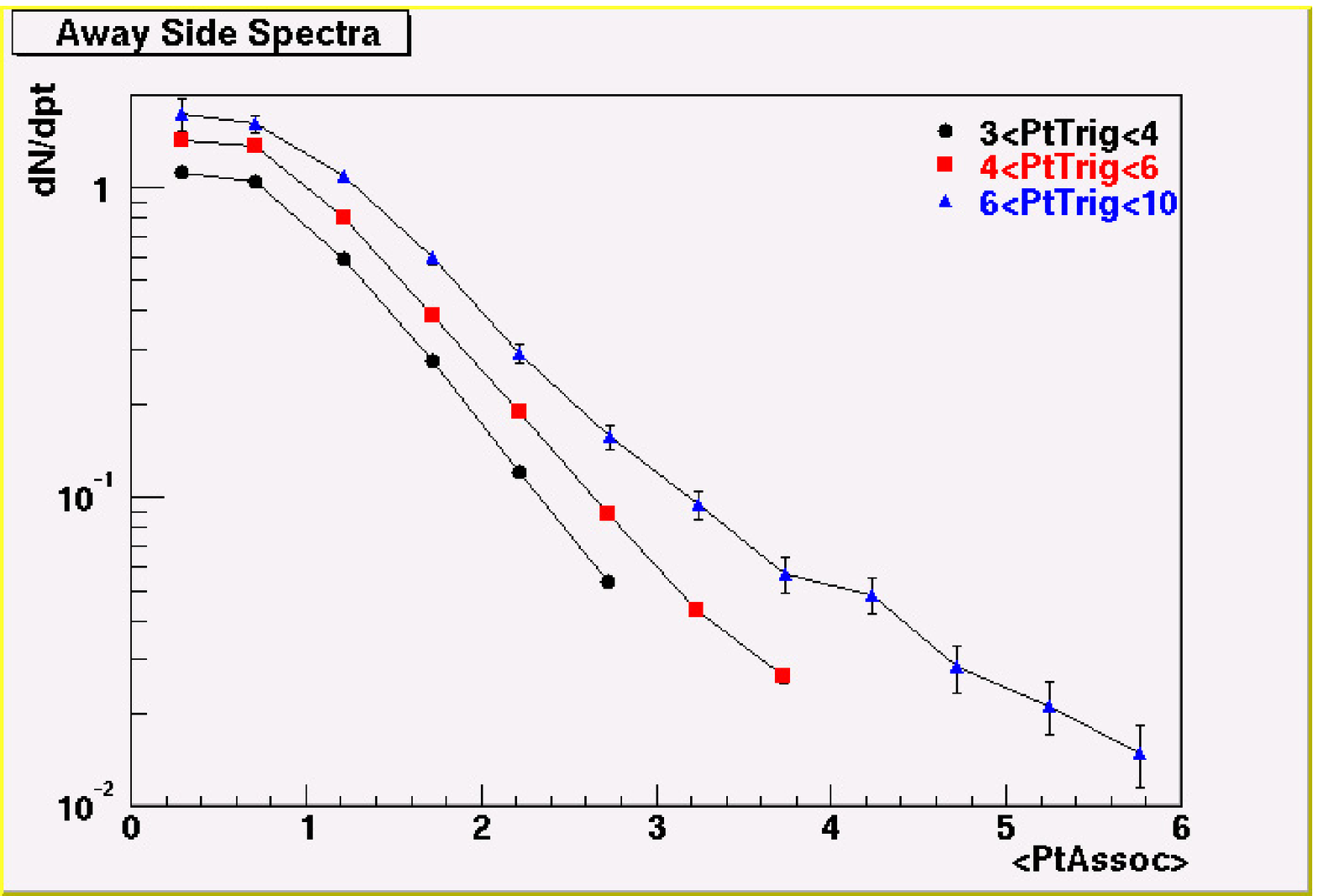}
			\end{minipage}
	\hfill
\begin{minipage}[t]{.49\textwidth}
	\includegraphics[width=1\textwidth]{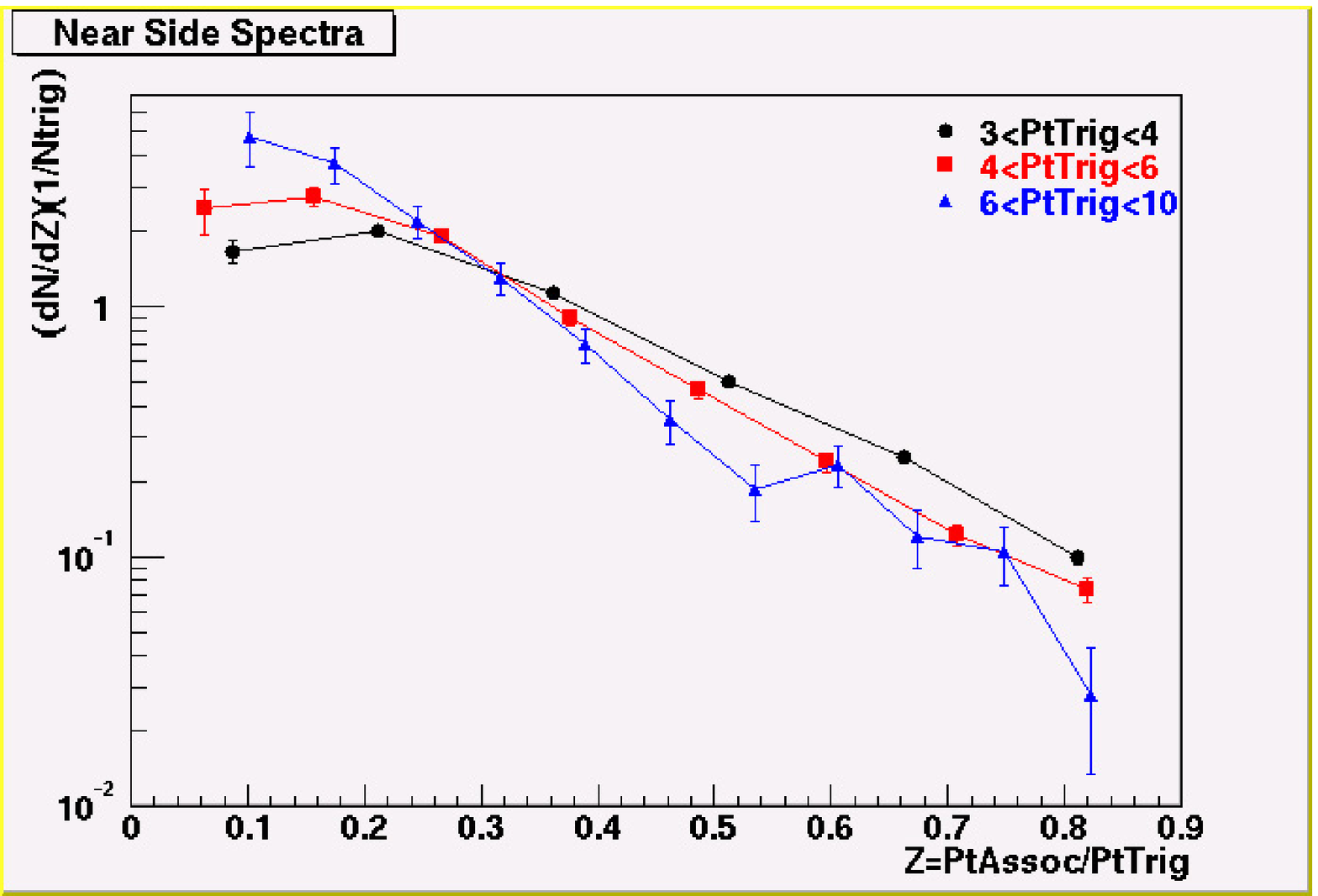}
	\includegraphics[width=1\textwidth]{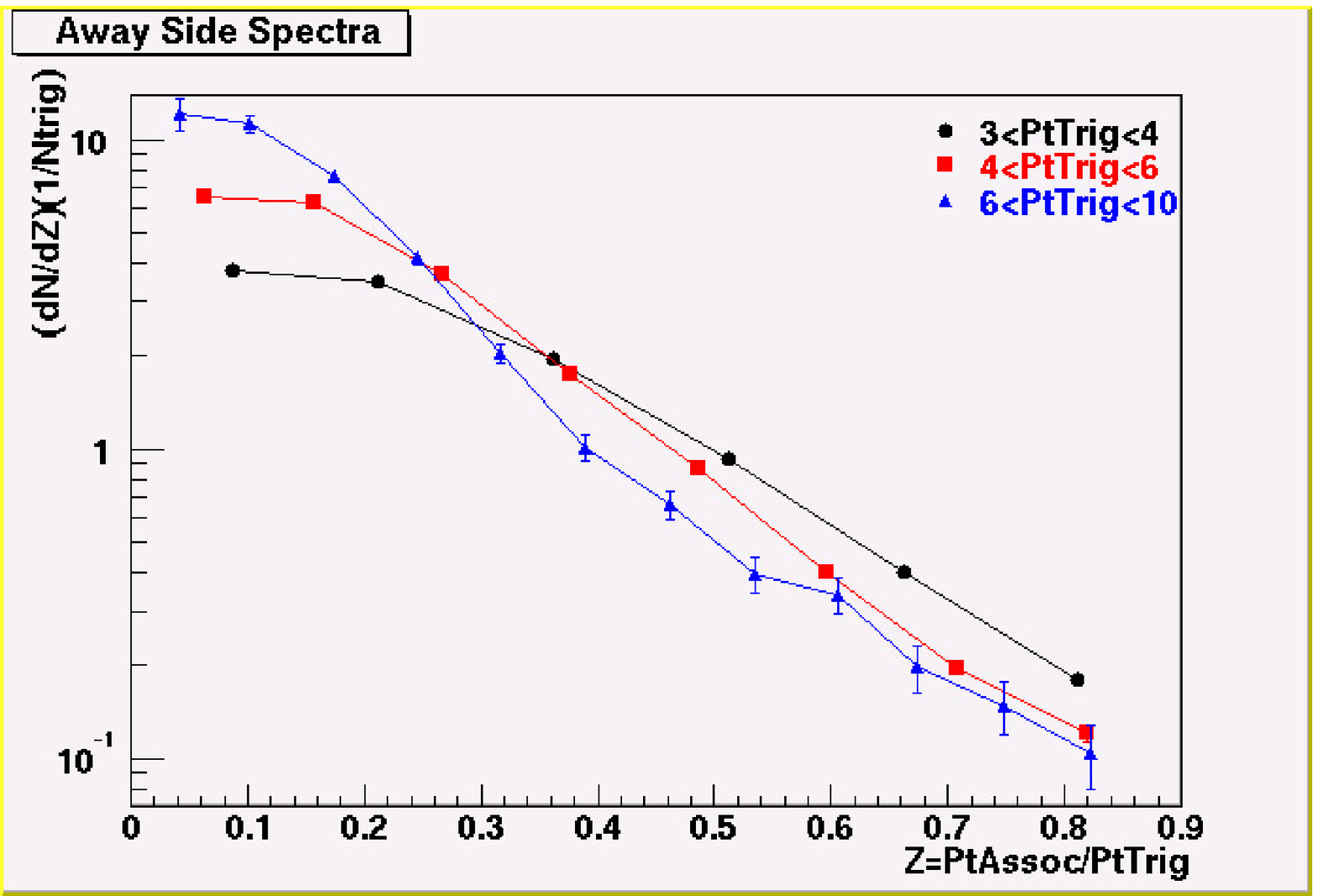}
			\end{minipage}
	\hfill
			\vspace*{-0.0cm}
	\caption{Associated particle spectra in d+Au collisions at $\sqrt{s_{NN}}=200$ GeV/c.  Left: $p_T$ spectra.  Right: $Z=p_{T}^{Assoc}/p_{T}^{Trig}$ spectra.  Top: Near-side spectra.  Bottom:  Away-side spectra.  Errors are statistical.}
	\label{fig:dAuSpect}	
\end{figure}

\begin{figure}[htb]
\centering
\includegraphics[width=0.6\textwidth]{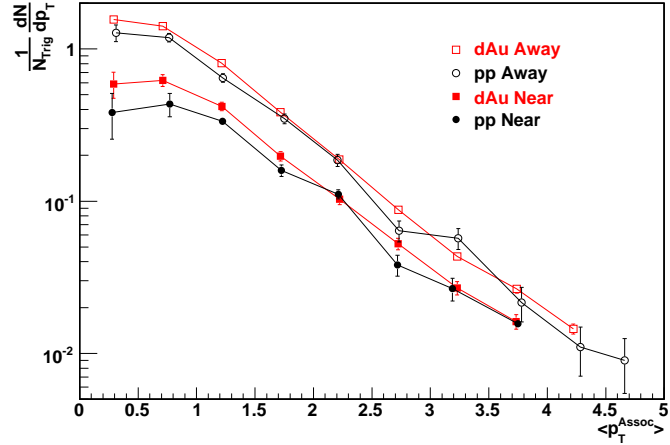}
\caption{Comparison of d+Au and {\it pp} spectra at $\sqrt{s_{NN}}=200$ GeV/c for $4<p_T^{Trig}<6$ GeV/c.  Both the near-side and away-side spectra are shown.  The $pp$ spectra is from \cite{fuqiang}.  Errors are statistical.}
\label{fig:spectcomp}
\end{figure}

We study the correlation functions versus associated particle $p_T$ and extract the correlated yields versus associated $p_T$.  Figure~\ref{fig:dAuSpect} shows the associated particle spectra.  The left column of plots show the near-side spectra and the away-side spectra on the top and bottom, respectively.  The spectra are shown for trigger particles of $3<p_{T}<4$ GeV/c, $4<p_{T}<6$ GeV/c, and $6<p_{T}<10$ GeV/c.  The spectra become harder as the trigger $p_T$ increases on both the near-side and away-side.   The right column of plots shows the $Z$-spectra where $Z=p_{T}^{Assoc}/p_{T}^{Trig}$.  The spectra are shown this way to present it in a format that is closer to the fragmentation function which is in terms of $z=p^{Assoc}/p^{Jet}$, where $p^{Jet}$ is the parton momentum (or energy) and and $p^{Assoc}$ is the fragment momentum projection along the parton momentum direction.  The $Z$ spectra are harder for the lower $p_T$ triggers, as seen in the figure.  The real fragmentation function in z is independent of the jet energy.  The reason for the change as a function of trigger $p_T$ is that the trigger particle momentum becomes a better proxy for the parton momentum as we increase $p_T$, our triggering on a high $p_T$ particle biases us towards jets where more of the parton energy is contained in the trigger momentum this bias increase with trigger $p_T$.  Figure~\ref{fig:spectcomp} shows the near-side and away-side spectra for both d+Au and $pp$ collisions.  The spectra are constant between the two collision systems showing no significant initial nuclear effects.

\begin{figure}[htb]
\hfill
\begin{minipage}[t]{.49\textwidth}
	\centering
		\includegraphics[width=1\textwidth]{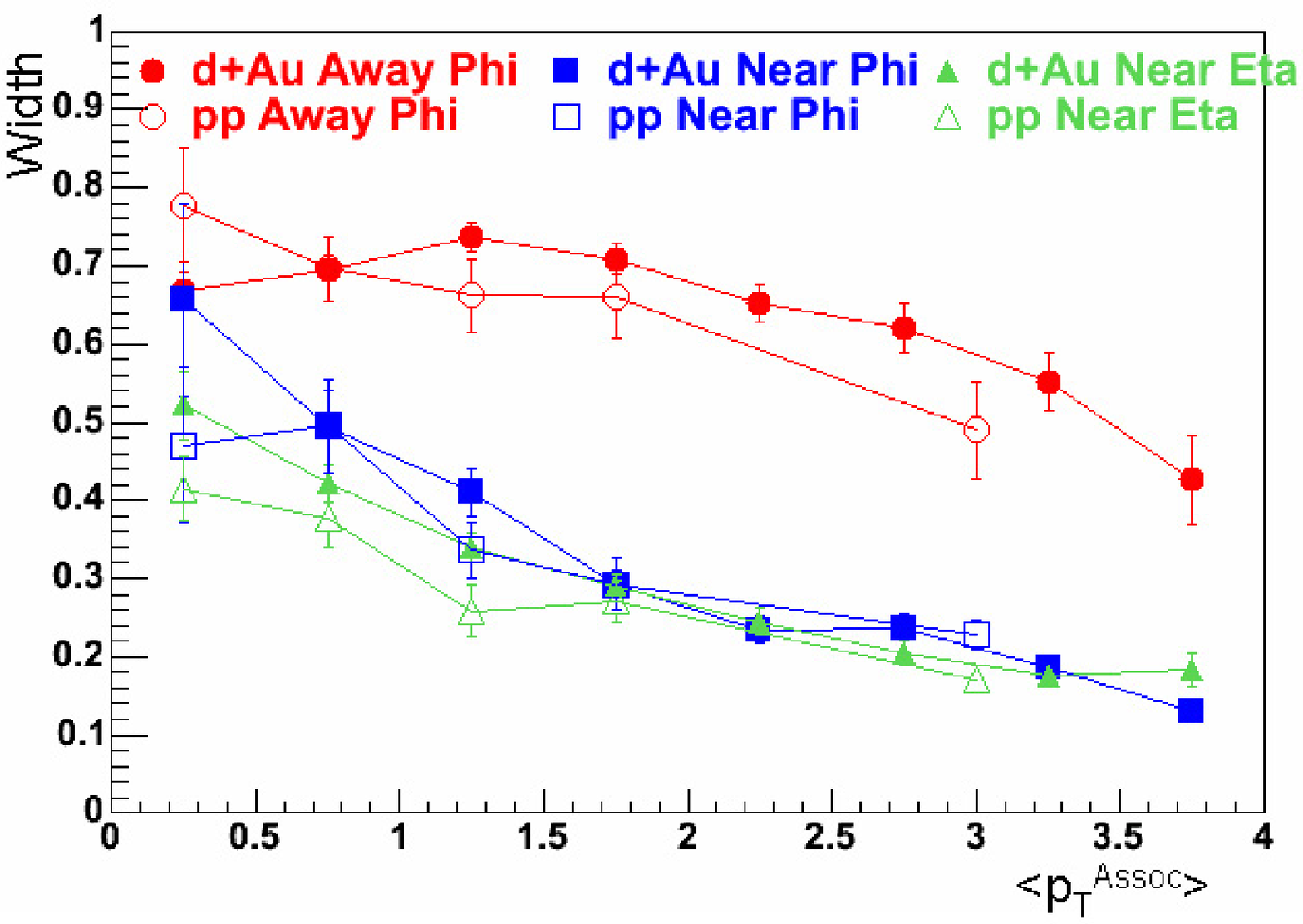}
			\end{minipage}
	\hfill
\begin{minipage}[t]{.49\textwidth}
	\includegraphics[width=1\textwidth]{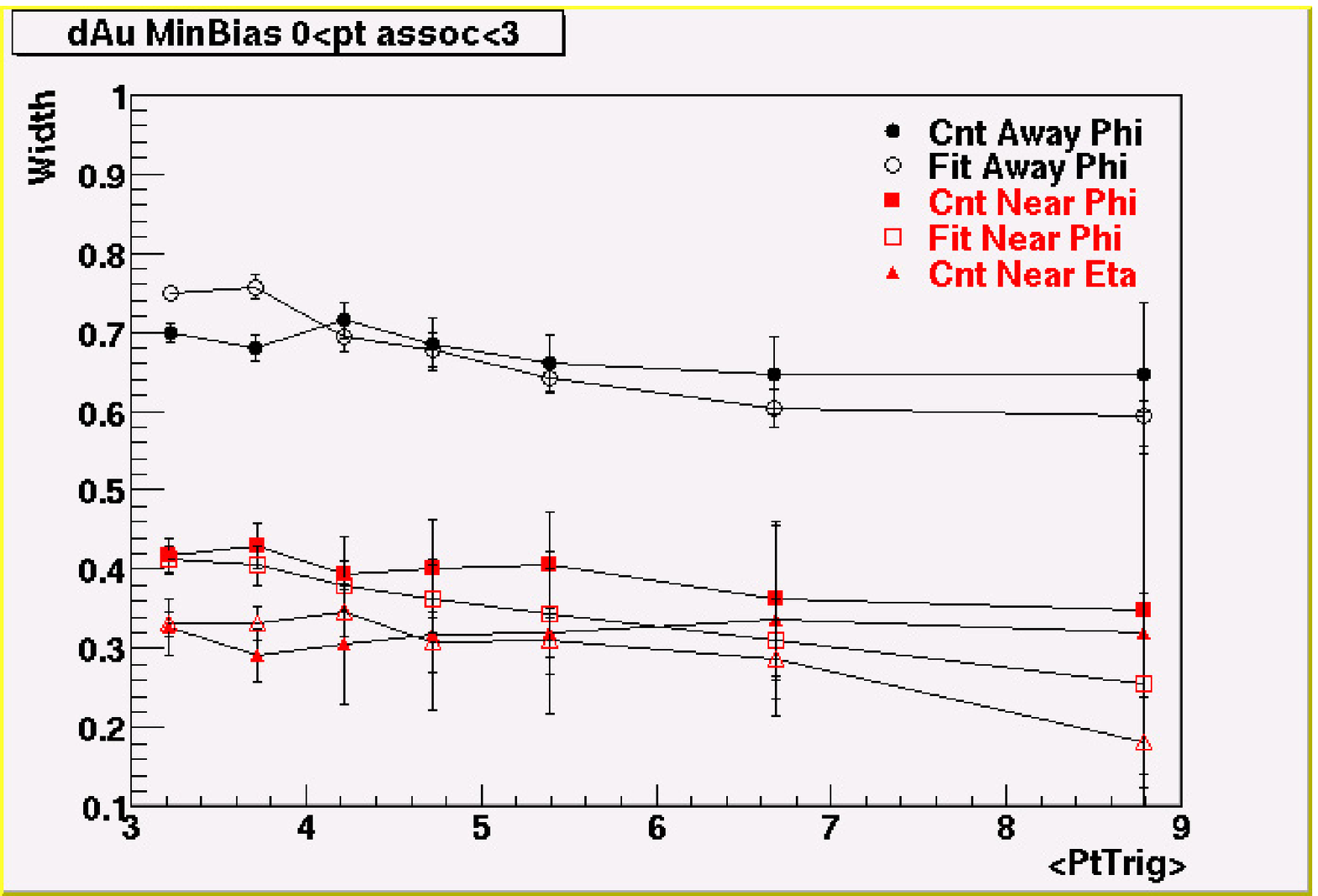}

			\end{minipage}
	\hfill
			\vspace*{-0.0cm}
	\caption{Jet-like correlation peak widths.  Left: RMS widths as a function of associated $p_{T}$.  Open symbols are for {\it pp}\cite{fuqiang} and filled symbols are for d+Au $\sqrt{s_{NN}}=200$ GeV/c.  Circles are for the away-side width in $\phi$.  Squares are for the near-side width in $\phi$.  Triangles are for the near-side width in $\eta$.  Right:  RMS width (filled) and Gaussian fit sigma (open) as a function of trigger $p_T$.  Errors are statistical.}
	\label{fig:dAuWidth}	
\end{figure}

The width of the jet-like correlations is also interesting.  Figure~\ref{fig:dAuWidth} (left) shows the RMS width as a function of associated particle $p_T$ for both {\it pp} and d+Au collisions for $4<p_{T}^{Trig}<6$ GeV/c.  Widths are shown in both $\eta$ and $\phi$ for the near-side and for $\phi$ on the away-side.  The width of the $\eta$ distribution on the away-side is not shown because the away-side is flat within our detector acceptance.  This is because the near-away and away-sides partons are not correlated in $\eta$ due to the colliding partons carrying different fractions of the longitudinal momentum to the collision giving the jet an overall longitudinal momenta of the colliding nucleons.  The d+Au widths show no significant changes from the width in {\it pp}.  The right panel shows the width as a function of trigger particle $p_{T}$ for associated particles of $0.15<p_{T}<3.0$ GeV/c.  The open symbols are the widths from Gaussian fits.  The filled symbols are from RMS.  The widths are consistent for the two measurement techniques because the peaks are very Gaussian.  In both figures, the widths are similar in $\eta$ and $\phi$ on the near-side.  This shows that the near-side emission is within a cone.  The near-side should not have same width as what would be calculated in fragmentation for widths respect to the jet-axis.  It should be broadened due to the use of the trigger particle as a proxy for the jet-axis.  The away-side has additional broadening due to two coupled reasons.  One is $k_{T}$ (initial parton transverse momentum).  This initial transverse momentum results in the near-side and away-side not being quite back-to-back in $\phi$ for a given di-jet.  Upon averaging over many di-jets this results in a broadening on the away-side.  The other reason is trigger bias, where the trigger particle tends to select the jet with higher transverse momentum, also due to the fact that the partons carry some initial transverse momentum. The backside partner jet has lower energy and thus the fragmentation cone is wider. In principle, one should be able to extract valuable information about the initial $k_T$ broadening from the width measurements in $pp$ and d+Au within a model framework.  One interesting note is that one would expect the $k_T$ broadening to be larger in d+Au than in pp because of initial multiple scatting in d+Au, however, our $pp$ and d+Au comparison result does not seem to show this is a significant effect.  

\subsection{Summary}

Jet-like correlations have been studied in d+Au collisions at $\sqrt{s_{NN}}=200$ GeV/c.  The correlations are consistent with jet-like correlations in {\it pp} collisions.  This implies that the modification to the away-side correlation observed in central Au+Au collisions is not an initial nuclear effect.  This also justifies the use of d+Au data for comparison to Au+Au collisions instead of {\it pp} collisions.  The benefit gained from this is the greatly increased statistics for d+Au collisions compared to {\it pp}.  Associated particle spectra and correlation widths are extracted for d+Au collisions and can be used for comparison to theoretical calculations.

\section{Au+Au Collisions}

Two-particle correlations have already been studied in Au+Au collisions at $\sqrt{s_{NN}}=200$ GeV at STAR \cite{disap,fuqiang}.  The publications used Au+Au data from the second year of RHIC running.  This thesis extends the measurements using data from the fourth year of RHIC running which has about an order of magnitude increased statistics.  In this thesis, we used the same trigger and associated $p_T$ cuts as are used in the  PHENIX publication\cite{phenix} to check for consistency between the two experiments.  

\begin{figure}[htb]
\centering
\includegraphics[width=0.6\textwidth]{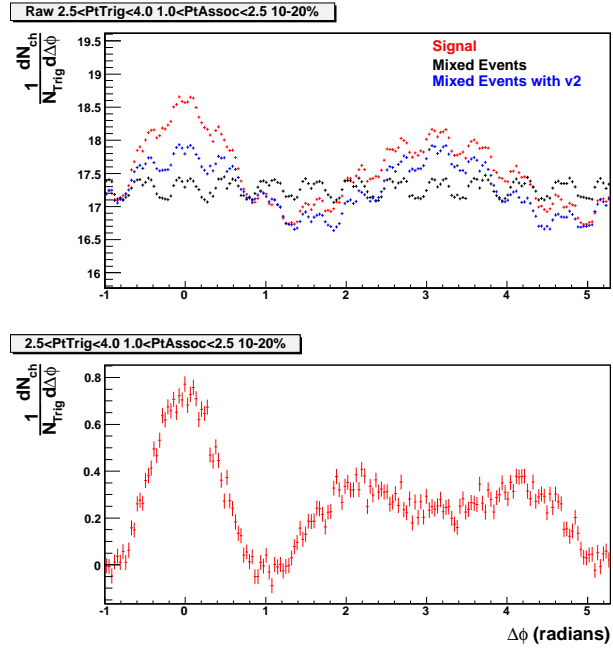}
\caption{Two-particle correlation functions for $2.5<p_{T}^{Trig}<4.0$ GeV/c and $1<p_{T}^{Assoc}<2.5$ GeV/c in 10-20\% Au+Au collisions at $\sqrt{s_{NN}}=200$ GeV/c.  Top:  Raw correlation function (red).  Background from event mixing (black).  Background from event mixing with elliptic flow added in pairwise by hand (blue).  Bottom:  Background subtraced 2-particle correlation function.  The background is normalized by ZYA1 before subtraction.  Errors are statistical.  Plots for all of the Au+Au centrality bins can be found in the appendix.}
\label{fig:Aubkg}
\end{figure}

The background subtraction method in Au+Au collisions is the same as that for the d+Au collisions except for the flow subtraction.  The elliptic flow has been added to the mixed events pairwise by hand as discussed in section 3.1.1.  The top panel of Figure~\ref{fig:Aubkg} shows the raw 2-particle correlation function in $\Delta\phi$ in red.  In black are the mixed events, where the associated particles are taken from different events of the same centrality as the trigger particle.  In blue are the mixed events with the elliptic flow correction added in pairwise.  The bumps and dips apparent in all three distributions are due to the TPC sector boundaries.  The bottom panel shows the background subtracted correlation function.  This is the raw 2-particle correlation function minus the normalized mixed event background with elliptic flow.  The mixed event with elliptic flow background is normalized such that the background subtracted correlation function is zero in the region $|\Delta\phi\pm1|<0.2$\cite{fuqiang} as was done in d+Au collisions.

\subsection{Comparison Between STAR and PHENIX}

\begin{figure}[htb]
\hfill
\begin{minipage}[t]{0.49\textwidth}
\centering
\includegraphics[width=1.0\textwidth]{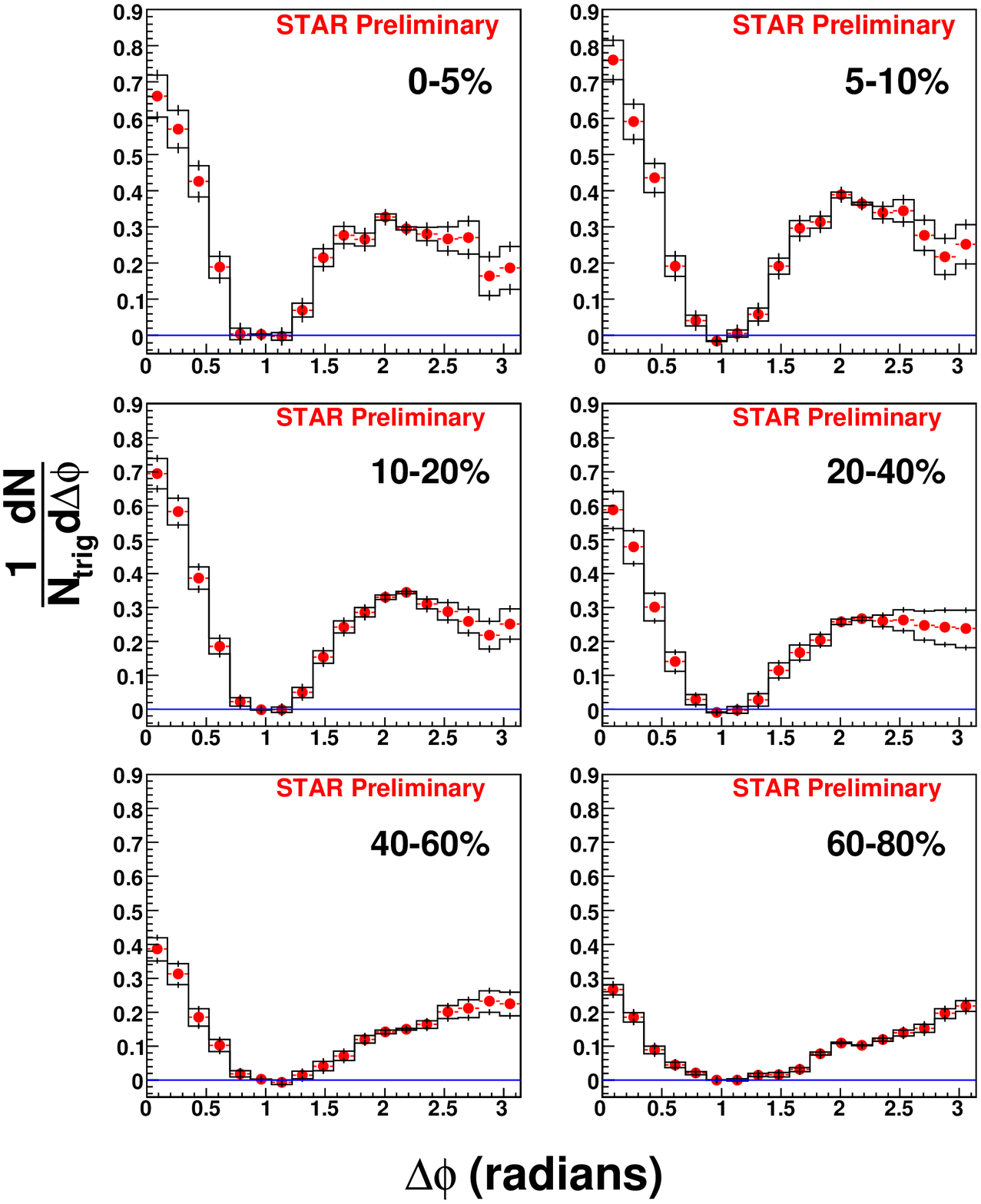}
\end{minipage}
\hfill
\begin{minipage}[t]{0.49\textwidth}
\centering
\includegraphics[width=1.0\textwidth]{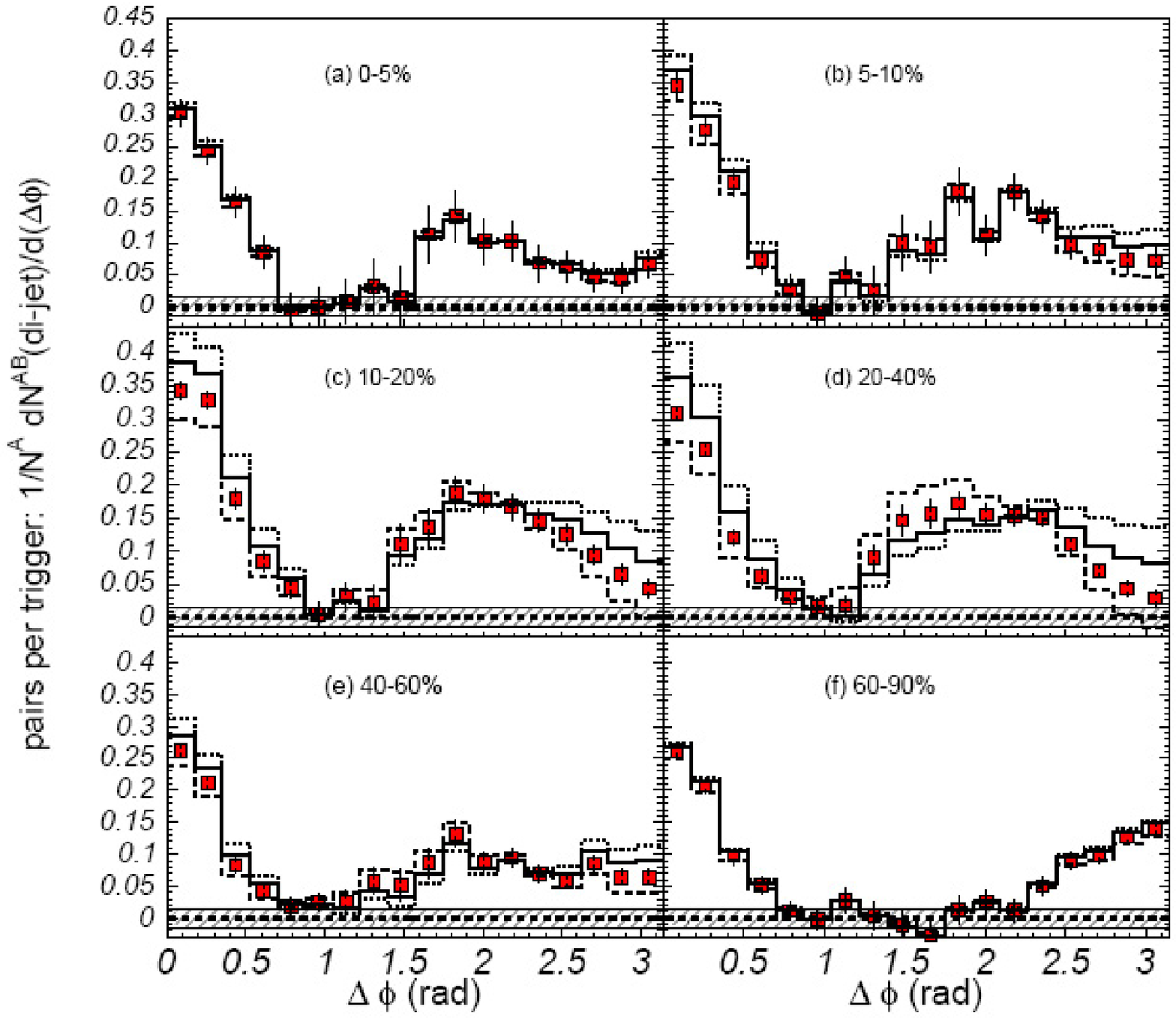}
\end{minipage}
\caption{Background subtracted correlation functions in Au+Au collisions at $\sqrt{s_{NN}}=200$ GeV/c for $2.5<p_{T}^{Trig}<4.0$ GeV/c and $1.0<p_{T}^{Assoc}<2.5$ GeV/c.  Left:  Our results with $|\eta^{Assoc}|<1.0$.  Right:  PHENIX results from \cite{phenix} with $|\eta^{Assoc}|<0.35$.  Error bars are statistical errors.  Histograms represent 1 sigma systematic uncertainty on the elliptic flow, except that the  top most histogram on the PHENIX results is -2 sigma systematic uncertainty.  Shaped bands on PHENIX results are the normalization uncertainty.}
\label{fig:AuAu}
\end{figure}

Figure~\ref{fig:AuAu} shows the background subtracted 2-particle correlation functions in Au+Au collisions at $\sqrt{s_{NN}}=200$ GeV/c using the high statistics year 4 data.  The correlation functions are for $2.5<p_{T}^{Trig}<4.0$ GeV/c and $1.0<p_{T}^{Assoc}<2.5$ GeV/c.  The left set of panels shows our results and the right panels show the published PHENIX results\cite{phenix}.  The magnitude of the peaks is higher in our results than the PHENIX results due to our larger acceptance of associated particles in $\eta$; our results have $|\eta^{Assoc}|<1.0$ while the PHENIX results are for $|\eta^{Assoc}|<0.35$.  This increases the magnitude of our away-side peaks by 1.0/0.35 since the away-side is evenly distributed in $\eta$\cite{fuqiang}.  Our near-side peak is also larger because of two reasons.  One is that the near-side peak is broad and our larger $\eta$ acceptance catches more of the associated particles.  The other is that we include more of the long range $\eta$ correlations (the ridge) that have been observed in 2-particle $\Delta\phi-\Delta\eta$ correlation functions\cite{fuqiang,dan,jorn}.  

\begin{figure}[htb]
\centering
\includegraphics[width=.6\textwidth]{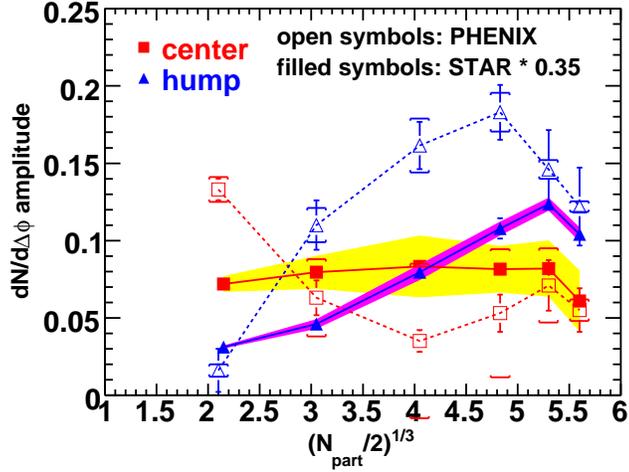}
\caption{Away-side correlation amplitudes in the hump ($\pi/3<|\Delta\phi-\pi|<4\pi/9$) and center ($|\Delta\phi-\pi|<\pi/9$) as a function of centrality, $(N_{part}/2)^{1/3}$ shown in blue and red respectively.  Our data is shown in filled symbols and has been scaled down by a factor of 0.35 to account for the $\eta$ acceptance.  The PHENIX data is shown in open symbols.  The bars shown the statistical errors and the systematic errors due to flow are shown in shaded bands and caps for our data and the PHENIX data, respectively.}
\label{fig:AuAu2}
\end{figure}   

We are most interested in comparing the shape of the away-side correlation between the two experiments.  Our results show a dip at $\Delta\phi=\pi$ in the three most central bins; however, this result is not systematically significant given the systematic errors (shown by the black histograms) due to the uncertainty on the elliptic flow result.  In our results, the dip is the strongest in the 0-5\% most central bin as would be expected if the dip is a medium-induced effect.  In the PHENIX results, the dip at $\Delta\phi=\pi$ is stronger and is systematicly significant with their uncertainty on elliptic flow (three histograms are $\pm1\sigma$ and $-2\sigma$).  The dip at $\pi$ in the PHENIX results is the strongest in mid-central collisions where the elliptic flow uncertianities are the largest.  This is illistrated by Fig.~\ref{fig:AuAu} where the centrality dependences of the hump ($\pi/3<|\Delta\phi-\pi|<4\pi/9$) and center ($|\Delta\phi-\pi|<\pi/9$) are shown for both our results and the PHENIX results.  The trends of the centrality dependences suggests that the discrepancy between our results and the PHENIX results may come from the systematics in the flow subtraction.  This figure also shows that the signal strength in the central region does not drop as centrality increases.  The double peaked structure in central Au+Au collisions results from an increased yield in the hump region.  

\begin{figure}[htb]
\hfill
\begin{minipage}[t]{0.49\textwidth}
\includegraphics[width=1.0\textwidth]{Plots/qmPhi_2.5Pt4.0_1.0pt2.5_6Cent.eps}
\end{minipage}
\hfill
\begin{minipage}[t]{0.49\textwidth}
\includegraphics[width=1.0\textwidth]{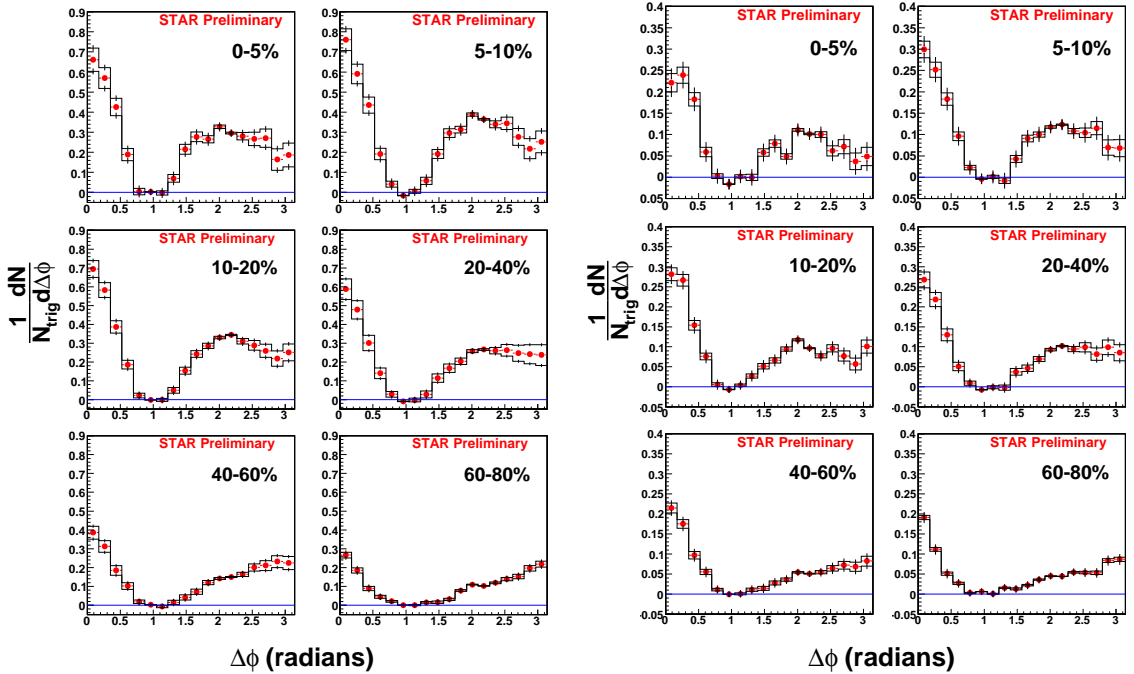}
\end{minipage}
\caption{Background subtracted correlation functions in Au+Au collisions at $\sqrt{s_{NN}}=200$ GeV/c for $2.5<p_{T}^{Trig}<4.0$ GeV/c and $1.0<p_{T}^{Assoc}<2.5$ GeV/c.  Left:  Our results with $|\eta^{Assoc}|<1.0$.  Right:  Our results with $|\eta^{Assoc}|<0.35$.  Error bars are statistical.  Histograms represent 1 sigma systematic uncertainty on the elliptic flow.}
\label{fig:AuAu3}
\end{figure}

A cross check has been done to see if there is any effect in the away-side shape due to the $\eta$ acceptance.  Figure~\ref{fig:AuAu3} shows a comparison of our results with $|\eta^{Assoc}|<1.0$ and $|\eta^{Assoc}|<0.35$.  No significant difference on the away-side shape is seen.

\section{Identified Trigger Particle Correlations}

\begin{figure}[htb]
\centering
\includegraphics[width=1.0\textwidth]{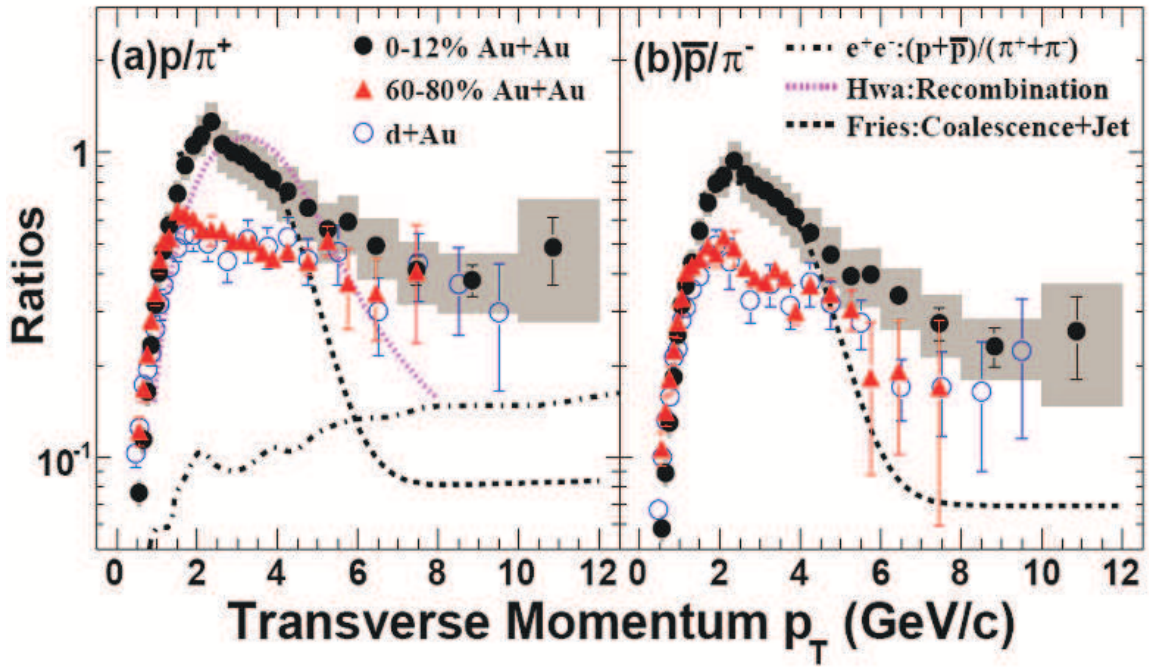}
\caption{(a) Proton to $\pi^{+}$ and (b) anti-proton to $\pi^{-}$ ratios for Au+Au collisions in two centralities and d+Au collisions at $\sqrt{s_{NN}}=200$ GeV/c from \cite{spectra}.}
\label{fig:bm}
\end{figure}   

An enhanced baryon to meson ratio at intermediate $p_T$ has been observed in central Au+Au collisions with respect to peripheral Au+Au, d+Au and $pp$ collisions\cite{pspectra,sspectra}.  Figure~\ref{fig:bm} shows the proton(anti-proton) to $\pi^{+}$ ($\pi^{-})$ in the left (right) panel for central Au+Au, peripheral Au+Au and d+Au.  This effect can be explained by coalescence and recombination models\cite{recomb3,recomb2,recomb}. In this model, the momentum comes from the sum of the quark momenta used to make the particle.  Since baryons are composed of three quarks and mesons of two the baryons are shifted to higher momentum than the mesons resulting in an enhanced baryon over meson ratio at intermediate $p_T$.  However, simple coalescence and recombination models do not contain any angular correlations, thus one would expect lower per trigger correlation strength for intermediate $p_T$ triggered baryons then triggered mesons.  In this section, we identify intermediate $p_T$ baryons and mesons and using them for trigger particles to do jet-correlation studies.  We attempt to identify and study any differences in the associated particle distributions to hopefully shed further light on the baryon/meson puzzle.

\subsection{Particle Identification in TPC at High $P_T$}

 \begin{figure}[htb]
	\centering
	\vspace*{0.0cm}
		\includegraphics[width=0.75\textwidth]{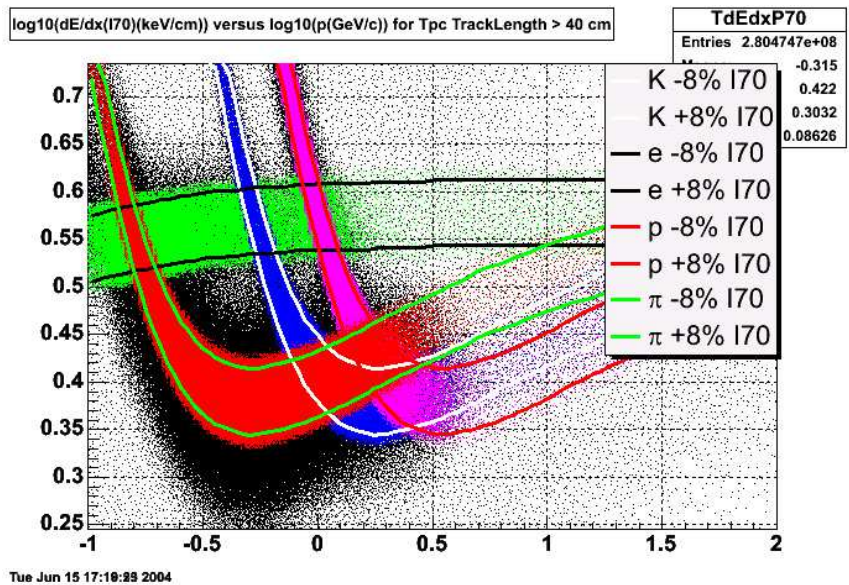}
			\vspace*{0.0cm}
	\caption{Ionization energy loss plotted as $log_{10}$(dE/dx) vs. $log_{10}$(p).   The lines represent $\pm\sigma$ ($\pm$8\% in relative dE/dx) bands for protons, kaon, electrons, and pions.  I70 stands for Bichsel's prediction for 30\% truncated mean for dE/dx.  Figure is from \cite{dedx}.}
	\label{fig:dedx}
\end{figure}

The STAR time projection chamber (TPC) was designed for identification of protons, anti-protons and charged pion and kaons at low $p_T$ using energy loss (dE/dx).  At around 1.5-2.0 GeV/c the energy loss is similar enough for all three particles that they cannot be distinguished.  However, particle identification can be performed for higher $p_T$ particles.  This is due to the relativistic rise of dE/dx.  The dE/dx distributions for different particle types is shown in figure~\ref{fig:dedx} as a function of the magnitude of momentum.  Pions can be distinguished from kaons and protons at around 2.5 GeV/c and above.  Protons and kaons can be distinguished from each other starting at about 3.0 GeV/c. 

\begin{figure}[htb]
	\centering
	\vspace*{0.0cm}
		\includegraphics[width=0.75\textwidth]{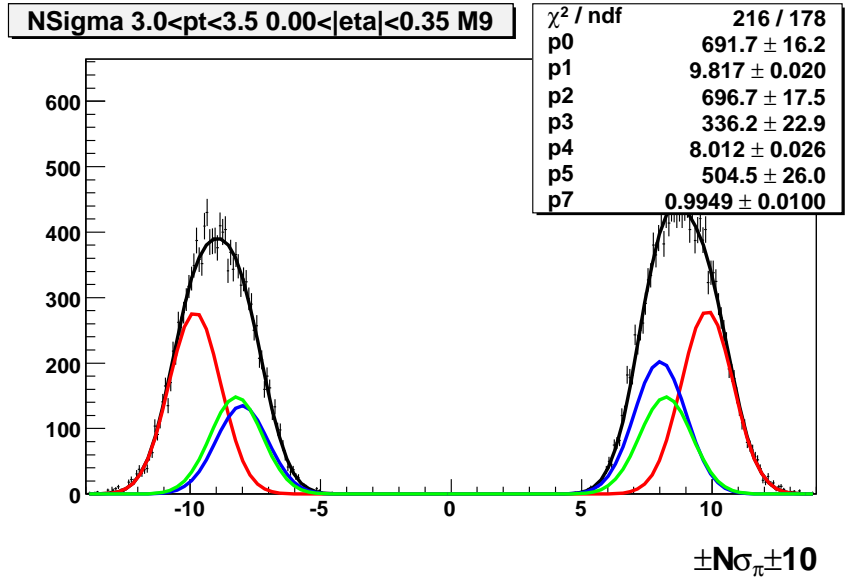}
			\vspace*{0.0cm}
	\caption{$N\sigma_{\pi}$ distribution for $3 < p_T < 3.5$ GeV/c and $| \eta | < $ 0.35 from 0-5\% Au+Au collsions at $\sqrt{s_{NN}}=200$ GeV/c.  Positive particles are plotted as $N\sigma_{\pi}$ plus 10, negative particles are plotted as -$N\sigma_{\pi}$ minus 10.  Curves are from fit, red is pion, green is kaon, blue is proton, and black is the sum of the three.  Errors are statistical.  Plots for other bins can be found in the appendix.}
	\label{fig:nsig}
\end{figure}

To separate the particles $N\sigma_{\pi}$ is used, where $N\sigma_{\pi}$ is the (dE/dx-predicted $\pi$ dE/dx)/resolution.  With this variable, if the data is well calibrated, the pions will form a Gaussian peak centered at zero with a $\sigma$ width of 1.  Figure~\ref{fig:nsig} shows an example $N\sigma_{\pi}$ distribution.  In this figure, positive and negative particles are separated by plotting $\pm N\sigma_{\pi}\pm10$ for positive and negative particles, respectively.  It is beneficial to bin the dE/dx distribution in $\eta$ in addition to $p_T$ and centrality.  This is because the energy loss is a function of the magnitude of the total velocity and with too large of an $\eta$ bin we are sampling a large range of total momentum magnitude within one $p_T$ bin, resulting in a reduced resolution.  We can than fit the $N\sigma_{\pi}$ distributions to six Gaussians, one each for $\pi^{+}$, $\pi^{-}$, $K^{+}$, $K^{-}$, p and $\bar{p}$.  However, this fit needs additional constraints due to overlap of the Gaussians, as seen in Fig.~\ref{fig:nsig}.  One set of constraints is the requirement of the same width and centroid (adjusted for the way it is plotted) for the positive and negative particles.   Since the protons and kaons are not well separated in this region from just energy loss more constraints are required to extract the protons.  We can use the Bethe-Block formula to calculate the energy loss.  This can be used to determine the kaon centroid position relative to the pion and proton.   From this calculation the kaon peak position is determined to be approximately 1/3 of the distance between the proton and pion peaks.  The calculated centroids cannot be simply fixed in the fit because the dE/dx calibration is not perfect.  It may also be necessary to use the measured $K^{0}_{s}$ yield for the charged kaon yield, with the assumption that $K^{0}_{s}$ has the same yield as $K^{+}$ and $K^{-}$.  An example fit is shown in Fig.~\ref{fig:nsig}.  This method of fitting the dE/dx distributions has been previously used to extract high $p_T$ identified particle spectra\cite{spectra}.

\begin{figure}[htb]
\hfill
\begin{minipage}[t]{0.49\textwidth}
\centering
	\includegraphics[width=1\textwidth]{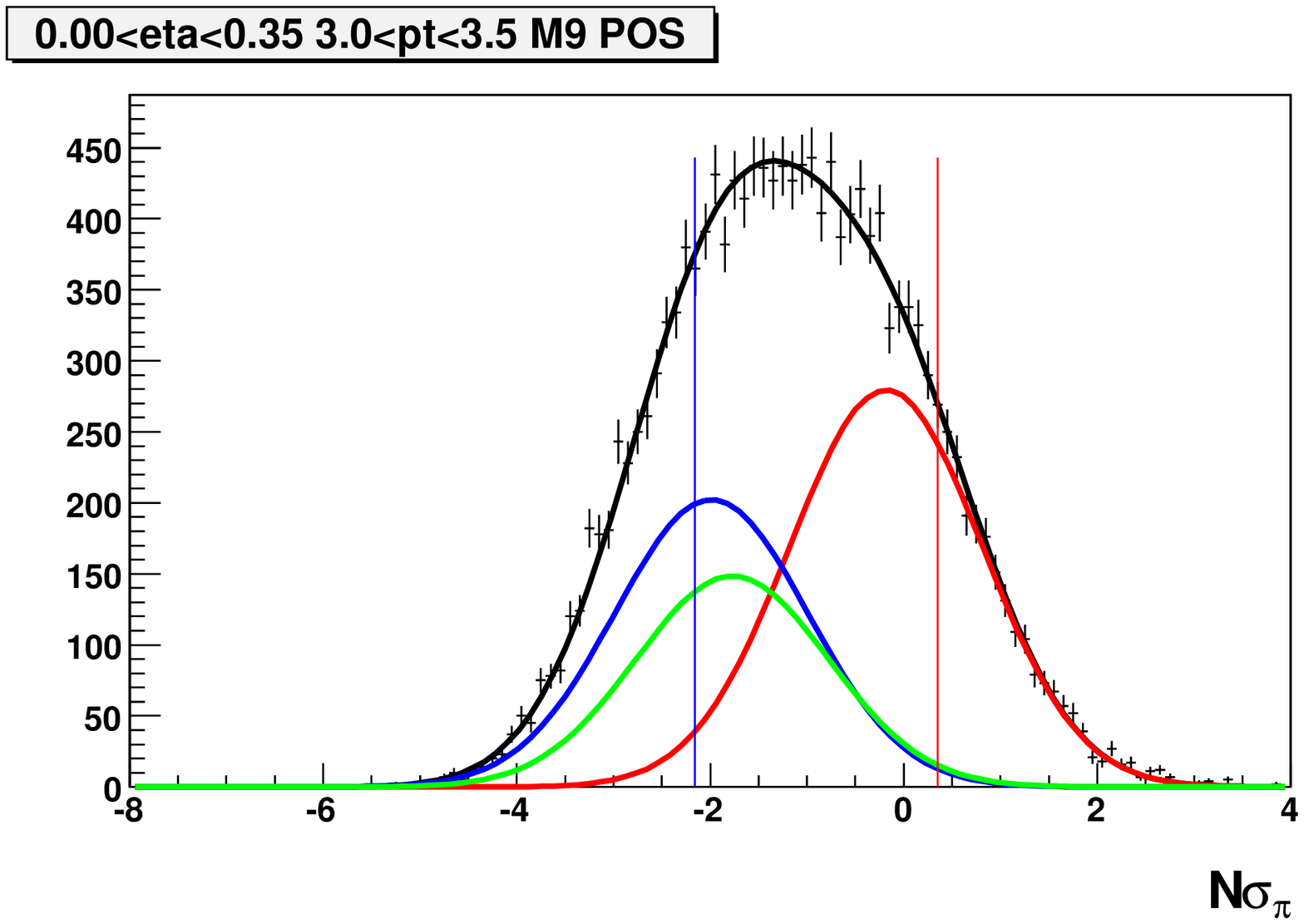}	
\end{minipage}
\hfill
\begin{minipage}[t]{0.49\textwidth}
\centering
	\includegraphics[width=1\textwidth]{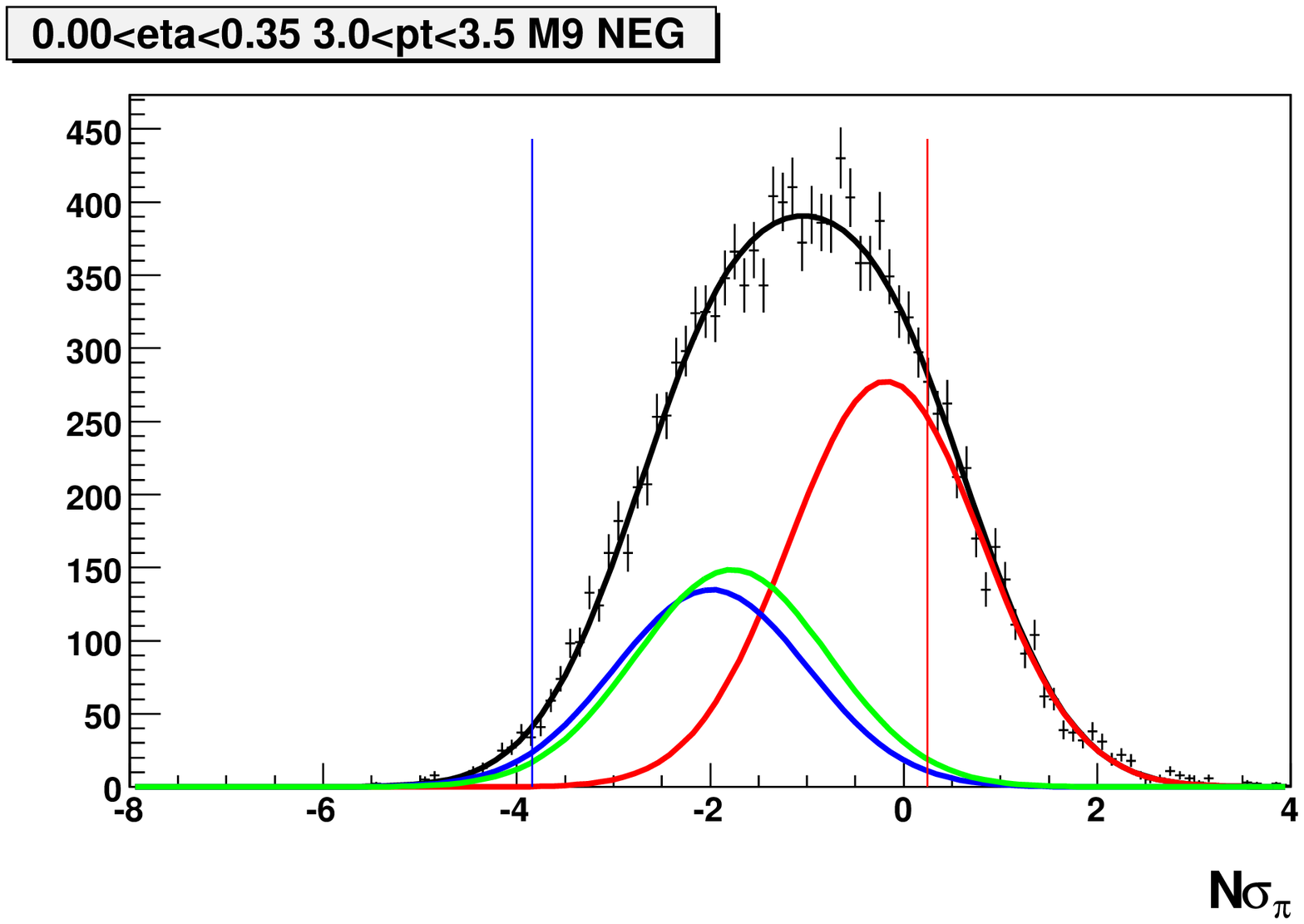}
\end{minipage}
\caption{$N\sigma_{\pi}$ distribution for $3 < p_T < 3.5$ GeV/c and $| \eta | < $ 0.35 from 0-5\% Au+Au collsions at $\sqrt{s_{NN}}=200$ GeV/c.  Left: Positive particles.  Right: Negative particles.  Curves are from fit, red is pion, green is kaon, blue is proton, and black is the sum of the other three.  The vertical lines show the position of the cuts.  $N\sigma_{\pi}$ greater than red line gives 95\% purity of charged pions.  $N\sigma_{\pi}$ less than blue line give 50\% purity of proton/anti-protons.  Errors are statistical.  Plots for other bins can be found in the appendix.}
\label{fig:pidcuts}
\end{figure}

Cuts can then be placed on the $N\sigma_{\pi}$ distribution to extract enriched samples of protons, anti-protons, and charged pions.  We can use our fit results to determine the purity of the selected particles from the cuts.  If there is large overlap of the Gaussians, we may have to sacrifice purity for statistics.  Figure~\ref{fig:pidcuts} shows an example $N\sigma_{\pi}$ distribution for the positive and negative particles with lines designating where cuts are to obtain 50\% purity for p/$\bar{p}$ and 95\% purity for $\pi^+/\pi^-$.  Particles with $N\sigma_{\pi}$ below the left line are taken to be protons and particles above the red line are taken to be $\pi^+$s with the given purities.  The purity is determined by integrating the curve, within the particle identification cut, obtained for the fit for the particlar particle type and dividing by the intergral of the curves for all particle types.  
   
\begin{figure}[htb]
	\centering
		\includegraphics[width=0.6\textwidth]{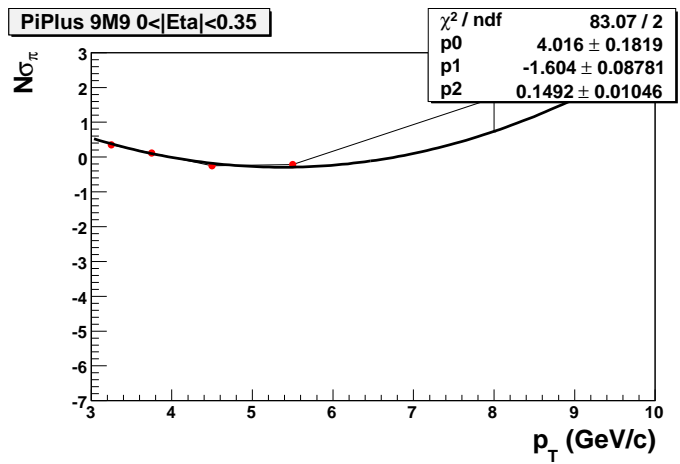}													
	\caption{Fits to the cuts on $N\sigma_{\pi}$ as a function of $p_{T}$ for $\pi^-$ with 95\% purity.  Plot is for Au+Au 0-5\% central collisions at $\sqrt{s_{NN}}=200$ GeV/c with $|\eta|<0.35$.  Errors are statistical.}
	\label{fig:fitcuts}	
\end{figure}

The position of these cuts can be improved upon then by fitting the cuts to a smooth curve as a function of $p_T$ for each $\eta$ bin.  An example fit is shown in Figures~\ref{fig:fitcuts}.  The cuts were computed out to higher $p_{T}$ but due to statistics only trigger particle of $3<p_{T}<4$ GeV/c were used.  The fits are to second degree polynomials.  Positive particles with $N\sigma_{\pi}$ greater than the $\pi^{+}$ fit and negative particles with $N\sigma_{\pi}$ greater than the $\pi^{-}$ fit are considered charged pions with 95\% purity.  Positive particles with $N\sigma_{\pi}$ less then the $p$ fit and negative particle with $N\sigma_{\pi}$ less than the $\bar{p}$ fit are considered protons/anti-protons with 50\% purity.

\subsection{Results and Discussions}

\begin{figure}[htbp]
\hfill
\begin{minipage}[t]{.32\textwidth}
	\centering
		\includegraphics[width=1\textwidth]{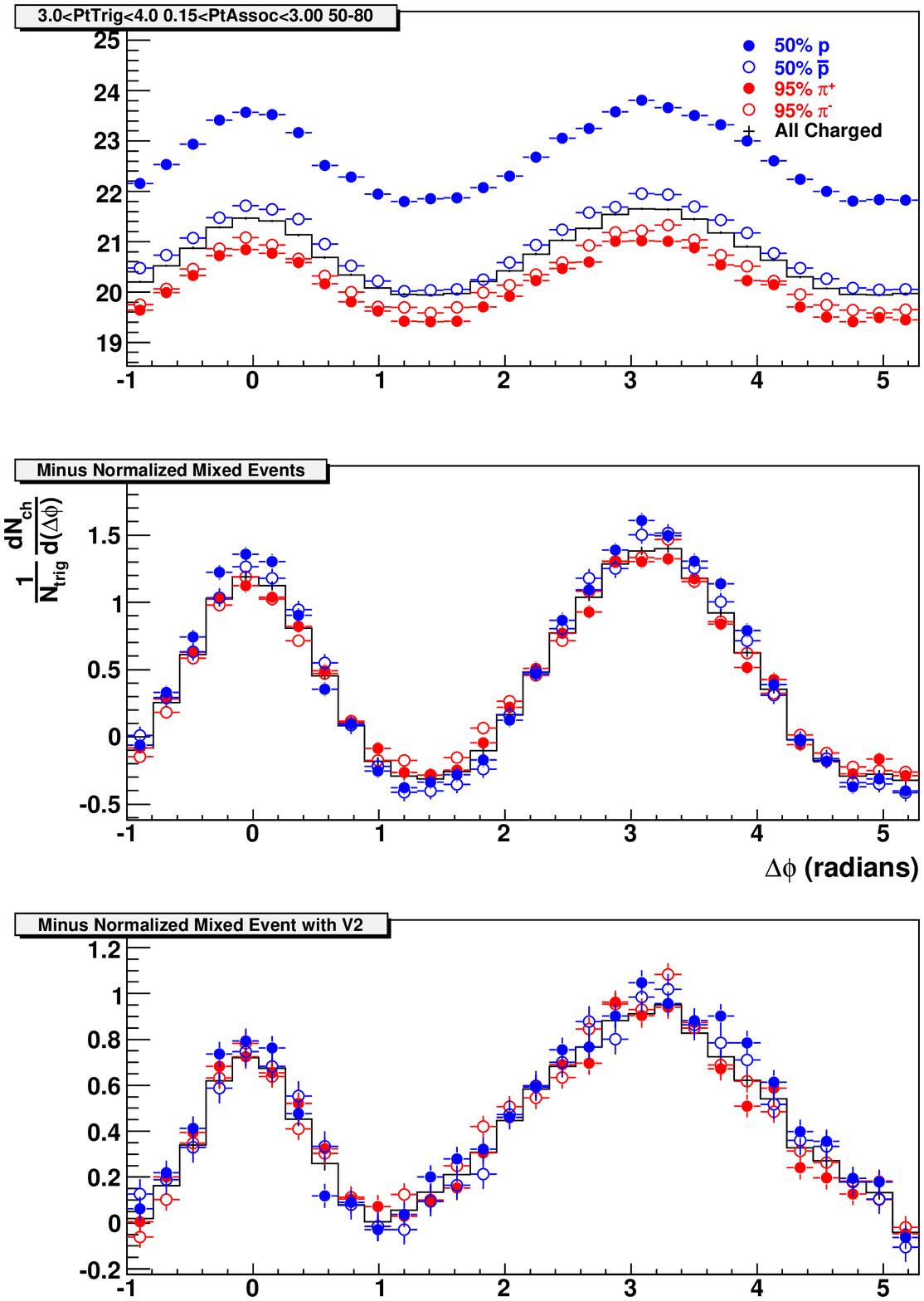}\\
		\vspace*{0.5cm}
		\includegraphics[width=1\textwidth]{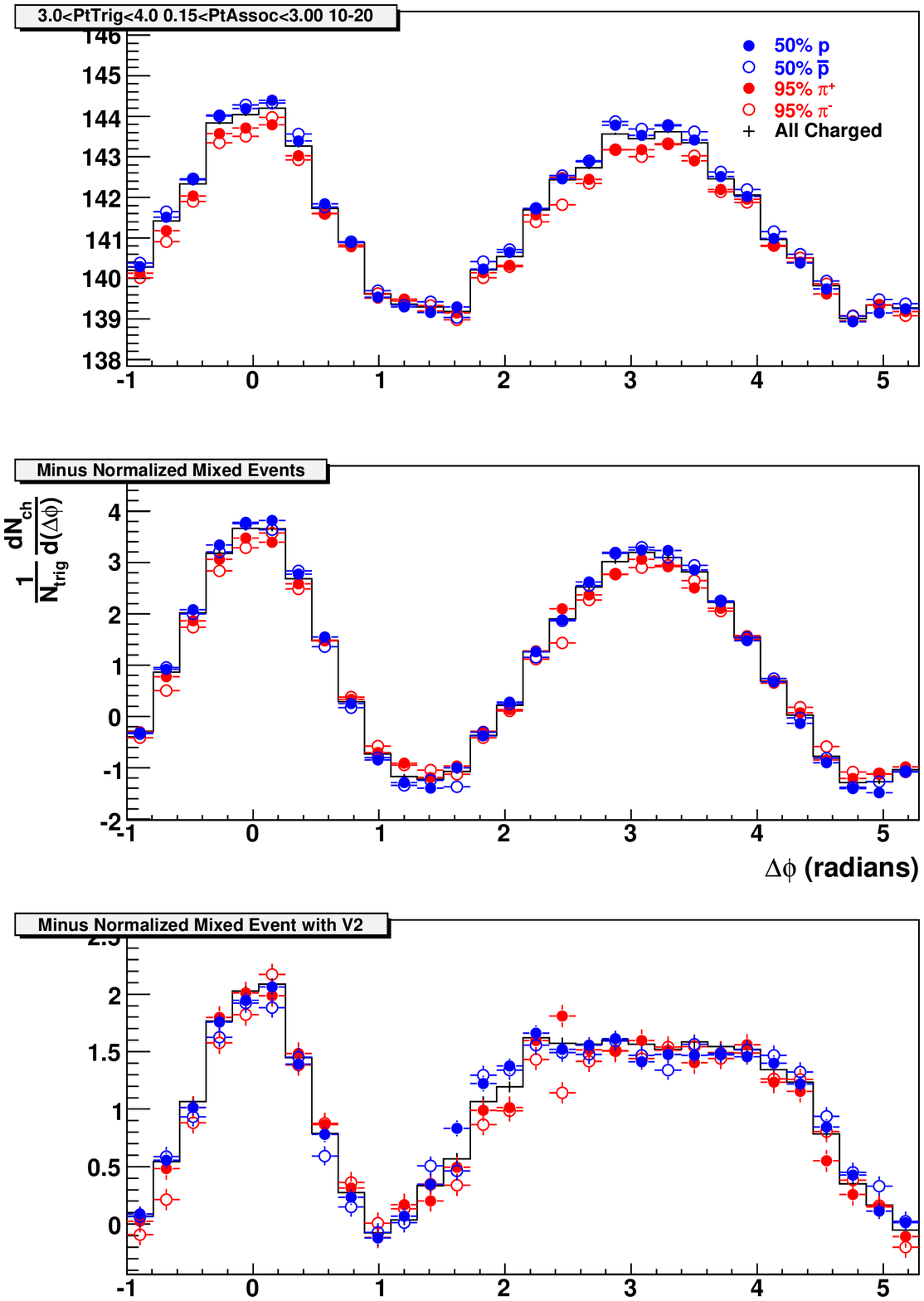}
\end{minipage}
\hfill
\begin{minipage}[t]{.32\textwidth}
	\centering
	\includegraphics[width=1\textwidth]{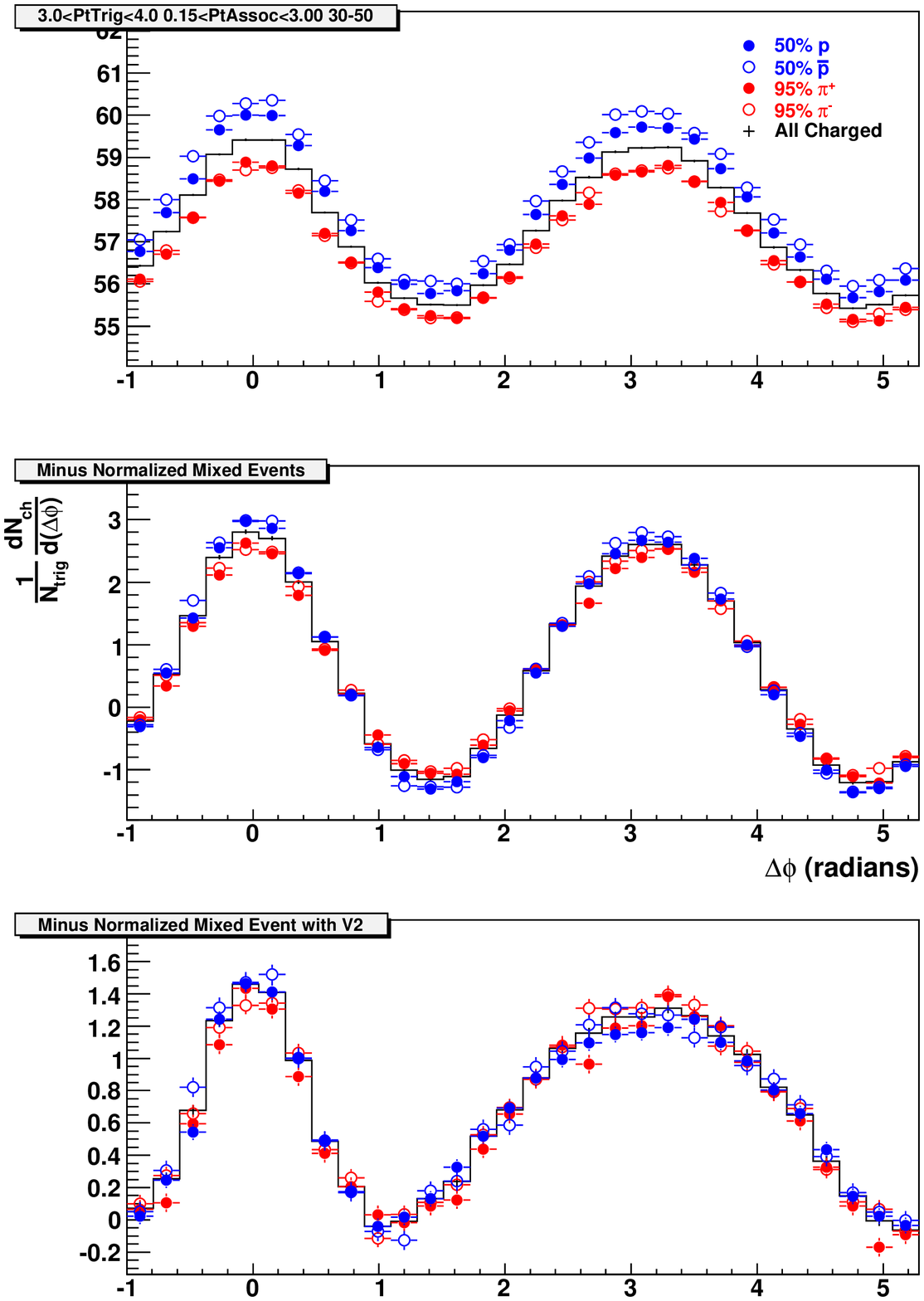}\\
		\vspace*{0.5cm}
	\includegraphics[width=1\textwidth]{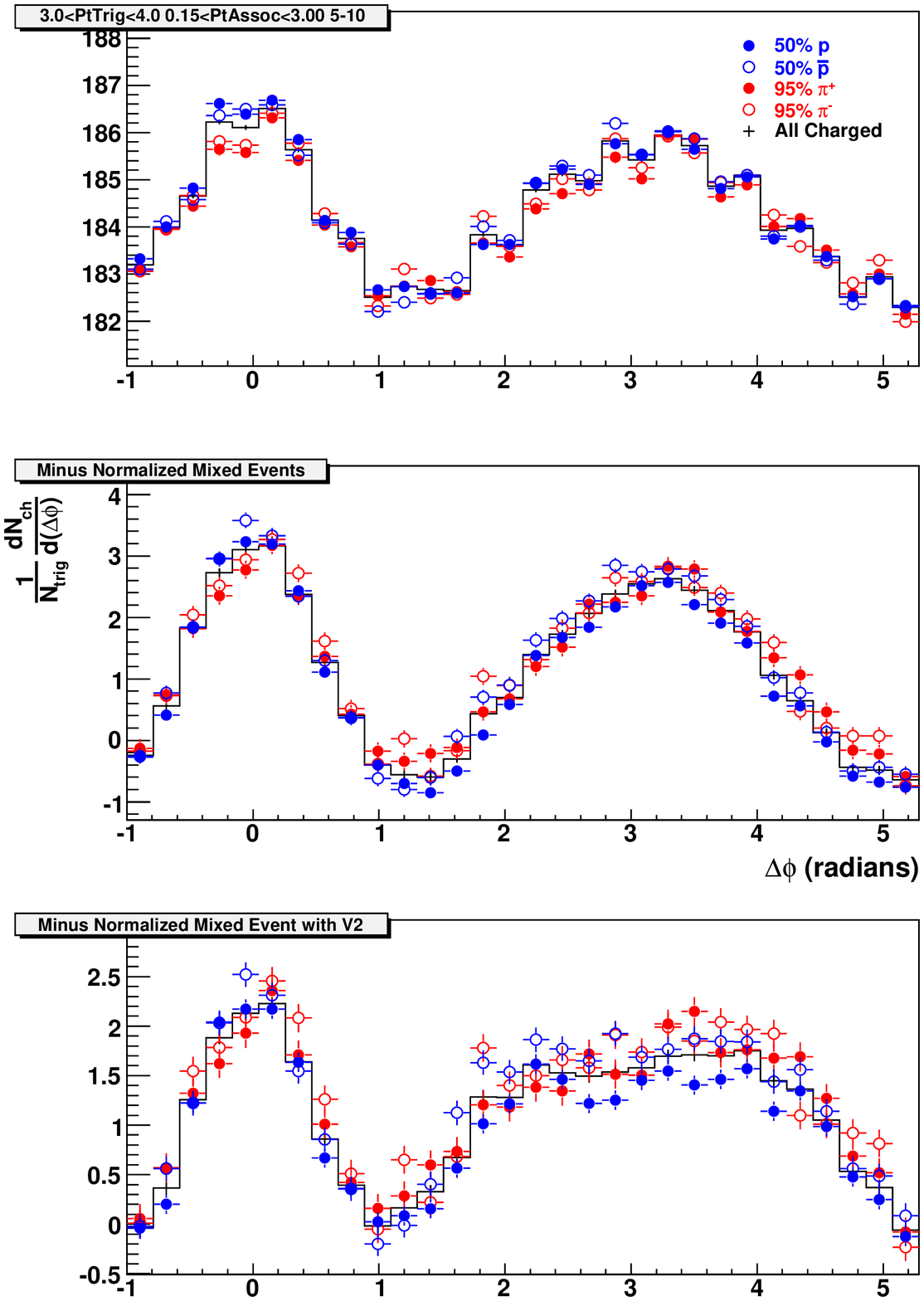}
\end{minipage}
\hfill
\begin{minipage}[t]{.32\textwidth}
	\centering
	\includegraphics[width=1\textwidth]{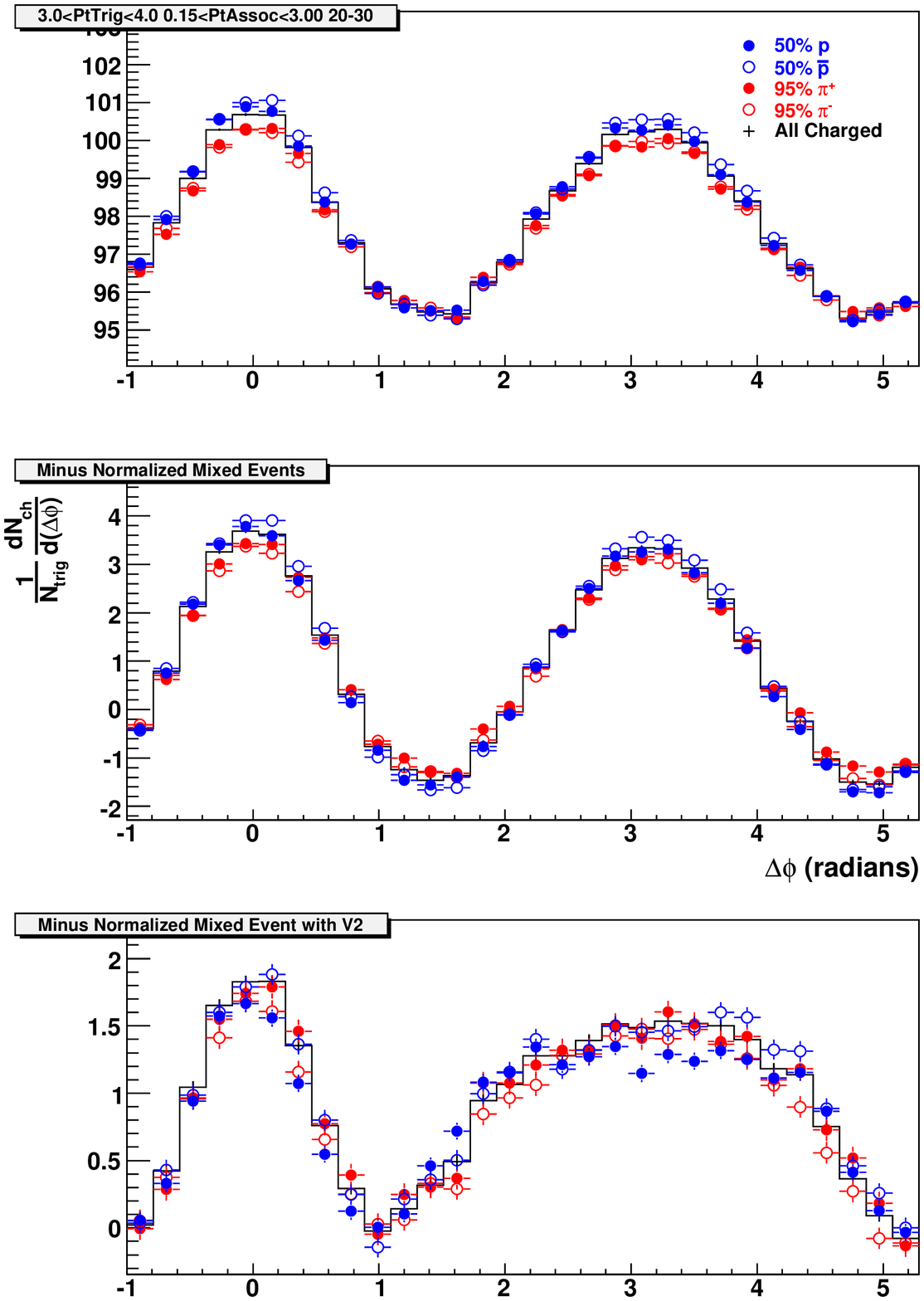}\\
		\vspace*{0.5cm}
	\includegraphics[width=1\textwidth]{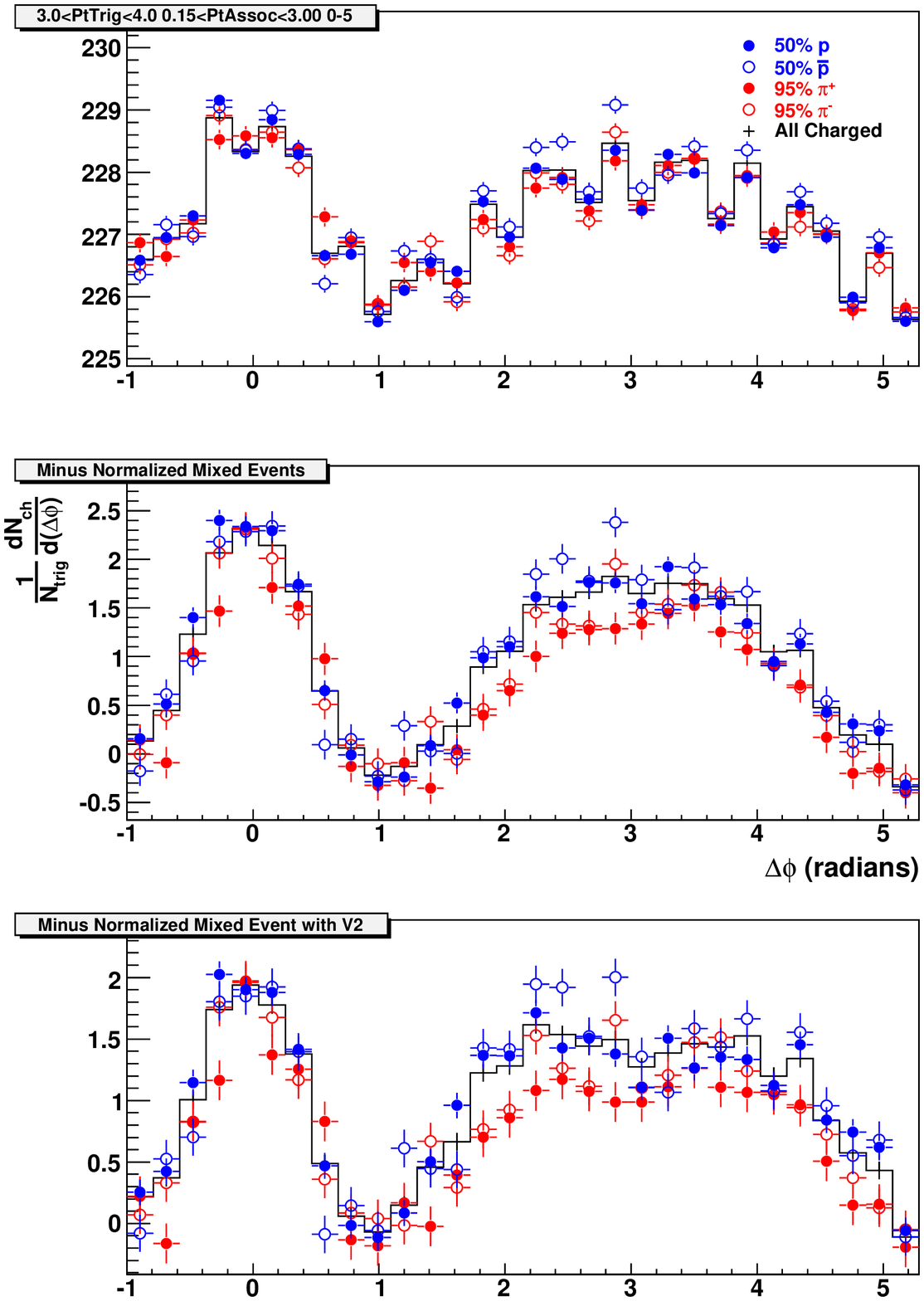}
\end{minipage}
\caption{Identified trigger particle correlations in Au+Au collisions at $\sqrt{s_{NN}}=200$ GeV/c.  Each group of 3 plots are (from top to bottom) raw signals, raw signal minus mixed events, and raw signal minus mixed events and $v_{2}$.  Groups from left to right, top to bottom are centralities 50-80\%, 30-50\%, 20-30\%, 10-20\%, 5-10\%, and 0-5\%.  All plots are for Au+Au collisions with $3<p_{T}^{Trig}<4$ GeV/c and $0.15<p_{T}^{Assoc}<3$ GeV/c.  Errors are statistical.}
\label{fig:pidcA0}
\end{figure}

\begin{figure}[htbp]
\hfill
\begin{minipage}[t]{.32\textwidth}
	\centering
		\includegraphics[width=1\textwidth]{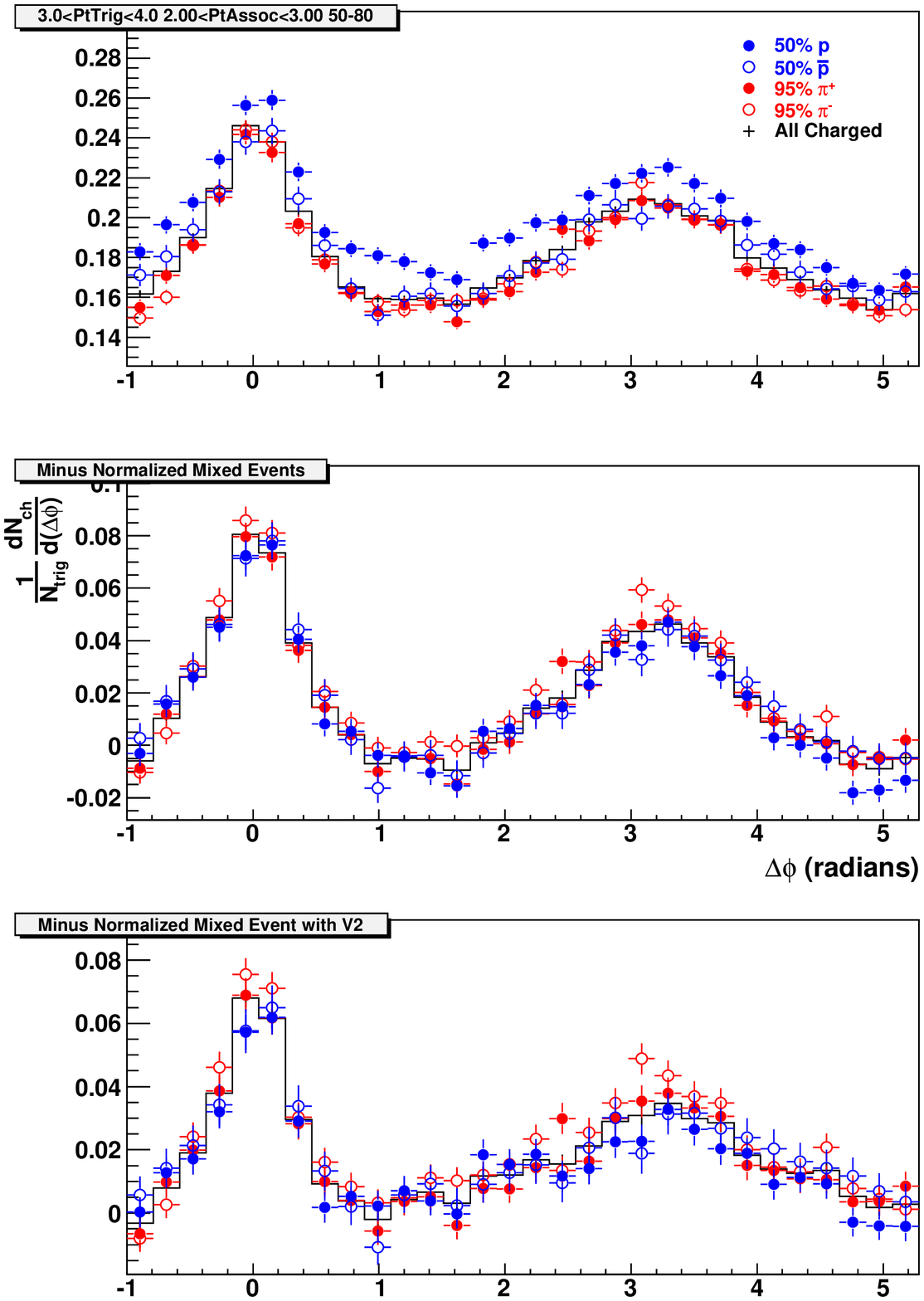}\\
		\vspace*{0.5cm}
		\includegraphics[width=1\textwidth]{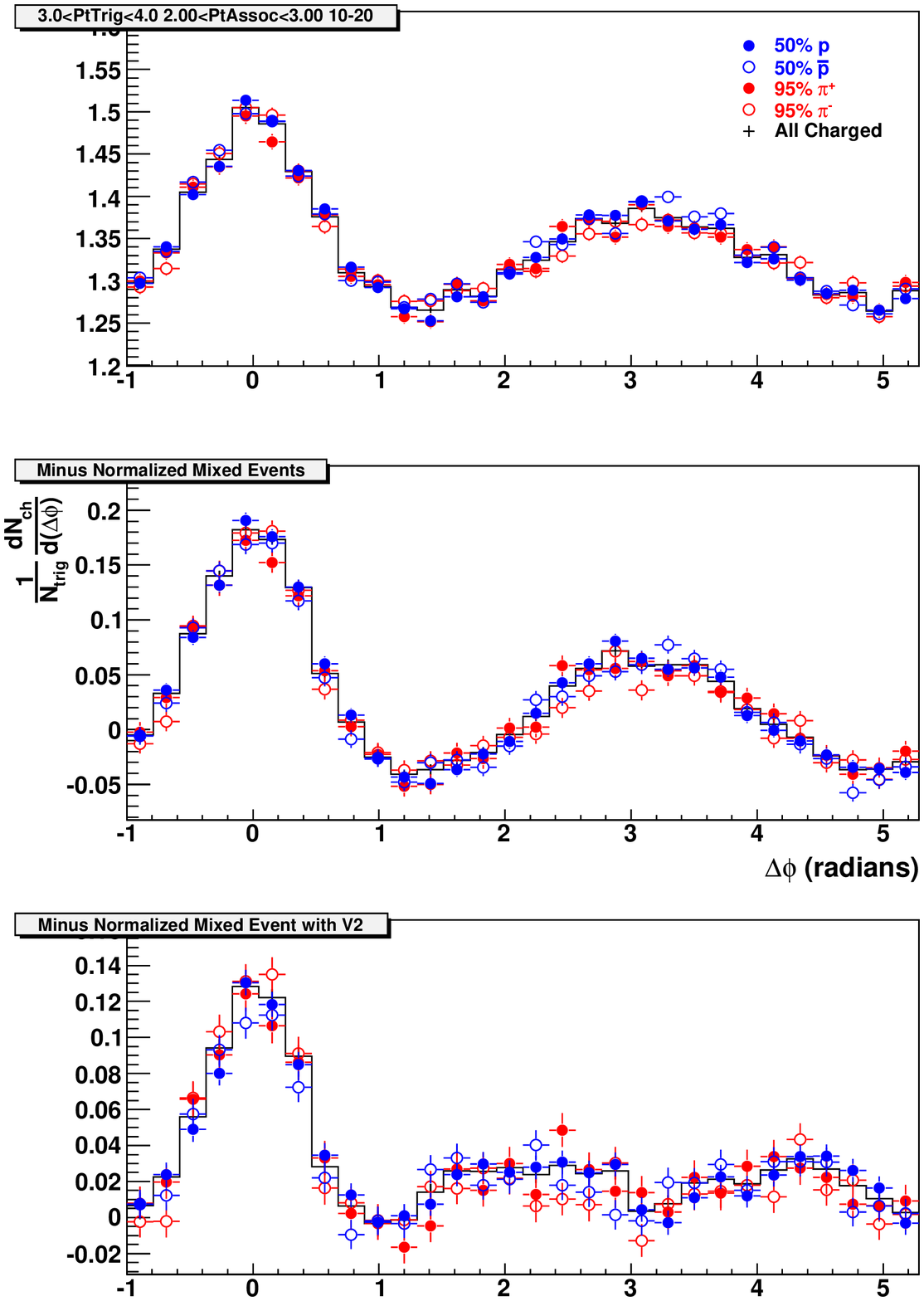}
\end{minipage}
\hfill
\begin{minipage}[t]{.32\textwidth}
	\centering
	\includegraphics[width=1\textwidth]{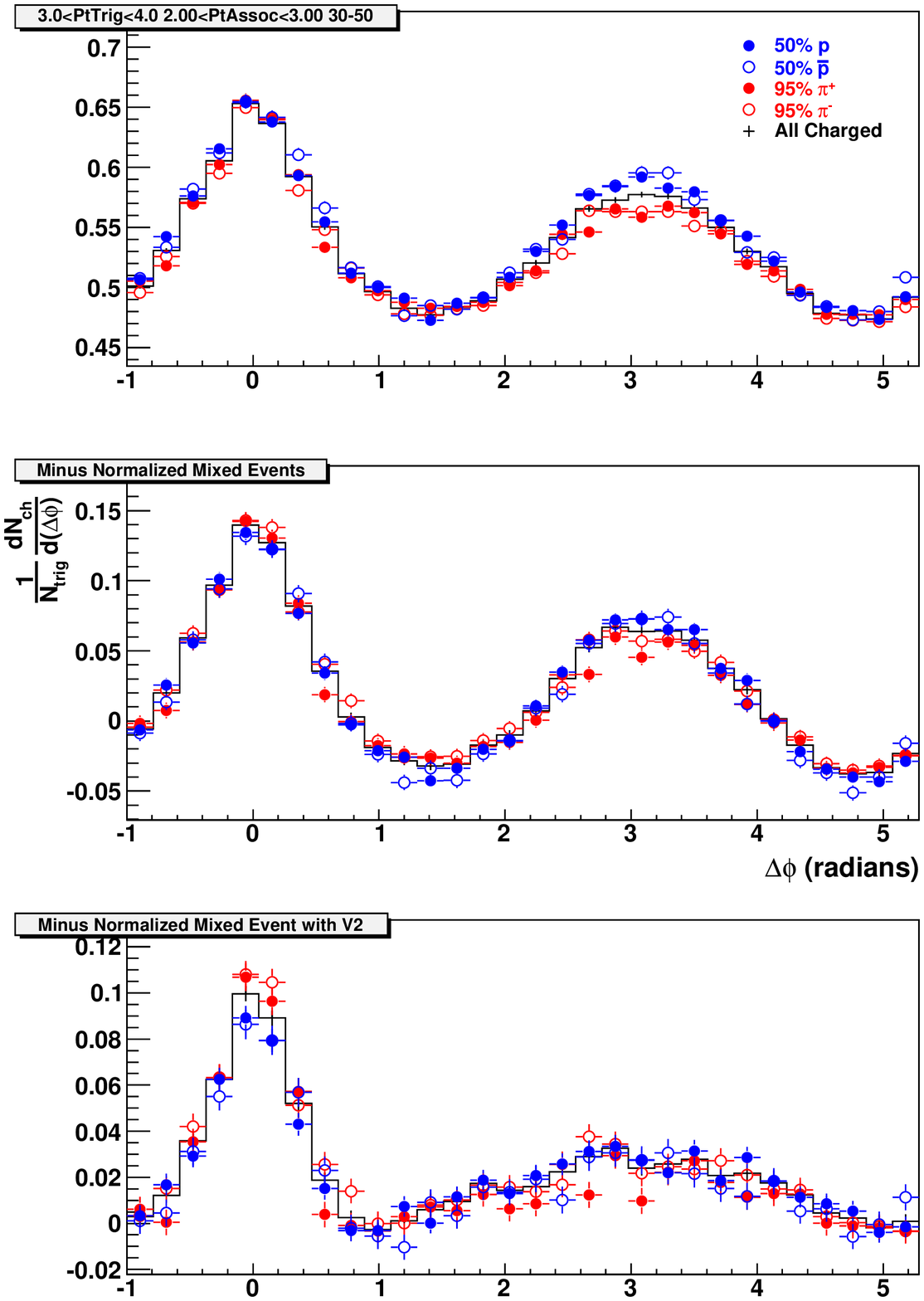}\\
		\vspace*{0.5cm}
	\includegraphics[width=1\textwidth]{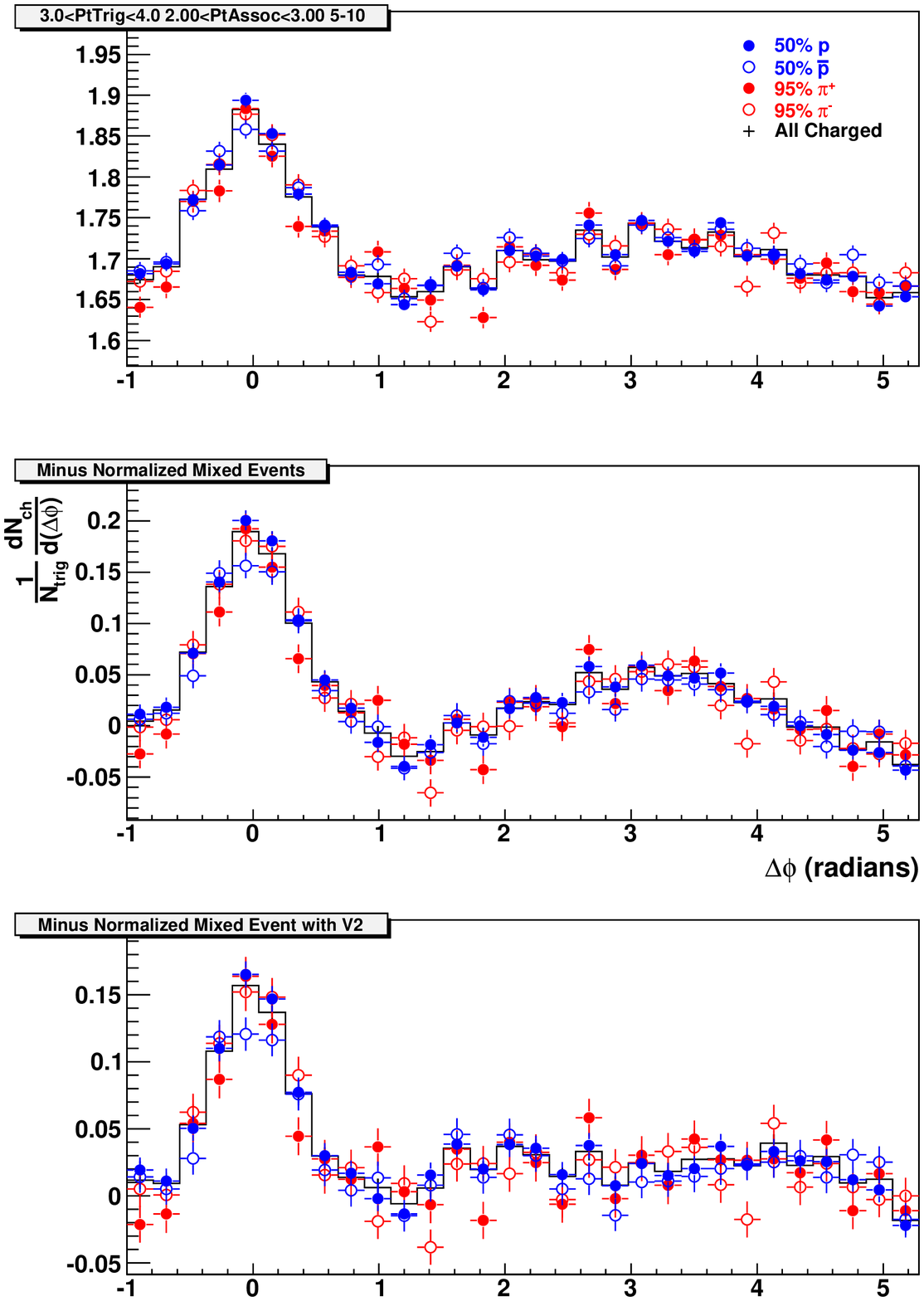}
\end{minipage}
\hfill
\begin{minipage}[t]{.32\textwidth}
	\centering
	\includegraphics[width=1\textwidth]{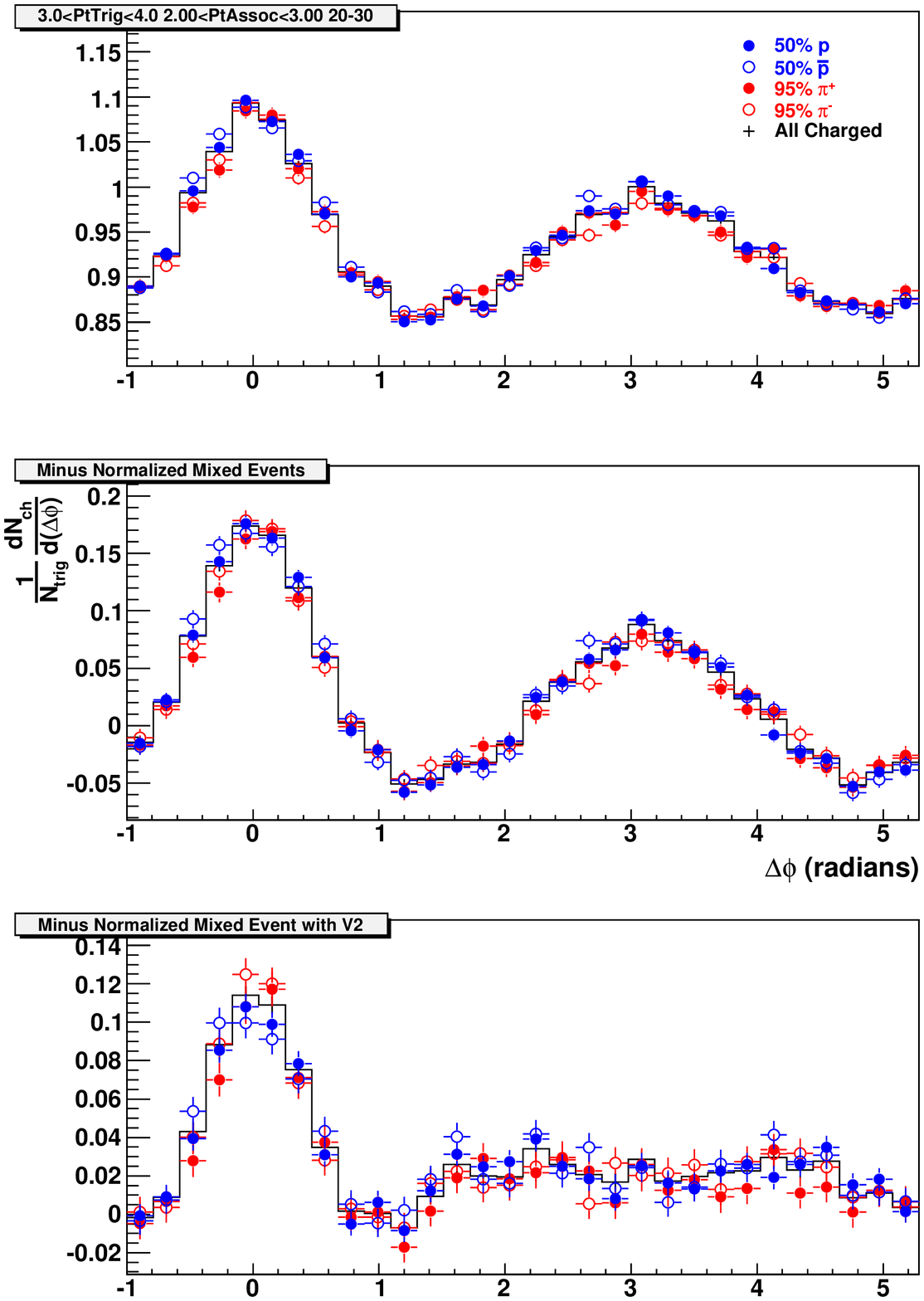}\\
		\vspace*{0.5cm}
	\includegraphics[width=1\textwidth]{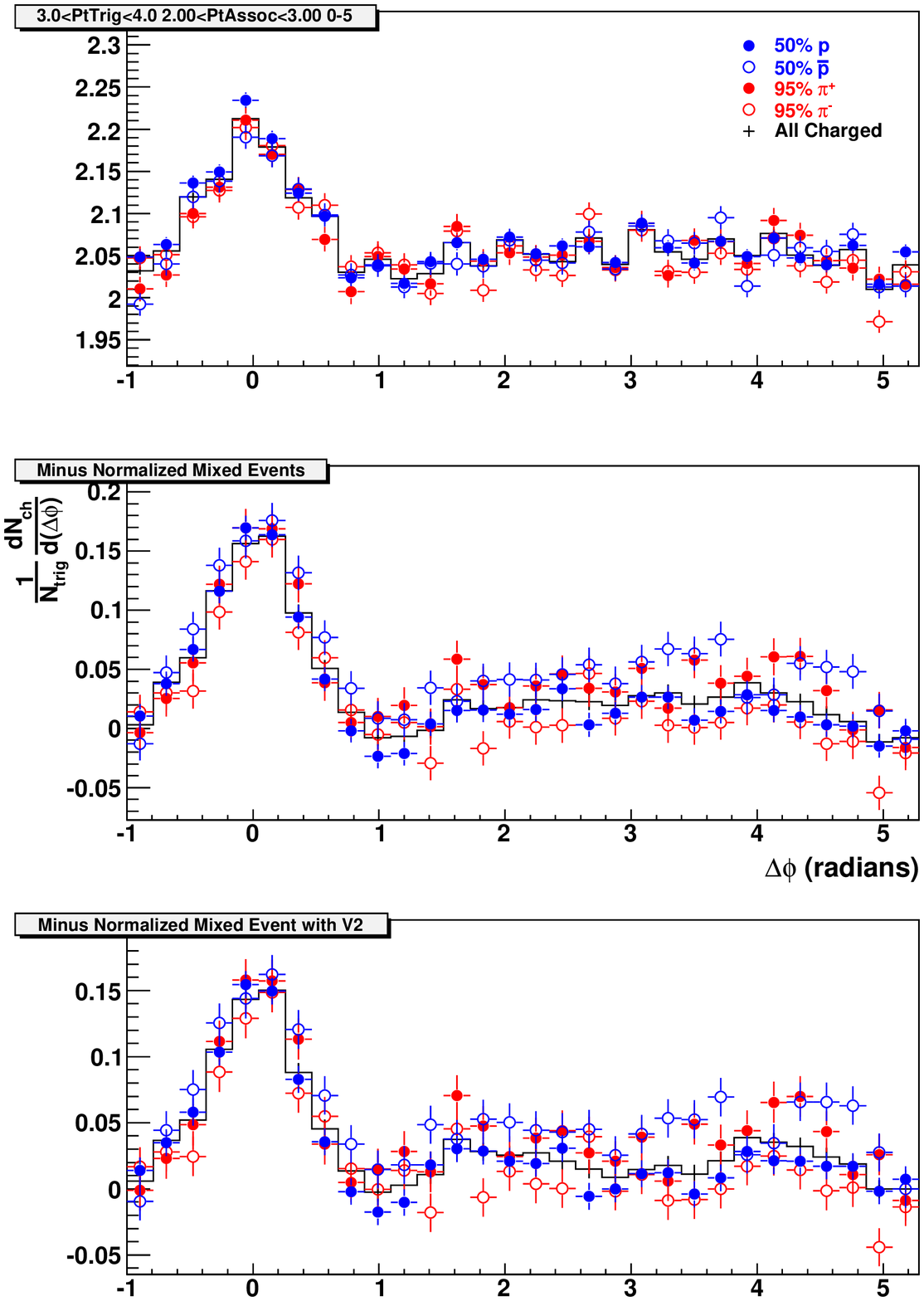}
\end{minipage}
\caption{Same as Figure 3.18 except for $2<p_{T}^{Assoc}<3$ GeV/c.}
\label{fig:pidcA2}
\end{figure}

\begin{figure}[htb]
\hfill
\begin{minipage}[t]{0.49\textwidth}
\centering
\includegraphics[width=.8\textwidth]{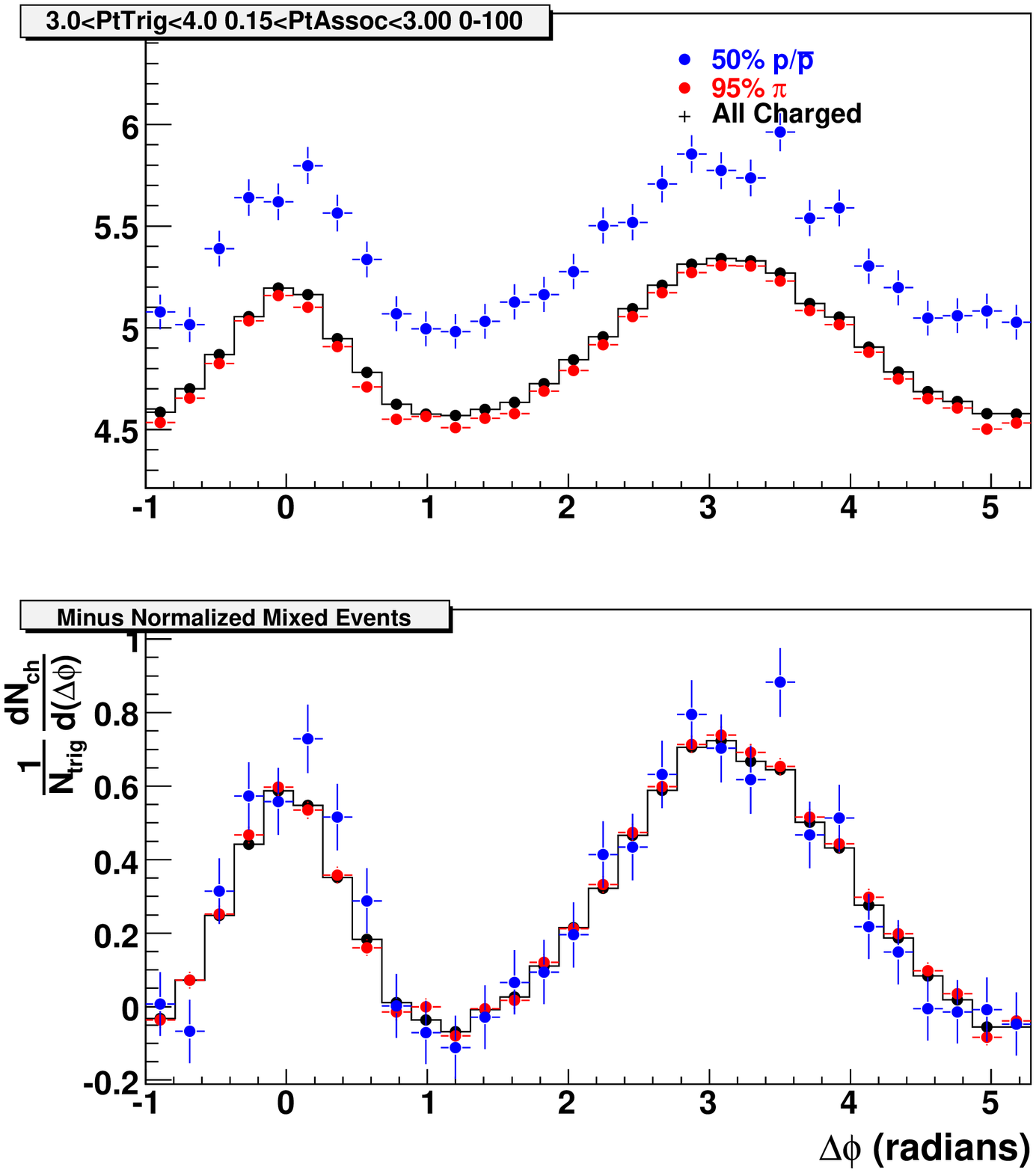}
\end{minipage}
\hfill
\begin{minipage}[t]{0.49\textwidth}
\centering
\includegraphics[width=.8\textwidth]{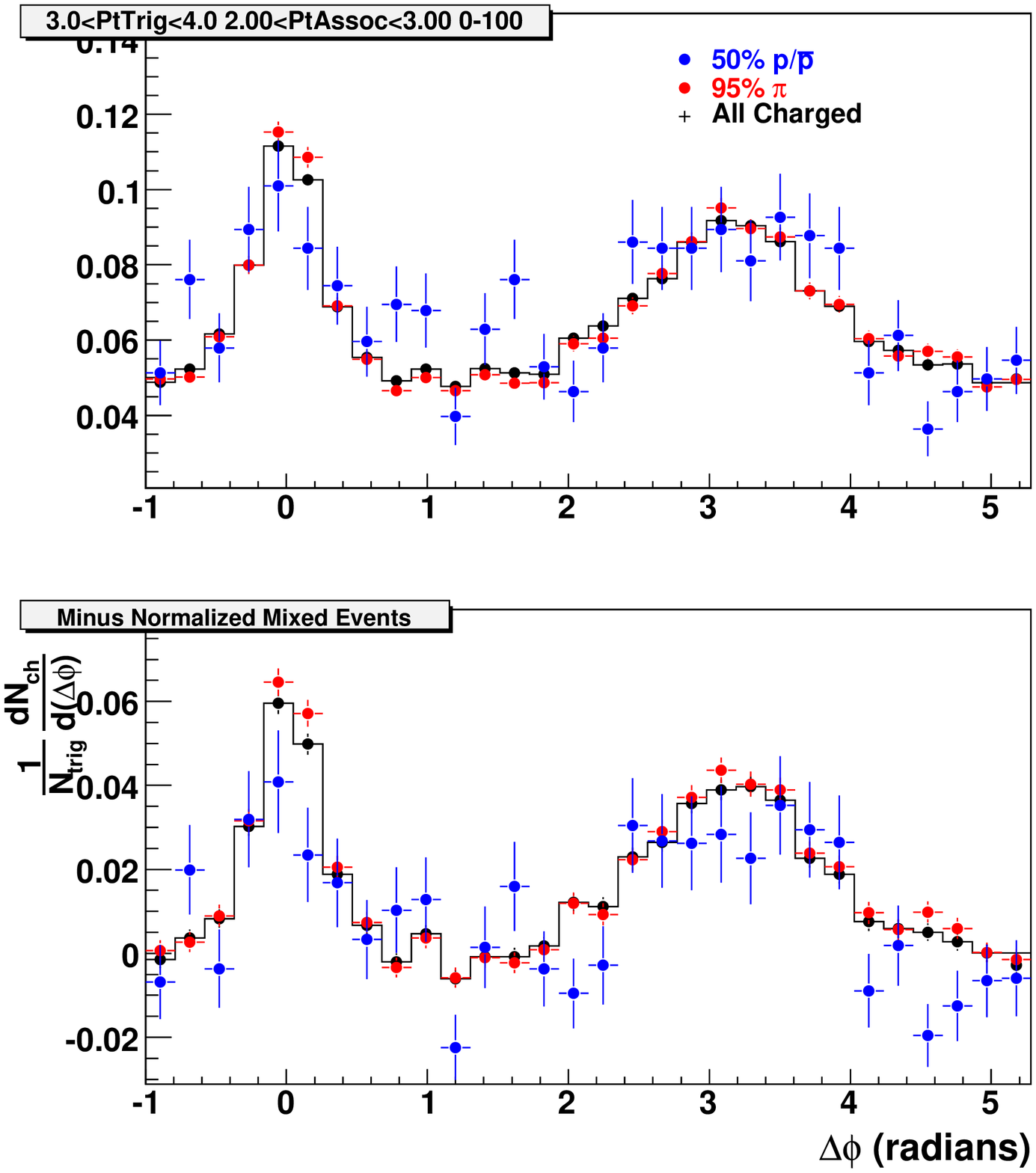}
\end{minipage}
\caption{Identified trigger particle correlations in d+Au collisions at $\sqrt{s_{NN}}=200$ GeV/c.  All plots are for $3<p_{T}^{Trig}<4$.  Left: $0.15<p_{T}^{Assoc}<3$.  Right:  $2<p_{T}^{Assoc}<3$.  Top:  Raw signals.  Bottom:  Raw signals minus normalized mixed events.  Errors are statistical.}
\label{fig:pidd1}
\end{figure}

Figure~\ref{fig:pidcA0} shows identified trigger particle correlations in Au+Au collisions for $3<p_{T}^{Trig}<4$ GeV/c and $0.15<p_{T}^{Assoc}<3$ GeV/c.  In this region, the correlations are dominated by the lower $p_{T}$ associated particles.  The raw signals are shown in the top panel of each set of three panels.  The difference in the background levels seen for the raw signal is due to different trigger biases for different types of triggers.  The second panel shows the raw signal minus mixed events. The mixed events are normalized such that the signal is ZYA1.  Once the difference level of the background is removed the correlations look similar.  Part of the correlation is not jet-like but is due to the correlation with the reaction plane.  The bottom panel shows the correlations with the elliptic flow modulated mixed events removed.  The flow is added into the mixed events pair-wise.  The flow is not the charged particle flow as used above, but is scaled such that the baryon/meson flow is 3/2.  This contains the assumption of quark number scaling to the elliptic flow which has been seen in data up to 2 GeV/c\cite{v2scale}.  The yields obtained from the Gaussian fits in the particle identification procedure are used to constrain the relative number of baryons and mesons.  After elliptic flow subtraction the jet-like correlations show no significant dependence on trigger particle type for all centrality bins.  Figure~\ref{fig:pidcA2} shows the identified trigger particle correlations in Au+Au collisions for $3<p_{T}^{Trig}<4$ and $2<p_{T}^{Assoc}<3$ GeV/c.  These higher $p_{T}$ associated results are similar to the lower $p_{T}$ associated results in that background subtracted jet-like correlations show no significant dependence on trigger particle type.  Figure~\ref{fig:pidd1} shows the identified trigger particle correlations in d+Au collisions for $3<p_{T}^{Trig}<4$ and $0.15<p_{T}^{Assoc}<4$ (left) and $2<p_{T}^{Assoc}<4$ GeV/c (right).  For the d+Au plots, the $p$ and $\bar{p}$ and the $\pi^{+}$ and $\pi^{-}$ have been combined due to low statistics.  Again for the identified trigger particle correlations in d+Au no significant differences are seen for the triggered baryons and mesons.

\begin{figure}[htb]
\hfill
\begin{minipage}[t]{0.32\textwidth}
\centering
\includegraphics[width=1\textwidth]{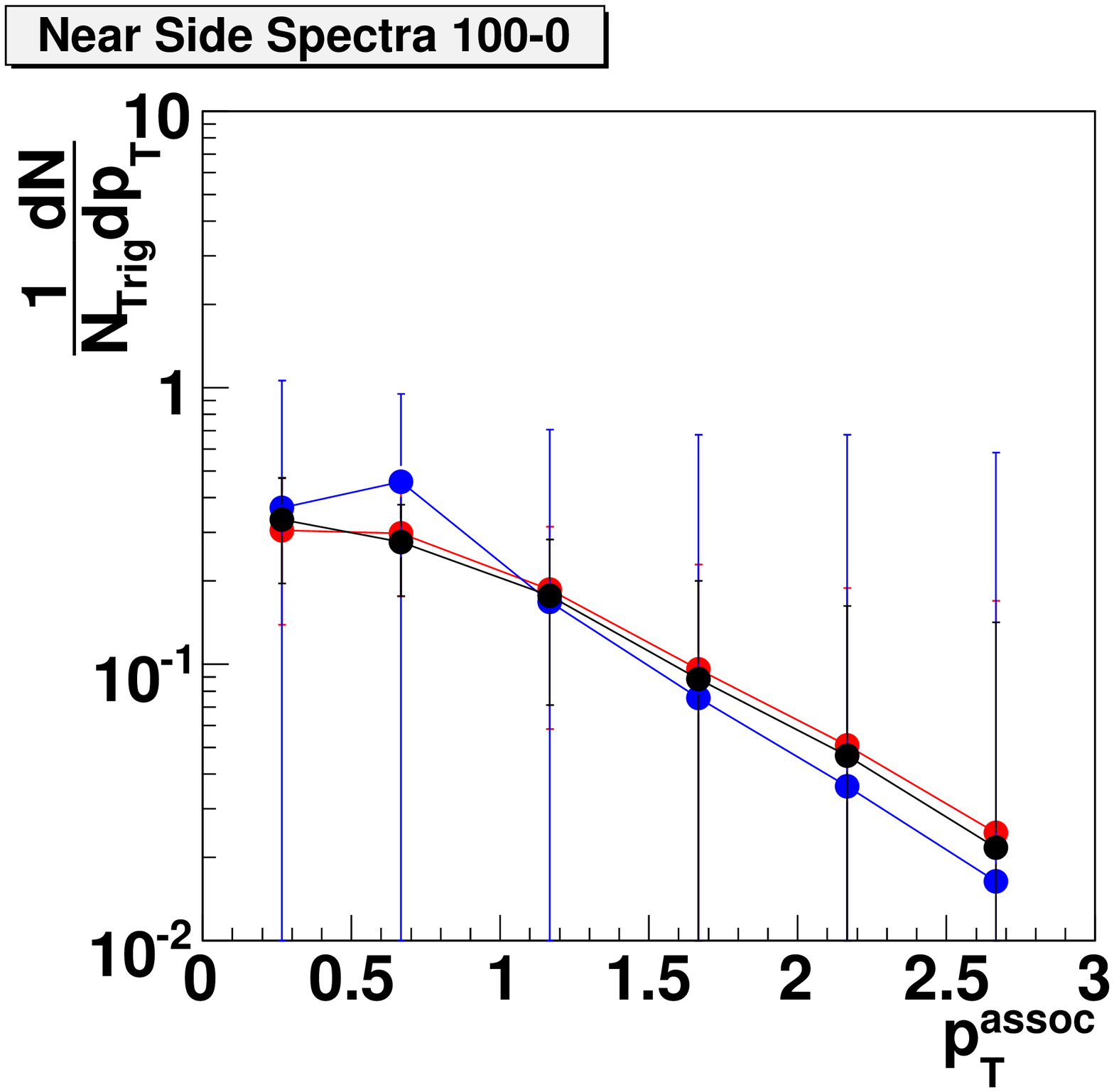}
\includegraphics[width=1\textwidth]{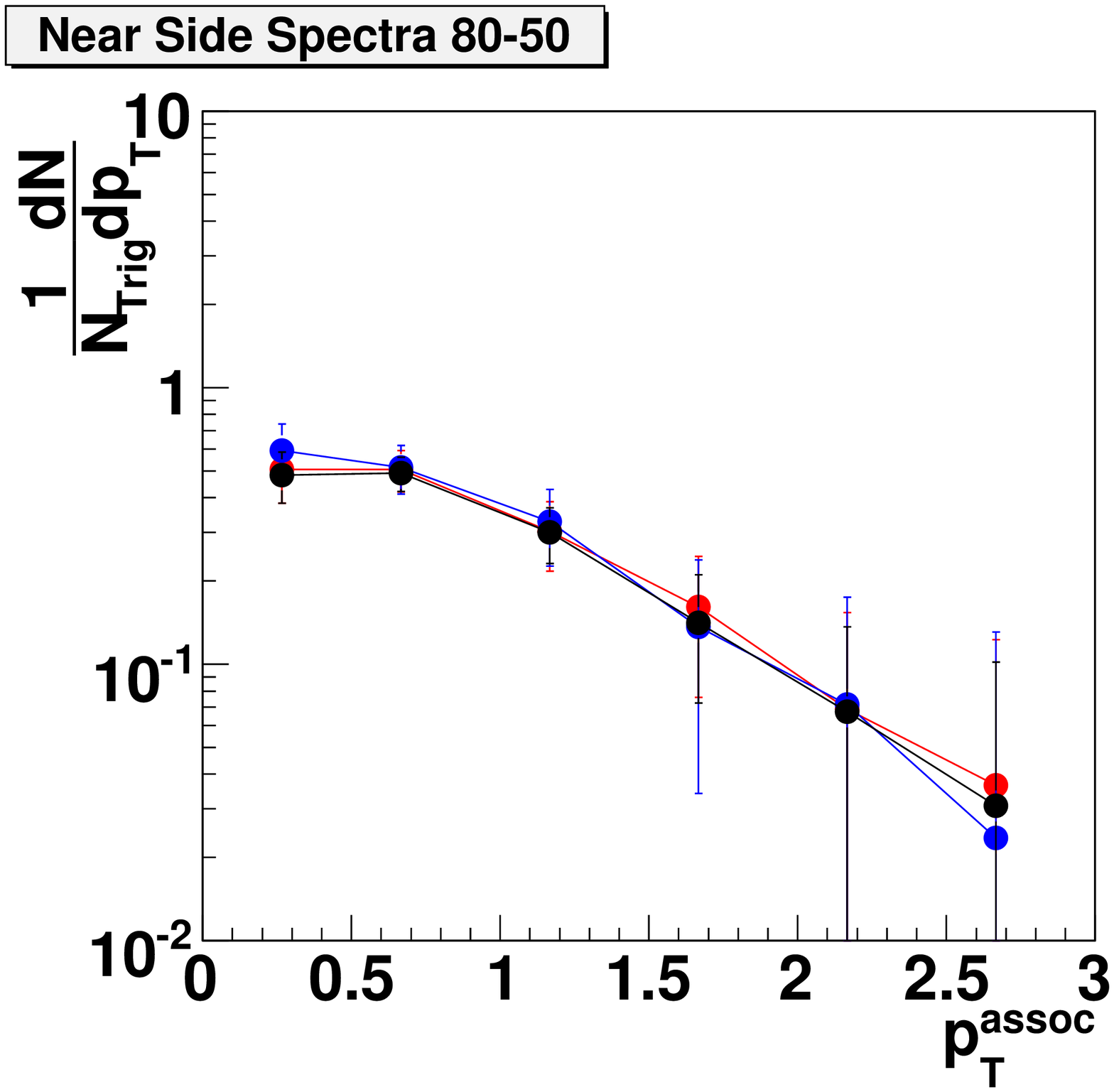}
\includegraphics[width=1\textwidth]{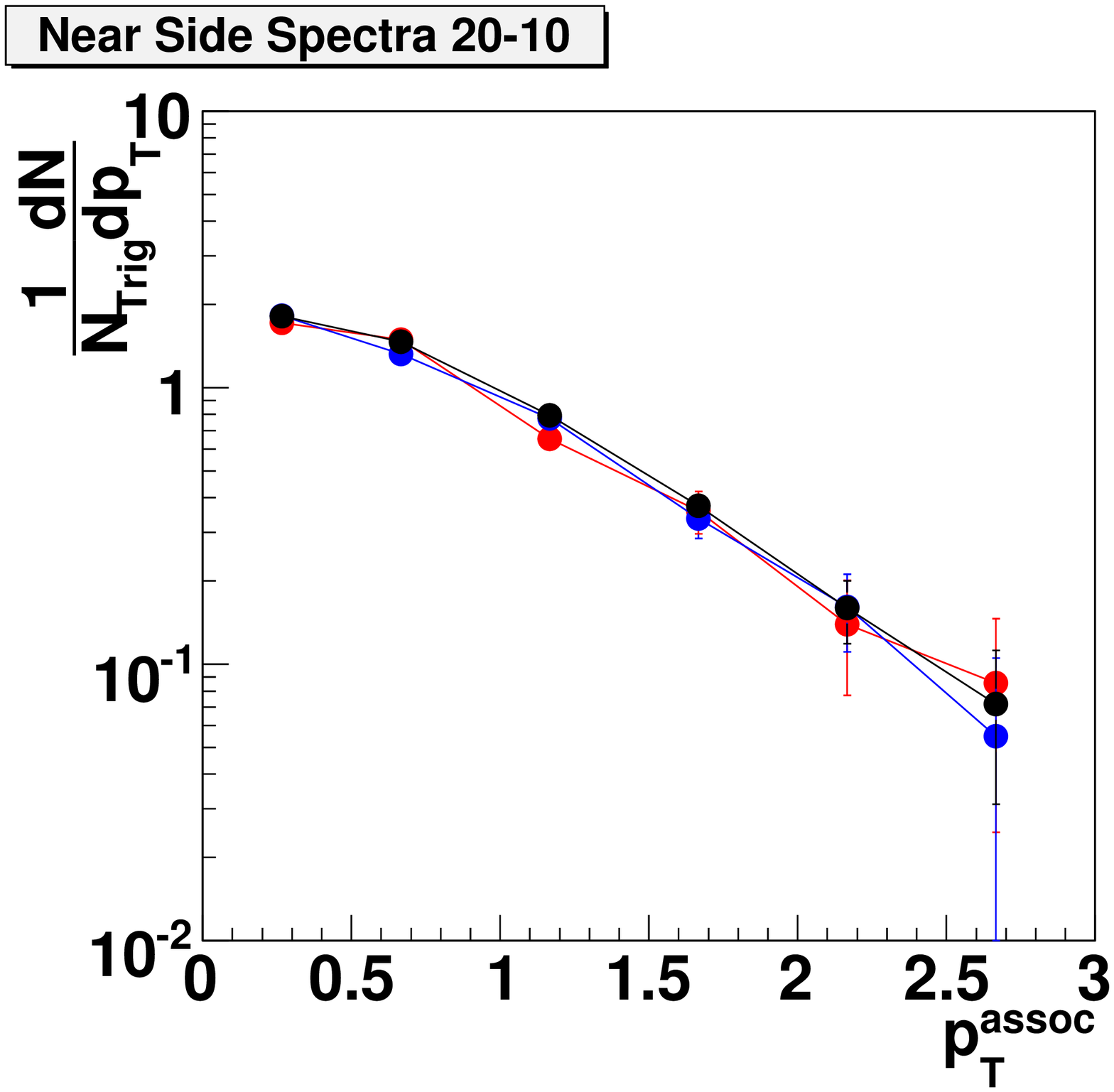}
\end{minipage}
\hfill
\begin{minipage}[t]{0.32\textwidth}
\centering
\includegraphics[width=1\textwidth]{Plots/spectblank.eps}
\includegraphics[width=1\textwidth]{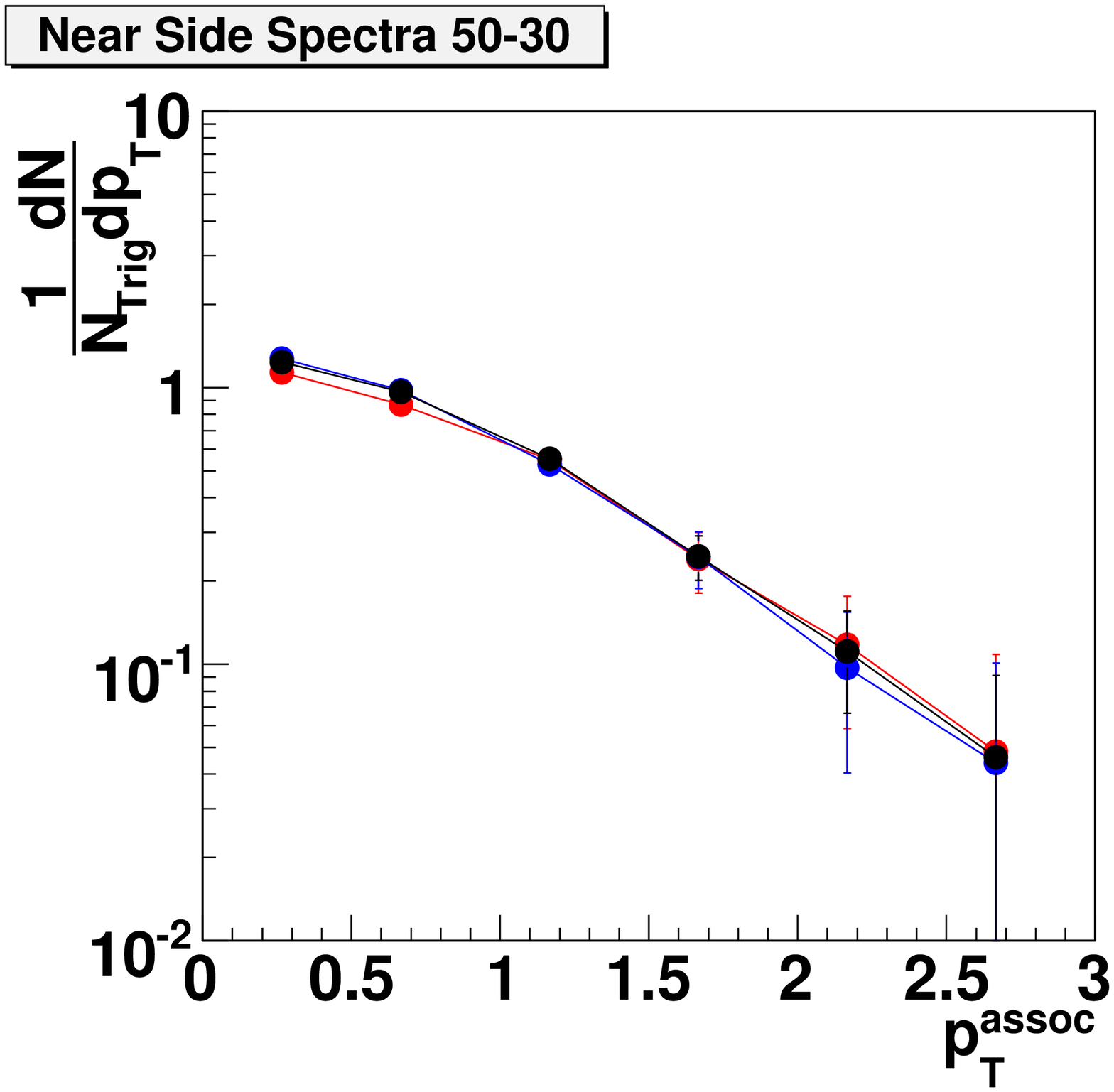}
\includegraphics[width=1\textwidth]{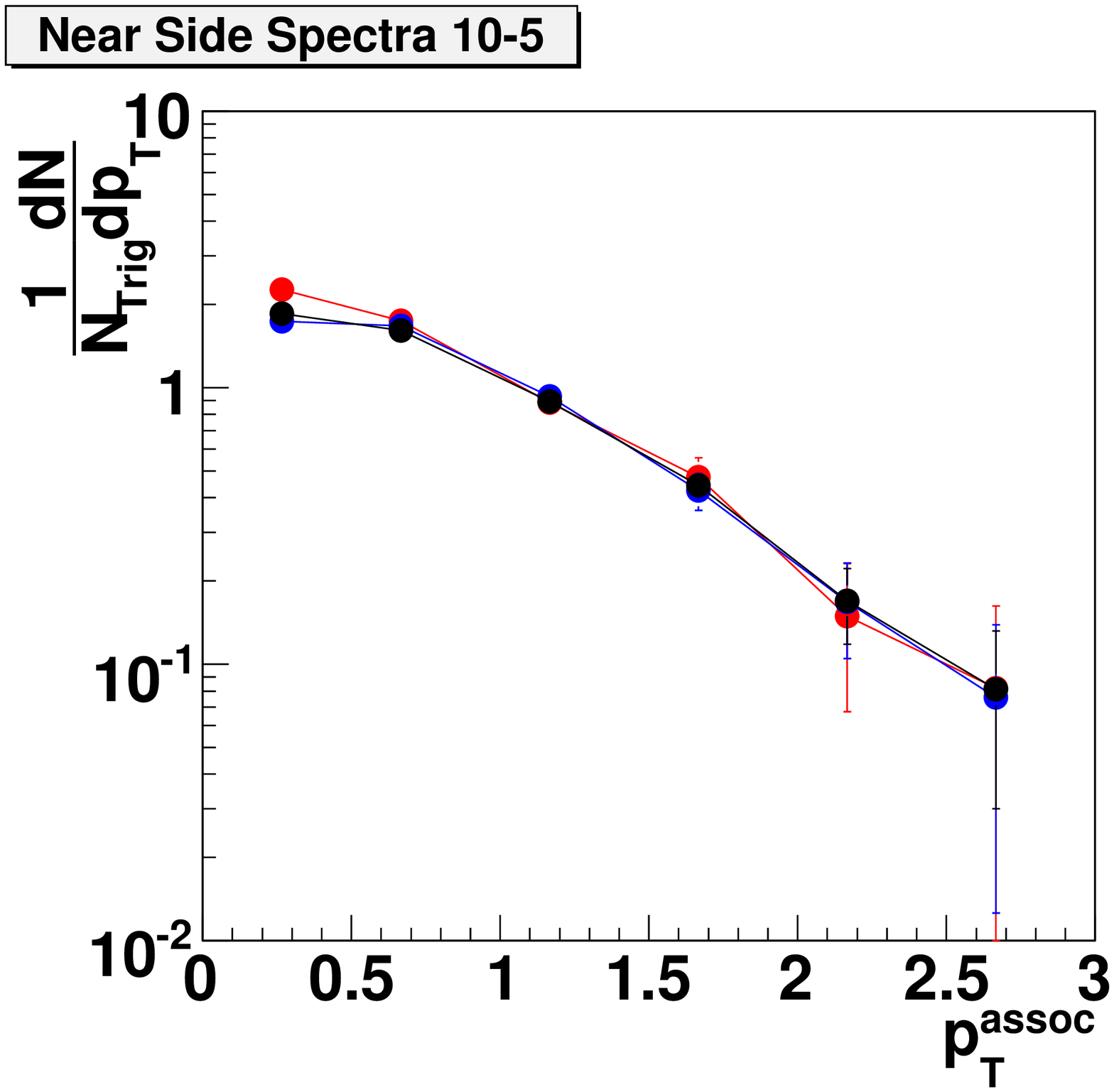}
\end{minipage}
\begin{minipage}[t]{0.32\textwidth}
\centering
\includegraphics[width=1\textwidth]{Plots/spectblank.eps}
\includegraphics[width=1\textwidth]{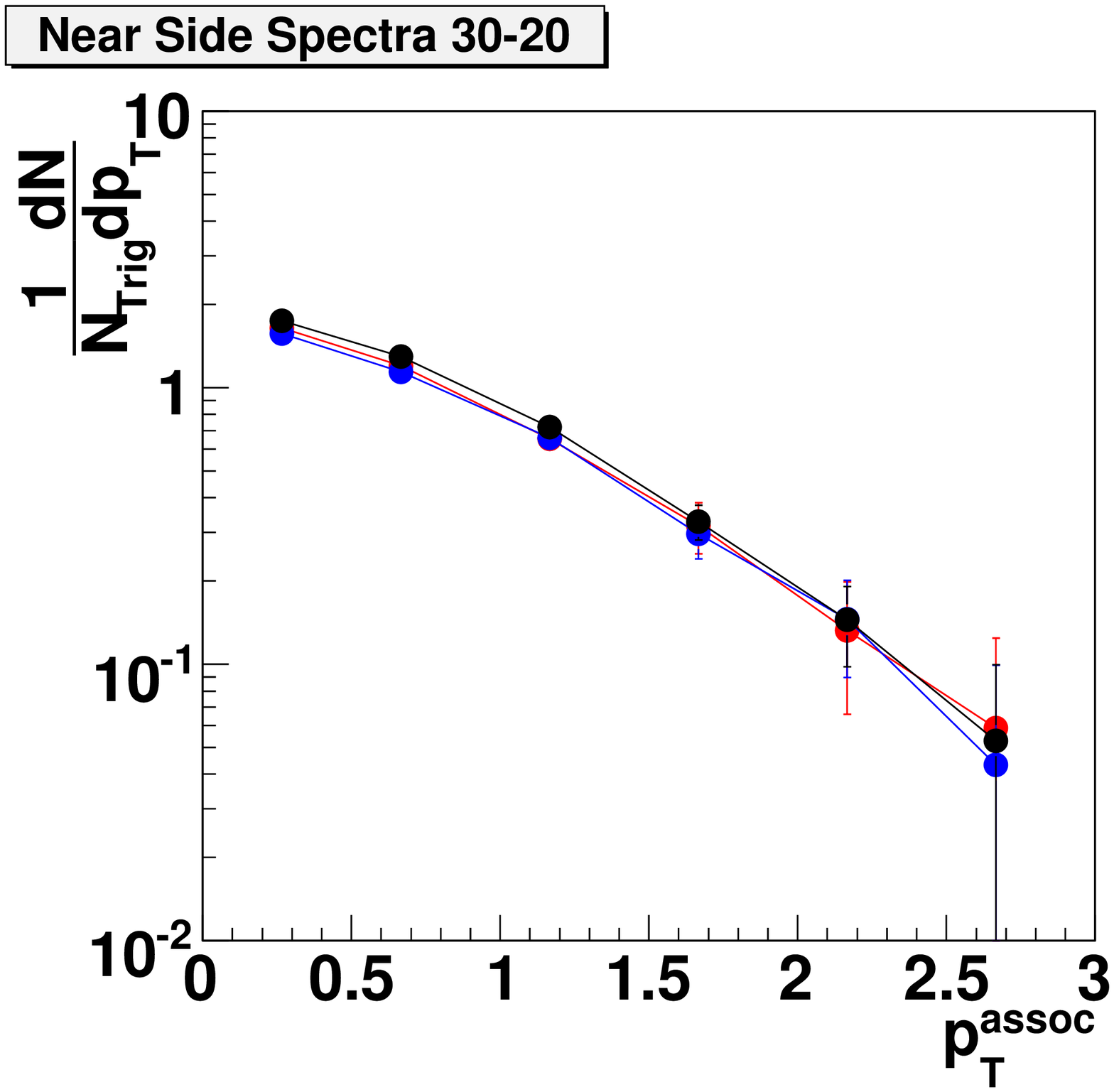}
\includegraphics[width=1\textwidth]{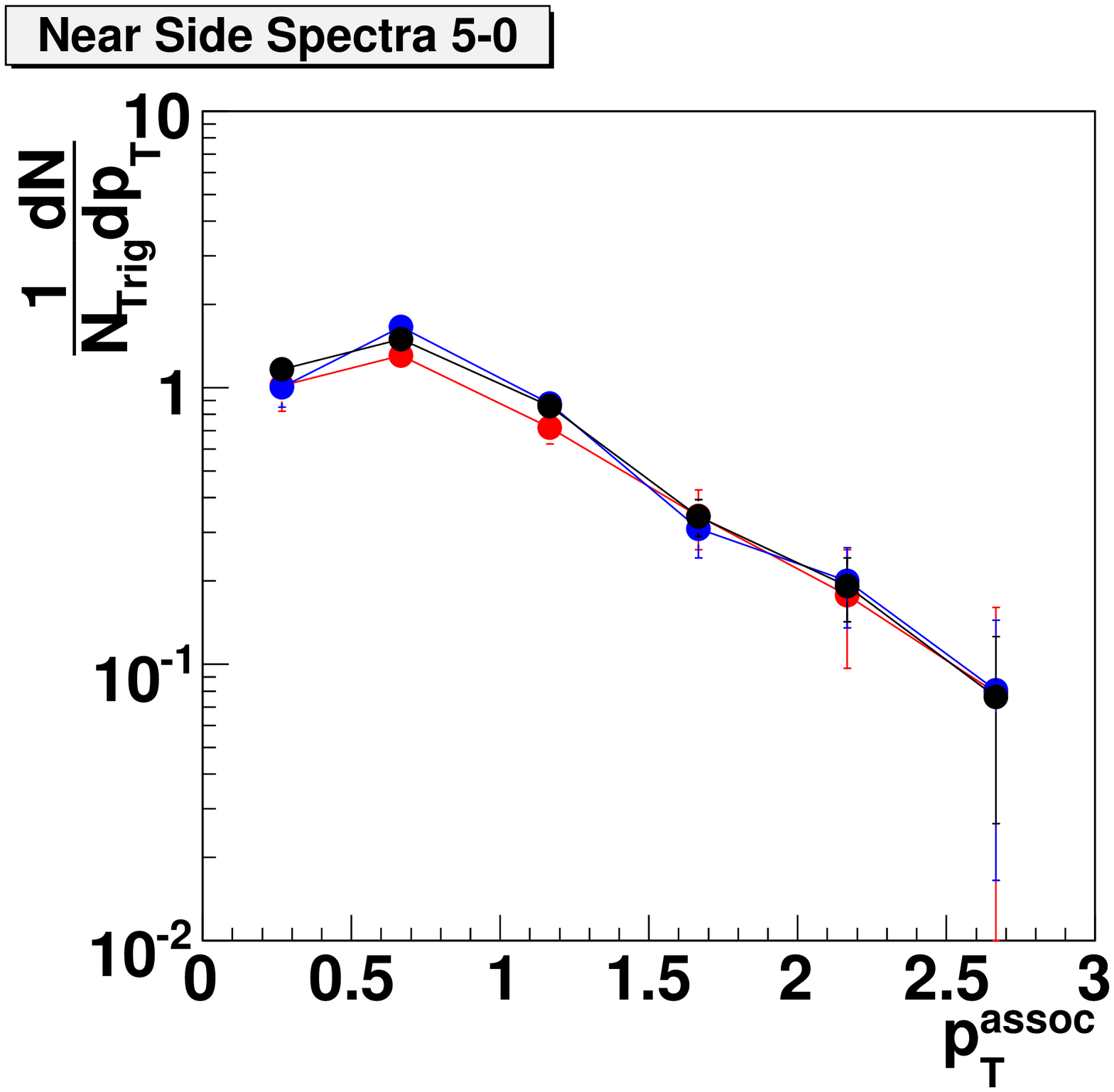}
\end{minipage}
\caption{Near side spectra for identified protons/anti-protons (blue), charged pions (red)  and all charged particles (black) in Au+Au collisions at $\sqrt{s_{NN}}=200$ GeV/c.  Panels are d+Au minimum bias and Au+Au centrality bins 50-80\%, 30-50\%, 20-30\%, 10-20\%, 5-10\%, and 0-5\% from left to right and top to bottom.  Errors are statistical.}
\label{fig:pidNspect}
\end{figure}

\begin{figure}[htb]
\hfill
\begin{minipage}[t]{0.32\textwidth}
\centering
\includegraphics[width=1\textwidth]{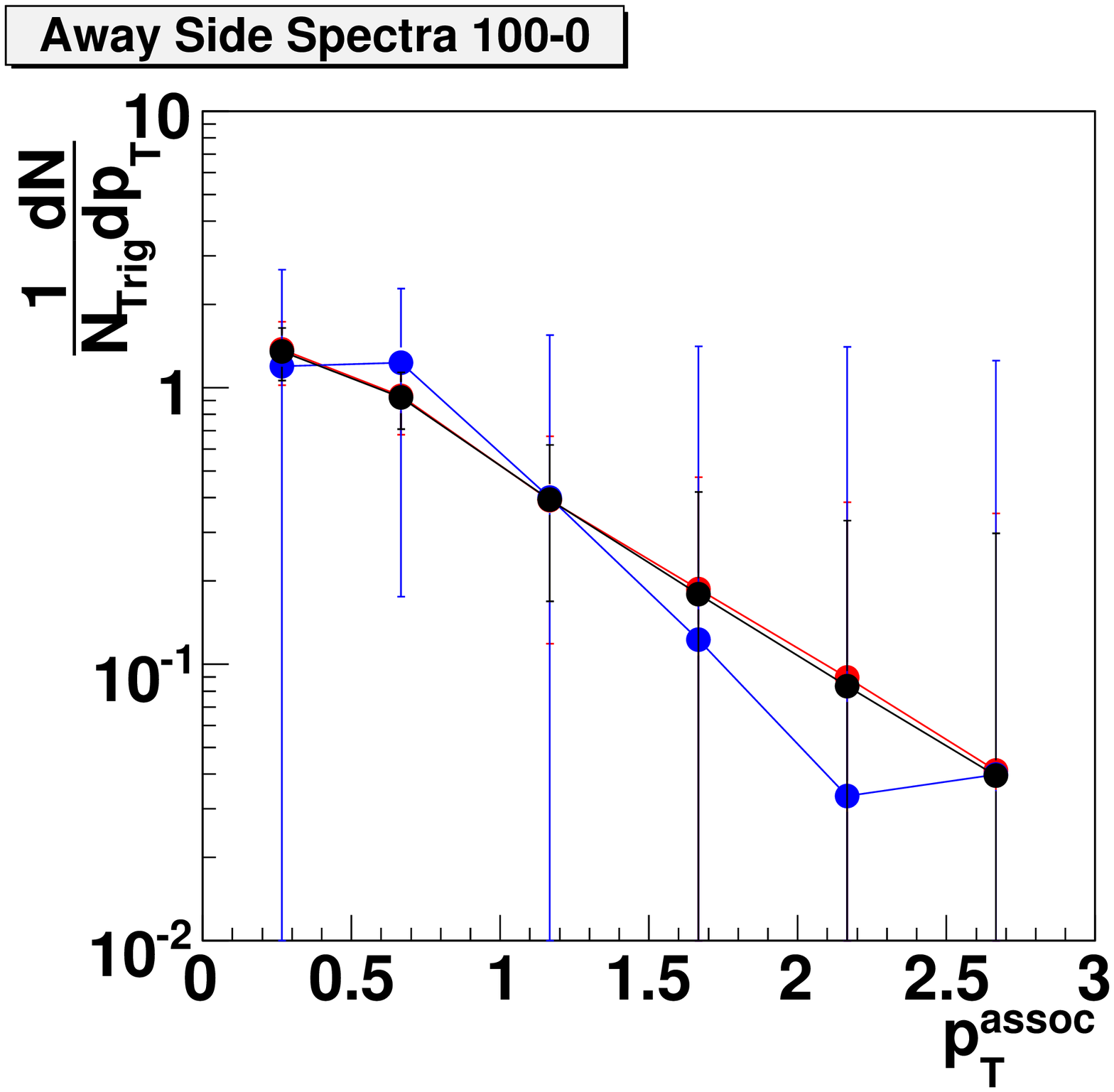}
\includegraphics[width=1\textwidth]{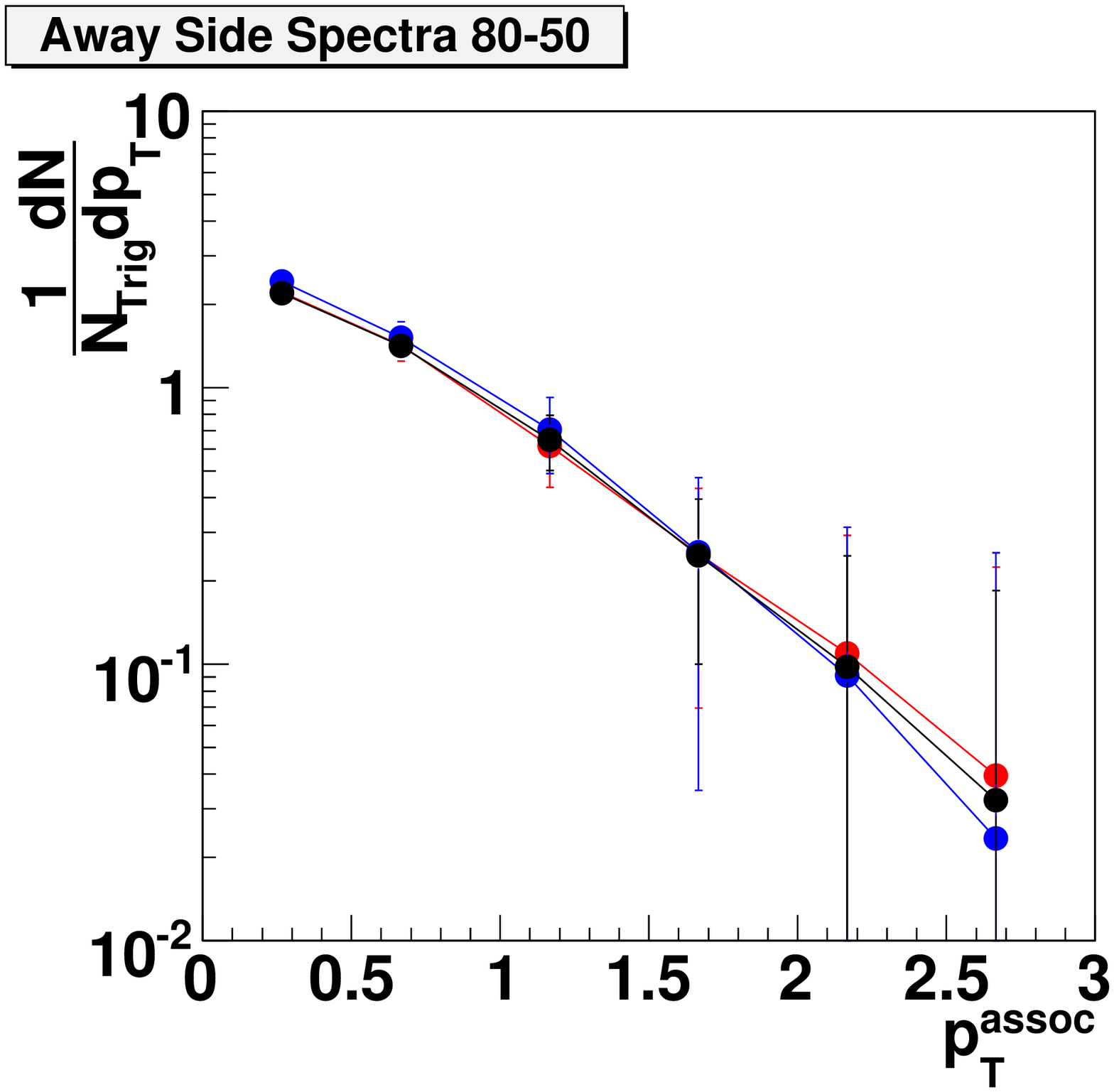}
\includegraphics[width=1\textwidth]{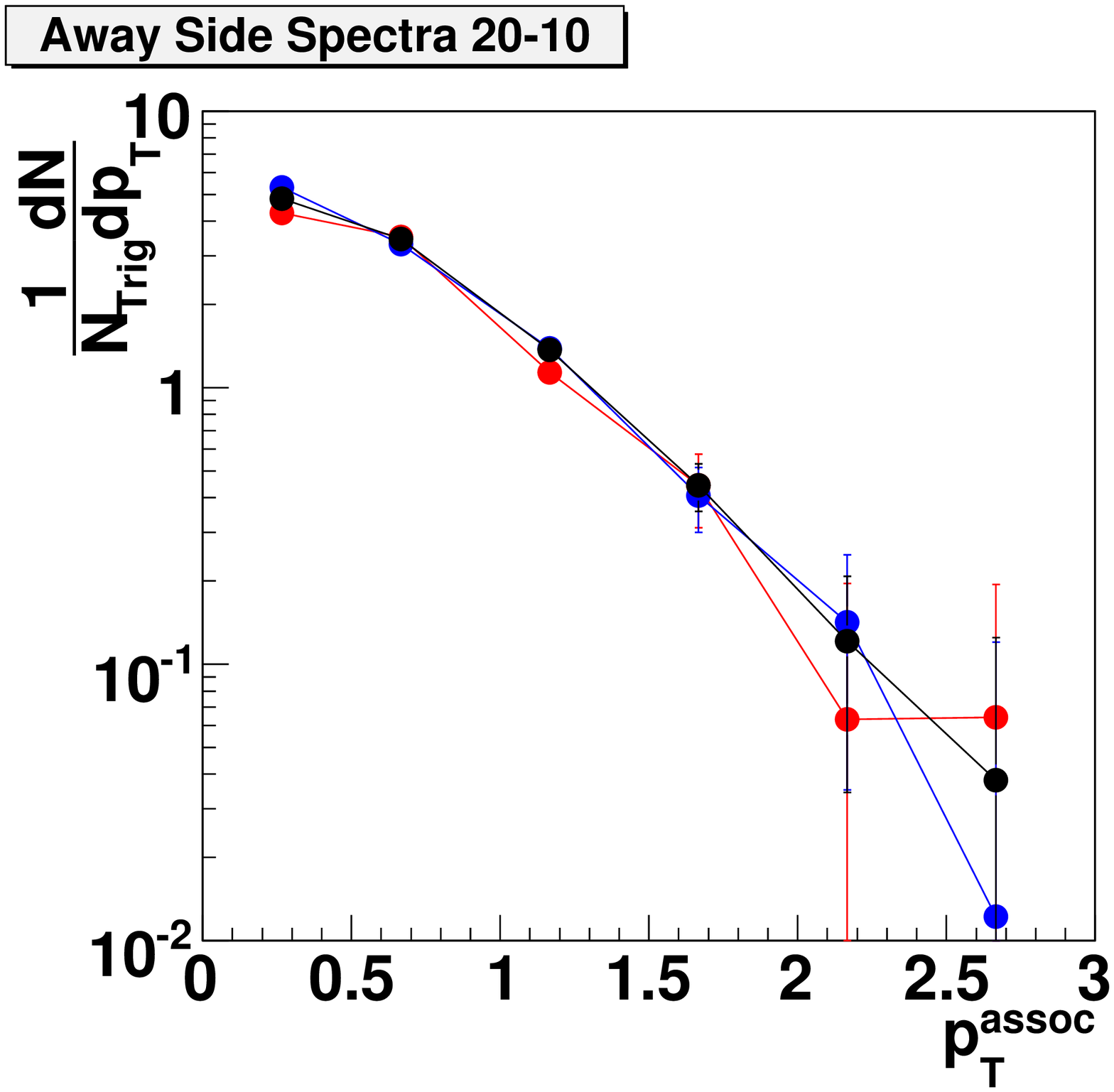}
\end{minipage}
\hfill
\begin{minipage}[t]{0.32\textwidth}
\centering
\includegraphics[width=1\textwidth]{Plots/spectblank.eps}
\includegraphics[width=1\textwidth]{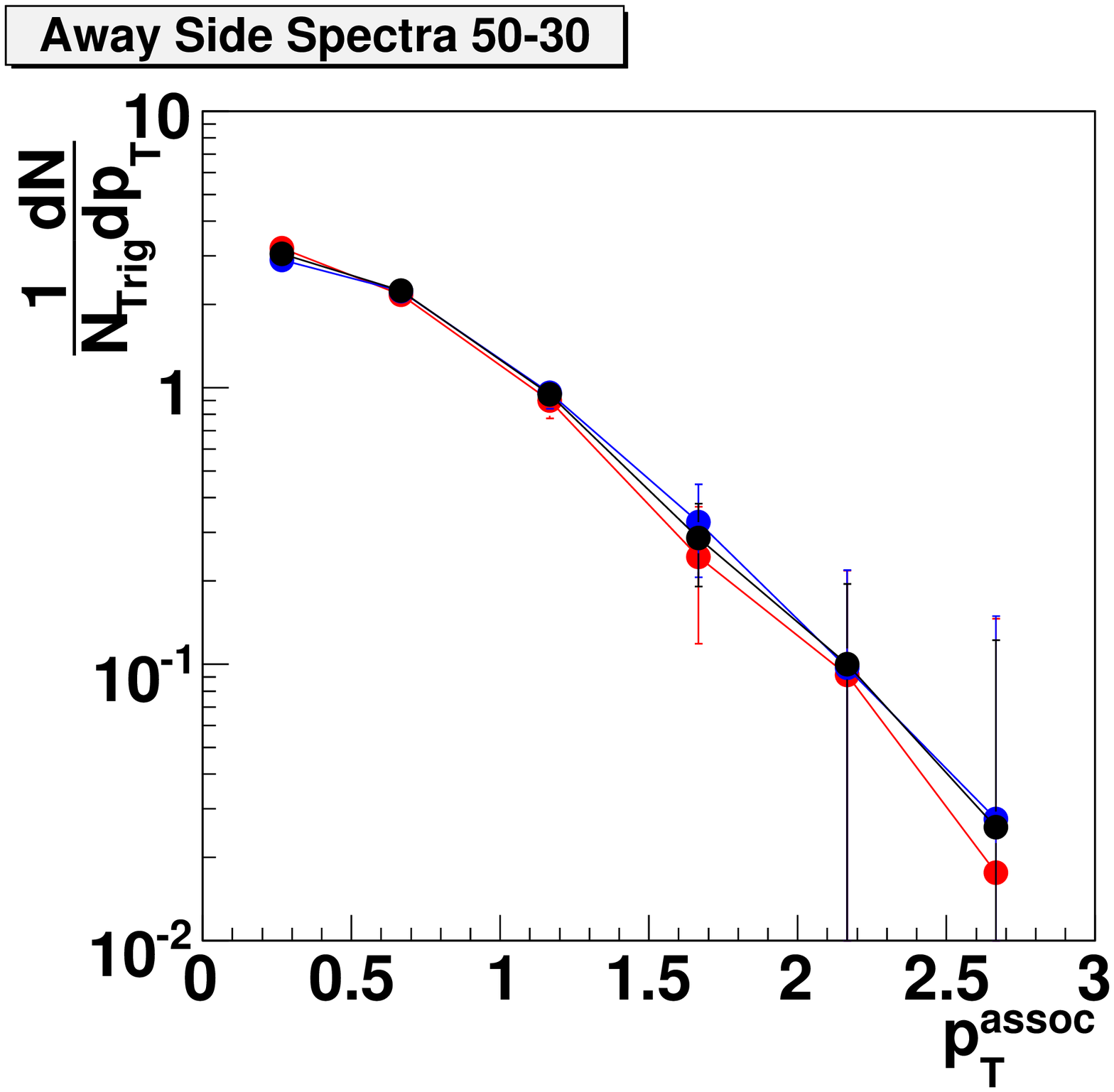}
\includegraphics[width=1\textwidth]{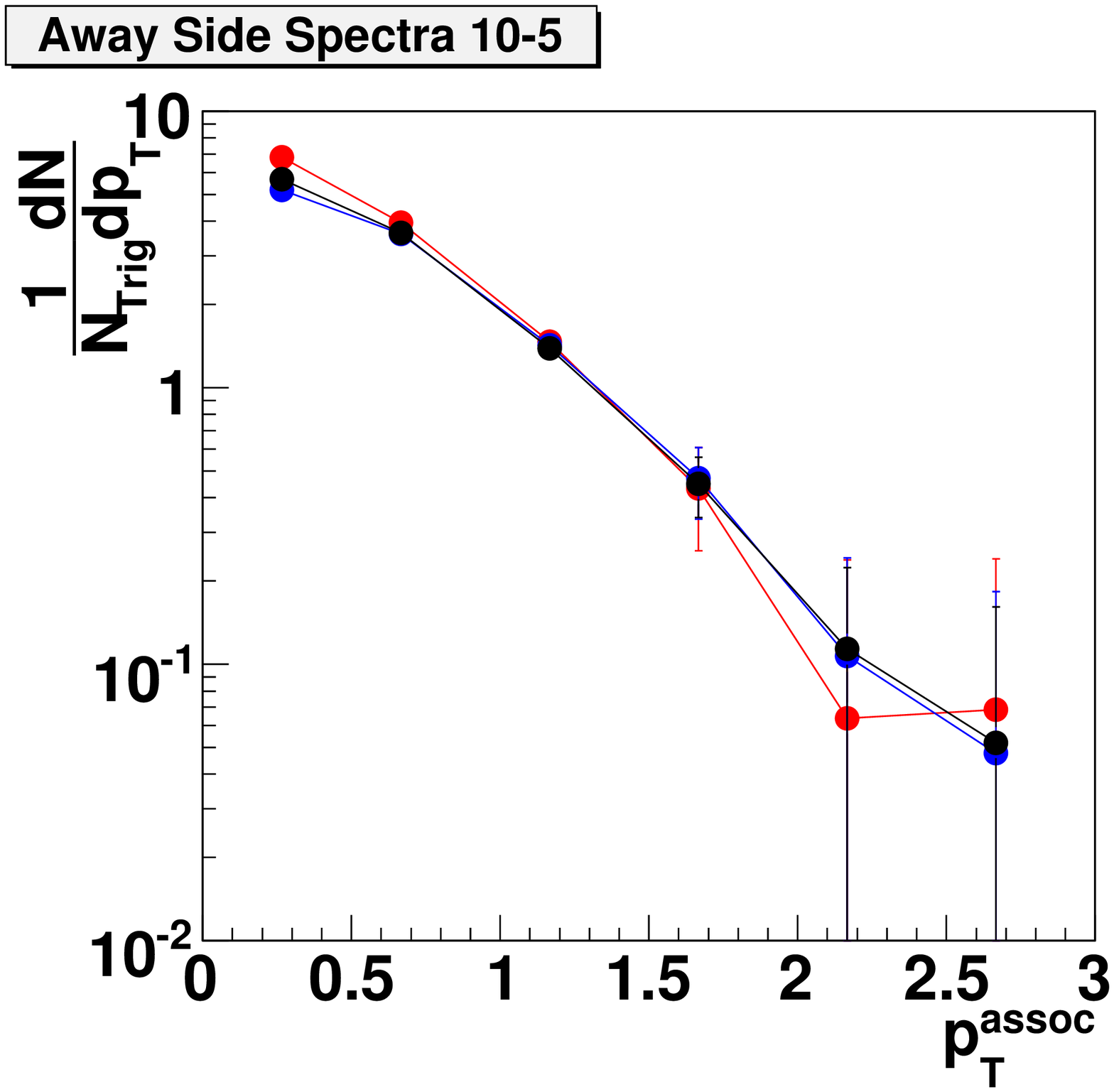}
\end{minipage}
\begin{minipage}[t]{0.32\textwidth}
\centering
\includegraphics[width=1\textwidth]{Plots/spectblank.eps}
\includegraphics[width=1\textwidth]{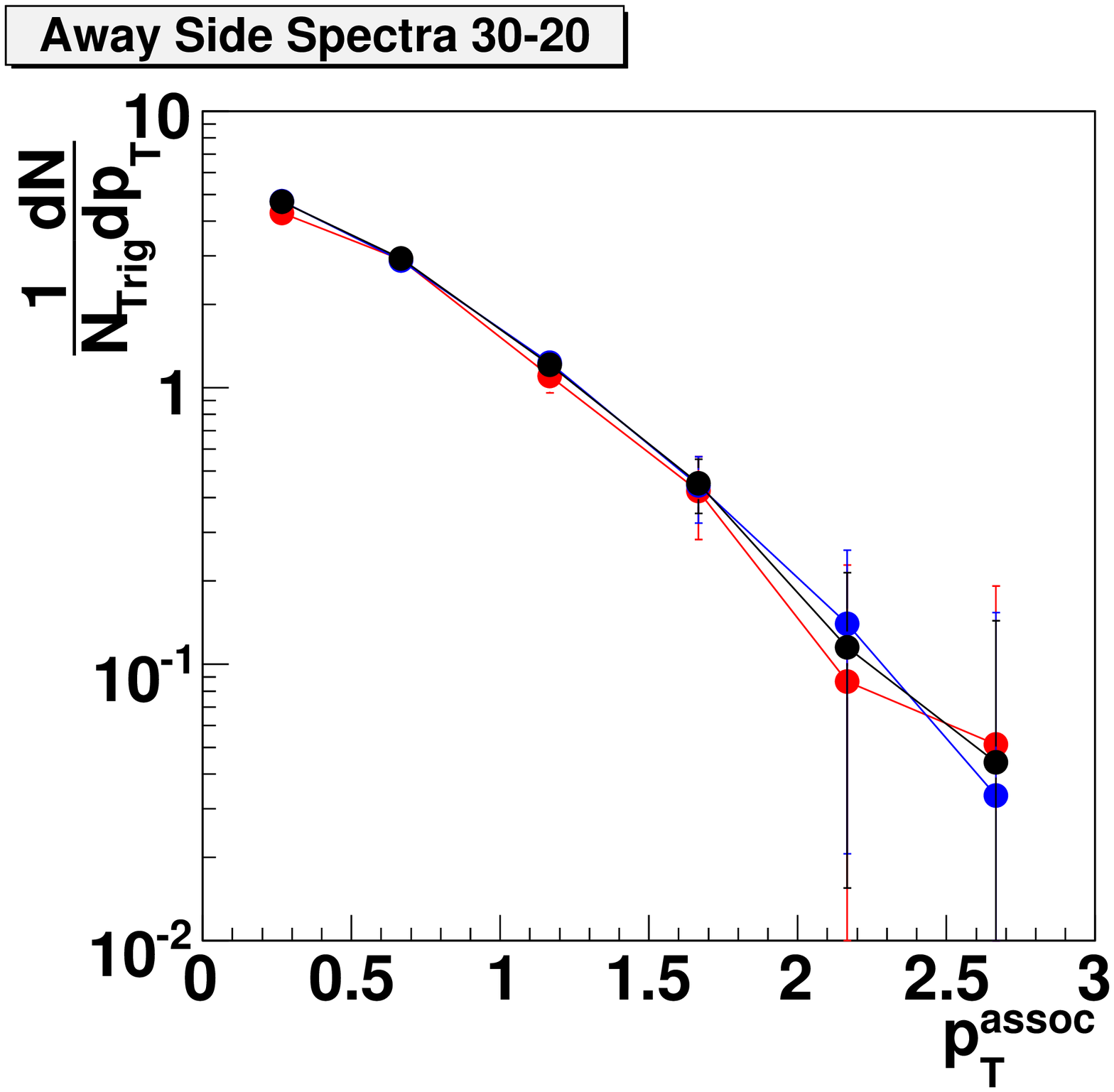}
\includegraphics[width=1\textwidth]{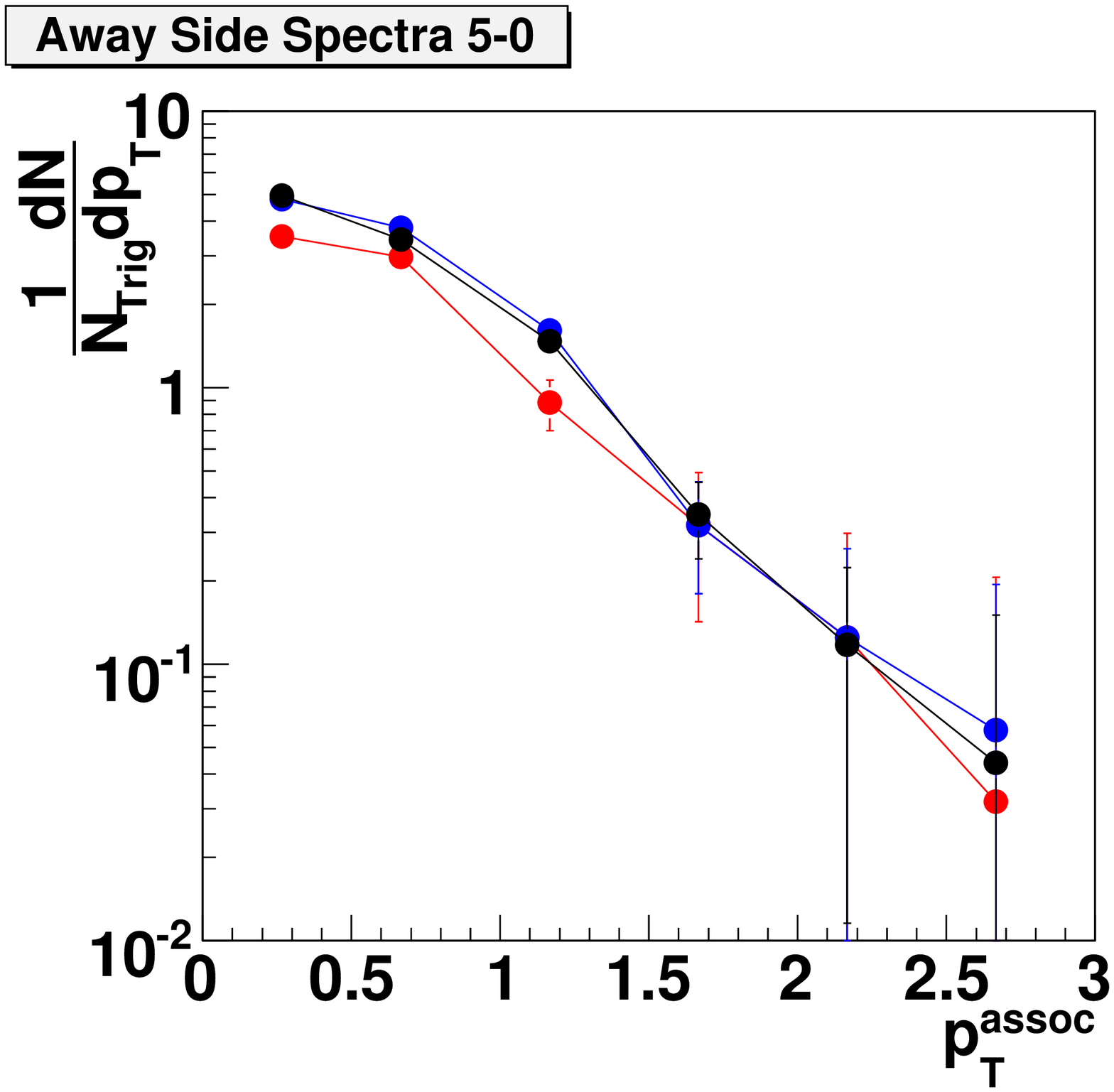}
\end{minipage}
\caption{Away side spectra for identified protons/anti-protons (blue), charged pions (red)  and all charged particles (black) in Au+Au collisions at $\sqrt{s_{NN}}=200$ GeV/c.  Panels are d+Au minimum bias and Au+Au centrality bins 50-80\%, 30-50\%, 20-30\%, 10-20\%, 5-10\%, and 0-5\% from left to right and top to bottom.  Errors are statistical.}
\label{fig:pidAspect}
\end{figure}

Figures~\ref{fig:pidNspect} and~\ref{fig:pidAspect} show the near-side and away-side associated particle spectra for triggered protons/anti-protons, charged pions, and all charged particles.  The associated particle spectra for both triggered protons/anti-protons and charged pions are consistent with the associated spectra for unidentified triggers.  This is true for all centralities in Au+Au collisions.

\subsection{Summary}

Jet-like correlations have been studied with identified trigger particles at $\sqrt{s_{NN}}=200$ GeV/c.  This study is motivated by the observed baryon/meson puzzle (the large baryon/meson ratio at intermediate $p_{T}$).  If more baryons are indeed formed from coalescence of thermal constituent quarks, then the per-trigger normalized jet-like correlation will be reduced for trigger baryons relative to that for trigger mesons.  This analysis is done for intermediate $p_{T}$ triggers for two reasons.  First, we are statistics limited to explore the $p_{T}^{Trig}$ space in more detail.  Second, this is the region of interest where the baryon/meson ratio peaks.  No significant difference is seen between jet-like correlations with identified baryons ($p$ and $\bar{p}$) and mesons ($\pi^{-}$ and $\pi^{+}$).  The increased baryon/meson ratio does not seem to have a significant effect on the associated particle production.
\chapter{THREE-PARTICLE JET-LIKE CORRELATIONS}

\section{Introduction and Motivation}
Two-particle azimuthal correlations that are reported in Chapter 3 of this thesis and elsewhere\cite{fuqiang,phenix} have shown a modified away-side structure in central Au+Au collisions, with respect to {\it pp}, d+Au, and peripheral Au+Au.  The away-side distribution is broadened or even double-humped in central Au+Au collisions. Different physics mechanisms have been suggested to explain this modification including: large angle gluon radiation\cite{gluon1,gluon2}, jets deflected by radial flow\cite{deflected} or preferential selection of particles due to path-length dependent energy loss\cite{path}, hydrodynamic conical flow generated by Mach-cone shock waves\cite{mach0,mach1,mach2,mach25,mach3,mach4,mach5,mach6}, and \v{C}erenkov gluon radiation\cite{cerenkov1,cerenkov2}.  Three-particle correlations can be used to differentiate the mechanisms with conical emission, Mach-cone and \v{C}erenkov gluon radiation, from other physics mechanisms. 

\begin{figure}[htb]
\centering
\includegraphics[width=.5\textwidth]{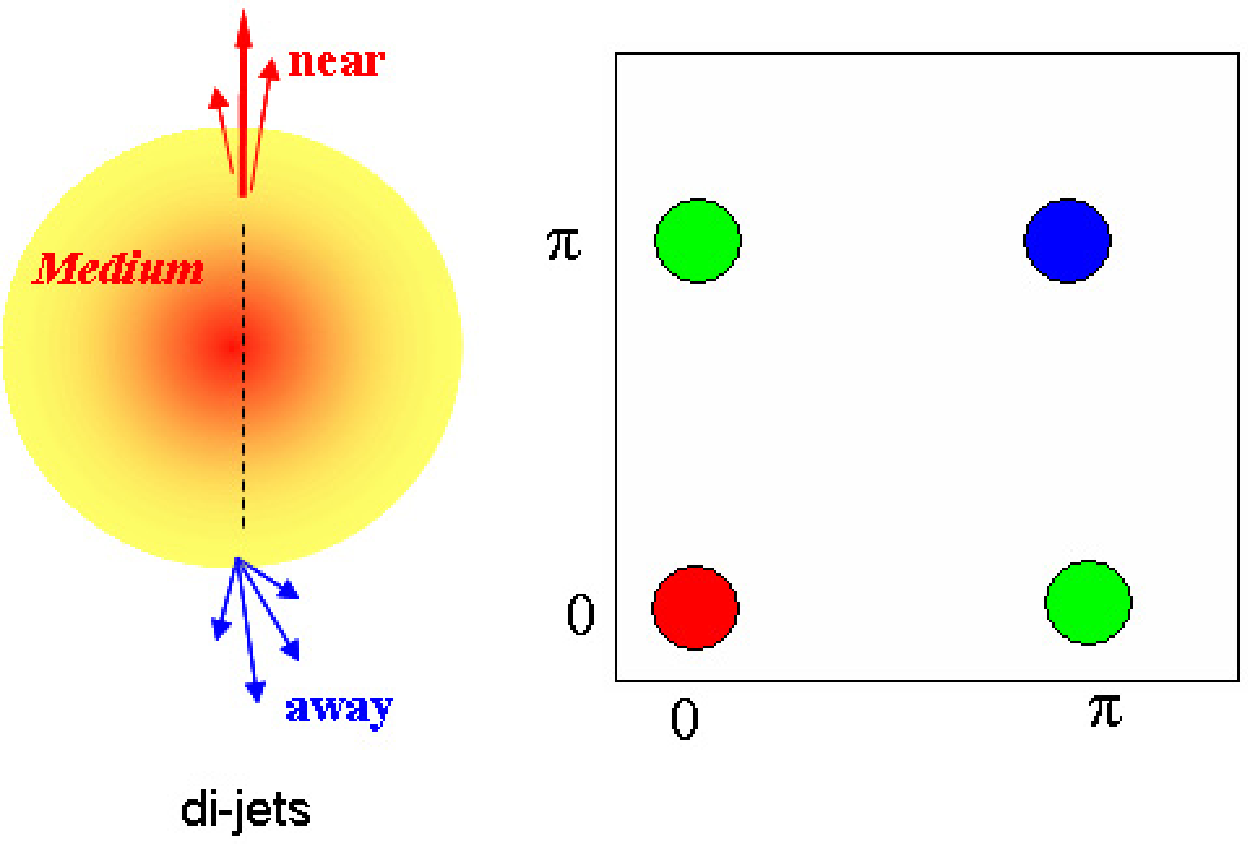}
\includegraphics[width=.5\textwidth]{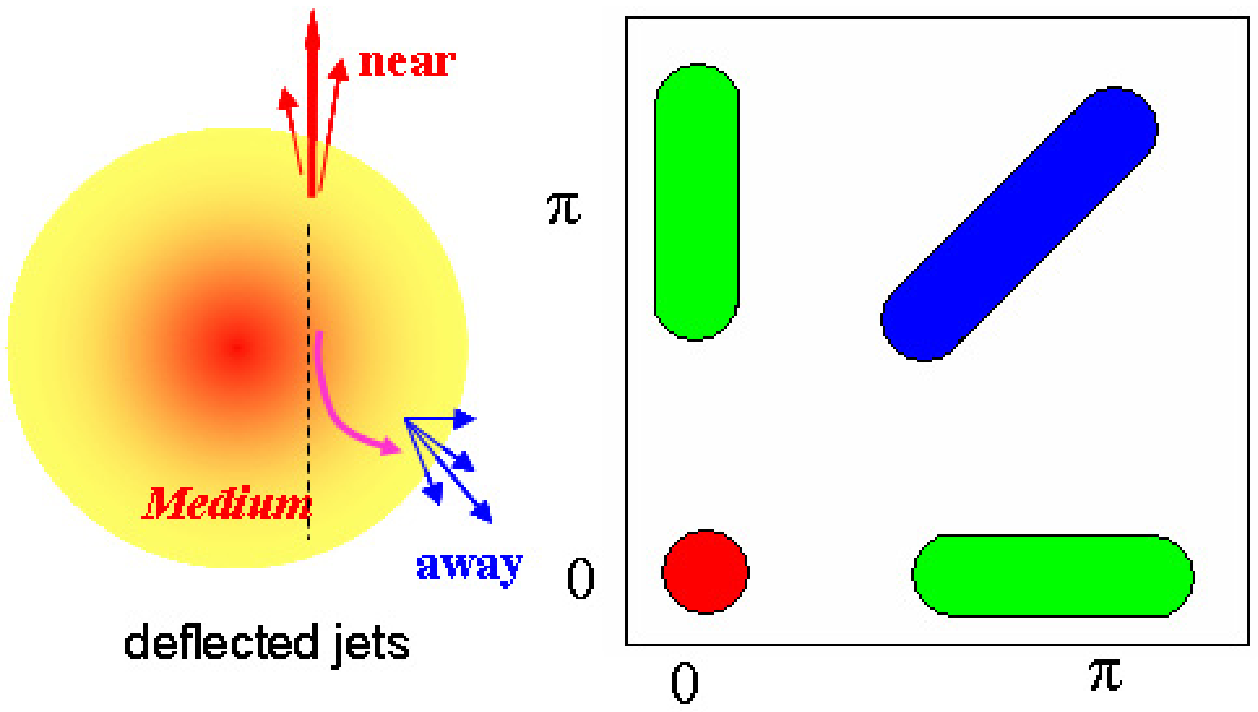}
\includegraphics[width=.5\textwidth]{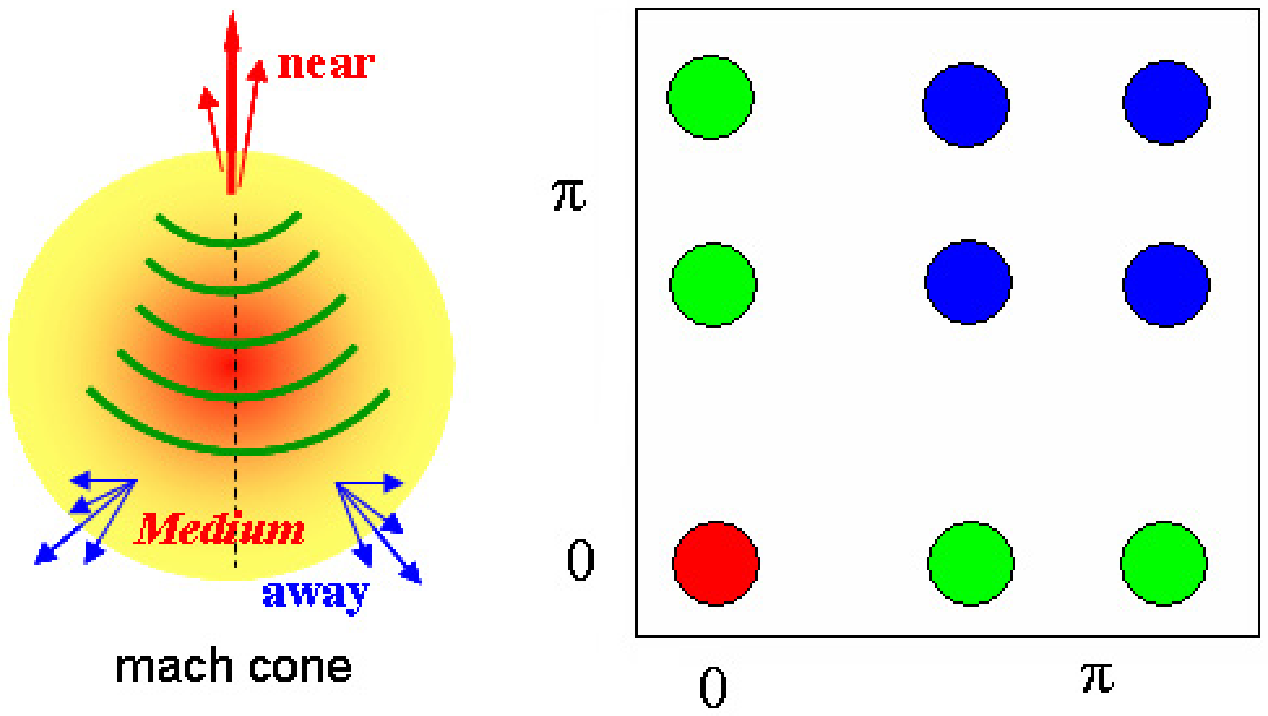}
\caption{Cartoon representations of the expected results for 3-particle azimuthal correlations for different physics mechanisms.  Top: Unmodified back-to-back di-jets.  Middle:  Jets deflected by radial flow or path length dependent energy loss.  Bottom:  Conical emission.}
\label{fig:toon}
\end{figure}

Figure~\ref{fig:toon} is a cartoon view of expected 3-particle correlation signals for different scenarios.  The top panel shows unmodified back-to-back di-jets.  In $\Delta\phi_1-\Delta\phi_2$ space, , where $\Delta\phi_{i}=\phi_{i}-\phi_{Trig}$, the 3-particle correlation will have four peaks, one at $(0,0)$ for instances where both associated particles are on the near-side, one at $(\pi,\pi)$ for instances where both associated particles are on the away-side and peaks at $(0,\pi)$ and $(\pi,0)$ for instances where one of the associated particles is on the near-side and the other is on the away-side.  The middle panel shows what happens in situations where the away-side jet is modified such that particles only come out to one side of $\pi$.  In a single event, this would result in an away-side peak on-diagonal but displaced from $\pi$; however we average over many events resulting in an on-diagonal structure.  The structure stays on-diagonal because even though the away-side particles are not at $\pi$ they are still close together.  This scenario can come about due to physics mechanisms such as: jets deflected by radial flow or path-length dependent energy loss.  The bottom panel shows the expected cartoon results for conical emission.  For conical emission particles can be off-set to both sides of $\pi$.  When both associated particle are off-set to the same side of $\pi$ we get on-diagonal peaks again.  If we take particles that are off-set to opposite sides of $\pi$ the particles are no long close together and we get peaks on the off-diagonal.  This gives us four peaks on the away-side.  The two off-diagonal peaks are only expected for conical emission and therefore will be our signature for conical emission.

\section{Analysis Procedure}

Results from this analysis have been publicly shown by STAR\cite{proc1,proc2,proc3,proc4,proc5}.  This chapter presents the details the analysis.  The details of the analysis techniques have been described in \cite{writeup}.  The analysis assumes that an event is composed, besides the trigger particle, of two components,  particles that are correlated with the trigger particle (other than the correlation due to anisotropic flow) and background particles from the bulk medium (only correlated with the trigger particle due to anisotropic flow).  Suppose the number of jet-correlated particles is $N_{jet}$ and the number of background particles is $N_{bkgd}$.  The total number of particles is then $N = N_{jet} + N_{bkgd}$.  The number of particle pairs is $N(N-1)$ and is composed of three parts: the number of background pairs $N_{bkgd}(N_{bkgd}-1)$, the number of jet-correlated pairs $N_{jet}(N_{jet}-1)$, and the number of cross pairs $N_{jet}N_{bkgd}+N_{bkgd}N_{jet}=2N_{jet}N_{bkgd}$.   Since we are interested in the angular correlation of the jet-correlated pairs which is an unknown, we must subtract the angular correlations of the background pairs and the cross pairs that are known.  The angular correlations of the background pairs and the cross pairs can be obtained because they are uncorrelated (except for the anisotropic flow correlation).  Therefore, by subtraction from the raw 3-particle correlation the angular correlations of the background pairs and the cross pairs, one can obtain the genuine 3-particle jet-correlation.
Lets introduce some equations to restate this more formally. For 2-particle correlations:
\begin{eqnarray}
J_{2}(\Delta\phi)&=&dN/d\Delta\phi, \nonumber \\
B_{2}(\Delta\phi)&=&dN_{bkgd}/d\Delta\phi, \nonumber \\
\hat{J_{2}}(\Delta\phi)&=&dN_{jet}/d\Delta\phi, \nonumber \\
\hat{J_{2}}(\Delta\phi)&=&J_{2}(\Delta\phi)-B_{2}(\Delta\phi),
\end{eqnarray}
where $\Delta\phi$ is the associated particle azimuthal angle relative to that of the trigger particle.  $J_{2}$ is a notation for the raw azimuthal distribution relative to the trigger particle azimuth for the entire event.  $B_{2}$ is a notation for the azimuthal distribution of the underlying background particles relative to the trigger particle azimuth.  $\hat{J_{2}}$ is the azimuthal distribution of the associated jet particles relative to the trigger particle azimuth.

The raw 3-particle azimuthal correlation between the trigger particle and two other particles in this notation is:
\begin{eqnarray}
J_{3}(\Delta\phi_{1},\Delta\phi_{2})&=&J_{2}(\Delta\phi_{1}) \otimes J_{2}(\Delta\phi_{2}) \nonumber \\ 
&=&[\hat{J_{2}}(\Delta\phi_{1})+B_{2}(\Delta\phi_{1})]\otimes[\hat{J_{2}}(\Delta\phi_{2})+B_{2}(\Delta\phi_{2})]
\end{eqnarray}
where $A\otimes B$ is not a simple product of $A$ and $B$ due to correlations, $\Delta\phi_{1}=\phi_{1}-\phi_{trig}$, and $\Delta\phi_{2}=\phi_{2}-\phi_{trig}$.  The raw 3-particle correlation can be divided into three parts: the 3-particle jet-correlations we are interested in,
\begin{equation}
\hat{J_{3}}(\Delta\phi_{1},\Delta\phi_{2})=\hat{J_{2}}(\Delta\phi_{1})\otimes\hat{J_{2}}(\Delta\phi_{2}),
\end{equation}
and two background terms.  One of the combinatoric backgrounds is between an associated jet particle and a background particle,
\begin{eqnarray}
\hat{J_{2}} \otimes B_{2} &=& \hat{J_{2}}(\Delta\phi_{1}) \otimes B_{2}(\Delta\phi_{2}) + B_{2}(\Delta\phi_{1}) \otimes \hat{J_{2}}(\Delta\phi_{2}) \nonumber \\
&=& \hat{J_{2}}(\Delta\phi_{1})B_{2}(\Delta\phi_{2})+B_{2}(\Delta\phi_{1})\hat{J_{2}}(\Delta\phi_{2}),
\label{eqn:hsbg}
\end{eqnarray}
which we term the {\it hard-soft} background.  Since the jet-correlated particle and the background particle are uncorrelated, the hard-soft background can be obtained from the simple product of the 2-particle jet-correlation function $\hat{J_{2}}$ and the background $B_{2}$.  The other background is the combinatoric  background between the two background particles,
\begin{equation}
B_{3}(\Delta\phi_{1},\Delta\phi_{2})=B_{2}(\Delta\phi_{1}) \otimes B_{2}(\Delta\phi_{2}),
\end{equation}
which we term the {\it soft-soft} background.  The soft-soft background can be obtained from inclusive events and will be discussed in greater detail below.  The final 3-particle correlation can thus be determined by,
\begin{equation}
\hat{J_{3}}(\Delta\phi_{1},\Delta\phi_{2})=J_{3}(\Delta\phi_{1},\Delta\phi_{2})-[\hat{J_{2}}(\Delta\phi_{1})B_{2}(\Delta\phi_{2})+B_{2}(\Delta\phi_{1})\hat{J_{2}}(\Delta\phi_{2})]-B_{3}(\Delta\phi_{1},\Delta\phi_{2}).
\end{equation}

\subsection{Flow Correction}
In heavy-ion collisions, particle emission is correlated to the reaction plane due to hydrodynamic collective flow of the bulk medium and the anisotropic overlap region between the colliding nuclei.  We have already examined this flow correlation in 2-particle correlation studies up to the second order harmonic.  Because the signal to noise ratio in 3-particle correlations is small, the fourth order harmonics of the flow correlation cannot be neglected.  The flow correlation, expressed in harmonics up to fourth order, is given by,
\begin{equation}
\frac{dN}{d\phi}=\frac{N}{2\pi}[1+2v_{2}\cos2(\phi-\Psi)+2v_{4}\cos4(\phi-\Psi)]
\label{2partflow}
\end{equation}
where $\phi$ is the azimuthal angle of the particle and $\Psi$ is the azimuthal angle of the reaction plane.  Due to symmetry the first and third (and all other odd) harmonic terms vanish at mid-rapidity for symmetric collisions systems. 
The distribution of the trigger-background particle pairs is given by
\begin{eqnarray}
\frac{d^{4}N}{d\Psi d\phi_{trig} d\phi d(\Delta\phi)}&=&\frac{N_{trig}}{2\pi}[1+2v_{2}^{trig}\cos 2(\phi_{trig}-\Psi)+2v_{4}^{trig}\cos 4(\phi_{trig}-\Psi)] \nonumber \\
&&\cdot \frac{N_{bkgd}}{2\pi}[1+2v_{2}\cos 2(\phi-2\Psi)+2v_{4}\cos(4\phi-4\Psi)] \nonumber \\
&&\cdot \frac{1}{2\pi}\delta(\Delta\phi-(\phi-\phi_{trig}))
\end{eqnarray}
for a given $\Psi$, $\phi_{trig}$, and $\phi$.  $N_{trig}$ is the number of trigger particles and $N_{bkgd}$ is the total number of background particles.  Integrating over $\Psi$, $\phi_{trig}$, and $\phi$ we obtain,
\begin{eqnarray}
\frac{dN}{d(\Delta\phi)}&=&\int_{0}^{2\pi} \int_{0}^{2\pi} \int_{0}^{2\pi}d\Psi d\phi d\phi_{trig} \frac{d^{4}N}{d\Psi d\phi d\phi_{trig}} \nonumber \\
&=&\frac{N_{trig}N_{bkgd}}{2\pi}[1+2v_{2}^{trig}v_{2}\cos(2\Delta\phi)+2v_{4}^{trig}v_{4}\cos(4\Delta\phi)].
\end{eqnarray}
When normalized per trigger particle, the anisotropic flow modulation is then,
\begin{equation}
B_{2}(\Delta\phi)=B_{1}[1+2v_{2}^{trig}v_{2}\cos(2\Delta\phi)+2v_{4}^{trig}v_{4}\cos(4\Delta\phi)]
\end{equation}
where $B_{1}=N_{bkgd}/2\pi$ is the average background density level.

The distribution of the trigger-background-background triplets is given by,
\begin{eqnarray}
\frac{d^{6}N}{d\Psi d\phi_{trig} d\phi_{1} d\phi_{2} d(\Delta\phi_{1}) d(\Delta\phi_{2})} &=&\frac{N_{trig}}{2\pi}[1+2v_{2}^{trig}\cos(2\phi_{trig}-2\Psi) + 2v_{4}^{trig}\cos(4\phi_{trig}-4\Psi)] \nonumber \\
&& \cdot \frac{N_{1}}{2\pi}[1+2v_{2}^{(1)}\cos(2\phi_{1}-2\Psi)+2v_{4}^{1}\cos(4\phi_{1}-4\Psi)] \nonumber \\
&& \cdot \frac{N_{2}}{2\pi}[1+2v_{2}^{(2)}\cos(2\phi_{2}-2\Psi)+2v_{4}^{2}\cos(4\phi_{2}-4\Psi)] \nonumber \\
&& \cdot \frac{1}{2\pi}\delta(\Delta\phi_{1}-(\phi_{1}-\phi_{trig})) \delta(\Delta\phi_{2}-(\phi_{2}-\phi_{trig}))
\end{eqnarray}
for a given set of $\Psi$, $\phi_{trig}$, $\phi_{1}$, and $\phi_{2}$.  After integrating over $\Psi$, $\phi_{trig}$, $\phi_{1}$, and $\phi_{2}$ we obtain,
\begin{eqnarray}
\frac{d^{2}N}{d(\Delta\phi_{1})d(\Delta\phi_{2})} &=& \int_{0}^{2\pi}\int_{0}^{2\pi}\int_{0}^{2\pi}\int_{0}^{2\pi}d\Psi d\phi_{trig} d\phi_{1} d\phi_{2} \frac{d^{6}N}{d\Psi d\phi_{trig} d\phi_{1} d\phi_{2} d(\Delta\phi_{1}) d(\Delta\phi_{2})} \nonumber \\
&=&\frac{N_{trig}N_{1}N_{2}}{(2\pi)^{2}} [1+2v_{2}^{trig}v_{2}^{(1)}\cos(2\Delta\phi_{1}) \nonumber \\
&& + 2v_{2}^{trig}v_{2}^{(2)}\cos(2\Delta\phi_{2}) + 2v_{2}^{(1)}v_{2}^{(2)}\cos(2\Delta\phi_{1}-2\Delta\phi_{2}) \nonumber \\
&& + 2v_{4}^{trig}v_{4}^{(1)}\cos(4\Delta\phi_{1}) + 2v_{4}^{trig}v_{4}^{(2)}\cos(4\Delta\phi_{2}) \nonumber \\
&& + 2v_{4}^{(1)}v_{4}^{(2)}\cos(4\Delta\phi_{1}-4\Delta\phi_{2}) + 2v_{2}^{trig}v_{2}^{(1)}v_{4}^{(2)}\cos(2\Delta\phi_{1}-4\Delta\phi_{2}) \nonumber \\
&& + 2v_{2}^{trig}v_{2}^{(2)}v_{4}^{(1)}\cos(4\Delta\phi_{1}-2\Delta\phi_{2}) + 2v_{2}^{(1)}v_{2}^{(2)}v_{4}^{trig}\cos(2\Delta\phi_{1}+2\Delta\phi_{2})]. \nonumber \\
\label{eqn:flow3}
\end{eqnarray}
Here $v_n^{i}$ is the $n^{th}$ order harmonic of the particle i (where i can be the trigger, or either one of the two associated particles).  In Eqn.~\ref{eqn:flow3} we have taken the two background particles from two separate sets with multiplicities $N_{1}$ and $N_{2}$ giving us $N_{1}N_{2}$ pairs.  For particles from the same set that becomes $N_{bkgd}(N_{bkgd}-1)$.  Normalized per trigger particle, the anisotropic flow modulation is then given by,
\begin{eqnarray}
B_{3}(\Delta\phi_{1},\Delta\phi_{2})&=&\frac{N_{bkgd}(N_{bkdg}-1)}{(2\pi)^{2}} [1+2v_{2}^{trig}v_{2}^{(1)}\cos(2\Delta\phi_{1}) \nonumber \\
&& + 2v_{2}^{trig}v_{2}^{(2)}\cos(2\Delta\phi_{2}) + 2v_{2}^{(1)}v_{2}^{(2)}\cos(2\Delta\phi_{1}-2\Delta\phi_{2}) \nonumber \\
&& + 2v_{4}^{trig}v_{4}^{(1)}\cos(4\Delta\phi_{1}) + 2v_{4}^{trig}v_{4}^{(2)}\cos(4\Delta\phi_{2}) \nonumber \\
&& + 2v_{4}^{(1)}v_{4}^{(2)}\cos(4\Delta\phi_{1}-4\Delta\phi_{2}) + 2v_{2}^{trig}v_{2}^{(1)}v_{4}^{(2)}\cos(2\Delta\phi_{1}-4\Delta\phi_{2}) \nonumber \\
&& + 2v_{2}^{trig}v_{2}^{(2)}v_{4}^{(1)}\cos(4\Delta\phi_{1}-2\Delta\phi_{2}) + 2v_{2}^{(1)}v_{2}^{(2)}v_{4}^{trig}\cos(2\Delta\phi_{1}+2\Delta\phi_{2})] \nonumber \\
\end{eqnarray}

\begin{figure}[htb]
\centering
\includegraphics[width=0.6\textwidth]{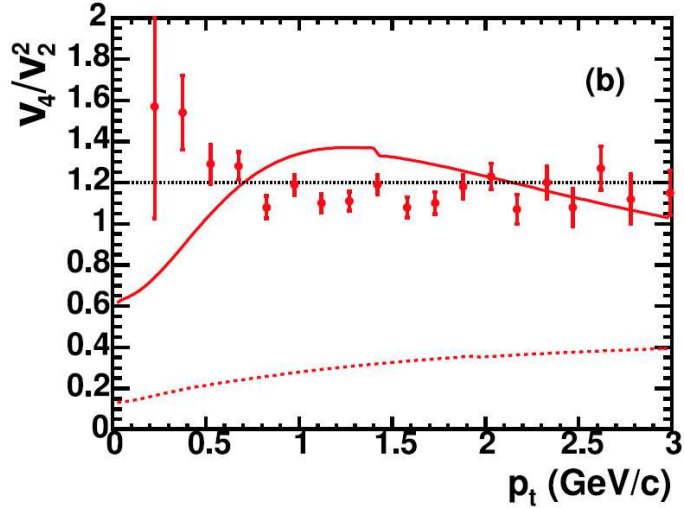}
\caption{Results for $v_{4}/v_{2}^{2}$\cite{flow}.  A fit to a constant in the $p_T$ range of $1<p_T<2$ GeV/c is used to parameterize $v_{4}$ in terms of $v_{2}^{2}$.  The black line shows a fit to a constant over the entire $p_T$ range.  The red curves are from blast wave fits.}
\label{fig:v4}
\end{figure}

As in 2-particle correlations we use the average $v_{2}$ from the reaction plane and 4-particle cumulant measurements.  Since we are now using the fourth order harmonics we also need $v_{4}$ values.  Figure~\ref{fig:v4} shows the measured $v_{4}/v_{2}^{2}$\cite{flow}.   We fit the measurement of $v_{4}/v_{2}^{2}$ to a constant in the $p_T$ range of 1<$p_T$<2 GeV/c, yielding $v_{4}=1.15v_{2}^{2}$.

\subsection{Acceptance and Efficiency Corrections}
\begin{figure}[htb]
\hfill
\begin{minipage}[t]{0.49\textwidth}
\centering
\includegraphics[width=1.0\textwidth]{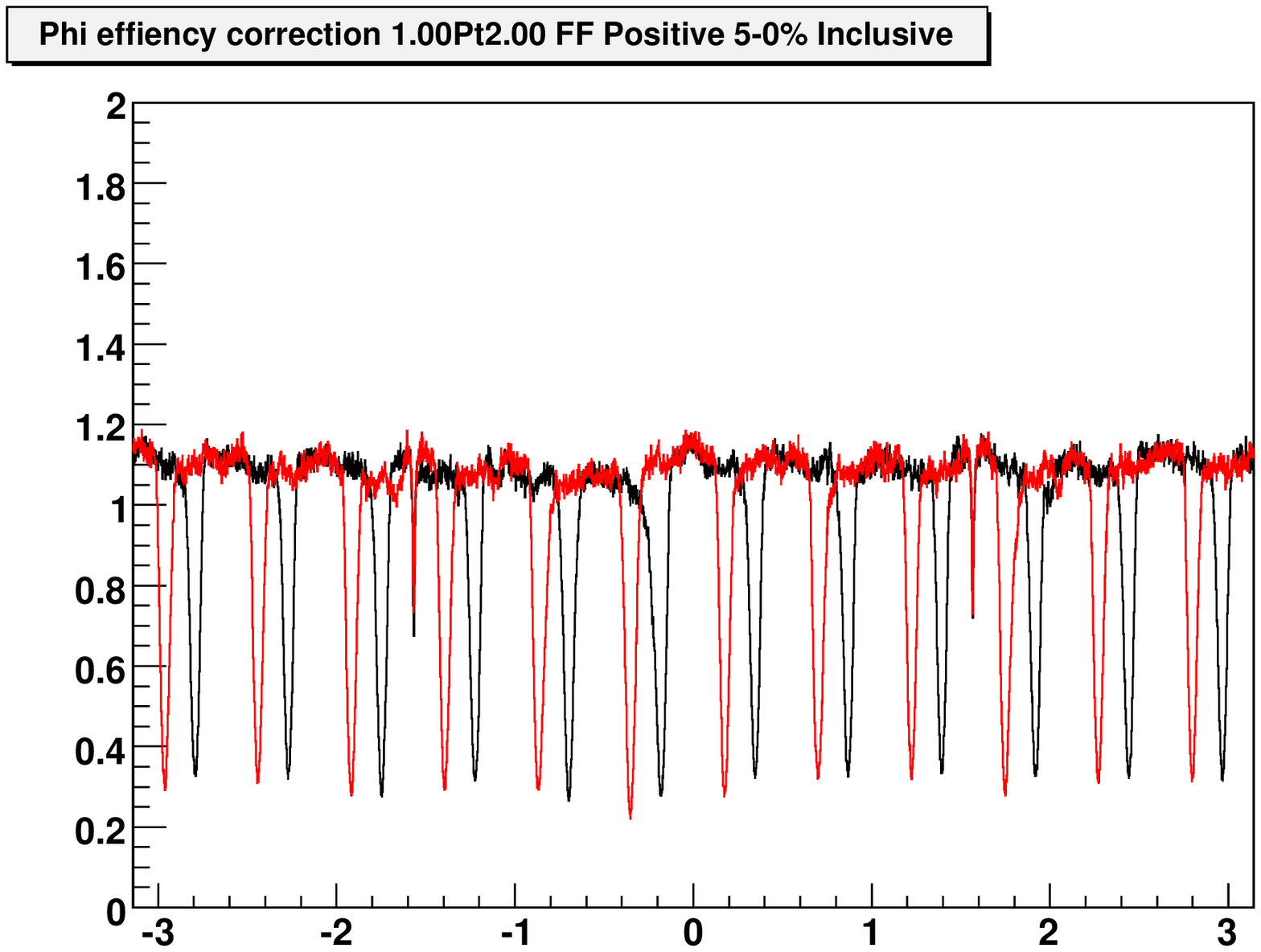}
\end{minipage}
\hfill
\begin{minipage}[t]{0.49\textwidth}
\centering
\includegraphics[width=1.0\textwidth]{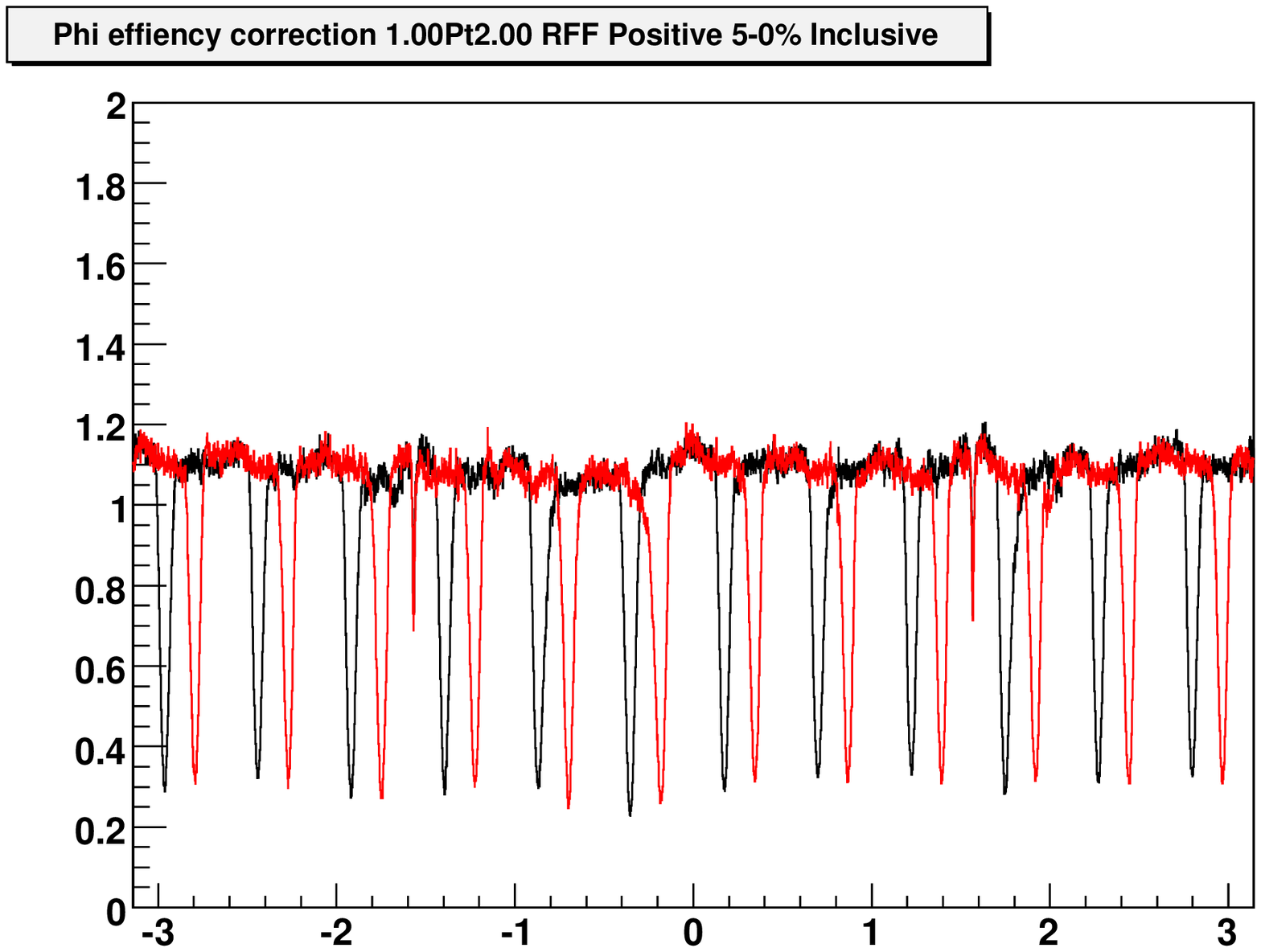}
\end{minipage}
\label{fig:acc}
\caption{TPC acceptance in $\phi$ for particles of $1<p_T<2$ GeV/c with $\pm$0.5 Tesla magnetic field.  Positive particles are shown in black and negative particles are shown in red.  Plots are for 0-5\% Au+Au collisions.  Left:  Magnetic field with postive polarity.  Right:  Magnetic field with negative polarity.  Errors are statistical.  Plots for other centralities and for $3<p_T<4$ GeV/c particles can be found in the appendix.}  
\end{figure}

All associated particles are corrected for the overall $\phi$-averaged detector efficiency as is discussed in 3.1.2.  The $\phi$ dependent efficiency (acceptance) is accounted for separately.  The $\phi$ dependence of the efficiency can be corrected for to first order by mixed events.  This is sufficient in 2-particle correlations (i.e. the 2-particle acceptance correction) but not for 3-particle correlation due to the low signal to noise ratio.  The lack of a $\phi$-dependent acceptance correction can also have an impact on the normalization in 3-particle correlations.  To improve this correction we use a single particle $\phi$ dependent acceptance correction.  This correction is done separately for each charge and magnetic field setting.  Figure~\ref{fig:acc} shows examples of the $\phi$ acceptance for two polarities of magnetic field settings for our associated particles in central Au+Au collisions.  The $\phi$ acceptance correction plots are binned in very narrow $\phi$ bins to make any binning effect negligible.  Particles are weighted by the reciprocal of the $\phi$ dependent acceptance.  The $\phi$ dependent acceptance is normalized such that the average is one, since we have already corrected for the number of particles with the overall  $\phi$-averaged detector efficiency.  Because high $p_T$ particles curve little in the STAR magnetic field, the trigger particle acceptance goes to zero at the TPC sector boundaries.  To avoid large corrections we do not use trigger particles that have less than a 10\% $\phi$ acceptance.  Since the quantities are per trigger this has no effect on our overall efficiencies.  The number of trigger particles is accumulated by the $\phi$-dependent correction factor, and is used in the final normalization of the per trigger correlation functions.  

\section{Background Construction}
We are using charged particles in the STAR TPC.  Correlations are between a trigger particle of $3<p_{T}<4$ GeV/c and two associated particle of $1<p_{T}<2$ GeV/c.  The data used in this thesis are from 2 million $pp$ events from the second year of RHIC running, 6.5 million d+Au events used from third year of RHIC running, and 12 million minimum bias and 19 million central triggered events from the fourth year of RHIC running.  The number of events listed are good events after vertex cuts.  The tracks used all have a distance of closest approach (DCA) to the primary vertex less than 2 cm.  They have at least 20 points fit by the track reconstruction software and at least 51\% of the maximum possible number of fit points (determined by the detector geometry for the particular track) to avoid split tracks.  All tracks are within $|\eta|<1.0$.  The events were divided in centrality bins based on the so-called reference multiplicity.  In Au+Au collisions, the reference multiplicity is the number of tracks with $DCA<1$ cm, number of fit points greater than 10, and $|\eta|<0.5$ (the STAR reference multiplicity).  In d+Au collisions, the reference multiplicity is the number of tracks with $DCA<2$, number of fit points great than 15 and $|\eta|<1.0$.  The centrality divisions are shown in Table~\ref{tab:cent}.  The analysis was performed in these individual centrality bins and then combined, weighted by the number of trigger particles, into larger centrality bins.  This was done to reduce non-Possion and multiplicity dependent effects that increase with the size of the centrality bin.

\begin{table}[hbt]
\centering
\caption{Centrality bins definitions used in this thesis in d+Au and Au+Au collisions.  The minimum and maximum are applied on the reference multiplicity. See text for the definition of reference multiplicity in d+Au and Au+Au collisions.}
\begin{tabular}{|l|l|l|} 
\hline
Centrality&Minimum&Maximum\\
\hline
d+Au 100-20\%&0&17\\
d+Au 20-10\%&18&26\\
d+Au 0-10\%&27&100\\
Au+Au 70-80\%&15&31\\
Au+Au 60-70\%&32&57\\
Au+Au 50-60\%&58&96\\
Au+Au 40-50\%&97&150\\
Au+Au 30-40\%&151&222\\
Au+Au 20-30\%&223&319\\
Au+Au 10-20\%&320&441\\
Au+Au 5-10\%&442&520\\
Au+Au 0-5\%&521&1000\\
\hline
\end{tabular}
\label{tab:cent}
\end{table}

\begin{figure}[htb]
\centering
\includegraphics[width=0.49\textwidth]{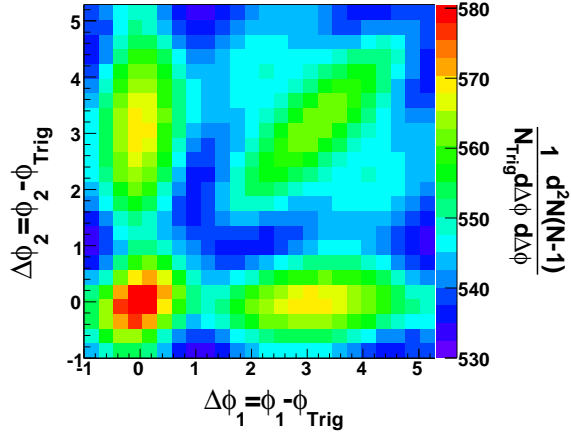}
\caption{Raw 3-particle correlation function for 0-12\% central Au+Au collisions.}
\label{fig:raw3}
\end{figure}

Figure~\ref{fig:raw3} shows an example raw 3-particle correlation function, $J_{3}(\Delta\phi_{1},\Delta\phi_{2})$, where $\Delta\phi_{i}=\phi_{i}-\phi_{Trig}$.  The raw 3-particle correlation function contains our desired jet-like 3-particle correlation and background terms.  These background terms are discussed in detail below.

\subsection{Hard-Soft Background}
\begin{figure}[htb]
\hfill
\begin{minipage}[t]{0.49\textwidth}
\centering
\includegraphics[width=1.0\textwidth]{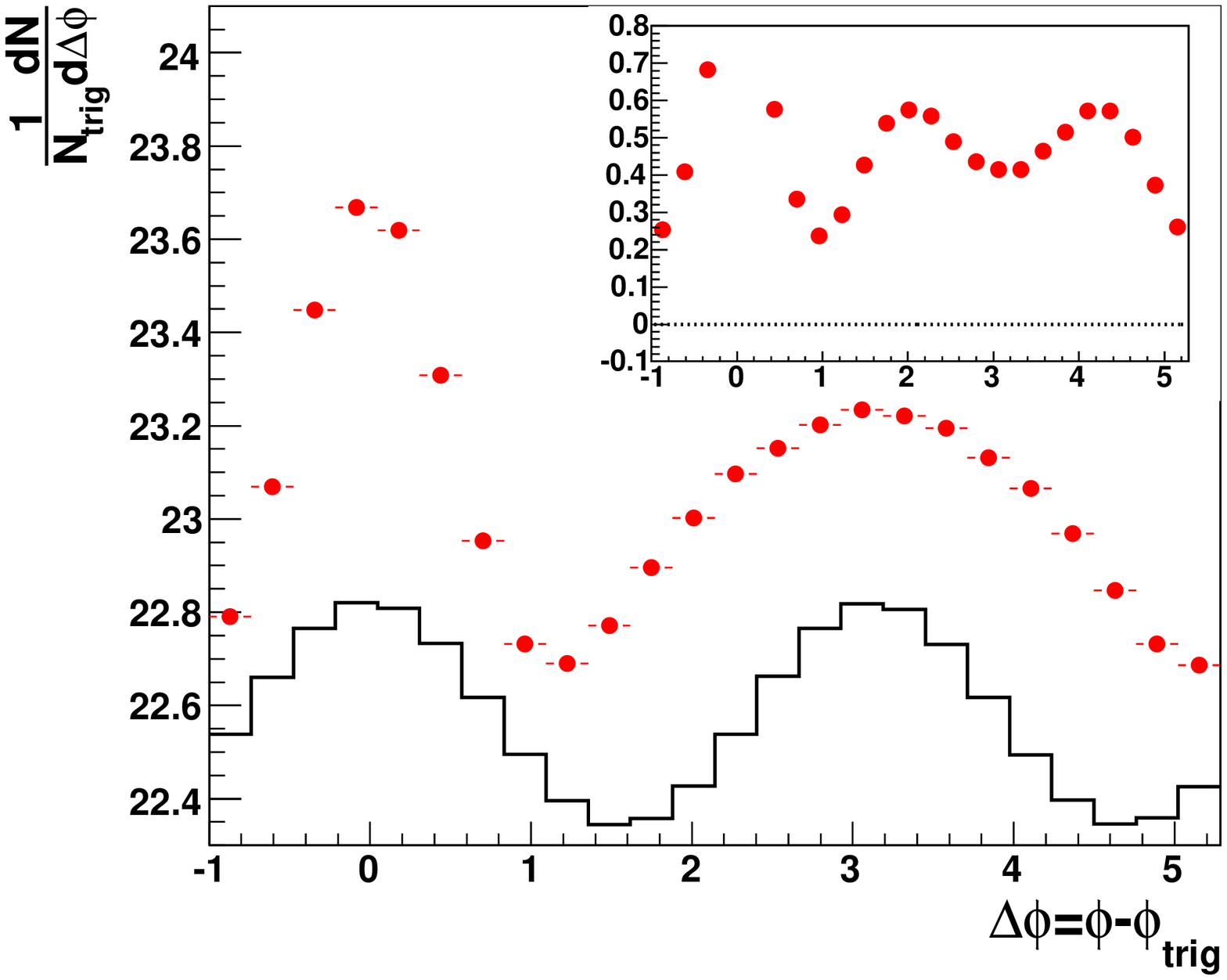}
\end{minipage}
\hfill
\begin{minipage}[t]{0.49\textwidth}
\centering
\includegraphics[width=1.0\textwidth]{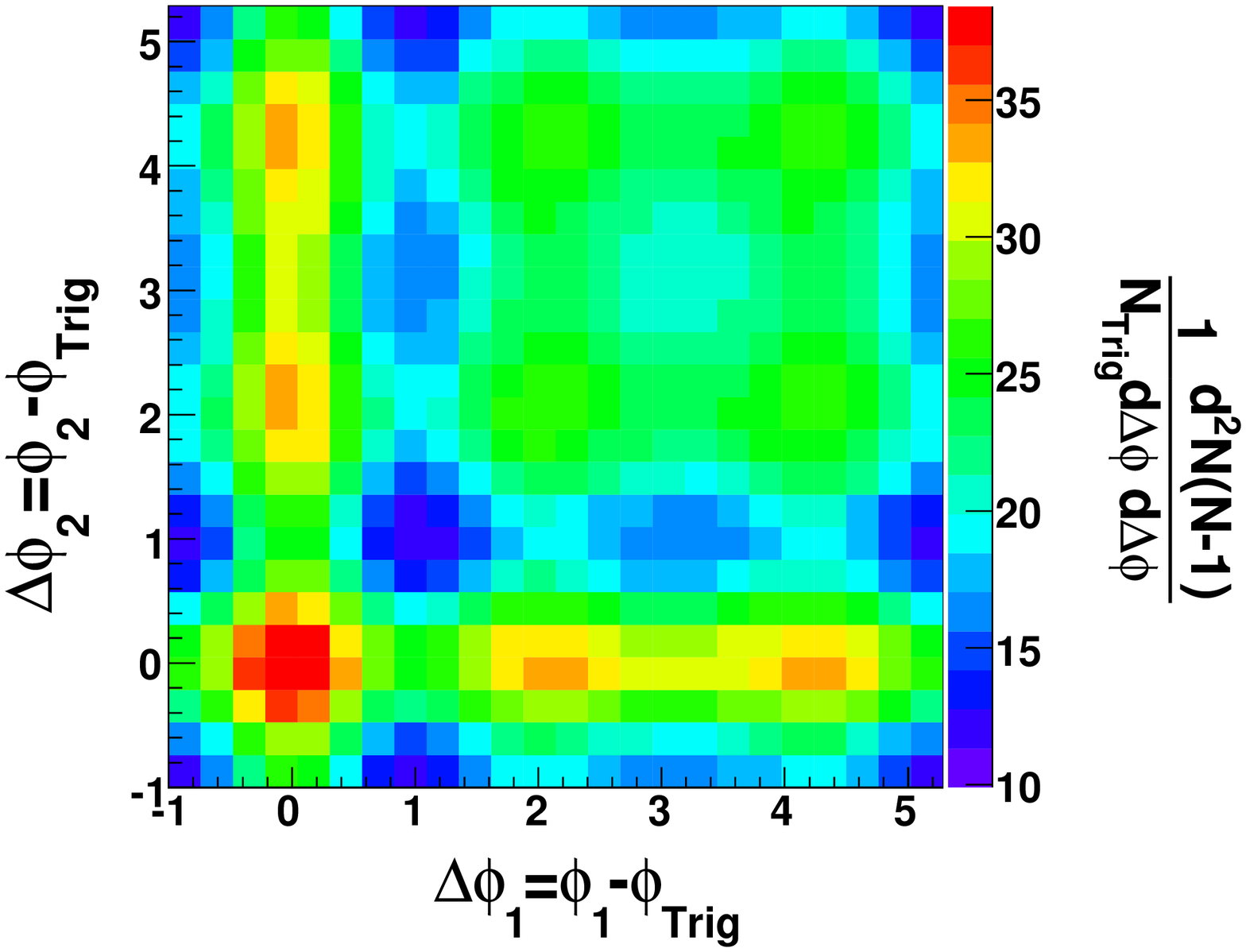}
\end{minipage}
\caption{Left:  Raw 2-particle correlation function is shown in red, $\hat{J_2}$.  Normalized background from mixed events with $v2$ and $v4$ added, $aB_{inc}F_{2}$, is shown in black.  Normalization is done by 3-particle ZYAM.  Background subtracted 2-particle correlation is shown in the minipanel, $\hat{J_2}$.  Right:  Hard-soft background term, $\hat{J_{2}} \otimes B_{2}$.  Plots are from 0-12\% central Au+Au collisions at $\sqrt{s_{NN}}=200$ GeV/c.  Errors are statistical.}
\label{fig:hs}
\end{figure}

The two-particle correlation function is given by (as discussed in Chapter 3),
\begin{equation}
\hat{J_{2}}(\Delta\phi)=J_{2}(\Delta\phi)-aB_{inc}F_{2}(\Delta\phi)
\end{equation}
where
\begin{equation}
F_{2}^{a,b}(\Delta\phi)=1+2v_{2}^{a}v_{2}^{b}cos(2\Delta\phi)+2v_{4}^{a}v_{4}^{b}cos(4\Delta\phi)
\end{equation}
is the 2-particle flow modulation up to the fourth harmonic and $B_{inc}=dN_{inc}/d(\Delta\phi)$.  Here our background event is an inclusive event, a minimum bias event of the same centrality.  The normalization factor $a$ is used to adjust the level of the background due to trigger bias effects.  In the 3-particle correlation analysis $a$ will not be normalized to 2-particle ZYA1 as is done in the 2-particle analysis (presented in Chapt. 3).  Here we will use the assumption the 3-particle jet-like correlation is positive definite and use 3-particle ZYAM (zero yield at minimum).  This can provide a better constraint on the 2-particle background since we have additional information in the 3-particle correlations.  The systematics on this assumption will later be discussed.  An example raw 2-particle correlation $J_{2}(\Delta\phi)$ is shown in figures~\ref{fig:hs} (left), in filled symbols.  The normalized background from mixed events is shown by the solid black line with flow, $aB_{inc}F_{2}(\Delta\phi)$.  The mixed events were constructed by mixing the trigger particle with inclusive events.  The flow, both $v_2$ and $v_4$, is added pairwise during the event mixing.  The $v_2$ is the average of the reaction plane and 4-particle cumulant measurements.  The $v_4$ is taken to be $1.15v_2^2$.  The minipanel shows the background subtracted 2-particle correlations, $\hat{J_{2}}$.  The right panel shows the hard-soft background term.  This term is given in equation~\ref{eqn:hsbg}.  This is a folding of the background subtracted 2-particle correlation and its background.  This background term removes 2-particle correlations, where one particle is correlated with the trigger and the other is only correlated via the reaction plane, from our 3-particle correlation.  In real data analysis this term becomes,
\begin{eqnarray}
J_{2} \otimes B_{2} &=& \hat{J_{2}}(\Delta\phi_{1})aB_{inc}[1+F_{2}^{Jet,(2)}(\Delta\phi_{2})] + \hat{J_{2}}(\Delta\phi_{2})aB_{inc}[1+F_{2}^{Jet,(1)}(\Delta\phi_{1})] \nonumber \\
&=& [J_{2}(\Delta\phi_{1})-aB_{inc}F_{2}^{Trig,(1)}(\Delta\phi_{1})]aB_{inc}[1+F_{2}^{Jet,(2)}(\Delta\phi_{2})] \nonumber \\
&& + [J_{2}(\Delta\phi_{2})-aB_{inc}F_{2}^{Trig,(2)}(\Delta\phi_{2})]aB_{inc}[1+F_{2}^{Jet,(1)}(\Delta\phi_{1})]. \nonumber \\
\end{eqnarray}
Then we fold the jet-like 2-particle correlation with the flow modulated background.  To obtain $v_2^{Jet}$ we make the assumption that the trigger-associated jet-like correlated pair are correlated with the reaction plane with the same flow as the trigger particle ($v_2^{Jet}=v_2^{Trig}$).  The systematics on this assumption will be discussed later.

\subsection{Soft-Soft Background}

\begin{figure}[htb]
\hfill
\begin{minipage}[t]{0.49\textwidth}
\centering
\includegraphics[width=1.0\textwidth]{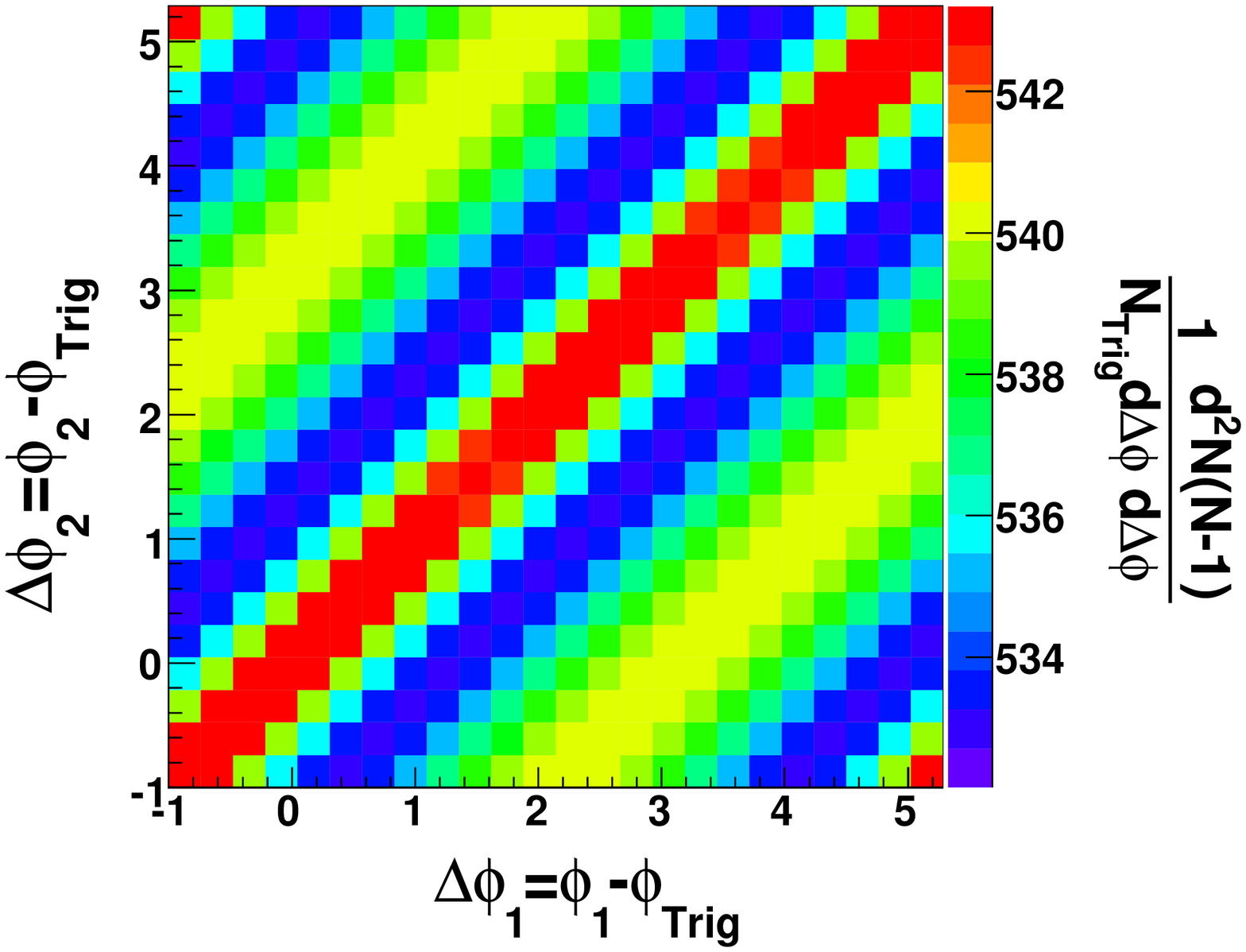}
\end{minipage}
\hfill
\begin{minipage}[t]{0.49\textwidth}
\centering
\includegraphics[width=1.0\textwidth]{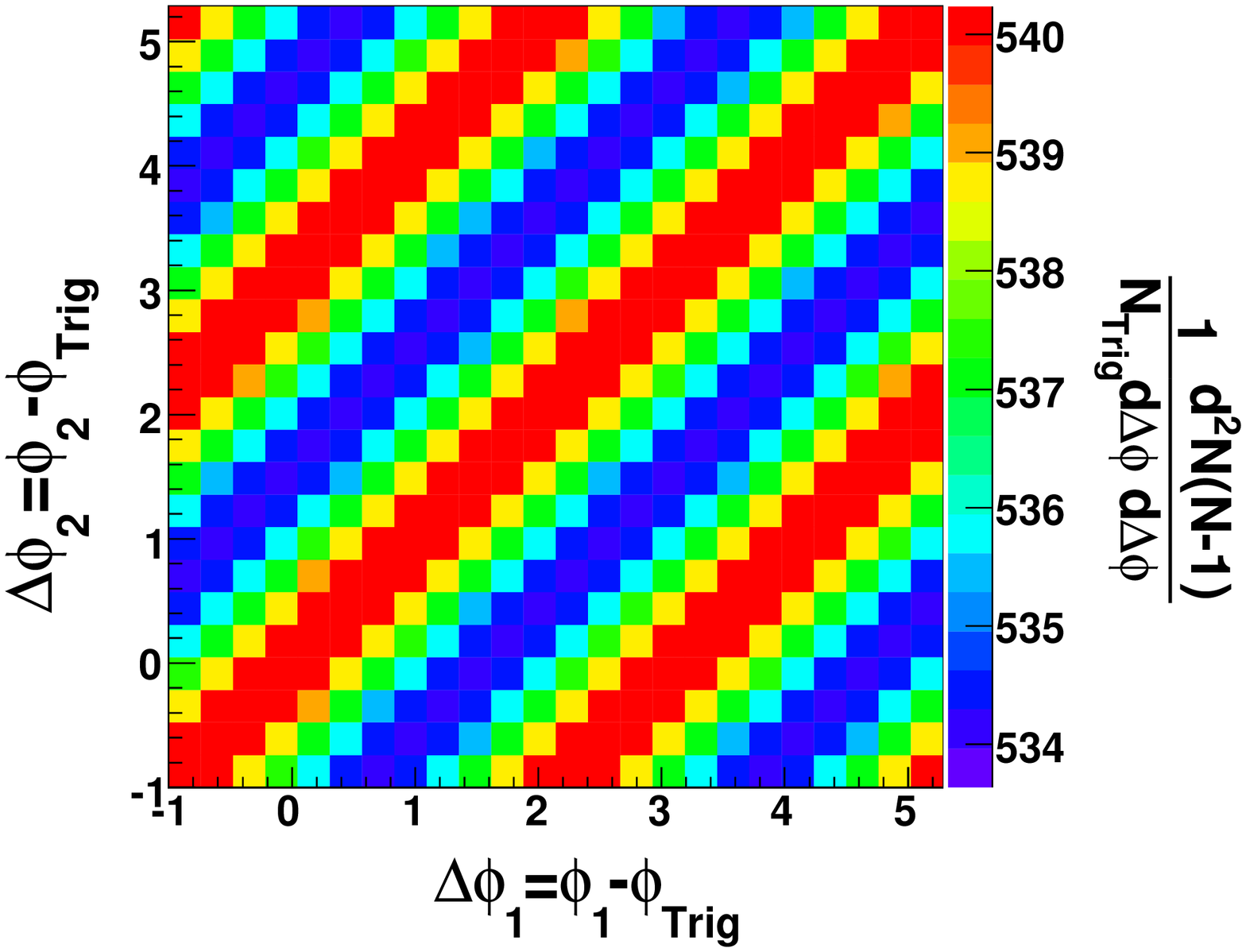}
\end{minipage}
\caption{Left:  Soft-soft background term, $a^{2}bJ_{2}^{inc}$.  Right:  Flow component of the soft-soft background term, $a^{2}bJ_{2}^{inc,flow}$.   Plots are from 0-12\% central Au+Au collisions at $\sqrt{s_{NN}}=200$ GeV/c.}
\label{fig:ss}
\end{figure}

Figure~\ref{fig:ss} shows the soft-soft background term.  This term contains all correlations between the two softer particles that are independent of the trigger particle.  In {\it pp} and d+Au collisions, this term is composed of minijets, background jets, and decays.  In Au+Au collisions, there is also a contribution from the flow correlation between the two softer particles.  This flow component is the dominant part of the soft-soft term in Au+Au collisions.  The soft-soft term is constructed by mixing a trigger particle from one event with pairs of softer particles from another (inclusive) event.   It is the inclusive event particle pair density with respect to a random trigger, 
\begin{equation}
J_{2}^{inc}(\Delta\phi_{1},\Delta\phi_{2})=\frac{d^{2}(N_{inc}(N_{inc}-1))}{d(\Delta\phi_{1})d(\Delta\phi_{2})}
\end{equation}
The flow component already contained in this term is,
\begin{eqnarray}
J_{2}^{inc,flow}(\Delta\phi_{1},\Delta\phi_{2})&=&\frac{N_{inc}(N_{inc}-1)}{(2\pi)^{2}}[1+2v_{2}^{(1)}v_{2}^{(2)}\cos2(\Delta\phi_{1}-\Delta\phi_{2}) \nonumber \\
&&+2v_{4}^{(1)}v_{4}^{(1)}\cos4(\Delta\phi_{1}-\Delta\phi_{2})]
\end{eqnarray}
 and is shown in Fig.~\ref{fig:ss} (right).  This was constructed by mixing the trigger particle with two different inclusive events and adding the flow in tripletwise.  There is a $\langle N_{inc}(N_{inc}-1)\rangle/\langle N^{(1)}_{inc}N^{(2)}_{inc}\rangle$ factor applied so that the number of associated pairs is the same as if they came from the same event.  The soft-soft background is scaled by $a^2 b$ where $a$ is the scaling factor used in the 2-particle correlations and accounts for triggering effects, as discussed in Chapt. 3, and $b$ accounts for non-Poisson effects.  The factor $a$ is determined by a 3-particle ZYAM assumption.  The factor $b$ is required because the event multiplicity distributions are not Poisson.  The factor $b$ is determined by the ratio of the deviation from Poisson of the triggered events to the inclusive event,
\begin{equation}
b=\frac{\frac{\langle N_{trig}(N_{trig}-1)\rangle}{\langle N_{trig}\rangle^{2}}}{\frac{\langle N_{inc}(N_{inc}-1)\rangle}{\langle N_{inc}\langle
^{2}}}.
\end{equation}
This method of obtaining $b$ makes the assumption that the underlying background and the triggered events have the same level of deviation from Poisson.
  
\subsection{Trigger Flow Correlation}

\begin{figure}[htb]
\hfill
\begin{minipage}[t]{0.32\textwidth}
\centering
\includegraphics[width=1.0\textwidth]{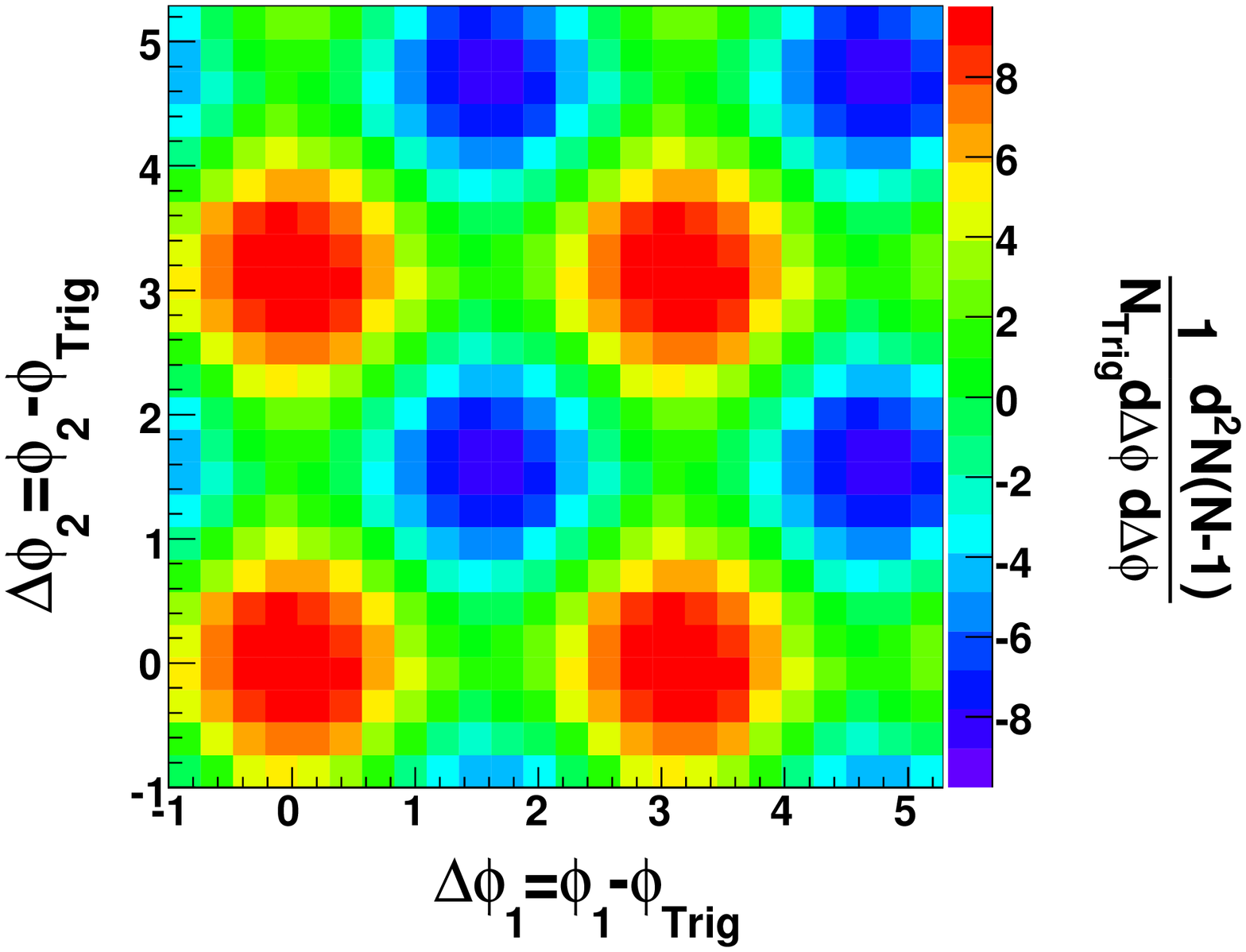}
\end{minipage}
\hfill
\begin{minipage}[t]{0.32\textwidth}
\centering
\includegraphics[width=1.0\textwidth]{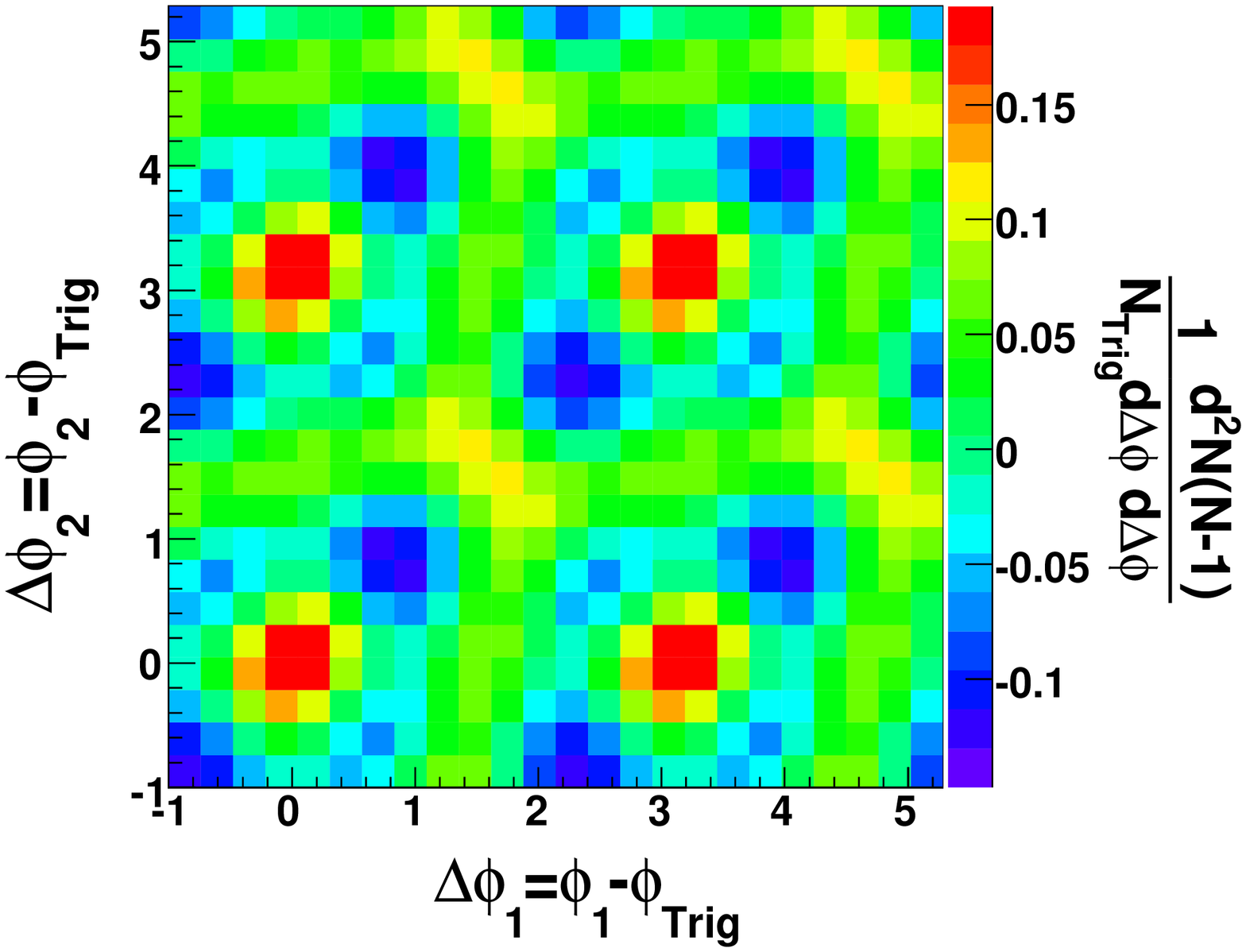}
\end{minipage}
\hfill
\begin{minipage}[t]{0.32\textwidth}
\centering
\includegraphics[width=1.0\textwidth]{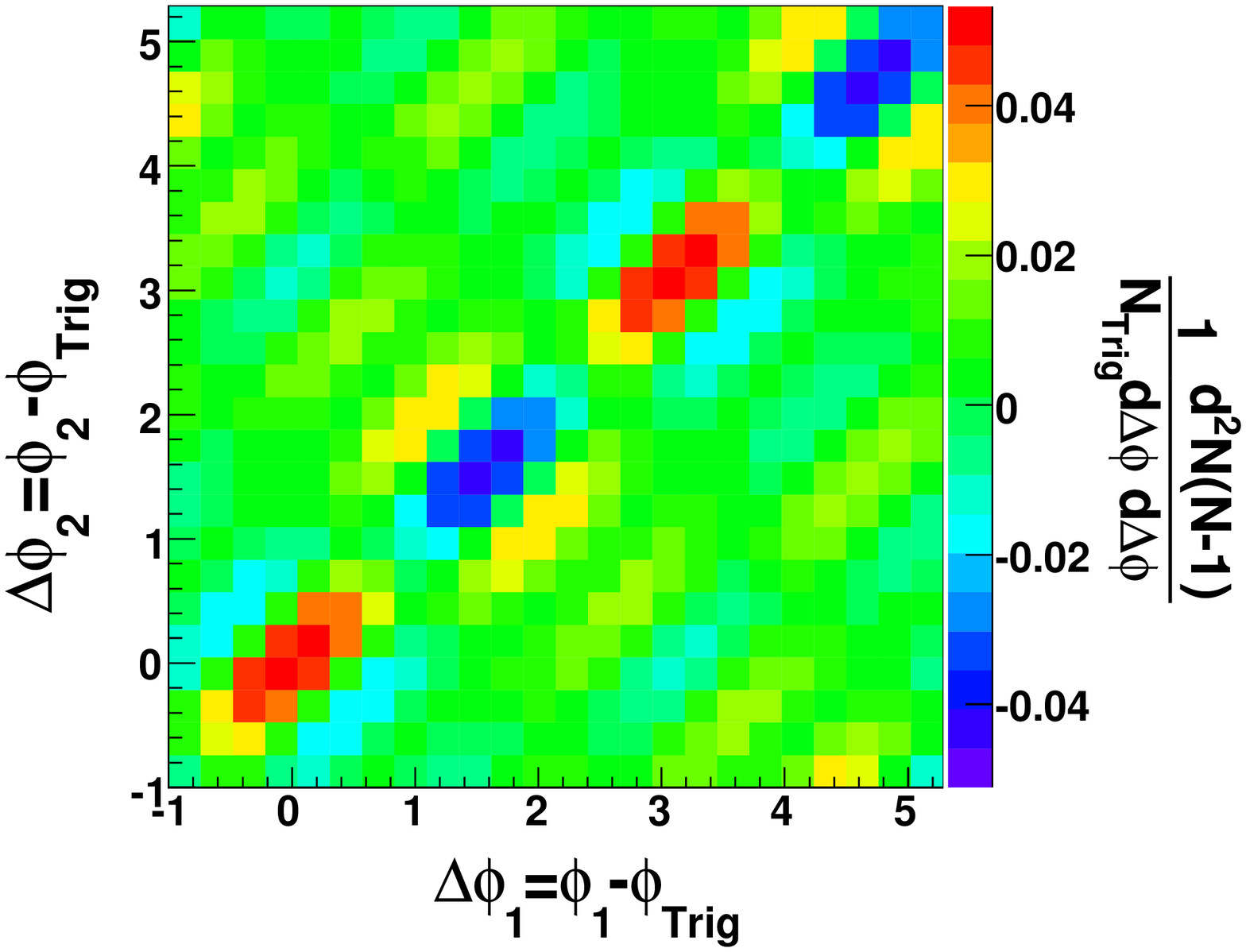}
\end{minipage}
\caption{Left:  Trigger flow correlation up to order $v_2^2$.  Center:  Trigger flow correlation of order $v_2^{4}$.  Right:  Trigger flow correlation with non-flow on soft-soft, $J_{3}^{TF}$.  Plots are from 0-12\% central Au+Au collisions at $\sqrt{s_{NN}}=200$ GeV/c.}
\label{fig:3flow}
\end{figure}

There are additional flow correlations that are not preserved in the event mixing and thus must be subtracted by hand.  There is the flow correlation between the trigger particle and the two associated particles, where both of the associated particles come from the underlying background in the triggered event.    The total trigger flow is,
\begin{equation}
J_{3}^{TF}=\frac{N_{inc}(N_{inc}-1)}{(2\pi)^{2}}F_{3}(\Delta\phi_{1},\Delta\phi_{2})
\end{equation}
with
\begin{eqnarray}
F_{3}(\Delta\phi_{1},\Delta\phi_{2})&=& F_{2}(\Delta\phi_{1}) + F_{2}(\Delta\phi_{2})\nonumber \\
&& + 2v_{2}^{trig}v_{2}^{(1)}v_{4}^{(2)}cos(2\Delta\phi_{1}-4\Delta\phi_{2})\nonumber \\
&& + 2v_{2}^{trig}v_{4}^{(1)}v_{2}^{(2)}cos(4\Delta\phi_{1}-2\Delta\phi_{2})\nonumber \\
&& + 2v_{4}^{trig}v_{2}^{(1)}v_{2}^{(2)}cos(2\Delta\phi_{1}+2\Delta\phi_{2}).
\end{eqnarray}
Figure~\ref{fig:3flow} (left) shows the flow correlation with the trigger particle up to order $v_{2}^{2}$.  The central panel shows the flow correlation with the trigger particle of order $v_{2}^{4}$ where $v_{4}$ assumed proportional to $v_{2}^{2}$.  The sum of these two panels is $J_{3}^{TF}(\Delta\phi_{1},\Delta\phi_{2})$.  These terms are constructed by mixing the trigger particle with associated particles from two different inclusive events.  The flow terms are added in tripletwise with the $v_2$ from the average of the reaction plane and 4-particle cumulant measurements and $v_4=1.15v_2^2$.  Since the associated particles come from two different inclusive events there is a factor of $\langle N_{inc}(N_{inc}-1)\rangle/\langle N^{(1)}_{inc}N^{(2)}_{inc}\rangle$ applied so that the average number of associated pairs is same as the average number of associated pairs when both associated particle come from the same event.  As with the soft-soft term, the trigger flow terms are scaled by $a^2 b$.

This would be exact if the particles were only correlated by flow.  However, there is an additional term required due to the flow correlation of the non-flow structure on the soft-soft term (such as minijets, decays and background jets) with the trigger particle.  We need to correct for this effect in the trigger flow.  To first order this can be approximated as,
\begin{equation}
F_{3}^{CF}=\left[\frac{J_{2}^{inc}(\Delta\phi_{1},\Delta\phi_{2})}{J_{2}^{inc,flow}(\Delta\phi_{1},\Delta\phi_{2})}-1 \right]F_{3}(\Delta\phi_{1},\Delta\phi_{2})
\end{equation}
An example of this term is shown in Fig.~\ref{fig:3flow} (right).  Since the ratio of the soft-soft term and the flow between the two soft particles is close to 1, this term is small and the first order correction is sufficient.

\subsection{Background Subtraction}

\begin{figure}[htb]
\hfill
\begin{minipage}[t]{0.49\textwidth}
\centering
\includegraphics[width=1.0\textwidth]{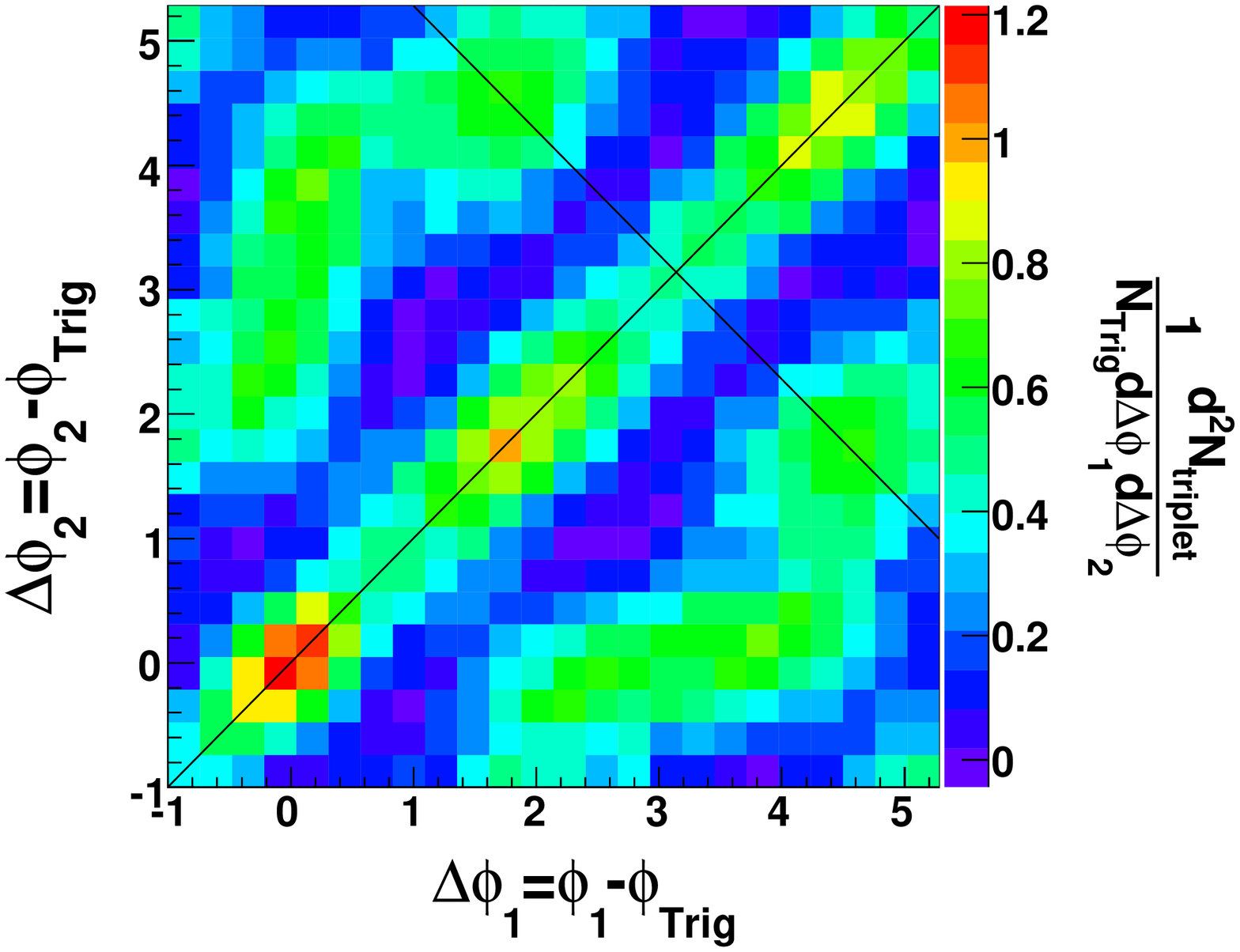}
\end{minipage}
\hfill
\begin{minipage}[t]{0.49\textwidth}
\centering
\includegraphics[width=1.0\textwidth]{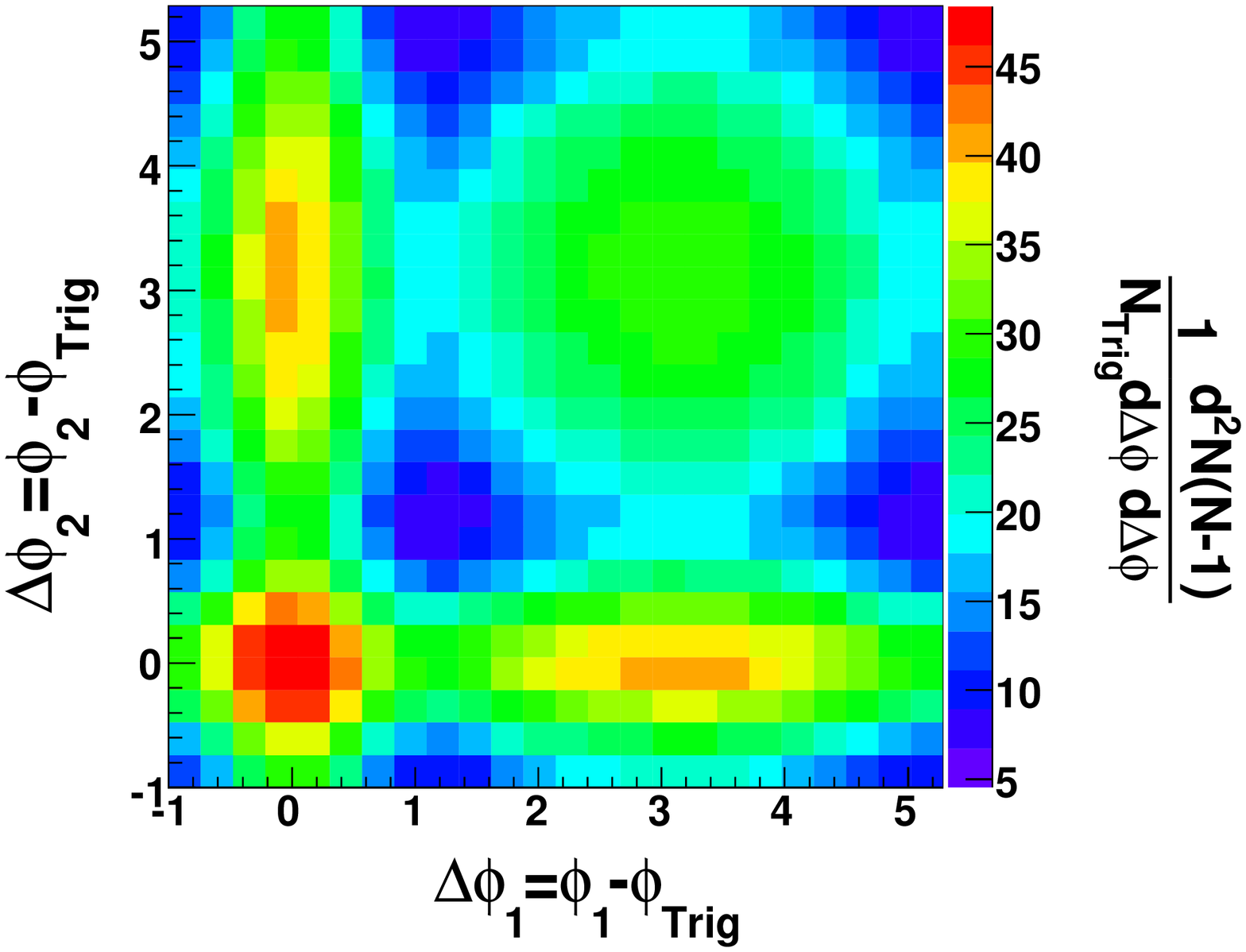}
\end{minipage}
\caption{Left:  Background subtracted 3-particle correlation, $\hat{J_3}$.  Right:  Sum of hard-soft and trigger flow background terms,  $J_{2} \otimes B_{2} + J_{3}^{TF}$.  Plots are from 0-12\% central Au+Au collisions at $\sqrt{s_{NN}}=200$ GeV/c.}
\label{fig:3sig}
\end{figure}

An example background subtracted 3-particle correlation is shown in Fig.~\ref{fig:3sig}, left.  This is obtained from the raw signal through the subtraction of the background terms discussed by,
\begin{eqnarray}
\hat{J_{3}}(\Delta\phi_{1},\Delta\phi_{2})&=& J_{3}(\Delta\phi_{1},\Delta\phi_{2})\nonumber \\
&& - a\hat{J_{2}}(\Delta\phi_{1})B_{inc}F_{2}^{Jet,(2)}(\Delta\phi_{2}) - a\hat{J_{2}}(\Delta\phi_{2})B_{inc}F_{2}^{Jet,(1)}(\Delta\phi_{1})\nonumber \\
&& - a^{2}b[J_{2}^{inc}(\Delta\phi_{1},\Delta\phi_{2})\nonumber \\
&& + \frac{\langle N(N-1)\rangle}{(2\pi)^2} \{F_{3}(\Delta\phi_{1},\Delta\phi_{2})+F_{3}^{CF}(\Delta\phi_{1},\Delta\phi_{2})\}]\nonumber \\
\end{eqnarray}
where $a$ is determined such that $\hat{J_{3}}(\Delta\phi_{1},\Delta\phi_{2})$ is ZYAM and .  To obtain the ZYAM, the lowest 10\% of the bins are used and $a$ is adjusted till these bins average zero.  The bins used for the minimum are recalculaed for each values of $a$.   The factor $b$ is there because the event multiplicity distributions are not Poisson.

One item of note in the background subtraction is the partial cancellation of the flow.  The flow term of $v_2^{Trig}v_2^{(i)}$ and $v_4^{Trig}v_4^{(i)}$, where $i=1,2$, partially cancel between the flow subtracted to form 2-particle jet-like correlation used to construct the hard-soft term and the terms in the trigger flow.  If the events are Poisson, then these terms will entirely cancel.  This partial cancellation gives a much smaller uncertainty on the sum of the hard-soft term and the trigger flow terms (Fig.~\ref{fig:3sig}, right) due to the uncertainty in the elliptic flow measurement than they have individually.  This allows us obtain a significant background subtracted 3-particle signal even in bins where the uncertainties on the hard-soft and trigger flow terms due to the uncertainty on the $v_2$ measurement are larger than our signal.  

\section{Results}
\subsection{Correlation Functions}

\begin{figure}[htb]
\hfill
\begin{minipage}[t]{0.32\textwidth}
\centering
\includegraphics[width=1.0\textwidth]{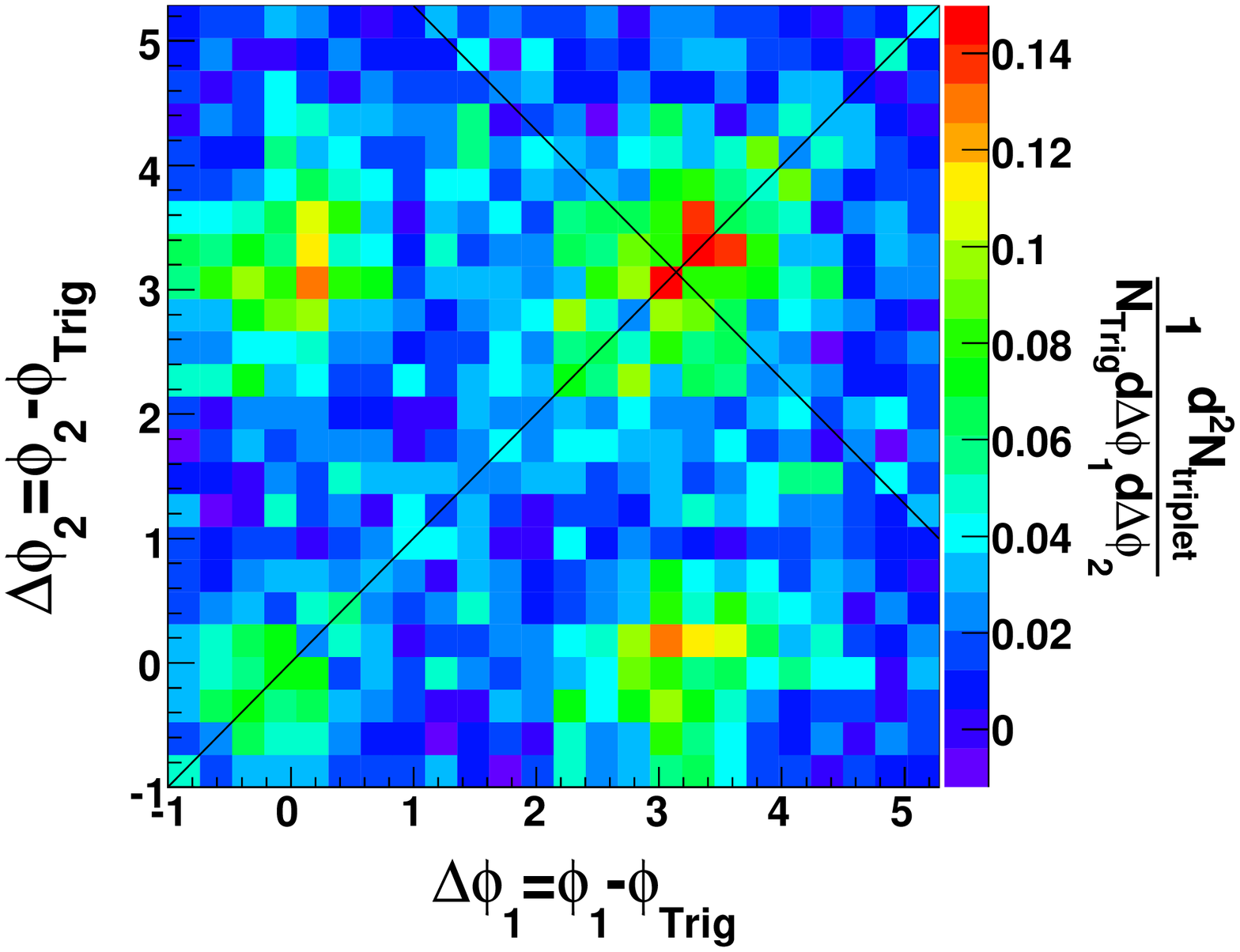}
\includegraphics[width=1.0\textwidth]{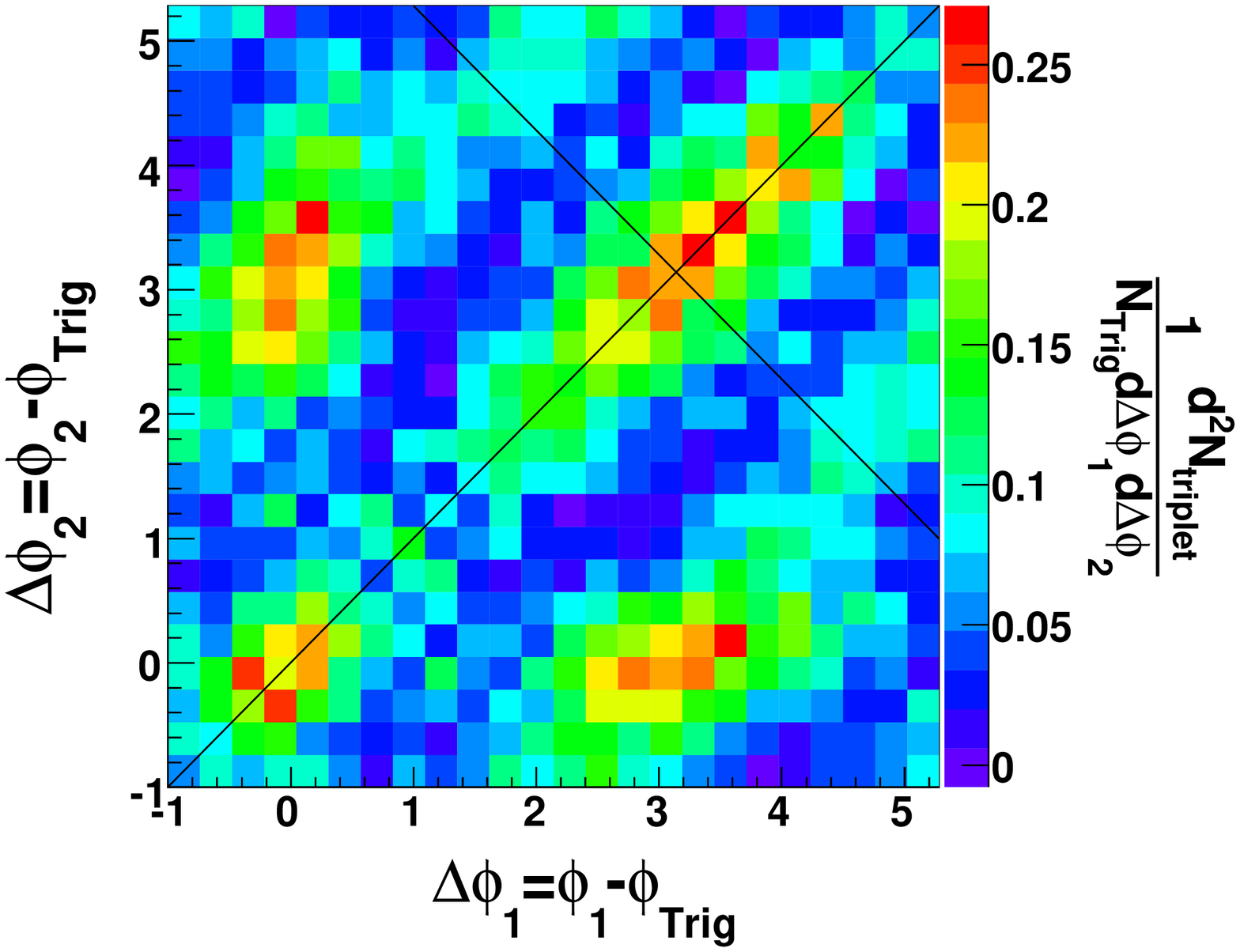}
\includegraphics[width=1.0\textwidth]{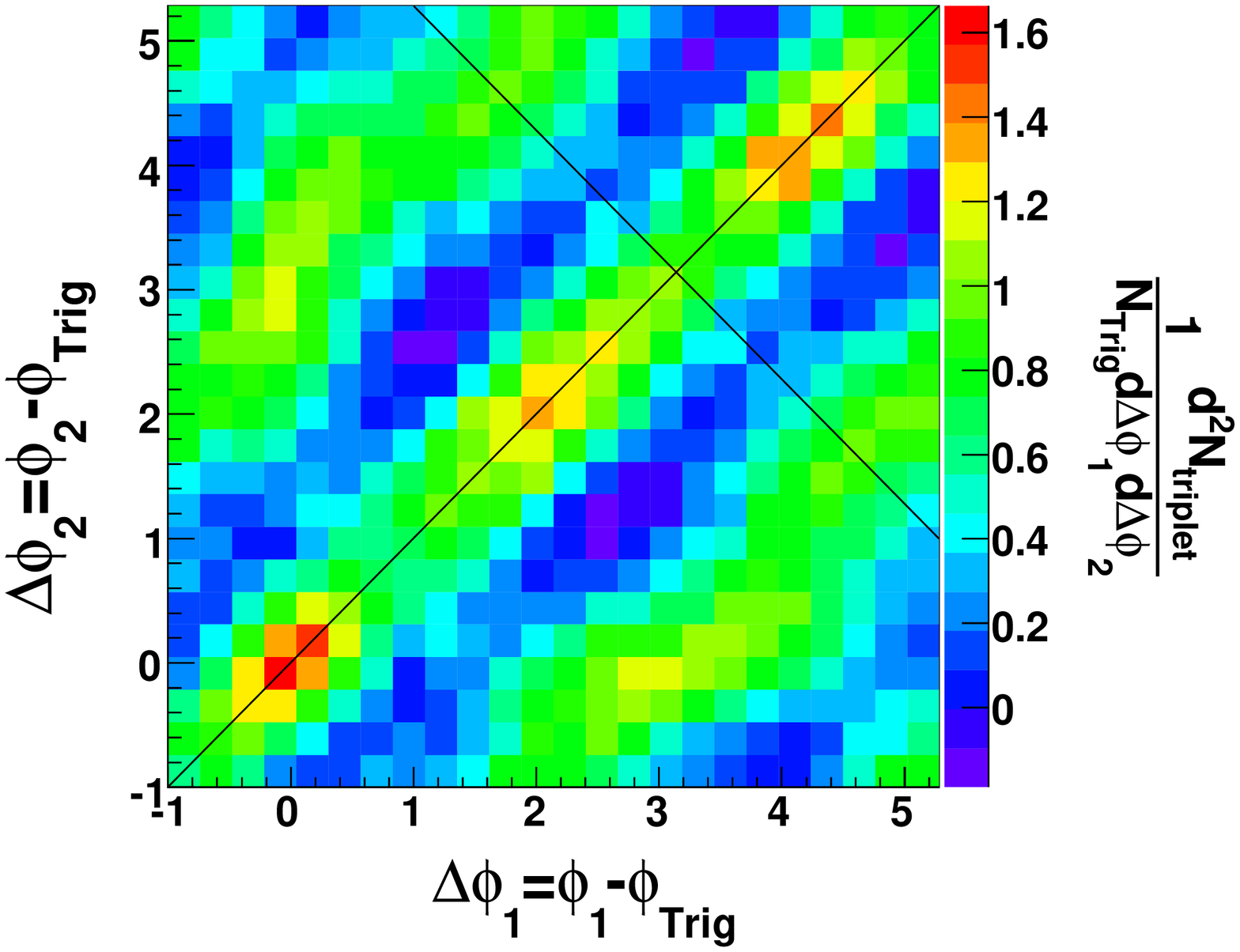}
\end{minipage}
\hfill
\begin{minipage}[t]{0.32\textwidth}
\centering
\includegraphics[width=1.0\textwidth]{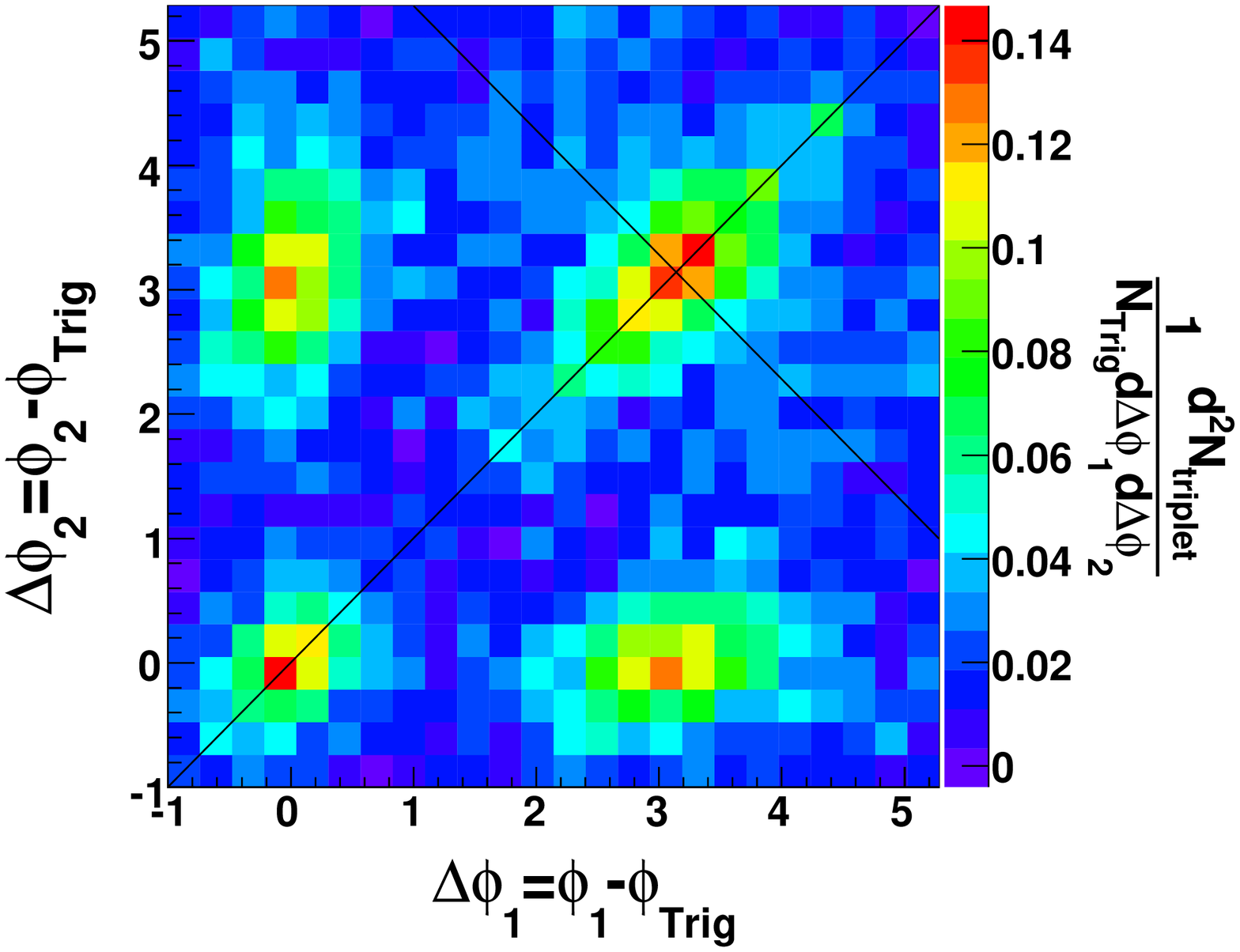}
\includegraphics[width=1.0\textwidth]{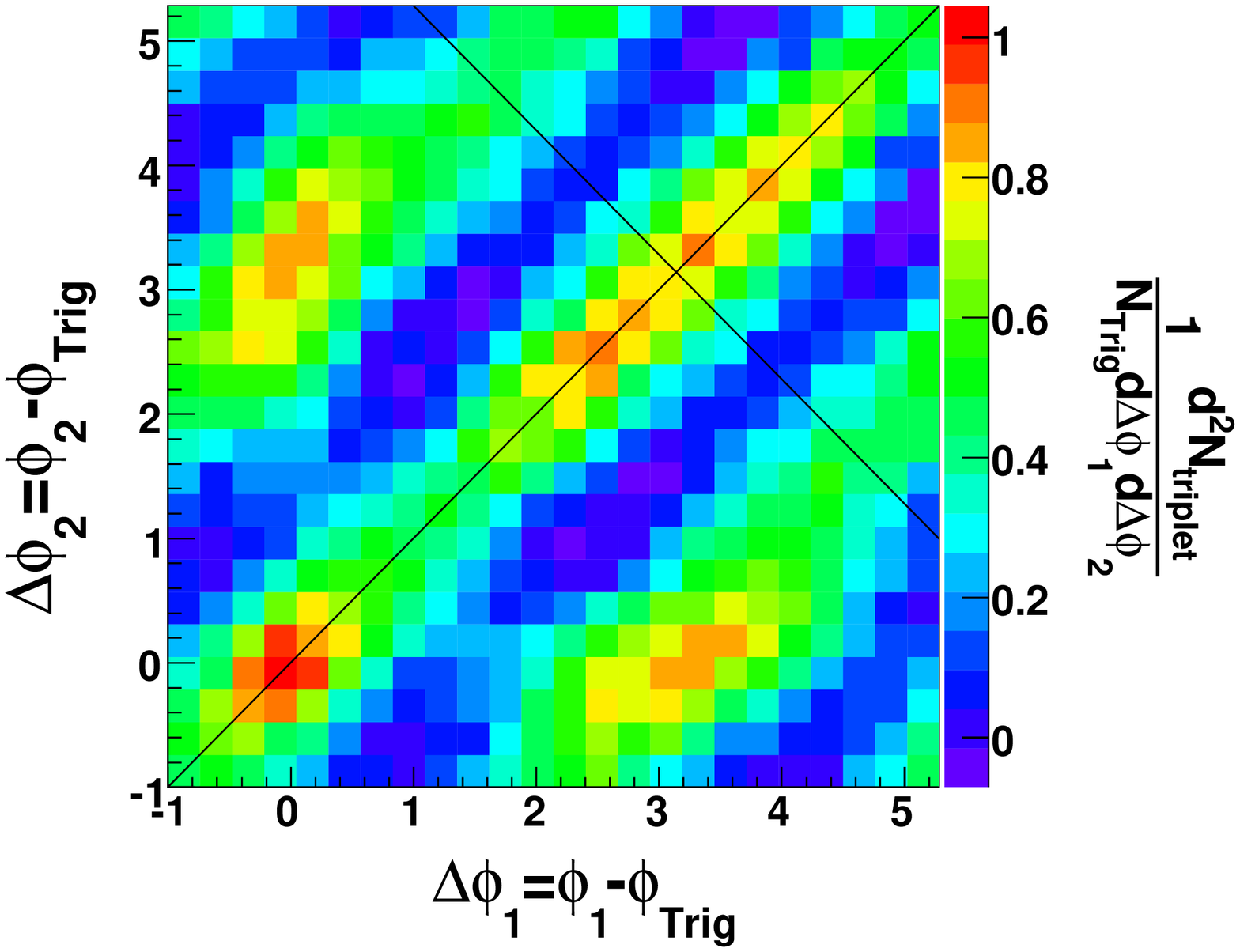}
\includegraphics[width=1.0\textwidth]{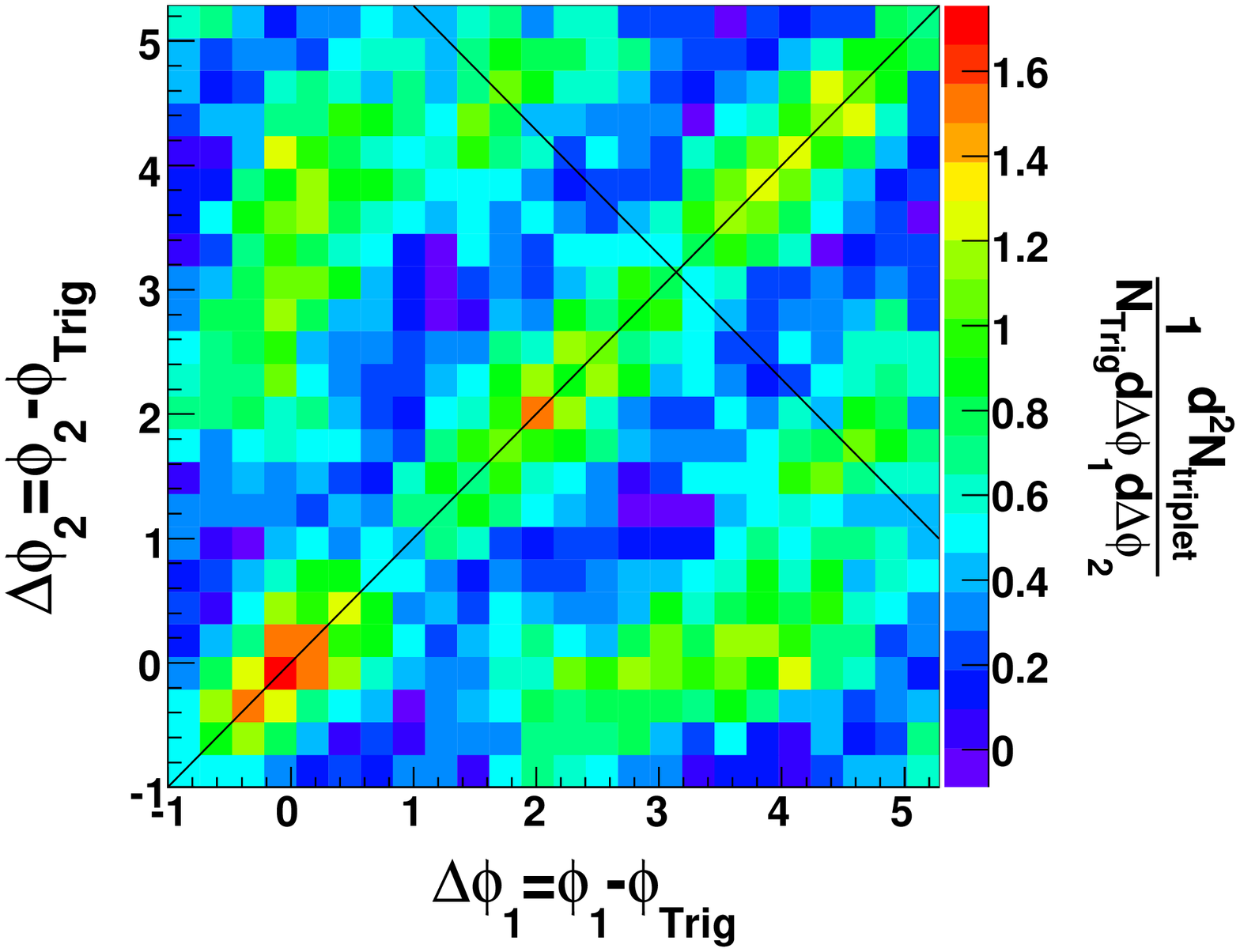}
\end{minipage}
\hfill
\begin{minipage}[t]{0.32\textwidth}
\centering
\includegraphics[width=1.0\textwidth]{Plots/Fig3Label_blank.eps}
\includegraphics[width=1.0\textwidth]{Plots/Fig3Label_blank.eps}
\includegraphics[width=1.0\textwidth]{Plots/Fig3Label_AuAuY4Cent_3.00Pt4.00_1.00pt2.00_6M9_dPhi2_.eps}
\end{minipage}
\caption{Background subtracted 3-particle correlations in (from right to left, top to bottom) {\it pp}, d+Au, Au+Au 50-80\%, Au+Au 30-50\%, Au+Au 10-30\%, Au+Au 0-10\% and ZDC triggered Au+Au 0-12\%. collisions at $\sqrt{s_{NN}}=200$ GeV/c}
\label{fig:final3}
\end{figure}

Figure~\ref{fig:final3} shows the background subtracted 3-particle correlations for {\it pp}, d+Au, four centralities of Au+Au from minimum bias data and ZDC triggered Au+Au.  The {\it pp} and d+Au background subtracted 3-particle correlations are very similar.  Both show four peaks.  A peak is at $(0,0)$ from instances where both associated particle are on the near-side.  Another peak is at $(\pi,\pi)$ from instances where both particles are on the away-side.  This peak is slightly elongated along the diagonal.  This elongation is qualitatively consistent with $k_{T}$ broadening which is due to nonzero initial total transverse momentum of the colliding partons.  This results in di-jets that are not exactly back-to-back.  This effect is enhanced by selecting high $p_T$ trigger particles because the selection preferentially picks up those parton scatterings that have a large total transverse momentum.  The $k_{T}$ is along the diagonal because it displaces the away-side from $\pi$ but all the jet particles on the away-side stay close together.  The peaks at $(0,\pi)$ and $(\pi,0)$ are also elongated for the same reason.  

\begin{figure}[htb]
\centering
\includegraphics[width=0.3\textwidth]{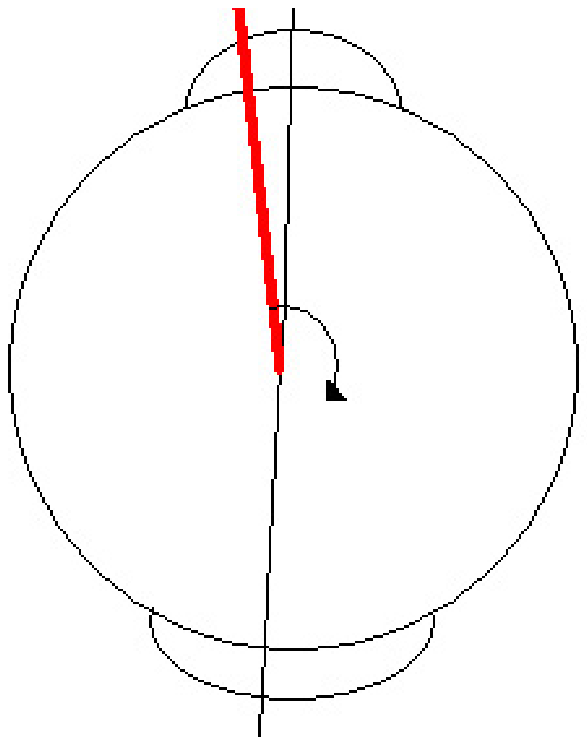}
\caption{Useful for visualizing the on-diagonal elongation and the tilting of the near-away peaks.  The black line is used to represent the jet-axis.  The red line is used to represent the trigger particle.  The bumps represent the near-side and away-side distribution.}
\label{fig:elong}
\end{figure}

In the peripheral Au+Au collisions, additional broadening is seen along the diagonal.  This additional elongation could be due to jets deflected by radial flow, path length dependent energy loss, or large angle gluon radiation.  In the 30-50\%  Au+Au centrality bin, potential peaks start to develop on the off-diagonal, suggesting a small contribution from conical emission.   The on-diagonal elongation continues into this bin.  There is also on-diagonal elongation of the near-side.  This is because we are using the trigger particle azimuthal angle as a proxy for the azimuthal angle of the jet-axis.  When the trigger paritcle is to one side of the jet-axis, we are more likly to get both associated particle to the other side of the jet-axis.  This also has the effect of slanting near-away peaks.  Figure~\ref{fig:elong} is a cartoon to help visualize this effect.  In the 10-30\% Au+Au bin, definite off-diagonal peaks are visible.  The 0-10\% Au+Au bin seems to have a simialar signal; however the statistics are very poor.  For increased statistics in central collisions we have taken enriched central data samples afforded by an online ZDC trigger.  This trigger provides a 0-12\% most central Au+Au collsions with an order of magnitude more statistics.  With this data sample, very distinct off-diagonal peaks can be seen, providing unambiguous evidence for conical emission.  

\begin{figure}[htb]
\hfill
\begin{minipage}{0.28\textwidth}
\centering
\includegraphics[width=1.0\textwidth]{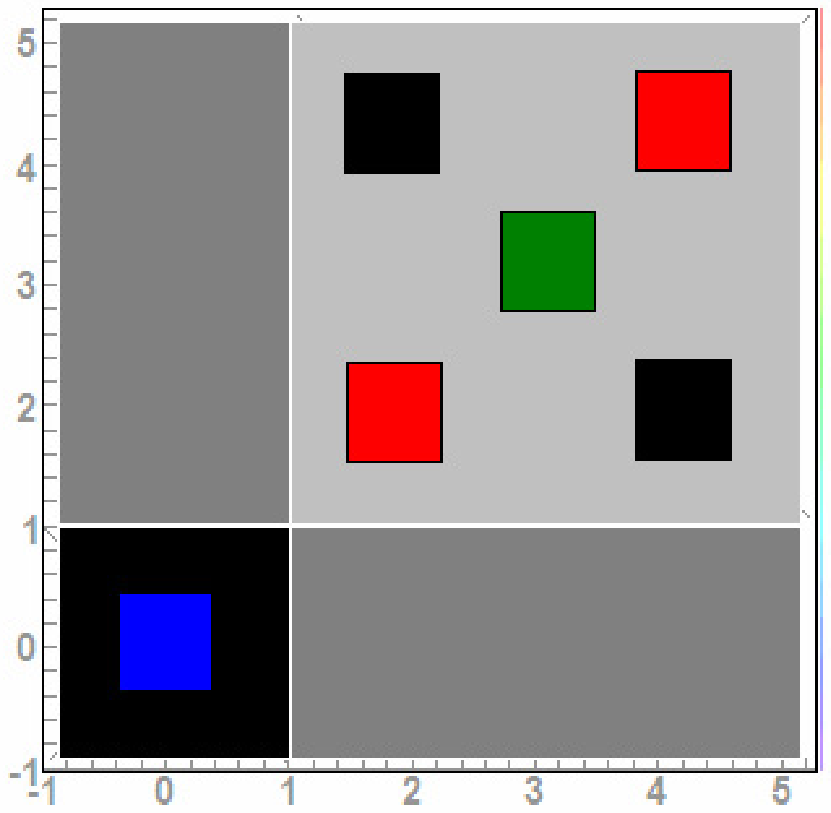}
\end{minipage}
\begin{minipage}{0.69\textwidth}
\centering
\includegraphics[width=1.0\textwidth]{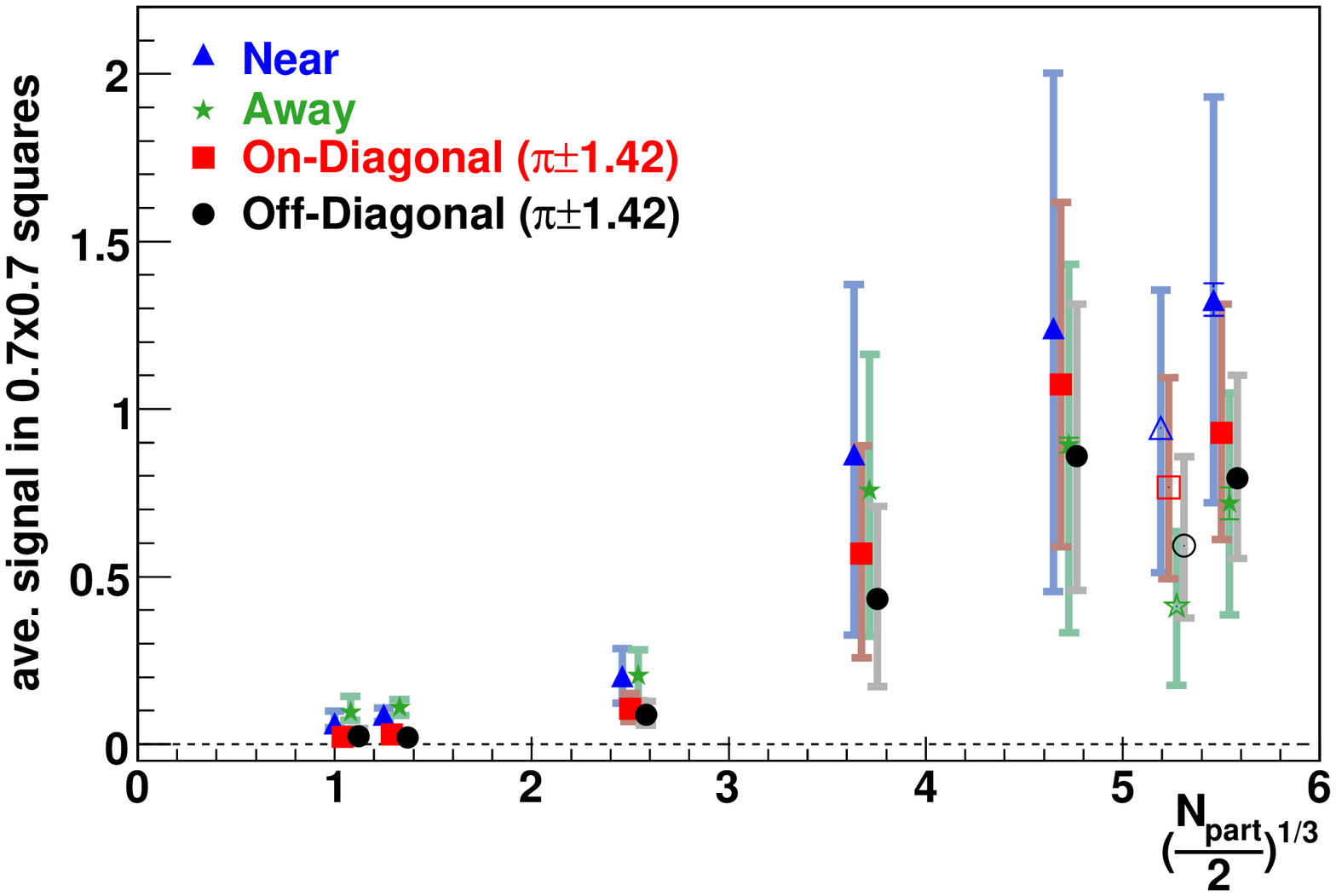}
\end{minipage}
\caption{Left:  Cartoon with approximate positions of squares used for the yields.  Right:  Average yield in 0.7x0.7 squares centered on the near-side $(0,0)$, the away-side $(\pi,\pi)$, on-diagonal $(\pi\pm1.42,\pi\pm1.42)$, and off-diagonal $(\pi\pm1.42,\pi\mp1.42)$.  $N_{part}$ is the number of participants.  The solid errors are statistical and the shaded are systematic.  The ZDC triggered 0-12\% most central Au+Au collions point is shifted to the left for clarity.  Collisions are at $\sqrt{s_{NN}}=200$ GeV/c.}
\label{fig:yieldcent}
\end{figure}

To examine the results quantitatively, we calculated the average yield, of the background subtracted 3-particle correlation, in boxes centered at regions of interest.  Figure~\ref{fig:yieldcent} (left) shows a cartoon with the locations of the boxes.  The right panel shows the average yields at these locations, the near-side $(0,0)$, the away-side $(\pi,\pi)$, on-diagonal $(\pi\pm1.42,\pi\pm1.42)$ and off-diagonal $(\pi\pm1.42,\pi\mp1.42)$ plotted as a function of centrality.  The near-side yield increases with centrality.  The away-side yield increases through {\it pp}, d+Au and peripheral Au+Au collisions but seems to level out in mid-central to central Au+Au collisions.  The on-diagonal and off-diagonal yields are consistent with zero in {\it pp}, d+Au, and peripheral Au+Au collisions.  In mid-central Au+Au collisions, they are significantly above zero.  The on-diagonal peak is consistently larger than the off-diagonal peak.  

\begin{figure}[htb]
\hfill
\begin{minipage}[t]{0.32\textwidth}
\centering
\includegraphics[width=1.0\textwidth]{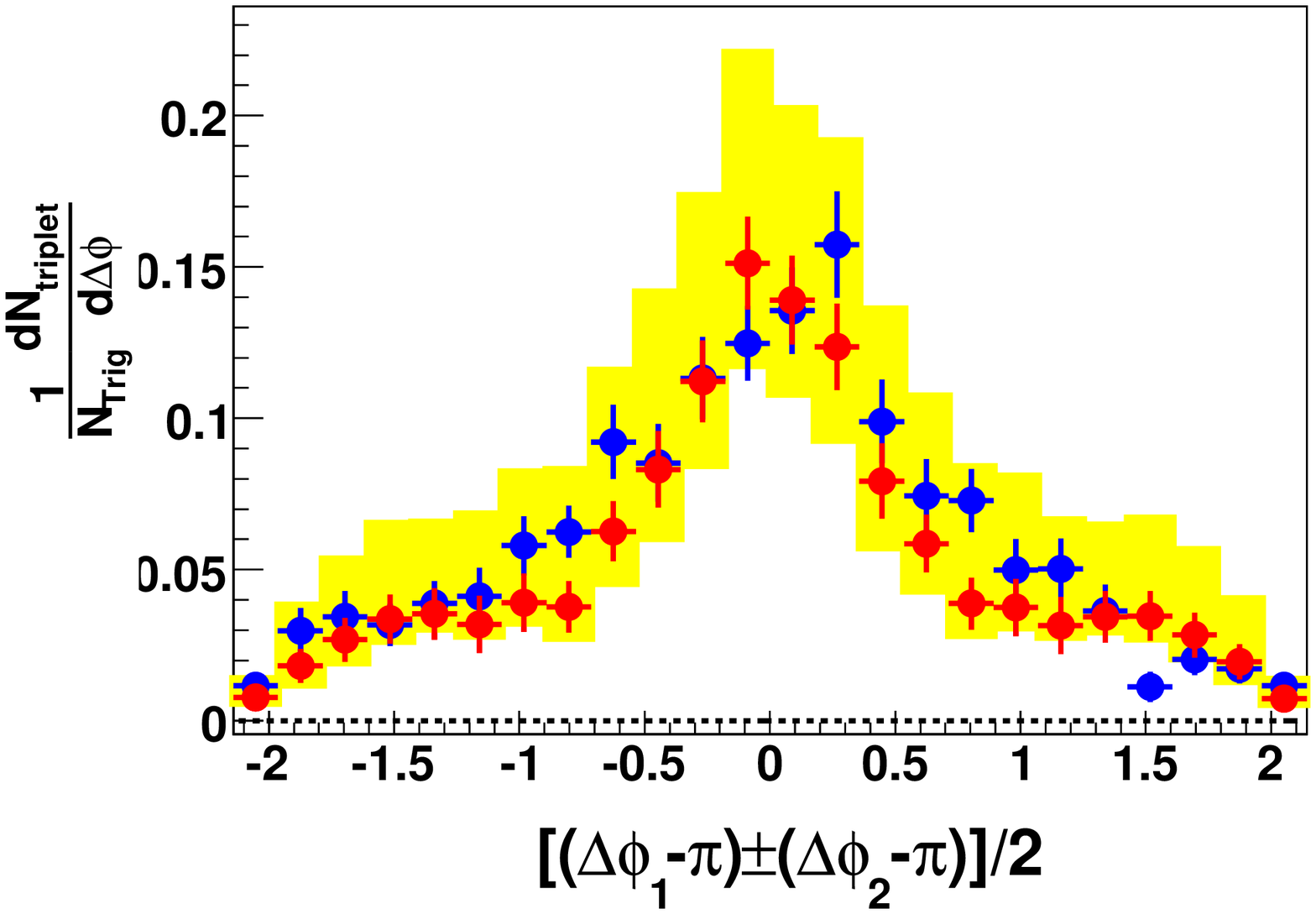}
\includegraphics[width=1.0\textwidth]{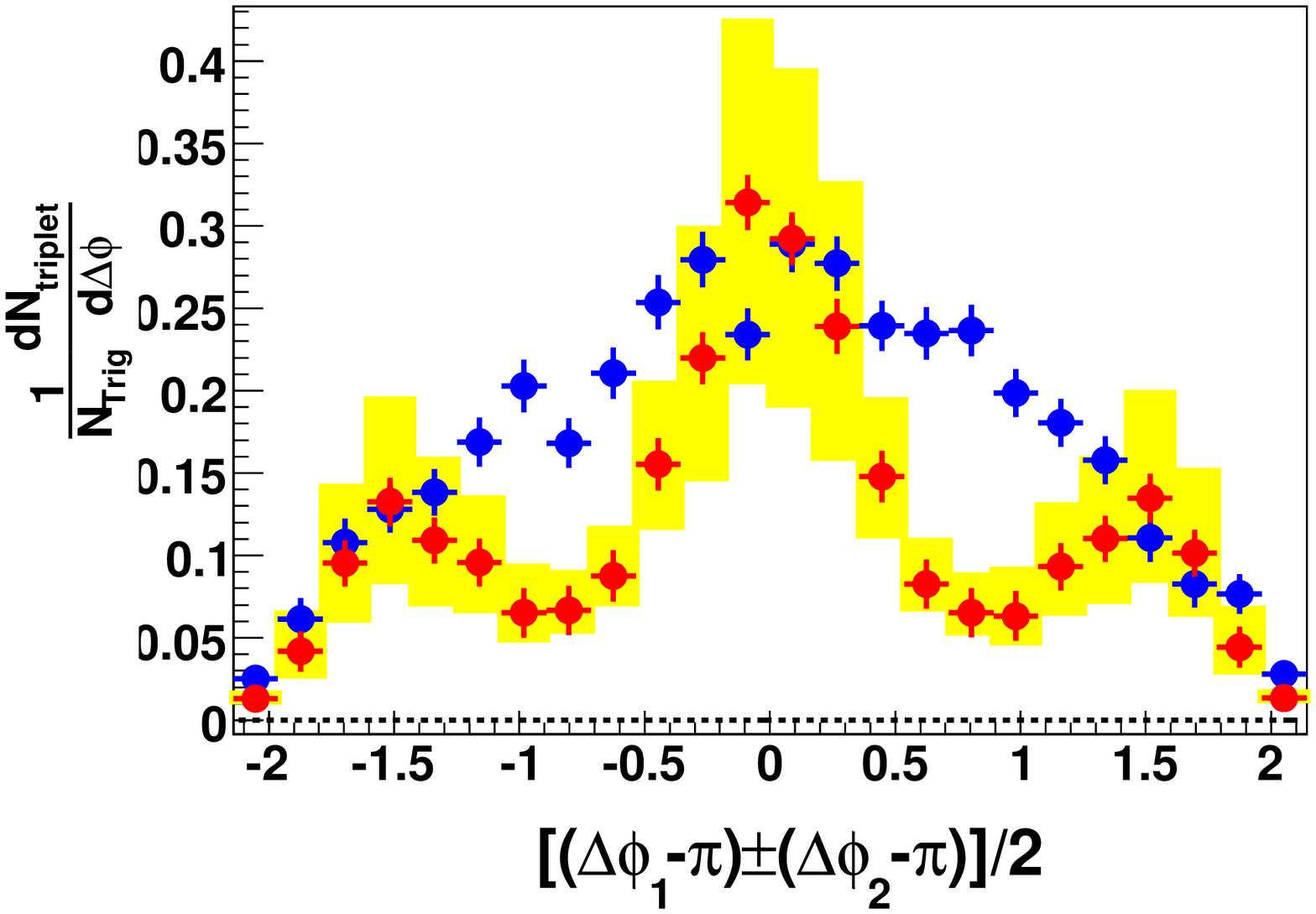}
\includegraphics[width=1.0\textwidth]{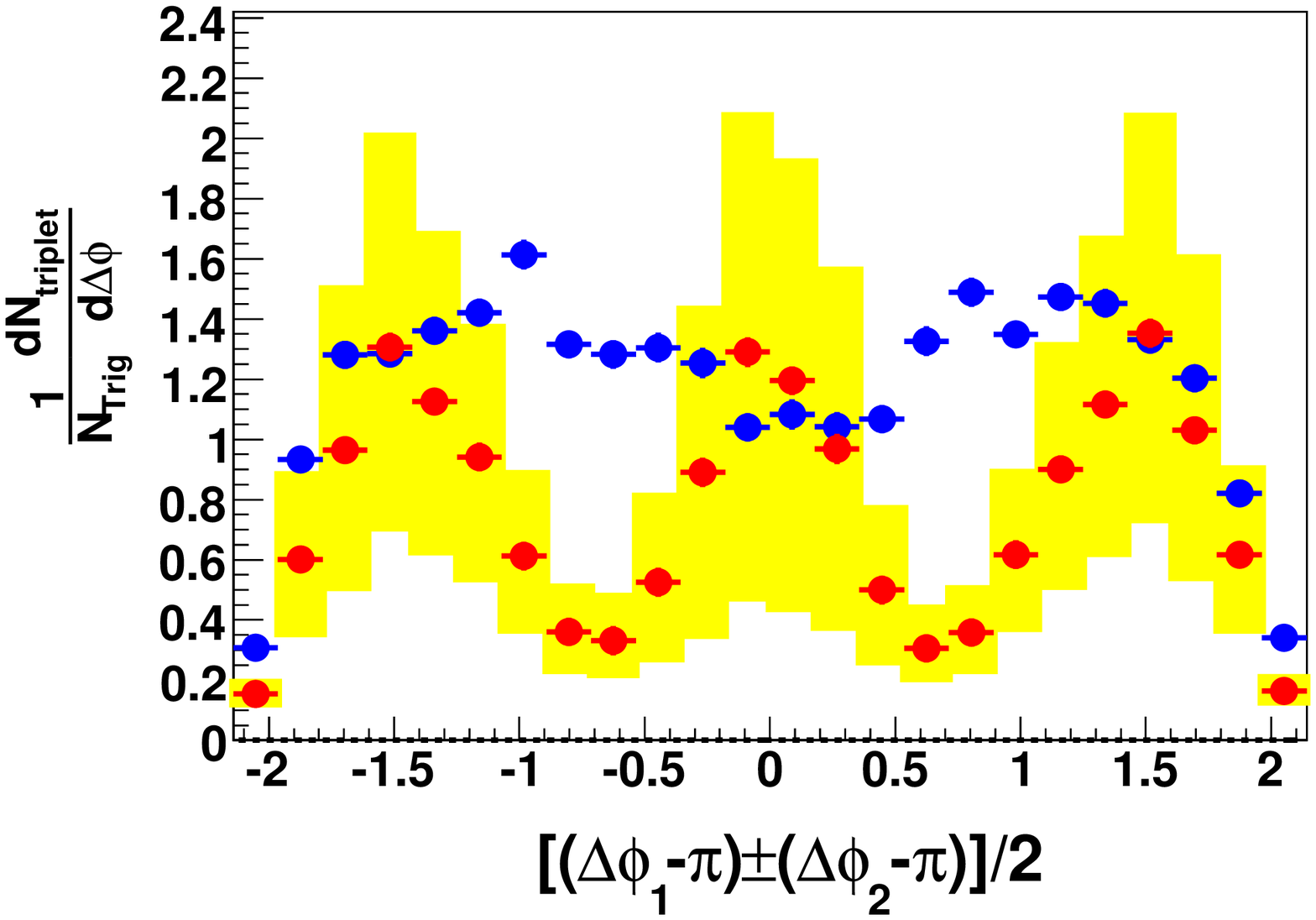}
\end{minipage}
\hfill
\begin{minipage}[t]{0.32\textwidth}
\includegraphics[width=1.0\textwidth]{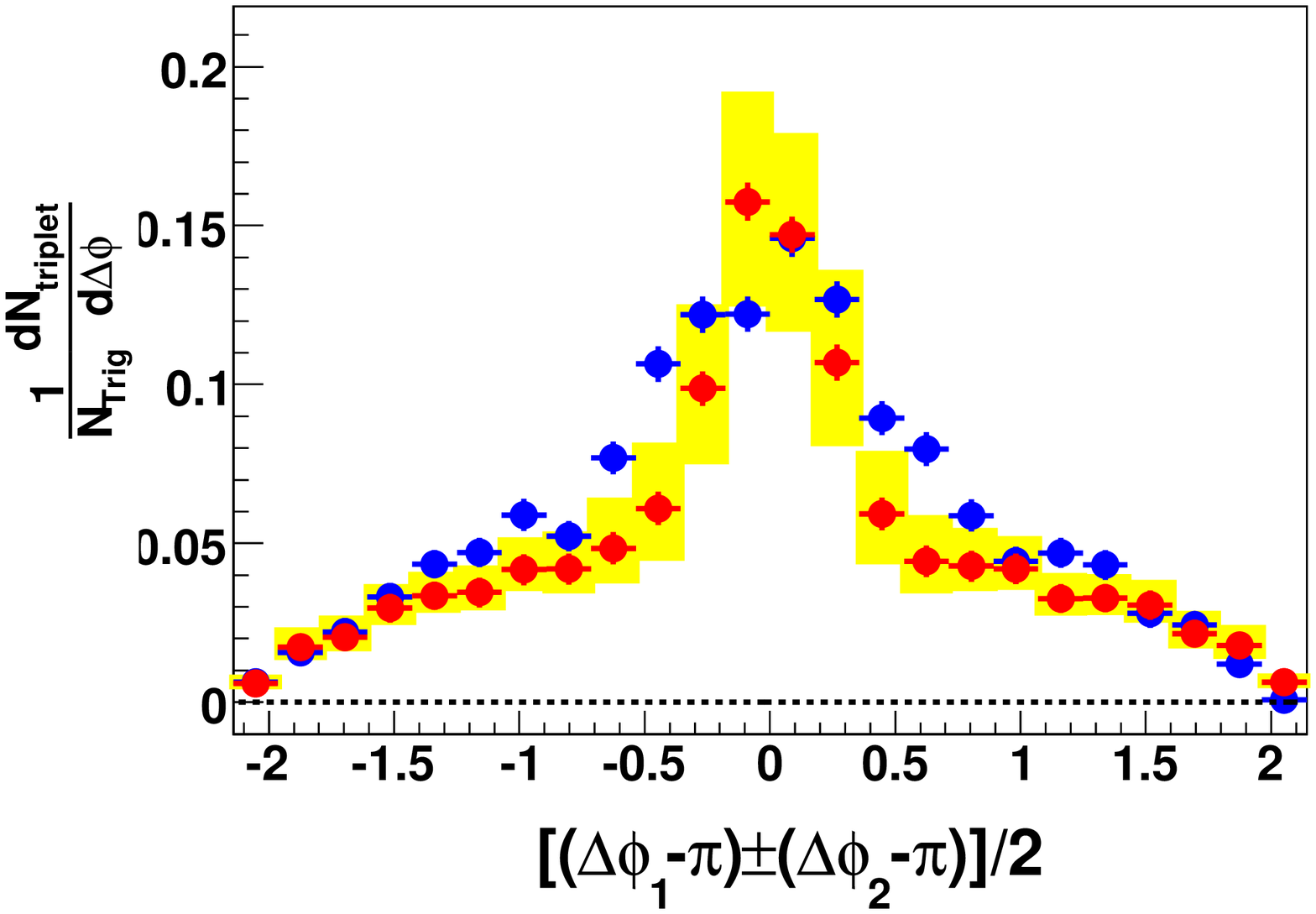}
\includegraphics[width=1.0\textwidth]{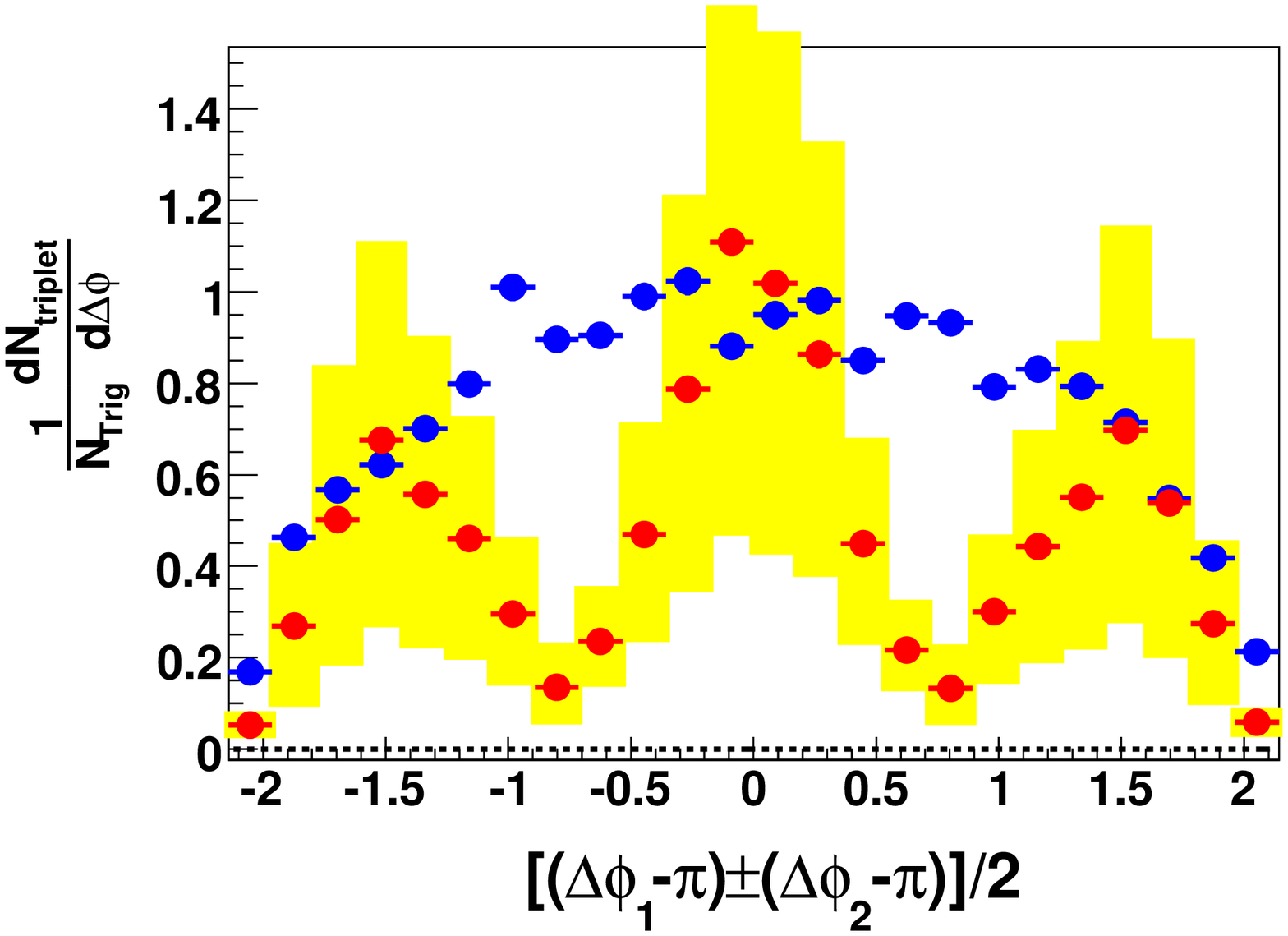}
\includegraphics[width=1.0\textwidth]{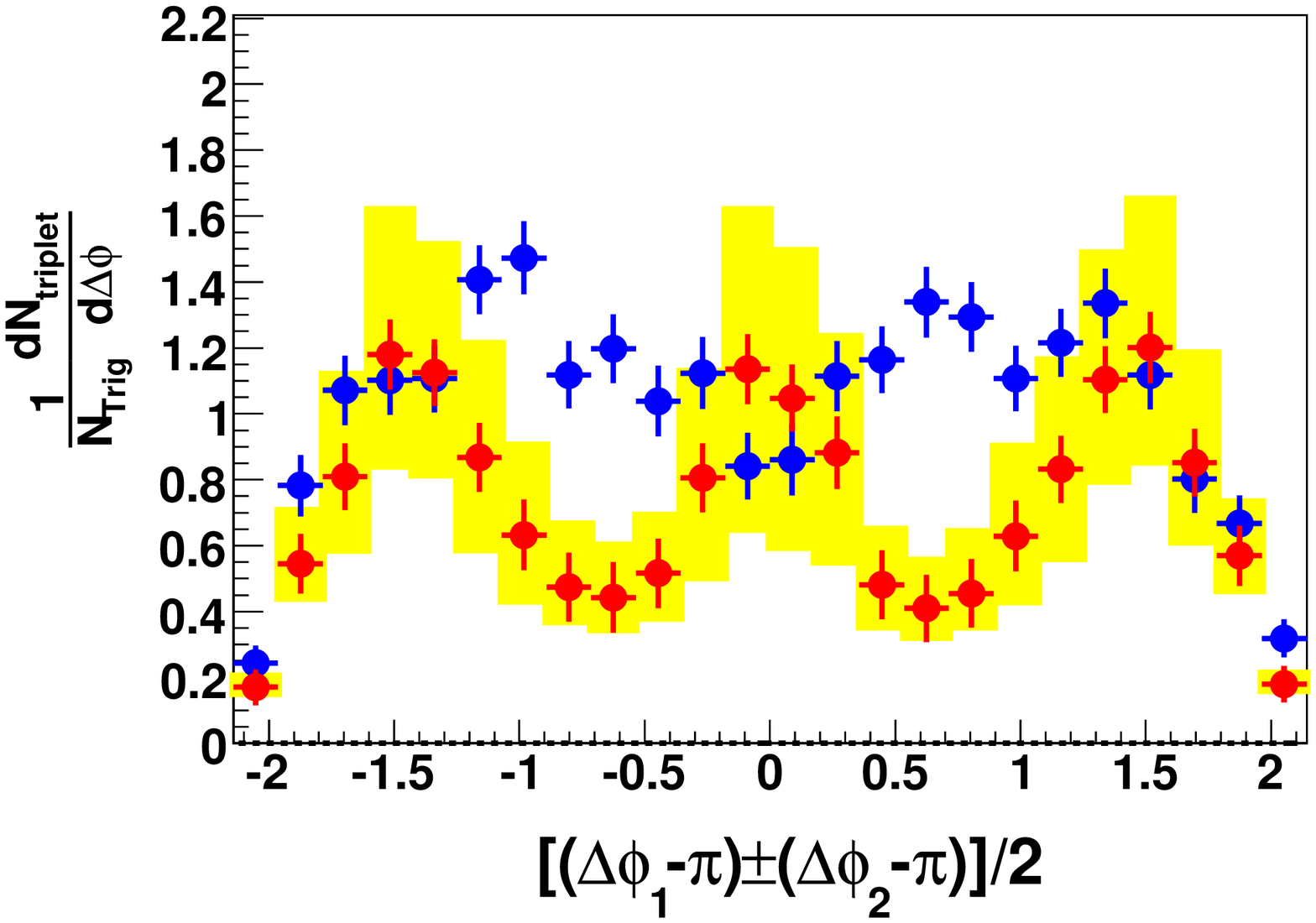}
\end{minipage}
\hfill
\begin{minipage}[t]{0.32\textwidth}
\includegraphics[width=1.0\textwidth]{Plots/ProjStripBlank.eps}
\includegraphics[width=1.0\textwidth]{Plots/ProjStripBlank.eps}
\includegraphics[width=1.0\textwidth]{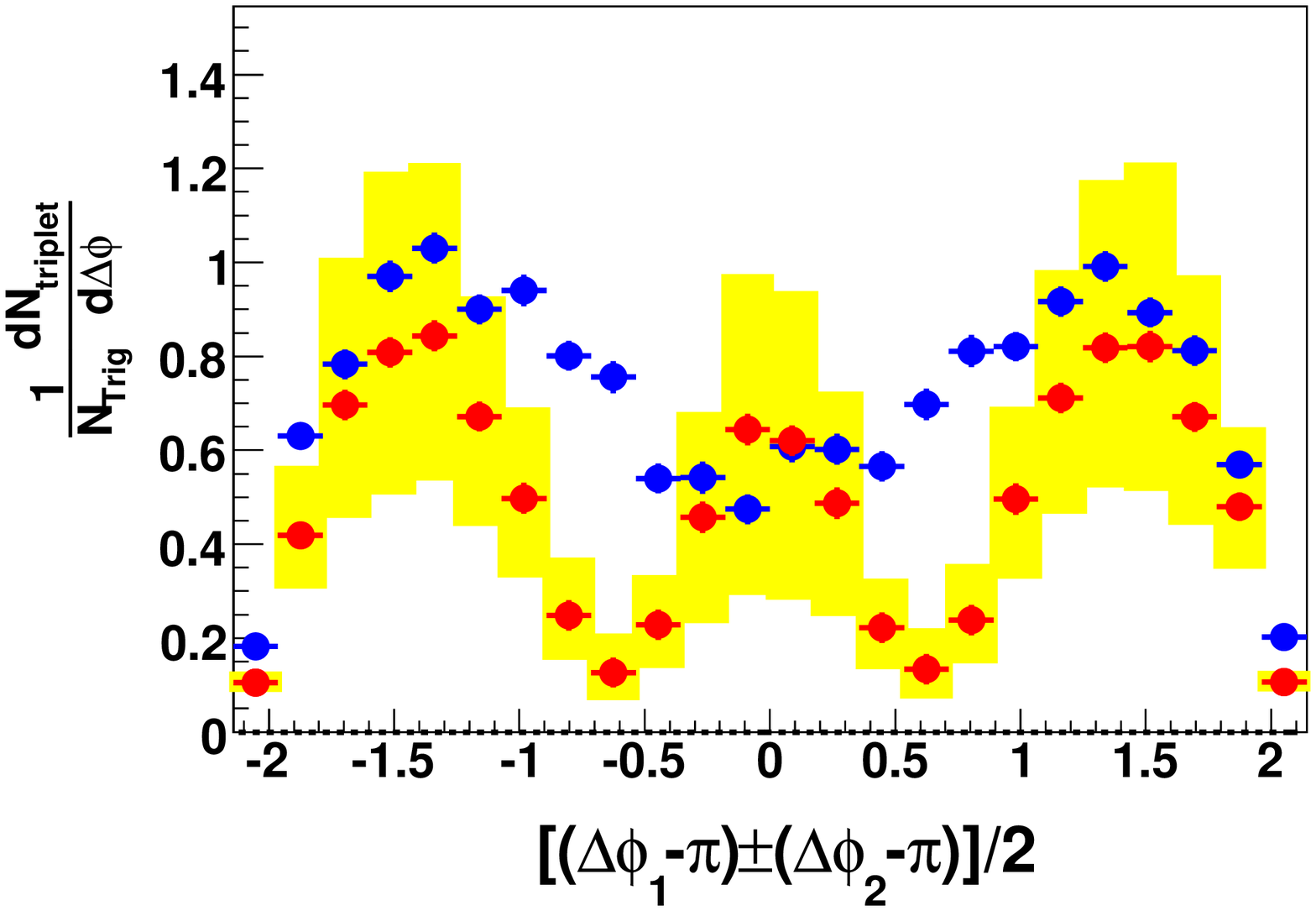}
\end{minipage}
\caption{Projections of strips of full width 0.7 radians on the away-side.  On-diagonal projections (blue) and off-diagonal projection (red) in centralities (from left to right, top to bottom) {\it pp}, d+Au, Au+Au 50-80\%, Au+Au 30-50\%, Au+Au 10-30\%, Au+Au 0-10\% and ZDC triggered Au+Au 0-12\% collisions at $\sqrt{s_{NN}}=200$ GeV/c.  Error bars are statistical errors.  Systematic errors on the off-diagonal projections are shown in the yellow band.  Dashed black lines are at zero.  The on-diagonal projections with systematic errors can be found in the appendix.}
\label{fig:proj}
\end{figure}

We obtain more quantitative information by studying the projections of the background subtracted 3-particle correlations.  We expect two side peaks in the off-diagonal projection for conical emission.  Figure~\ref{fig:proj} shows on-diagonal and off-diagonal projections of the background subtracted 3-particle correlations shown in Fig~\ref{fig:final3}.  These projections are done by projecting strips of width 0.7 radians about the diagonal (off-diagonal) to the diagonal (off-diagonal) for particles on the away-side, a diagram showing the projected regions is shown in Fig.~/ref{fig:ProjDiagram}.  Since we have finite binning and we are not projecting along one of the axes, bin width effects will appear in these projections if we do not take care of them.  This is removed by sampling each individual $(\Delta\phi_{1},\Delta\phi_{2})$ bin randomly 100 times (i.e. $1/100^{th}$ of the bin content is projected at 100 random positions within each bin).  This removes the jagged effects due to finite bin width but does leave us with bin smearing of up to half a bin width.  

\begin{figure}[htb]
\centering
\includegraphics[width=0.3\textwidth]{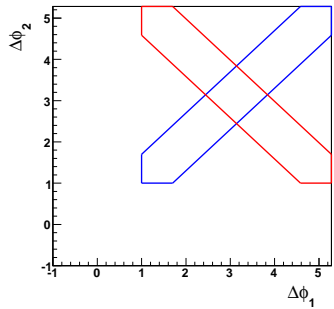}
\caption{Regions that are projected for the on-diagonal and off-diagonal projections are shown in blue and red respectively.}
\label{fig:ProjDiagram}
\end{figure}

As seen from Figure~\ref{fig:proj}, in {\it pp} and d+Au collisions the on-diagonal (blue) and off-diagonal (red with yellow systematic error boxes) projections look very similar, both are single peaked.  The only difference is the on-diagonal projection is wider.  This is likely due to $k_{T}$ broadening as previously discussed.  The most peripheral Au+Au bin (50-80\%) has a very broad on-diagonal peak.  The off-diagonal projection shows a central peak and two symmetric side peaks\footnote{The 3-particle correlation functions are symmetric with respect to $\Delta\phi_{1}$ and $\Delta\phi_{2}$ by construction so the off-diagonal projections are always symmetric, except for effects from the randomization and the bins are not set such that $\pi$ is a division.  The on-diagonal projection has no such forced symmetry.  The statistical errors in the projections do not have a correction for the double counting of associated pairs.}.  These side-peaks are not very significant, perhaps indicating only a small contribution from conical emission to this centrality.  The next centrality bin (30-50\% most) shows an on-diagonal peak that is very broad and consistent with being flat on top.  The off-diagonal projection shows side peaks that are larger than the 50-80\% centrality bin, however the systematic errors are also larger and the peaks are only about two sigma effect.  In the 10-30\% Au+Au centrality bin, the on-diagonal projection has become dipped at $[\Delta\phi_1+\Delta\phi_2]/2=\pi$.  The off-diagonal projection shows significant side peaks, evidence for conical emission.  The 0-10\% most central Au+Au collisons look very similar although the statistical errors are much larger.  The 0-12\% ZDC triggered central Au+Au data again looks similar (but with smaller errors) with a dip in the on-diagonal projection, a centeral peak and two symmetric side peaks in the off-diagonal projection.  These side peaks are consistent with conical emisson in the central Au+Au data.  

\begin{figure}[htb]
\hfill
\begin{minipage}[t]{0.32\textwidth}
\centering
\includegraphics[width=1.0\textwidth]{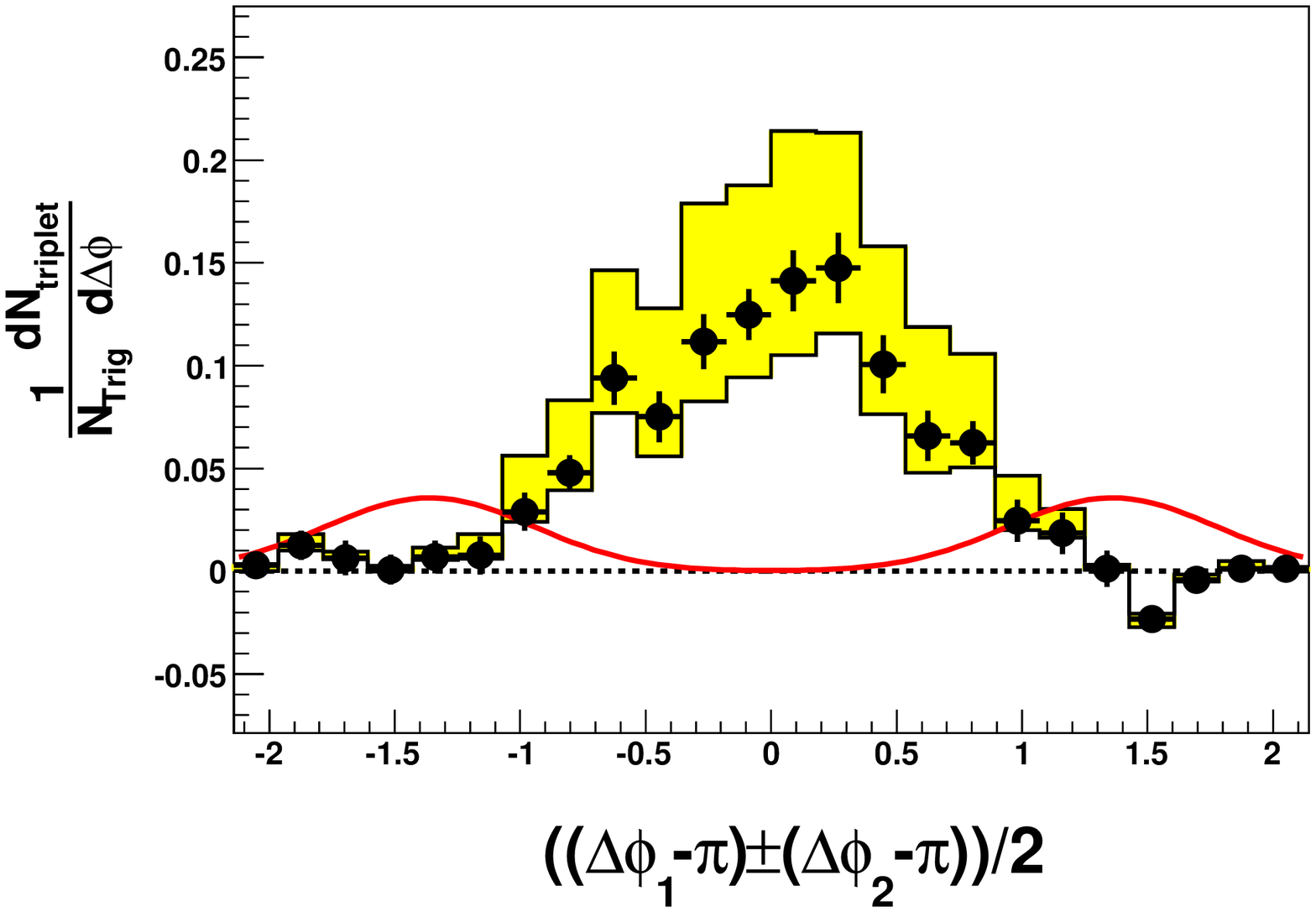}
\includegraphics[width=1.0\textwidth]{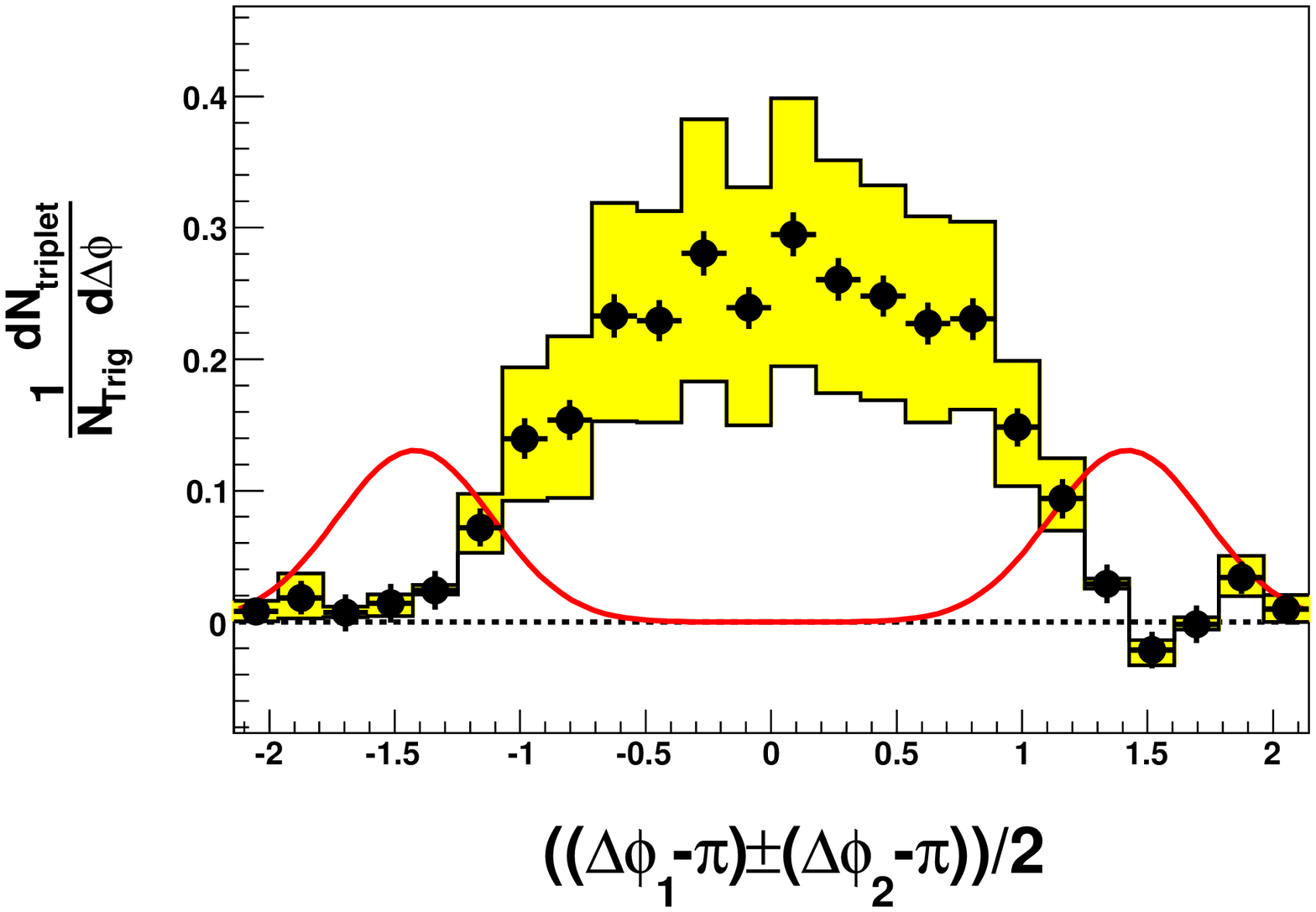}
\includegraphics[width=1.0\textwidth]{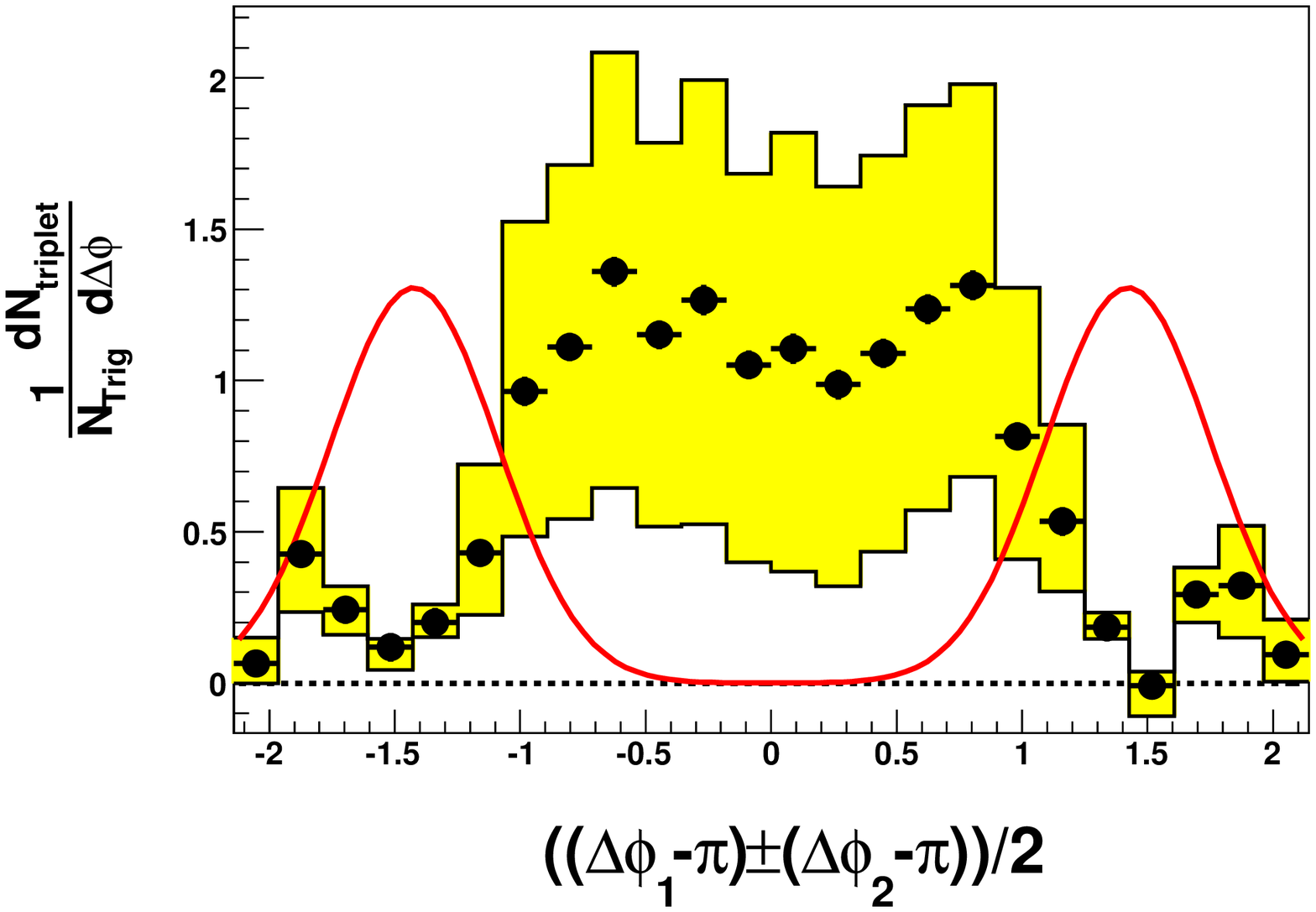}
\end{minipage}
\hfill
\begin{minipage}[t]{0.32\textwidth}
\includegraphics[width=1.0\textwidth]{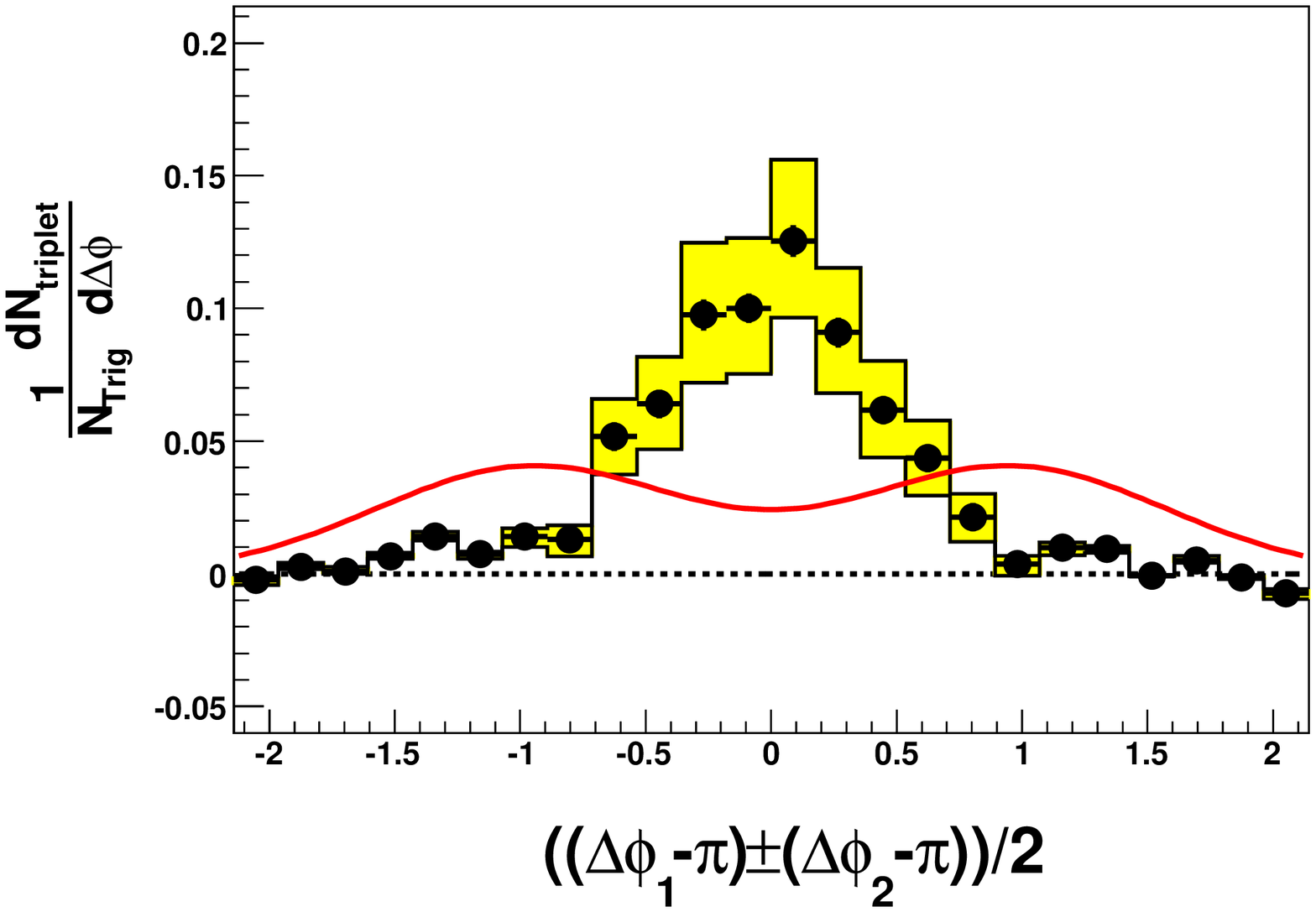}
\includegraphics[width=1.0\textwidth]{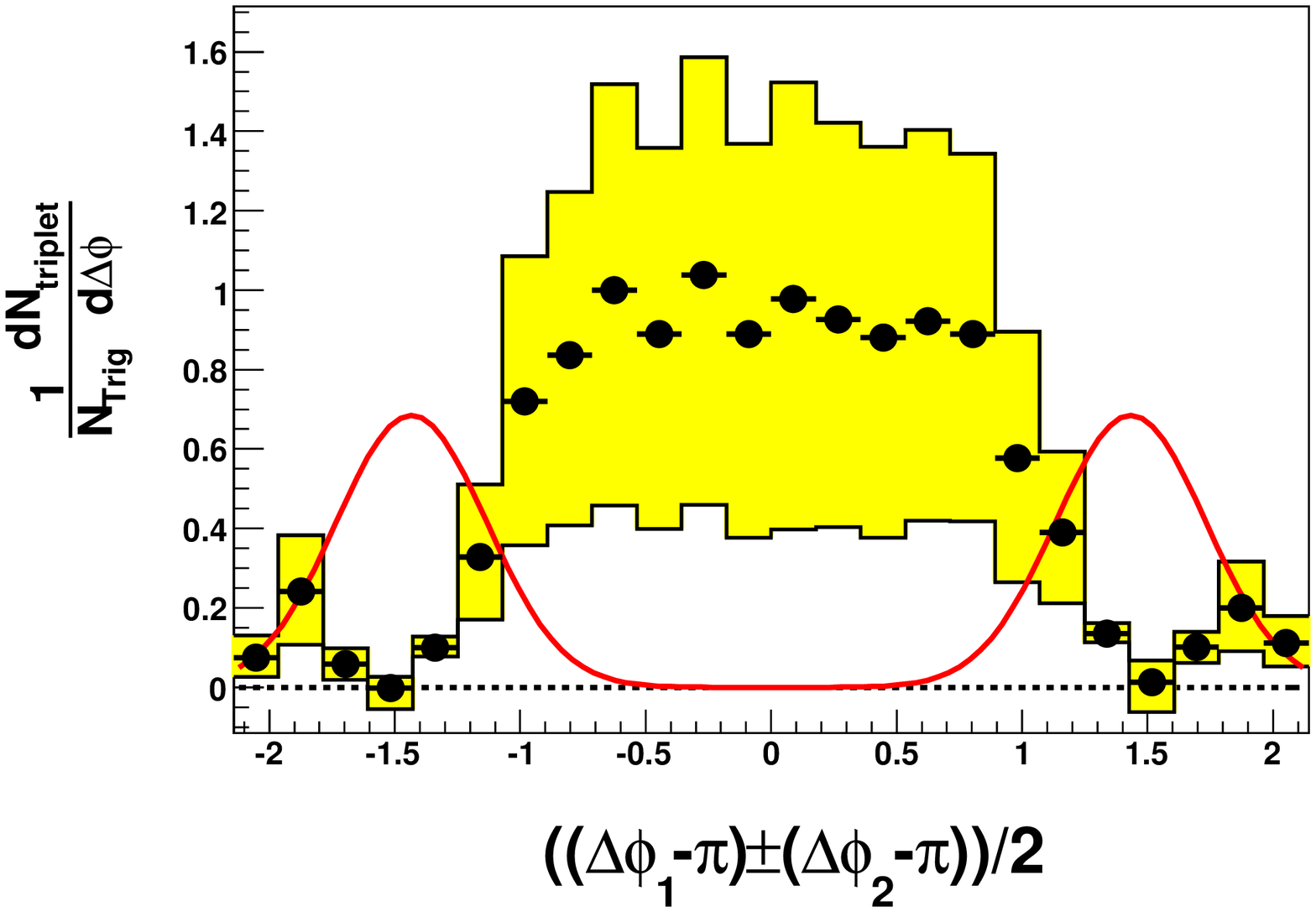}
\includegraphics[width=1.0\textwidth]{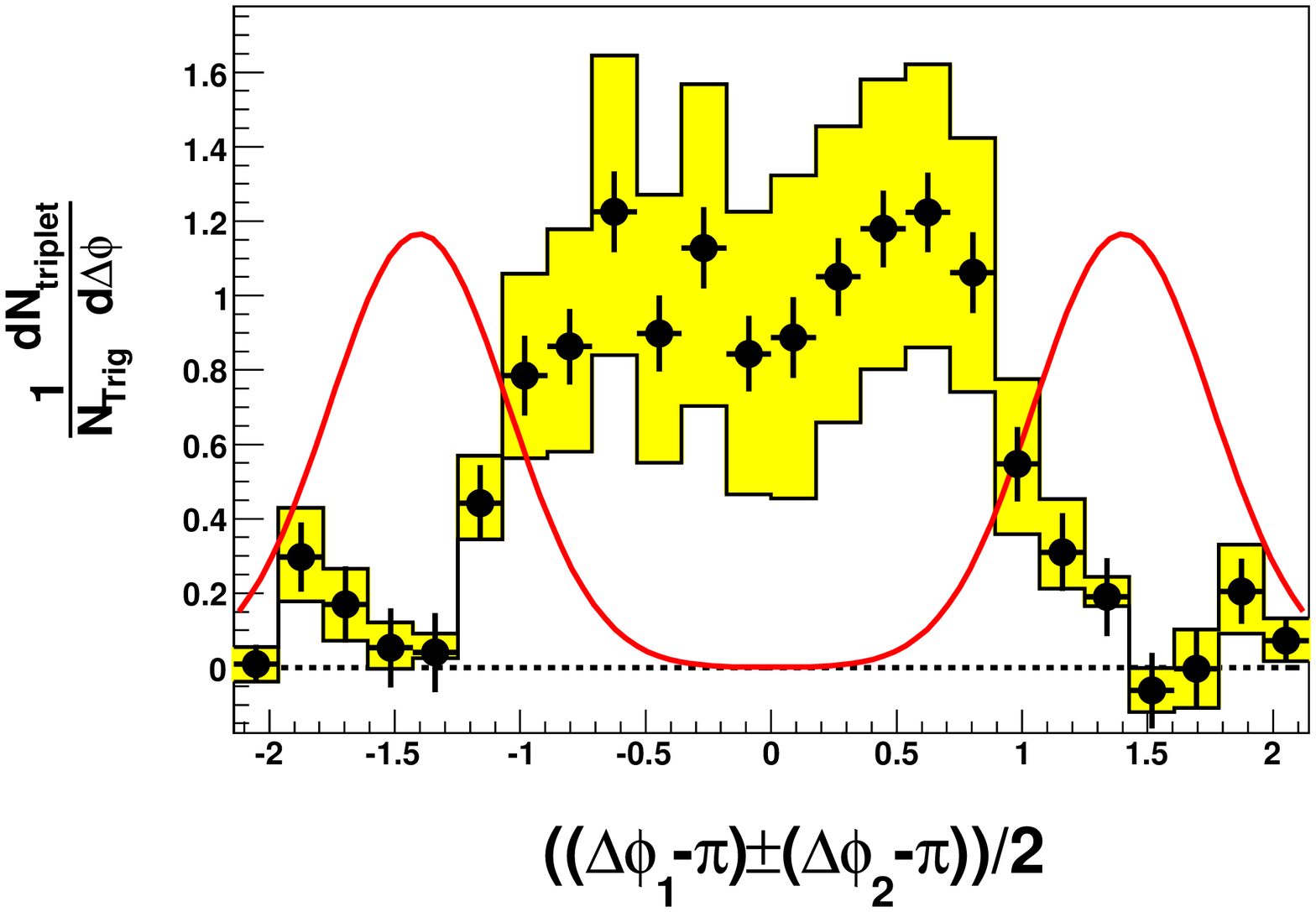}
\end{minipage}
\hfill
\begin{minipage}[t]{0.32\textwidth}
\includegraphics[width=1.0\textwidth]{Plots/ProjStripBlank.eps}
\includegraphics[width=1.0\textwidth]{Plots/ProjStripBlank.eps}
\includegraphics[width=1.0\textwidth]{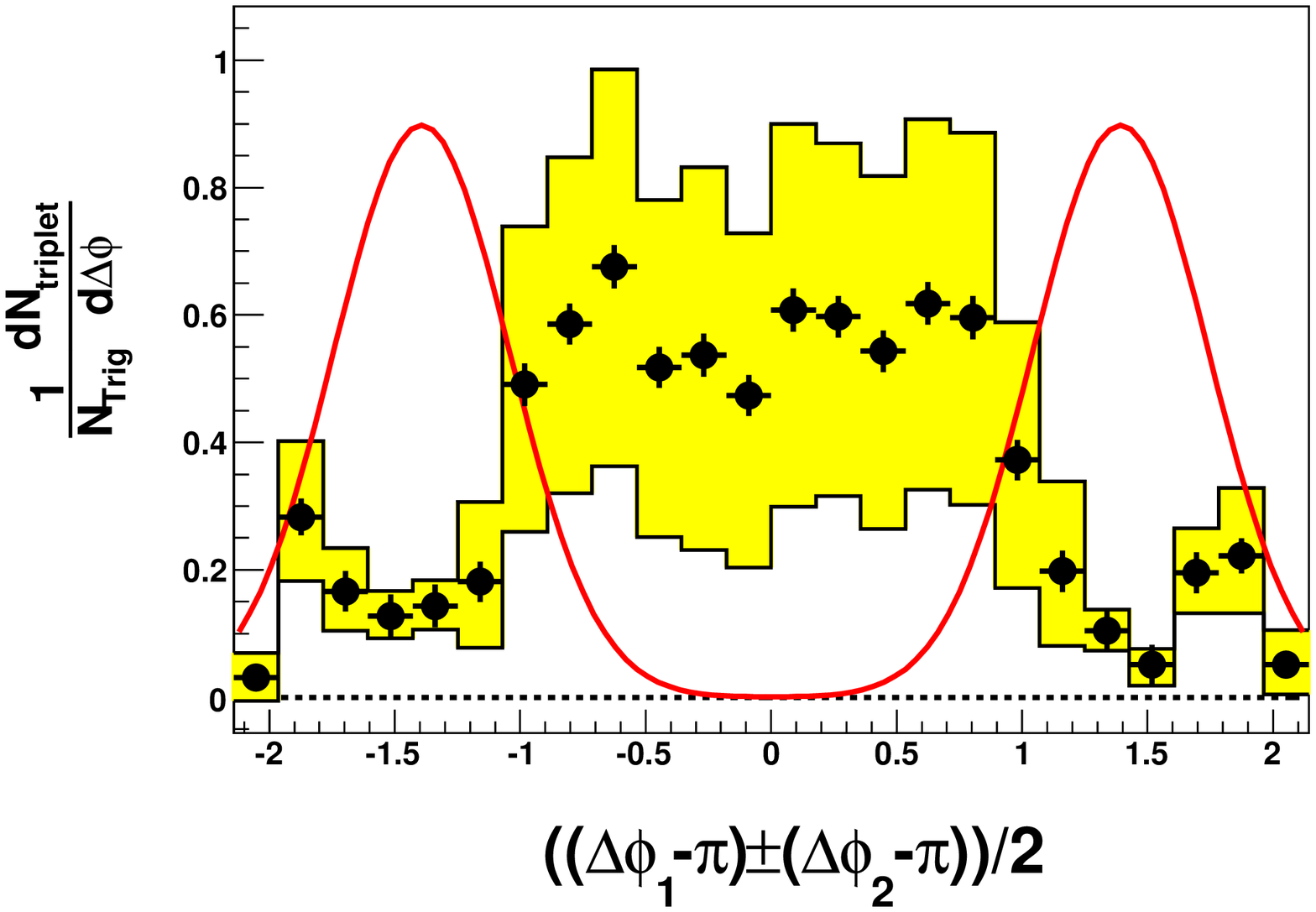}
\end{minipage}
\caption{Difference between in the on-diagonal projection and the side Gaussians from fit to the off-diagonal projection.  Projections are in strips of full width 0.7 radians on the away-side in centralities (from left to right, top to bottom) {\it pp}, d+Au, Au+Au 50-80\%, Au+Au 30-50\%, Au+Au 10-30\%, Au+Au 0-10\% and ZDC triggered Au+Au 0-12\% collisions at $\sqrt{s_{NN}}=200$ GeV/c.  Error bars as statistical from the on-diagonal projection only.  Systematic errors are shown in the yellow band.  Dashed black lines are at zero.}
\label{fig:projdiff}
\end{figure}

There is extra signal on the on-diagonal projections when compared to the off-diagonal projections.  This is shown by the difference in the projections in Fig.~\ref{fig:projdiff}.  The difference plotted is the on-diagonal projection minus the side Gaussians obtained from the fit to the off-diagonal projection.  In $pp$ and d+Au, there is little signal in the off-diagonal projections in the region of the side peaks so this difference is very similar to the on-diagonal projection.  In central and mid-central Au+Au collisions, there is a central wide peak that is rather flat.  There is also some small contribution on the outside edges.  The peaks seen in the difference could be due to many different physics mechanism such as: large angle gluon radiation, jets deflected by radial flow or preferential selection of particles due to path-length dependent energy loss, hydrodynamic conical flow generated by Mach-cone shock waves that couple with flow\cite{mach6}, and/or jets for which the away-side jet undergoes relatively little medium modification. 

\subsection{Emission Angle and Associated $P_T$ Dependence}

\begin{figure}[htb]
\centering
\includegraphics[width=0.75\textwidth]{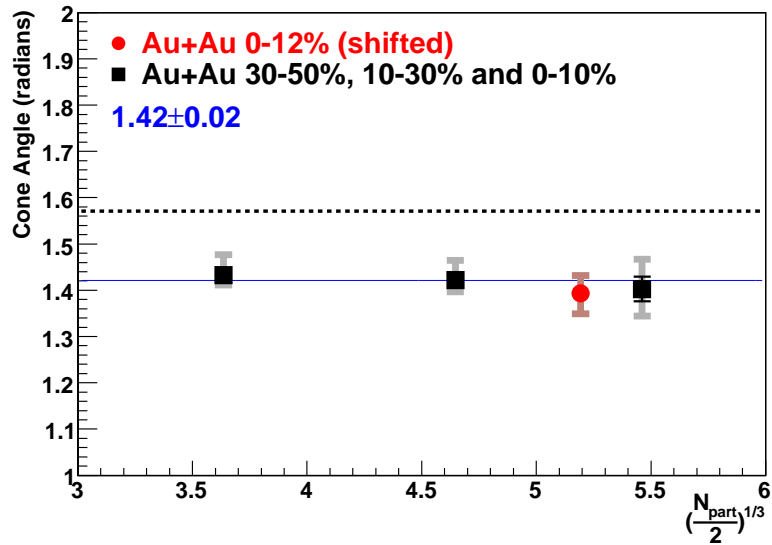}
\caption{Angles from Gaussian fits to the off-diagonal projections.  The statistical errors (solid) are the fit errors.  Systematic errors are shaded.  The blue line is a fit of the points to a constant, yielding $1.42\pm0.02$ (fit error using quadrature sum of statistical and systematic errors on the points).  $N_{part}$ is the number of participants.  The ZDC triggered 0-12\% most central Au+Au point is shifted to the left for clarity.  The dashed line is at $\pi/2$.  The collisions are at $\sqrt{s_{NN}}=200$ GeV/c.}
\label{fig:angle}
\end{figure}

We can extract the conical emission angle by fitting the off-diagonal projections to a central Gaussian and symmetric side Gaussians. Table~\ref{tab:angle} gives the angles obtained from the fits for mid-central and central Au+Au collisions.  They are also plotted in figure~\ref{fig:angle}.  If the emission angle is the Mach-cone angle, we may use this angle to extract the average speed of sound of the medium, $c_s$, from the obtained Mach-cone angle $\theta^M$, by,  
\begin{equation}
cos(\theta^{M})=c_{s}/c
\end{equation}
This is the average speed of sound, averaging over the entire time evolution of heavy ion collisions, which may undergo several stages:  QGP, phase transition, and hadronic stages.  Expected sound velocities for different phases are listed in table~\ref{tab:cs}.  This equation gives $c_{s}=0.15c$ for $\theta^{M}=1.42$ radians.    One should be cautious about this number because it neglects effects of hydrodynamics (other than Mach-cone) and expansion.   The relation between the emission angle in data and the speed of sound in the medium could be very complicated and is a subject of on-going theoretical work.

\begin{table}[hbt]
\centering
\caption{Angles from Gaussian fits to the off-diagonal projections.}
\begin{tabular}{|l|l|l|l|} 
\hline
Centrality&Angle&Statistical Error&Systematic Error\\
\hline
AuAu 30-50\%&1.43&$\pm0.01$&$^{+0.04}_{-0.03}$\\
AuAu 10-30\%&1.42&$\pm0.01$&$^{+0.04}_{-0.03}$\\
AuAu  0-10\%&1.40&$\pm0.03$&$^{+0.07}_{-0.05}$\\
AuAu ZDC 0-12\%&1.39&$\pm0.01$&$^{+0.04}_{-0.04}$\\
\hline
\end{tabular}
\label{tab:angle}
\end{table}

\begin{table}[hbt]
\centering
\caption{Expected speeds of sound in different phases of matter~\cite{cs1,cs2}.}
\begin{tabular}{|l|l|}
\hline
Phase&Speed of Sound\\
\hline
QGP& $\frac{1}{\sqrt{3}}c$\\
Mixed Phase& 0\\
Resonance Gas& $0.47c$\\
\hline
\end{tabular}
\label{tab:cs}
\end{table}

It is important to investigate the emission angle as a function of associated particle $p_{T}$.  For conical emission from a Mach-cone the angle is expect to remain constant as a function of associated particle $p_{T}$, since the speed of sound is an intrinsic property of the medium and independent of the partonic momentum.  Simple \v{C}erenkov radiation models predict a sharply decreasing angle as a function of associated particle $p_{T}$\cite{cerenkov2}.  This $p_T$ dependence can be used to distinguish conical emission from Mach-cone and \v{C}erenkov radiation.

\begin{figure}[htb]    
\hfill
\begin{minipage}{0.32\textwidth}
\centering
\includegraphics[width=1.0\textwidth]{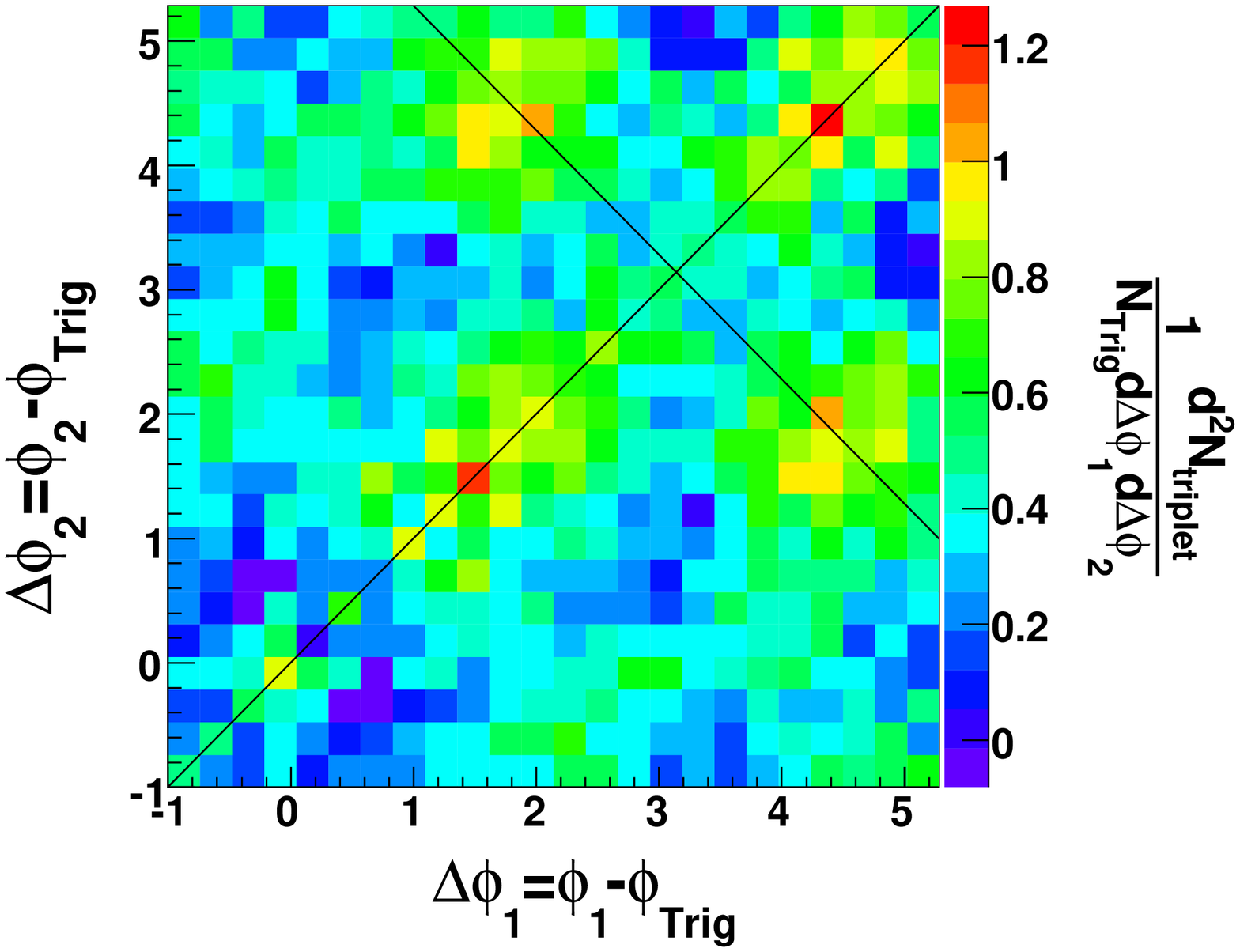}
\includegraphics[width=1.0\textwidth]{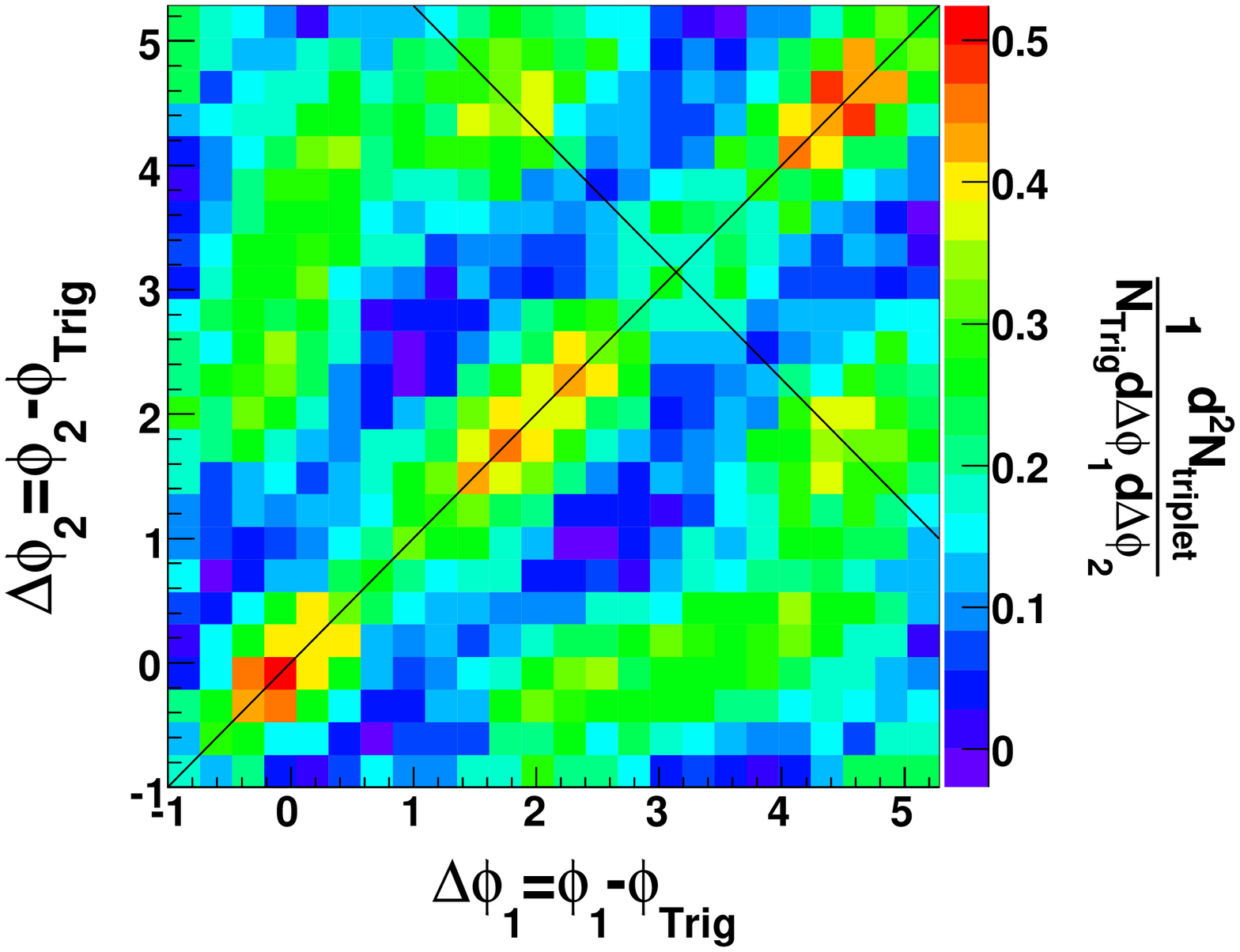}
\end{minipage}
\hfill
\begin{minipage}{0.32\textwidth}
\centering
\includegraphics[width=1.0\textwidth]{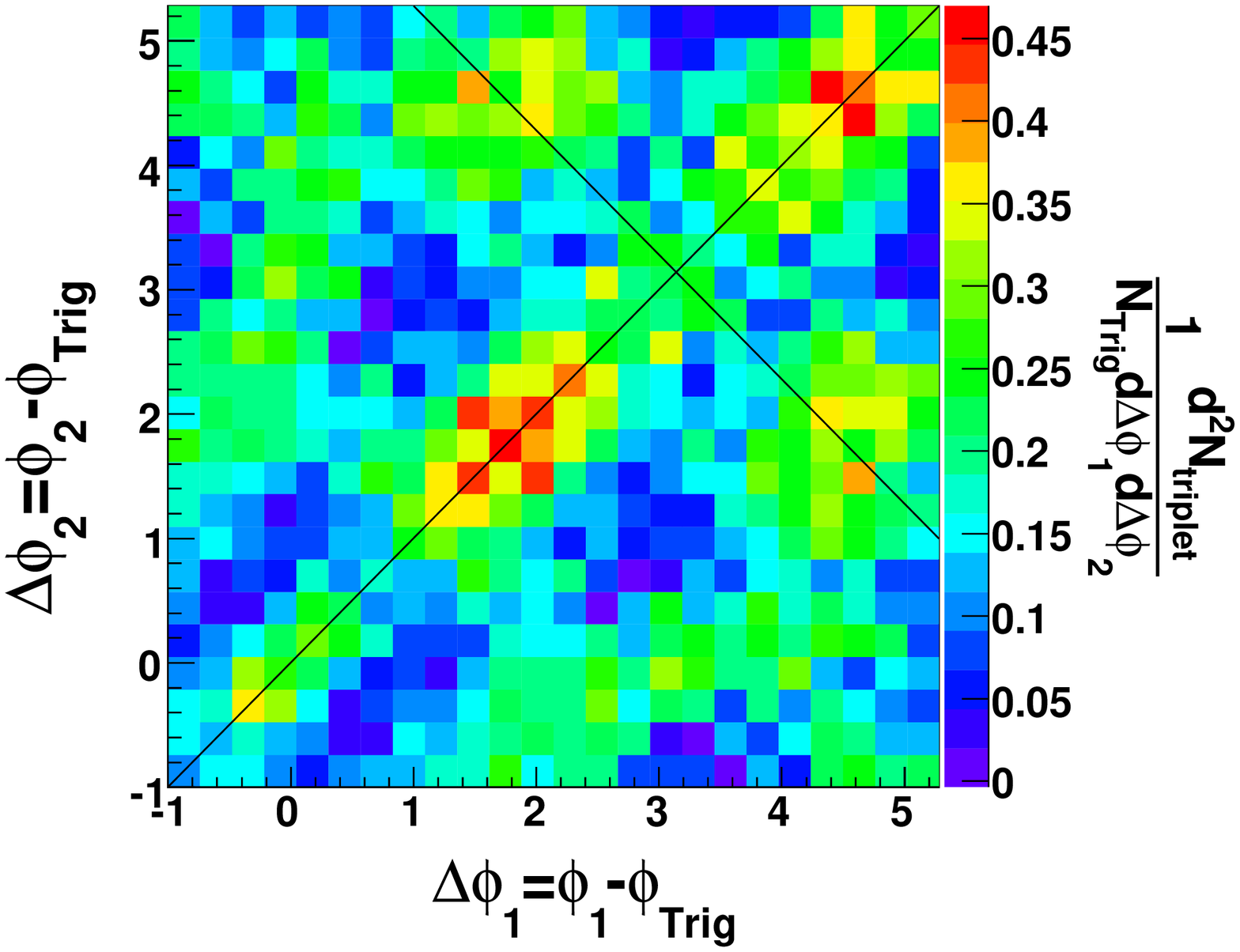}
\includegraphics[width=1.0\textwidth]{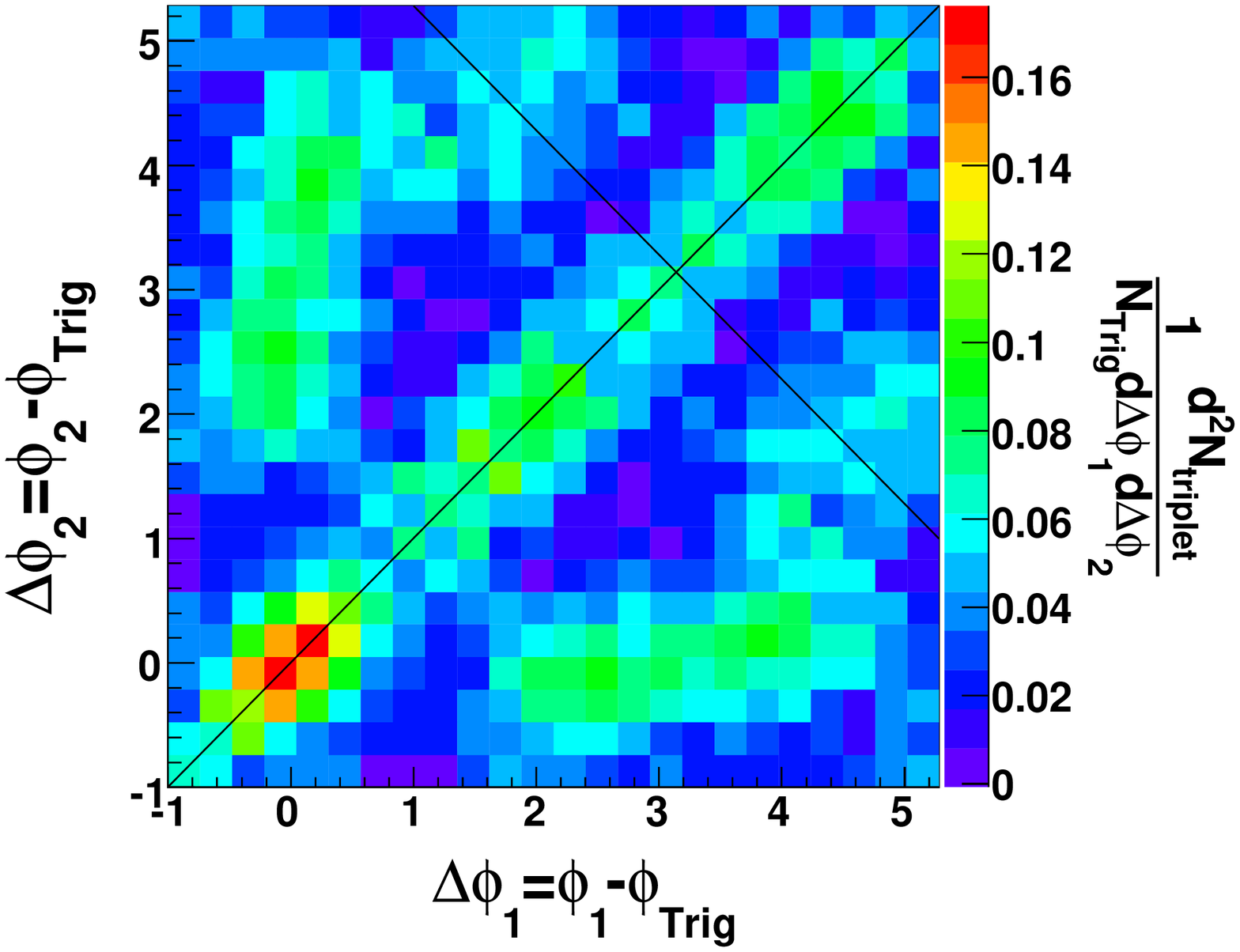}
\end{minipage}
\hfill
\begin{minipage}{0.32\textwidth}
\centering
\includegraphics[width=0.75\textwidth]{Plots/blank.eps}
\includegraphics[width=1.0\textwidth]{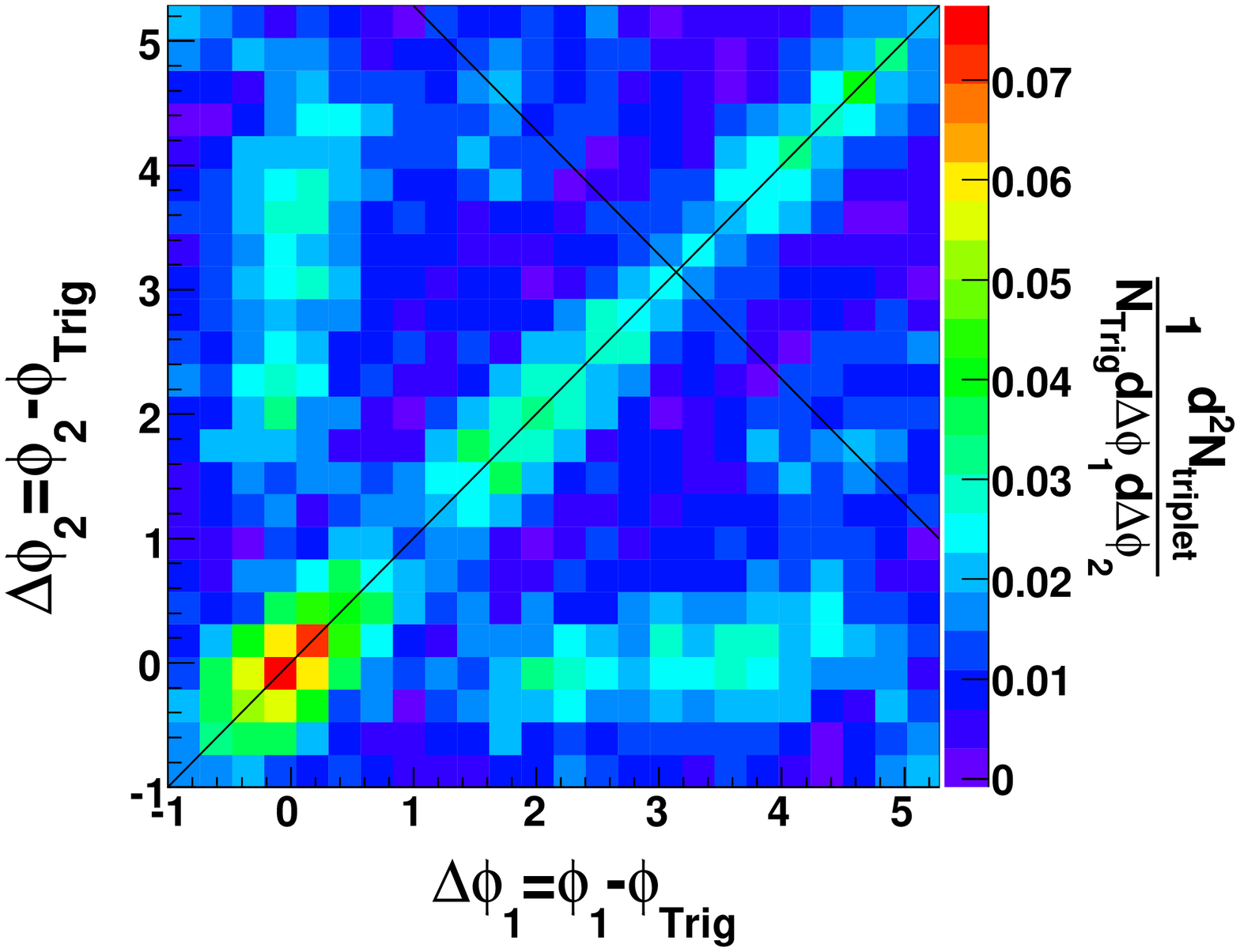}
\end{minipage}
\caption{Background subtracted 3-particle correlations for ZDC triggered 0-12\% most central Au+Au collisions at $\sqrt{s_{NN}}=200$ GeV/c.  Trigger particle is $3<p_{T}<4$ GeV/c.  Associated particle transverse momentum from left to right top to bottom is $0.5<p_{T}<0.75$, $0.75<p_{T}<1$, $1<p_{T}<1.5$, $1.5<p_{T}<2$, and $2<p_{T}<3$ GeV/c.}
\label{fig:pt1}
\end{figure}

\begin{figure}[htb]    
\hfill
\begin{minipage}{0.32\textwidth}
\centering
\includegraphics[width=1.0\textwidth]{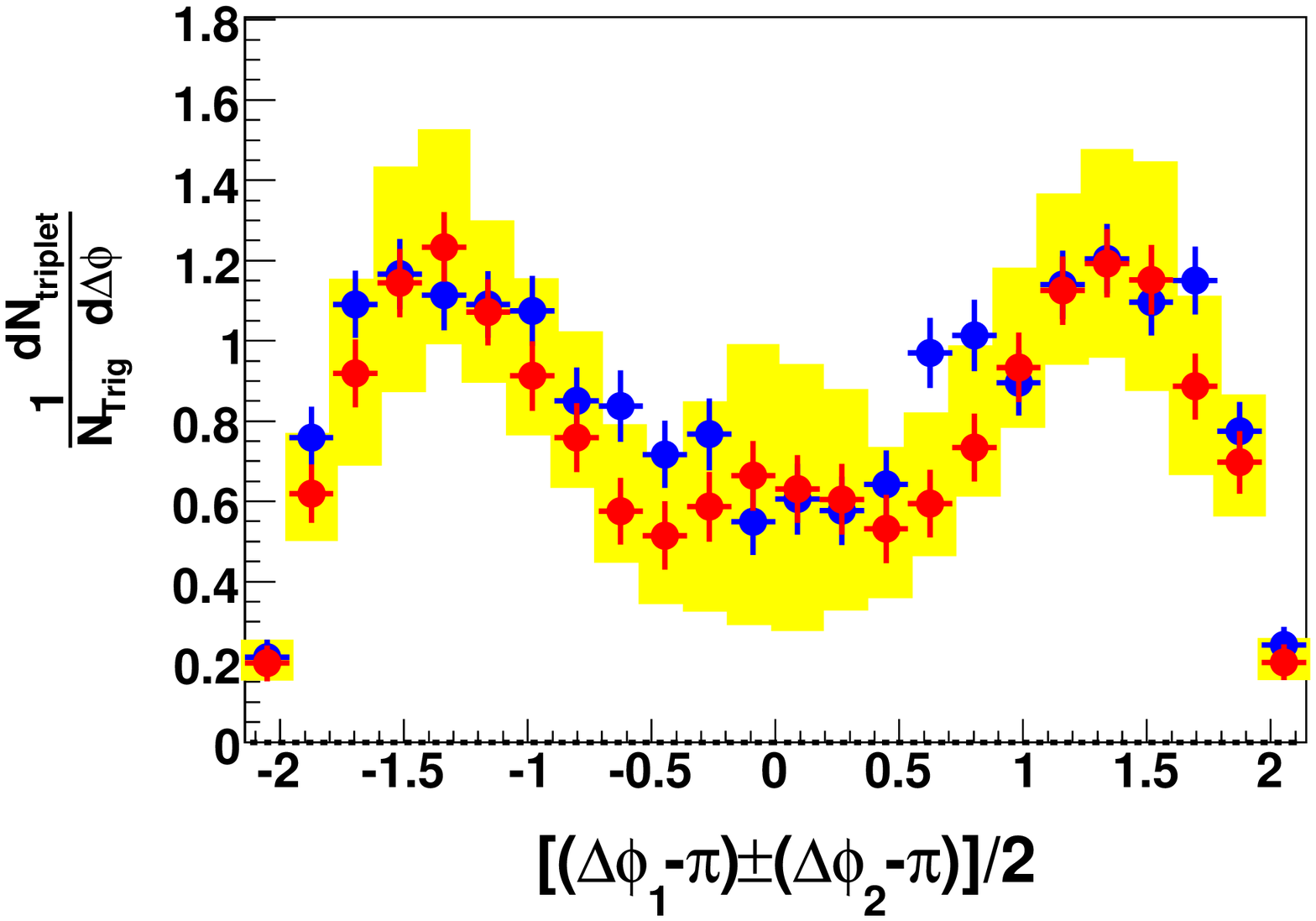}
\includegraphics[width=1.0\textwidth]{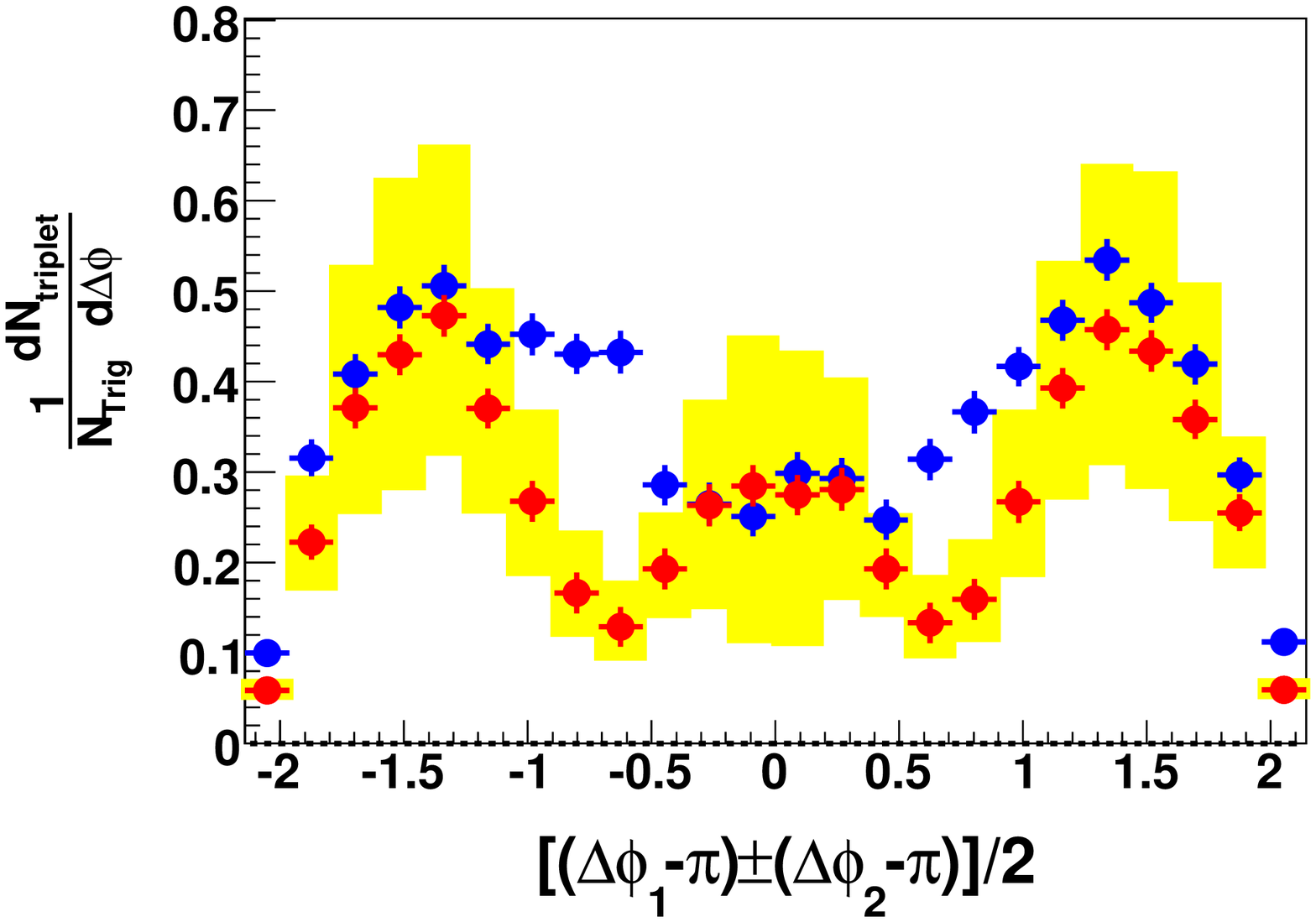}
\end{minipage}
\hfill
\begin{minipage}{0.32\textwidth}
\centering
\includegraphics[width=1.0\textwidth]{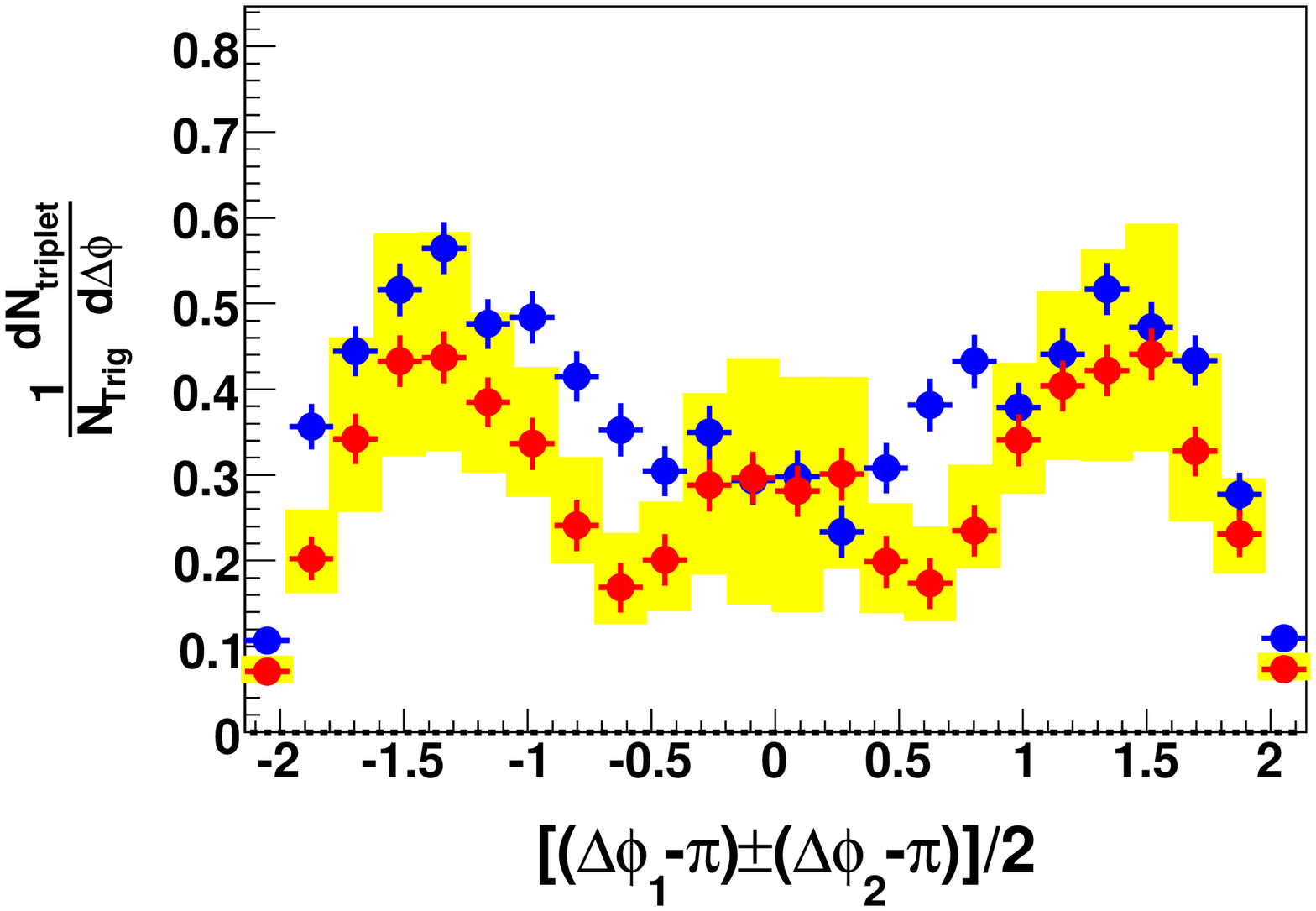}
\includegraphics[width=1.0\textwidth]{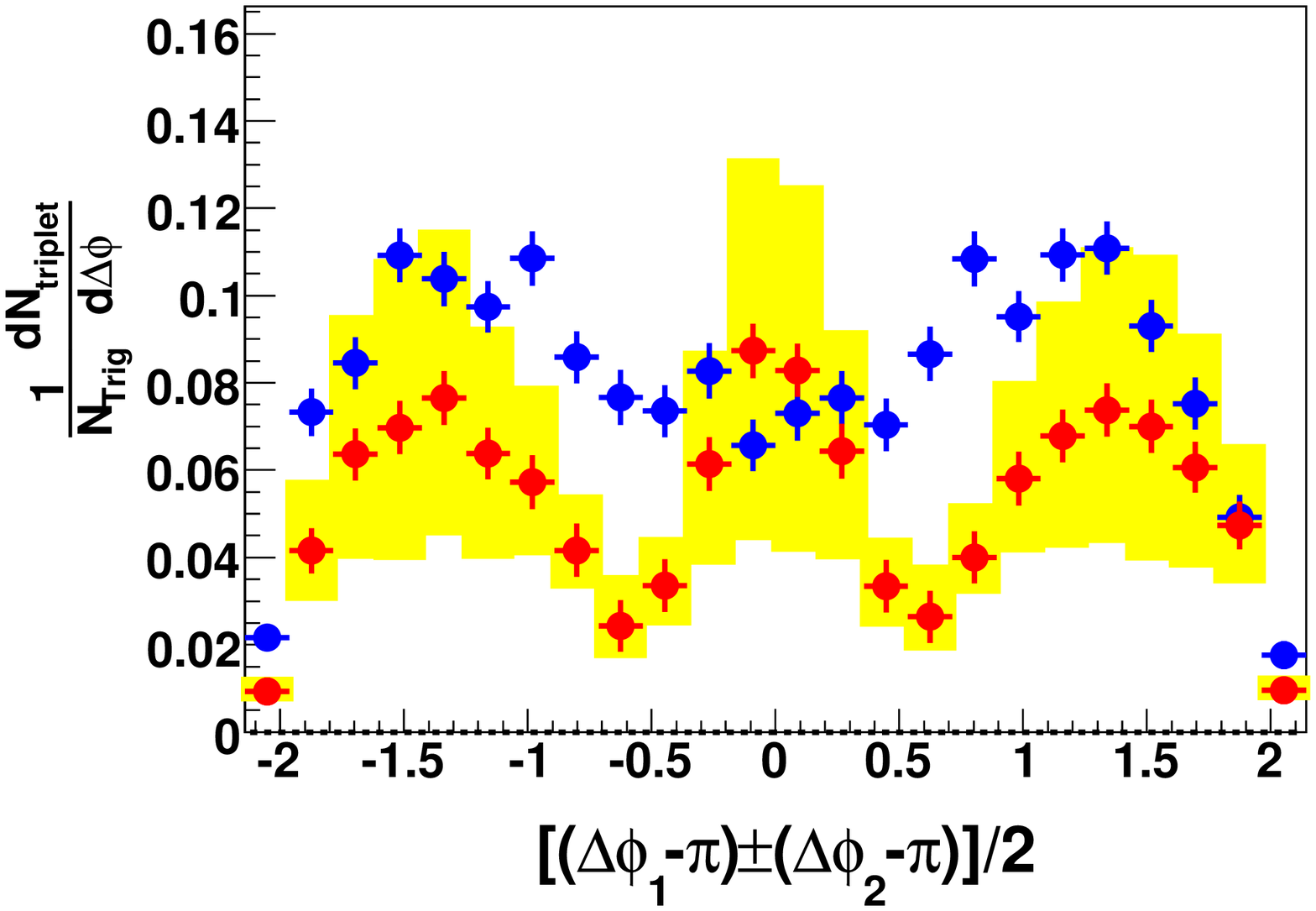}
\end{minipage}
\hfill
\begin{minipage}{0.32\textwidth}
\centering
\includegraphics[width=1.0\textwidth]{Plots/ProjStripBlank.eps}
\includegraphics[width=1.0\textwidth]{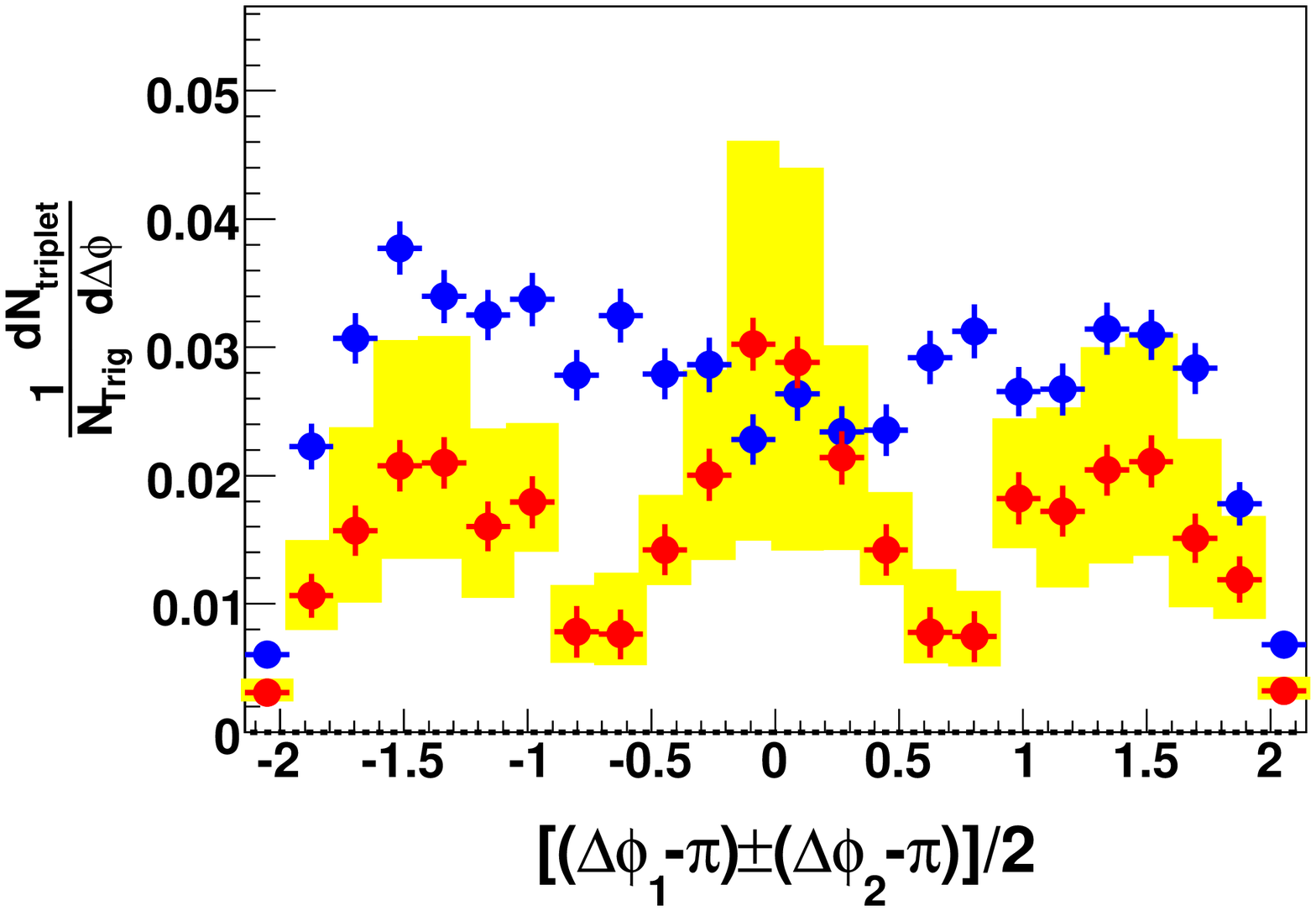}
\end{minipage}
\caption{Away-side projections of background subtracted 3-particle correlations for ZDC triggered 0-12\% most central Au+Au collisions at $\sqrt{s_{NN}}=200$ GeV/c.  The on-diagonal projection is in blue and off-diagonal projection in red.  Yellow bands represent the systematic error on the off-diagonal projection.  Trigger particle is $3<p_{T}<4$ GeV/c.  Associated particle transverse momentum from left to right top to bottom is $0.5<p_{T}<0.75$, $0.75<p_{T}<1$, $1<p_{T}<1.5$, $1.5<p_{T}<2$, and $2<p_{T}<3$ GeV/c  The projections are of strips of full width of 0.7 radians.}
\label{fig:pt3}
\end{figure}

\begin{figure}[htb]    
\hfill
\begin{minipage}{0.32\textwidth}
\centering
\includegraphics[width=1.0\textwidth]{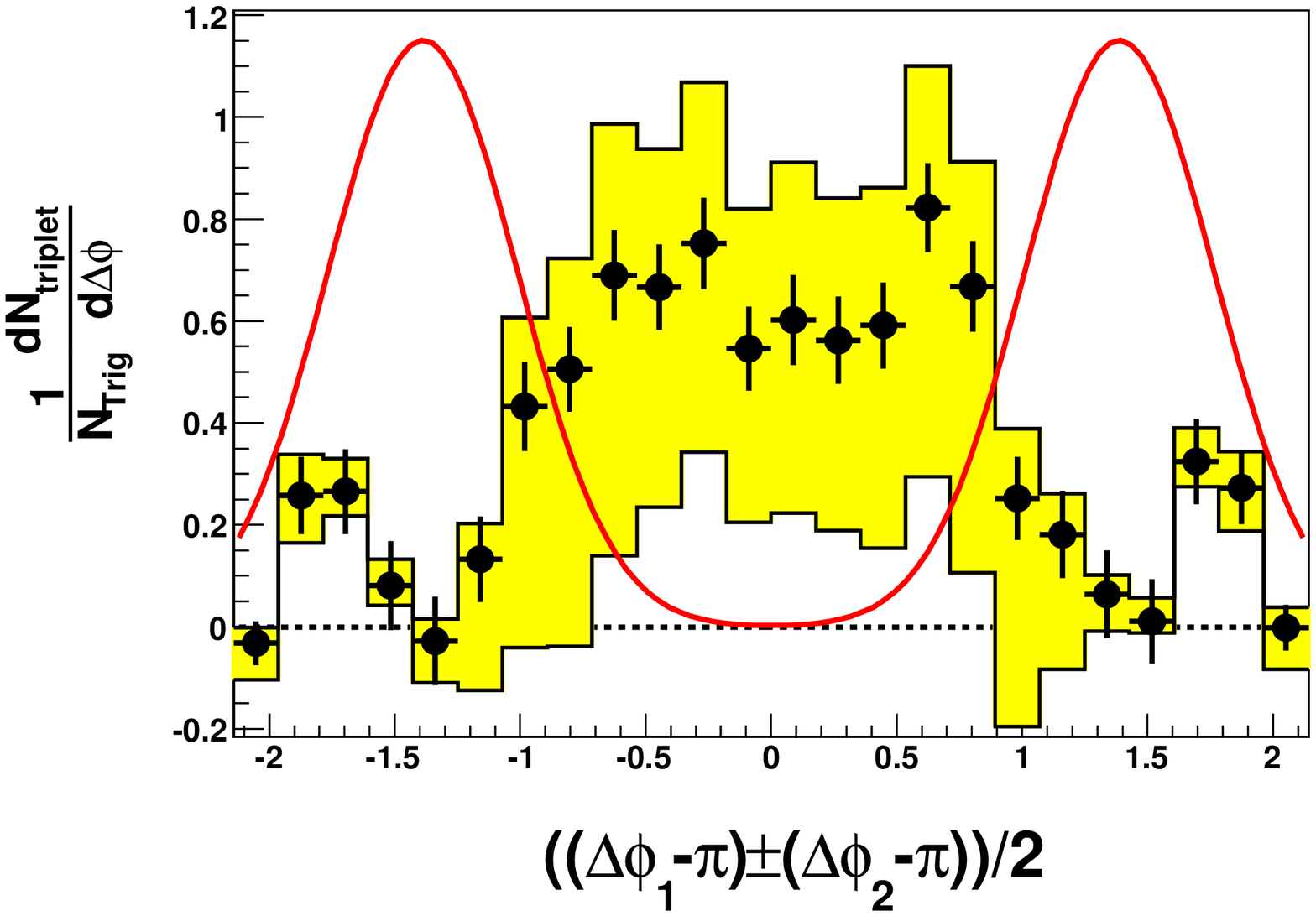}
\includegraphics[width=1.0\textwidth]{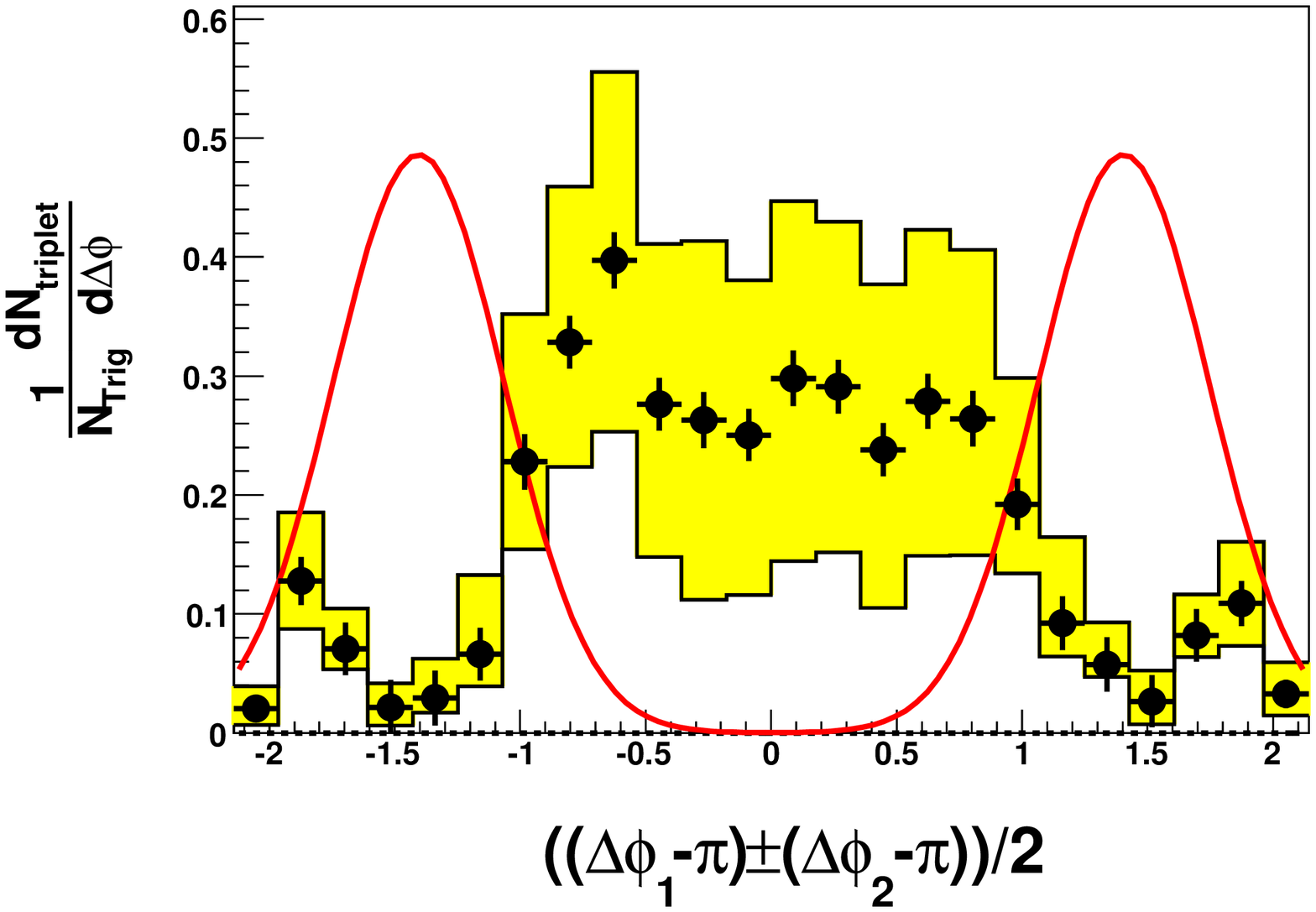}
\end{minipage}
\hfill
\begin{minipage}{0.32\textwidth}
\centering
\includegraphics[width=1.0\textwidth]{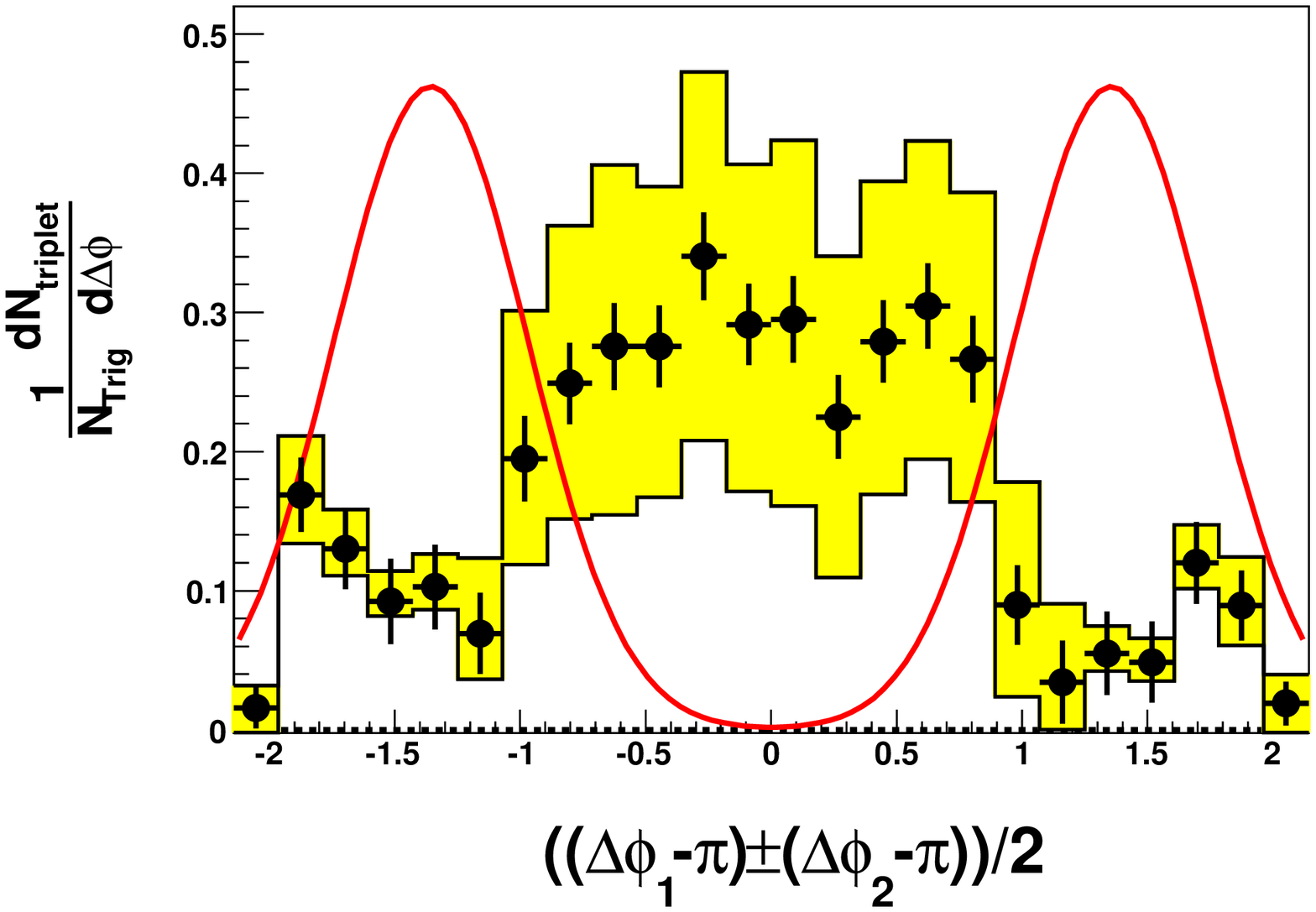}
\includegraphics[width=1.0\textwidth]{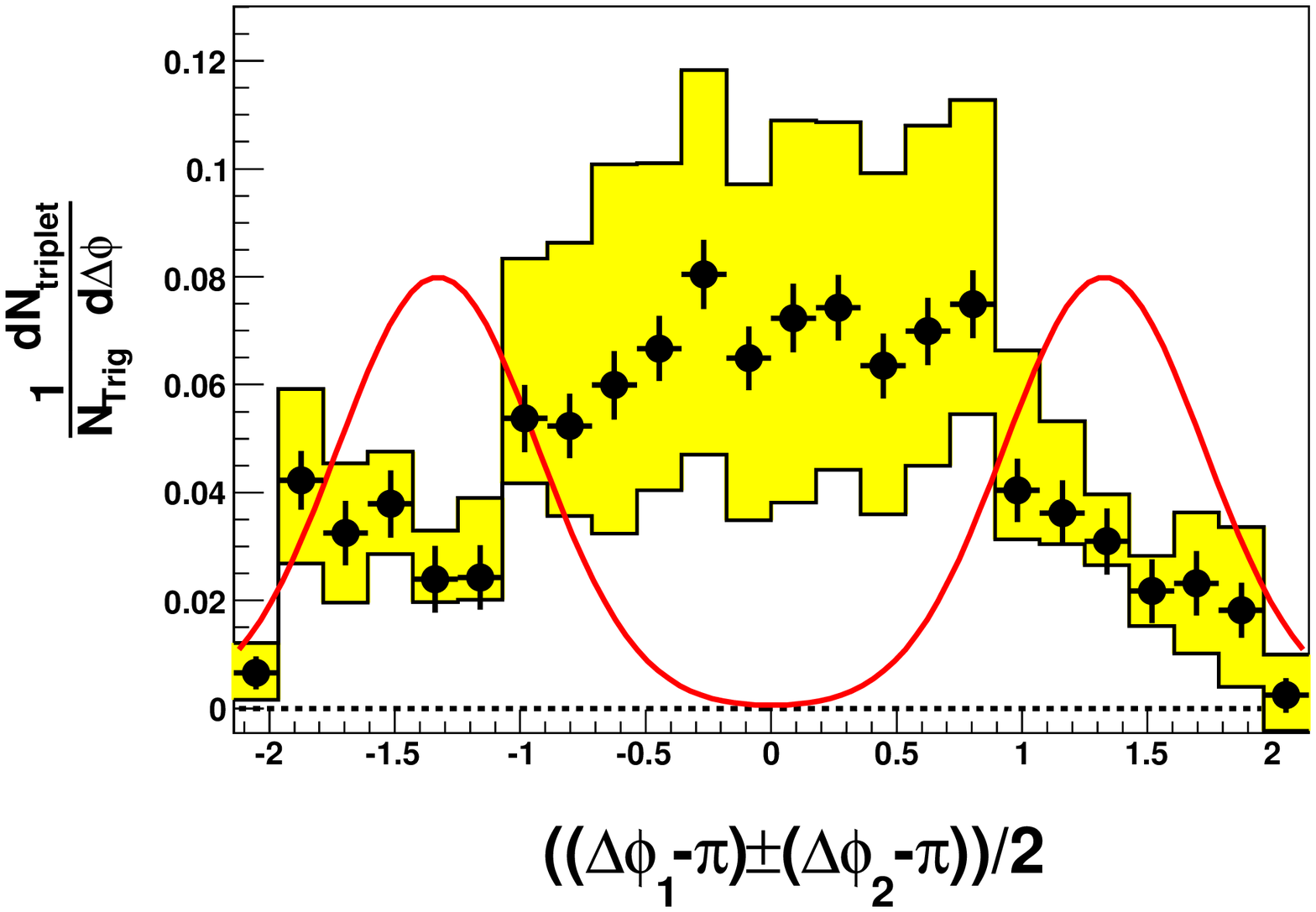}
\end{minipage}
\hfill
\begin{minipage}{0.32\textwidth}
\centering
\includegraphics[width=1.0\textwidth]{Plots/ProjStripBlank.eps}
\includegraphics[width=1.0\textwidth]{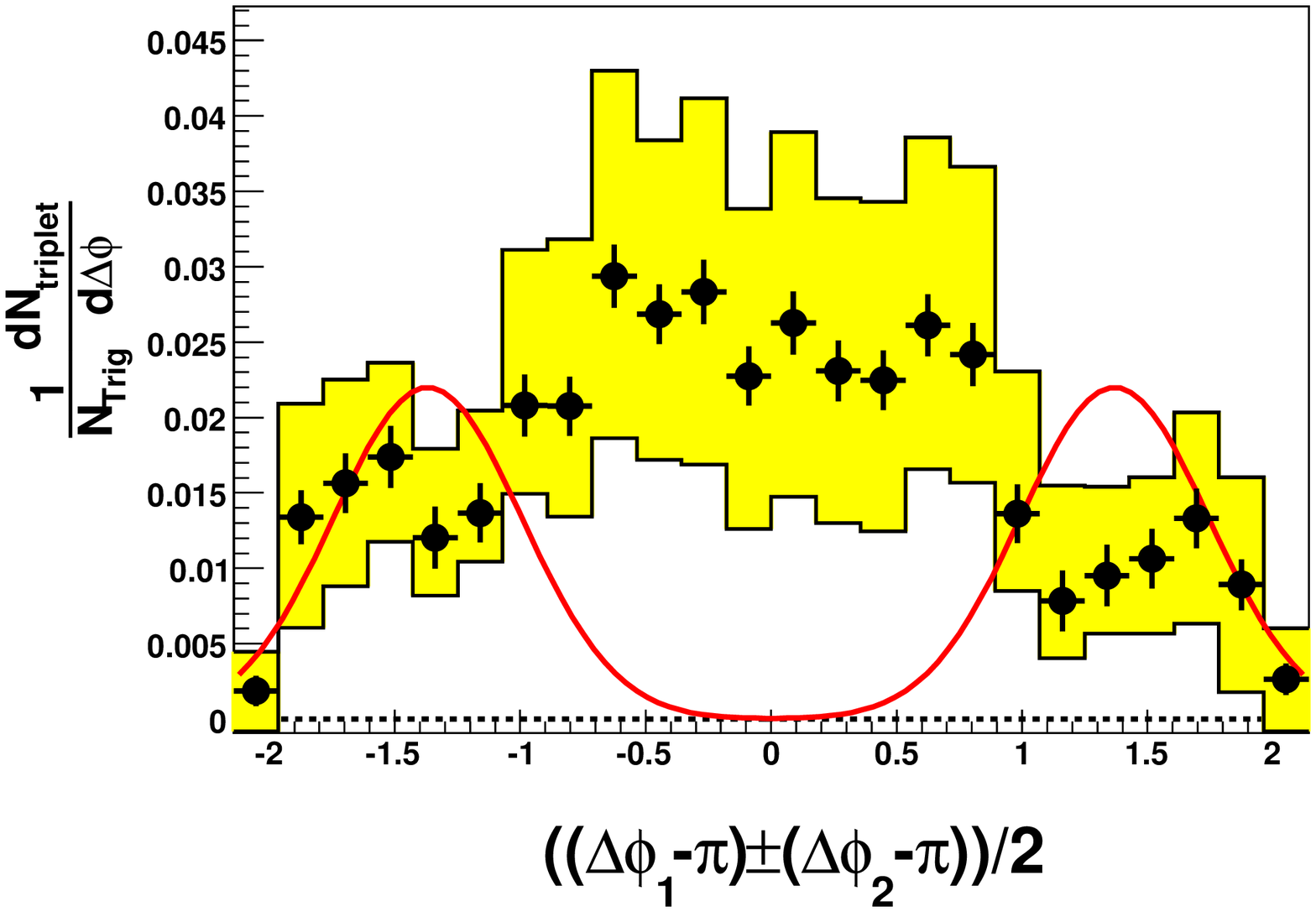}
\end{minipage}
\caption{On-diagonal projections with the side Gaussians from the off-diagonal projection removed.  The red curves are the Gaussians that were subtracted.  Yellow bands represent the systematic error on the off-diagonal projection.  Trigger particle is $3<p_{T}<4$ GeV/c.  Associated particle transverse momentum from left to right top to bottom is $0.5<p_{T}<0.75$, $0.75<p_{T}<1$, $1<p_{T}<1.5$, $1.5<p_{T}<2$, and $2<p_{T}<3$ GeV/c  The projections are of strips of full width of 0.7 radians.  Plots are for 0-12\% ZDC triggered Au+Au collisions at $\sqrt{s_{NN}}=200$ GeV/c.}
\label{fig:pt4}
\end{figure}

\begin{figure}[htb]
\centering
\includegraphics[width=0.6\textwidth]{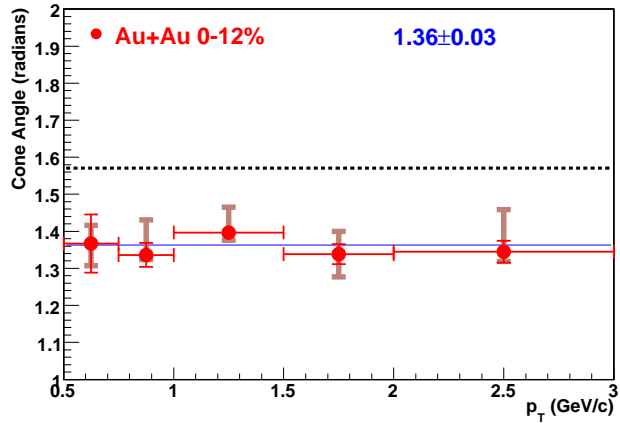}
\caption{Angles from fits to off-diagonal projections as a function of associated particle $p_{T}$ for ZDC triggered 0-12\% most central Au+Au collisions at $\sqrt{s_{NN}}=200$ GeV/c.  Trigger particle is $3<p_{T}<4$ GeV/c.  The solid errors are statistical and the shaded are systematic.  The blue line is a fit to a constant, yielding yielding $1.36\pm0.03$ (fit error using quadrature sum of statistical and systematic errors on the points).}
\label{fig:pt2}
\end{figure}

Figure~\ref{fig:pt1} shows background subtracted 3-particle correlations in ZDC triggered 0-12\% Au+Au collisions for different associated particle $p_{T}$ bins.  The $p_{T}$ dependent 3-particle correlation plots were not corrected for $\phi$ acceptance on a seperate basis for the different magnetic field settings.  However, the results for the $1<p_{T}<2$ GeV/c associated particles have very little dependence on whether or not this correction was done for both field setting together or seperately for the different fields.  The magnitude of the off-diagonal peaks decreases with increasing associated particle $p_{T}$ as expected.  Figure~\ref{fig:pt3} shows the on-diagonal and off-diagonal projections of the plots in Fig.~\ref{fig:pt1}.  The on-diagonal projections with the conical emission from the fit to the off-diagonal projection removed are shown in Fig.~\ref{fig:pt4}.  From Figures~\ref{fig:pt3} and~\ref{fig:pt4} it can be seen that the relative difference between the on-diagonal signal and the off-diagonal signal increases with associated particle $p_T$.  Figure~\ref{fig:pt2} shows the angle from the Gaussian fits to the off-diagonal projections as a function of associated particle $p_{T}$.  The angle is consistent with remaining constant as a function of associated particle $p_{T}$ as expectd for Mach-cone emission.  It is inconsistent with a sharply decreasing angle as a function of associated particle $p_{T}$ as predicated by simple \v{C}erenkov radiation\cite{cerenkov2}.

\subsection{Widths}

\begin{figure}[htb]
\hfill
\begin{minipage}[t]{0.49\textwidth}
\centering
\includegraphics[width=1.0\textwidth]{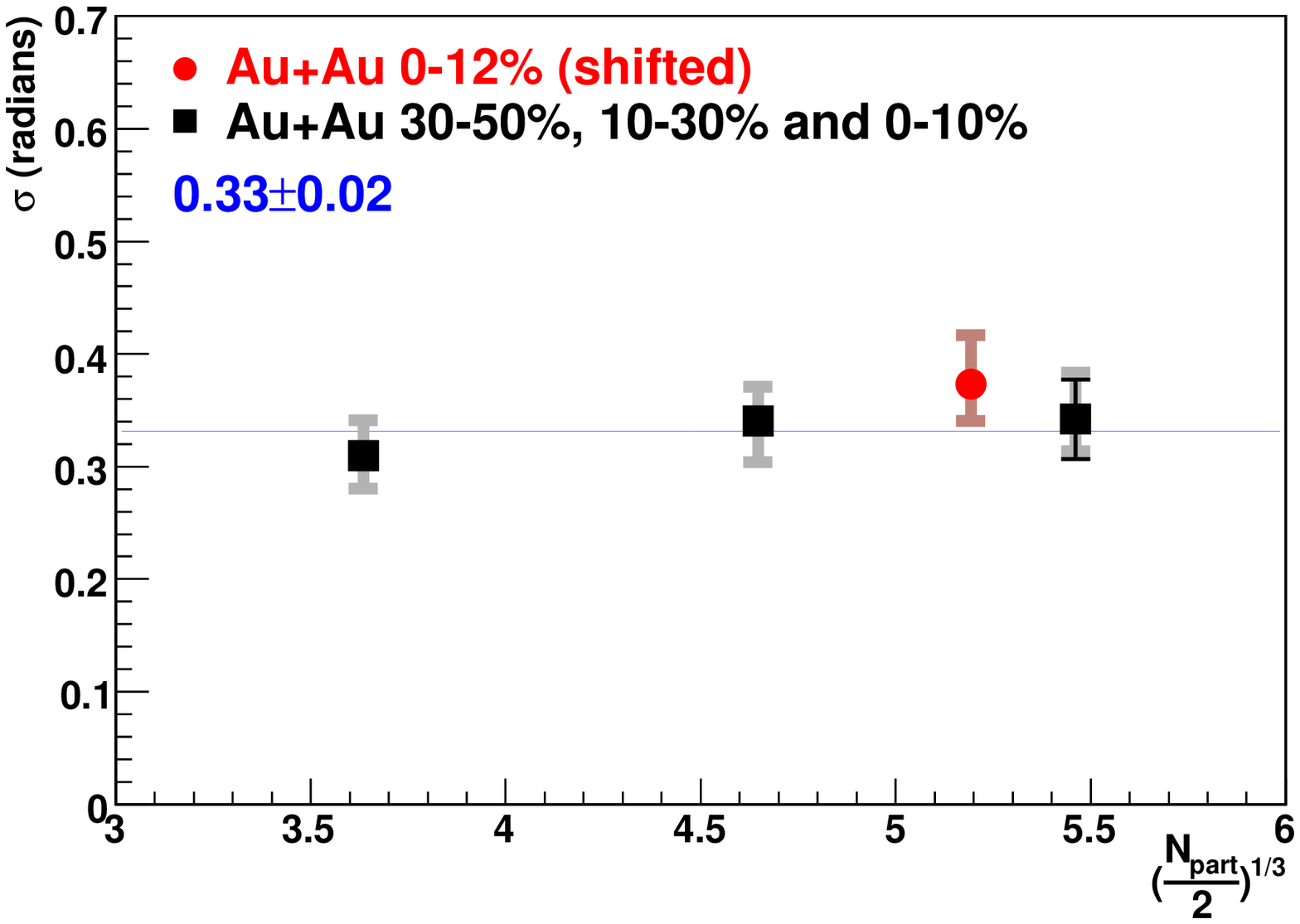}
\includegraphics[width=1.0\textwidth]{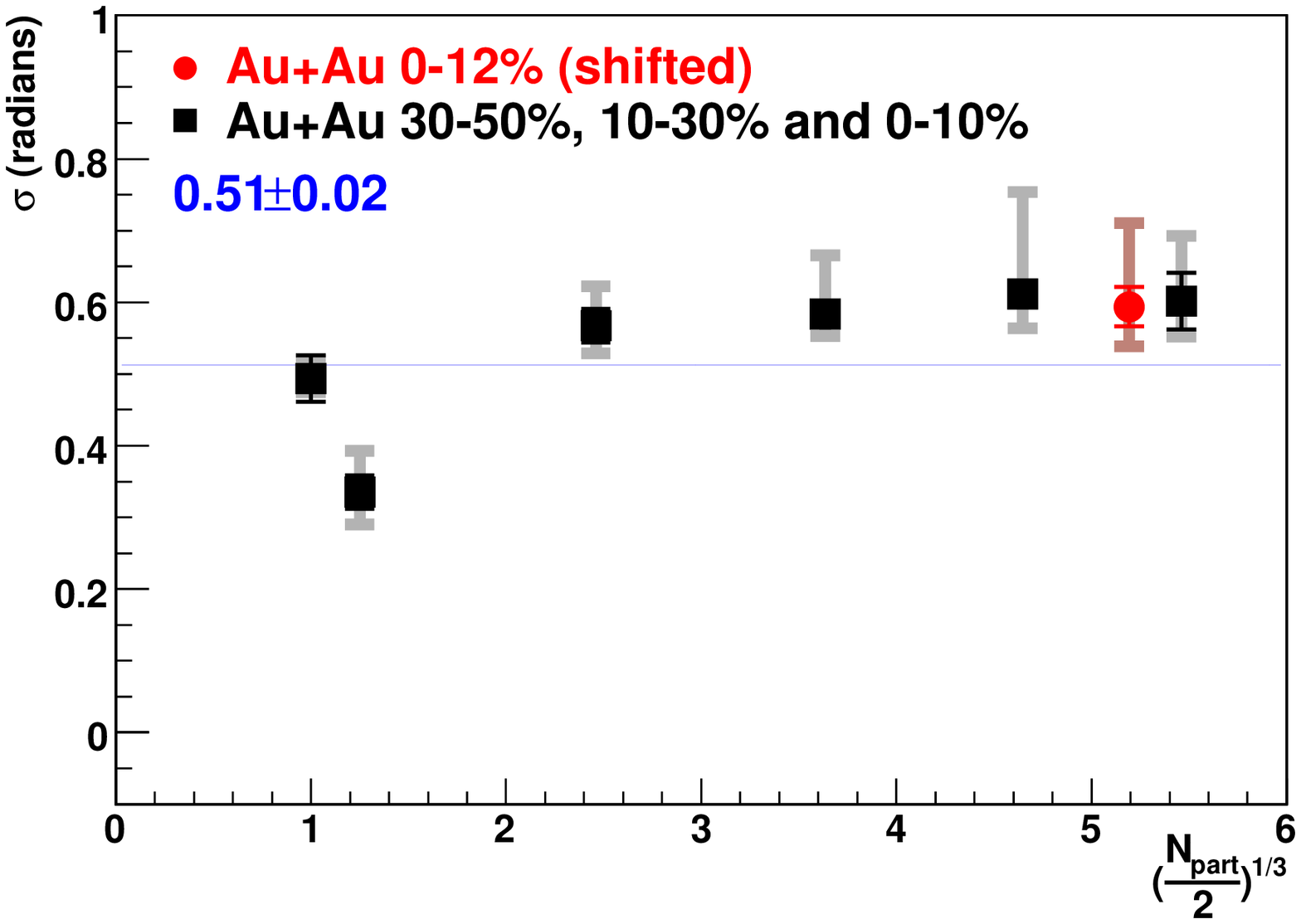}
\end{minipage}
\hfill
\begin{minipage}[t]{0.49\textwidth}
\centering
\includegraphics[width=1.0\textwidth]{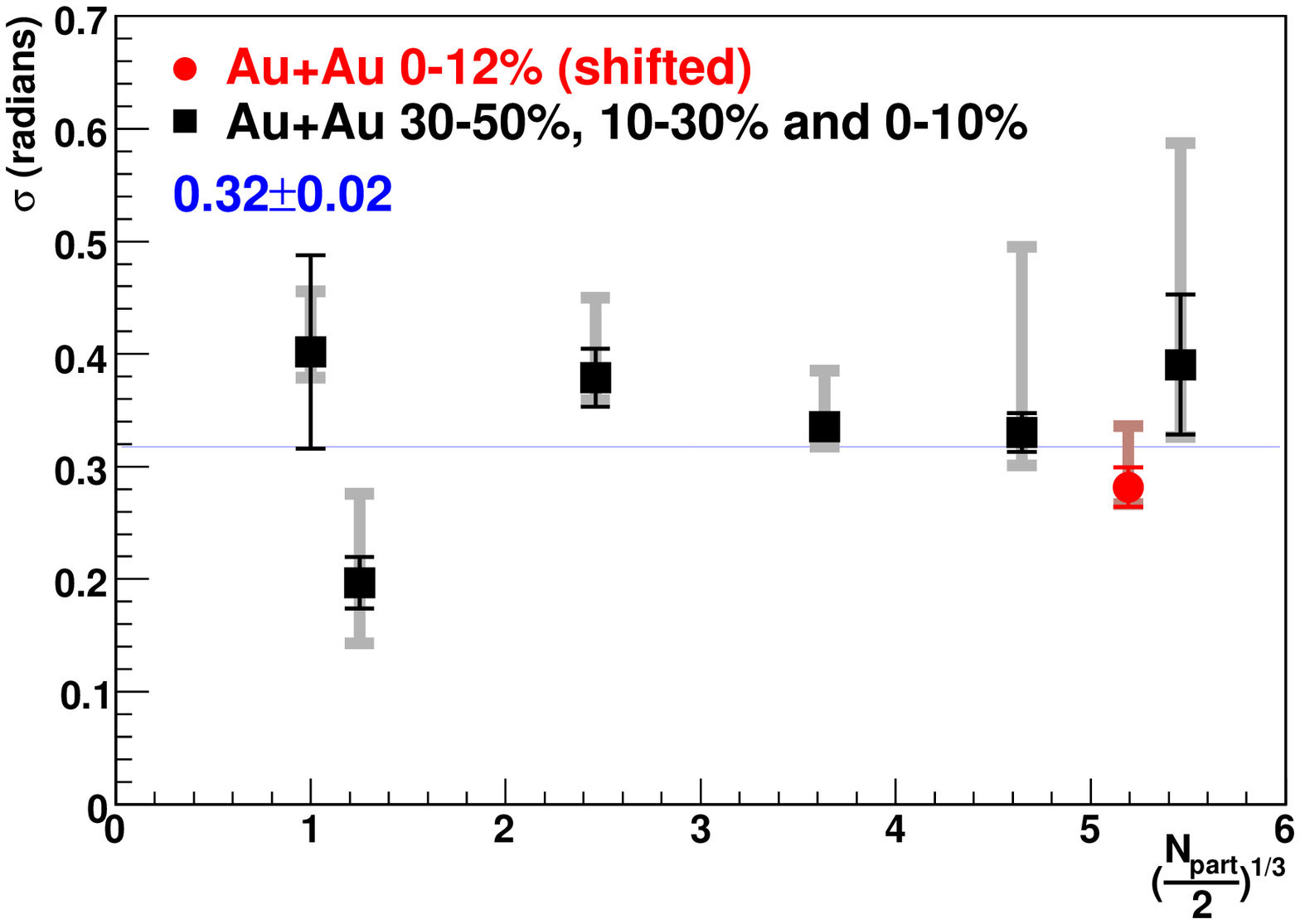}
\includegraphics[width=1.0\textwidth]{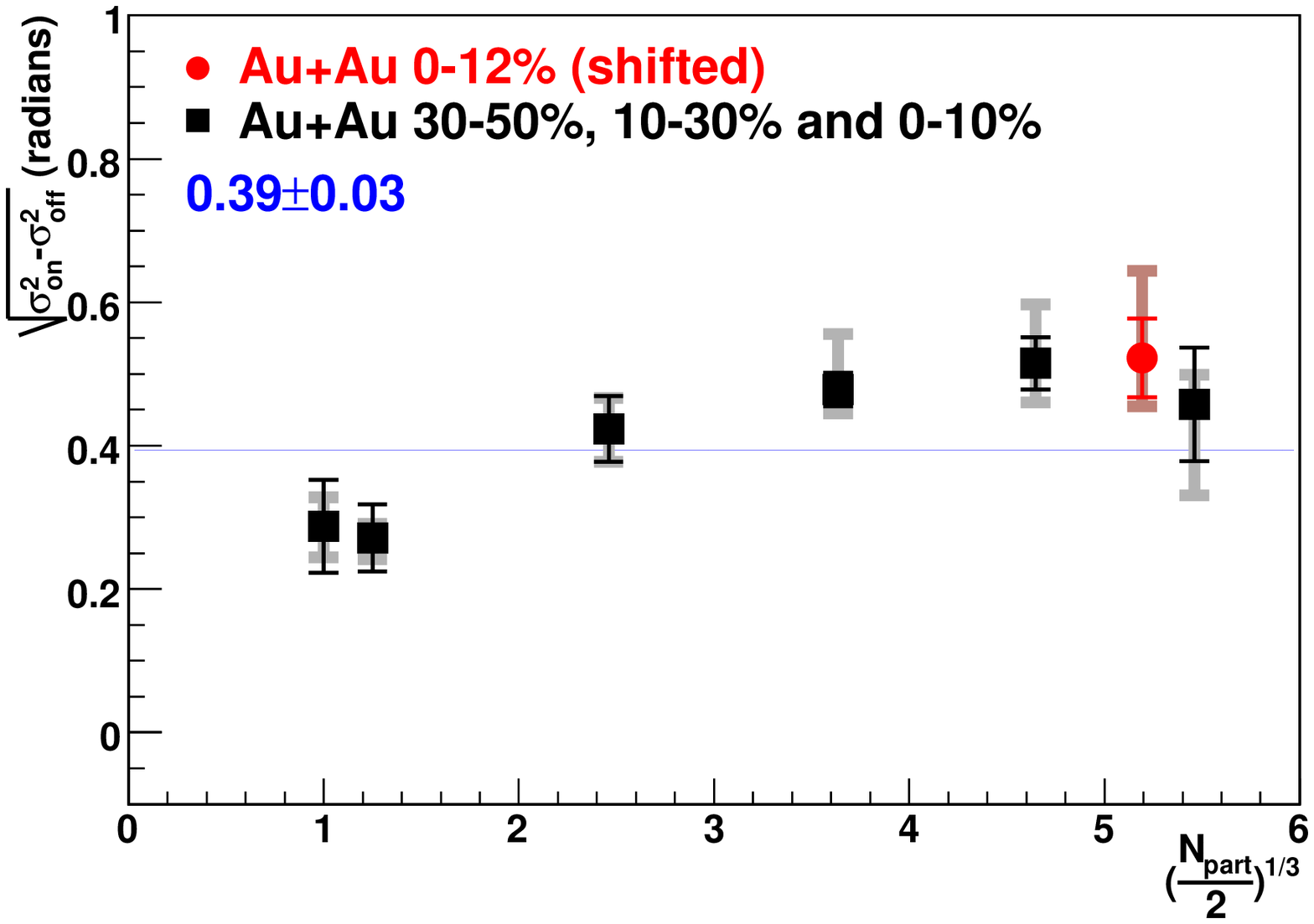}
\end{minipage}
\caption{Centrality dependence of the width ($\sigma$ of Gaussians) from fits to away-side projections of the background subtracted 3-particle correlations.  Top Left:  Width of the side Gaussians in the off-diagonal projection.  Top Right:  Width of the central peak in the off-diagonal projection.  Bottom Left:  Width of the central peak in the on-diagonal projection after subtraction of the side Gaussians from the fit to the off-diagonal projection.  Bottom Right:  Difference in the widths of the central peaks done as $\sqrt{\sigma_{on-diagonal}^{2}-\sigma_{off-diagonal}^{2}}$.  The numbers indicate the constant fit results.  The solid errors are statistical and the shaded are systematic.}
\label{fig:centw1}
\end{figure}

The widths of the conical emission peaks are shown as a function of centrality in Fig.~\ref{fig:centw1}, top left.  The widths are constant with being independent of centrality or increasing with centrality.  An increase in the width with centrality could point towards more dispersion of the cone through more interactions with the medium in central collisions.  The top right and bottom left plots show the widths of the central away-side peak in the off-diagonal and on-diagonal projections, respectively.  There is not significant centrality dependence of central peak in the off-diagonal projection.  The width of the central peak in the on-diagonal projections is the width of the peak shown in Fig.~\ref{fig:projdiff}.  This peak appears to increase in width as the centrality increases.  The interpretation of the broadening is complicated as previously discussed and could come from many different physics mechanisms.  The bottom right plot shows the difference in the widths of the on-diagonal projection and the off-diagonal projection.  This difference increases with centrality.  Again the interpretation is complicated but the broadening of the on-diagonal central peak relative to the off-diagonal central peak could be due to interactions between away-side jet and the medium.

\begin{figure}[htb]
\hfill
\begin{minipage}[t]{0.49\textwidth}
\centering
\includegraphics[width=1.0\textwidth]{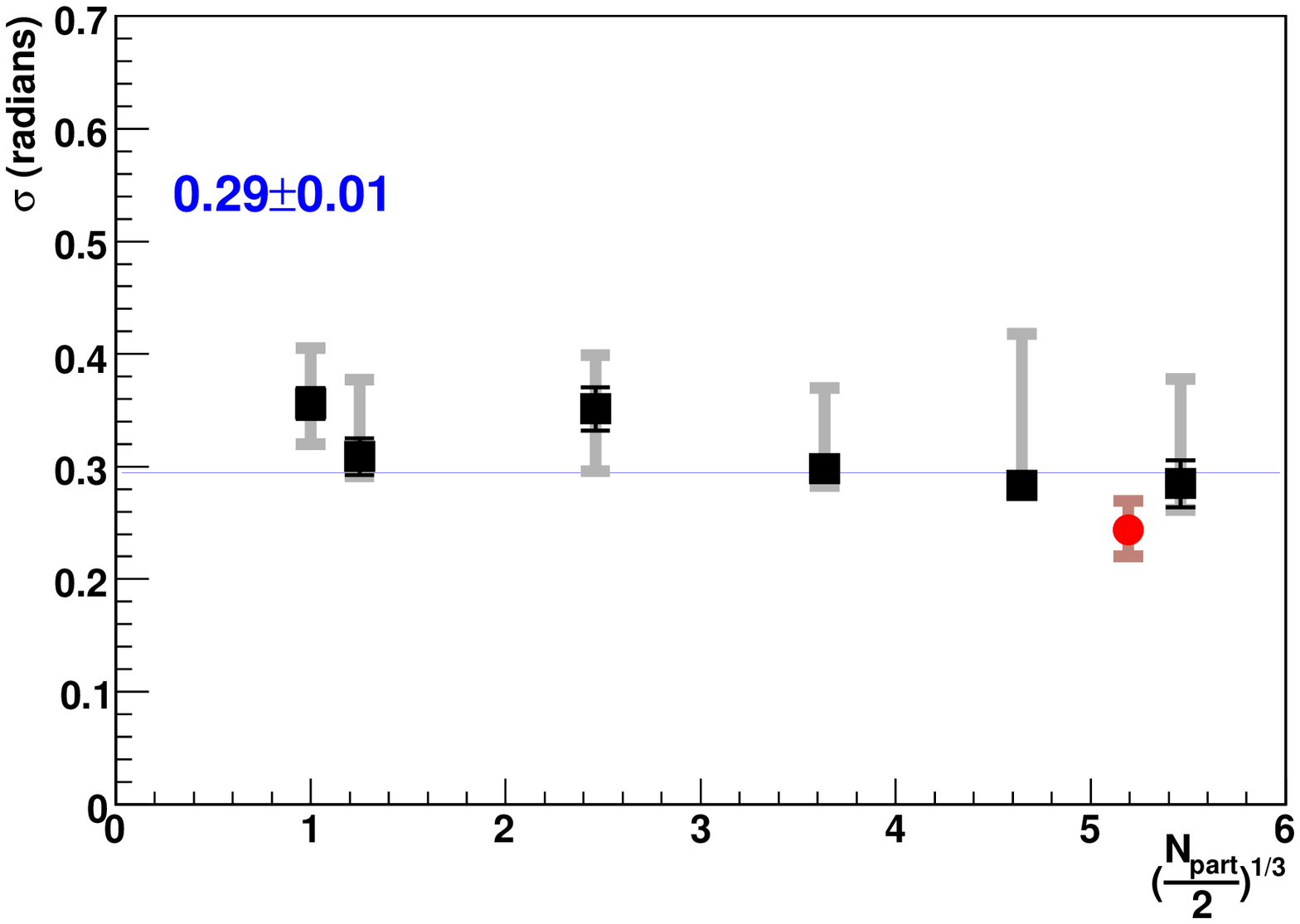}
\includegraphics[width=1.0\textwidth]{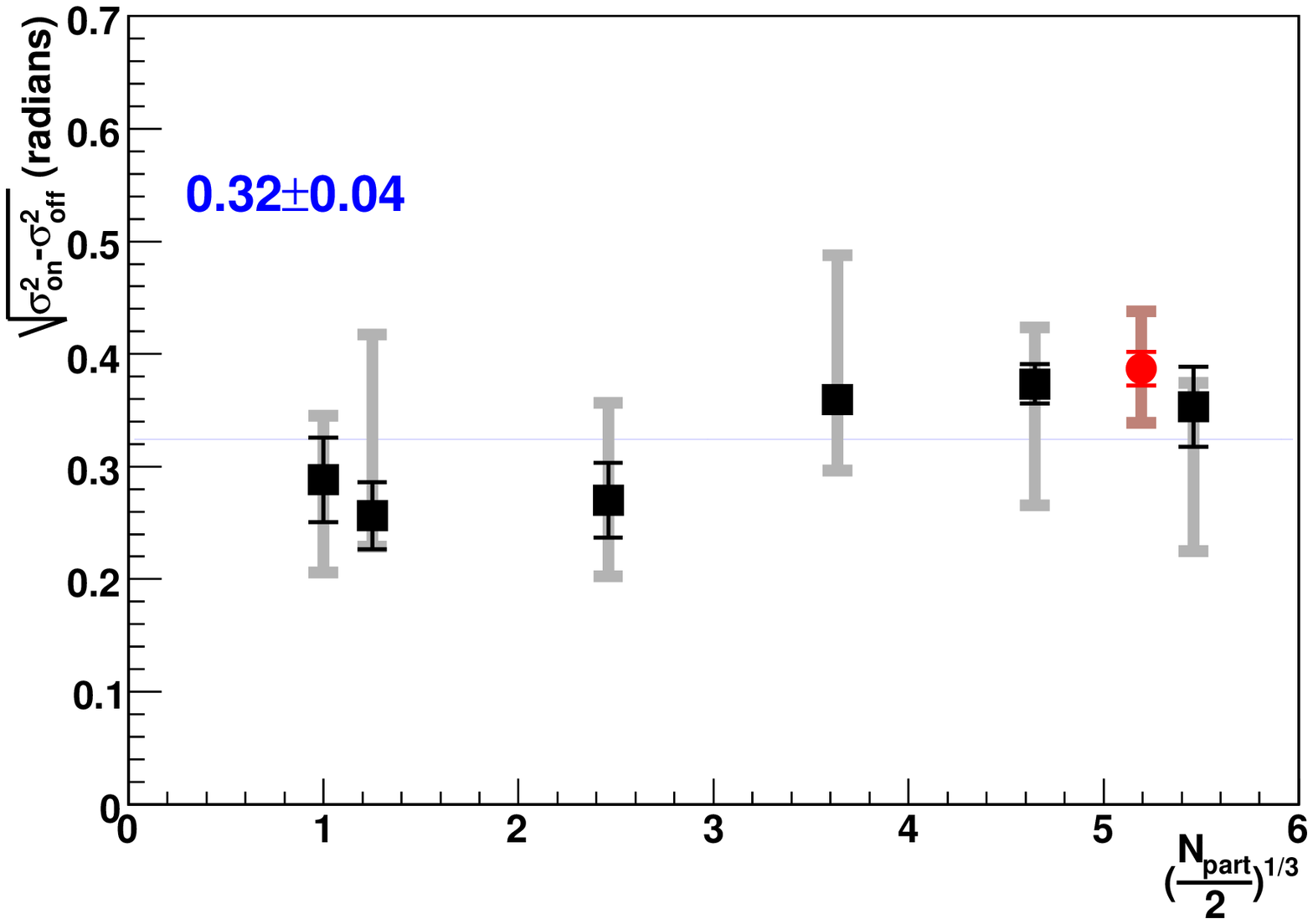}
\end{minipage}
\hfill
\begin{minipage}[t]{0.49\textwidth}
\centering
\includegraphics[width=1.0\textwidth]{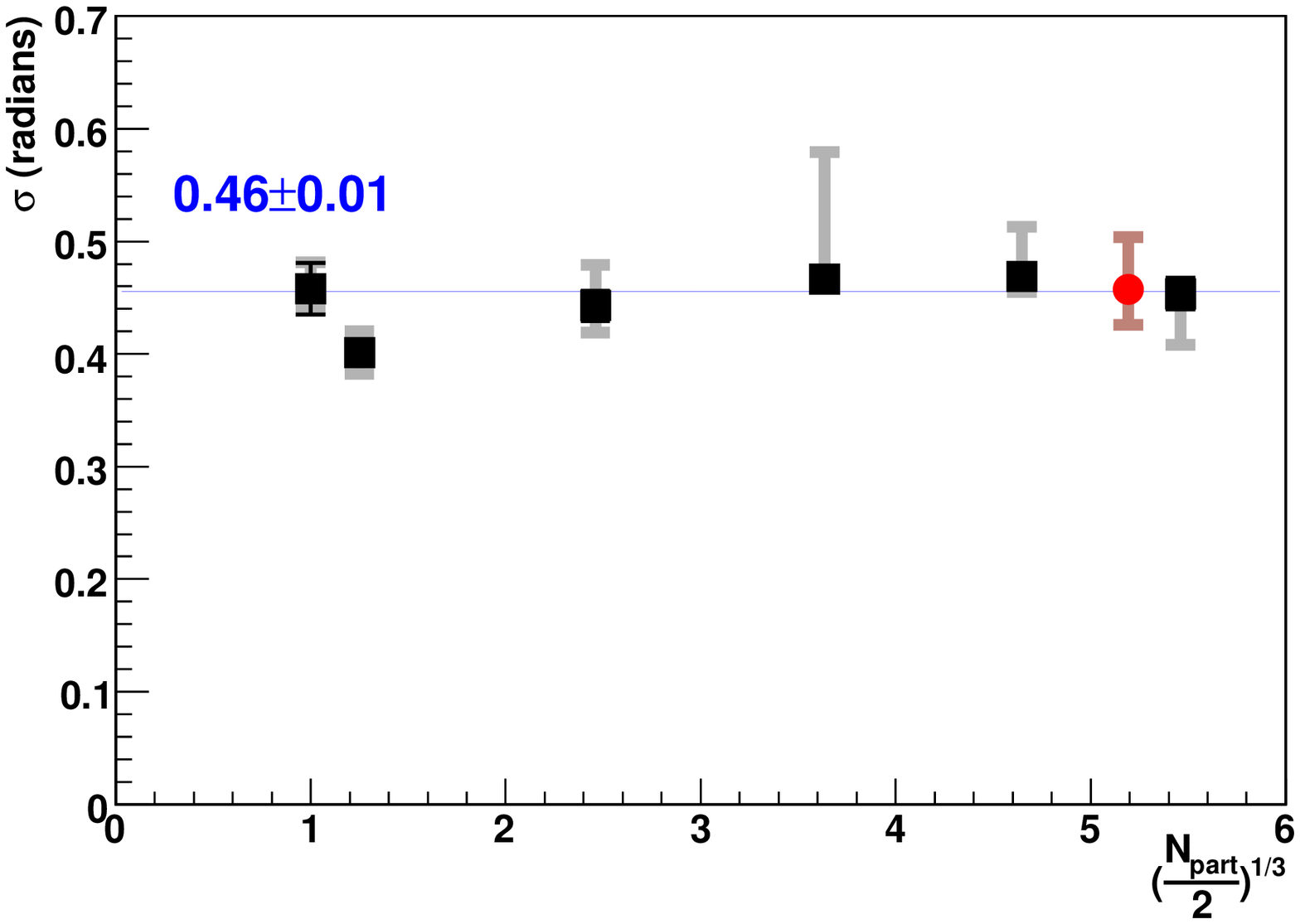}
\includegraphics[width=1.0\textwidth]{Plots/ProjStripBlank.eps}
\end{minipage}
\caption{Centrality dependence of the width ($\sigma$ of Gaussians) from fits to near-side projections of the background subtracted 3-particle correlations.  Top Left:  Off-diagonal peak width.  Top Right:  On-diagonal peak width.  Bottom:  Difference in the peak widths as $\sqrt{\sigma_{on-diagonal}^{2}-\sigma_{off-diagonal}^{2}}$.   The numbers indicate the constant fit results.  The solid errors are statistical and the shaded are systematic.}
\label{fig:centw2}
\end{figure}

Fig.~\ref{fig:centw2} shows the centrality dependence of widths of the near-side peak, in both the off-diagonal and on-diagonal projection, top left and right, respectively.  The near-side peak width is consistent with being independent of centrality.  This is not unexpected if most of the trigger particles come from surface emission.  The on-diagonal width is wider than the off-diagonal width because we are using the trigger particle as a proxy for the jet-axis.  If the trigger particle is to one side of the jet-axis than both associated particles are more likely to be to one side of the trigger particle (which puts them on-diagonal) and more likely to be further from the trigger particle than the jet-axis (which results in the broadening).  The difference in the widths is shown in the bottom panel.  This difference is consistent with no centrality dependence.

\begin{figure}[htb]
\hfill
\begin{minipage}[t]{0.49\textwidth}
\centering
\includegraphics[width=1.0\textwidth]{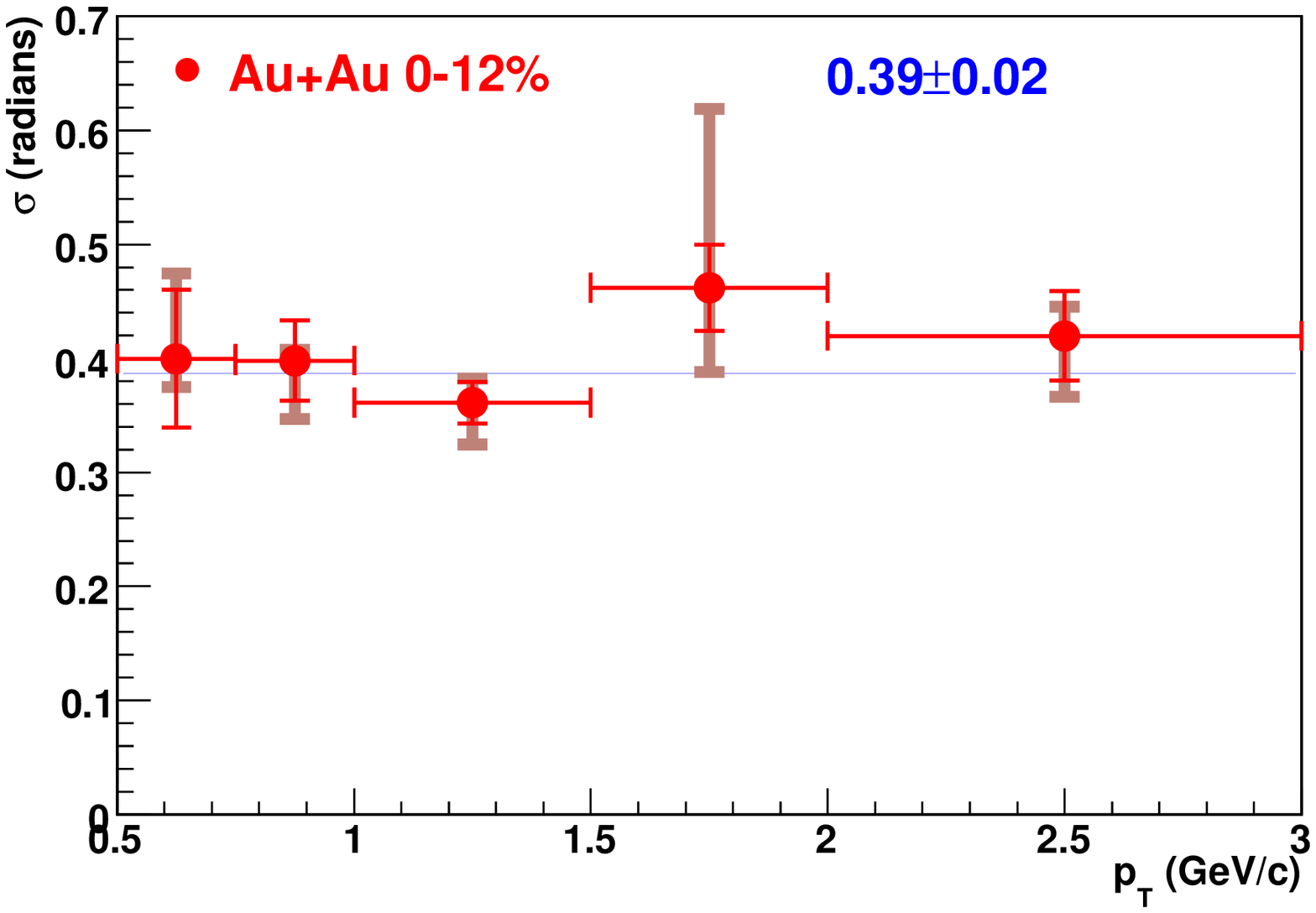}
\end{minipage}
\hfill
\begin{minipage}[t]{0.49\textwidth}
\centering
\includegraphics[width=1.0\textwidth]{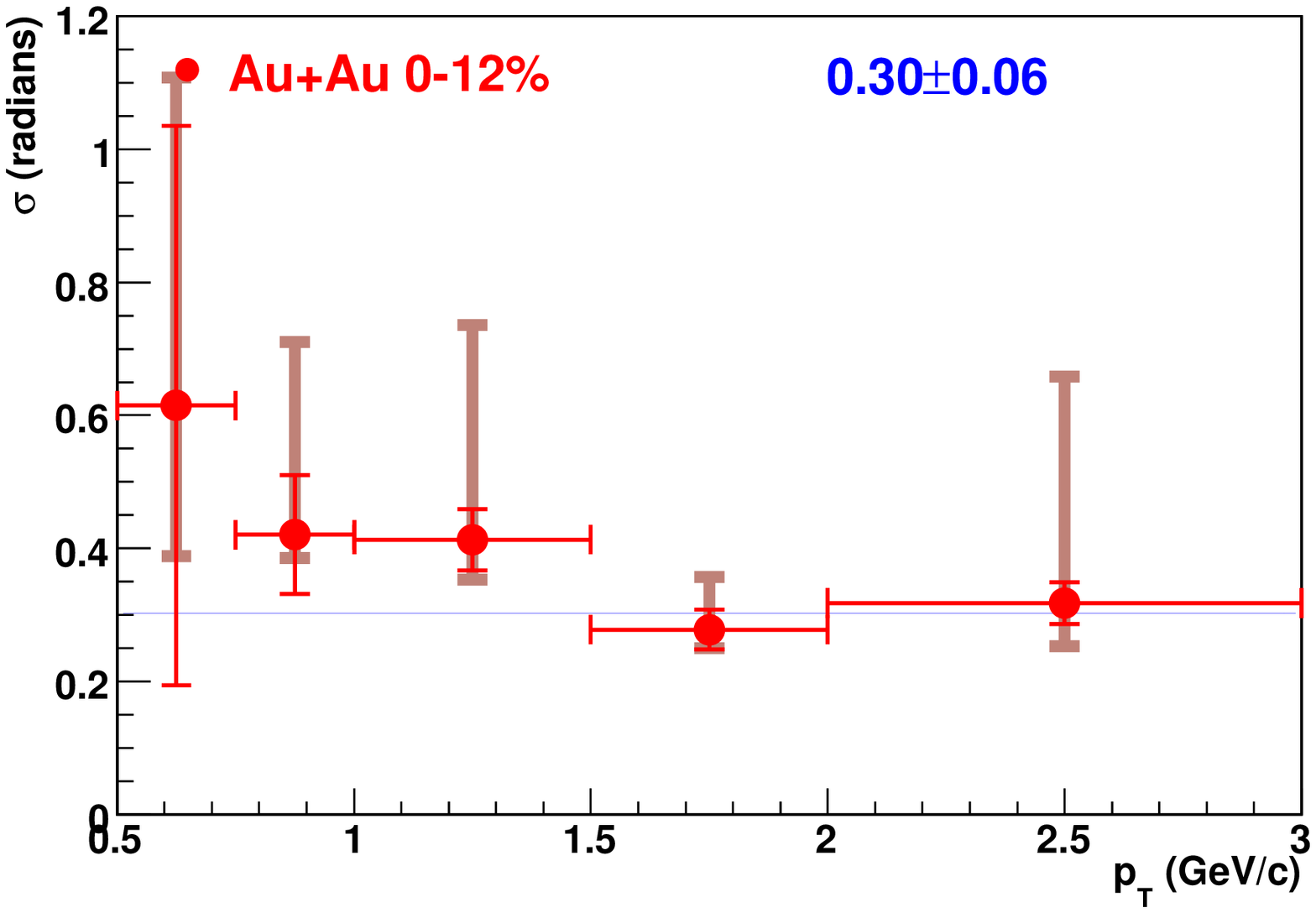}
\end{minipage}
\caption{Width ($\sigma$ of Gaussians) of away-side peaks from the off-diagonal projections of the background subtracted 3-particle correlations as a function of associated particle $p_T$.  Left:  Width of the side Gaussians.  Right:  Width of the central Gaussian.  Plots are for 0-12\% ZDC triggered Au+Au at $\sqrt{s_{NN}}=200$ GeV/c with $3<p_{T}^{Trig}<4$ GeV/c.    The numbers indicate the constant fit results.  The solid errors are statistical and the shaded are systematic.}
\label{fig:ptw1}
\end{figure}

Figure~\ref{fig:ptw1} shows the widths of the away-side peaks as a function of associated particle $p_T$.  Although the errors are large, the conical emission peaks (left) seem to have associated particle $p_T$ independent widths.  The width of the central peak (right) is consistent with decreasing with associated $p_T$.  

\begin{figure}[htb]
\hfill
\begin{minipage}[t]{0.49\textwidth}
\centering
\includegraphics[width=1.0\textwidth]{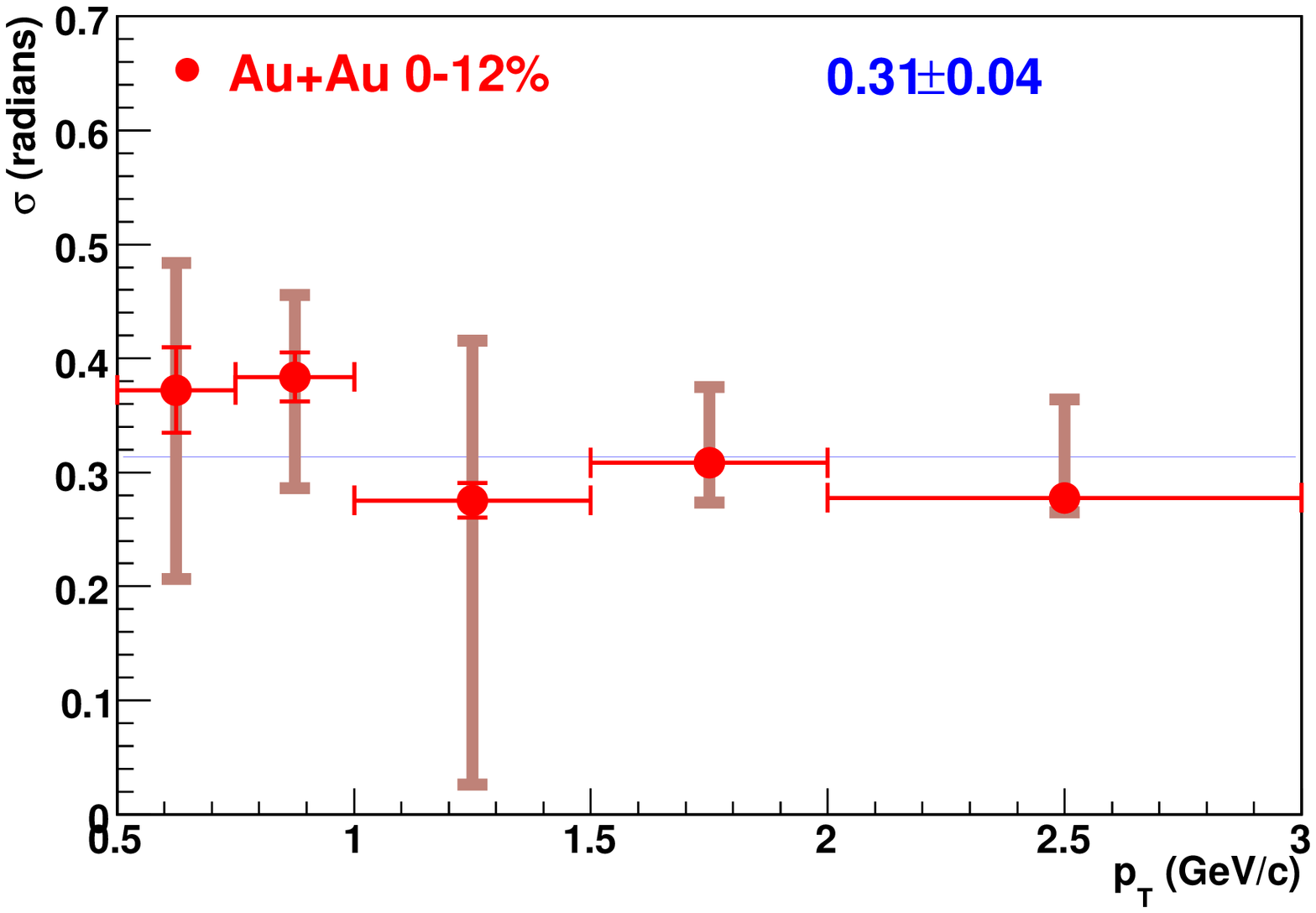}
\includegraphics[width=1.0\textwidth]{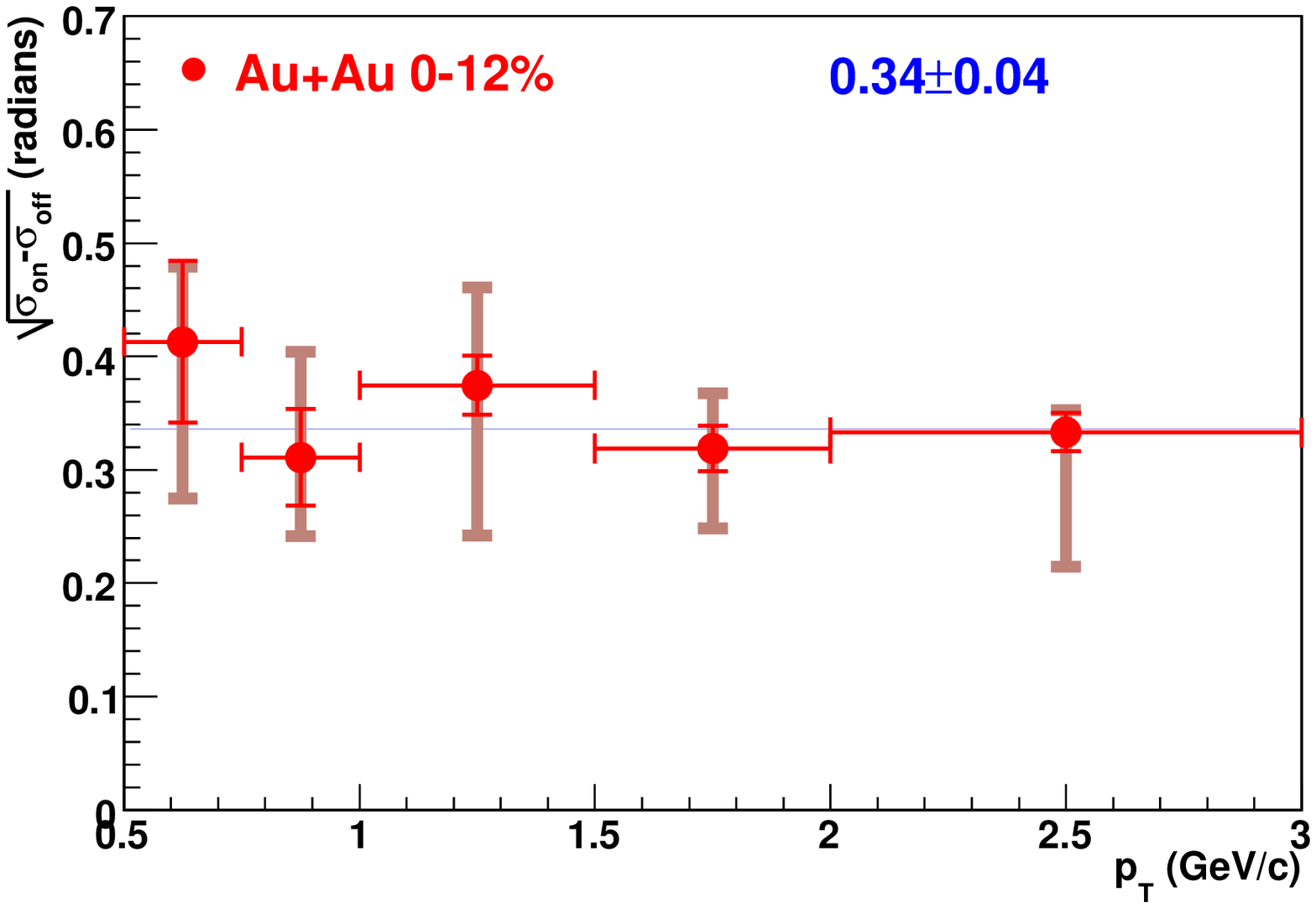}
\end{minipage}
\hfill
\begin{minipage}[t]{0.49\textwidth}
\centering
\includegraphics[width=1.0\textwidth]{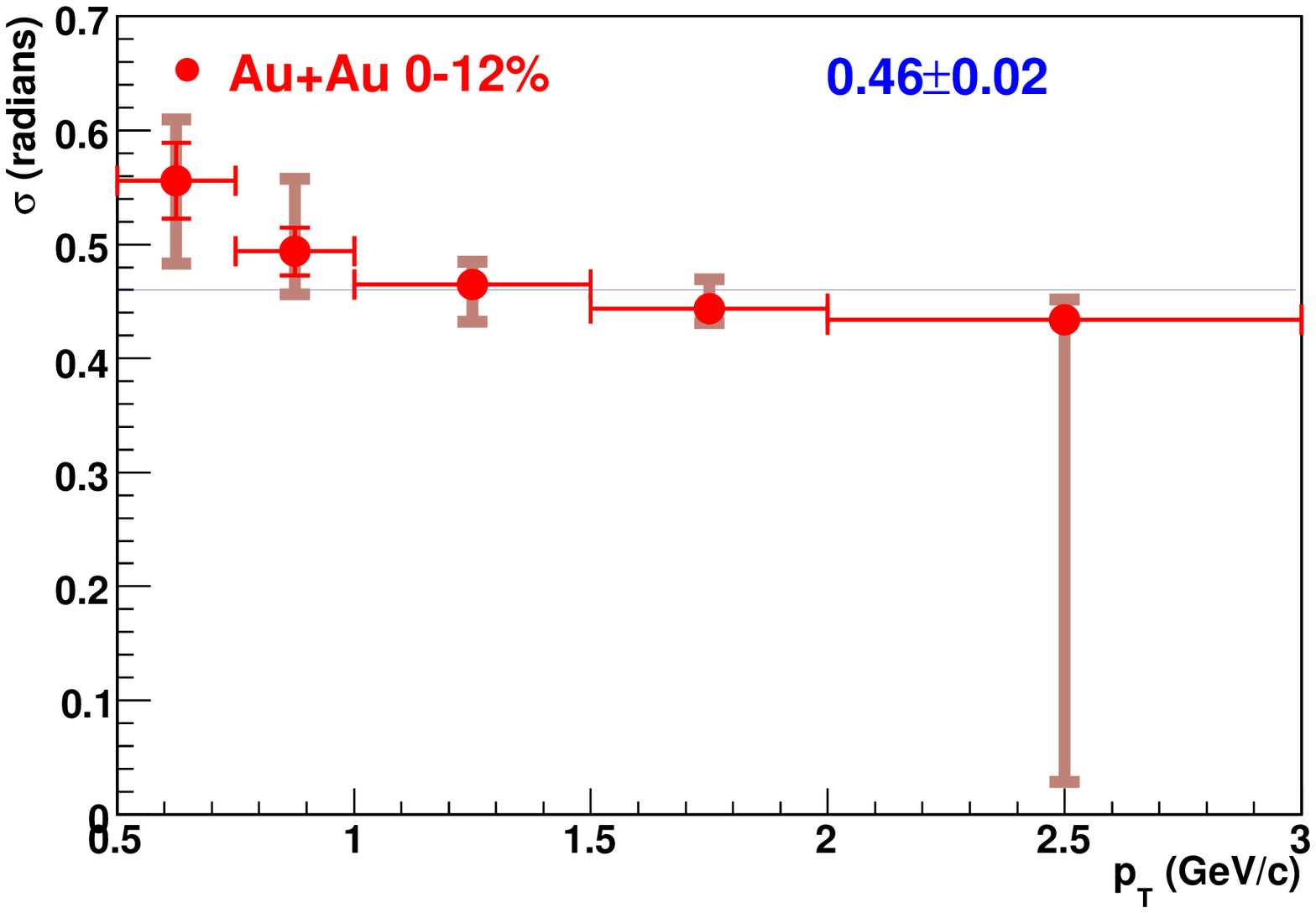}
\includegraphics[width=1.0\textwidth]{Plots/ProjStripBlank.eps}
\end{minipage}
\caption{Width ($\sigma$ of Gaussians) of near-side peaks from projections of the background subtracted 3-particle correlations as a function of associated particle $p_T$.  Top Left:  Off-diagonal peak width.  Top Right:  On-diagonal peak width.  Bottom:  Difference in the peak widths as $\sqrt{\sigma_{on-diagonal}^{2}-\sigma_{off-diagonal}^{2}}$.   Widths are from Au+Au 0-12\% ZDC triggered data at $\sqrt{s_{NN}}=200$ GeV/c.  The numbers indicate the constant fit results.  The solid errors are statistical and the shaded are systematic.}
\label{fig:ptw2}
\end{figure}

Figure~\ref{fig:ptw2} shows the widths of the near-side peak as a function of associated particle $p_T$.  The widths of the on-diagonal projection appear to decrease with associated particle $p_T$.  This is probably because the high $p_T$ associated particles will be more aligned with the jet-axis\footnote{This is because the average momentum perpendicular to the jet-axis ($j_T$) is independent of the associated particle $p_T$.  For higher $p_T$ particles this is a smaller fraction of their transverse momentum so it leads to a smaller angular deviation from the jet-axis.}.  The off-diagonal near-side projection widths are consistent with either a decrease with associated particle $p_T$ or independent of associated particle $p_T$.  There is no significant change in the difference of the widths.

\subsection{Trigger $P_T$ Dependence}

\begin{figure}[htb]
\hfill
\begin{minipage}{0.45\textwidth}
\centering
\includegraphics[width=1.0\textwidth]{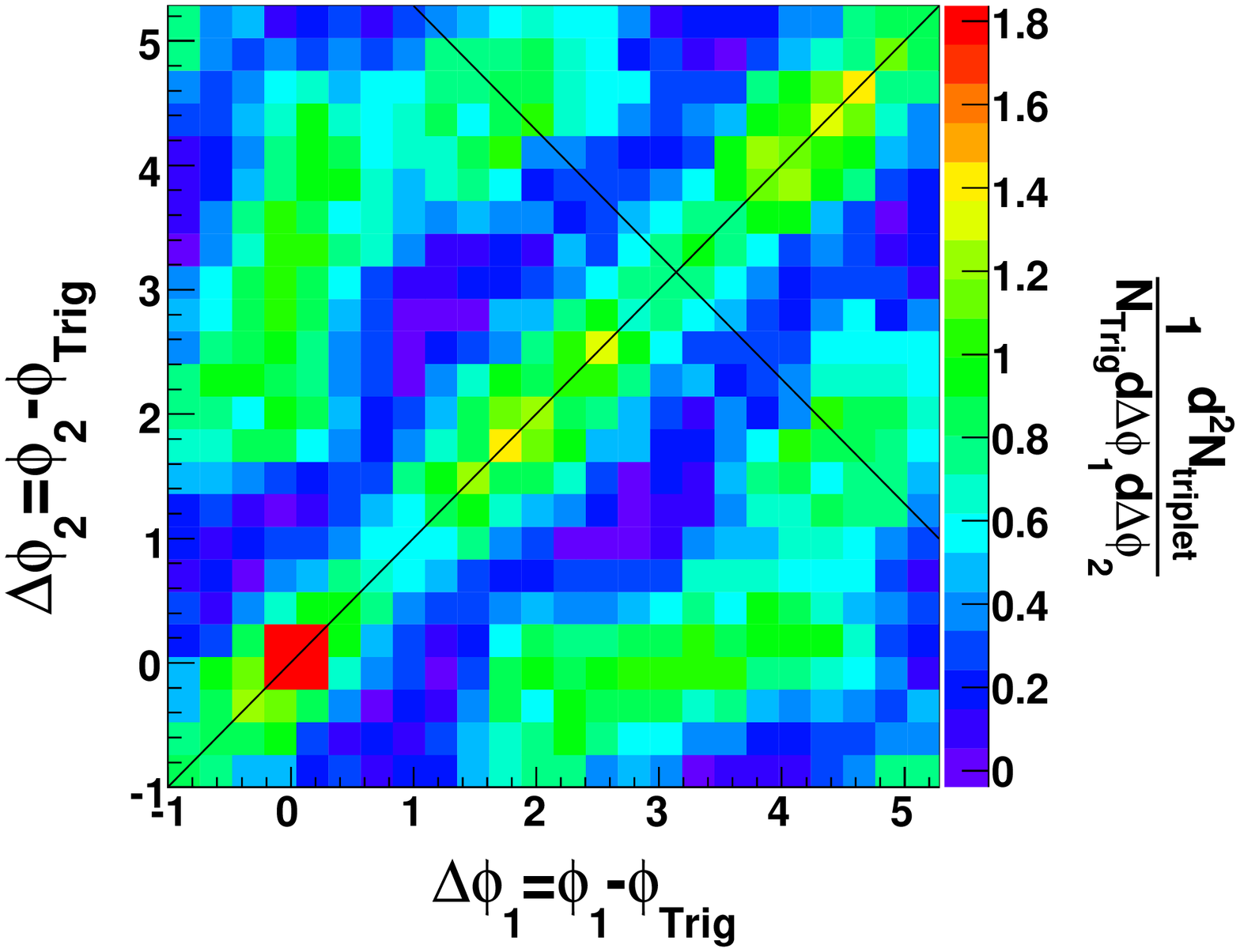}
\includegraphics[width=1.0\textwidth]{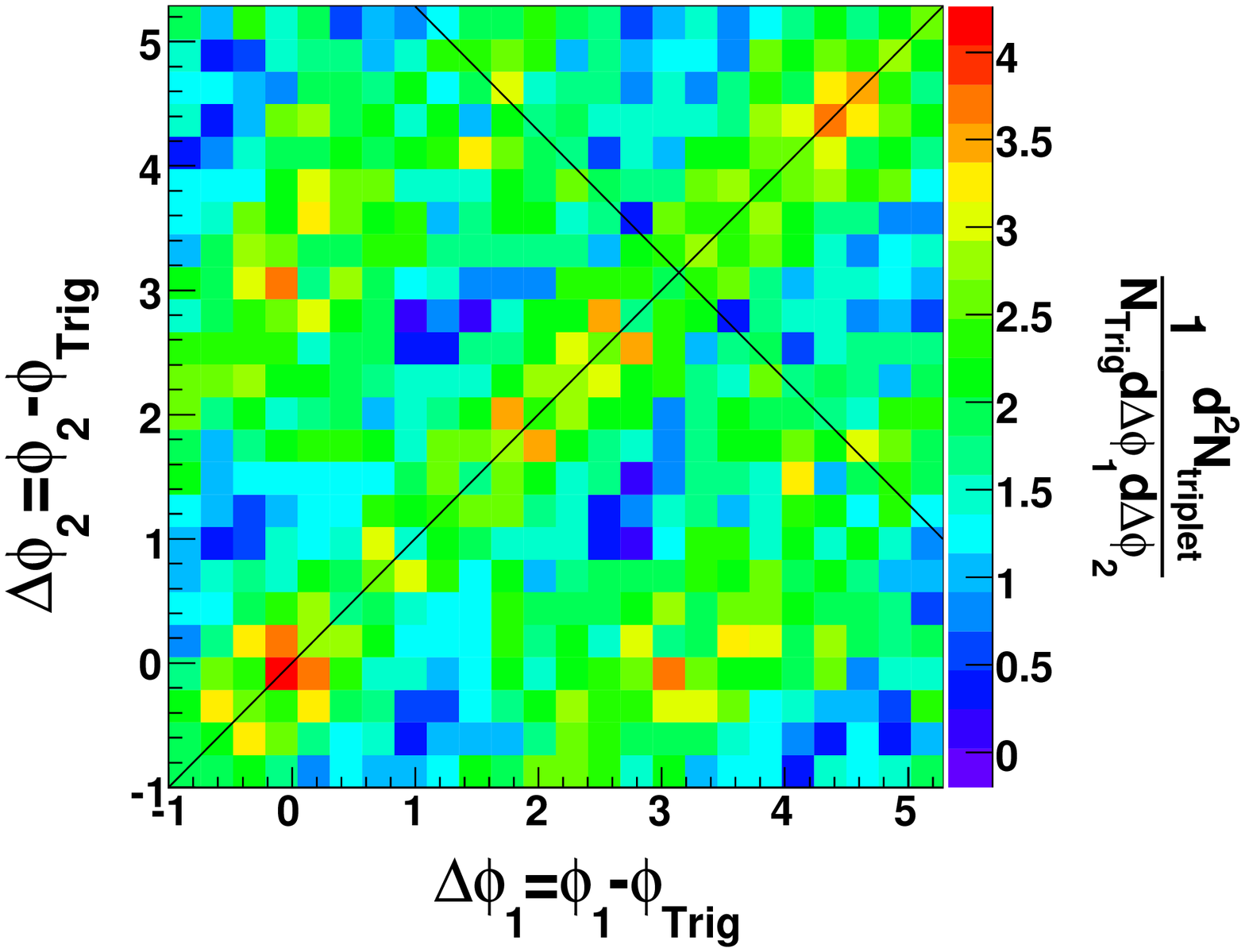}
\end{minipage}
\hfill
\begin{minipage}{0.45\textwidth}
\centering
\includegraphics[width=1.0\textwidth]{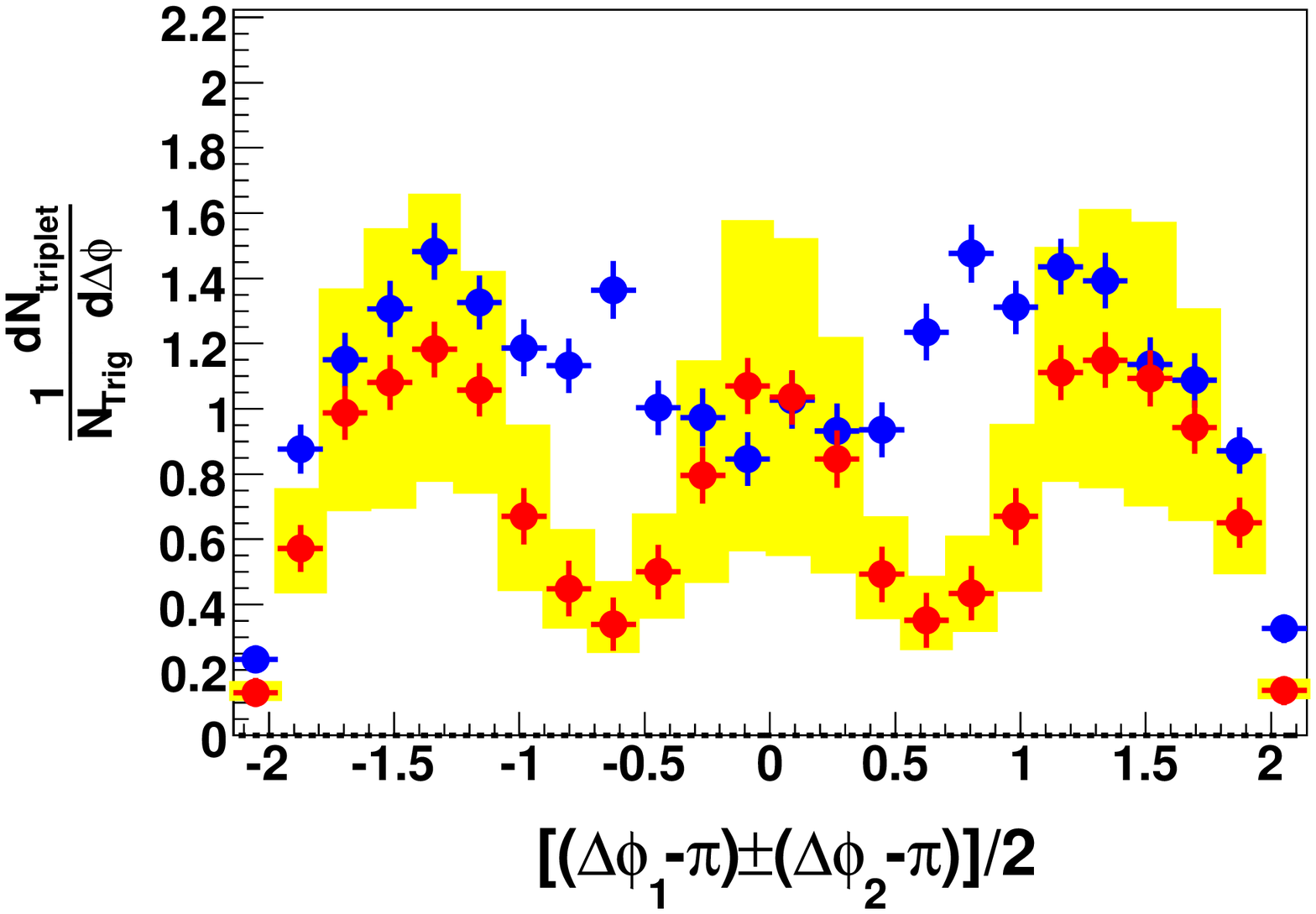}
\includegraphics[width=1.0\textwidth]{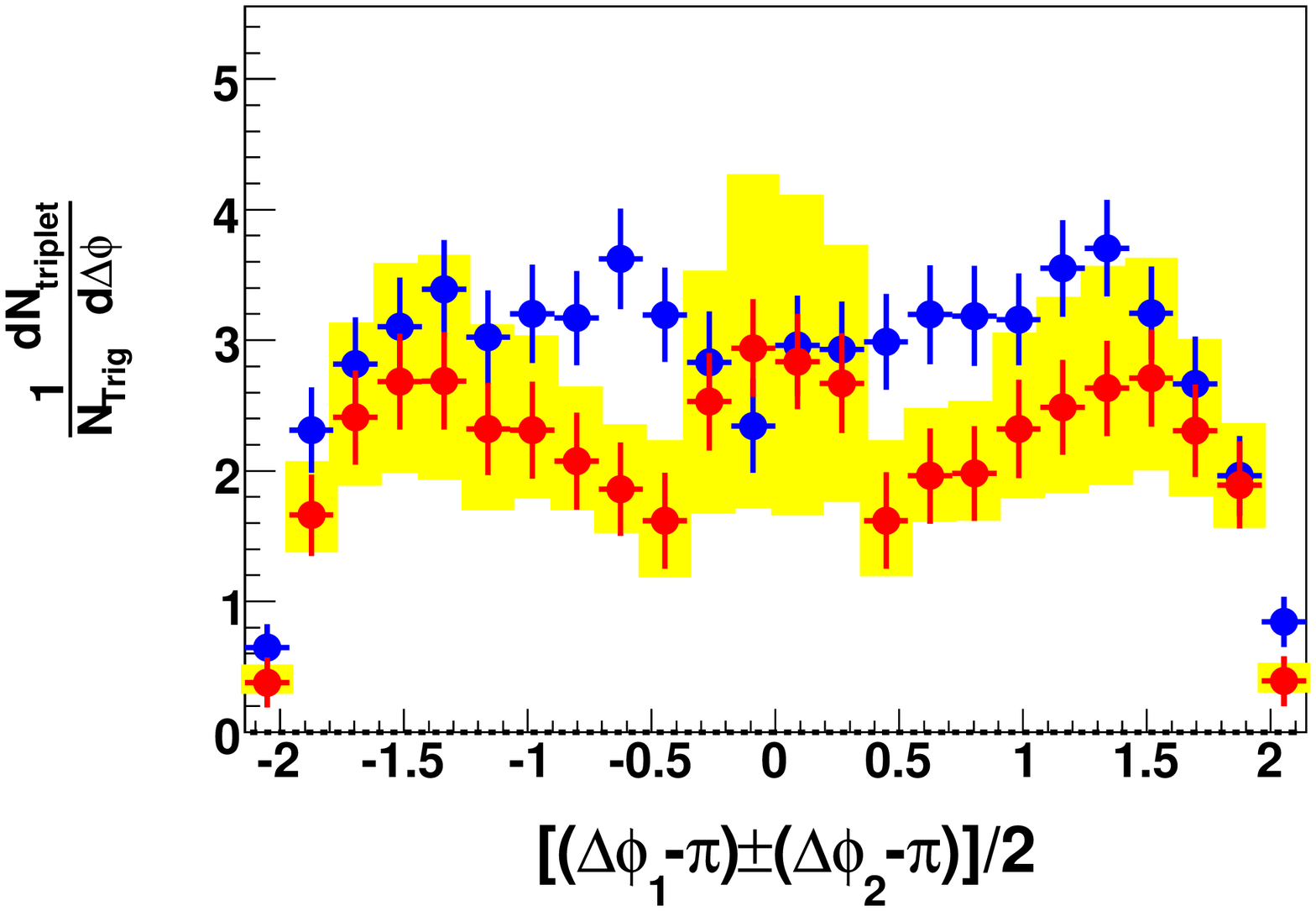}
\end{minipage}
\caption{Three-particle correlations with trigger particles of $4<p_{T}^{trig}<6$ (top) and $6<p_{T}^{Trig}<10$ (bottom) with associated particles of $1<p_{T}<2$ GeV/c for 0-12\% ZDC triggered Au+Au collisions at $\sqrt{s_{NN}}=200$ GeV/c.  Left:  Background subtracted 3-particle correlation.  Right:  On-diagonal projection (blue) and off-diagonal (red).  Error bars are the statistical error. Yellow band is the systematic error on the off-diagonal projection.}
\label{fig:TPt}
\end{figure}

\begin{figure}[htb]
\hfill
\begin{minipage}{0.49\textwidth}
\centering
\includegraphics[width=1.0\textwidth]{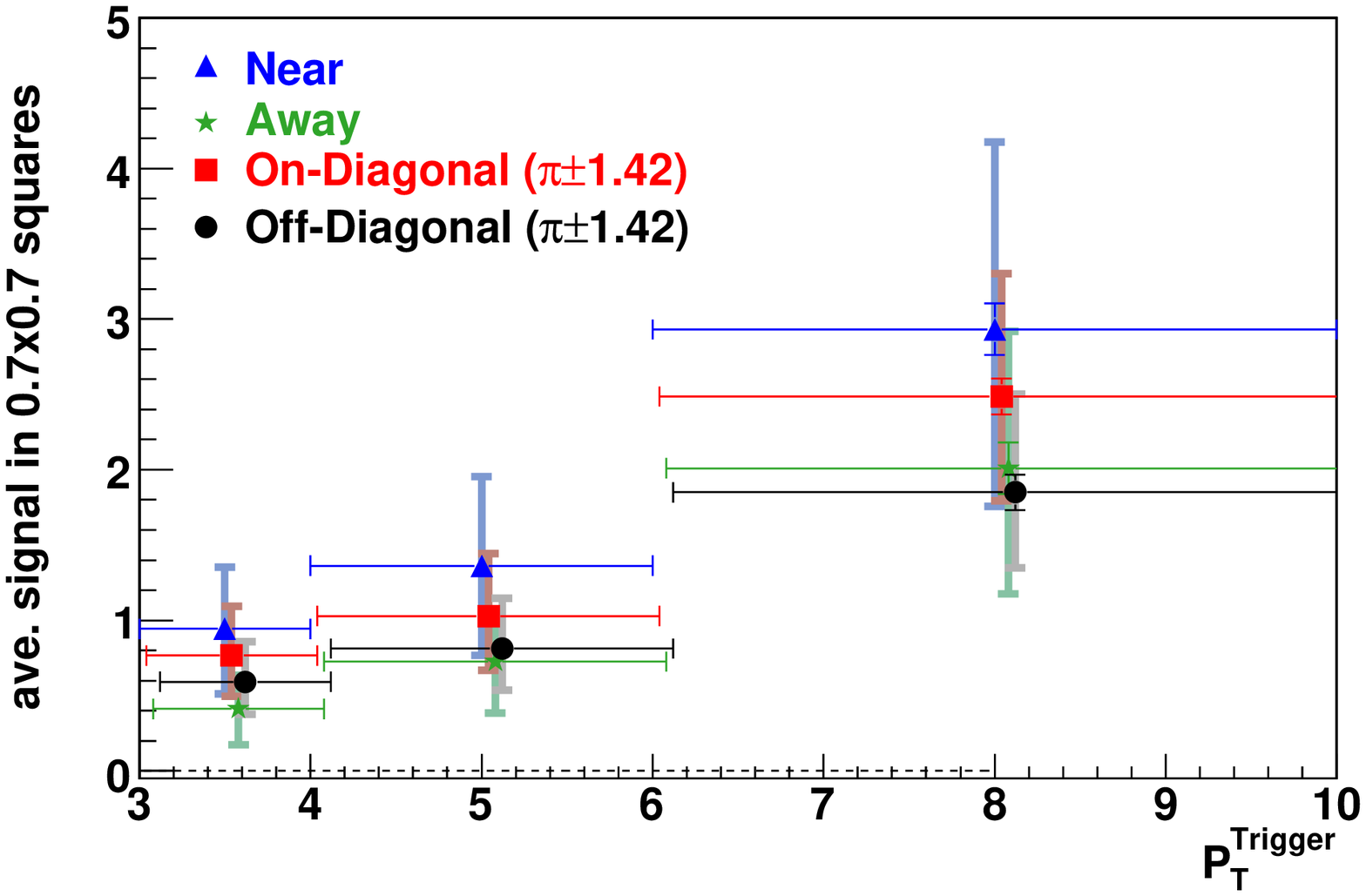}
\end{minipage}
\hfill
\begin{minipage}{0.49\textwidth}
\centering
\includegraphics[width=1.0\textwidth]{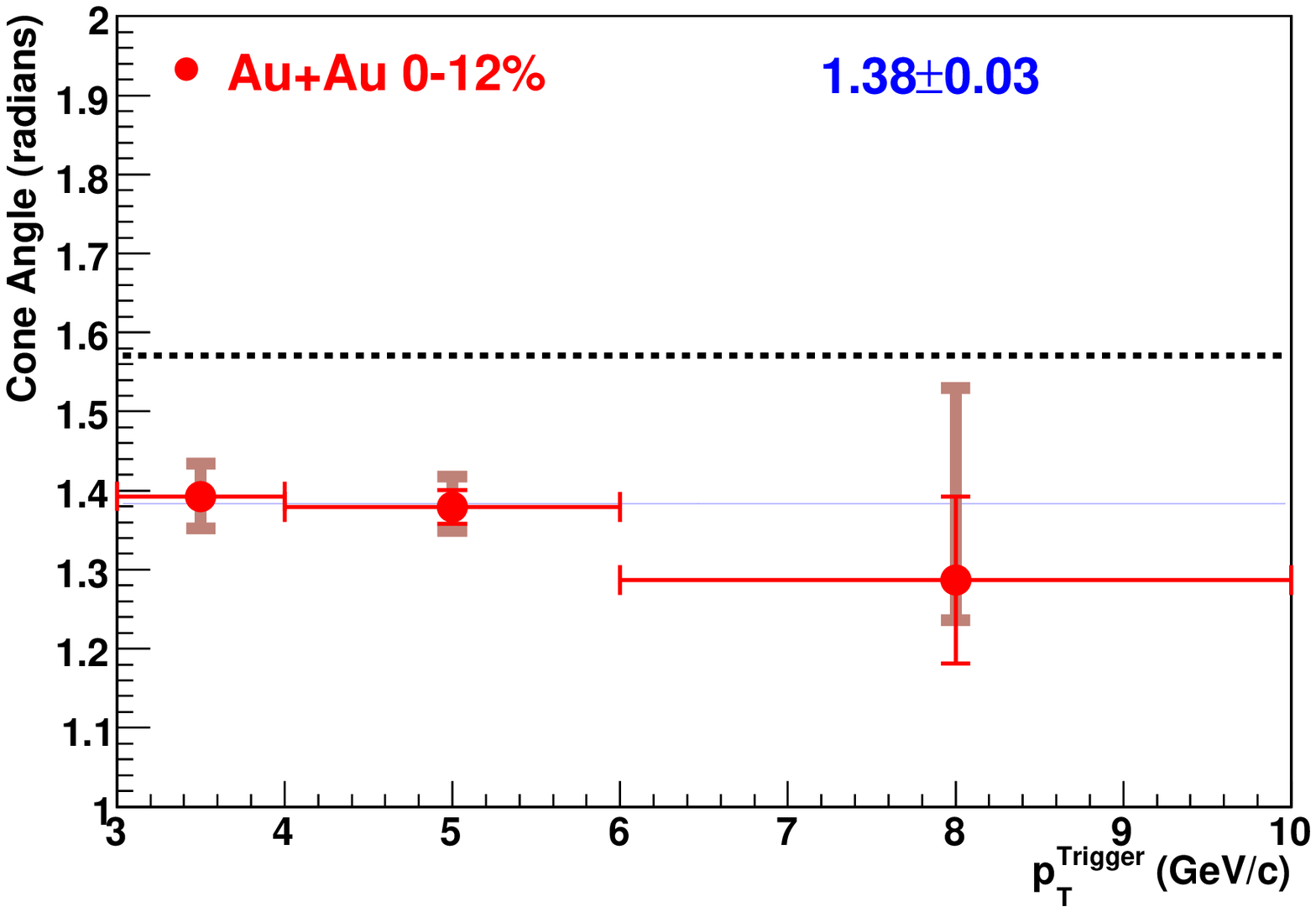}
\end{minipage}
\caption{Left:  Average yield in 0.7x0.7 squares centered on the near-side $(0,0)$, the away-side $(\pi,\pi)$, on-diagonal $(\pi\pm1.42,\pi\pm1.42)$, and off-diagonal $(\pi\pm1.42,\pi\mp1.42)$ as a function of trigger particle $p_T$.  Right:   Angles from fits to off-diagonal projections.  The solid errors are statistical and the shaded are systematic.  Plots are for 0-12\% ZDC triggered Au+Au collisions at $\sqrt{s_{NN}}=200$ GeV/c.}
\label{fig:TPt2}
\end{figure}

We can also look at how the 3-particle correlation changes with trigger particle $p_{T}$.  Figure~\ref{fig:TPt} (left) shows the background subtracted 3-particle correlation for triggers of $4<p_{T}<6$ GeV/c and $6<p_T<10$ GeV/c.  The result looks very similar to that for $3<p_{T}<4$ GeV/c.  There does appear to be a little less on-diagonal broadening on both the near-side and the away-side for the higher $p_{T}$ trigger.  This is expected because the azimuthal angle of the higher $p_{T}$ trigger is a better proxy for the jet-axis.  Figure~\ref{fig:TPt} also shows the on-diagonal and off-diagonal projections.  The yields and the angle of the side peaks in the off-diagonal projection are shown in Fig.~\ref{fig:TPt2}.  The yields increase with increasing trigger particle $p_{T}$.  The away-side yield increases more quickly then the off-diagonal peak yield.  This is consistent with more of the away-side jet punching though the center for higher energy jets.  The angle of the side peaks in the off-diagonal projection is consistent with no trigger particle $p_T$ dependence.  

\section{Systematic Uncertainties}

Systematic error bars have been shown in different plots in the results section.  This section discusses what goes into the systematic errors.  The systematics have been rigorously studied.  There are two dominant sources of systematic uncertainties.  These are normalization and uncertainty on the flow measurement.   Other sources of systematic error include $v_{2}$ of the jet, effect of flow fluctuations, the parameterization of $v_{4}$, and finite multiplicity bin width effects.  There is also an overall 10\% systematic uncertianity from the uncertianity in the efficiency.  In the following sections, we discuss the sources of systematic uncertainty in detail.

\subsection{Background Normalization}

We have two factors that are used in our background normalization.  Both of these parameters have been explored in parameter space.  The default values have been chosen based on our best knowledge of jet-like correlations at RHIC

\begin{figure}[htbp]
\centering
\hfill
\begin{minipage}{0.16\textwidth}
\includegraphics[width=1.0\textwidth]{Plots/blank2.eps}
\end{minipage}
\hfill
\begin{minipage}{0.20\textwidth}
\includegraphics[width=1.0\textwidth]{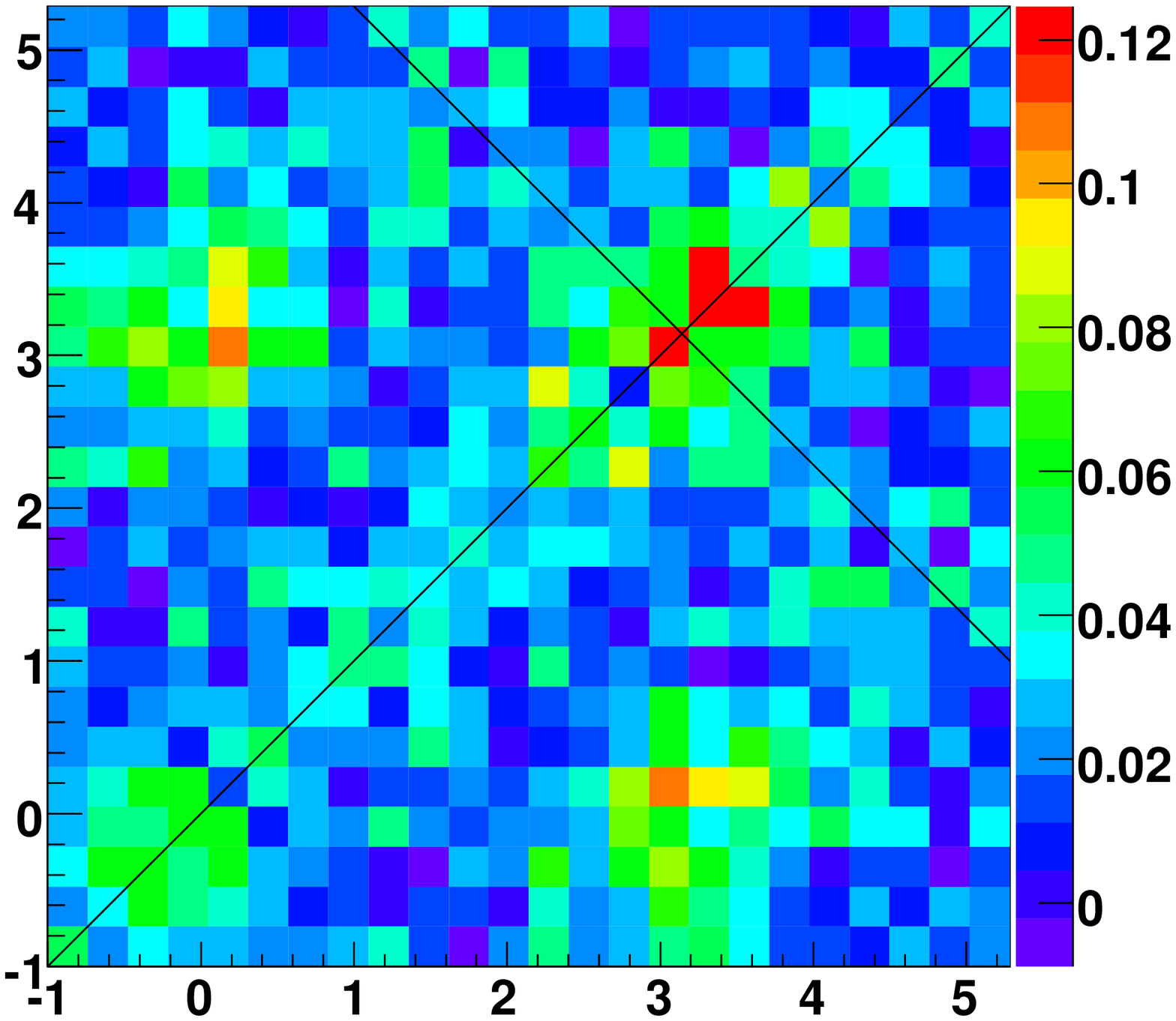}
\includegraphics[width=1.0\textwidth]{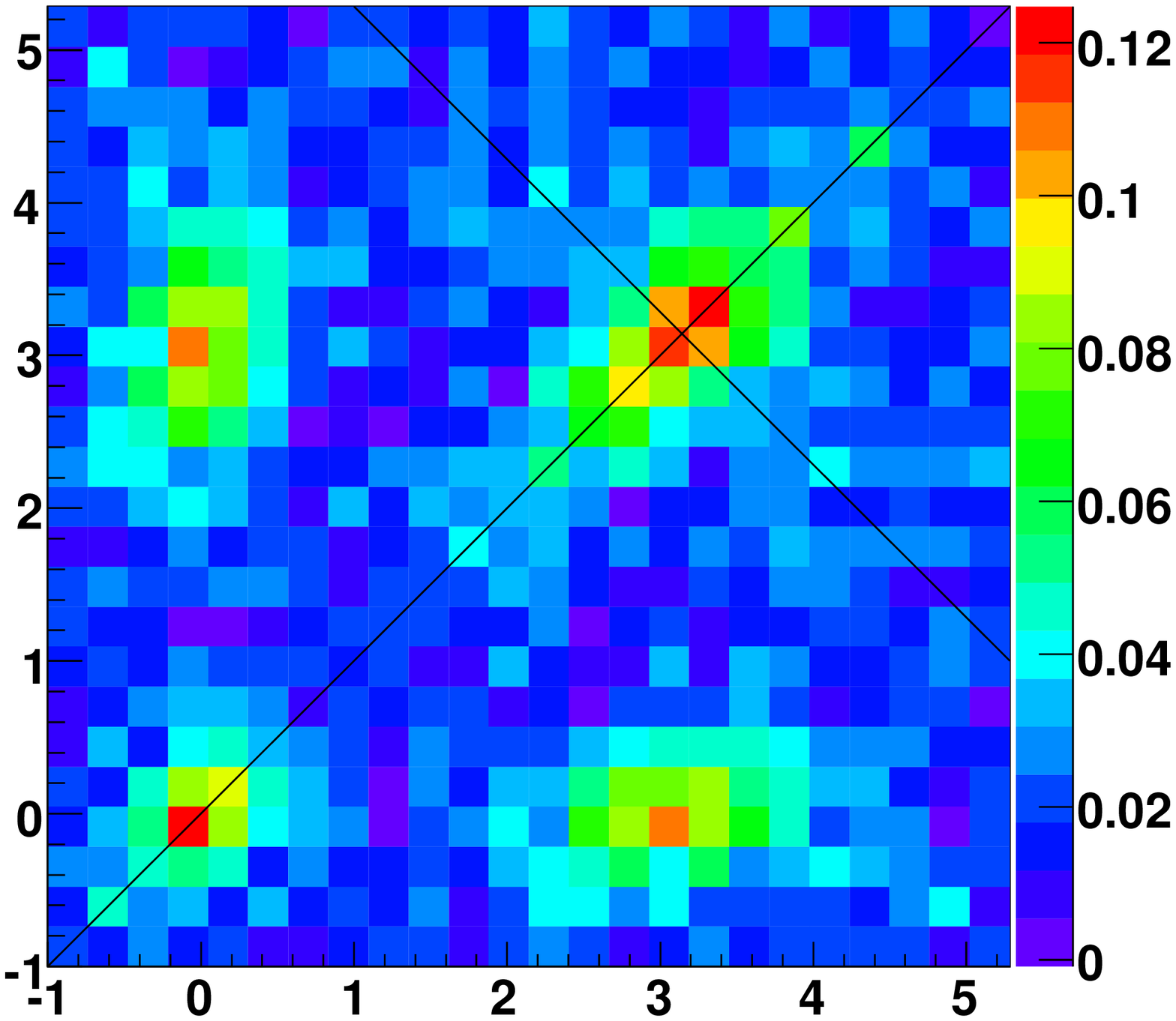}
\includegraphics[width=1.0\textwidth]{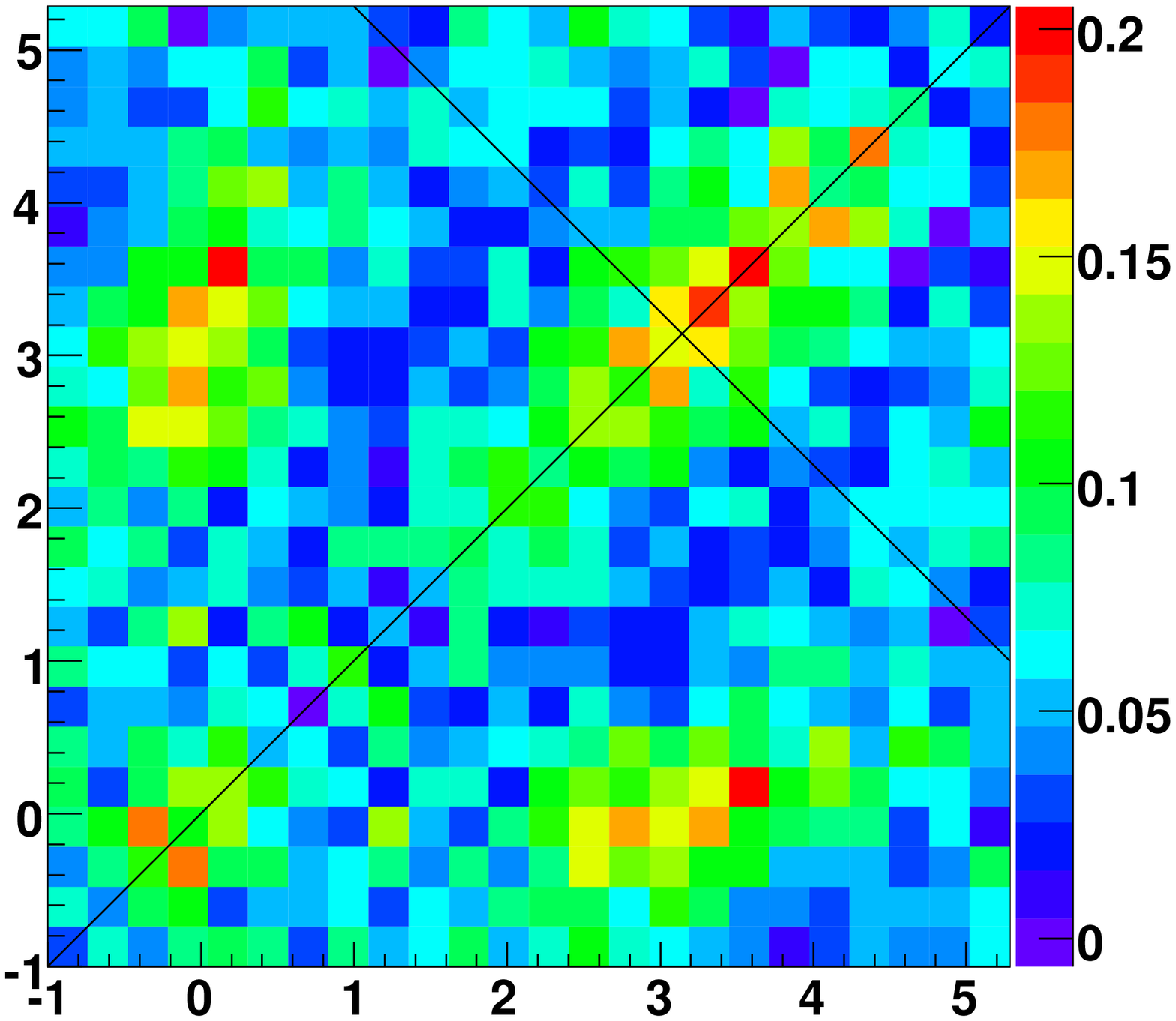}
\includegraphics[width=1.0\textwidth]{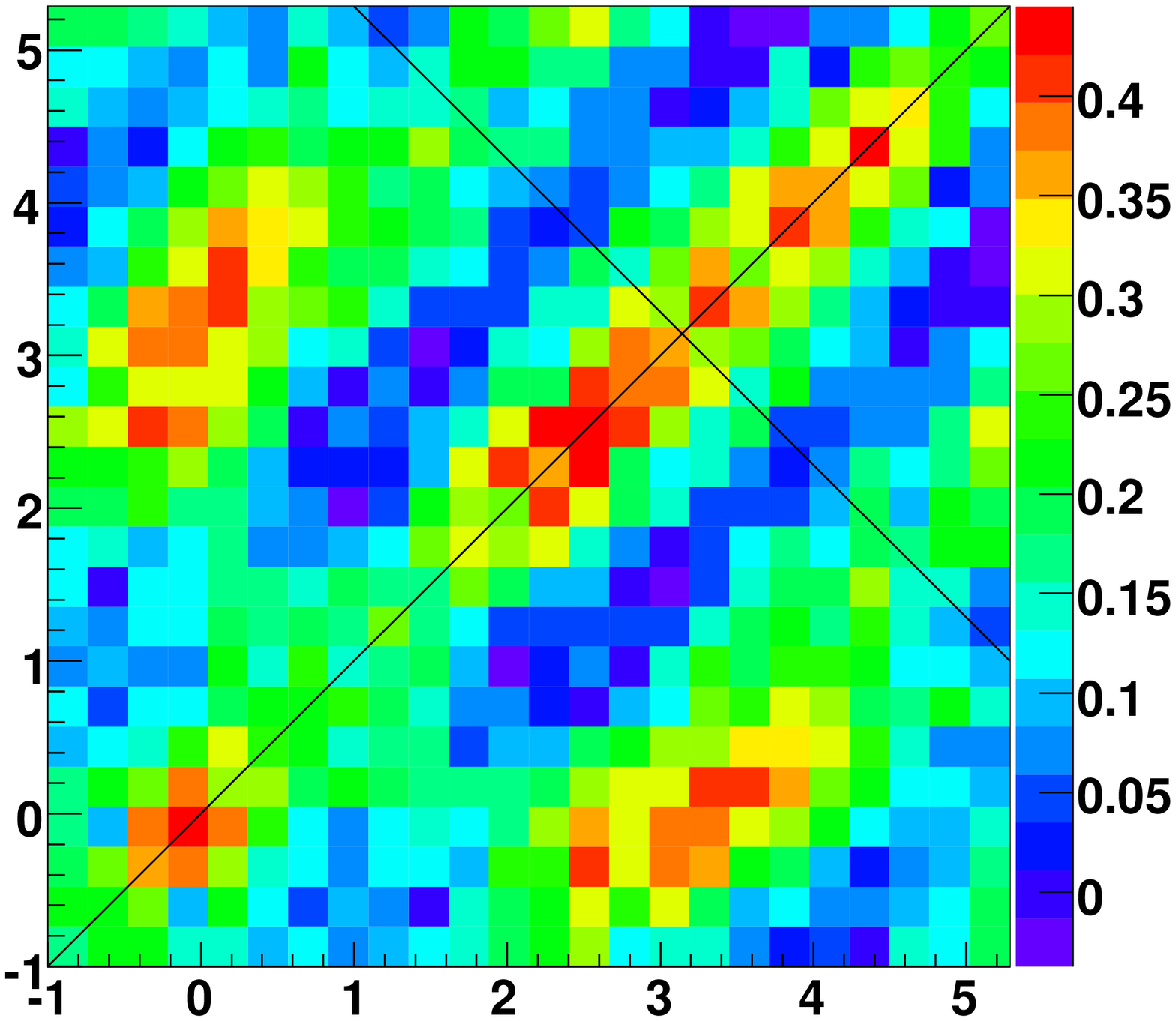}
\includegraphics[width=1.0\textwidth]{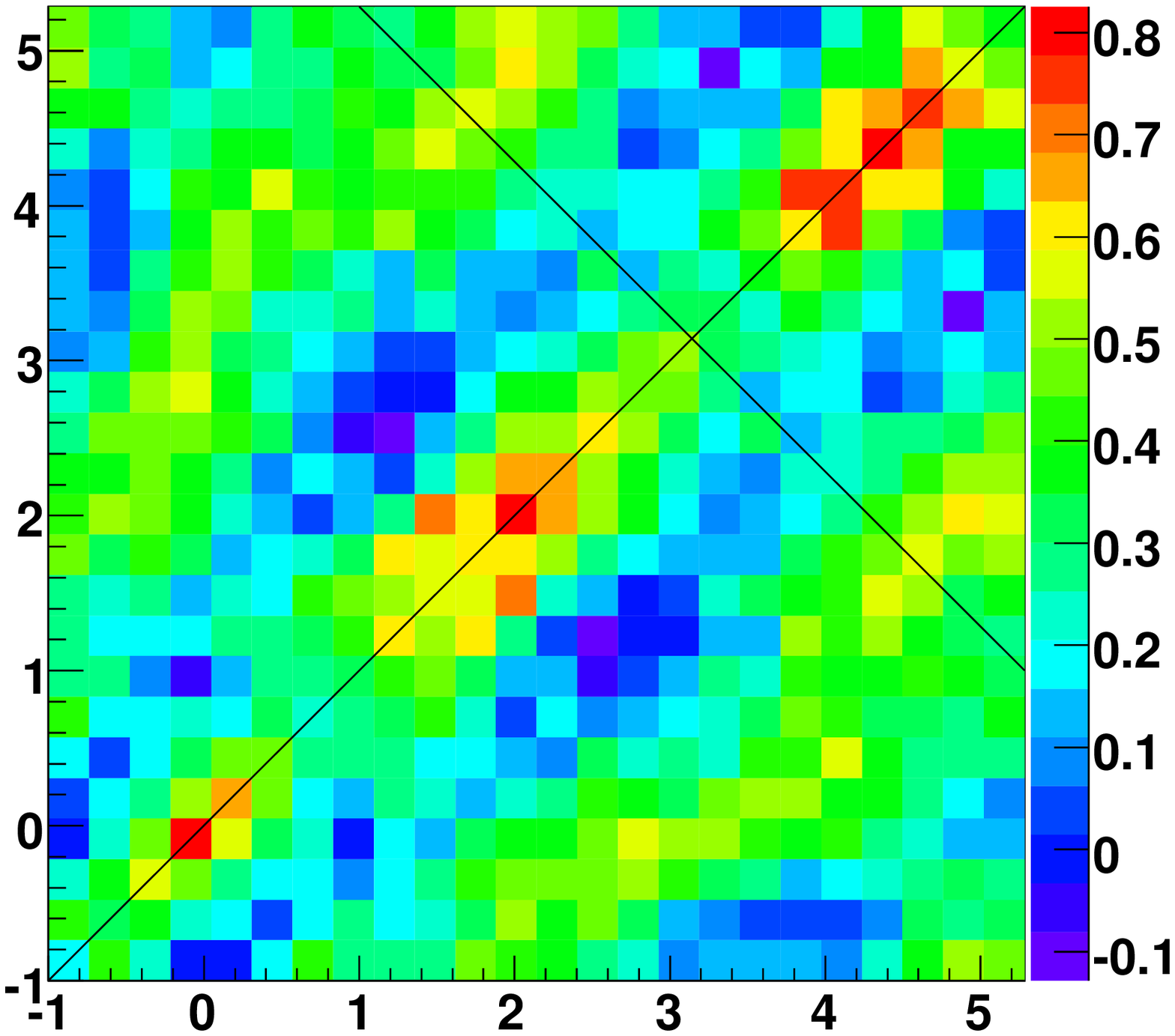}
\includegraphics[width=1.0\textwidth]{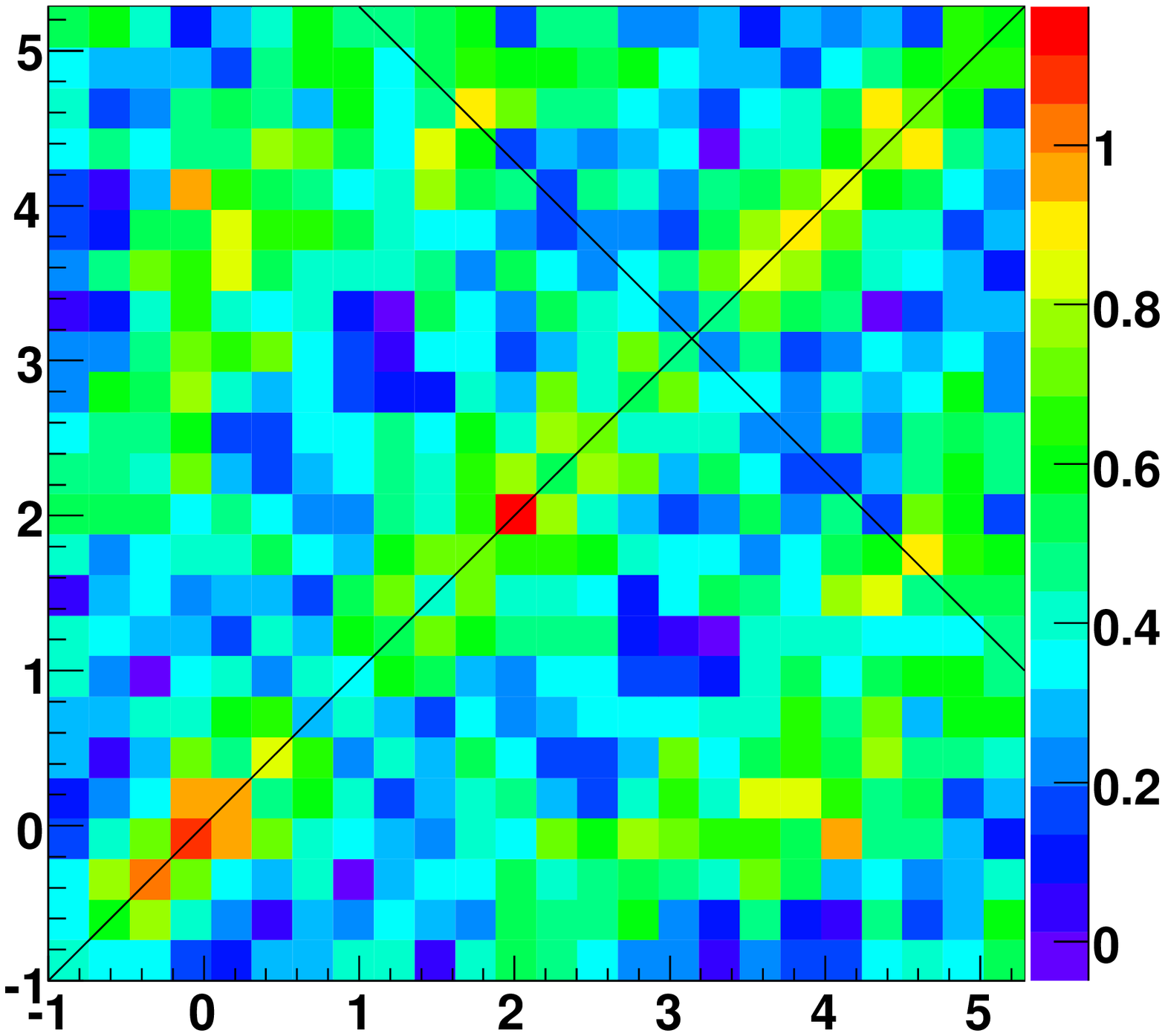}
\includegraphics[width=1.0\textwidth]{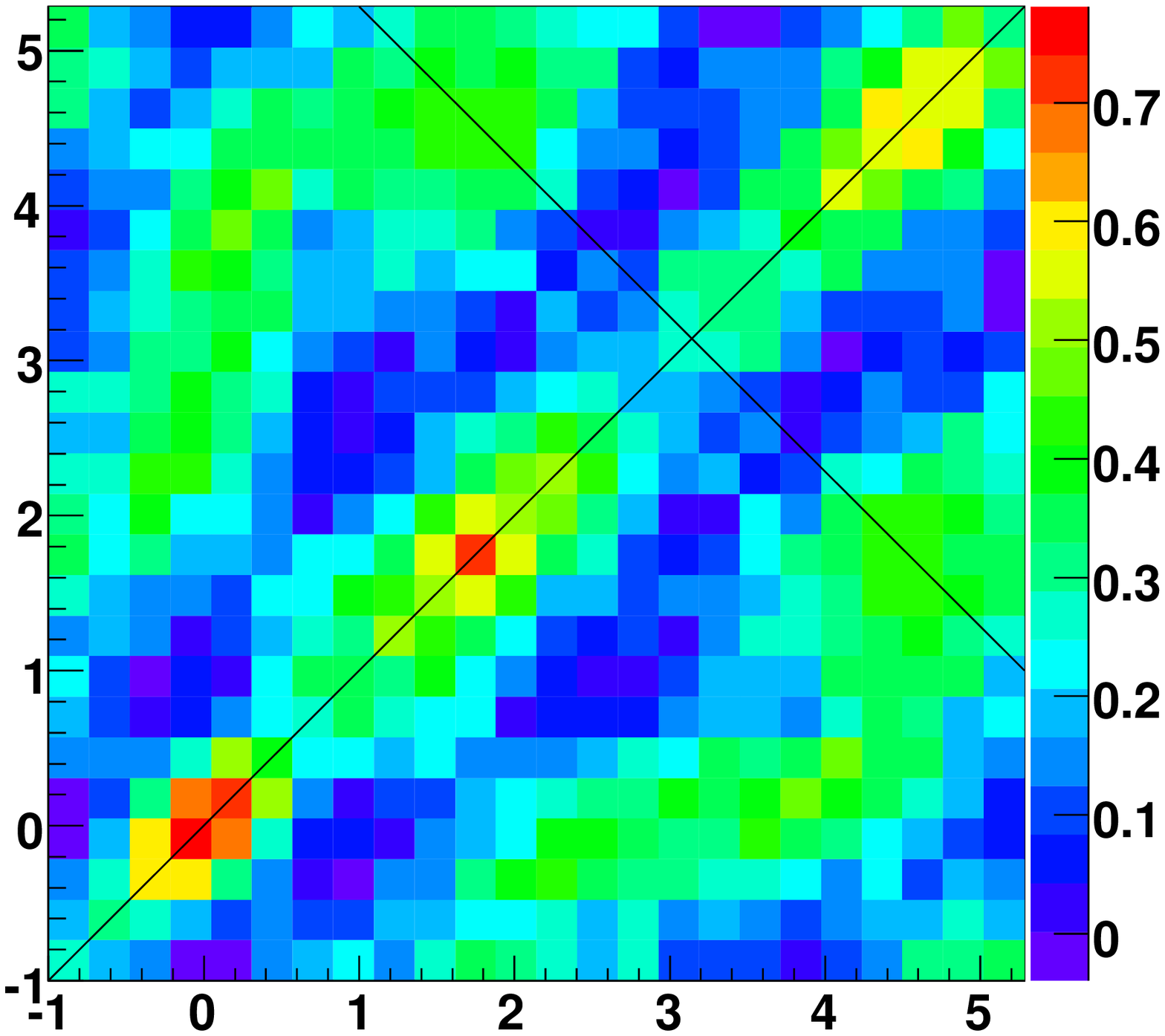}
\end{minipage}
\hfill   
\begin{minipage}{0.20\textwidth}
\includegraphics[width=1.0\textwidth]{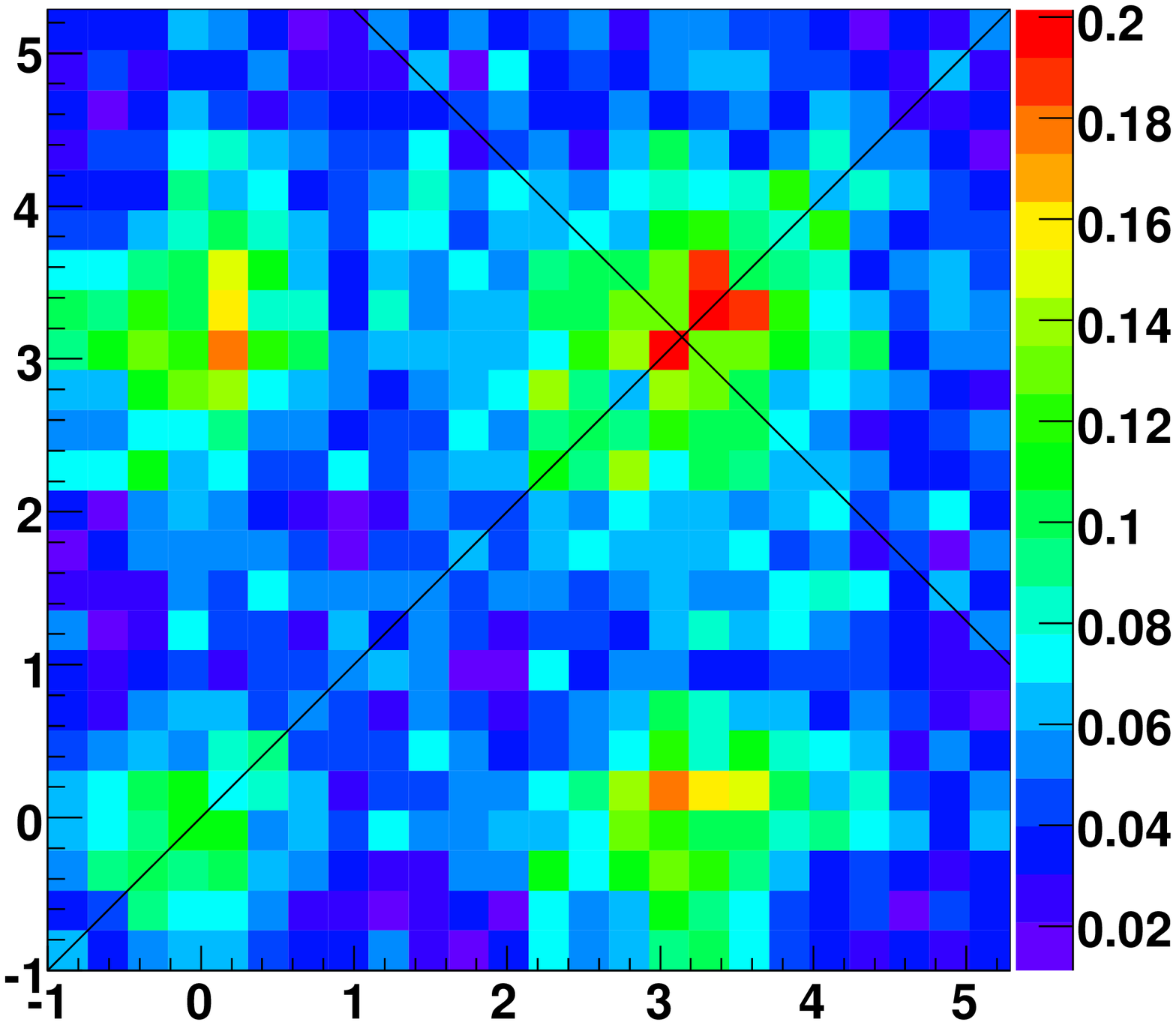}
\includegraphics[width=1.0\textwidth]{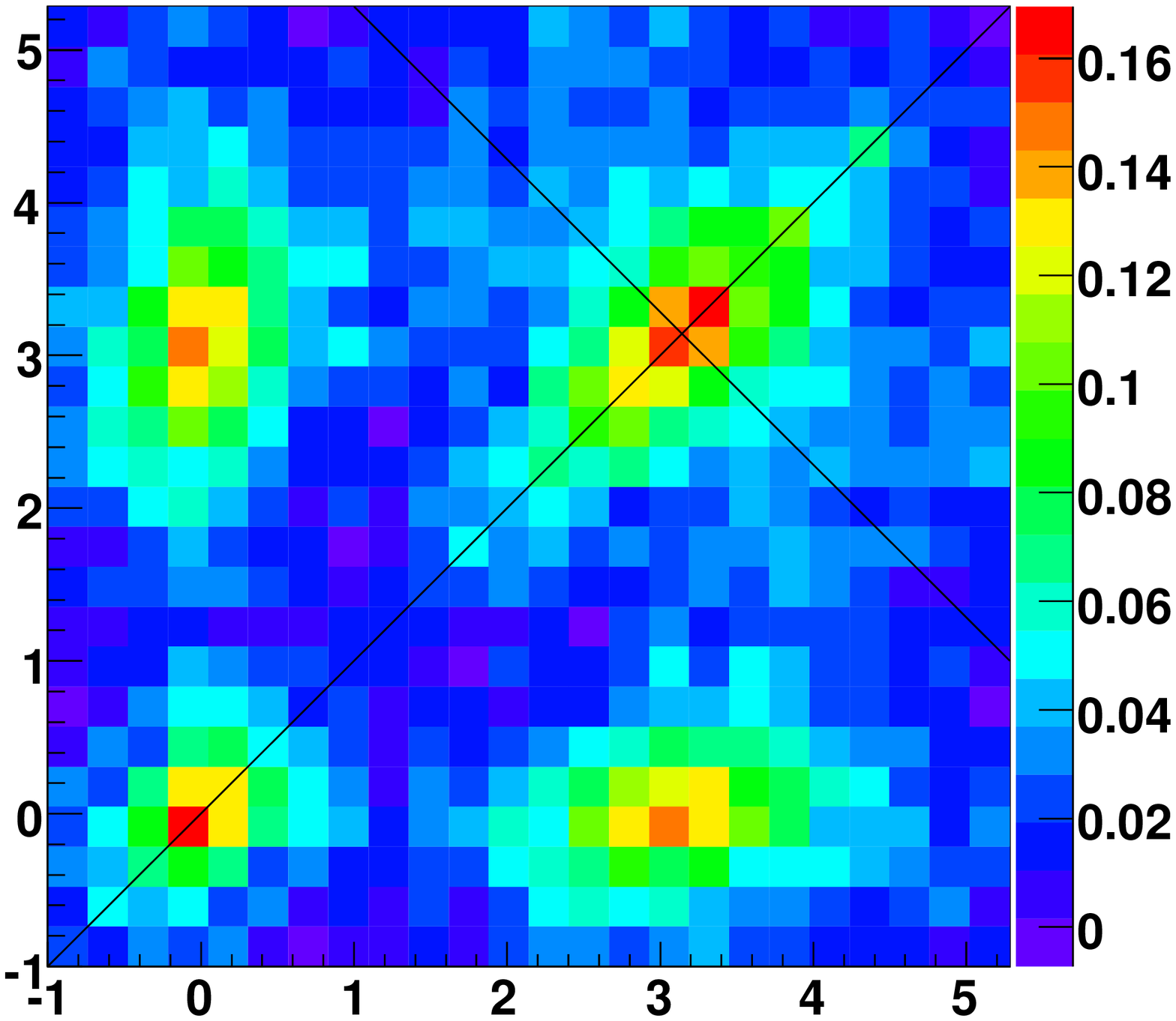}
\includegraphics[width=1.0\textwidth]{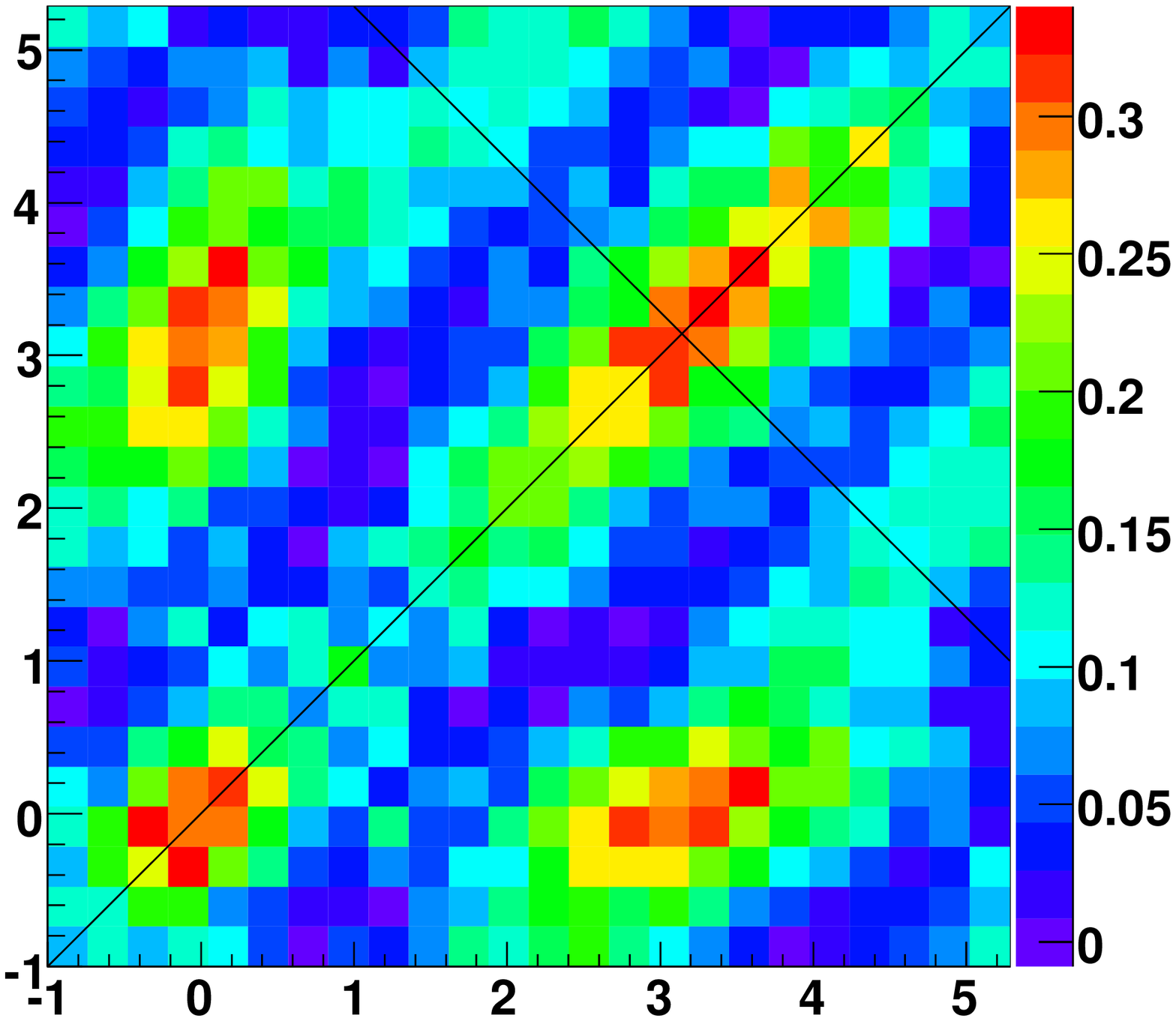}
\includegraphics[width=1.0\textwidth]{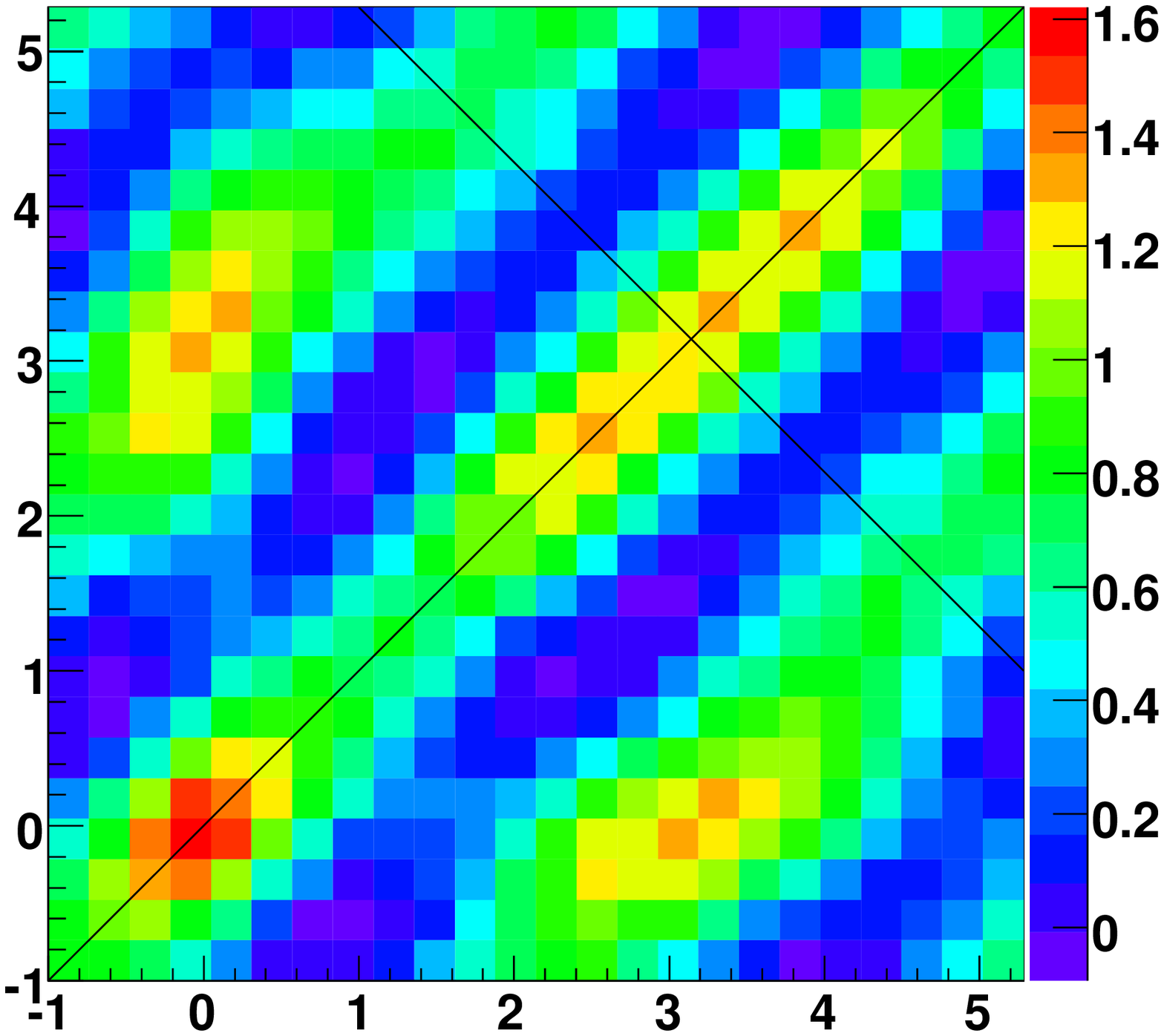}
\includegraphics[width=1.0\textwidth]{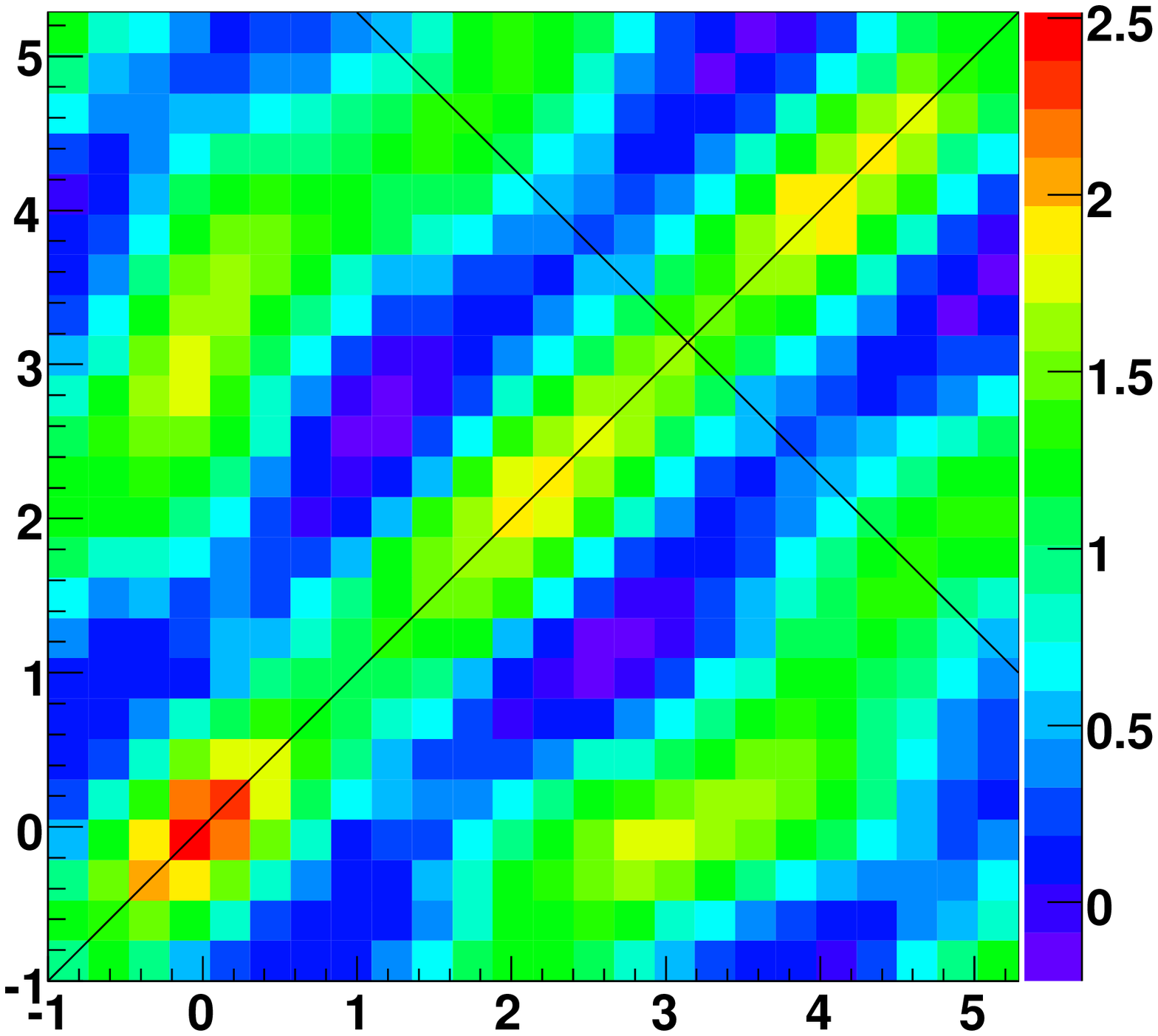}
\includegraphics[width=1.0\textwidth]{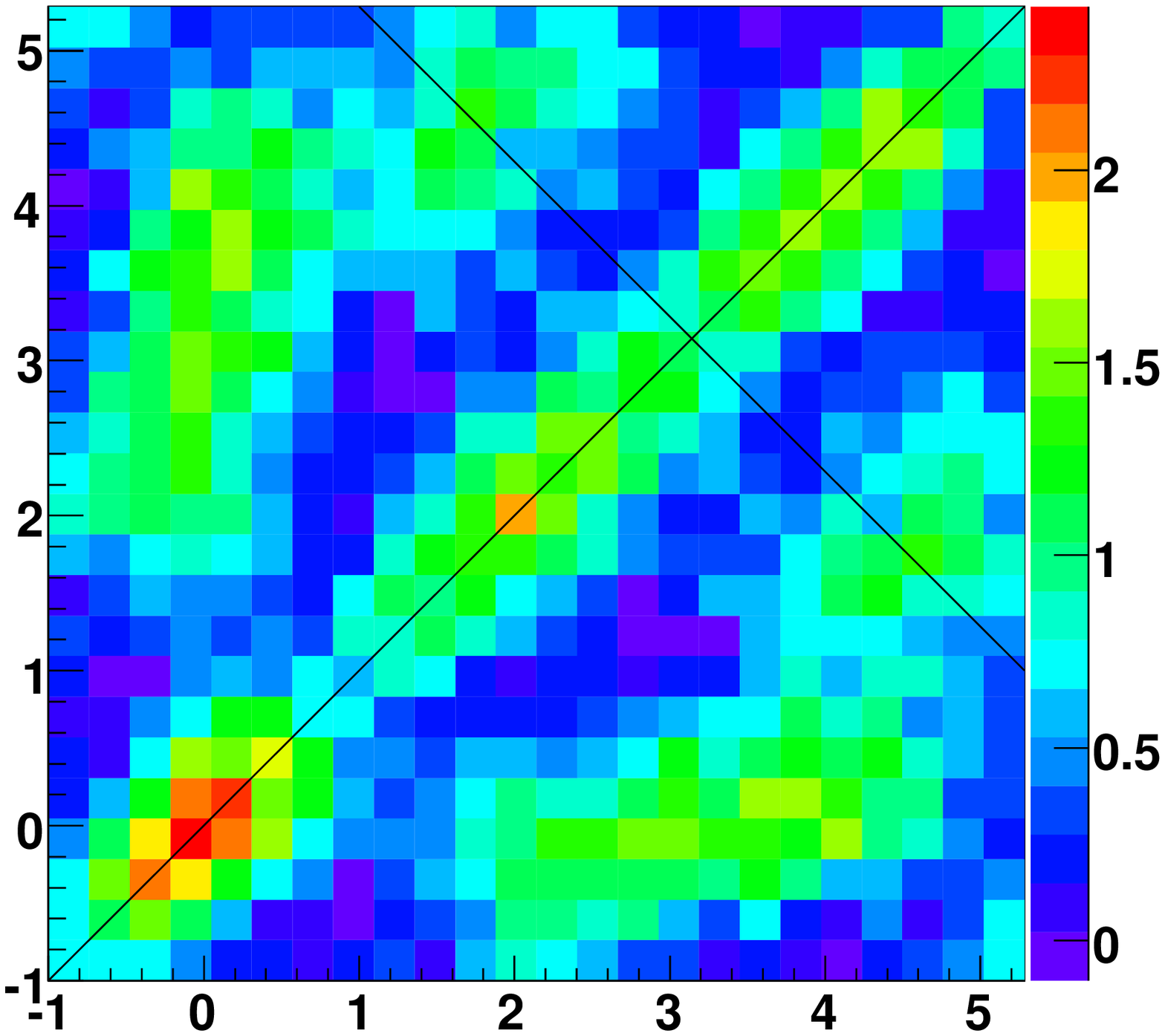}
\includegraphics[width=1.0\textwidth]{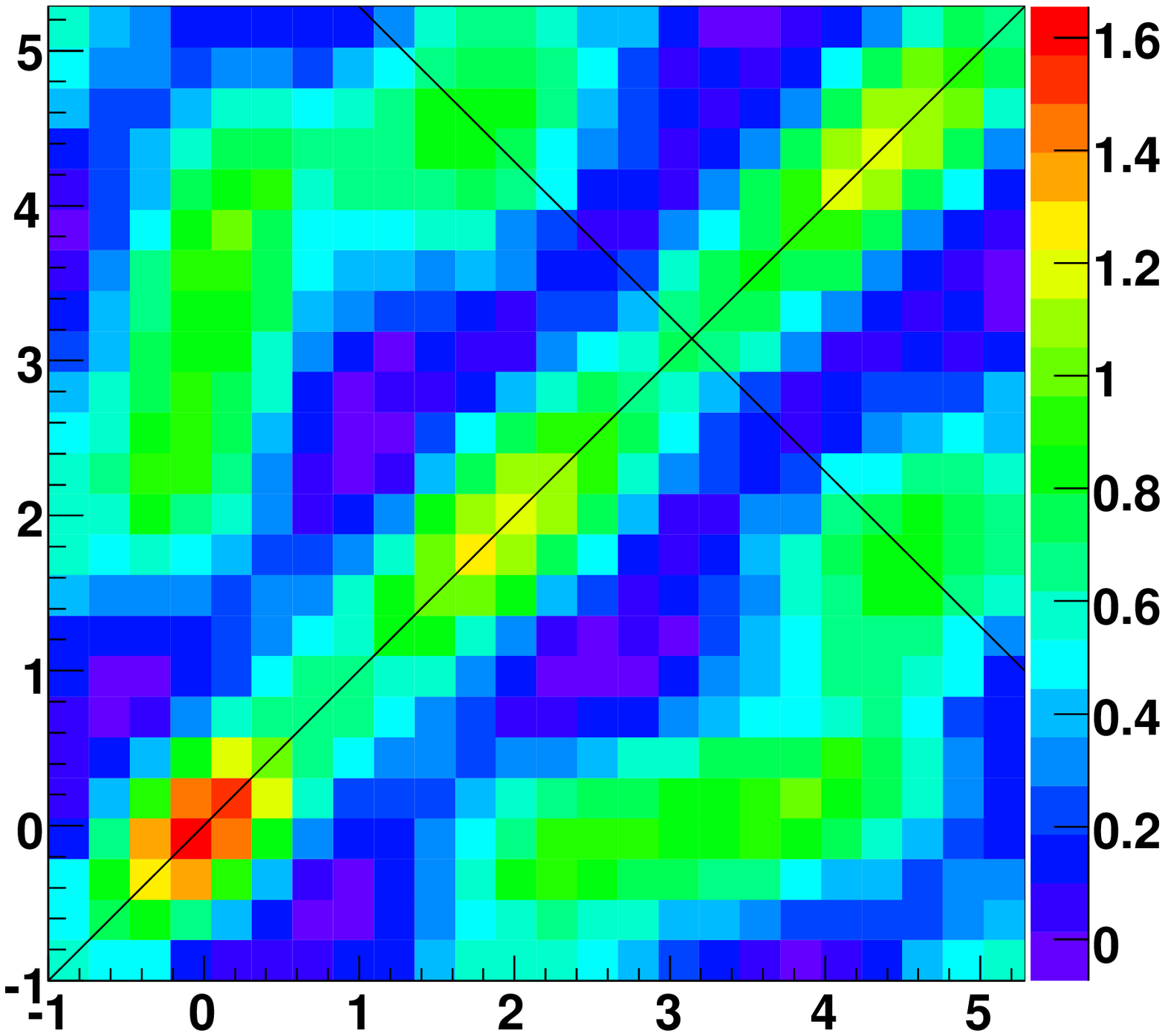}
\end{minipage}
\hfill
\begin{minipage}{0.20\textwidth}
\includegraphics[width=1.0\textwidth]{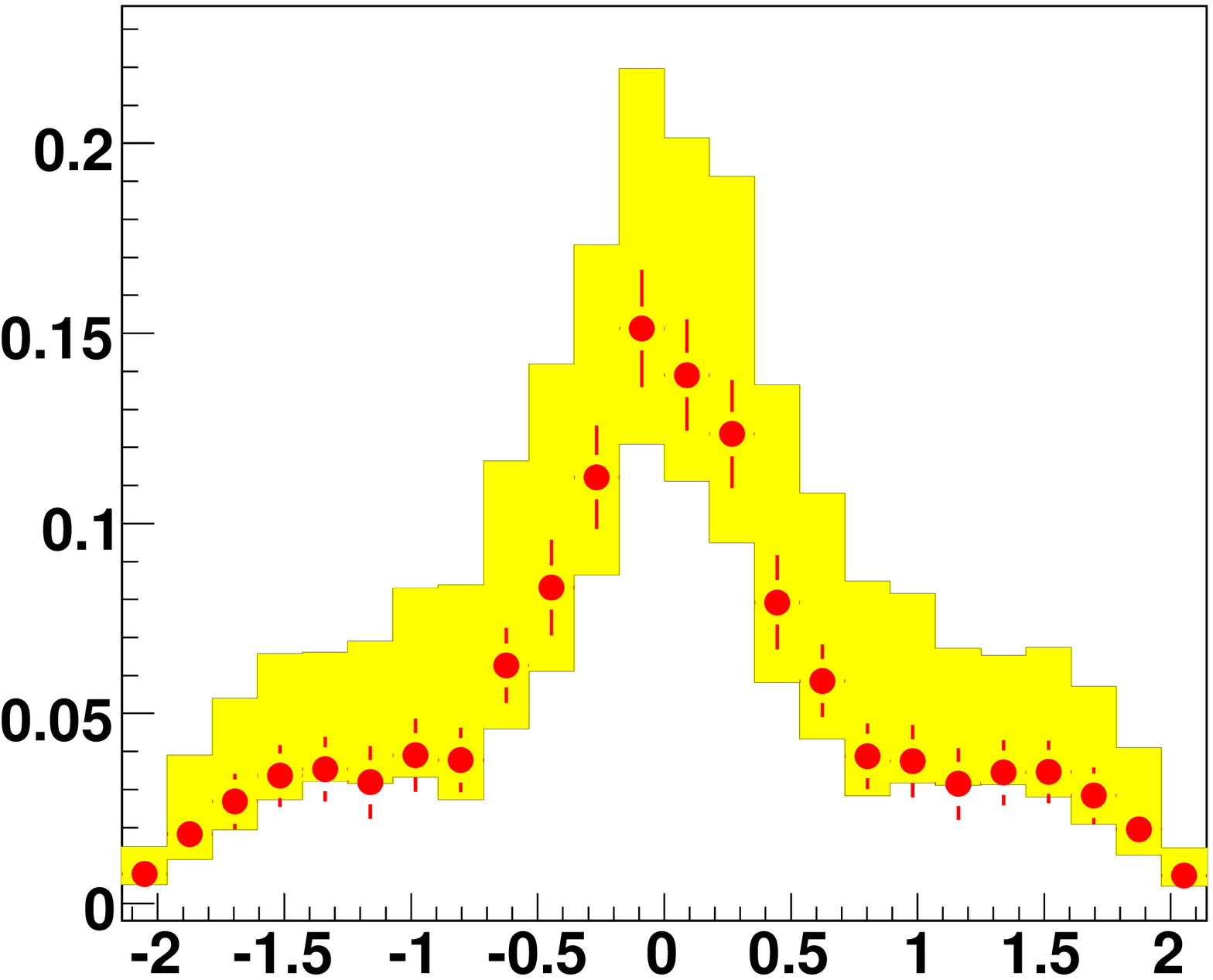}
\includegraphics[width=1.0\textwidth]{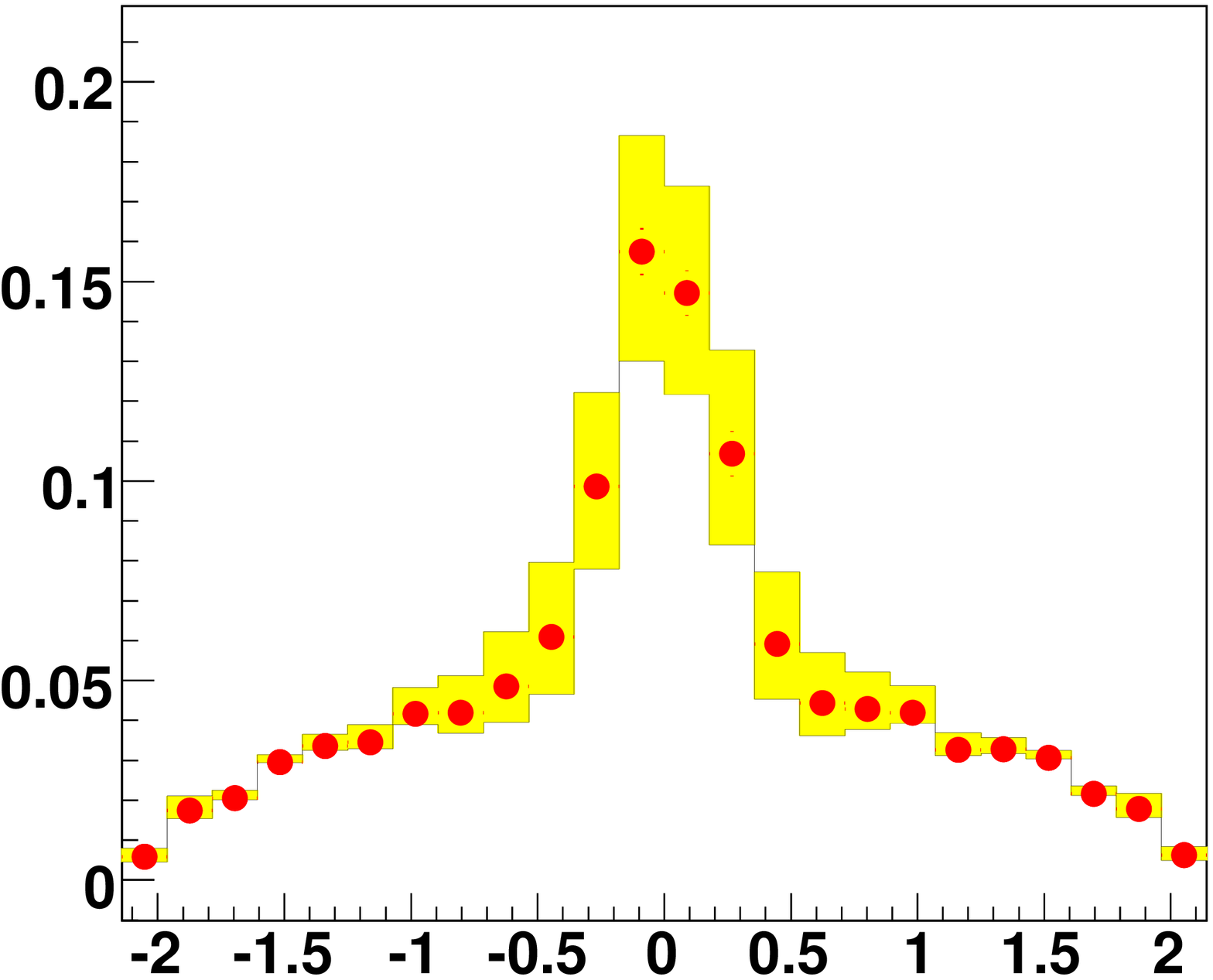}
\includegraphics[width=1.0\textwidth]{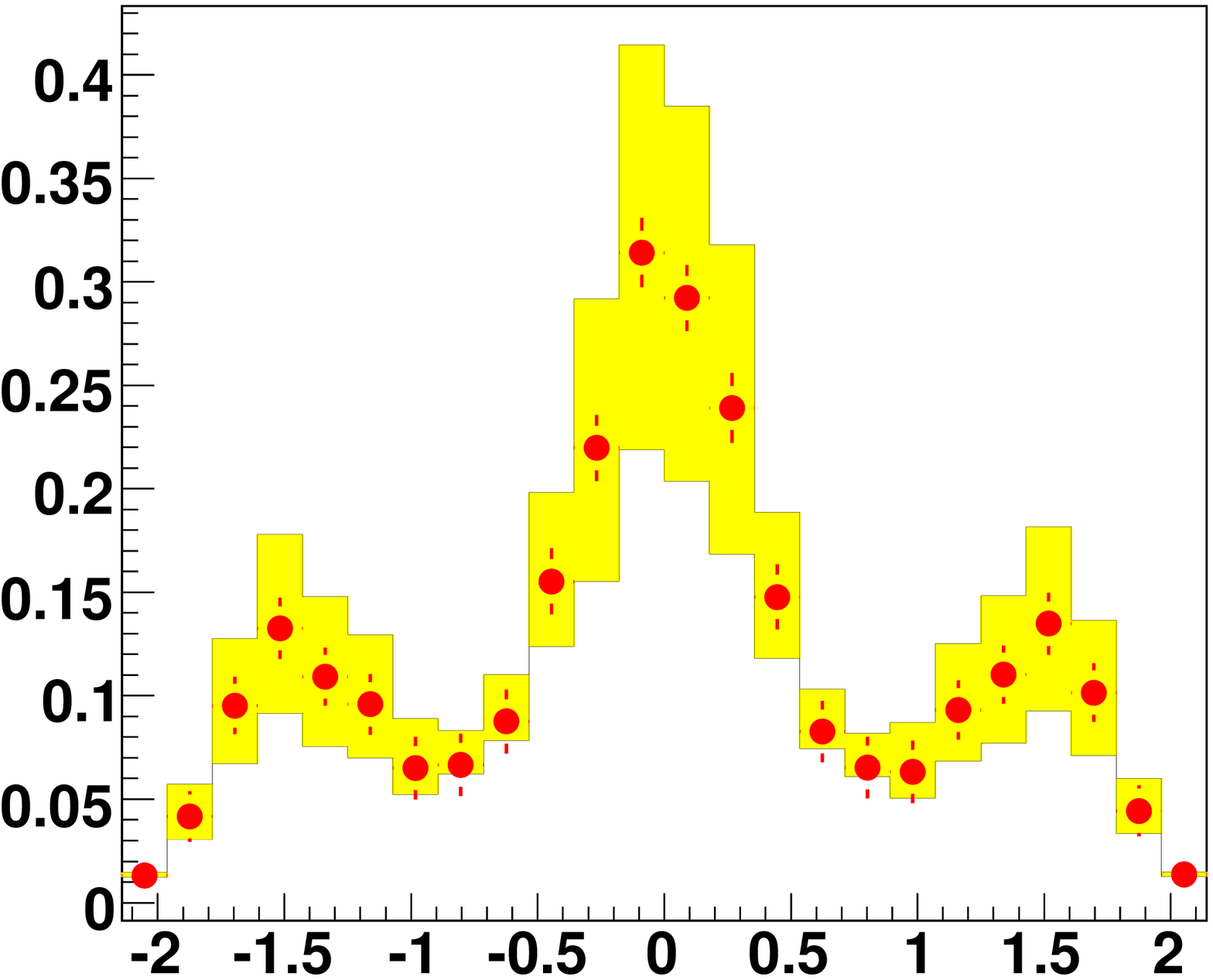}
\includegraphics[width=1.0\textwidth]{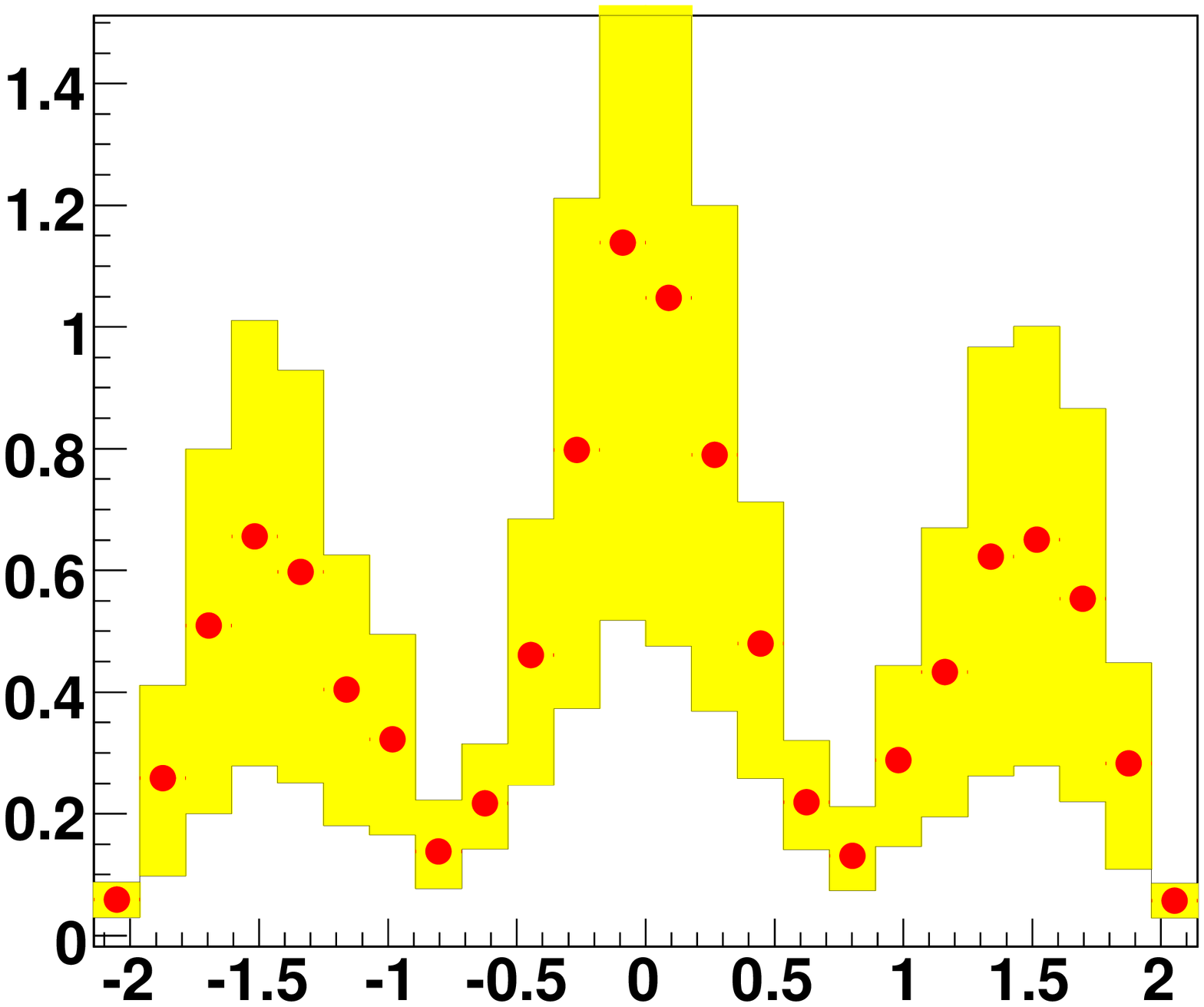}
\includegraphics[width=1.0\textwidth]{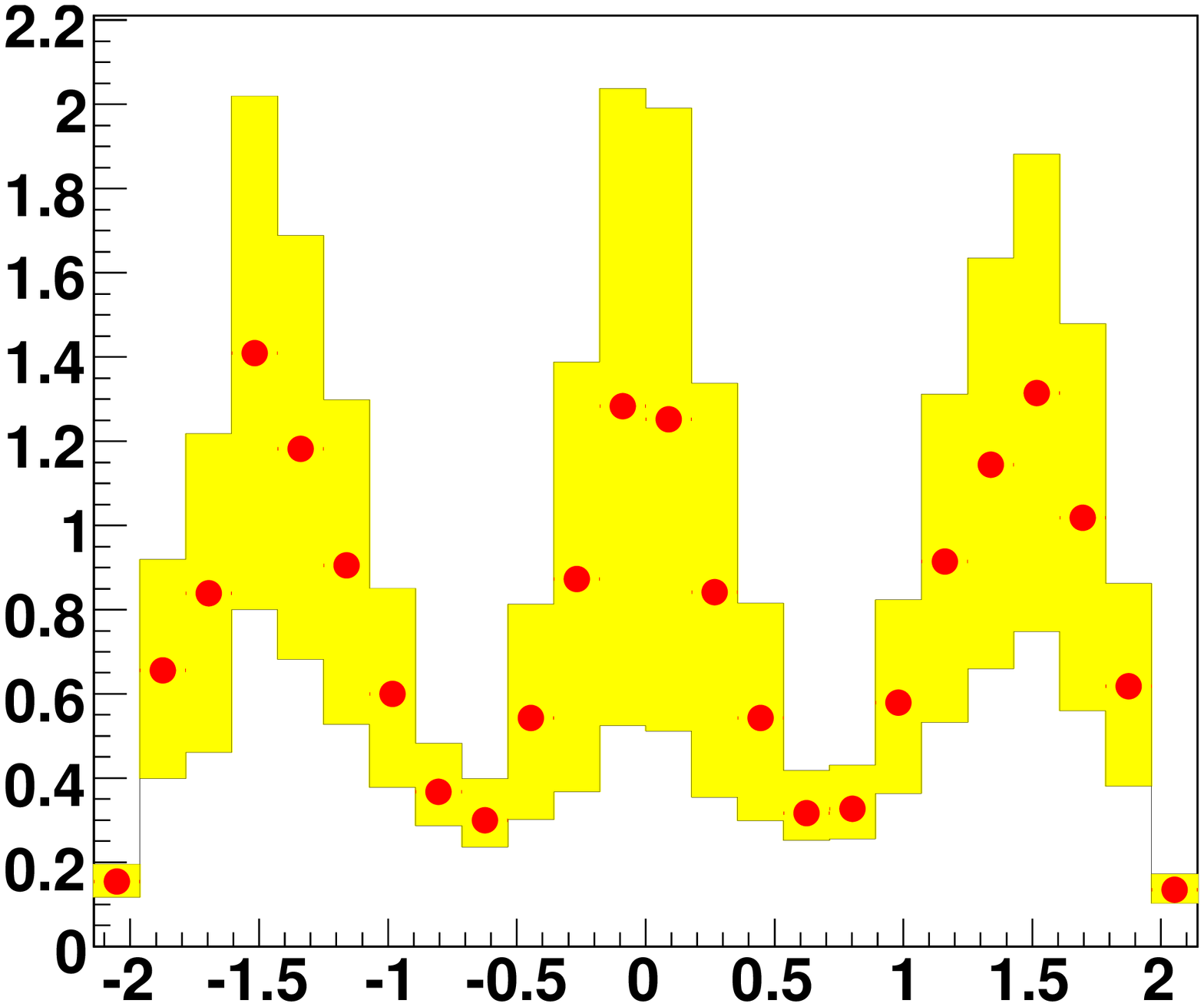}
\includegraphics[width=1.0\textwidth]{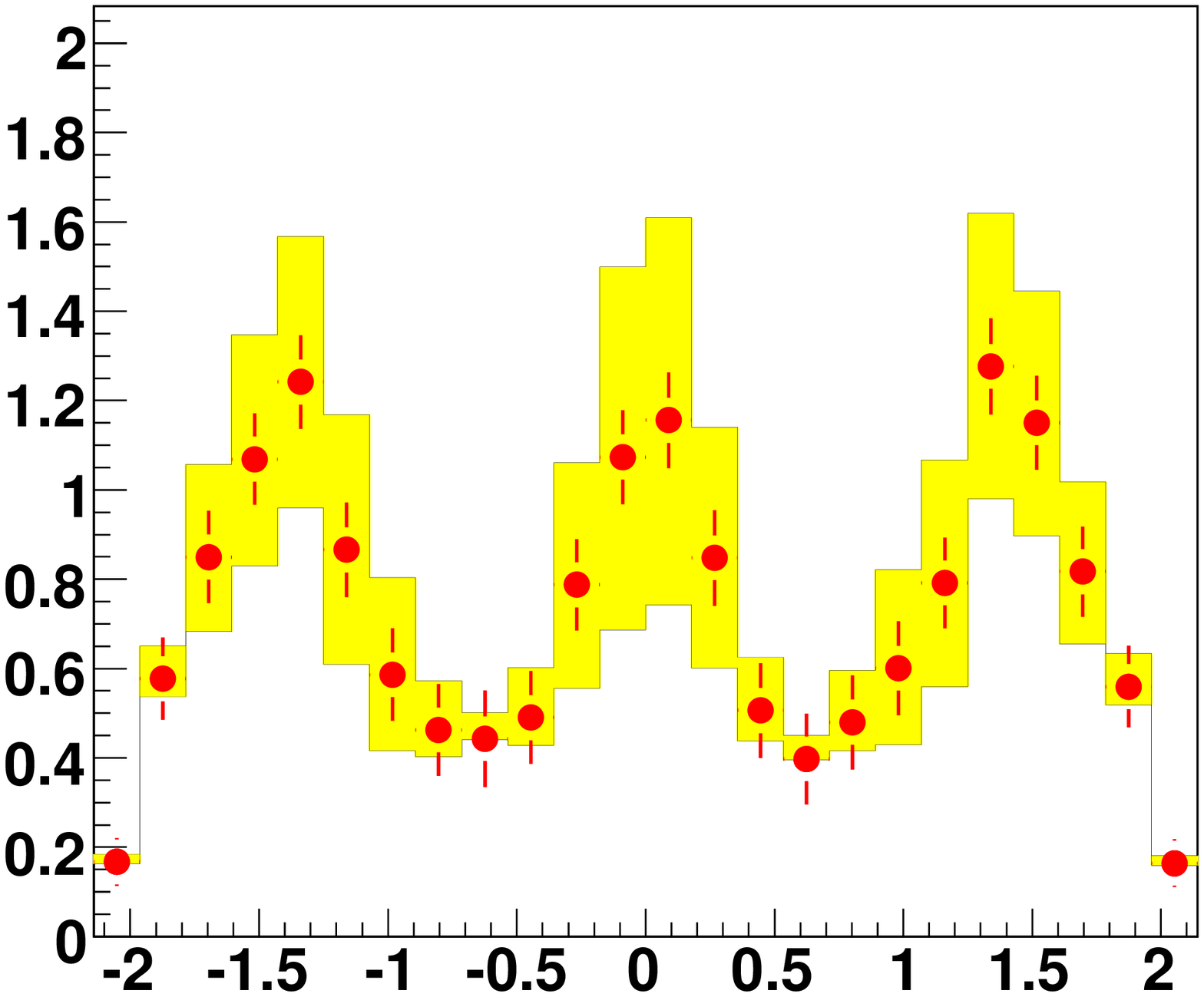}
\includegraphics[width=1.0\textwidth]{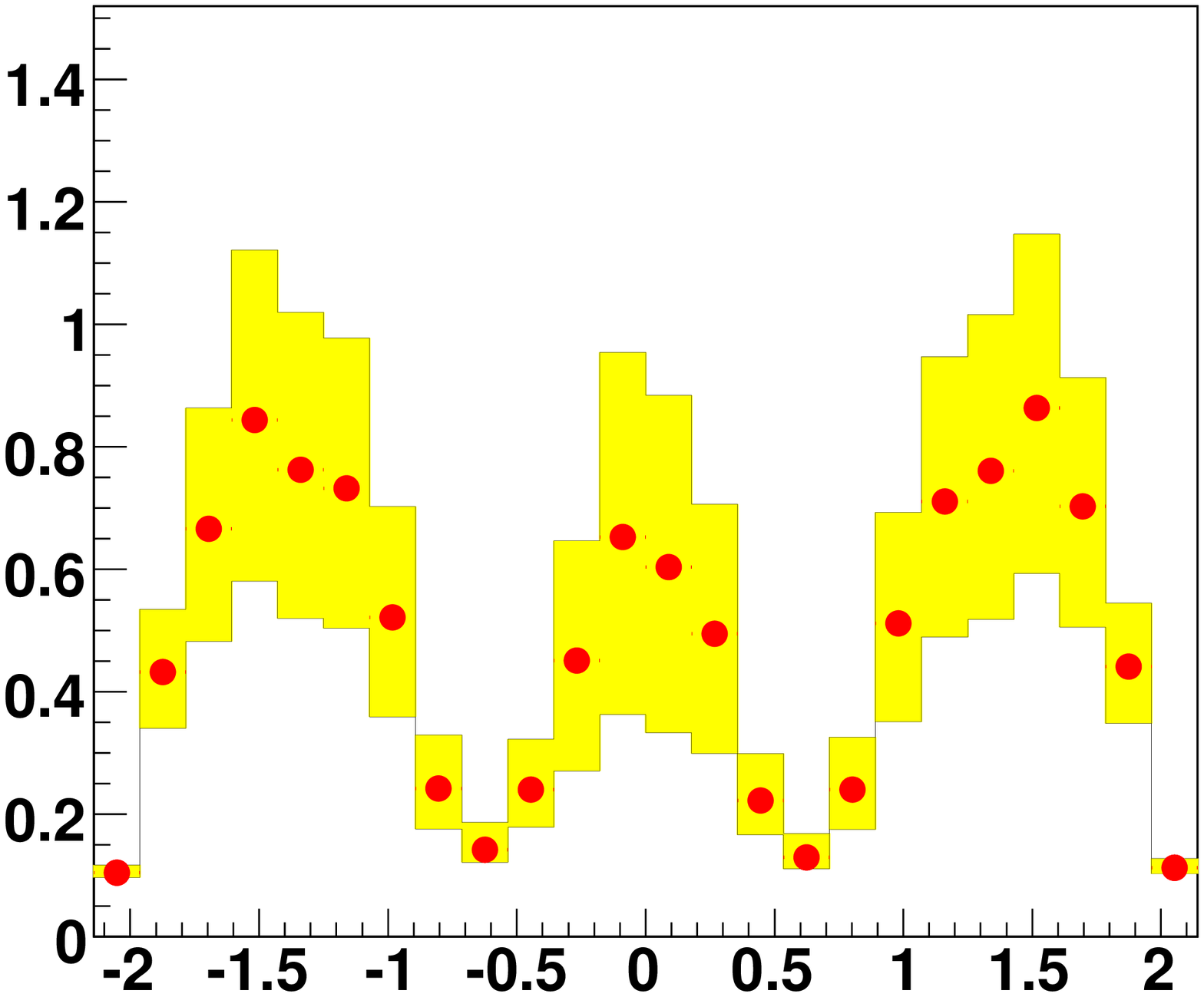}
\end{minipage}
\hfill
\begin{minipage}{0.16\textwidth}
\includegraphics[width=1.0\textwidth]{Plots/blank2.eps}
\end{minipage}
\caption{Background subtracted 3-particle correlations for the uncertainty on normalization factor $a$.  Left:  Upper limit on $a$ where it is obtained from 2-particle ZYA1.  Center:  Lower limit on $a$  where it is the default from 3-particle ZYAM minus the difference between the upper limit and the default.  Right:  Off-diagonal projection from default with systematic uncertainty from the uncertainty on $a$ shown in shaded band.  From top to bottom plots are {\it pp}, d+Au, Au+Au 50-80\%, Au+Au 30-50\%, Au+Au 10-30\%, Au+Au 0-10\%, and ZDC triggered Au+Au 0-12\% collisions at $\sqrt{s_{NN}}=200$ GeV/c.}
\label{fig:a}
\end{figure}

The normalization factor $a$ was at first chosen to be the same as is used in 2-particle correlations.  The assumption used was that the 2-particle correlated yield was zero at $\Delta\phi=1$ (the minimum of the background subtracted 2-particle correlation in Au+Au collisions falls at 1).  This was always known as a lower limit on the yield and an upper limit on $a$ but in 2-particle correlations there is not enough information to do any better.  This is an upper limit because our signal represents the number of particles (in a particular $p_{T}$ range) associated with a high $p_T$ trigger particle which is positive definite.  This can now be improved upon in 3-particle correlations.  The 3-particle correlation signal should be positive definite for the same reason, the number of pairs associated with a high $p_T$ trigger particle should be positive definite.  We therefore determine $a$ using the assumption that the 3-particle correlation signal is zero yield at minimum.  This is done by selecting the 10\% lowest bins (58 of $24 \times 24$) and adjusting $a$ until the average content of these bins is zero.  The lowest bins were redetermined for each adjusted value of $a$.  For the systematic error assigned to our normalization factor $a$, we use our known upper limit, the value from 2-particle ZYA1.  The difference between the default value and the value obtained from the 2-particle ZYA1 is considered to be our uncertainty.  Therefore for our lower limit we subtract this difference from the default value.  Figure~\ref{fig:a} shows the background subtracted 2-particle correlations when $a$ is obtained from 2-particle ZYA1 (our upper systematic limit) and from when $a$ is twice the default value minus the 2-particle ZYA1 value (our lower systematic limit) in the left and right panels, respectively.  Table~\ref{tab:ab} shows the default values of $a$ for each centrality bin along with the systematic errors applied.

\begin{table}[hbtp]
\centering
\caption{Values for normalization factors $a$ and $b$ with systematic errors.  ZDC triggered Au+Au collisions are divided into centrality bins as if they were minimum bias events.  The cross section numbers in quotes correspond to the cross sections as if events were minimum bias but not the actual cross sections for these events.}
\begin{tabular}{|l|r|r|r|}
\hline
Collision&Centrality&$a$&$b$\\
\hline
pp&0-100\%&$0.687\pm1.040$&$0.81+0.60-0.82$\\
d+Au&20-100\%&$1.001\pm0.039$&$1.264+0.016-0.022$\\
d+Au&10-20\%&$0.652\pm0.177$&$1.342+0.030-0.049$\\
d+Au&0-10\%&$0.748\pm0.093$&$1.339+0.011-0.018$\\
Au+Au&70-80\%&$0.946\pm0.140$&$0.980+0.015-0.027$\\
Au+Au&60-70\%&$0.964\pm0.081$&$0.9930+0.0036-0.0059$\\
Au+Au&50-60\%&$0.971\pm0.053$&$0.9965+0.0011-0.0019$\\
Au+Au&40-50\%&$0.959\pm0.053$&$0.9982+0.0008-0.0012$\\
Au+Au&30-40\%&$0.956\pm0.051$&$0.9991+0.0005-0.0008$\\
Au+Au&20-30\%&$0.969\pm0.035$&$0.9996+0.0004-0.0006$\\
Au+Au&10-20\%&$0.980\pm0.021$&$0.99998+0.00019-0.00026$\\
Au+Au&5-10\%&$0.976\pm0.020$&$1.00003+0.00011-0.00019$\\
Au+Au&0-5\%&$0.986\pm0.013$&$1.00023+0.00008-0.00012$\\
Au+Au ZDC&``20-30\%''&$0.944\pm0.057$&$0.99940+0.00070-0.00084$\\
Au+Au ZDC&``10-20\%''&$0.980\pm0.022$&$0.99963+0.00010-0.00012$\\
Au+Au ZDC&``5-10\%''&$0.982\pm0.015$&$1.00001+0.00005-0.00009$\\
Au+Au ZDC&0-5\%&$0.994\pm0.005$&$1.00021+0.00003-0.00005$\\
\hline
\end{tabular}
\label{tab:ab}
\end{table}

In the beginning of this analysis, the normalization factor $b$ was not applied, the equivalent of $b=1$.  It was obvious in d+Au collisions that this factor was necessary, for without this factor there was a large pedestal in the background subtracted 3-particle correlation signal.  This factor is necessary because the multiplicity distributions are non-Poisson.  If the events were Poisson than $\langle N(N-1) \rangle = \langle N \rangle^2$.  From data we know this is not true.  We did not require the events to be Poisson in our analysis, but assumed that $\langle N(N-1) \rangle / \langle N \rangle^2$ is the same for the triggered events, inclusive events and the underlying background.  This is a less stringent requirement.  With this assumption the soft-soft and trigger flow terms are scaled by $a^2$.  However, we have found that $\langle N(N-1) \rangle / \langle N \rangle^2$ is not the same for the triggered events and the inclusive events so we introduce $b$ as a correction, where  

\begin{equation}
b=\frac{\frac{\langle N_{trig}(N_{trig}-1) \rangle}{\langle N_{trig}\rangle^{2}}}{\frac{\langle N_{inc}(N_{inc}-1)\rangle}{\langle N_{inc}\rangle^{2}}}.
\end{equation}
We are therefore assuming that the underlying background deviates from Poisson similar to the deviation in triggered events.

\begin{figure}[htbp]
\hfill
\begin{minipage}{0.16\textwidth}
\includegraphics[width=1.0\textwidth]{Plots/blank2.eps}
\end{minipage}
\hfill
\begin{minipage}{0.20\textwidth}
\includegraphics[width=1.0\textwidth]{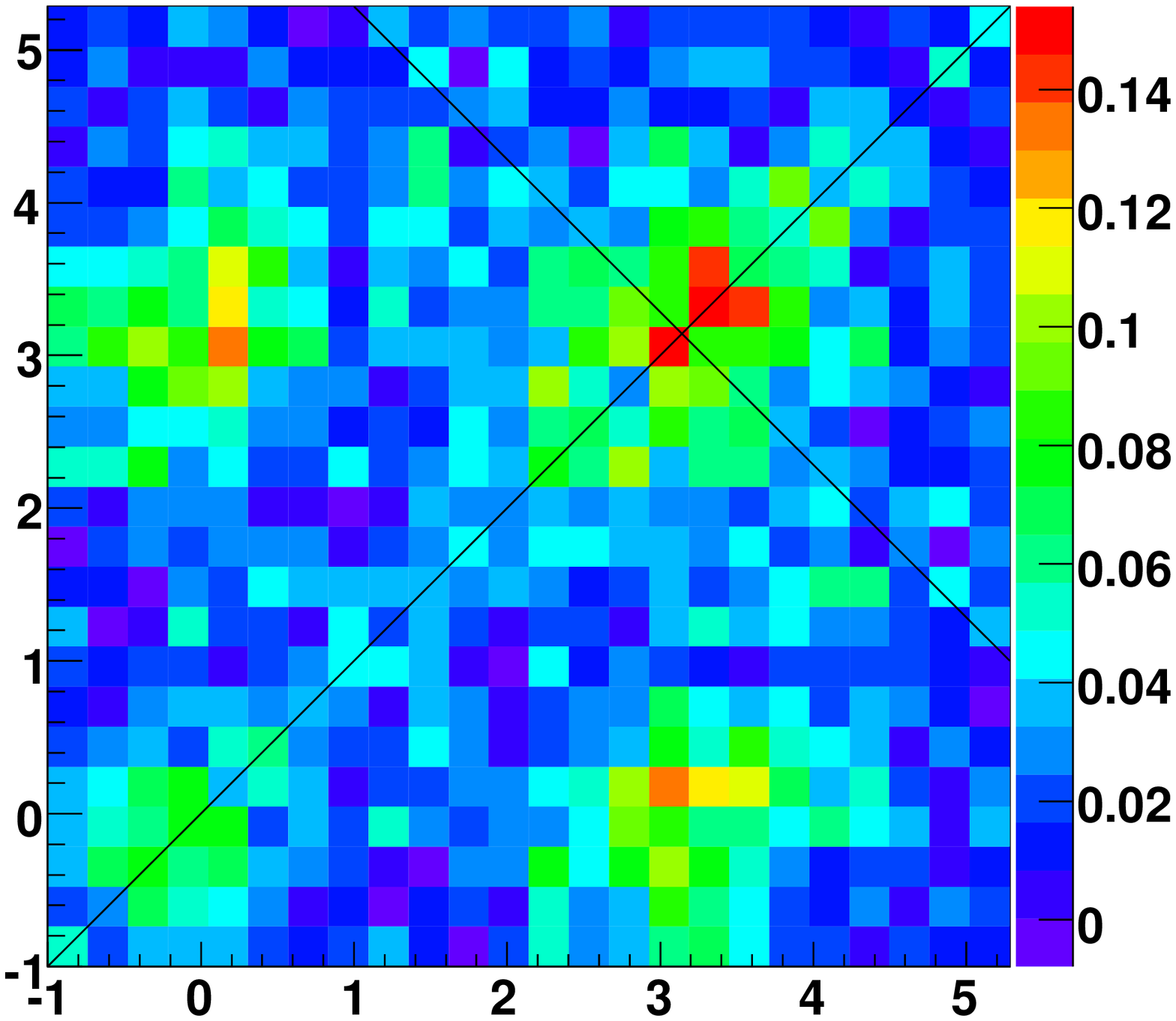}
\includegraphics[width=1.0\textwidth]{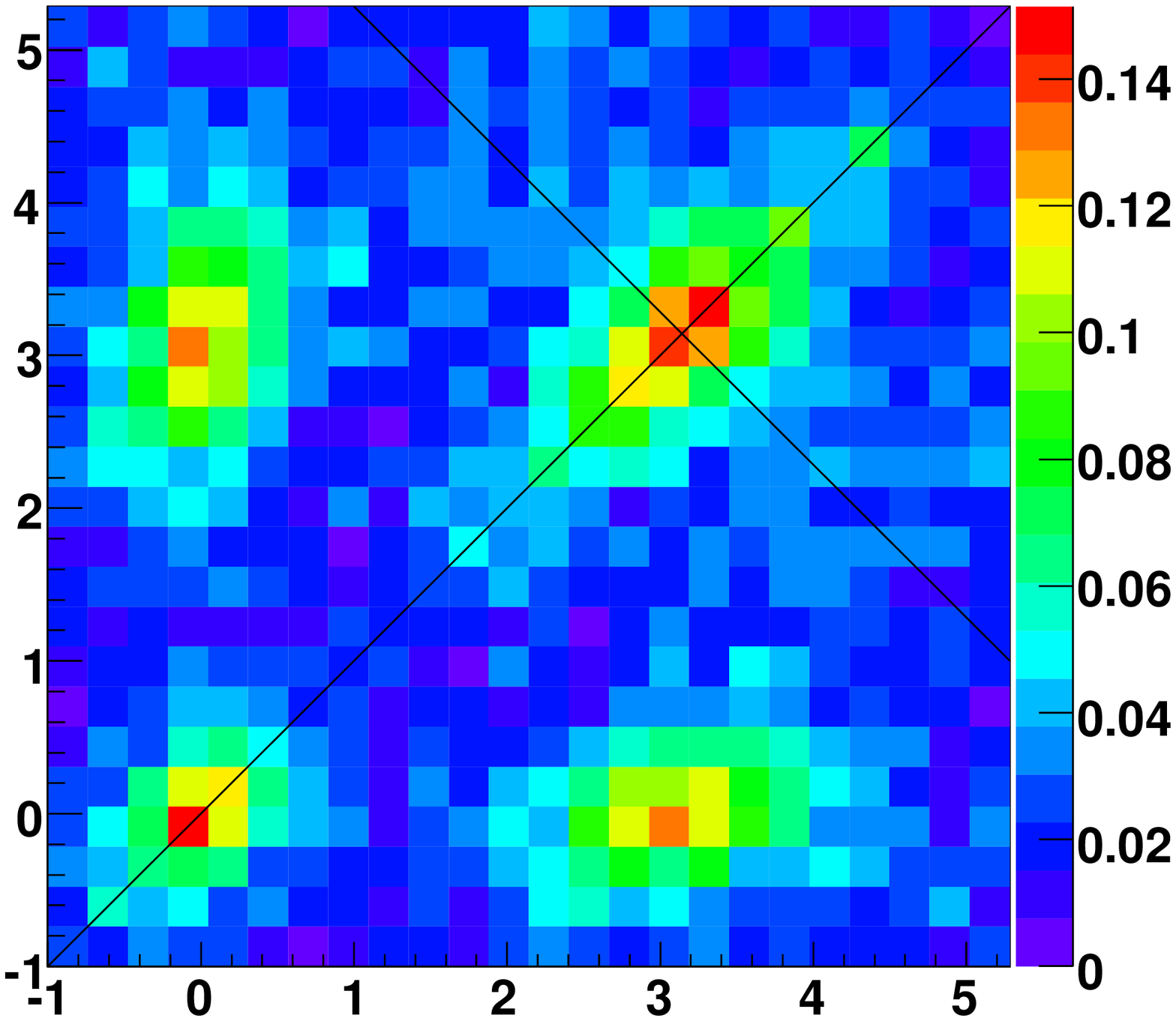}
\includegraphics[width=1.0\textwidth]{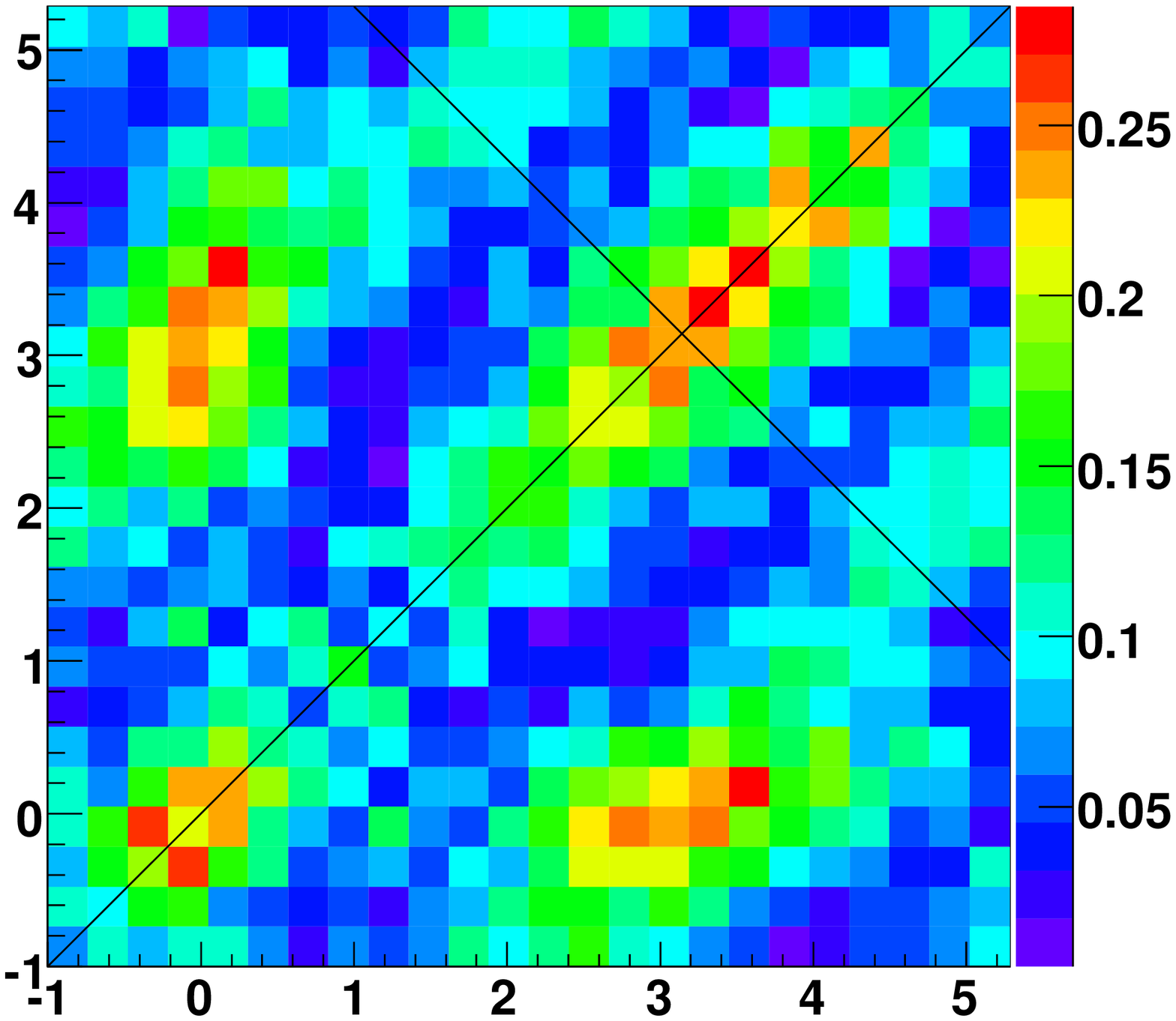}
\includegraphics[width=1.0\textwidth]{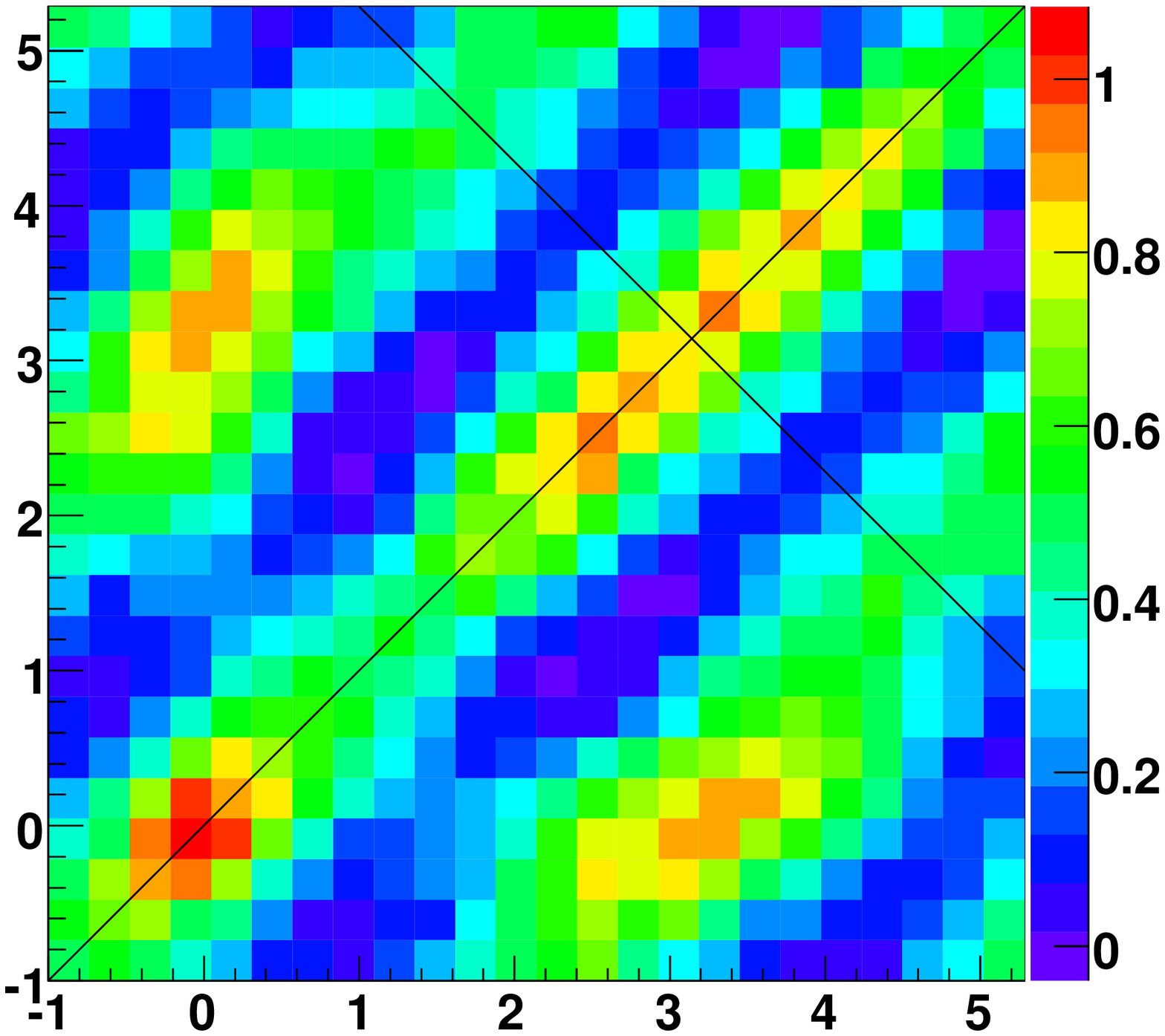}
\includegraphics[width=1.0\textwidth]{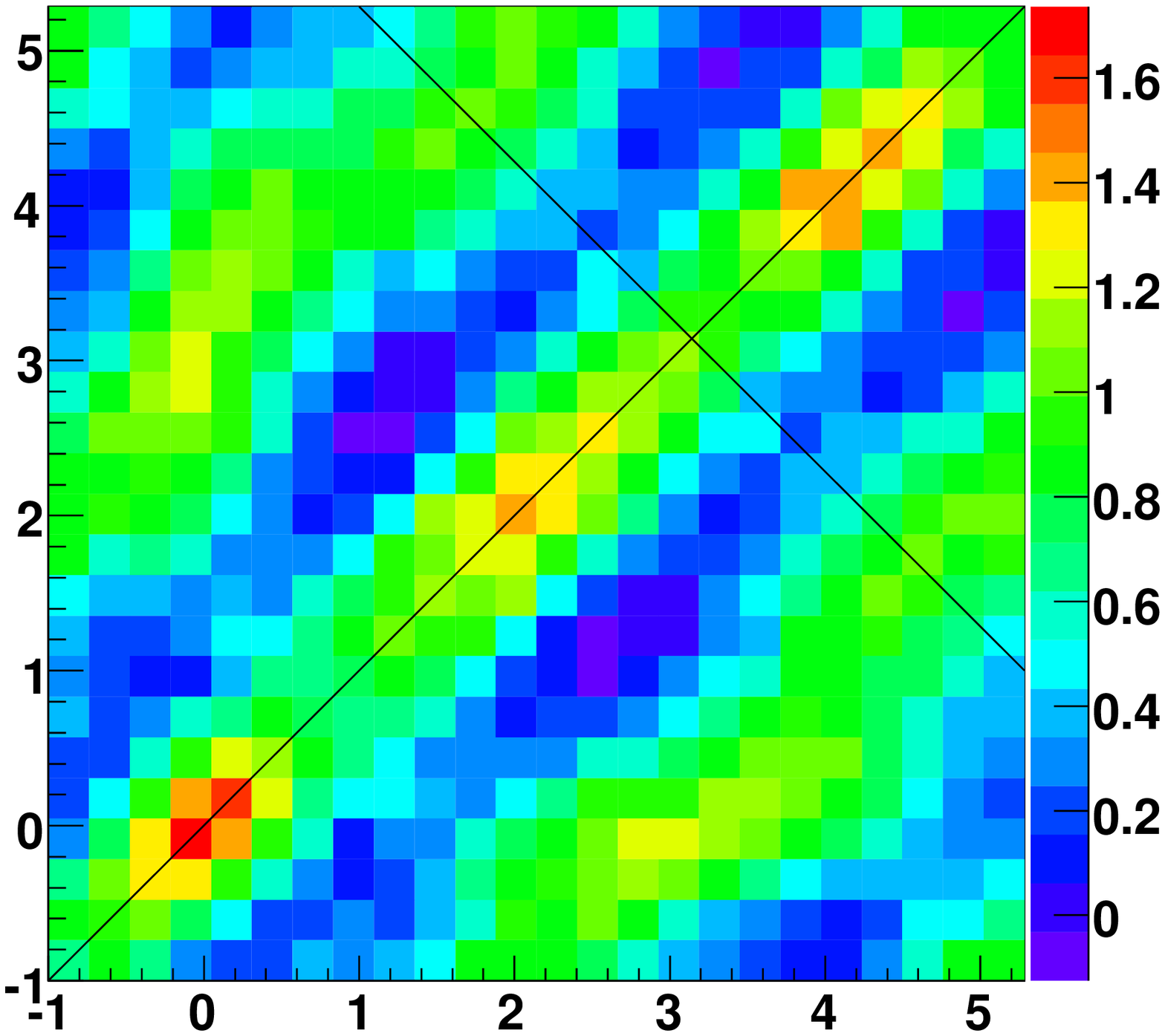}
\includegraphics[width=1.0\textwidth]{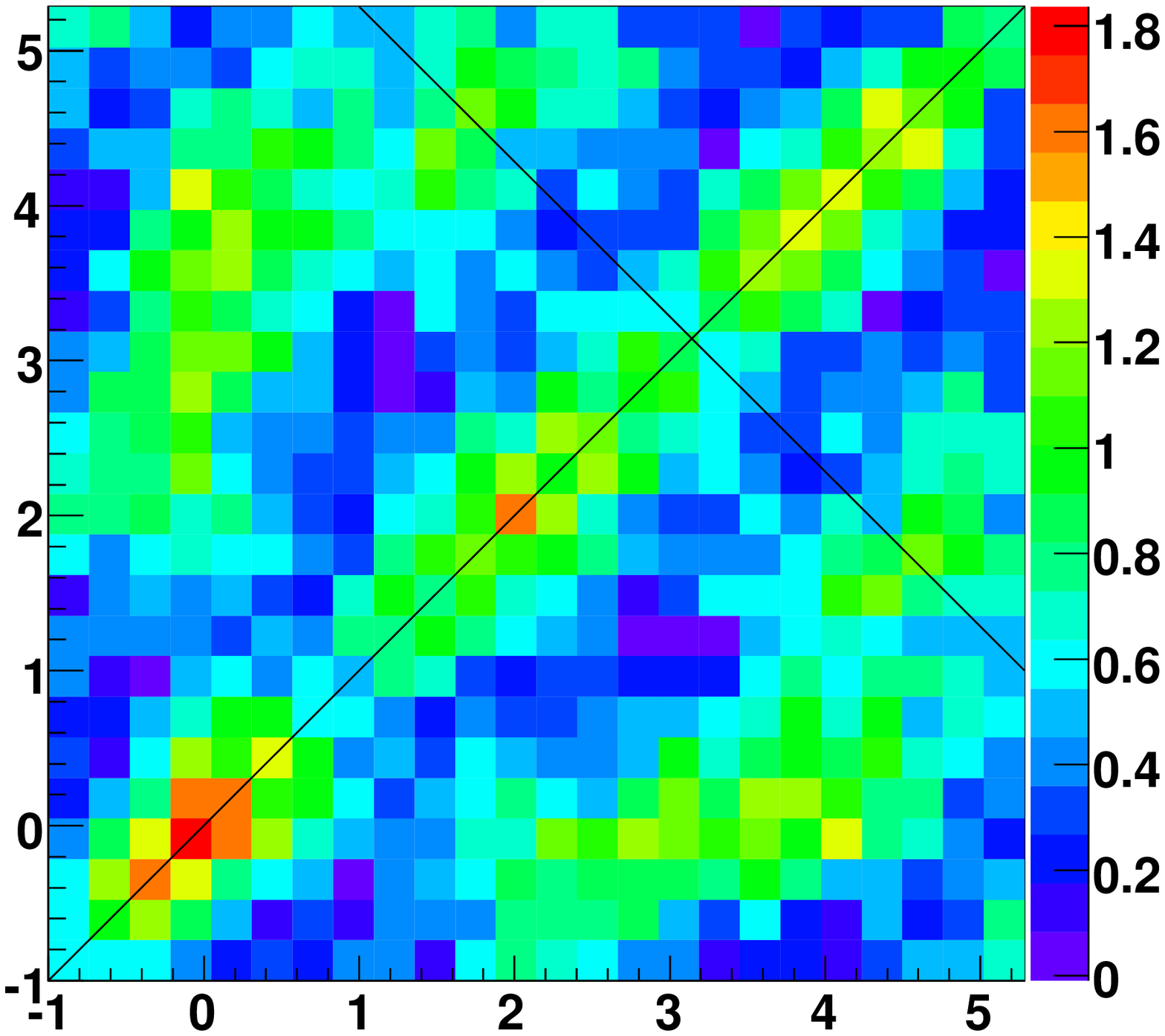}
\includegraphics[width=1.0\textwidth]{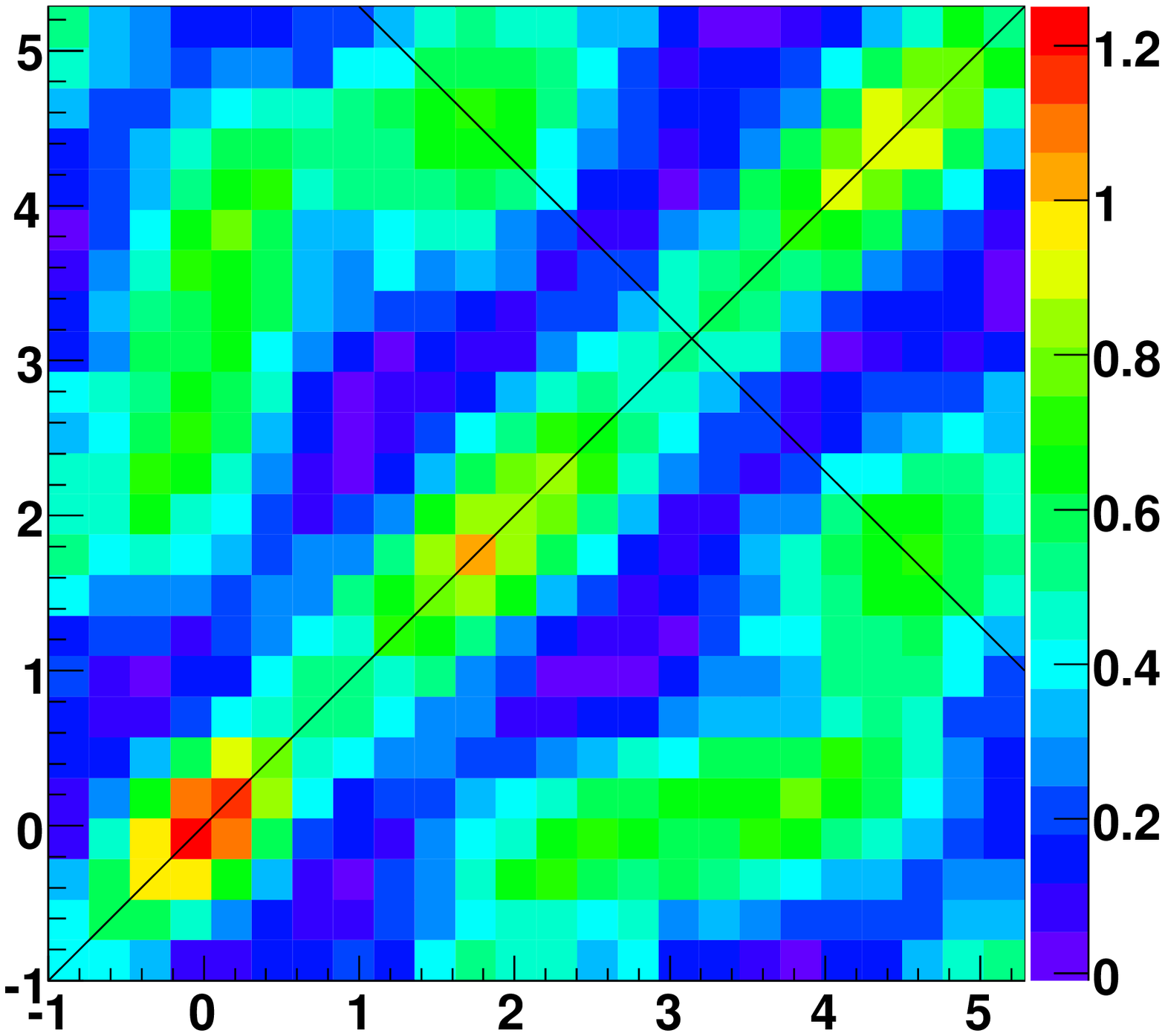}
\end{minipage}
\hfill   
\begin{minipage}{0.20\textwidth}
\includegraphics[width=1.0\textwidth]{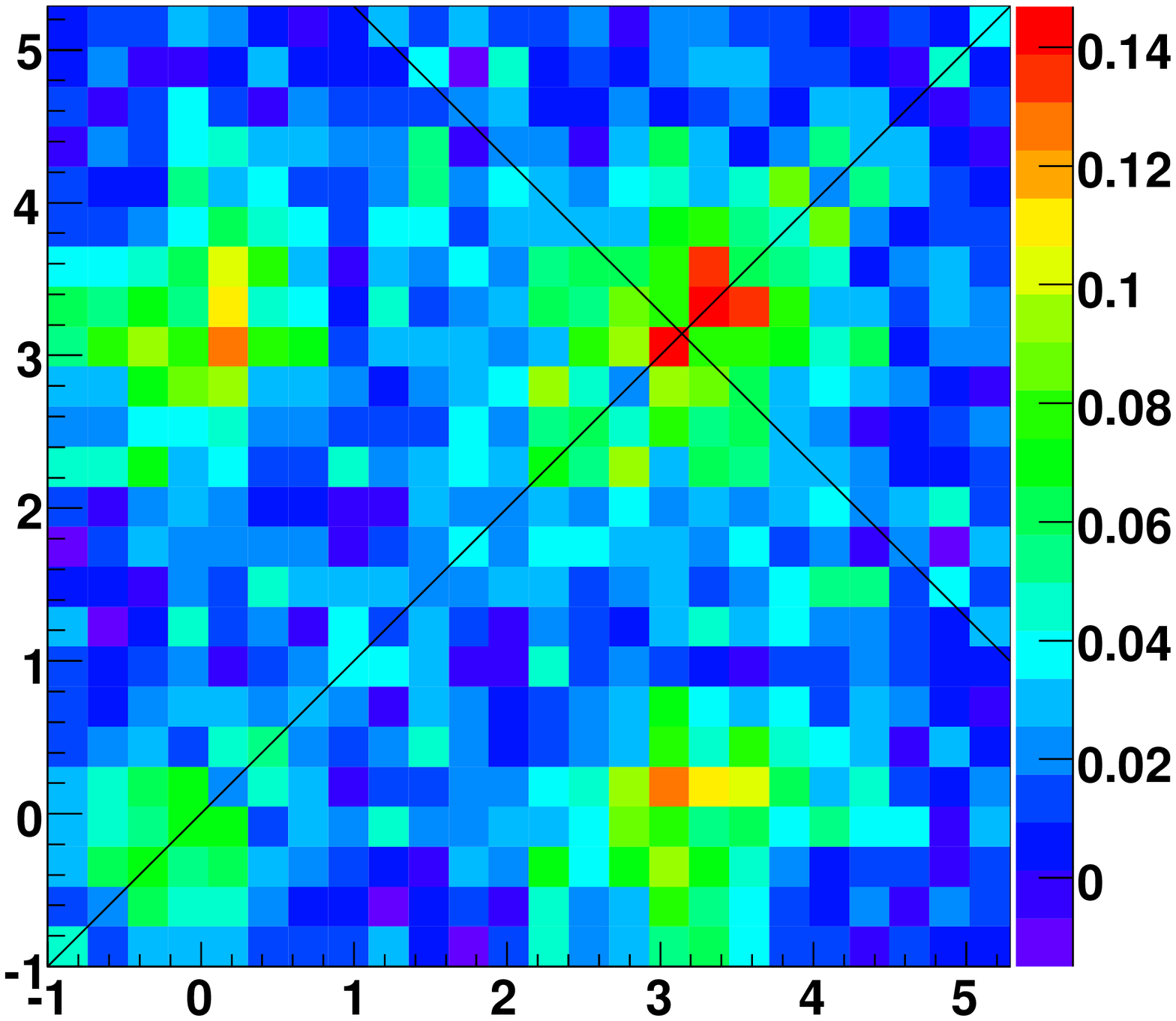}
\includegraphics[width=1.0\textwidth]{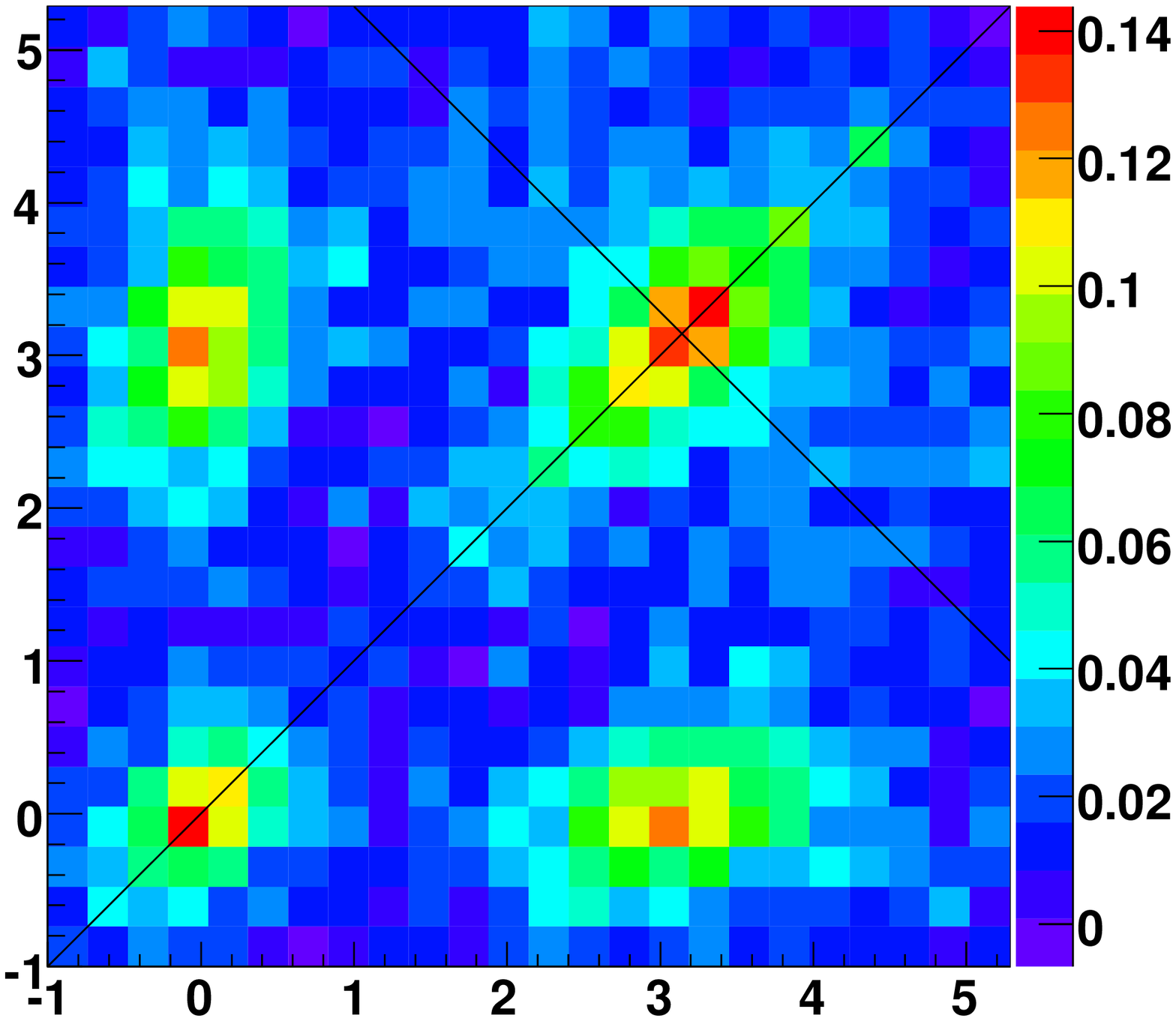}
\includegraphics[width=1.0\textwidth]{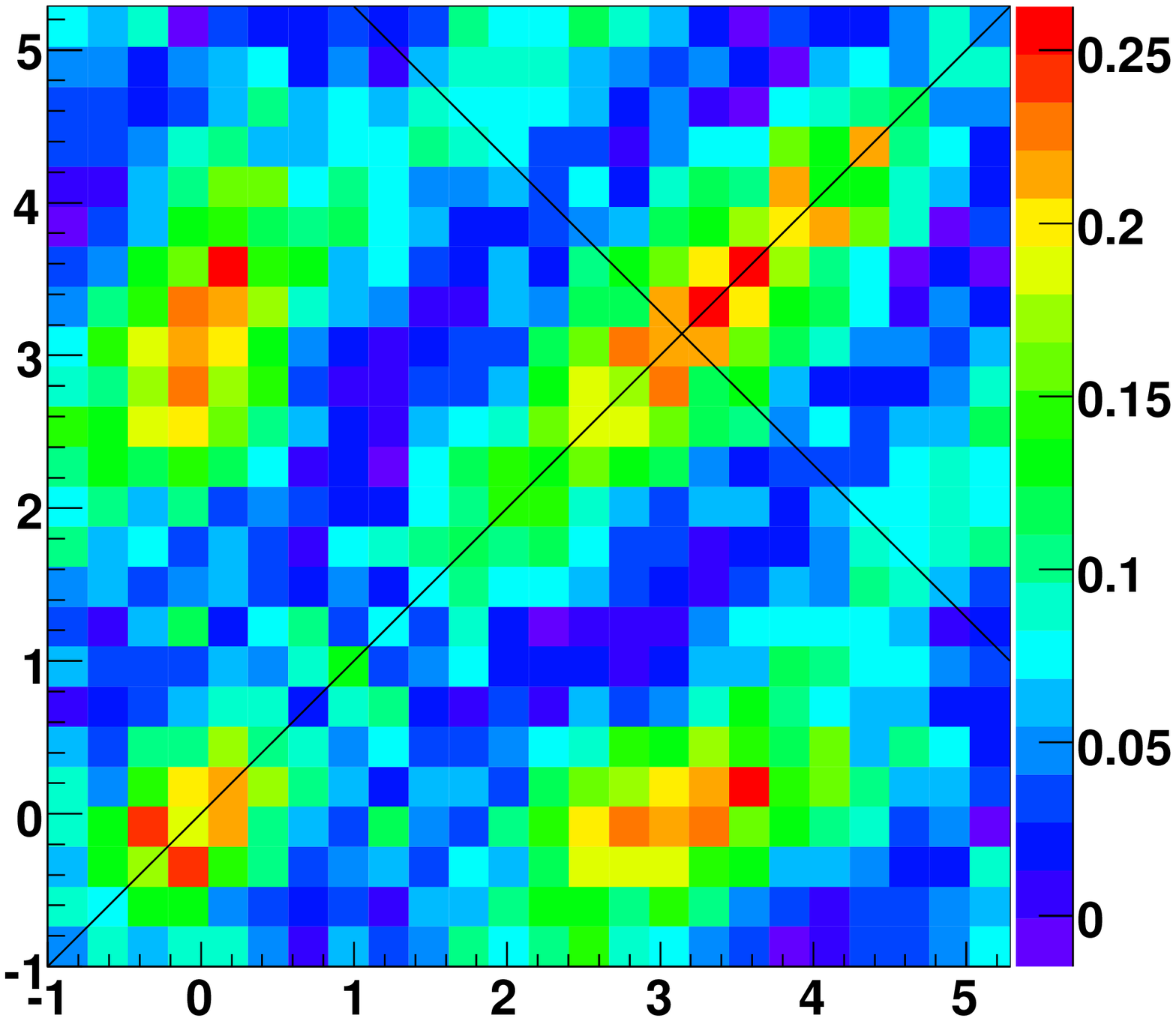}
\includegraphics[width=1.0\textwidth]{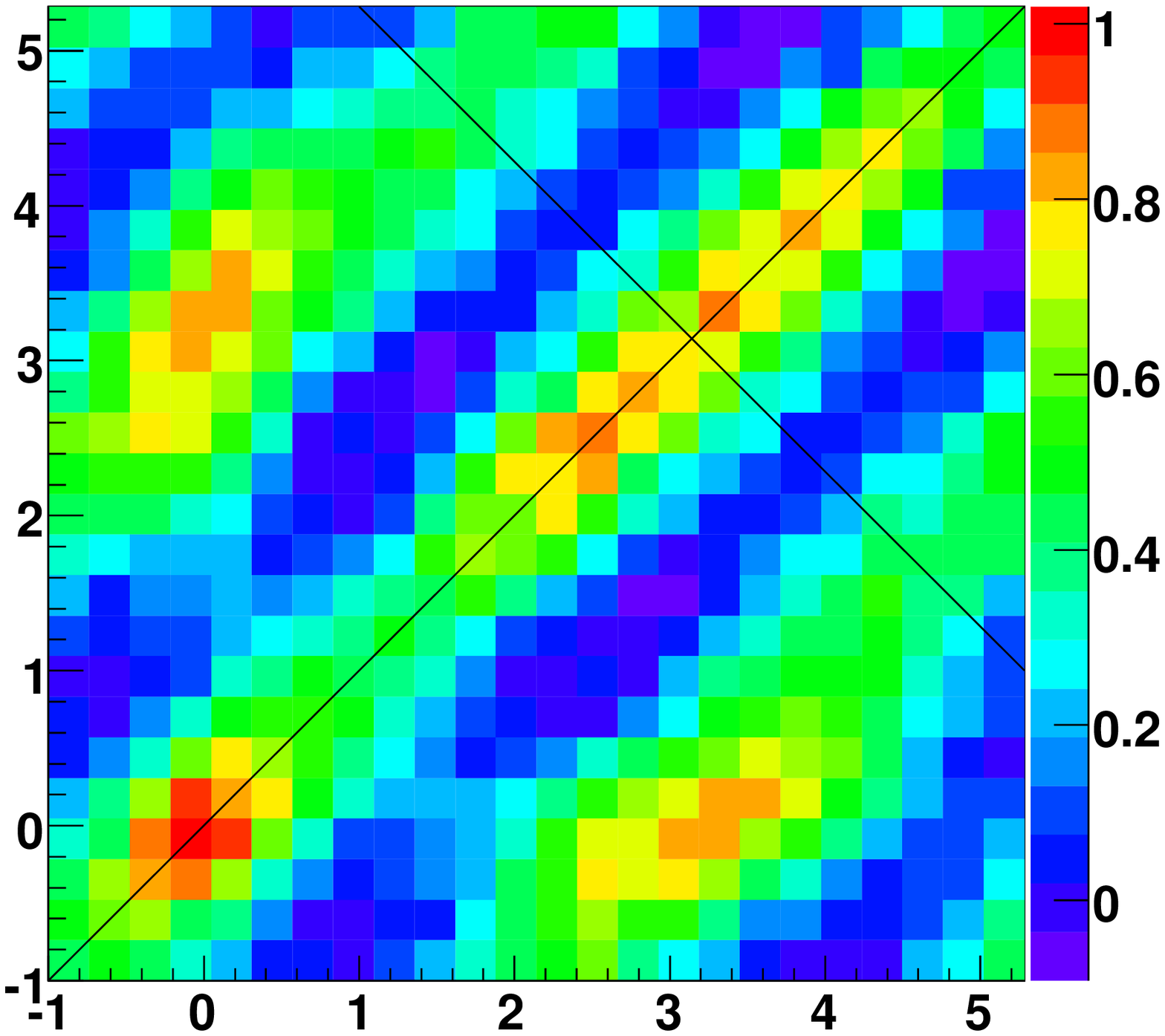}
\includegraphics[width=1.0\textwidth]{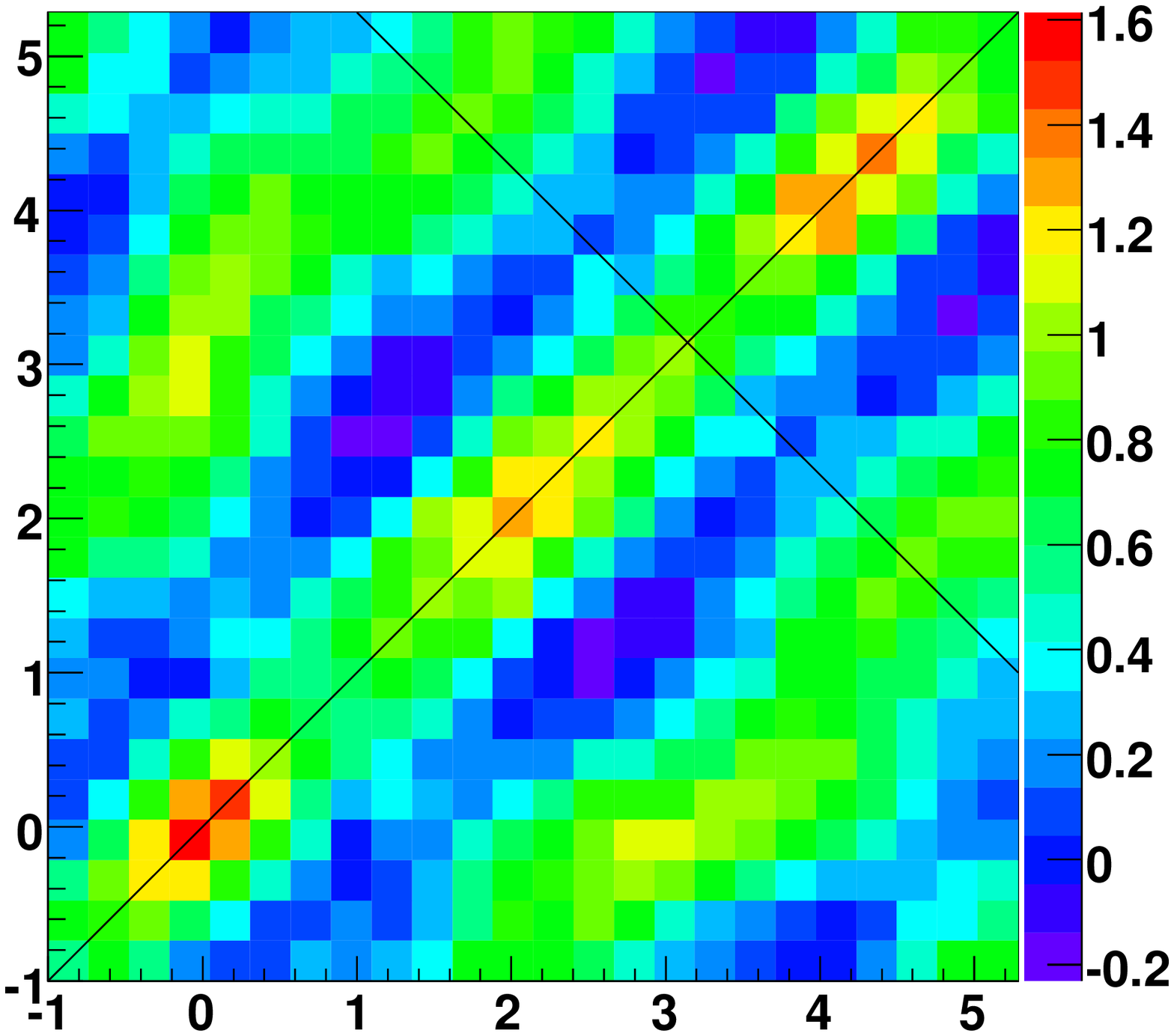}
\includegraphics[width=1.0\textwidth]{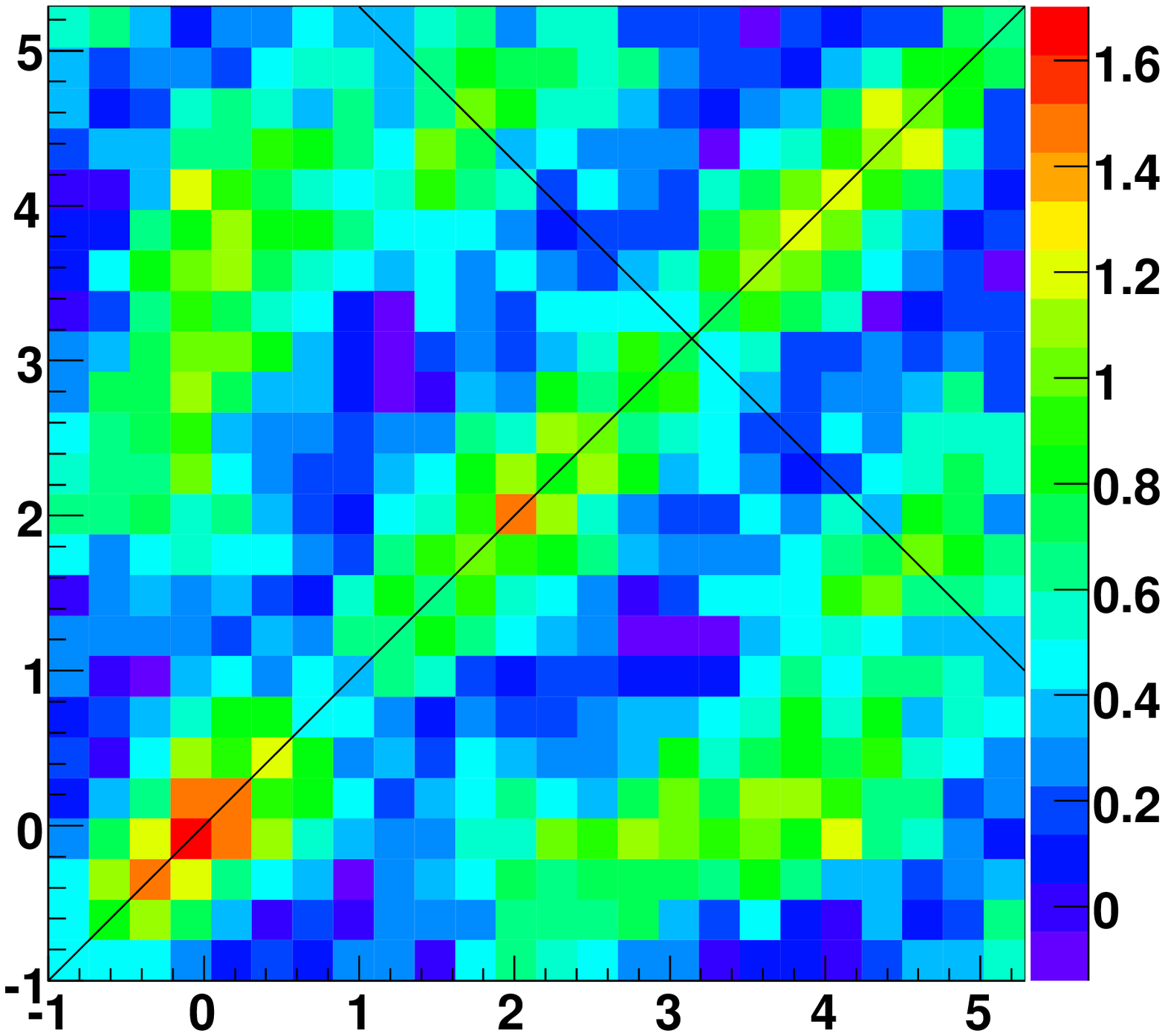}
\includegraphics[width=1.0\textwidth]{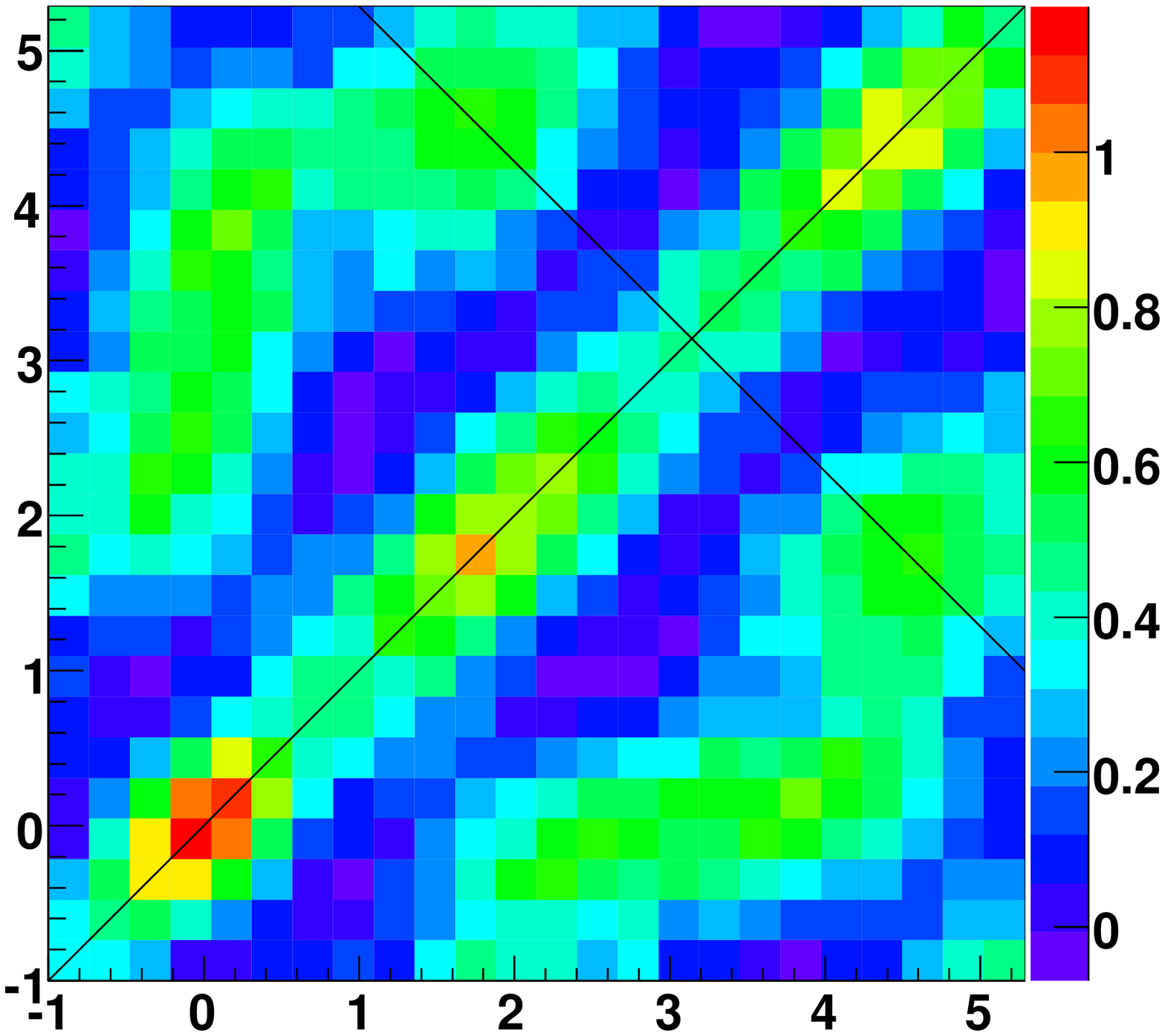}
\end{minipage}
\hfill
\begin{minipage}{0.20\textwidth}
\includegraphics[width=1.0\textwidth]{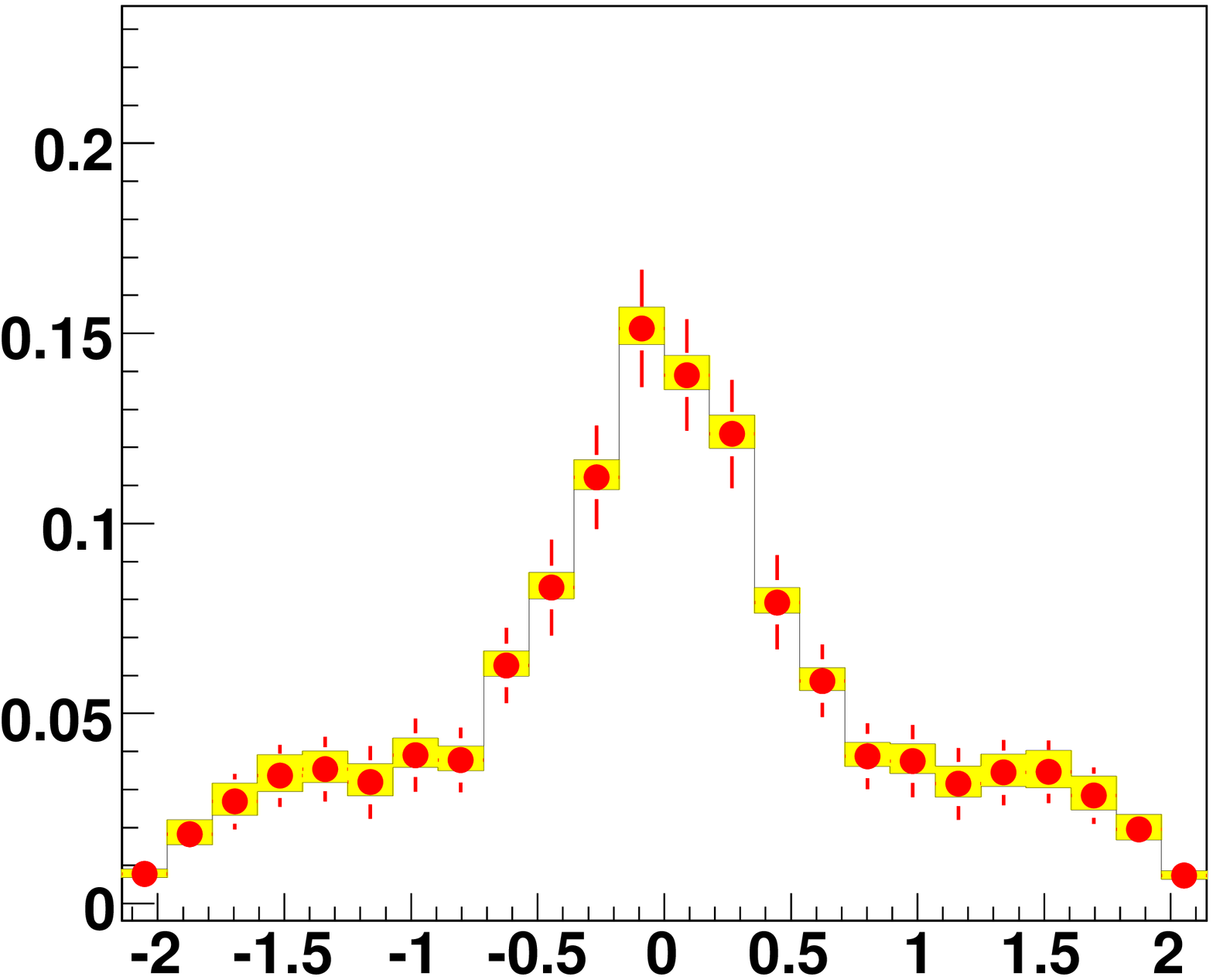}
\includegraphics[width=1.0\textwidth]{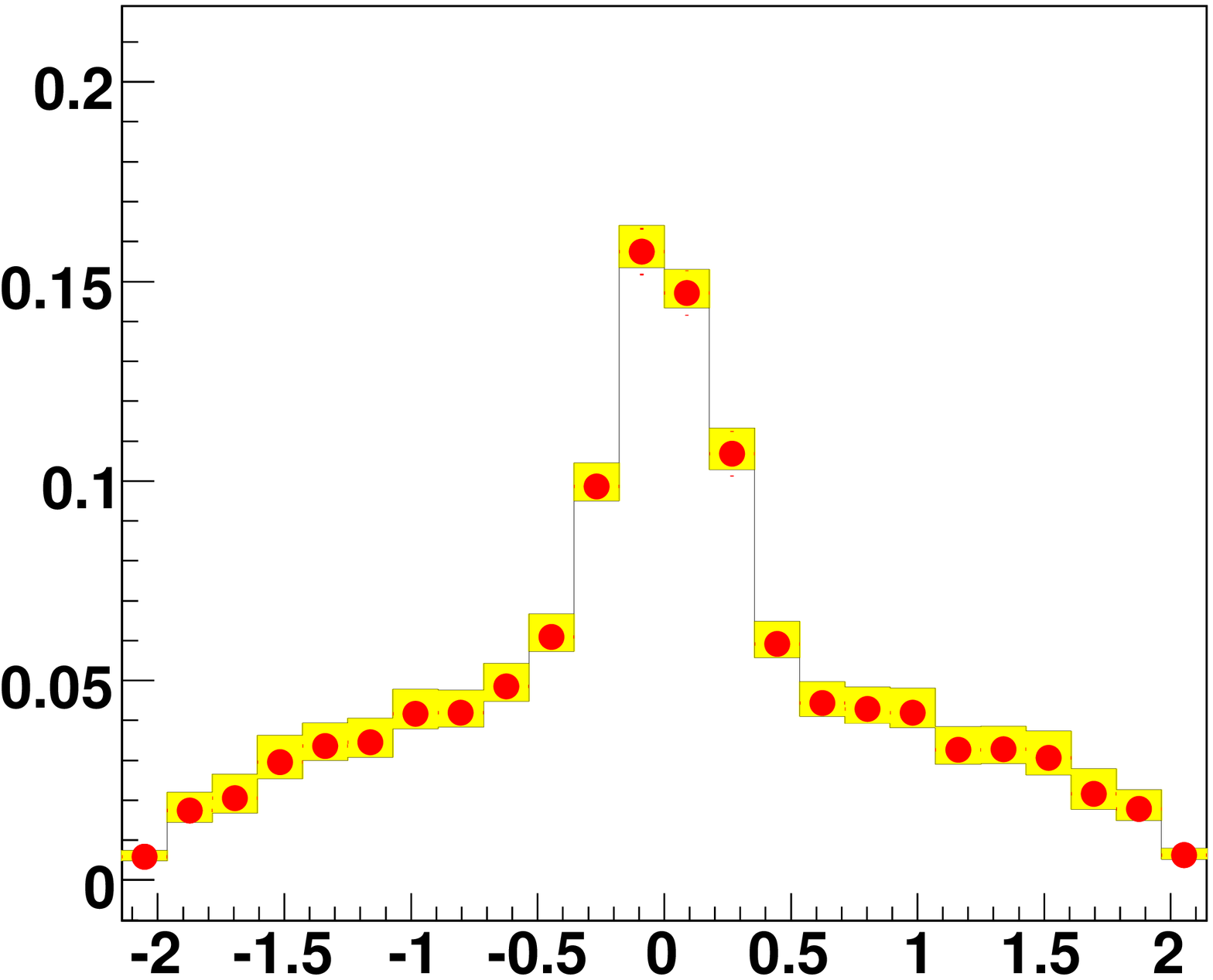}
\includegraphics[width=1.0\textwidth]{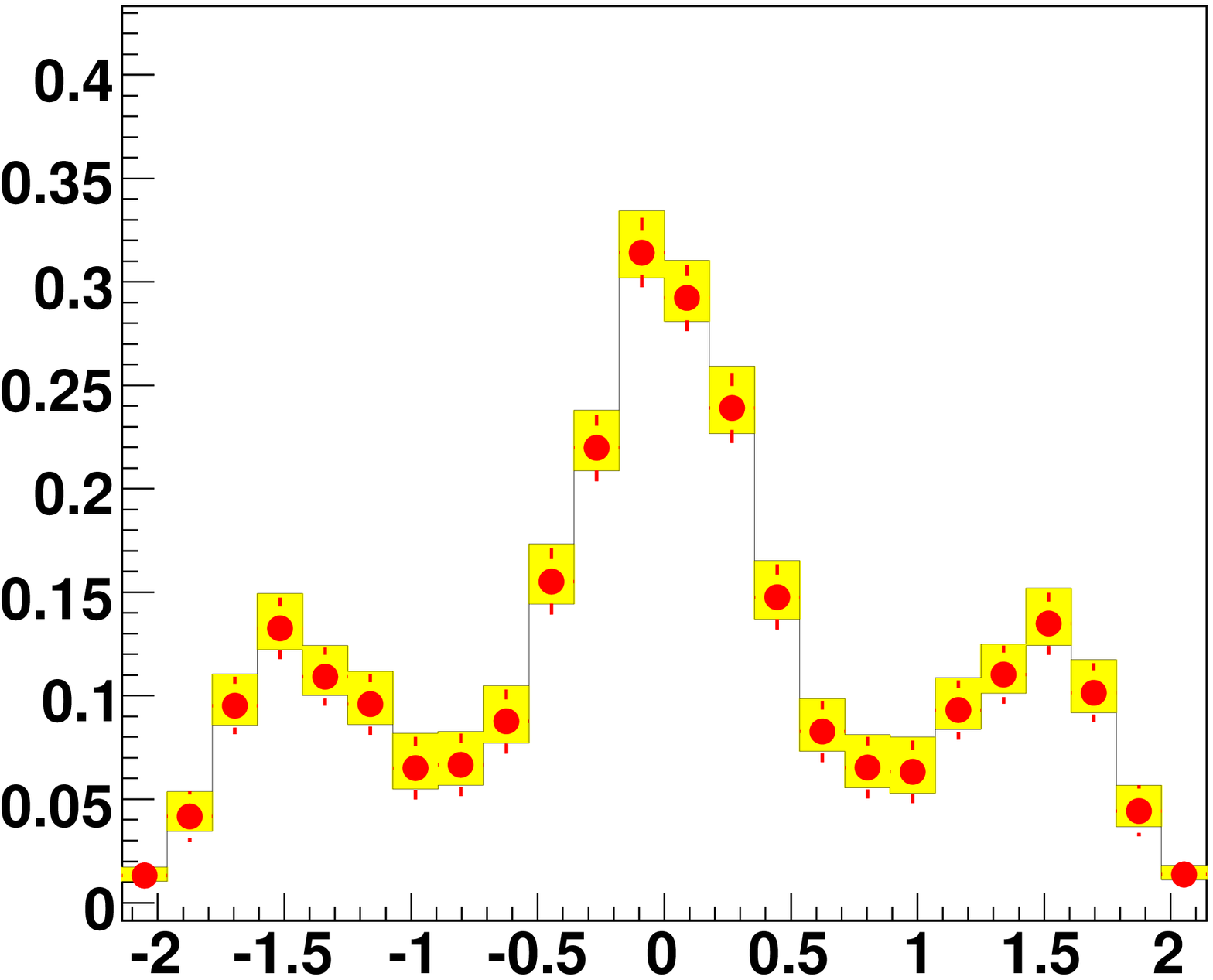}
\includegraphics[width=1.0\textwidth]{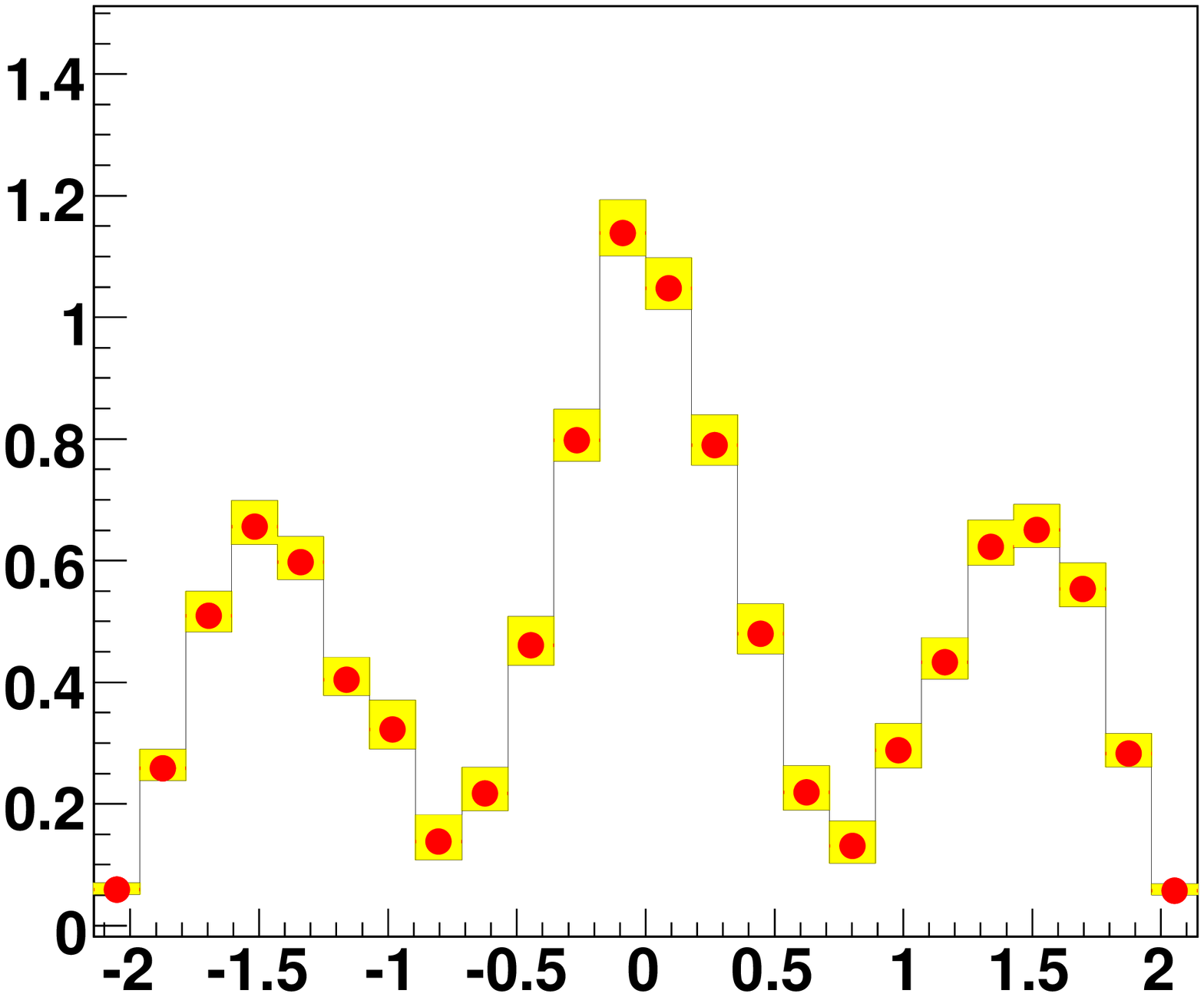}
\includegraphics[width=1.0\textwidth]{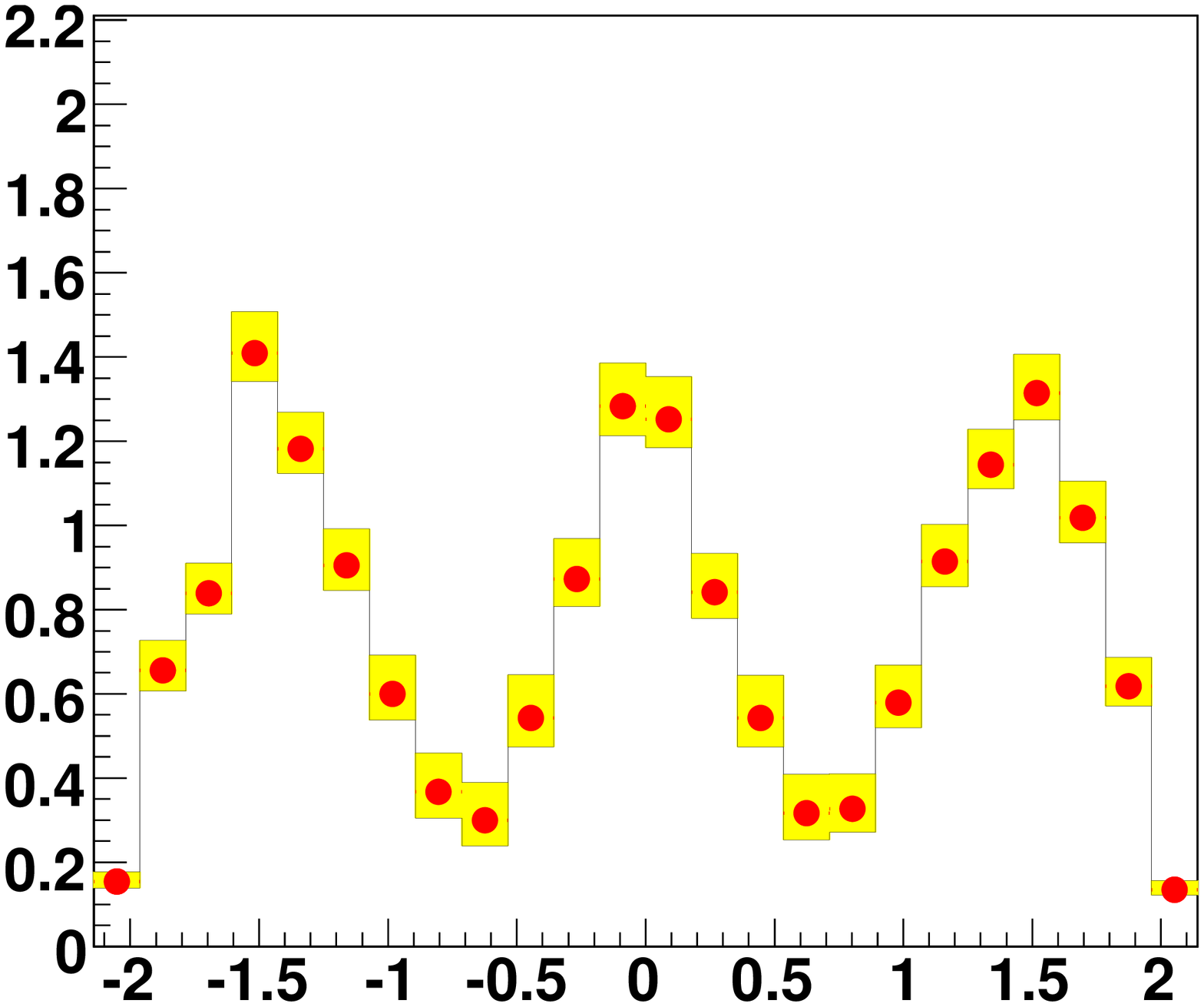}
\includegraphics[width=1.0\textwidth]{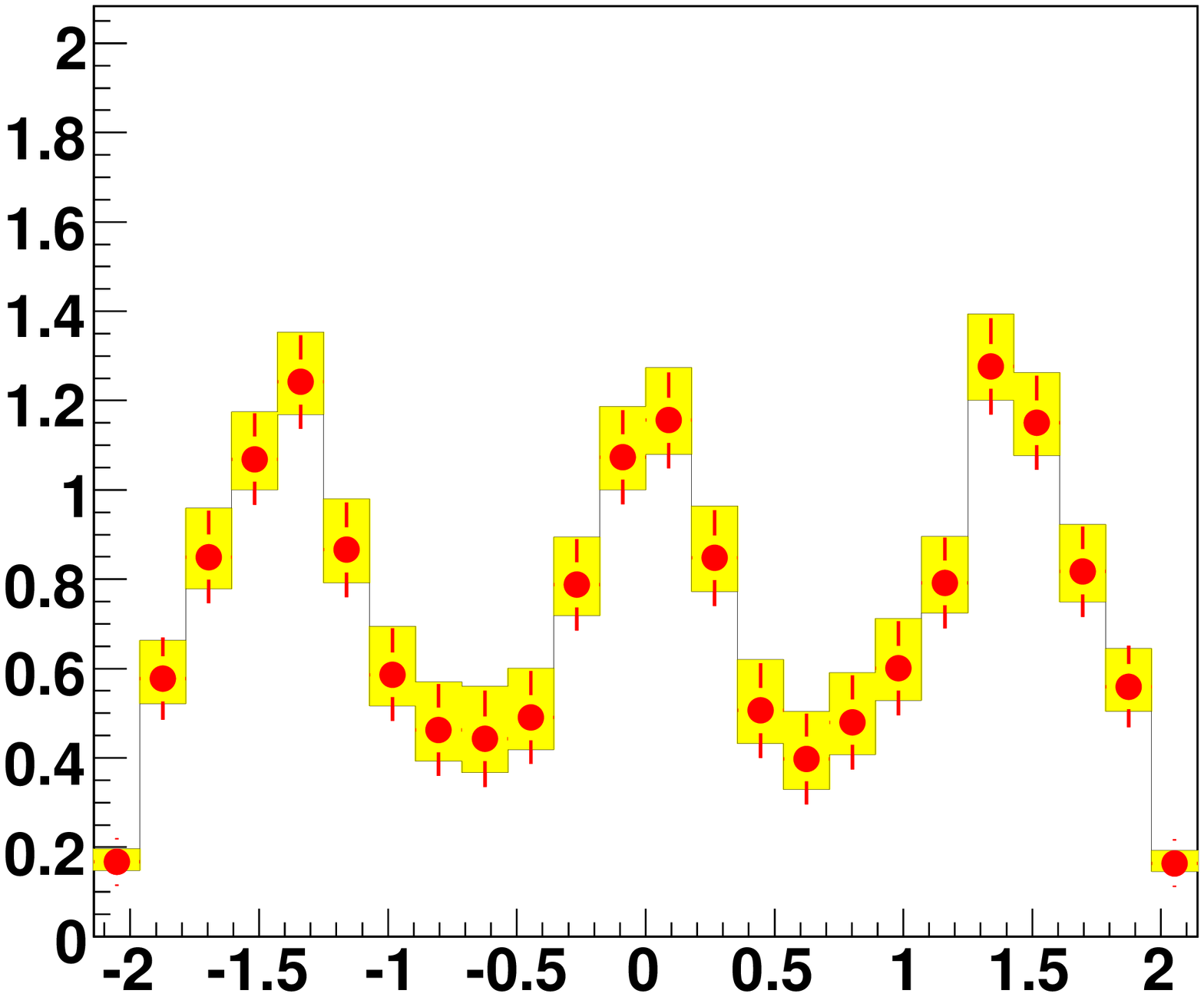}
\includegraphics[width=1.0\textwidth]{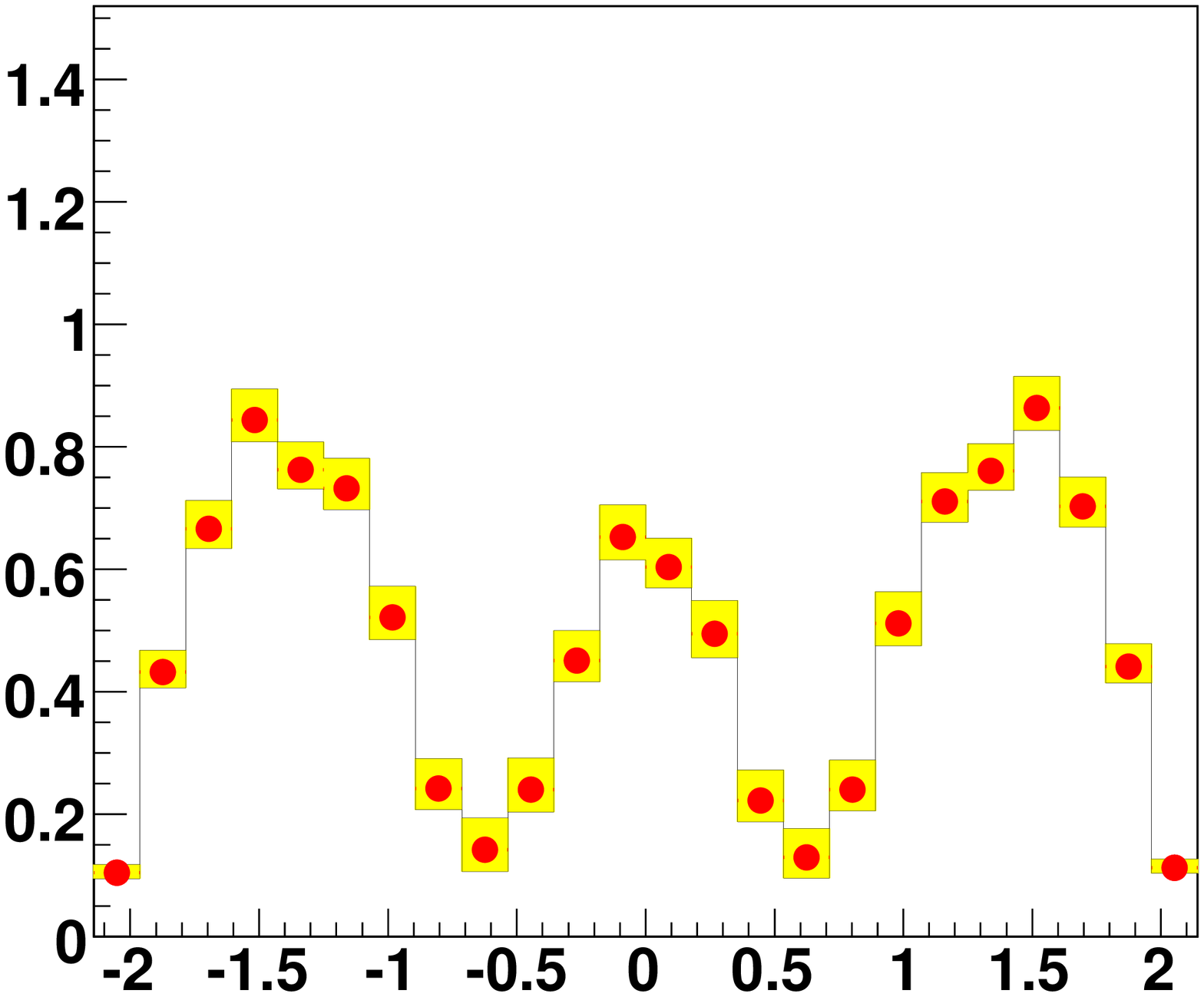}
\end{minipage}
\hfill
\begin{minipage}{0.16\textwidth}
\includegraphics[width=1.0\textwidth]{Plots/blank2.eps}
\end{minipage}
\caption{Background subtracted 3-particle correlations for the uncertainty on normalization factor $b$.  Left:  $b$ from 5\% of bins for ZYAM.  Center:  $b$ from 15\% of the bins for ZYAM.  Right:  Off-diagonal away-side projection from default with systematics uncertainty from the uncertainty on $b$ shown in shaded band.  From top to bottom plots are {\it pp}, d+Au, Au+Au 50-80\%, Au+Au 30-50\%, Au+Au 10-30\%, Au+Au 0-10\%, and ZDC triggered Au+Au 0-12\% collisions at $\sqrt{s_{NN}}=200$ GeV/c.}
\label{fig:b}
\end{figure}

The uncertainty applied to $b$ comes from the uncertainty in the 3-particle ZYAM.  This takes care of our uncertainty on ZYAM due to the number of bins used and gives us a relative change between $a$ and $b$.  We use the default value of $a$ that was determined by the 3-particle ZYAM, using 10\% of the bins, with the default value of $b$.  We then change the number of bins used in the ZYAM from 10\% to 5\% and 15\%.  Figure~\ref{fig:b} shows the background subtracted 3-particle correlations when $b$ is obtained from the 3-particle ZYAM using the default $a$ for 5\% and 15\% of the bins.  Table~\ref{tab:ab} shows the default values for $b$ and the systematic uncetainties applied.  The uncertainties listed are only from this change in the number of bins.  The change in $b$ that is fully correlated with the uncertainty in $a$ is not listed (i.e. how much $b$ changes to perseve 3-particle ZYAM for variations of $a$ within its uncertainty).

\subsection{Flow Correlation}

There are several systematics associated with flow correlations.  The dominant systematic from flow is the uncertainty on the flow measurement.  The other sources of systematic error due to flow are much smaller.  These include systematics associated with using the trigger particle flow for the jet flow, effect of $v_{2}$ fluctuations, and the trigger particle $v_{4}$.    

\begin{figure}[htbp]
\hfill
\begin{minipage}{0.07\textwidth}
\includegraphics[width=1.0\textwidth]{Plots/blank2.eps}
\end{minipage}
\hfill
\begin{minipage}{0.25\textwidth}
\includegraphics[width=1.0\textwidth]{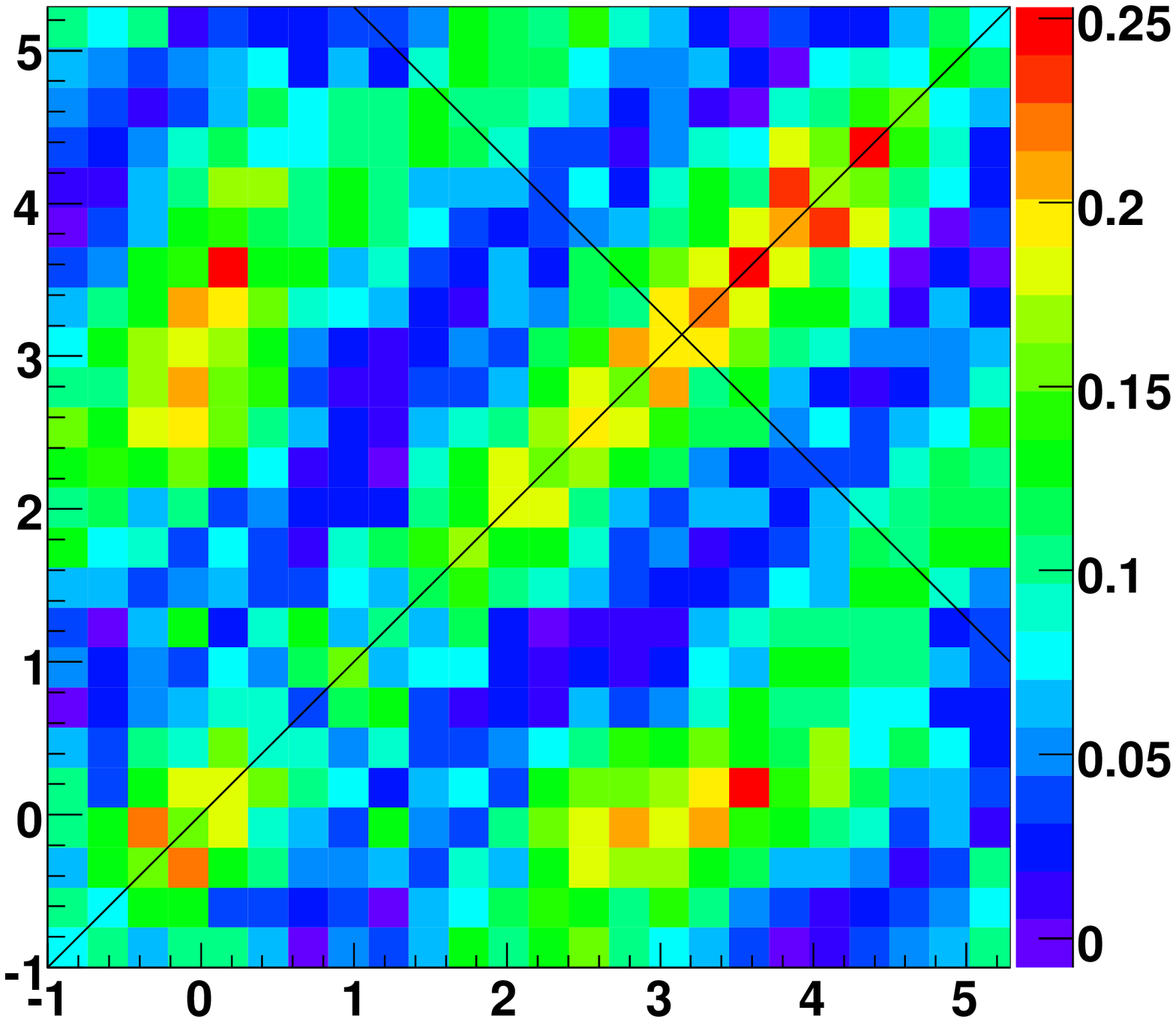}
\includegraphics[width=1.0\textwidth]{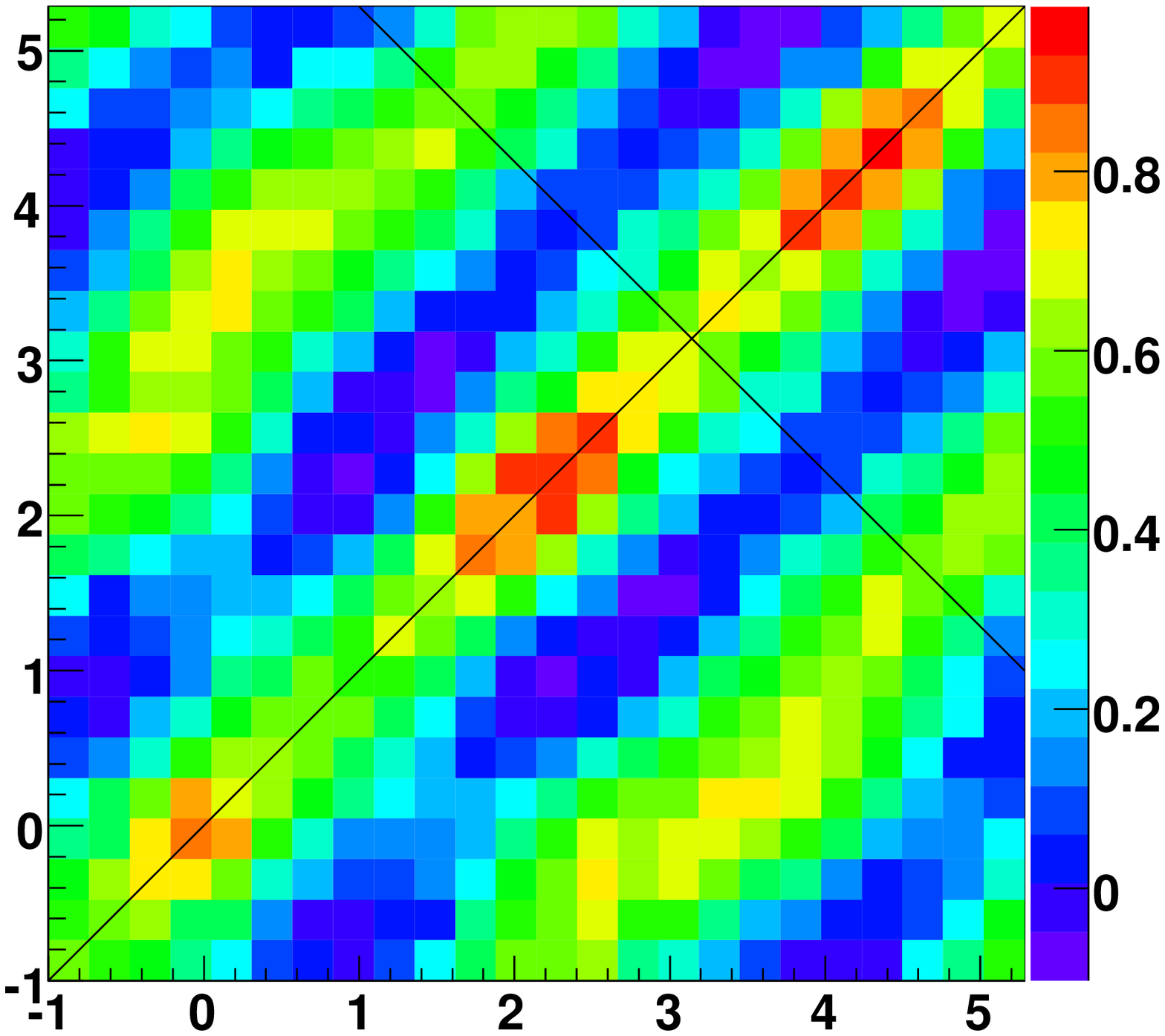}
\includegraphics[width=1.0\textwidth]{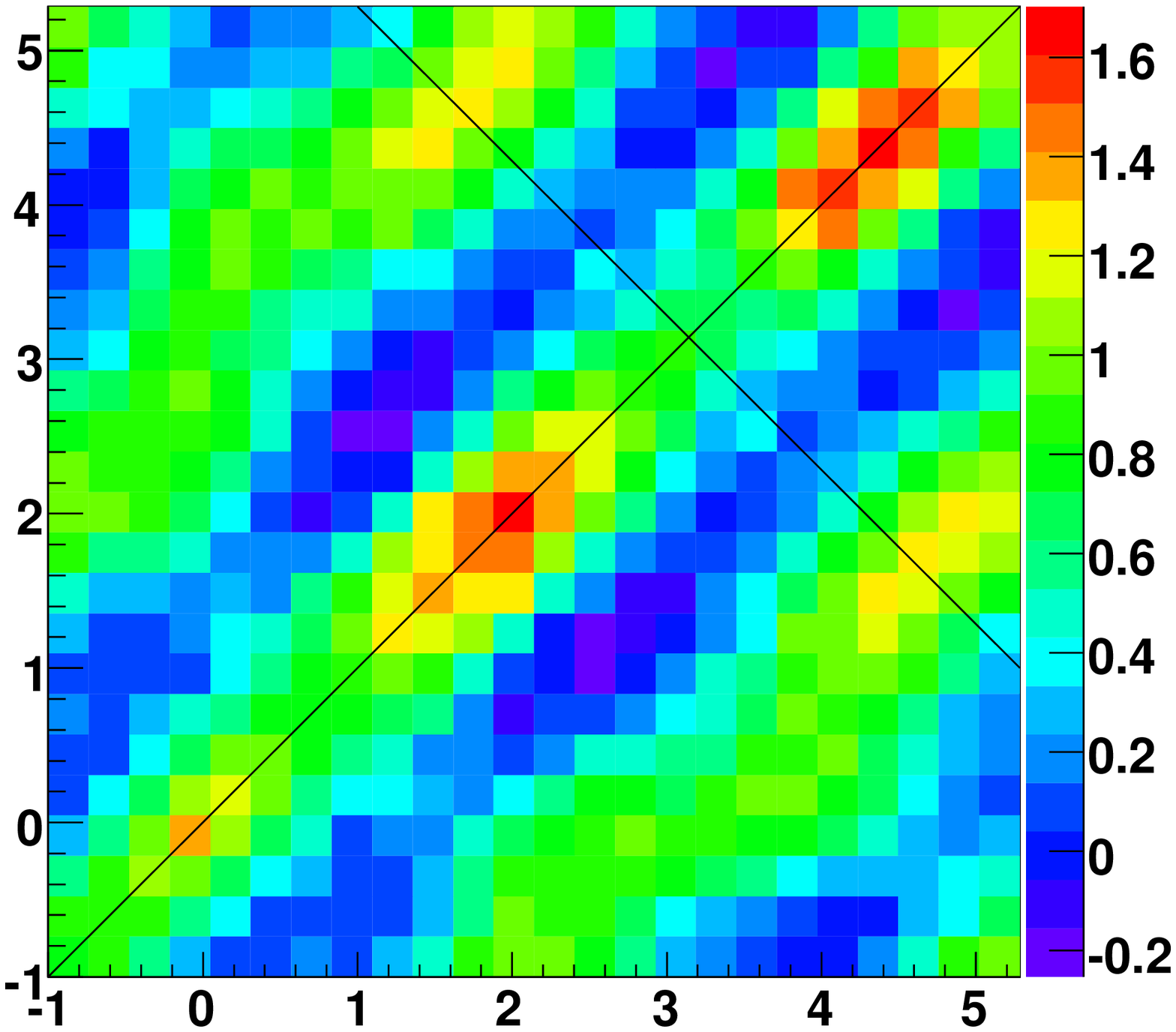}
\includegraphics[width=1.0\textwidth]{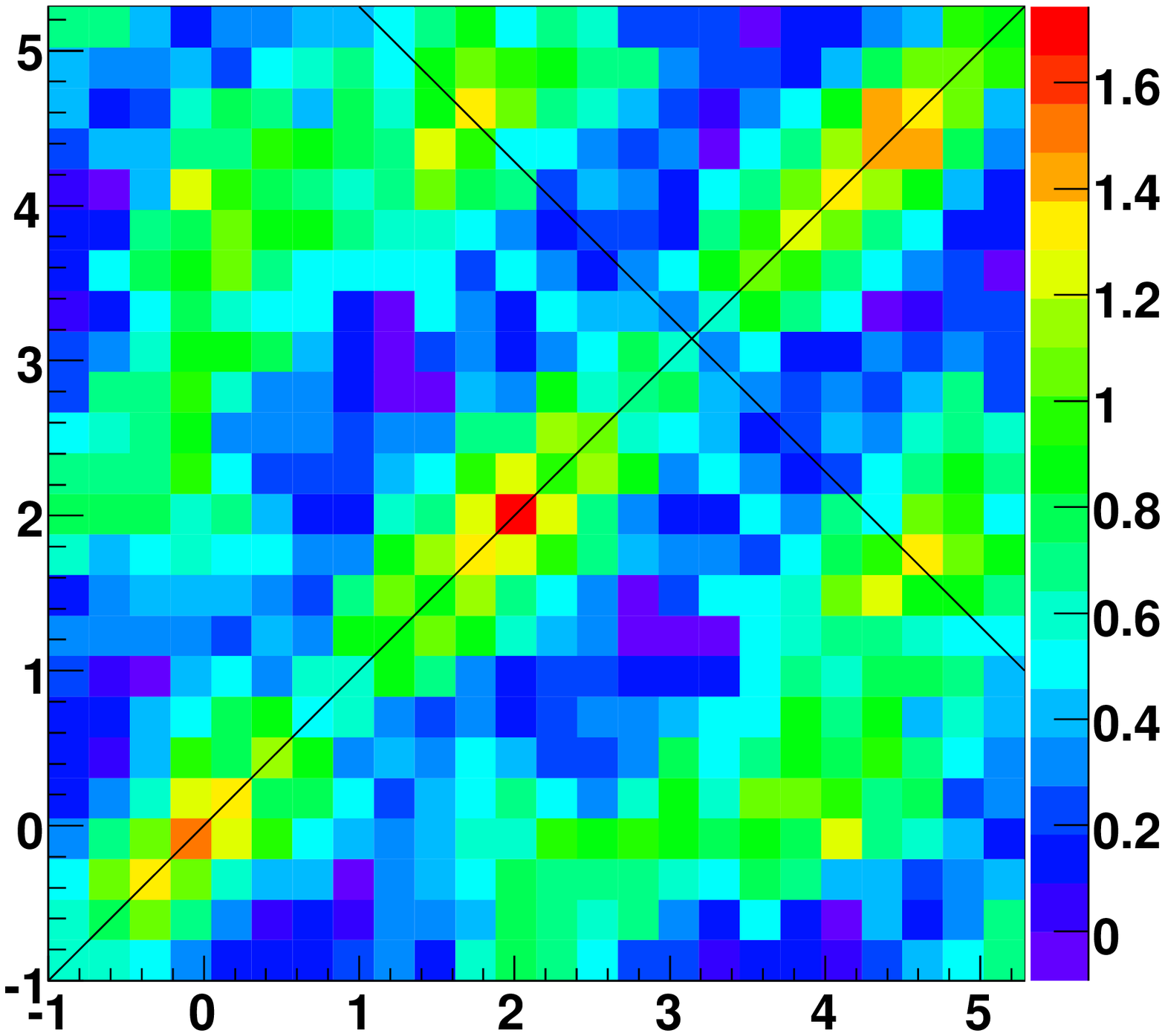}
\includegraphics[width=1.0\textwidth]{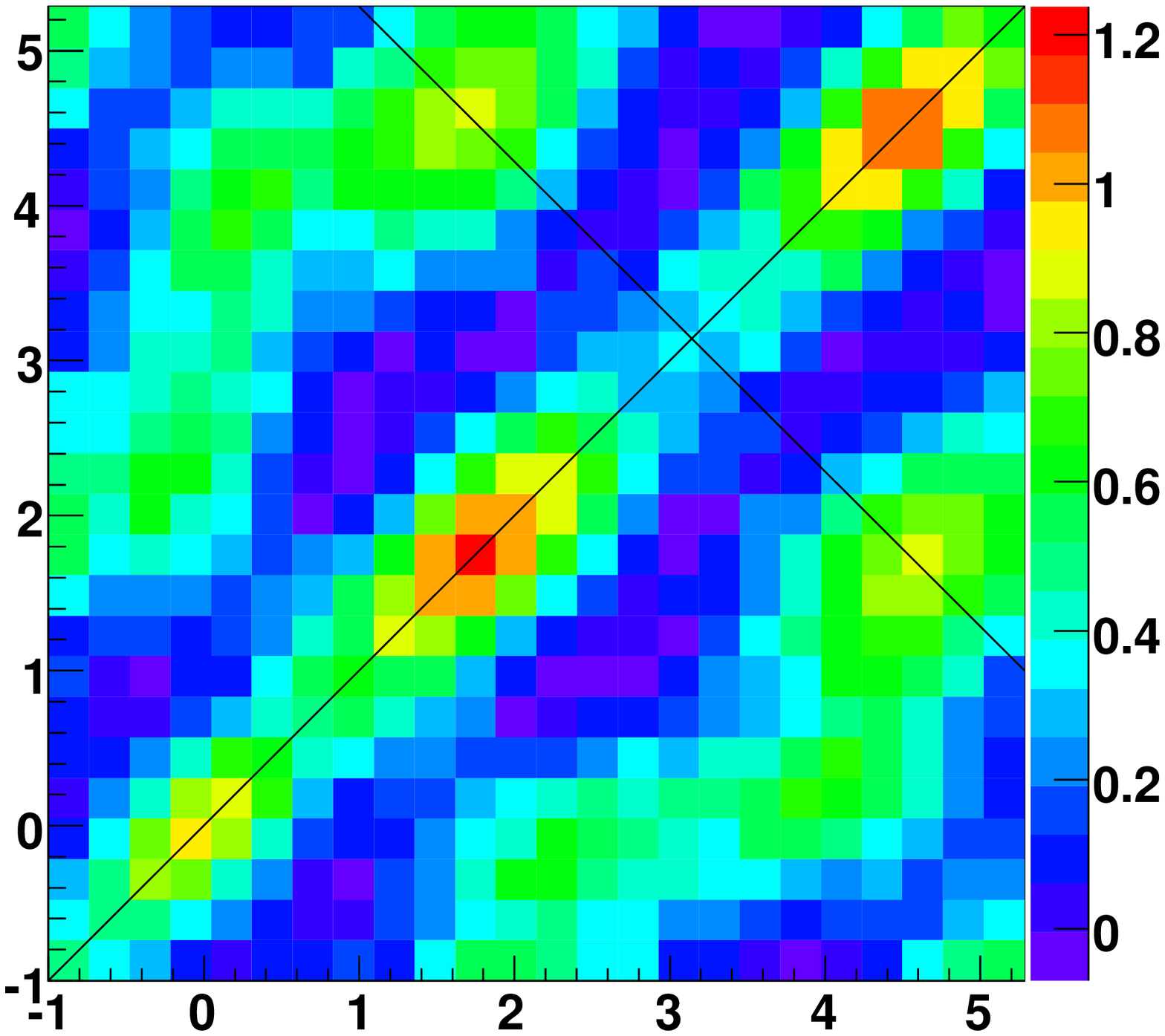}
\end{minipage}
\hfill   
\begin{minipage}{0.25\textwidth}
\includegraphics[width=1.0\textwidth]{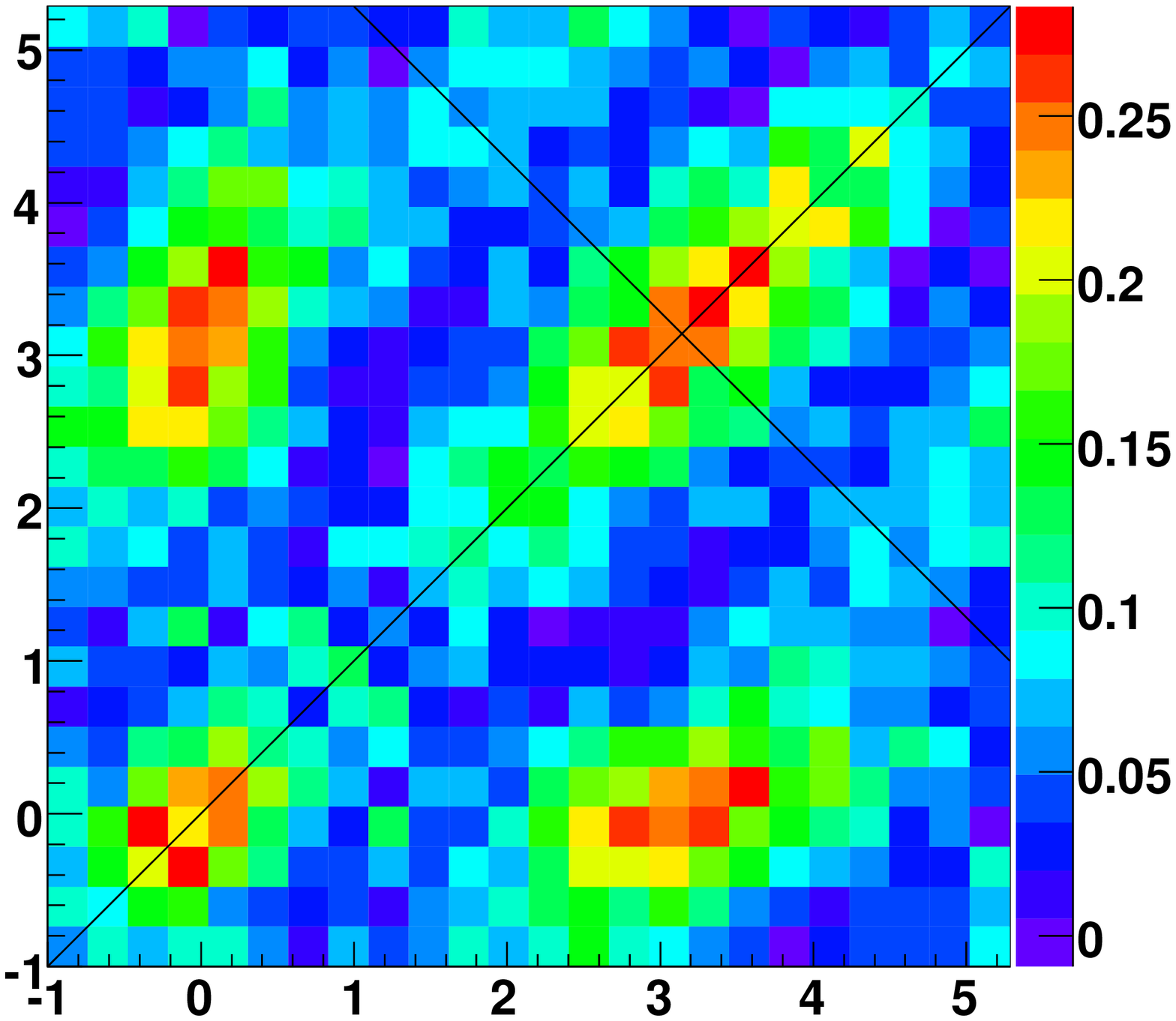}
\includegraphics[width=1.0\textwidth]{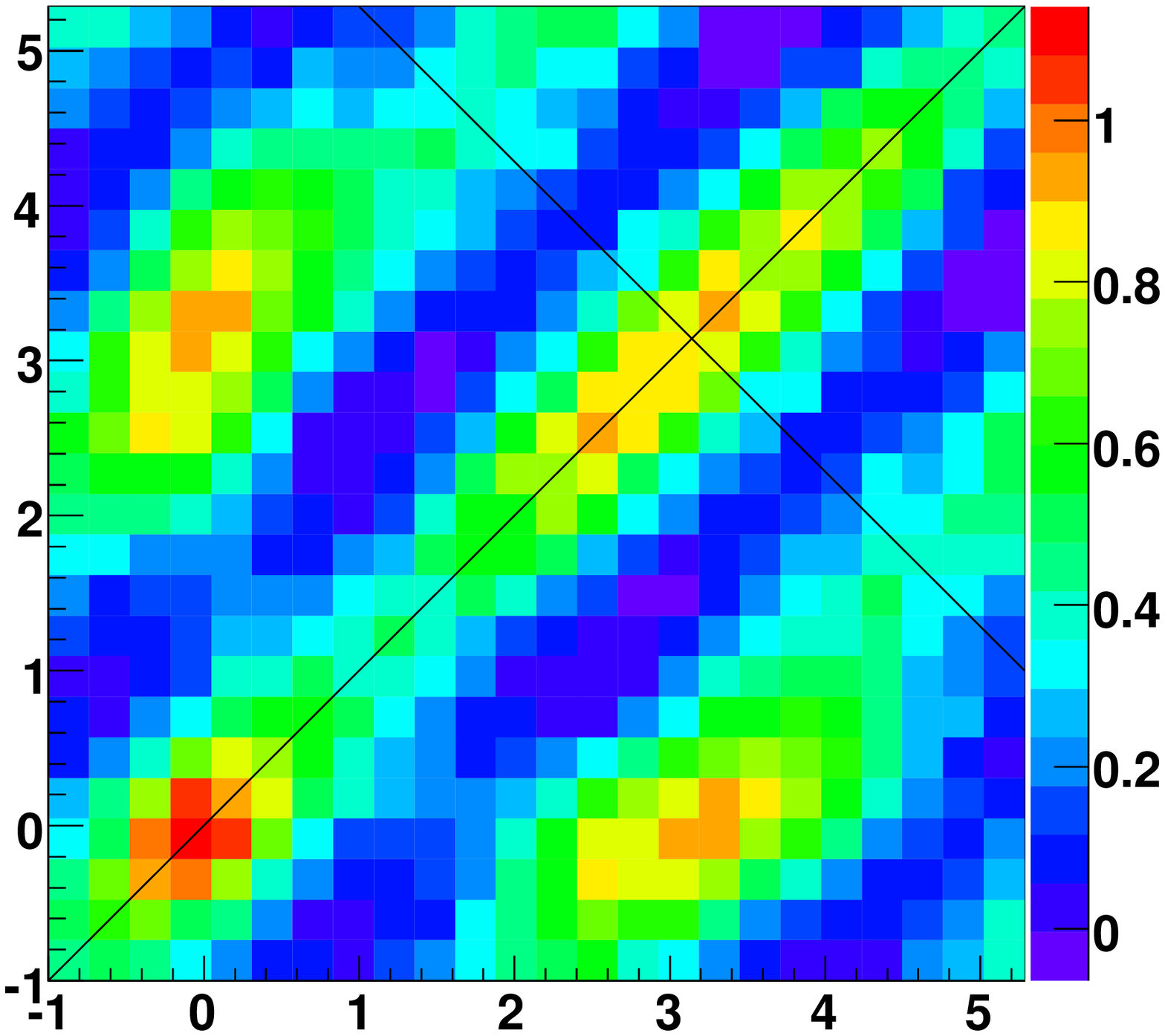}
\includegraphics[width=1.0\textwidth]{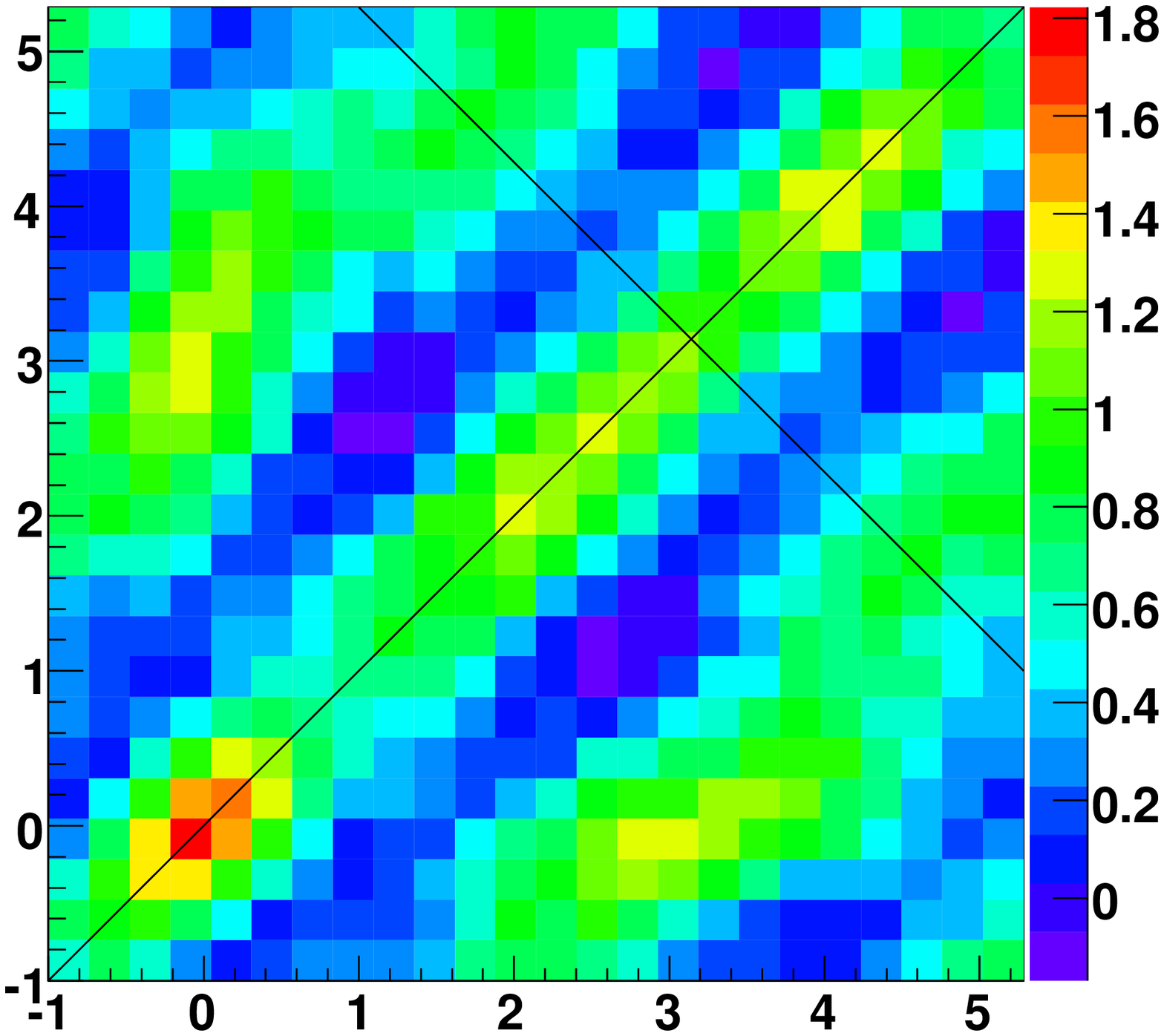}
\includegraphics[width=1.0\textwidth]{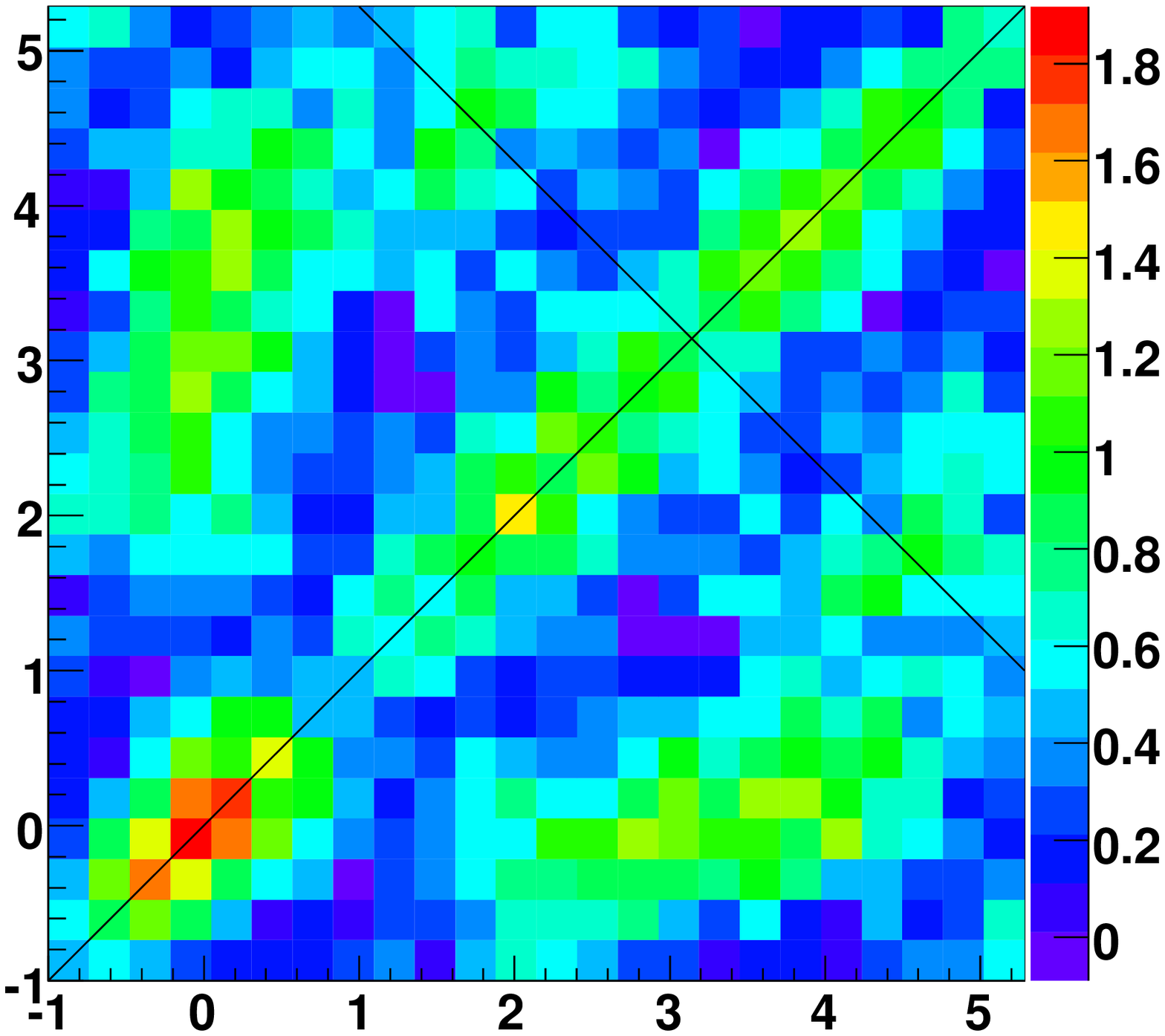}
\includegraphics[width=1.0\textwidth]{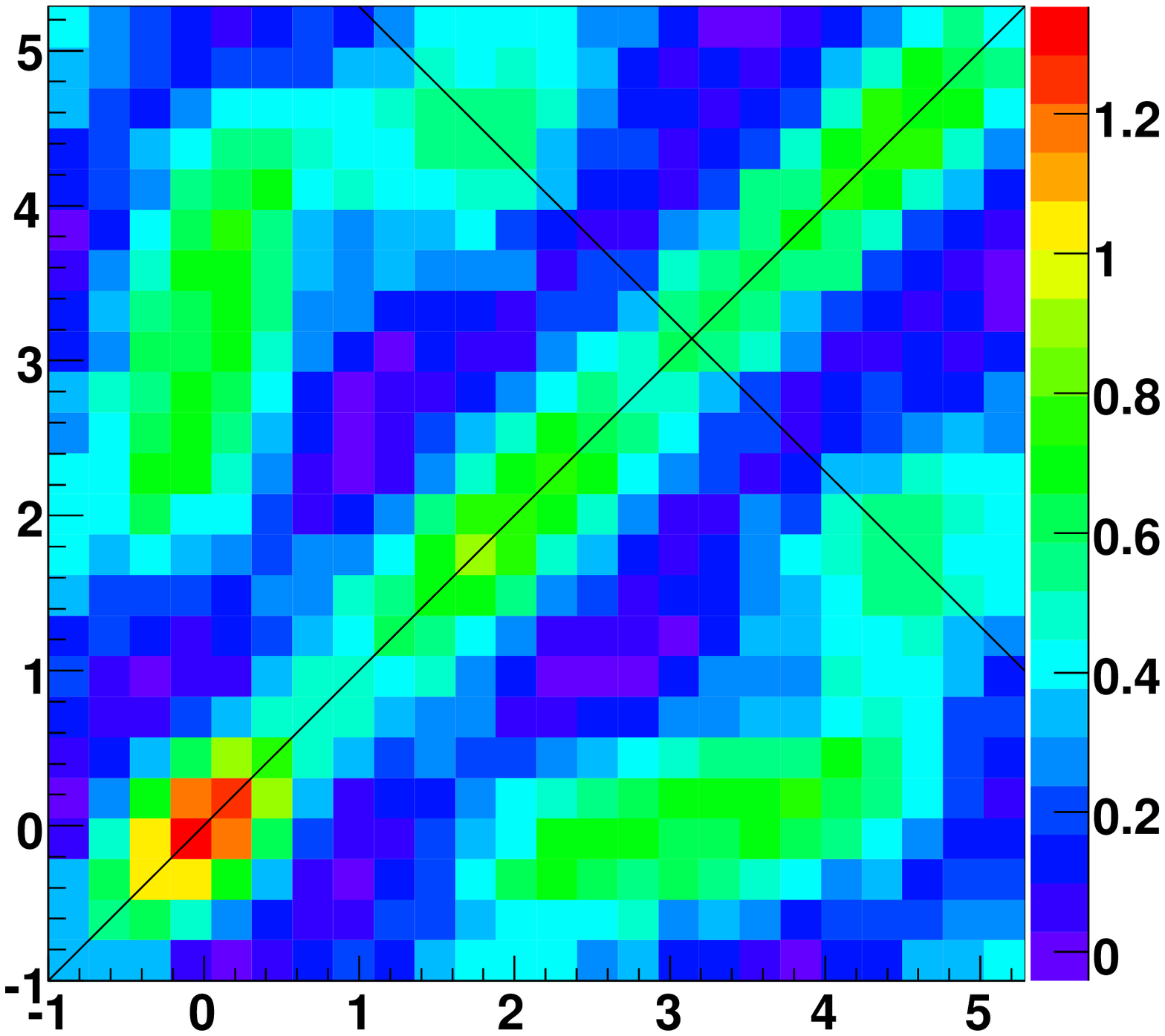}
\end{minipage}
\hfill
\begin{minipage}{0.25\textwidth}
\includegraphics[width=1.0\textwidth]{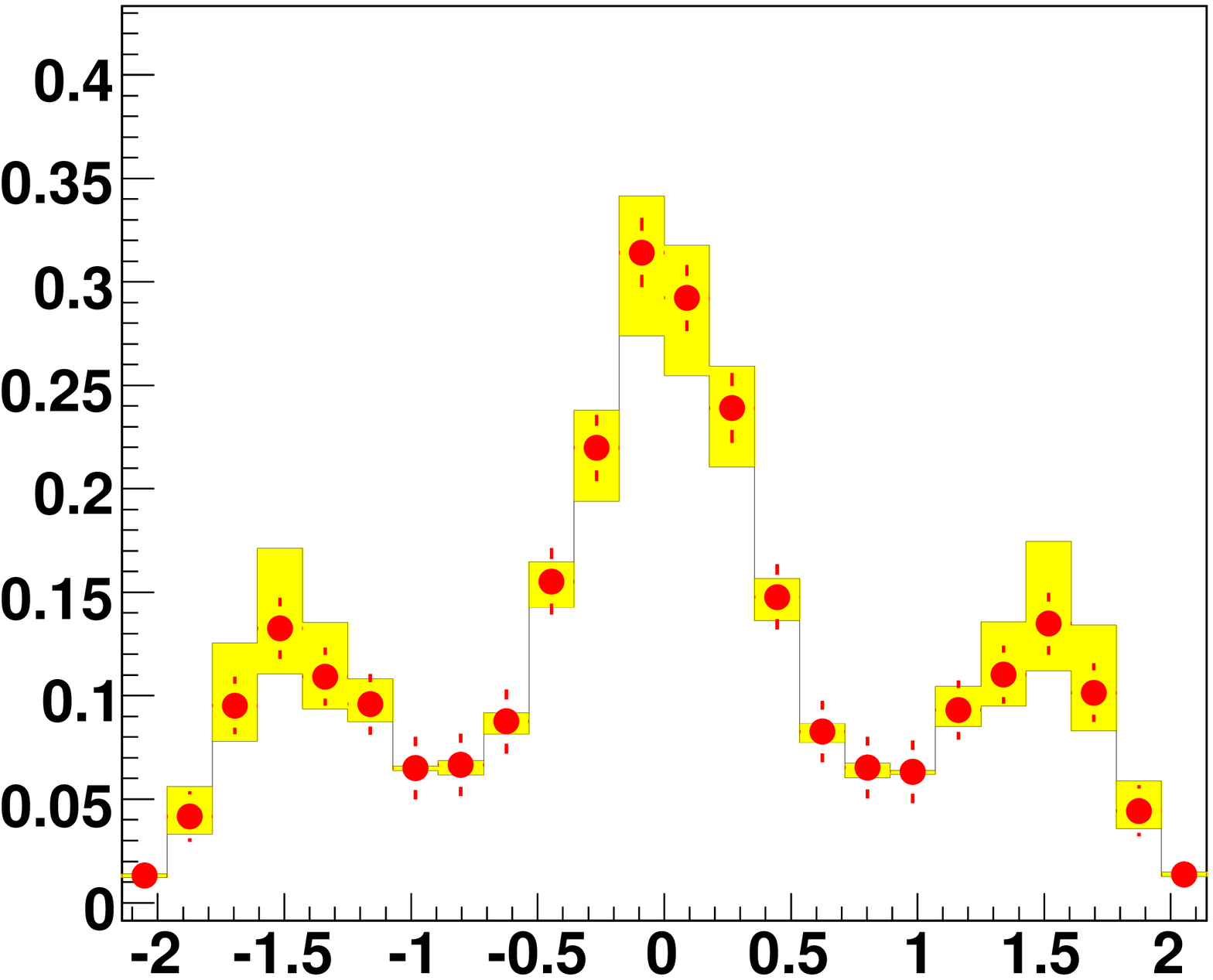}
\includegraphics[width=1.0\textwidth]{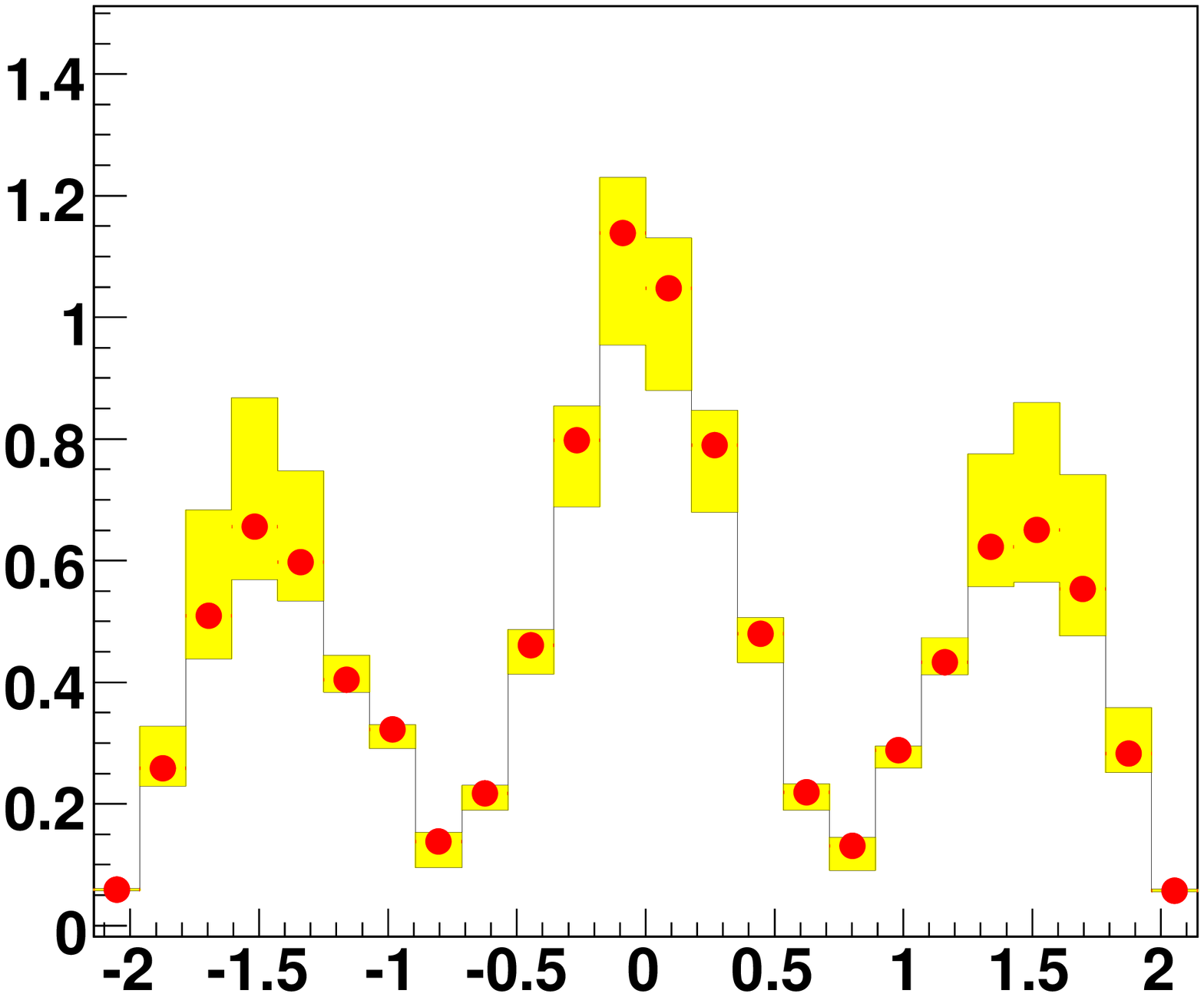}
\includegraphics[width=1.0\textwidth]{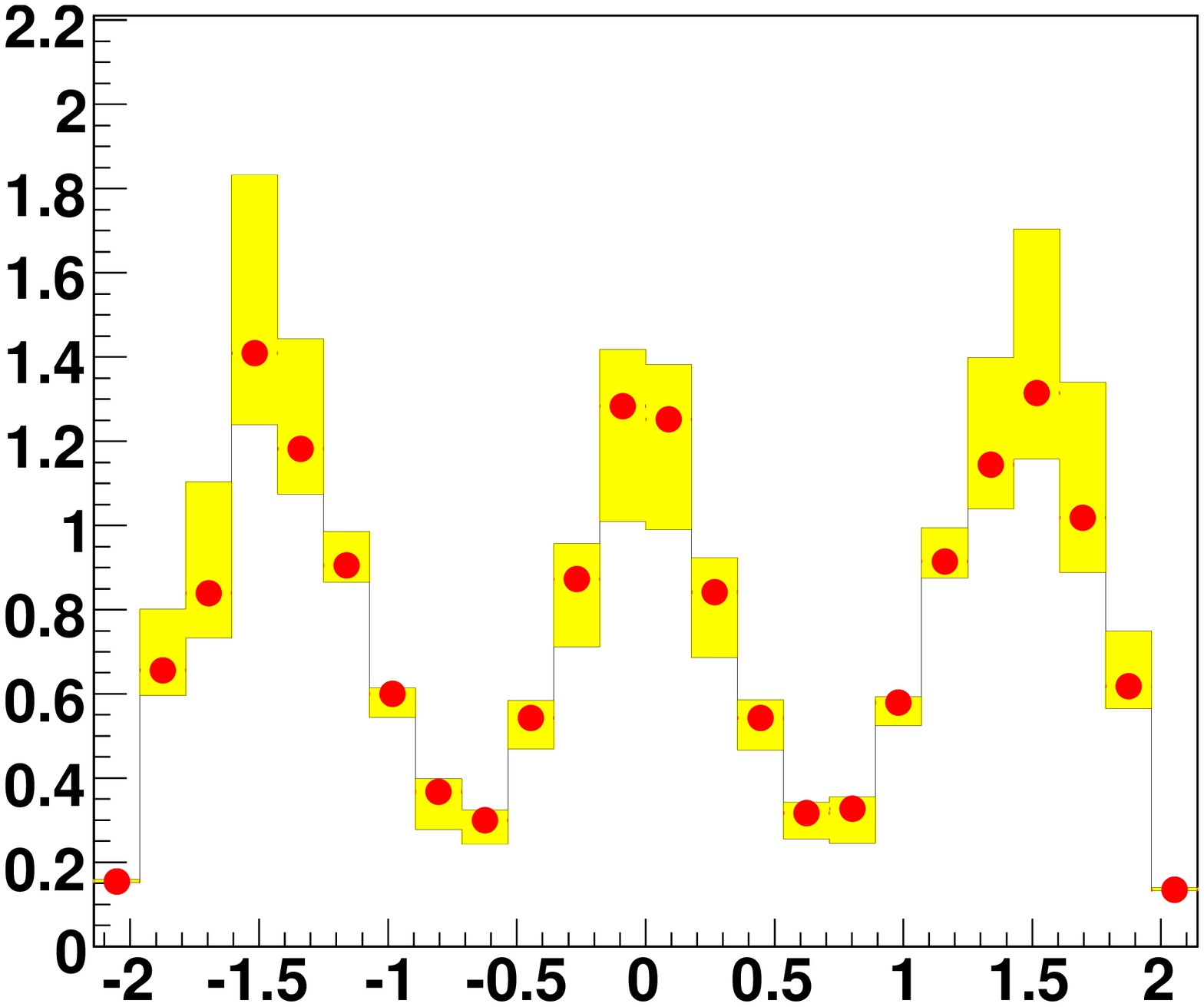}
\includegraphics[width=1.0\textwidth]{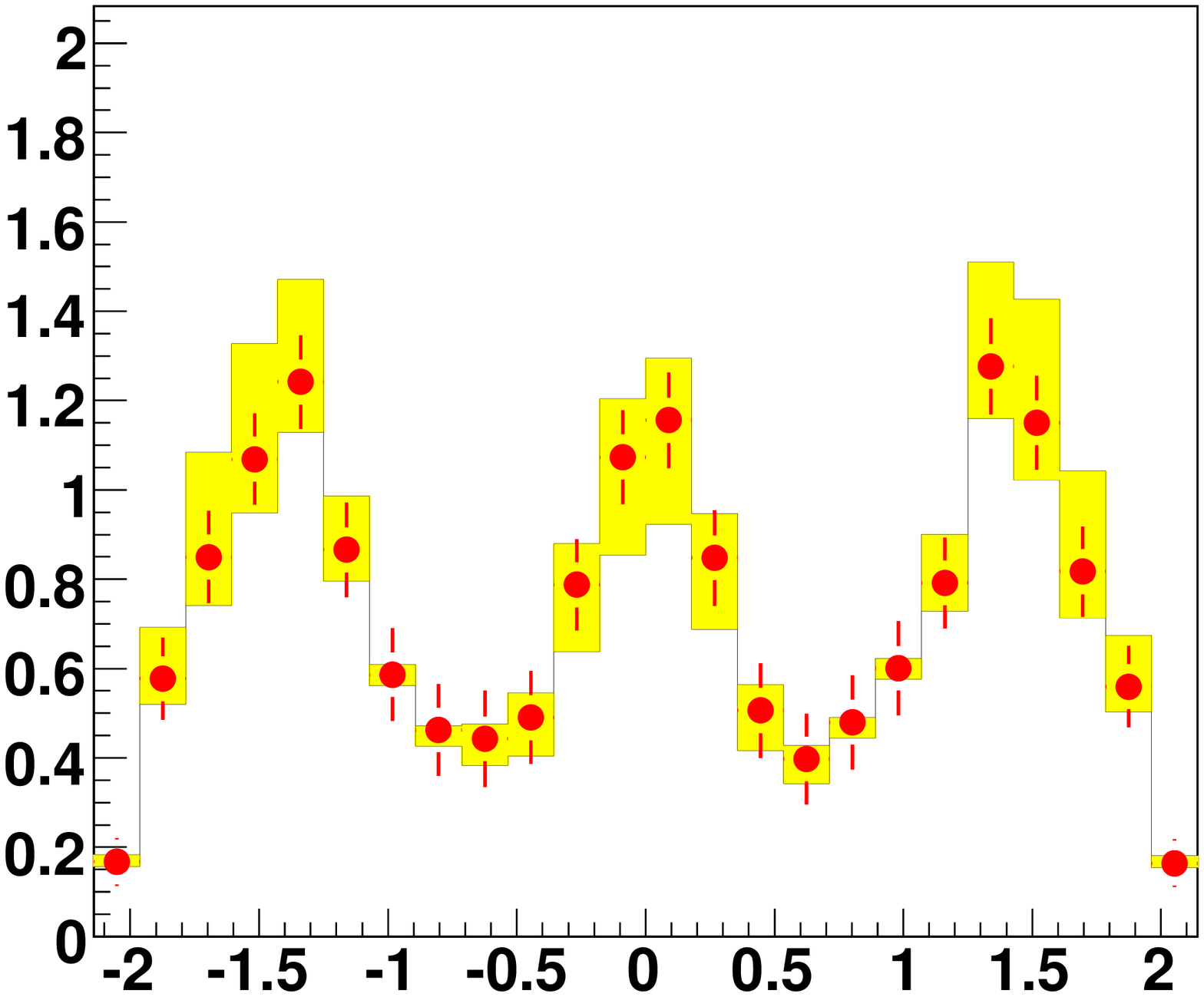}
\includegraphics[width=1.0\textwidth]{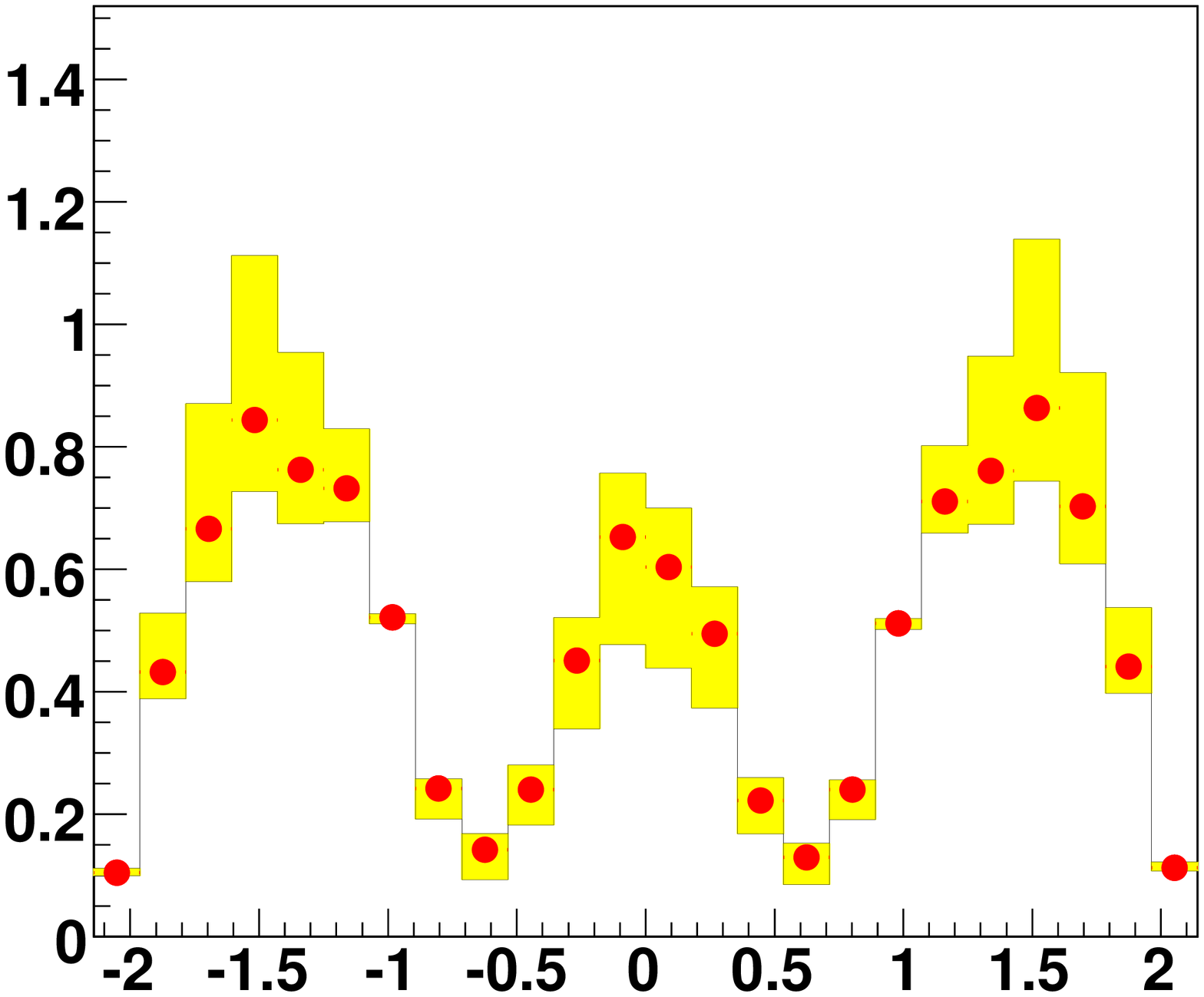}
\end{minipage}
\hfill
\begin{minipage}{0.07\textwidth}
\includegraphics[width=1.0\textwidth]{Plots/blank2.eps}
\end{minipage}
\caption{Background subtracted 3-particle correlations for uncertainty on $v_{2}$.  Left:  $v_{2}$ from the 2-particle cumulant measurement.  Center:  $v_{2}$ from the 4-particle cumulant method.  Right:  Off-diagonal away-side projection from default with systematic uncertainty from the uncertainty on $v_{2}$ shown in shaded band.  From top to bottom plots are Au+Au 50-80\%, Au+Au 30-50\%, Au+Au 10-30\%, Au+Au 0-10\%, and ZDC triggered Au+Au 0-12\% collisions at $\sqrt{s_{NN}}=200$ GeV/c.}
\label{fig:v2sys}
\end{figure}

There are several measurements of elliptic flow: the reaction plane method, the modified reaction plane method, the 2-particle cumulant and the 4-particle cumulant.  The different measurements give different results because the measured $v_2$ have different sensitivity to other effect such as non-flow.  The reaction plane method over estimates the $v_2$ due to non-flow.  The modified reaction plane method has reduced sensitivity from non-flow at high $p_T$, and will not over estimate the flow as much as the standard reaction plane method.  The 2-particle $v_2$ over estimates the flow due to non-flow and $v_2$ fluctuations.  The 4-particle cumulant has little sensitivity to non-flow and will underestimate the flow in the presence of $v_2$ fluctuations.
For our default $v_{2}$ value we use the average of the modified reaction plane and the 4-particle measurement.  For our systematics we use the 4-particle cumulant $v_{2}$ for the lower $v_{2}$ value and the 2-particle cumulant $v_{2}$ value is used for the upper limit.    The 2-particle cumulant is used as the upper limit instead of the modified reaction plane because the 2-particle cumulant includes the systematic uncertainty due to $v_2$ fluctuations.  In the most central and the most peripheral bins we do not have measurements for the 4-particle cumulant $v_{2}$.  The 4-particle cumulant $v_{2}$ is estimated from extrapolation to be about 50\% of the reaction plane $v_{2}$ (see Fig.~\ref{fig:flowratio}).  Figure~\ref{fig:v2sys} shows the background subtracted 3-particle correlations using the 2-particle cumulant $v_{2}$ and 4-particle cumulant $v_{2}$.  

\begin{figure}[htbp]
\hfill
\begin{minipage}{0.24\textwidth}
\includegraphics[width=1.0\textwidth]{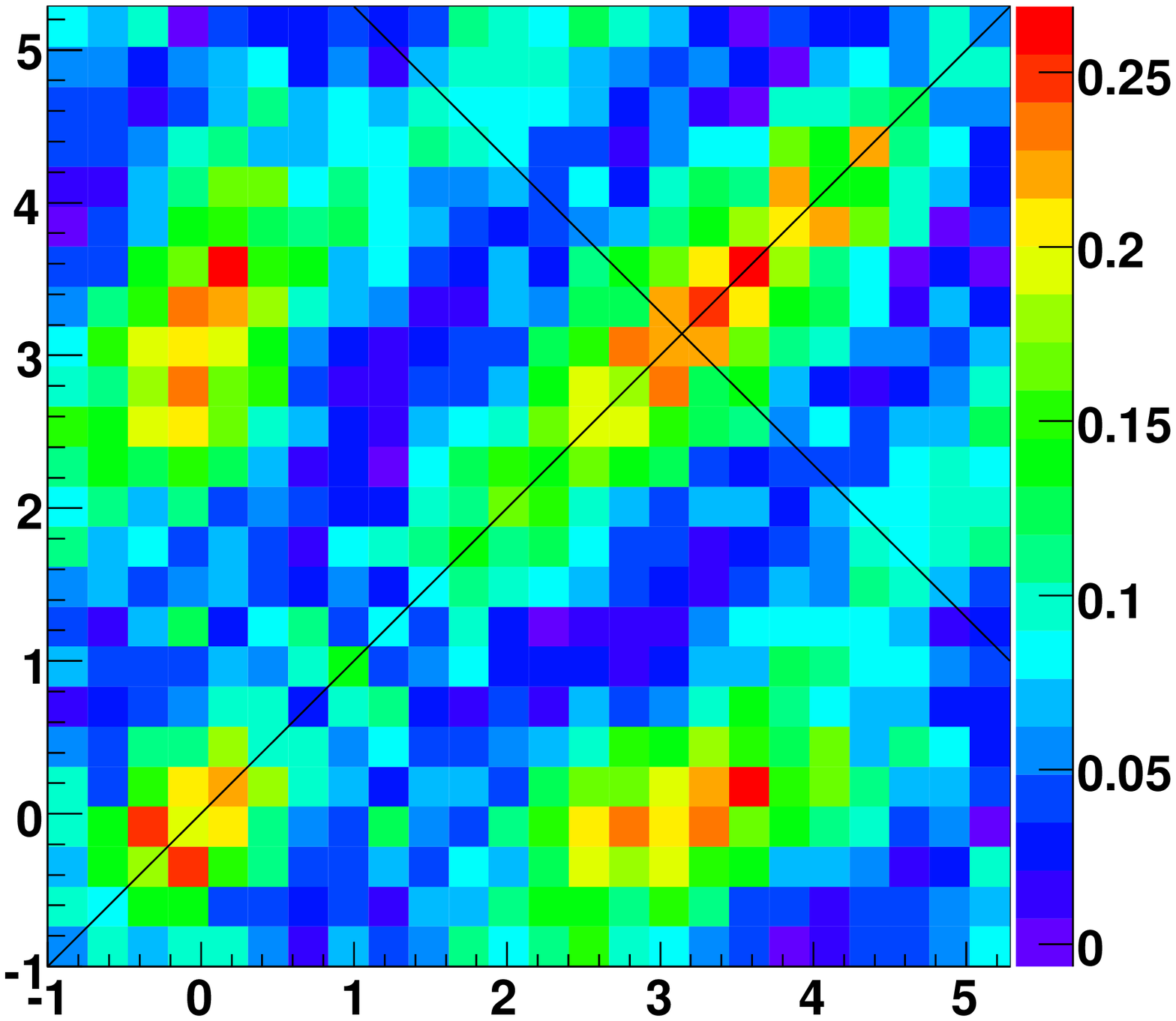}
\includegraphics[width=1.0\textwidth]{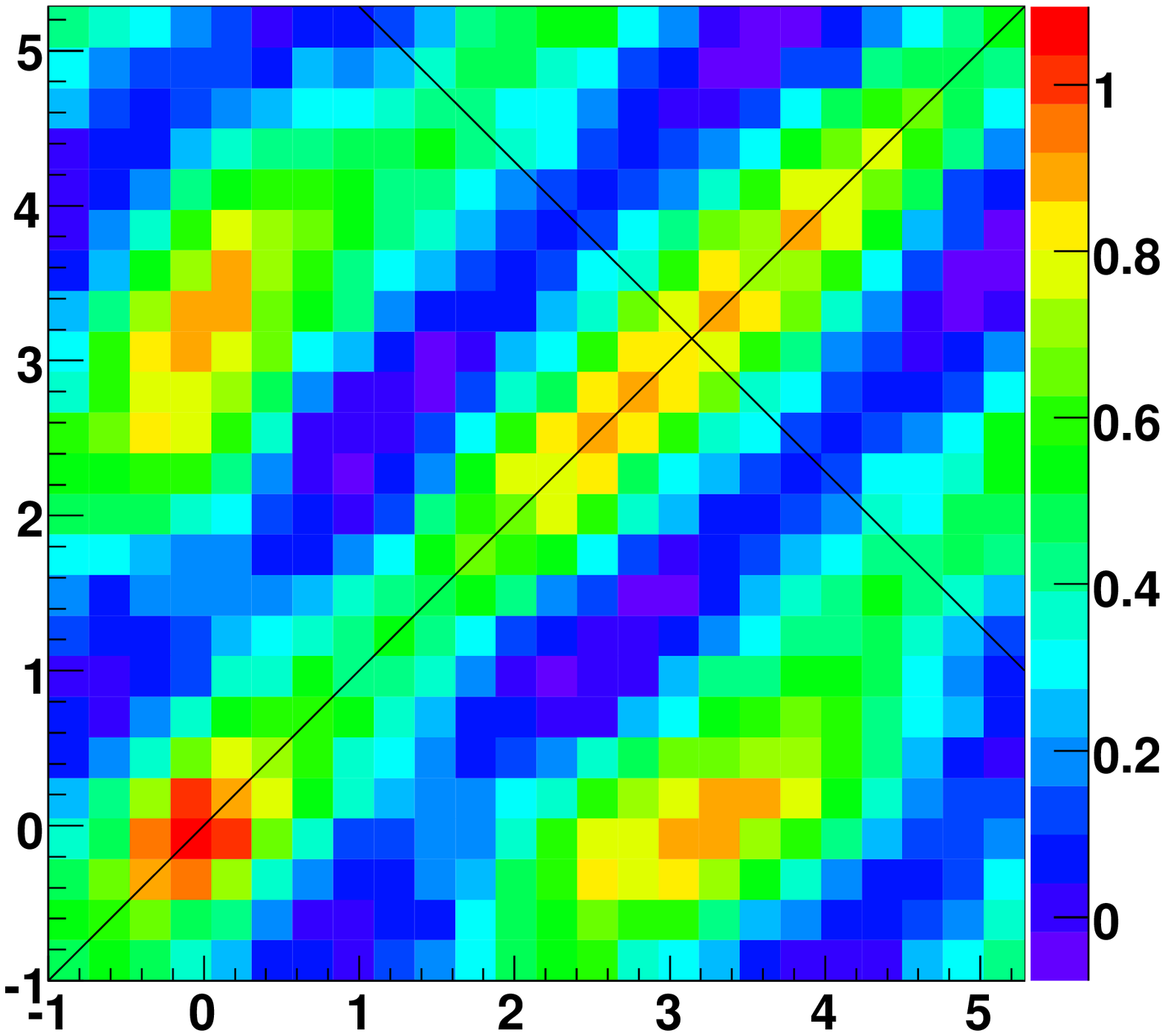}
\includegraphics[width=1.0\textwidth]{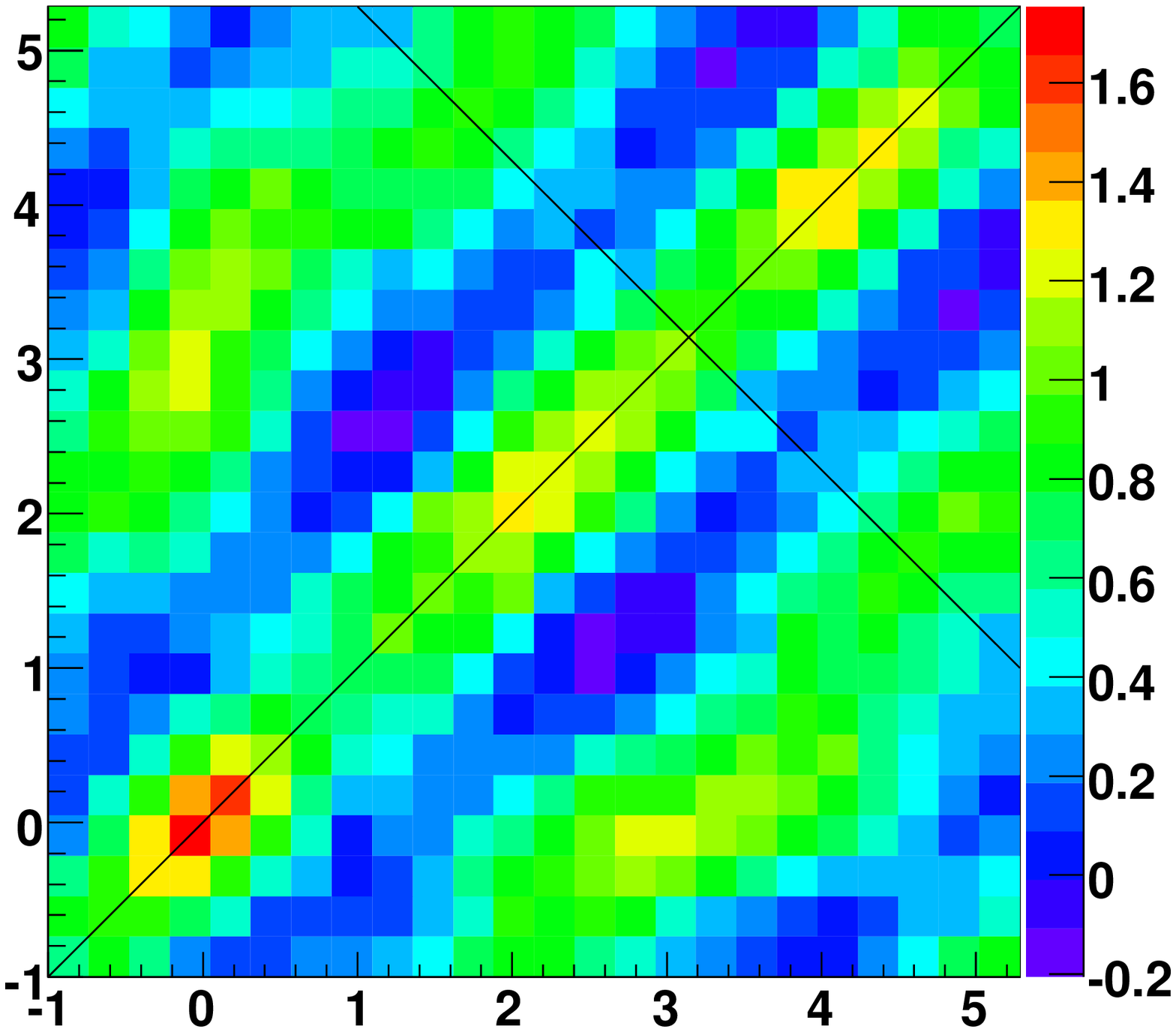}
\includegraphics[width=1.0\textwidth]{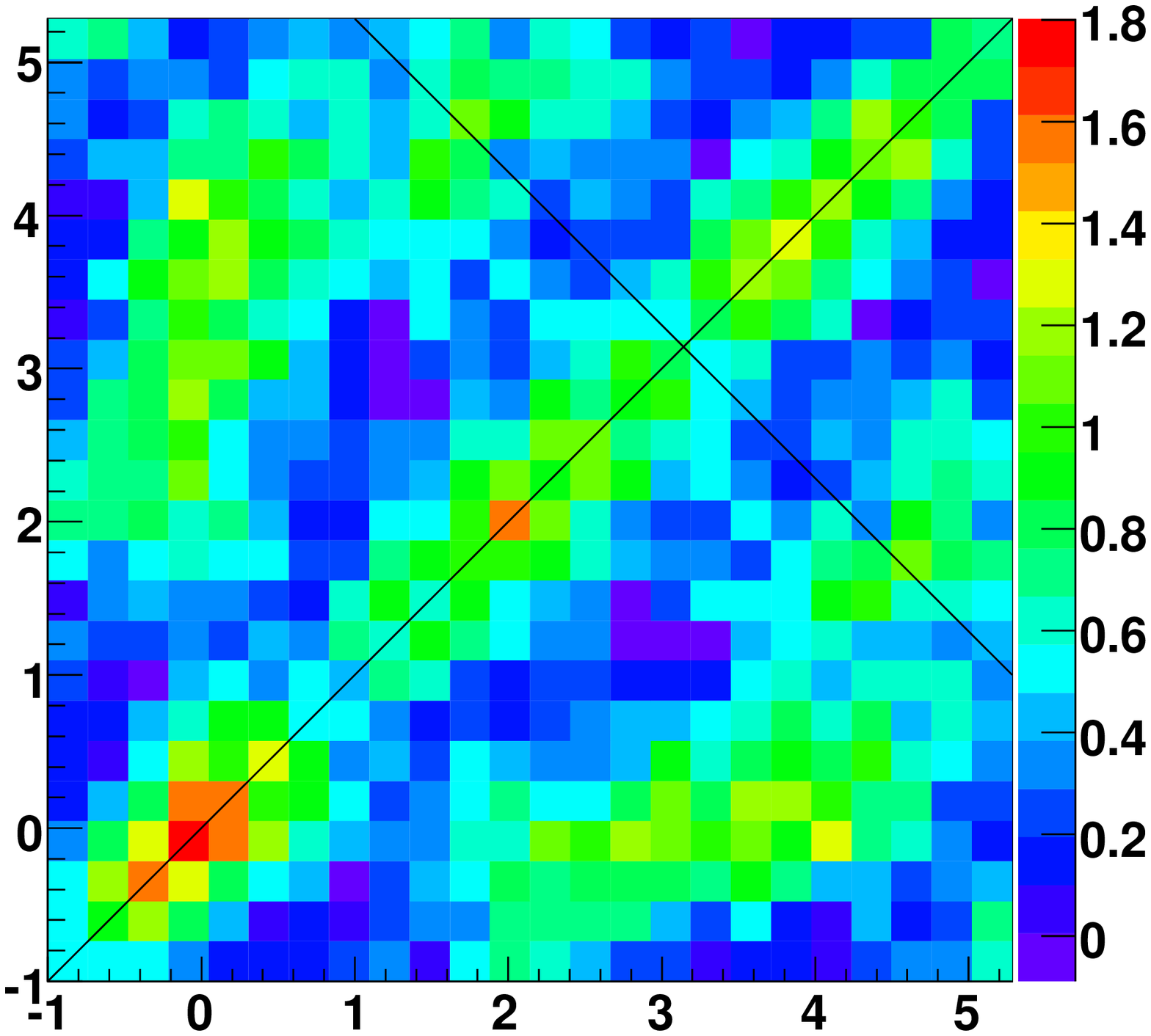}
\includegraphics[width=1.0\textwidth]{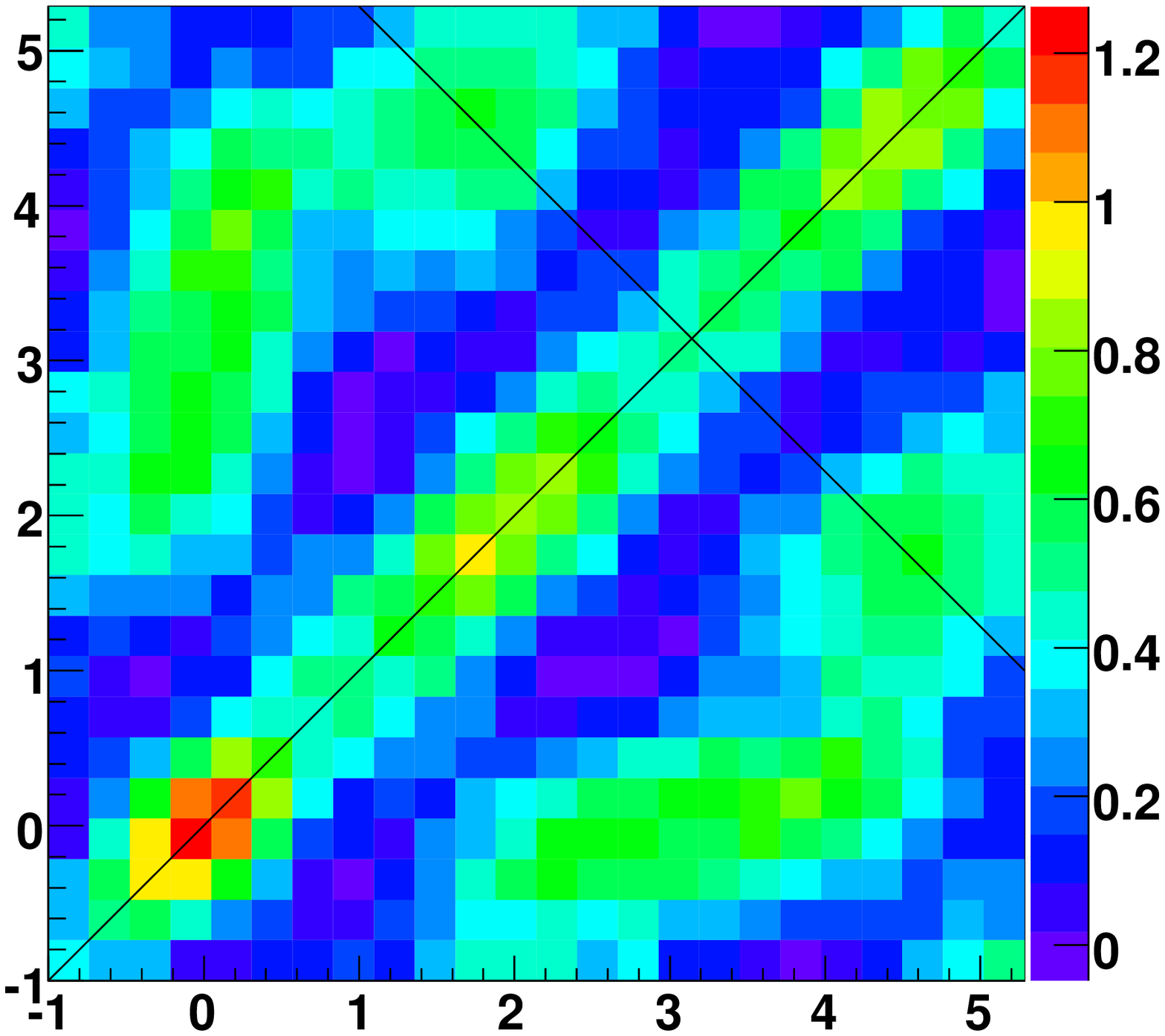}
\end{minipage}
\hfill   
\begin{minipage}{0.24\textwidth}
\includegraphics[width=1.0\textwidth]{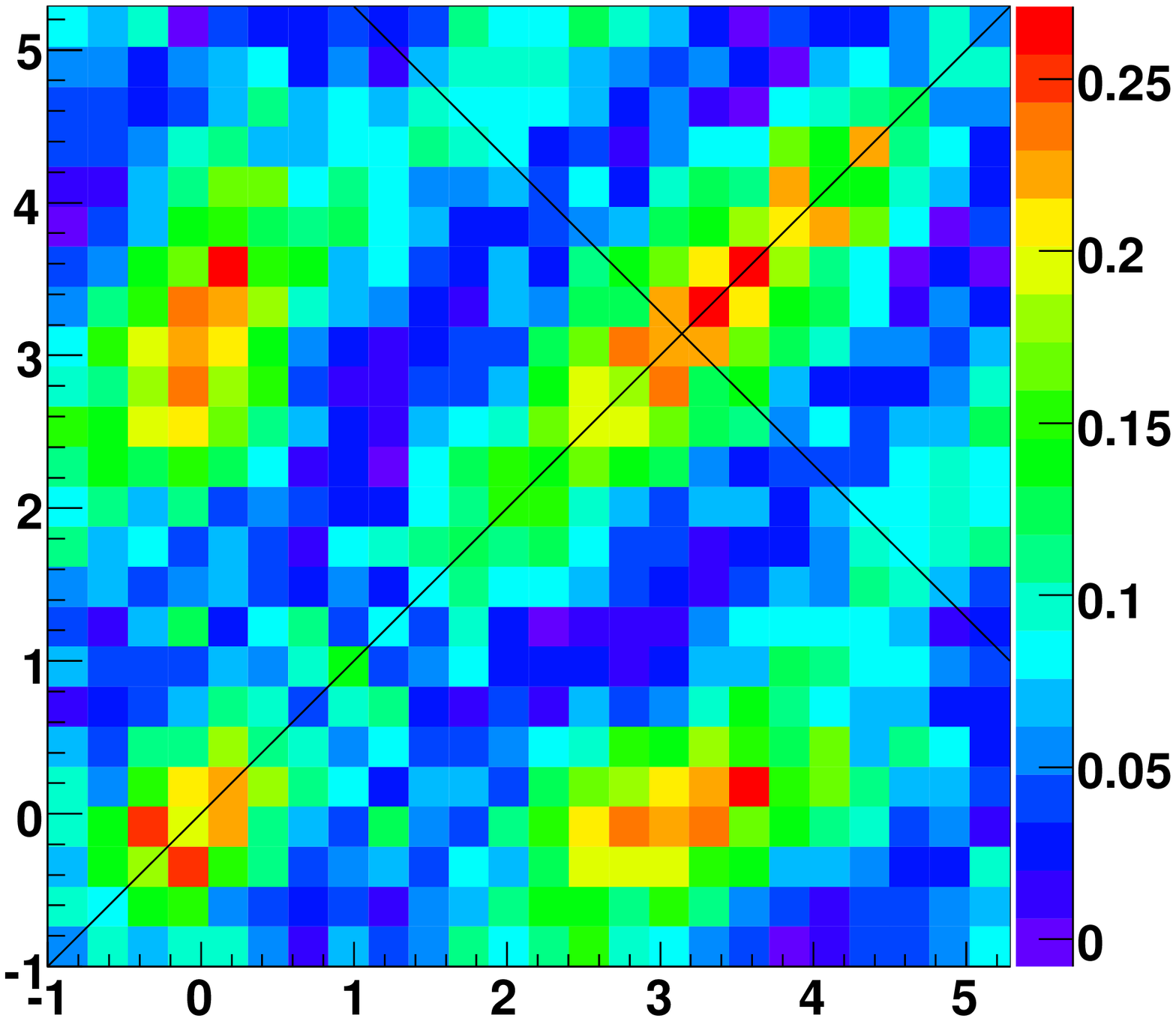}
\includegraphics[width=1.0\textwidth]{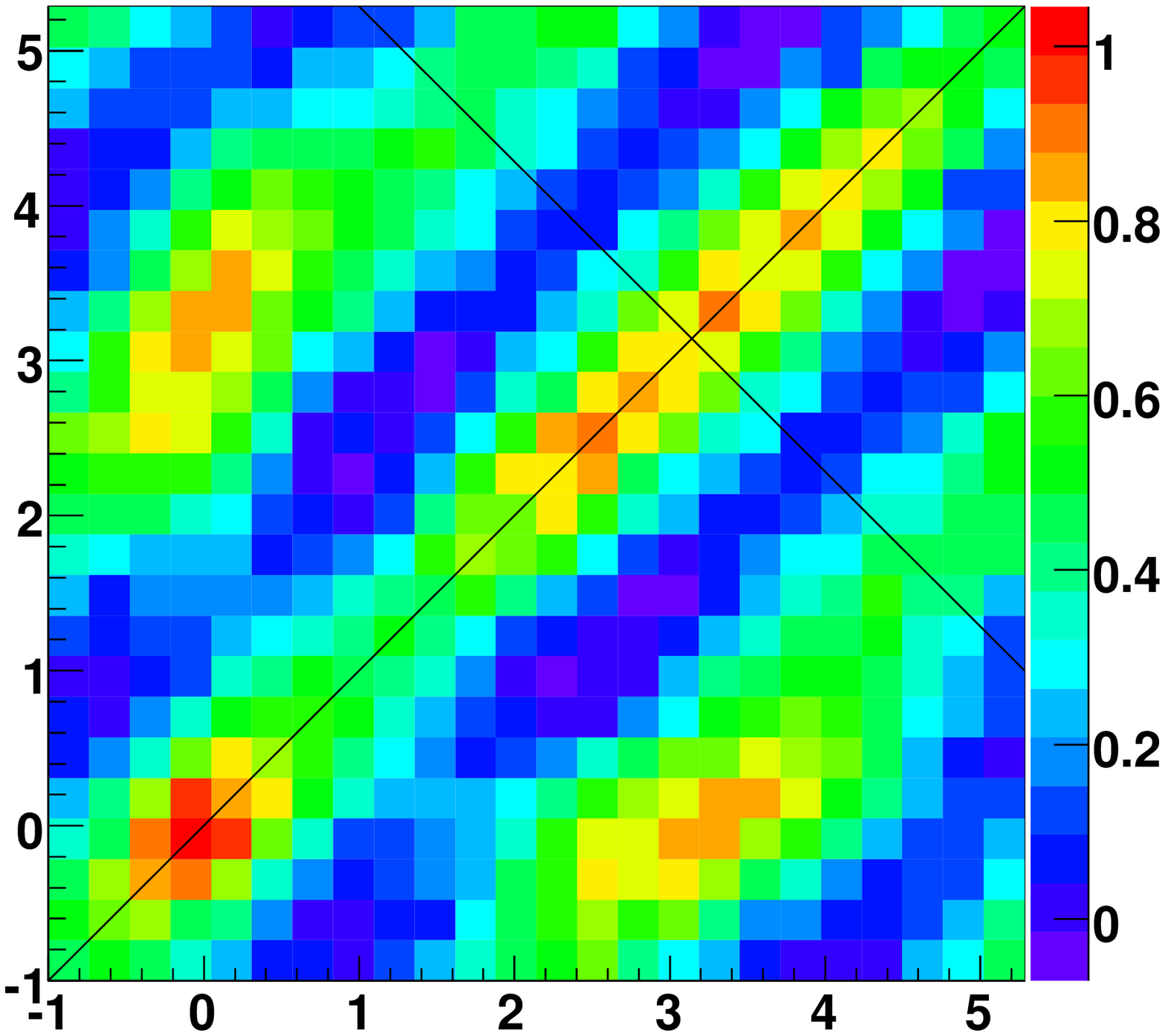}
\includegraphics[width=1.0\textwidth]{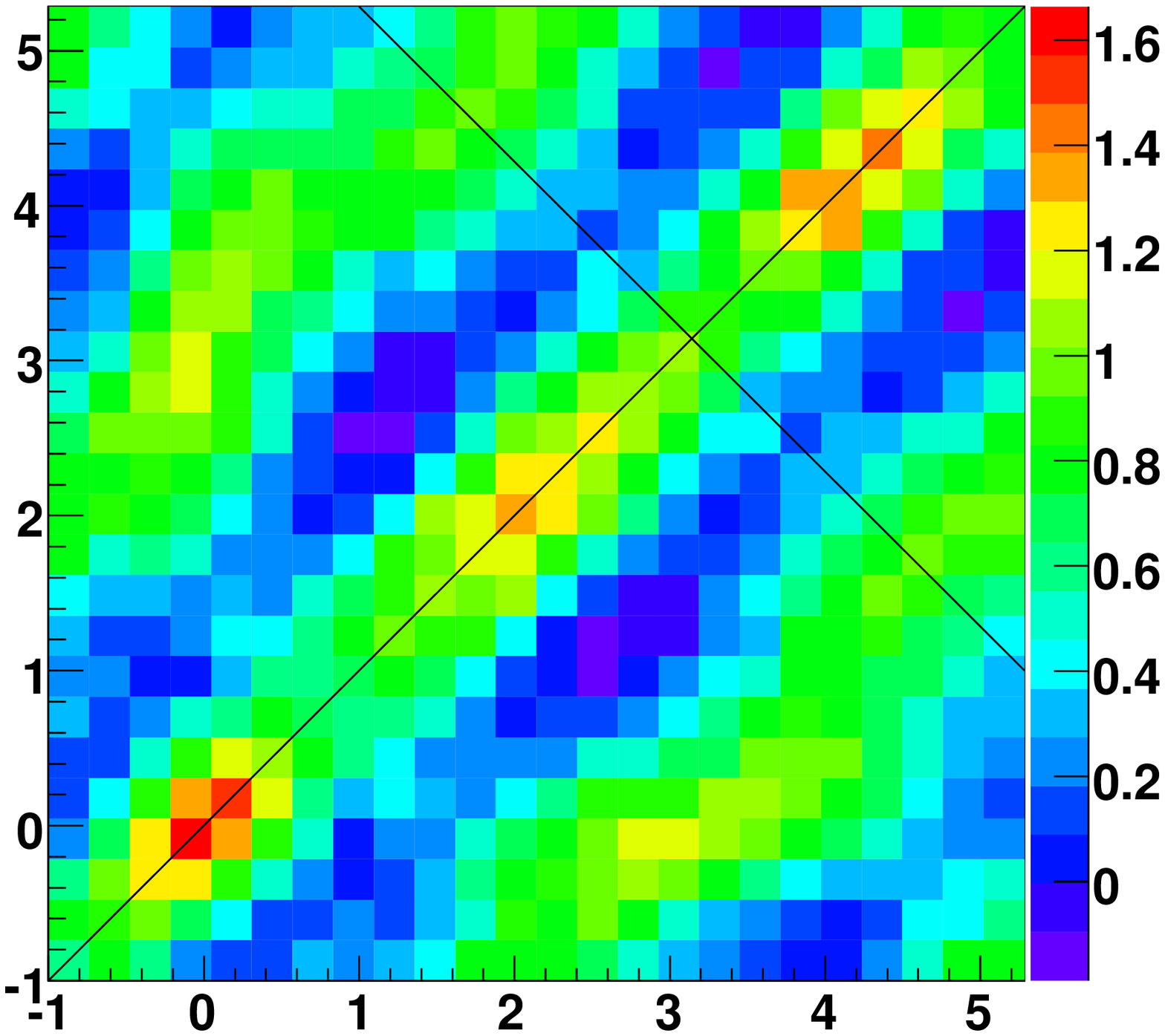}
\includegraphics[width=1.0\textwidth]{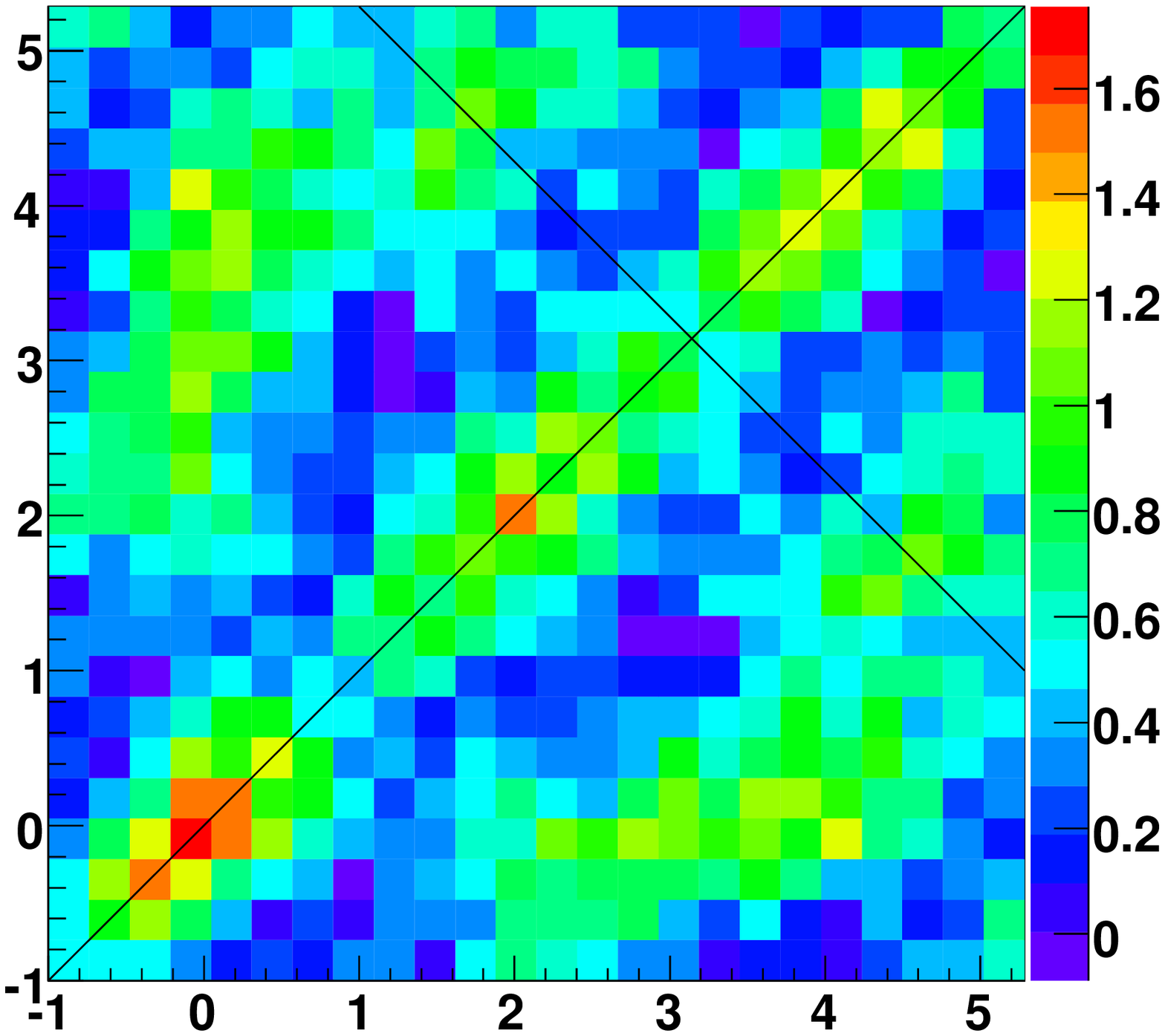}
\includegraphics[width=1.0\textwidth]{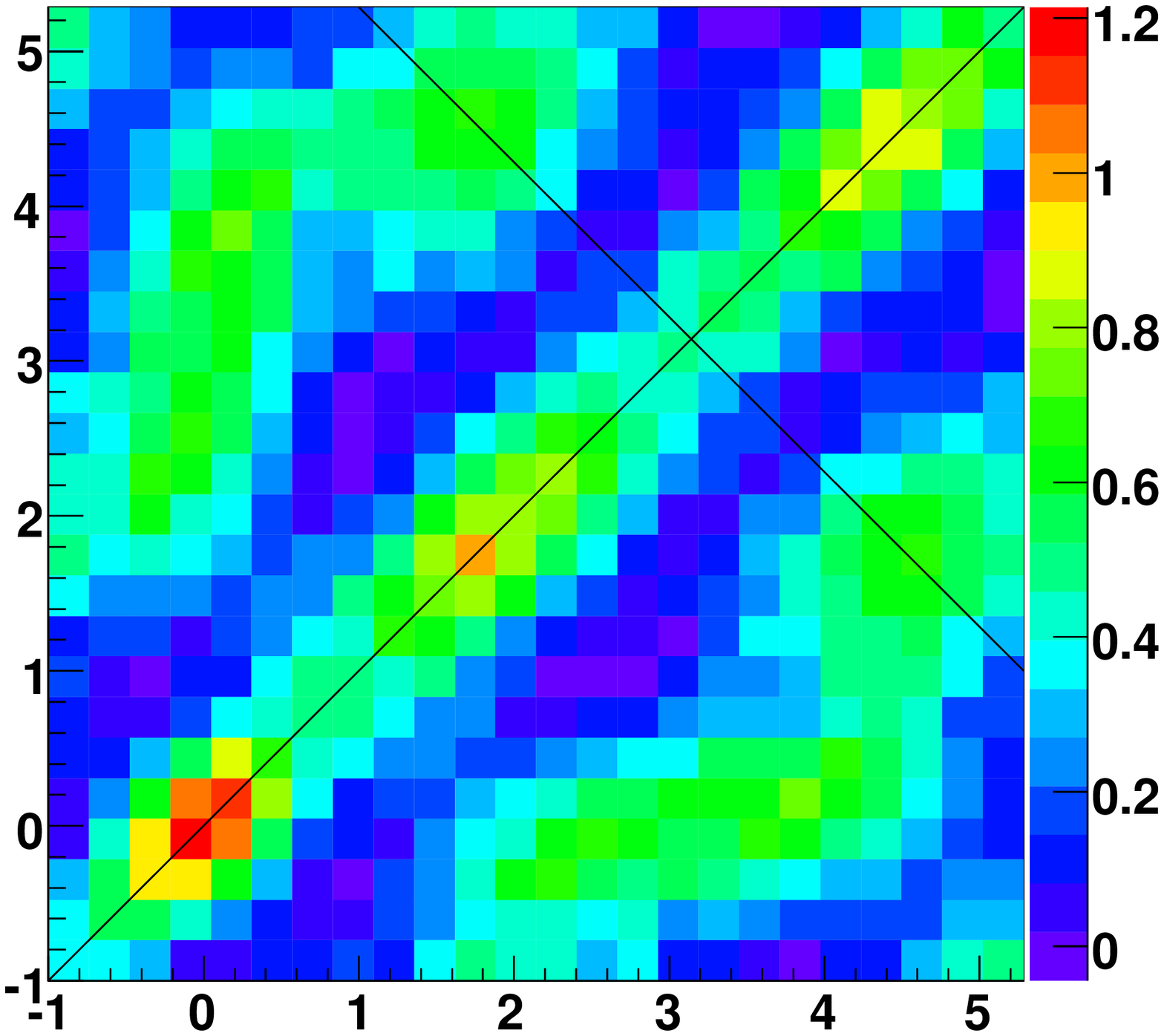}
\end{minipage}
\hfill
\begin{minipage}{0.24\textwidth}
\includegraphics[width=1.0\textwidth]{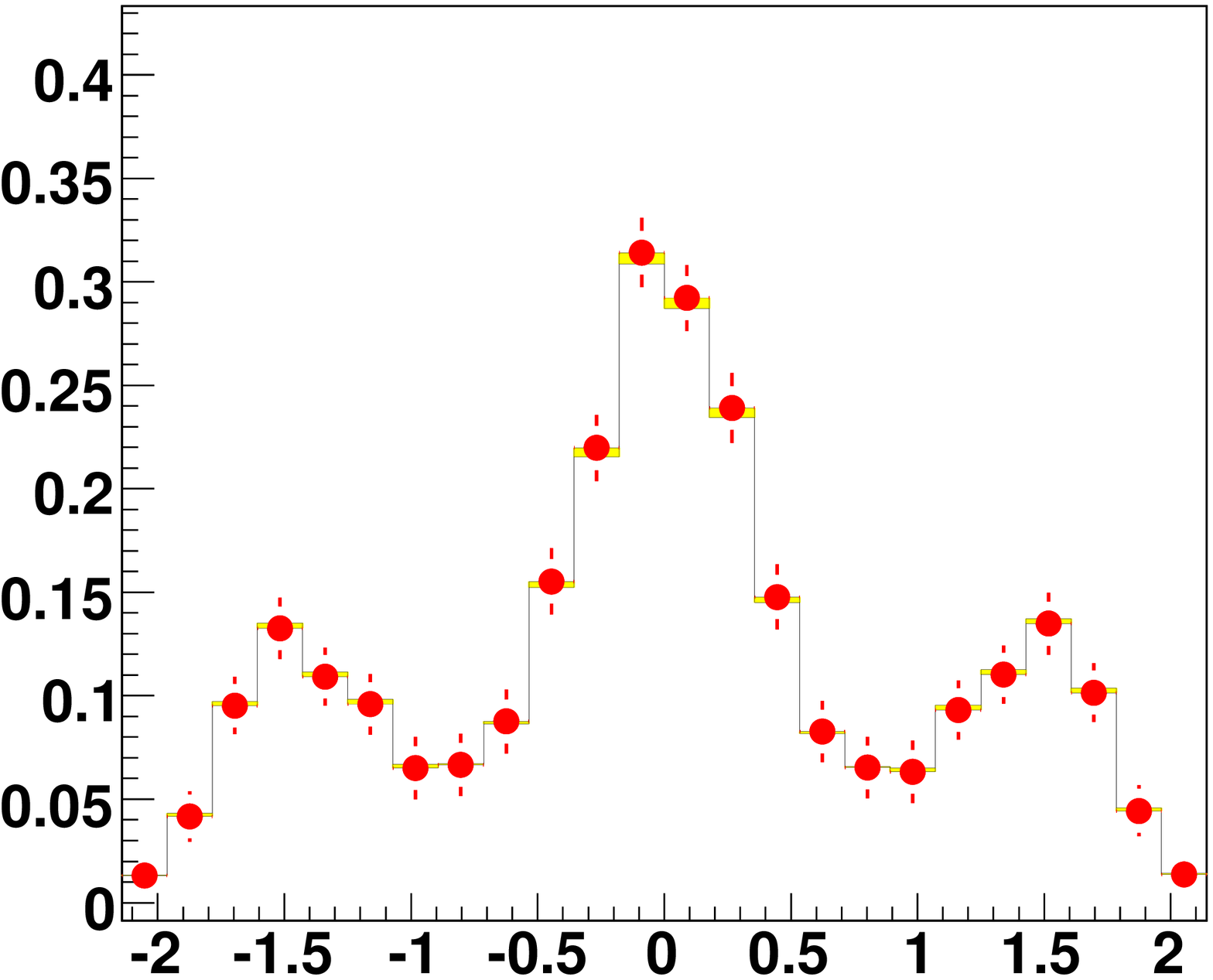}
\includegraphics[width=1.0\textwidth]{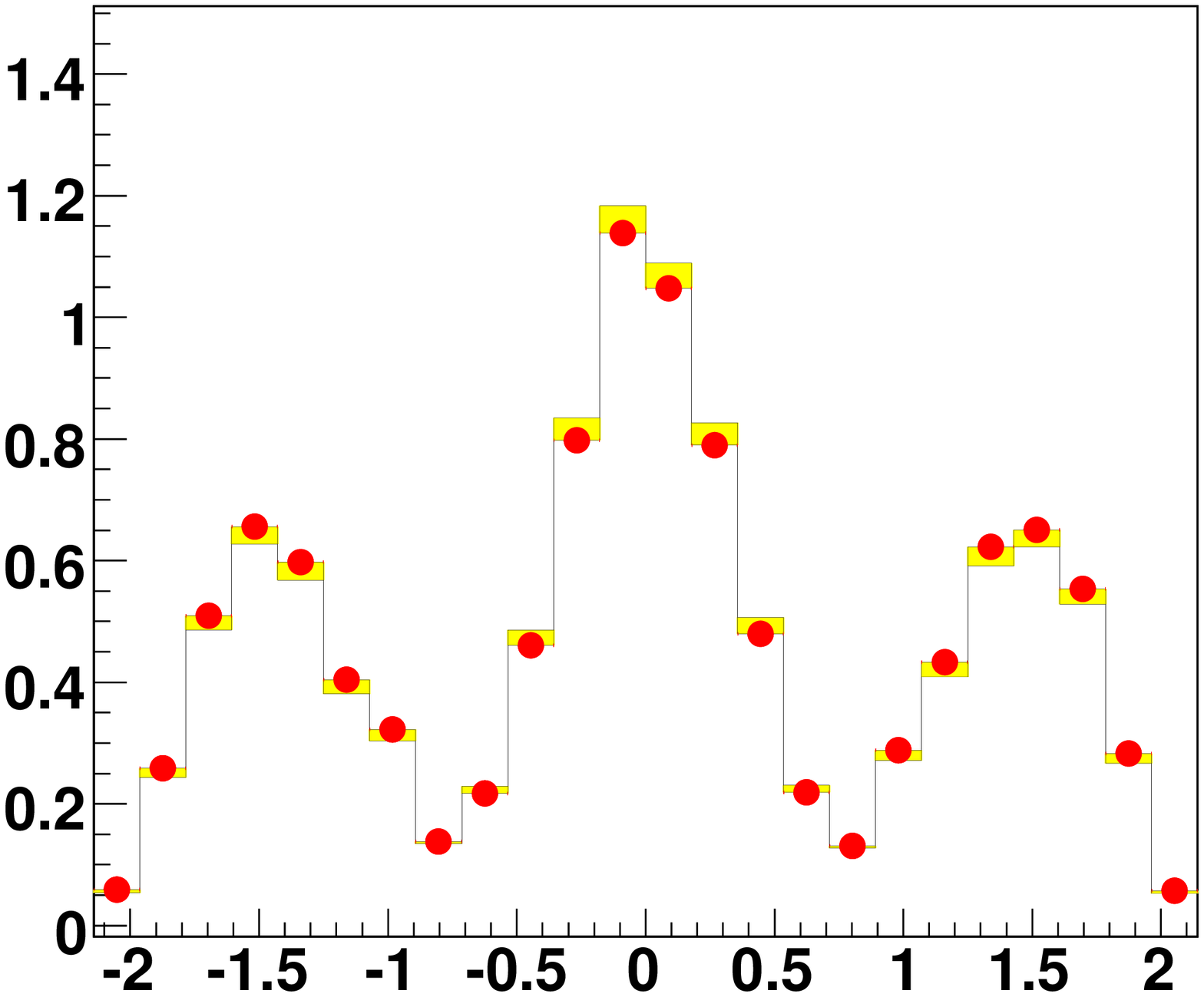}
\includegraphics[width=1.0\textwidth]{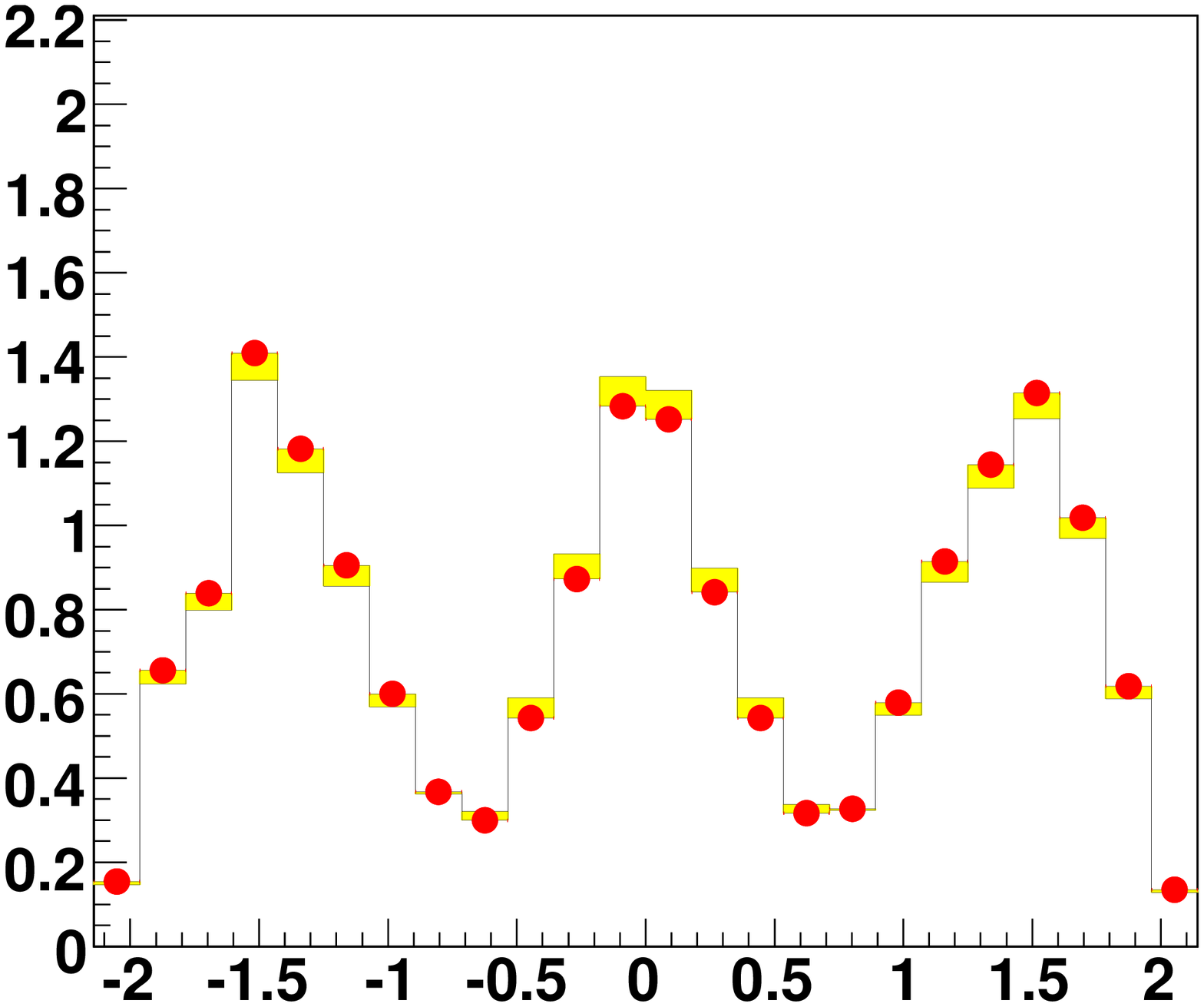}
\includegraphics[width=1.0\textwidth]{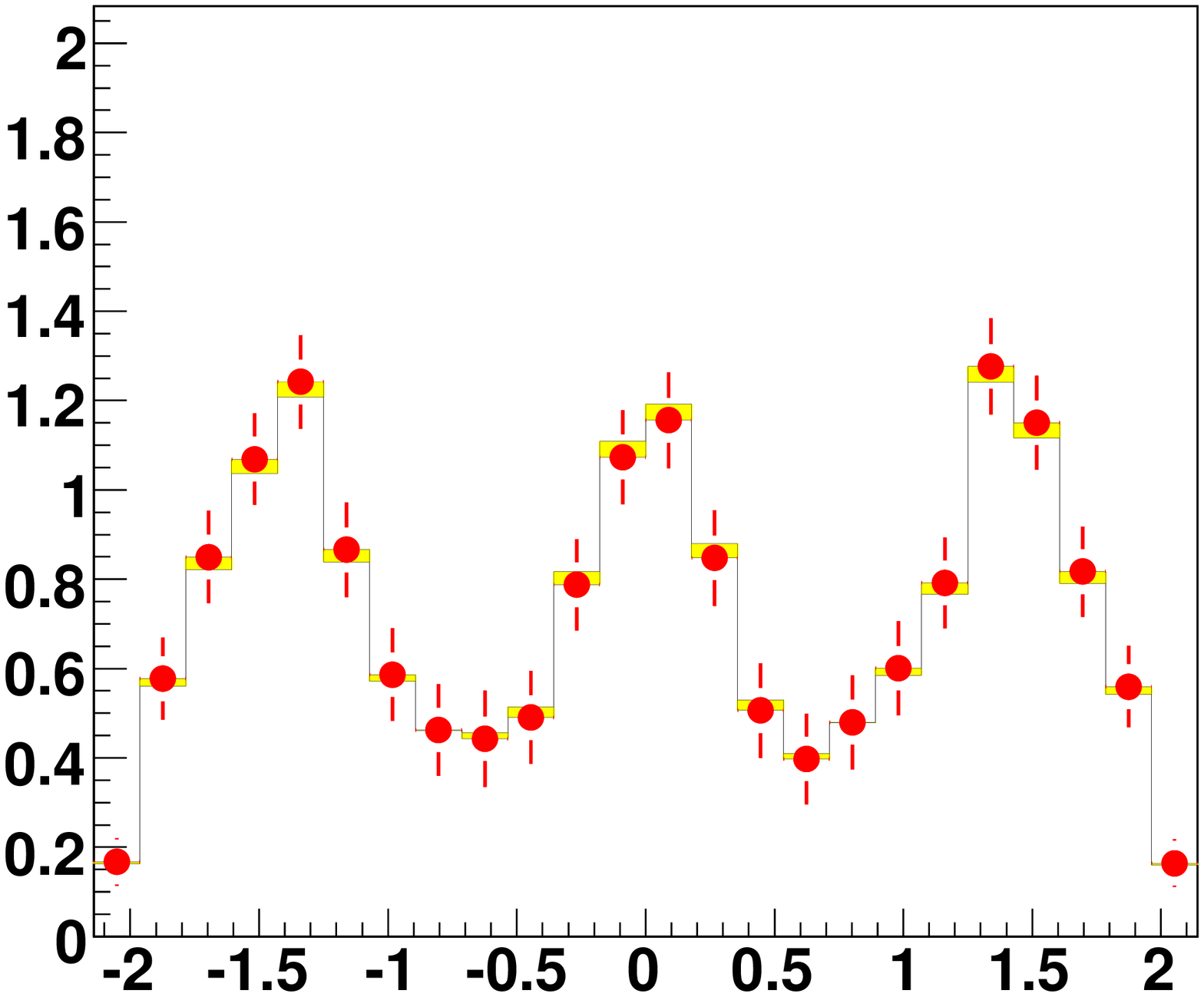}
\includegraphics[width=1.0\textwidth]{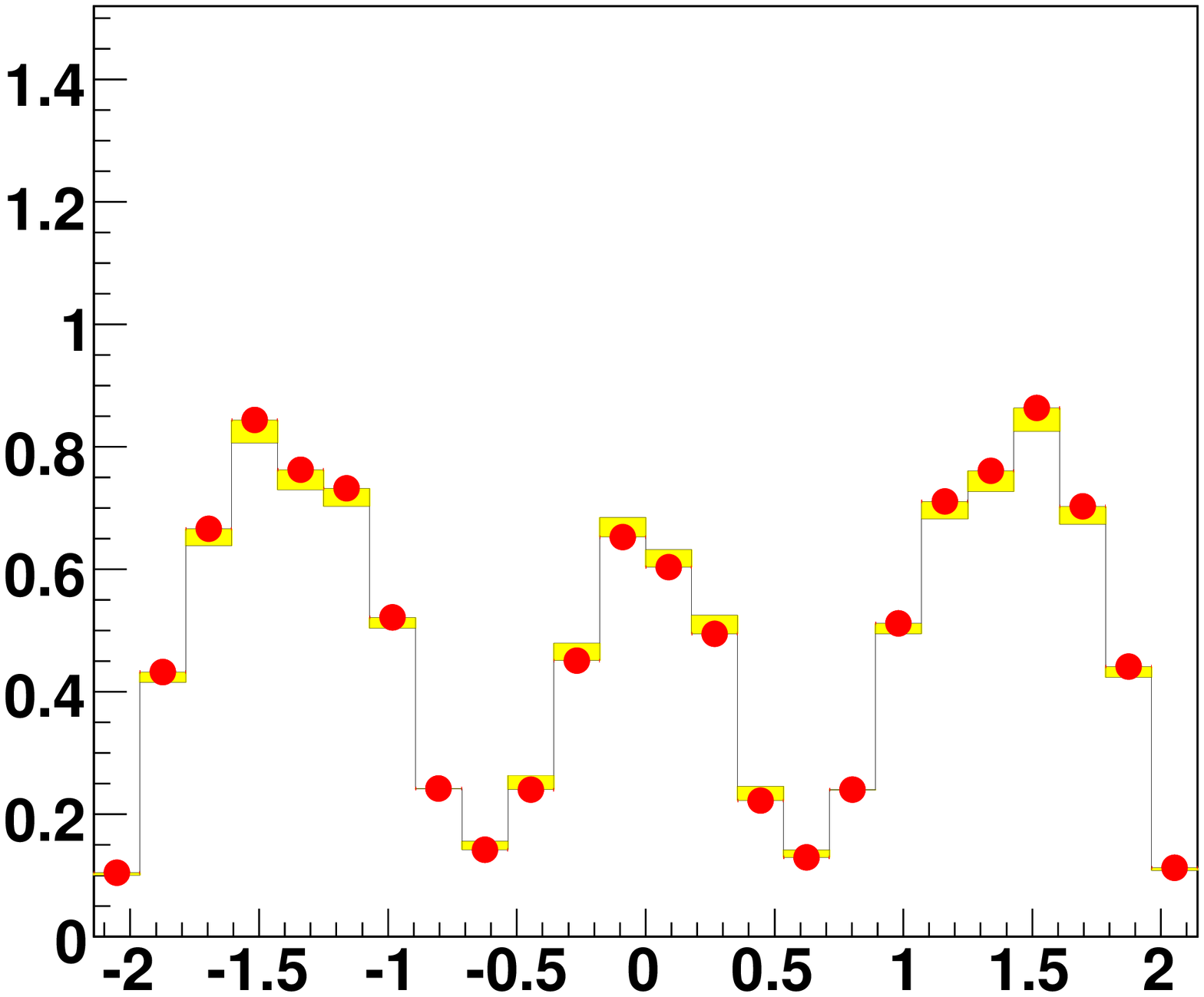}
\end{minipage}
\hfill
\begin{minipage}{0.24\textwidth}
\includegraphics[width=1.0\textwidth]{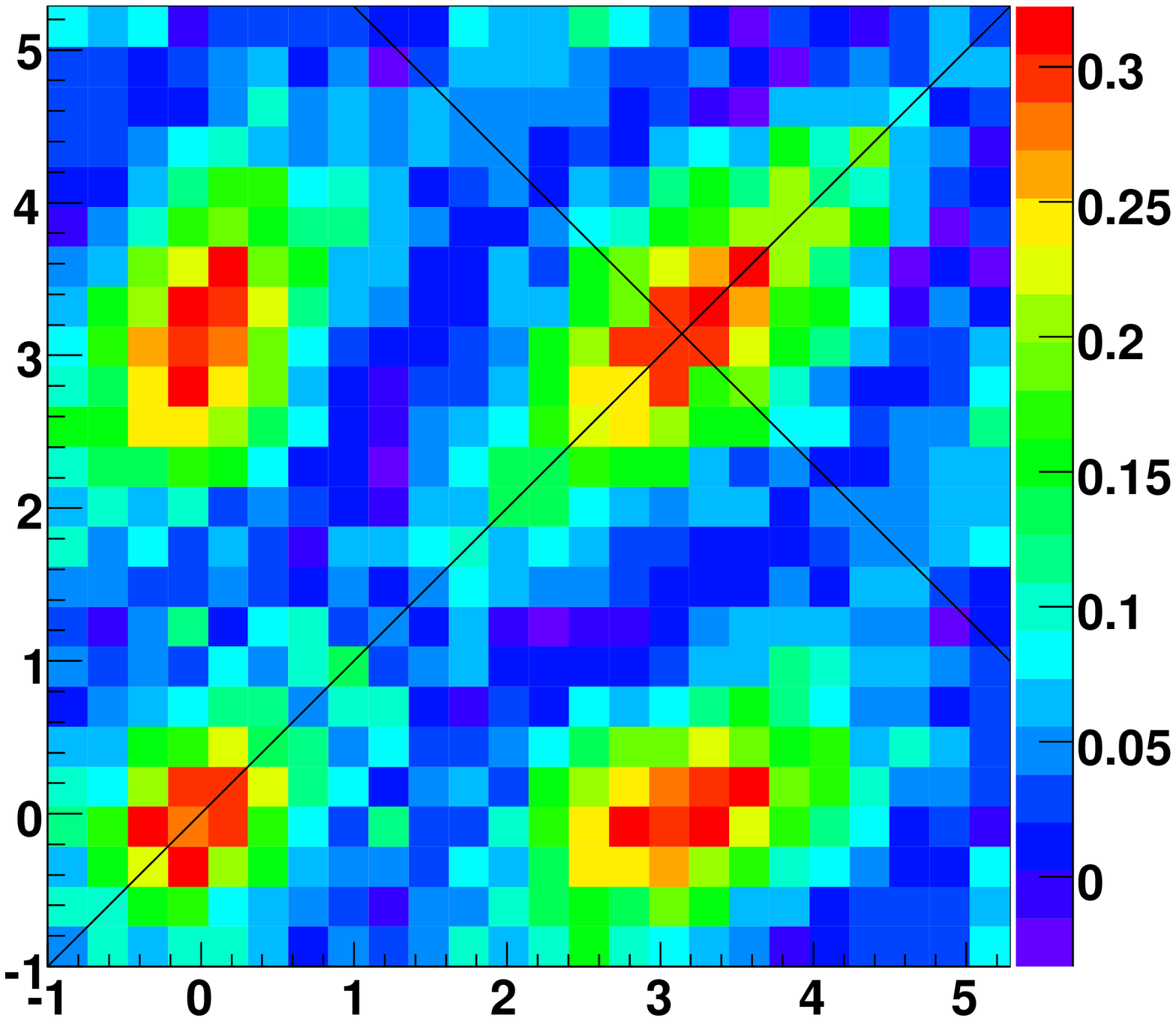}
\includegraphics[width=1.0\textwidth]{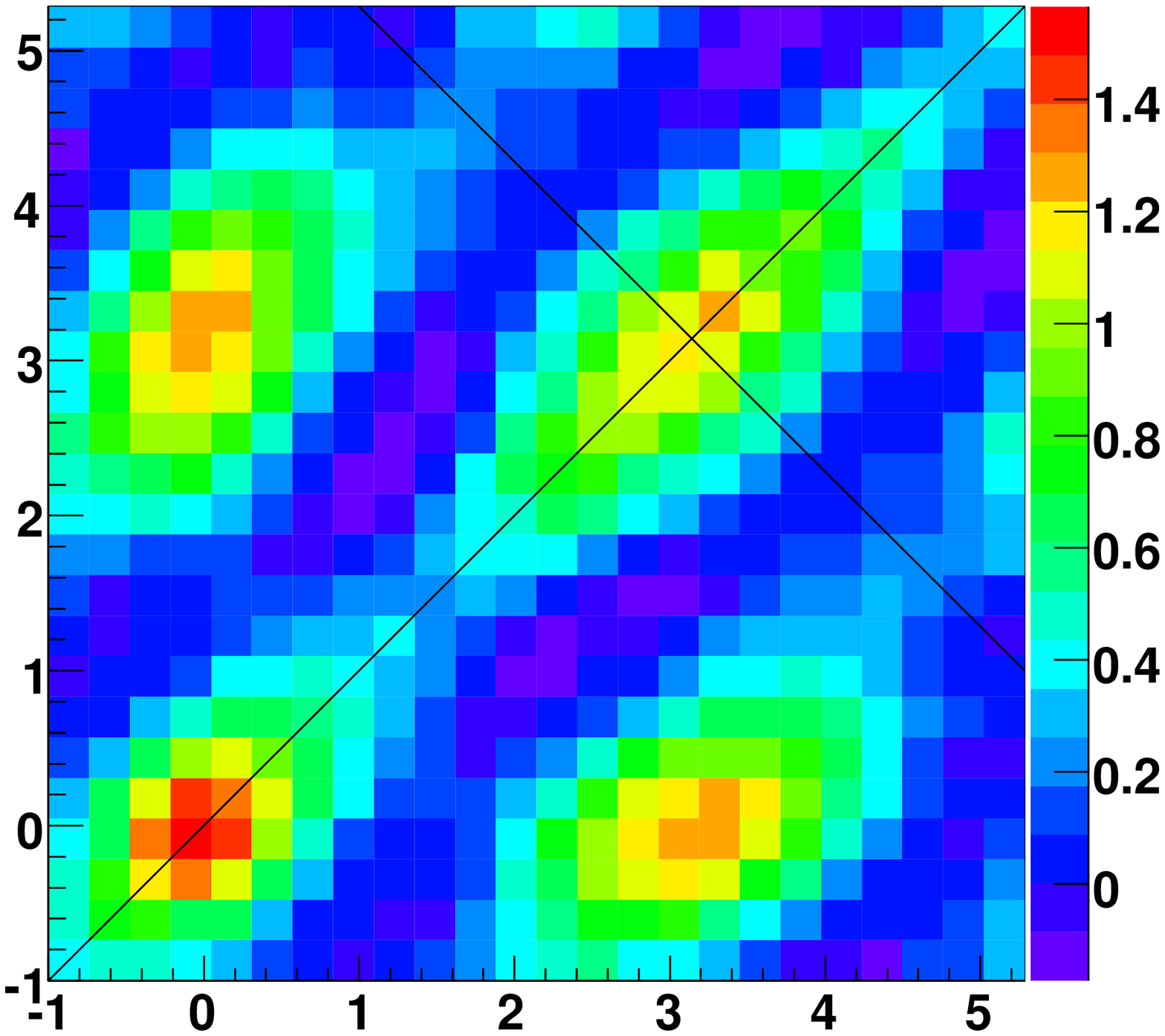}
\includegraphics[width=1.0\textwidth]{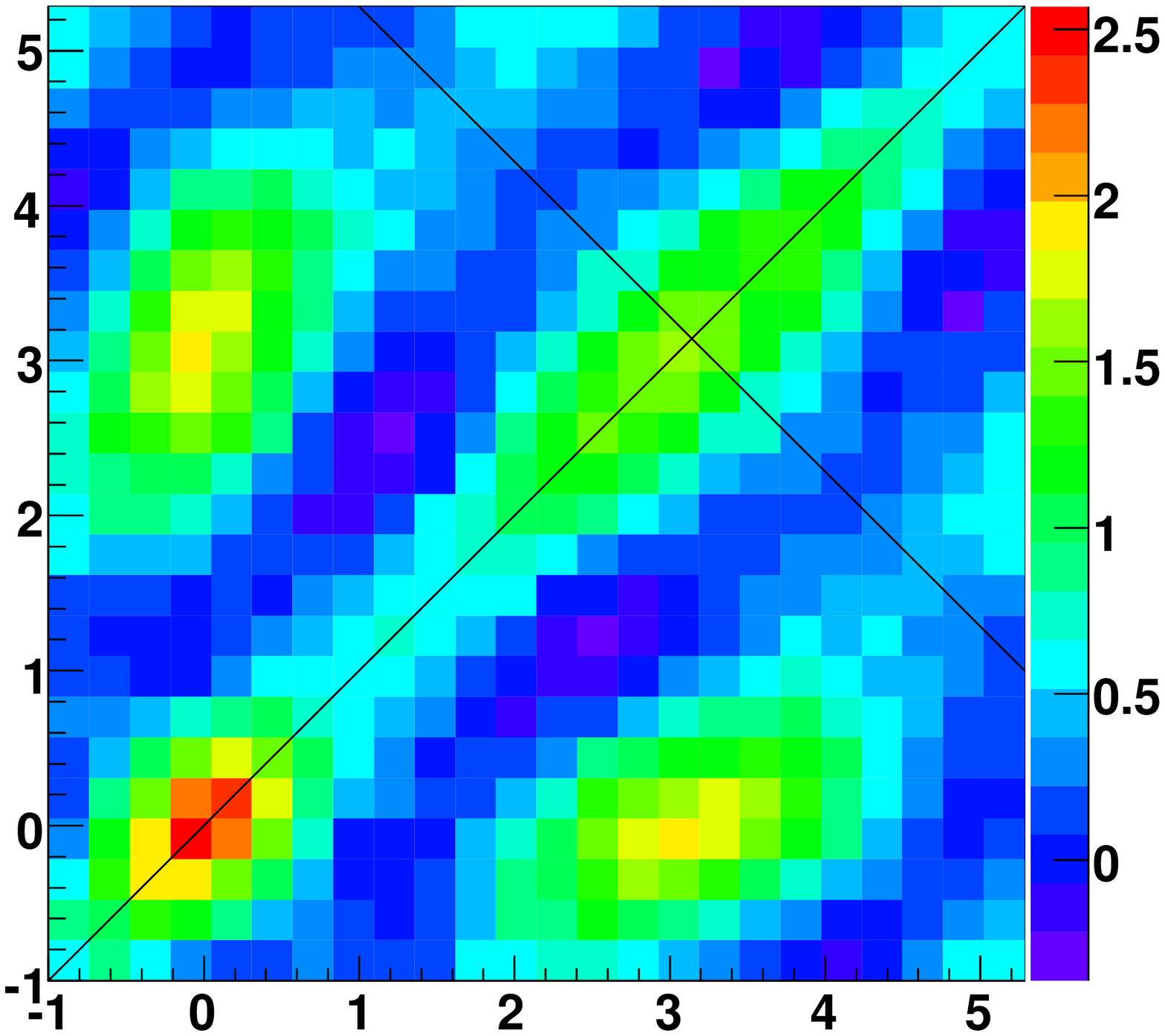}
\includegraphics[width=1.0\textwidth]{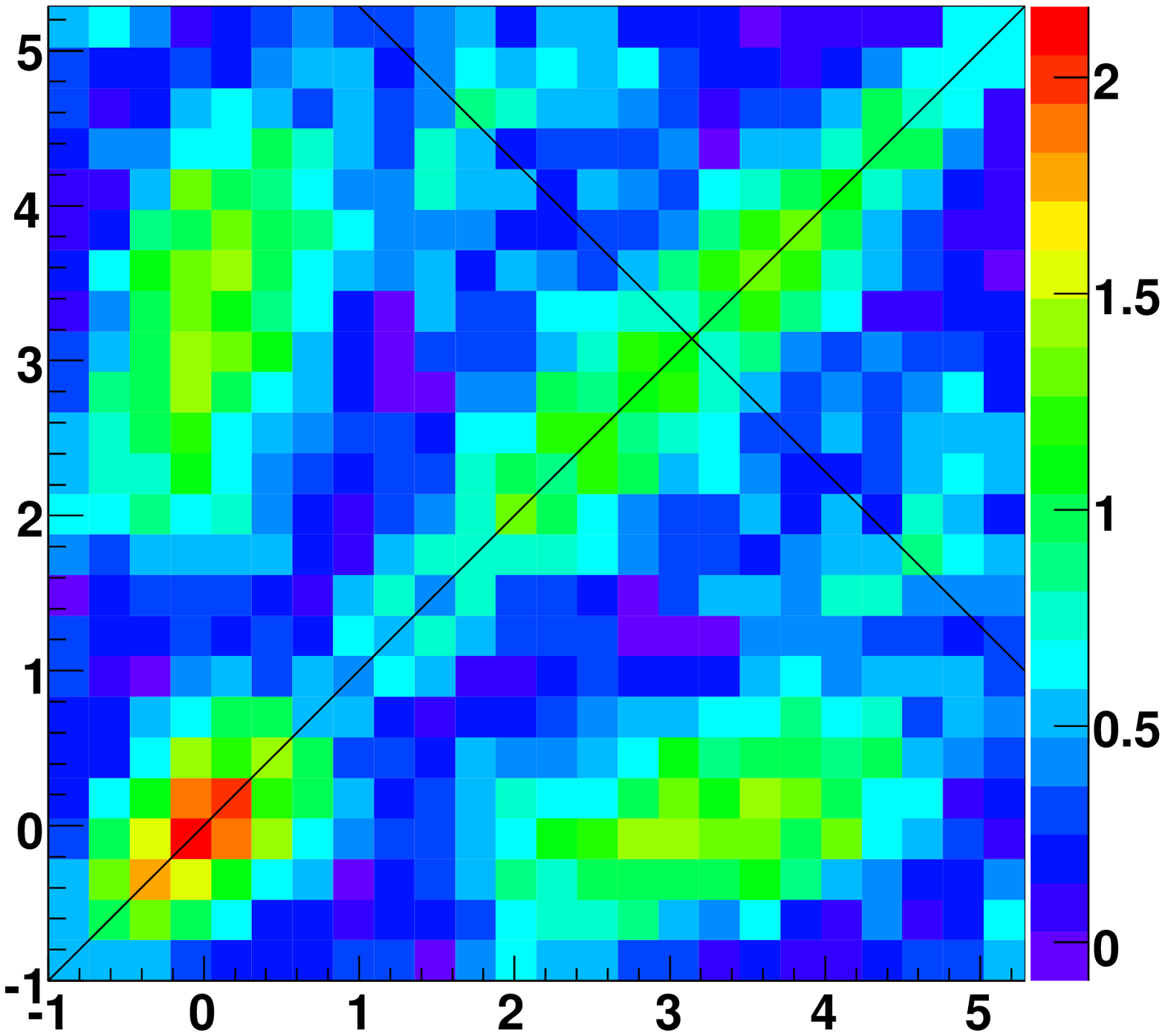}
\includegraphics[width=1.0\textwidth]{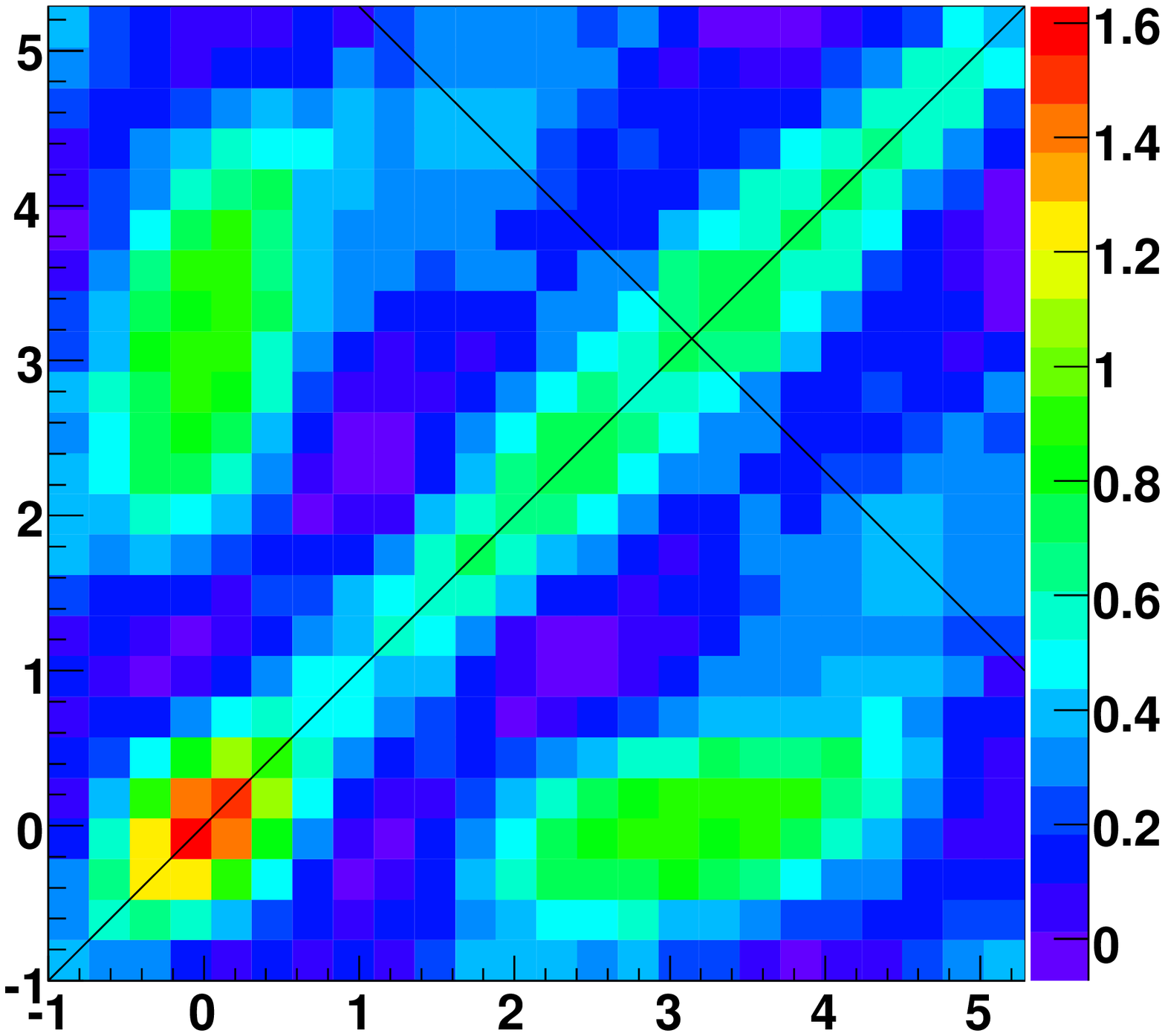}
\end{minipage}
\caption{Background subtracted 3-particle correlations for uncertainty on the jet-flow.  Left:  Jet-flow from 5 GeV/c particles.  Left Center:  Jet-flow same as trigger particle, default.  Right Center:  Off-diagonal away-side projection from default with systematic uncertainty from the uncertainty on jet flow shown in shaded band.  Right:  No jet-flow subtraction.  From top to bottom plots are Au+Au 50-80\%, Au+Au 30-50\%, Au+Au 10-30\%, Au+Au 0-10\%, and ZDC triggered Au+Au 0-12\% collisions at $\sqrt{s_{NN}}=200$ GeV/c.}
\label{fig:jetflow}
\end{figure}

One of the other flow systematics is on the jet-flow.  By default the trigger-associated pair is assumed to have the same azimuthal anisotropy  as the trigger particle.  To assess the effect of this, we assume that the azimuthal anisotropy of the trigger-associated pair may be more like the azimuthal anisotropy of a particle with $p_T=p_T^{Trig}+p_T^{assoc}$.  We use the $v_2$ of 5 GeV/c particles which is approximately the total $p_T$ of a trigger particle of $3<p_T<4$ GeV/c and an associated particle of $1<p_T<2$ GeV/c to estimate the systematic uncertainty due to jet-flow.  The jet-flow is put into the analysis in the construction of the hard-soft term by folding the flow modulated background with jet-flow value.  The background subtracted 3-particle correlation using the 5 GeV/c particle flow is shown in Fig.~\ref{fig:jetflow} along with the default jet-flow for comparison.  As an additional cross check, the background subtracted 3-particle correlation was examined without the jet-flow subtraction.  This was done as a check to see if observing a conical emission signal was dependent on whether on not this subtraction was performed.   It is not included in the systematic error bars.  The signal is reduced but still present without this subtraction as shown in Fig.~\ref{fig:jetflow}, right.  

The estimate of this systematic uncertainty on the azimuthal anisotropy of the trigger associated pair can be improved in the future by using 2-particle correlations for trigger particles with different orientations with respect to the reaction plane.  The 2-particle correlation can be divided into bins depending on the trigger particle orientation with respect to the reaction plane.  This would provide the reaction plane dependence of the trigger associated pair.  The hard-soft term could be constructed for each of these bins and summed together with a number of trigger particle weighting.  This would provide a better estimate of the azimuthal anisotropy of the trigger associated pair than what is currently used.  Once this has been performed, since it is a better estimate for the jet-flow, this should probably be used for the default value and the current value should be used for the assessment of the systematic uncertainty.

\begin{figure}[htbp]
\hfill
\begin{minipage}{0.07\textwidth}
\includegraphics[width=1.0\textwidth]{Plots/blank2.eps}
\end{minipage}
\hfill
\begin{minipage}{0.25\textwidth}
\includegraphics[width=1.0\textwidth]{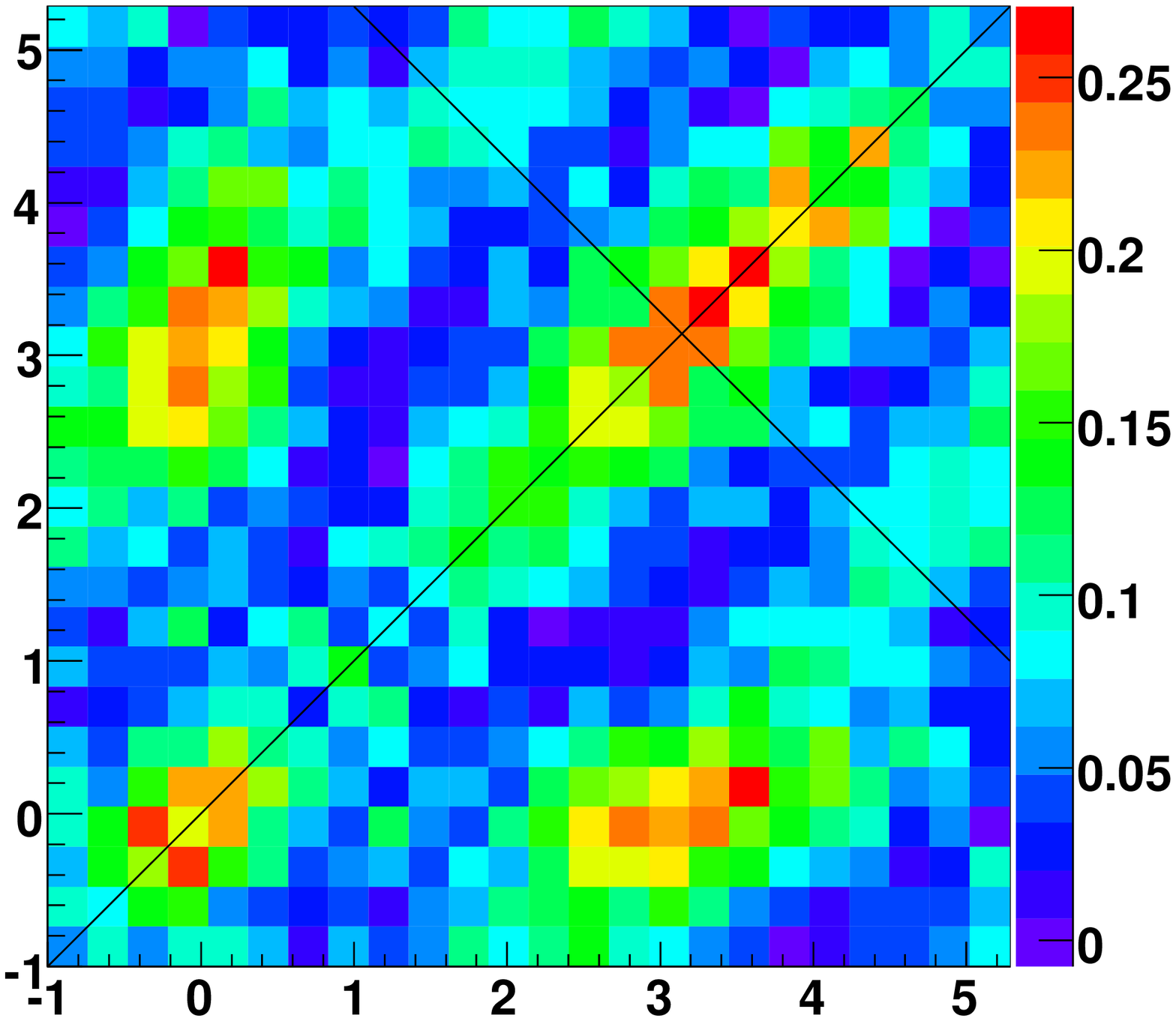}
\includegraphics[width=1.0\textwidth]{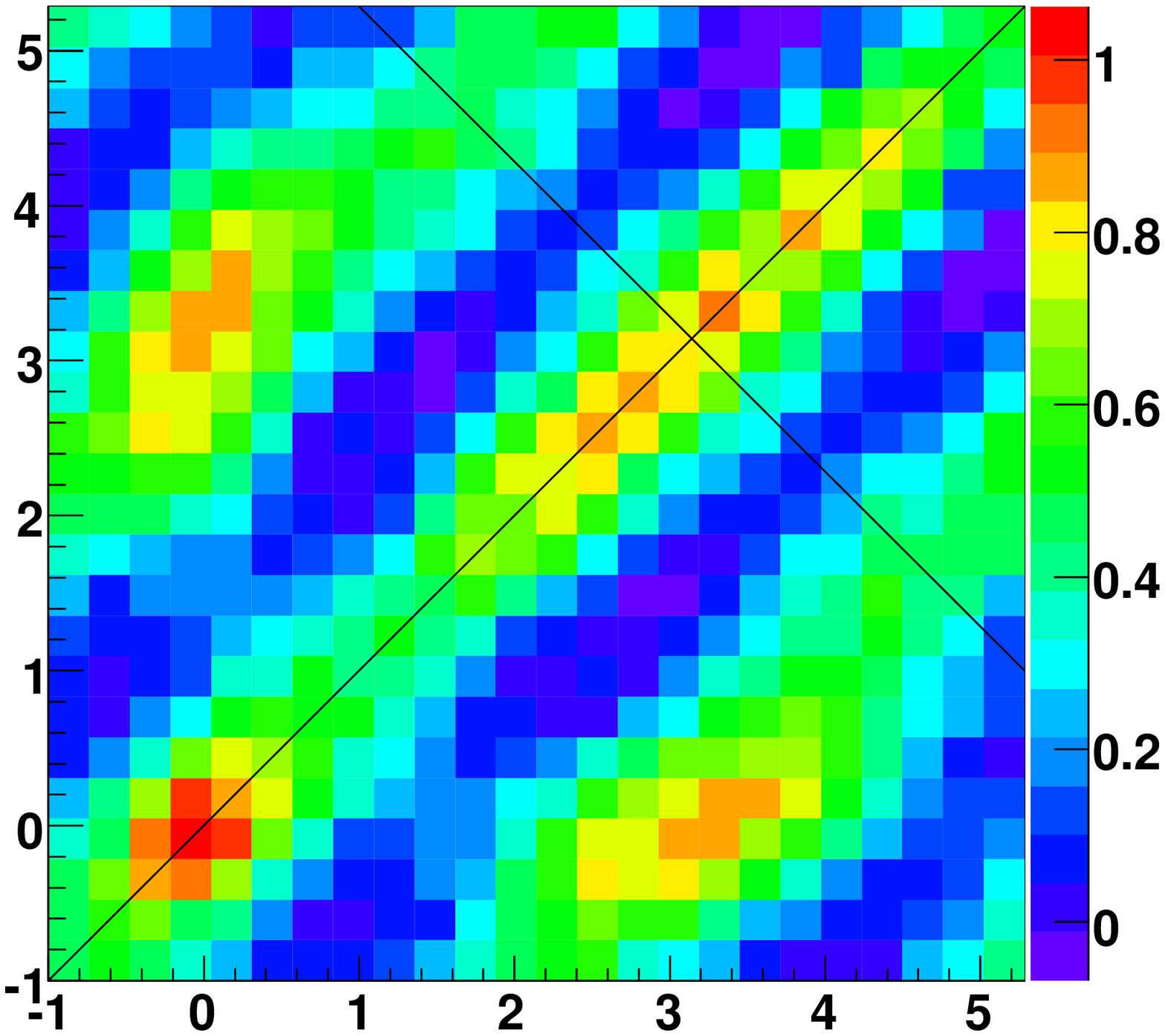}
\includegraphics[width=1.0\textwidth]{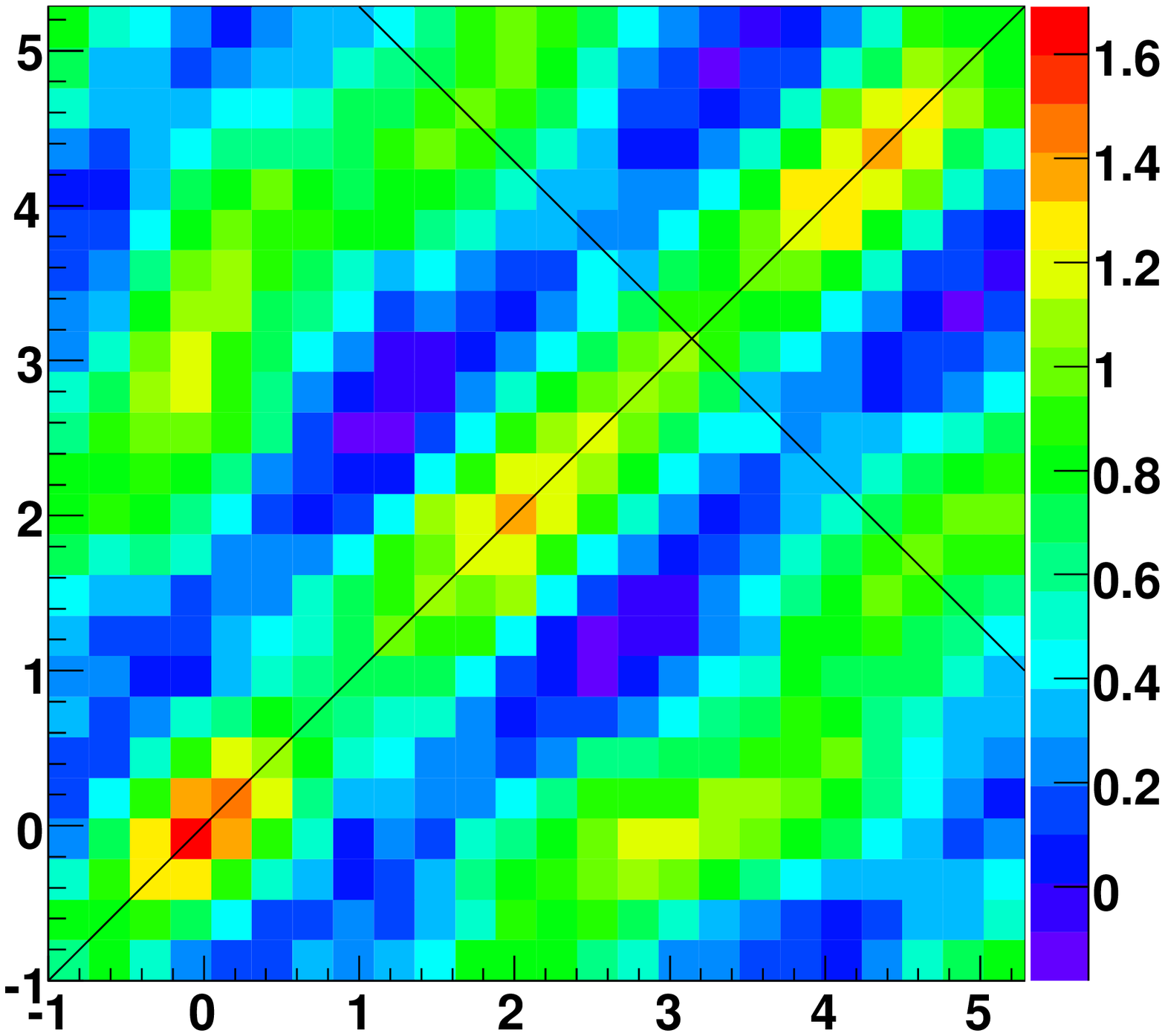}
\includegraphics[width=1.0\textwidth]{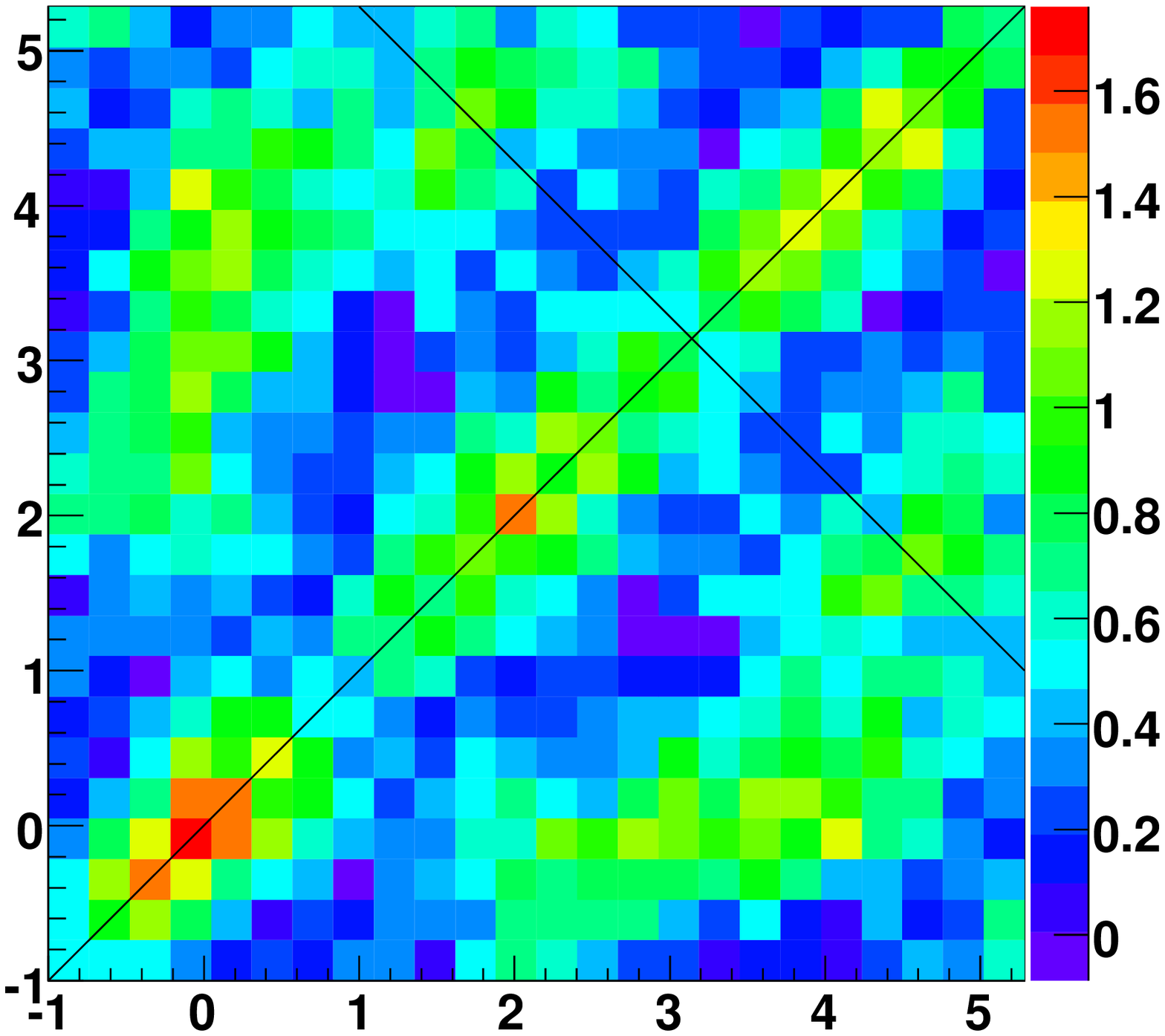}
\includegraphics[width=1.0\textwidth]{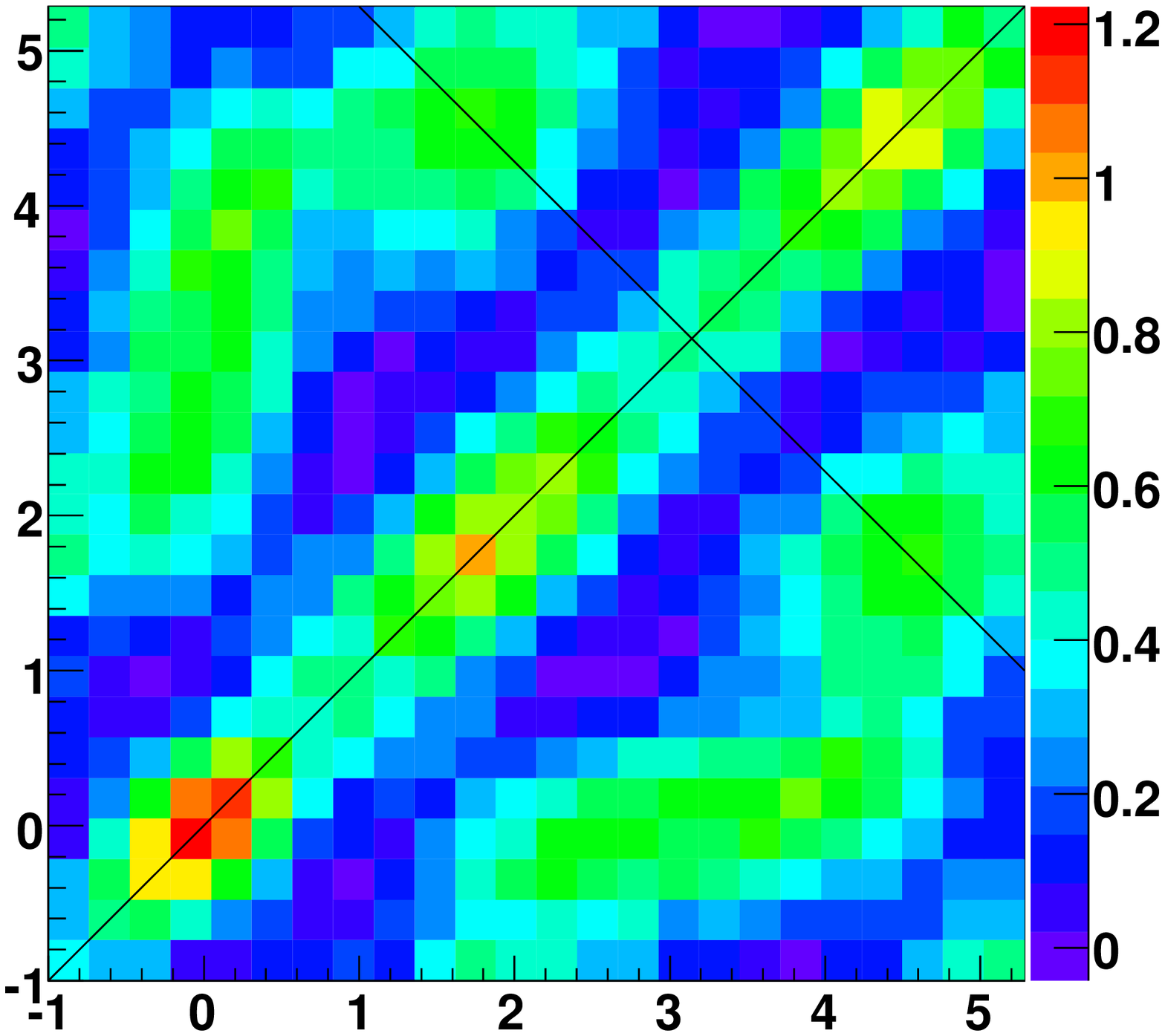}
\end{minipage}
\hfill   
\begin{minipage}{0.25\textwidth}
\includegraphics[width=1.0\textwidth]{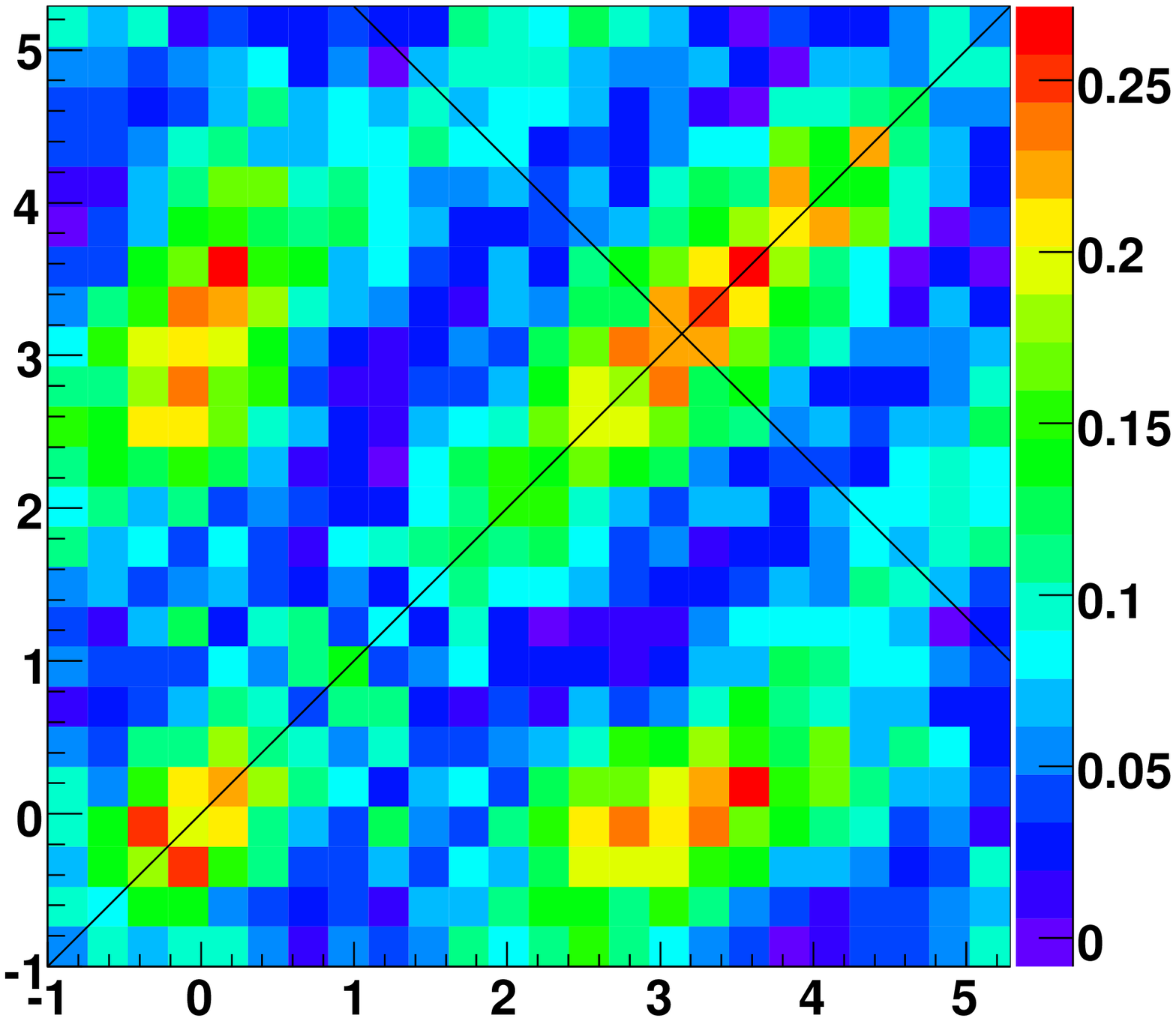}
\includegraphics[width=1.0\textwidth]{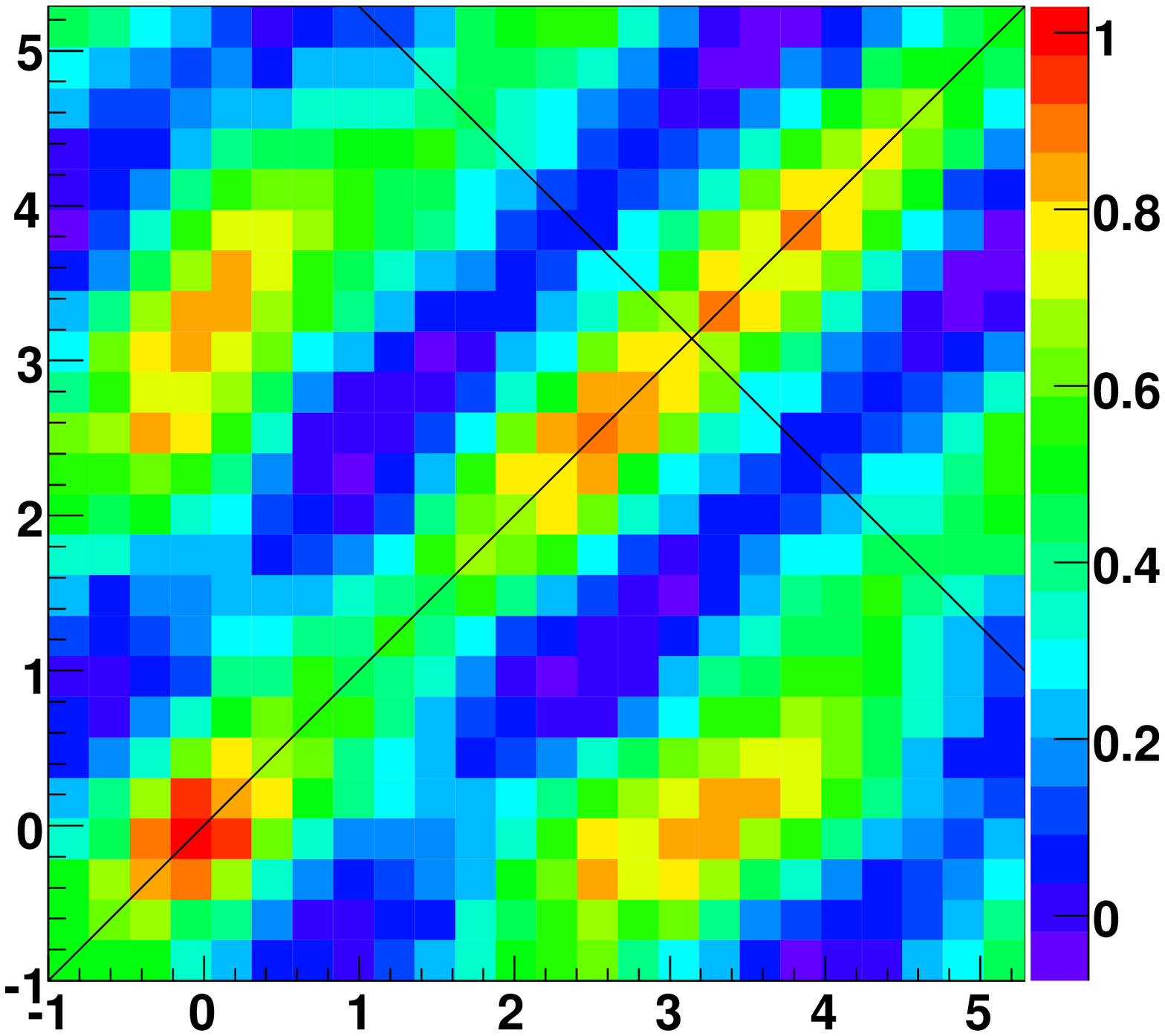}
\includegraphics[width=1.0\textwidth]{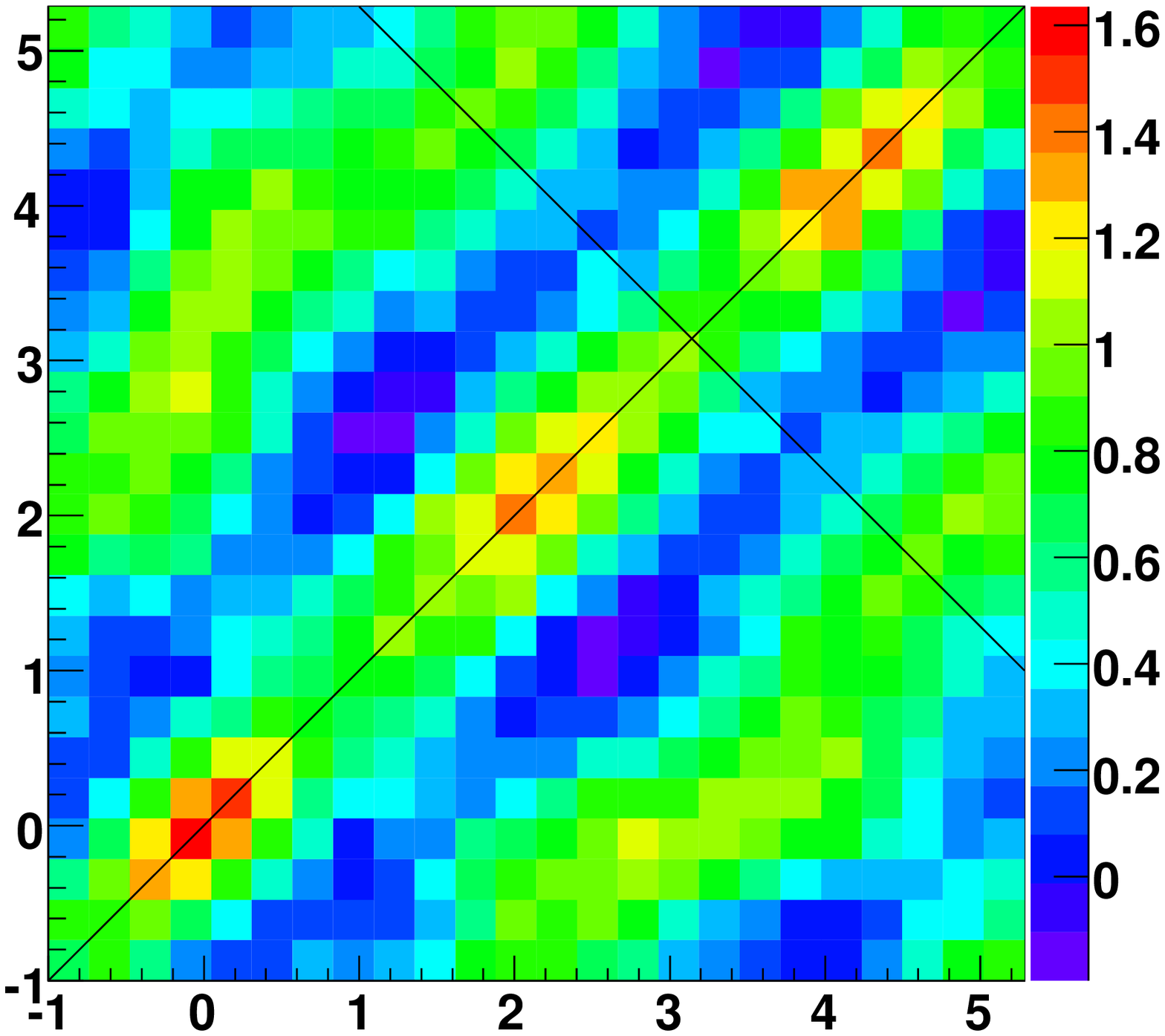}
\includegraphics[width=1.0\textwidth]{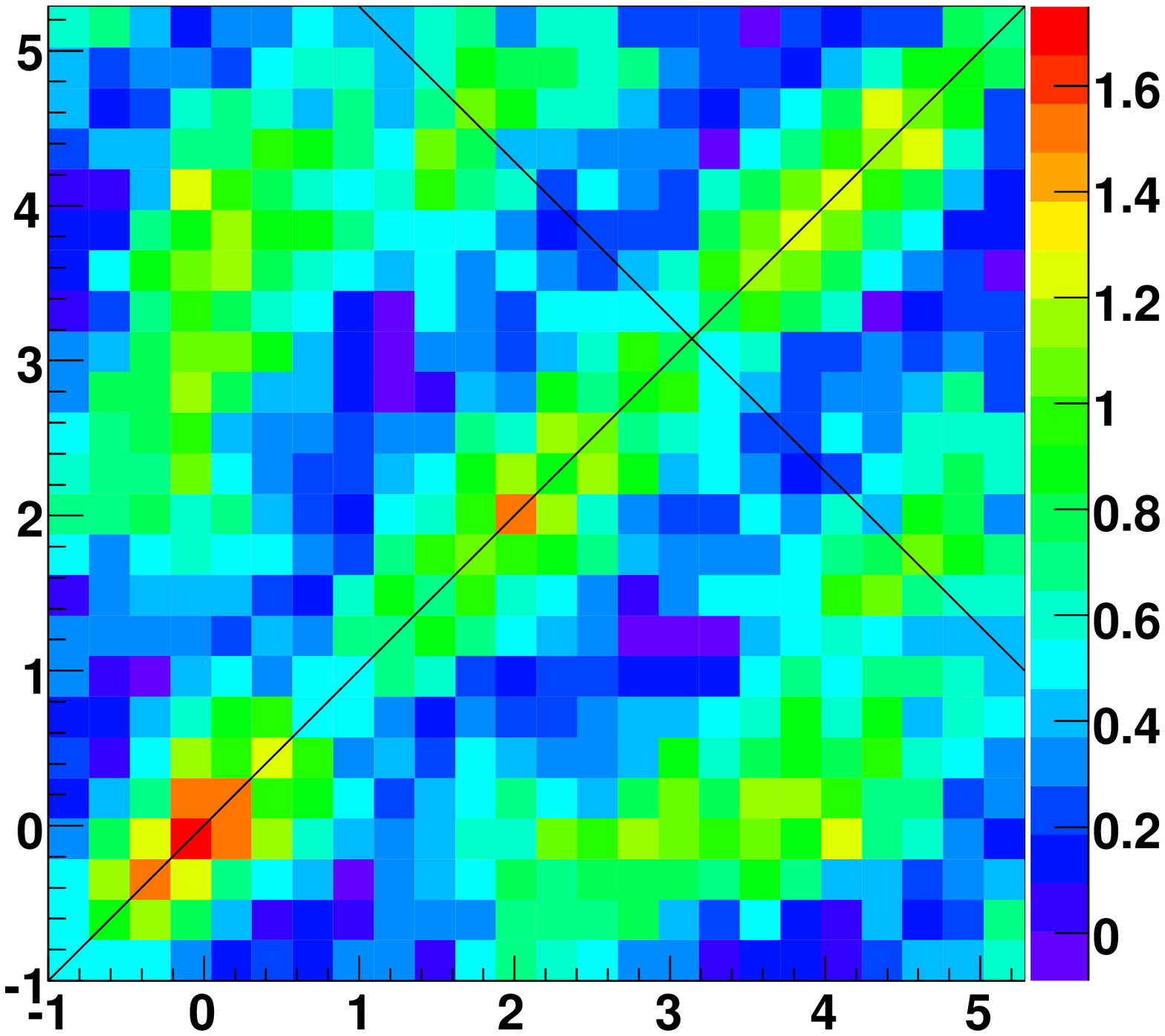}
\includegraphics[width=1.0\textwidth]{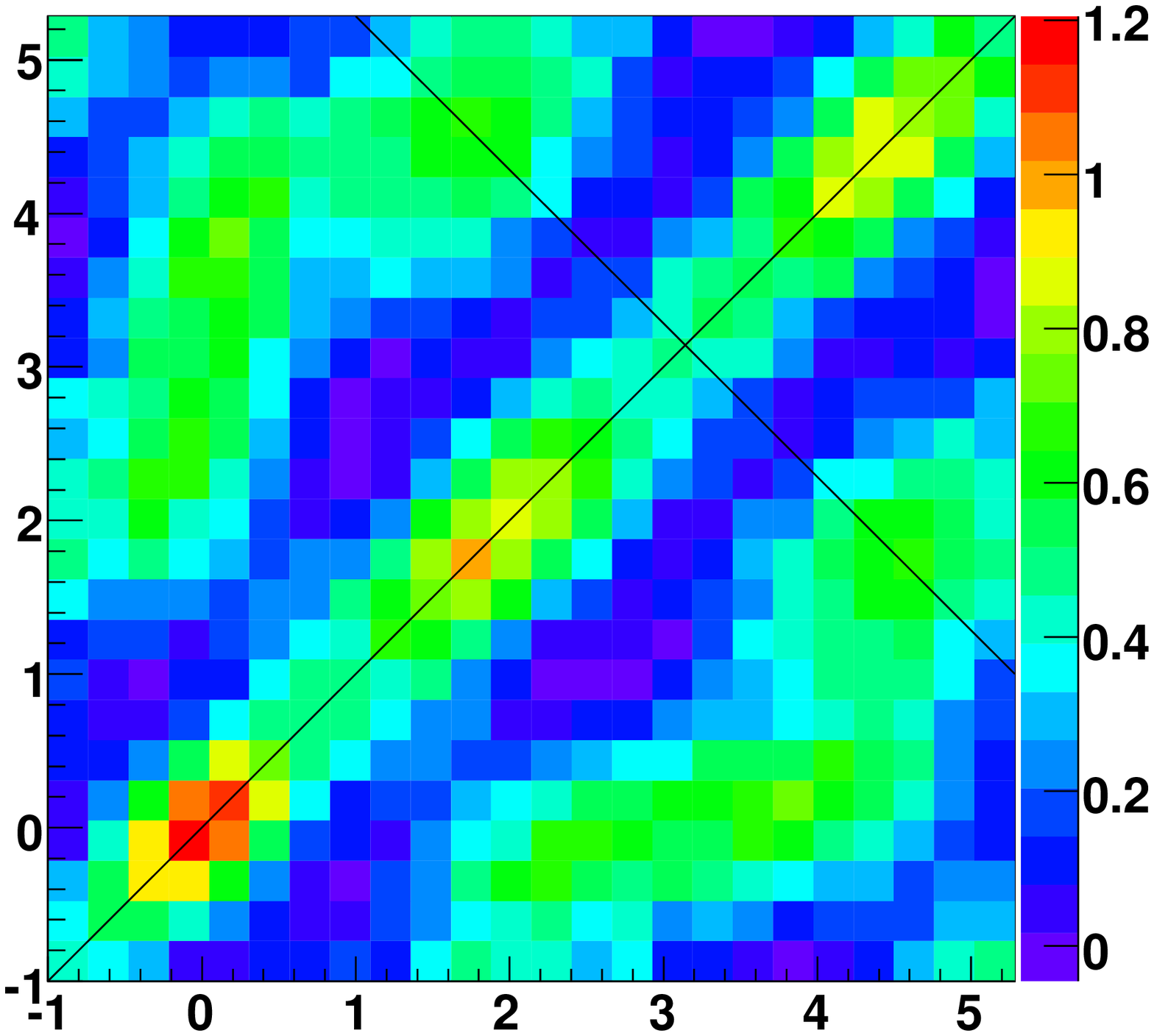}
\end{minipage}
\hfill
\begin{minipage}{0.25\textwidth}
\includegraphics[width=1.0\textwidth]{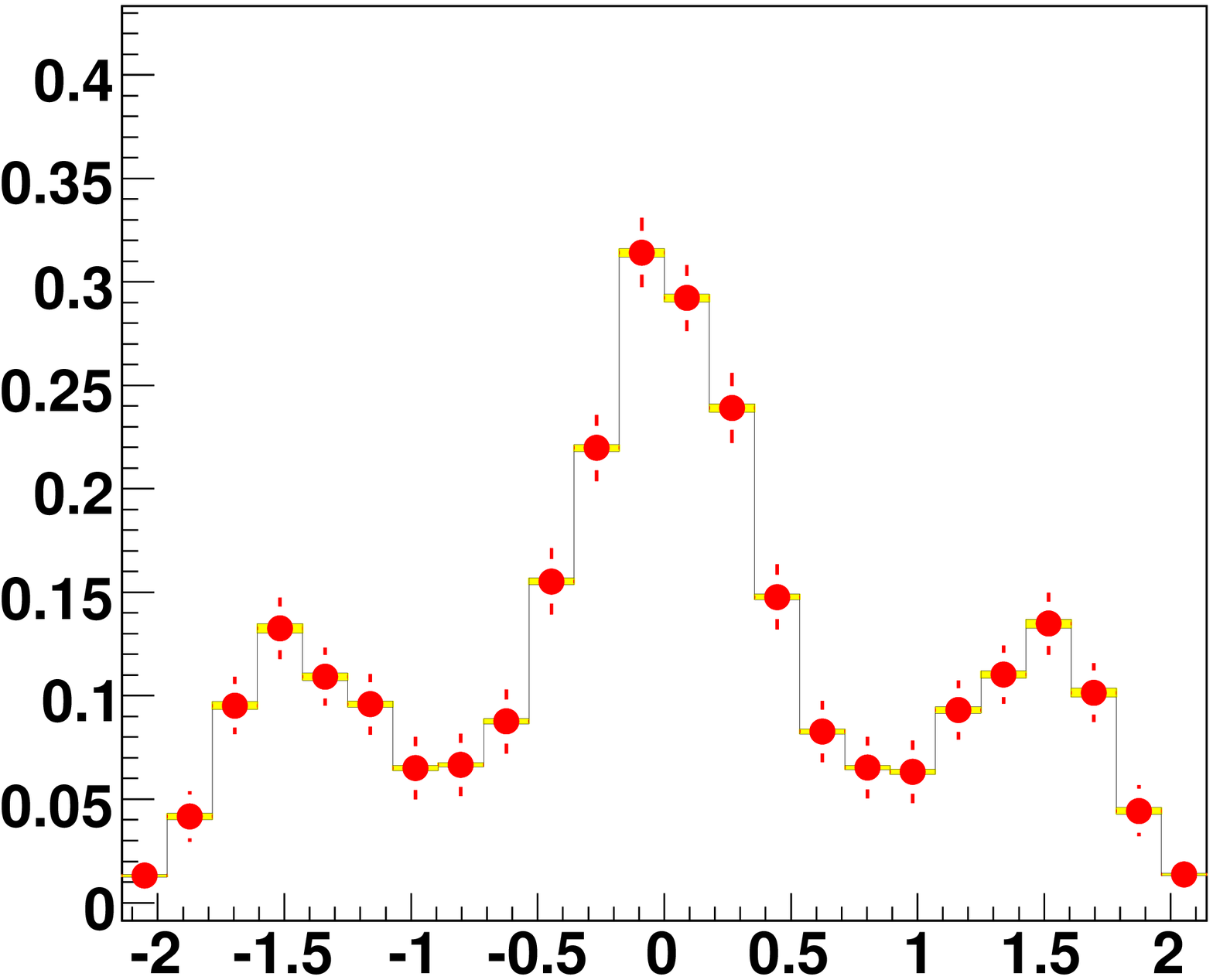}
\includegraphics[width=1.0\textwidth]{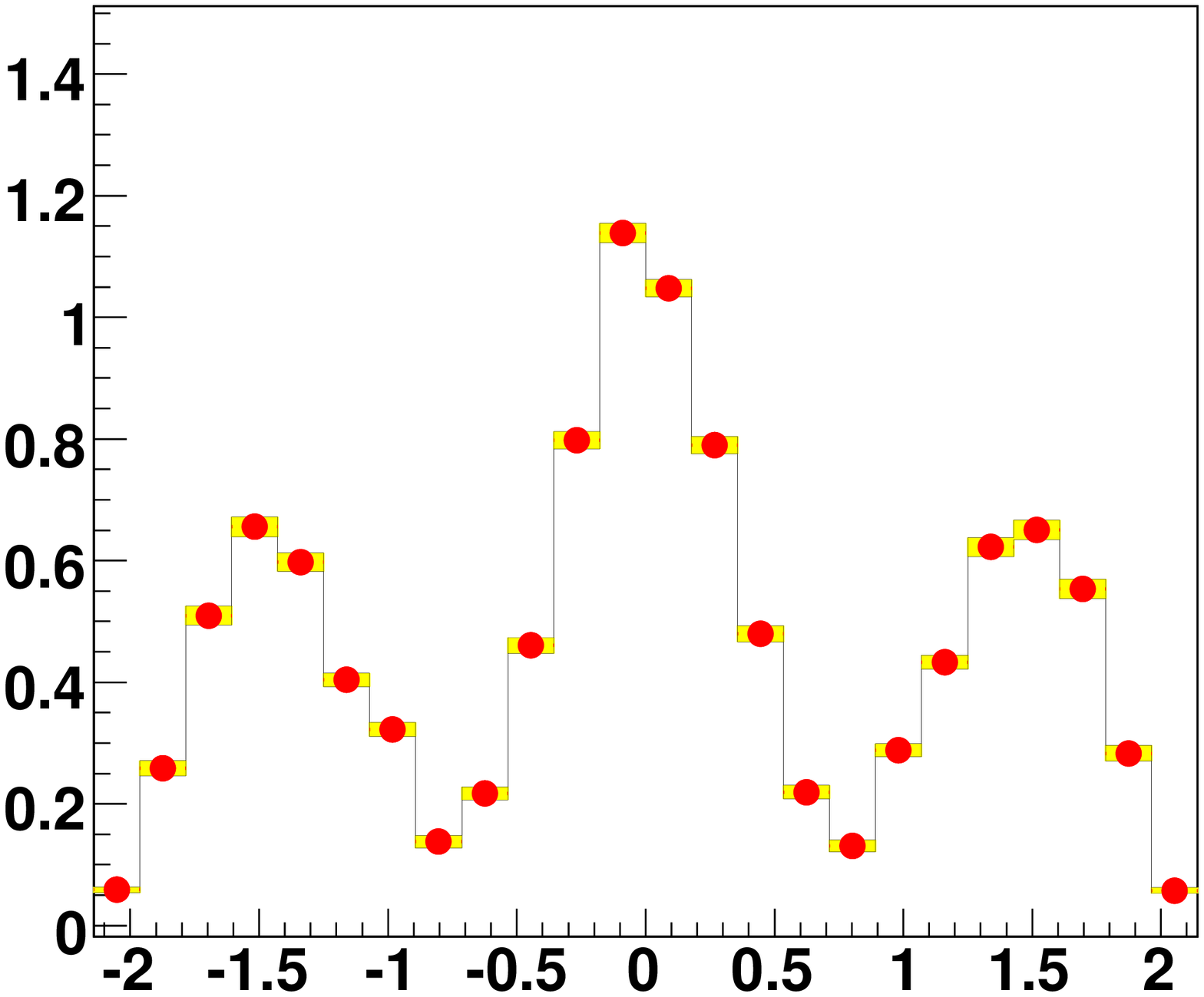}
\includegraphics[width=1.0\textwidth]{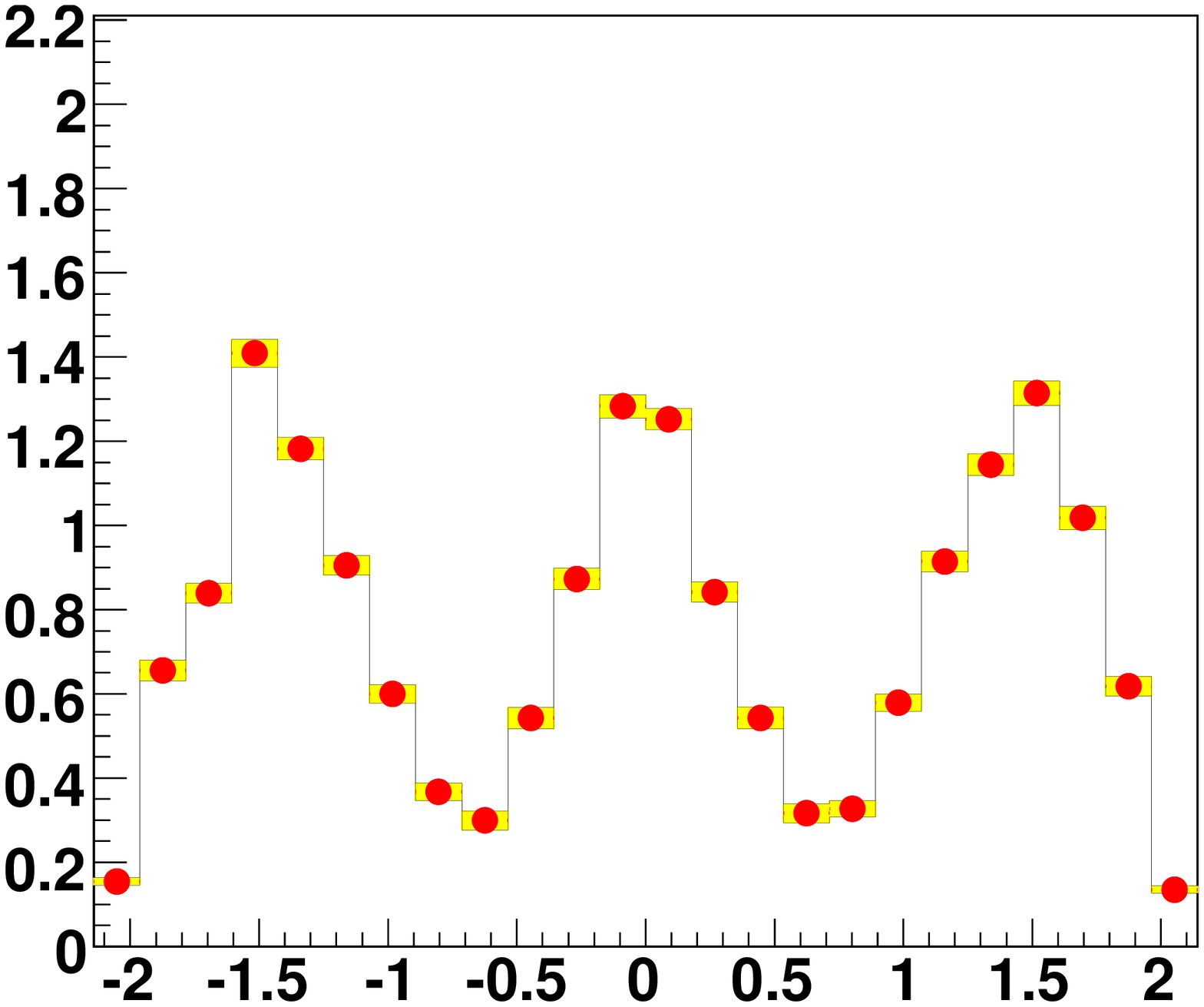}
\includegraphics[width=1.0\textwidth]{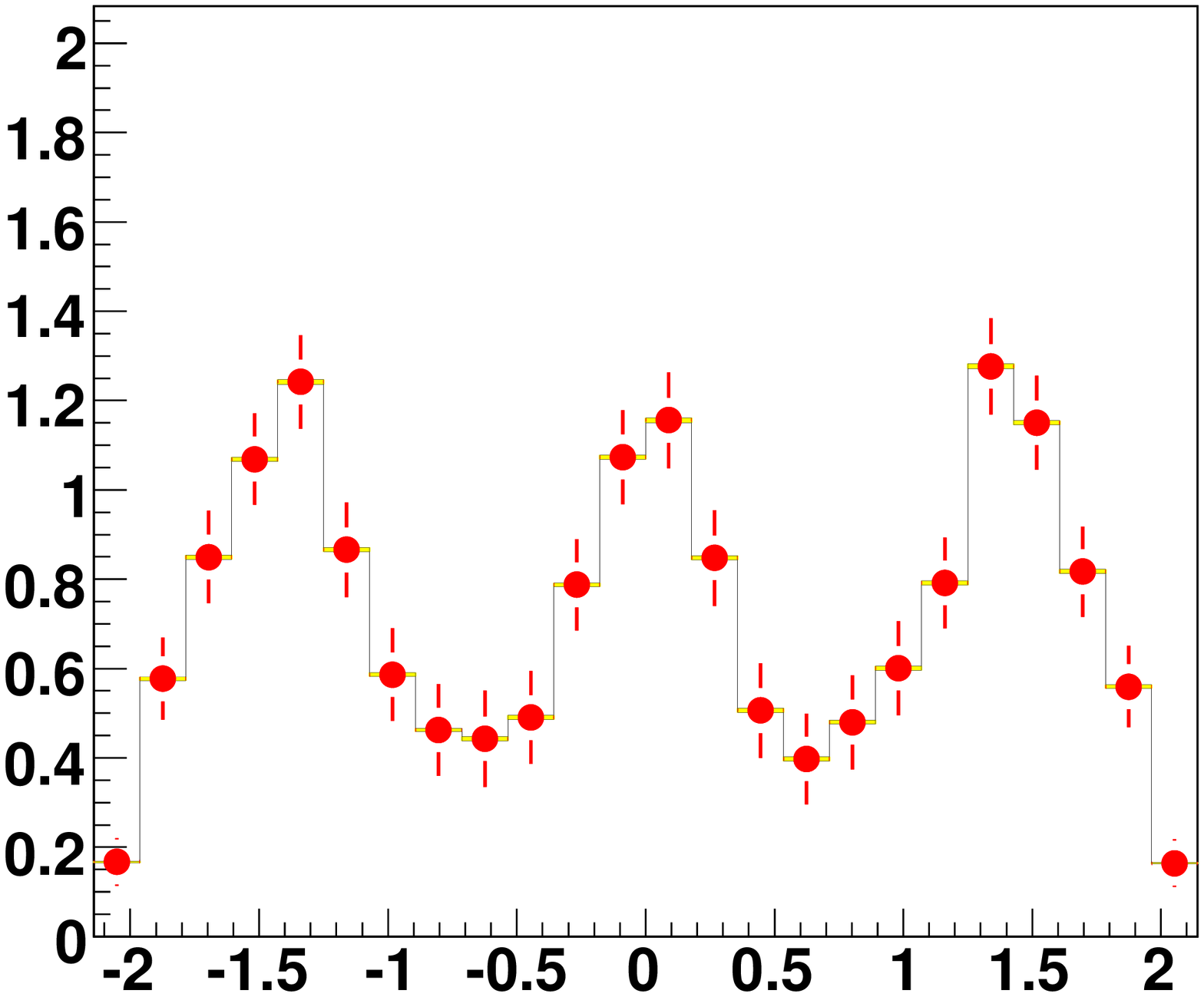}
\includegraphics[width=1.0\textwidth]{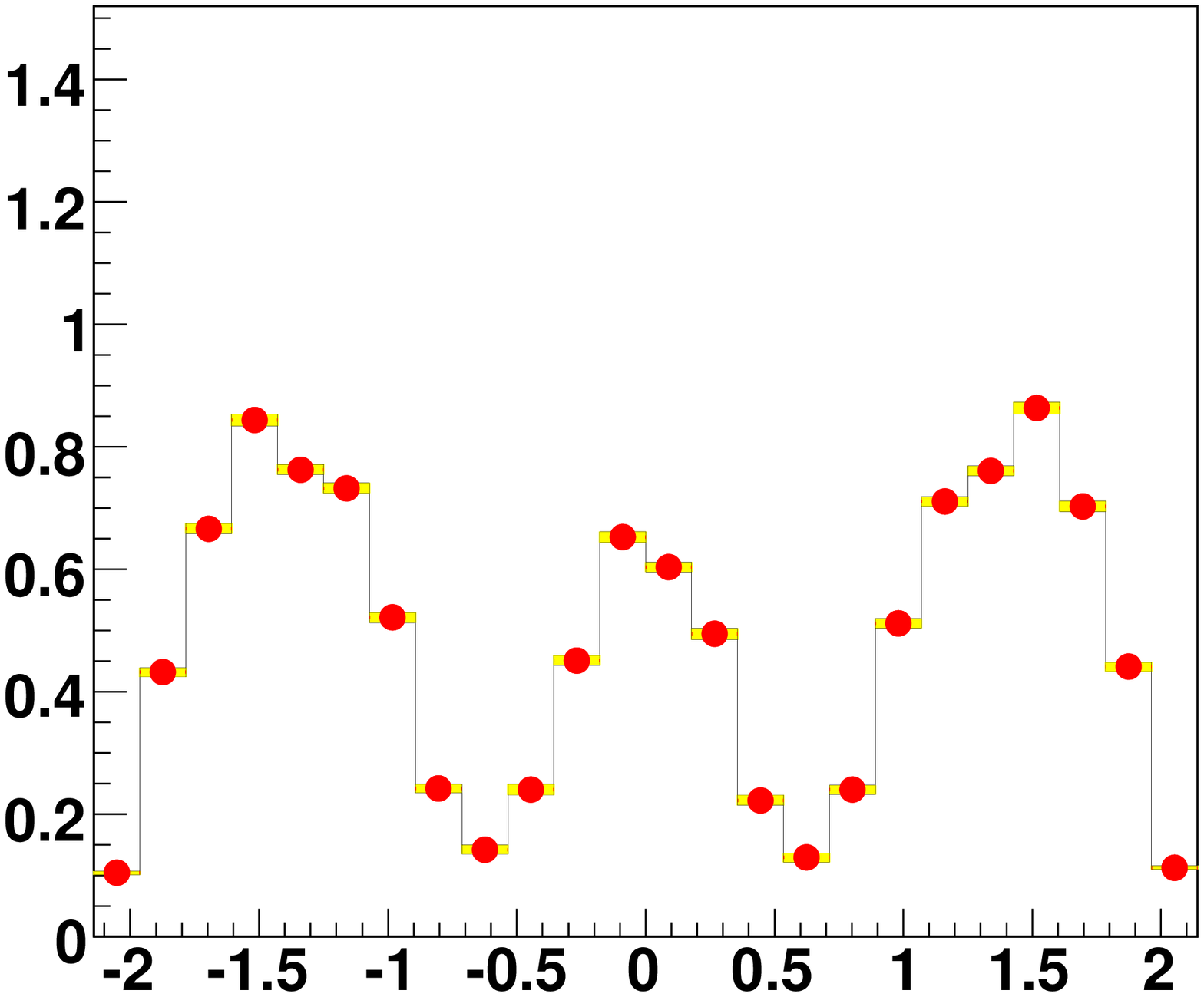}
\end{minipage}
\hfill
\begin{minipage}{0.07\textwidth}
\includegraphics[width=1.0\textwidth]{Plots/blank2.eps}
\end{minipage}
\caption{Background subtracted 3-particle correlations for uncertainty on the trigger particle $v_{4}$.  Left:  Trigger particle $v_{4}$ decreased by 20\%.  Center:  Trigger particle $v_{4}$ increased by 20\%.  Right:  Off-diagonal away-side projection from default with systematic uncertainty from the uncertainty on trigger particle $v_{4}$ shown in shaded band.  From top to bottom plots are Au+Au 50-80\%, Au+Au 30-50\%, Au+Au 10-30\%, Au+Au 0-10\%, and ZDC triggered Au+Au 0-12\% collisions at $\sqrt{s_{NN}}=200$ GeV/c.}
\label{fig:v4sys}
\end{figure}

Another uncertainty on the flow is from the parameterization of the $v_{4}$.  The $v_{4}$ was fit to the ratio of the $v_{4}/v_{2}^{2}$ for $1<p_{T}<2$ GeV/c.  This was not done for our trigger particles of $3<p_{T}<4$ GeV/c because the $v_{4}$ data does not go out to this high in $p_{T}$.  To account for the uncertainty on the trigger particle $v_{4}$ due to this a $\pm$20\% uncertianity on the trigger particle $v_4$ has been applied.  Figure~\ref{fig:v4sys} shows the background subtracted 3-particle correlations with the trigger particle $v_{4}$ decreased by 20\% and increased by 20\%.

There is an additional uncertainty on the elliptic flow in the ZDC triggered data.  There have been no flow measurements made for the ZDC data so we have used the minimum bias flow values.  These have been applied such that a ZDC event uses the $v_2$ value from the centrality bin that it would fall into if it were a minimum bias event.  A cross check has been applied by parameterizing the $v_2$ values as a function of multiplicity and using the value from the parameterization.  Due to first order cancellation of the $v_2*v_2$ and $v_4*v_4$ this was found to have a negligible impact on the final result. 

\subsection{Finite Multiplicity Bin Systematics}

\begin{figure}[htbp]
\hfill
\begin{minipage}{0.07\textwidth}
\includegraphics[width=1.0\textwidth]{Plots/blank2.eps}
\end{minipage}
\hfill
\begin{minipage}{0.25\textwidth}
\includegraphics[width=1.0\textwidth]{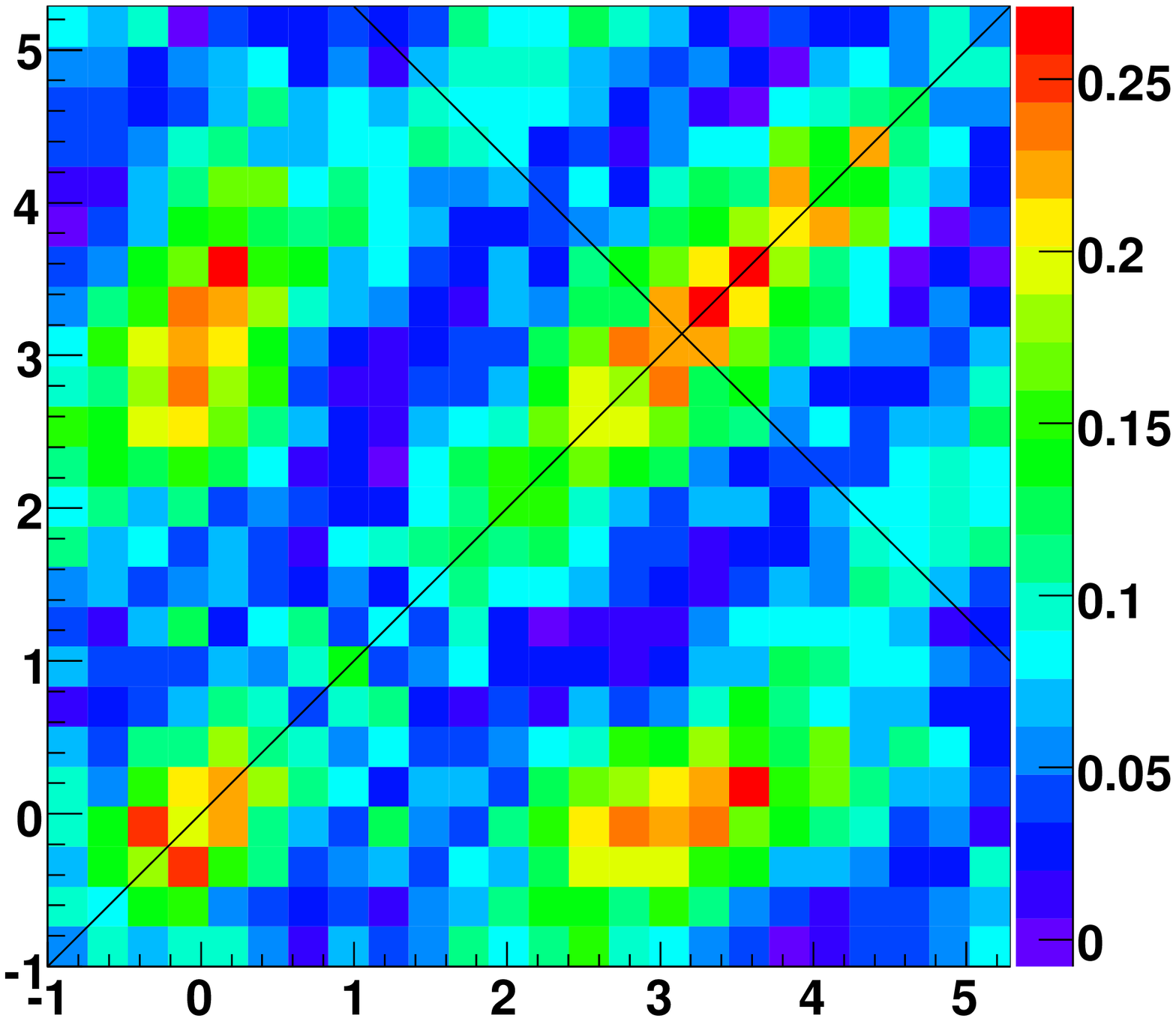}
\includegraphics[width=1.0\textwidth]{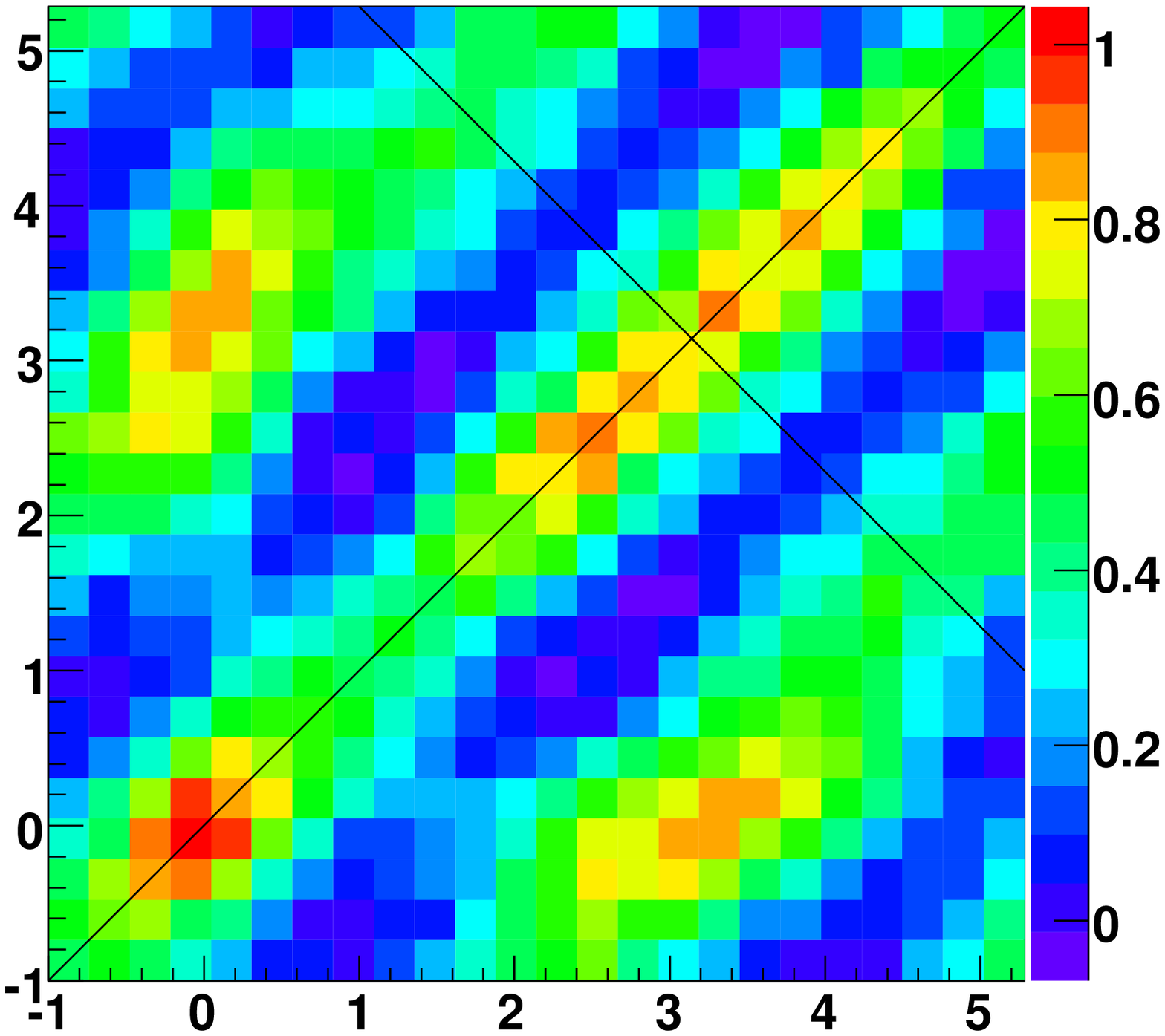}
\includegraphics[width=1.0\textwidth]{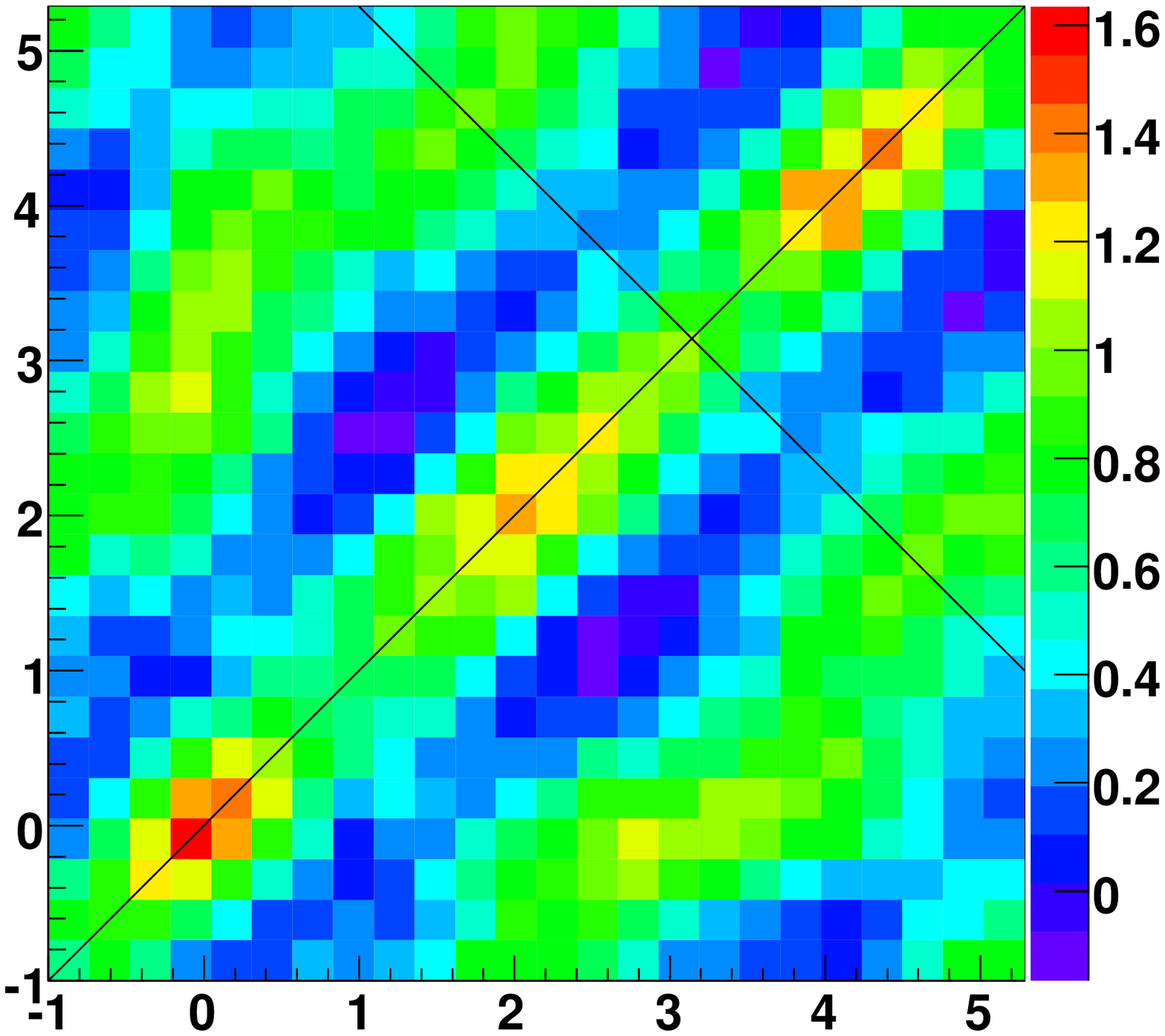}
\includegraphics[width=1.0\textwidth]{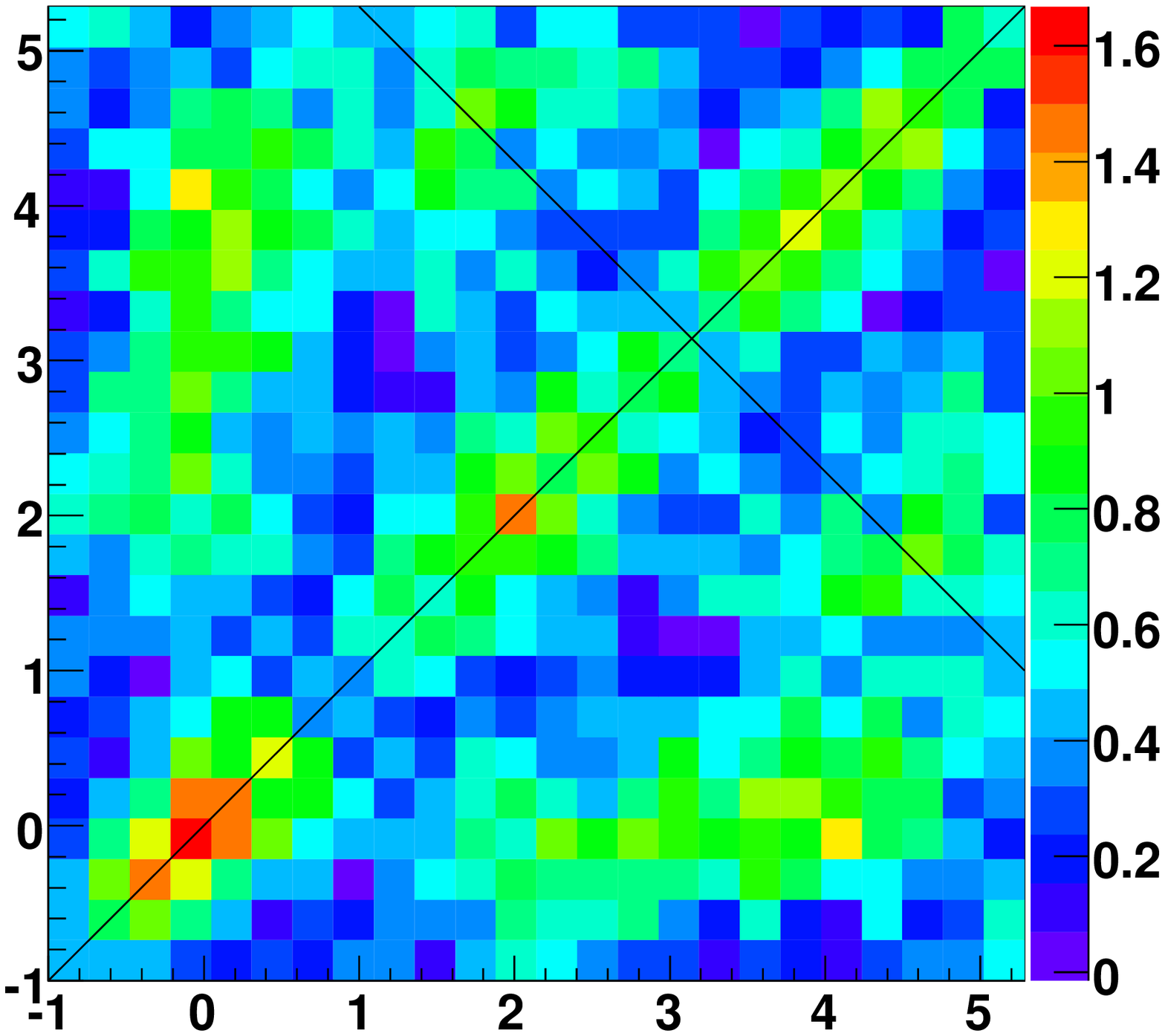}
\includegraphics[width=1.0\textwidth]{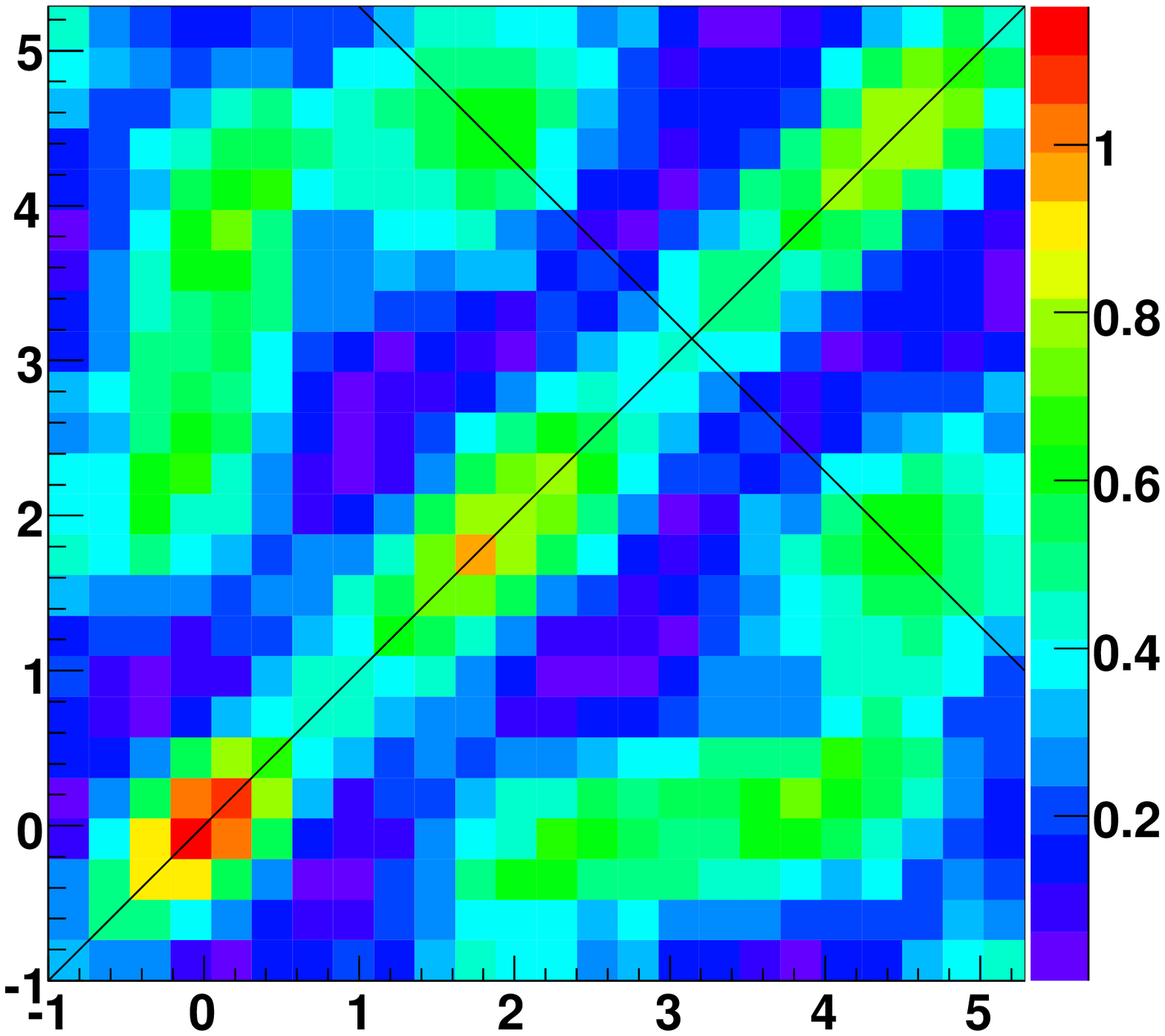}
\end{minipage}
\hfill   
\begin{minipage}{0.25\textwidth}
\includegraphics[width=1.0\textwidth]{Plots/Fig3Panel_AuAuY4MB_3.00Pt4.00_1.00pt2.00_1M3_E1.eps}
\includegraphics[width=1.0\textwidth]{Plots/Fig3Panel_AuAuY4MB_3.00Pt4.00_1.00pt2.00_4M5_E1.eps}
\includegraphics[width=1.0\textwidth]{Plots/Fig3Panel_AuAuY4MB_3.00Pt4.00_1.00pt2.00_6M7_E1.eps}
\includegraphics[width=1.0\textwidth]{Plots/Fig3Panel_AuAuY4MB_3.00Pt4.00_1.00pt2.00_8M9_E1.eps}
\includegraphics[width=1.0\textwidth]{Plots/Fig3Panel_AuAuY4Cent_3.00Pt4.00_1.00pt2.00_6M9_E1.eps}
\end{minipage}
\hfill
\begin{minipage}{0.25\textwidth}
\includegraphics[width=1.0\textwidth]{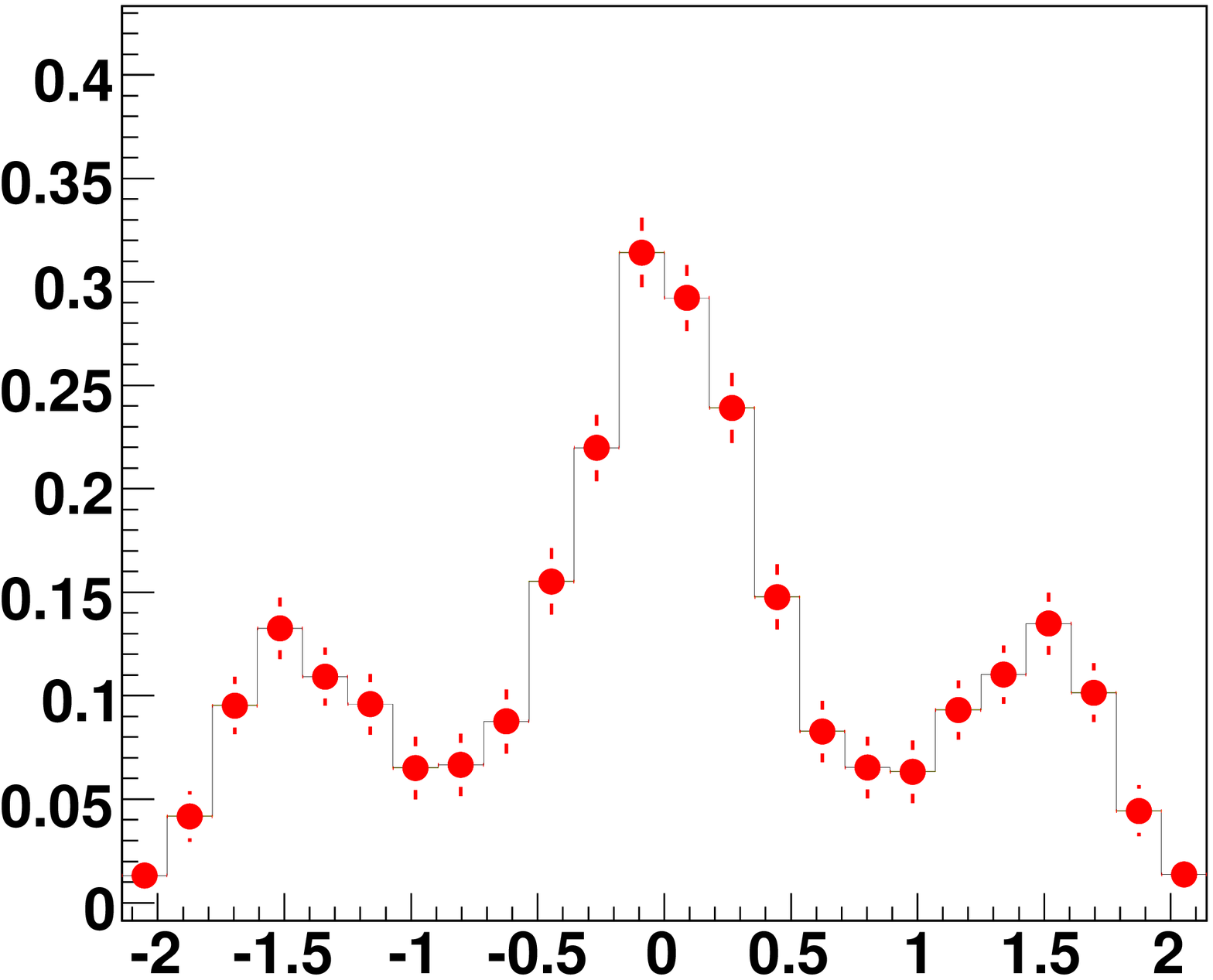}
\includegraphics[width=1.0\textwidth]{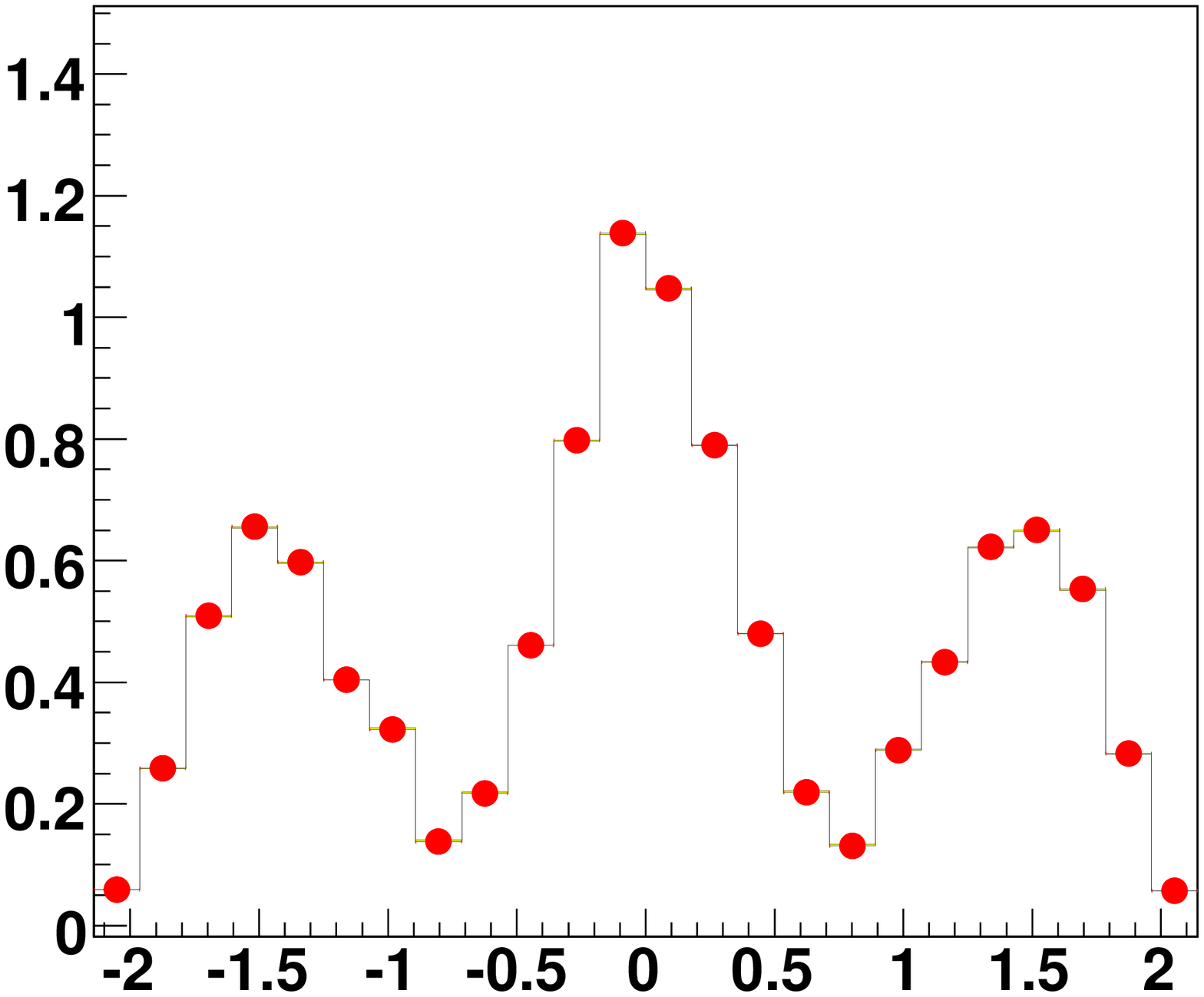}
\includegraphics[width=1.0\textwidth]{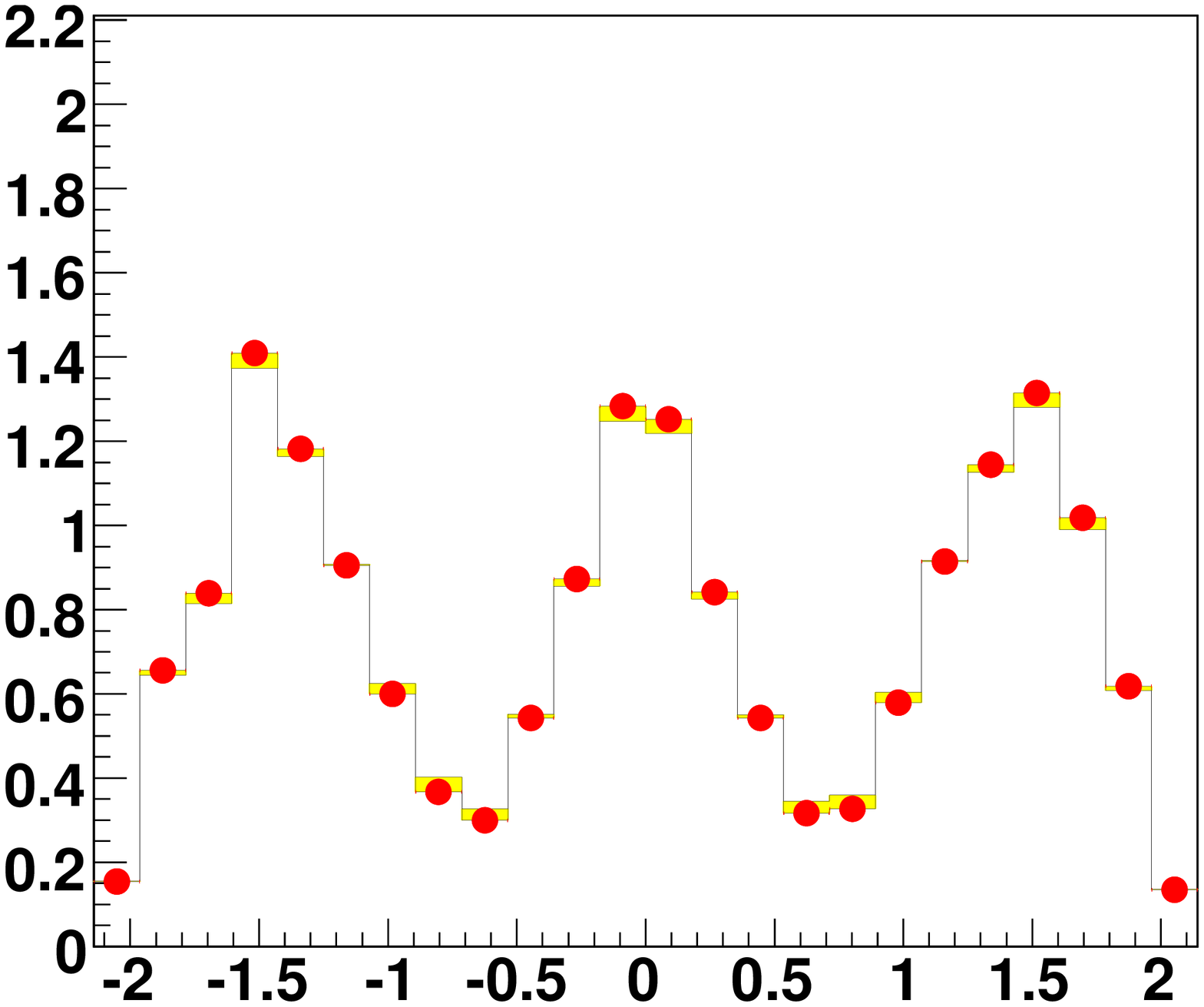}
\includegraphics[width=1.0\textwidth]{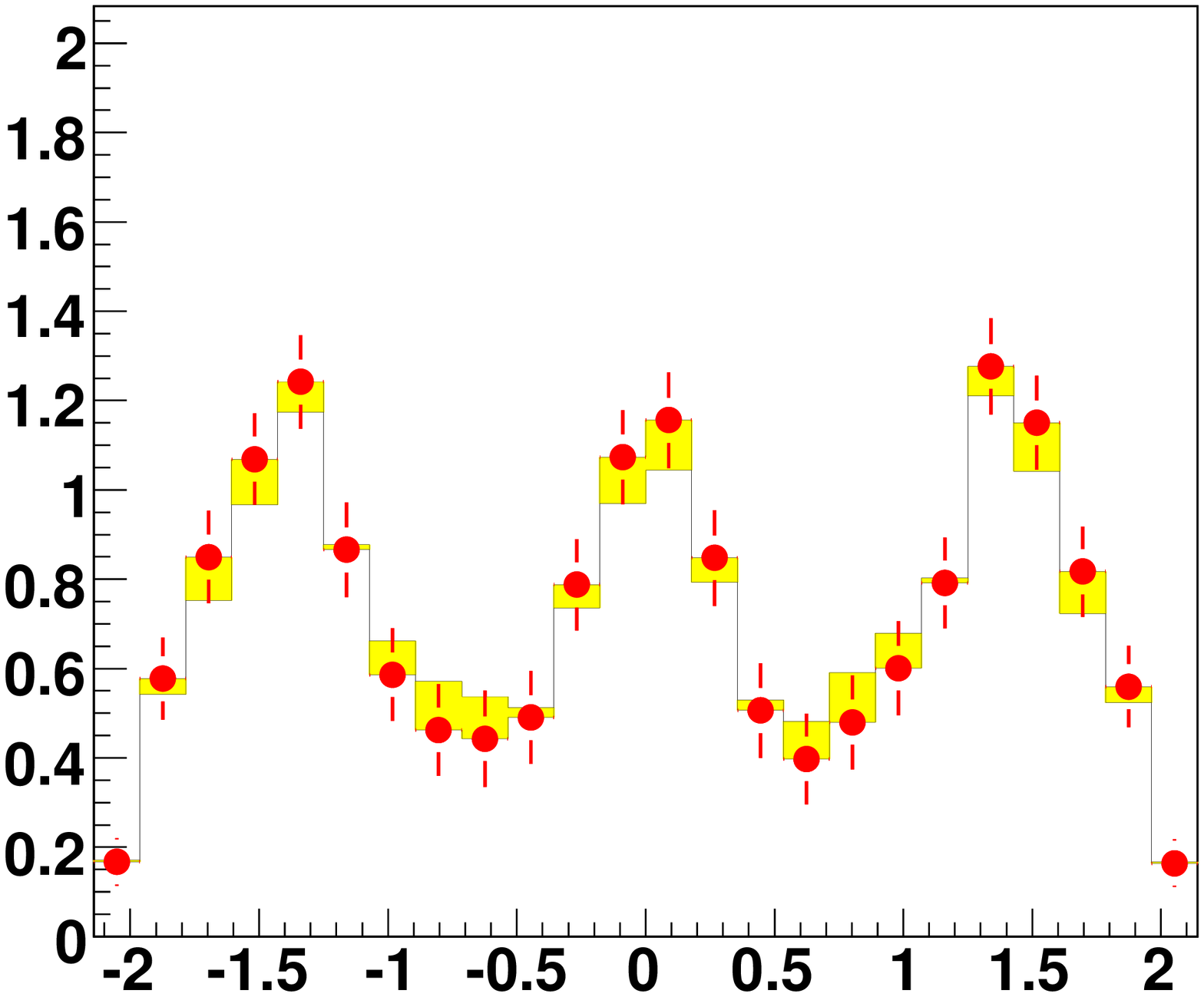}
\includegraphics[width=1.0\textwidth]{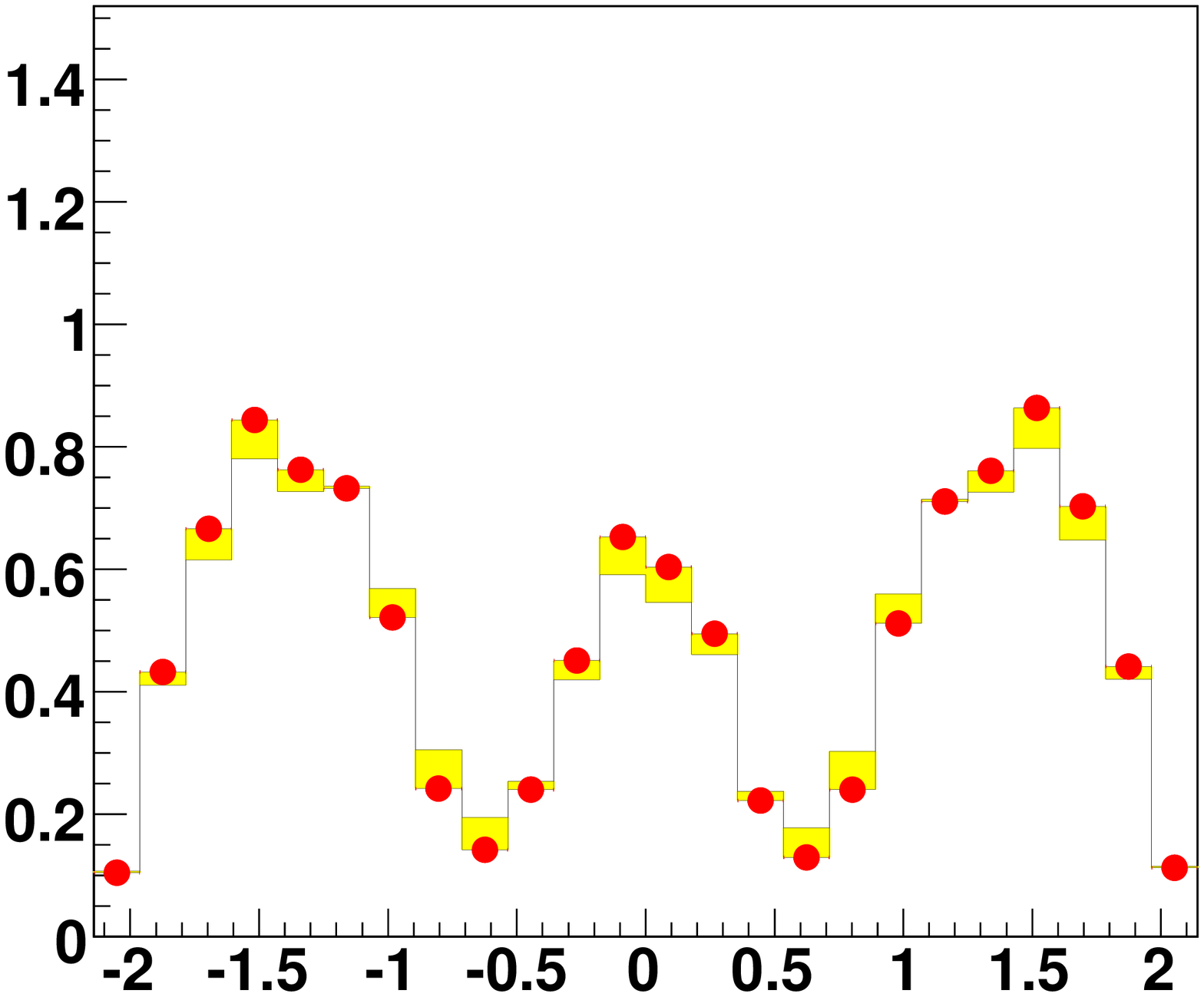}
\end{minipage}
\hfill
\begin{minipage}{0.07\textwidth}
\includegraphics[width=1.0\textwidth]{Plots/blank2.eps}
\end{minipage}
\caption{Background subtracted 3-particle correlations for uncertainty on the soft-soft term.  Left:  Soft-soft term constructed from weighed sum of soft-soft terms for individual multiplicities.  Center:  Default.  Right:  Off-diagonal away-side projection from default with systematic uncertainty from the uncertainty on soft-soft shown in shaded band.  From top to bottom plots are Au+Au 50-80\%, Au+Au 30-50\%, Au+Au 10-30\%, Au+Au 0-10\%, and ZDC triggered Au+Au 0-12\% collisions at $\sqrt{s_{NN}}=200$ GeV/c.}
\label{fig:sssys}
\end{figure}

There can be effects on the soft-soft and hard-soft background terms from doing the analysis in finite multiplicity bins.  In the soft-soft background term, the pedestal and the $\Delta\phi_1$, $\Delta\phi_2$ dependent structure scale differently with multiplicity.  As an assessment of the uncertainty in the soft-soft term from using a finite width centrality bin due to this difference in scaling, the soft-soft term was constructed in individual multiplicity bins.  The individual soft-soft terms were summed together in a weighted sum to make the soft-soft term for the entire multiplicity bin.  They were weighted by an estimated underlying event multiplicity distribution.  The underlying event multiplicity distribution was estimated from the trigger event multiplicity distribution minus the number of jet-like correlated particles counted in the multiplicity.  The number of jet-like correlated particles was estimated from a background subtracted 2-particle correlation using trigger particles with the same cuts as in the analysis and associated particles with the reference multiplicity cuts.  However, due to the way we have stored the subset of the data we process the number of fit points cut was 15 instead of 10.  This estimate had to be rounded to the nearest integer since the multiplicity distributions are integer.  Figure~\ref{fig:sssys} shows the the effect of using the soft-soft term constructed this way on the background subtracted 3-particle correlation results.

\begin{figure}[htb]
\centering
\includegraphics[width=0.75\textwidth]{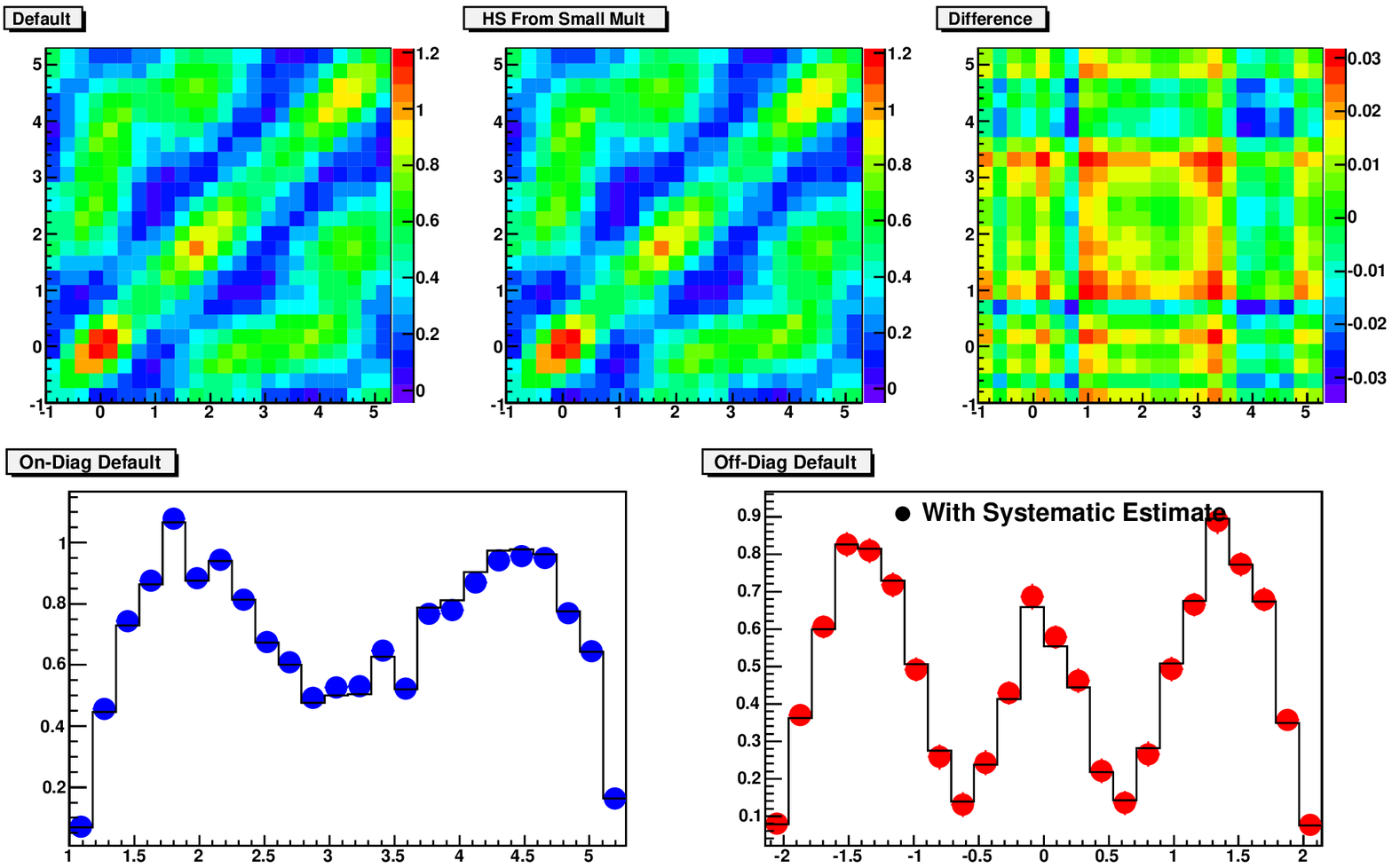}
\caption{Background subtracted 3-particle correlations for uncertianity on the hard-soft on the hard-soft term from bin width effects in ZDC triggered Au+Au 0-12\%.  Top Left: Default.  Top Center:  With hard-soft from finer multiplicity bins.  Top Right:  Difference.  Bottom:  On-diagonal, left, and off-diagonal projections, right.  The points are from the default and the histograms are from the finer mulitplicity binned hard-soft term.}
\label{fig:hssys}
\end{figure}

The hard-soft term can also have effects due to the finite multiplicity bin width.  Optimally the term would be created on an event-by-event basis; however, this cannot be done.  To check the magnitude of the effect, the ZDC trigger Au+Au data was dividing into multiplicity bins $1/10^{th}$ the size.  This effect should be the largest in central collisions where the largest change in the hard-soft term was 5\% that of the peaks, as shown in Fig.~\ref{fig:hssys}.  This could not be done in bins $1/10^{th}$ the size of the default bins for all centralities and data sets due to statistics, so an overall $\pm$5\% systematic uncertianity has been applied.  This systematic can be improved in the future by dividing the each of the centrality bins into as many subbins as the statistics will allow.  This will take into account the structure of this correction on $\Delta\phi_1$, $\Delta\phi_2$ space.

\section{Summary}
Three-particle azimuthal correlations have been measured to discover the physics mechanism(s) behind the broadened and maybe even double-peaked away-side structure in 2-particle correlations in central Au+Au collisions.  Three-particle correlations have the power to distinguish physics mechanisms with conical emission (hydrodynamic conical flow generated by Mach-cone shock waves and \v{C}erenkov gluon radiation) from other physics mechanisms.  This analysis is designed to extract the jet-like 3-particle correlations,  by treating the event as composed of  two components, particles jet-like correlated with the trigger particle and particles not jet-like correlated with the trigger particle (but correlated via anisotropic flow).  We assume the background subtracted 3-particle correlations are positive definite by normalizing the background via 3-particle ZYAM.  The systematics on this analysis have been studied in great detail.

The 3-particle azimuthal correlations have been studied for {\it pp}, d+Au, and different centralities of Au+Au collisions (for both minimum bias and ZDC triggered central data) for a trigger particle of $3<p_{T}^{Trig}<4$ GeV/c with two associated particles of $1<p_{T}^{Assoc}<2$ GeV/c.  We observed significant off-diagonal peaks, the signature of conical emission, in the mid-central and central Au+Au collisions at about 1.42 radians from $\pi$.  In ZDC triggered central Au+Au collisions, the associated particle $p_{T}$ dependence has been studied.  No strong $p_{T}$ dependence of the emission angle has been observed, suggesting the observed conical emission is due to Mach-cone shock waves, not \v{C}erenkov gluon radiation.  The average speed of sound of the medium is extracted to be $c_s=0.15c$ in a simple-minded Mach-cone scenario.  Further and more refined theoretical studies are urgently needed to assess the effects of hydrodynamic expansion and different scenarios of QGP-hadron phase transition and to connect our measurement to the properties of the medium and its equation of state.


%
%

\chapter{CONCLUSIONS AND RECOMMENDATIONS}

Jet quenching--suppression of high $p_T$ yields and correlations--observed at RHIC constitutes important evidence for the discovery of the created medium being a nearly perfect fluid.  In particular, jet-like correlations both at high $p_T$ and low $p_T$ have provided a valuable tool to study the properties of the medium.  This thesis contributes to the enterprise of jet-like correlation studies at RHIC.  Particularly, the 3-particle correlations, presented in this thesis, are the first such measurement and should provide new and unique insights into the medium at RHIC.

\section{Two-Particle Correlations of Charged Hadrons}

Jet-like correlations of charged hadrons have been studied in d+Au collisions.  This is the first d+Au jet-like correlation analysis to go down to very low associated $p_{T}$.  The correlations were found to be consistent with previous {\it pp} results.  This demonstrates that it is the final state in Au+Au collisions, not the initial state of Au nuclei, that is responsible for the modifications to jet-like correlations observed in central Au+Au collisions.  This also lets us use the d+Au data besides $pp$ as a reference for the Au+Au data to gain the benefit of the increased statistics in d+Au.  The widths of the correlation functions and the spectra of the correlated yields have been analyzed for d+Au collisions and can be used for comparison to theoretical calculations.

Jet-like correlations of charged hadrons have also been studied in Au+Au collisions.  This was done using Au+Au data from the fourth year of RHIC running which gave an increase in statistics of about an order of magnitude from the second year data.  This analysis was done using the same trigger and associated $p_T$ ranges $1.0<p_{T}^{Assoc}<2.5<p_{T}^{Trig}<4$ GeV/c as for the  published PHENIX results~\cite{phenix}.  The PHENIX data show a significant dip at $\pi$ in the 2-particle azimuthal correlations that is strongest in mid-central Au+Au collisions.  Our results also show a dip at $\pi$; however, the dip in our results is not as strong.  In our results, the dip is strongest in the most central collisions, but is not very significant given the systematic uncertainty on the elliptic flow measurement.

\section{Two-Particle Correlations with Identified Trigger Particles}

Jet-like correlations with identified trigger particles have been studied in d+Au collisions and Au+Au collisions in different centralities.  Proton-proton collisions did not have adequate statistics for this study.  The interest in studying identified correlation is due to the baryon/meson puzzle, the large baryon over meson ratio at intermediate $p_{T}$ in central Au+Au collisions.  The analysis was done for intermediate $p_{T}$ trigger particles where the large baryon/meson ratio is observed and where we have good statistics and particle identification capability.  Trigger particles are identified as $p$, $\bar{p}$, $\pi^{+}$, and $\pi^{-}$ by the relativistic rise of dE/dx.  No significant difference is observed between correlations with identified baryon and meson triggers.  The large baryon/meson ratio does not seem to have a large effect on the particles correlated with the baryons and mesons.  This imposes a serious challenge to the initial coalescence and recombination models which successfully and elegantly explained the large baryon/meson ratio at intermediate $p_T$ and the constituent quark scaling of the elliptic flow. 

For this analysis to be finalized, a detailed study of the systematics needs to be carried out.  Since the $v_{2}$ correlation will partially cancel in the difference between baryon and meson triggered correlations, the systematic effects of flow subtraction will not be very large.  The systematic effects due to background normalization should cancel to first order and be negligible in the difference.  A study needs to be done on the purities.  The purities depend on the measured $K^0_{s}$ to charged $\pi$ ratio.  This has two systematics associated with it.  The first is the assumption that the yields for $K^{+}=K^{-}=K^{0}_{s}$.  However, this systematic effect should be negligible if one sums $p+\bar{p}$ and $\pi^++\pi^-$ as $K^{+}+K^{-}=2K^{0}_{s}$ to a good approximation.  The second is the systematics due to the error in the $K^{0}_{s}$ measurement.  

Other similar analyses can be done that are of interest.  One is to use trigger particles identified through V0 reconstruction such as $\Lambda$,  and $K^{0}_{s}$.  This analysis is already on-going in STAR.  Another is to use a higher $p_{T}$ hadron trigger and use the identified particles as our associated particles.  This would allow one to look for baryon and meson dependences on the associated particles.  It would also be beneficial to look at both identified trigger and associated particles, at both low and high $p_T$, to see if there are differences in baryon-baryon, baryon-meson, meson-baryon, and meson-meson correlations.  Using charged hadrons as the trigger and identifying the associated particles may be an analysis that can be done with the current data.  Both analyses (and the current analysis) would benefit from increased statistics.  PHENIX has done a good job in identified particle correlations using their time of flight detector for particle identification.  Their results are currently limited to intermediate $p_T$.  The $p_T$ dependences of the identified correlations will be important to shed light on some of the puzzles.  There will be improvements to particle identification in STAR with the inclusion of the time of flight detector that will greatly benefit all of the identified particle analyses.

\section{Three-Particle Correlations}

Jet-like 3-particle azimuthal correlations have been studied in {\it pp}, d+Au, and different centralities of Au+Au collisions.  This analysis was done to investigate the physics mechanism(s) behind the broadened or double-peaked away-side correlation structure observed in central Au+Au collisions.  These physics mechanisms include: large angle gluon radiation, jet deflection by radial flow or preferential selection of particles due to path-length dependent energy loss, hydrodynamic conical flow generated by Mach-cone shock waves, and QCD \v{C}erenkov gluon radiation.  Three-particle correlations can identify conical emission which can be generated by Mach-cone shock waves or \v{C}erenkov gluon radiation, but not by other proposed physics mechanisms.  The analysis was carried out treating the triggered event as composed of particles jet-like correlated with the trigger and background particles to extract the genuine 3-particle jet-like correlations.  The combinatorial background was normalized assuming the 3-particle jet-like correlation is positive definite using ZYAM (zero yield at minimum).   The systematics have been studied in great detail.

We have observed significant off-diagonal peaks--the signature for conical emission--in mid-central and central Au+Au collisions.  These peaks are found to be at about $\pi\pm1.42$ radians from the trigger particle (i.e. emission angle).  We have studied the associated particle $p_{T}$ dependence and found the emission angle to have no significant dependence on the associated particle $p_{T}$.  This suggests that the observed conical emission is due to Mach-cone shock waves and not due to \v{C}erenkov gluon radiation, which predicts an angle that is sharply decreasing with associated particle $p_{T}$.  

Several recommendations can be made for future studies of 3-particle correlations.  They include:
\begin{itemize}
\item Trigger particle $p_T$ dependence.  This thesis has attempted this study but could not carry it to very high $p_T$ due to statistics.  By increasing the trigger particle $p_T$ one can probe different relative strengths of conical emission and back-to-back emission.  Also correlations with higher $p_T$ trigger benefit from an increased signal to noise ratio due to an increase in the signal.  Given enough statistics, potentially from an online high $p_T$ or high $E_T$ trigger, the increased signal to noise ratio could greatly reduce the systematic errors.
\item Three-particle correlations with identified associated particles (such as $p$ and $\pi$).  This can provide an important check on Mach-cone emission since protons and pions are expected to have different associated particle $p_T$ dependence on the emission strength.
\item Three-particle correlations with two trigger particles and one associated particle to probe different jet samples by varying the $p_T$ of the two triggers.
\item Trigger particle species dependence to look for differences between light quark triggers, heavy quark triggers, and non-photonic electron triggers.
\item The 3-particle correlation study in this thesis has concentrated on the away side.  There is also an interesting phenomenon on the near side.  The near-side ridge--particles that are jet-like correlated with the trigger particle in $\phi$ but not in $\eta$--can be studied through 3-particle $\Delta\eta$-$\Delta\eta$ correlations.
\end{itemize}
In addition, 4-particle azimuthal correlations can be explored.  This could be done using two trigger particles and two associated particles.  The second trigger particle would allow for the determination of the $\eta$ of the away-side jet-axis (which is not correlated with the near-side).  The two associated particles can then be studied in a polar coordinate system determined by the second trigger particle.  This is a more natural coordinate system to study conical emission.  Such an analysis would be very complicated due to a large number of background terms that can each be rather complicated.


%
%
%



%
%


\appendix
\renewcommand{\thechapter}{A}
\refstepcounter{chapter}
\makeatletter
\renewcommand{\theequation}{\thechapter.\@arabic\c@equation}
\makeatother
\chapter*{APPENDIX}
    \addcontentsline{toc}{chapter}{APPENDIX}

\begin{figure}[H]
\hfill
\begin{minipage}[t]{.4\textwidth}
	\centering
		\includegraphics[width=1\textwidth]{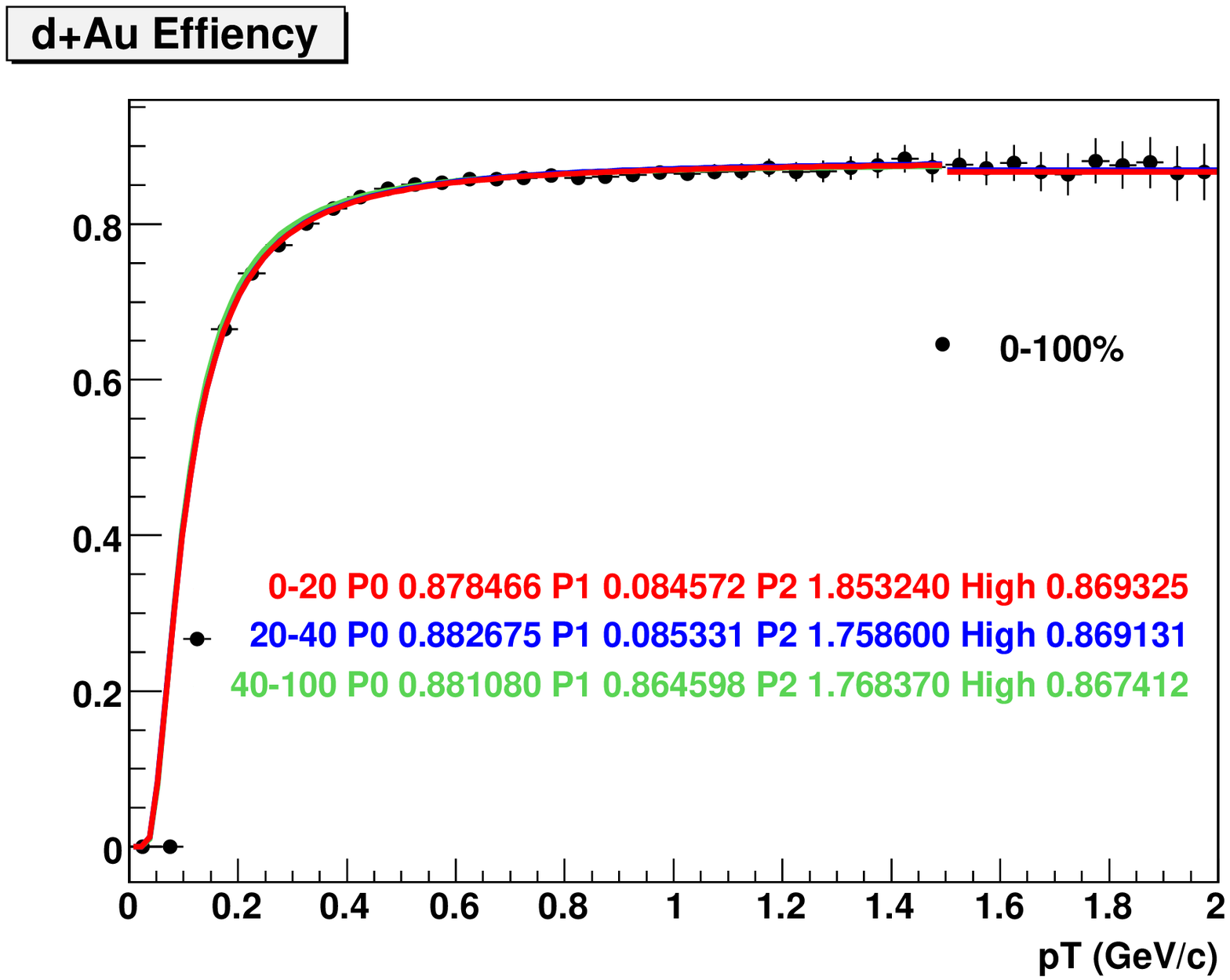}
		\includegraphics[width=1\textwidth]{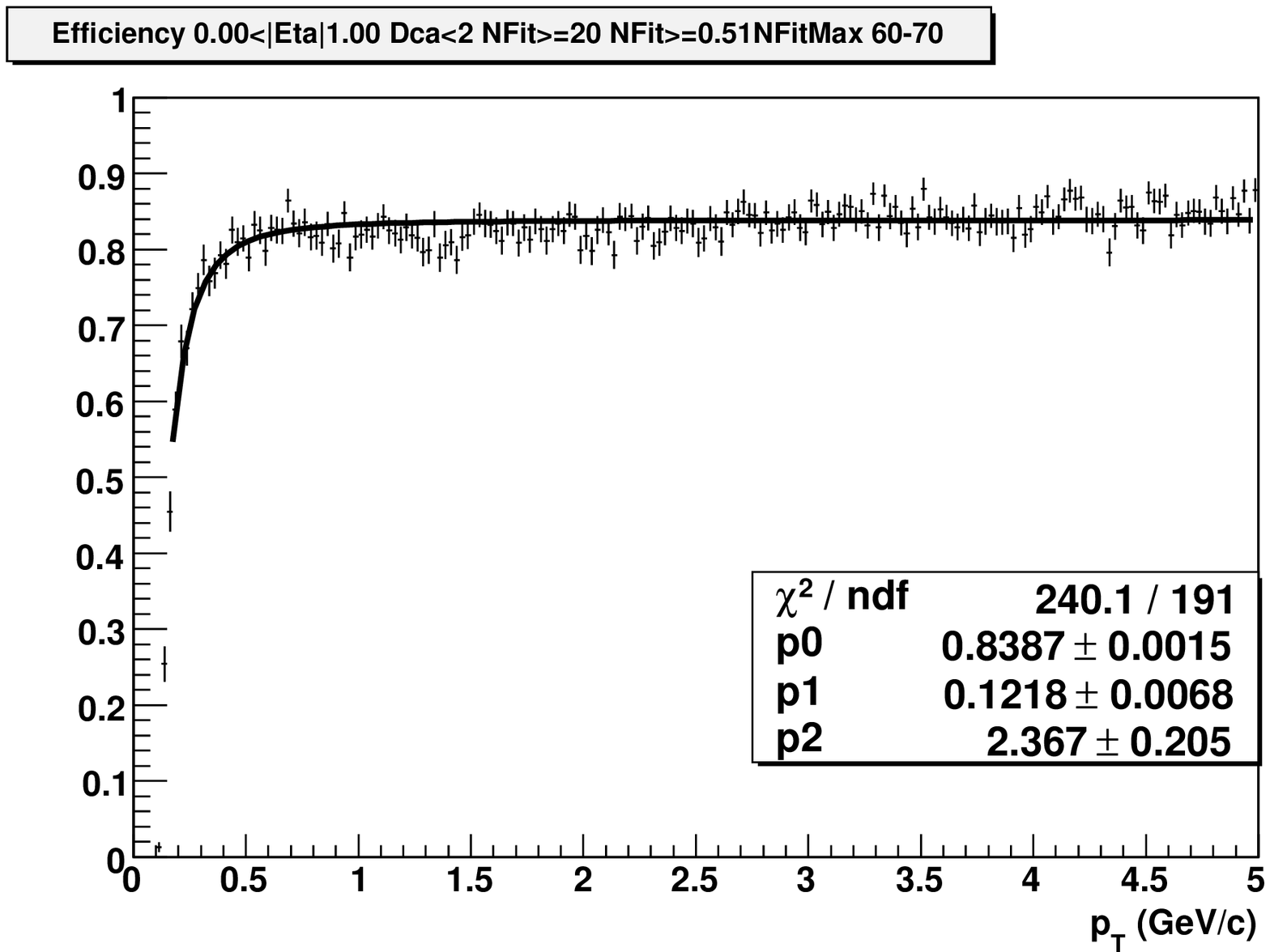}
		\includegraphics[width=1\textwidth]{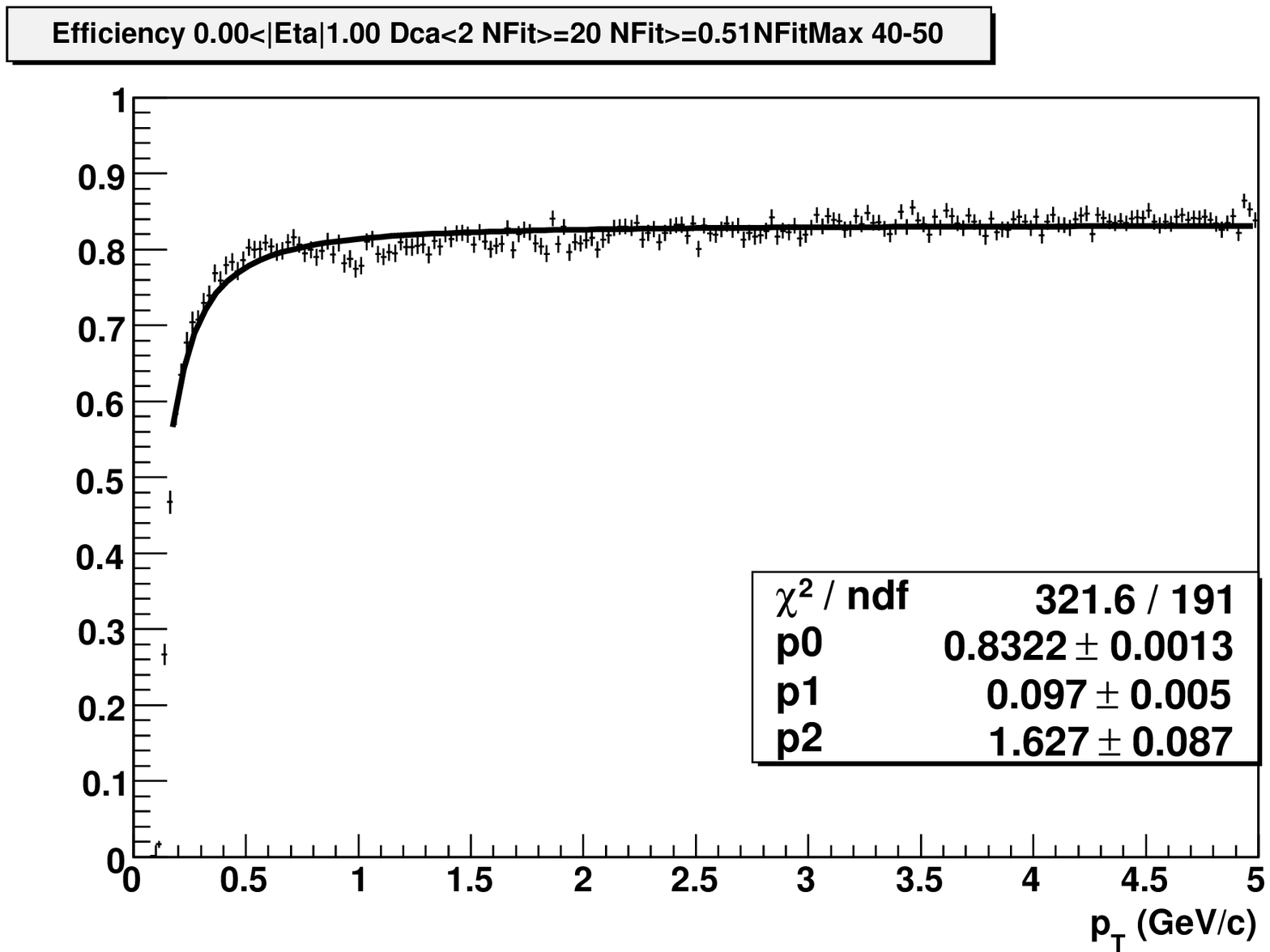}
		\includegraphics[width=1\textwidth]{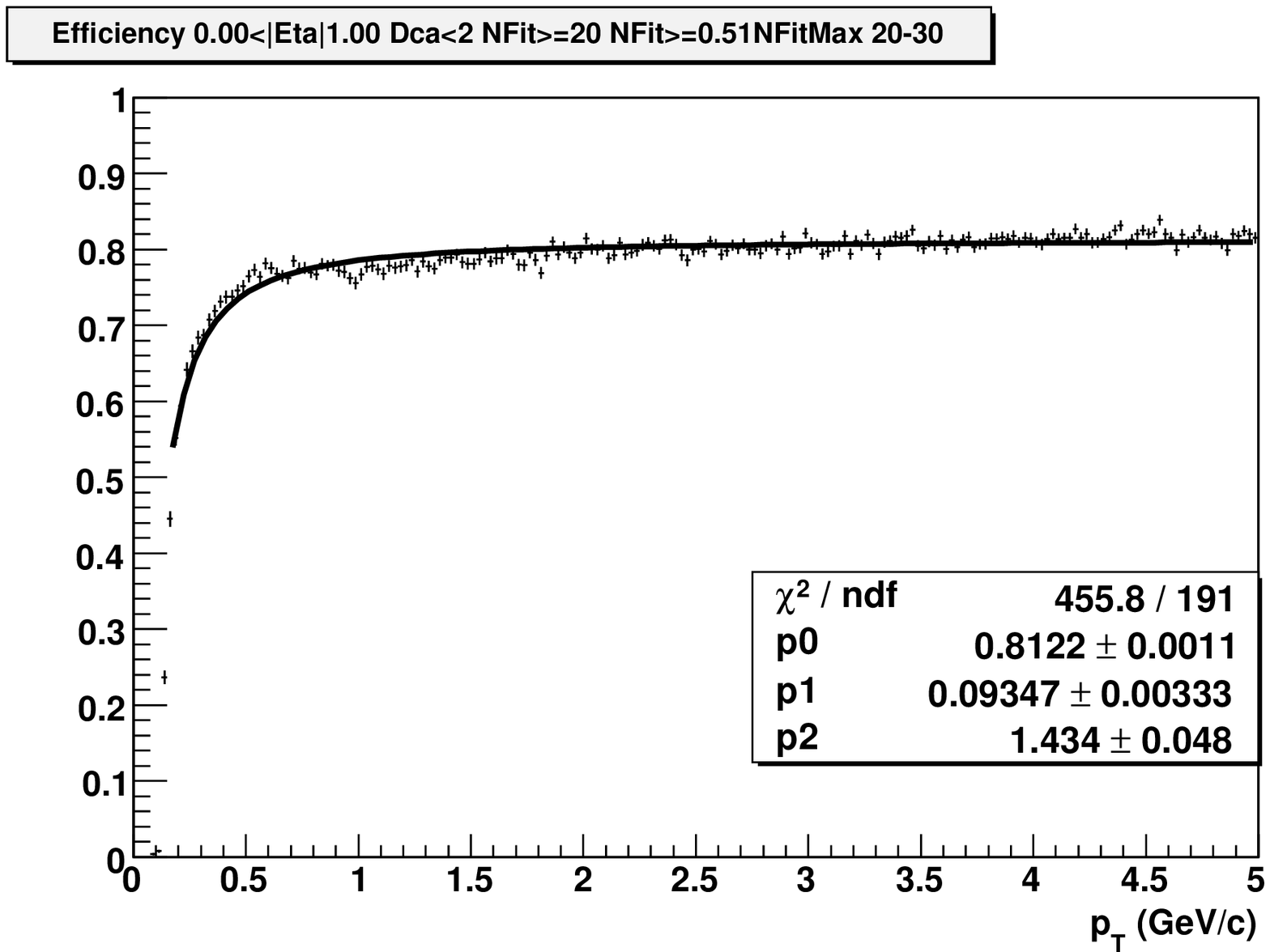}
		\includegraphics[width=1\textwidth]{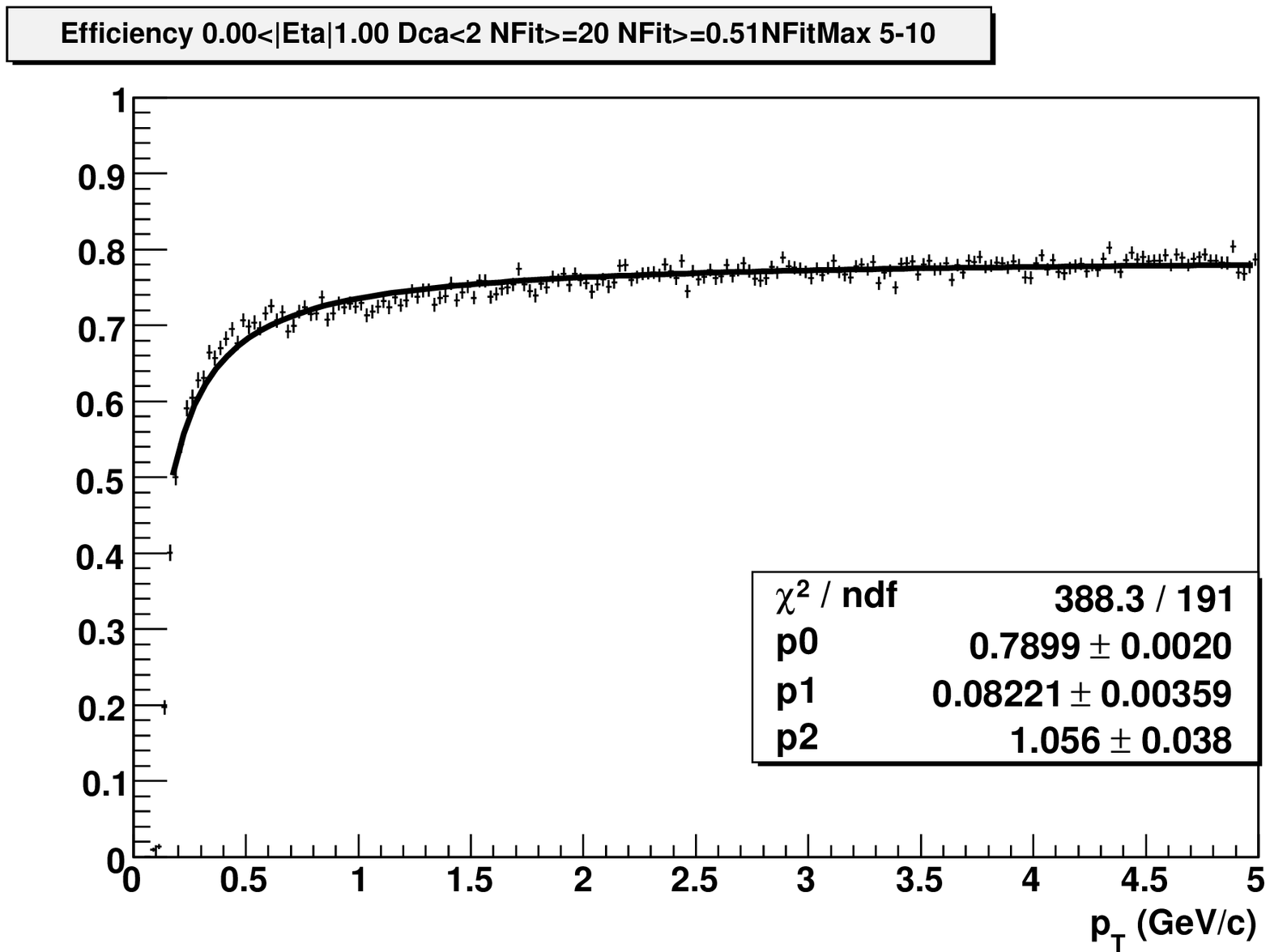}
			\end{minipage}
	\hfill
\begin{minipage}[t]{.4\textwidth}
	\includegraphics[width=1\textwidth]{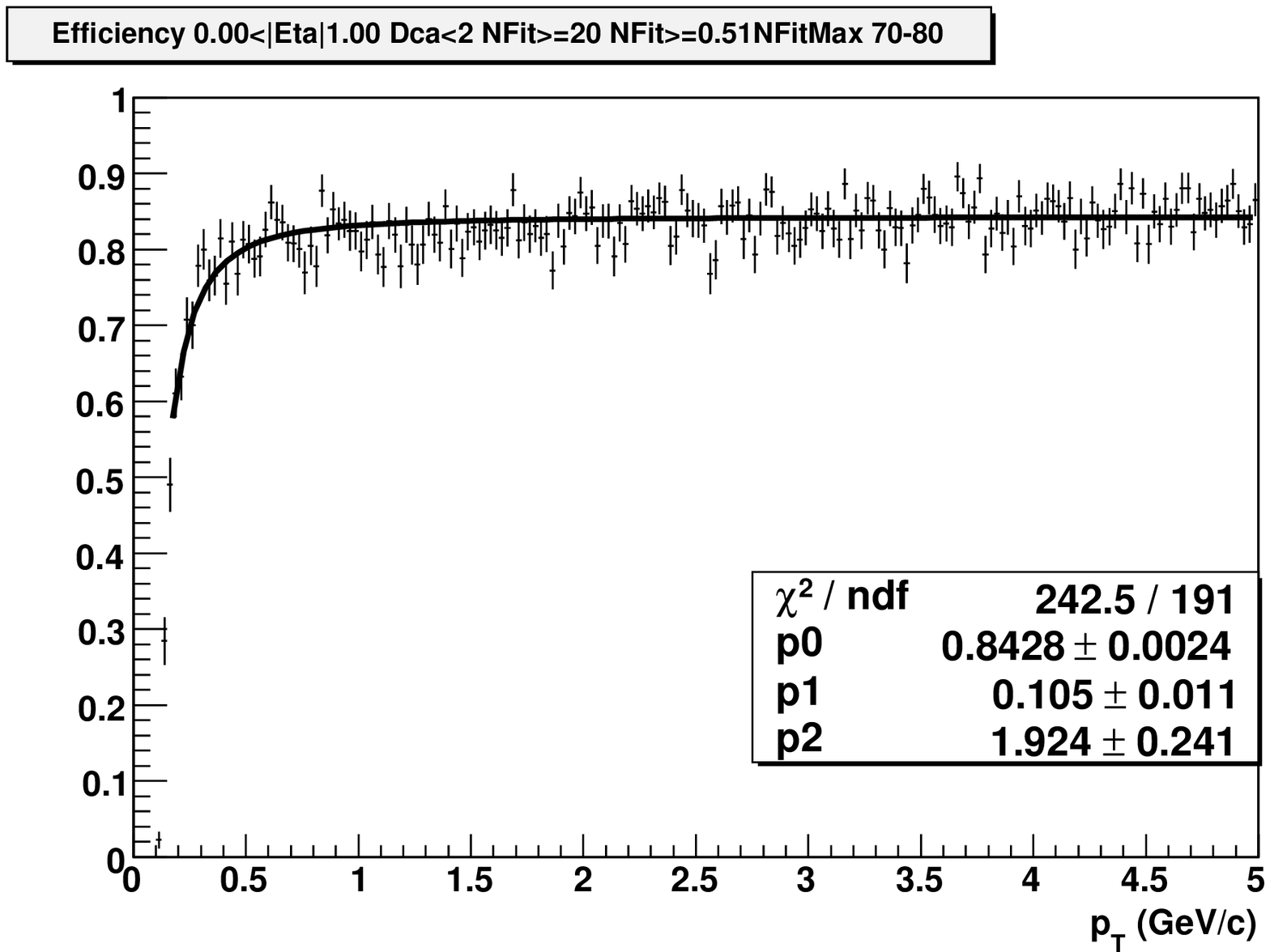}
	\includegraphics[width=1\textwidth]{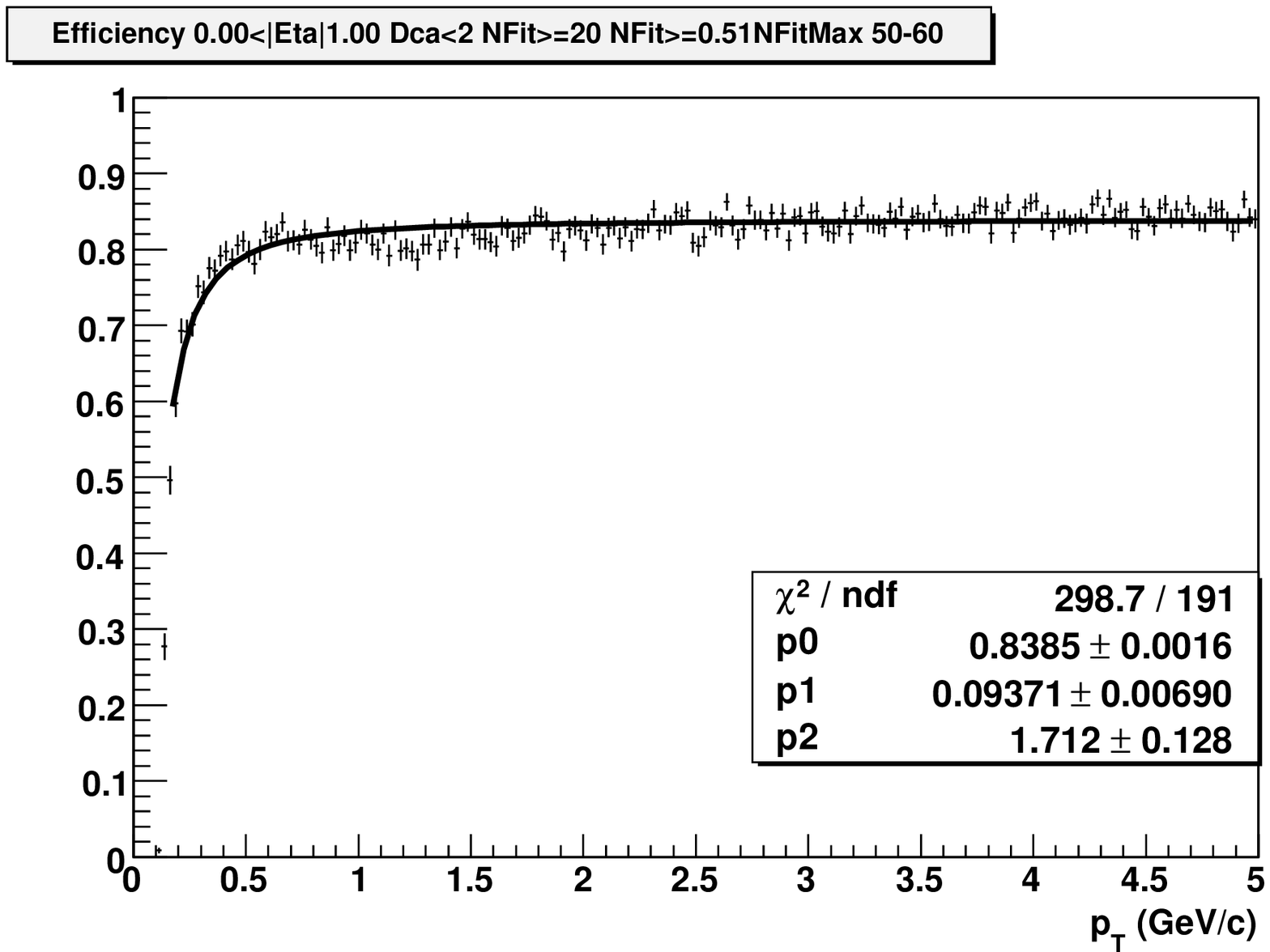}
	\includegraphics[width=1\textwidth]{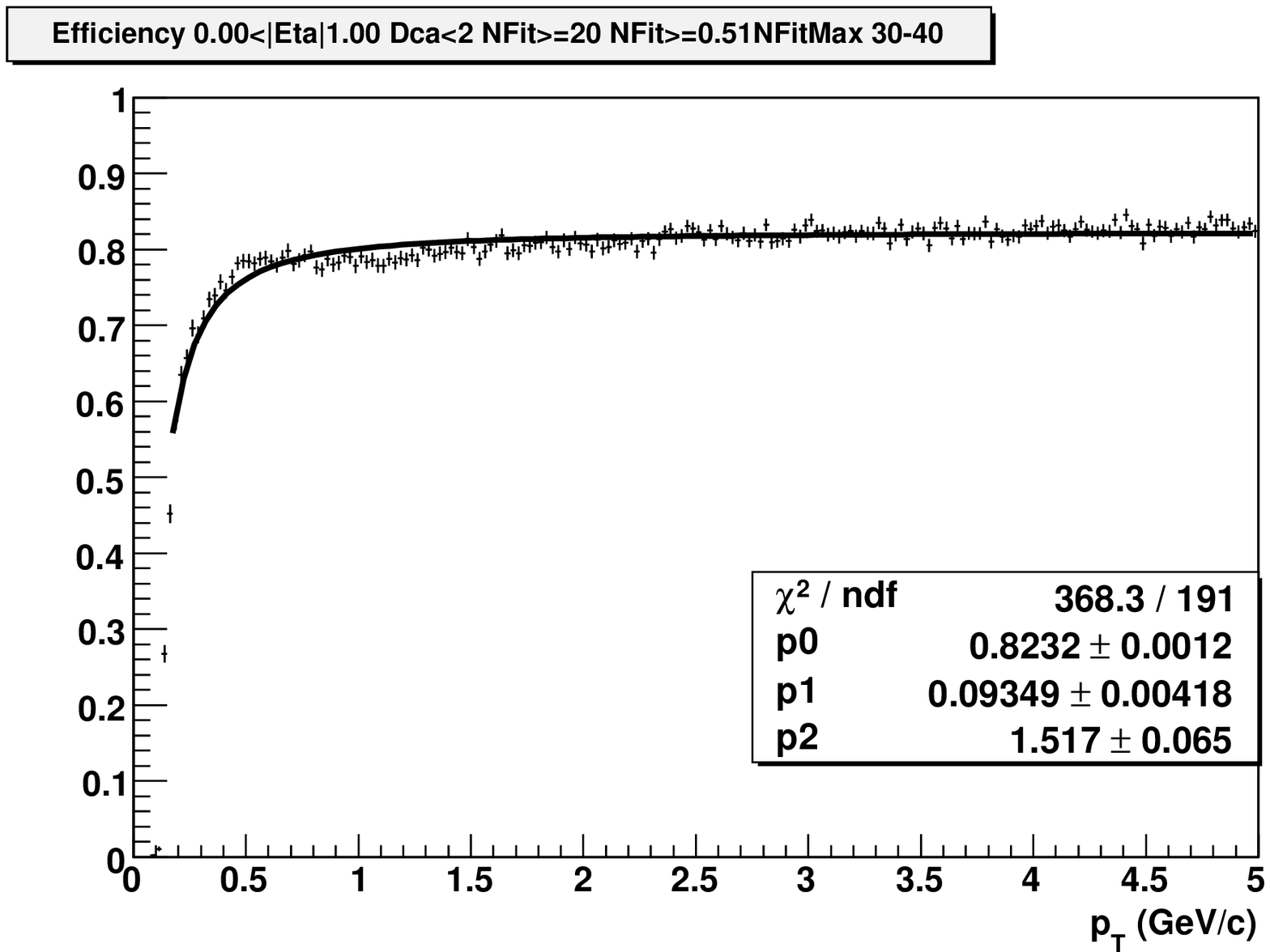}
	\includegraphics[width=1\textwidth]{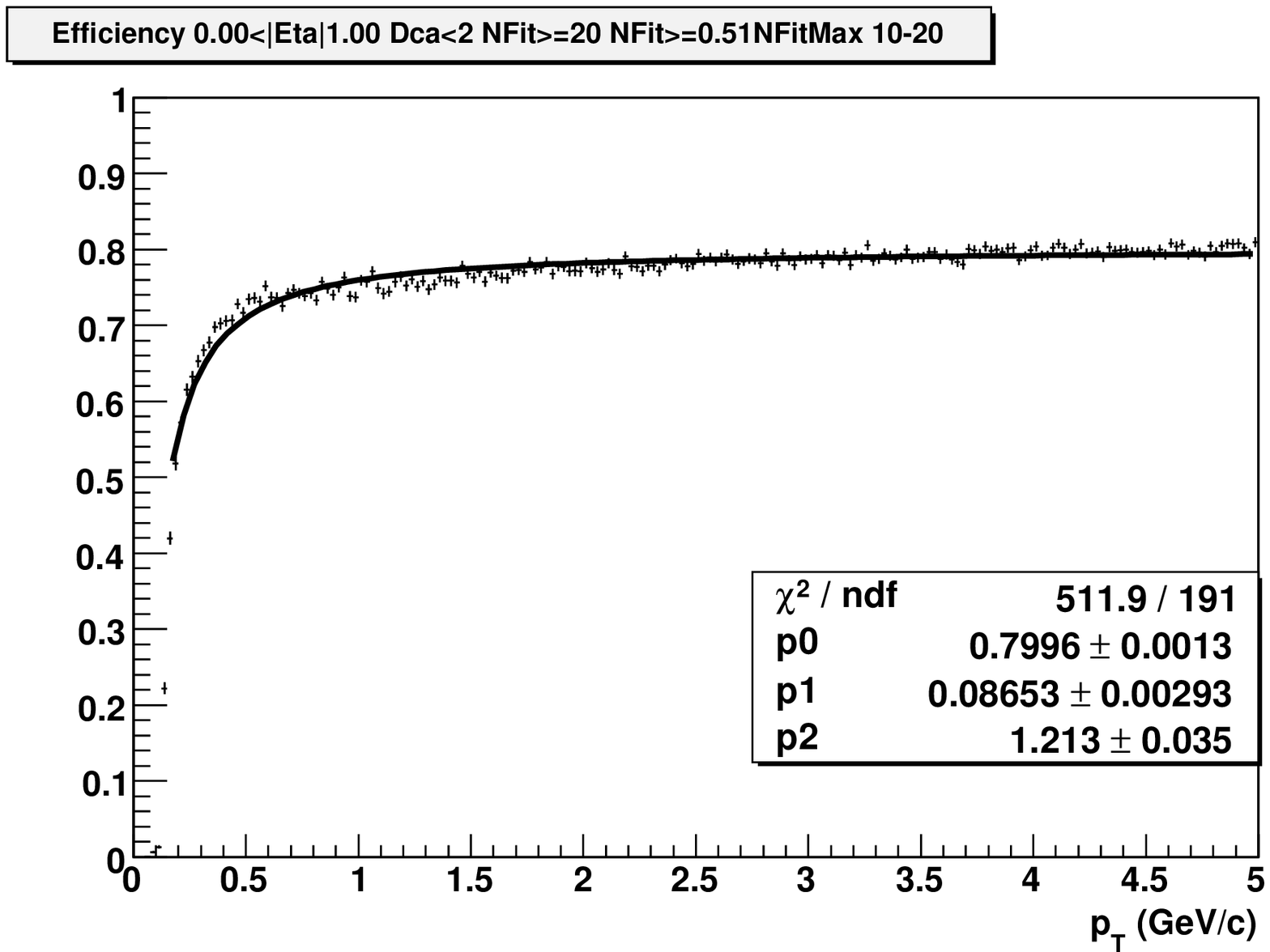}
		\includegraphics[width=1\textwidth]{Plots/Y4_0.0E1.0_D2_F20_M9.eps}
			\end{minipage}
	\hfill
			\vspace*{-0.0cm}
	\caption{Detector efficiency for charged pions for 3 centralities of d+Au collisions (top left) and 9 centralities of Au+Au collisions as a function of $p_T$ at $\sqrt{s_{NN}}=200$ GeV.}
	\label{fig:Eff2}	
\end{figure}

Figure~\ref{fig:Eff2} shows the $\phi$-averaged detector efficiency for 3 centralities of d+Au collisions and 9 centralities of Au+Au collisions.  The efficienies are fit to, 
\begin{equation}
	p0 e^{-(\frac{p1}{p_T})^{p2}}.
\label{eqn:effeqn}
\end{equation}
The inverse of the fit is used to correct for the number of particles at the single particle level.	

\begin{figure}[H]
\hfill
\begin{minipage}[t]{.18\textwidth}
	\centering
		\includegraphics[width=1\textwidth]{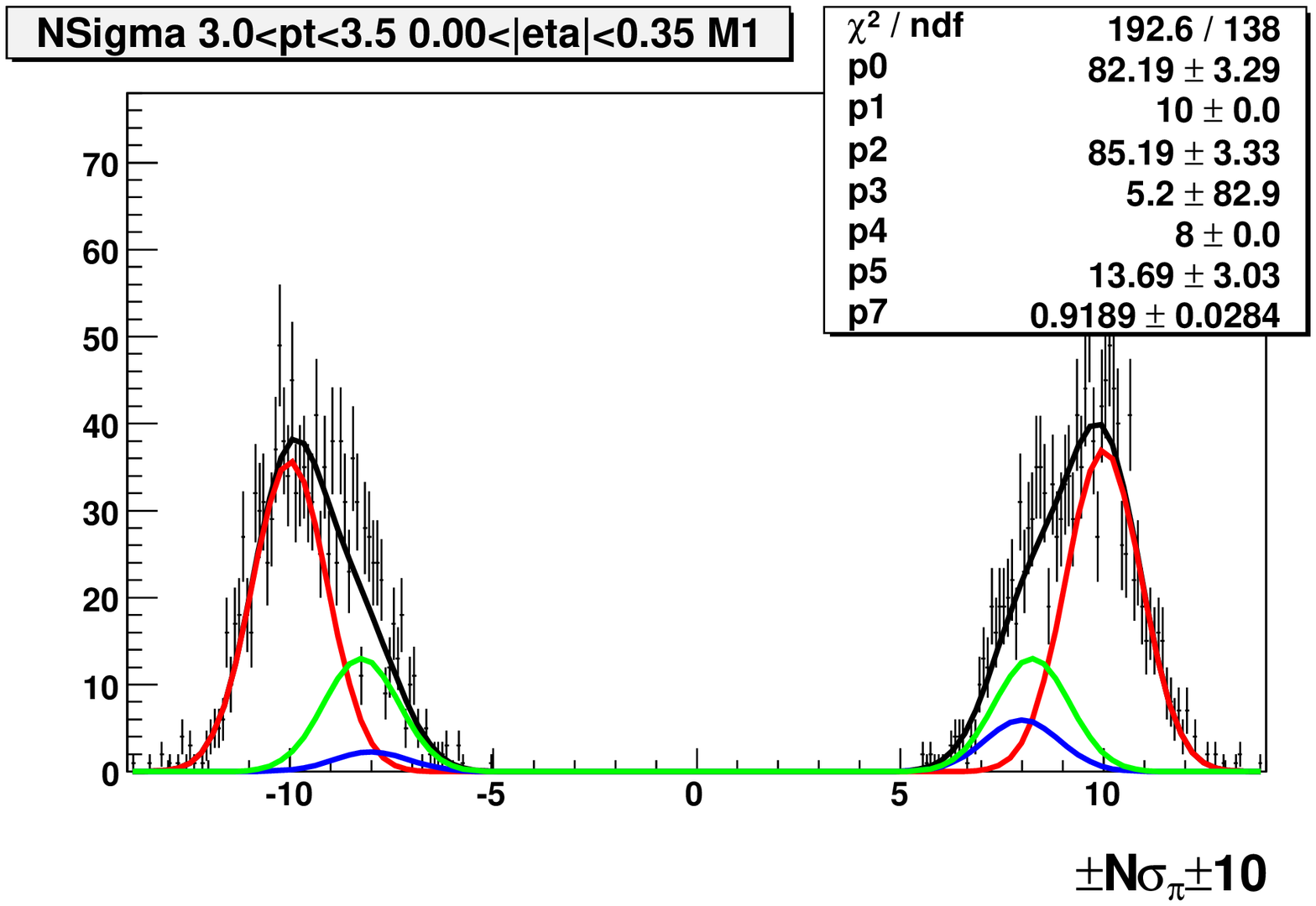}
		\includegraphics[width=1\textwidth]{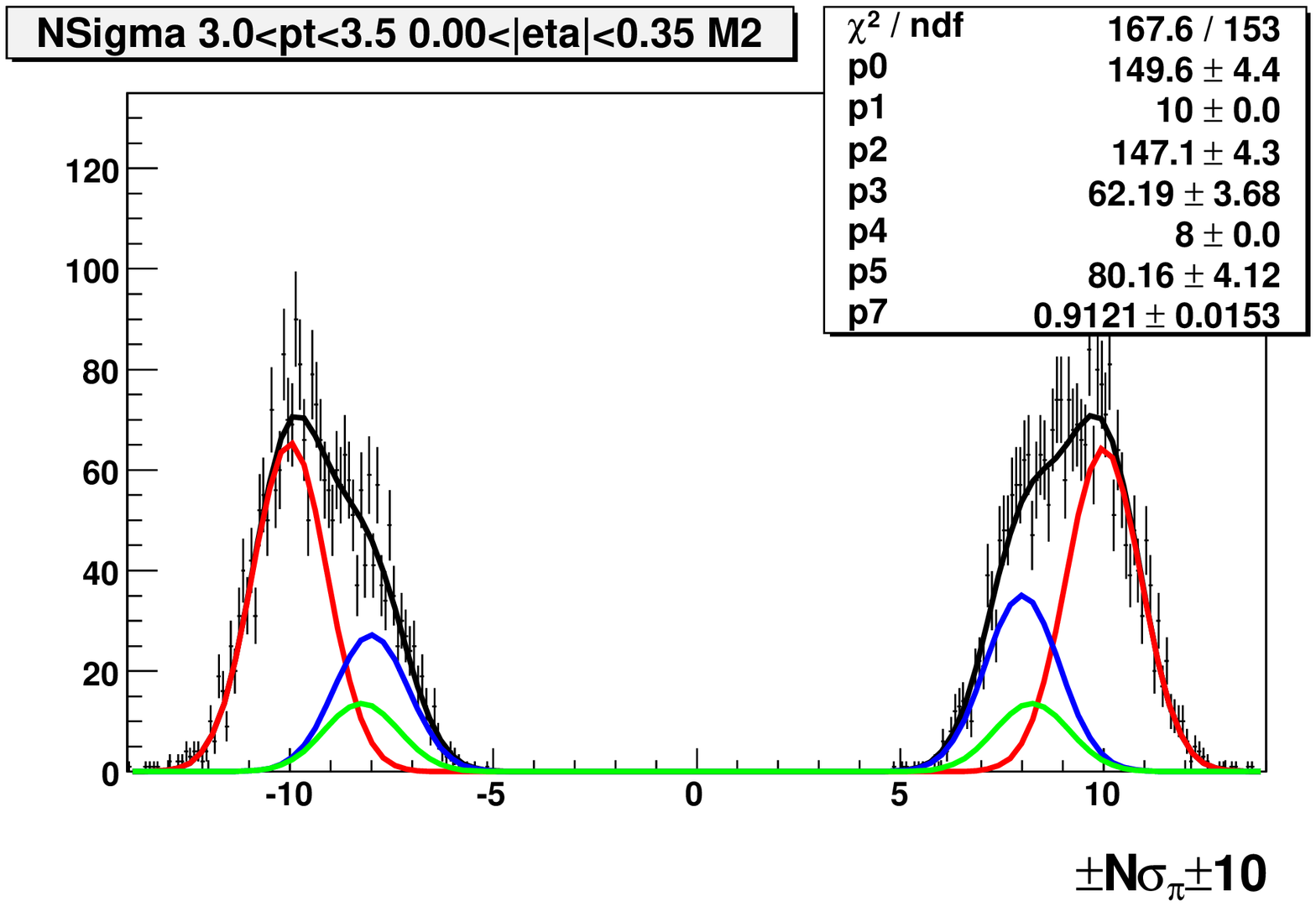}
		\includegraphics[width=1\textwidth]{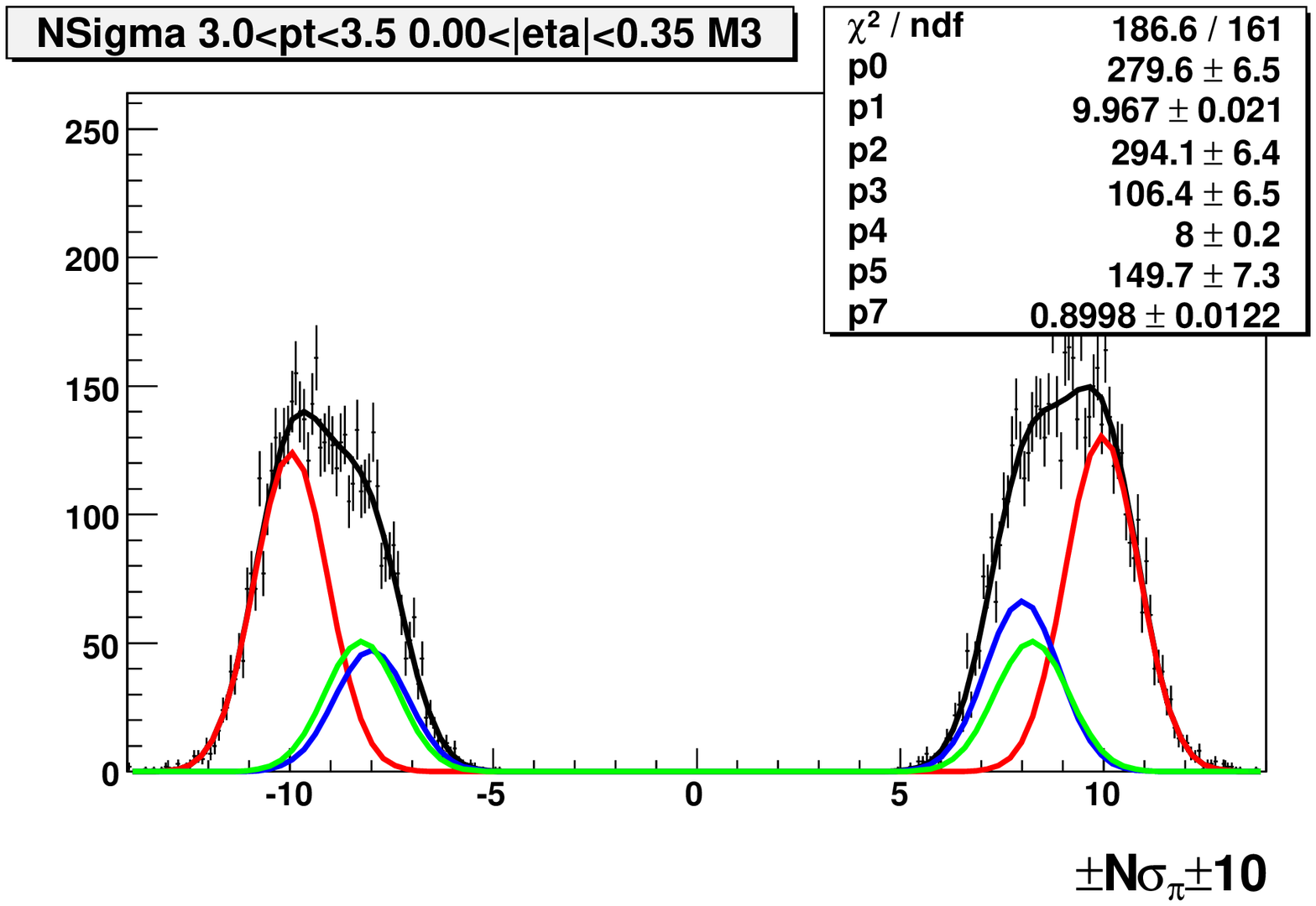}
		\includegraphics[width=1\textwidth]{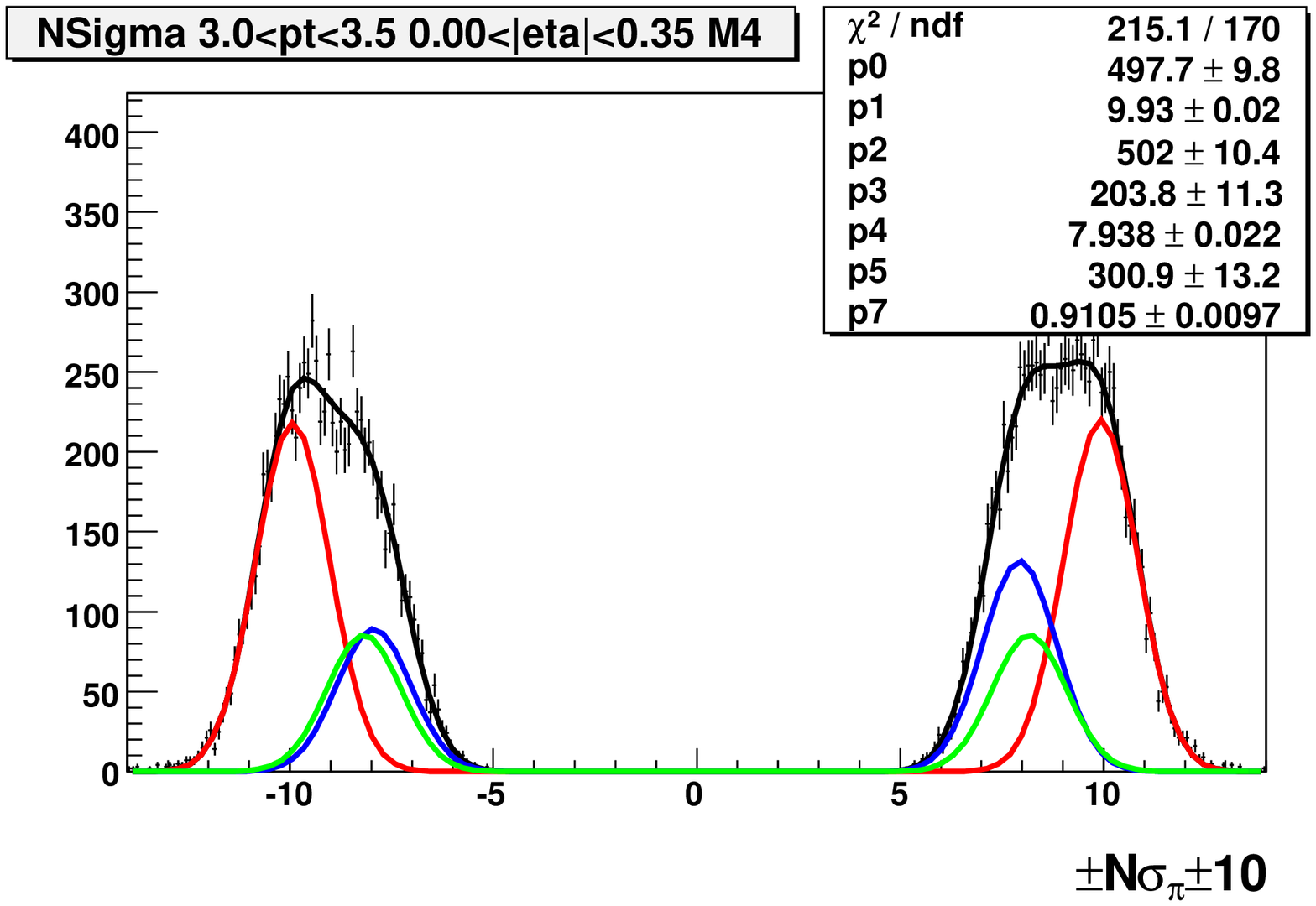}
		\includegraphics[width=1\textwidth]{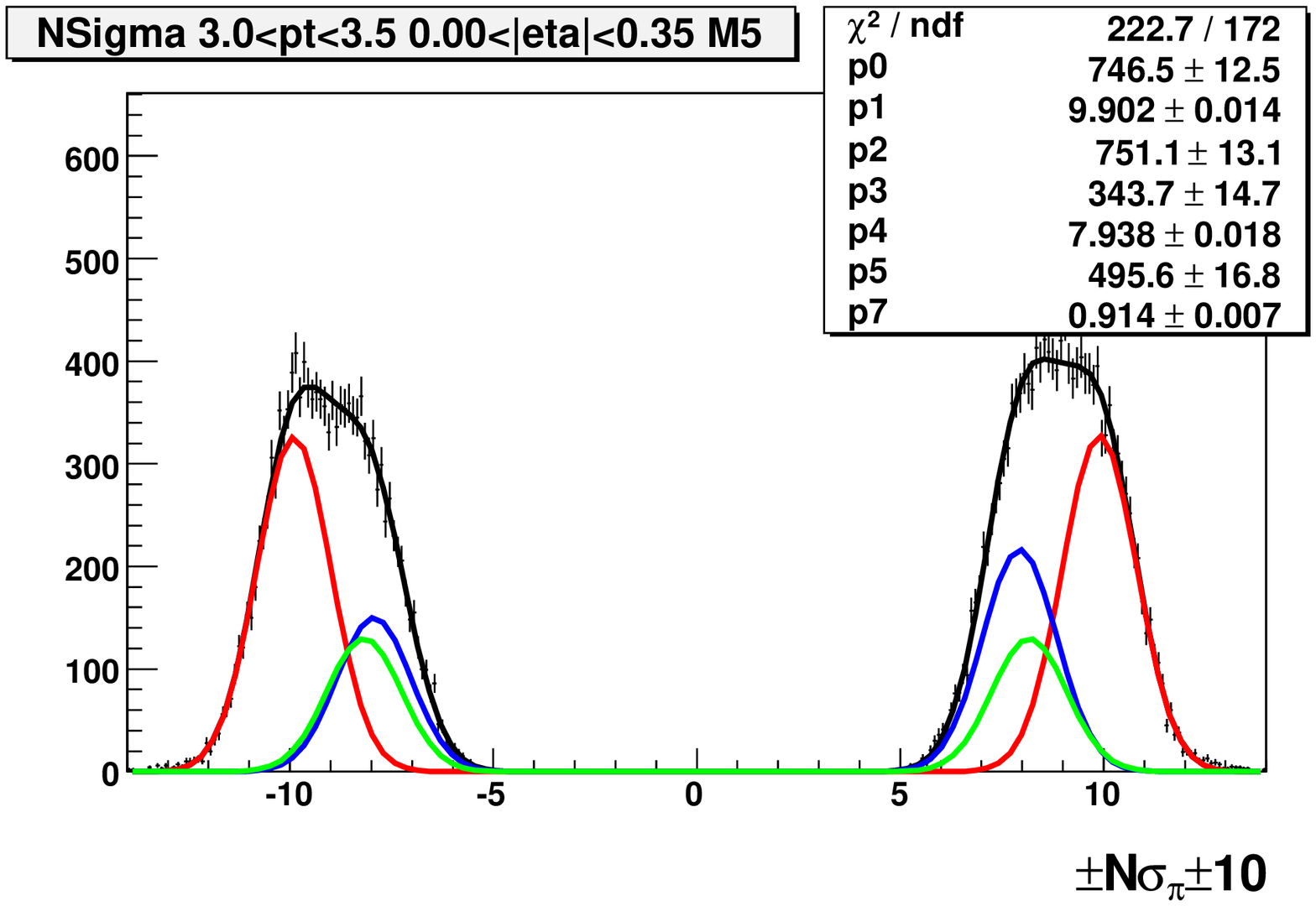}
		\includegraphics[width=1\textwidth]{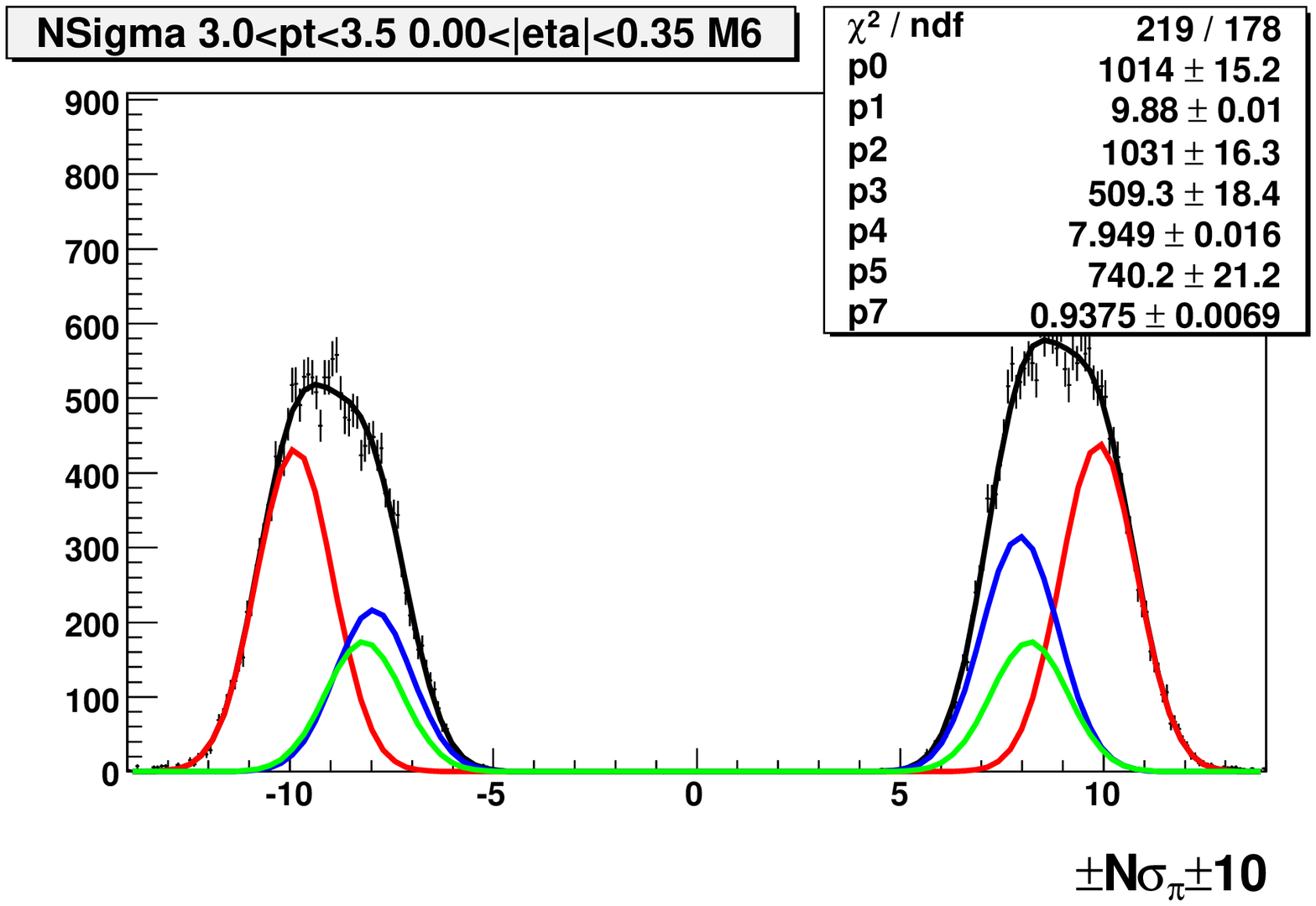}
		\includegraphics[width=1\textwidth]{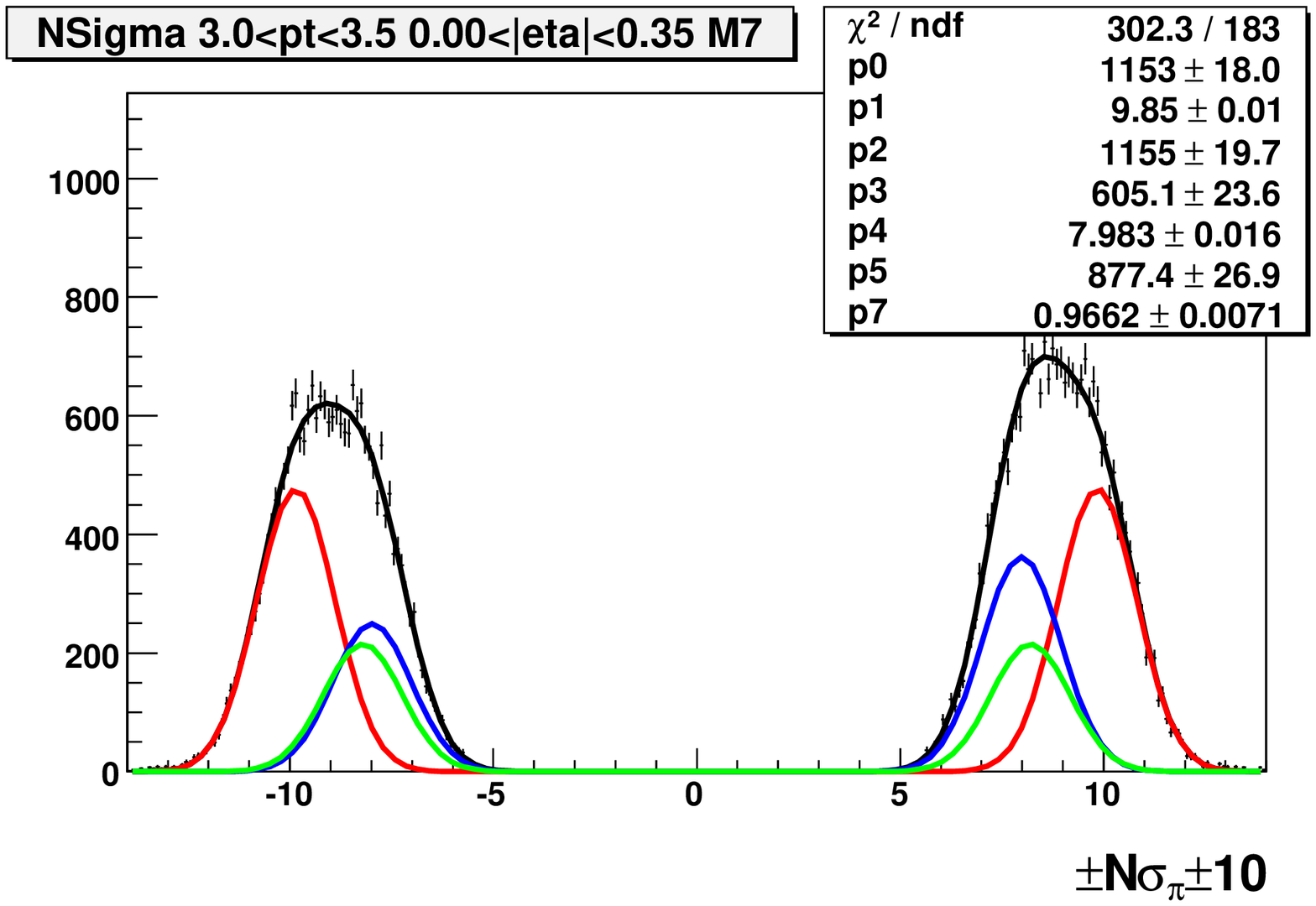}
		\includegraphics[width=1\textwidth]{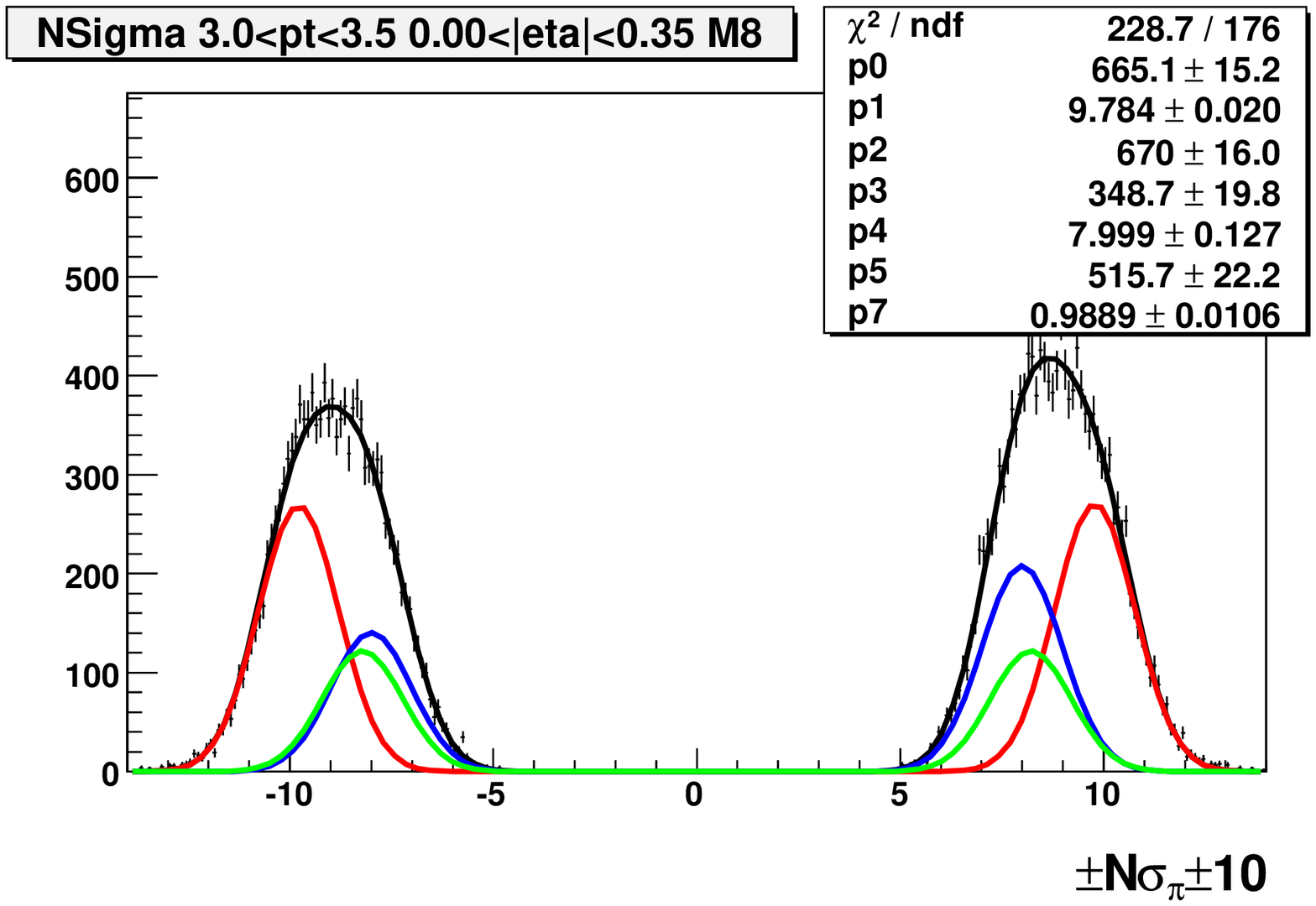}
		\includegraphics[width=1\textwidth]{Plots/fit_009.eps}									
			\end{minipage}
\hfill
\begin{minipage}[t]{.18\textwidth}
	\centering
		\includegraphics[width=1\textwidth]{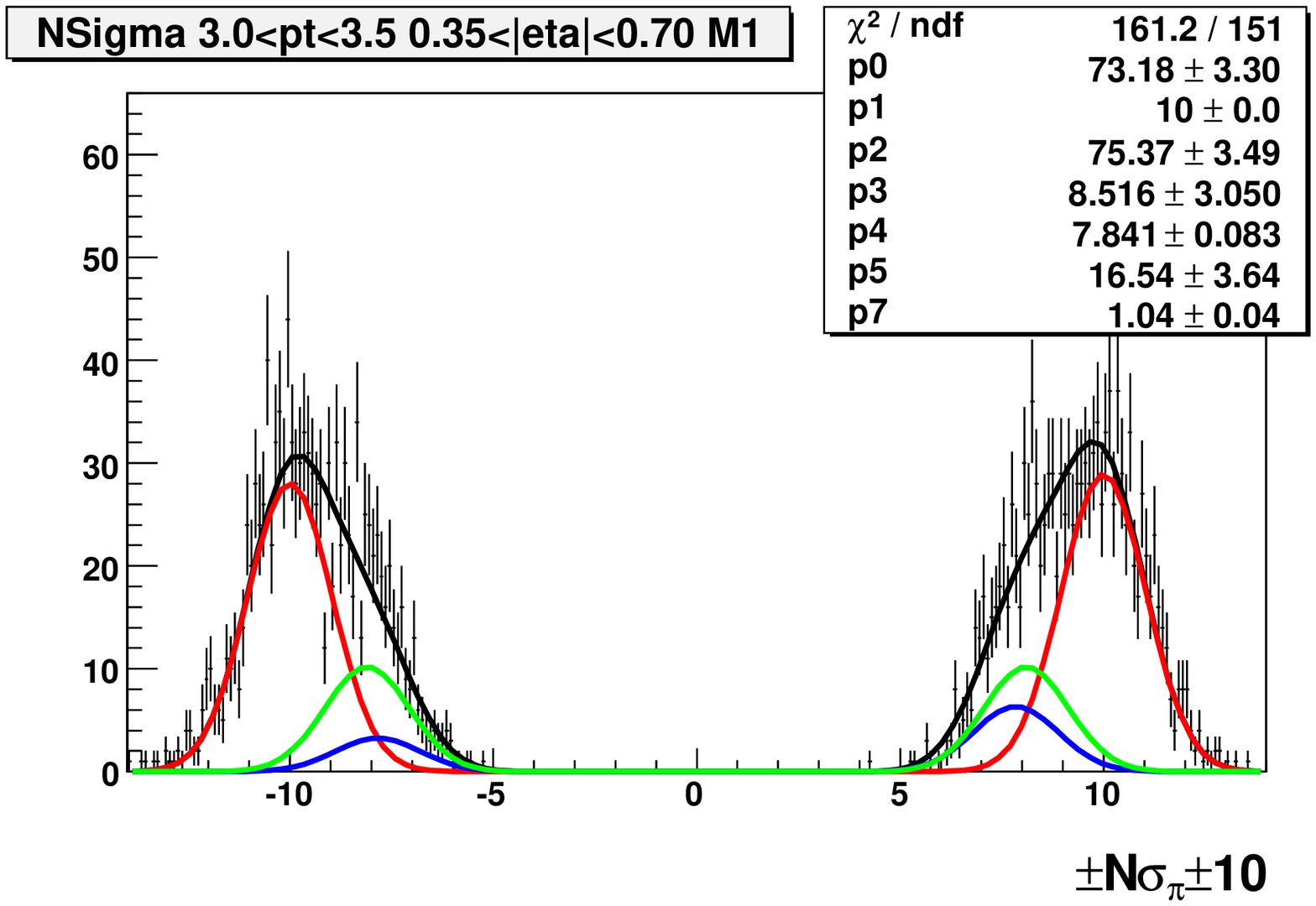}
		\includegraphics[width=1\textwidth]{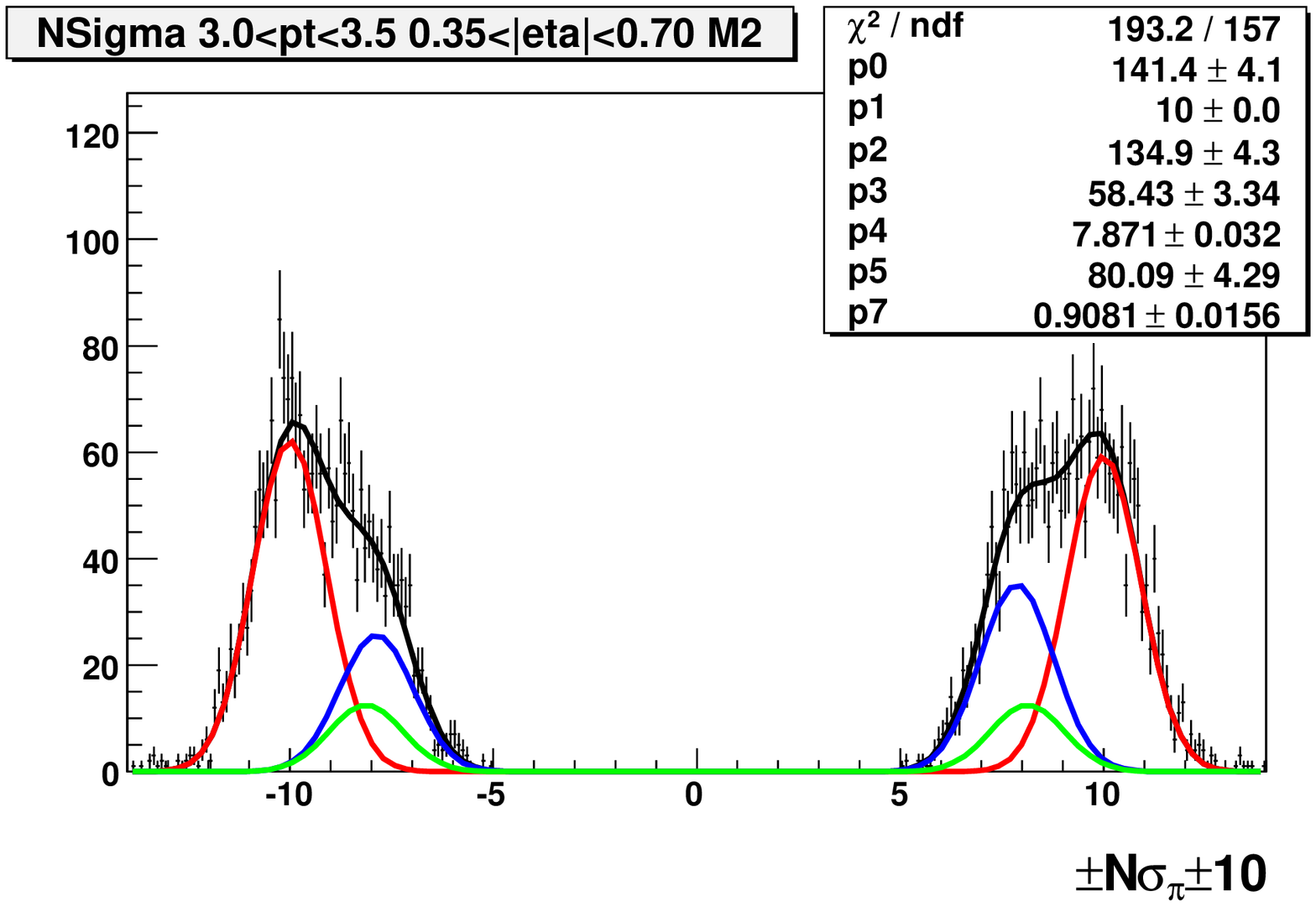}
		\includegraphics[width=1\textwidth]{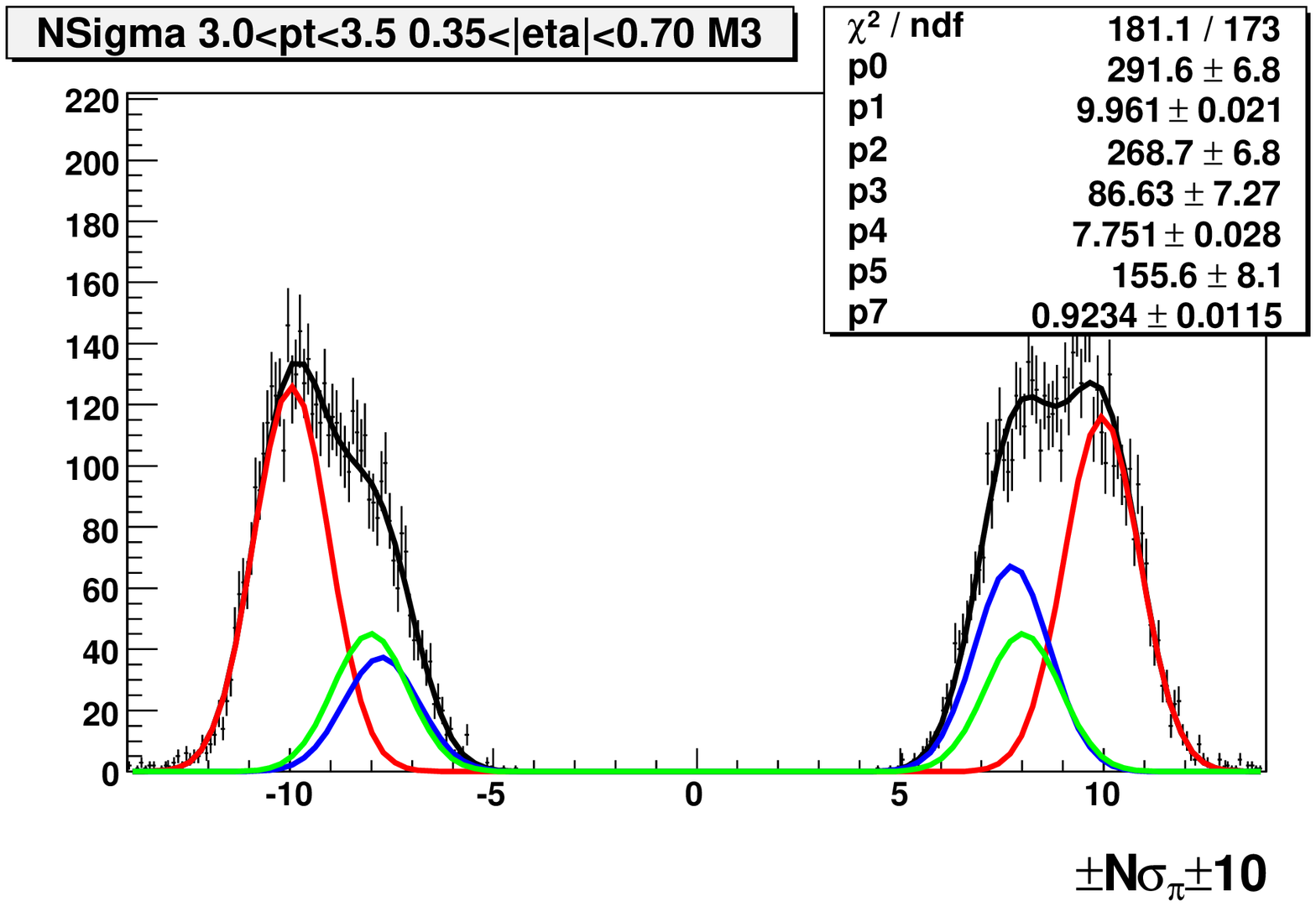}
		\includegraphics[width=1\textwidth]{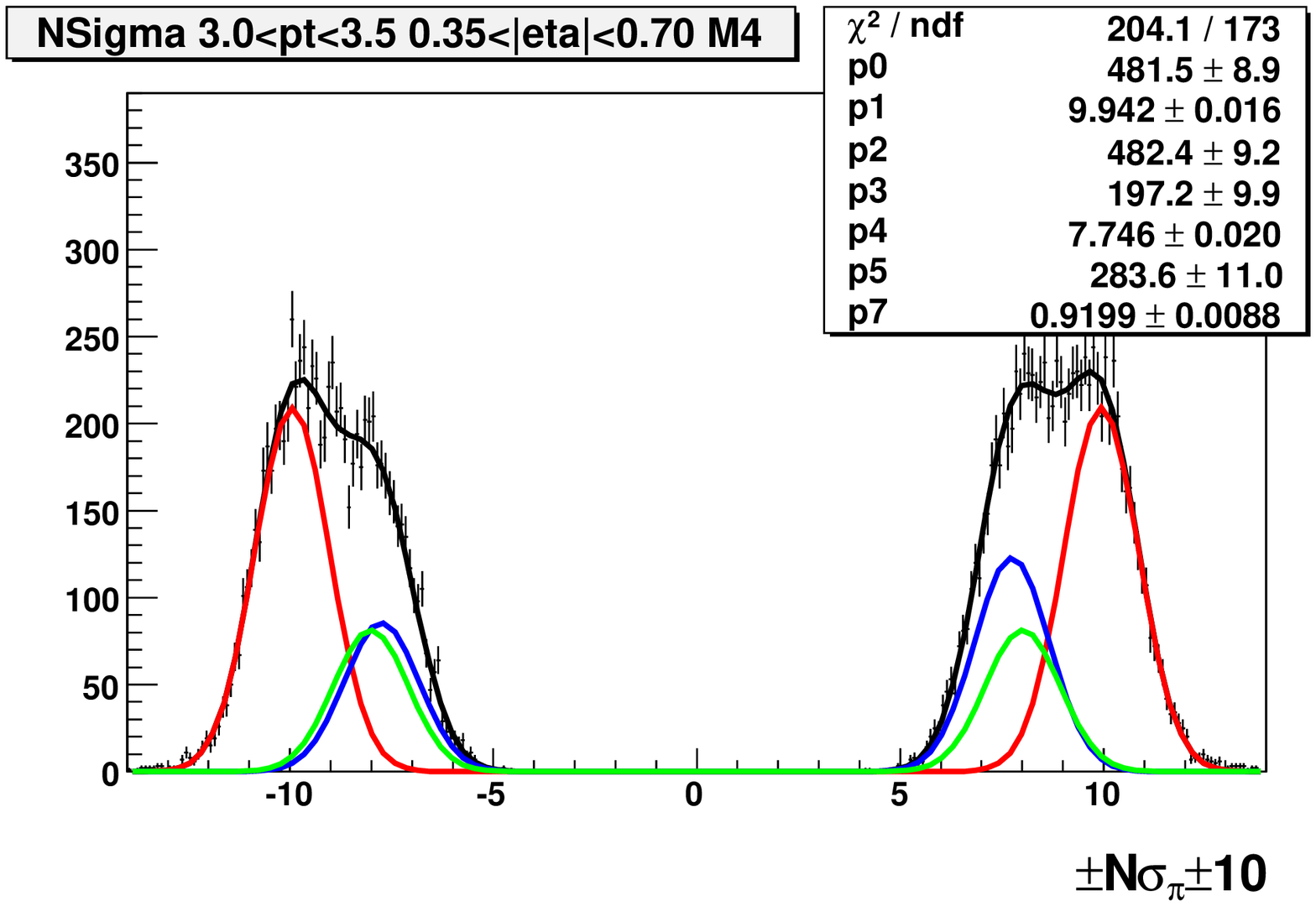}
		\includegraphics[width=1\textwidth]{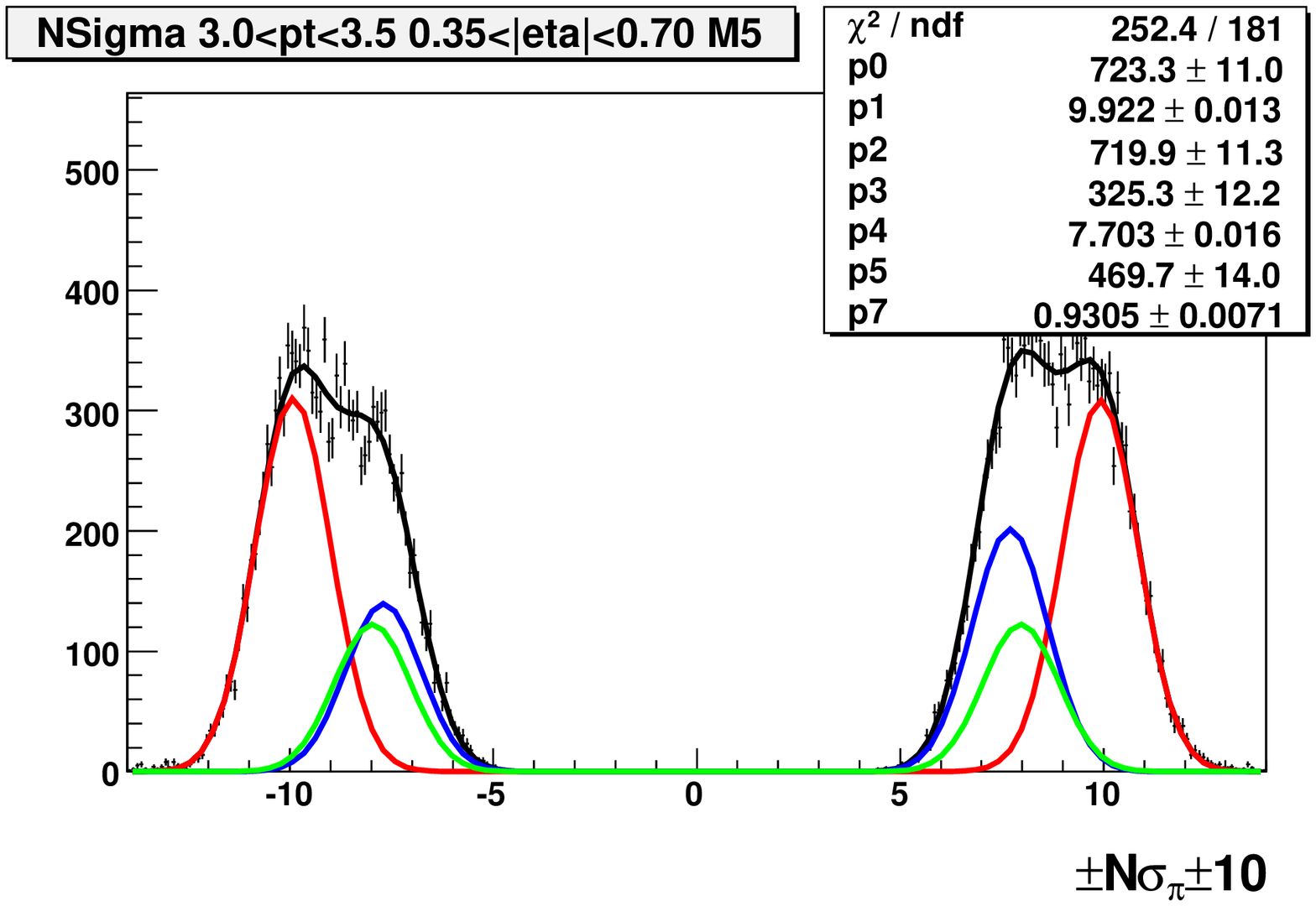}
		\includegraphics[width=1\textwidth]{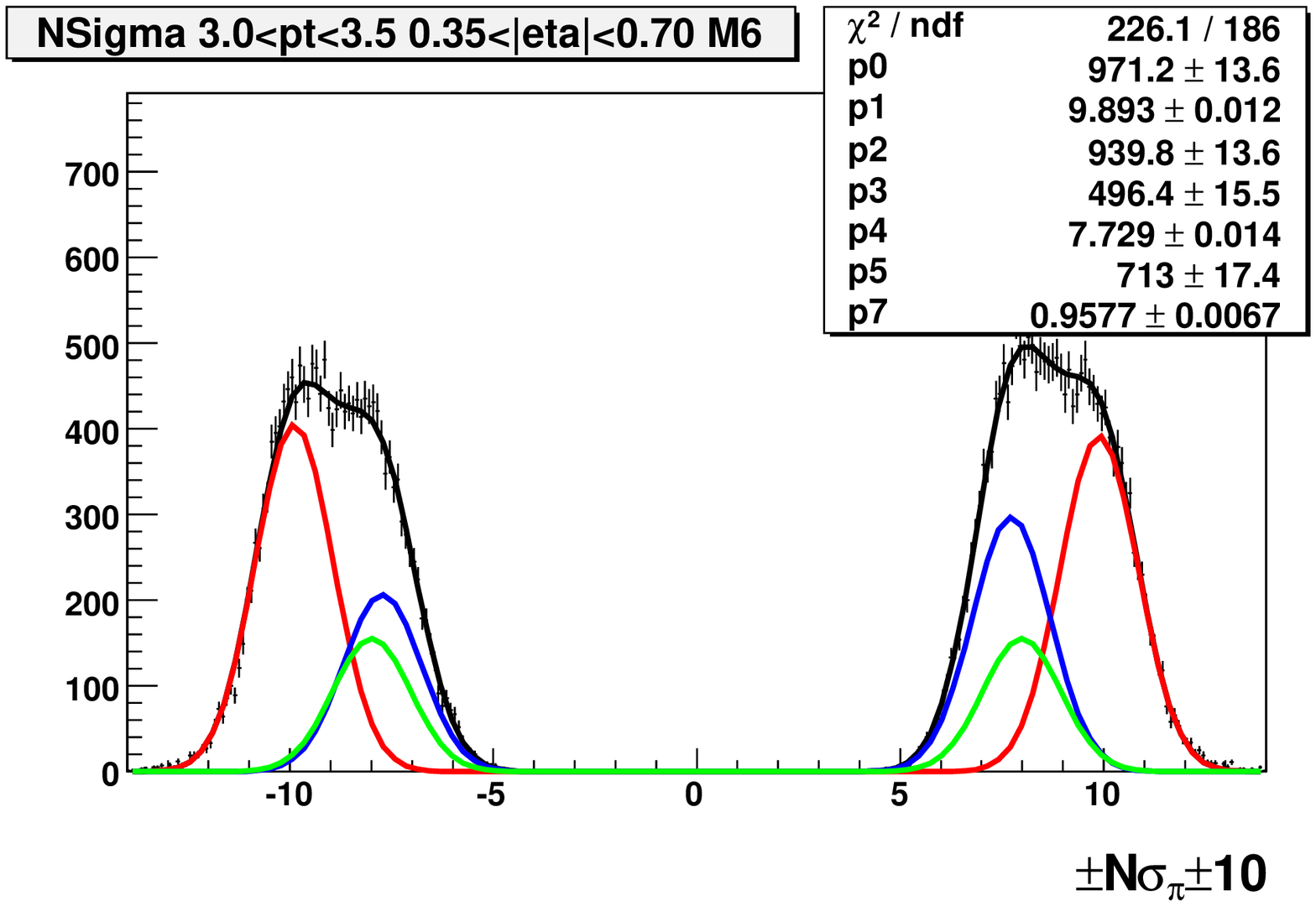}
		\includegraphics[width=1\textwidth]{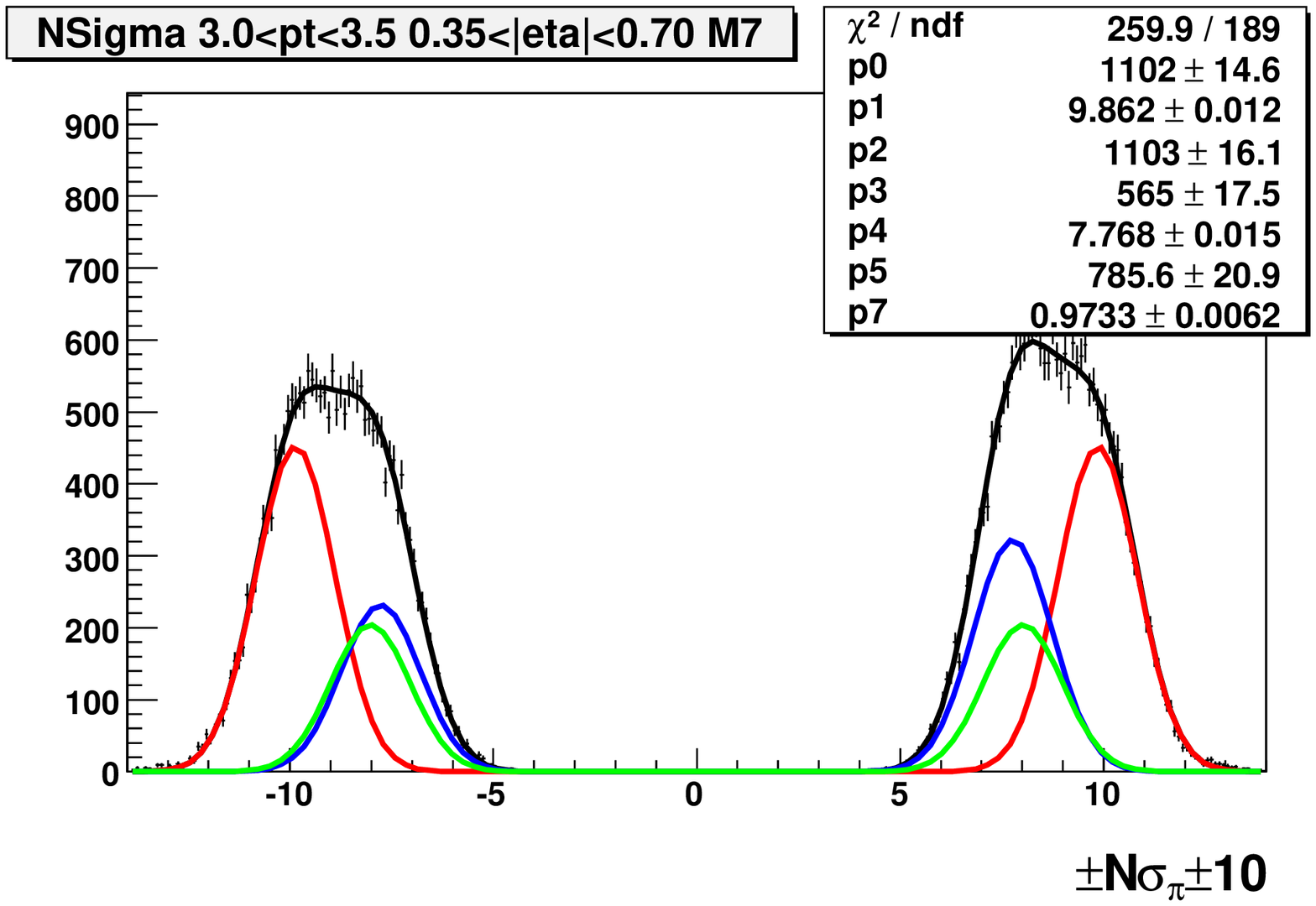}
		\includegraphics[width=1\textwidth]{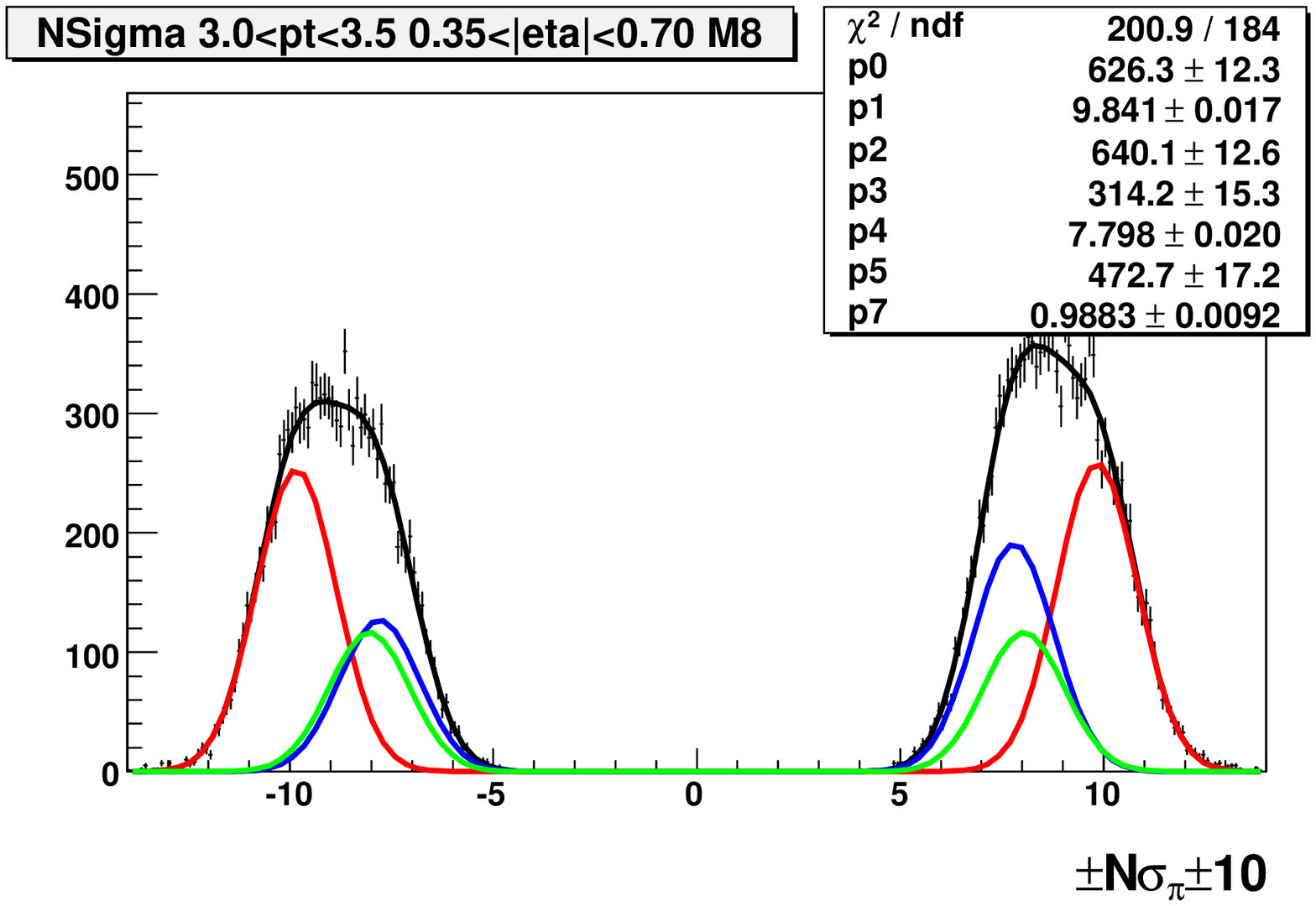}
		\includegraphics[width=1\textwidth]{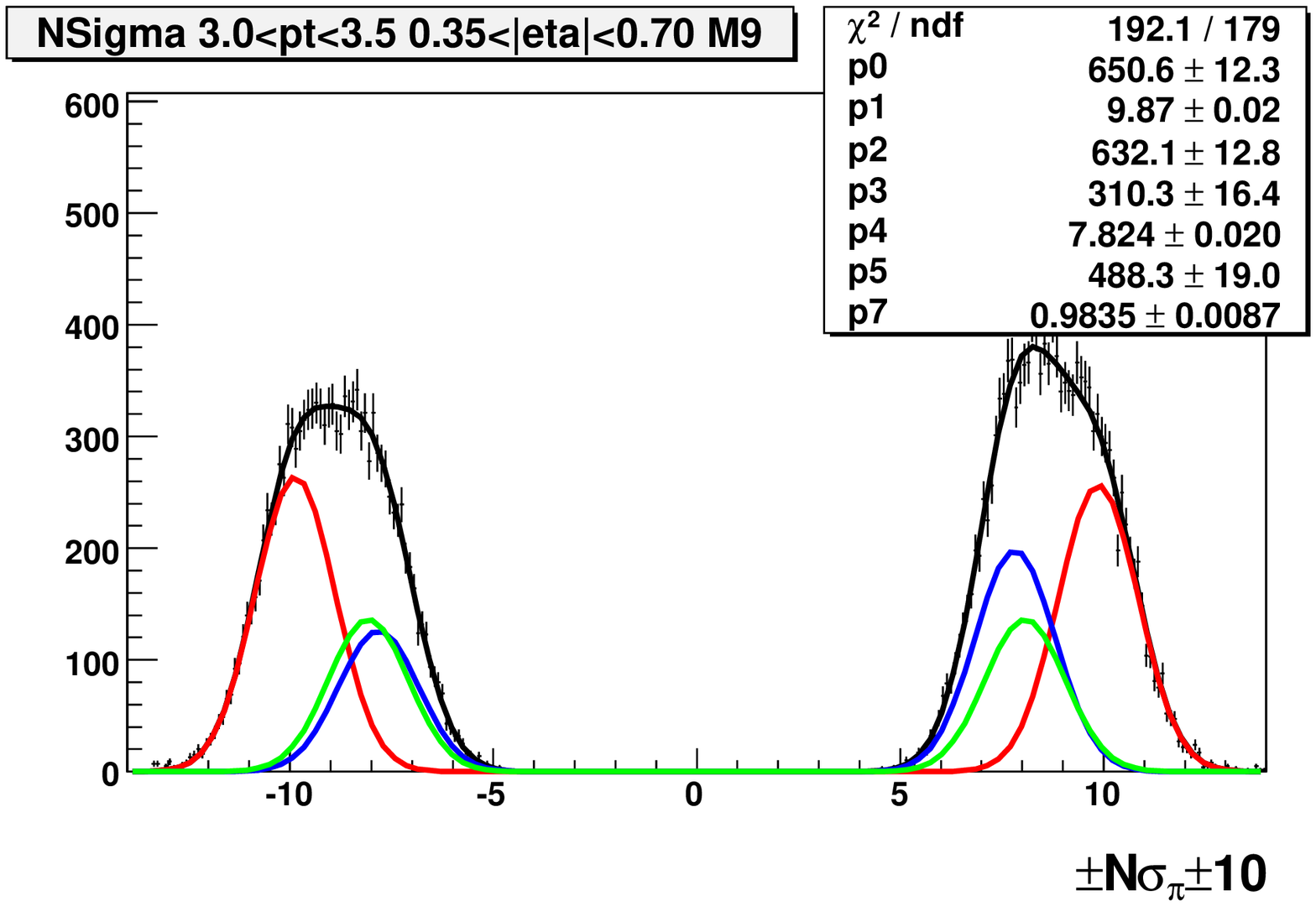}									
			\end{minipage}
\hfill
\begin{minipage}[t]{.18\textwidth}
	\centering
		\includegraphics[width=1\textwidth]{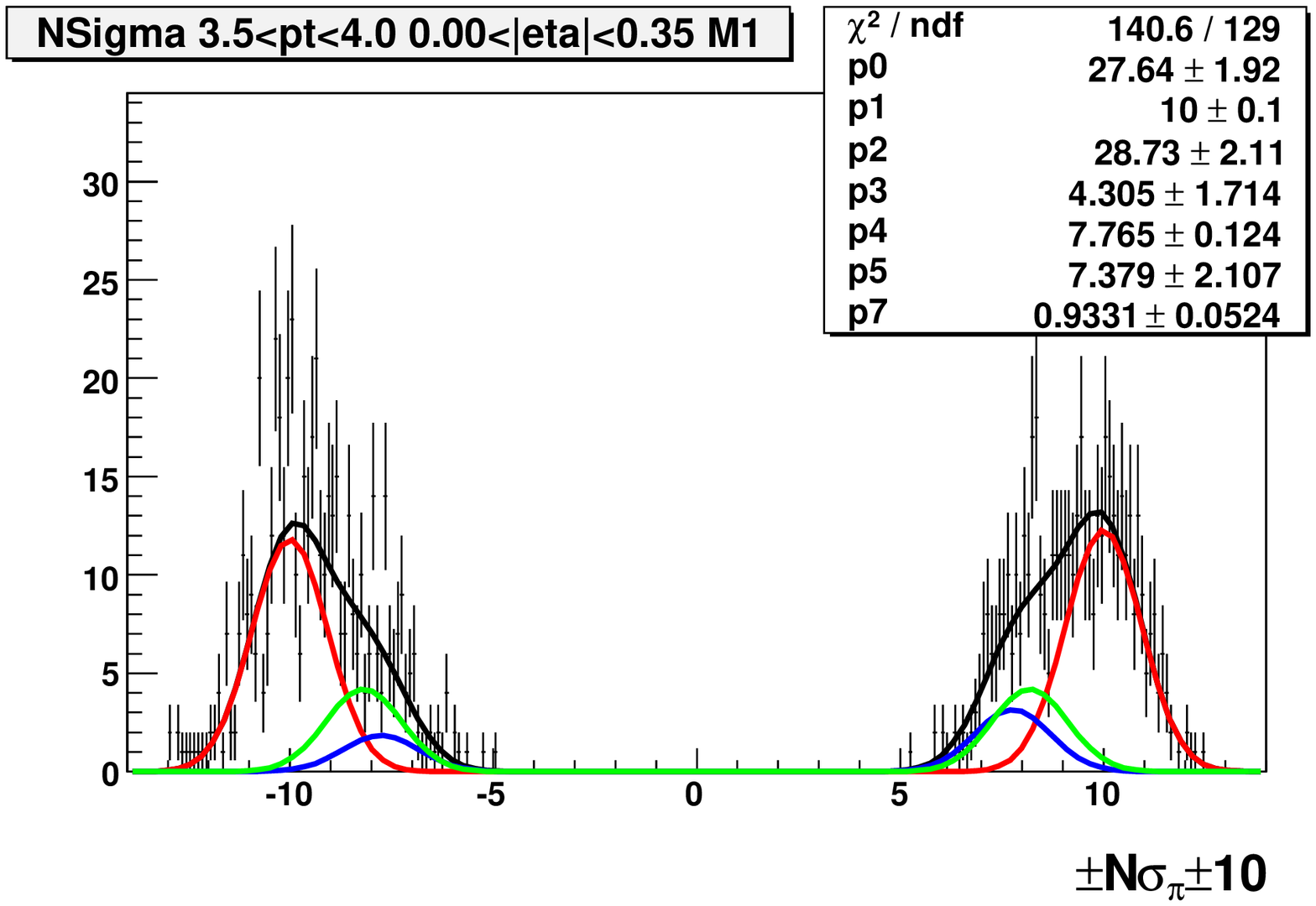}
		\includegraphics[width=1\textwidth]{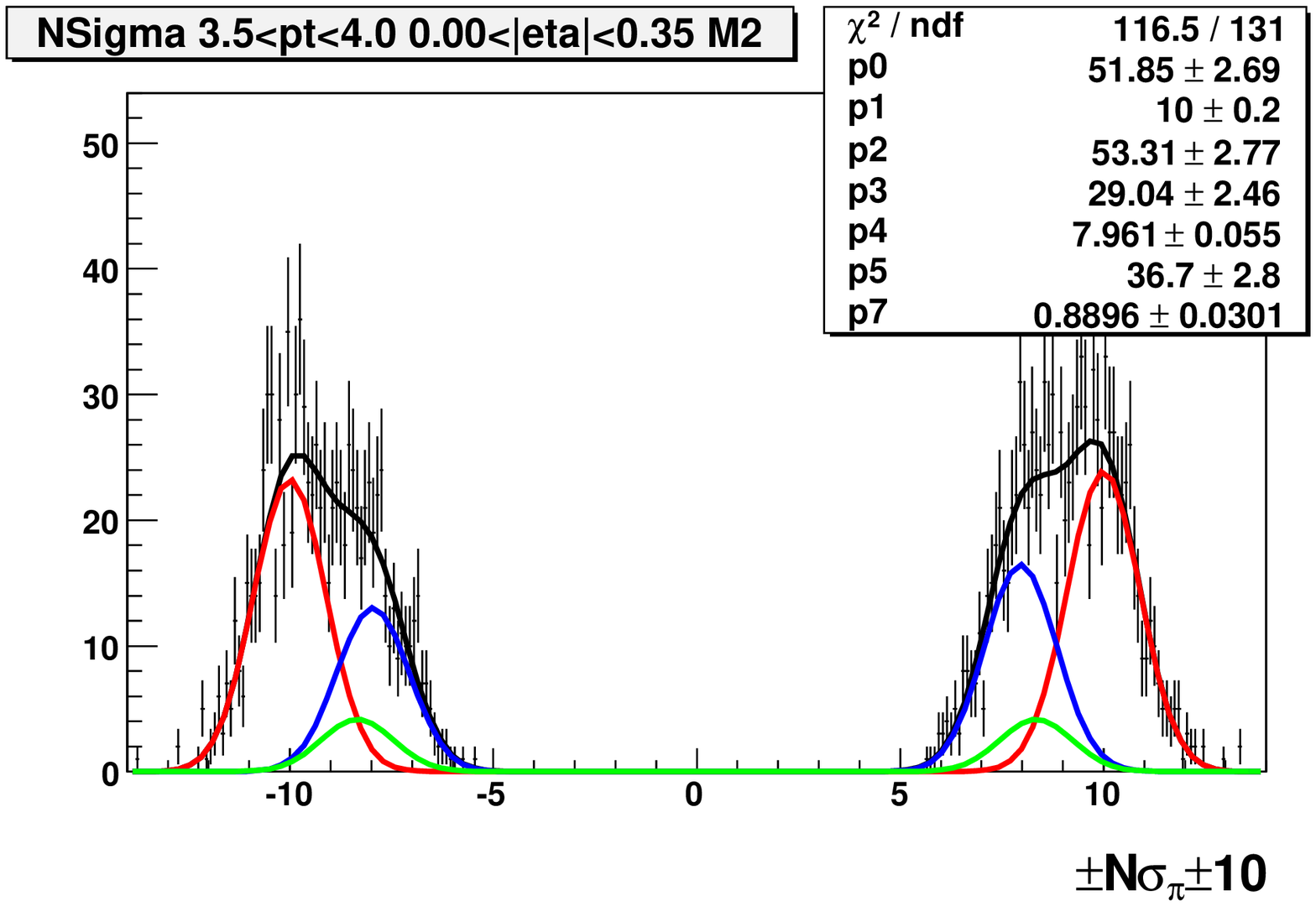}
		\includegraphics[width=1\textwidth]{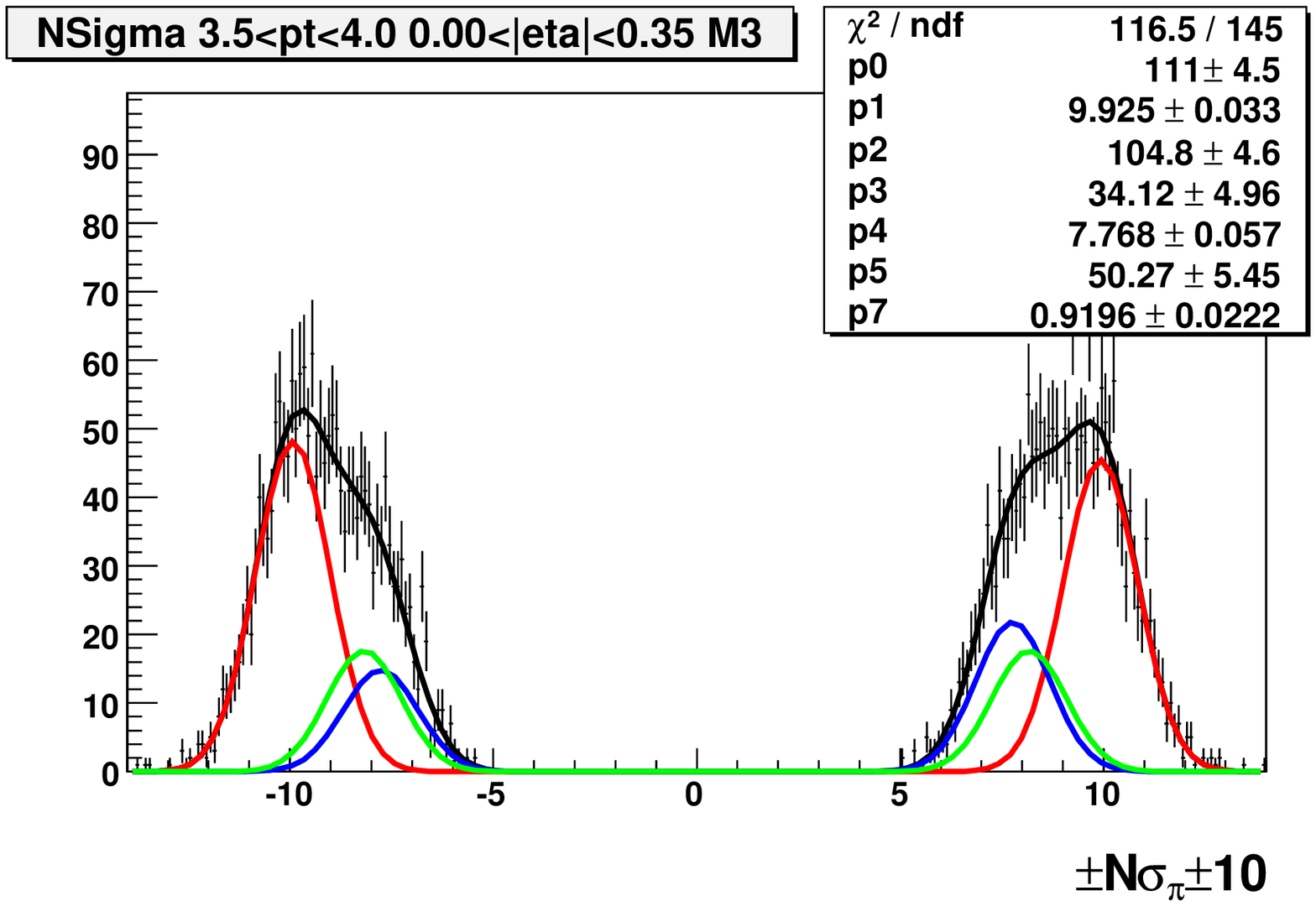}
		\includegraphics[width=1\textwidth]{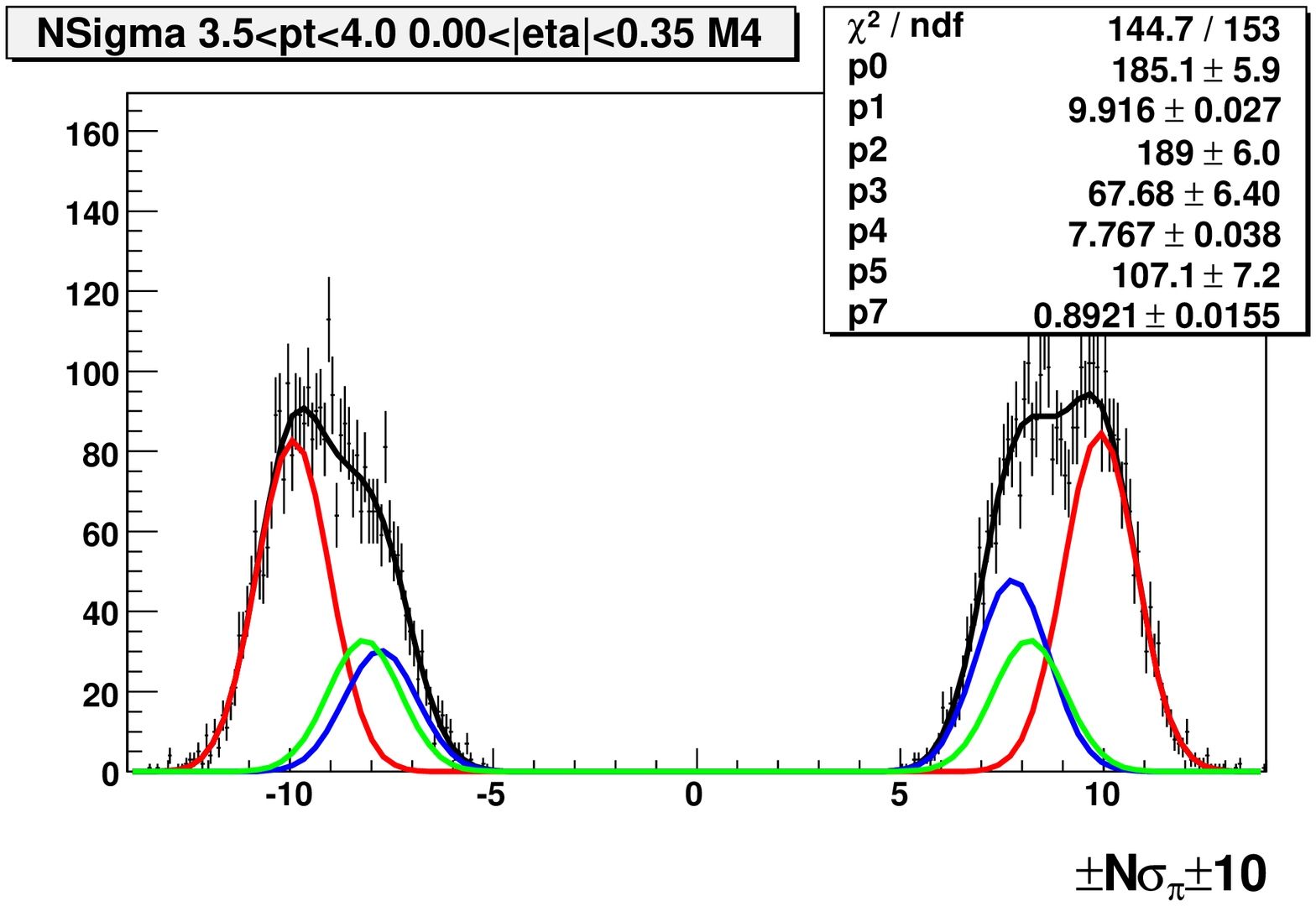}
		\includegraphics[width=1\textwidth]{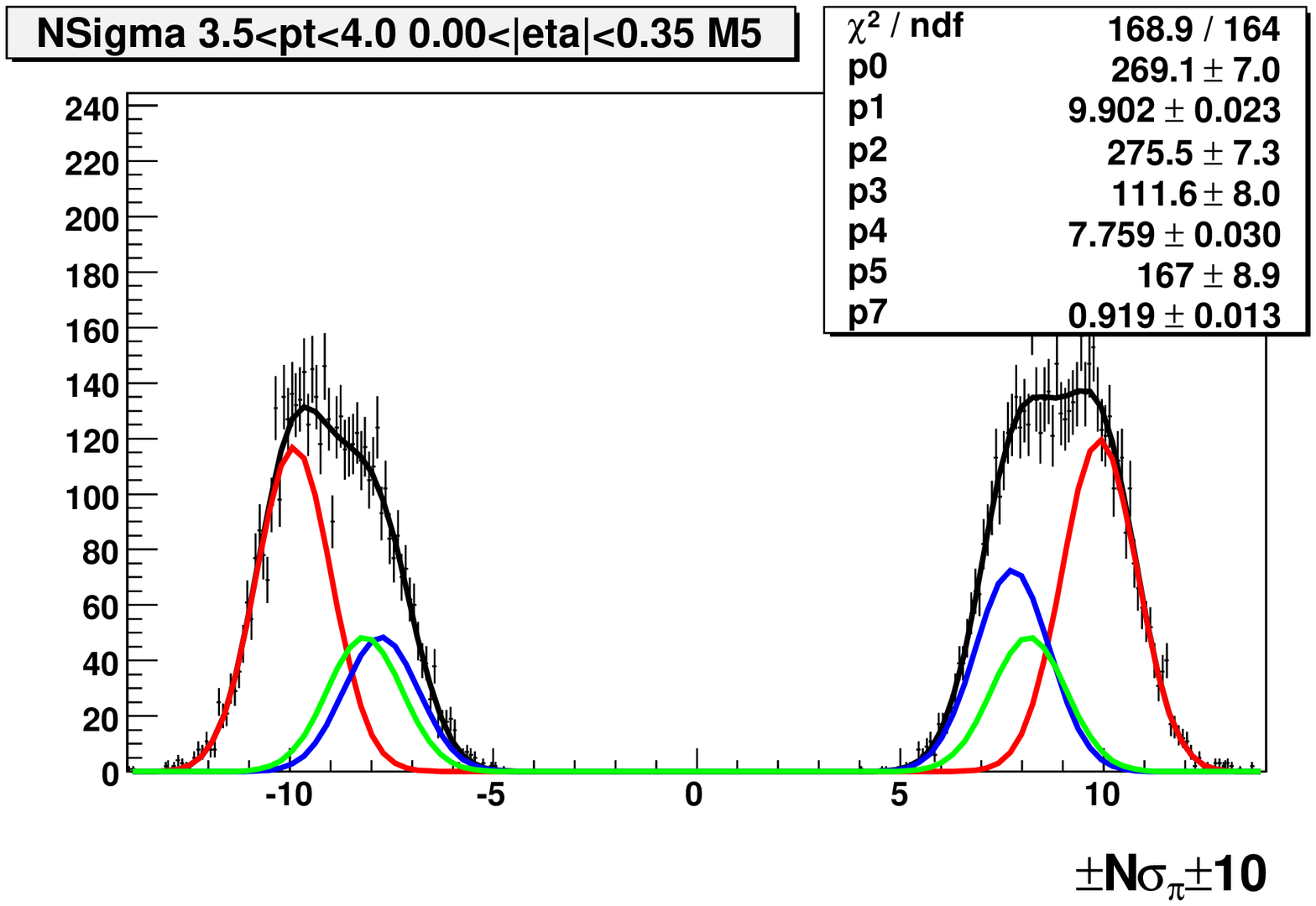}
		\includegraphics[width=1\textwidth]{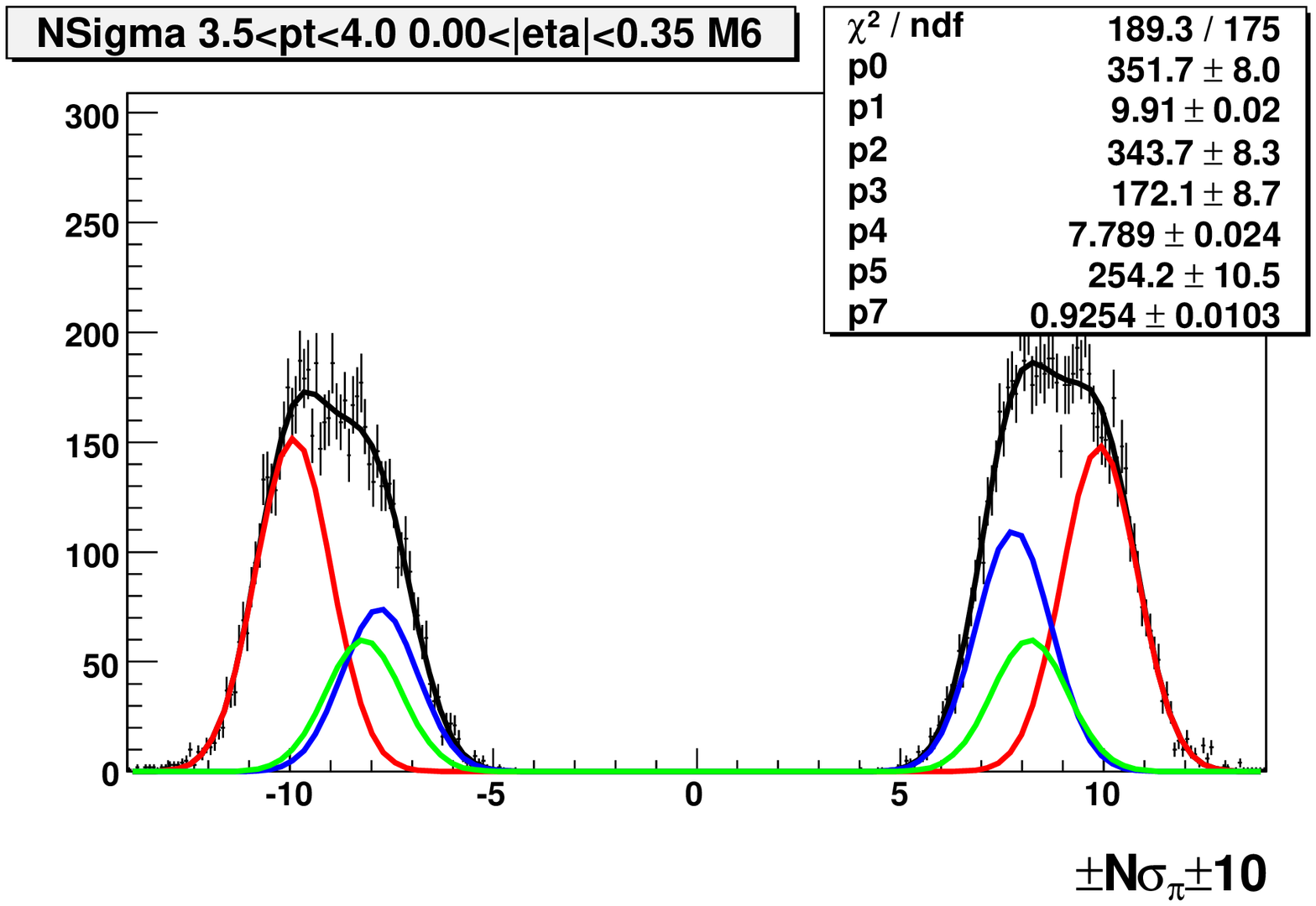}
		\includegraphics[width=1\textwidth]{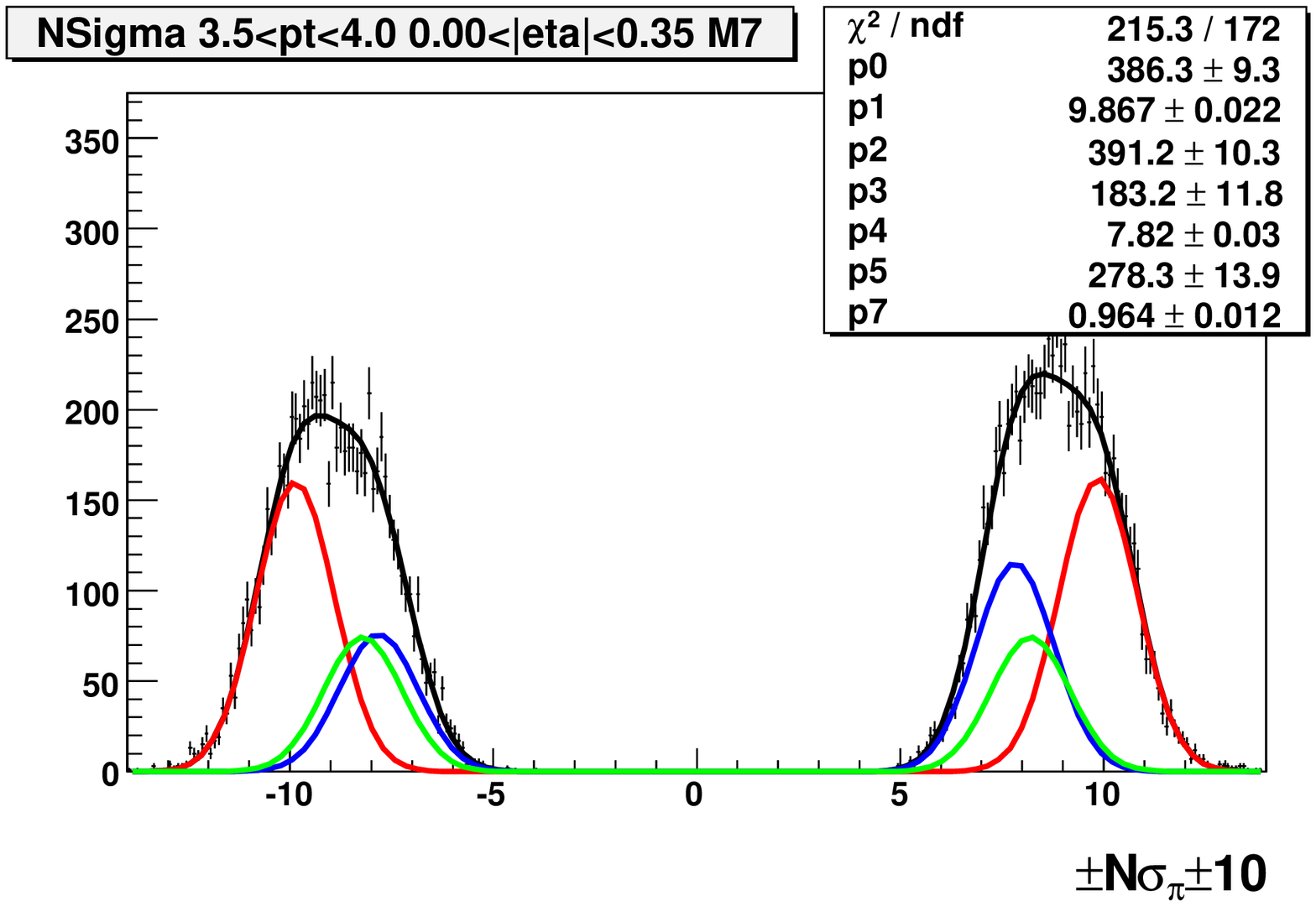}
		\includegraphics[width=1\textwidth]{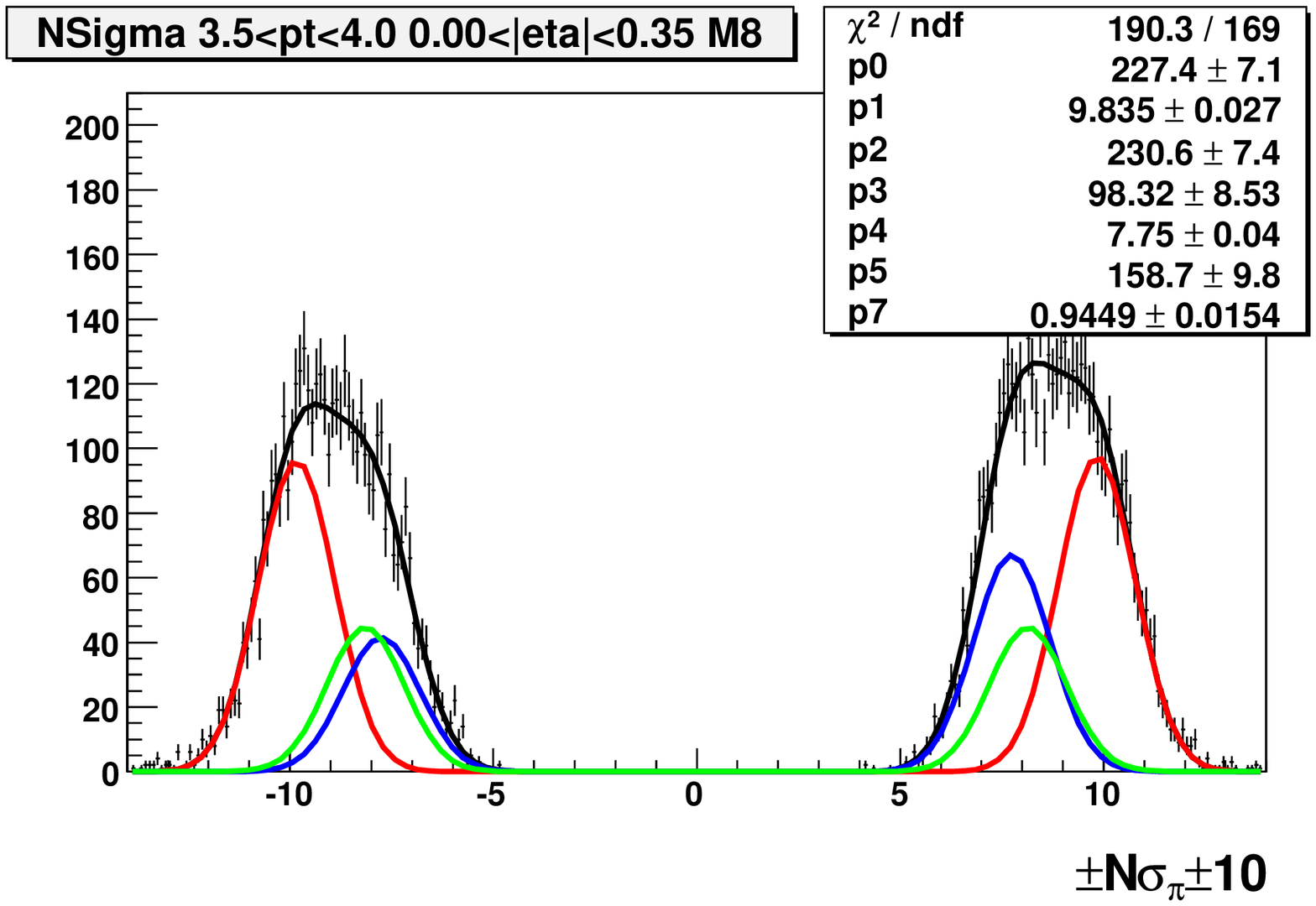}
		\includegraphics[width=1\textwidth]{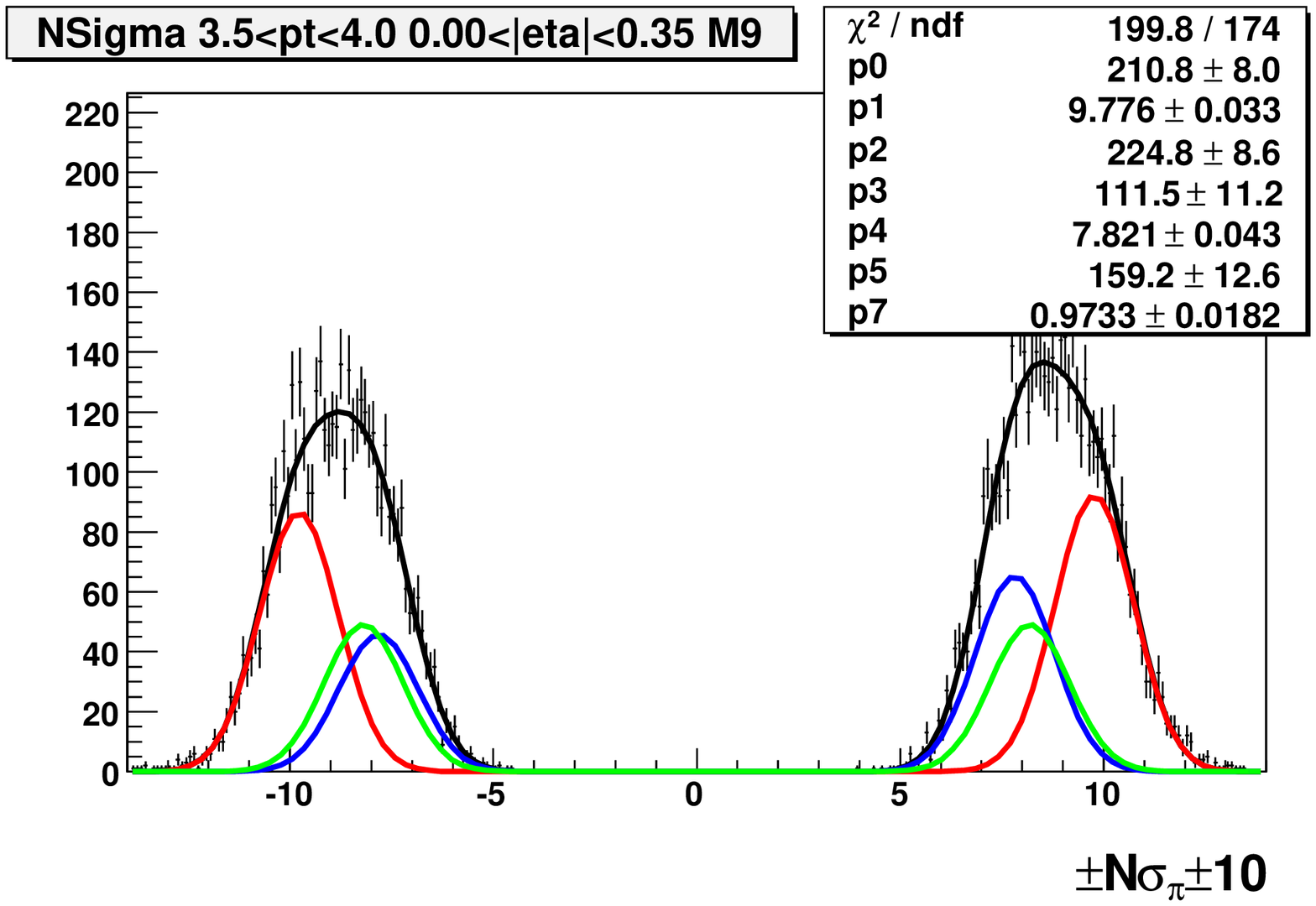}
										
			\end{minipage}
\hfill
\begin{minipage}[t]{.18\textwidth}
	\centering
		\includegraphics[width=1\textwidth]{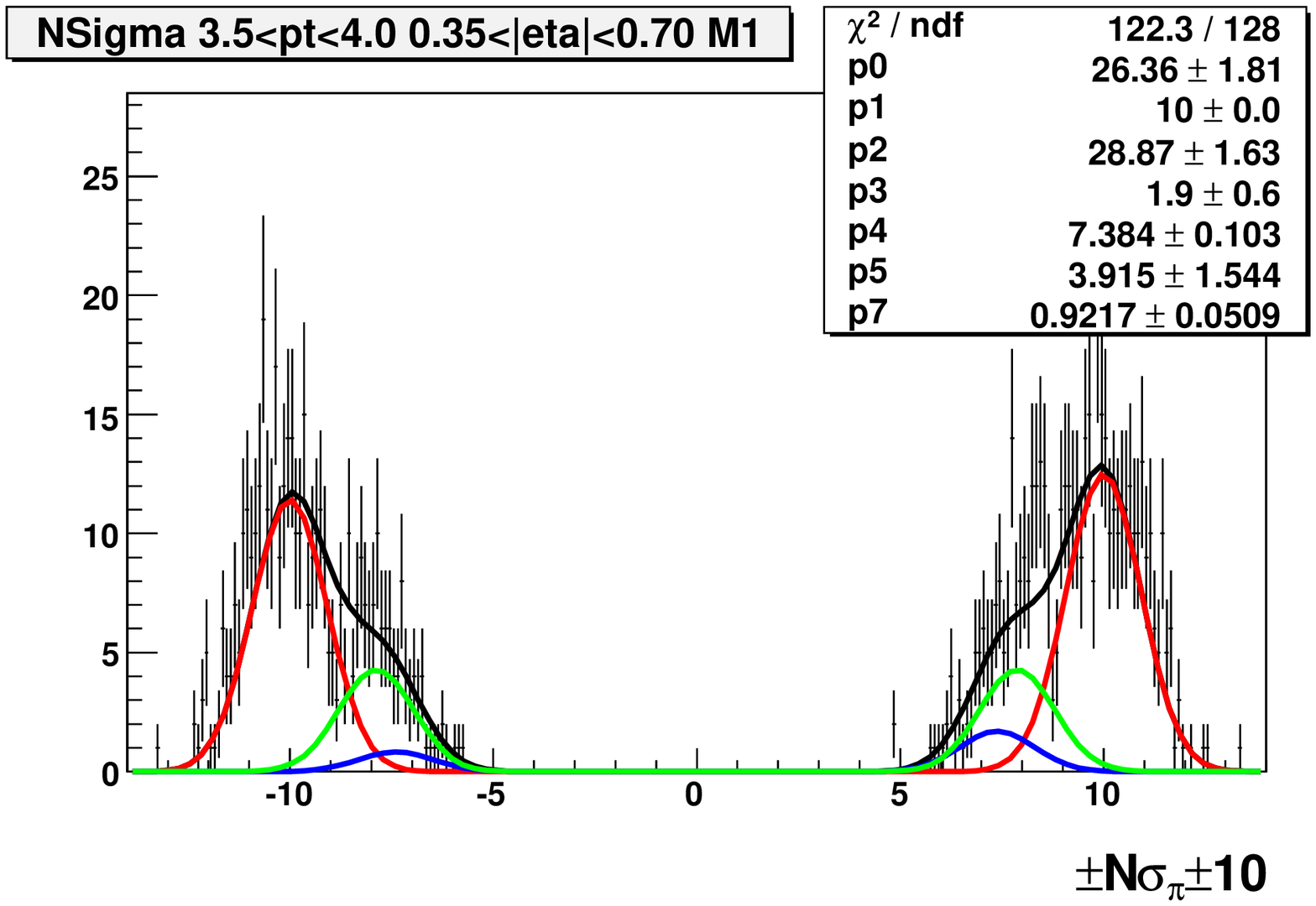}
		\includegraphics[width=1\textwidth]{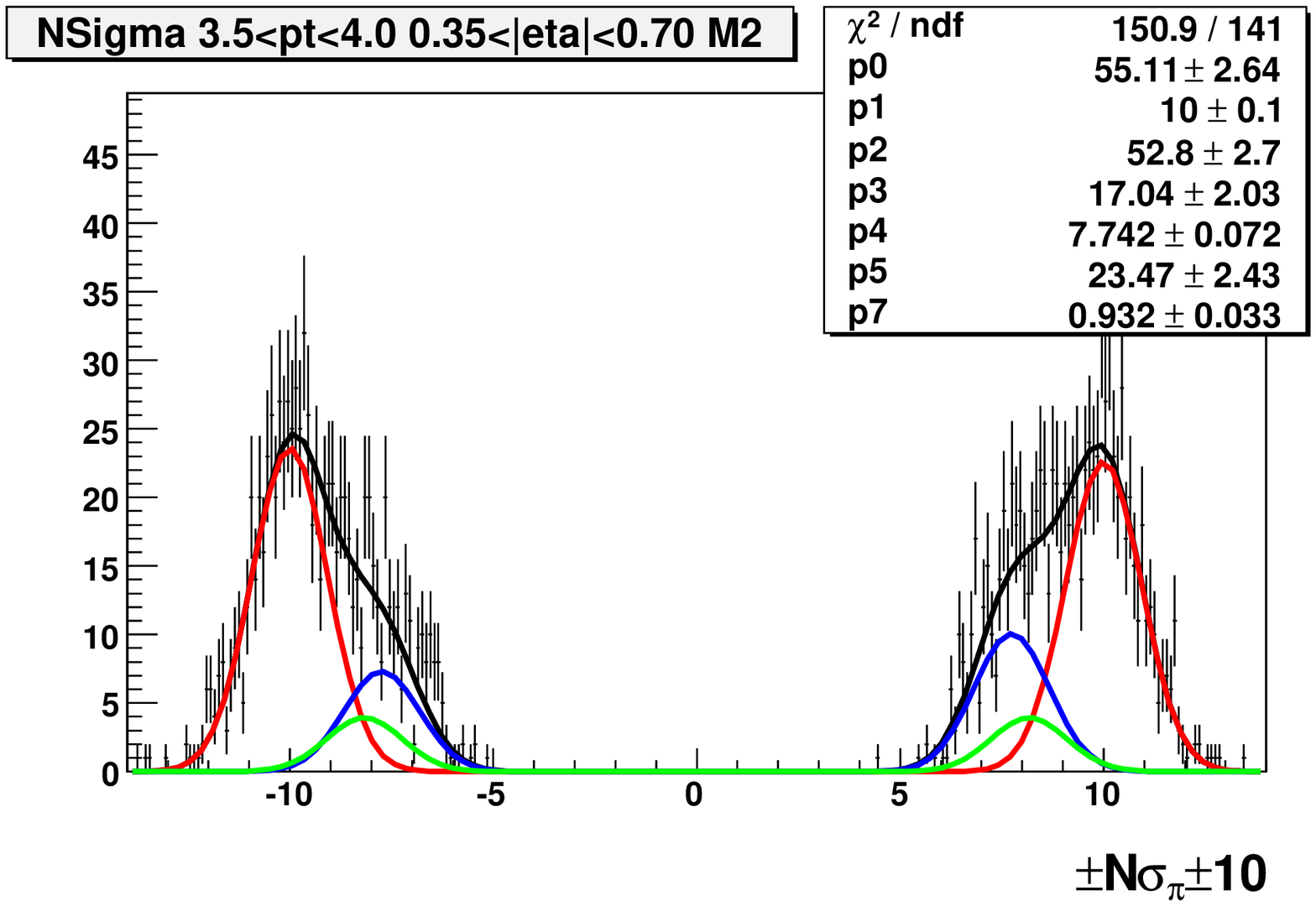}
		\includegraphics[width=1\textwidth]{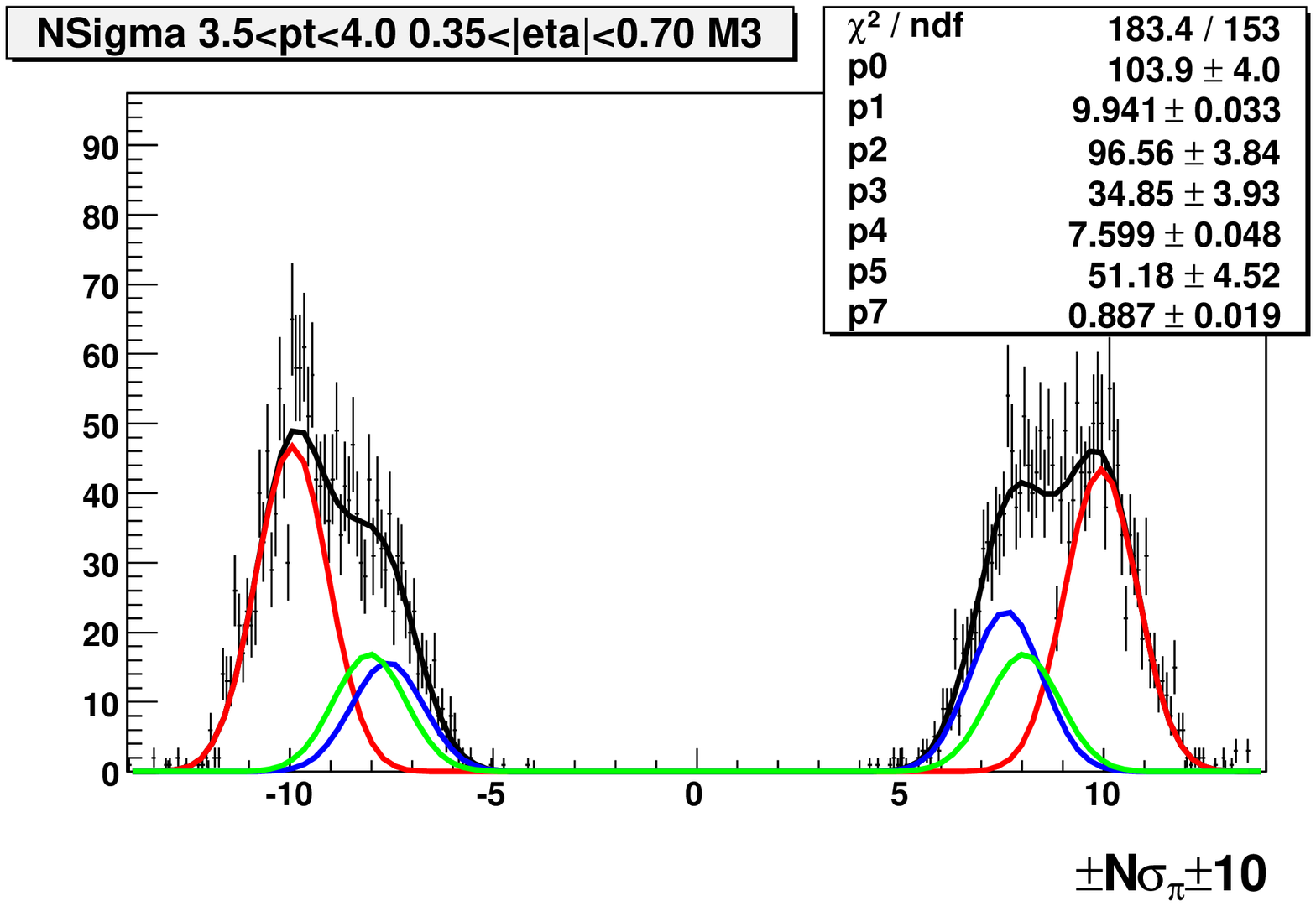}
		\includegraphics[width=1\textwidth]{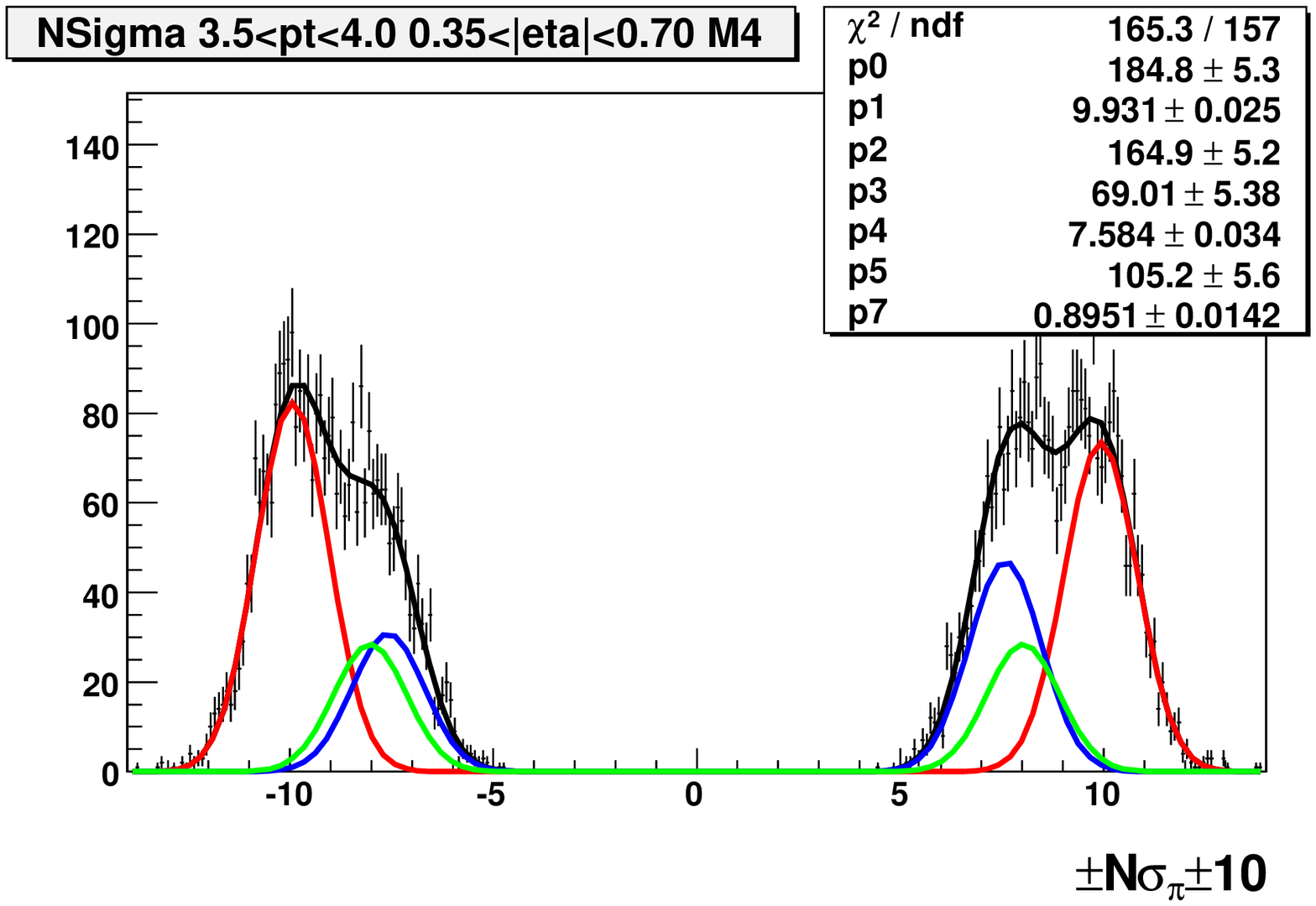}
		\includegraphics[width=1\textwidth]{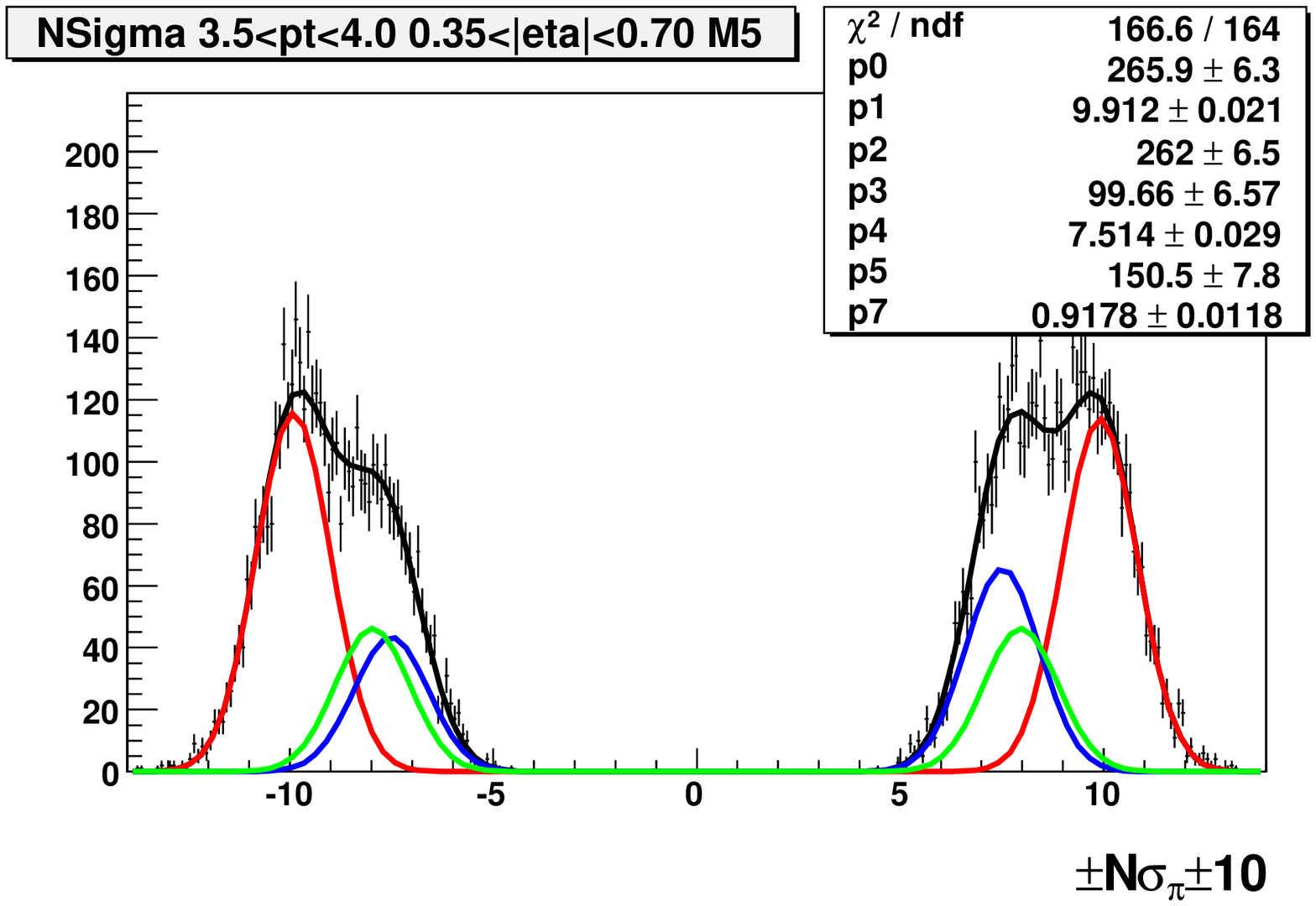}
		\includegraphics[width=1\textwidth]{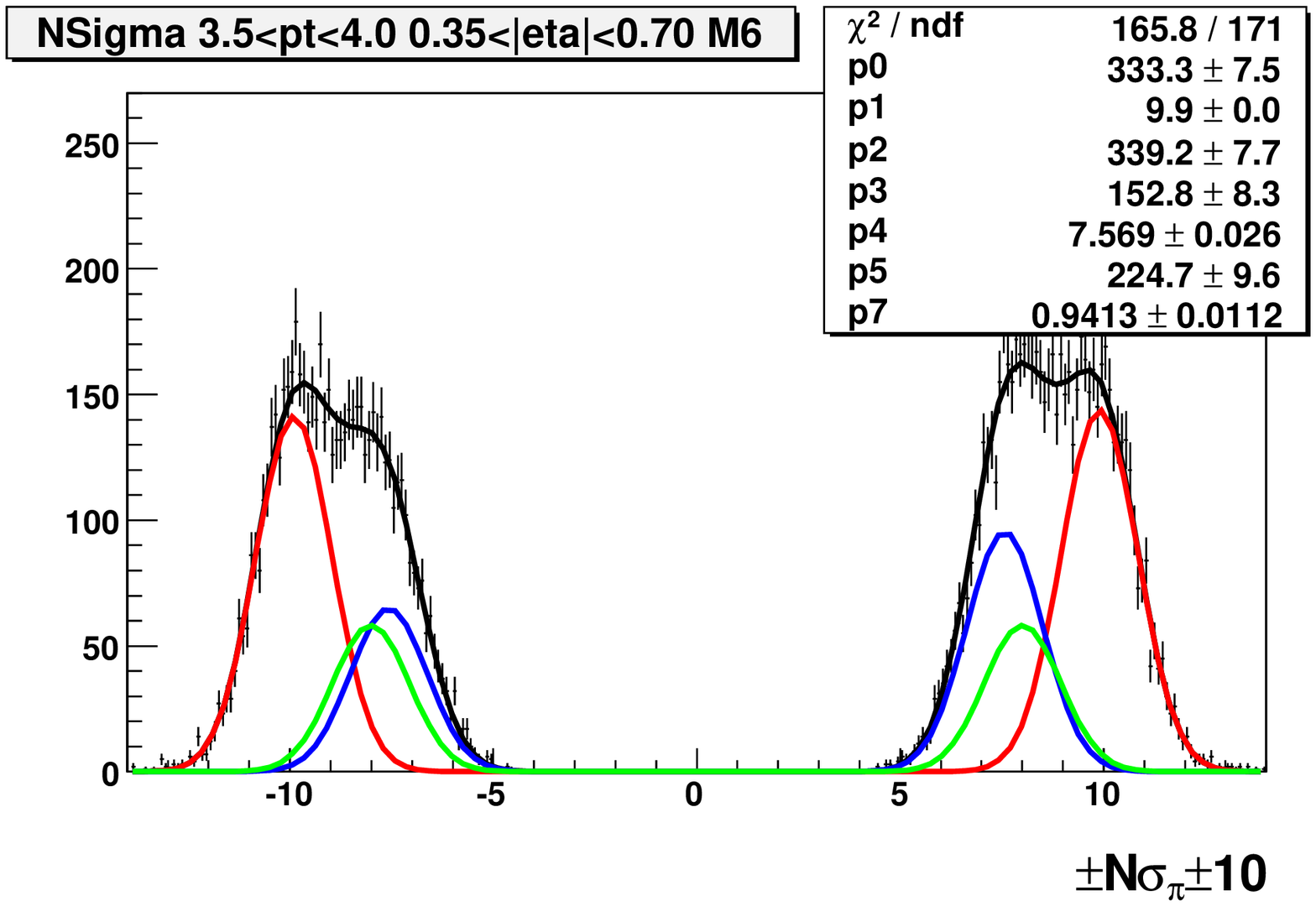}
		\includegraphics[width=1\textwidth]{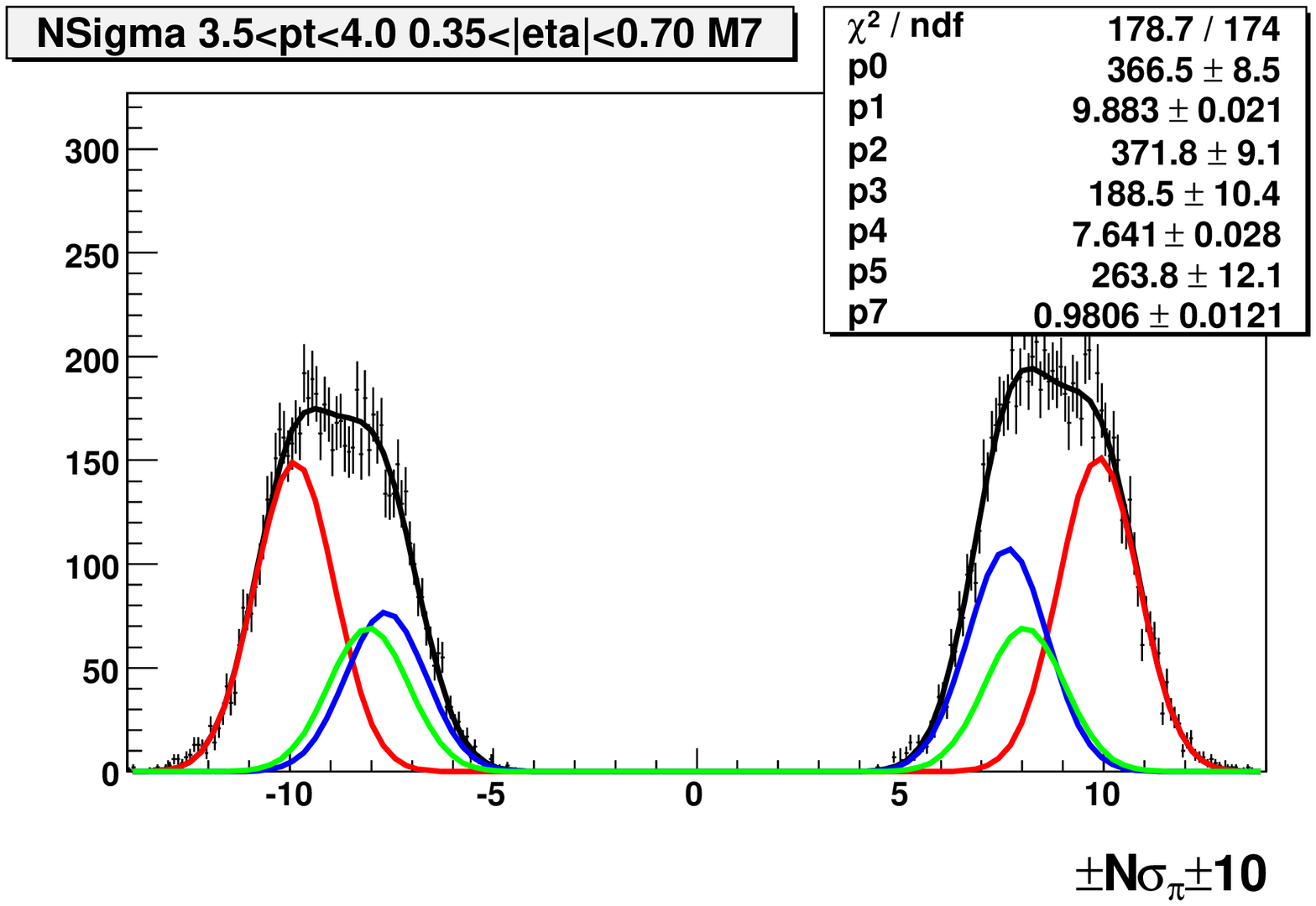}
		\includegraphics[width=1\textwidth]{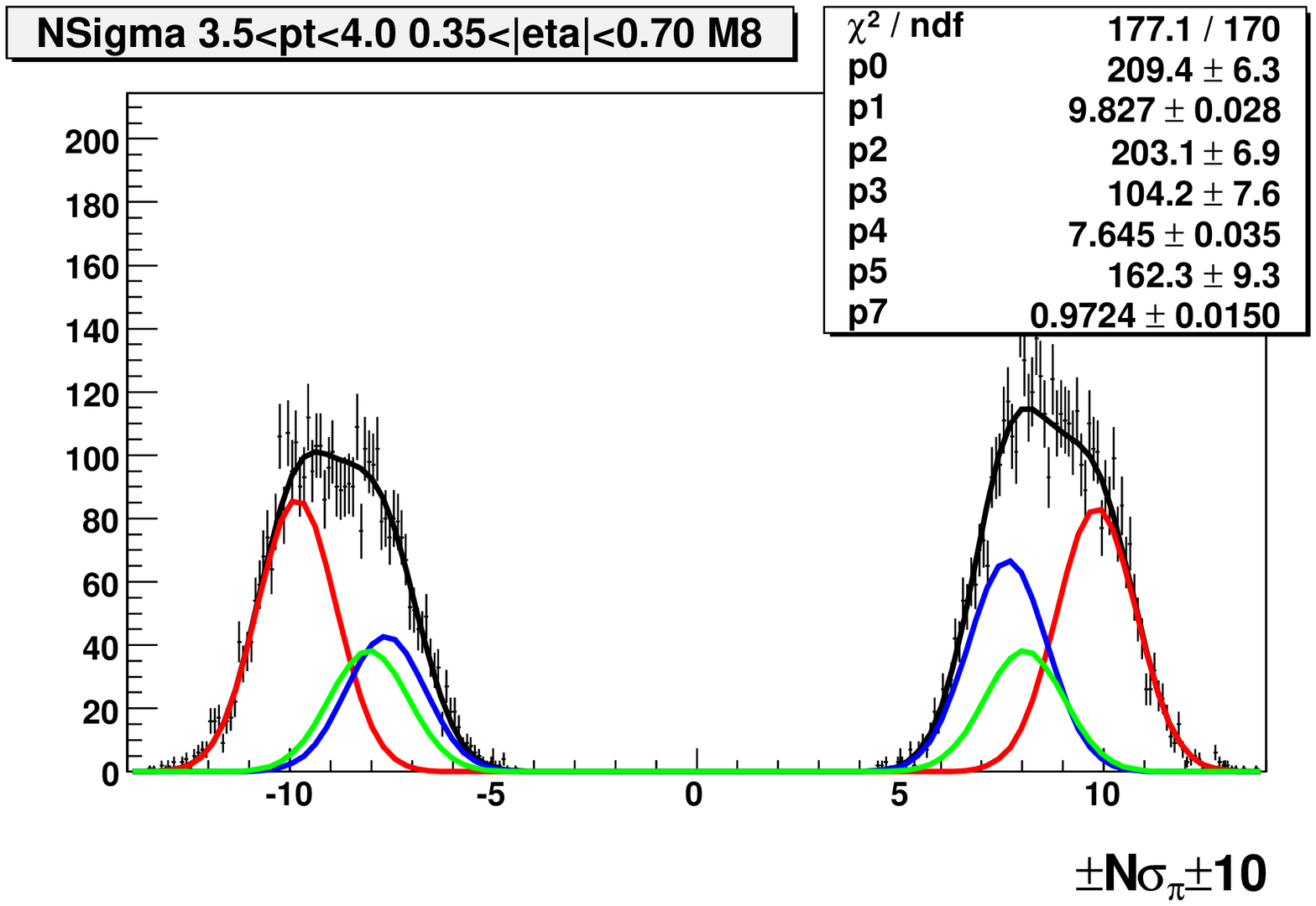}
		\includegraphics[width=1\textwidth]{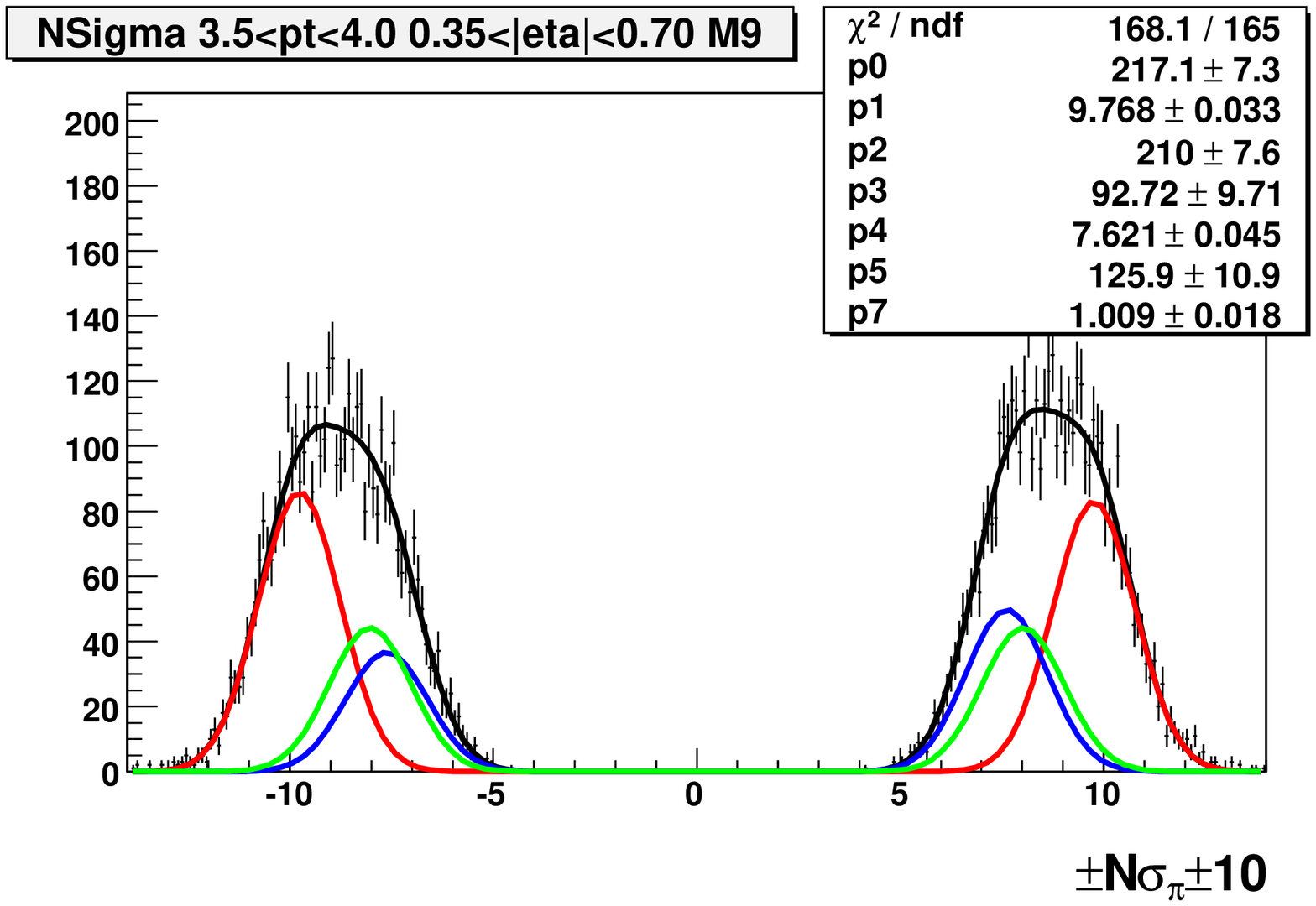}
		
			\end{minipage}						
	\caption{$N\sigma_{\pi}$ distributions with fits in Au+Au collisions at $\sqrt{s_{NN}}=200$ GeV/c.  Curves are from fit with red for pion, blue for proton and green for kaon.  Rows are the centrality bins 70-80\%, 60-70\%, 50-60\%, 40-50\%, 30-40\%, 20-30\%, 10-20\%, 5-10\% and 0-5\% from top to bottom.  Columns are for different $p_{T}$ and $\eta$ cuts.  Left: $|\eta|<0.35$ and $3<p_{T}<3.5$ GeV/c.  Left center:  $0.35<|\eta|<0.75$ and $3<p_{T}<3.5$ GeV/c.  Right center: $|\eta|<0.35$ and $3.5<p_{T}<4$ GeV/c.   Right:  $0.35<|\eta|<0.75$ and $3.5<p_{T}<4$ GeV/c.  Fit parameters are p0=$\pi^{-}$ yield, p1=$\pi$ centroid+10, p2=$\pi^{+}$ yield, p3=$\bar{p}$ yield, p4=$p/\bar{p}$ centroid+10, p5=$p$ yield, and p7=width}
	\label{fig:pidfit}	
\end{figure}

\begin{figure}[H]
\hfill
\begin{minipage}[t]{.23\textwidth}
	\centering
		\includegraphics[width=1\textwidth]{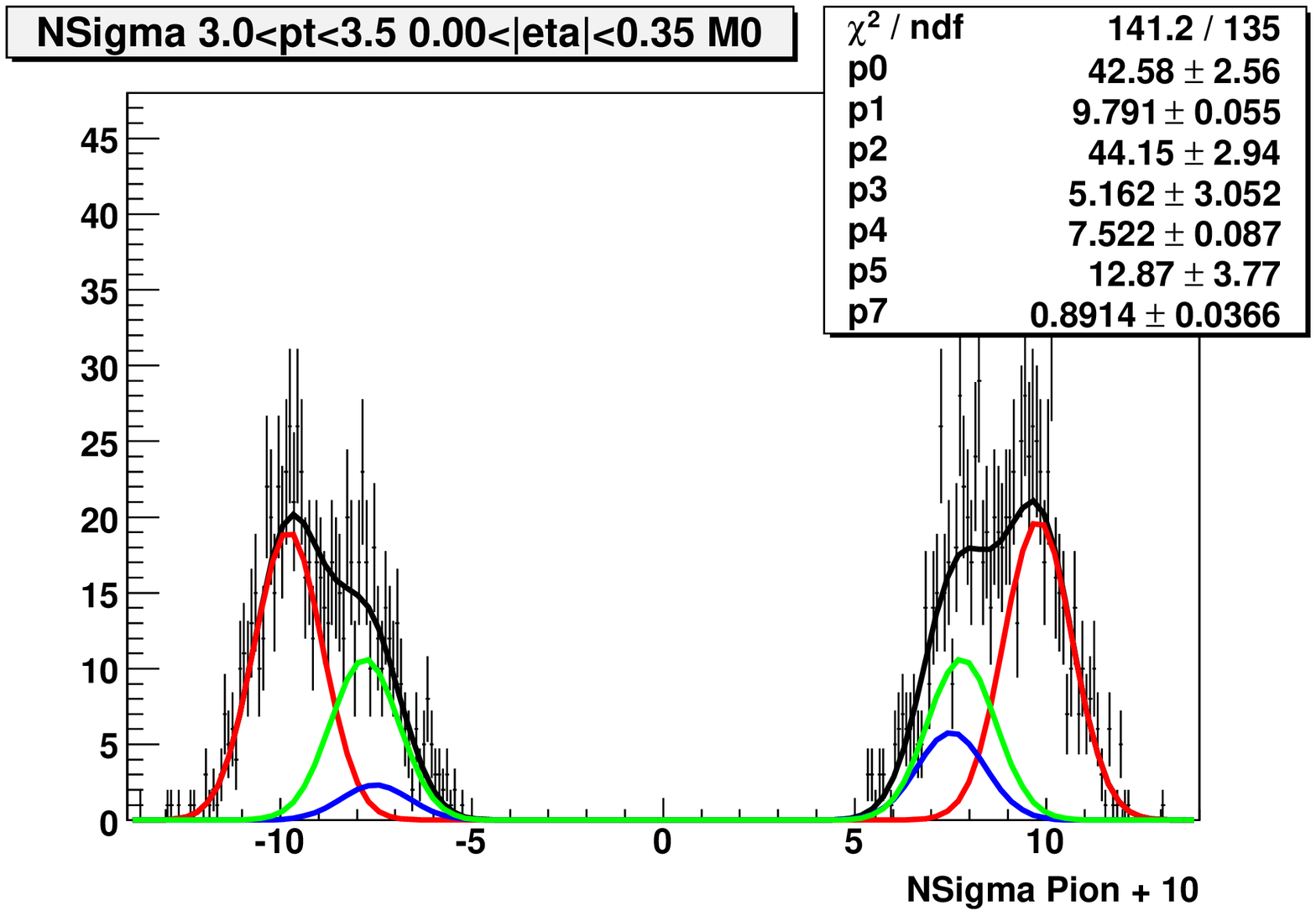}
		\includegraphics[width=1\textwidth]{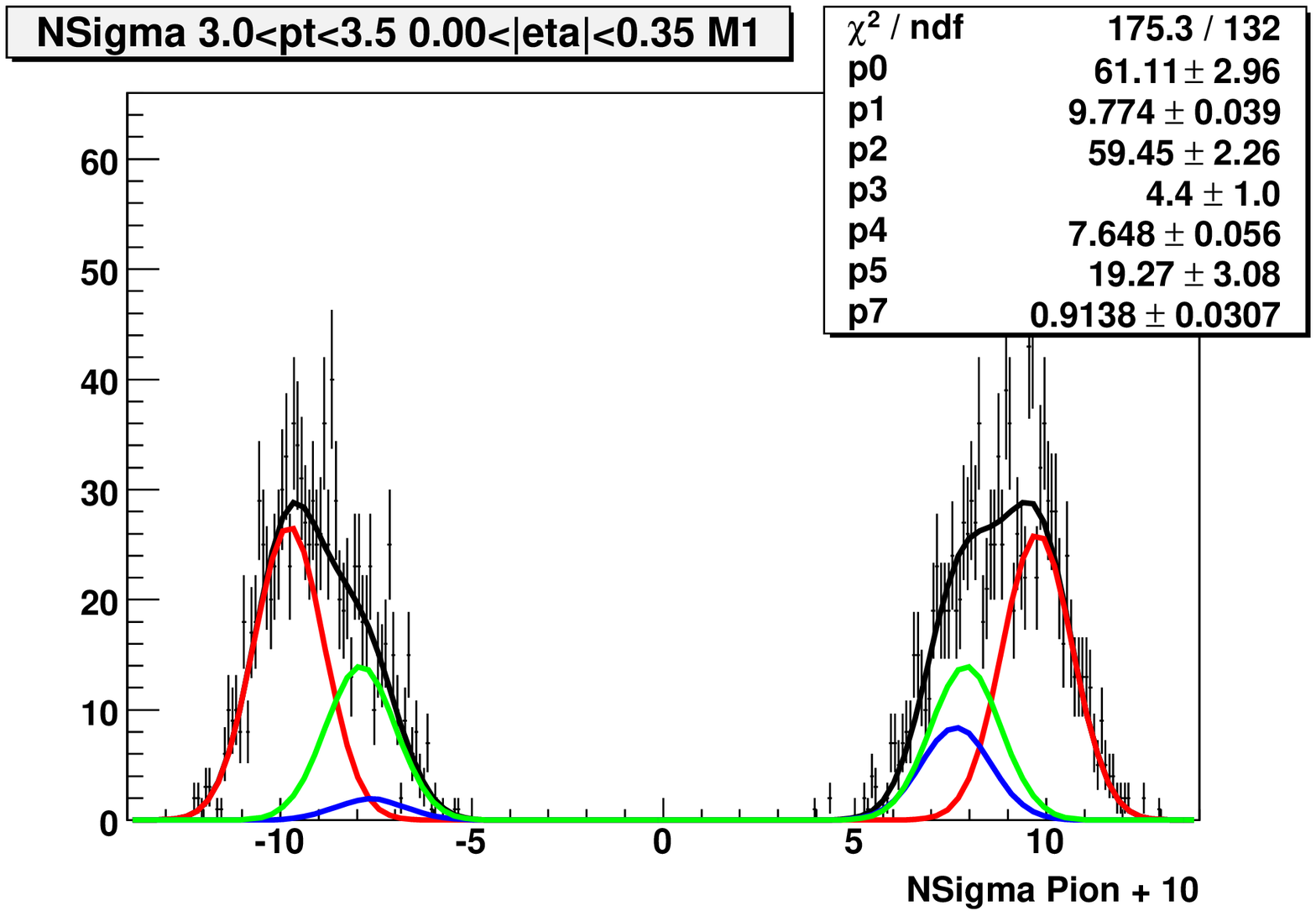}
		\includegraphics[width=1\textwidth]{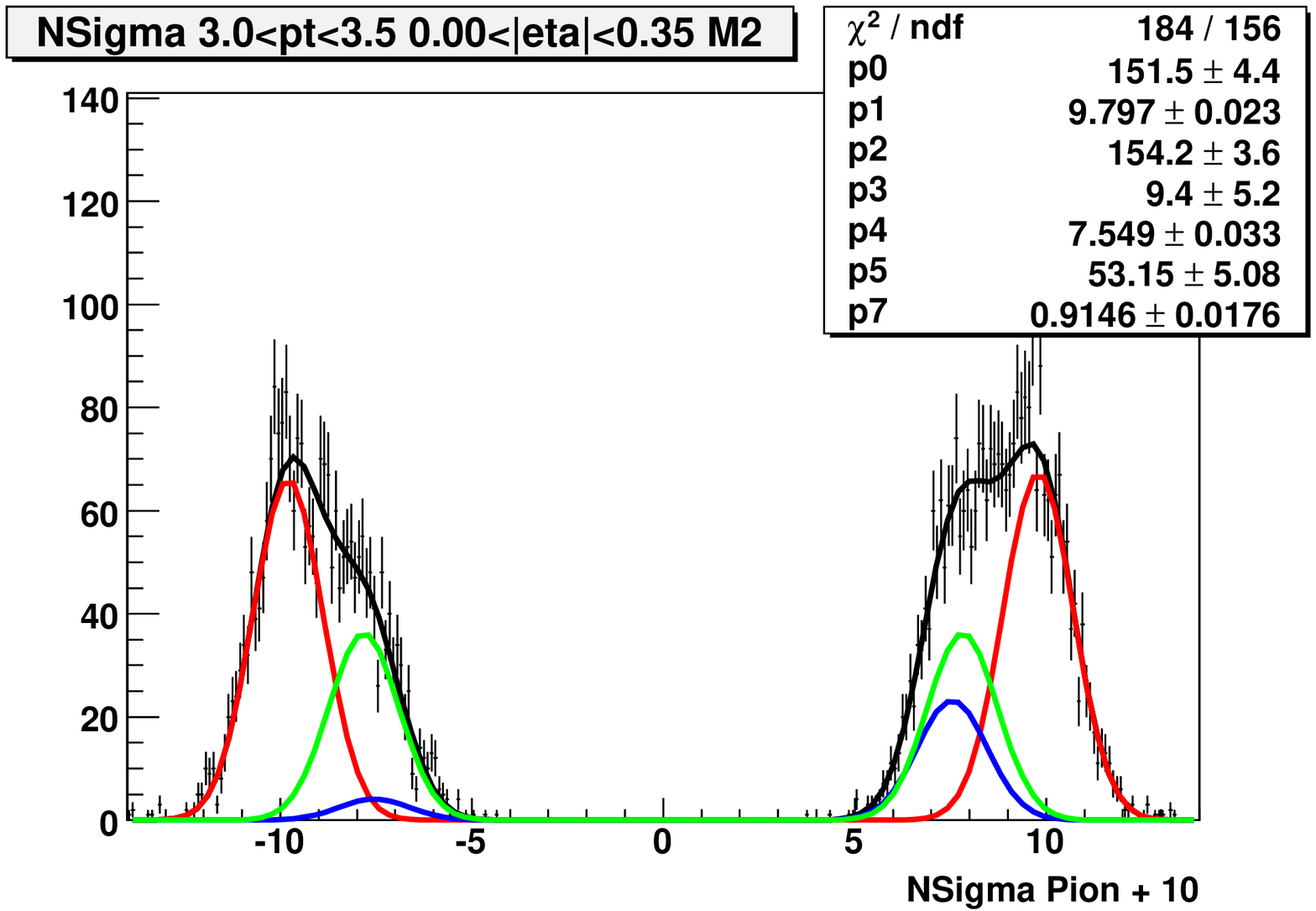}					
			\end{minipage}
\hfill
\begin{minipage}[t]{.23\textwidth}
	\centering
		\includegraphics[width=1\textwidth]{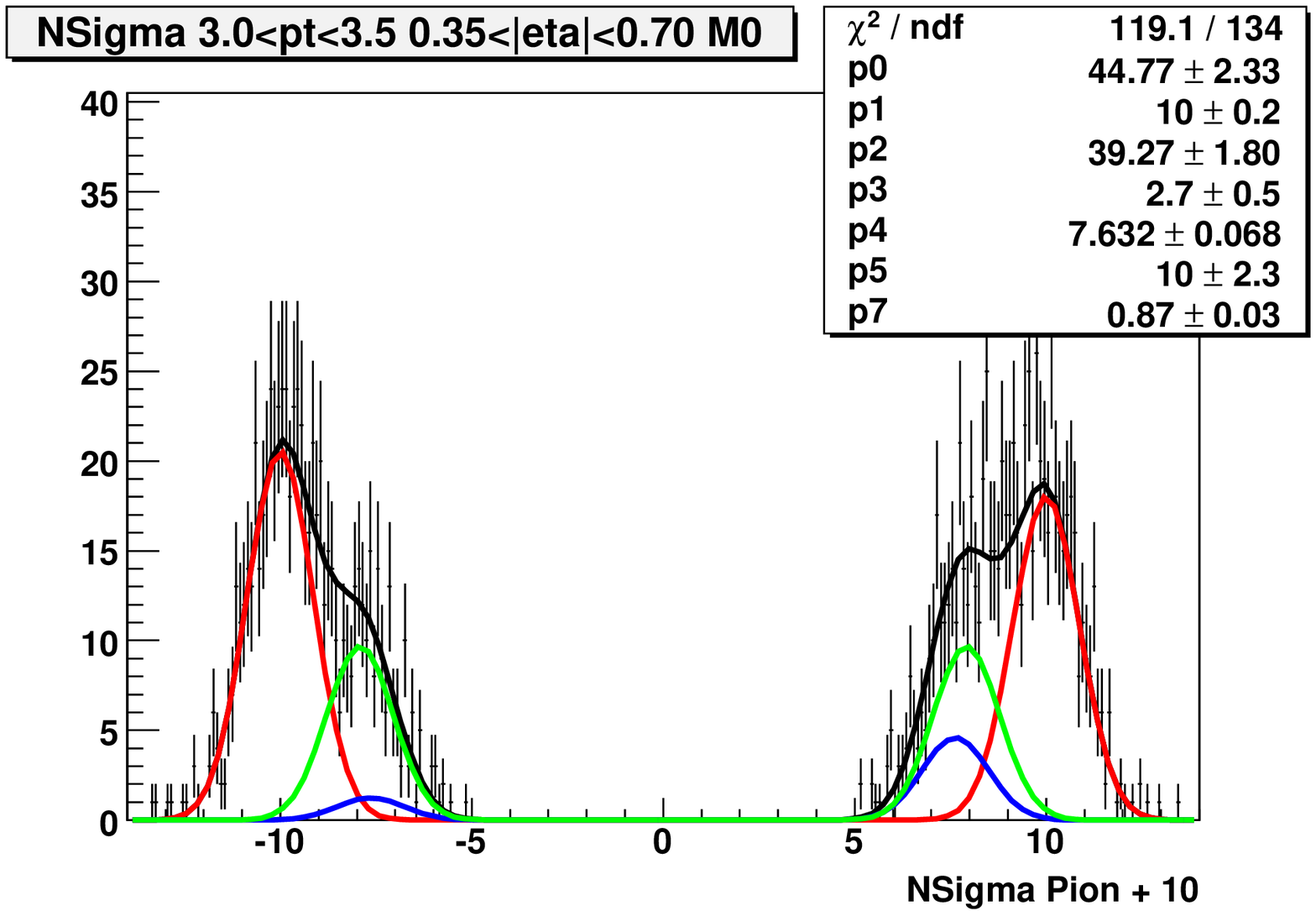}
		\includegraphics[width=1\textwidth]{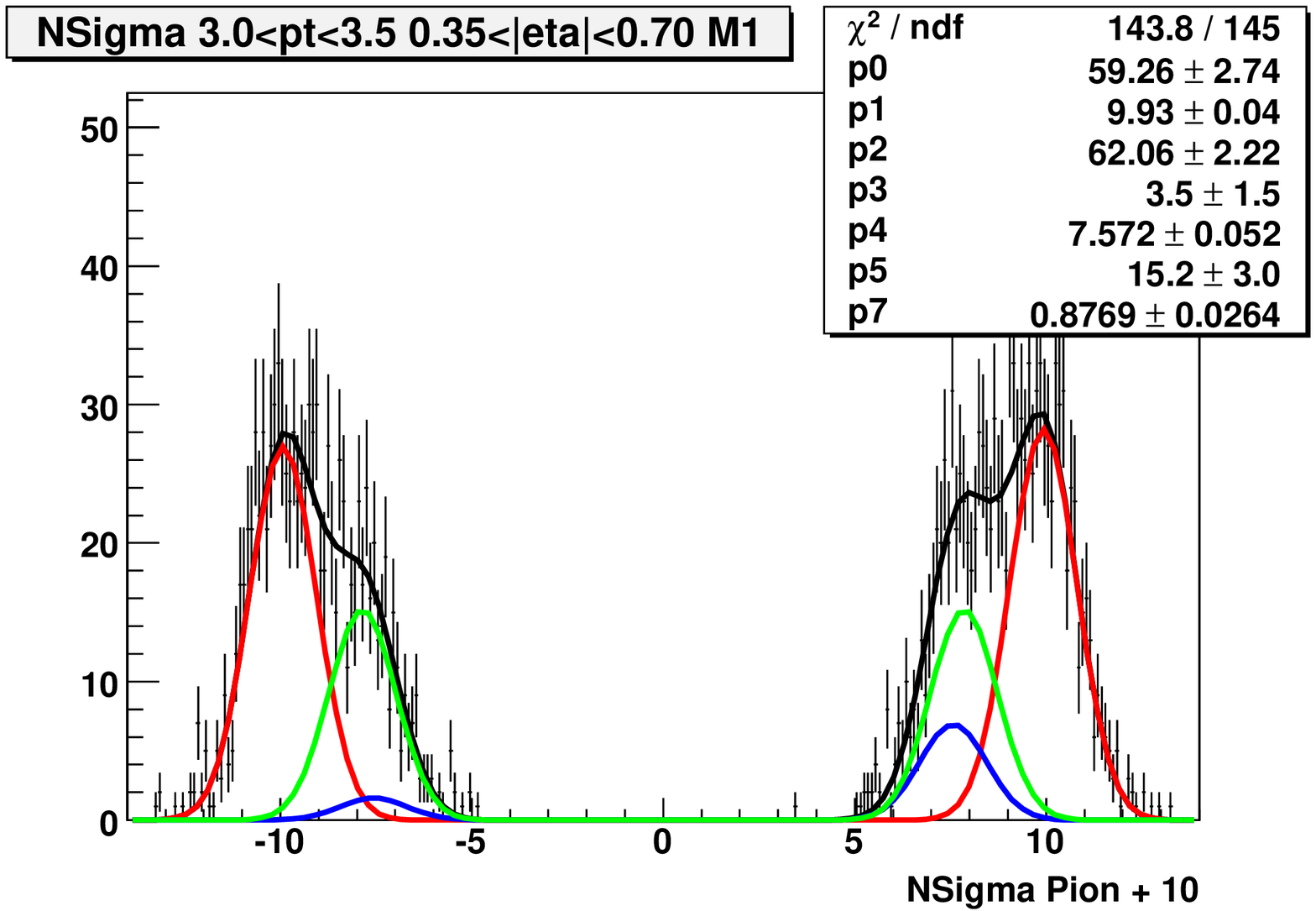}
		\includegraphics[width=1\textwidth]{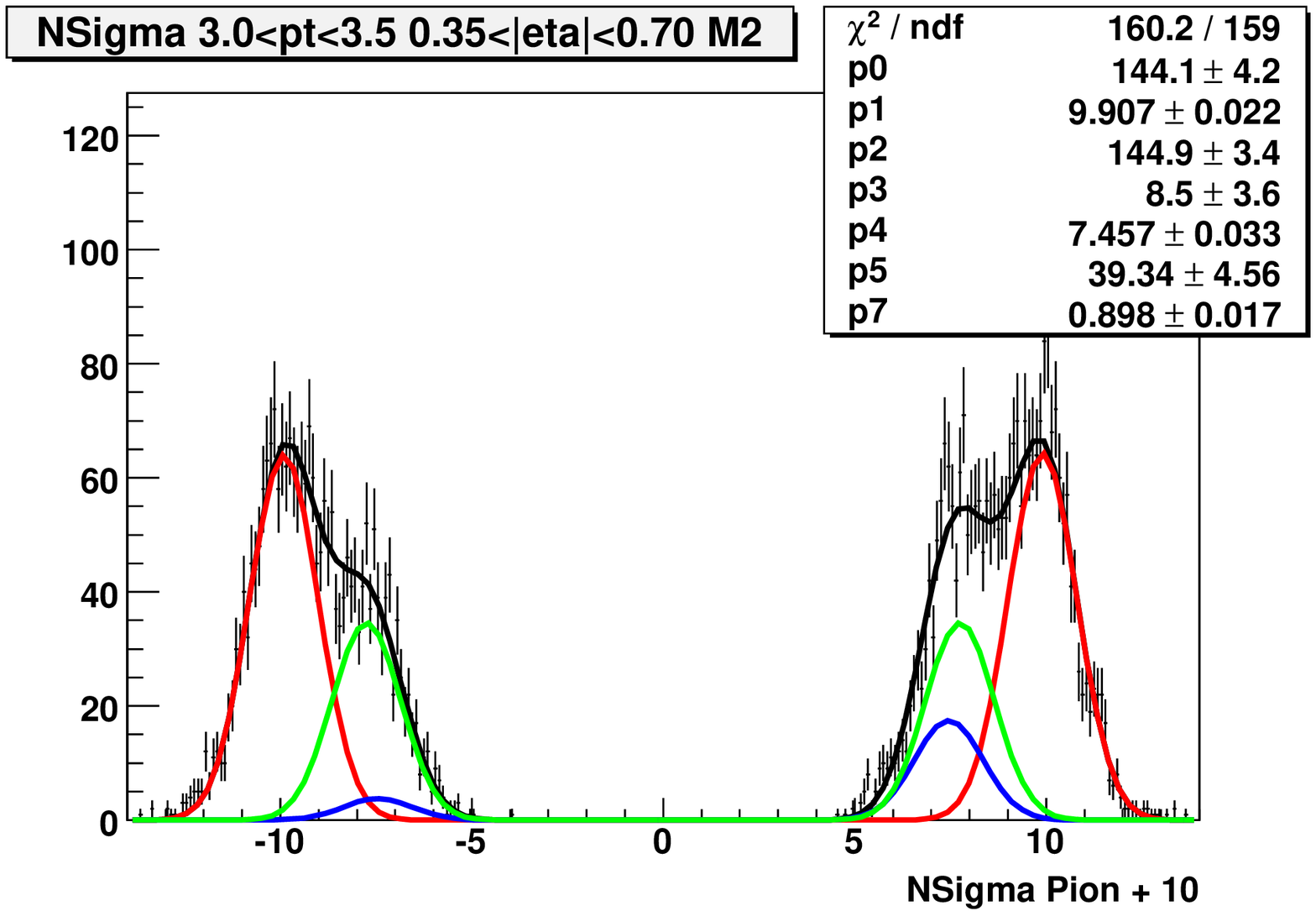}				
			\end{minipage}
\hfill
\begin{minipage}[t]{.23\textwidth}
	\centering
		\includegraphics[width=1\textwidth]{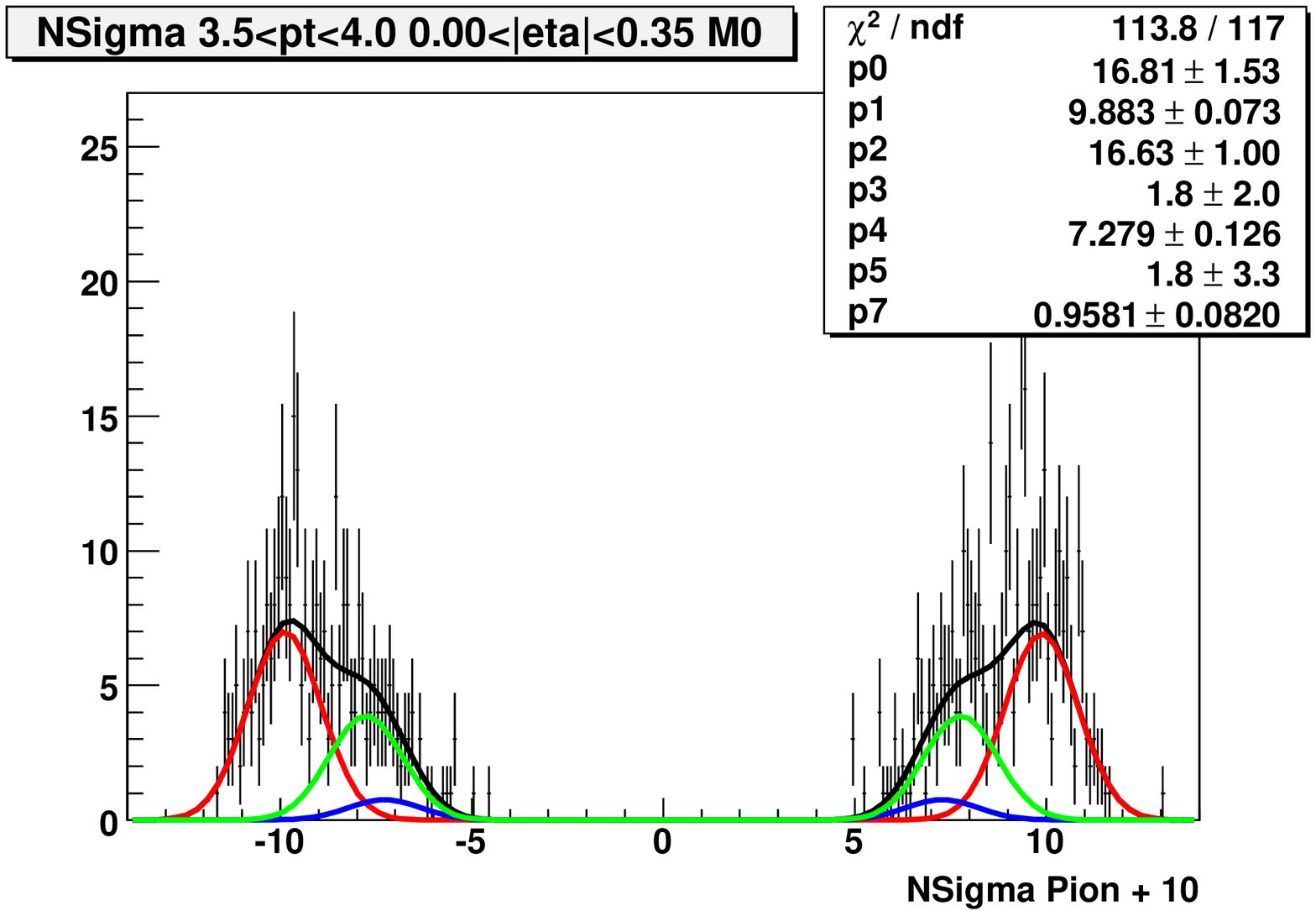}
		\includegraphics[width=1\textwidth]{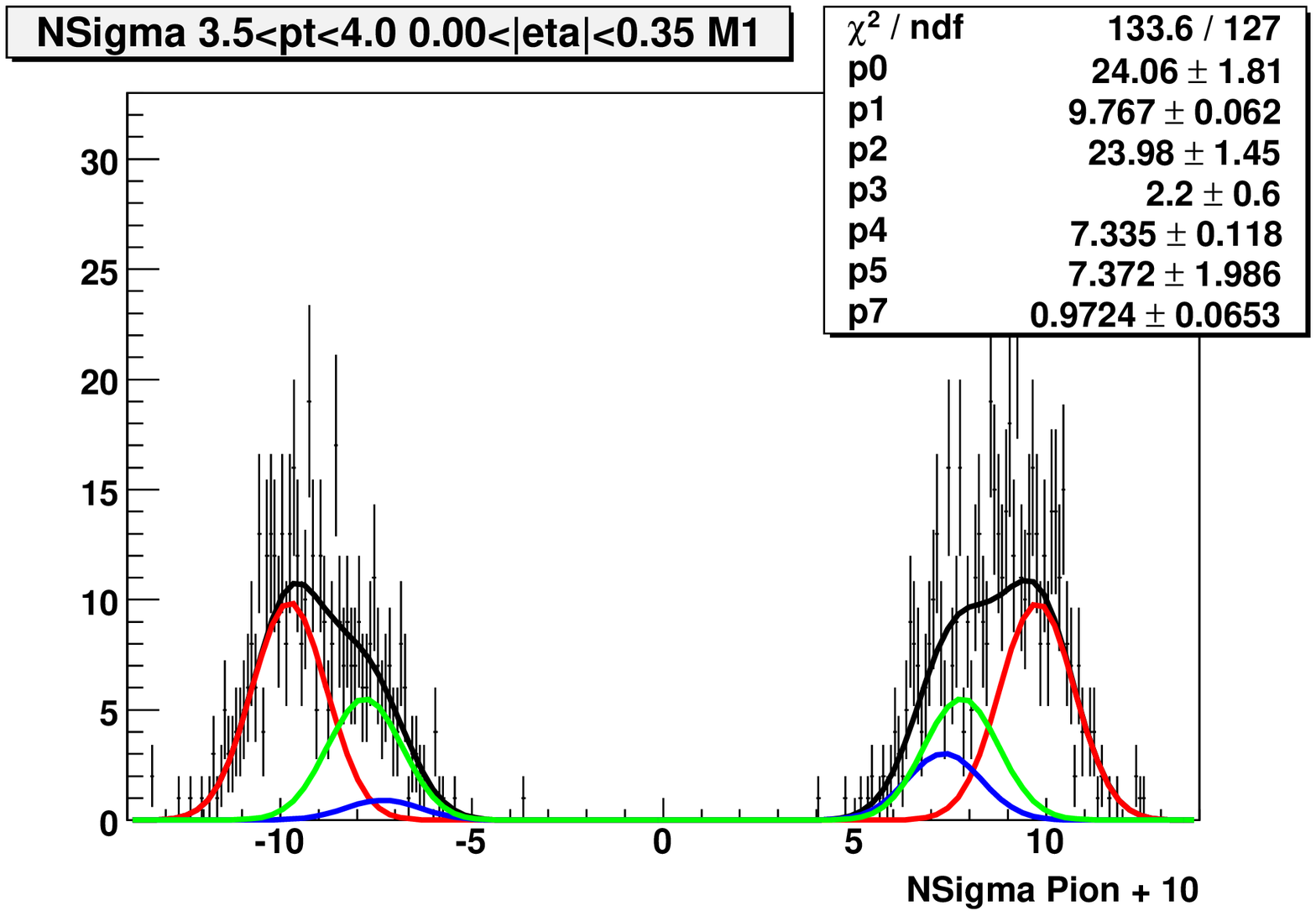}
		\includegraphics[width=1\textwidth]{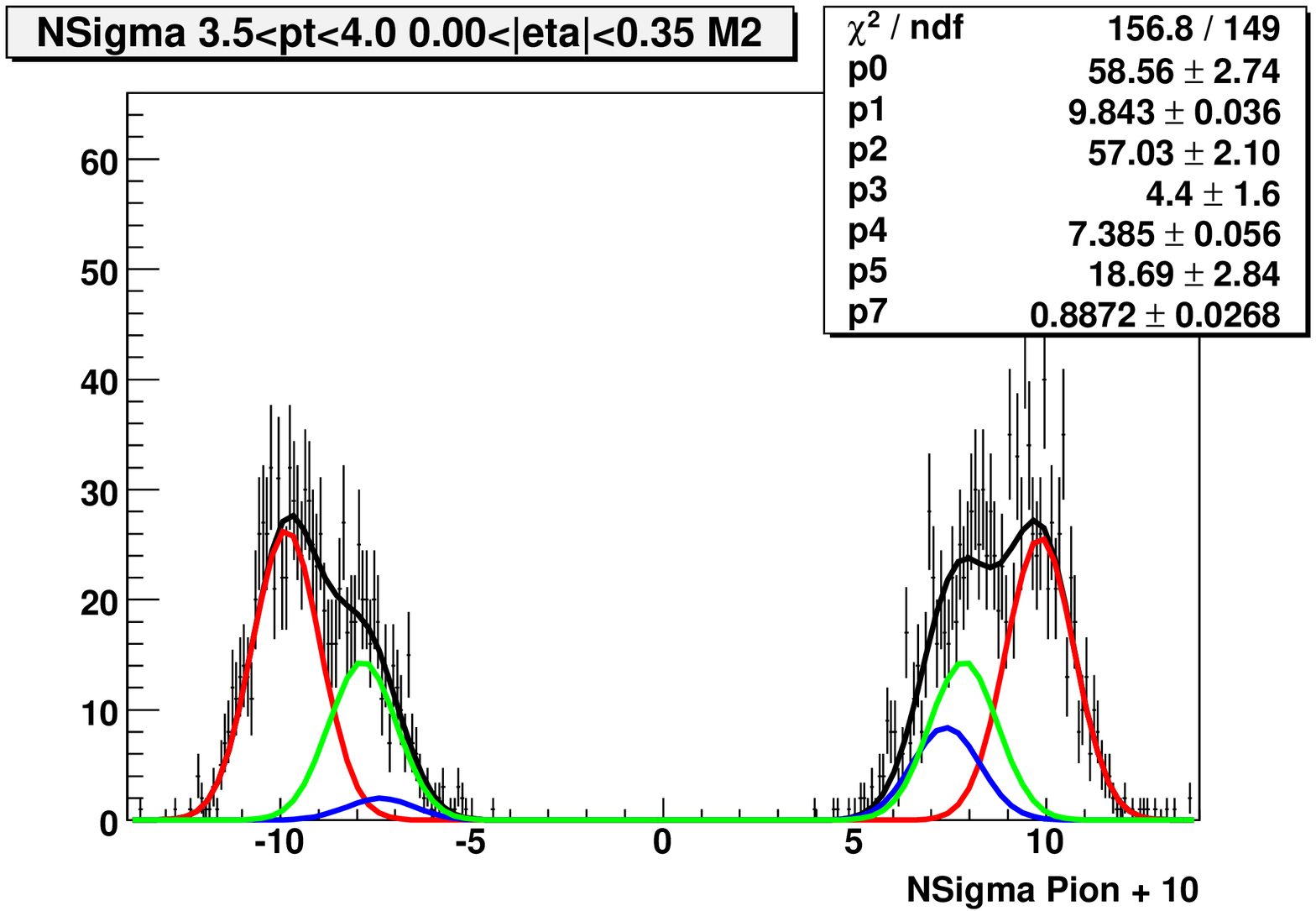}
	
			\end{minipage}
\hfill
\begin{minipage}[t]{.23\textwidth}
	\centering
		\includegraphics[width=1\textwidth]{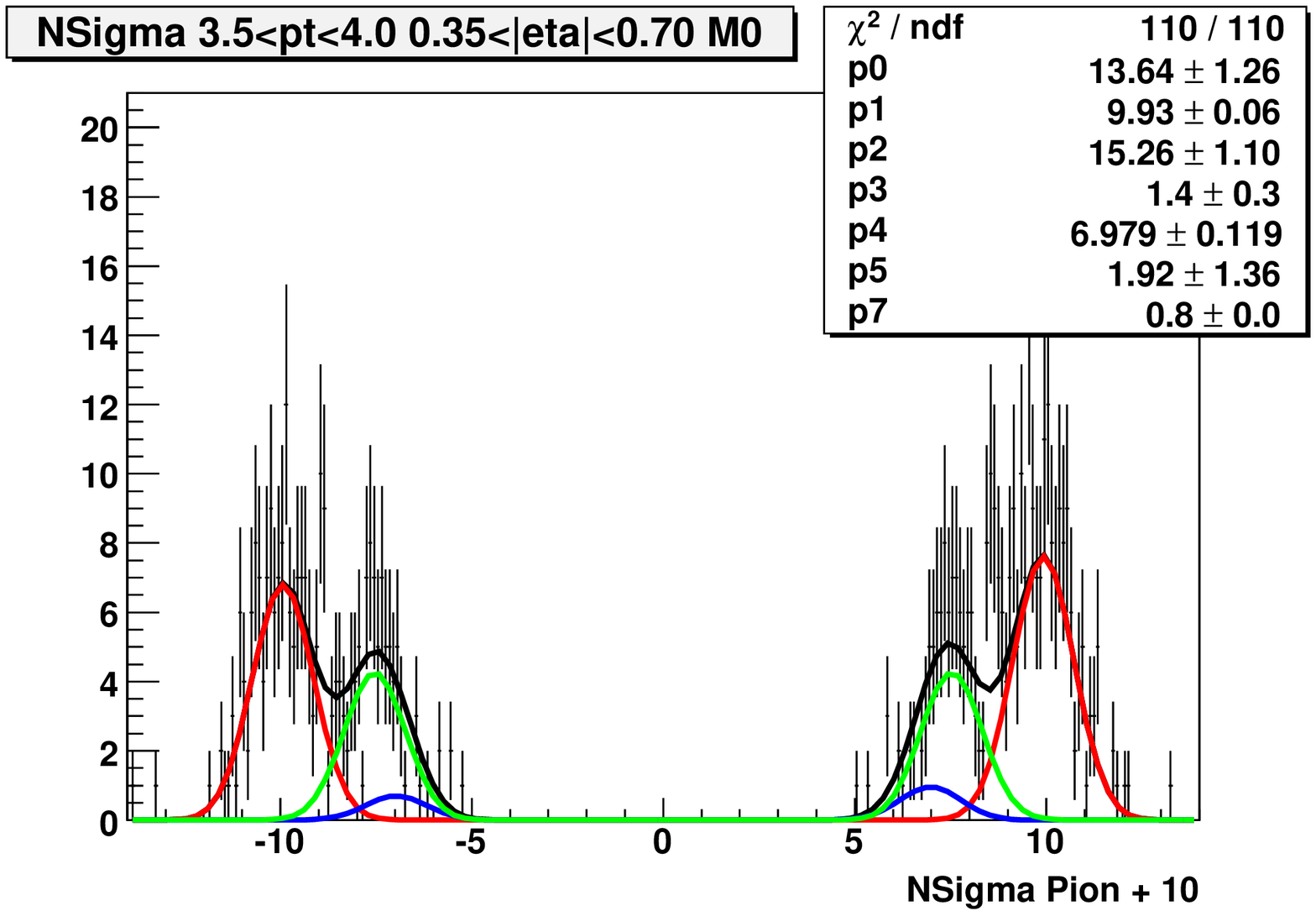}
		\includegraphics[width=1\textwidth]{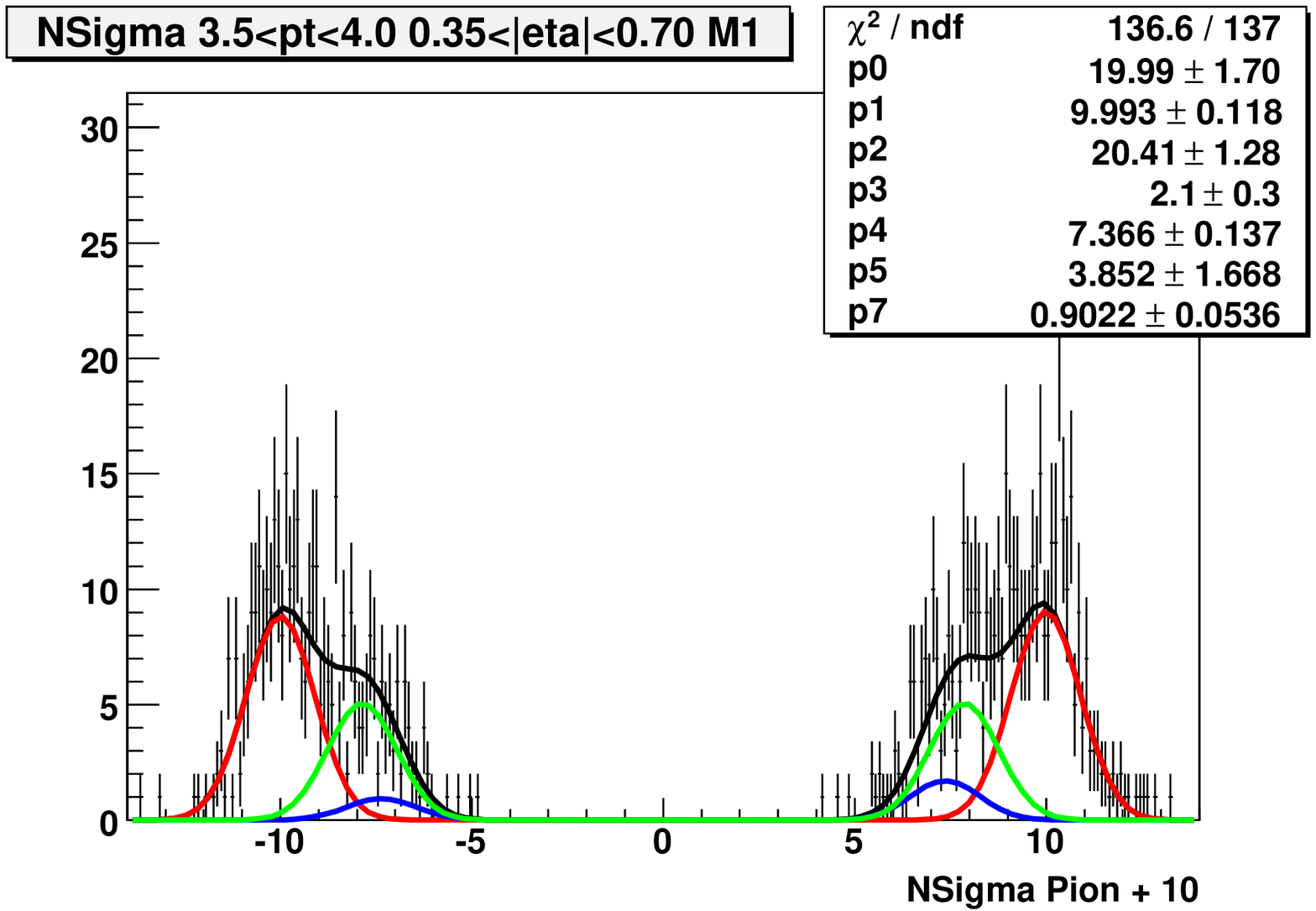}
		\includegraphics[width=1\textwidth]{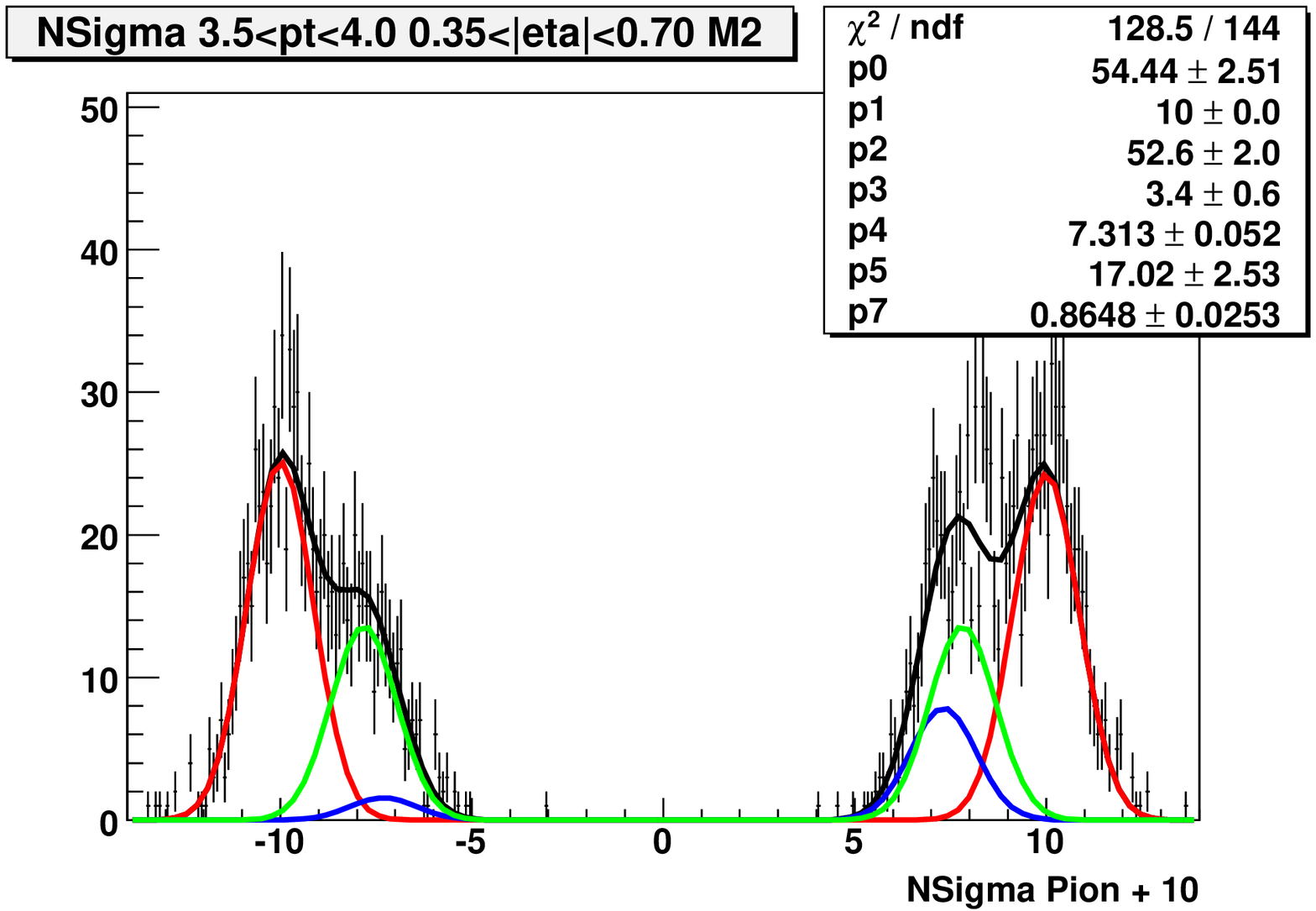}
			\end{minipage}						
	\caption{Same as Fig. 6.2 but for d+Au.  Rows are the centrality bins 100-40\%, 20-40\% and 0-20\% from top to bottom.}
	\label{fig:dAupidfit}	
\end{figure}

Figures~\ref{fig:pidfit} and~\ref{fig:dAupidfit} show $N\sigma_{\pi}$ distributions in Au+Au and d+Au respectively.  The distributions are fit to 6 Gaussians, one each for $p$, $\bar{p}$, $\pi^+$, $\pi^-$, $K^+$, $K^-$.  The yields for $\pi^-$, $\pi^+$, $\bar{p}$, and $p$ are p0, p2, p3, p5, respectively.  The positive and negative particles share the same centroid.  The centroids are given by p1 minus 10 and p4 minus 10 for $\pi^{\pm}$ and $p/\bar{p}$, respectively.  All peaks share a common width given by p7.  The $K^{\pm}$ yields are fixed by the measured $K_s^0$ to $\pi$ ratio.  The $K^{\pm}$ are fixed to $2/3$ of the distance between the pion and proton centroid (closer to the proton's centroid) which was determined from the Bethe-Bloch formula.

\begin{figure}[H]
\hfill
\begin{minipage}[t]{.19\textwidth}
	\centering
		\includegraphics[width=1\textwidth]{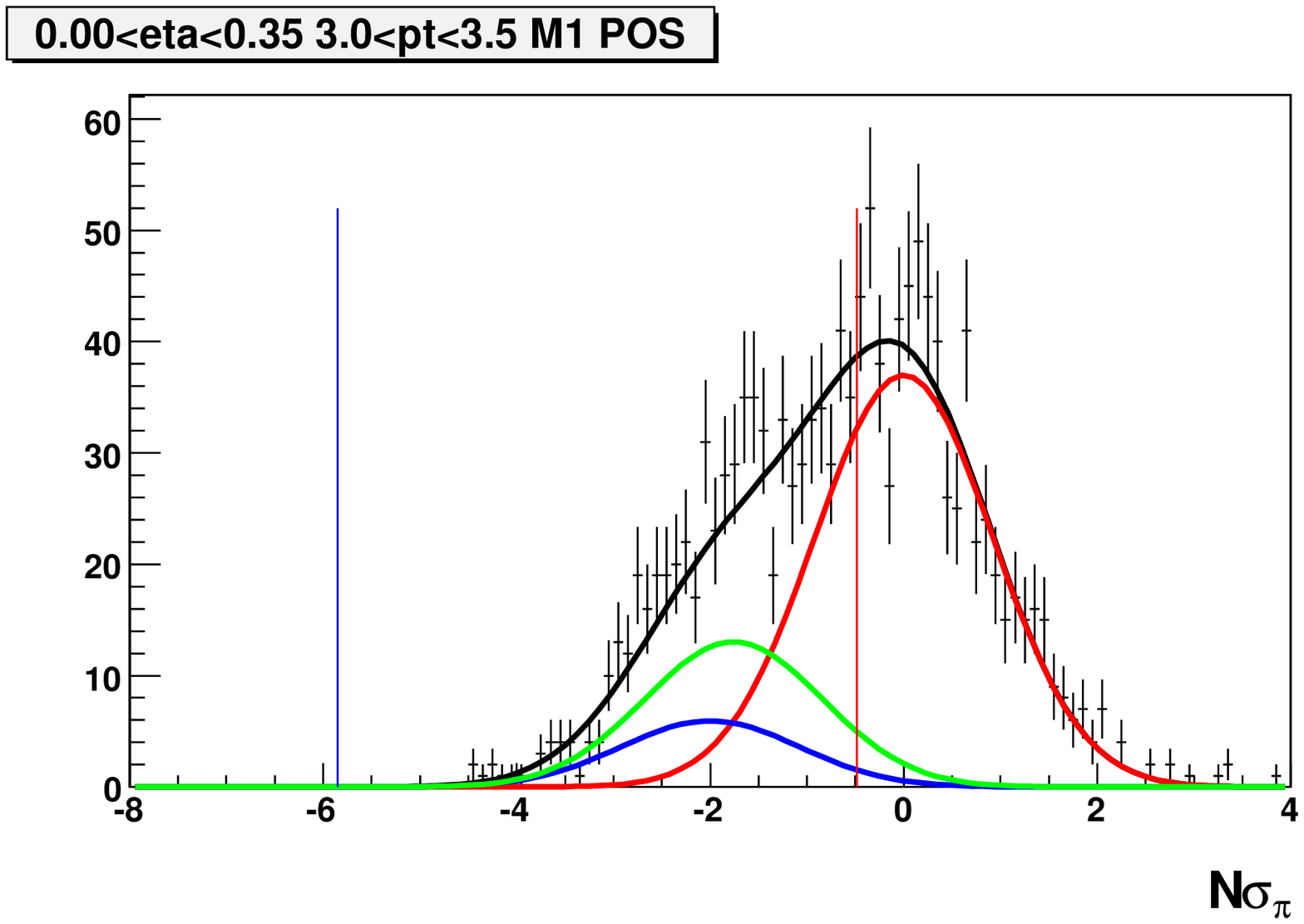}
		\includegraphics[width=1\textwidth]{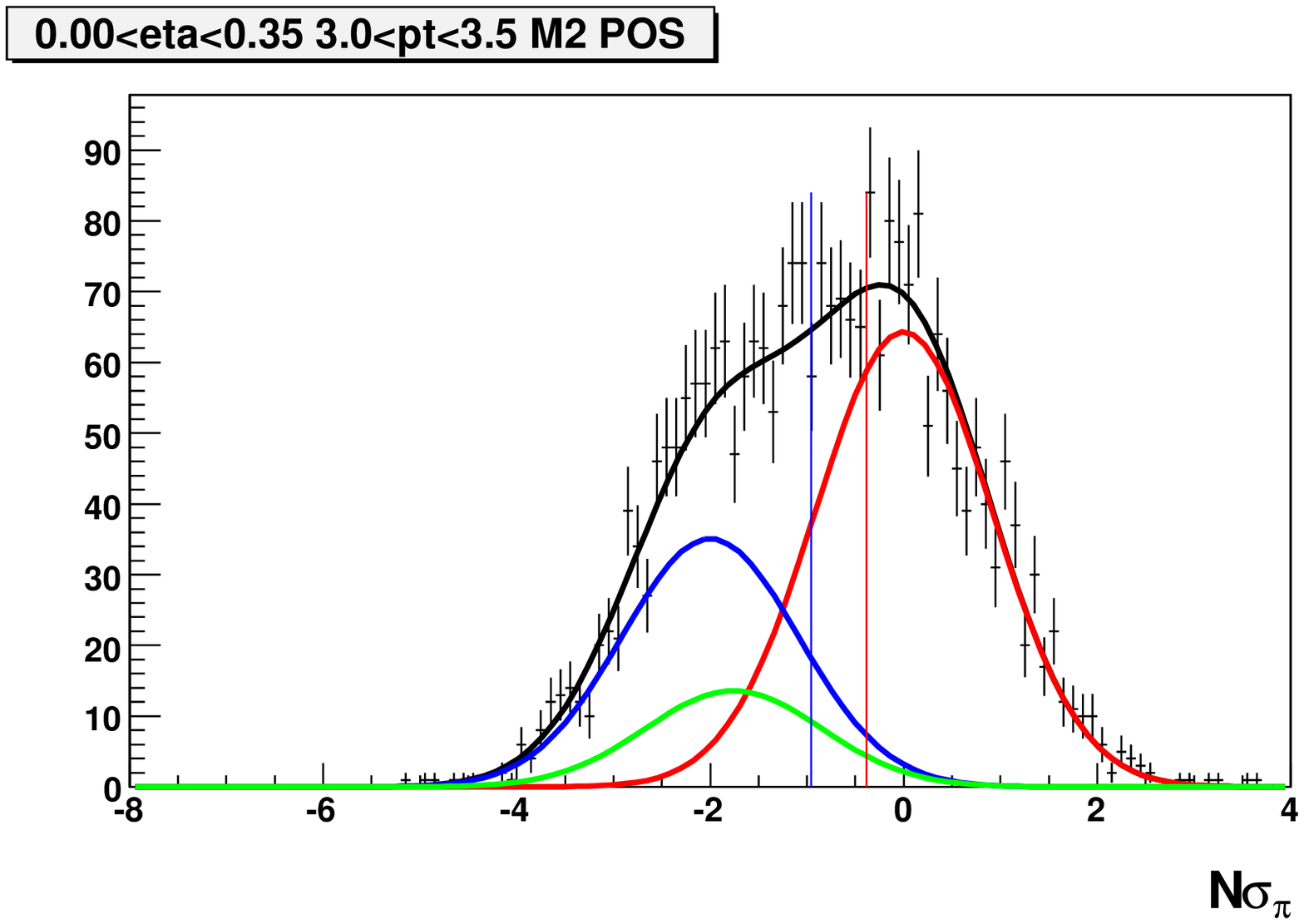}
		\includegraphics[width=1\textwidth]{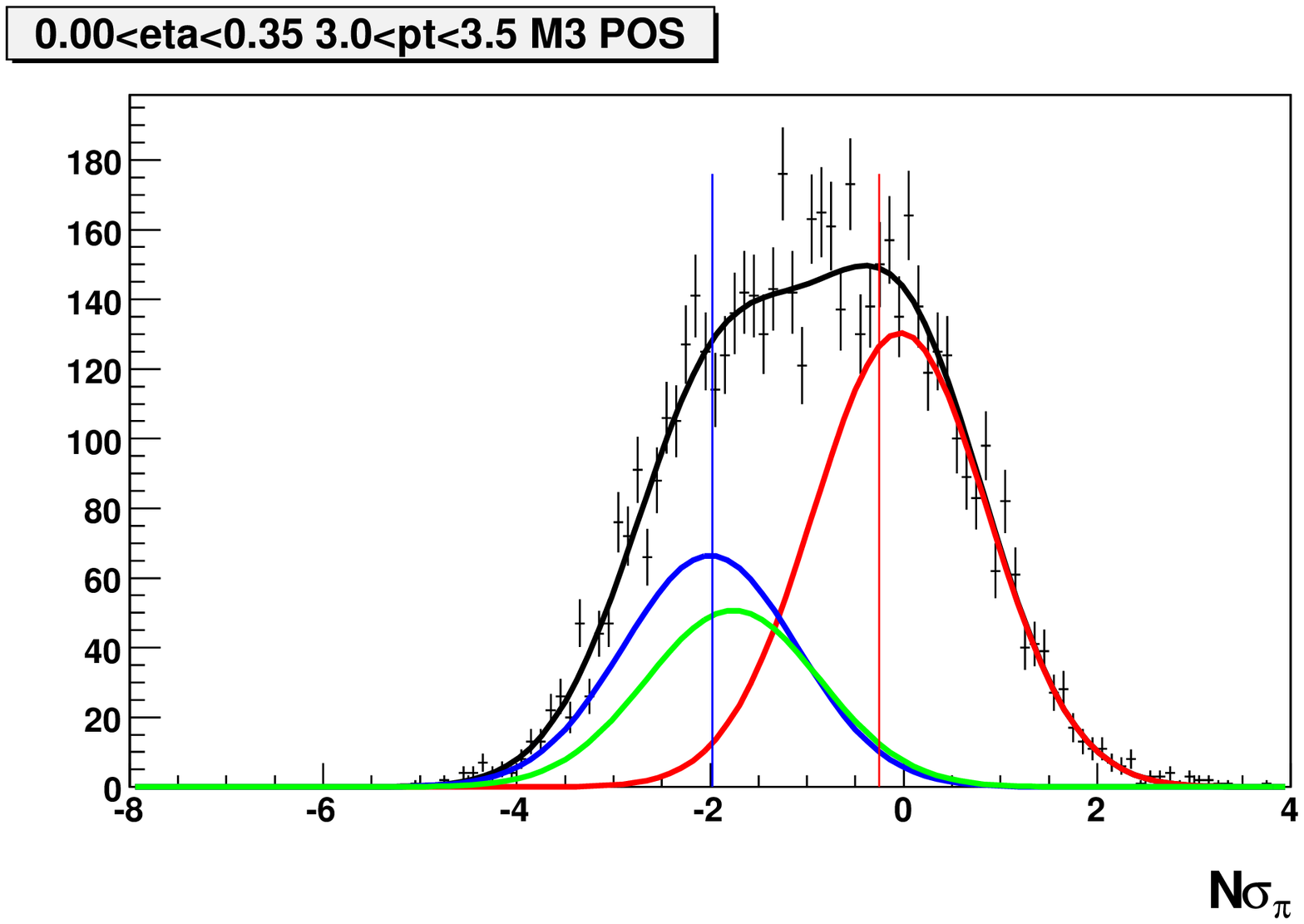}
		\includegraphics[width=1\textwidth]{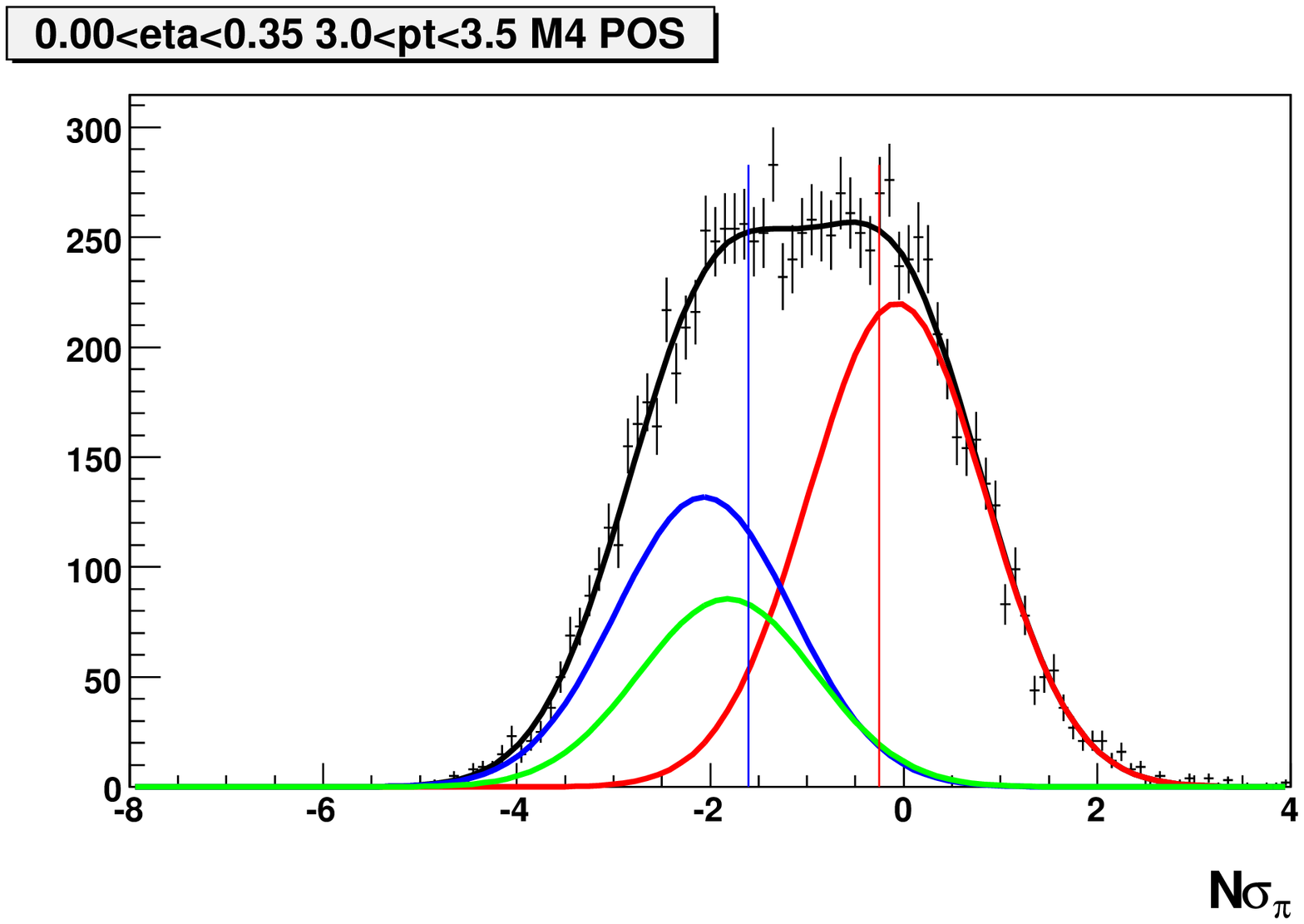}
		\includegraphics[width=1\textwidth]{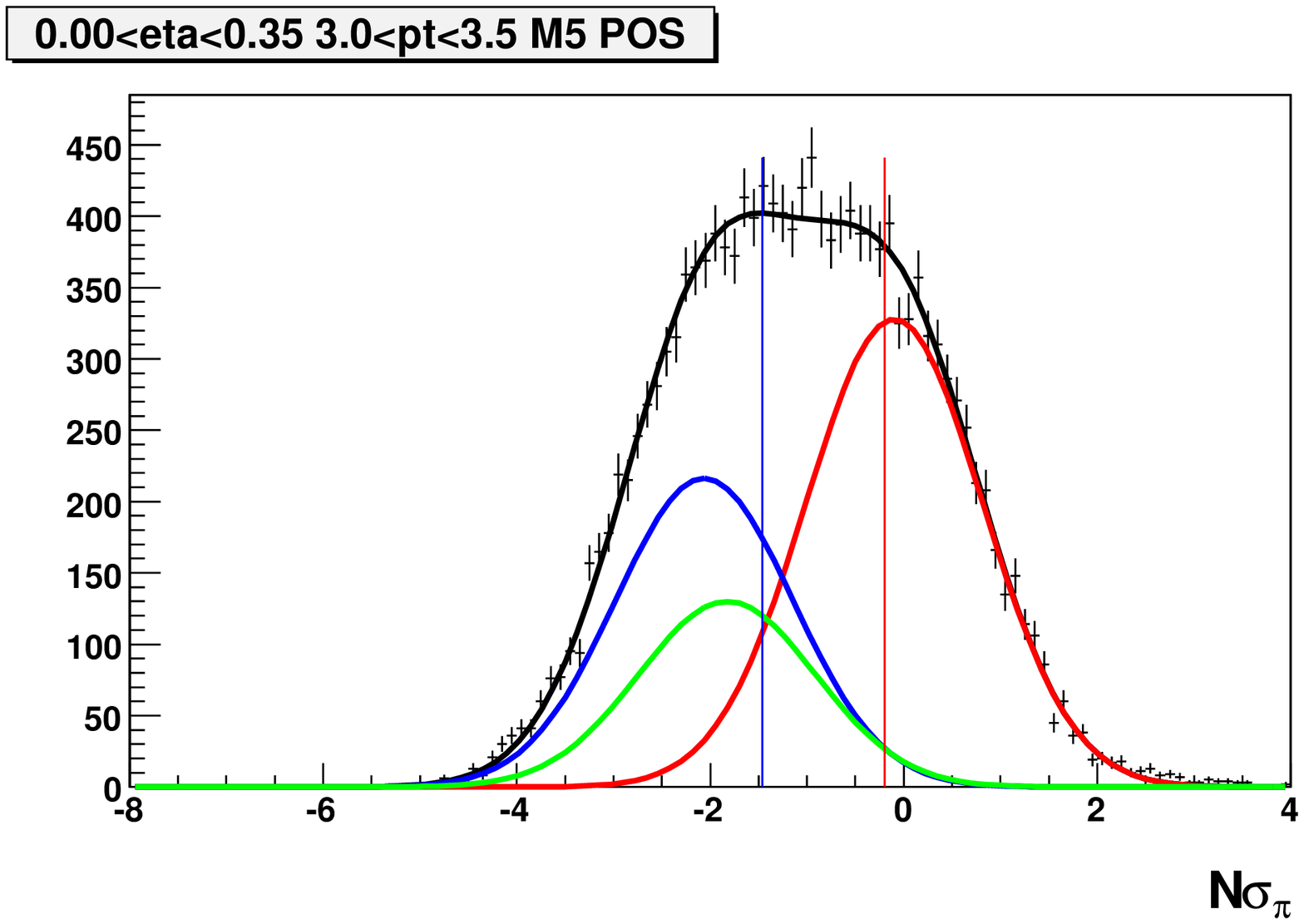}
		\includegraphics[width=1\textwidth]{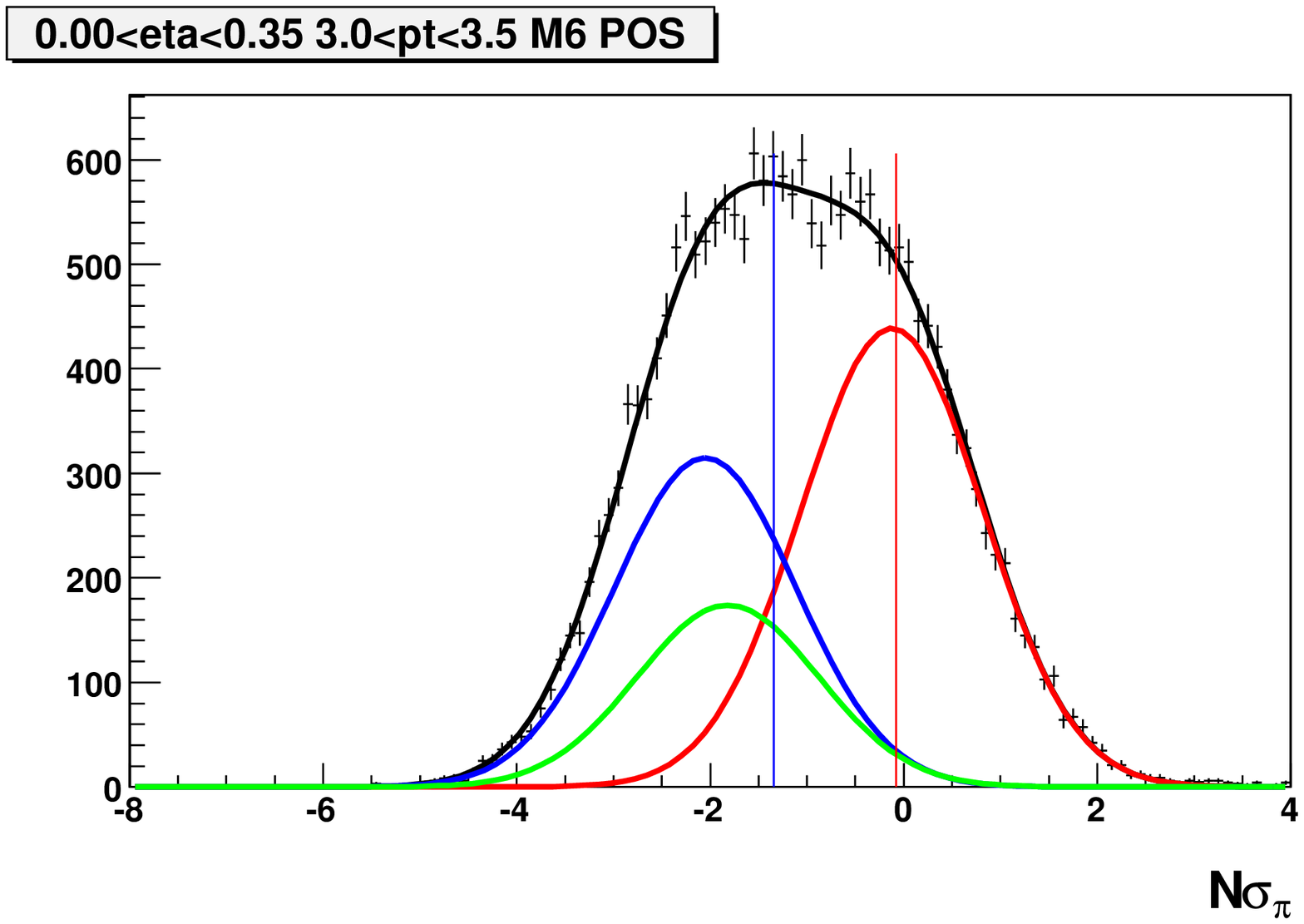}
		\includegraphics[width=1\textwidth]{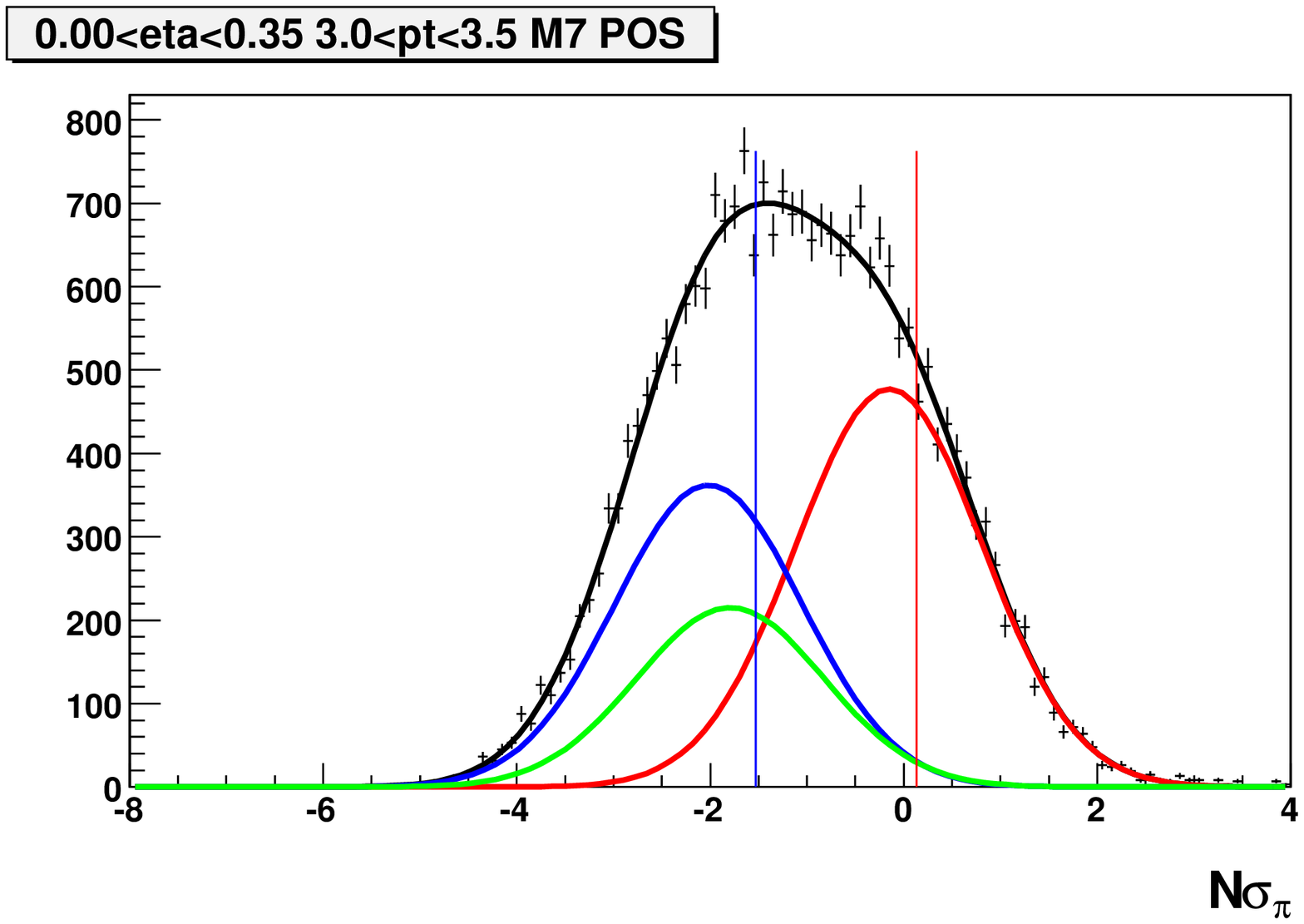}
		\includegraphics[width=1\textwidth]{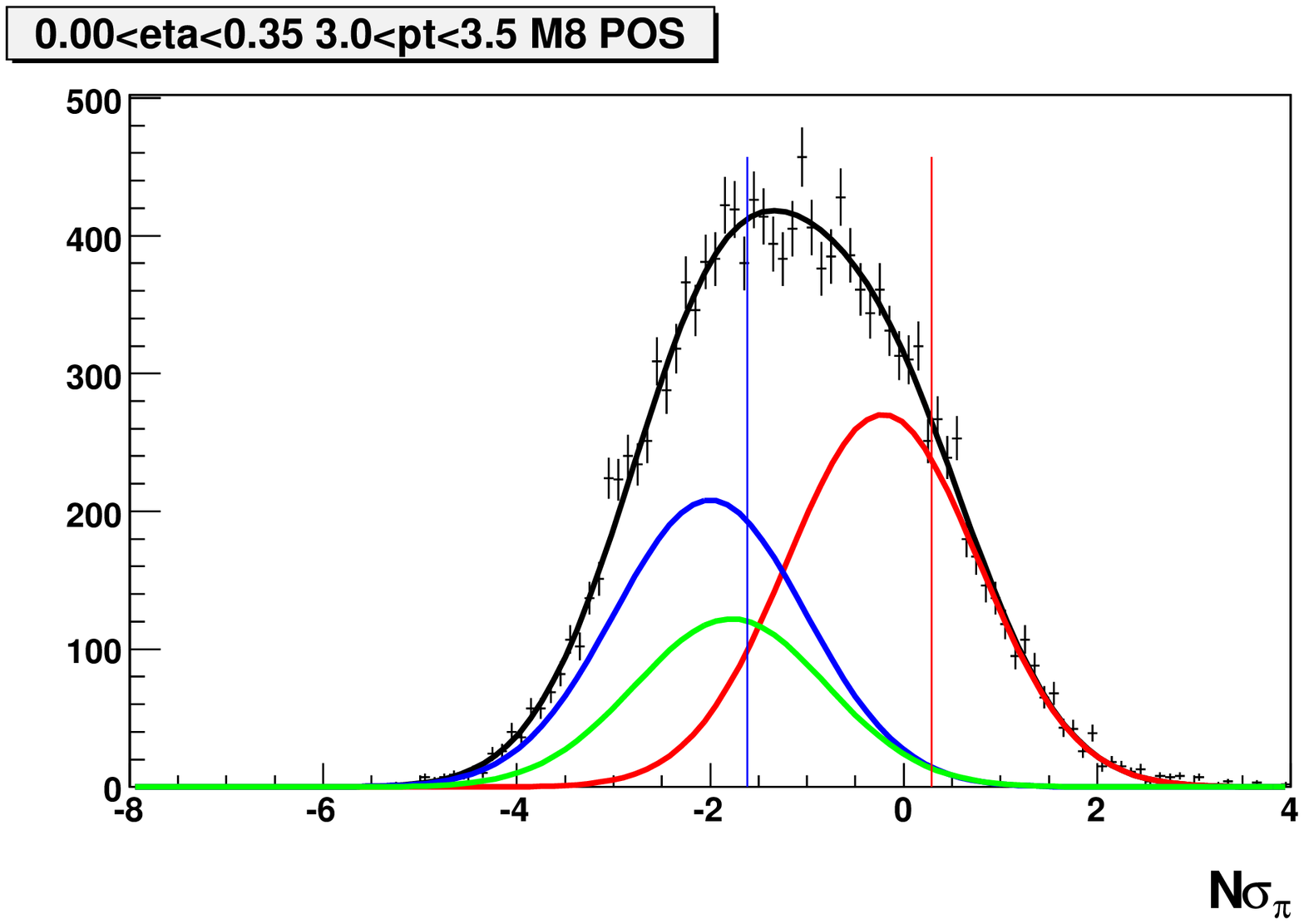}
		\includegraphics[width=1\textwidth]{Plots/cutP_009.eps}									
			\end{minipage}
\hfill
\begin{minipage}[t]{.19\textwidth}
	\centering
		\includegraphics[width=1\textwidth]{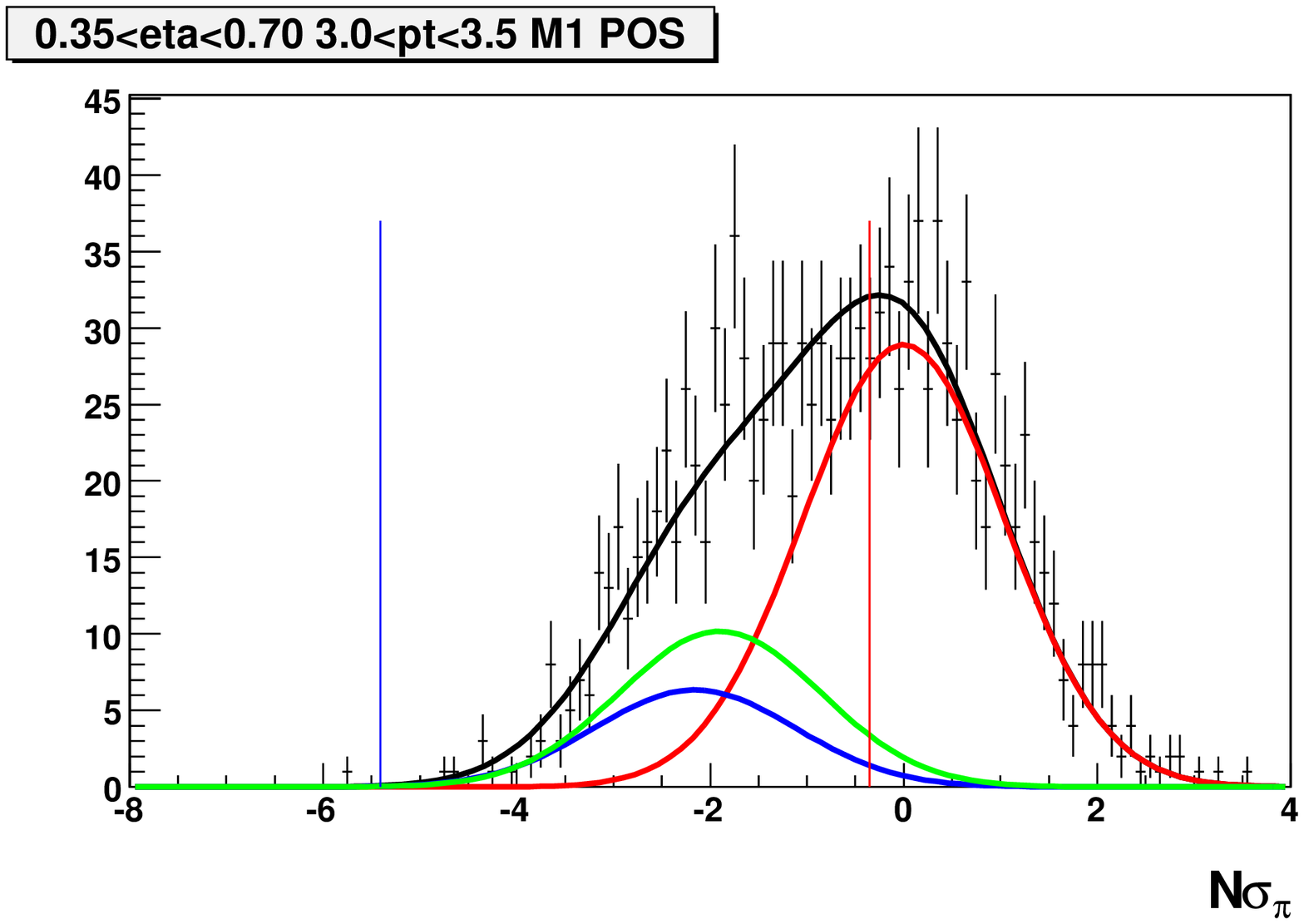}
		\includegraphics[width=1\textwidth]{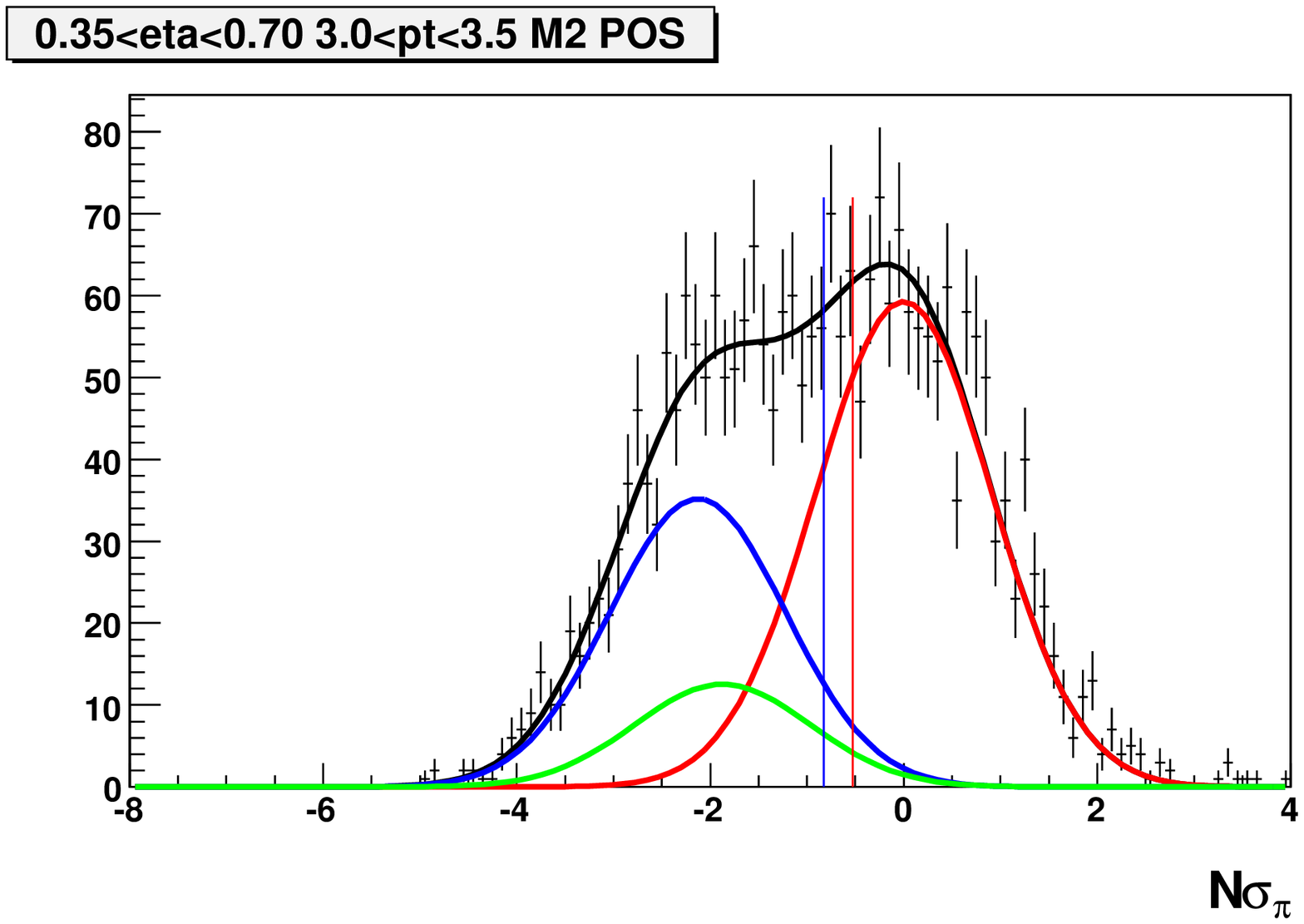}
		\includegraphics[width=1\textwidth]{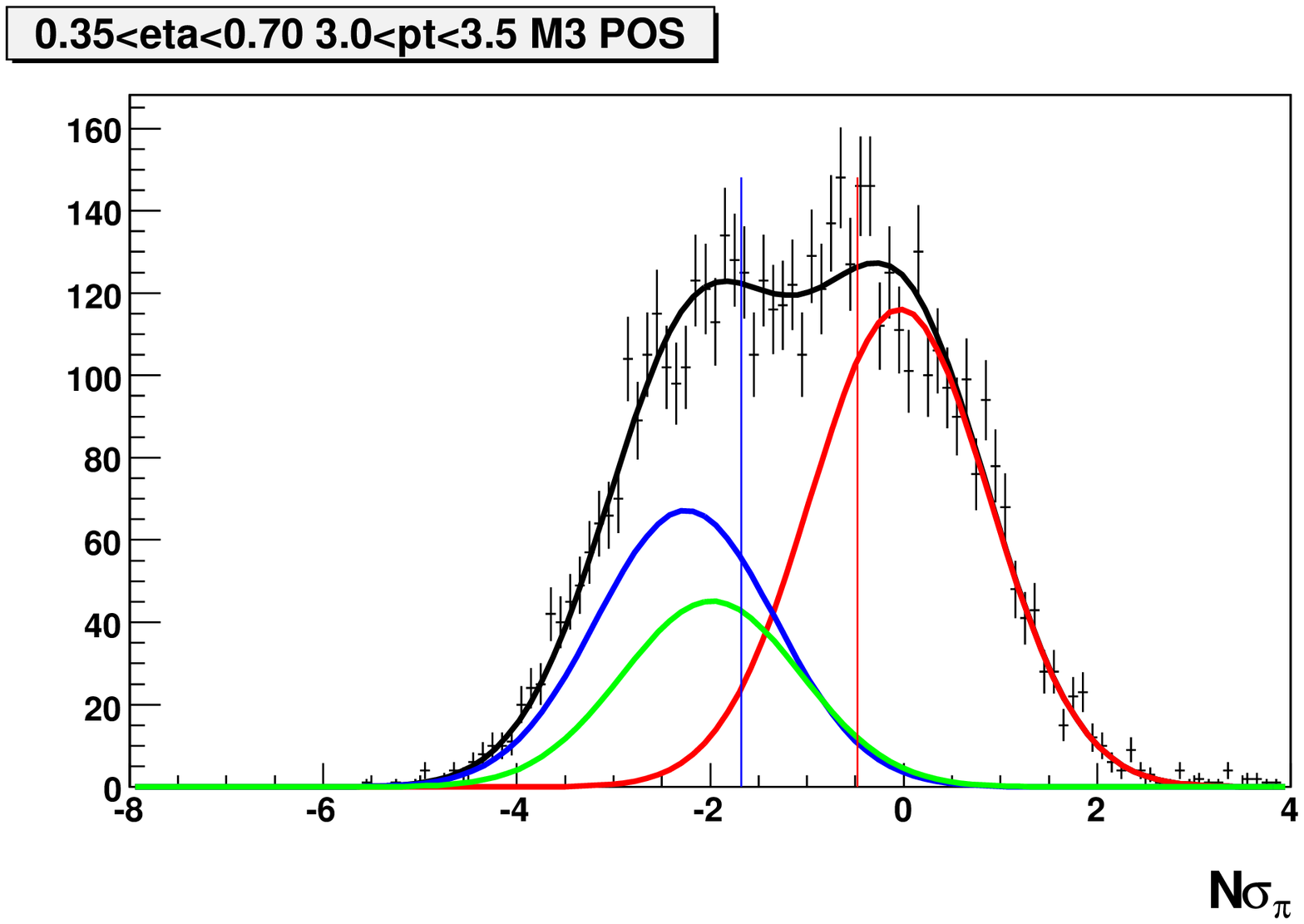}
		\includegraphics[width=1\textwidth]{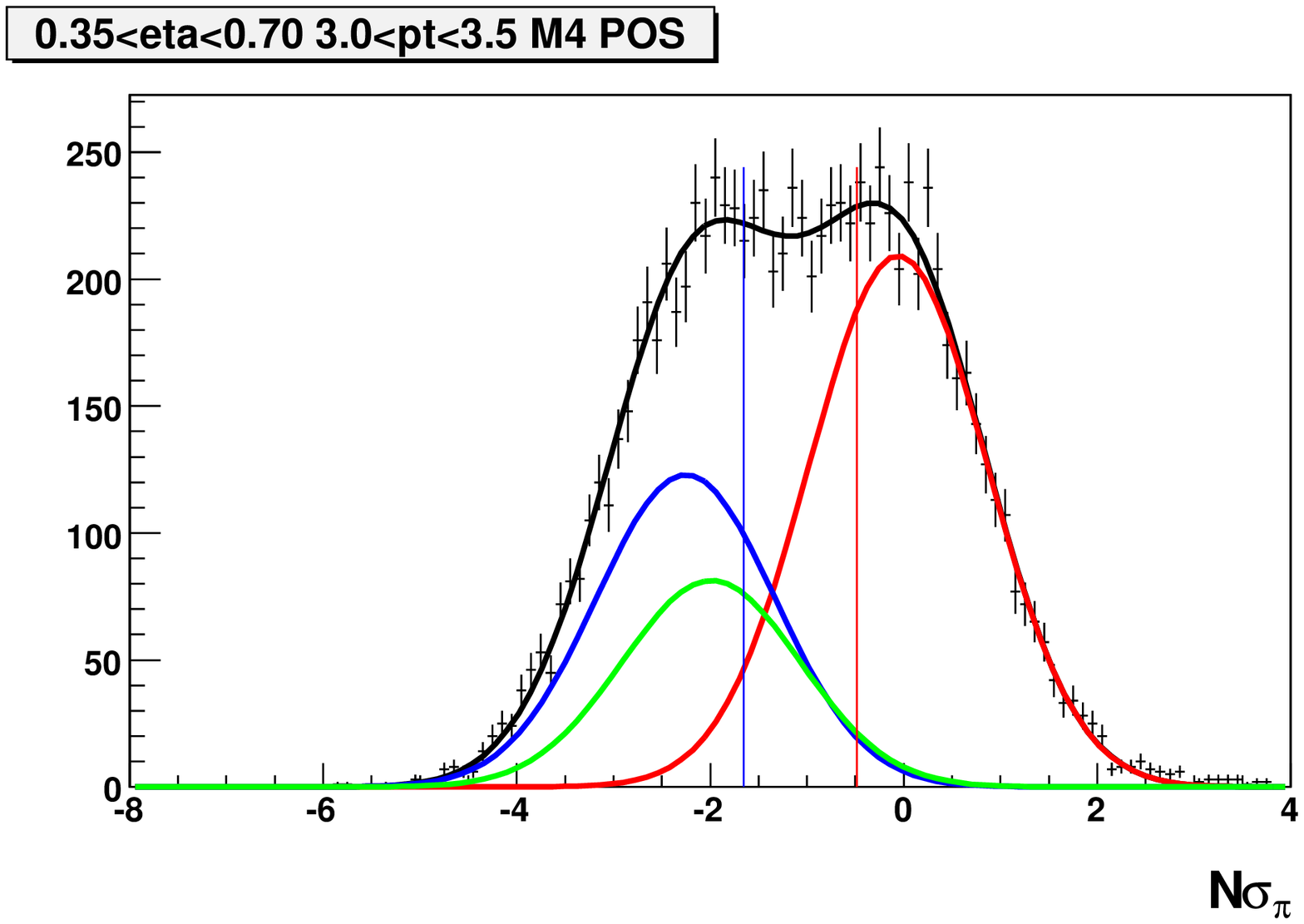}
		\includegraphics[width=1\textwidth]{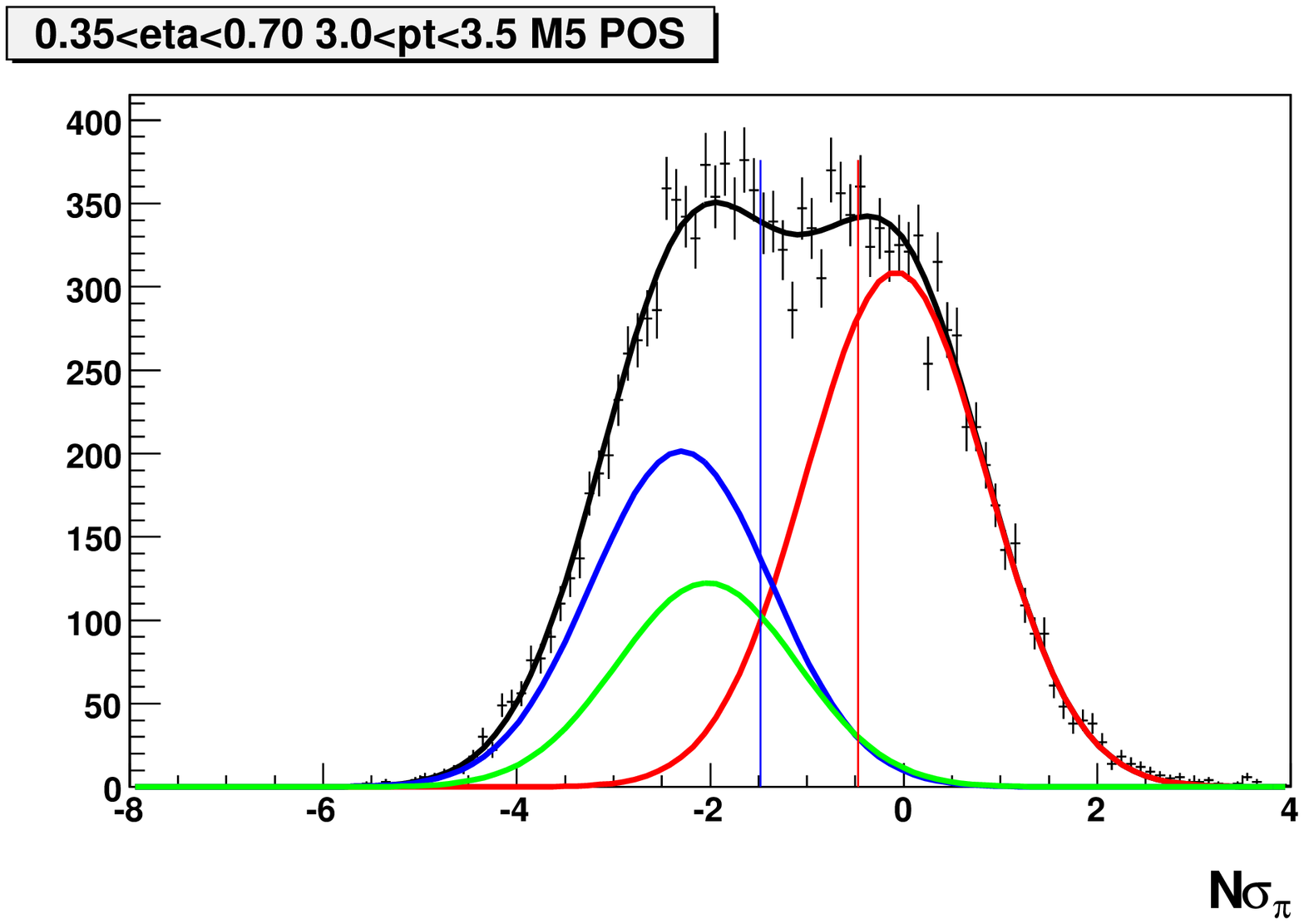}
		\includegraphics[width=1\textwidth]{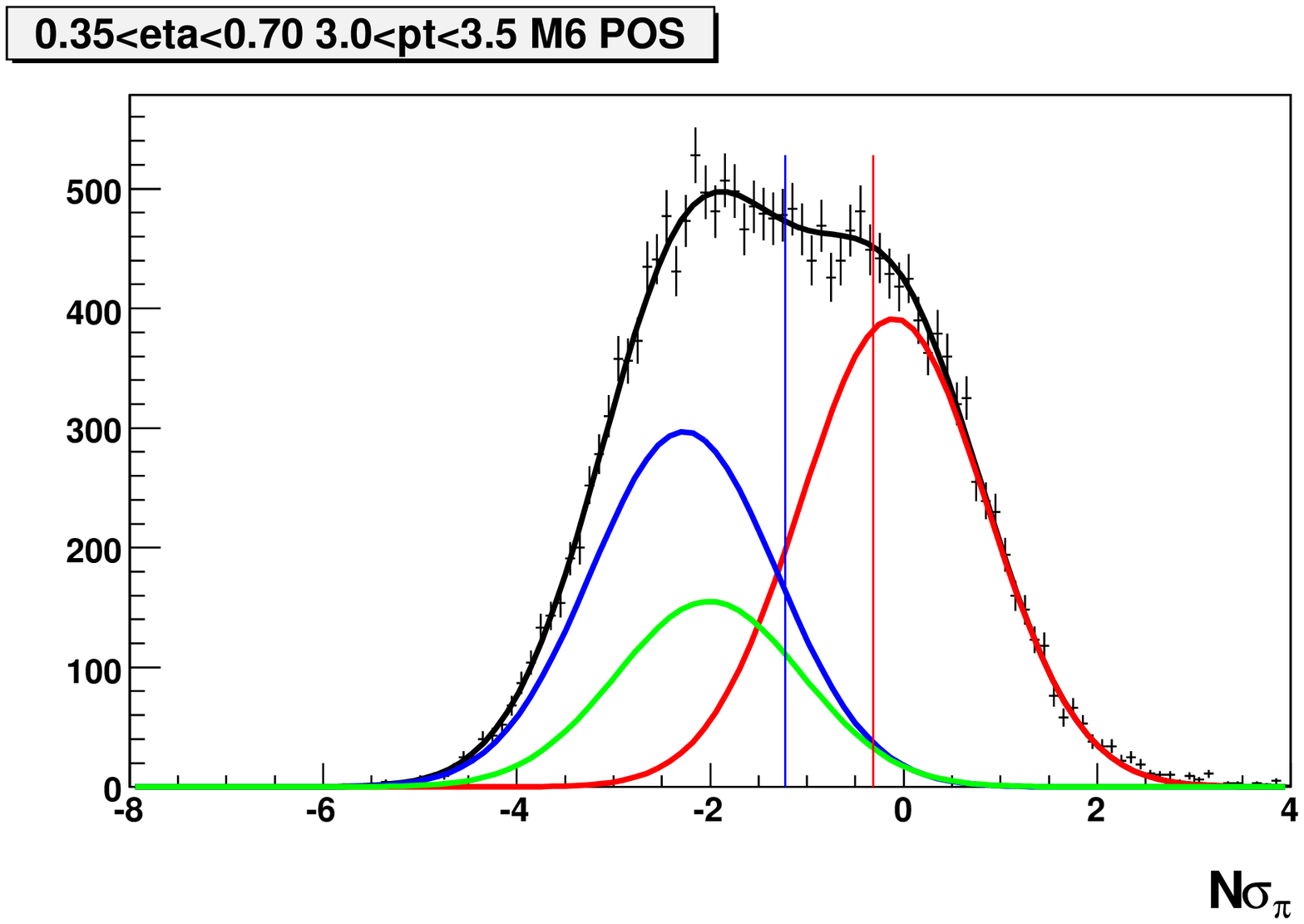}
		\includegraphics[width=1\textwidth]{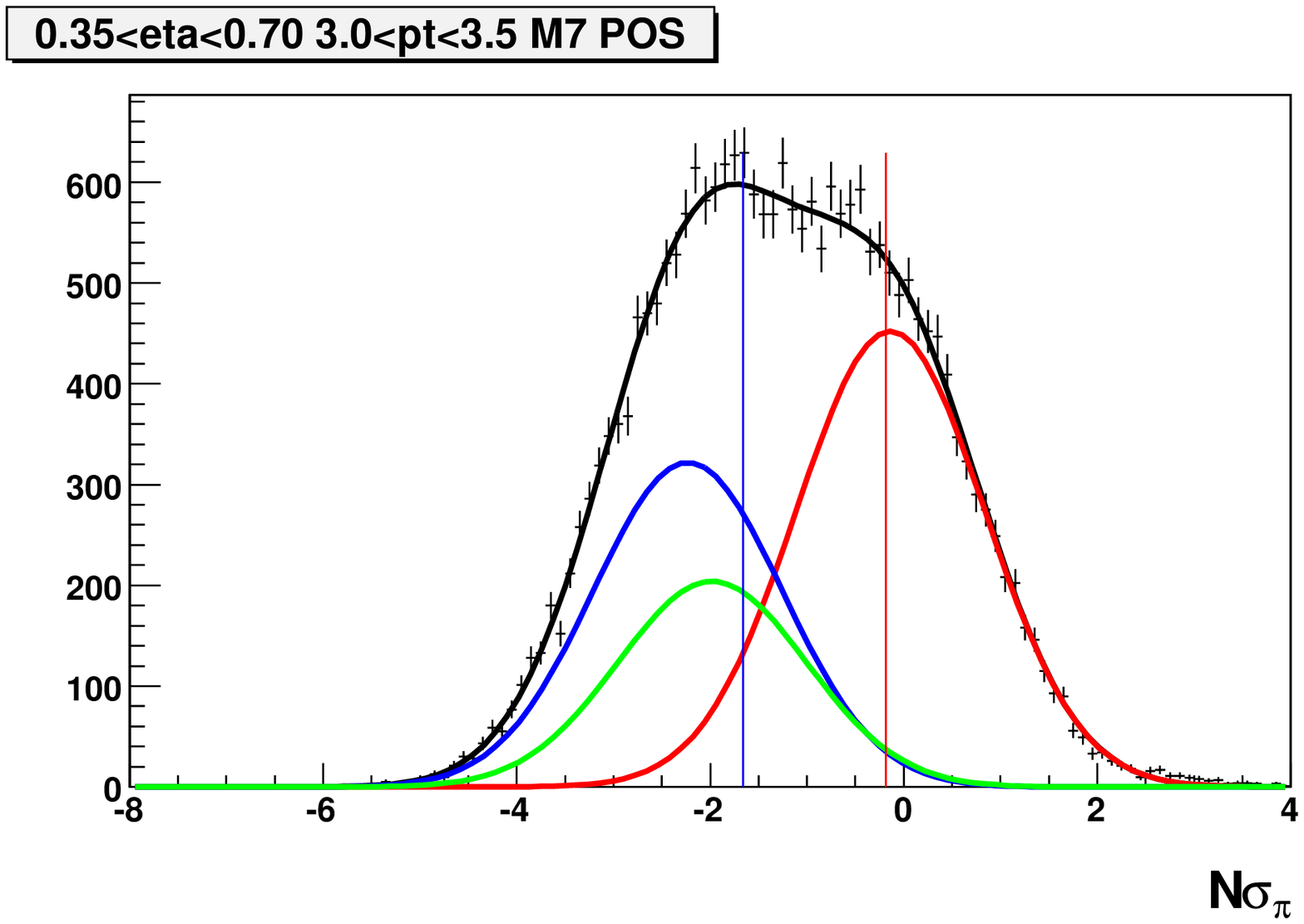}
		\includegraphics[width=1\textwidth]{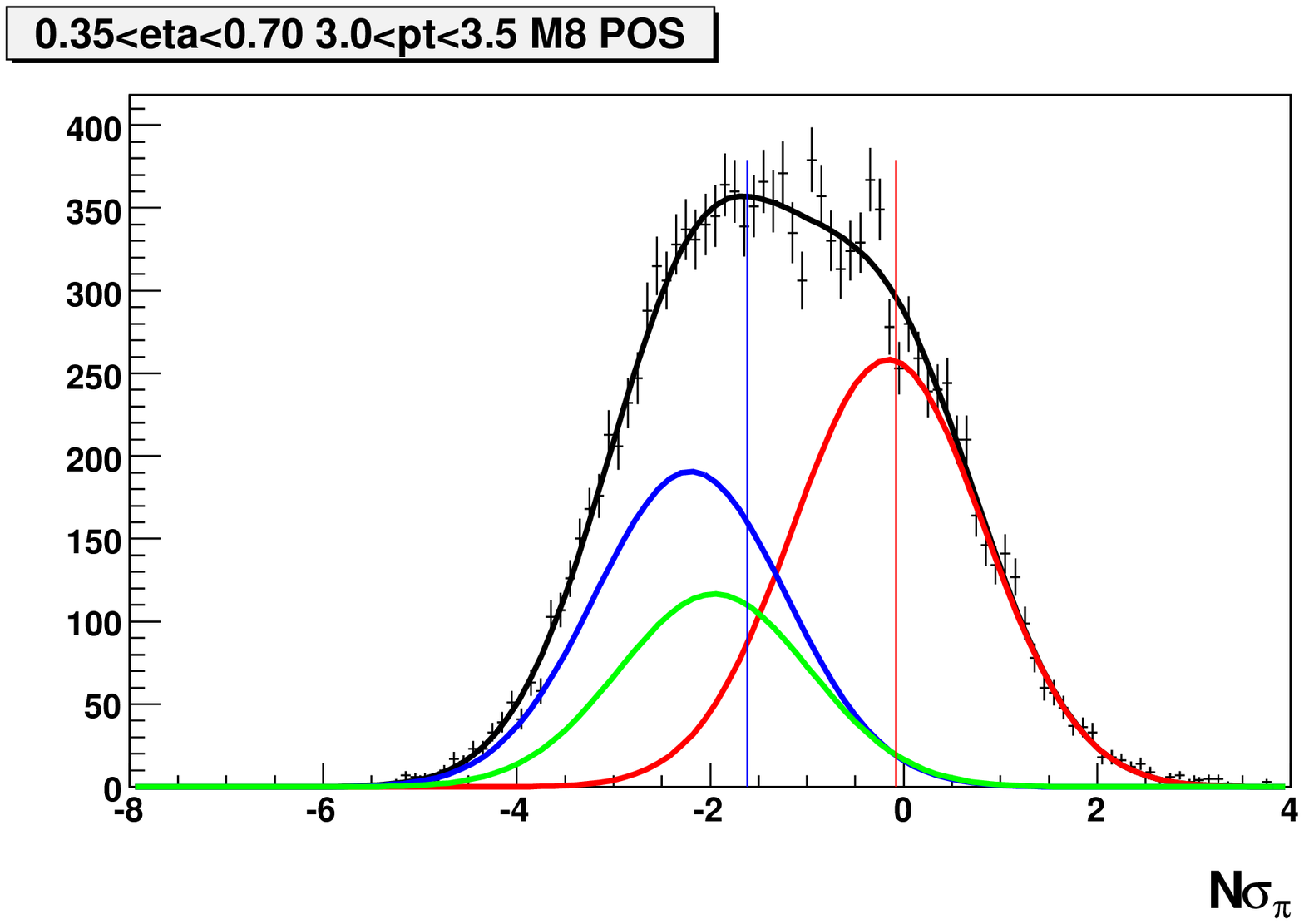}
		\includegraphics[width=1\textwidth]{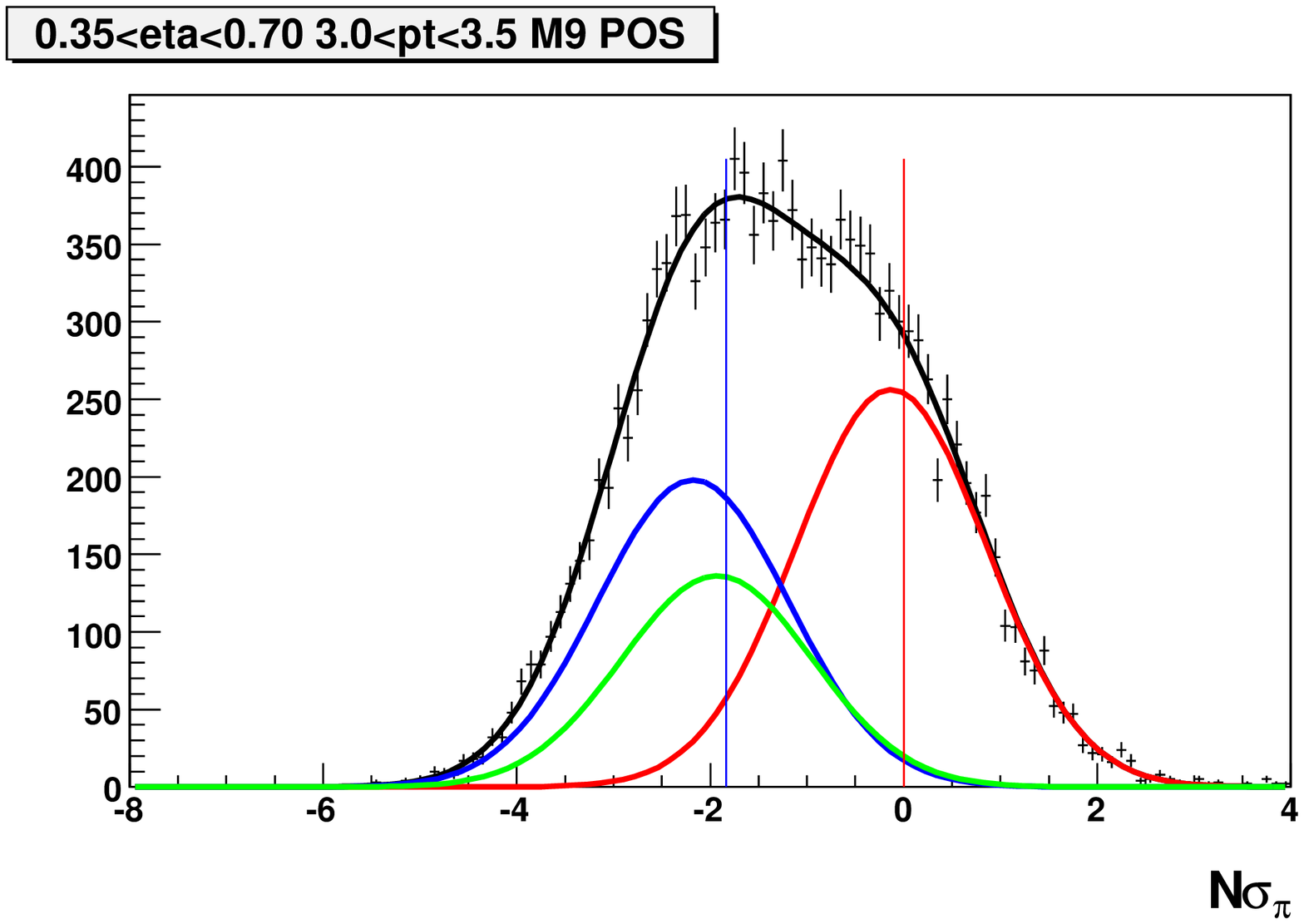}									
			\end{minipage}
\hfill
\begin{minipage}[t]{.19\textwidth}
	\centering
		\includegraphics[width=1\textwidth]{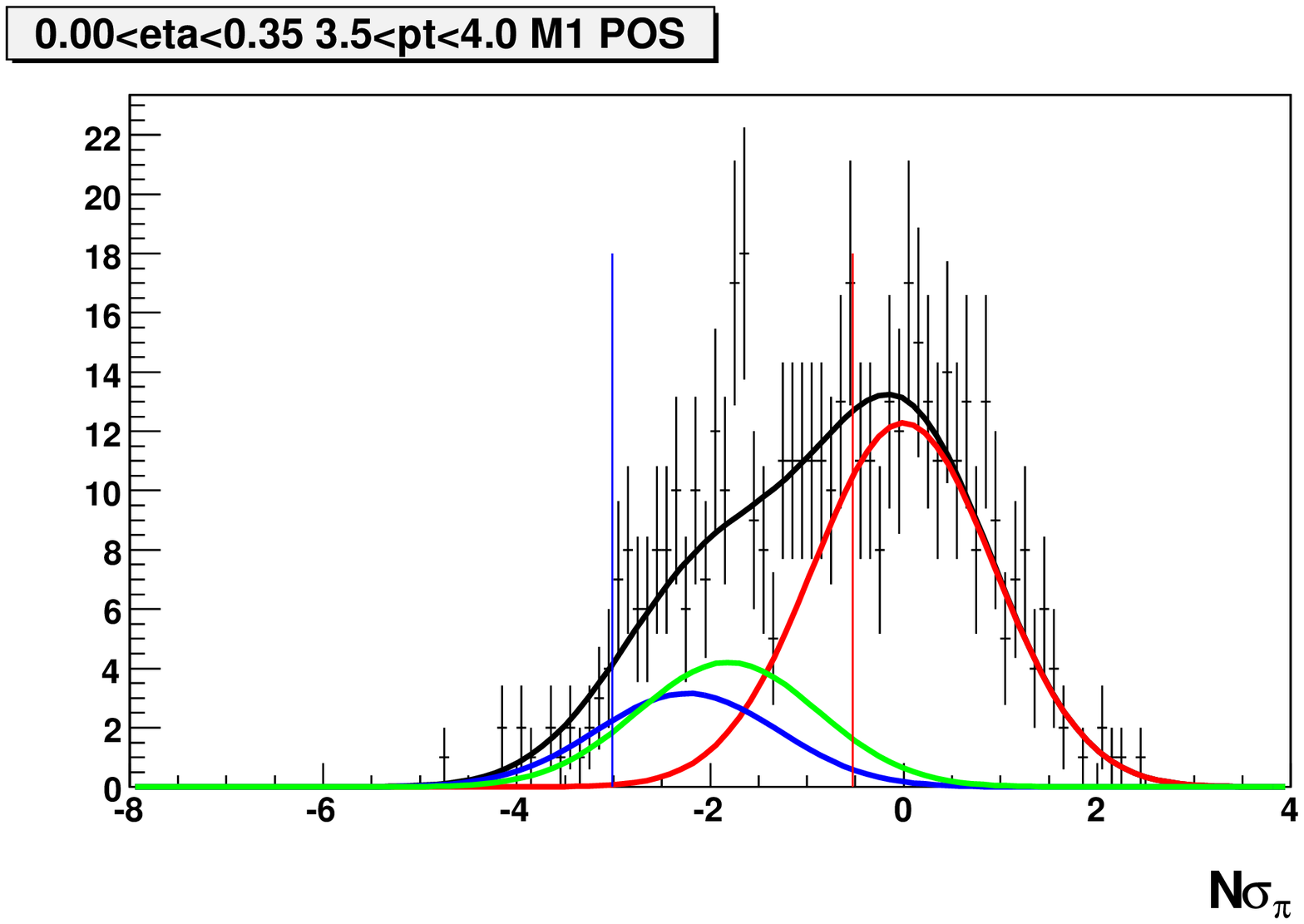}
		\includegraphics[width=1\textwidth]{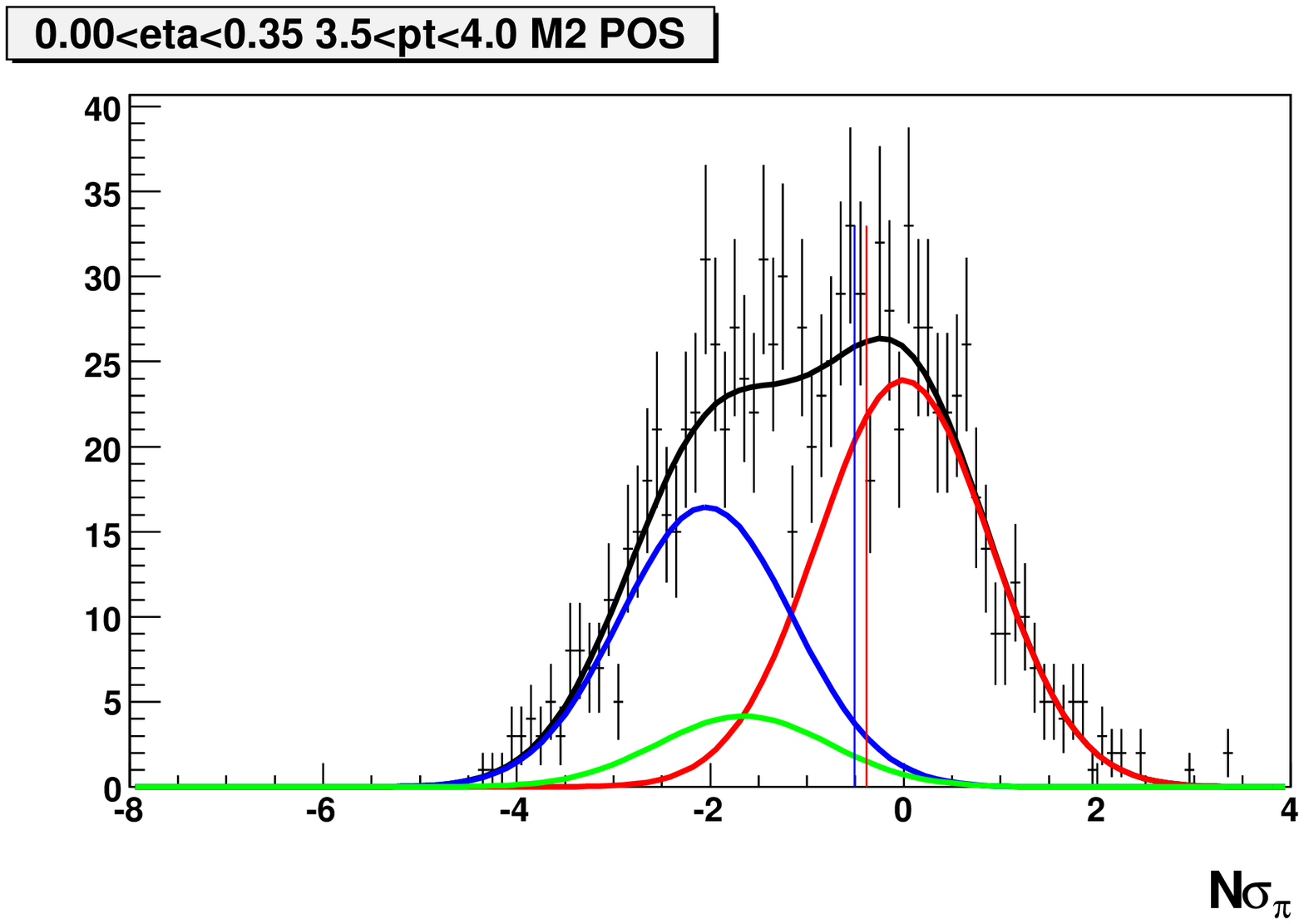}
		\includegraphics[width=1\textwidth]{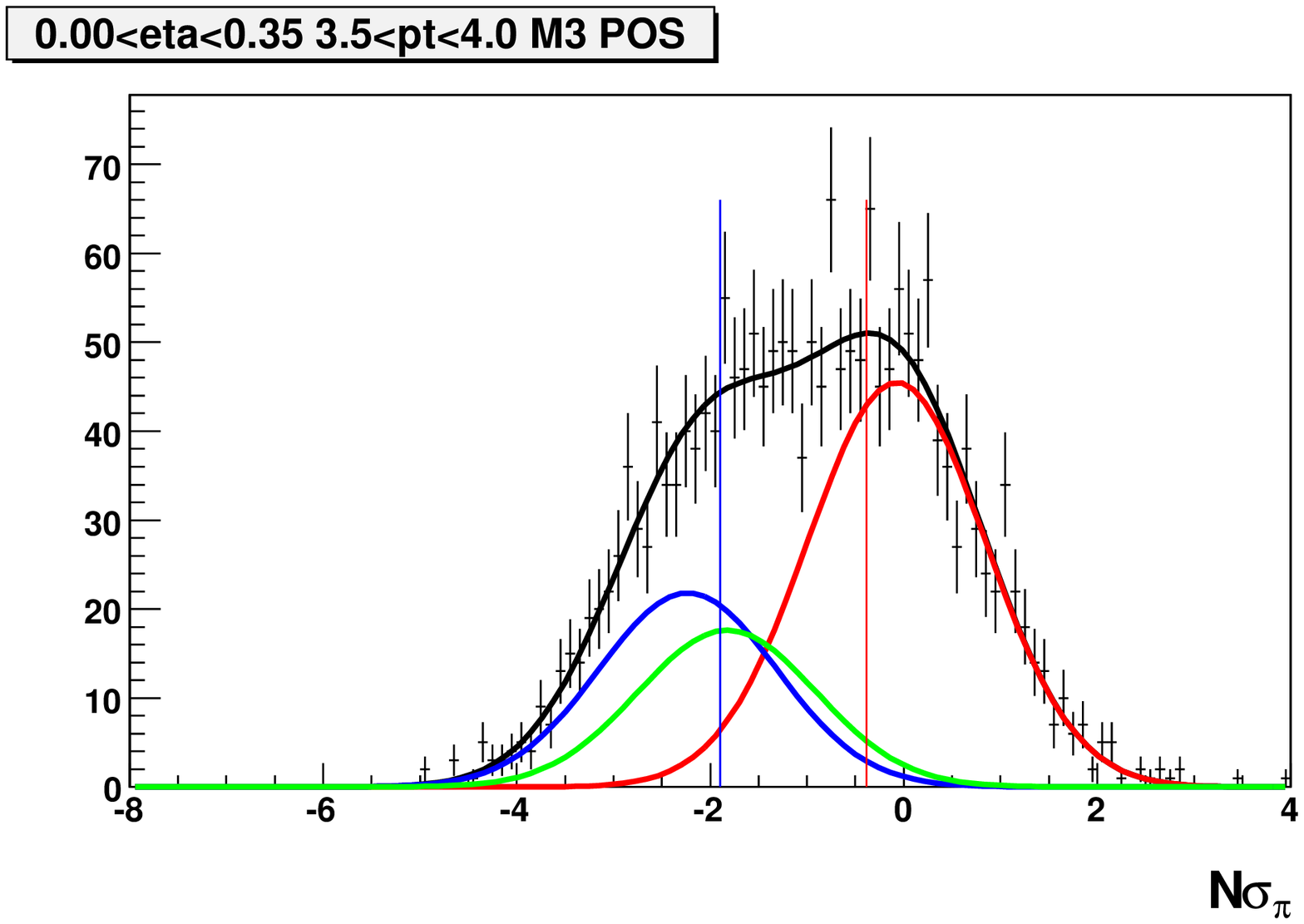}
		\includegraphics[width=1\textwidth]{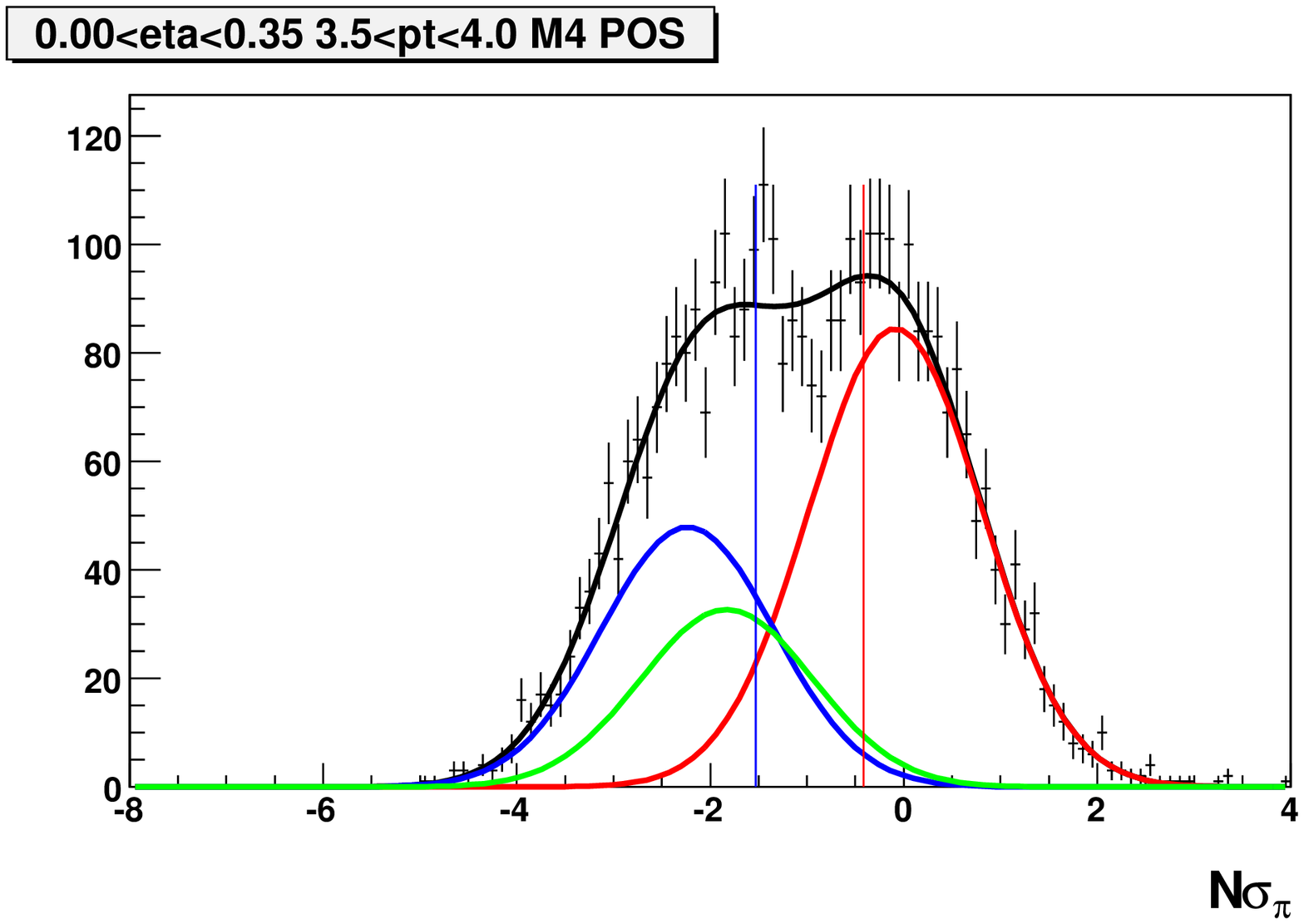}
		\includegraphics[width=1\textwidth]{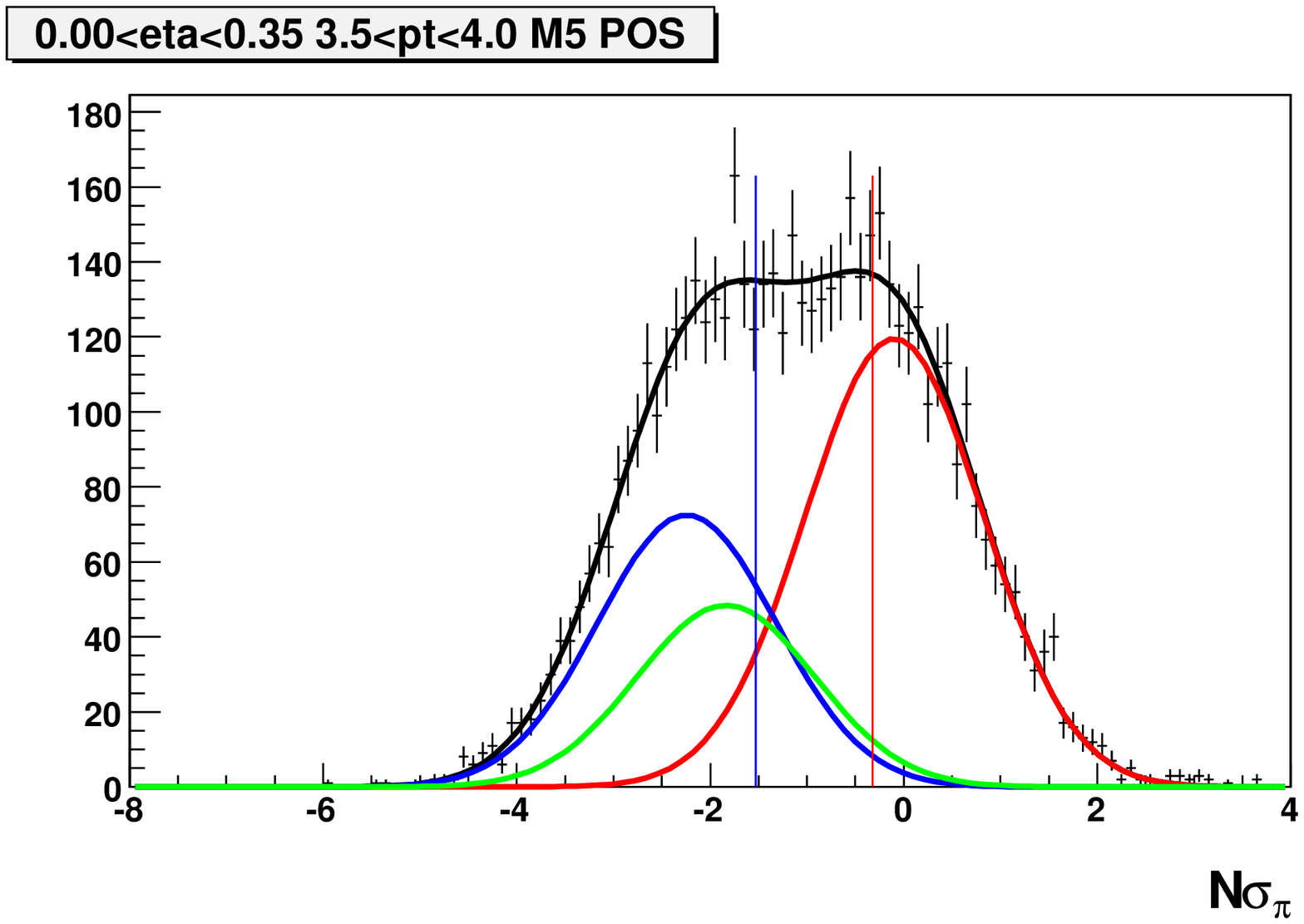}
		\includegraphics[width=1\textwidth]{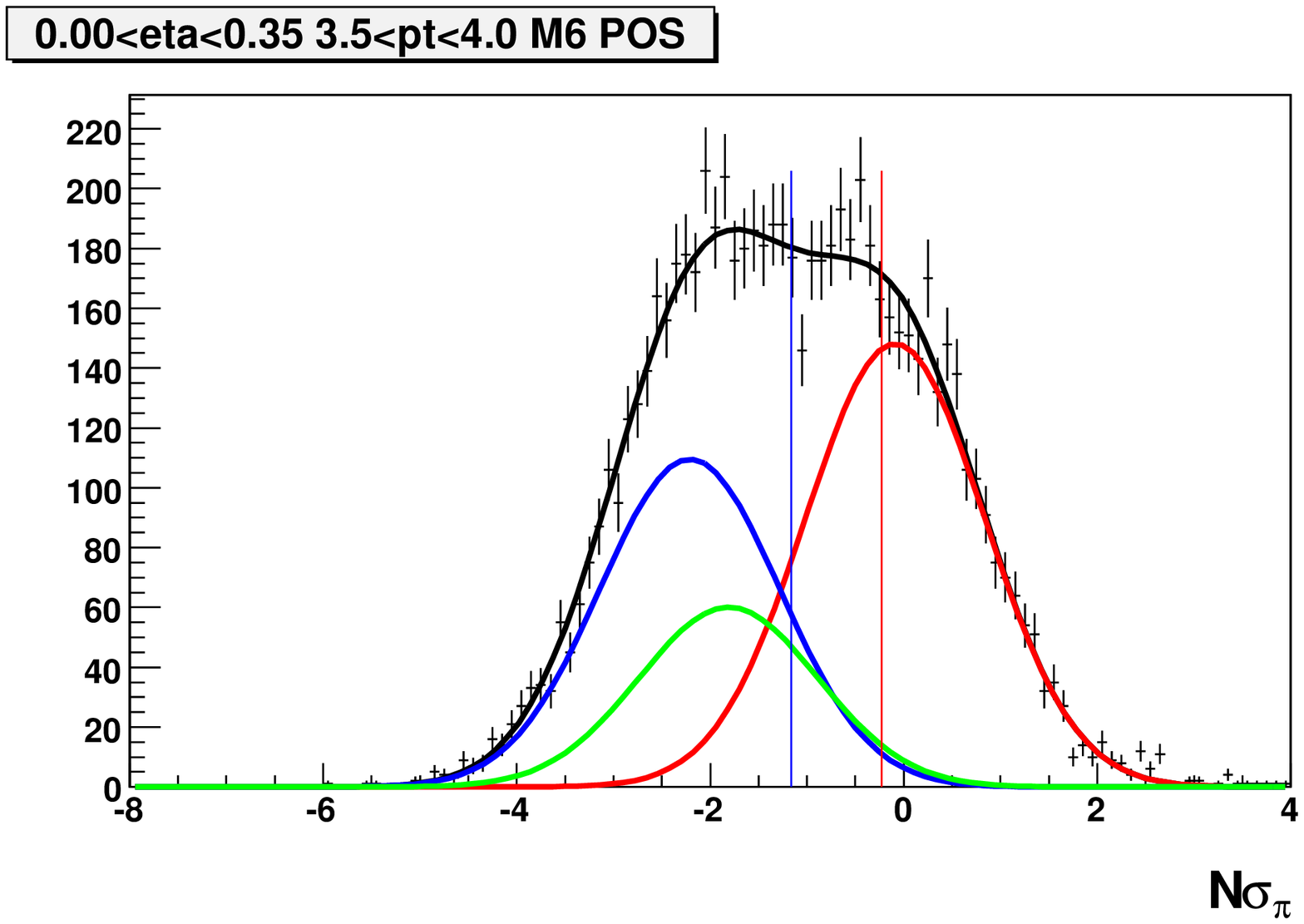}
		\includegraphics[width=1\textwidth]{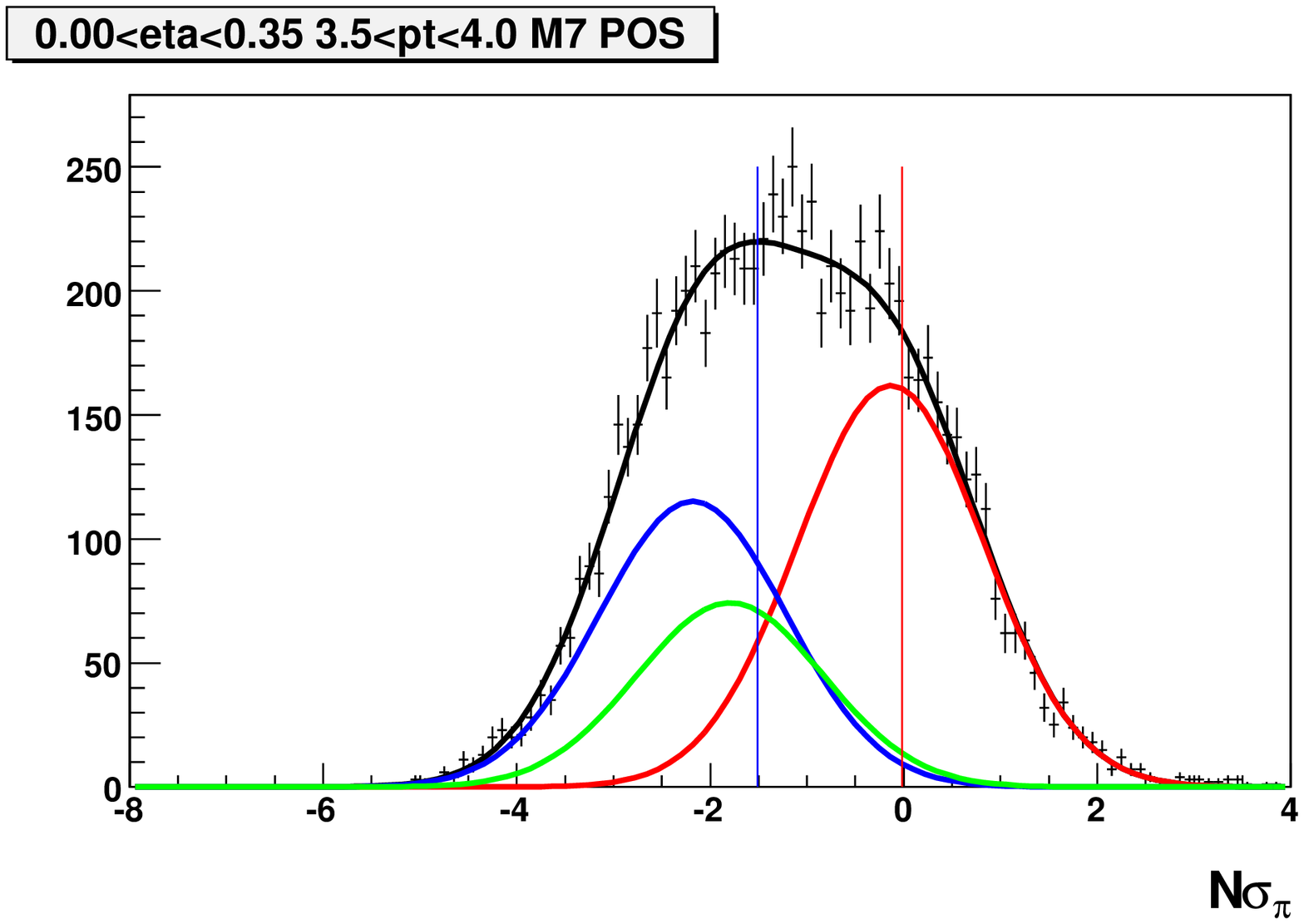}
		\includegraphics[width=1\textwidth]{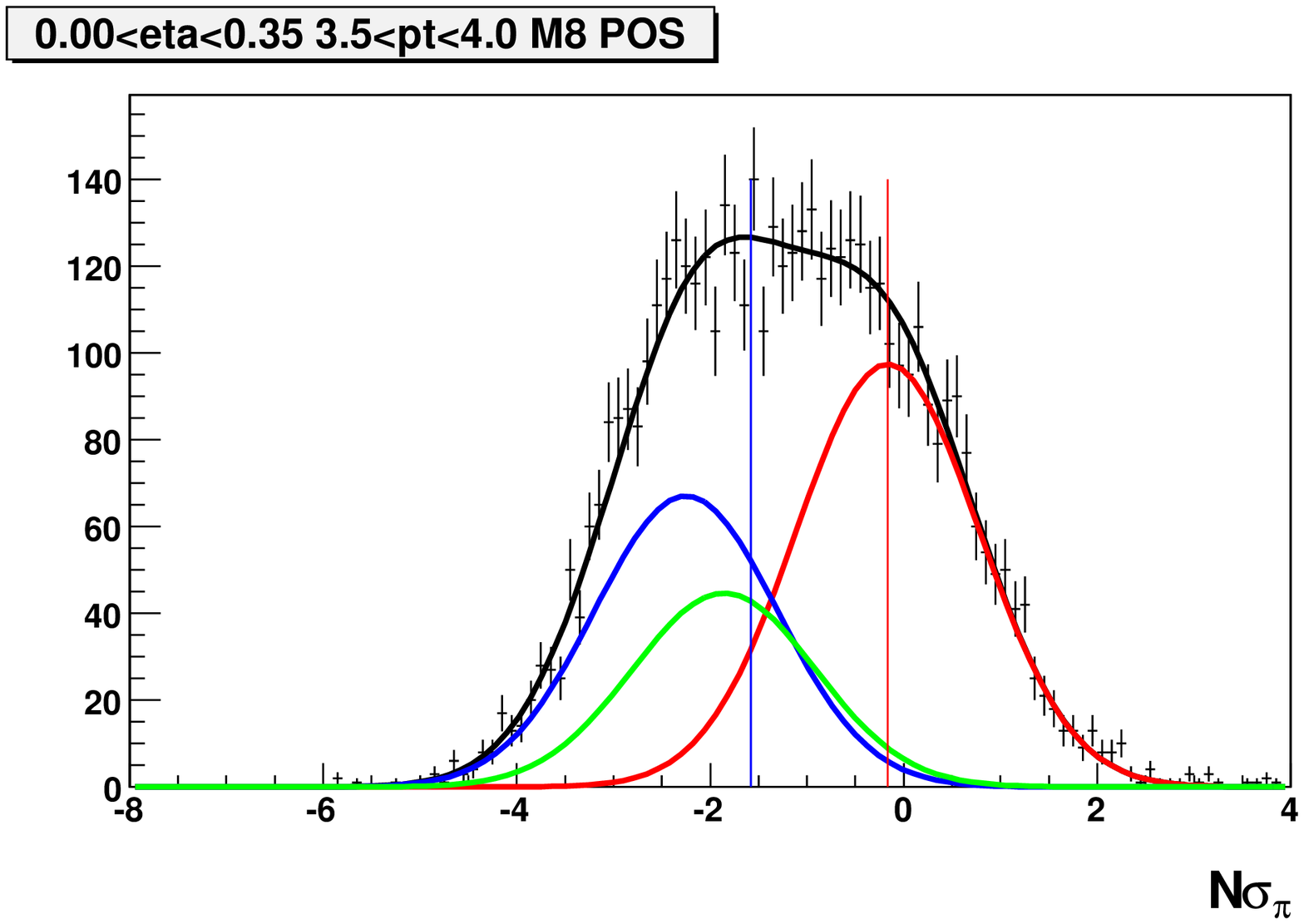}
		\includegraphics[width=1\textwidth]{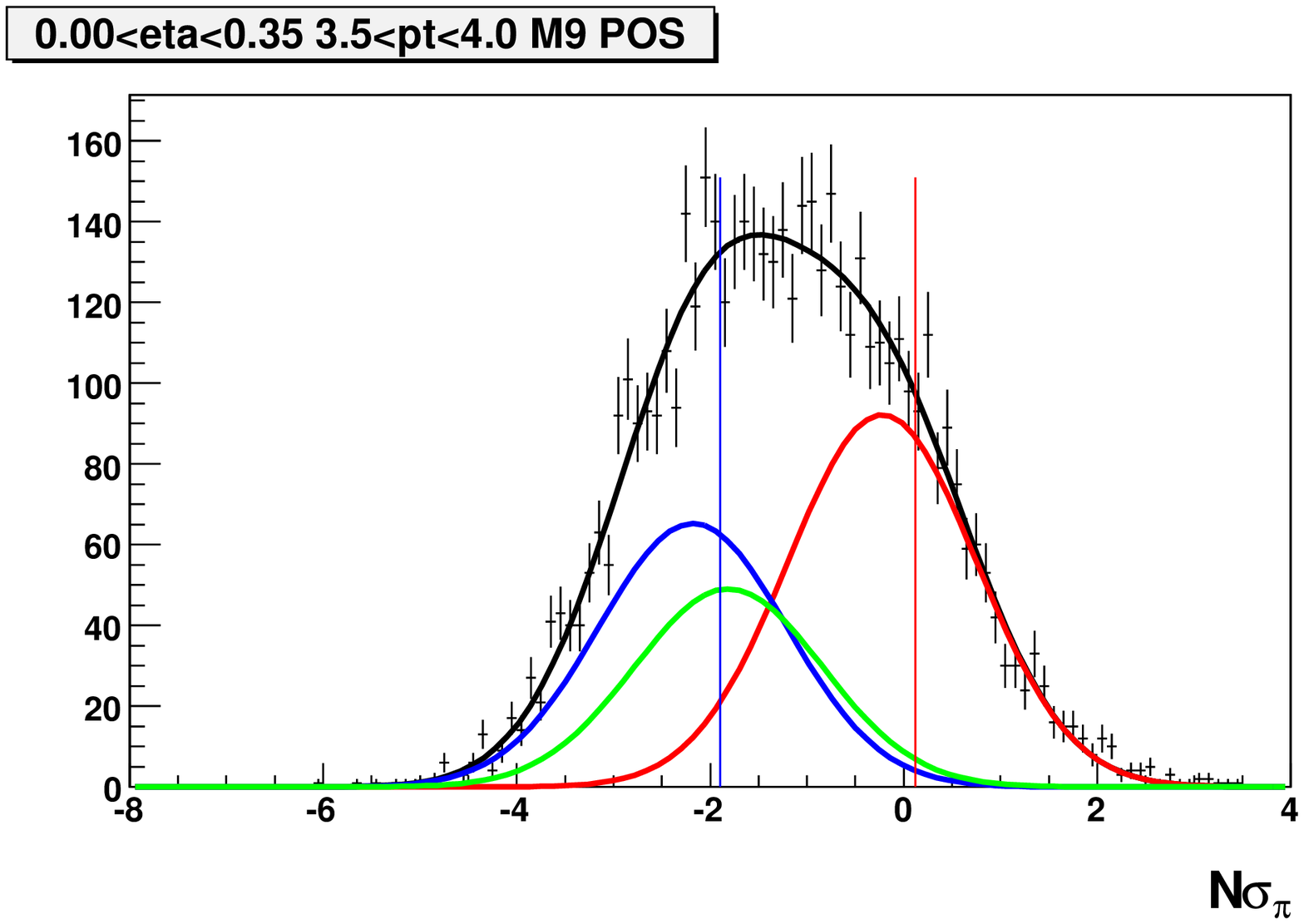}
										
			\end{minipage}
\hfill
\begin{minipage}[t]{.19\textwidth}
	\centering
		\includegraphics[width=1\textwidth]{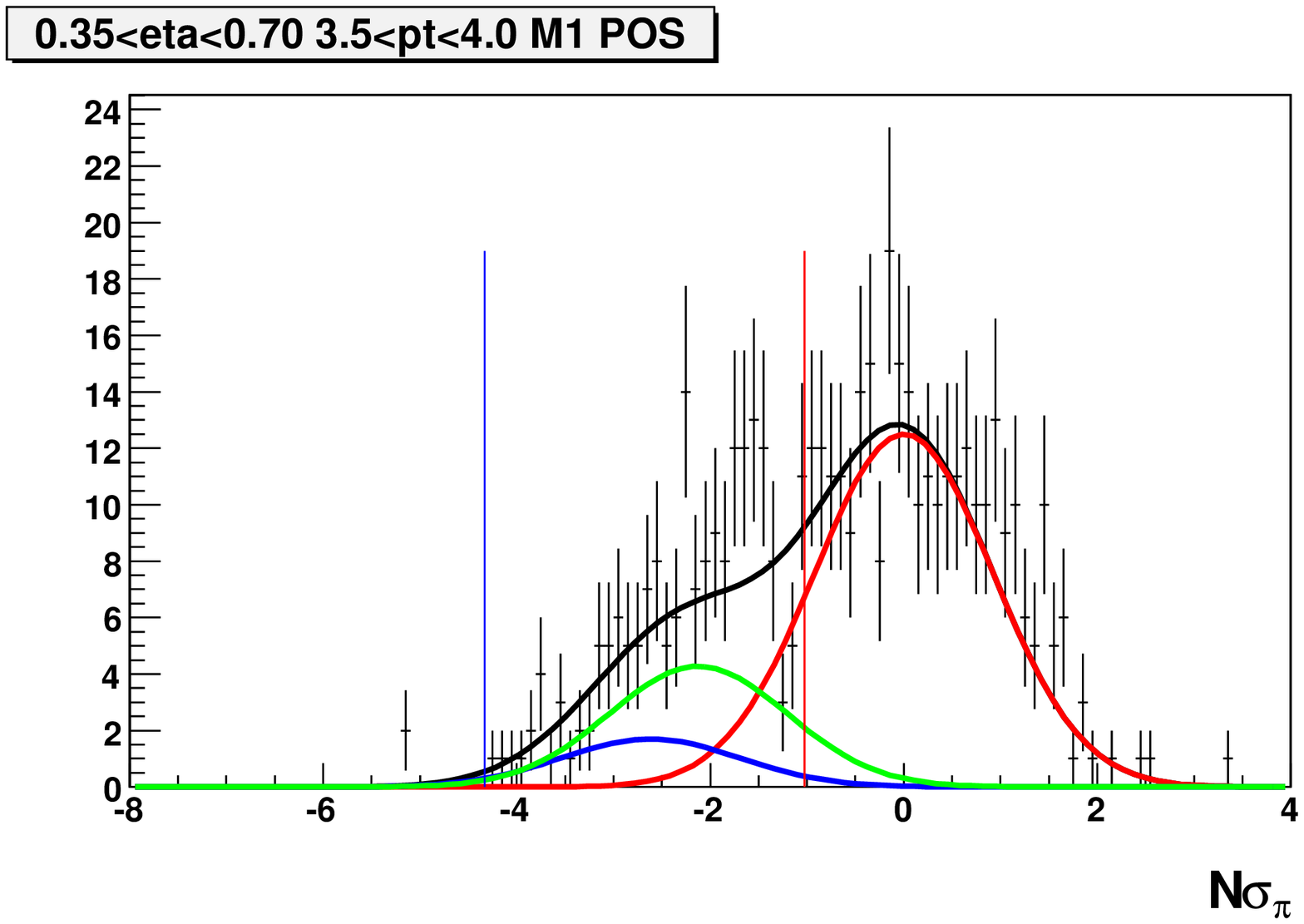}
		\includegraphics[width=1\textwidth]{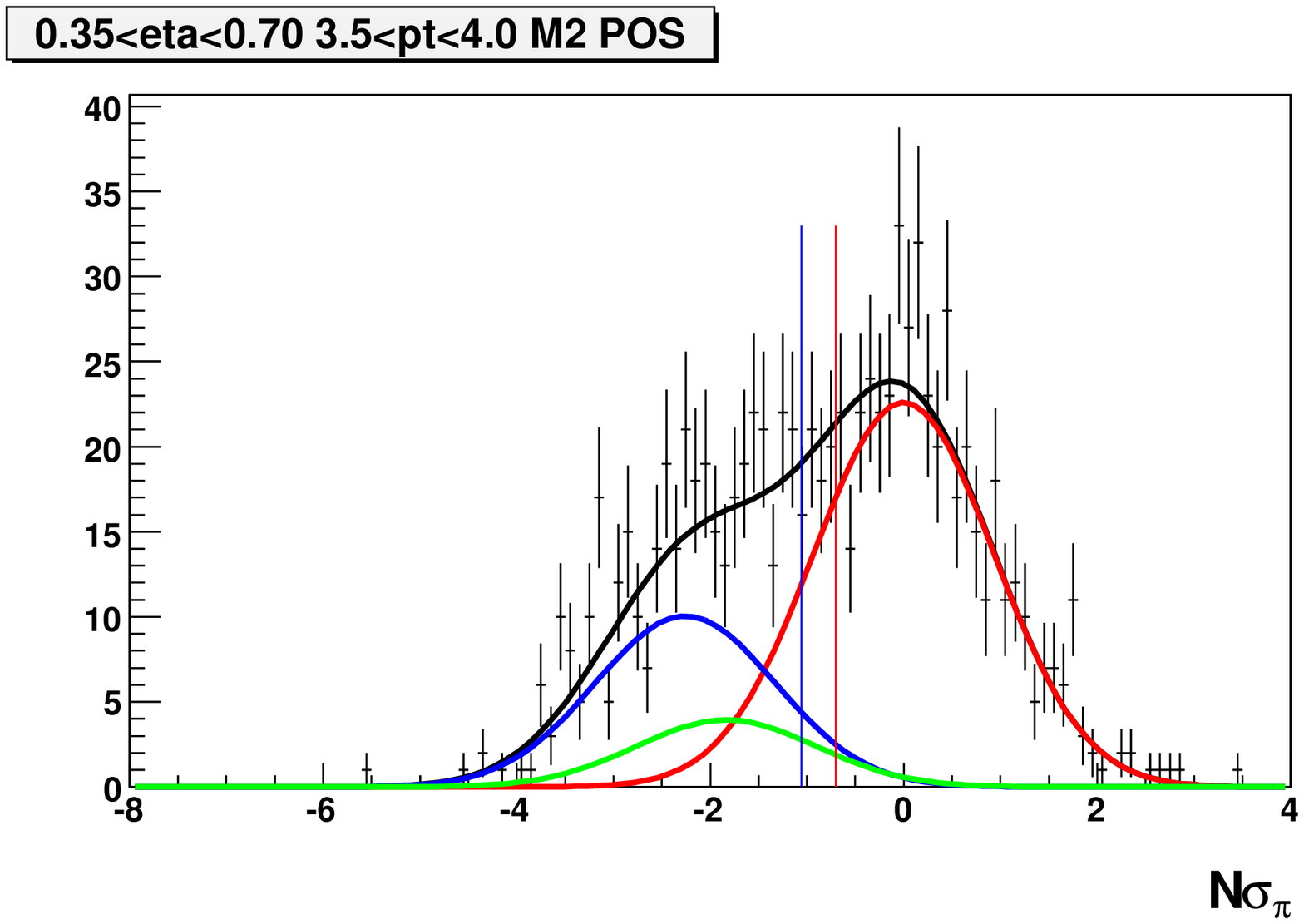}
		\includegraphics[width=1\textwidth]{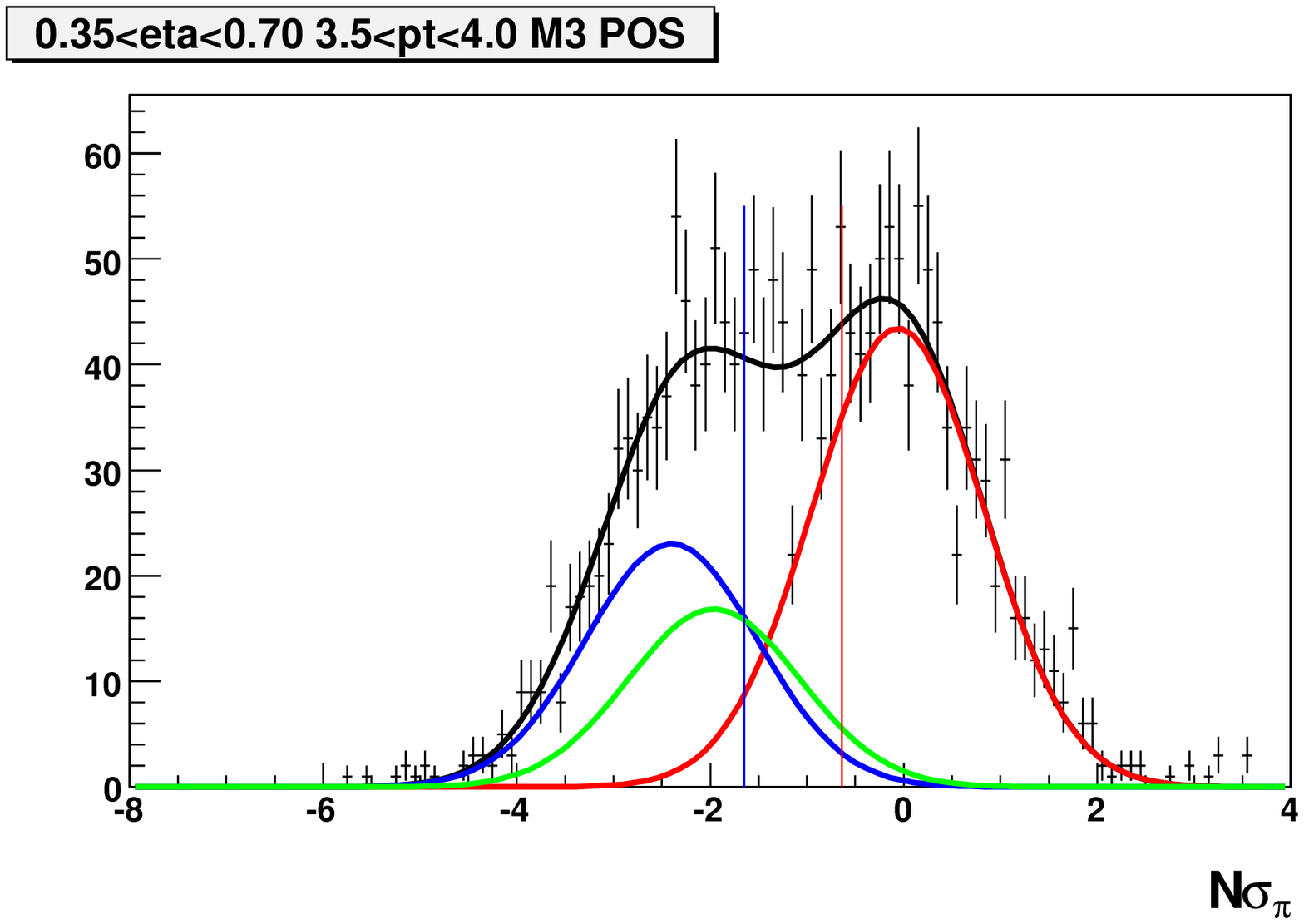}
		\includegraphics[width=1\textwidth]{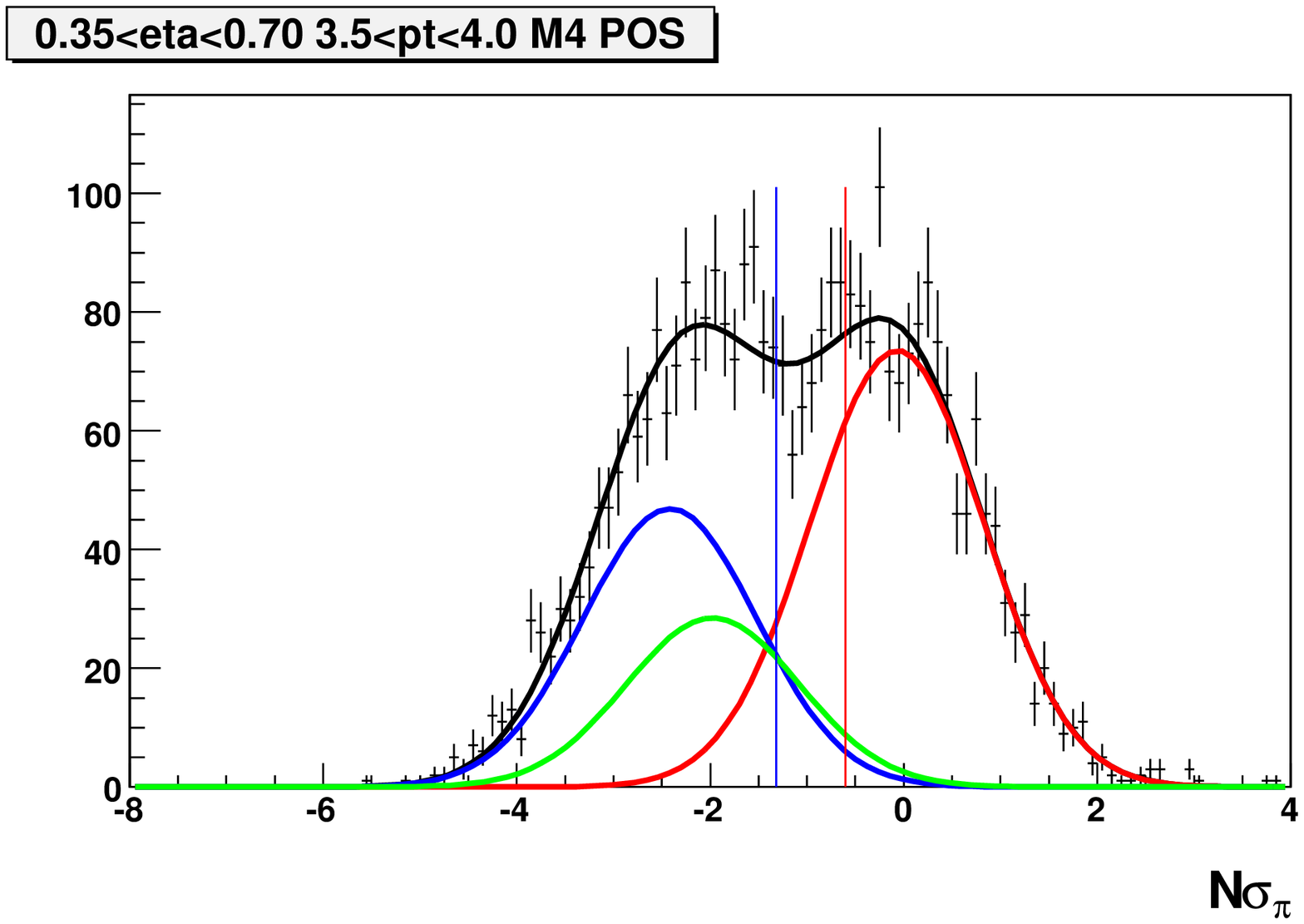}
		\includegraphics[width=1\textwidth]{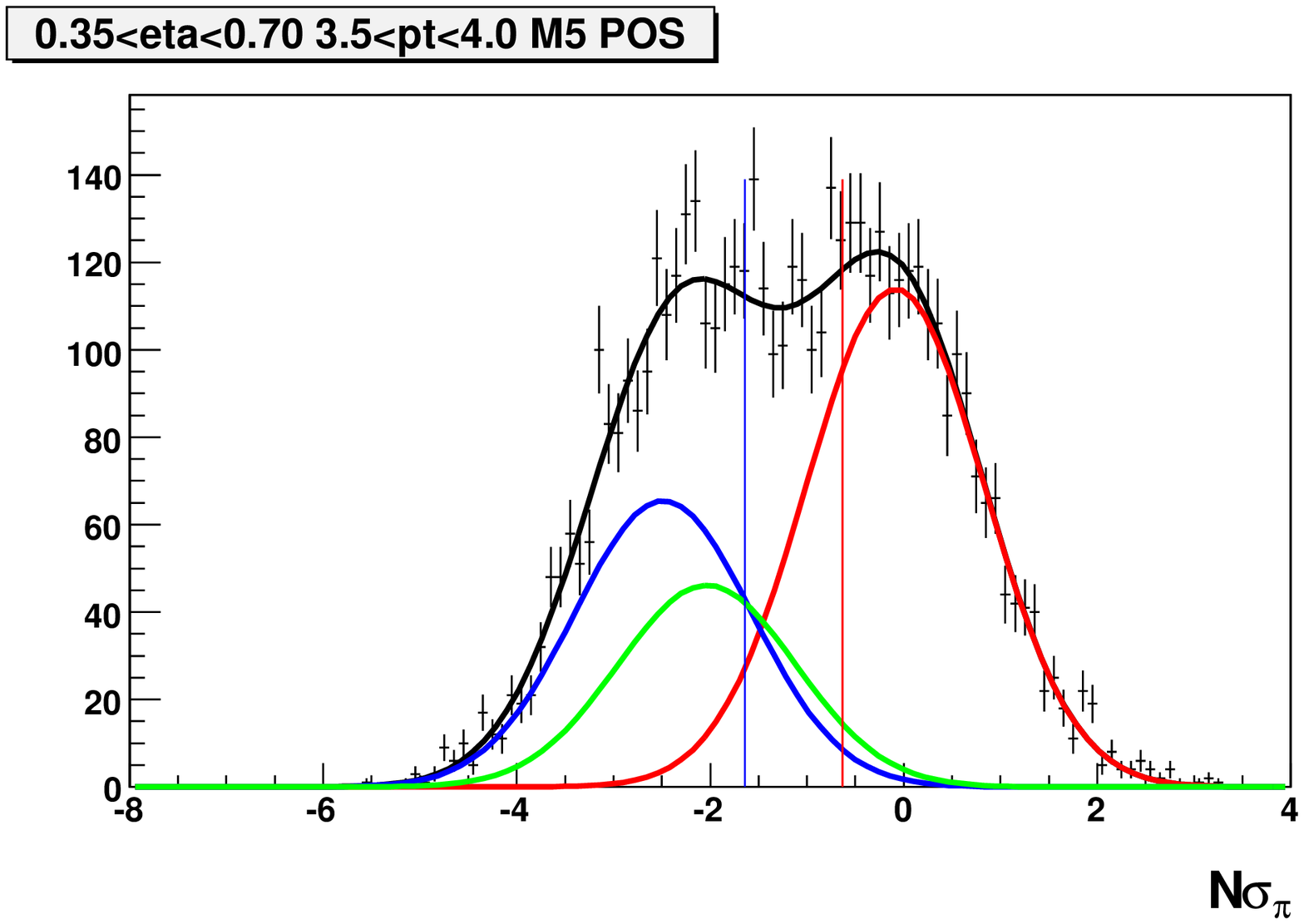}
		\includegraphics[width=1\textwidth]{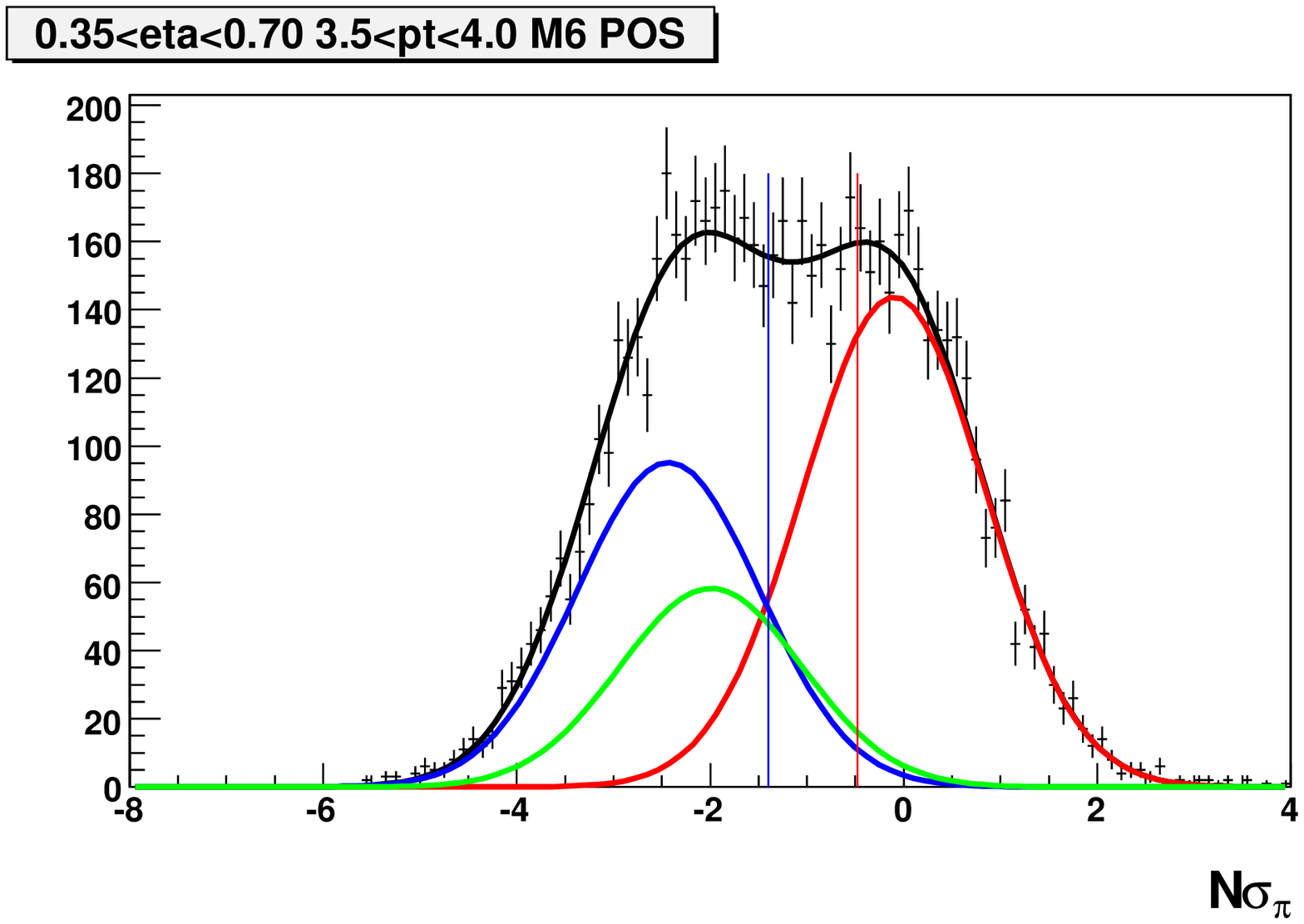}
		\includegraphics[width=1\textwidth]{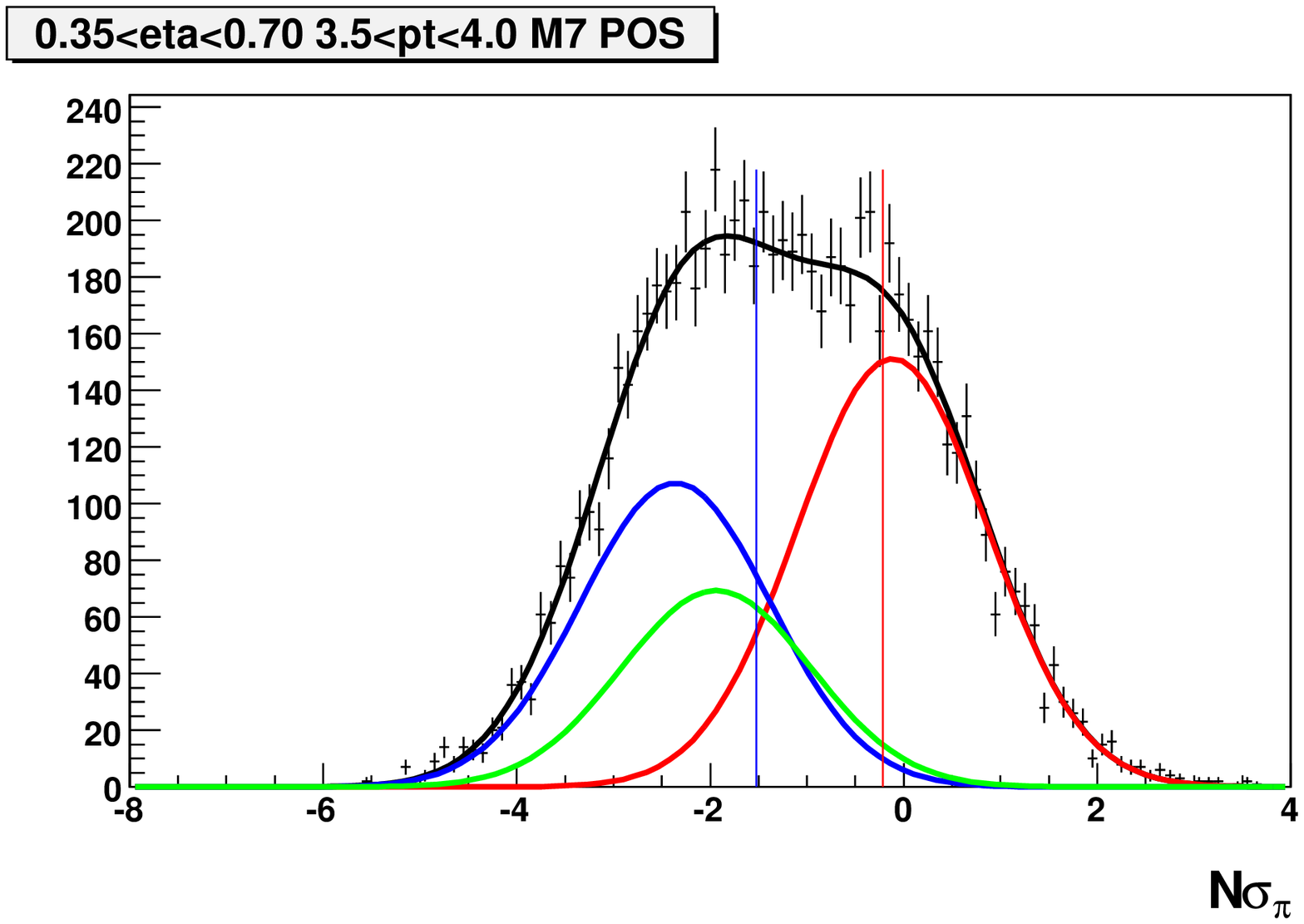}
		\includegraphics[width=1\textwidth]{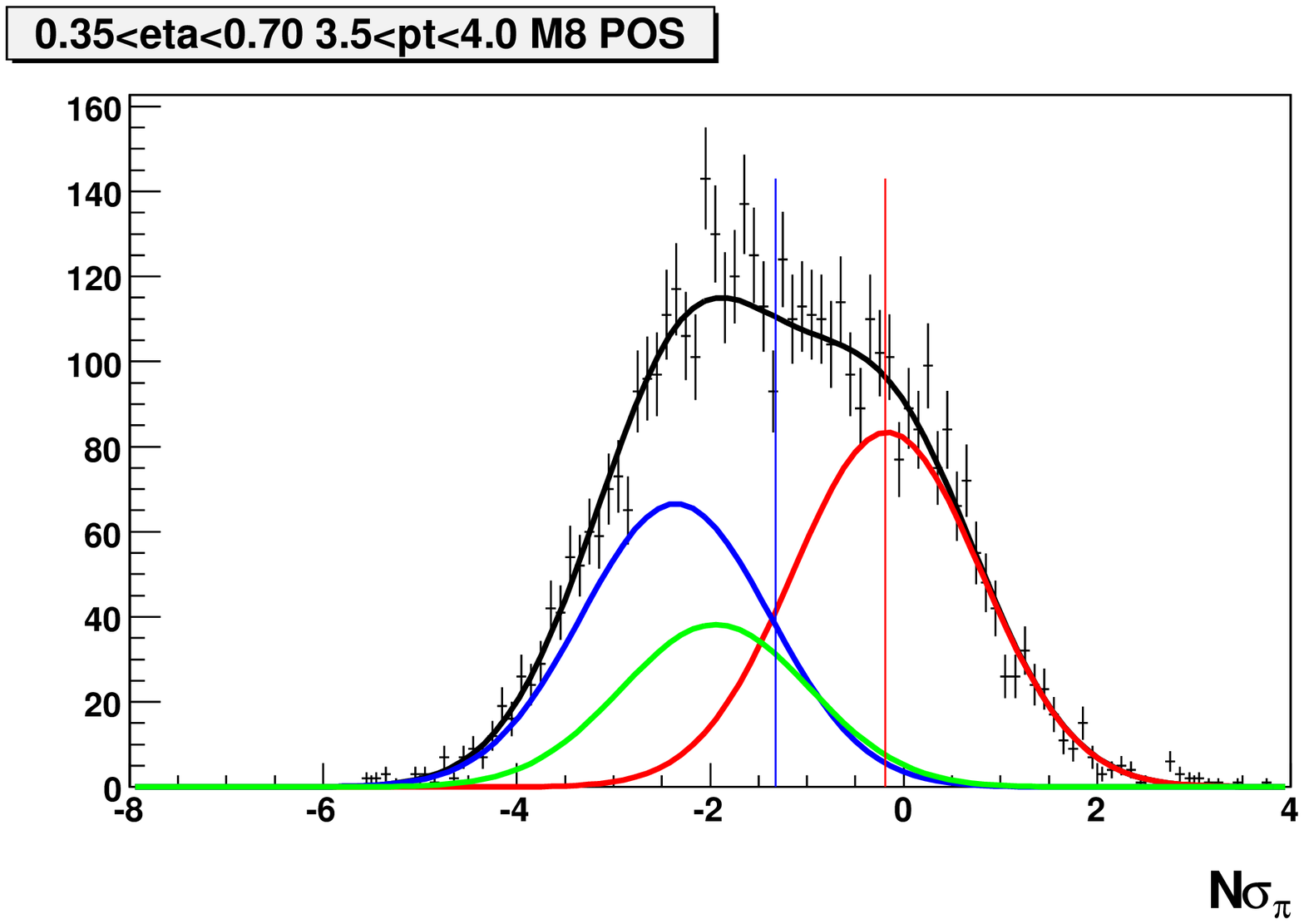}
		\includegraphics[width=1\textwidth]{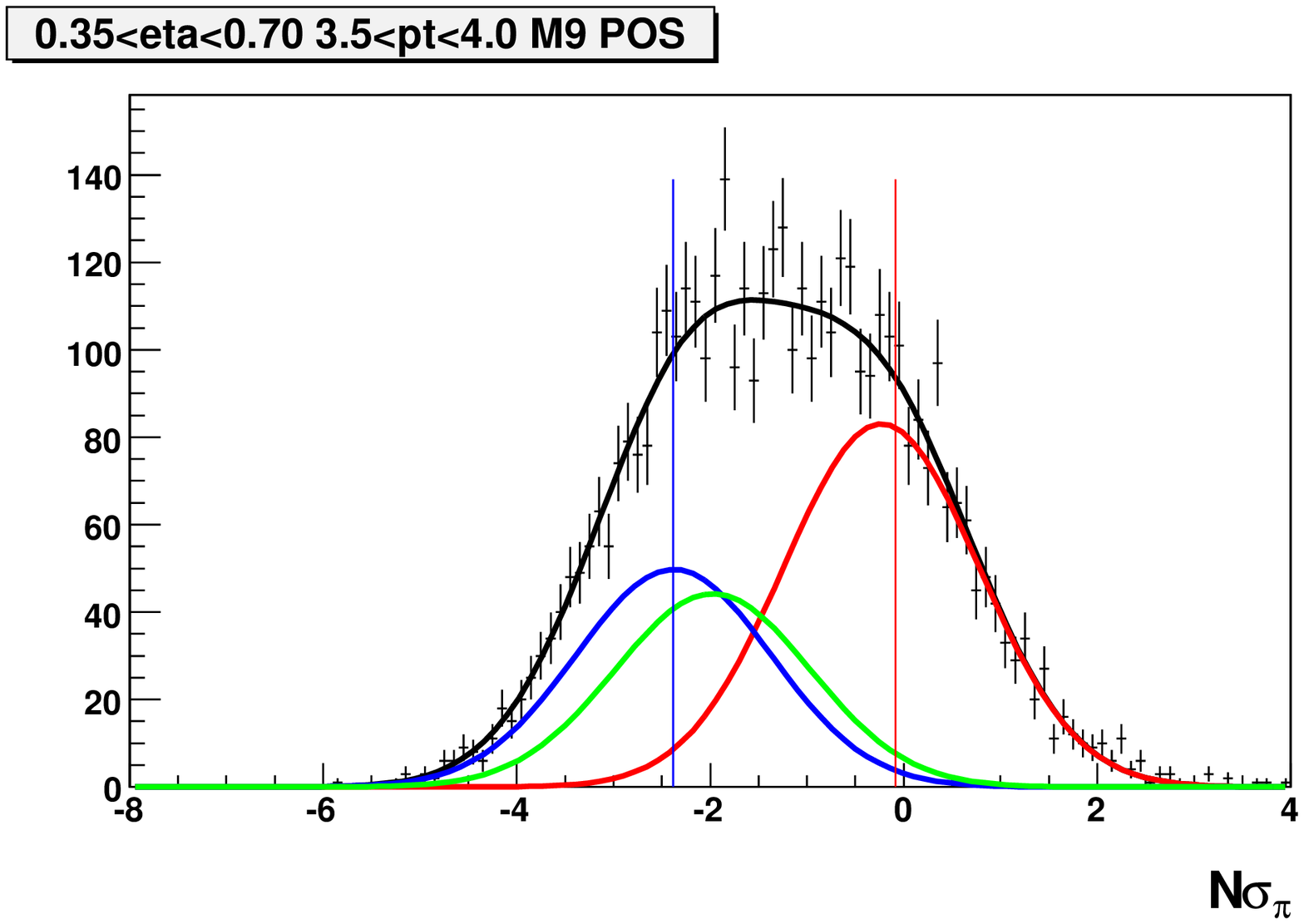}
		
			\end{minipage}						
	\caption{$N\sigma_{\pi}$ distributions with fits and cuts for positive particles in Au+Au collisions at $\sqrt{s_{NN}}=200$ GeV/c.  Curves are from fit with red for pion, blue for proton and green for kaon.  Particles to the left of the left line are 50\% protons and particles to the right of the right line are 95\% pion.  Rows are the centrality bins 70-80\%, 60-70\%, 50-60\%, 40-50\%, 30-40\%, 20-30\%, 10-20\%, 5-10\% and 0-5\% from top to bottom.  Columns are for different $p_{T}$ and $\eta$ cuts.  Left: $|\eta|<0.35$ and $3<p_{T}<3.5$ GeV/c.  Left center:  $0.35<|\eta|<0.75$ and $3<p_{T}<3.5$ GeV/c.  Right center: $|\eta|<0.35$ and $3.5<p_{T}<4.0$ GeV/c.  Right:  $0.35<|\eta|<0.75$ and $3.5<p_{T}<4.0$ GeV/c.}
	\label{fig:pidcutsP}	
\end{figure}

\begin{figure}[H]
\hfill
\begin{minipage}[t]{.19\textwidth}
	\centering
		\includegraphics[width=1\textwidth]{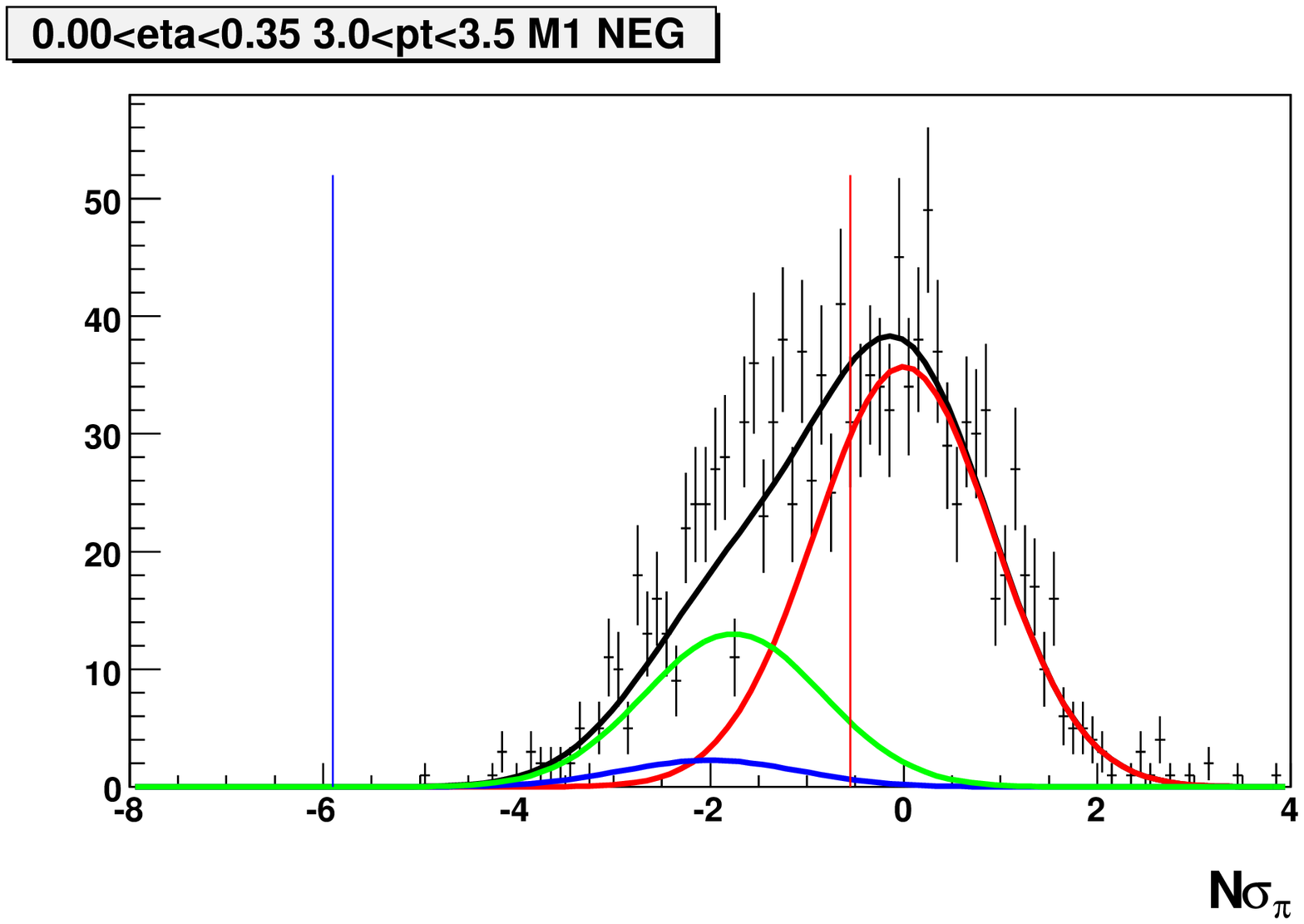}
		\includegraphics[width=1\textwidth]{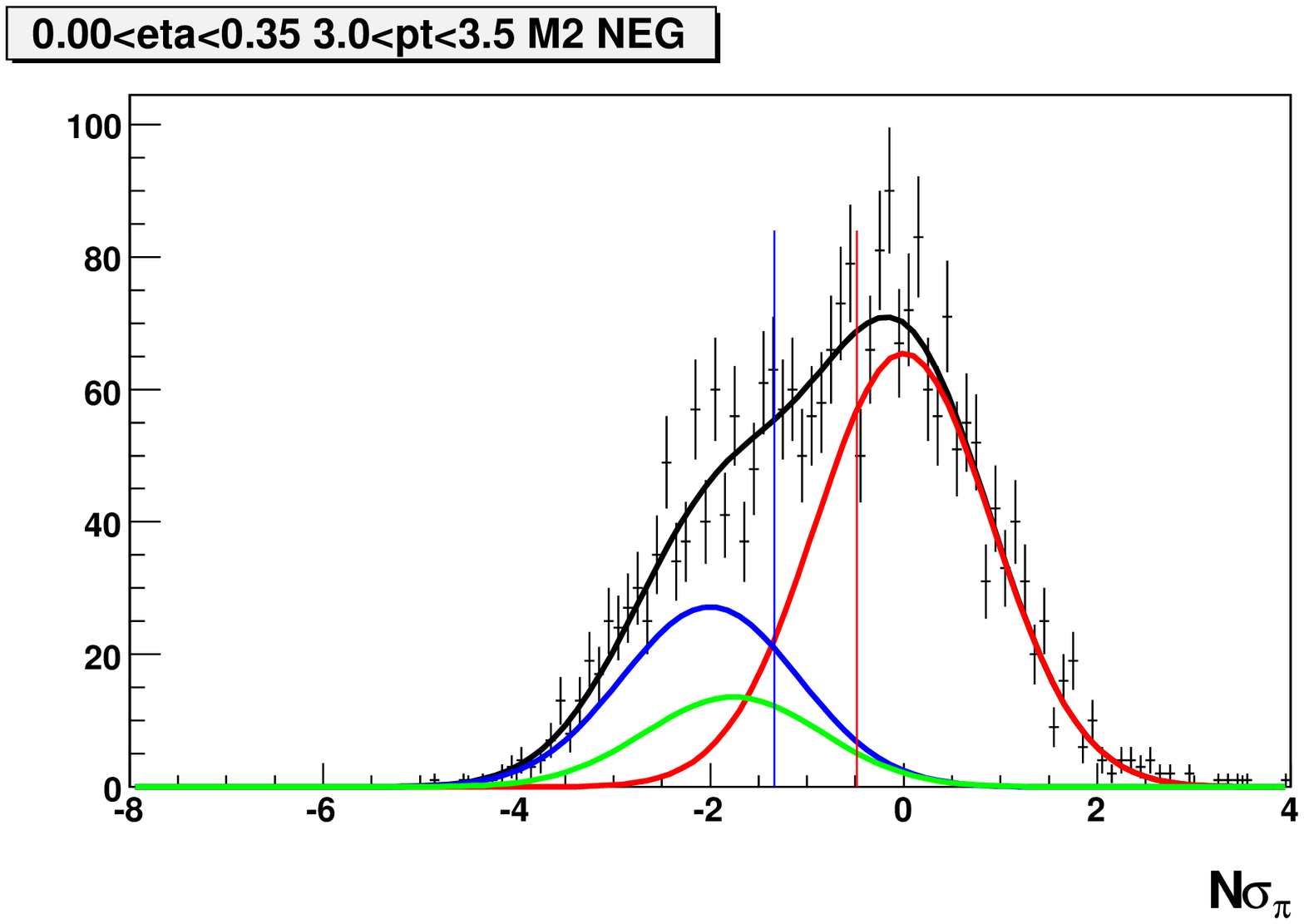}
		\includegraphics[width=1\textwidth]{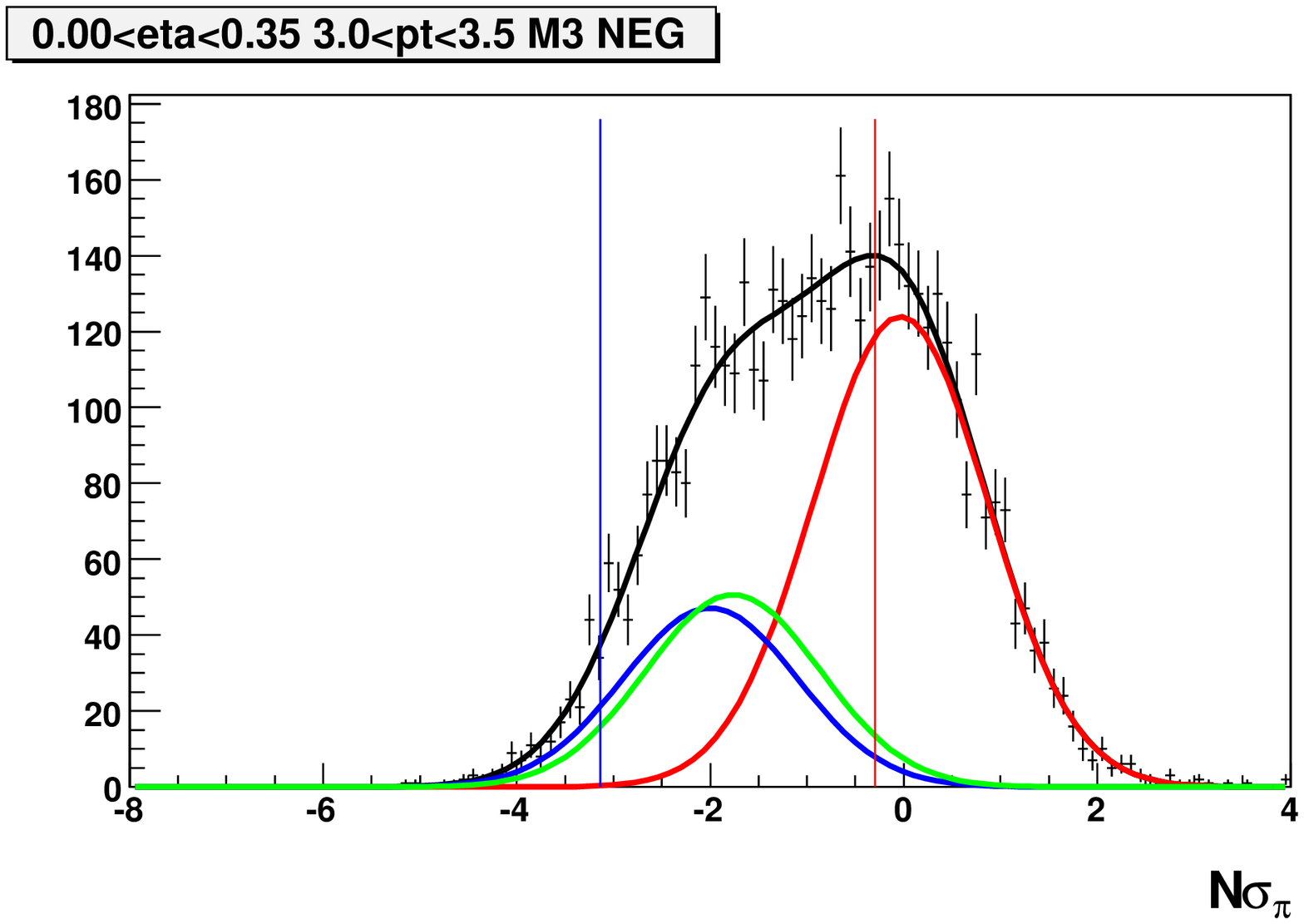}
		\includegraphics[width=1\textwidth]{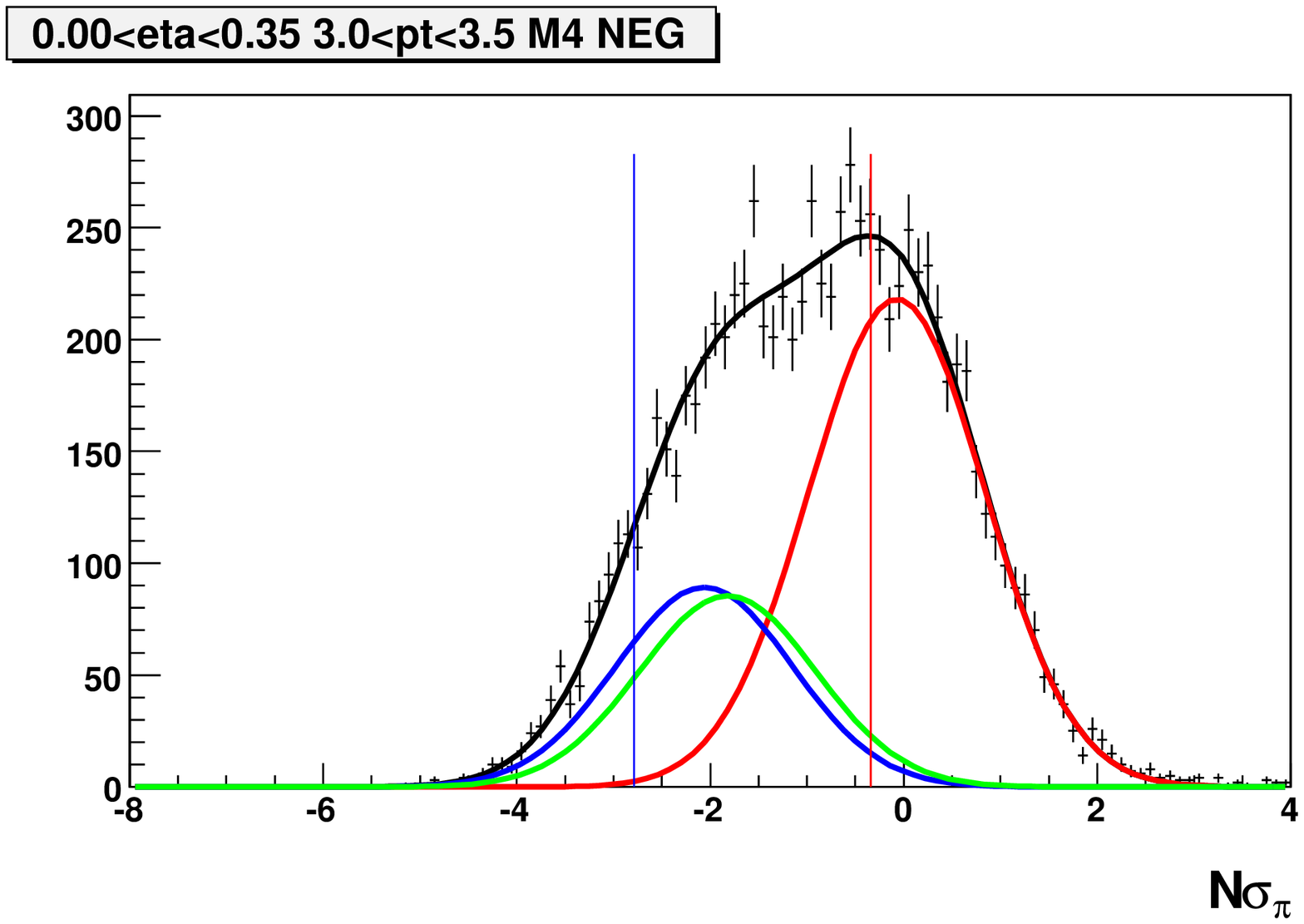}
		\includegraphics[width=1\textwidth]{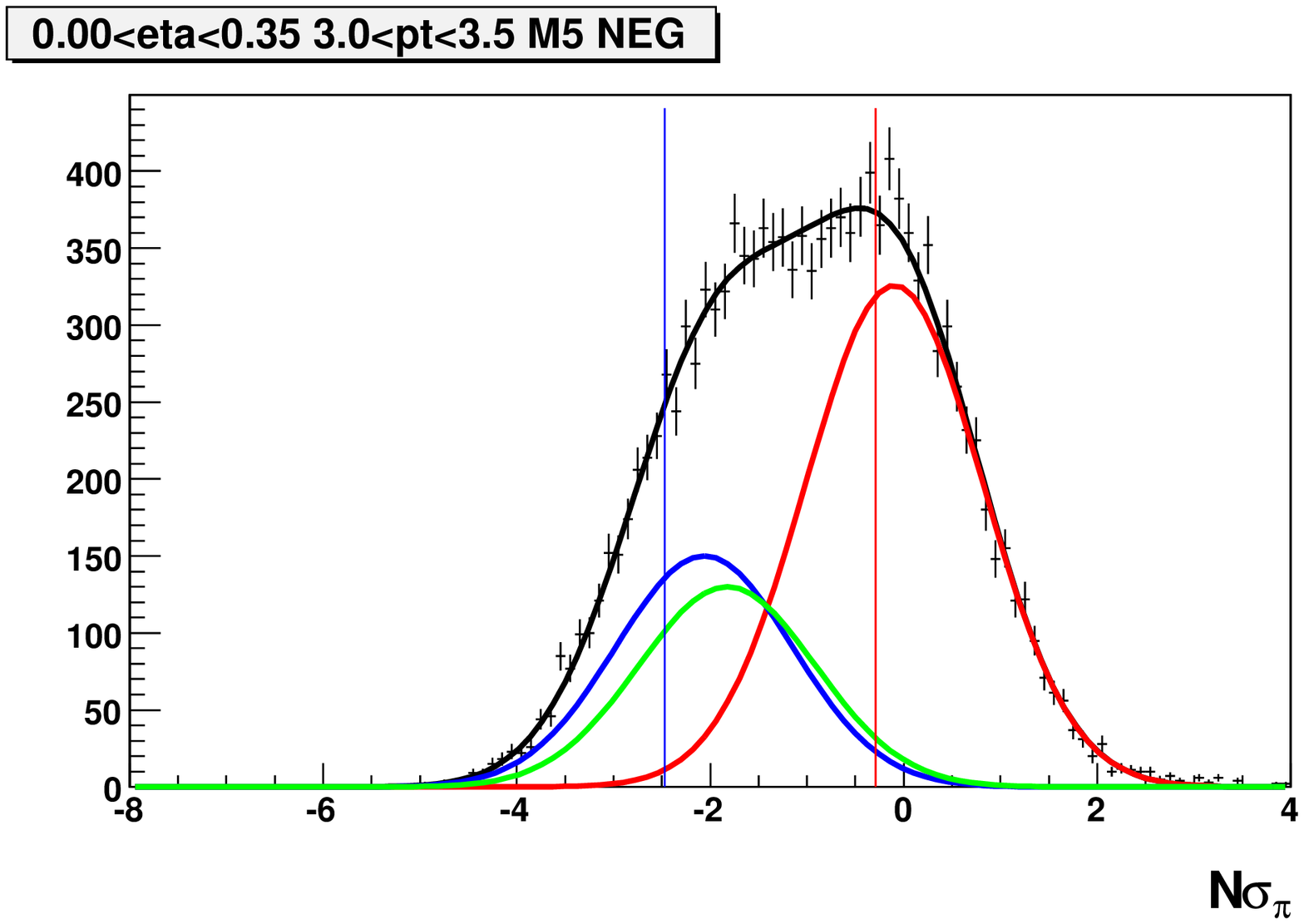}
		\includegraphics[width=1\textwidth]{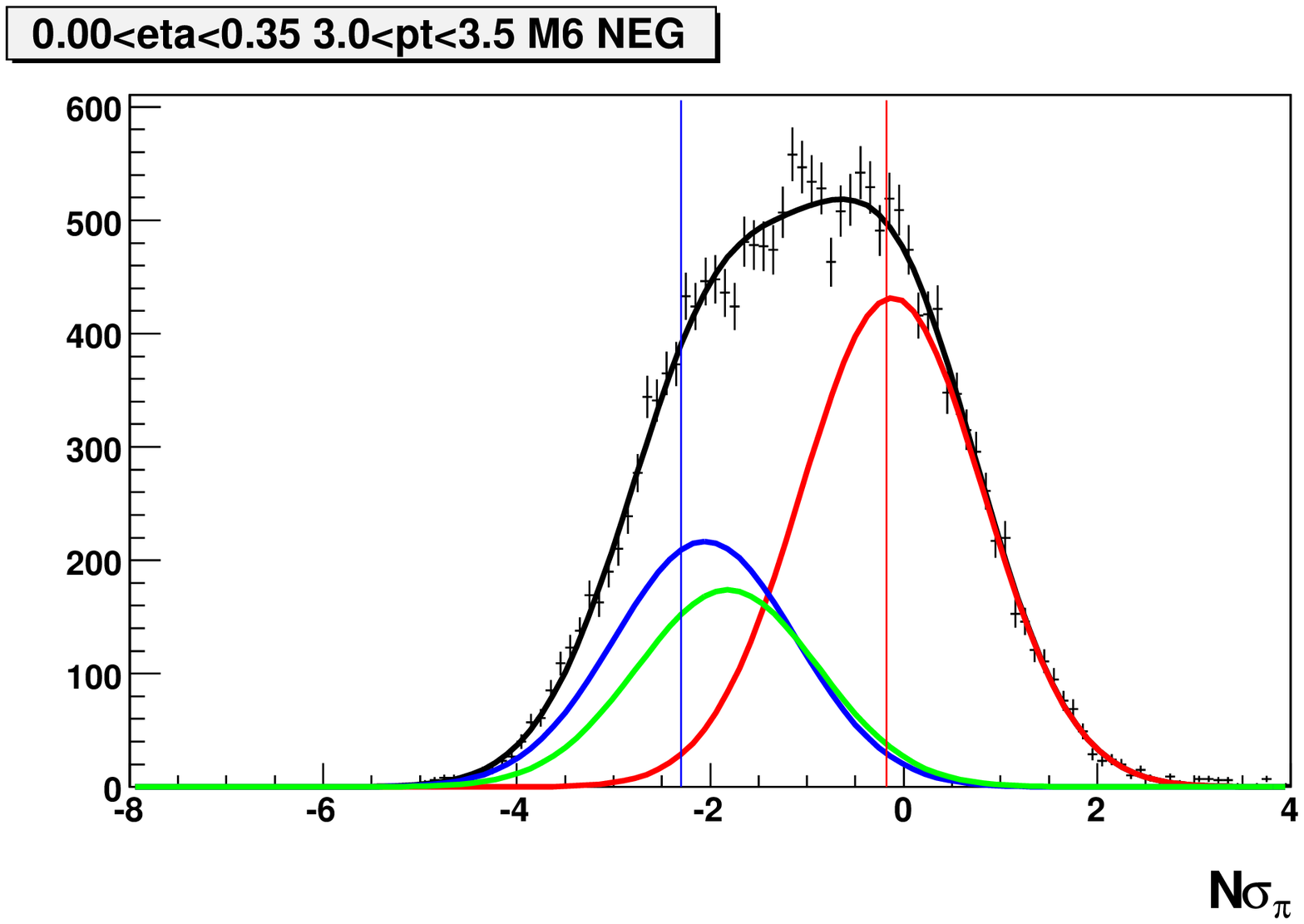}
		\includegraphics[width=1\textwidth]{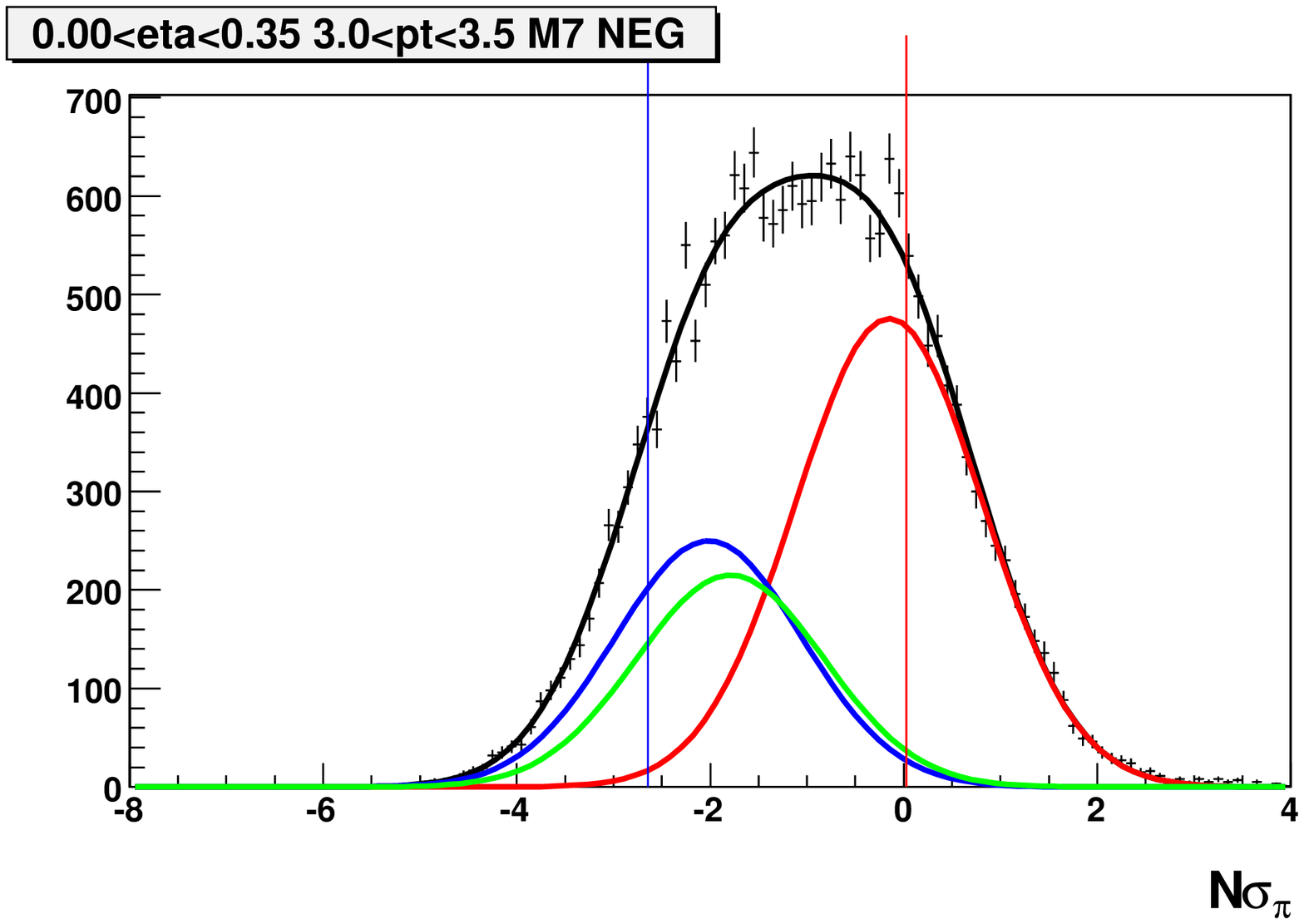}
		\includegraphics[width=1\textwidth]{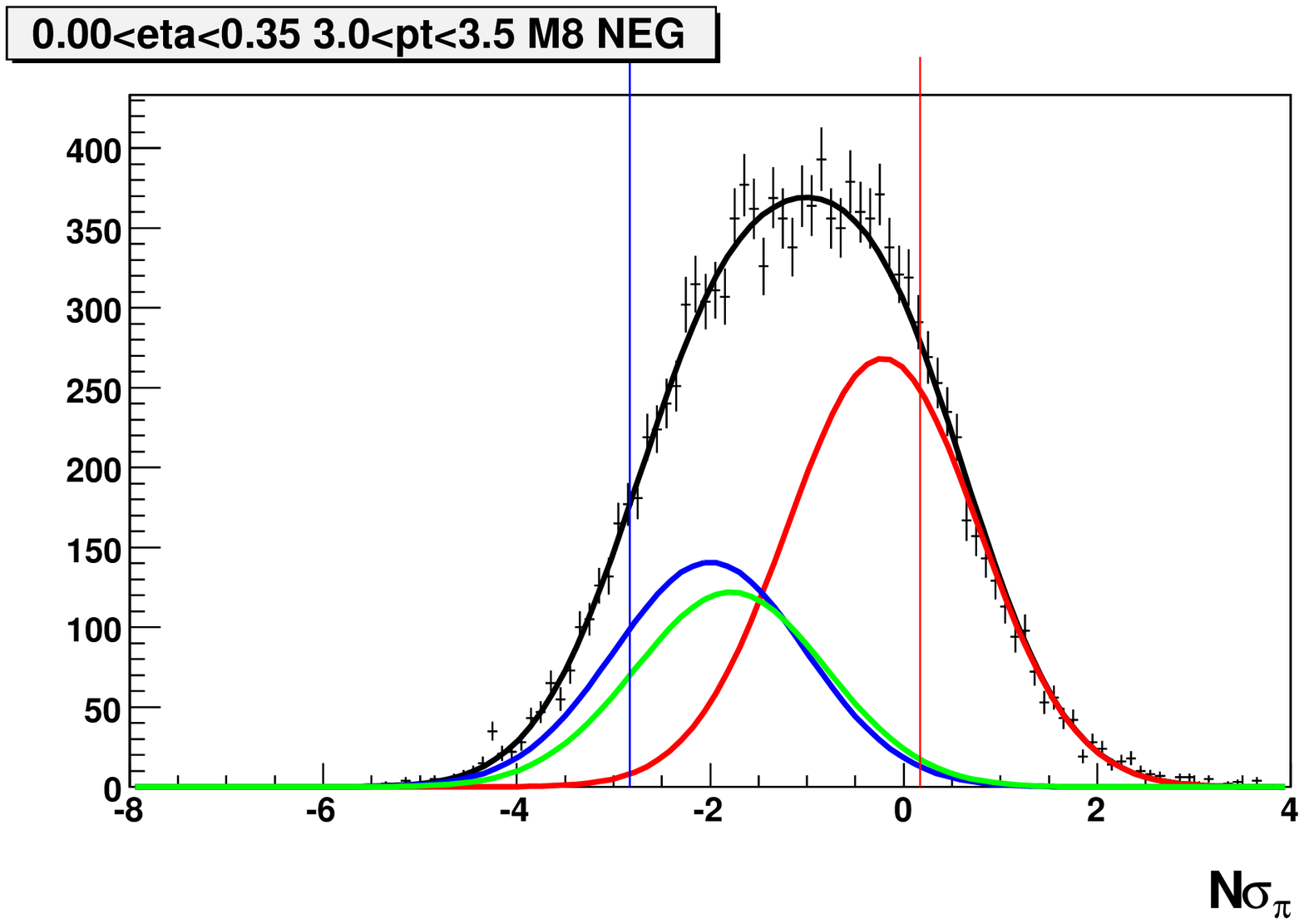}
		\includegraphics[width=1\textwidth]{Plots/cutN_009.eps}									
			\end{minipage}
\hfill
\begin{minipage}[t]{.19\textwidth}
	\centering
		\includegraphics[width=1\textwidth]{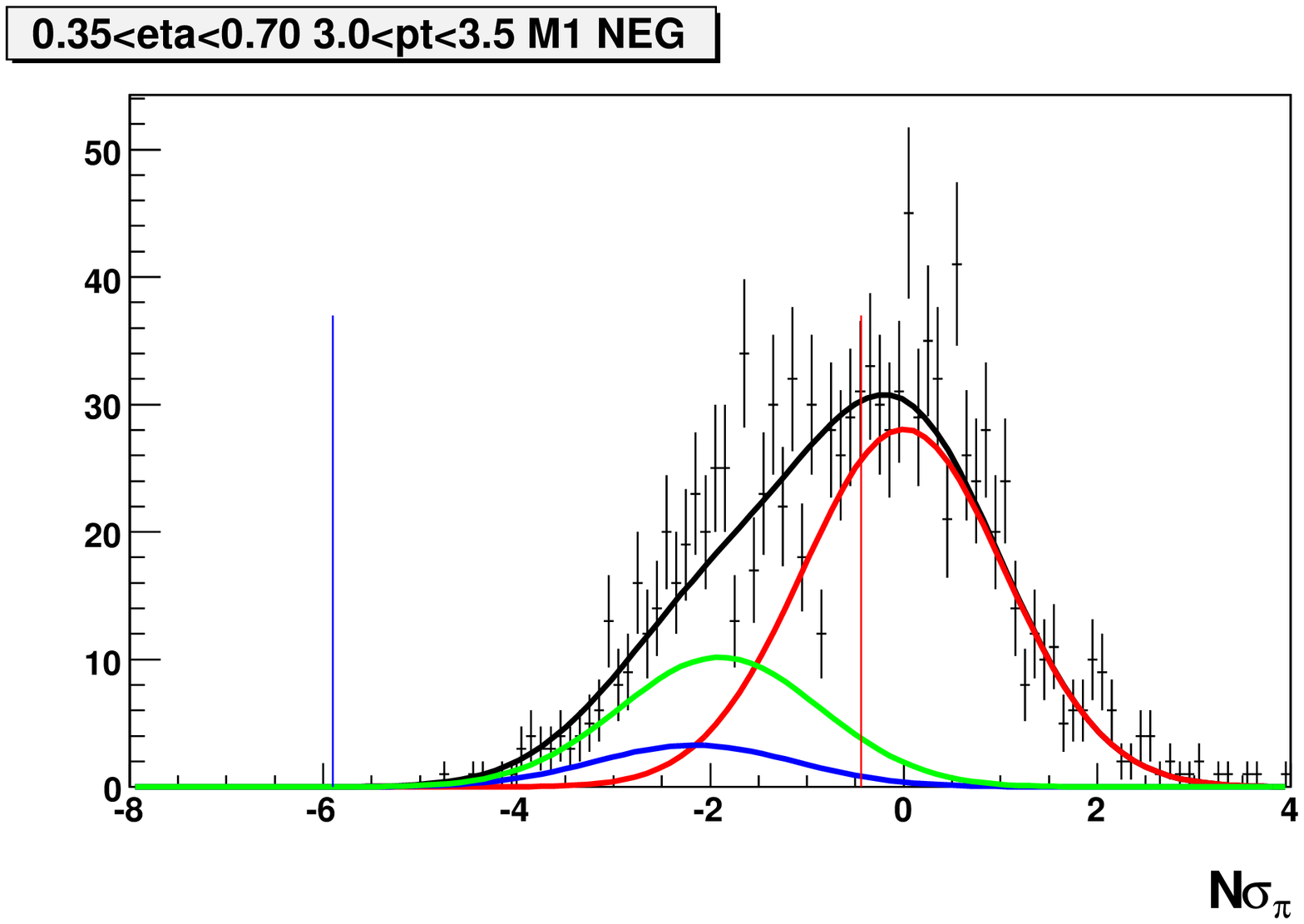}
		\includegraphics[width=1\textwidth]{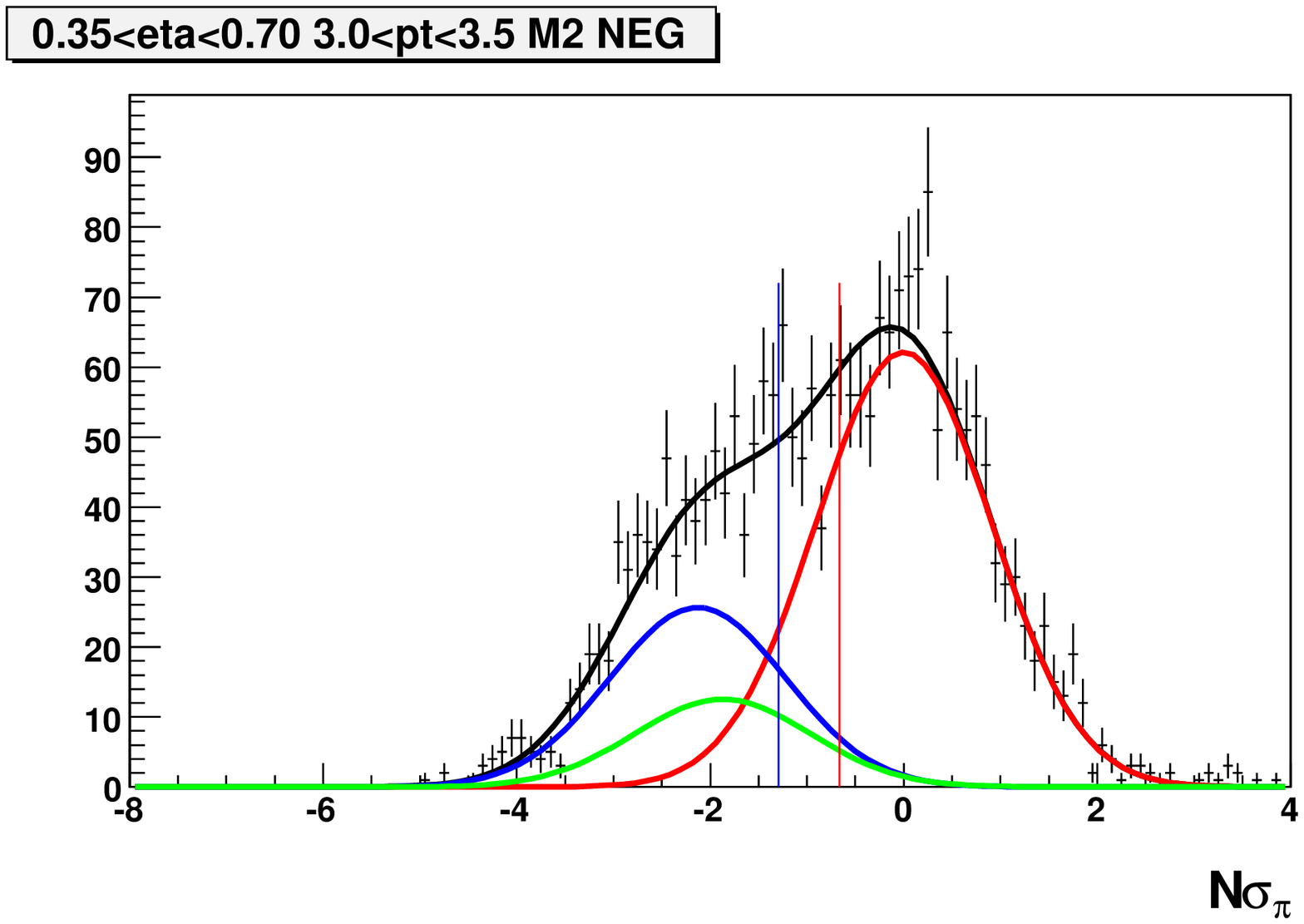}
		\includegraphics[width=1\textwidth]{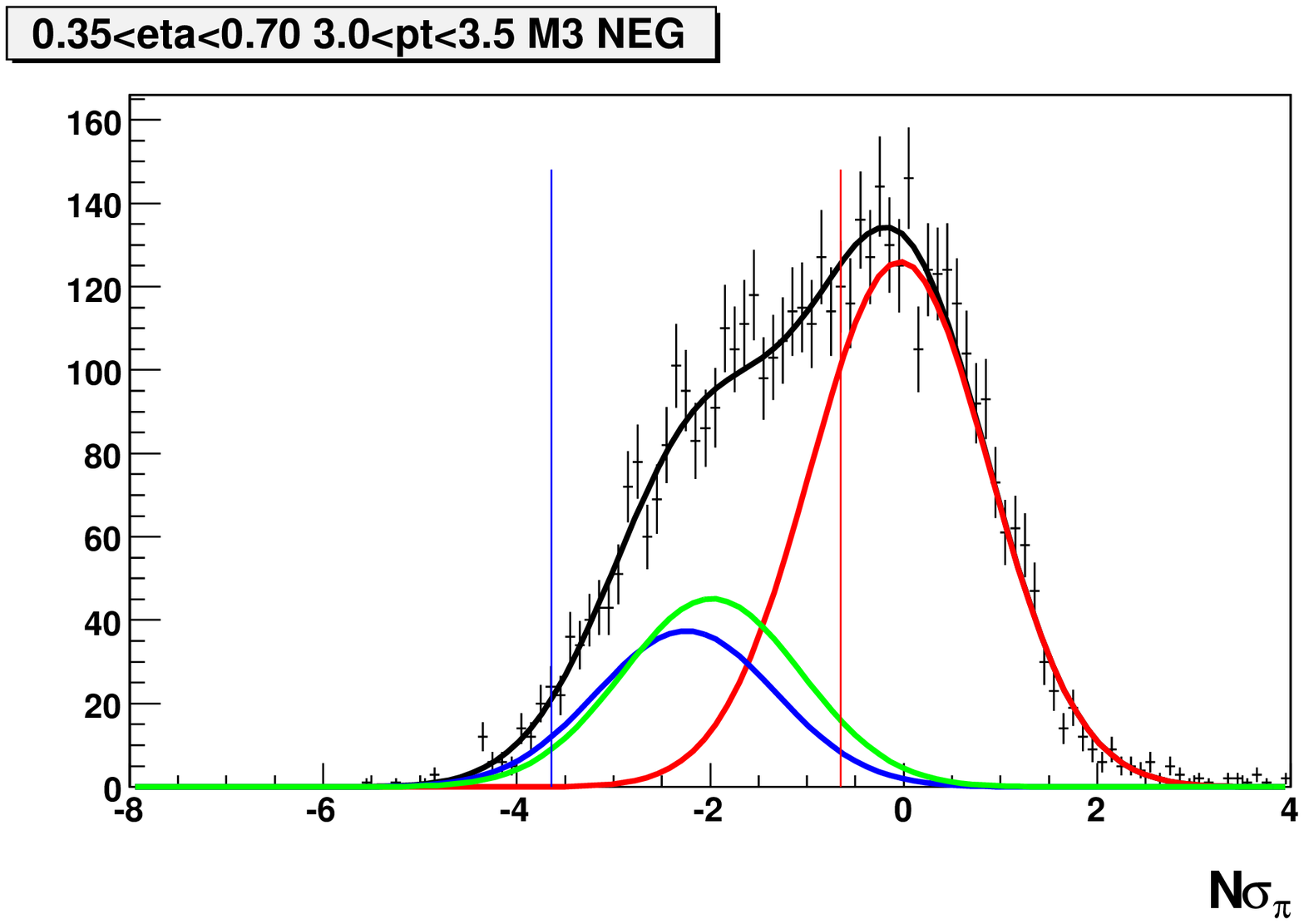}
		\includegraphics[width=1\textwidth]{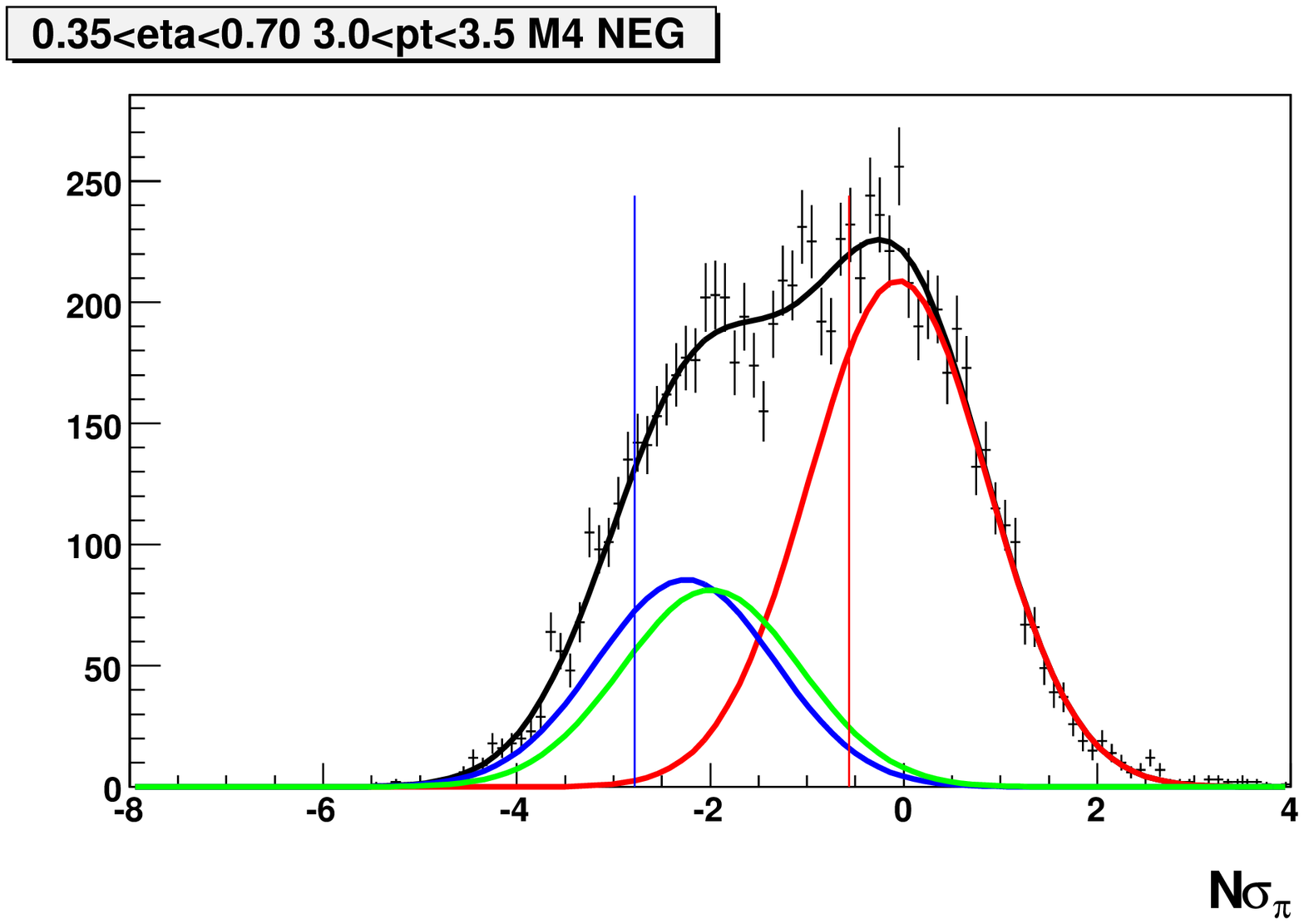}
		\includegraphics[width=1\textwidth]{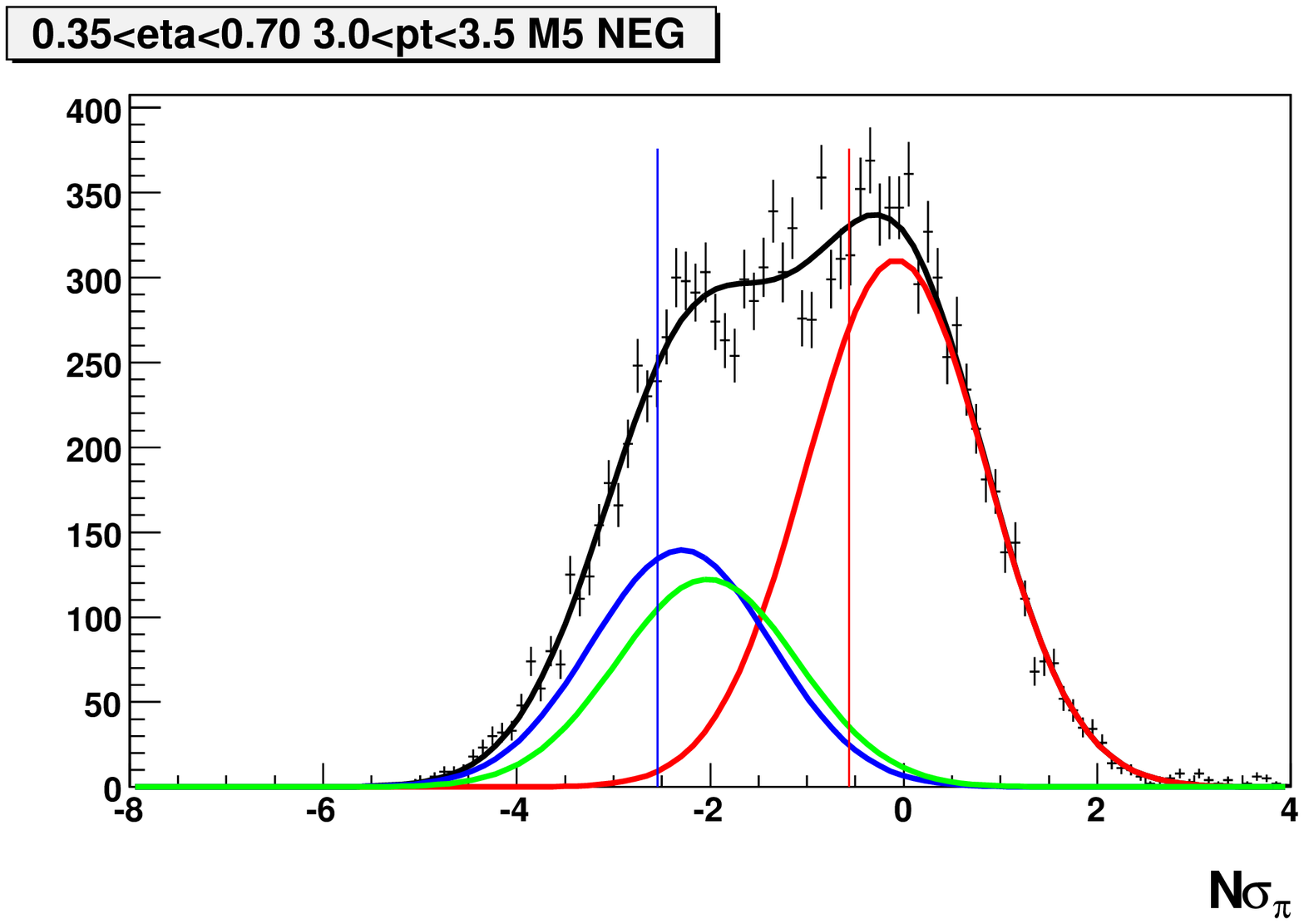}
		\includegraphics[width=1\textwidth]{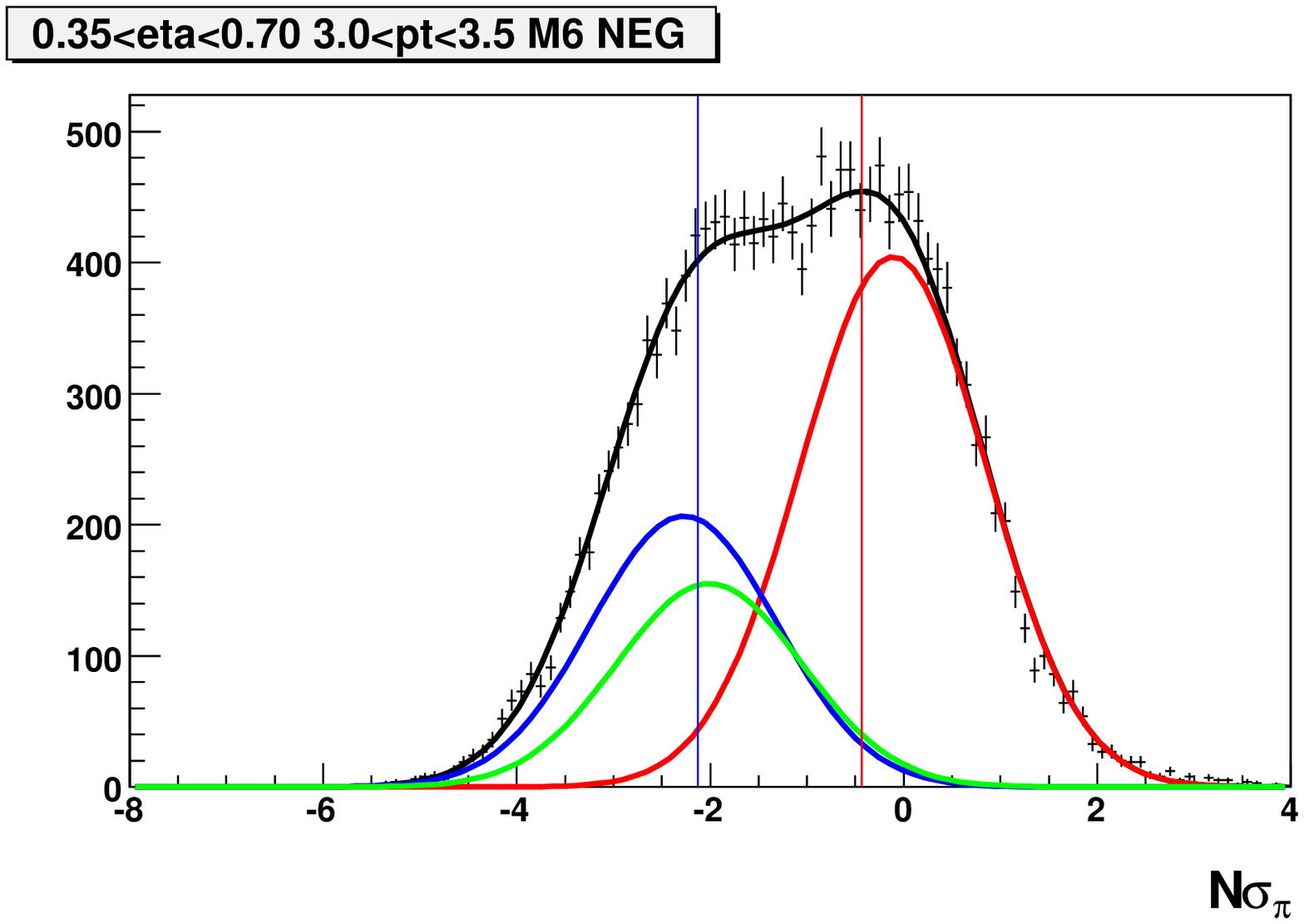}
		\includegraphics[width=1\textwidth]{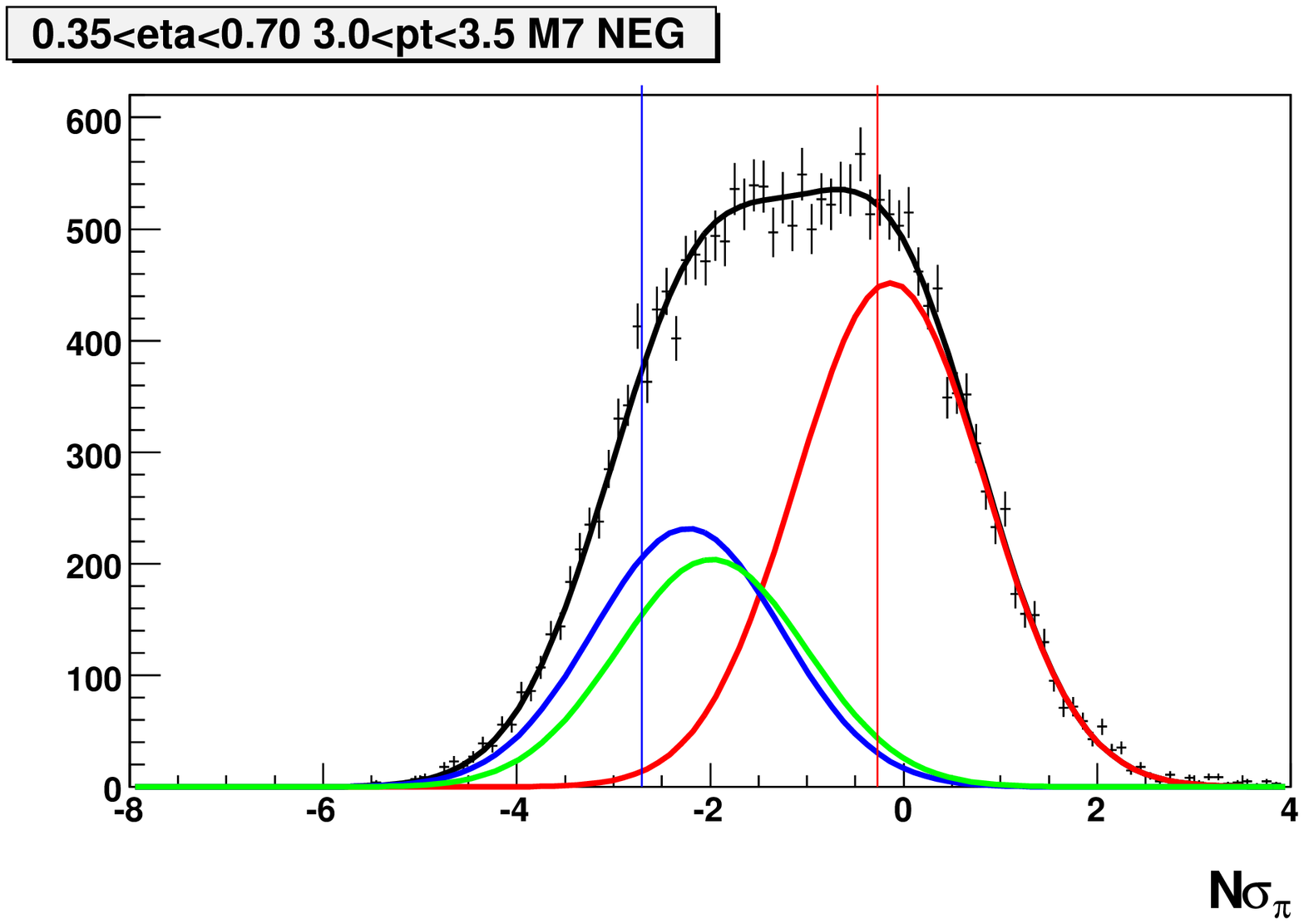}
		\includegraphics[width=1\textwidth]{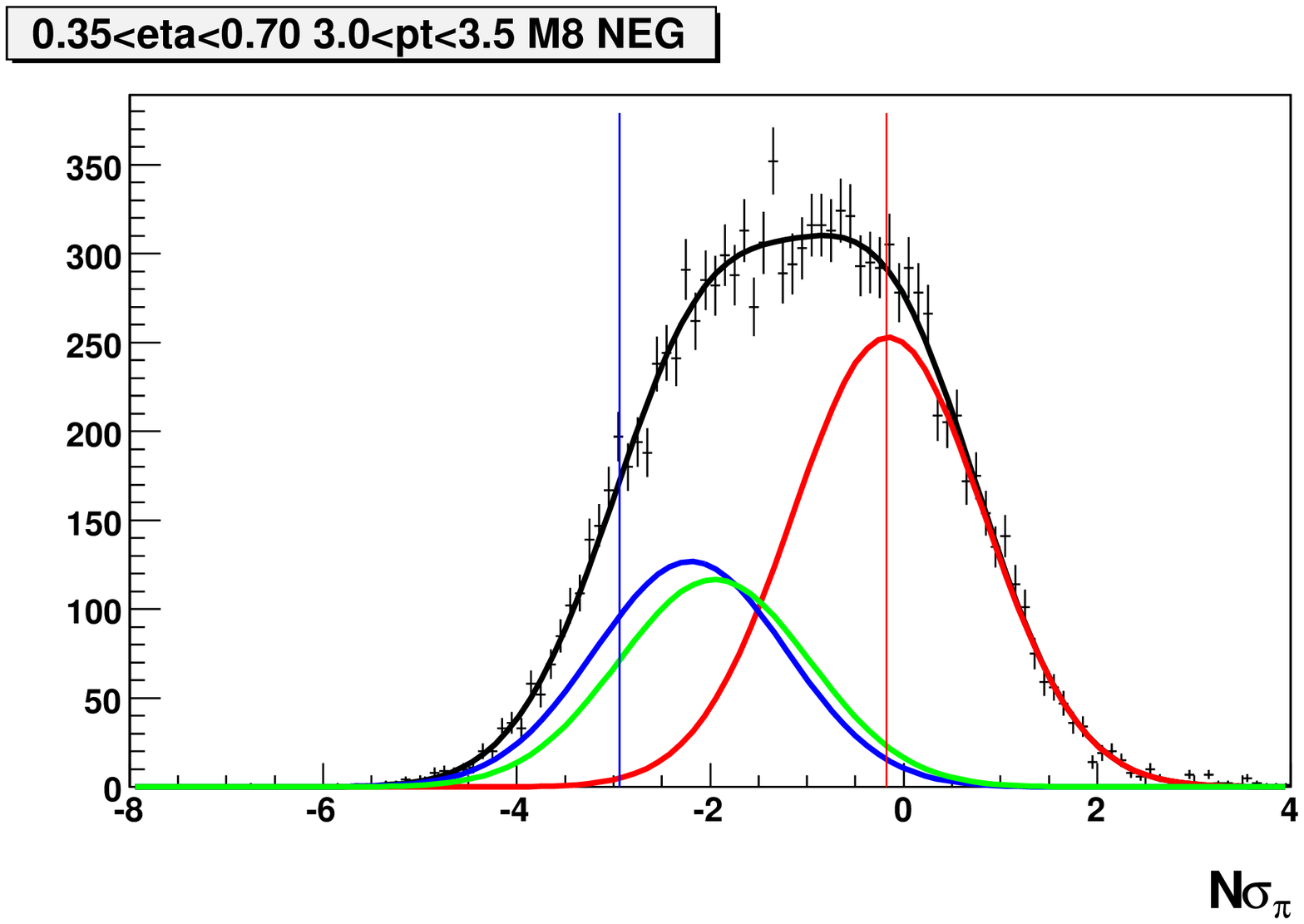}
		\includegraphics[width=1\textwidth]{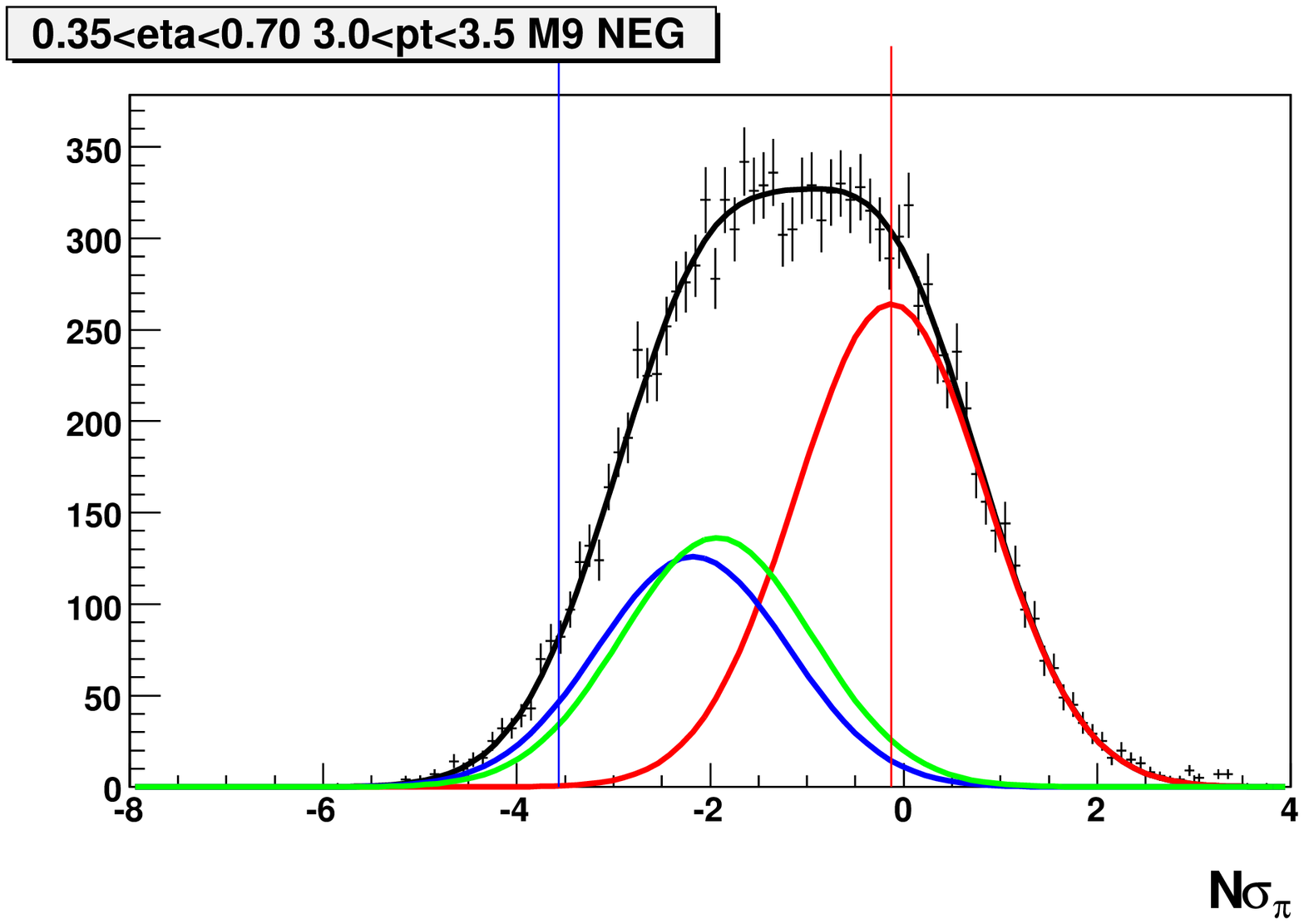}									
			\end{minipage}
\hfill
\begin{minipage}[t]{.19\textwidth}
	\centering
		\includegraphics[width=1\textwidth]{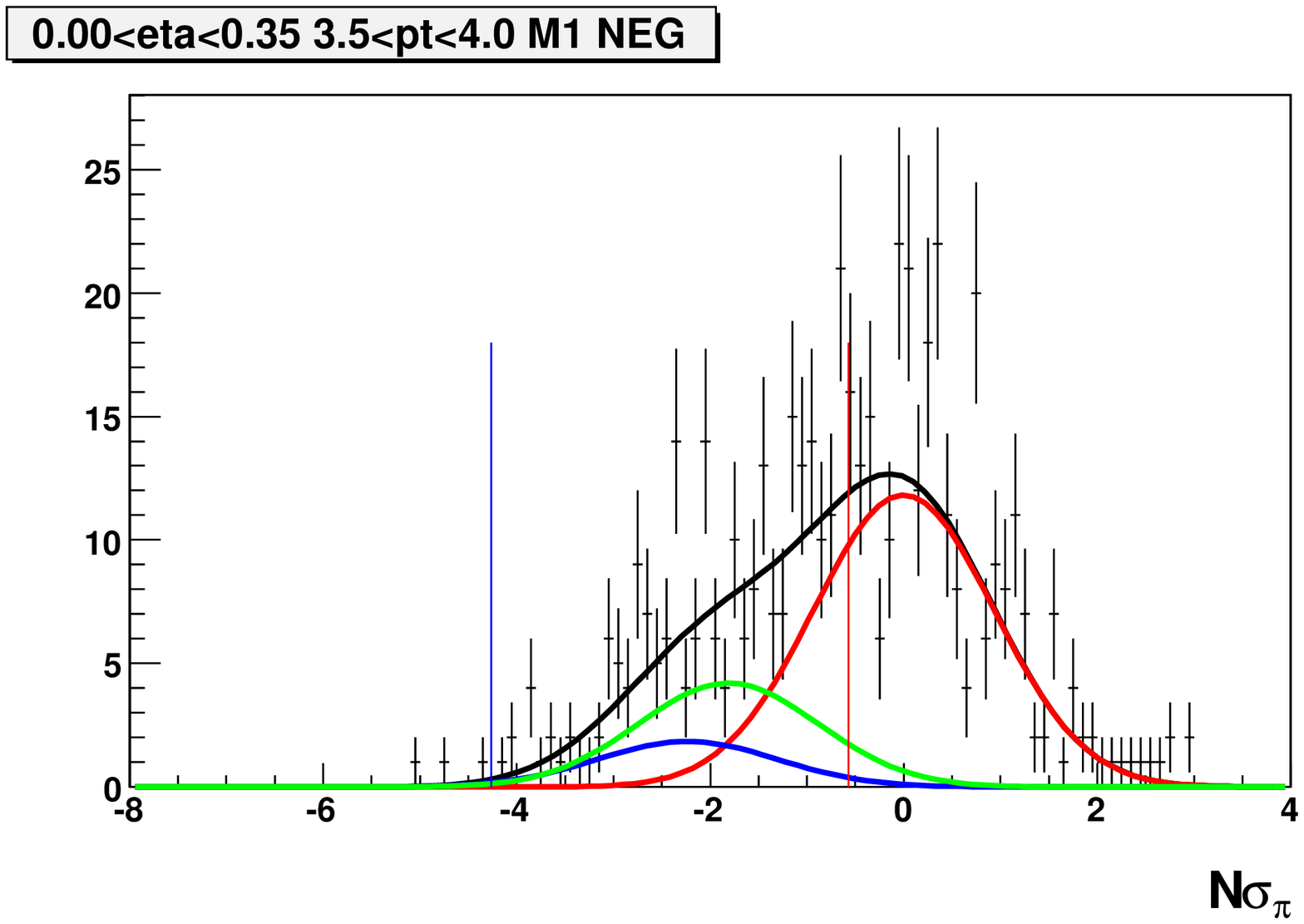}
		\includegraphics[width=1\textwidth]{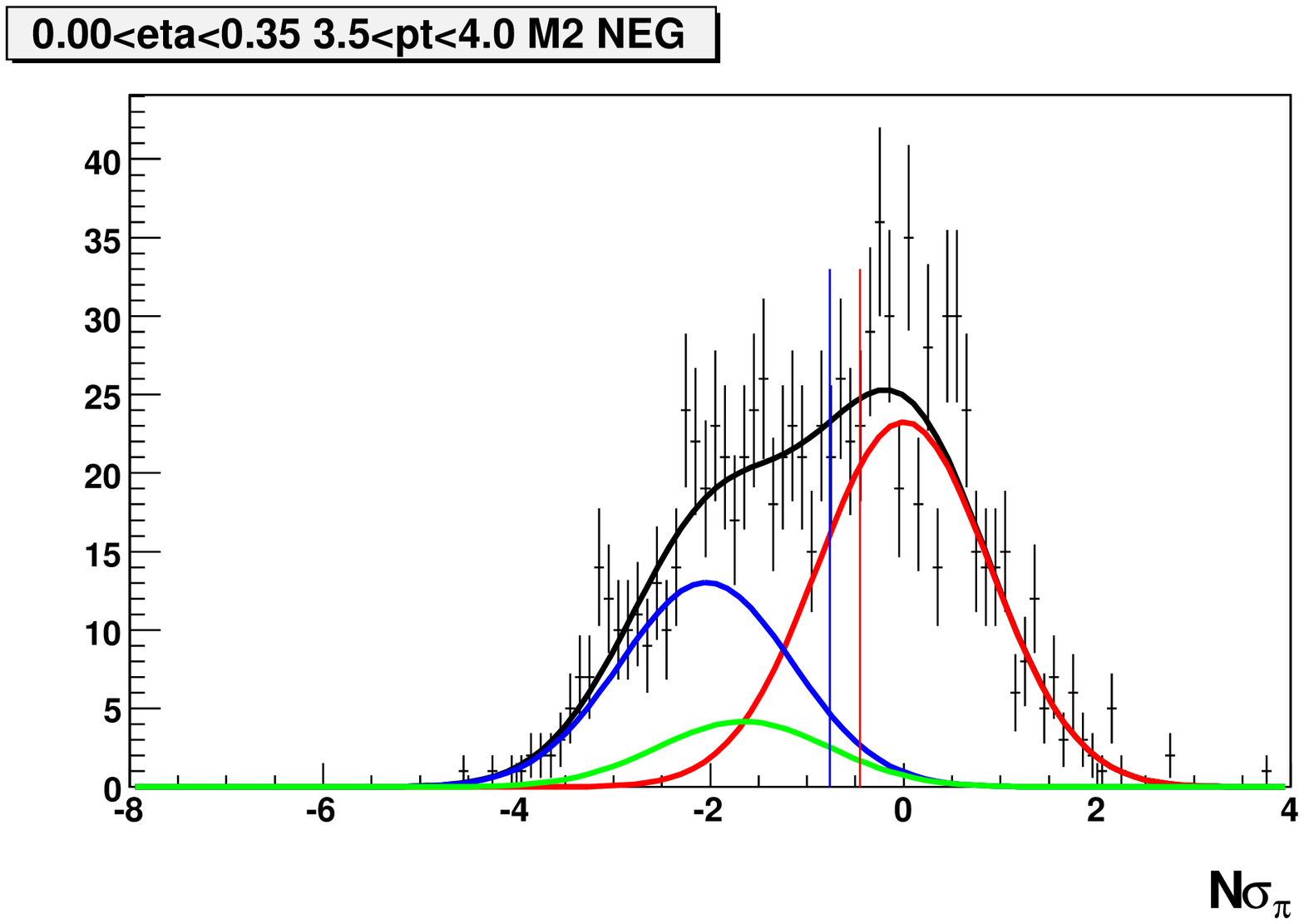}
		\includegraphics[width=1\textwidth]{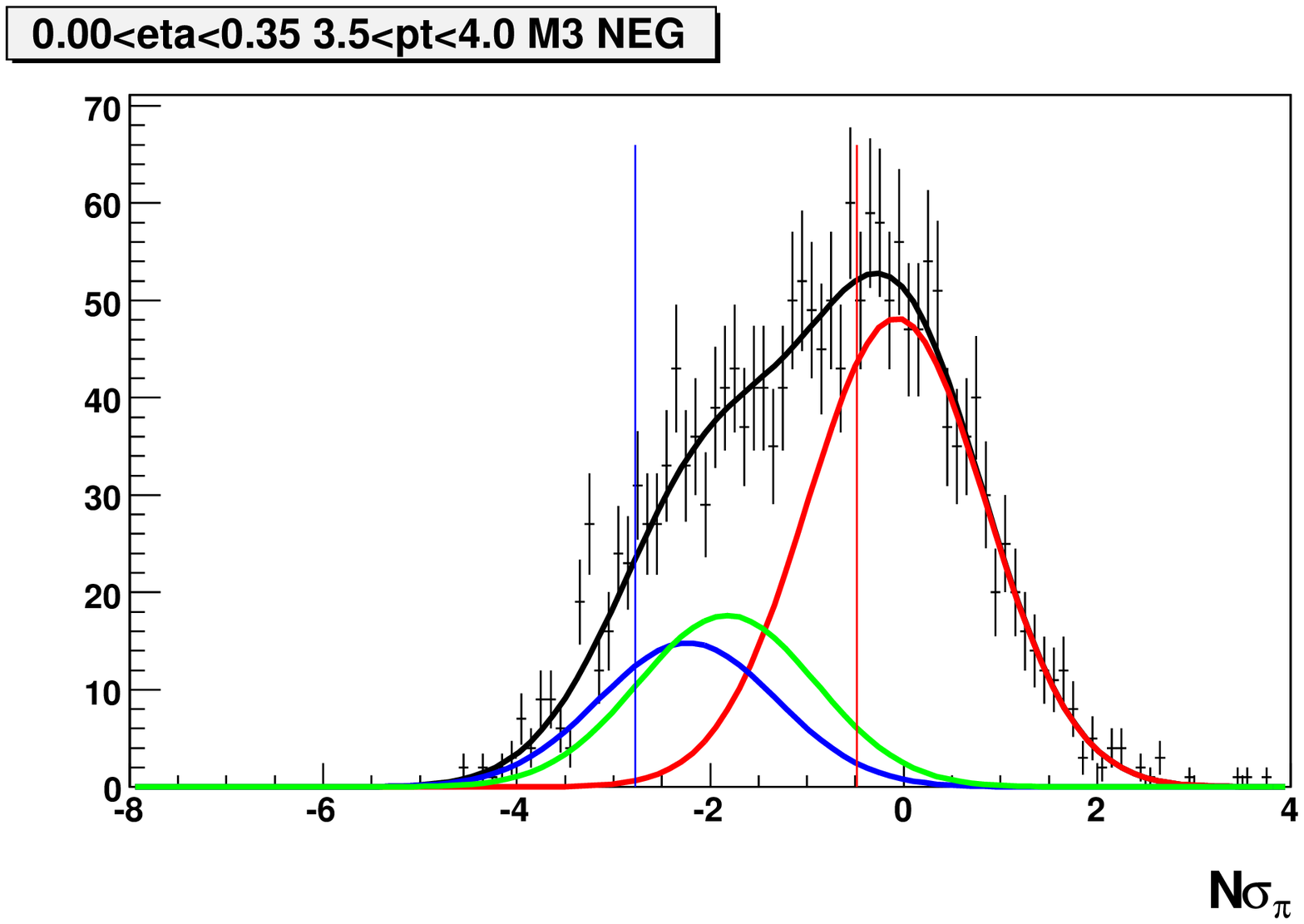}
		\includegraphics[width=1\textwidth]{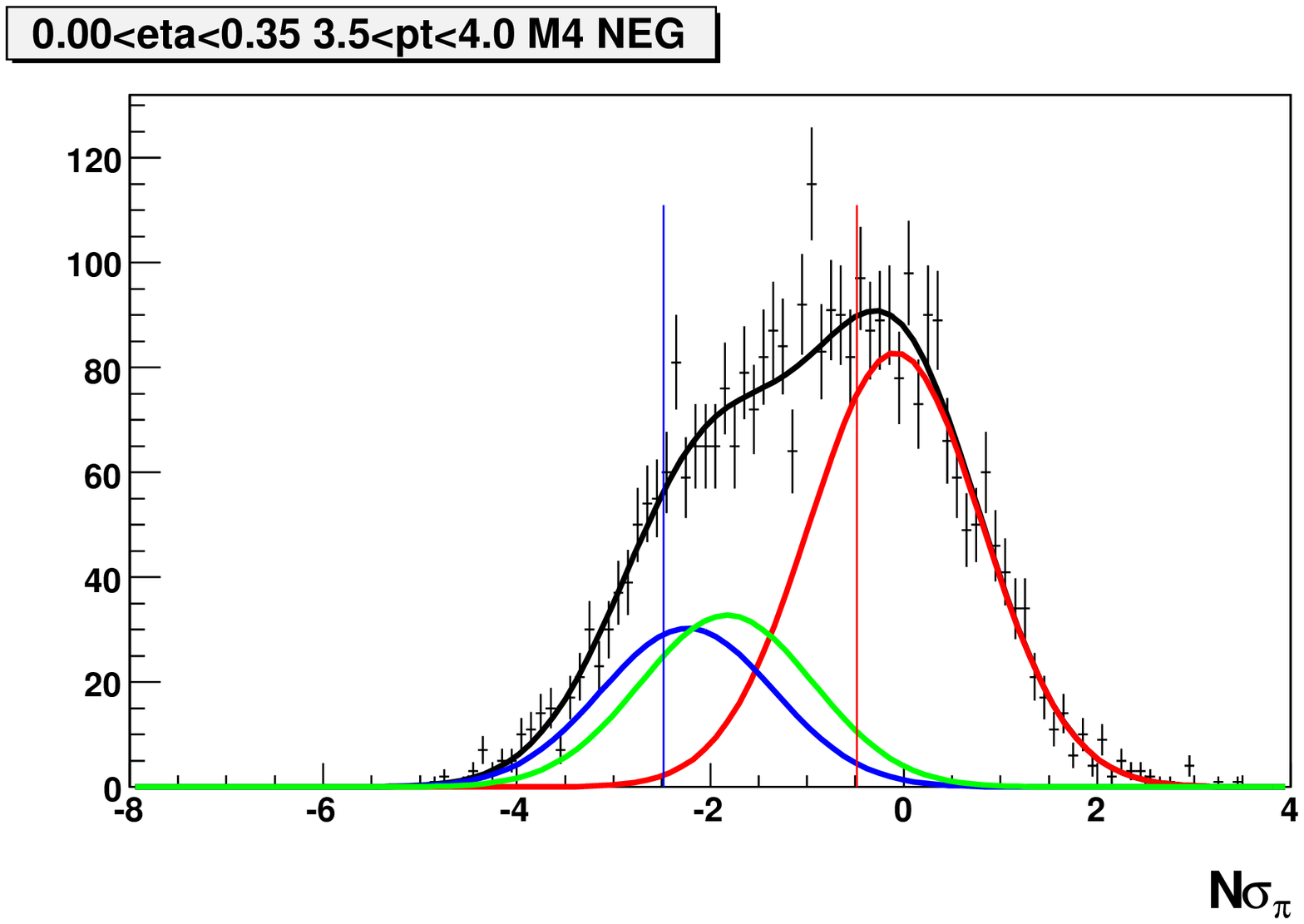}
		\includegraphics[width=1\textwidth]{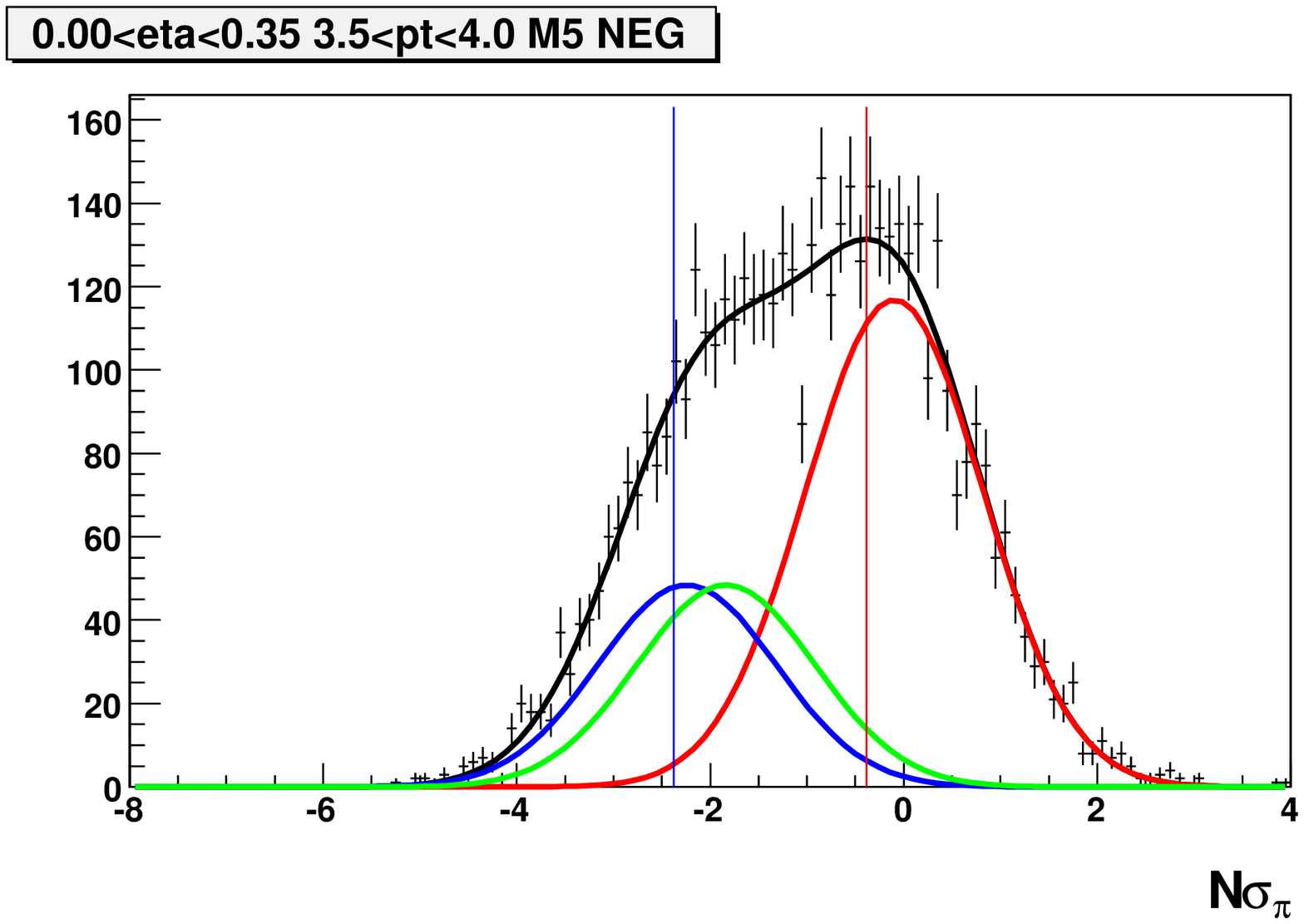}
		\includegraphics[width=1\textwidth]{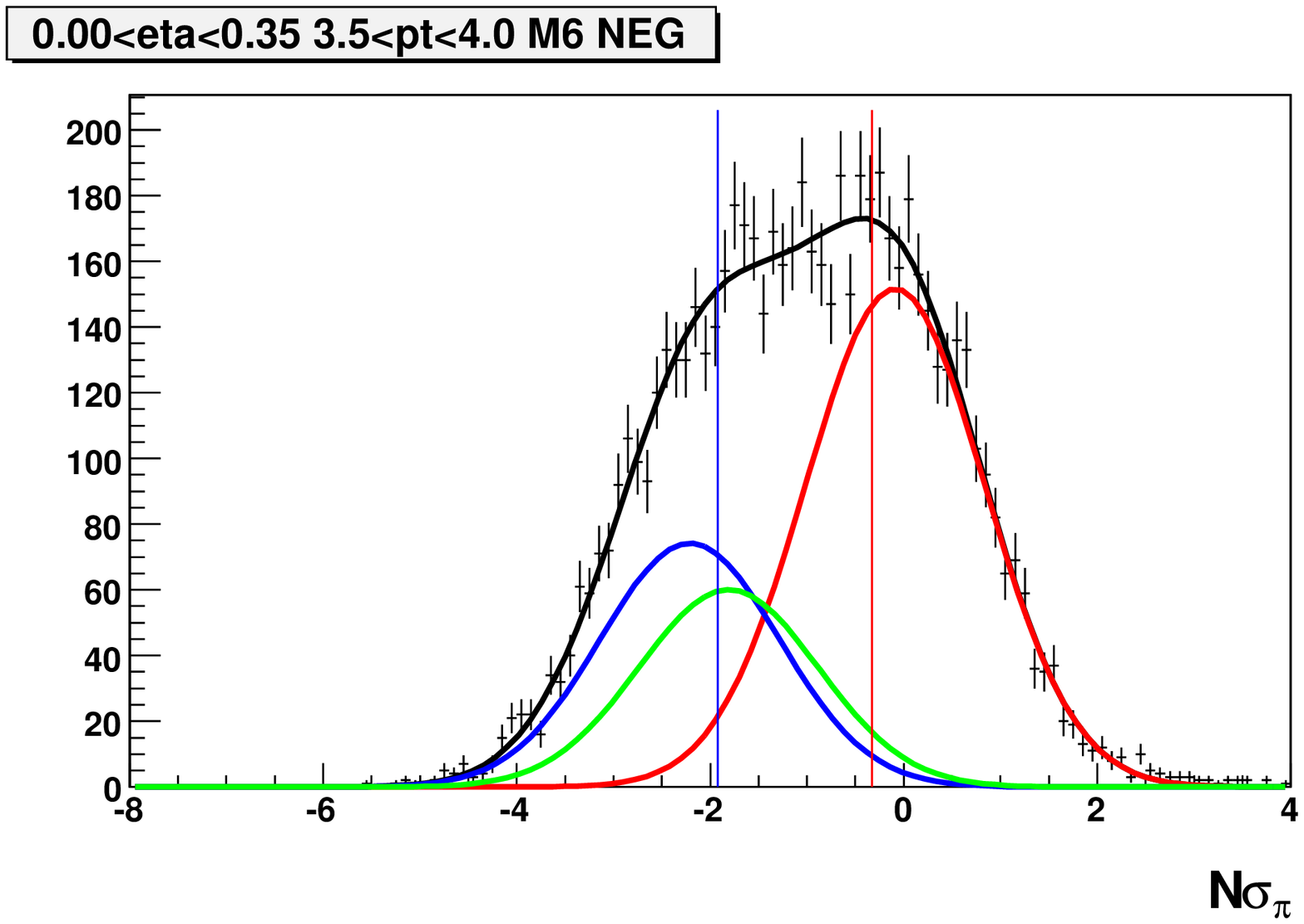}
		\includegraphics[width=1\textwidth]{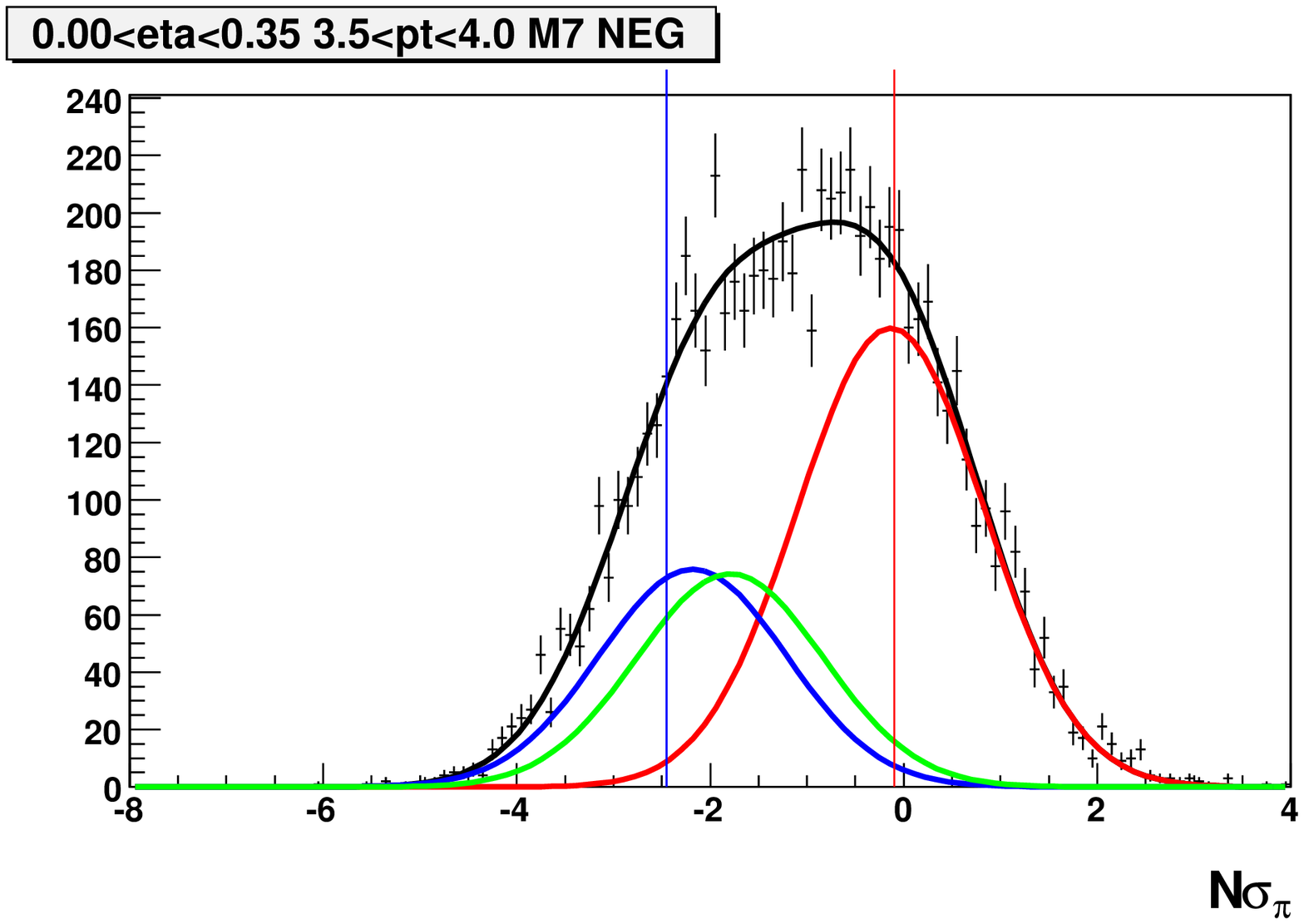}
		\includegraphics[width=1\textwidth]{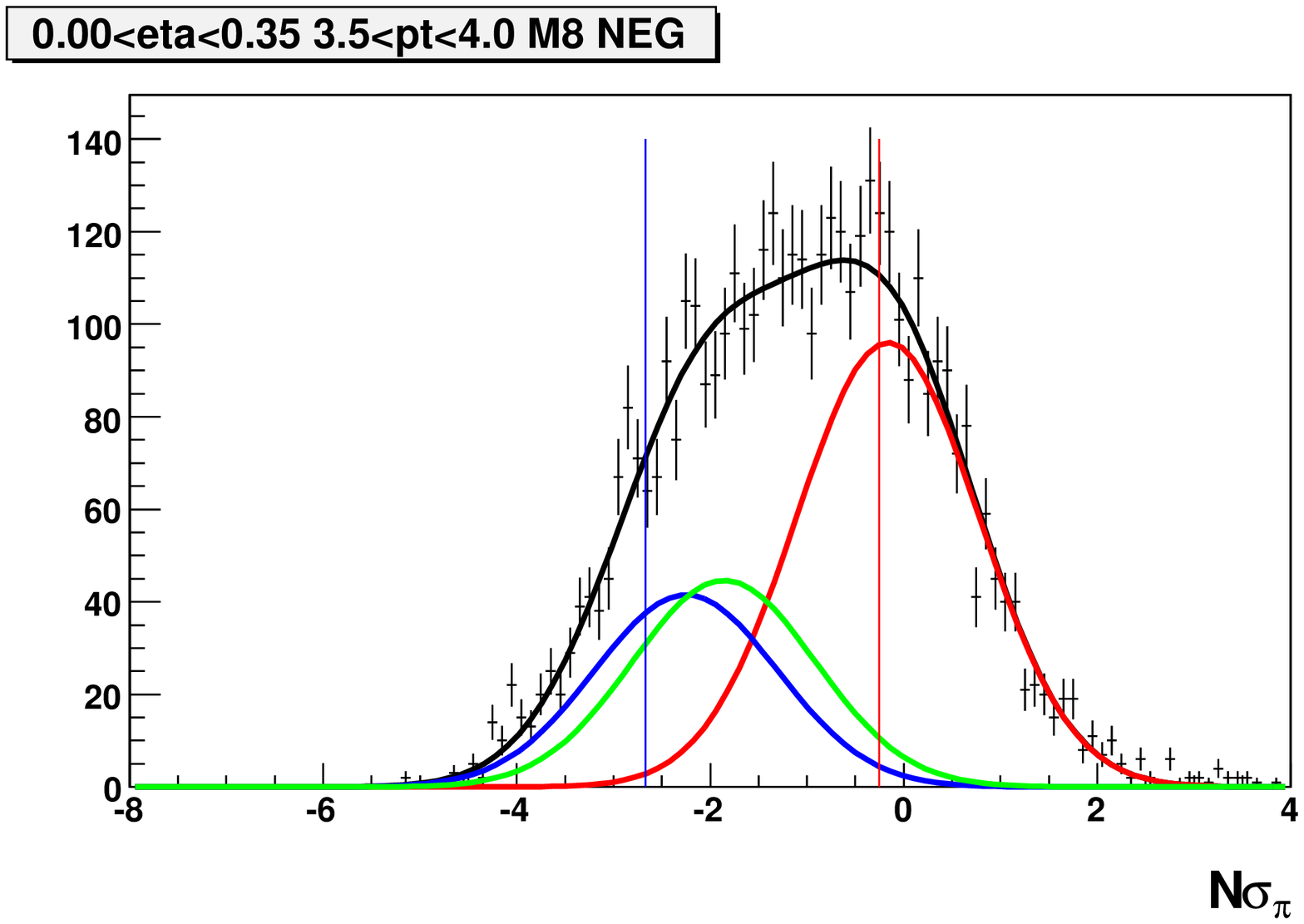}
		\includegraphics[width=1\textwidth]{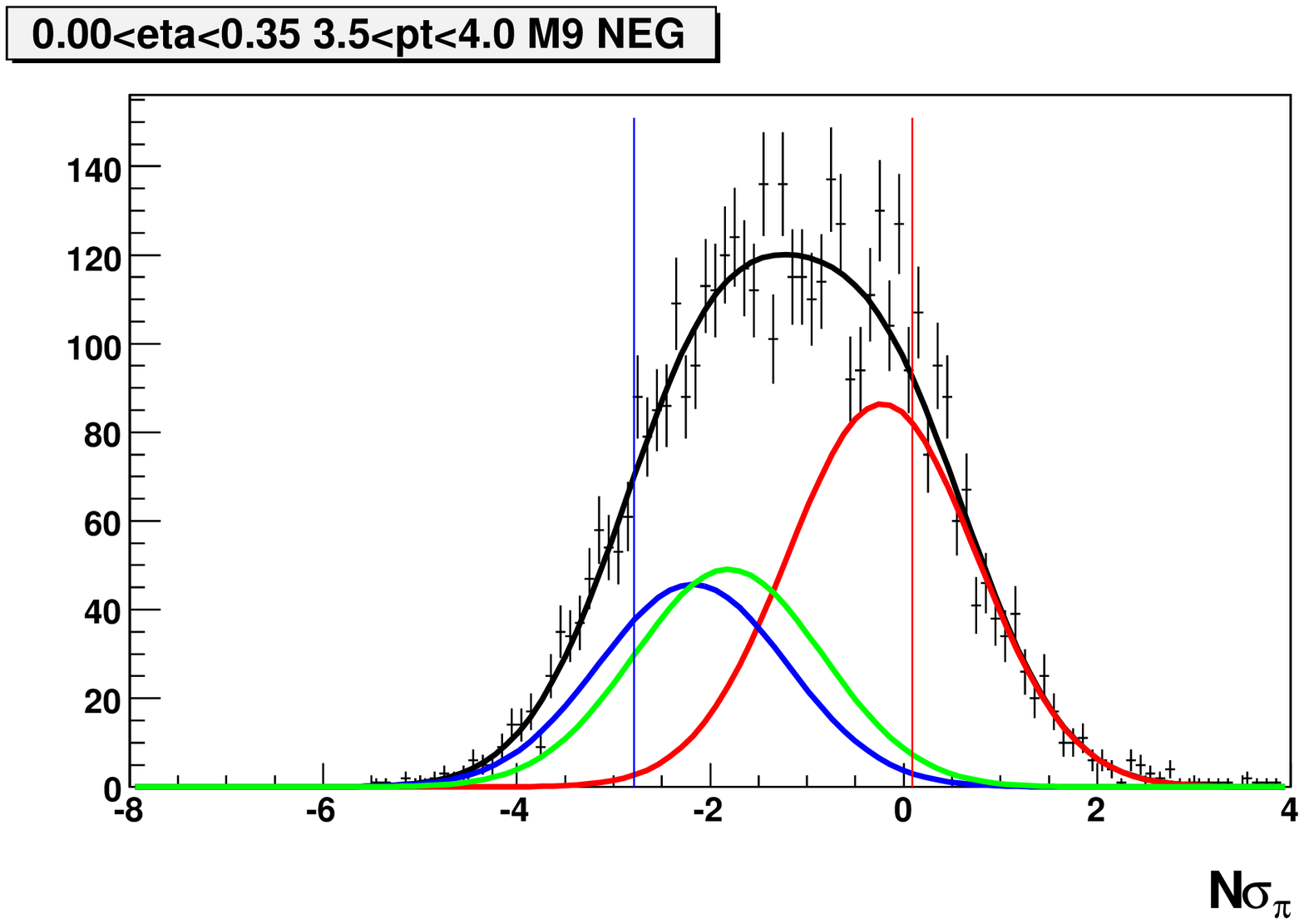}
										
			\end{minipage}
\hfill
\begin{minipage}[t]{.19\textwidth}
	\centering
		\includegraphics[width=1\textwidth]{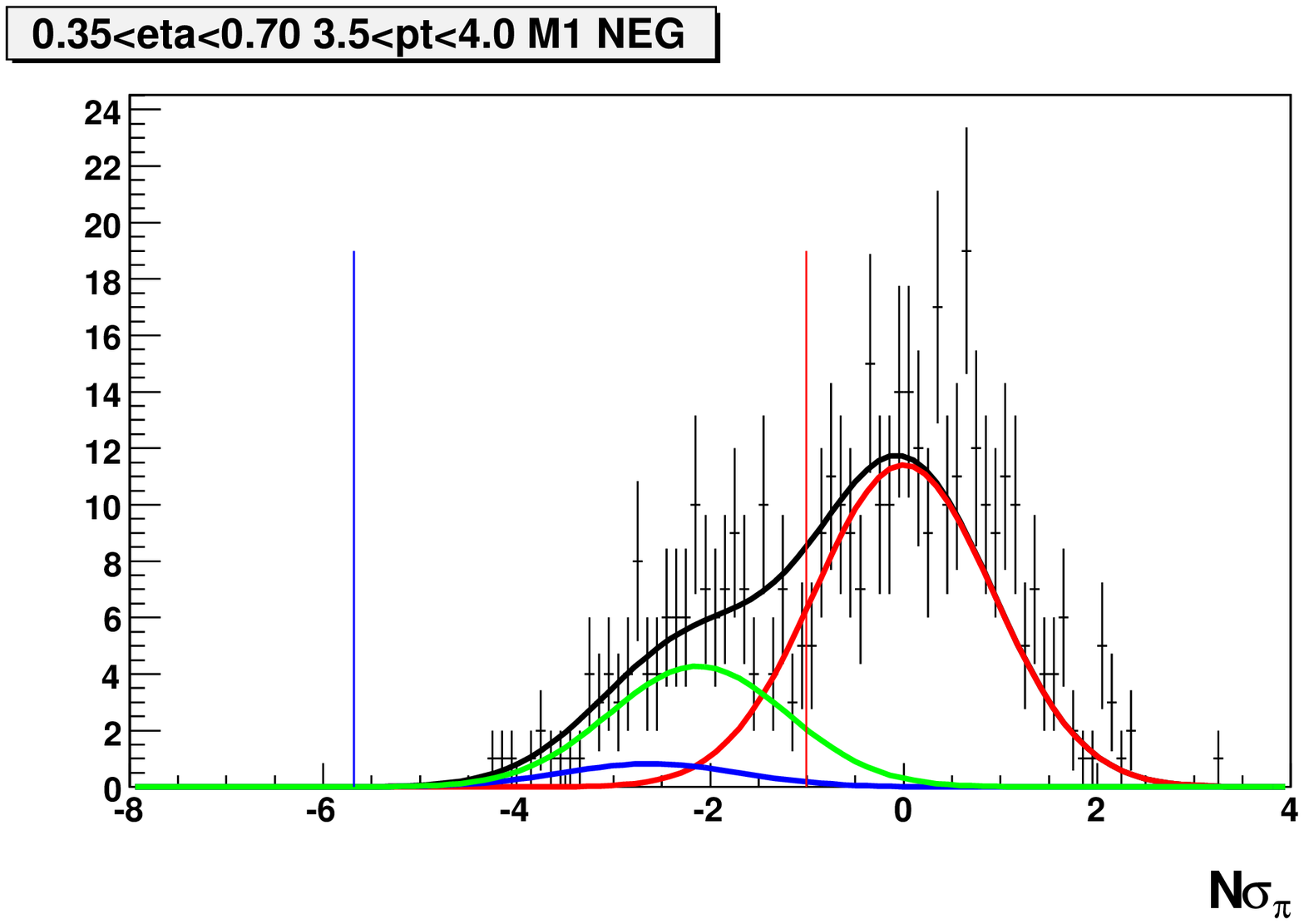}
		\includegraphics[width=1\textwidth]{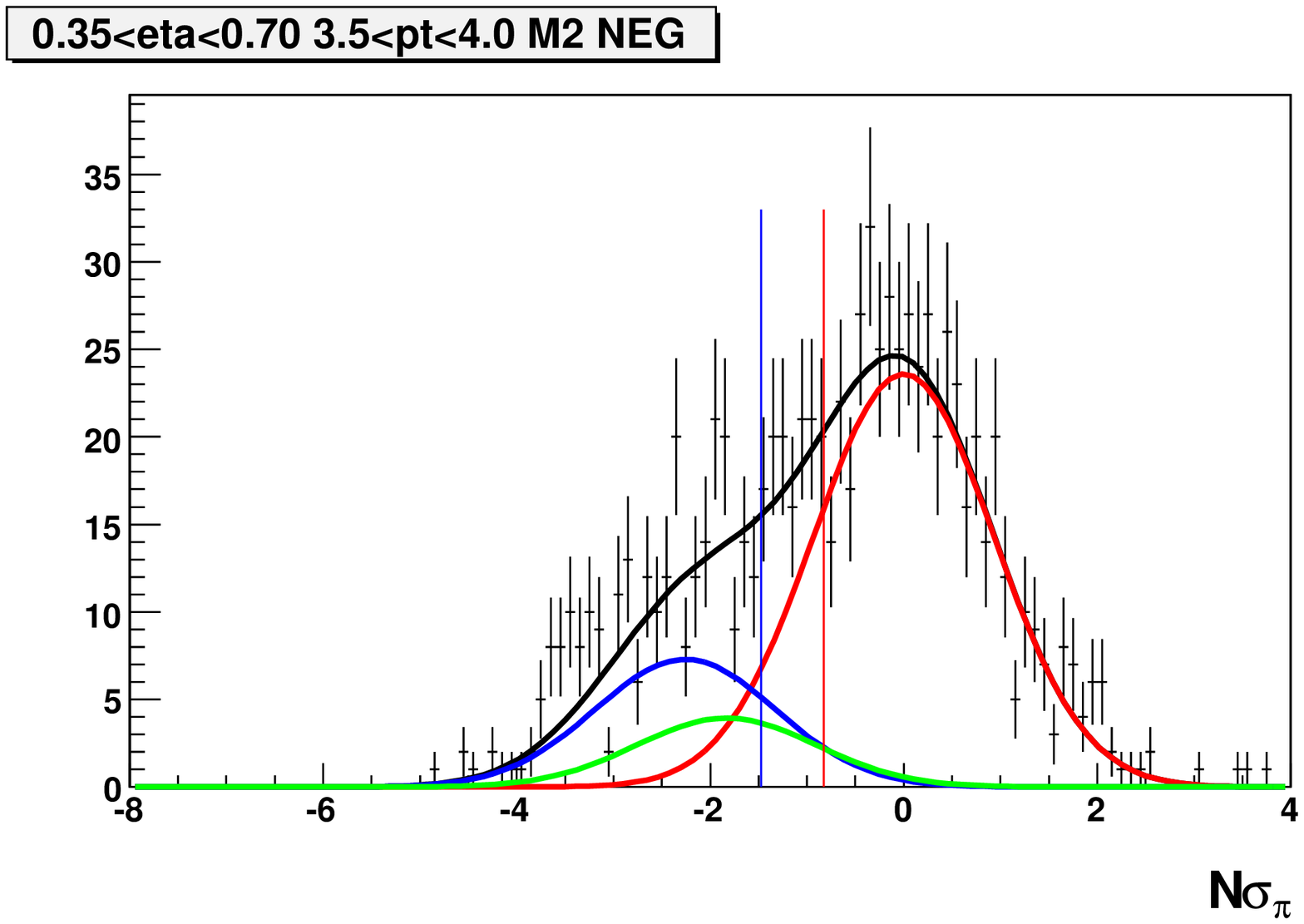}
		\includegraphics[width=1\textwidth]{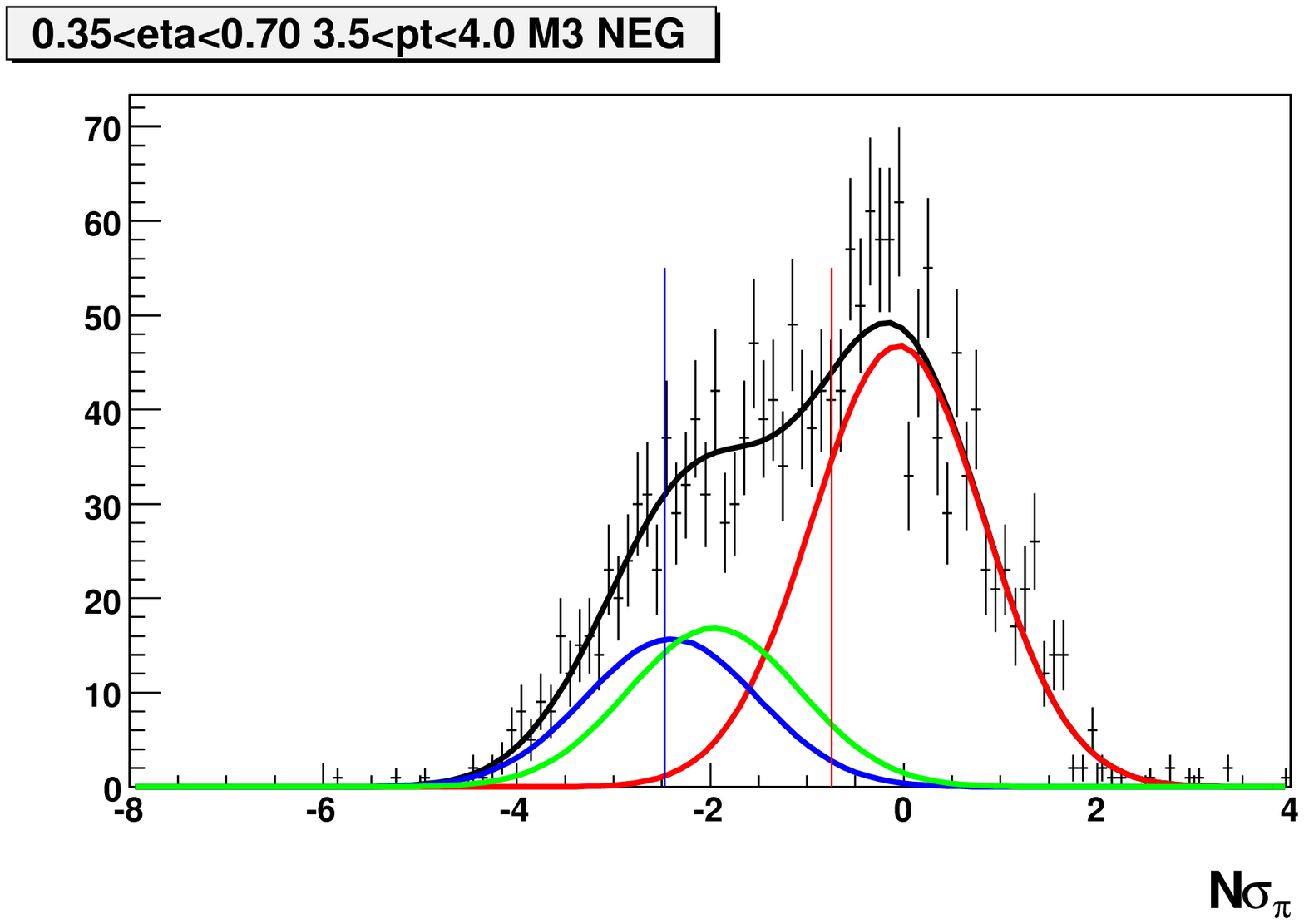}
		\includegraphics[width=1\textwidth]{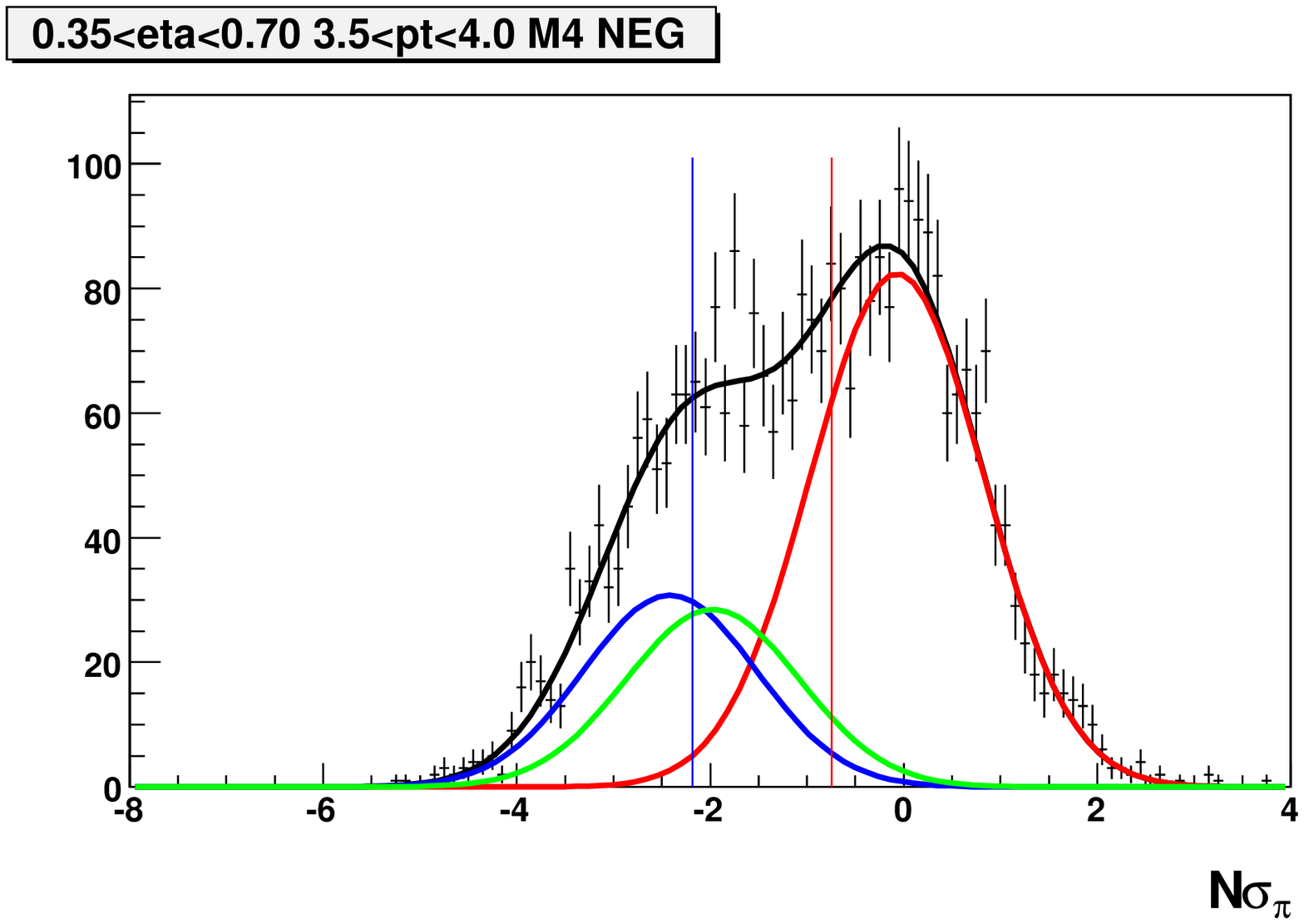}
		\includegraphics[width=1\textwidth]{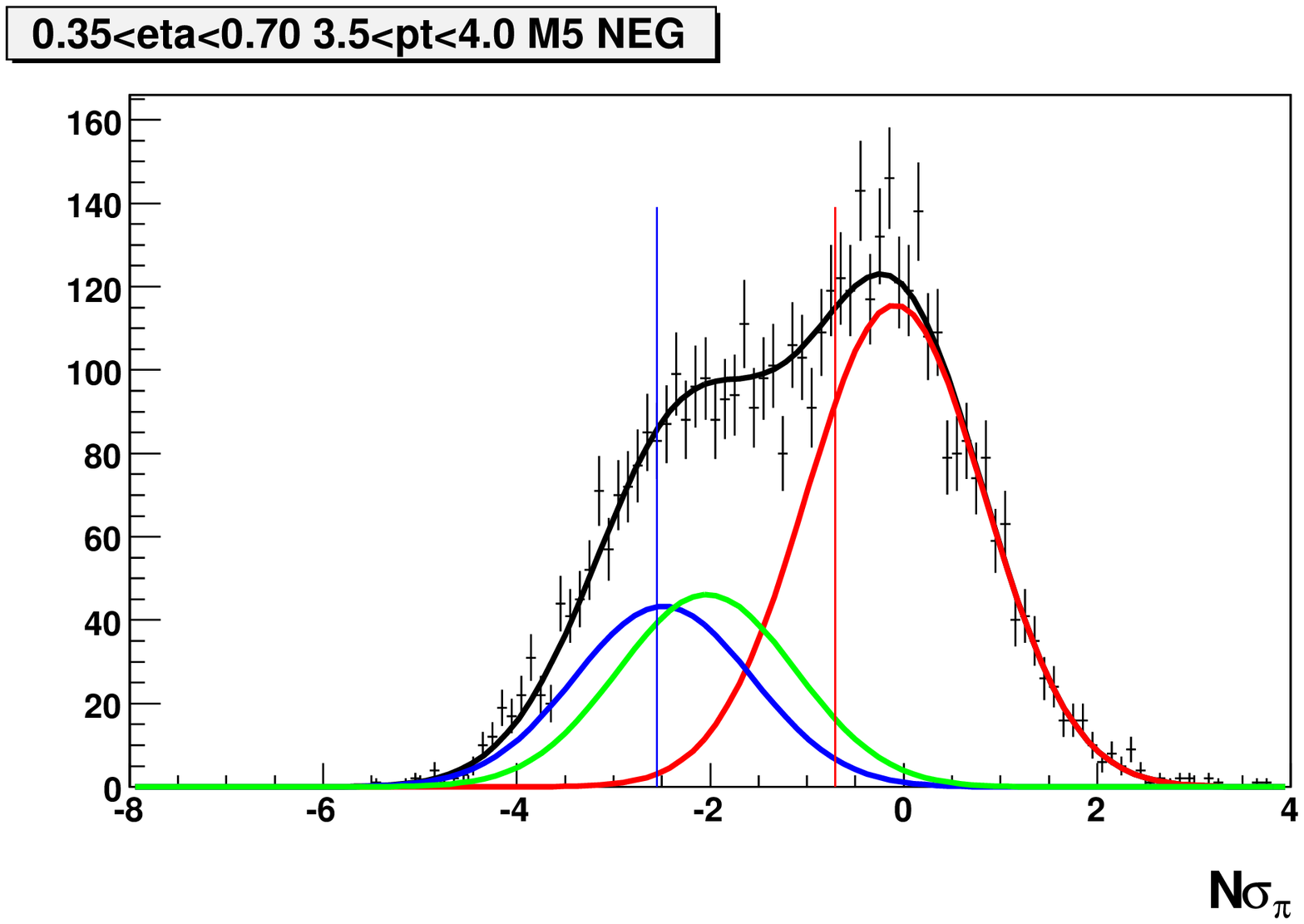}
		\includegraphics[width=1\textwidth]{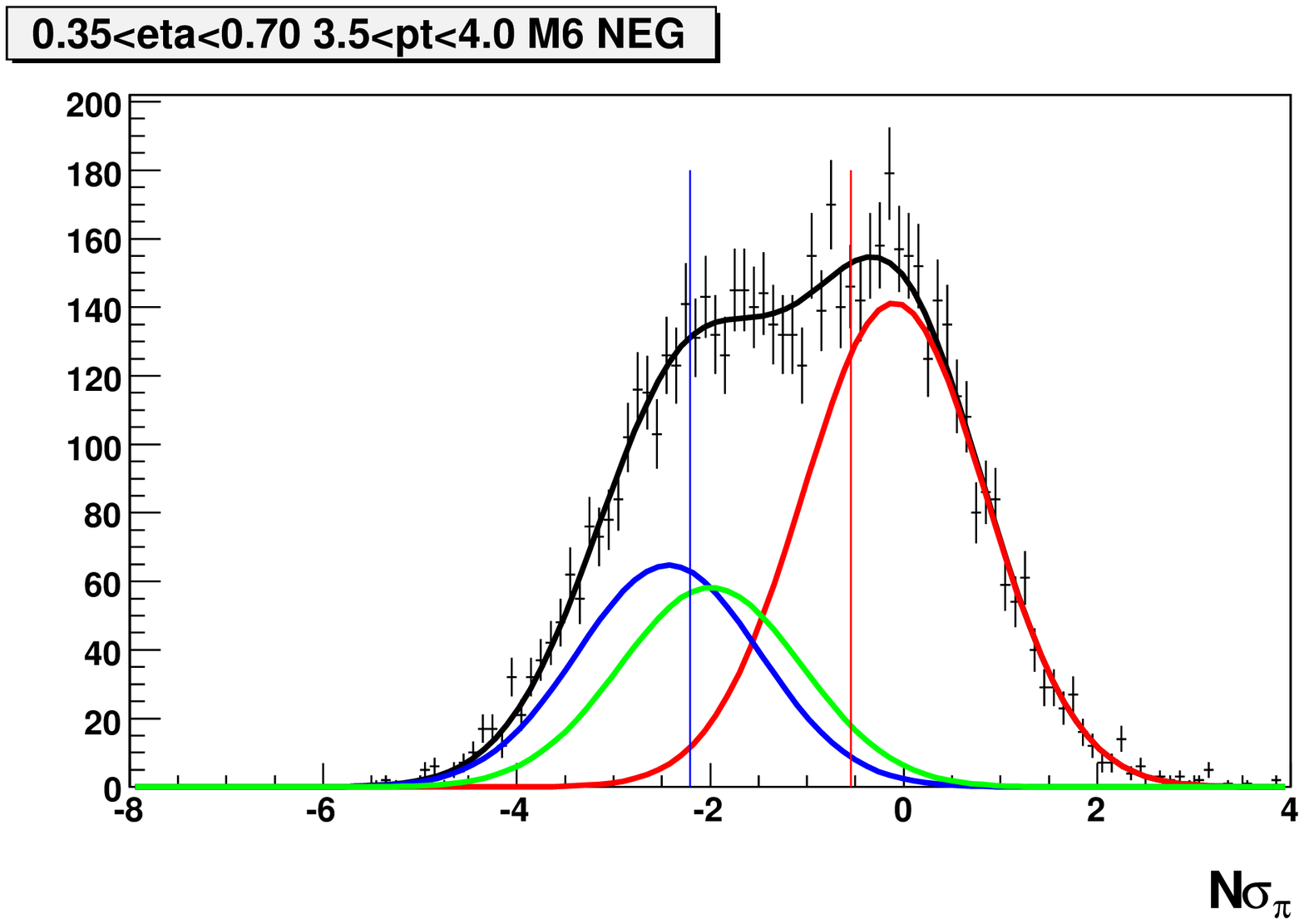}
		\includegraphics[width=1\textwidth]{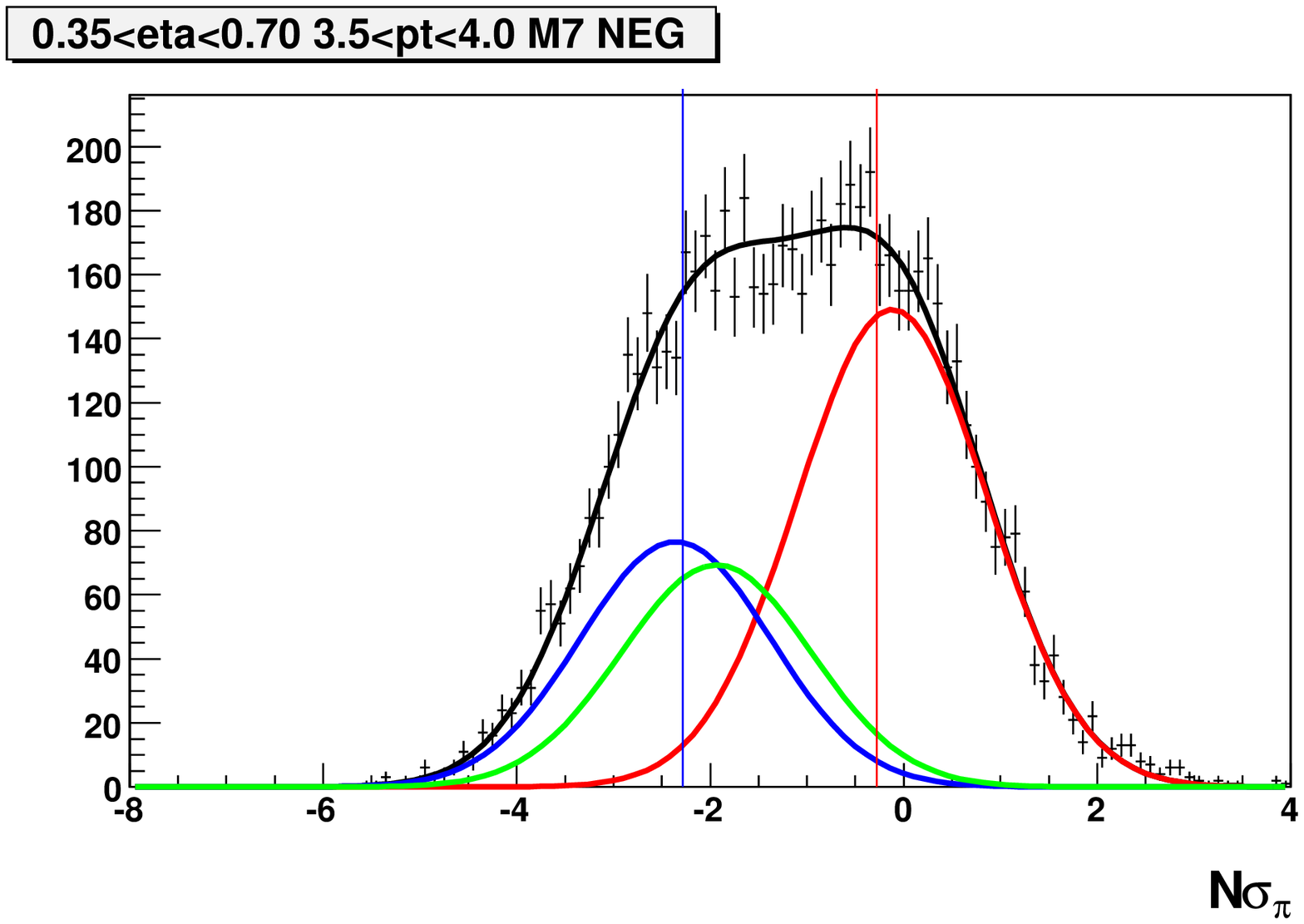}
		\includegraphics[width=1\textwidth]{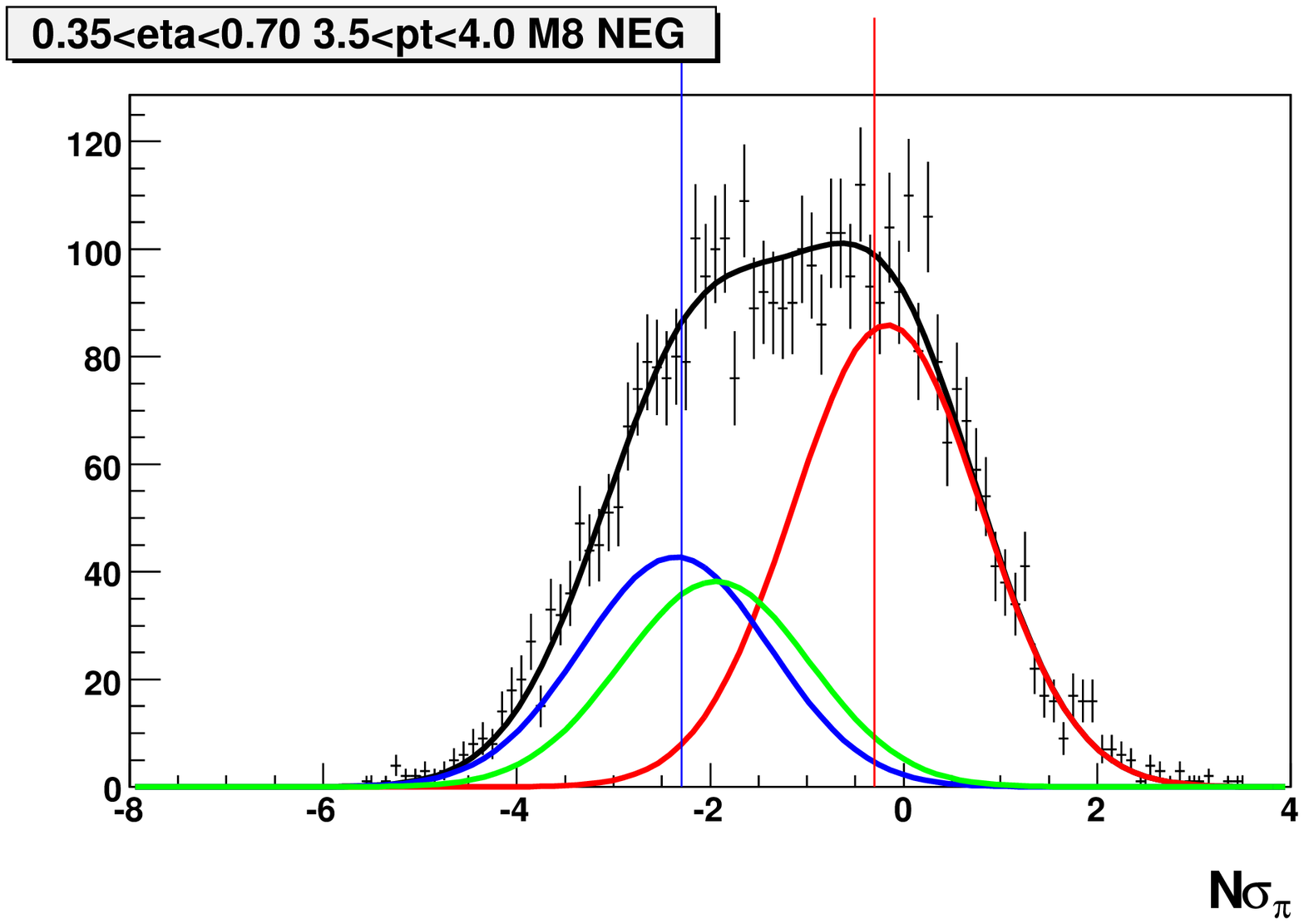}
		\includegraphics[width=1\textwidth]{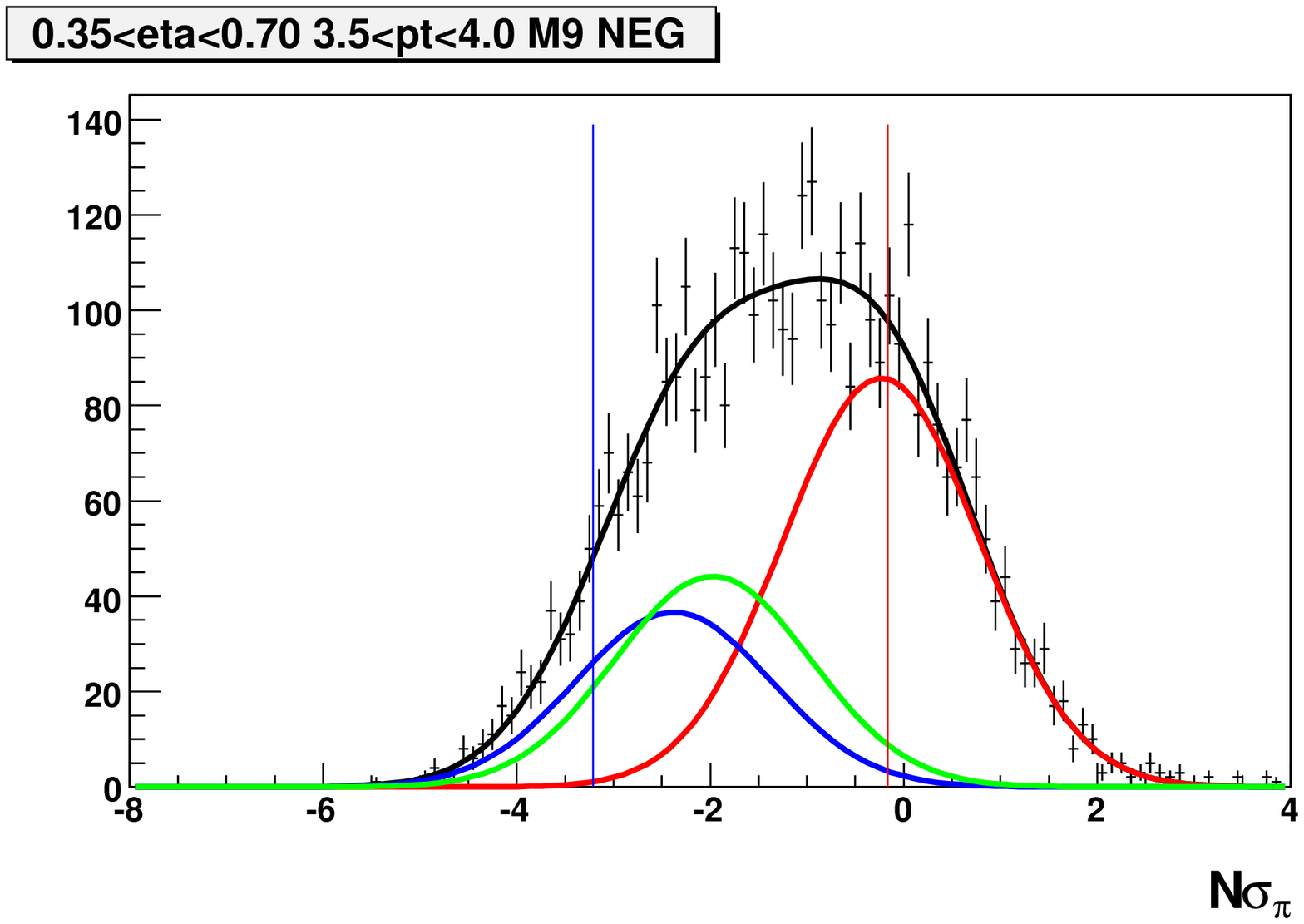}
		
			\end{minipage}	
					
	\caption{Same as Fig. 6.4 but for negative particles.}
	\label{fig:pidcutsN}	
\end{figure}

\begin{figure}[H]
\hfill
\begin{minipage}[t]{.23\textwidth}
	\centering
		\includegraphics[width=1\textwidth]{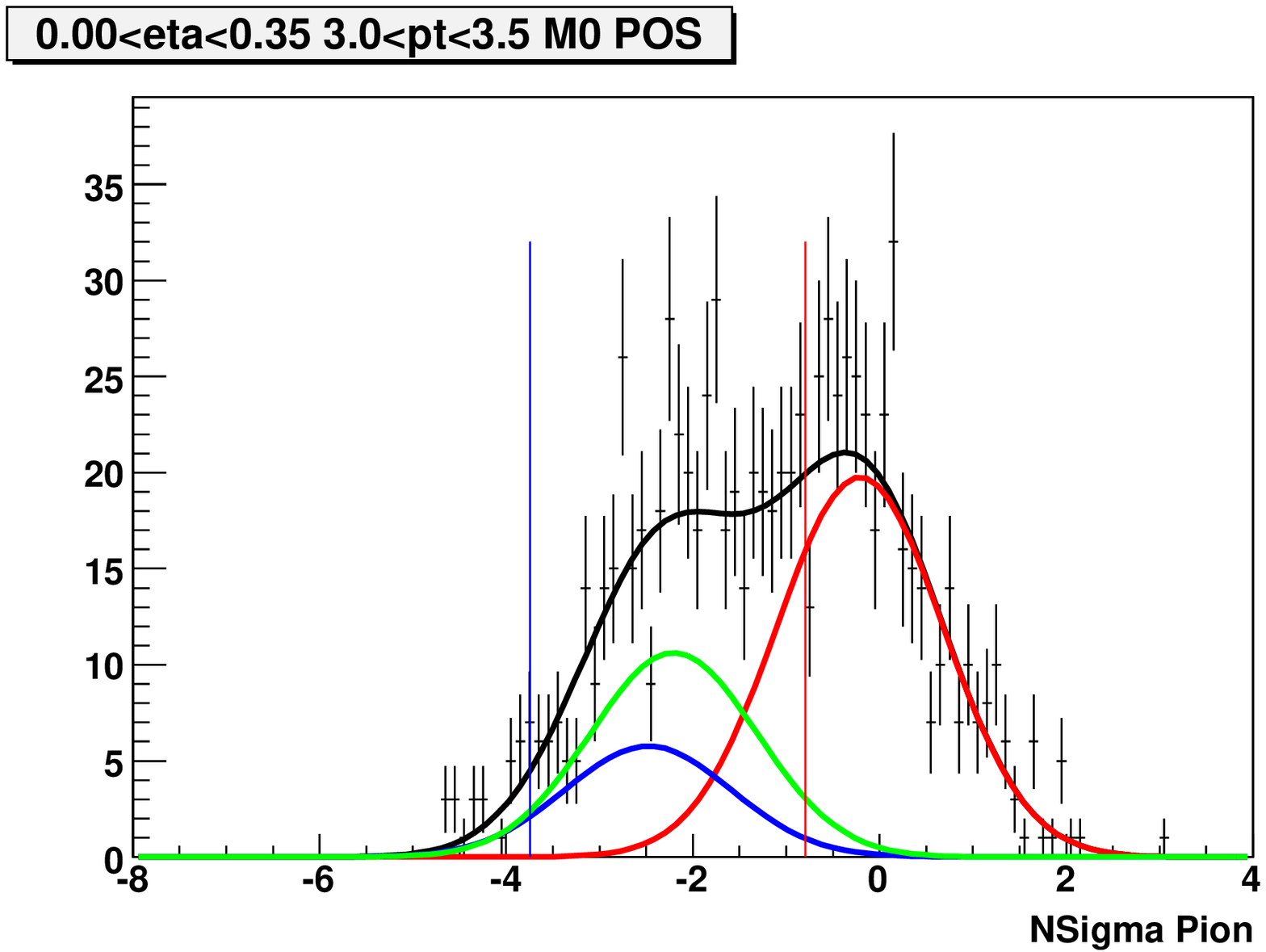}
		\includegraphics[width=1\textwidth]{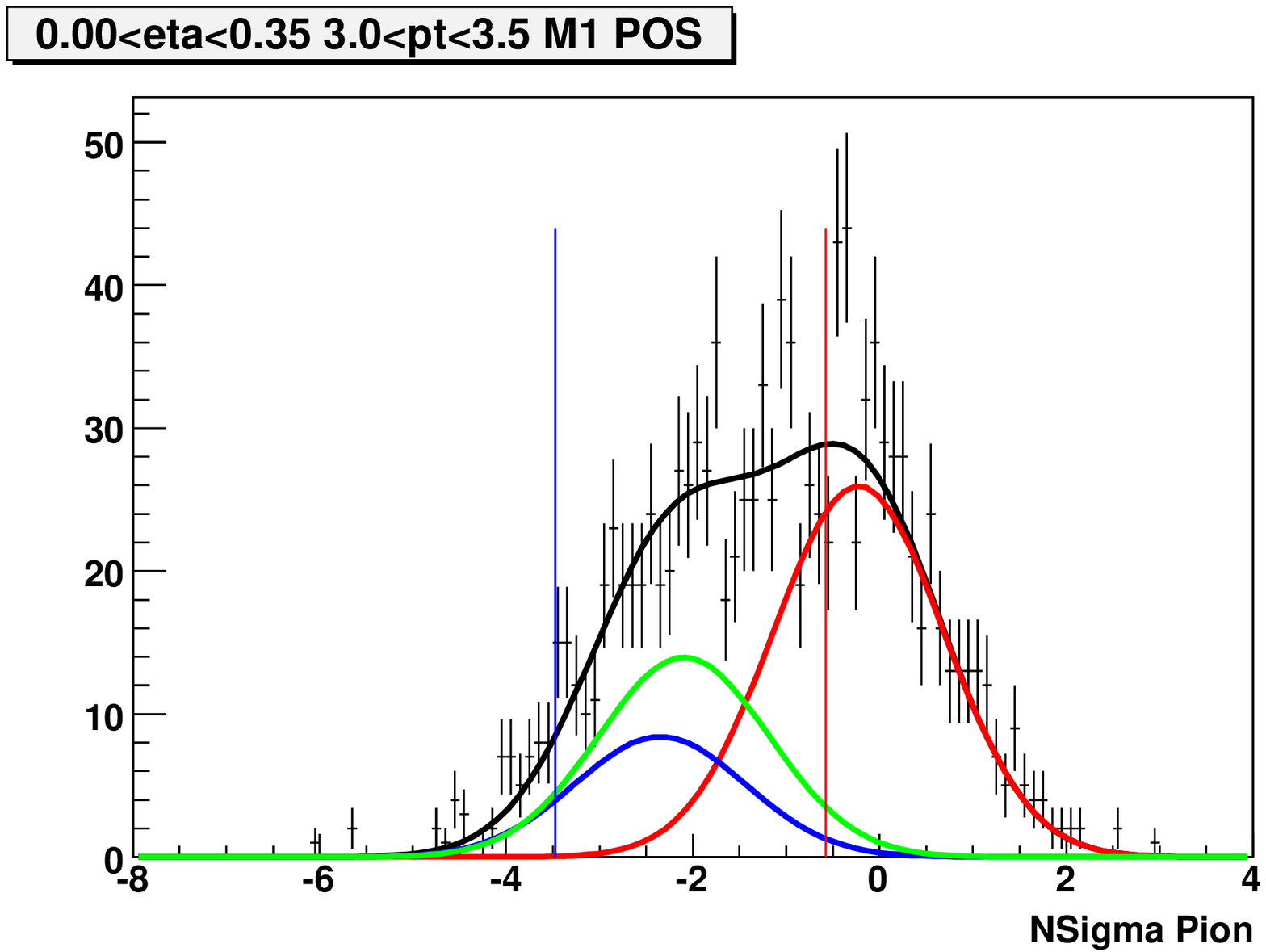}
		\includegraphics[width=1\textwidth]{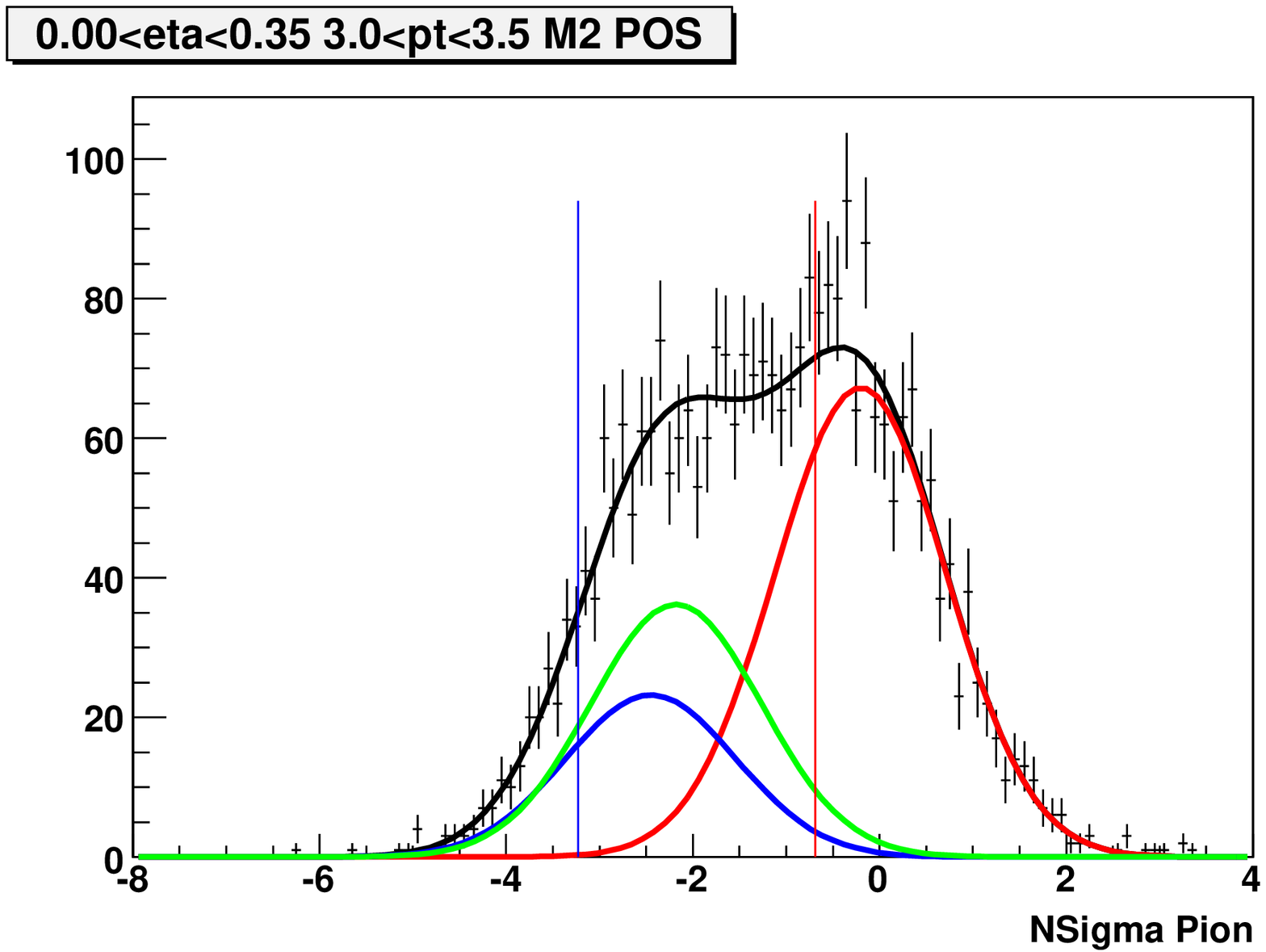}								
			\end{minipage}
\hfill
\begin{minipage}[t]{.23\textwidth}
	\centering
		\includegraphics[width=1\textwidth]{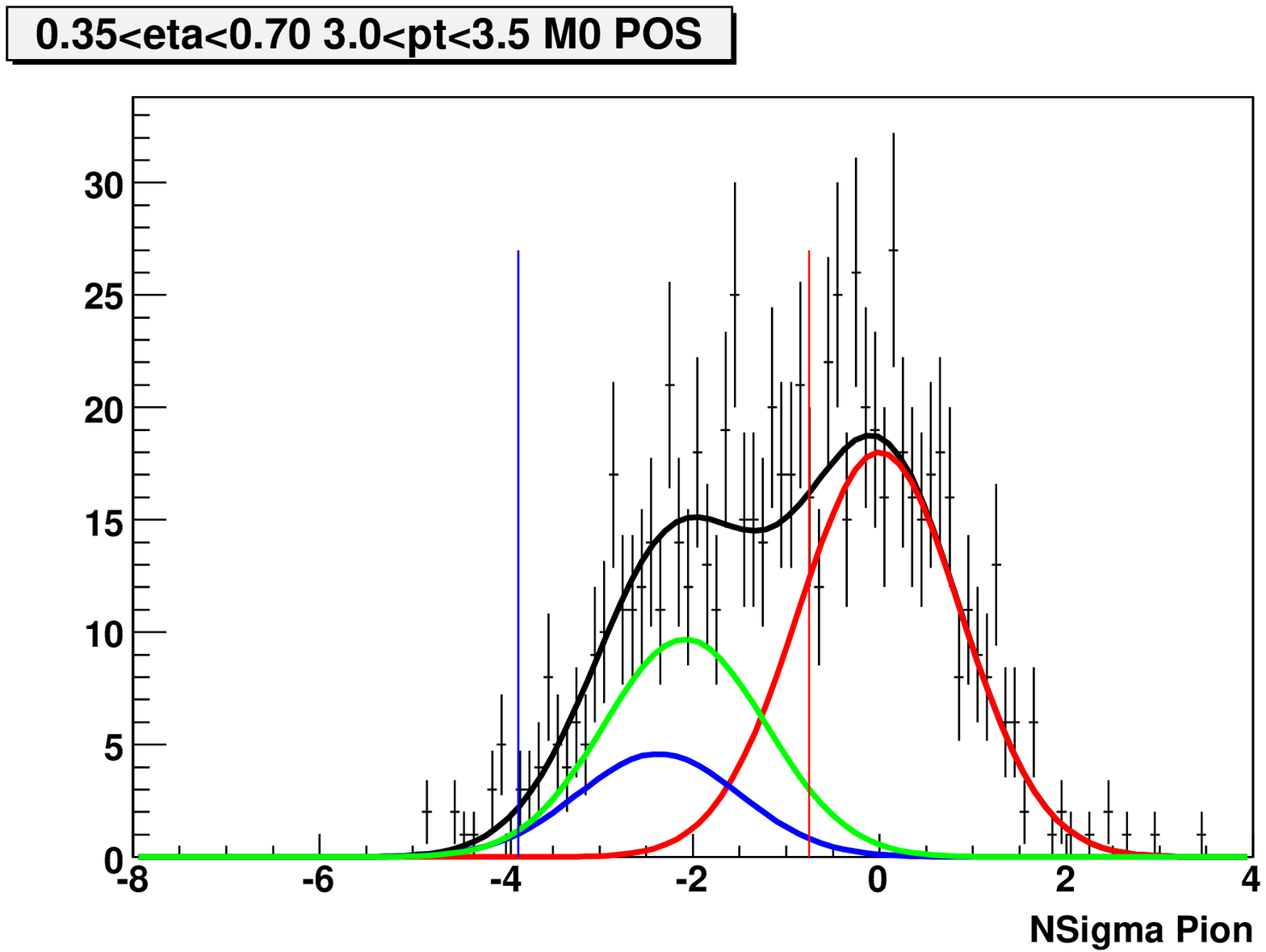}
		\includegraphics[width=1\textwidth]{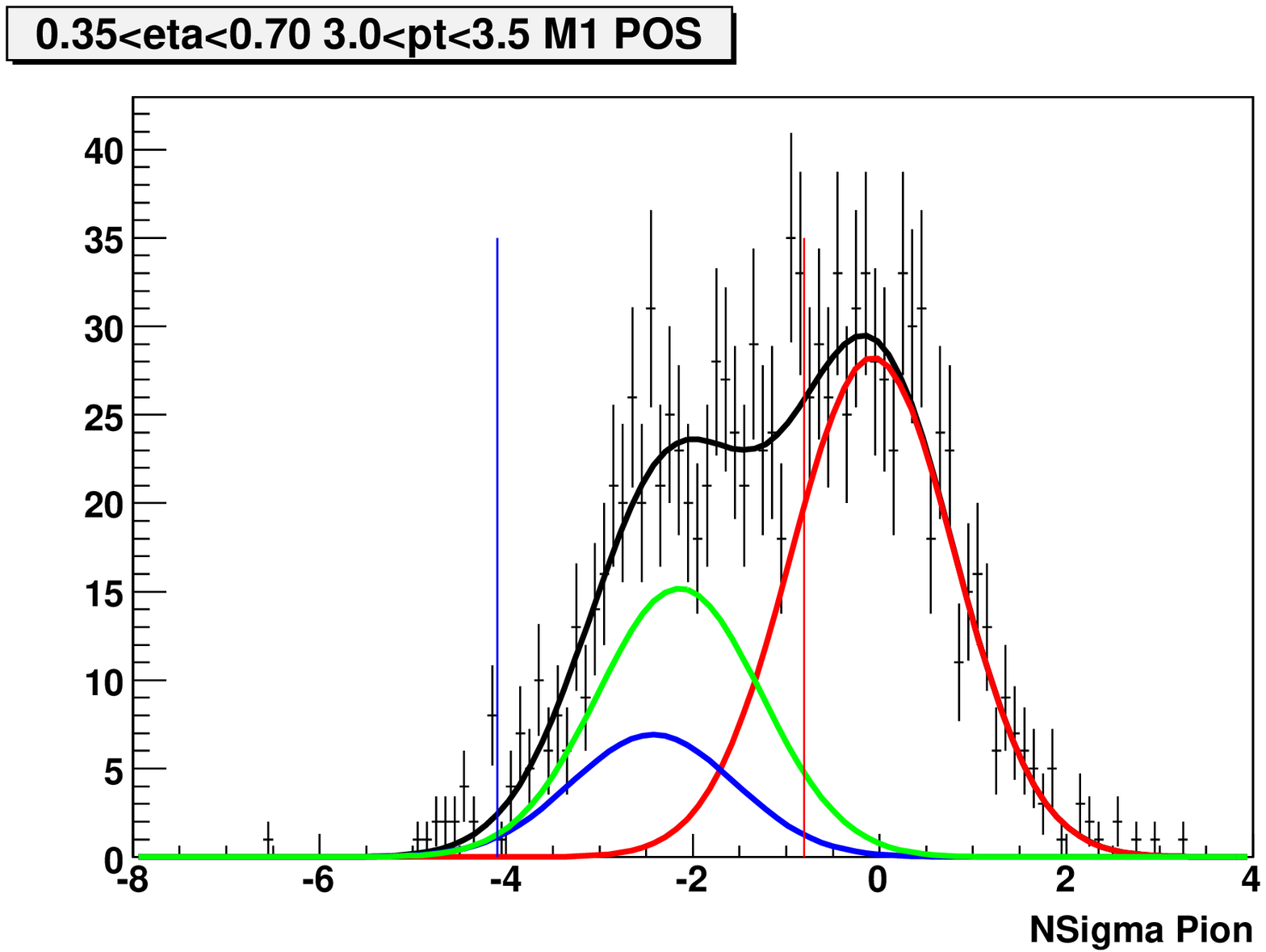}
		\includegraphics[width=1\textwidth]{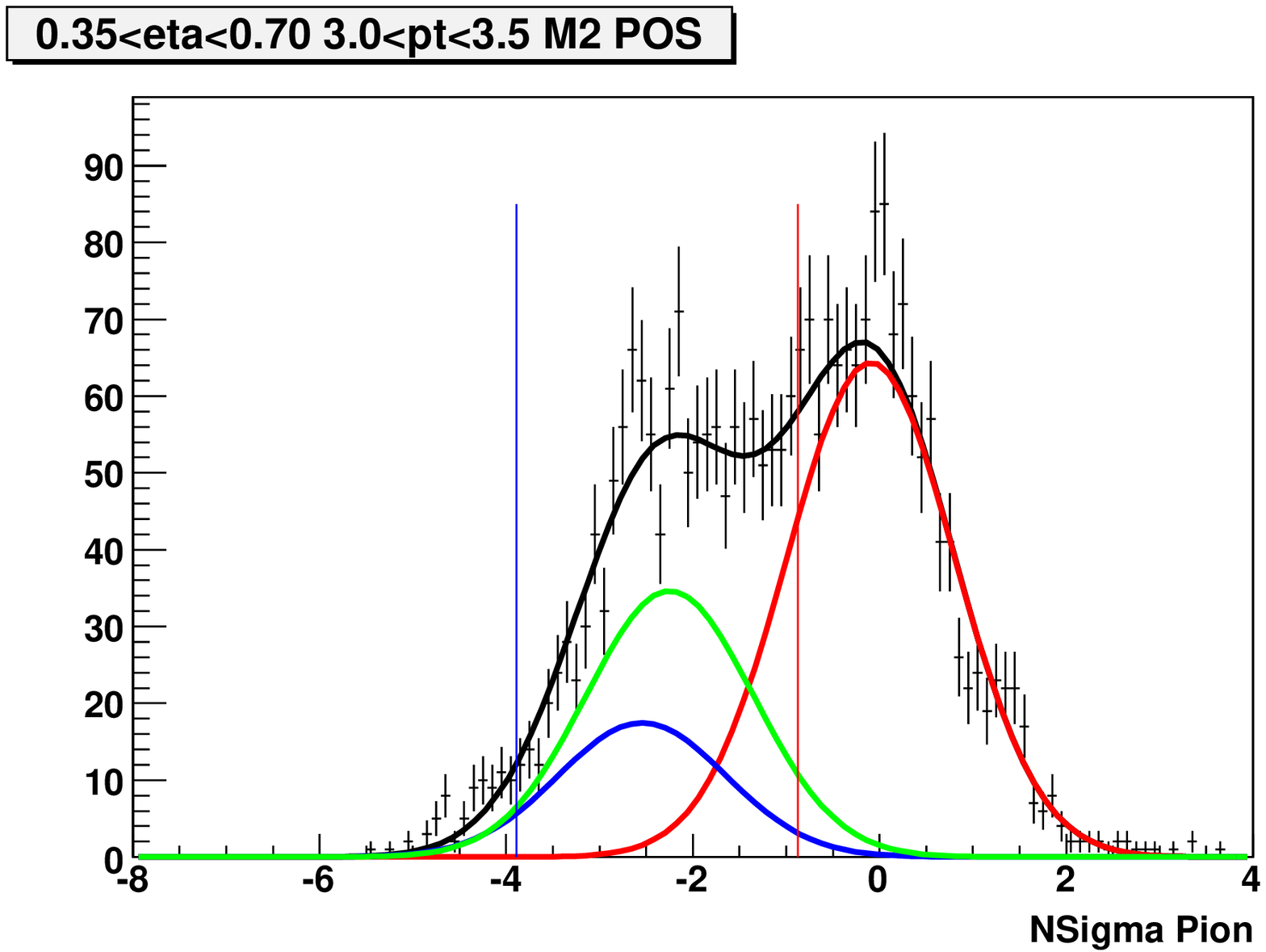}								
			\end{minipage}
\hfill
\begin{minipage}[t]{.23\textwidth}
	\centering
		\includegraphics[width=1\textwidth]{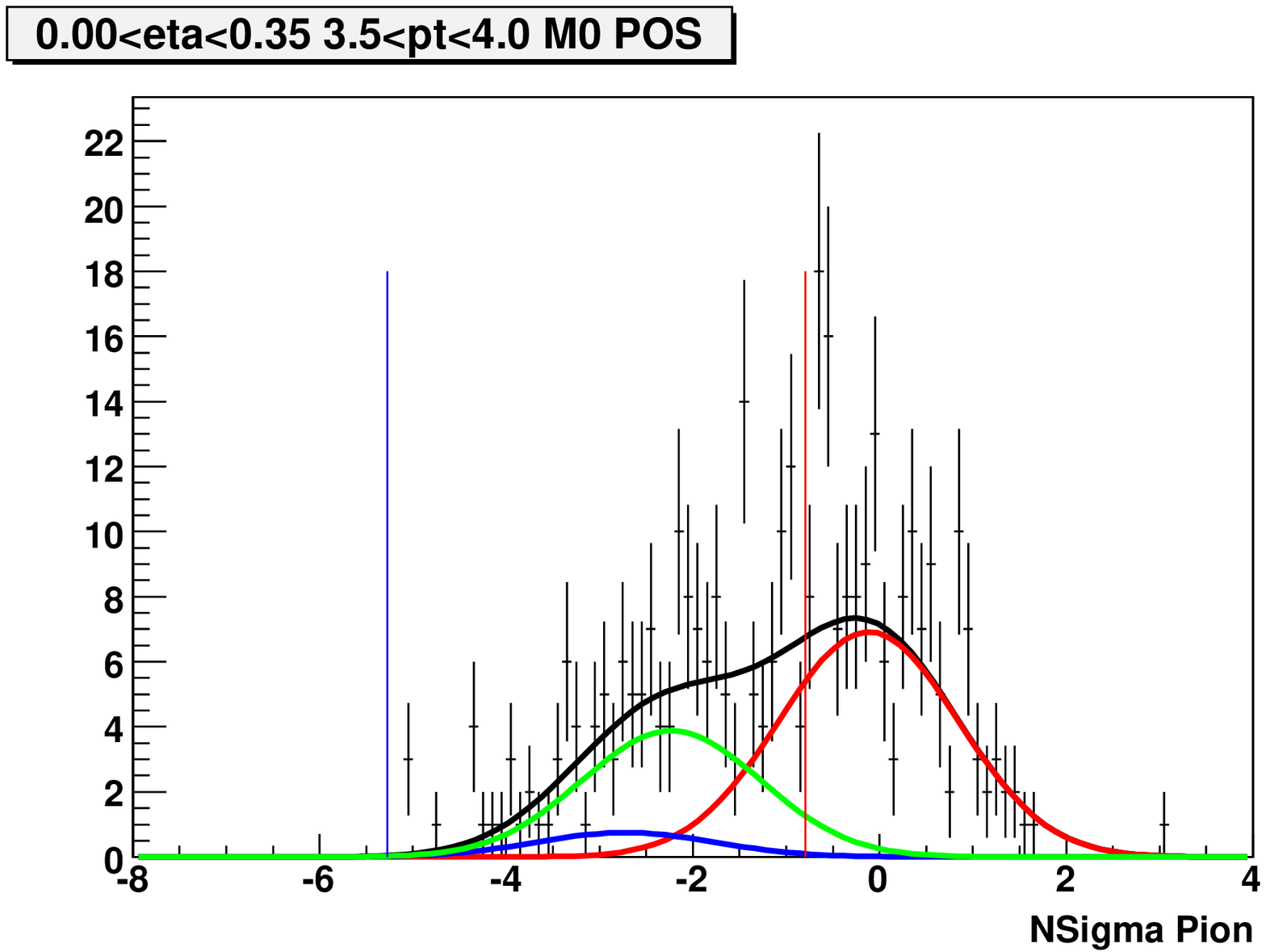}
		\includegraphics[width=1\textwidth]{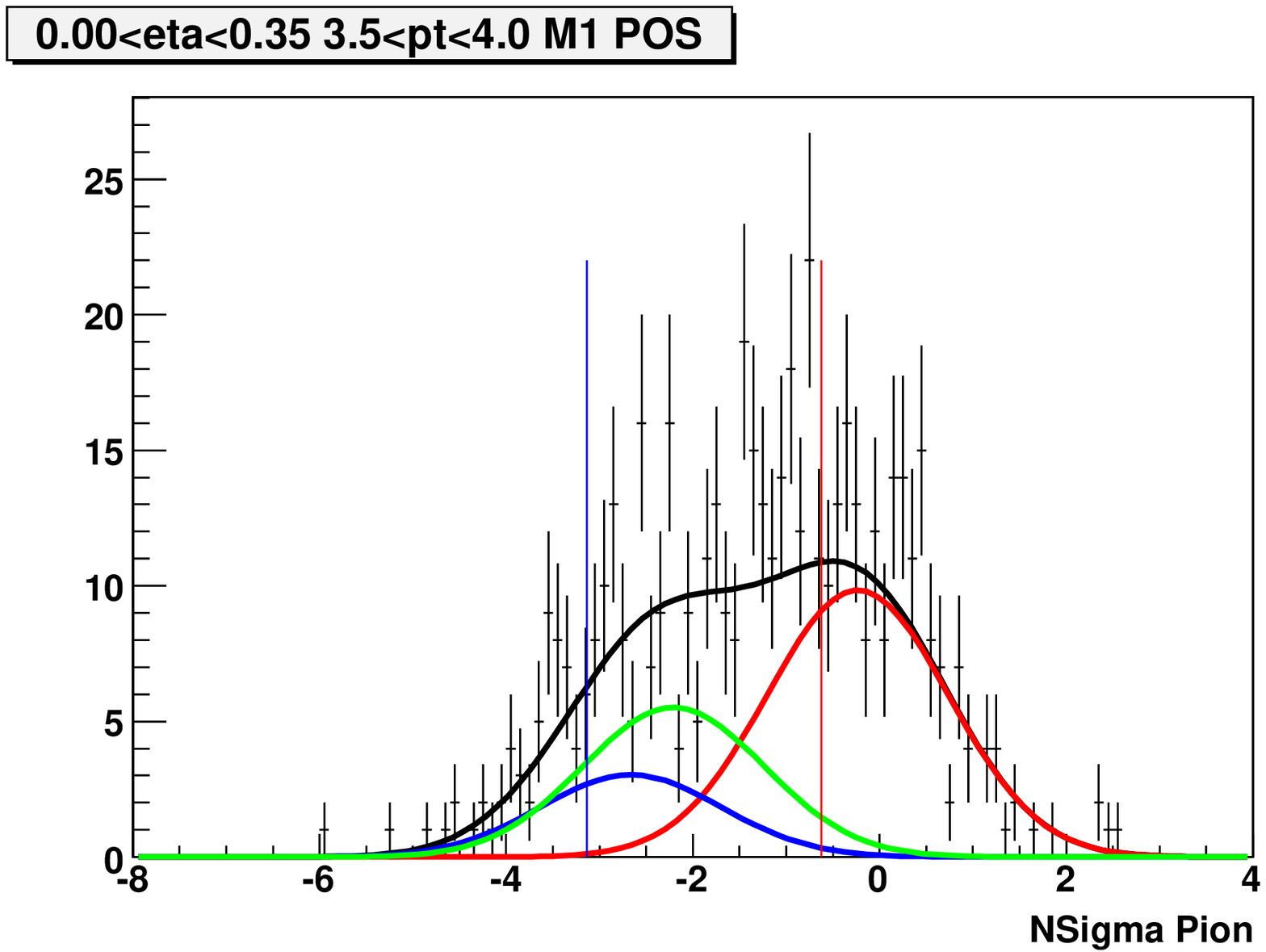}
		\includegraphics[width=1\textwidth]{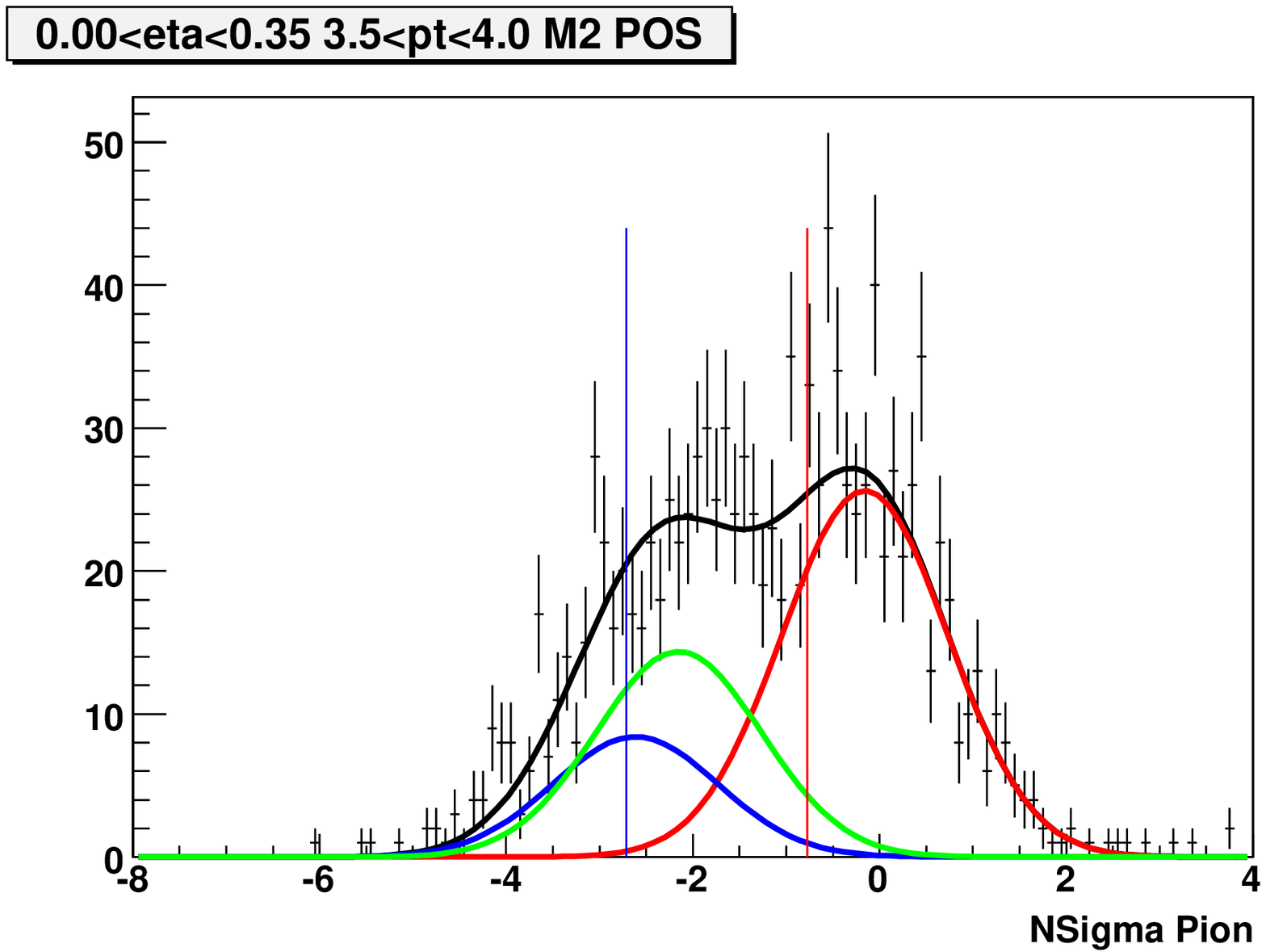}										
			\end{minipage}
\hfill
\begin{minipage}[t]{.23\textwidth}
	\centering
		\includegraphics[width=1\textwidth]{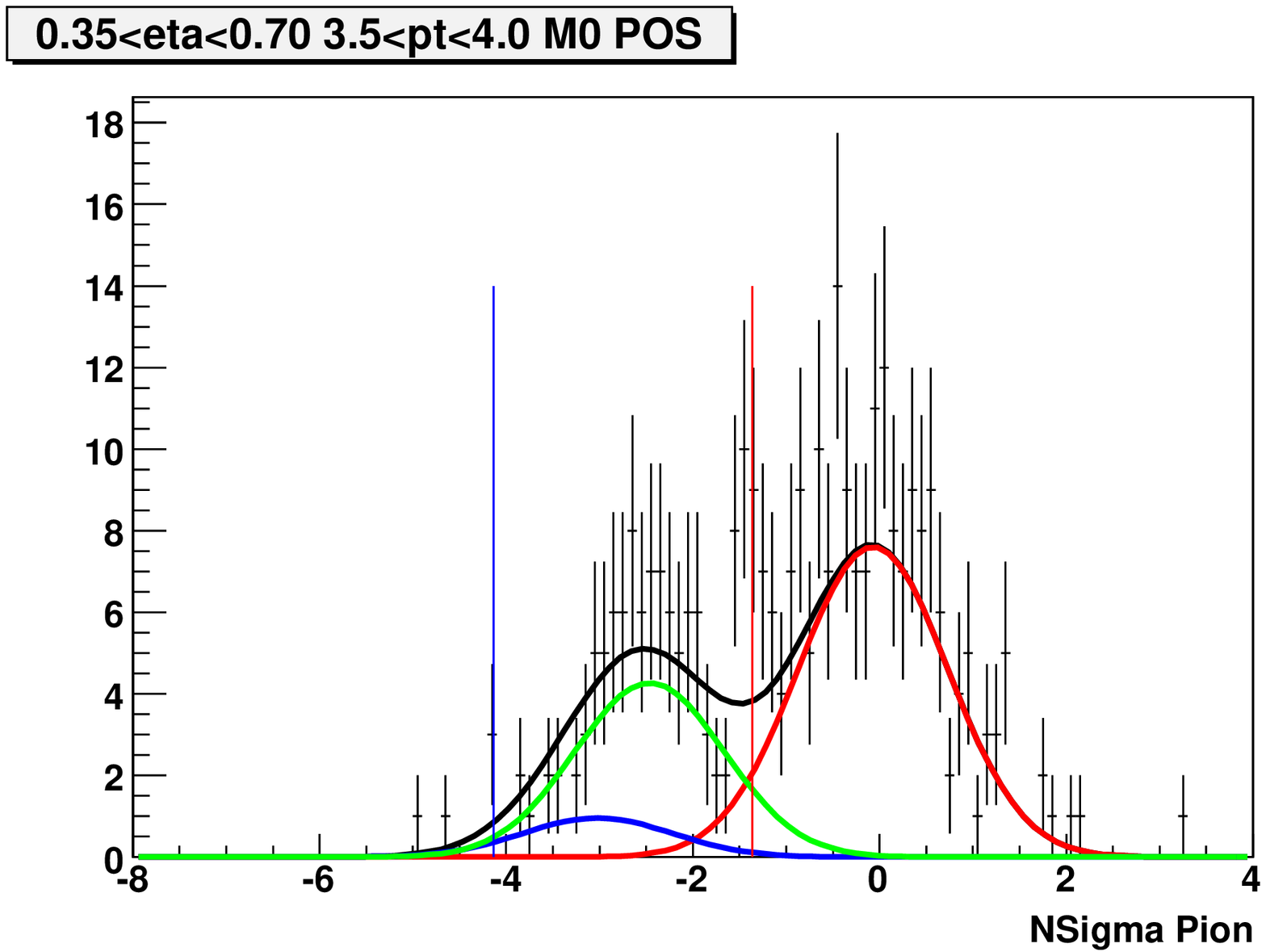}
		\includegraphics[width=1\textwidth]{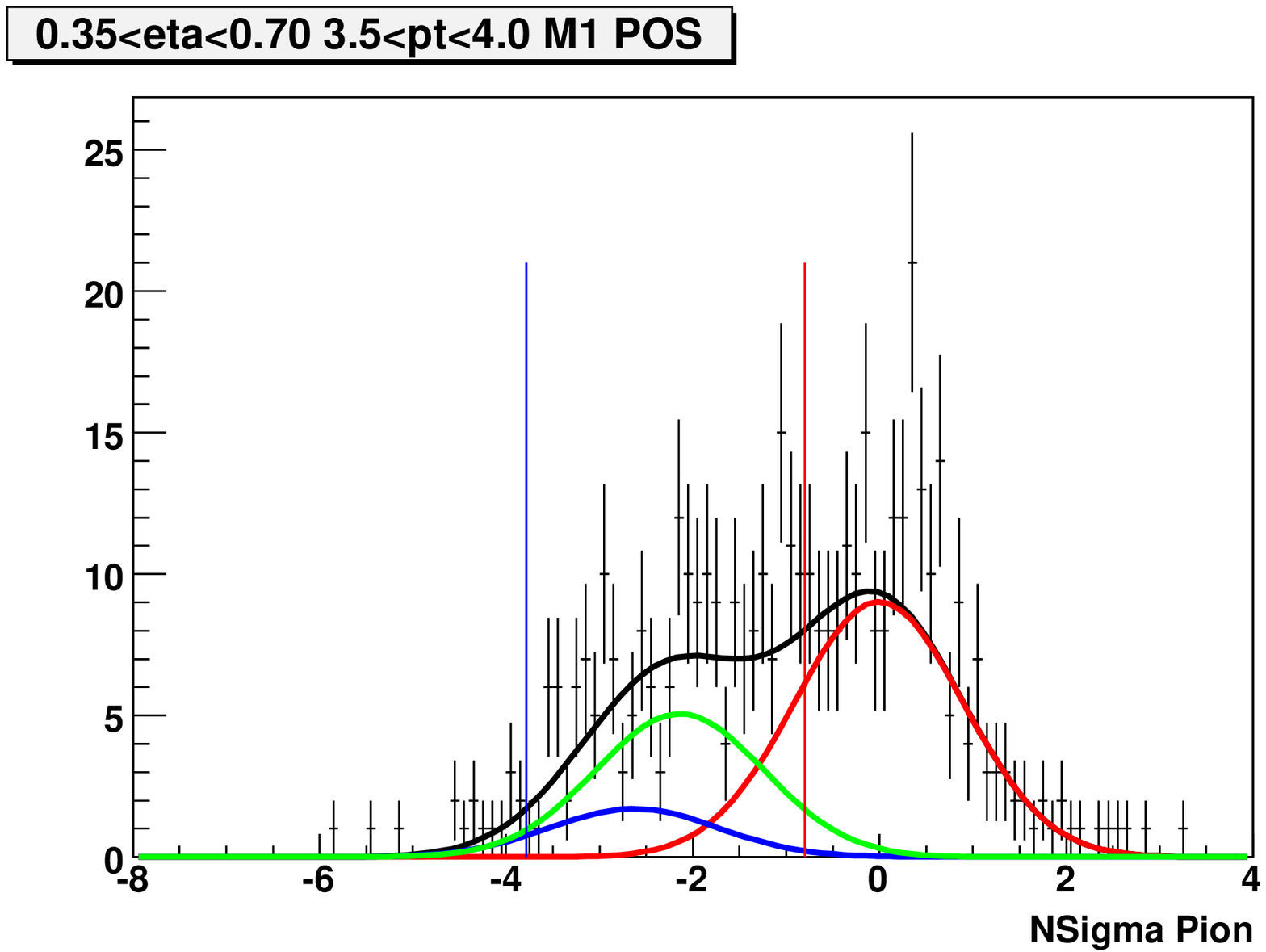}
		\includegraphics[width=1\textwidth]{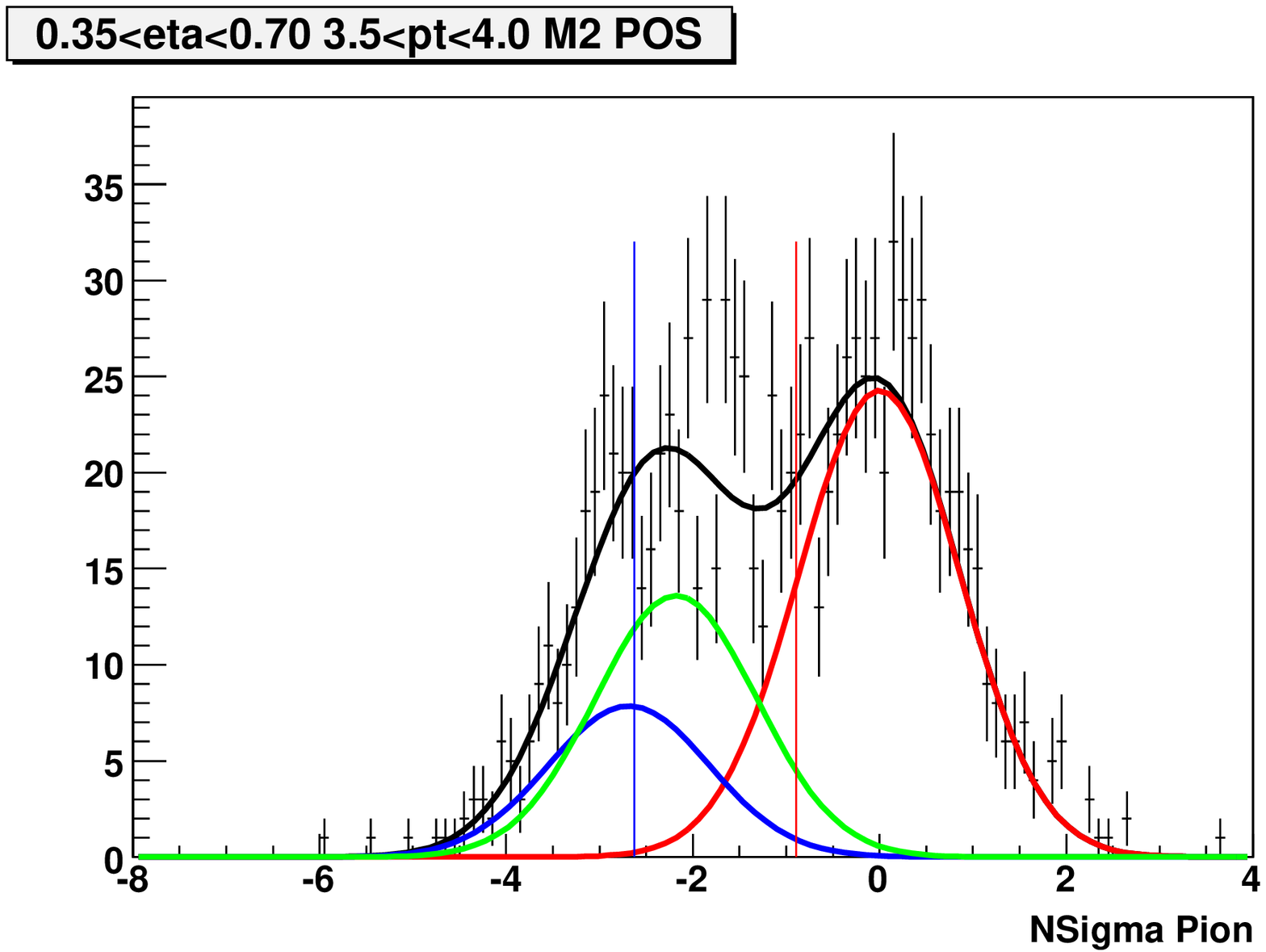}
		
			\end{minipage}						
	\caption{Same as Fig 6.4 but for d+Au.  Rows are the centrality bins 40-100\%, 20-40\% and 0-20\% from top to bottom.}
	\label{fig:dAupidcutsP}	
\end{figure}

\begin{figure}[H]
\hfill
\begin{minipage}[t]{.23\textwidth}
	\centering
		\includegraphics[width=1\textwidth]{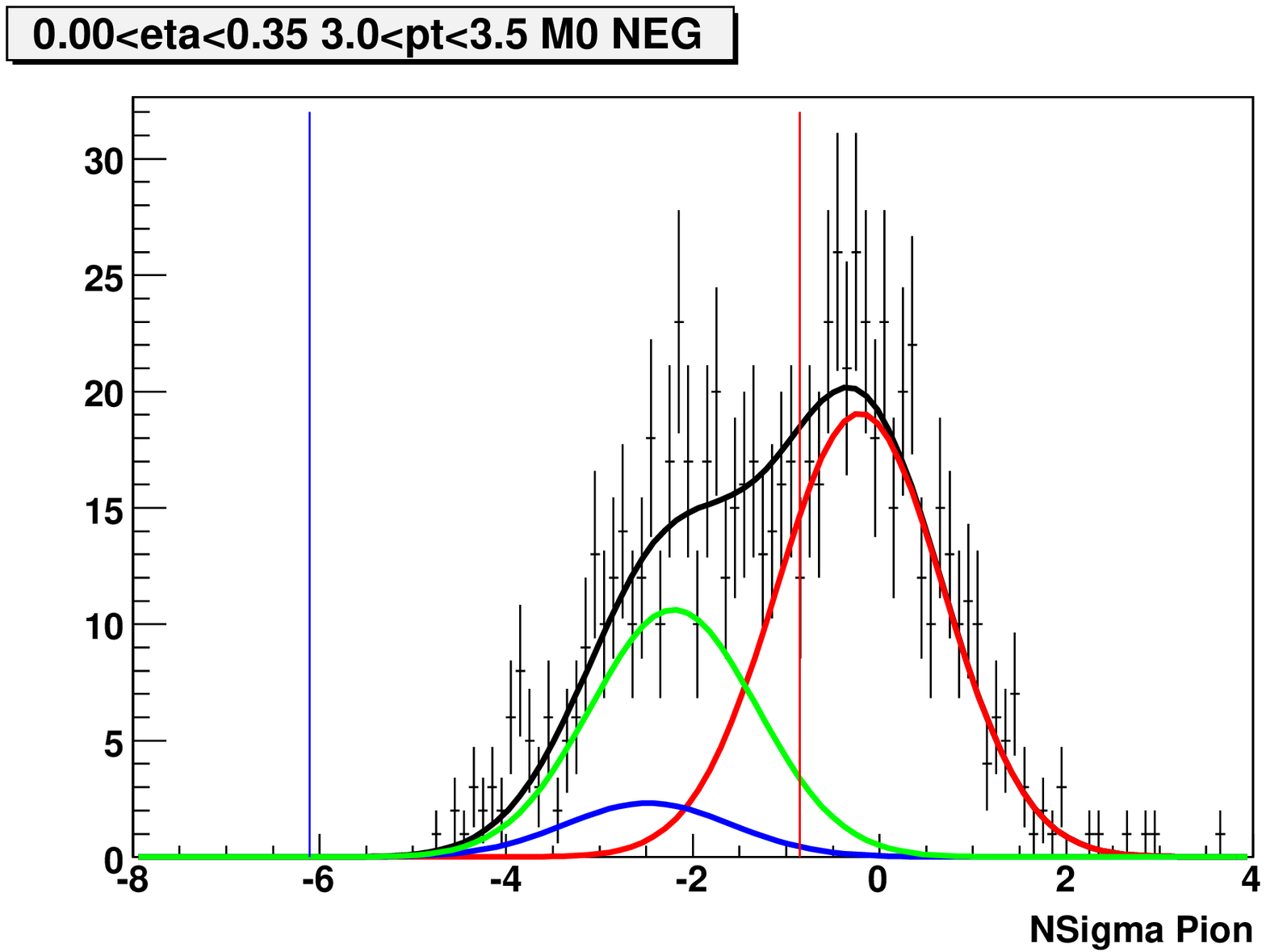}
		\includegraphics[width=1\textwidth]{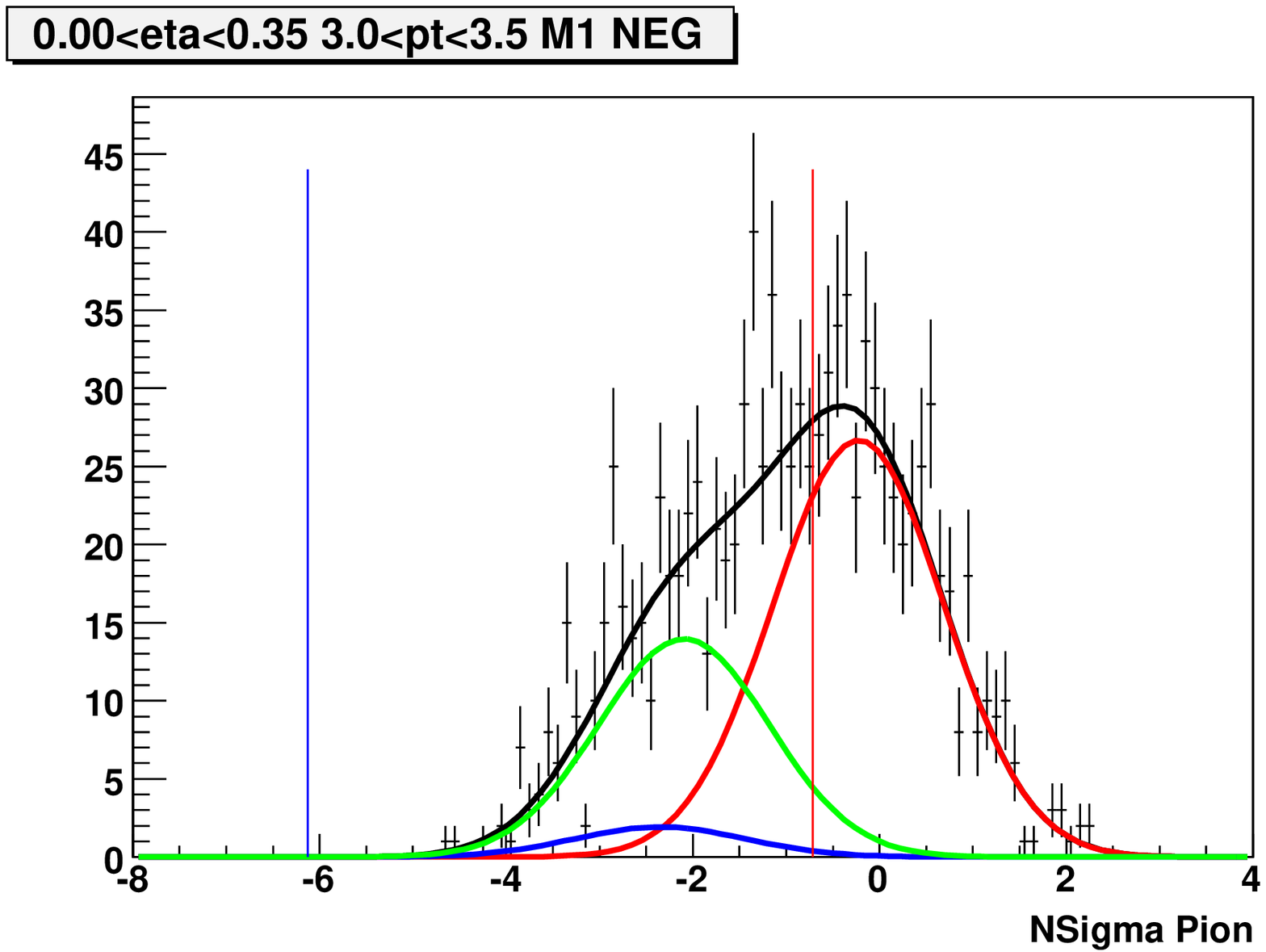}
		\includegraphics[width=1\textwidth]{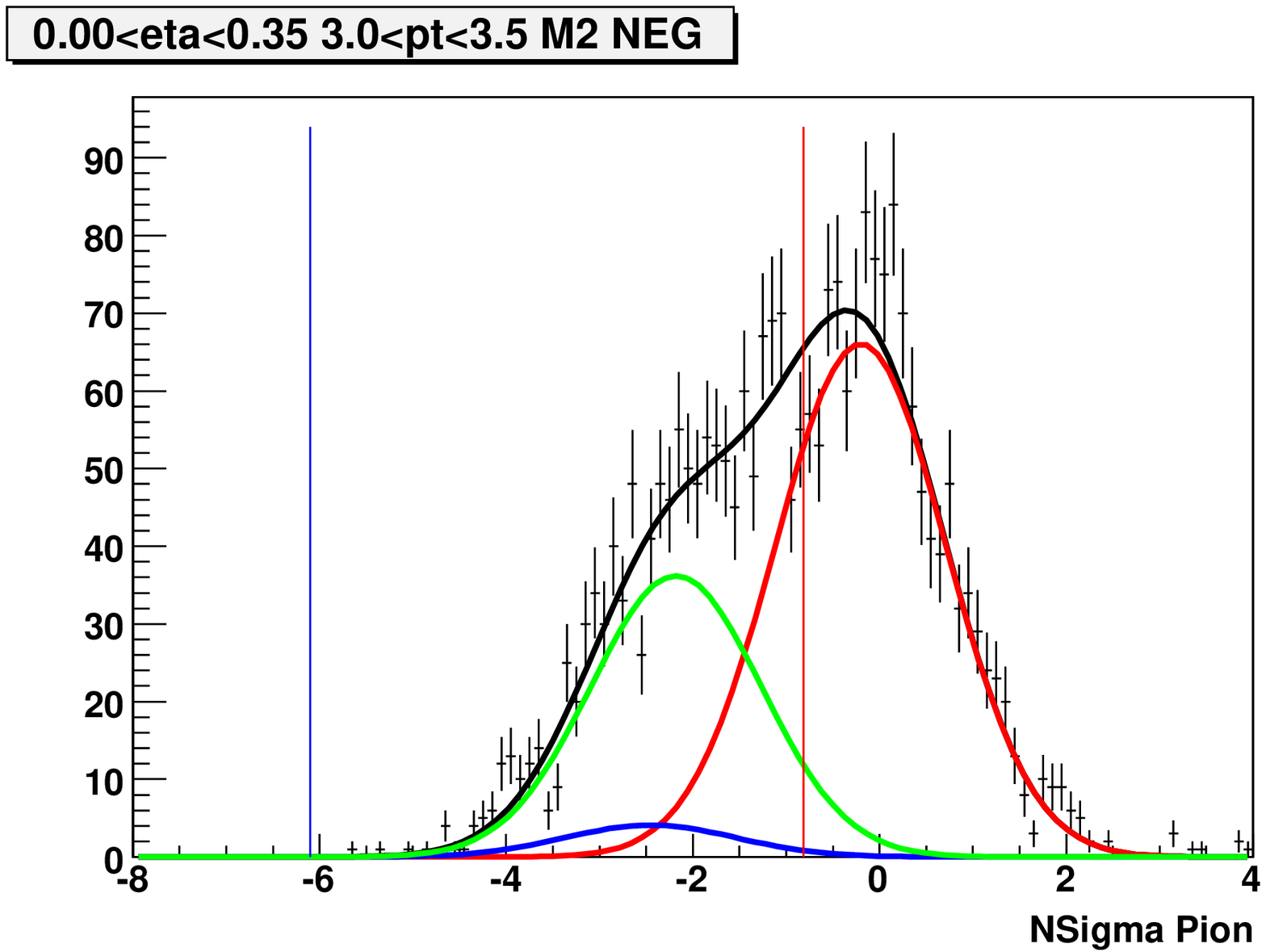}				
			\end{minipage}
\hfill
\begin{minipage}[t]{.23\textwidth}
	\centering
		\includegraphics[width=1\textwidth]{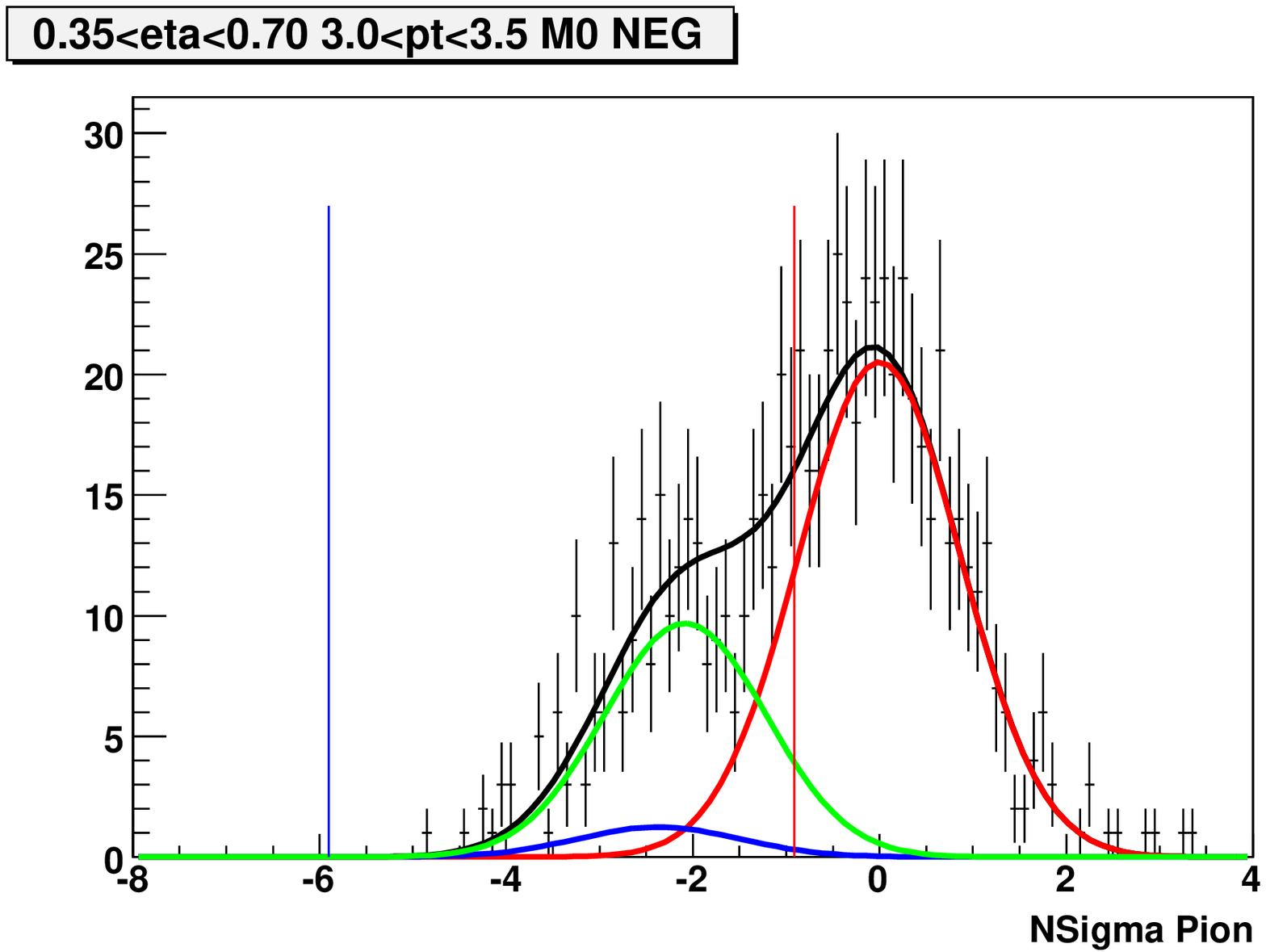}
		\includegraphics[width=1\textwidth]{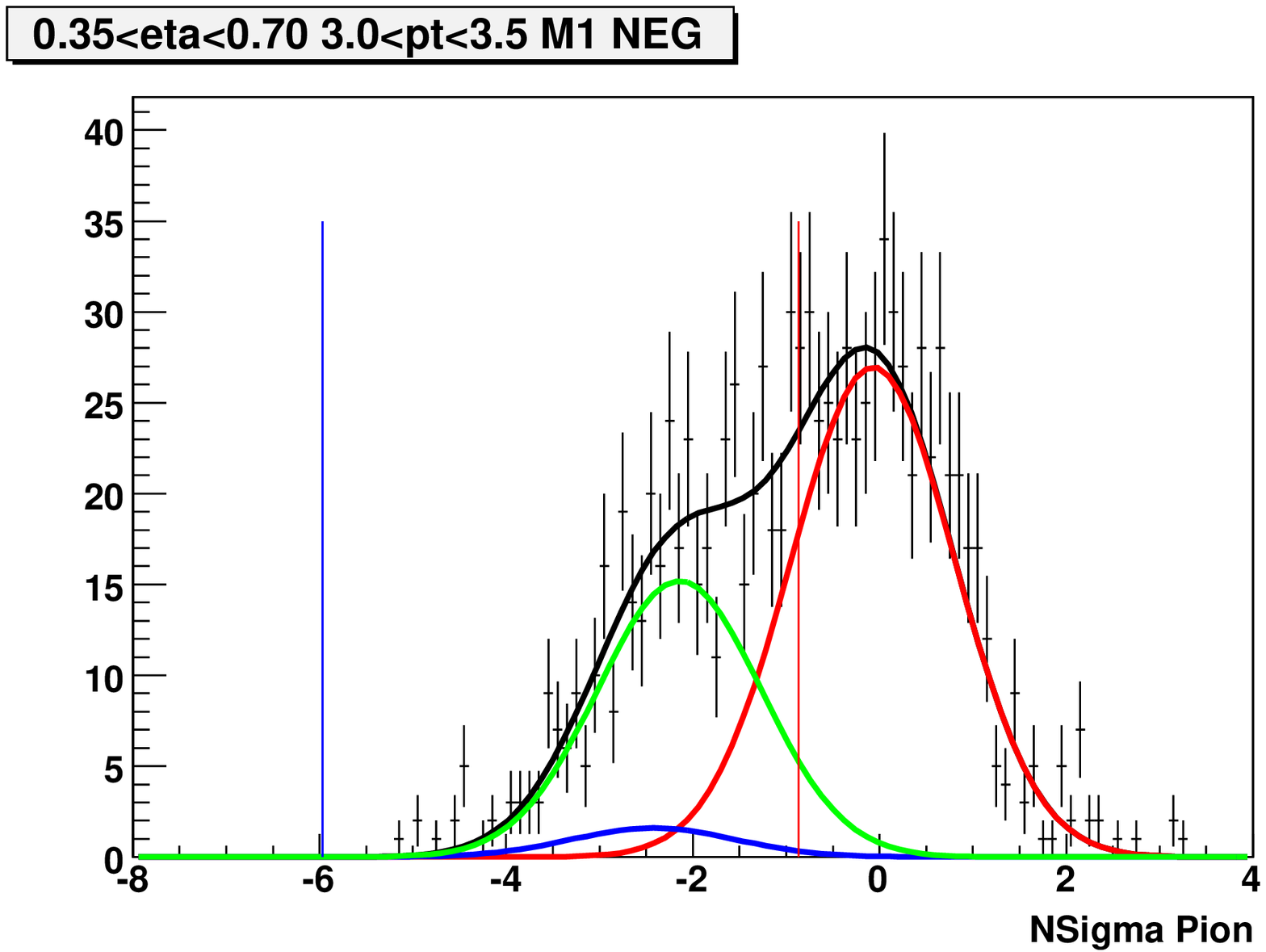}
		\includegraphics[width=1\textwidth]{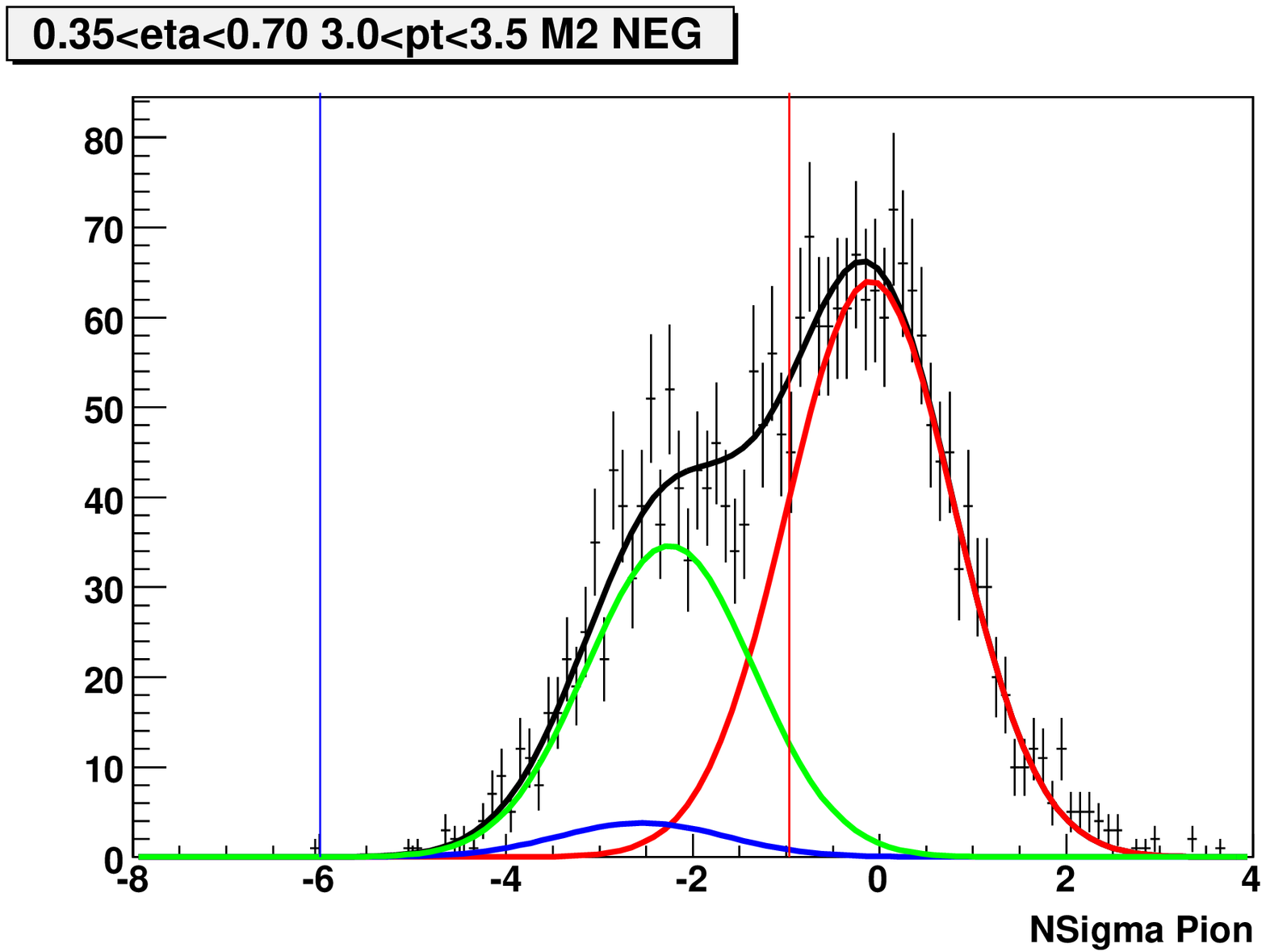}						
			\end{minipage}
\hfill
\begin{minipage}[t]{.23\textwidth}
	\centering
		\includegraphics[width=1\textwidth]{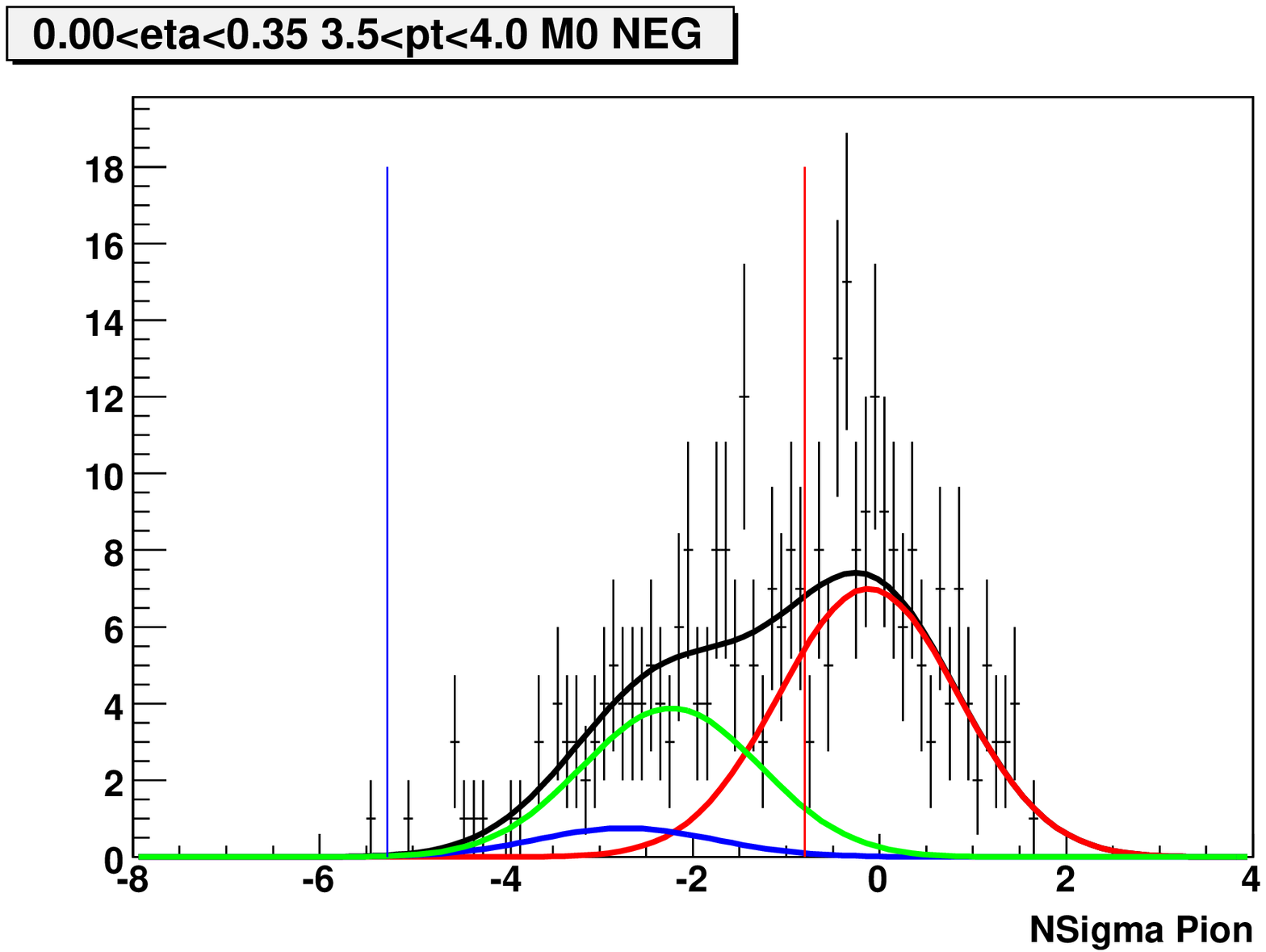}
		\includegraphics[width=1\textwidth]{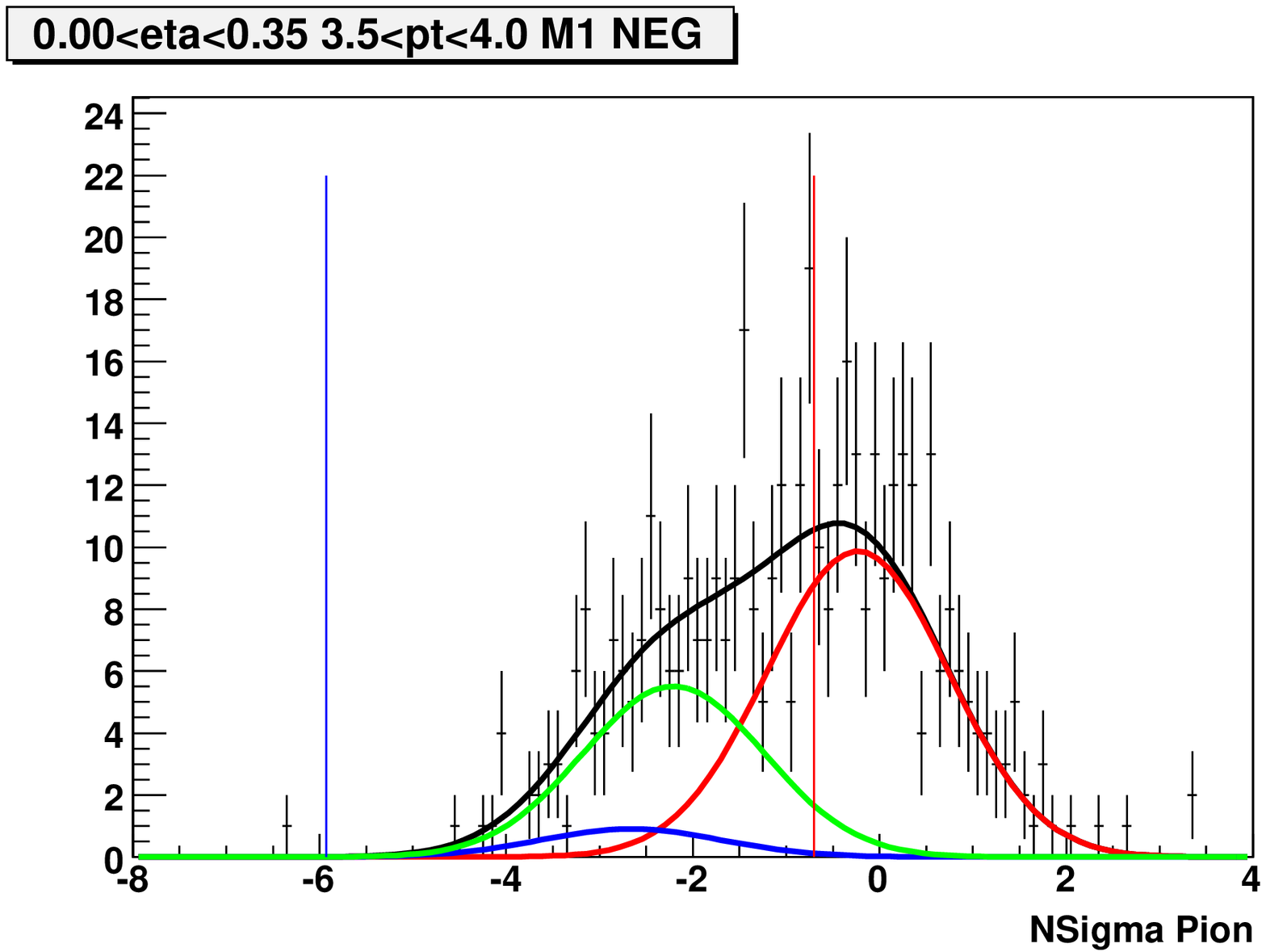}
		\includegraphics[width=1\textwidth]{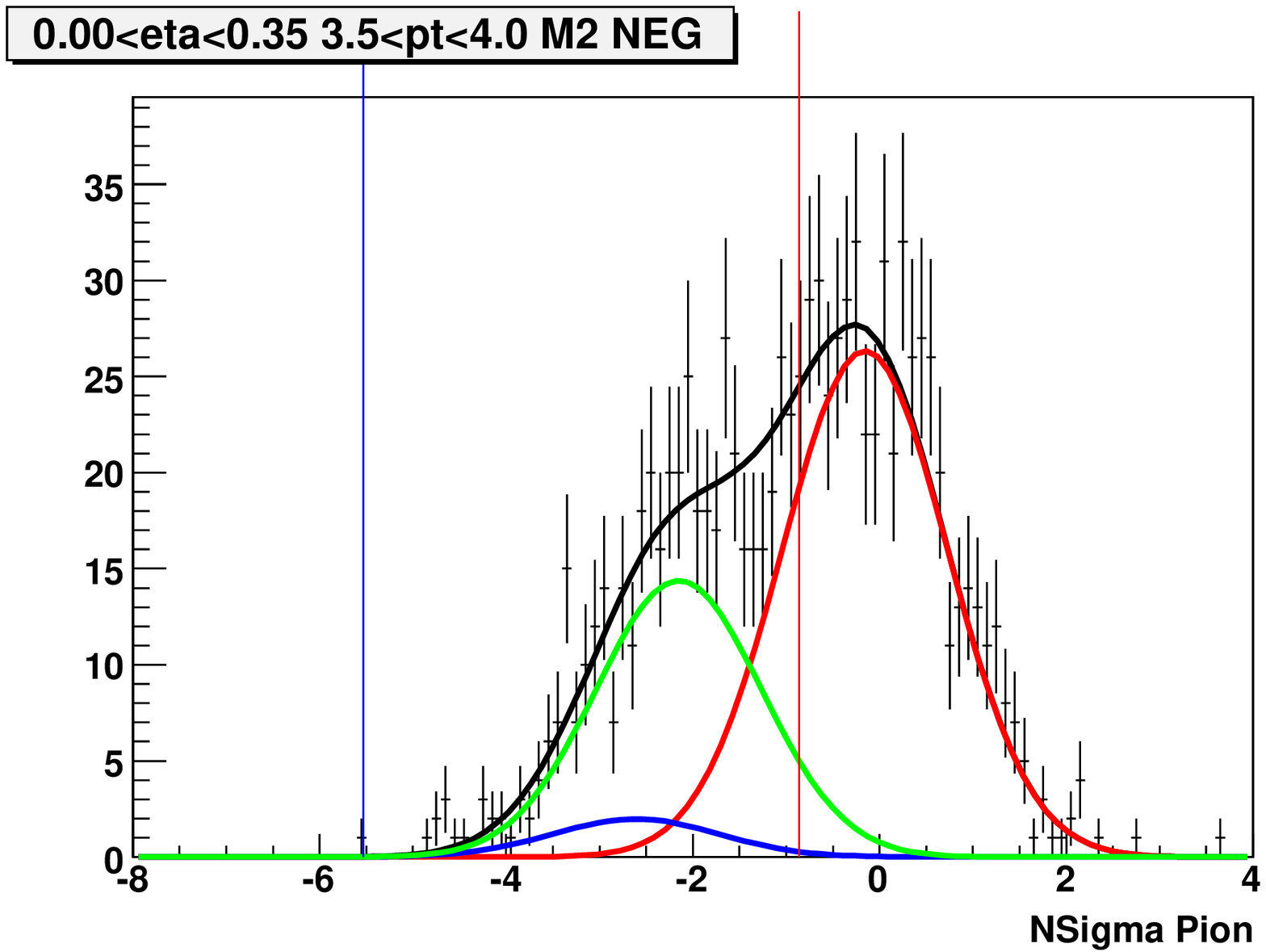}										
			\end{minipage}
\hfill
\begin{minipage}[t]{.23\textwidth}
	\centering
		\includegraphics[width=1\textwidth]{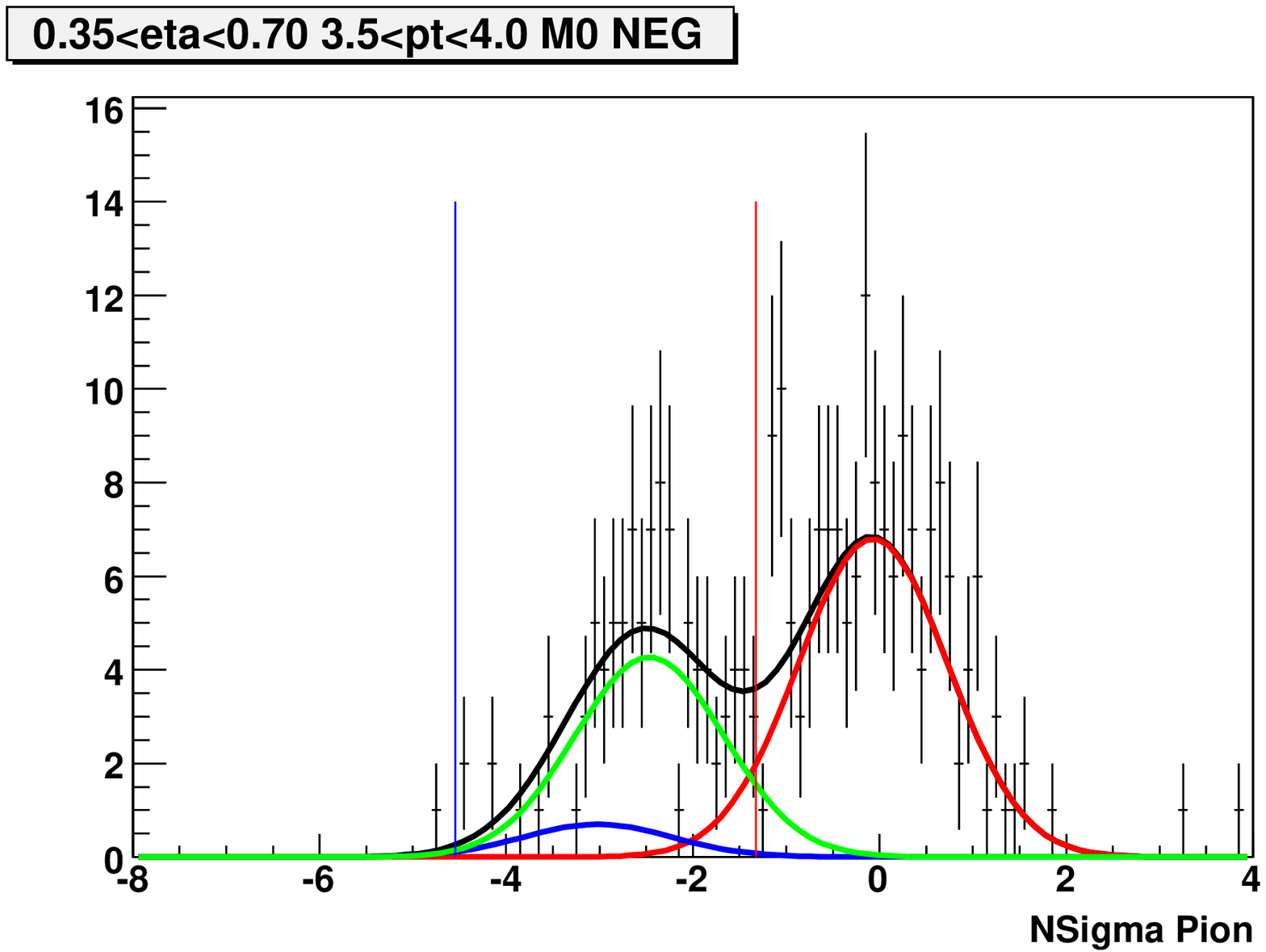}
		\includegraphics[width=1\textwidth]{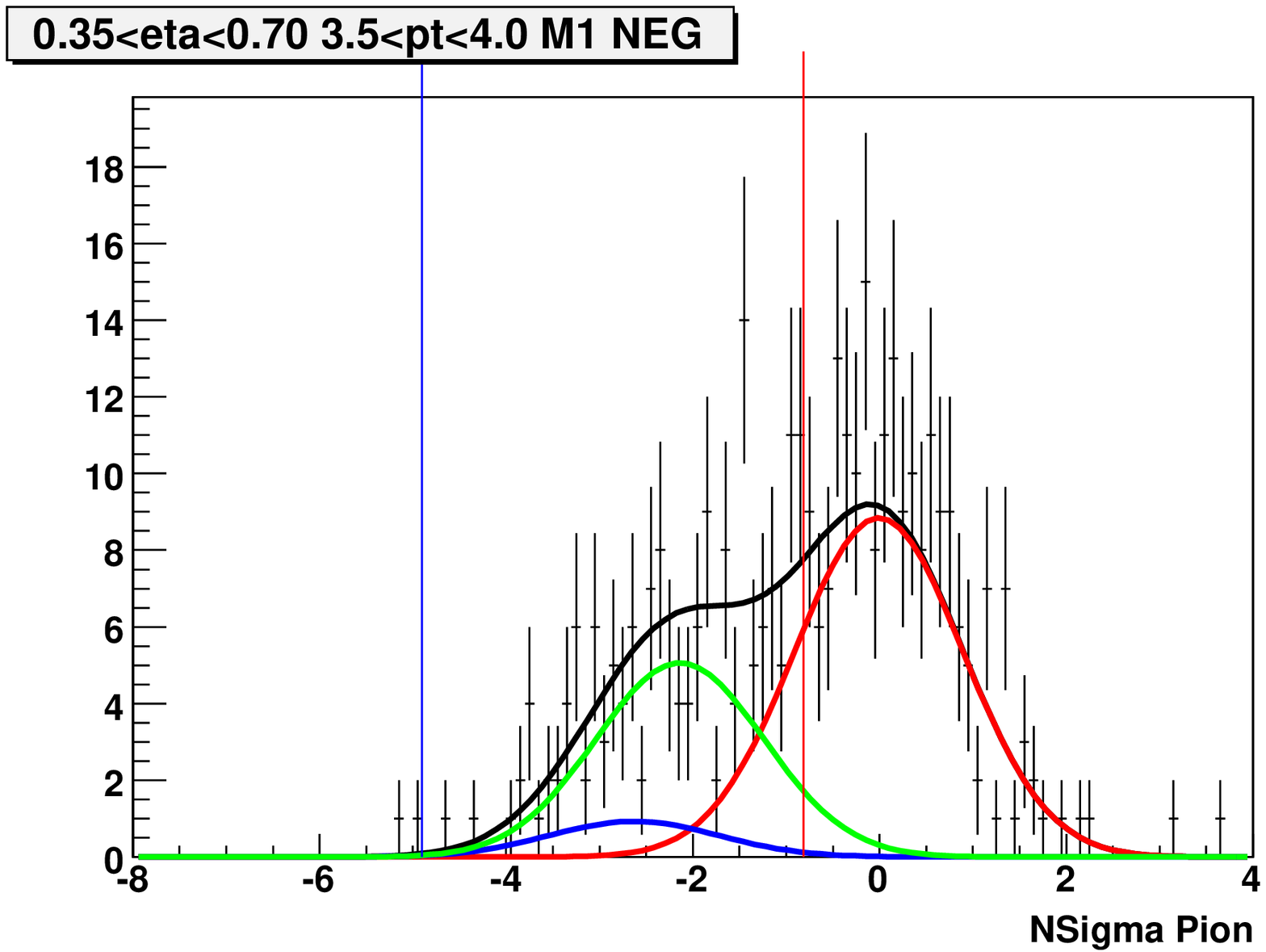}
		\includegraphics[width=1\textwidth]{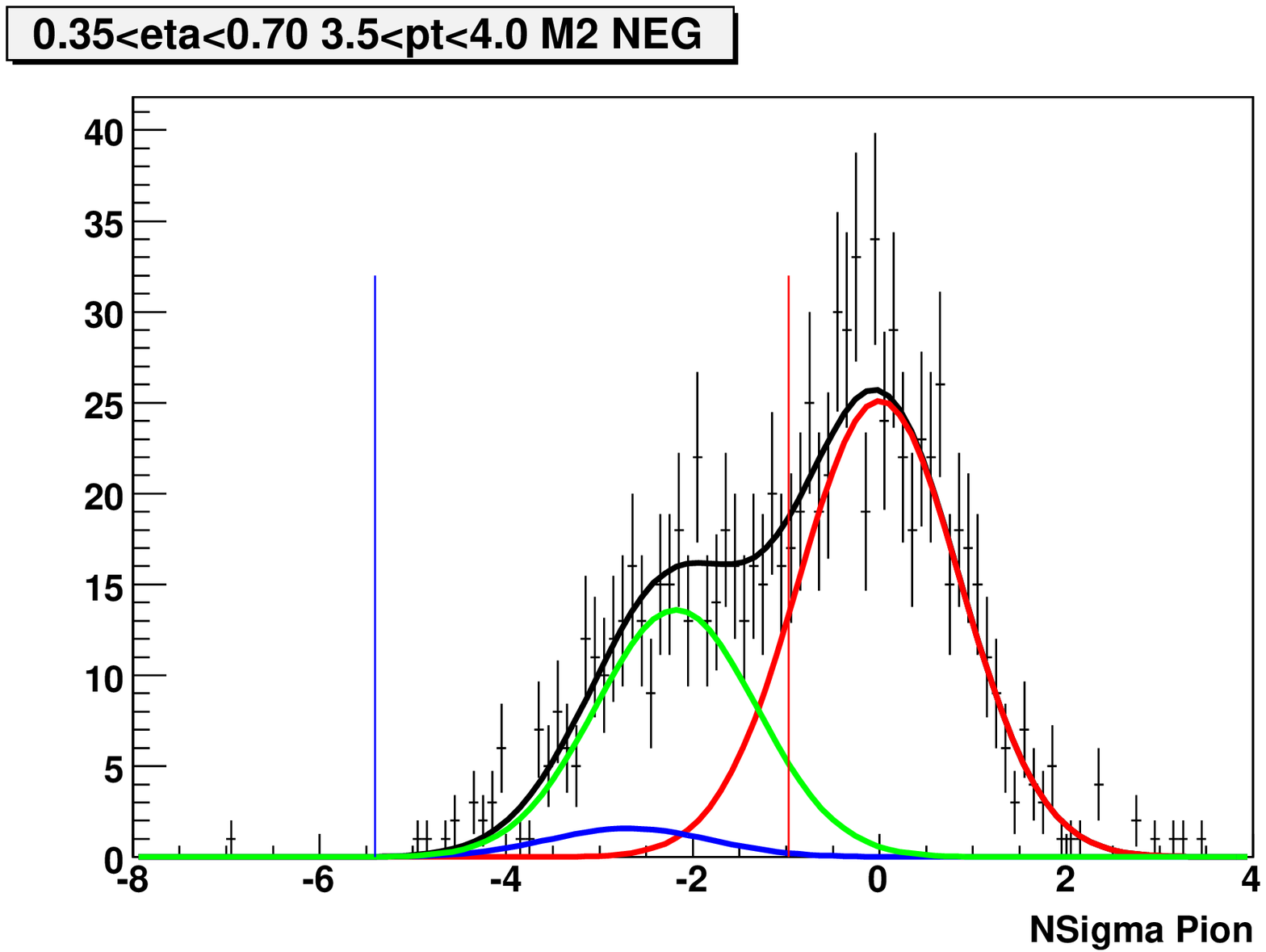}
			\end{minipage}	
					
	\caption{Same as Fig. 6.6 but for negative particles.}
	\label{fig:dAupidcutsN}	
\end{figure}

Figures~\ref{fig:pidcutsP}-\ref{fig:dAupidcutsN} show the $N\sigma_{\pi}$ distributions for positive and negative particles seperately with line representing cuts.  Particles with $N\sigma_{\pi}$ greater than the right line are $\pi^{\pm}$ with a 95\% purity.  Particles with $N\sigma_{\pi}$ less than the left line are $p(\bar{p})$ with a 50\% purity.  The curves are from the fits shown in Figures~\ref{fig:pidfit} and~\ref{fig:dAupidfit}.

\begin{figure}[H]
\hfill
\begin{minipage}[t]{.2\textwidth}
	\centering
		\includegraphics[width=1\textwidth]{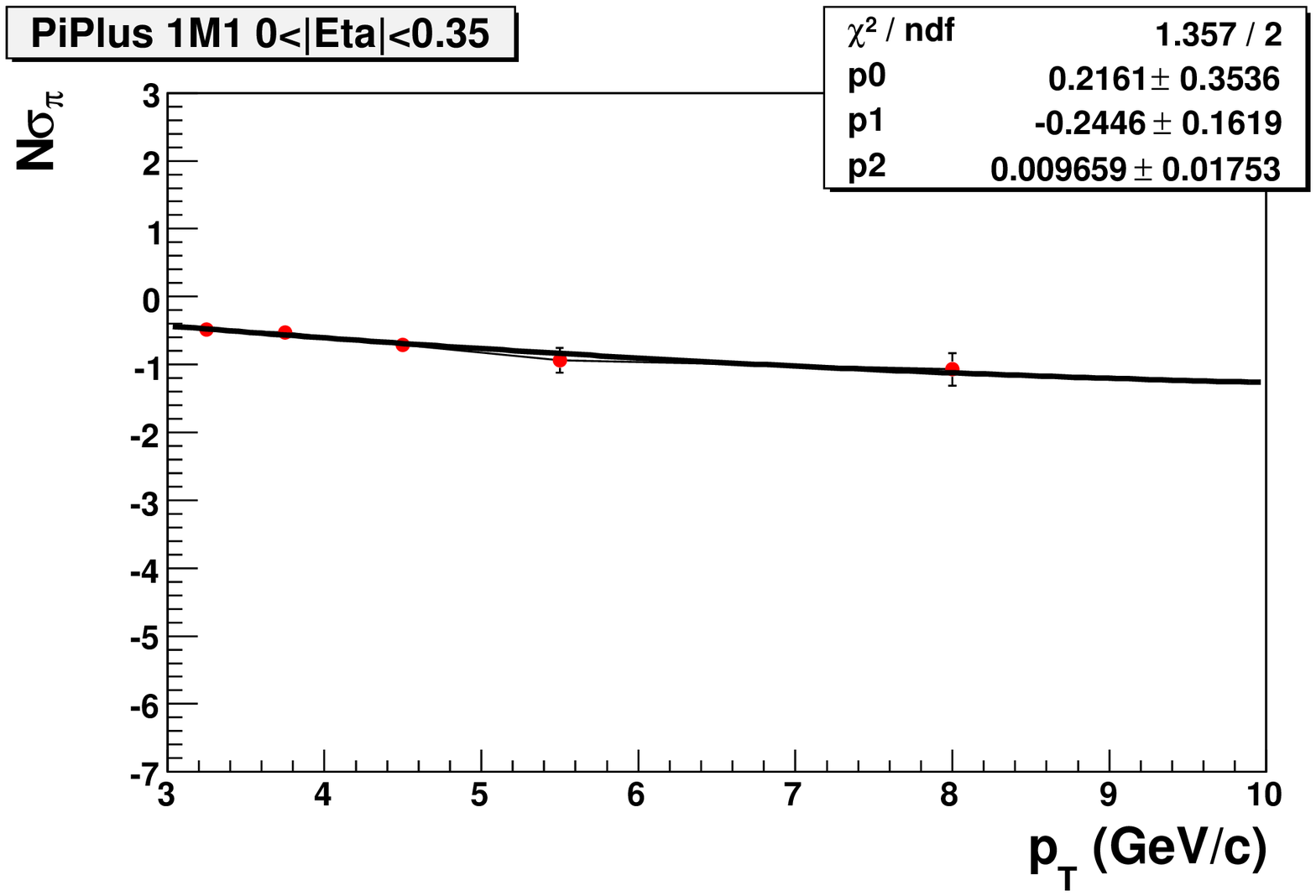}
		\includegraphics[width=1\textwidth]{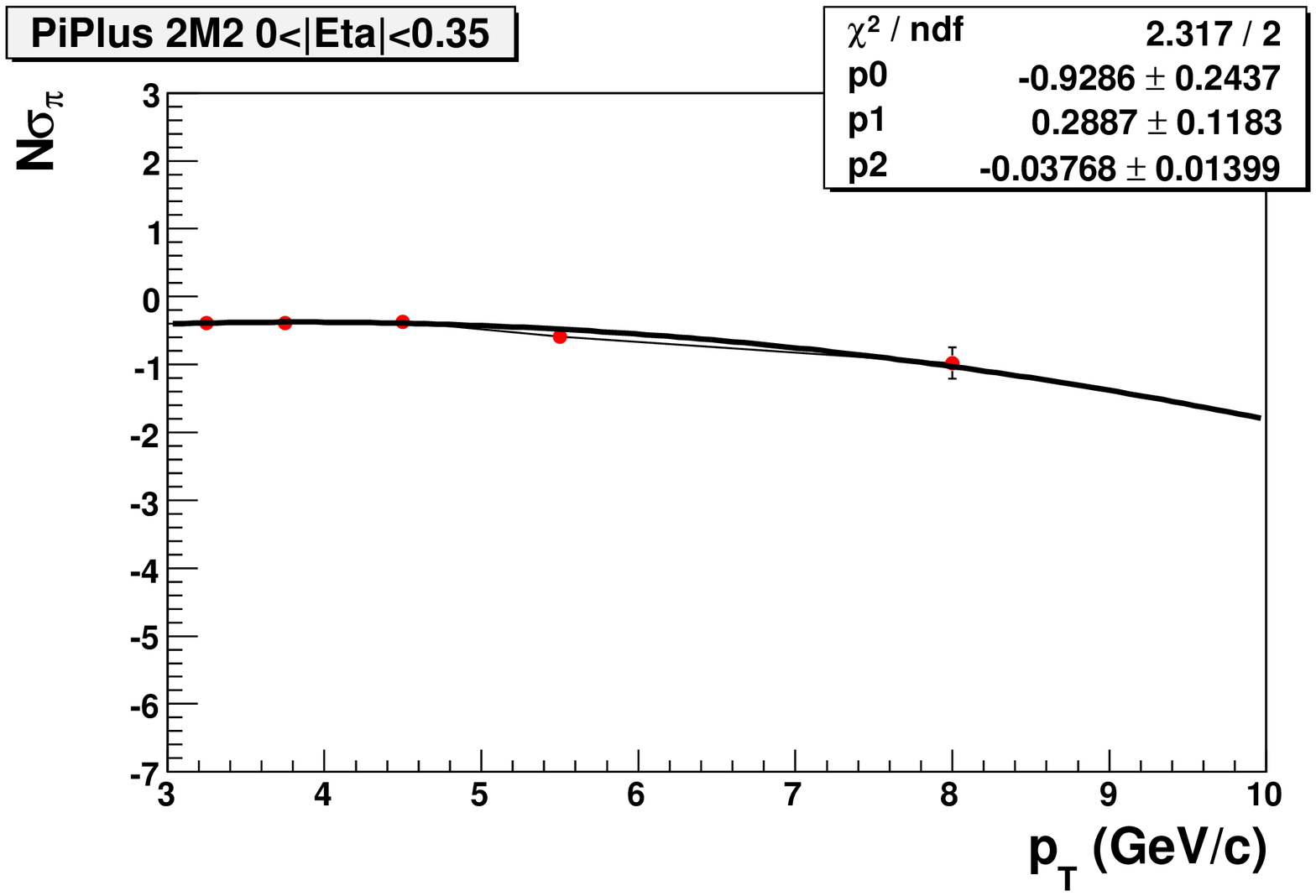}
		\includegraphics[width=1\textwidth]{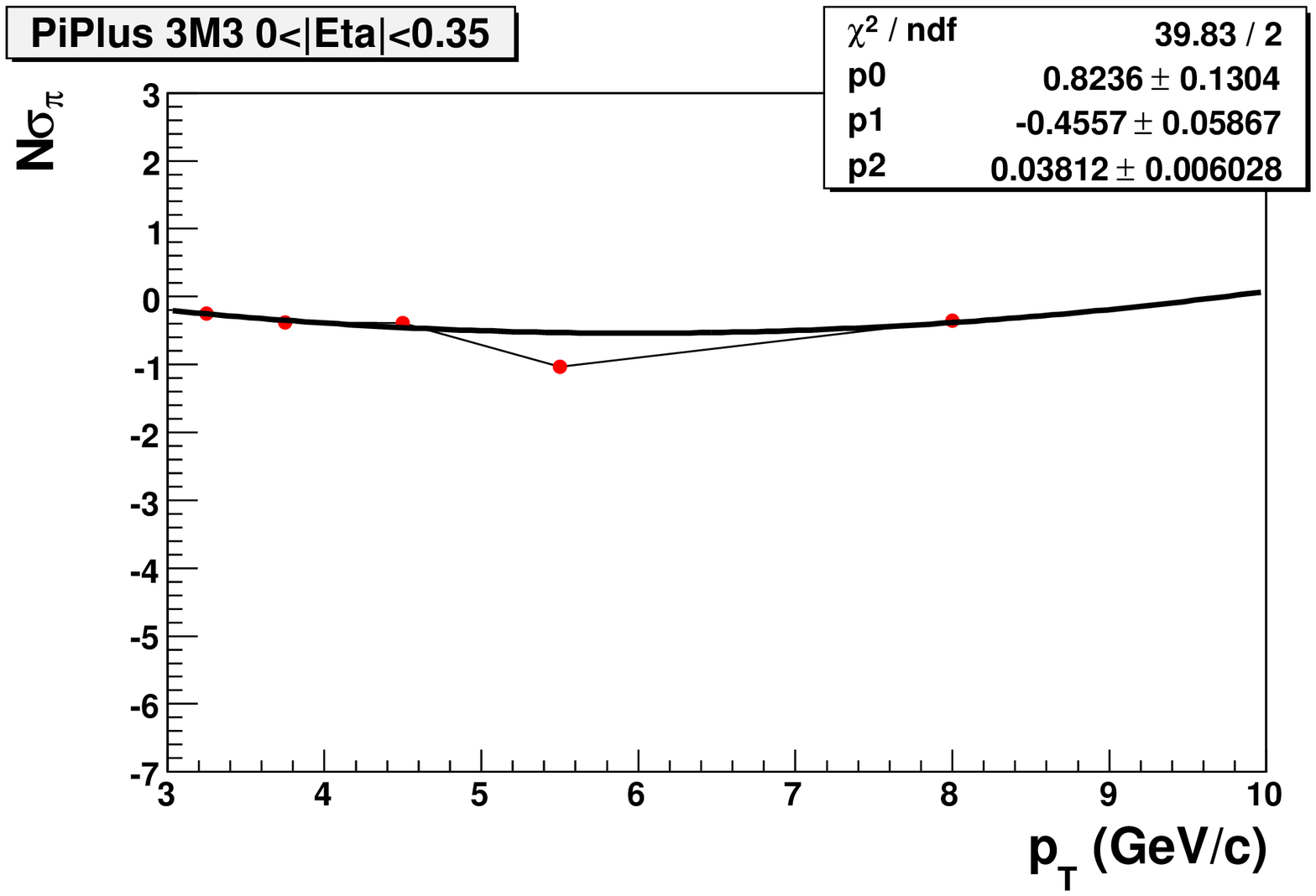}
		\includegraphics[width=1\textwidth]{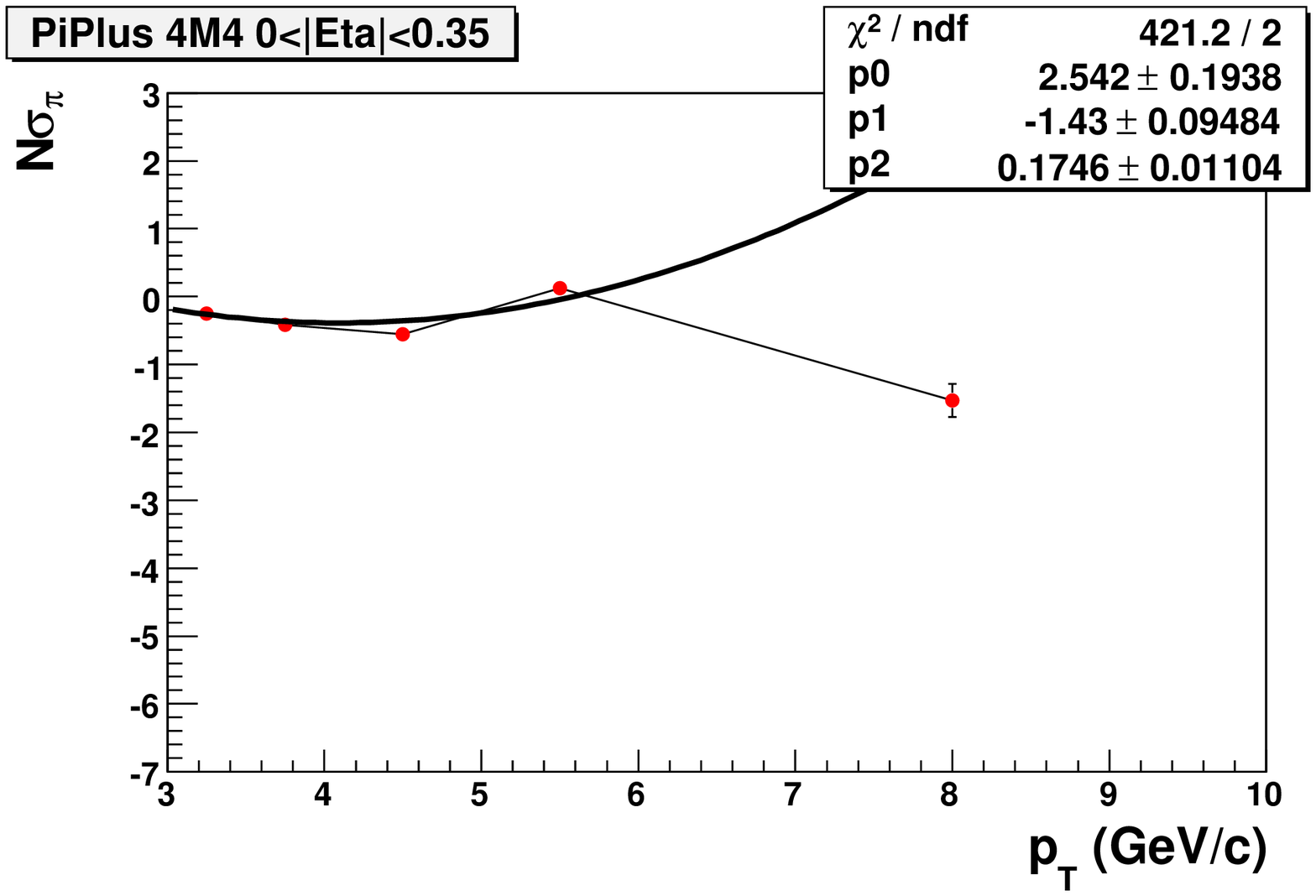}
		\includegraphics[width=1\textwidth]{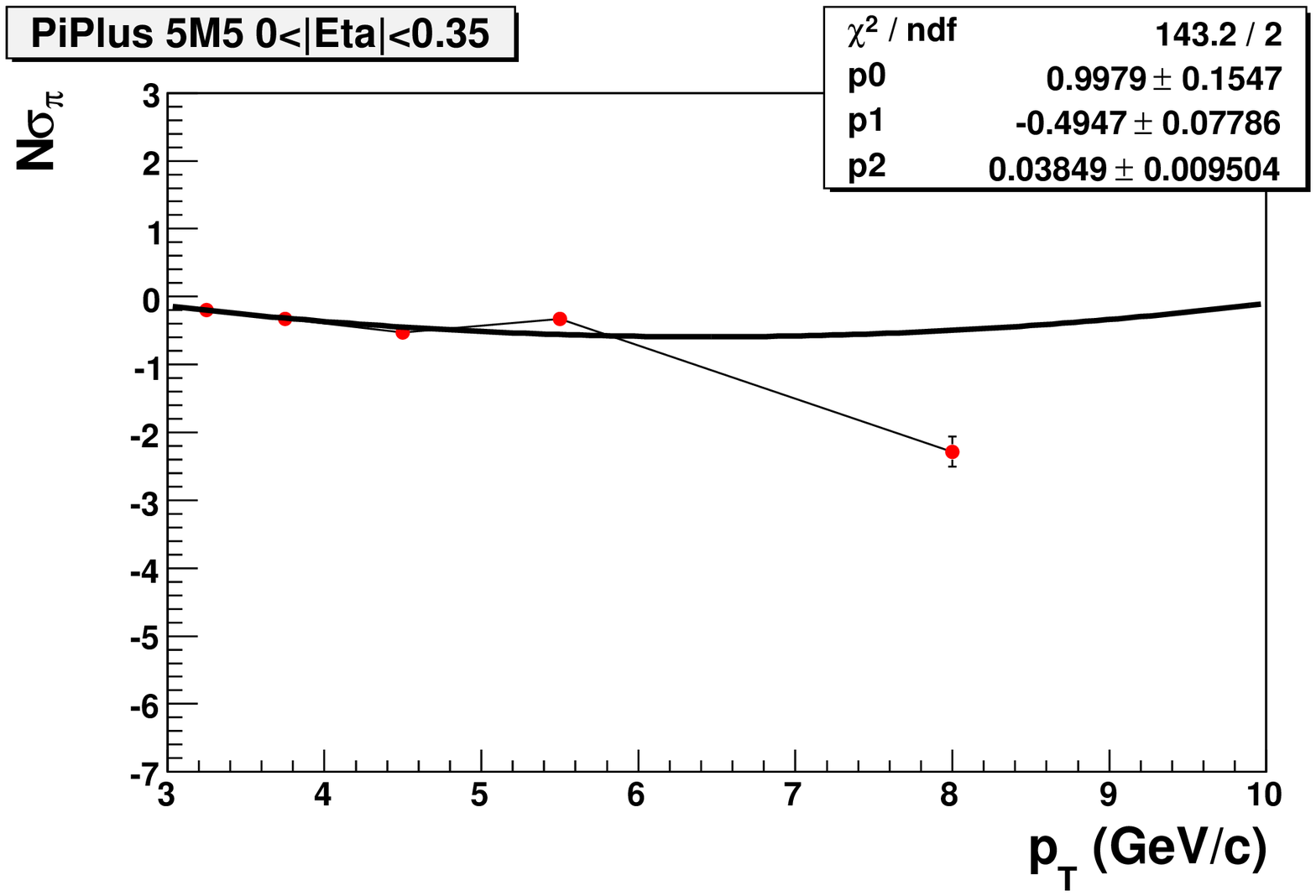}
		\includegraphics[width=1\textwidth]{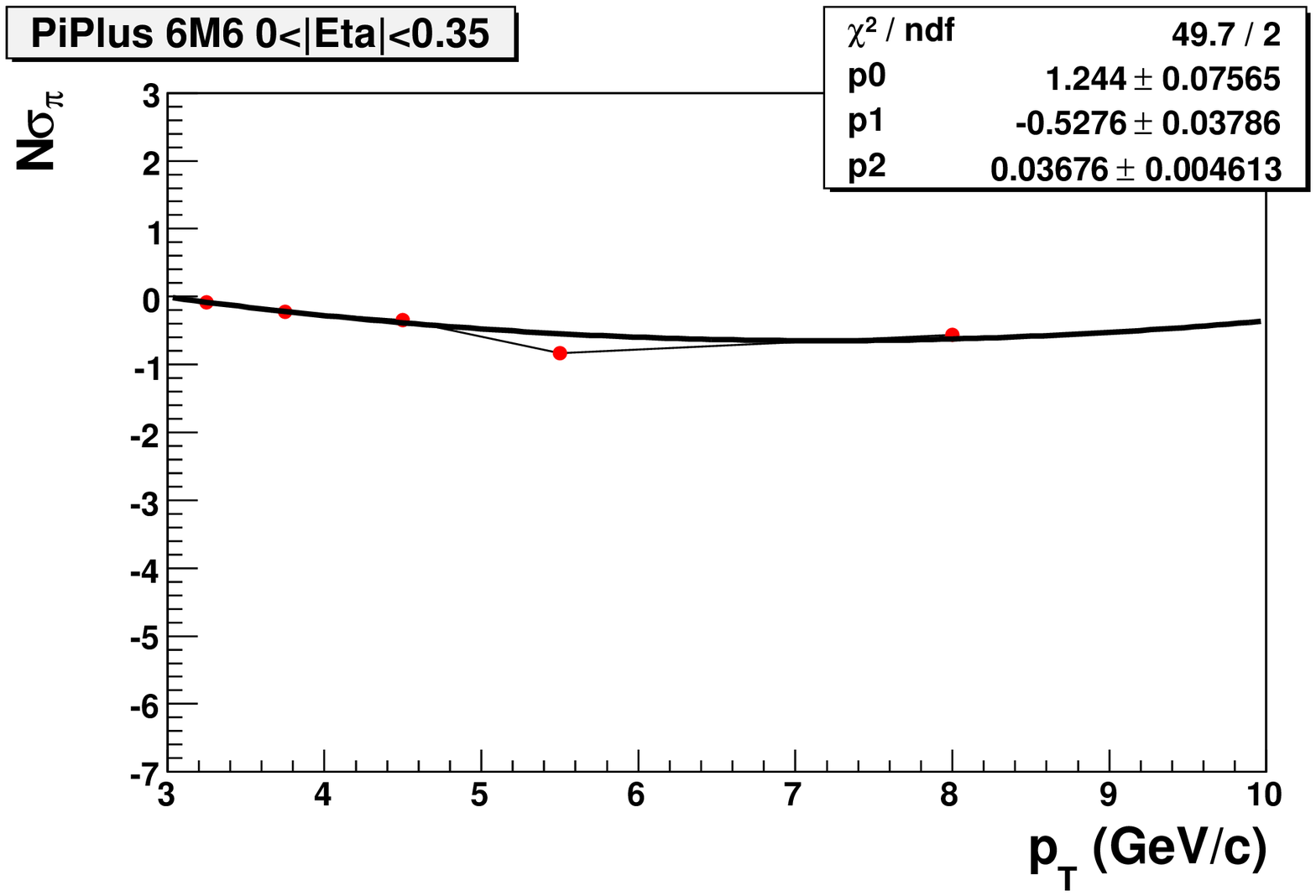}
		\includegraphics[width=1\textwidth]{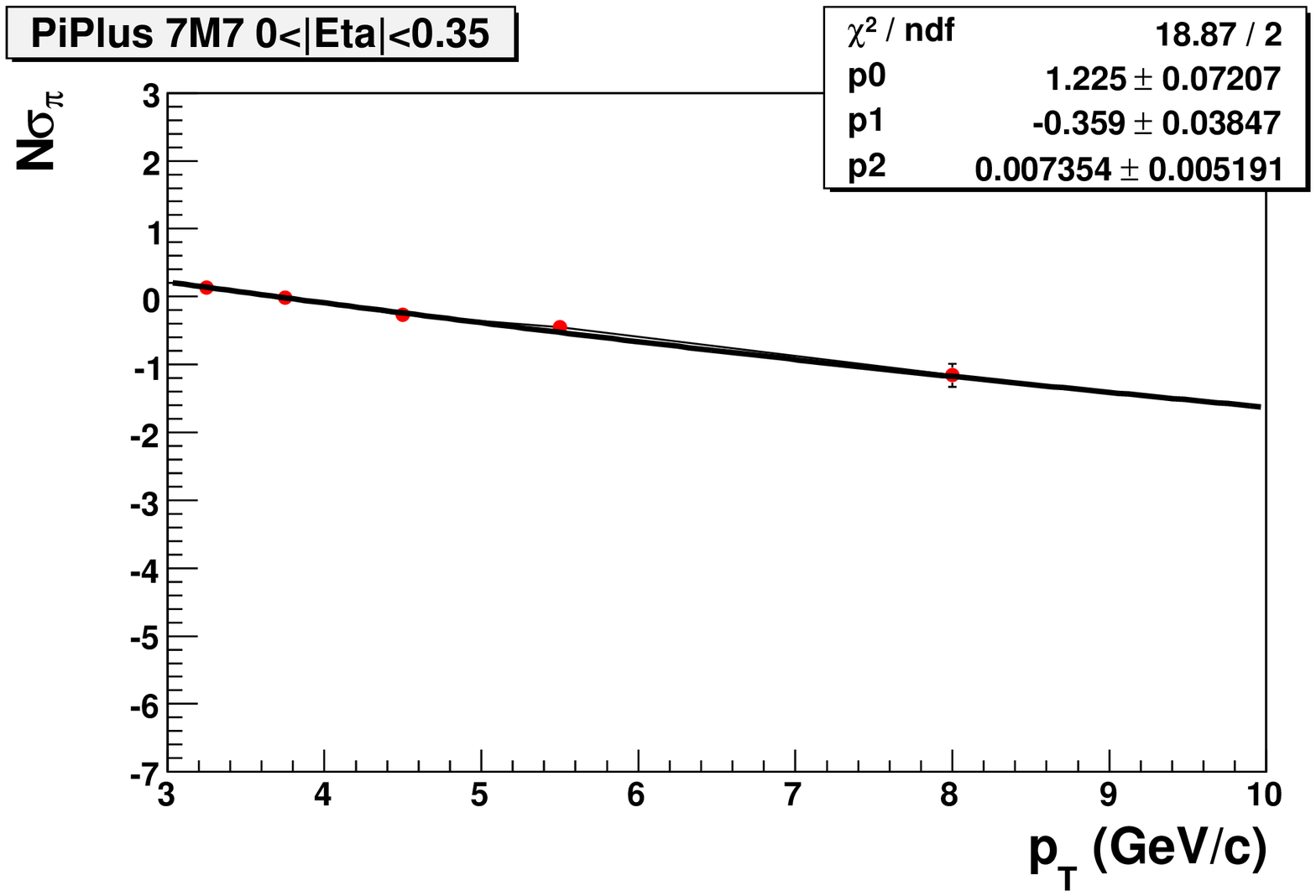}
		\includegraphics[width=1\textwidth]{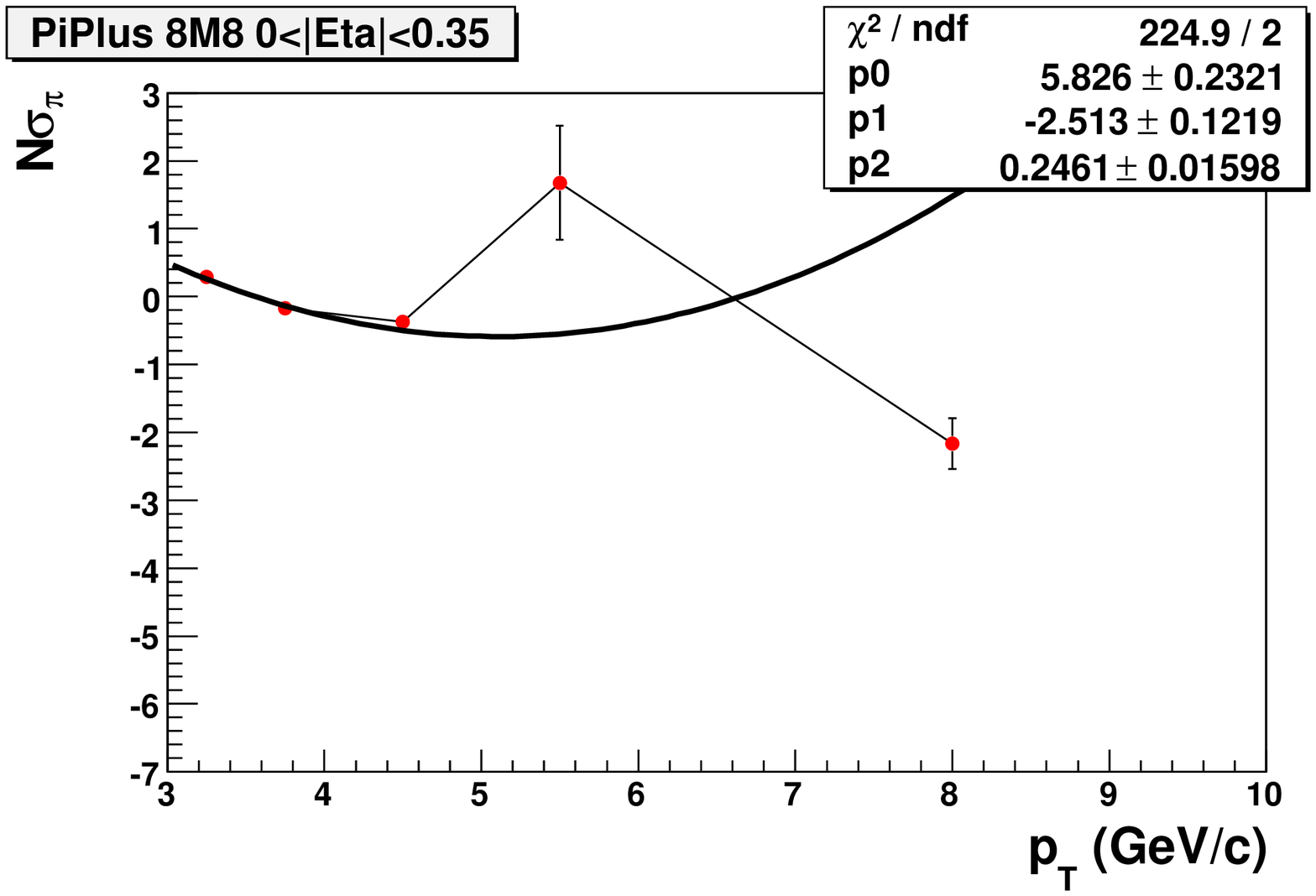}
		\includegraphics[width=1\textwidth]{Plots/PiPlus90.eps}									
			\end{minipage}
\hfill
\begin{minipage}[t]{.2\textwidth}
	\centering
		\includegraphics[width=1\textwidth]{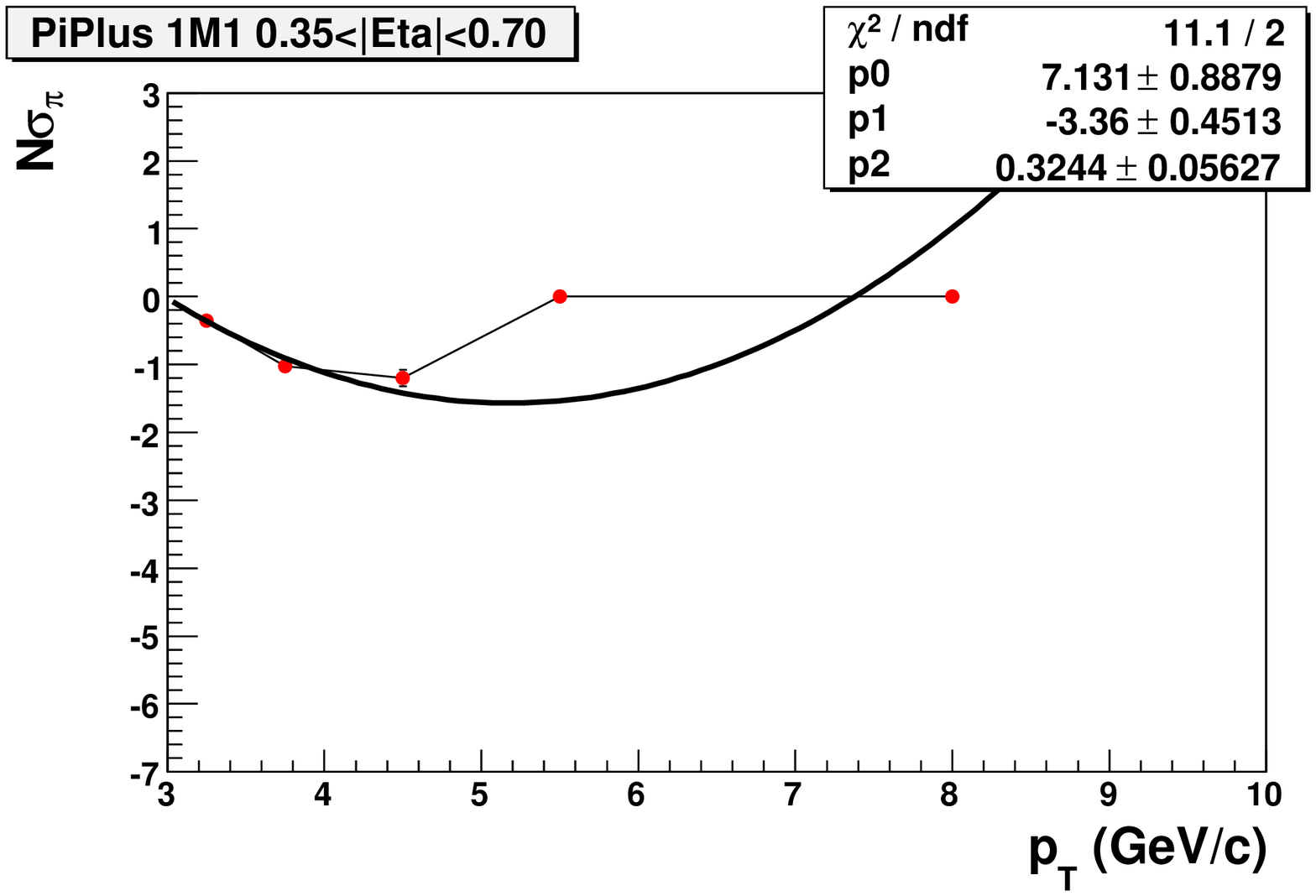}
		\includegraphics[width=1\textwidth]{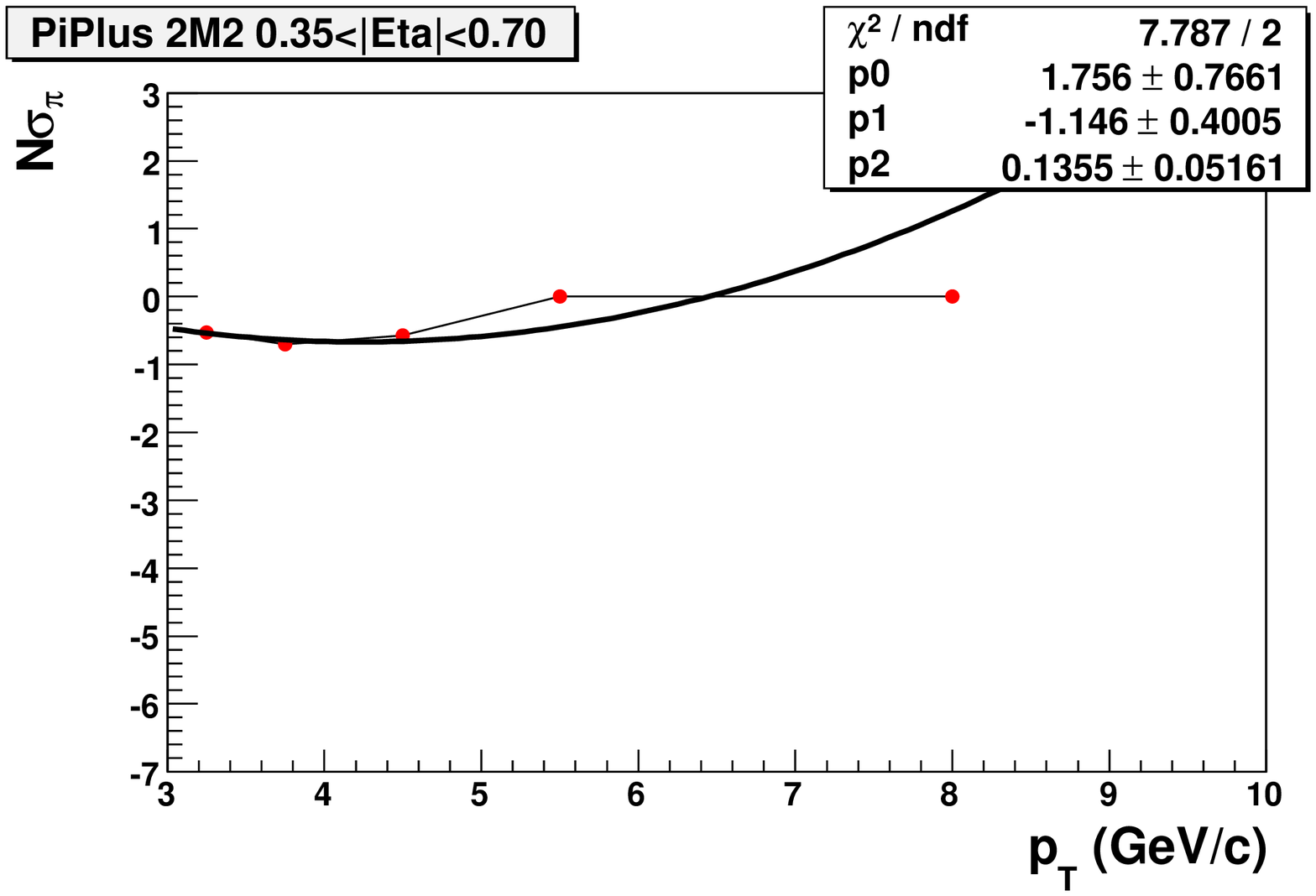}
		\includegraphics[width=1\textwidth]{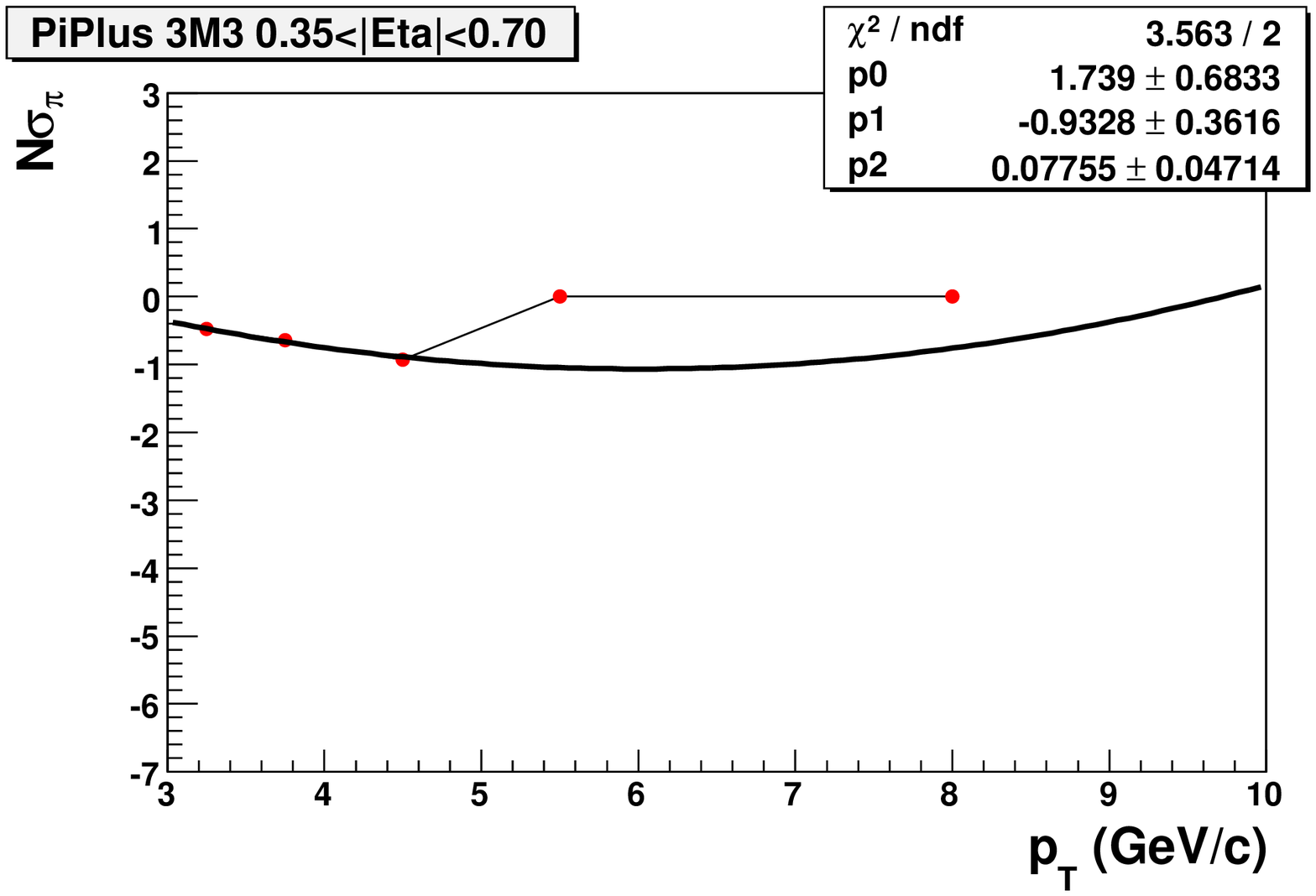}
		\includegraphics[width=1\textwidth]{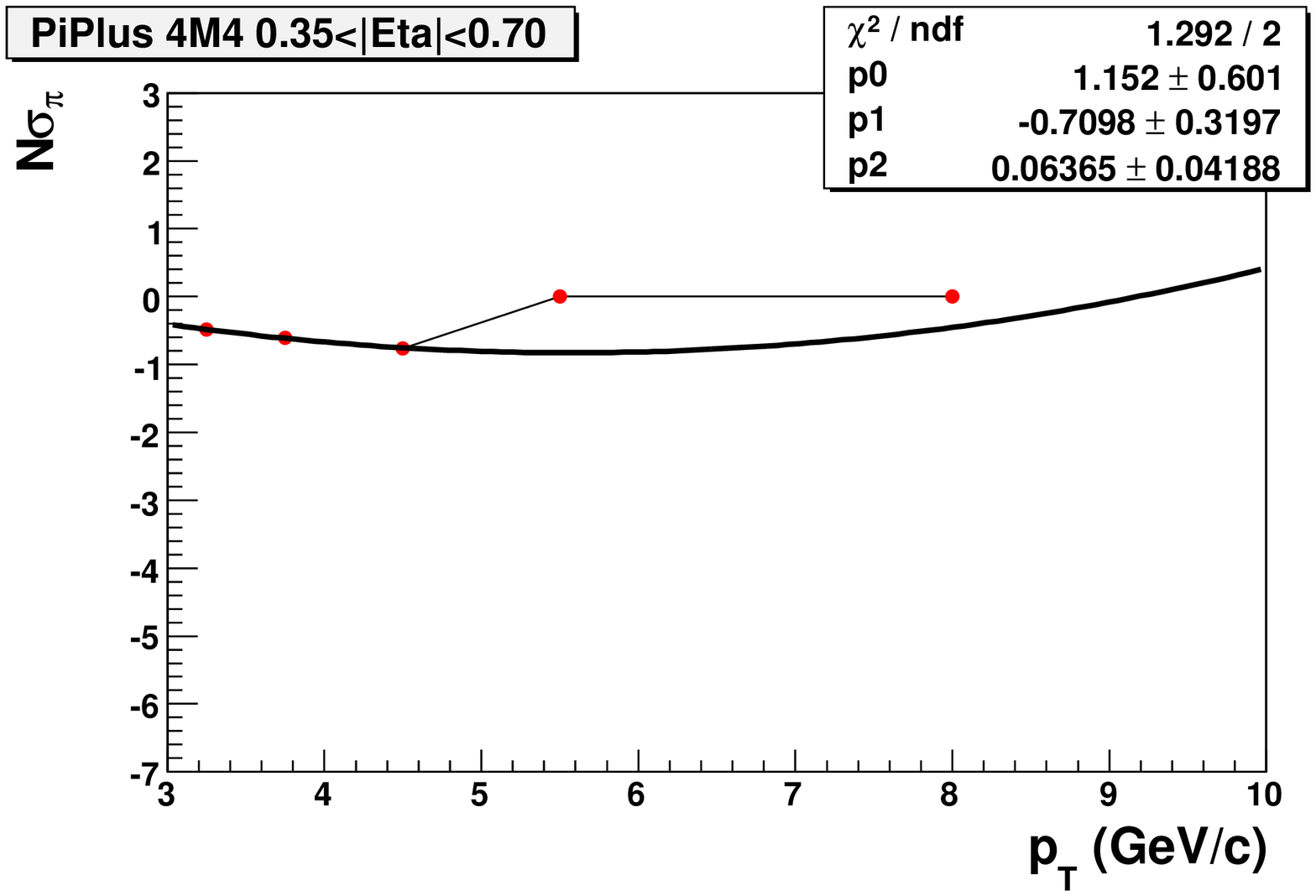}
		\includegraphics[width=1\textwidth]{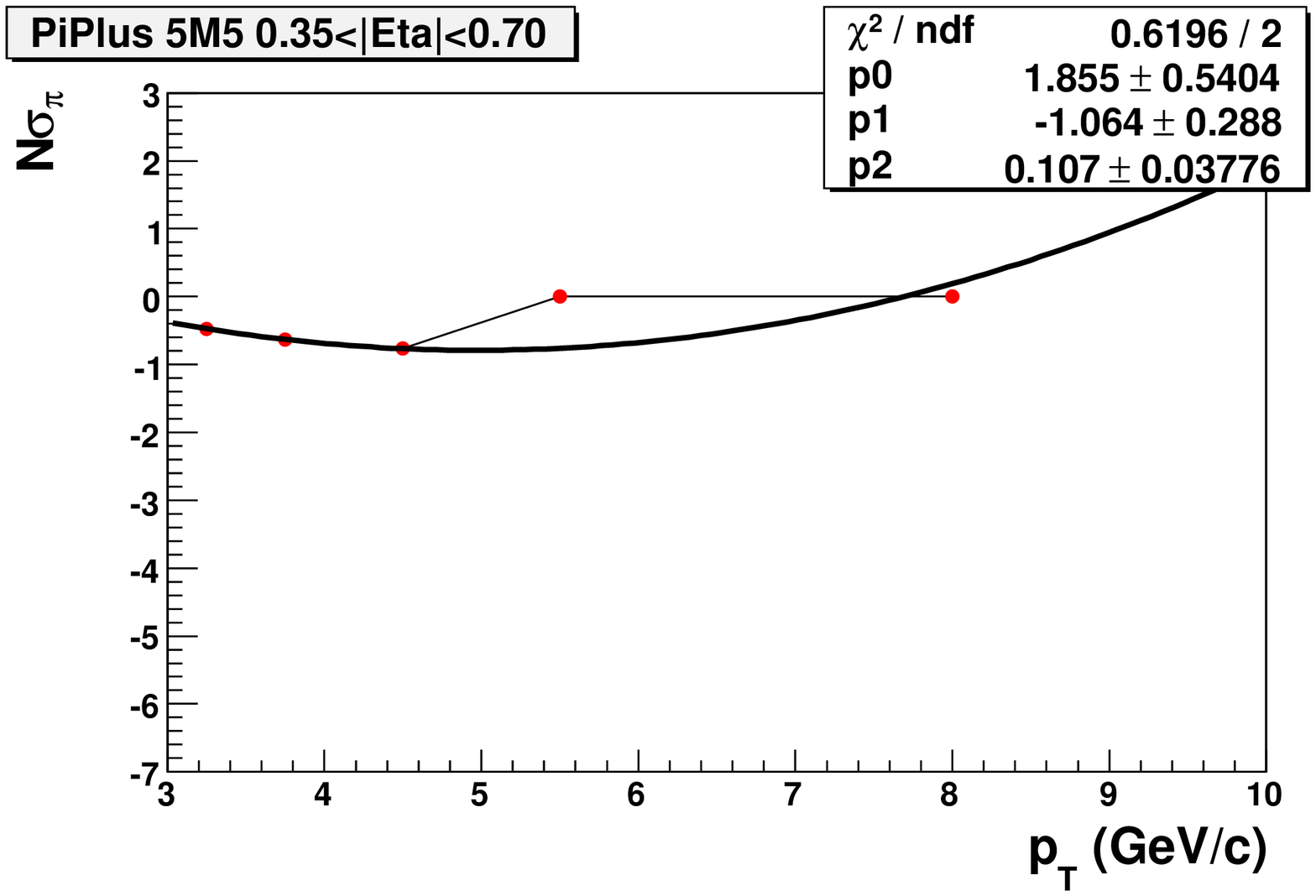}
		\includegraphics[width=1\textwidth]{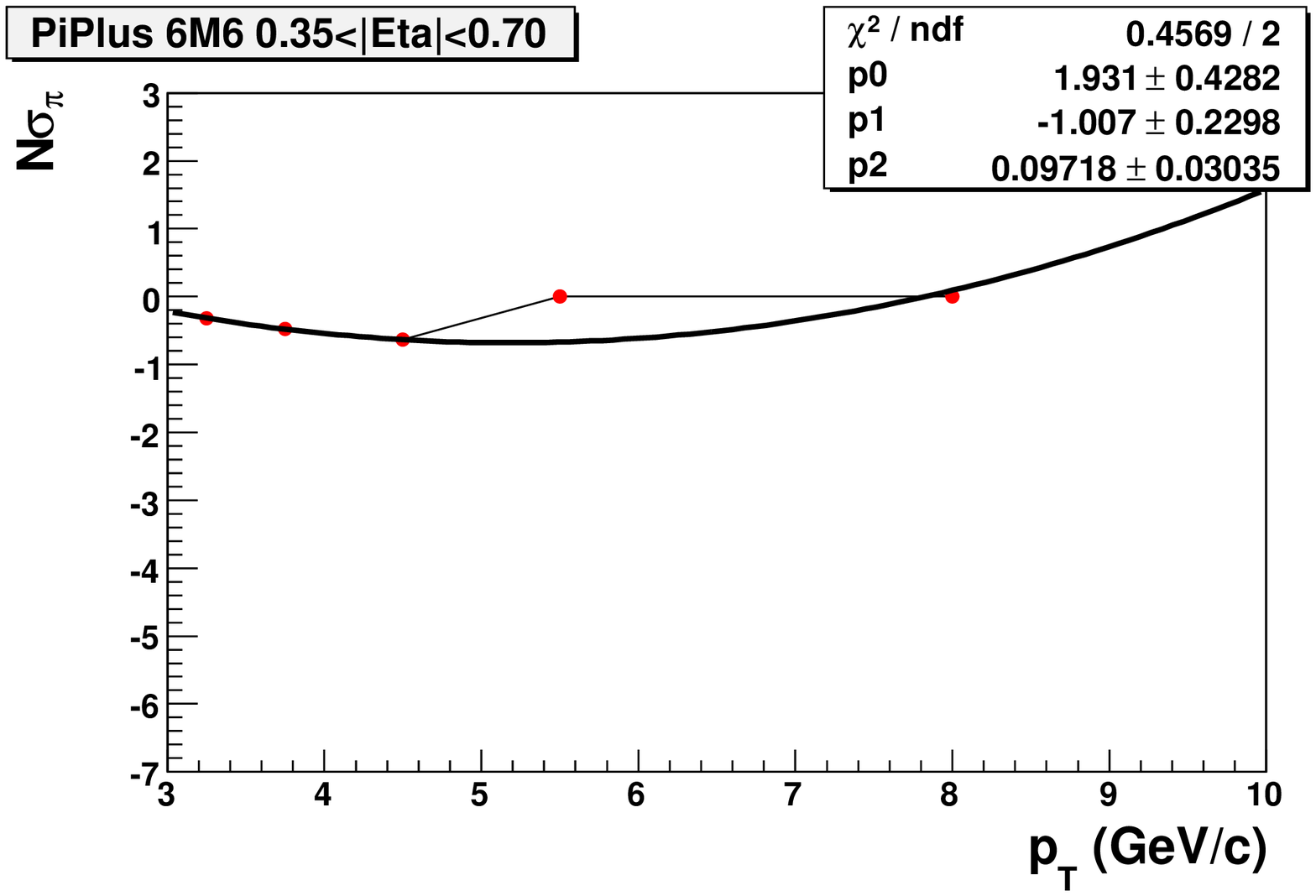}
		\includegraphics[width=1\textwidth]{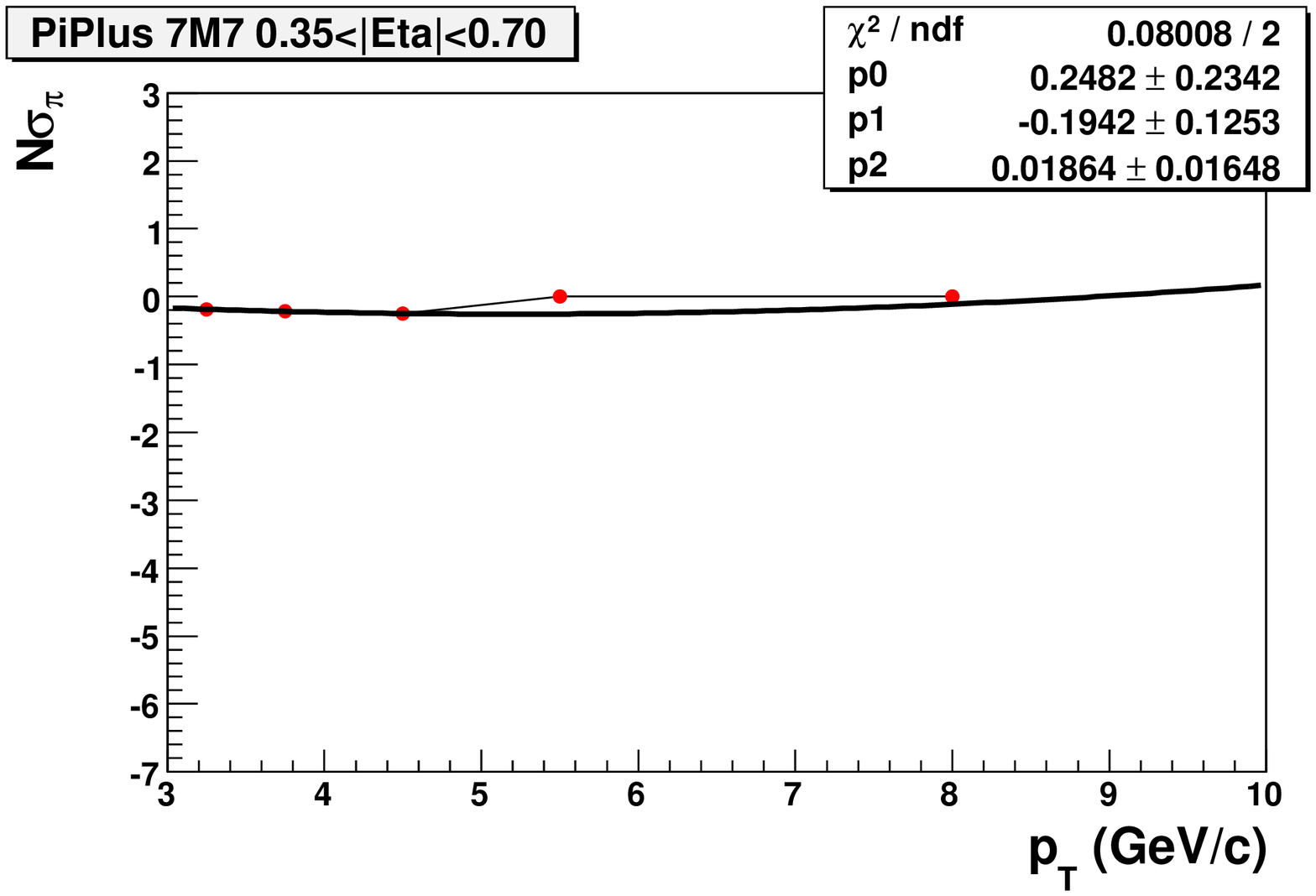}
		\includegraphics[width=1\textwidth]{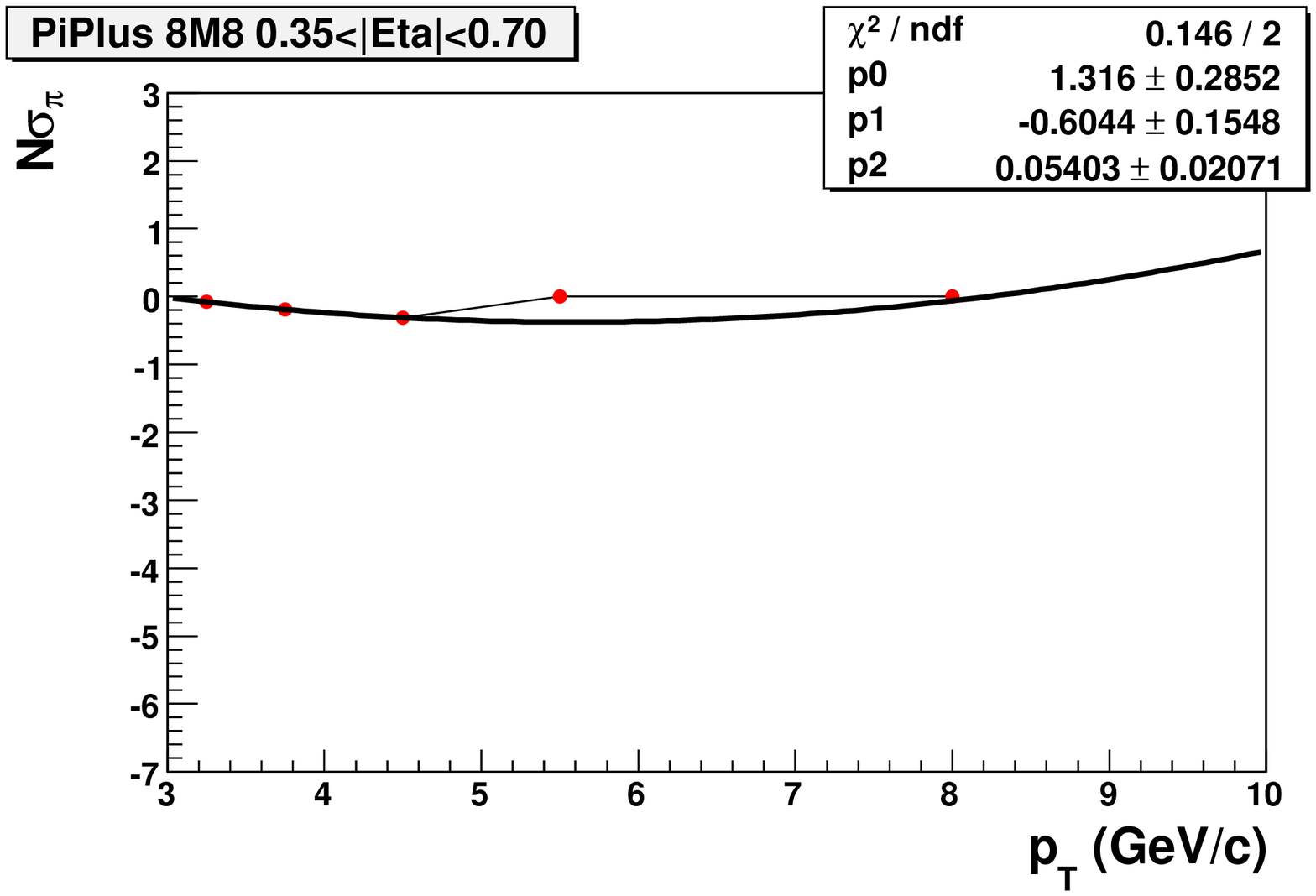}
		\includegraphics[width=1\textwidth]{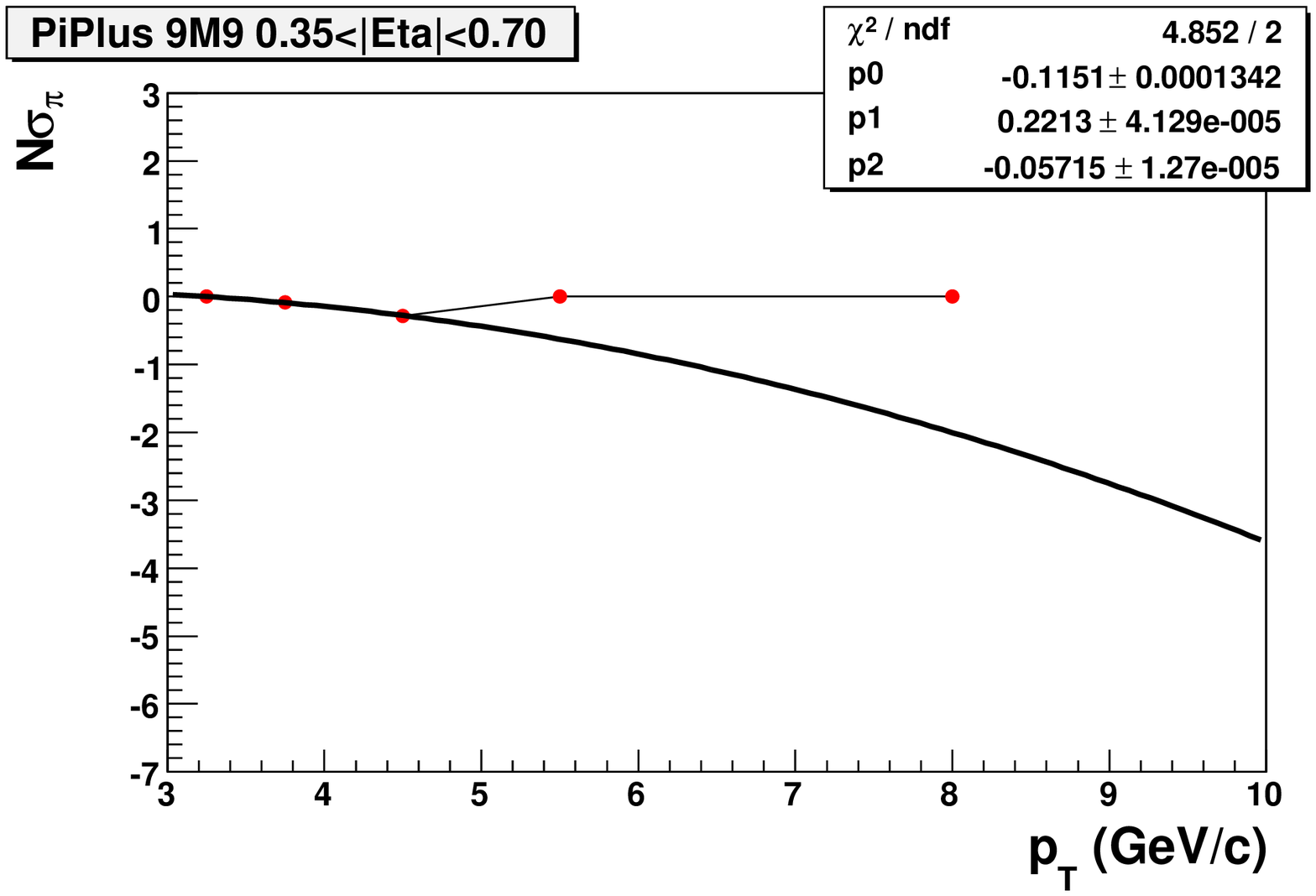}									
			\end{minipage}
\hfill
\begin{minipage}[t]{.2\textwidth}
	\centering
		\includegraphics[width=1\textwidth]{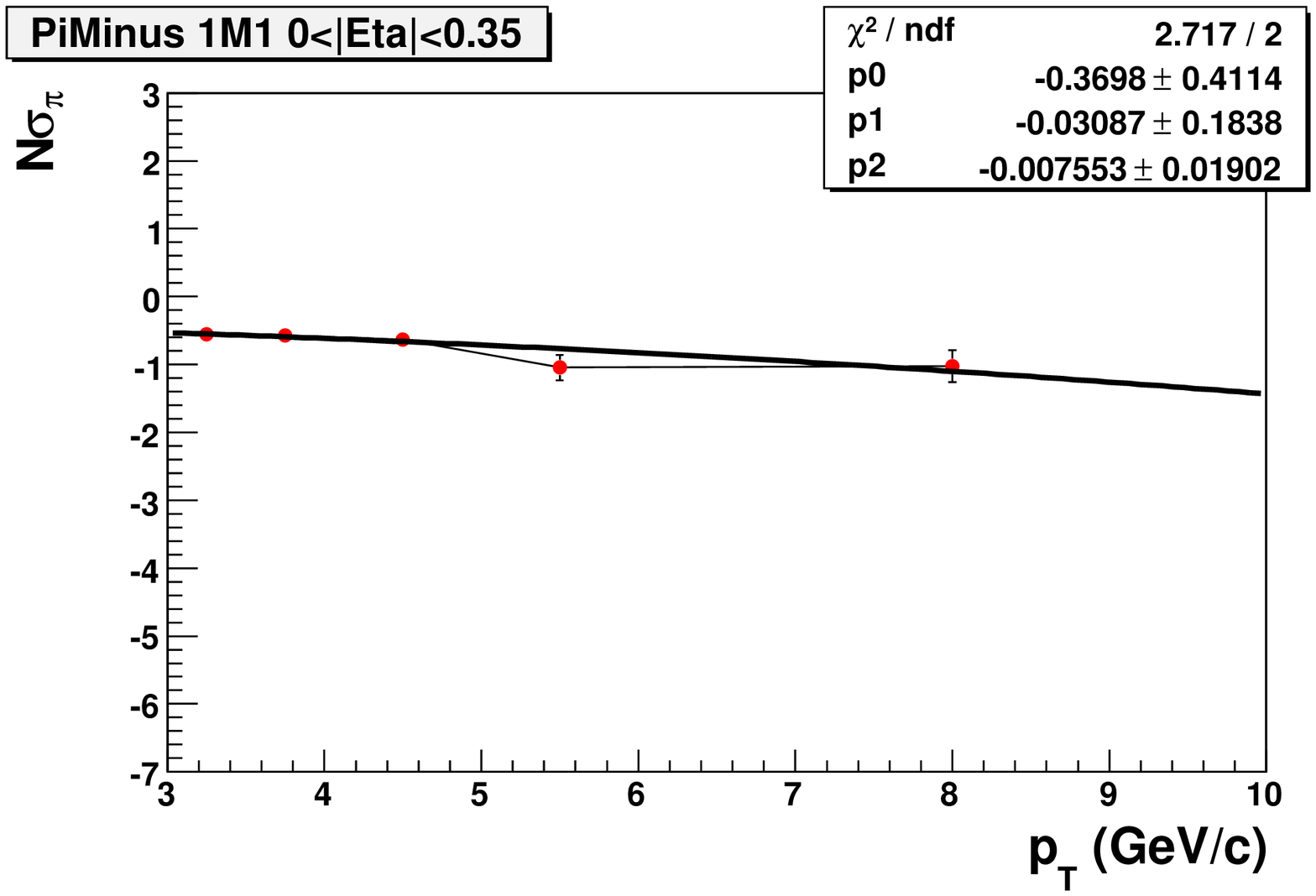}
		\includegraphics[width=1\textwidth]{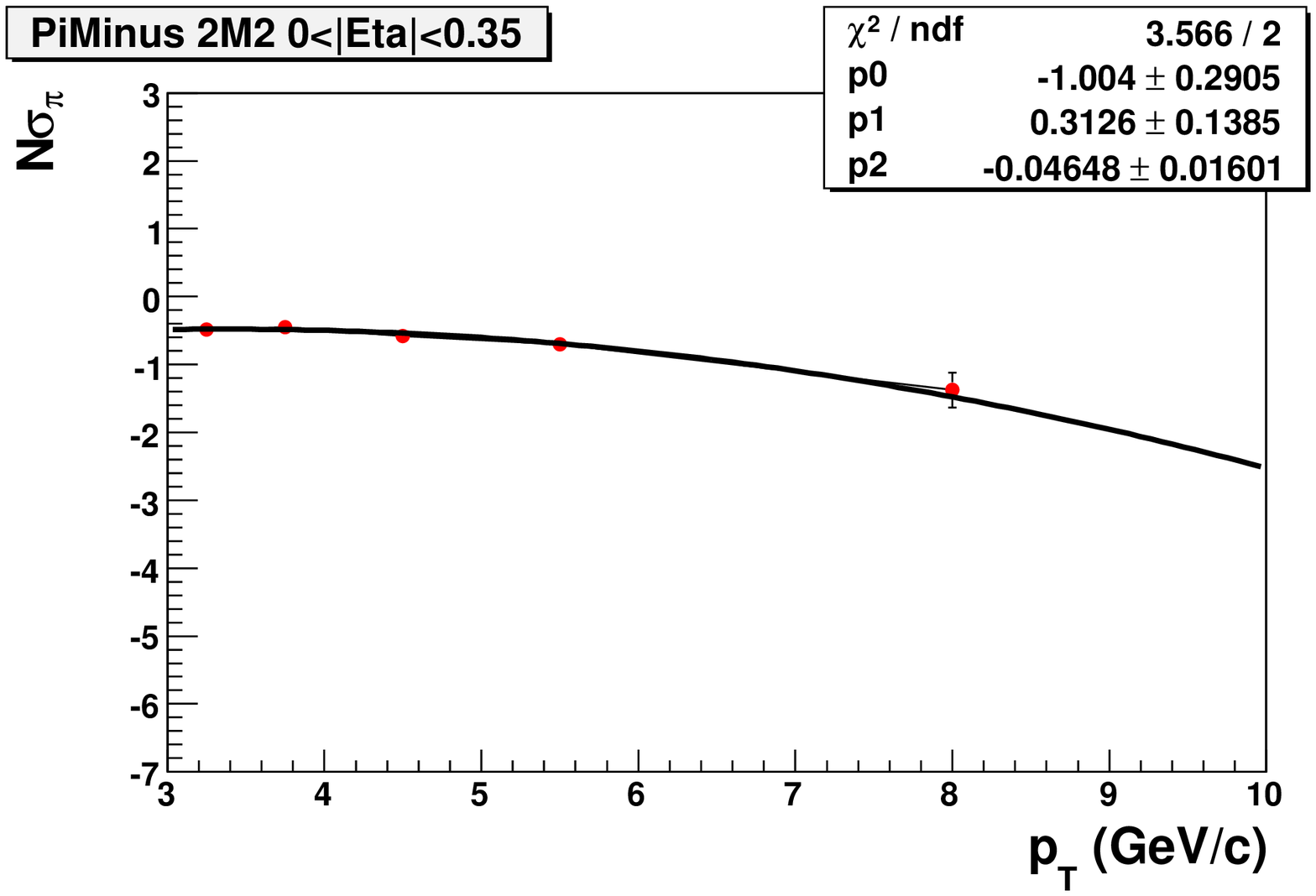}
		\includegraphics[width=1\textwidth]{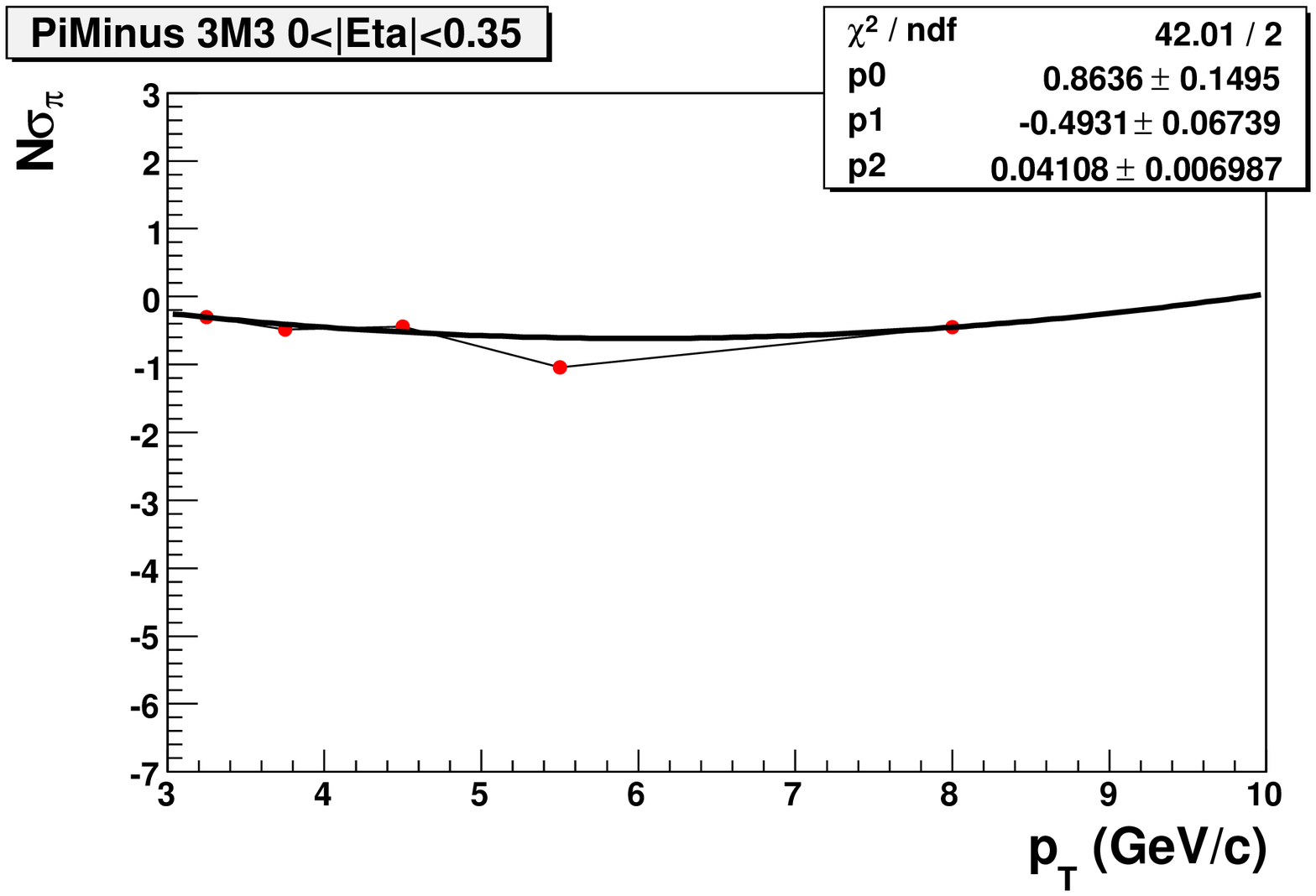}
		\includegraphics[width=1\textwidth]{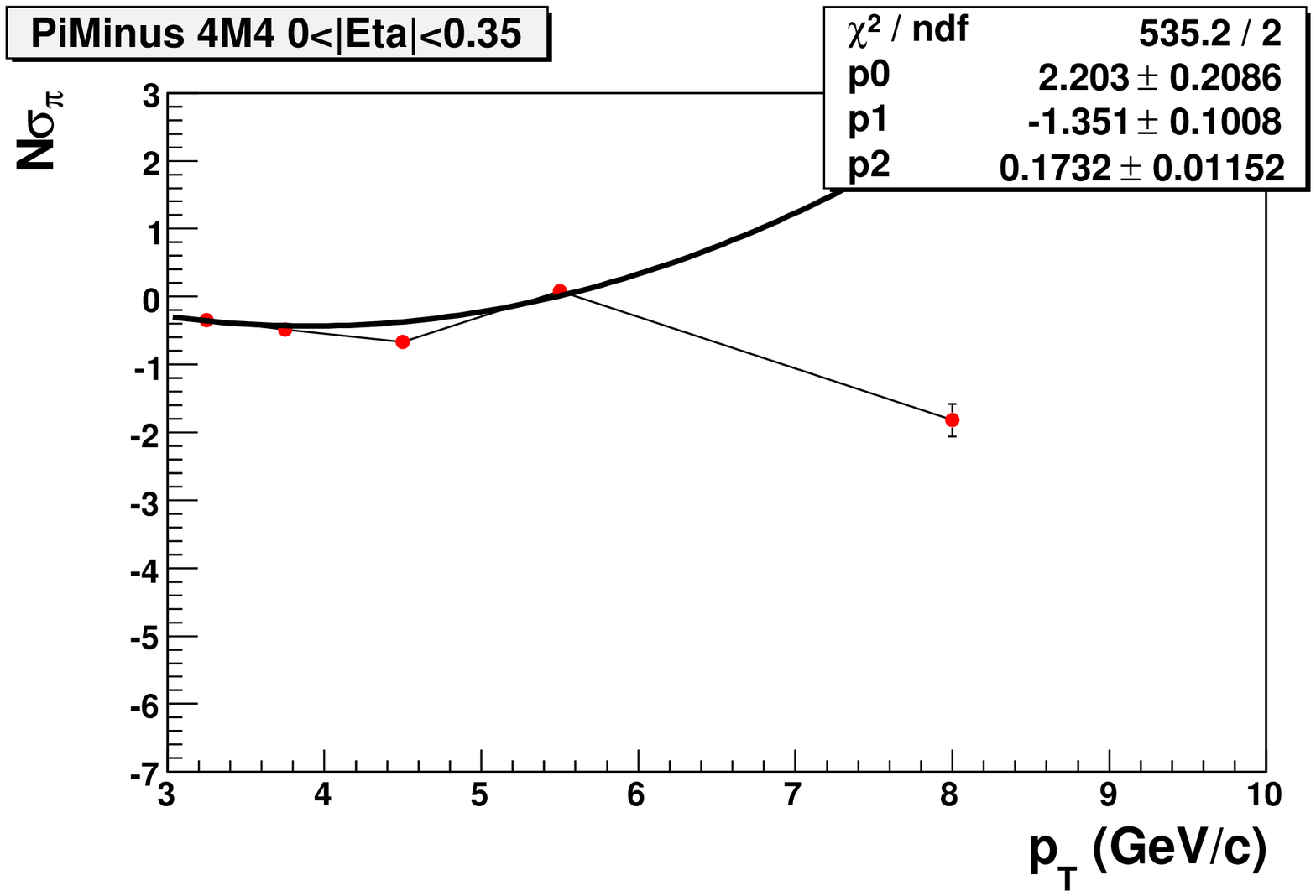}
		\includegraphics[width=1\textwidth]{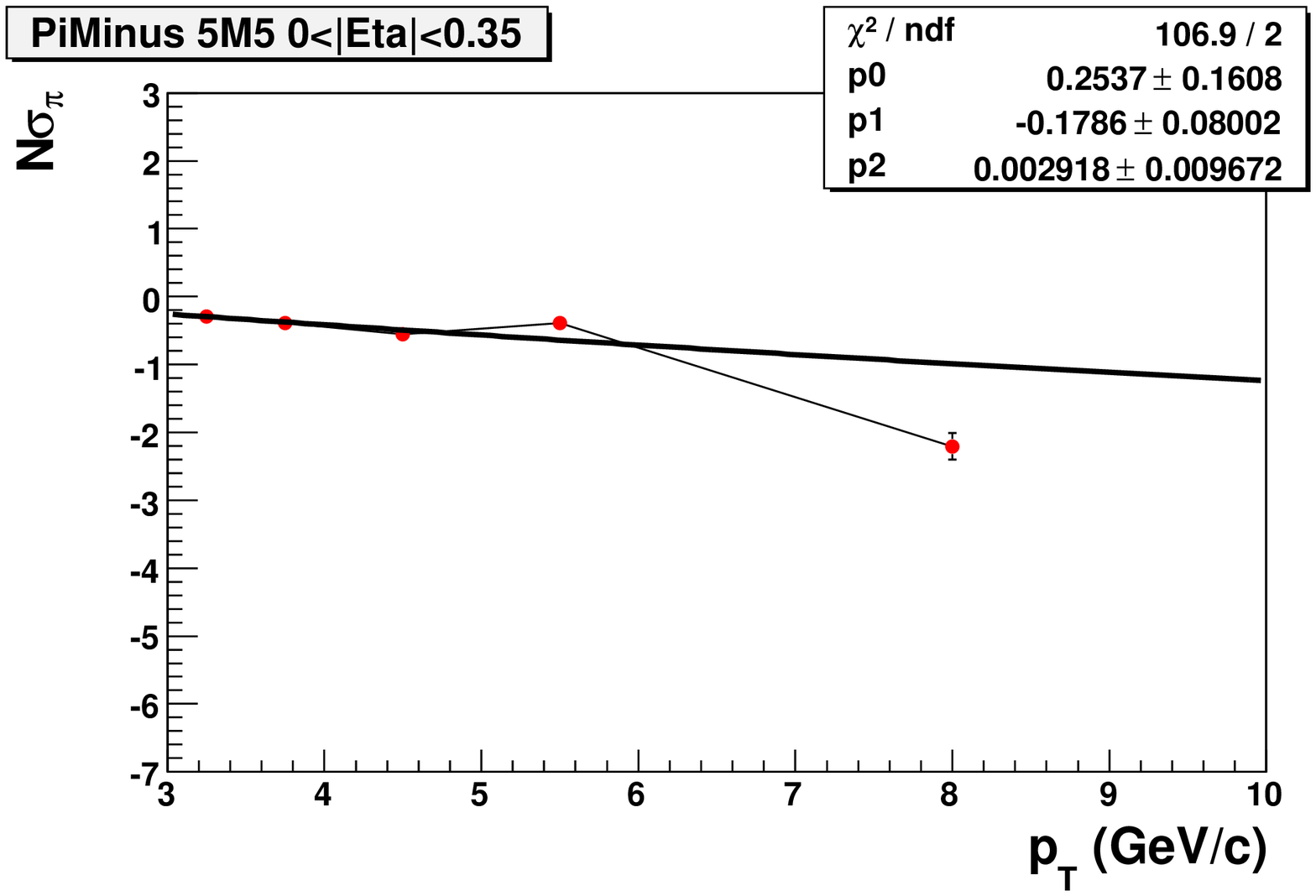}
		\includegraphics[width=1\textwidth]{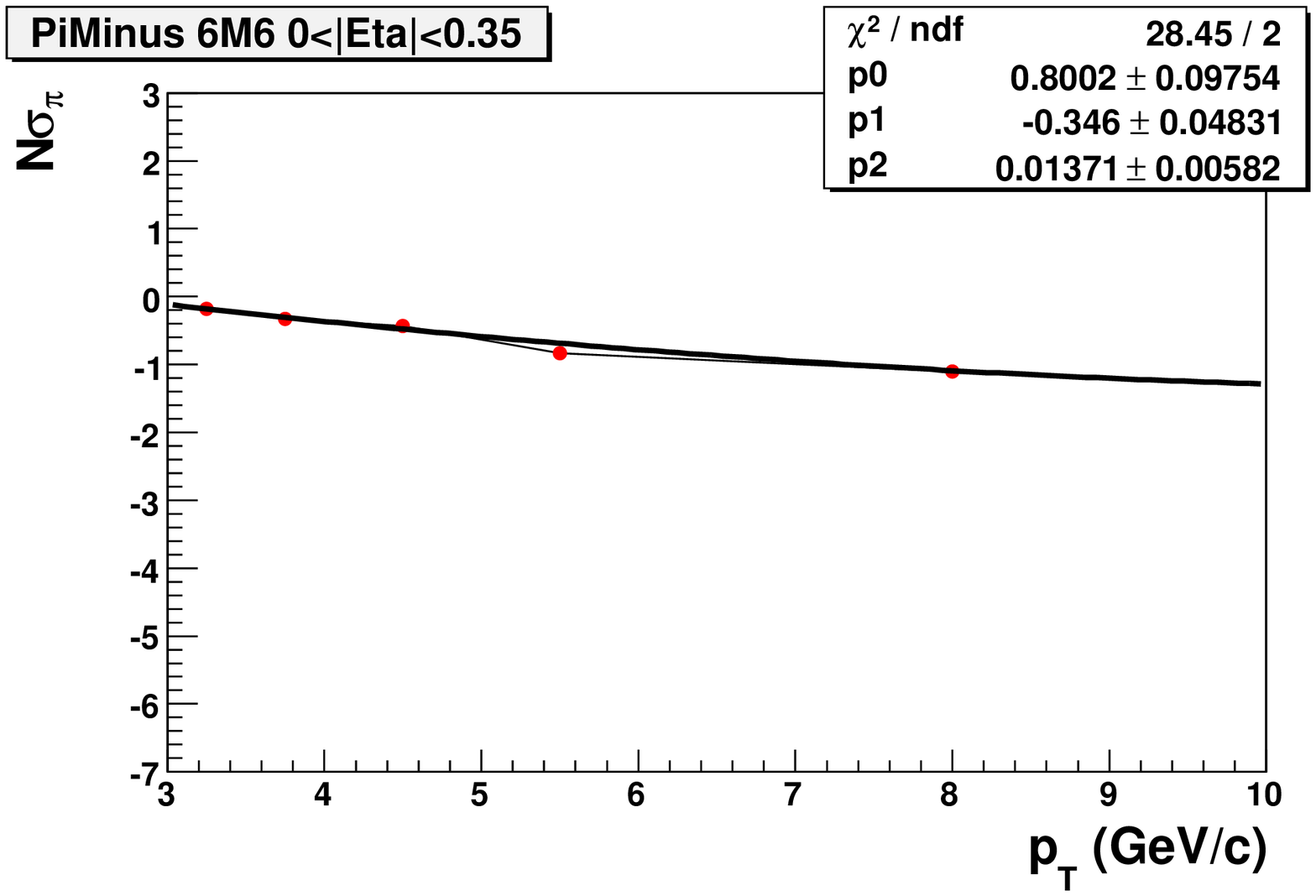}
		\includegraphics[width=1\textwidth]{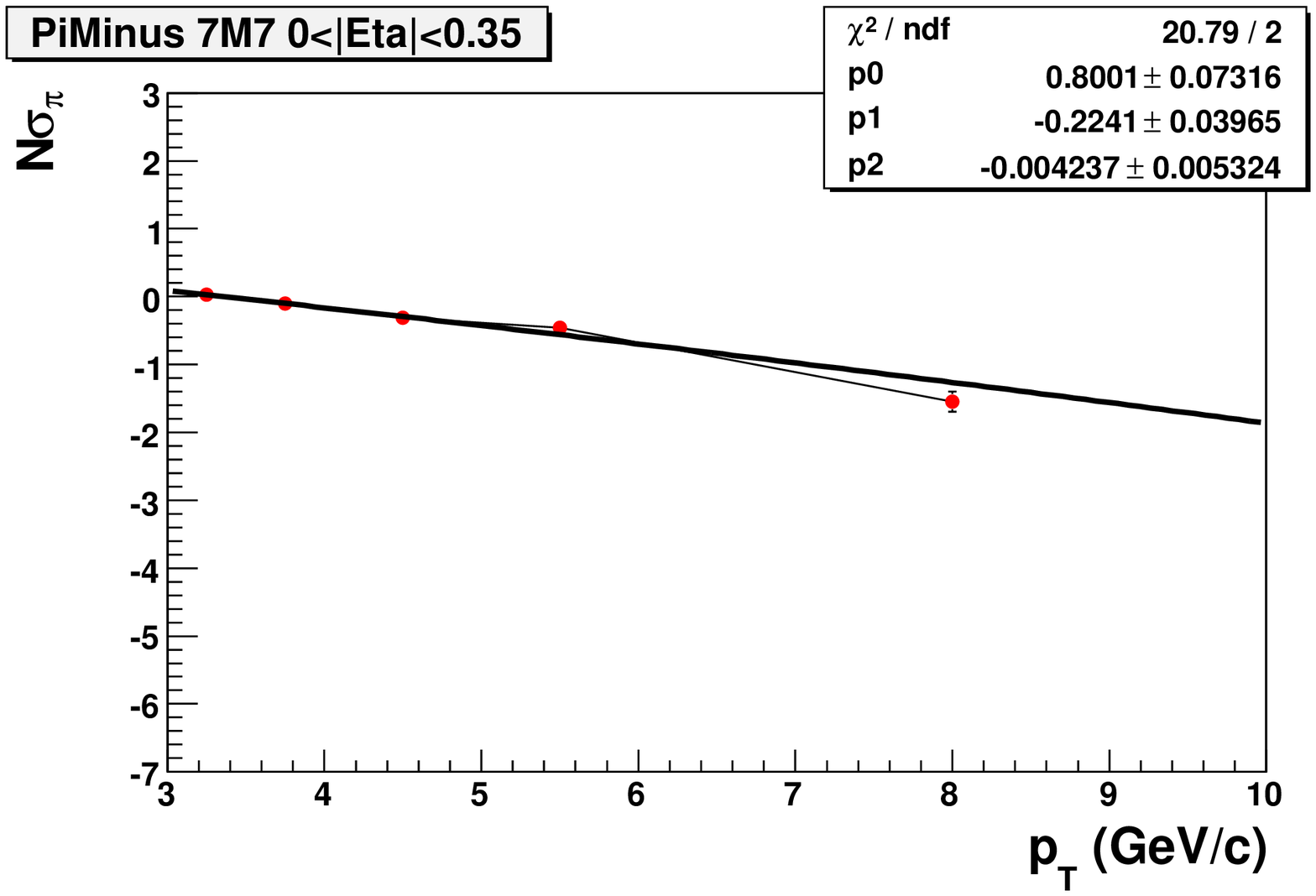}
		\includegraphics[width=1\textwidth]{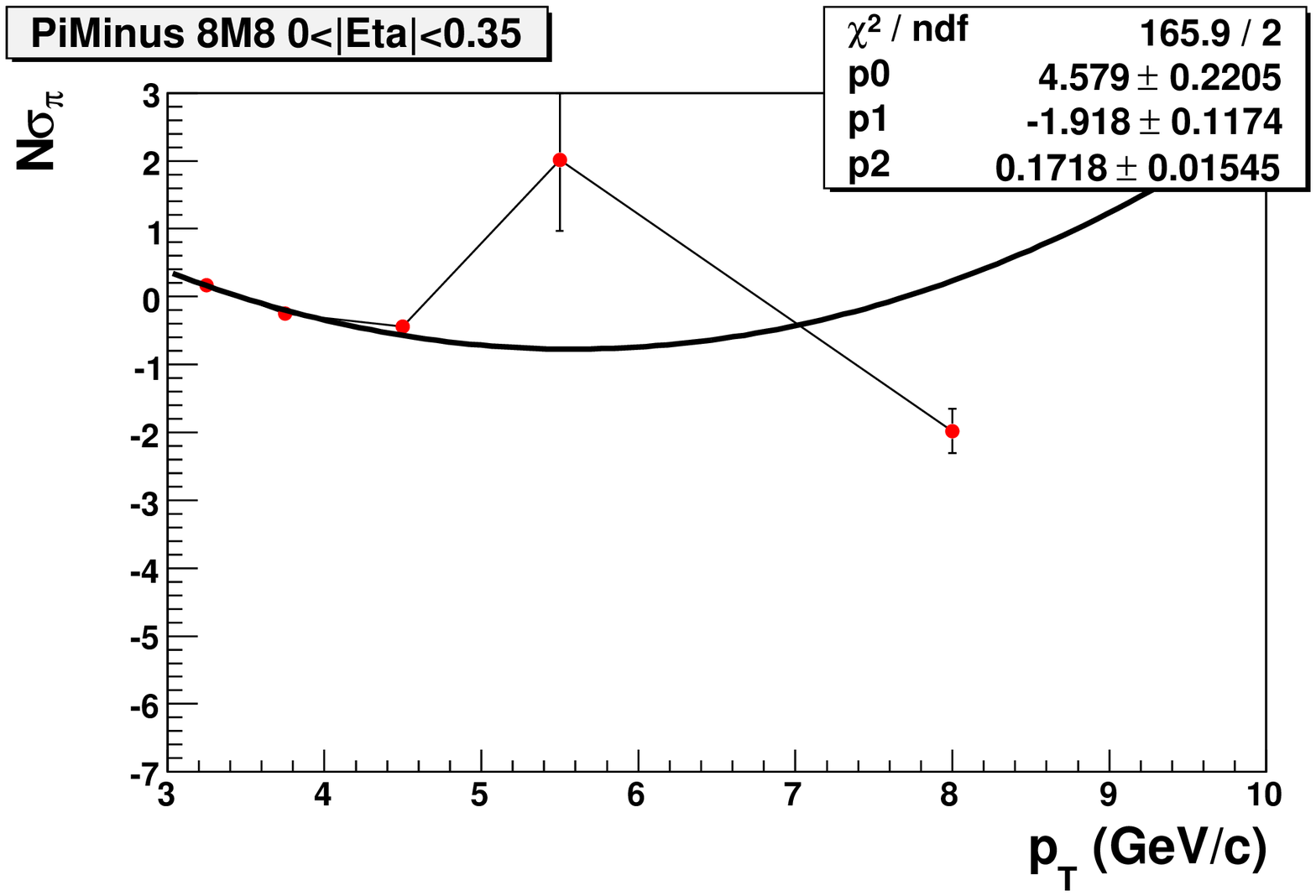}
		\includegraphics[width=1\textwidth]{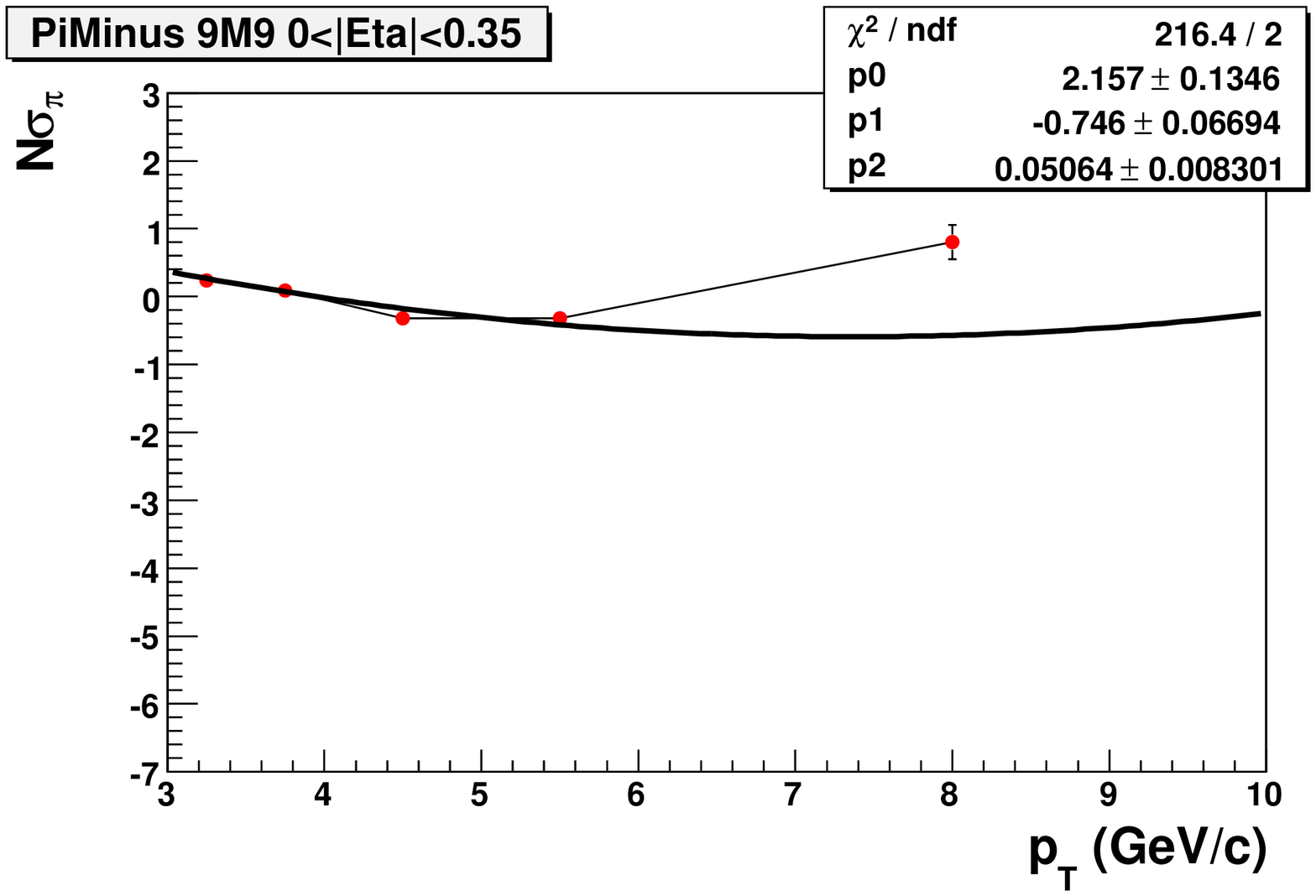}
										
			\end{minipage}
\hfill
\begin{minipage}[t]{.2\textwidth}
	\centering
		\includegraphics[width=1\textwidth]{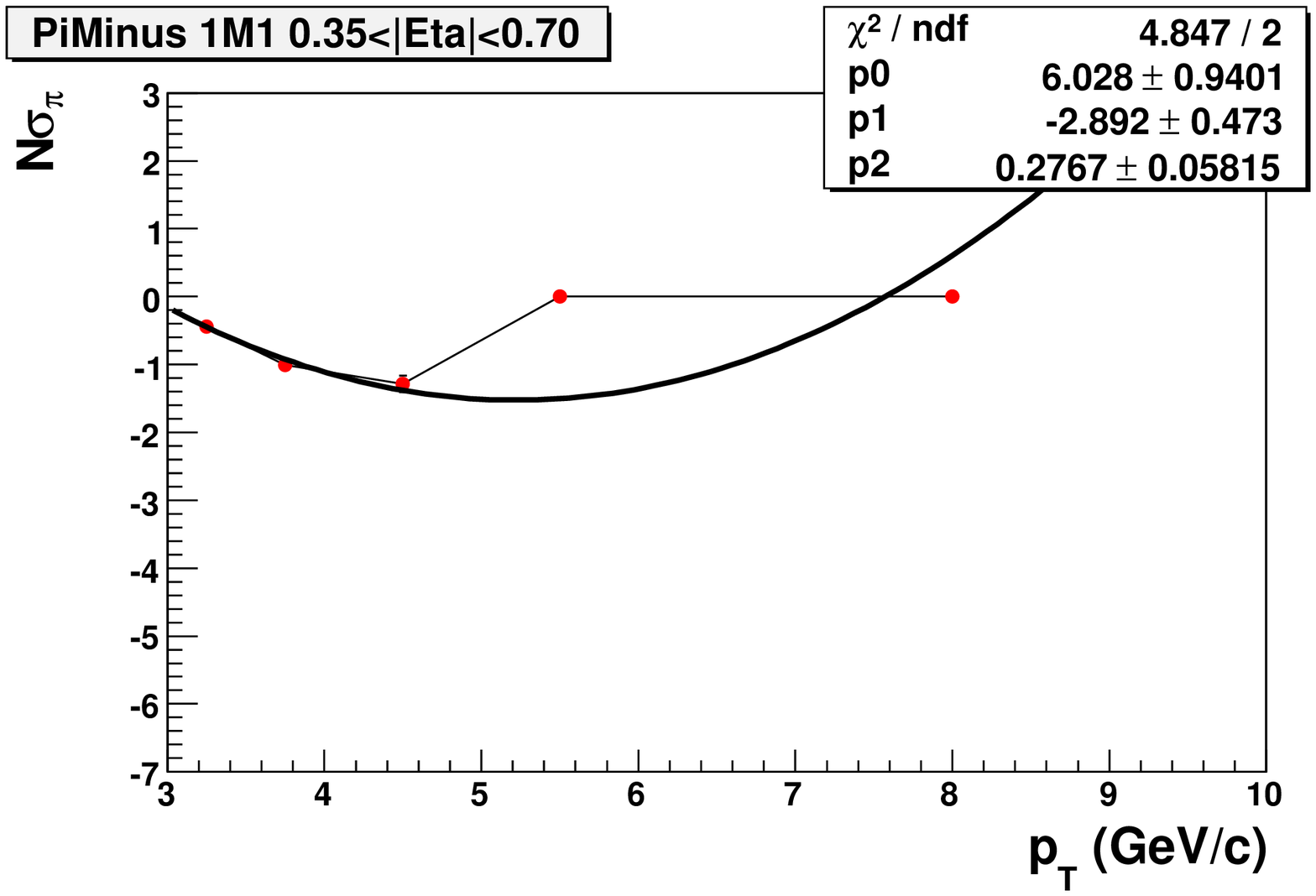}
		\includegraphics[width=1\textwidth]{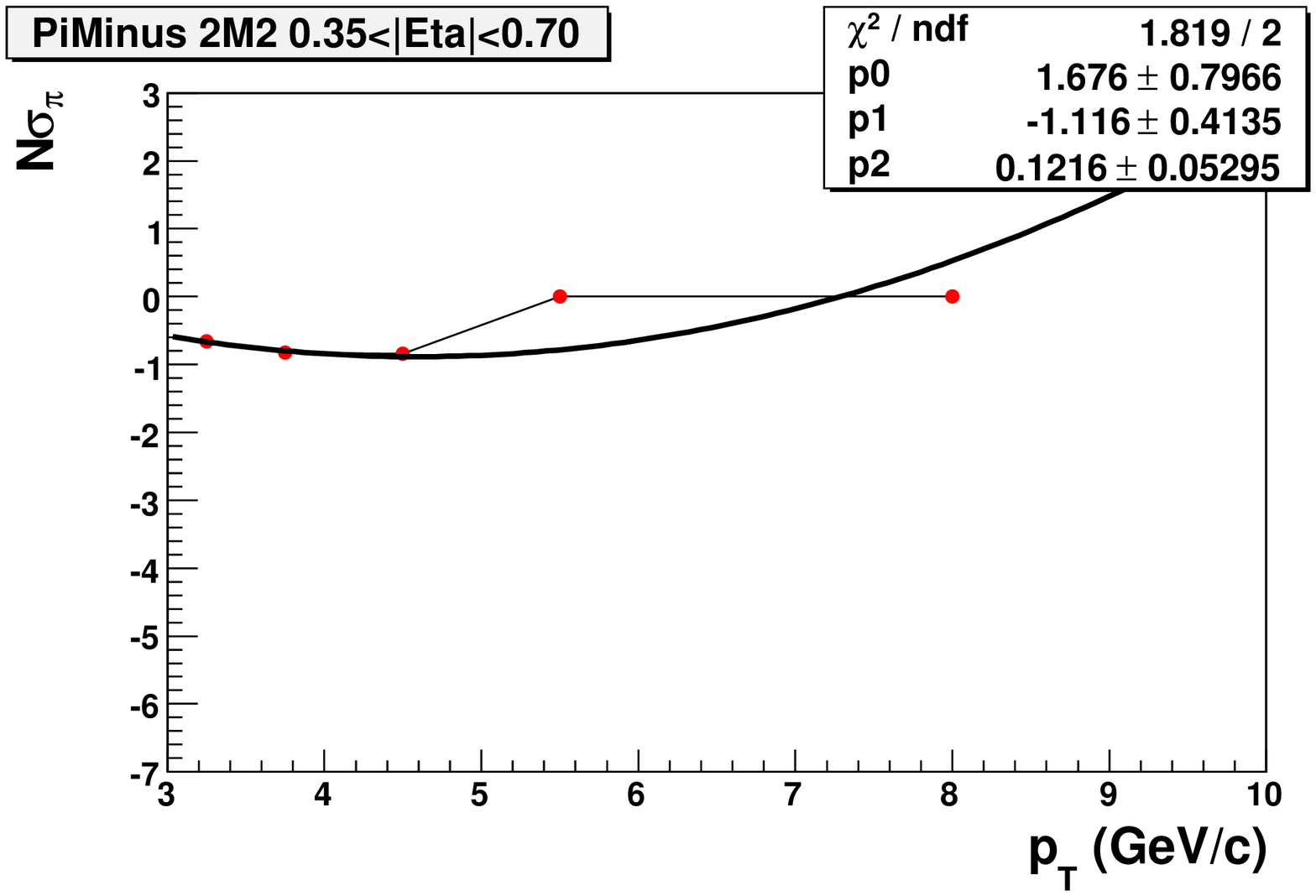}
		\includegraphics[width=1\textwidth]{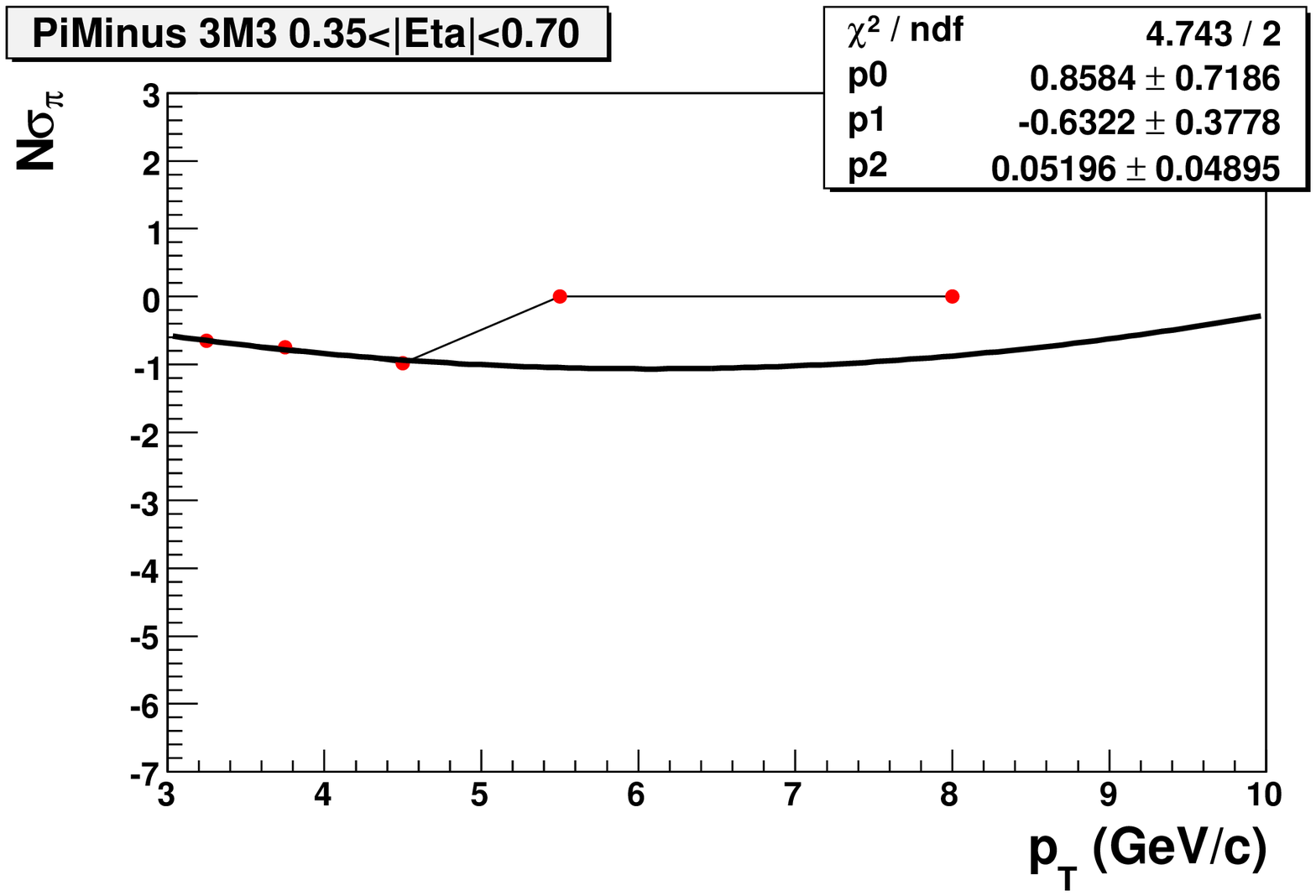}
		\includegraphics[width=1\textwidth]{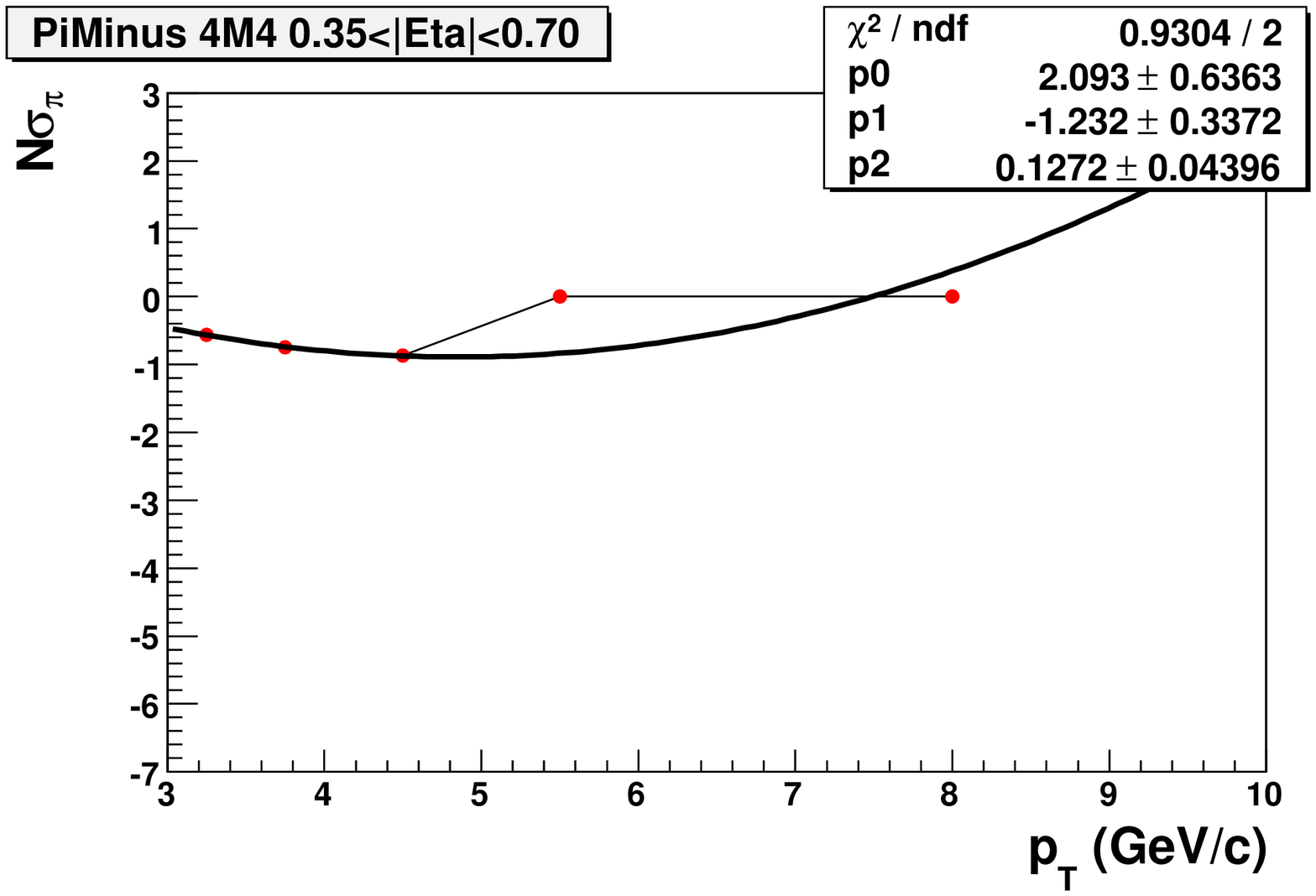}
		\includegraphics[width=1\textwidth]{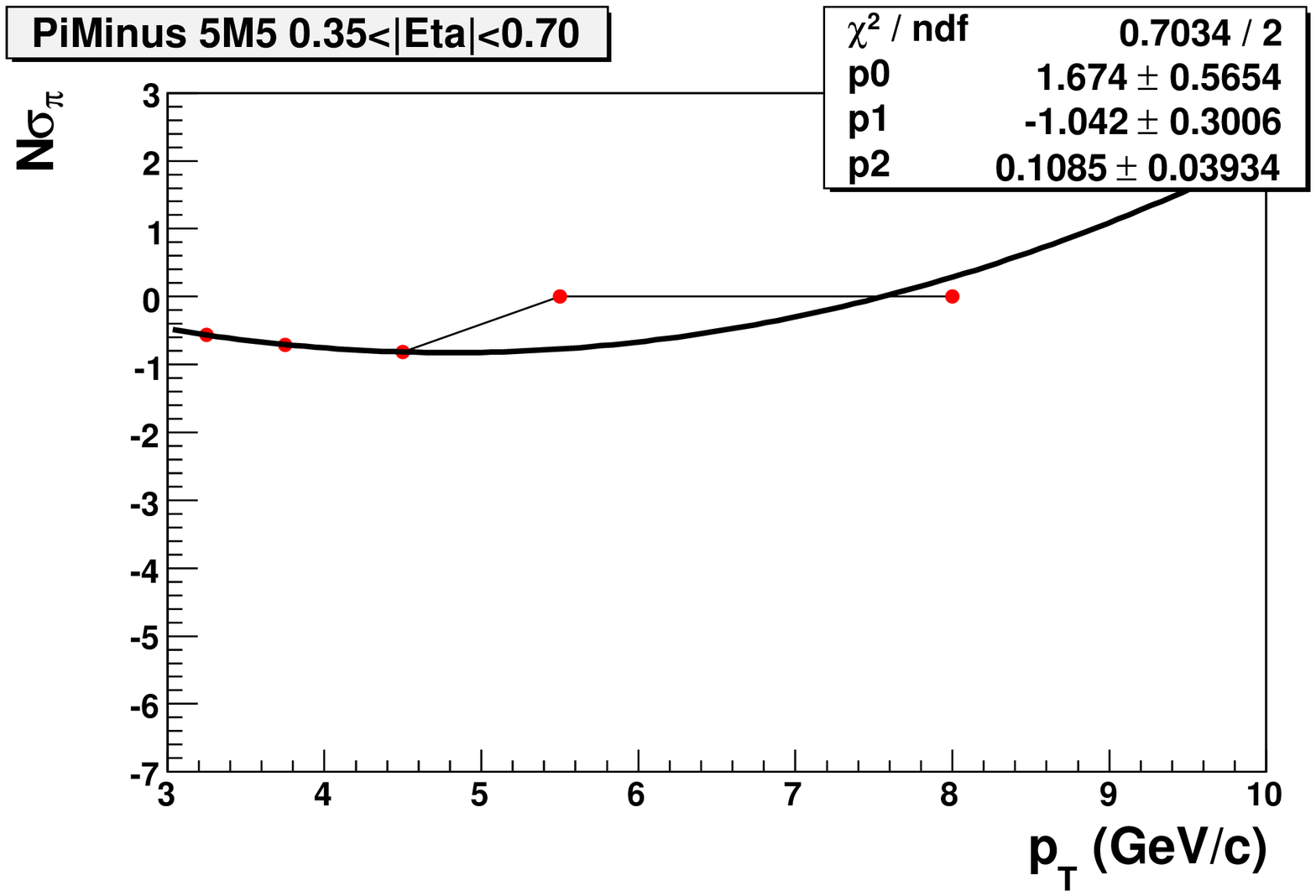}
		\includegraphics[width=1\textwidth]{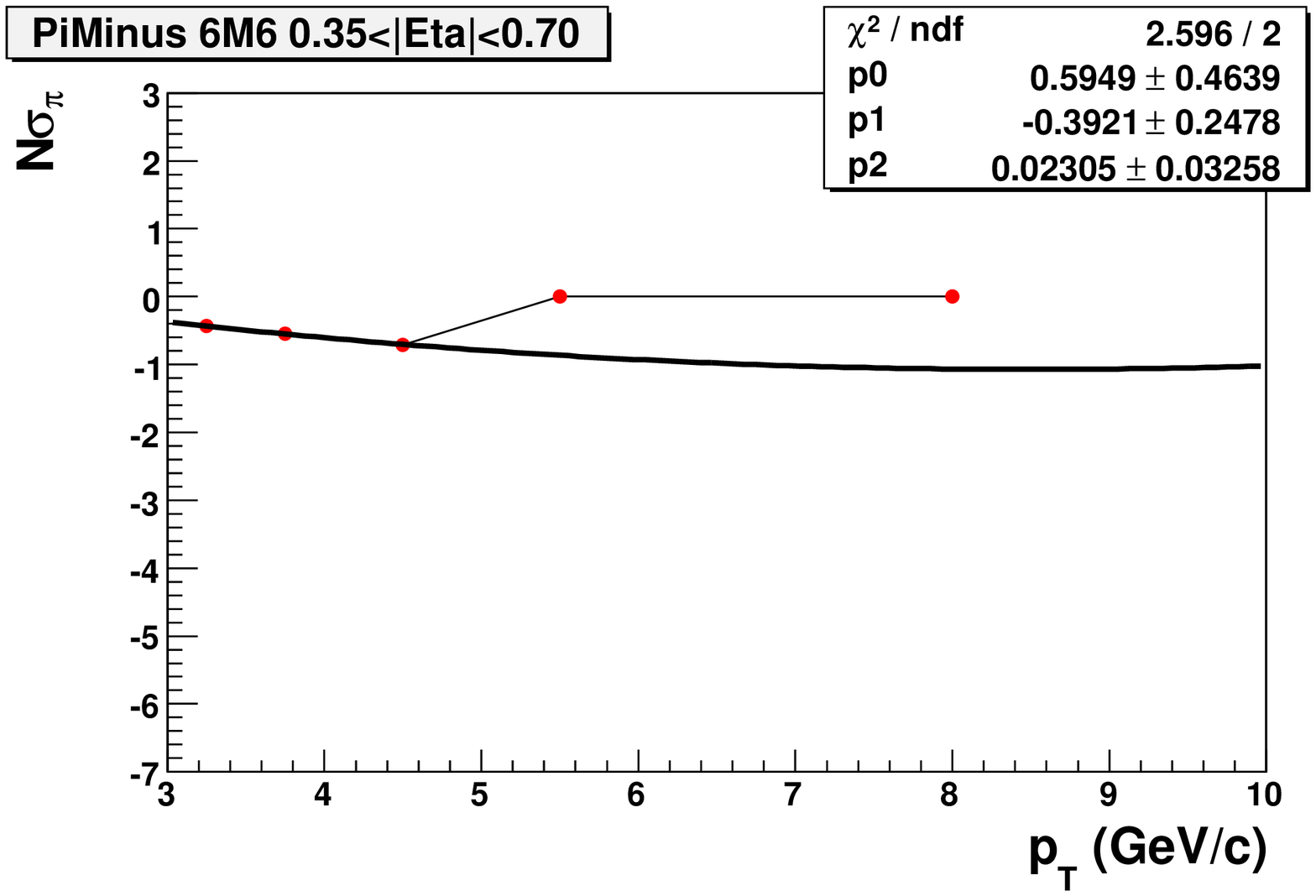}
		\includegraphics[width=1\textwidth]{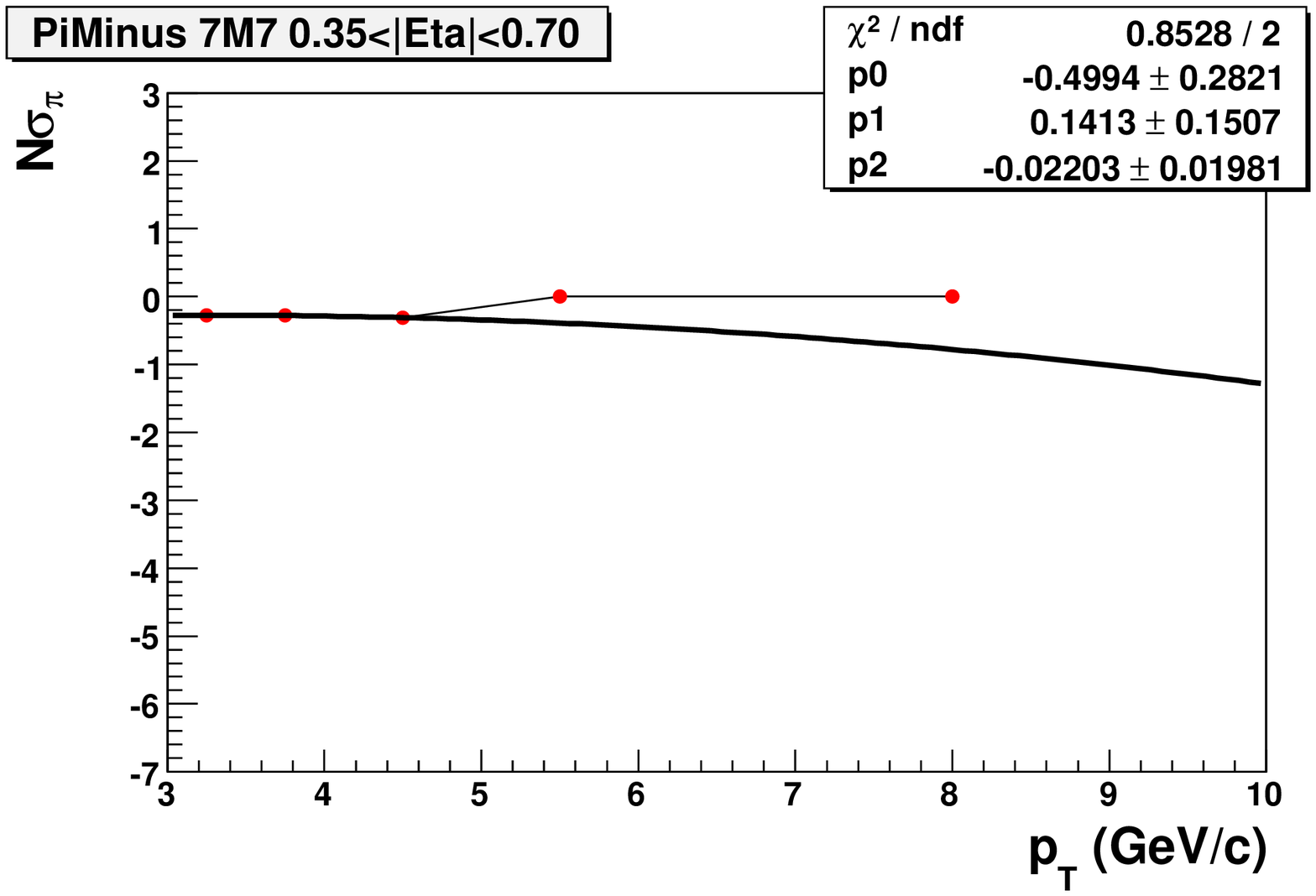}
		\includegraphics[width=1\textwidth]{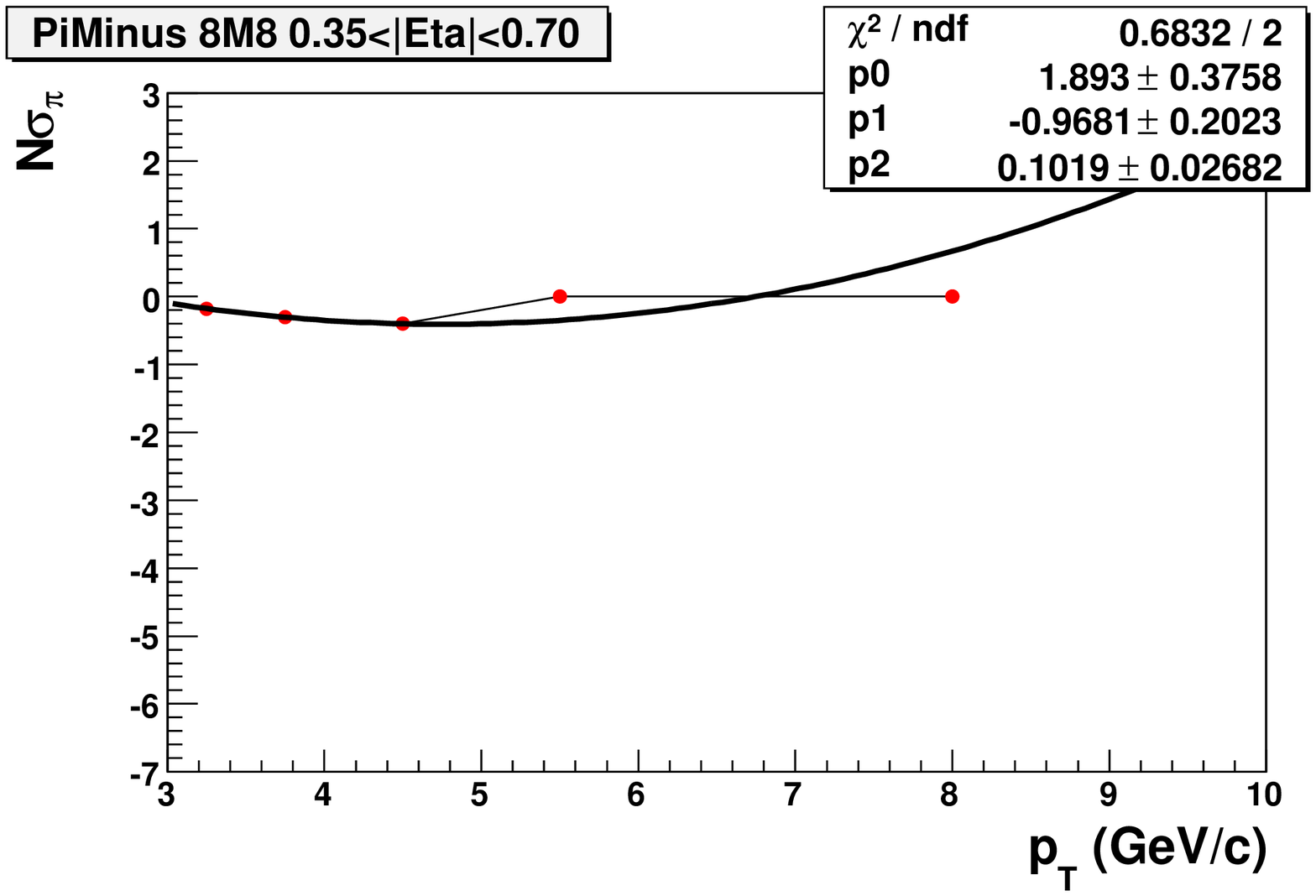}
		\includegraphics[width=1\textwidth]{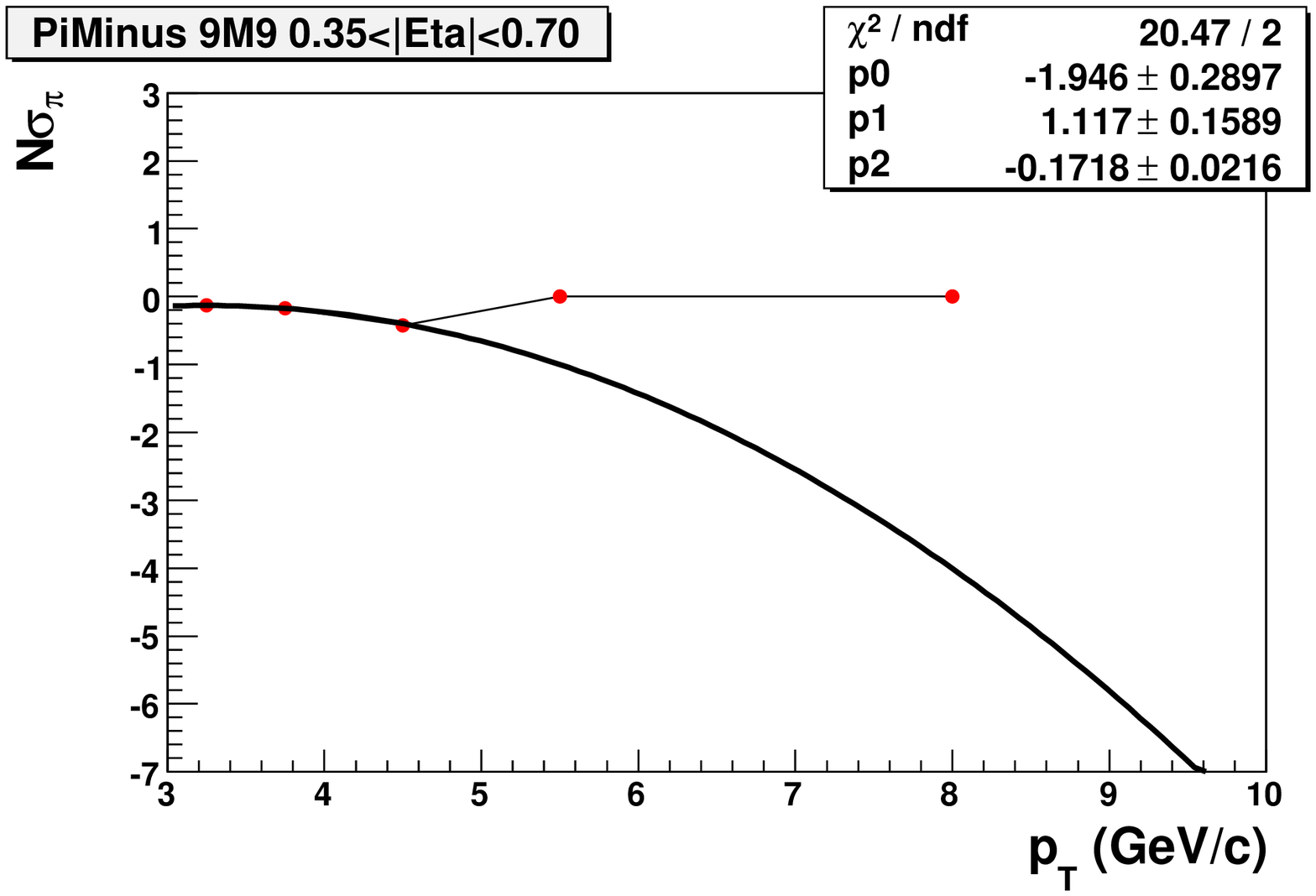}
		
			\end{minipage}	
					
	\caption{Fits to the cuts on $N\sigma_{\pi}$ as a function of $p_{T}$ for Au+Au collisions at $\sqrt{s_{NN}}=200$ GeV/c.  Rows correspond to the centrality bins 70-80\%, 60-70\%, 50-60\%, 40-50\%, 30-40\%, 20-30\%, 10-20\%, 5-10\% and 0-5\% from top to bottom.  Columns are for differents charges and $\eta$ cuts of pions.  Left: $\pi^{+}$ with $|\eta|<0.35$.  Left center:  $\pi^{+}$ with $0.35<|\eta|<0.7$.  Right center: $\pi^{-}$ with $|\eta|<0.35$.  Right:  $\pi^{-}$ with $0.35<|\eta|<0.7$.}
	\label{fig:fitcutspi}	
\end{figure}

\begin{figure}[H]
\hfill
\begin{minipage}[t]{.2\textwidth}
	\centering
		\includegraphics[width=1\textwidth]{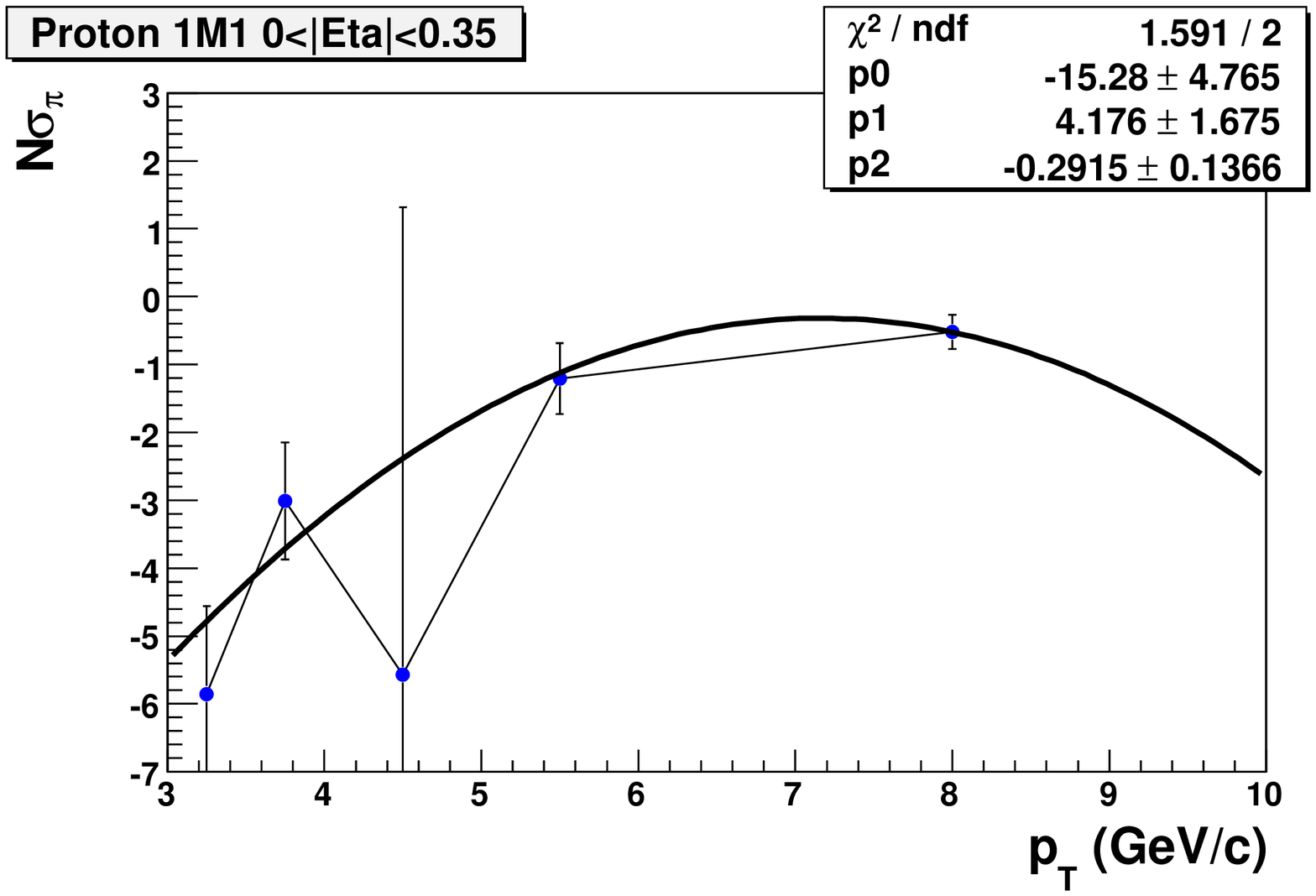}
		\includegraphics[width=1\textwidth]{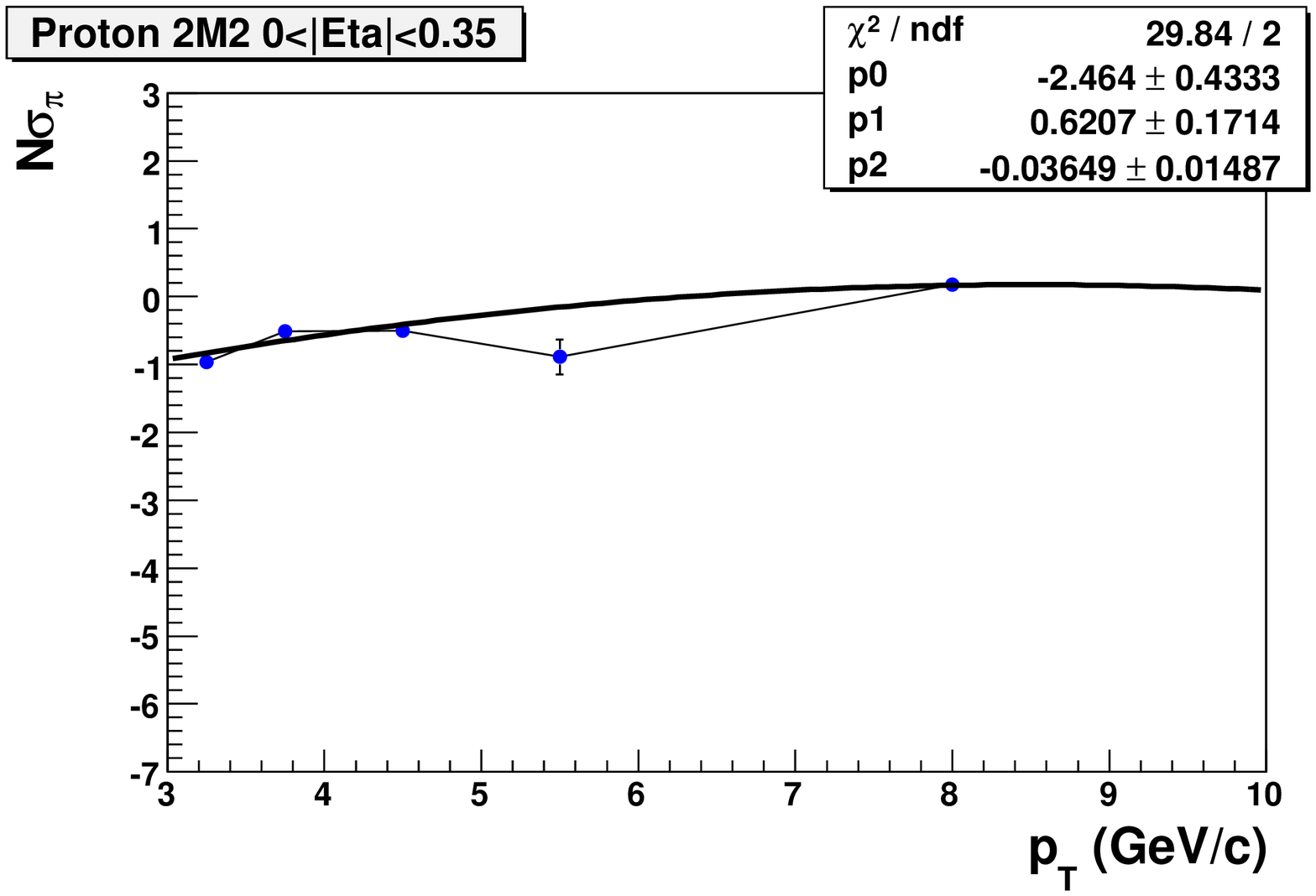}
		\includegraphics[width=1\textwidth]{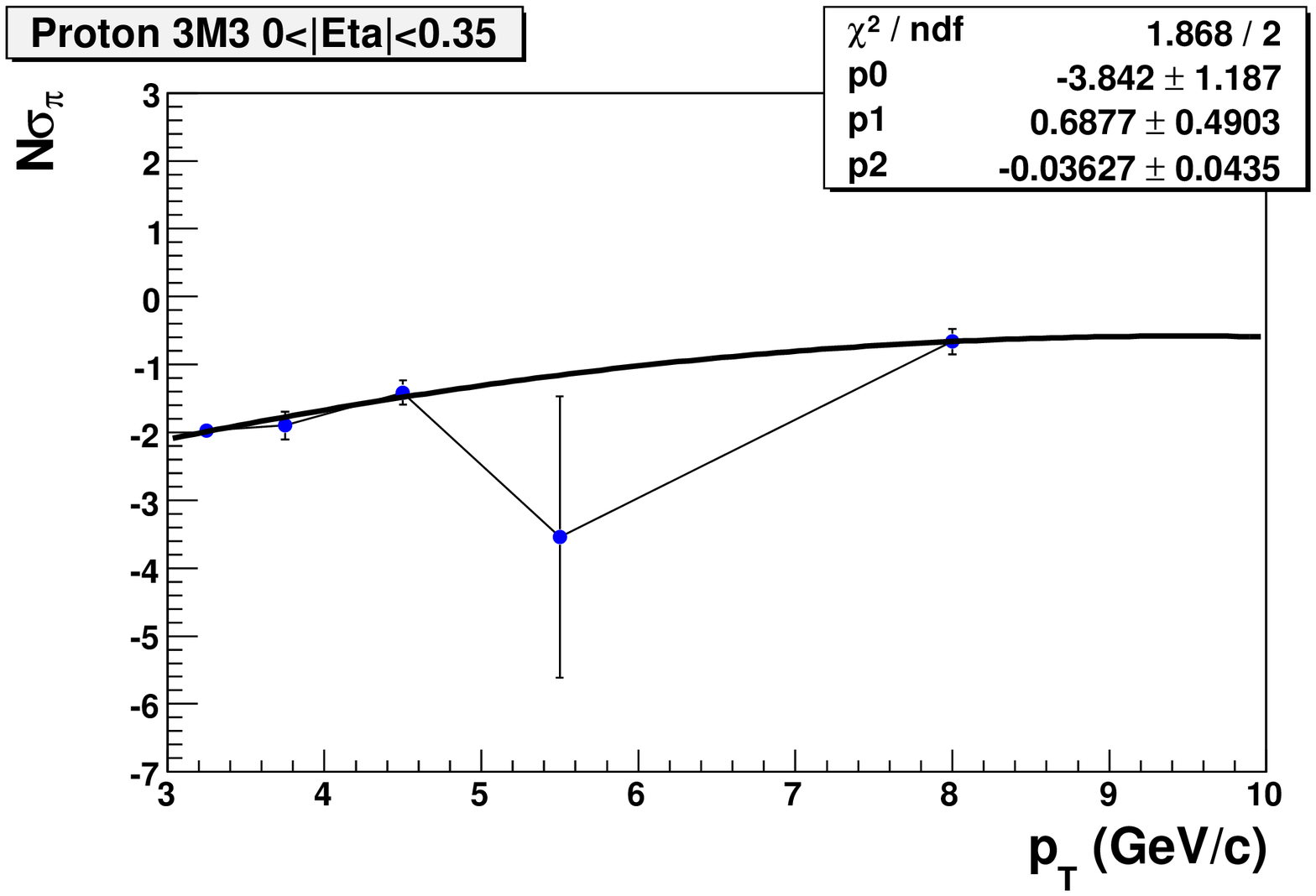}
		\includegraphics[width=1\textwidth]{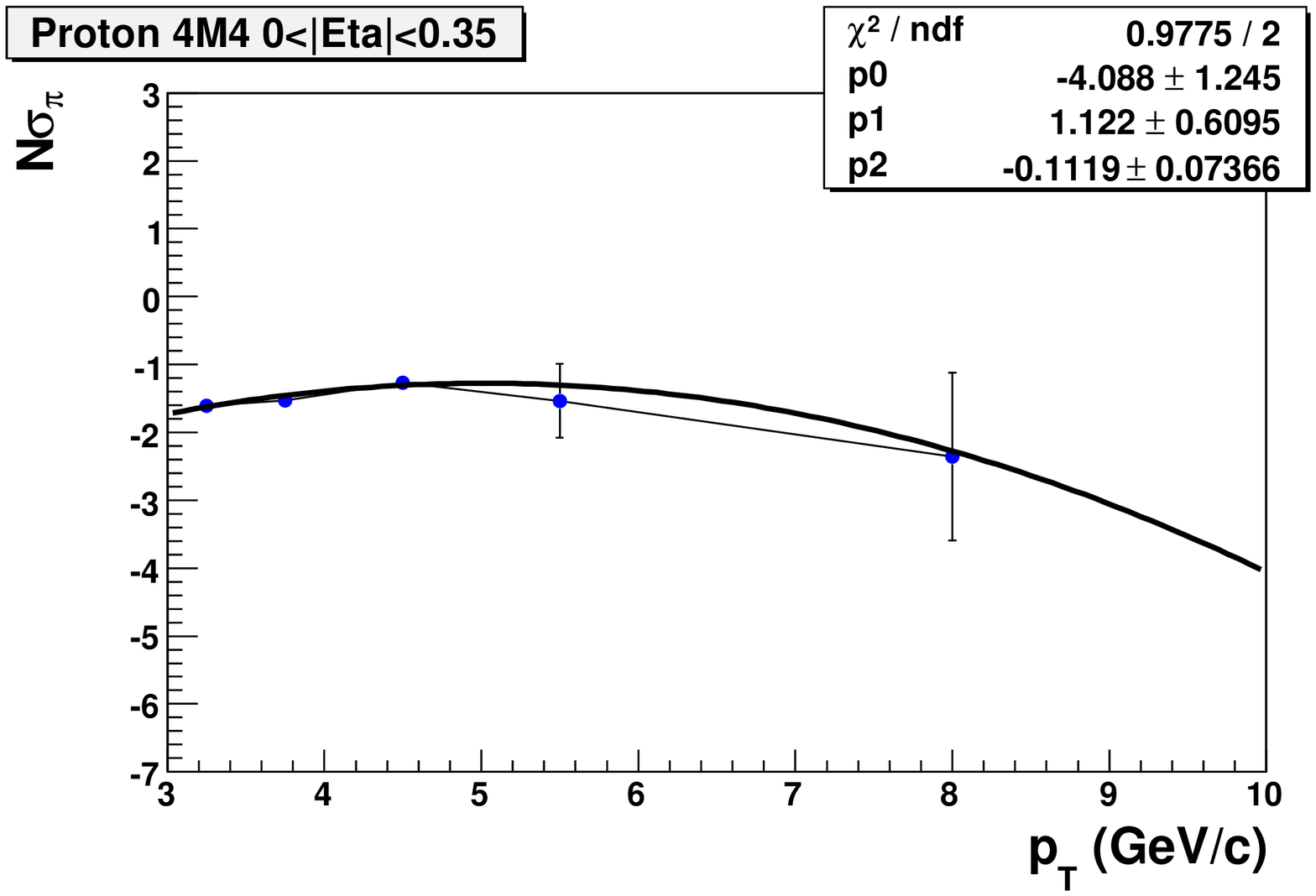}
		\includegraphics[width=1\textwidth]{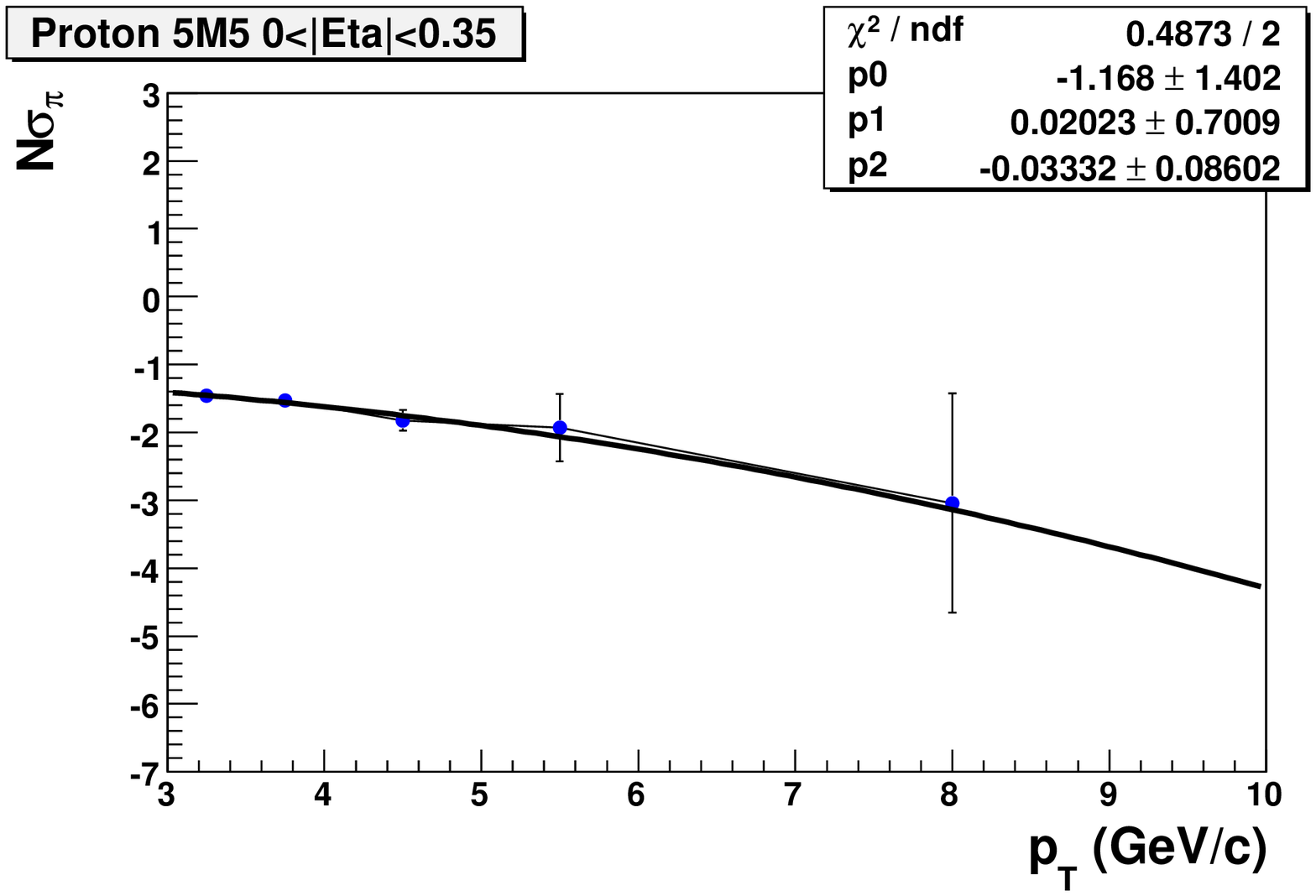}
		\includegraphics[width=1\textwidth]{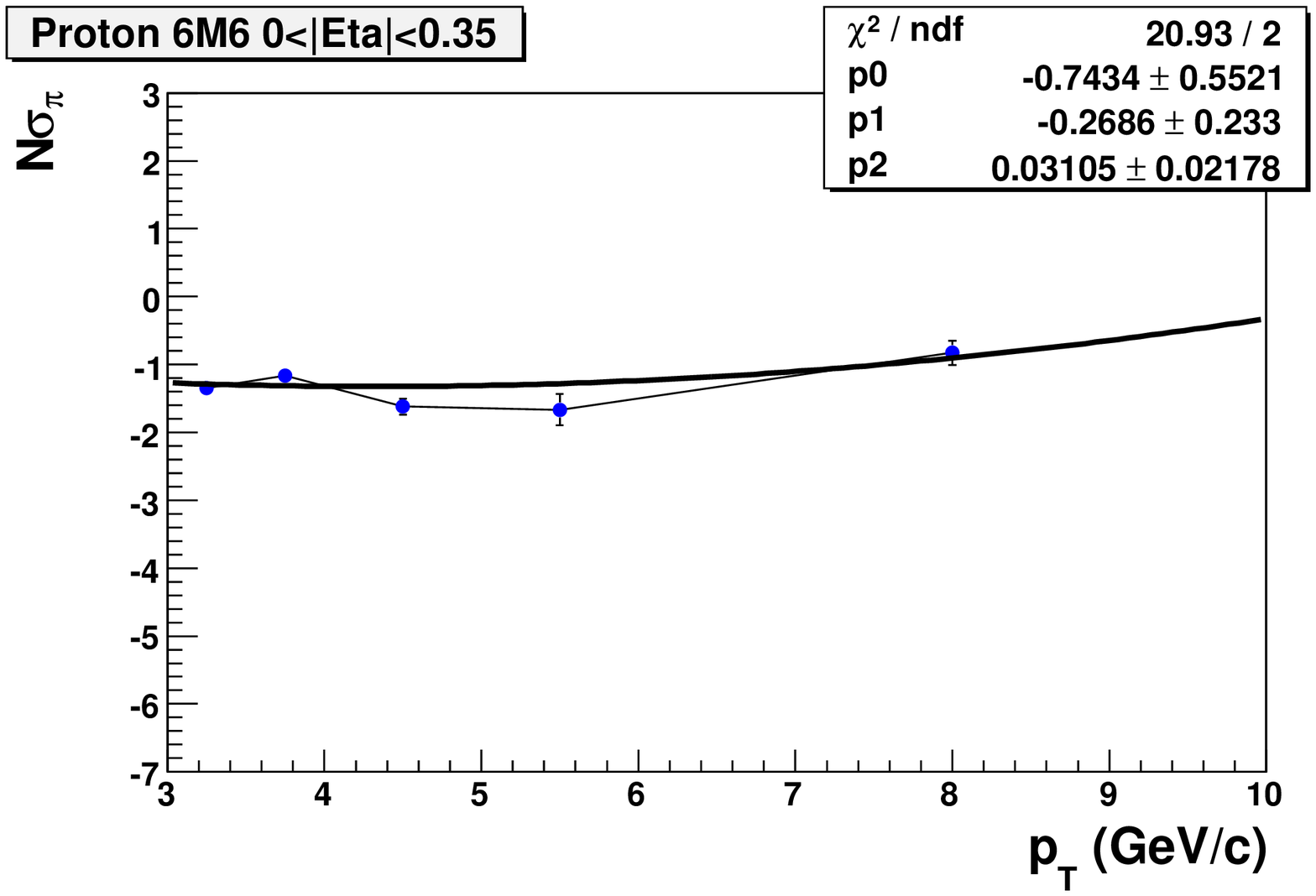}
		\includegraphics[width=1\textwidth]{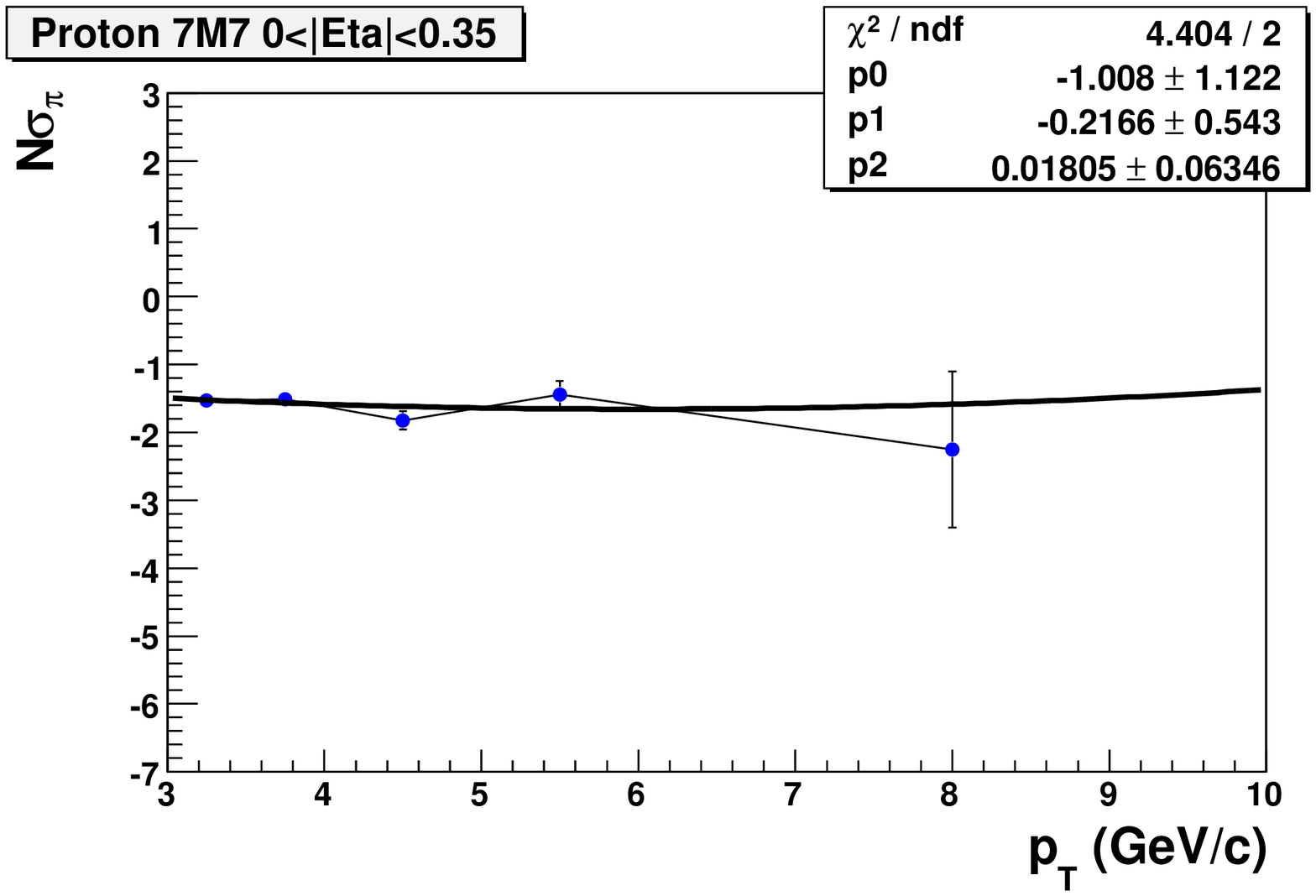}
		\includegraphics[width=1\textwidth]{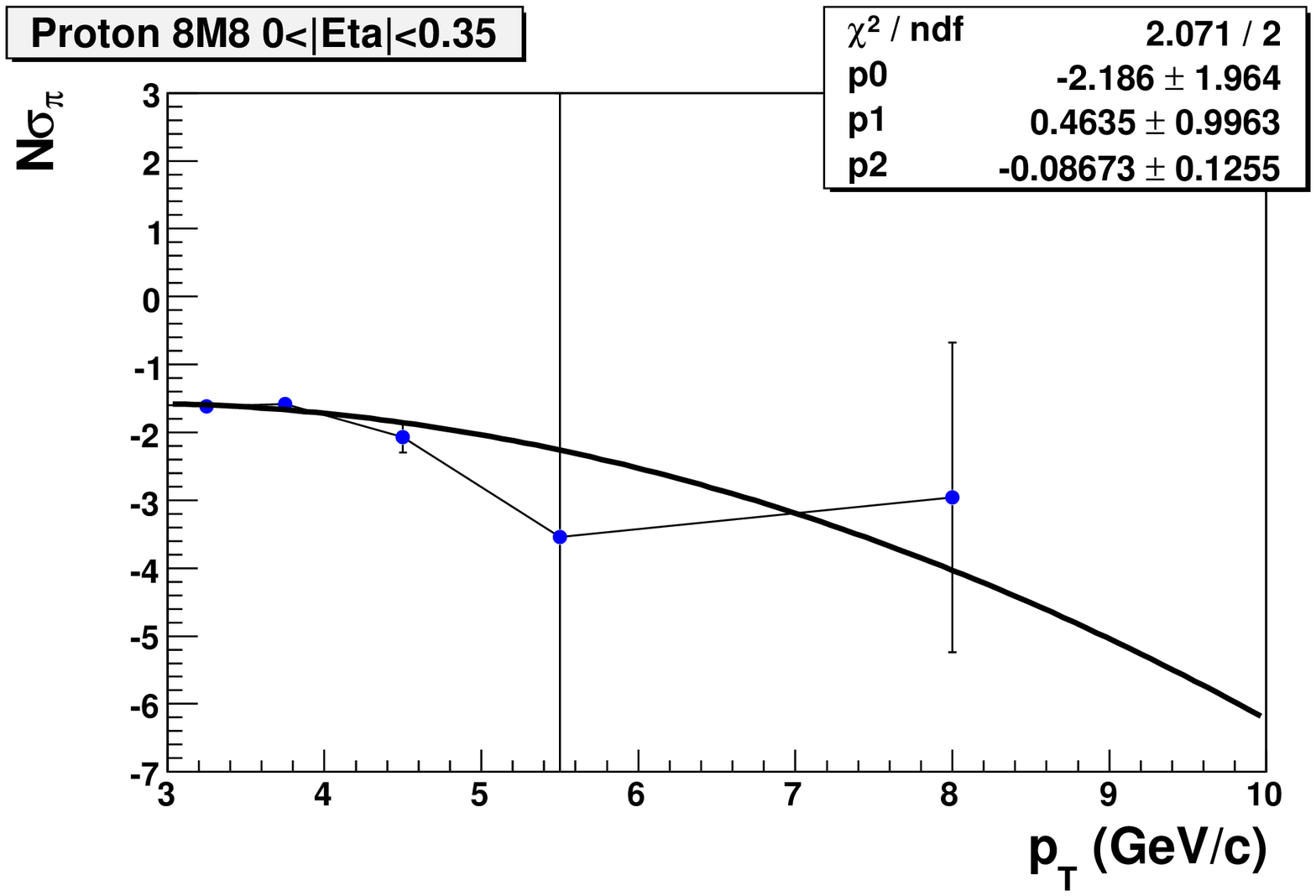}
		\includegraphics[width=1\textwidth]{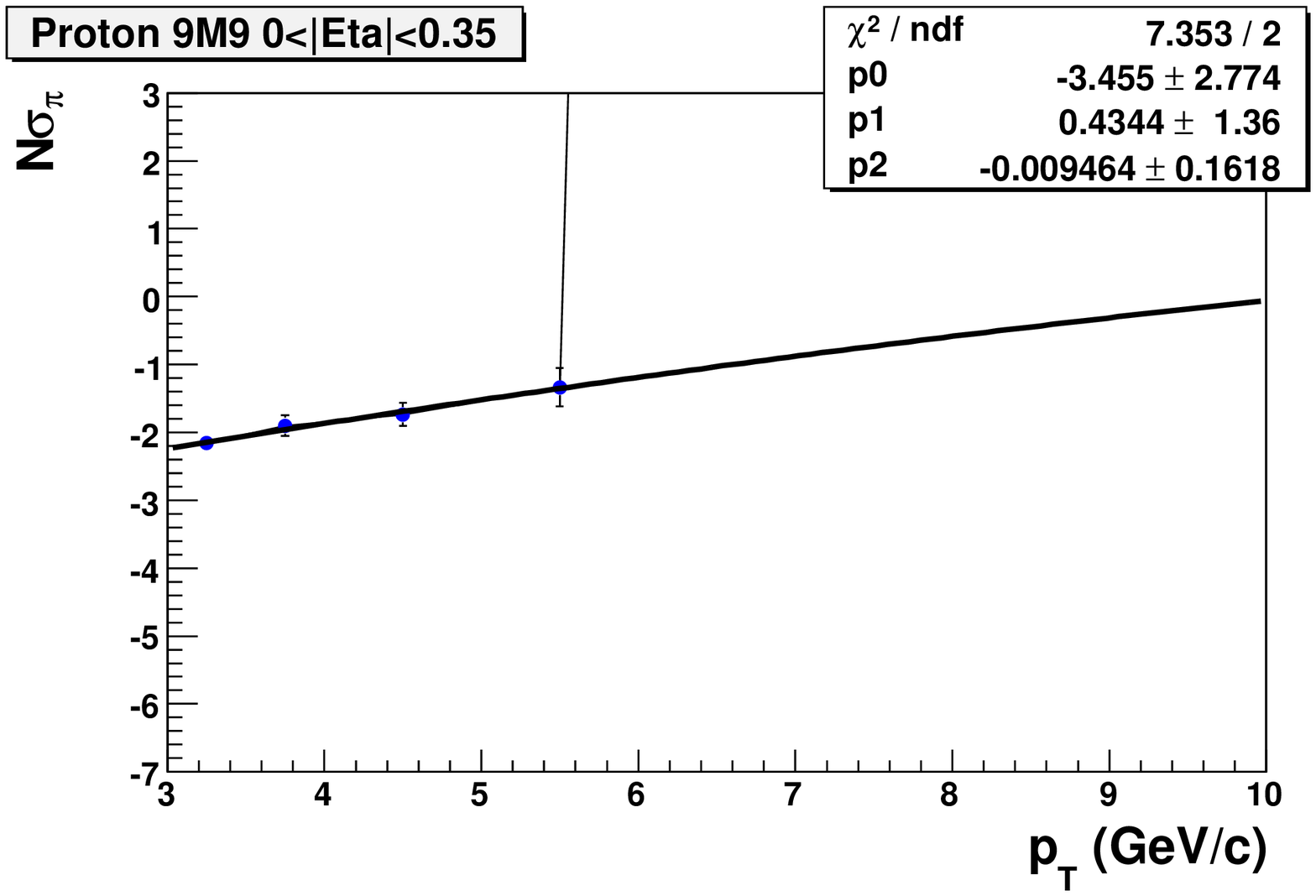}									
			\end{minipage}
\hfill
\begin{minipage}[t]{.2\textwidth}
	\centering
		\includegraphics[width=1\textwidth]{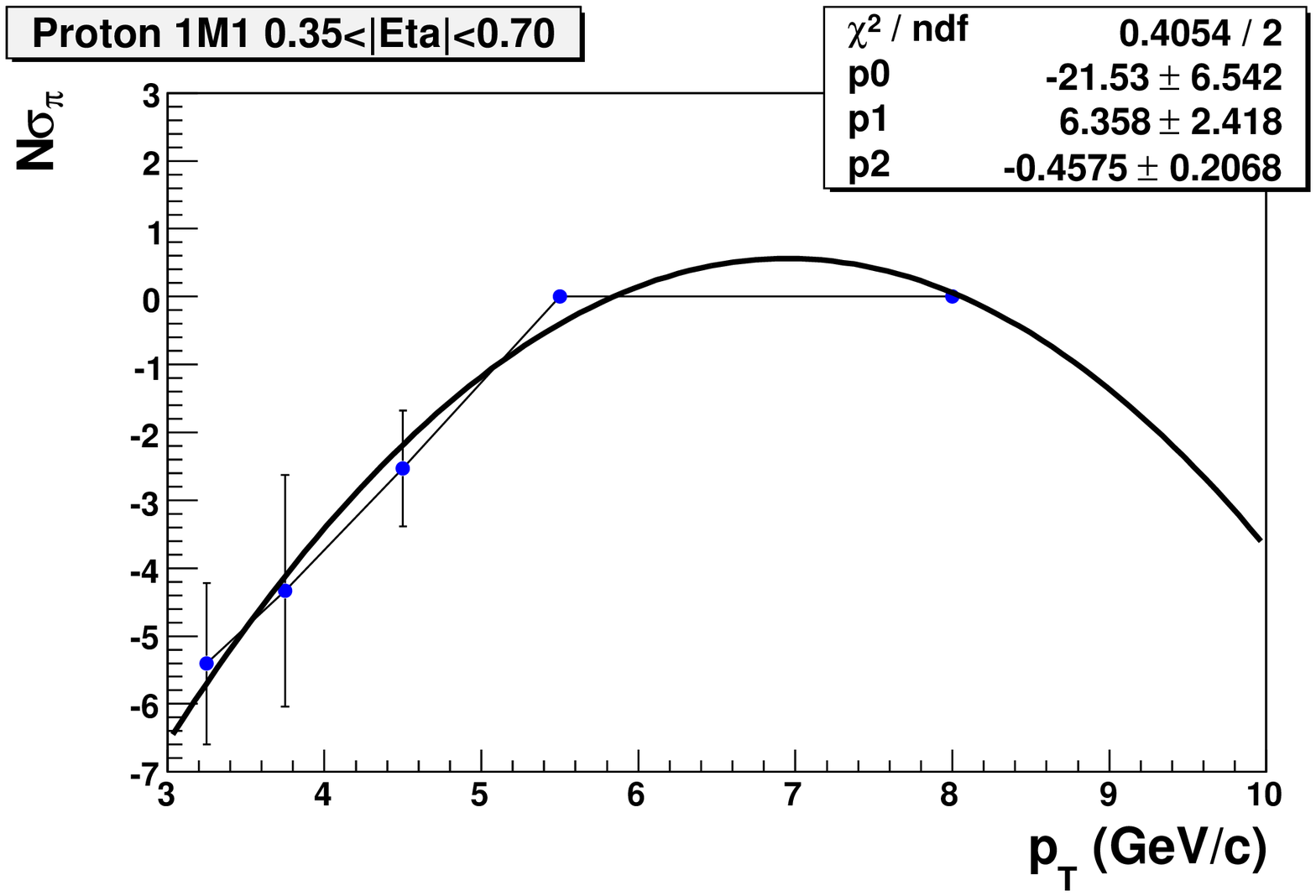}
		\includegraphics[width=1\textwidth]{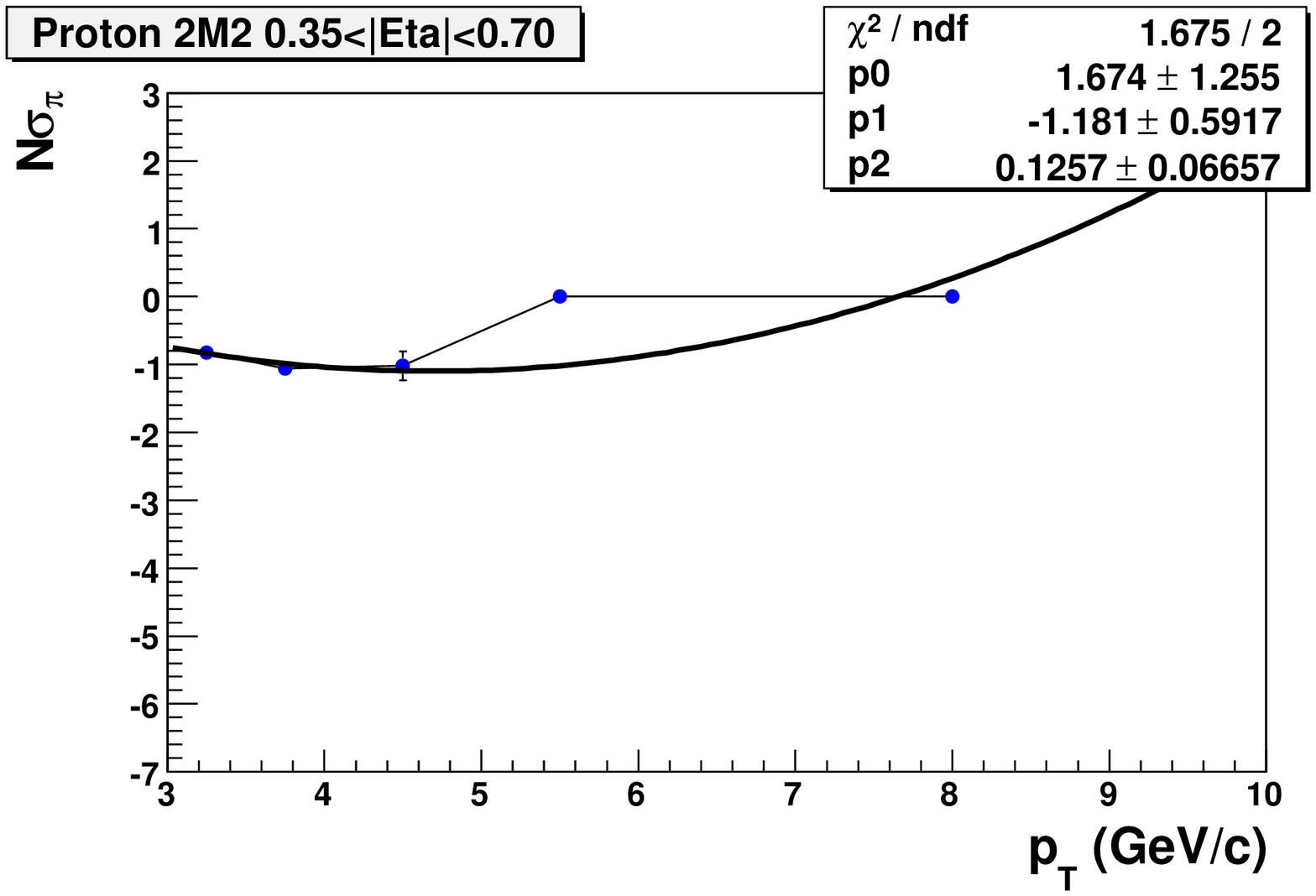}
		\includegraphics[width=1\textwidth]{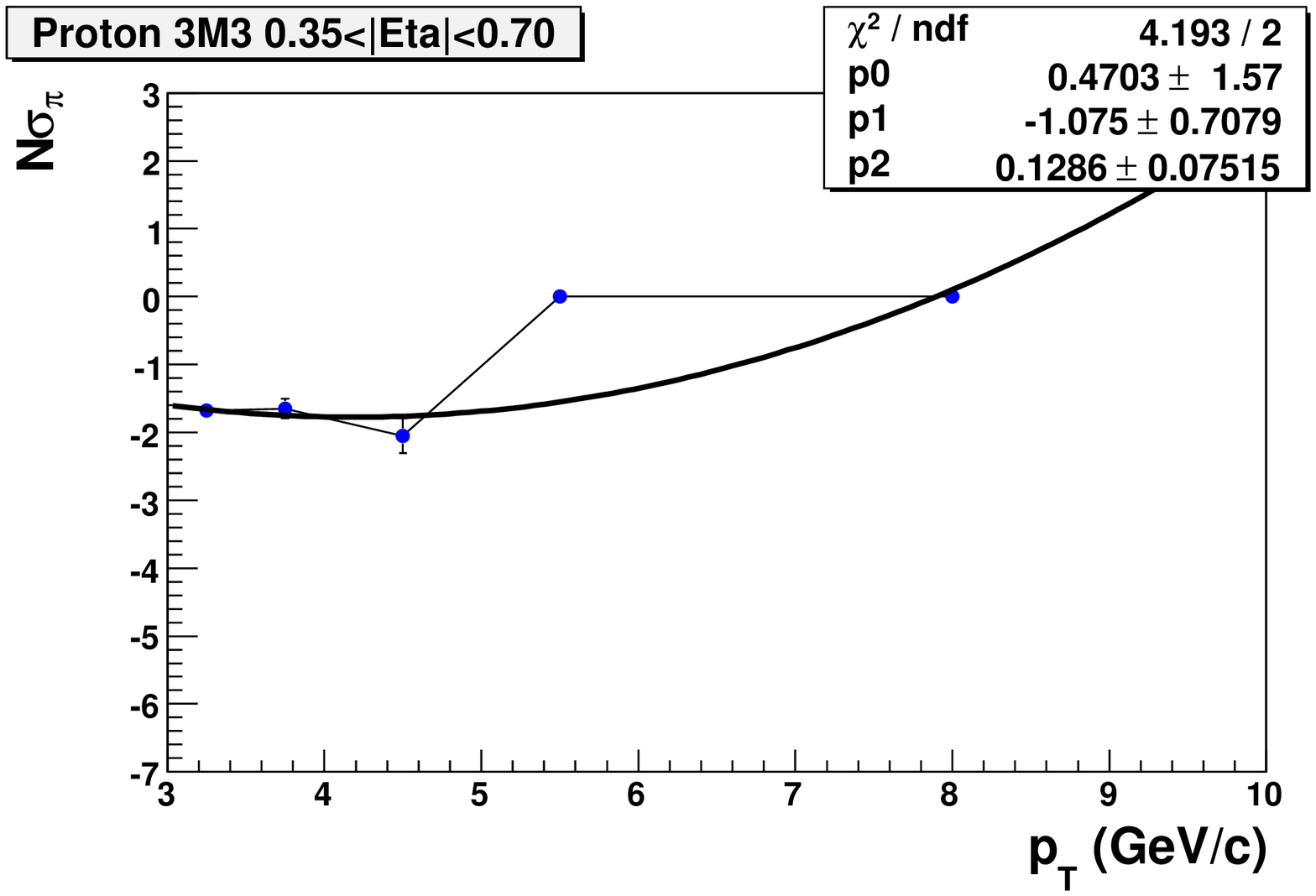}
		\includegraphics[width=1\textwidth]{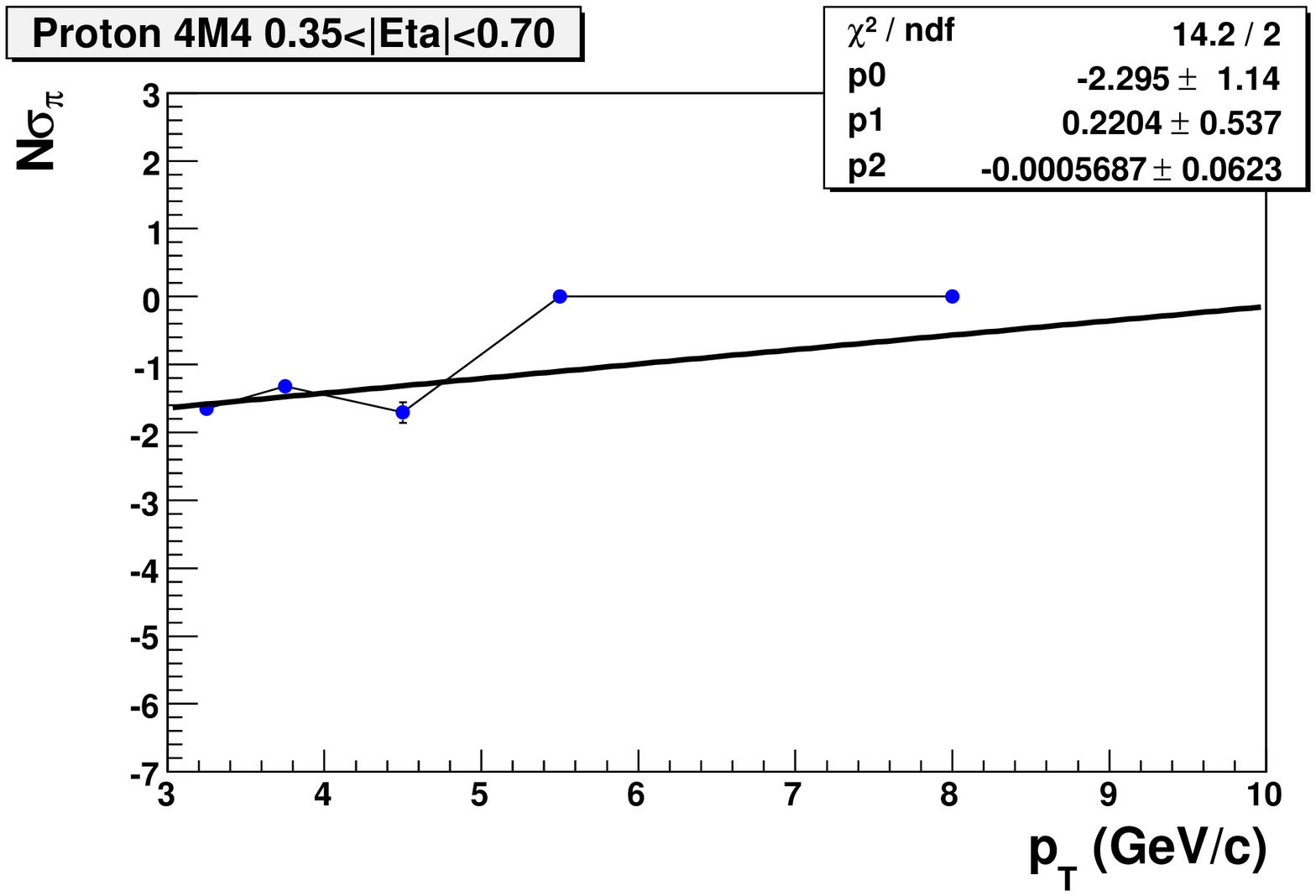}
		\includegraphics[width=1\textwidth]{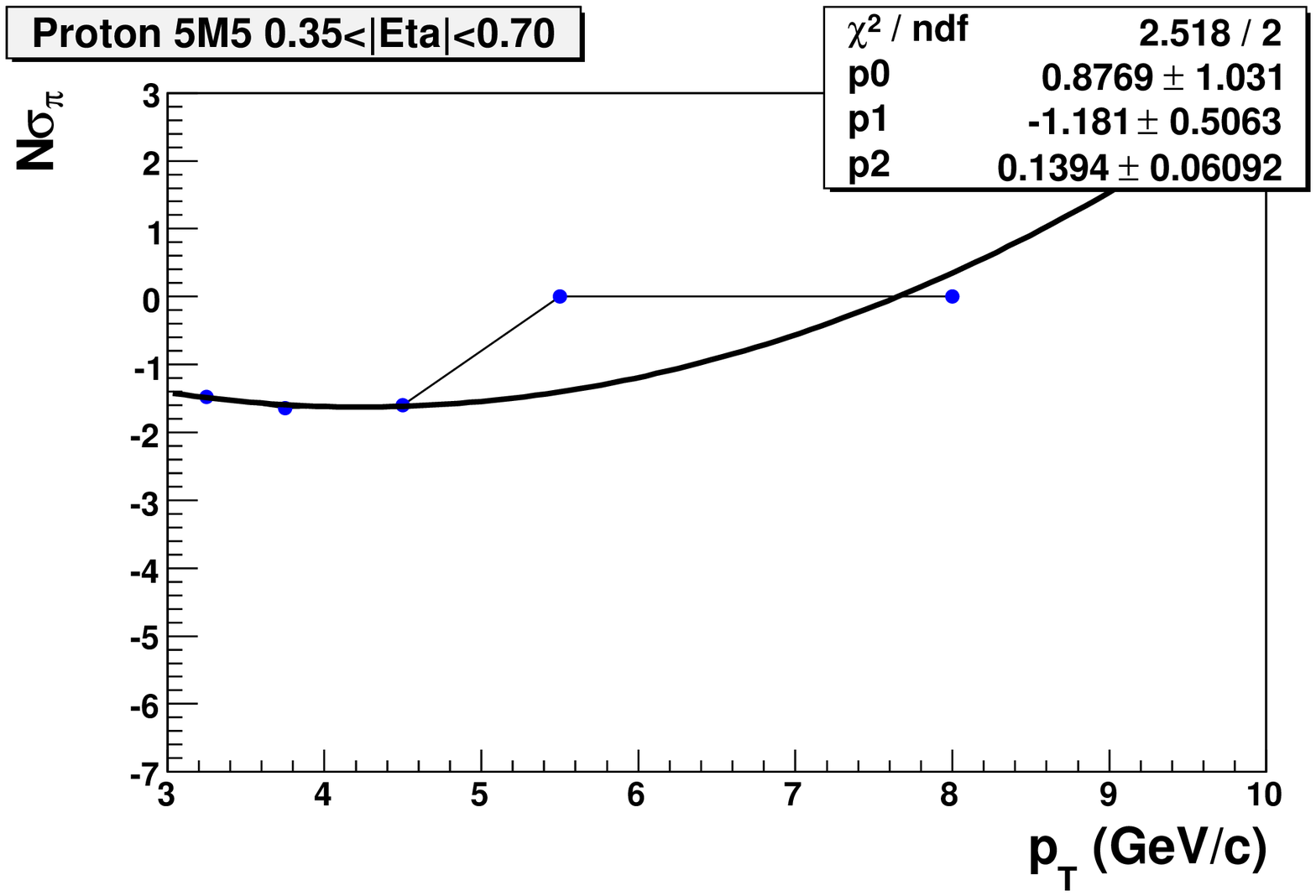}
		\includegraphics[width=1\textwidth]{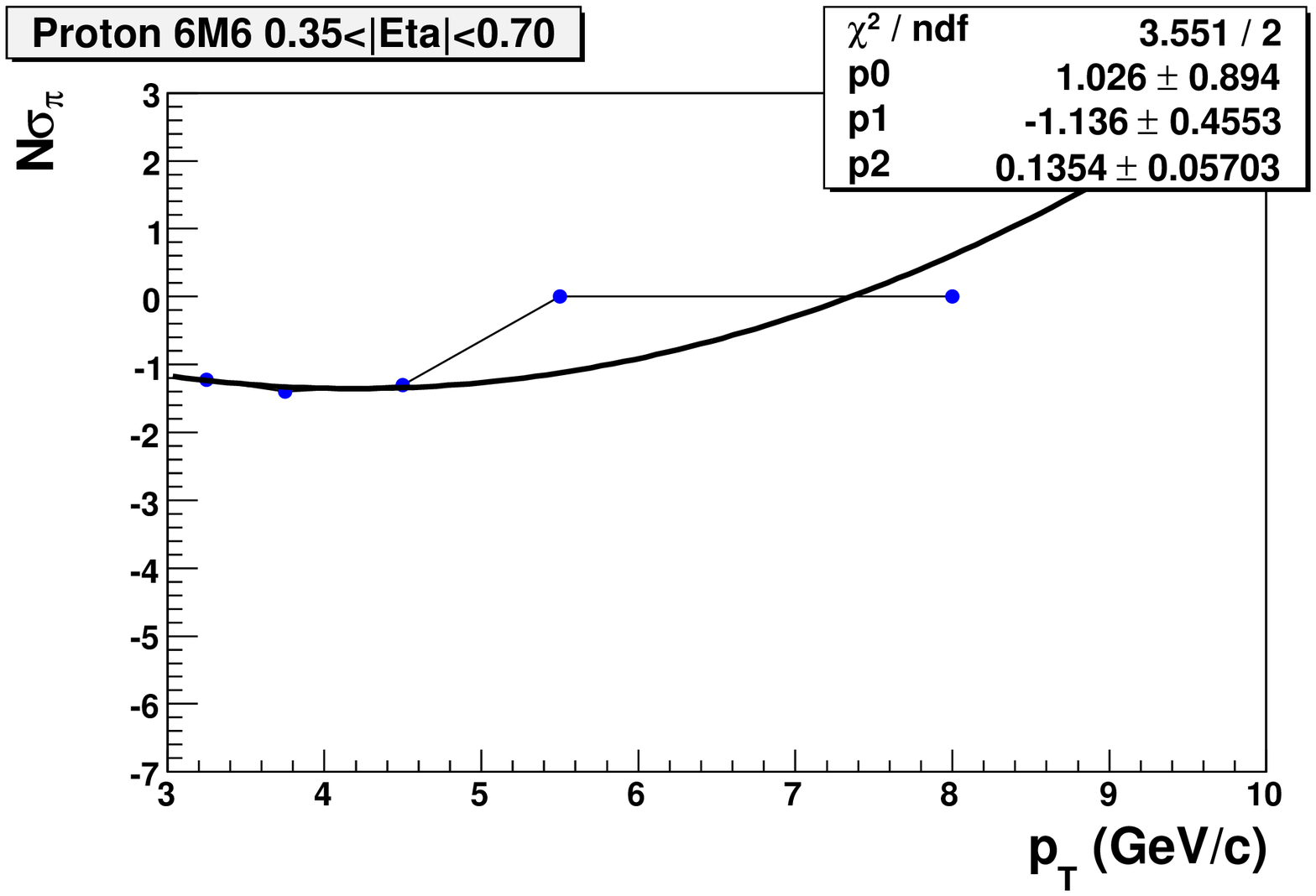}
		\includegraphics[width=1\textwidth]{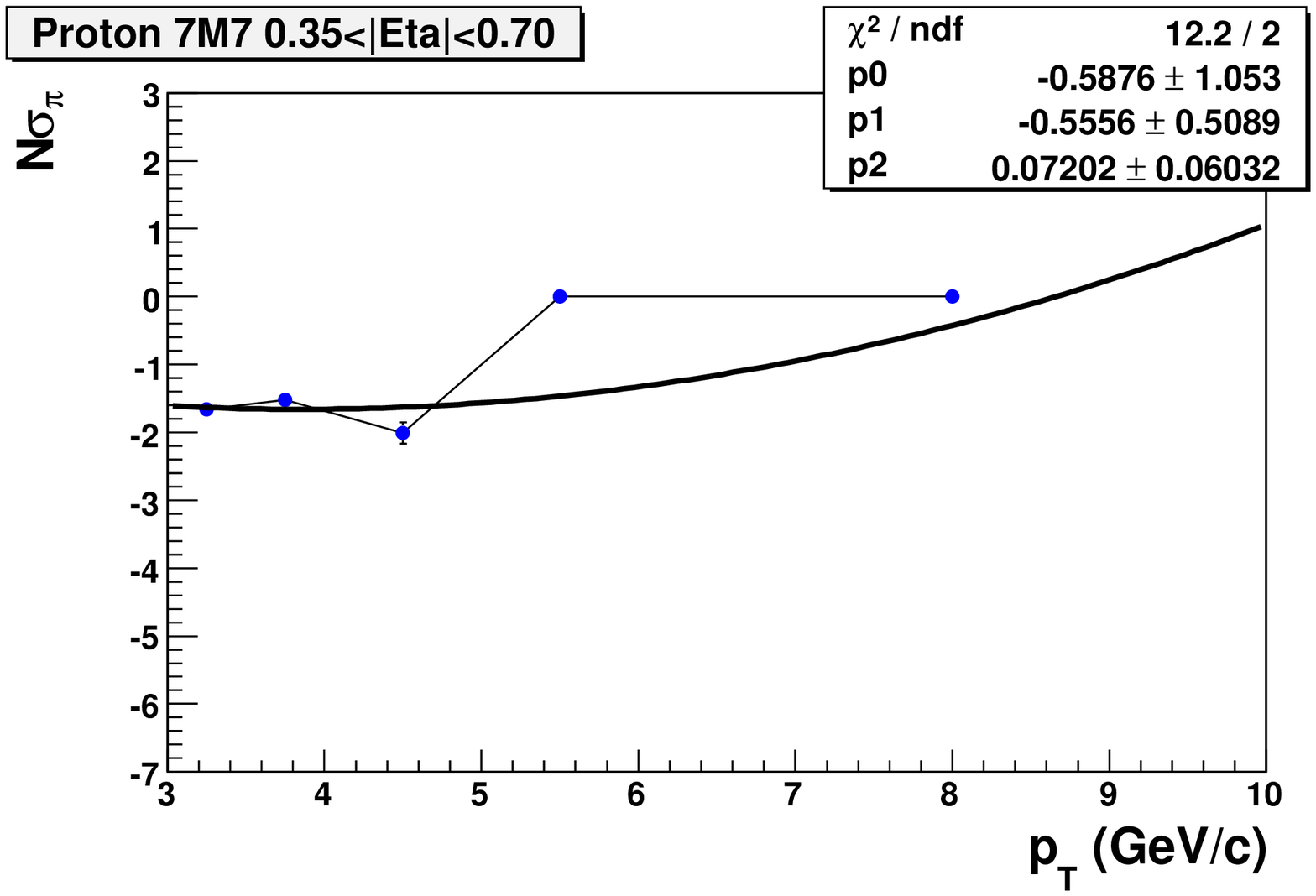}
		\includegraphics[width=1\textwidth]{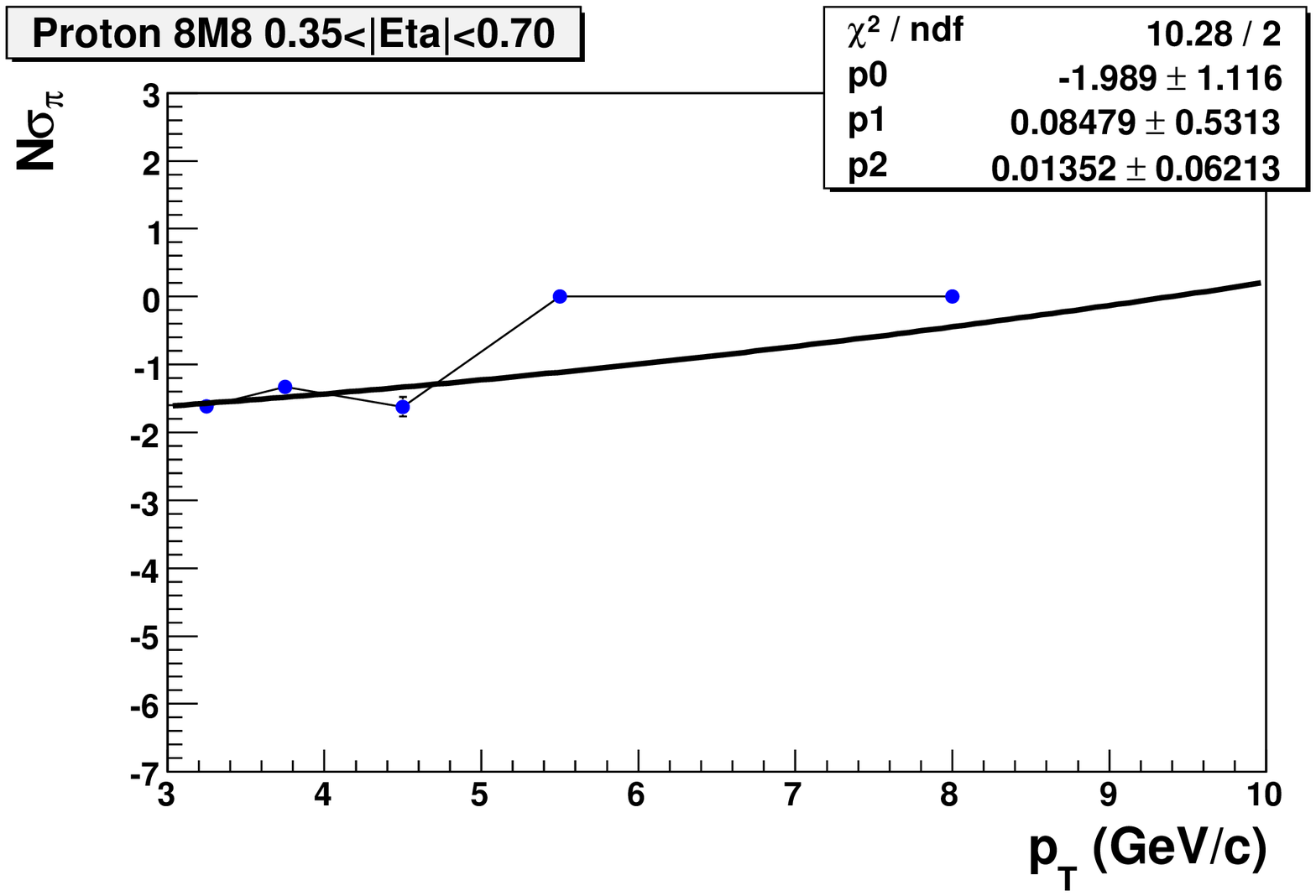}
		\includegraphics[width=1\textwidth]{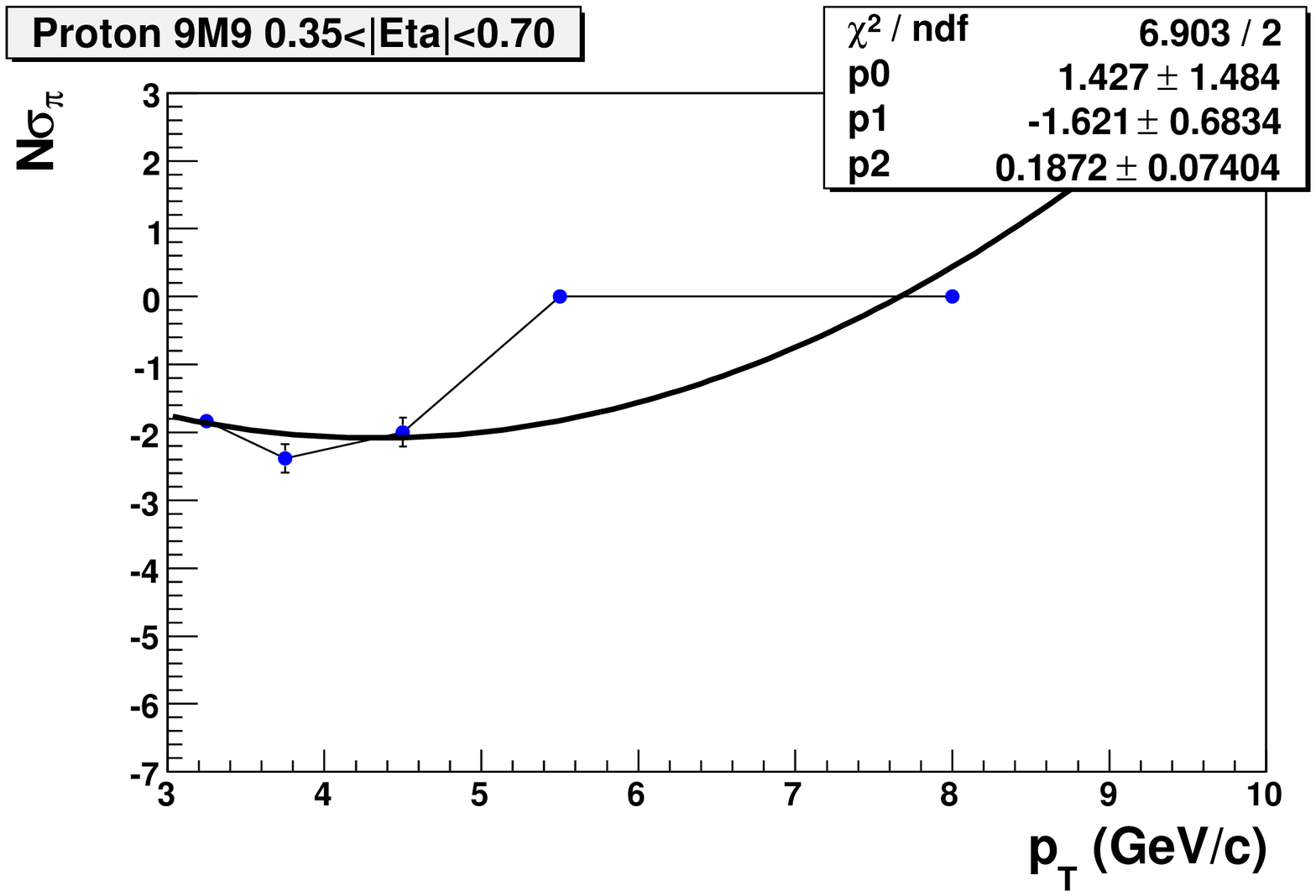}									
			\end{minipage}
\hfill
\begin{minipage}[t]{.2\textwidth}
	\centering
		\includegraphics[width=1\textwidth]{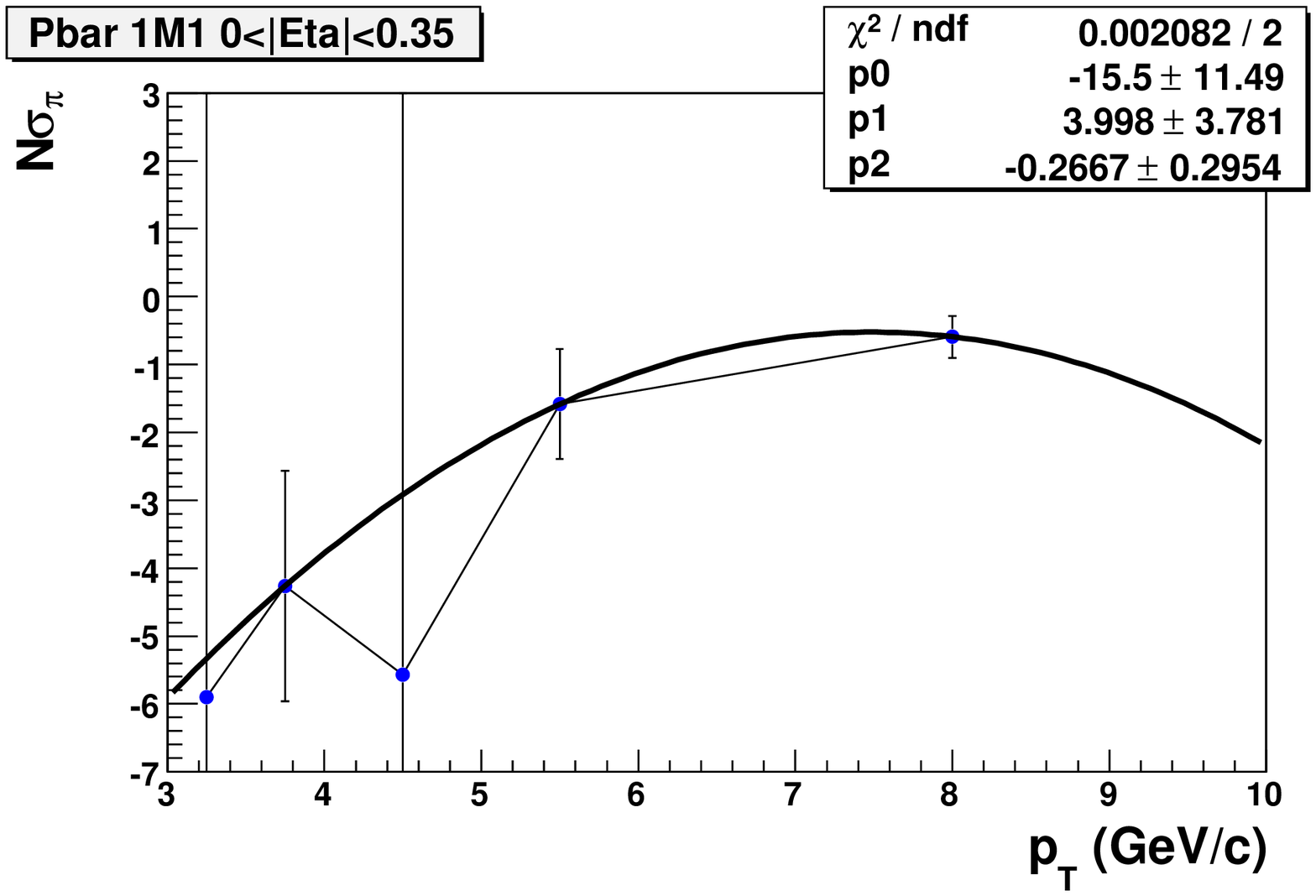}
		\includegraphics[width=1\textwidth]{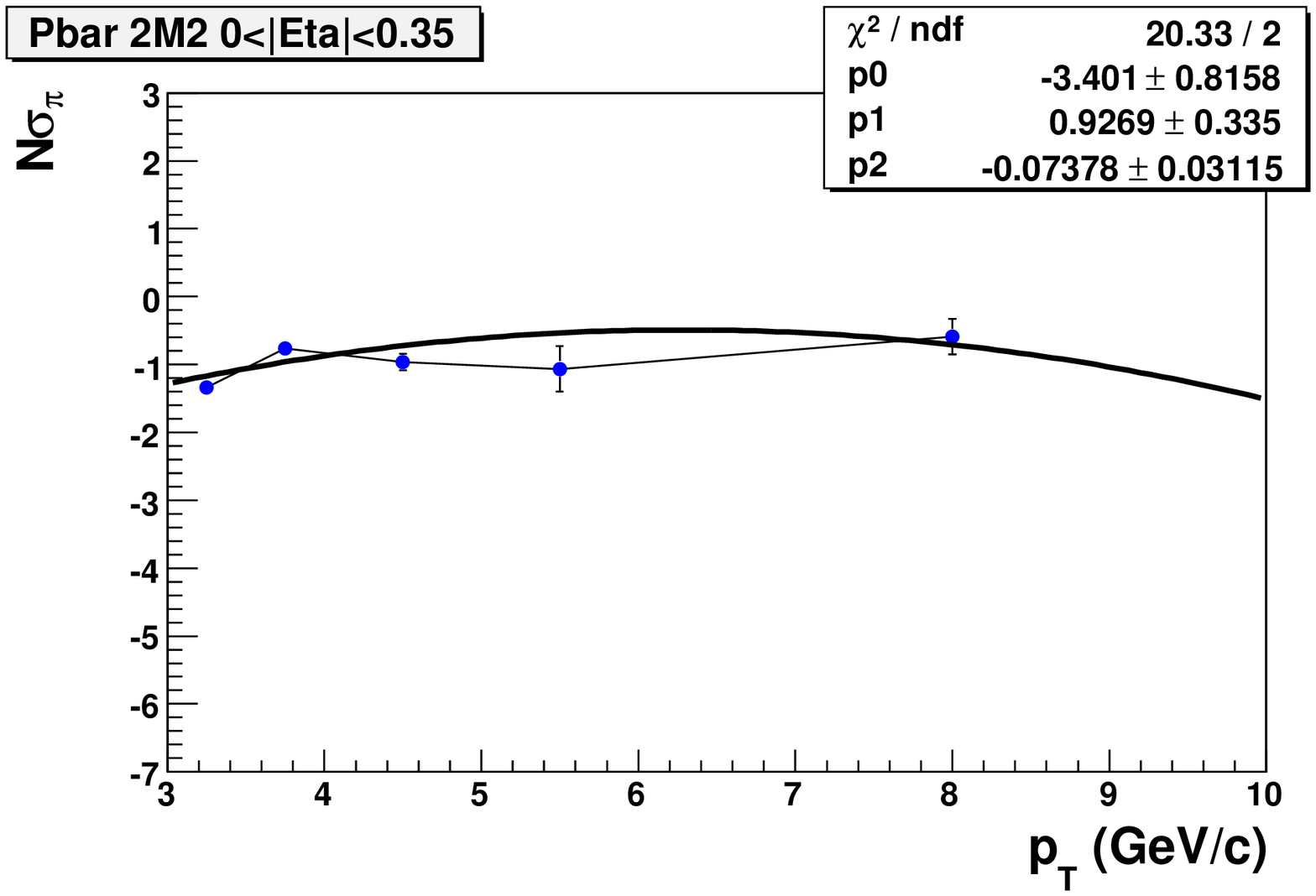}
		\includegraphics[width=1\textwidth]{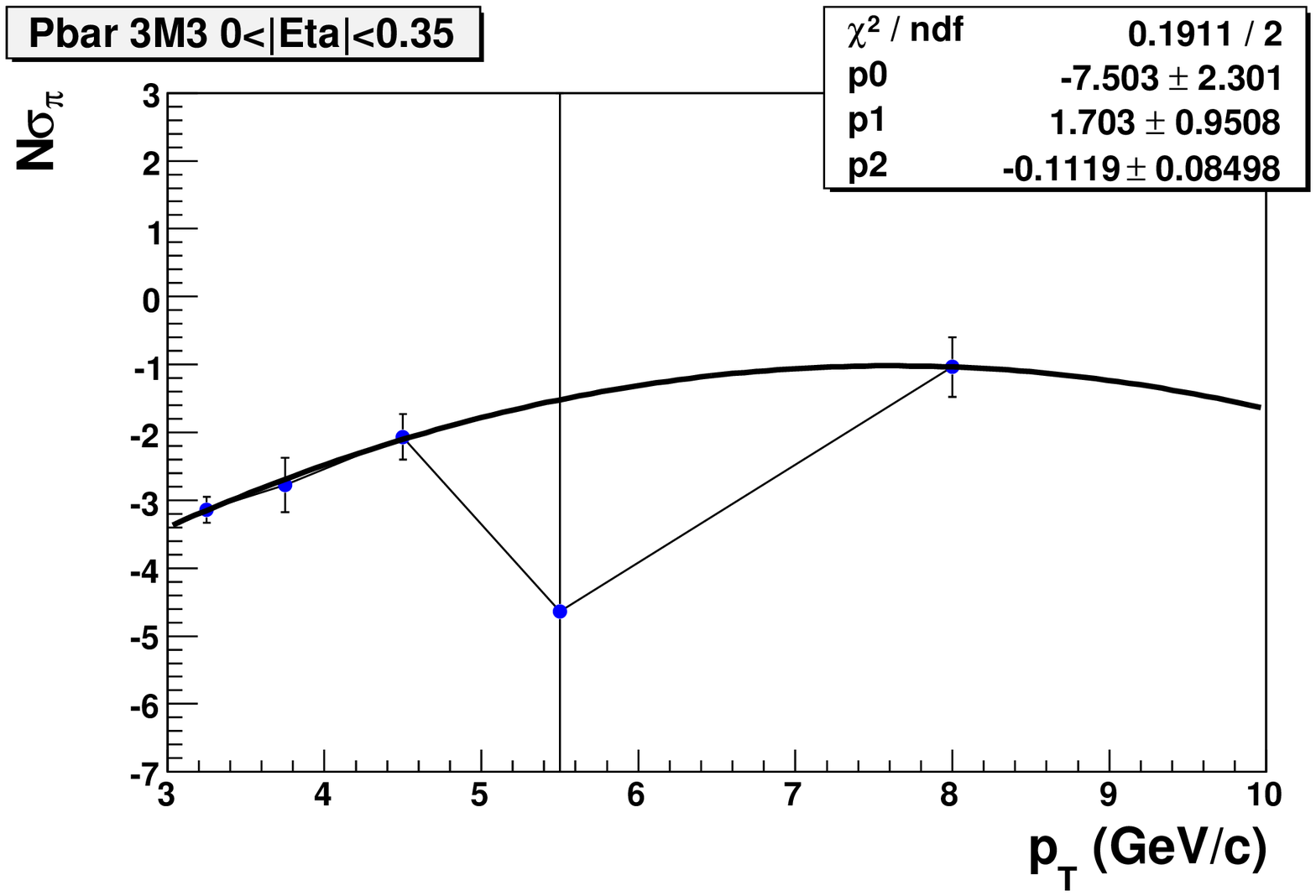}
		\includegraphics[width=1\textwidth]{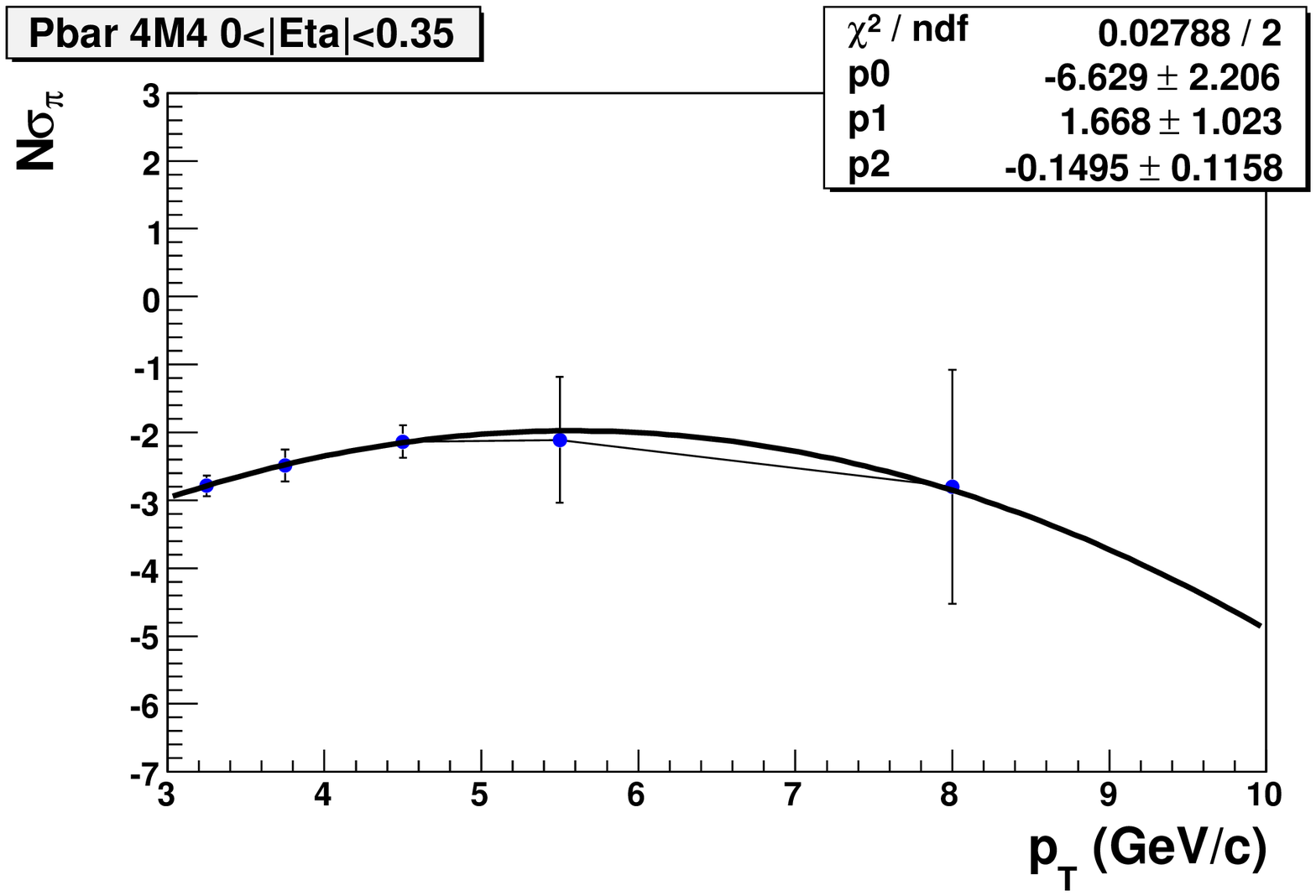}
		\includegraphics[width=1\textwidth]{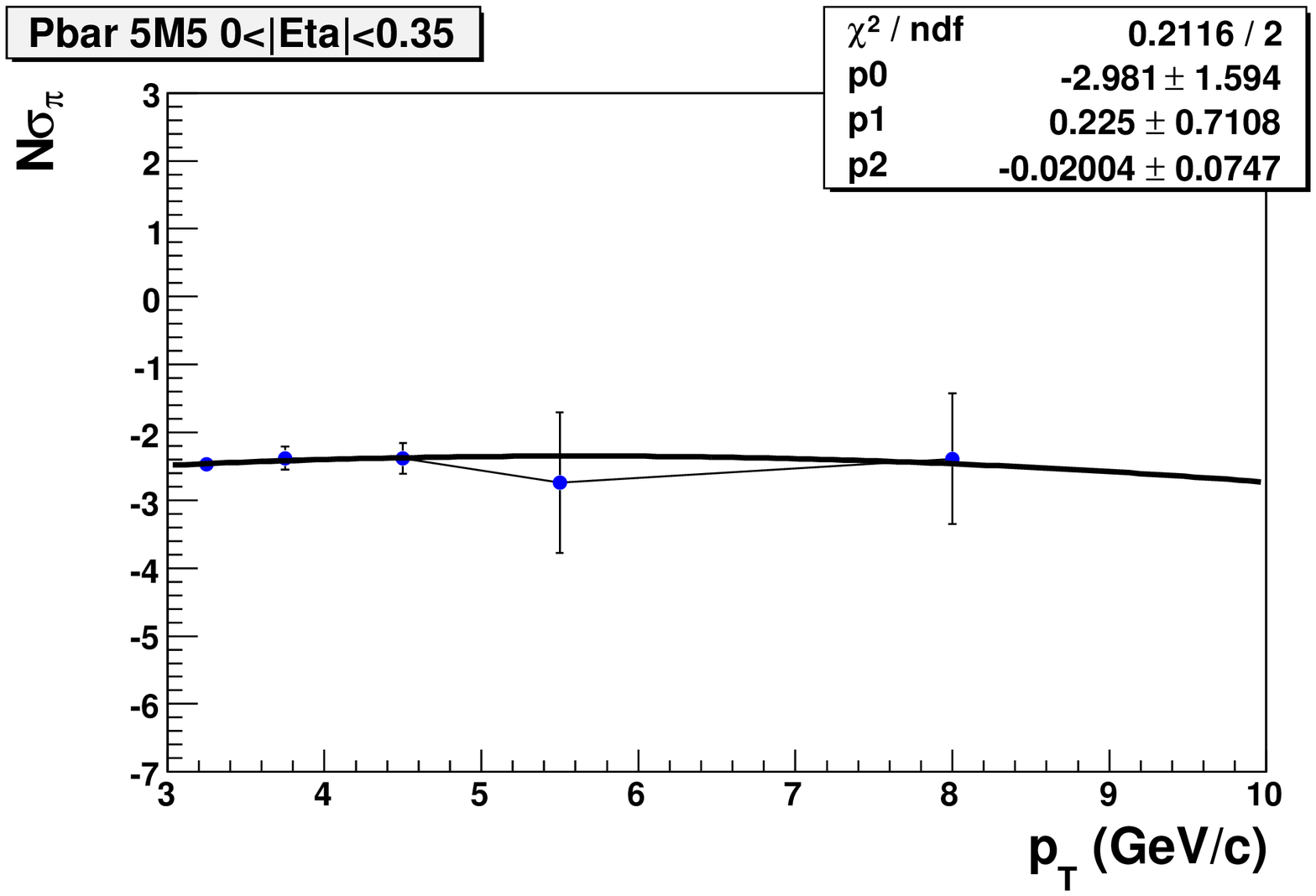}
		\includegraphics[width=1\textwidth]{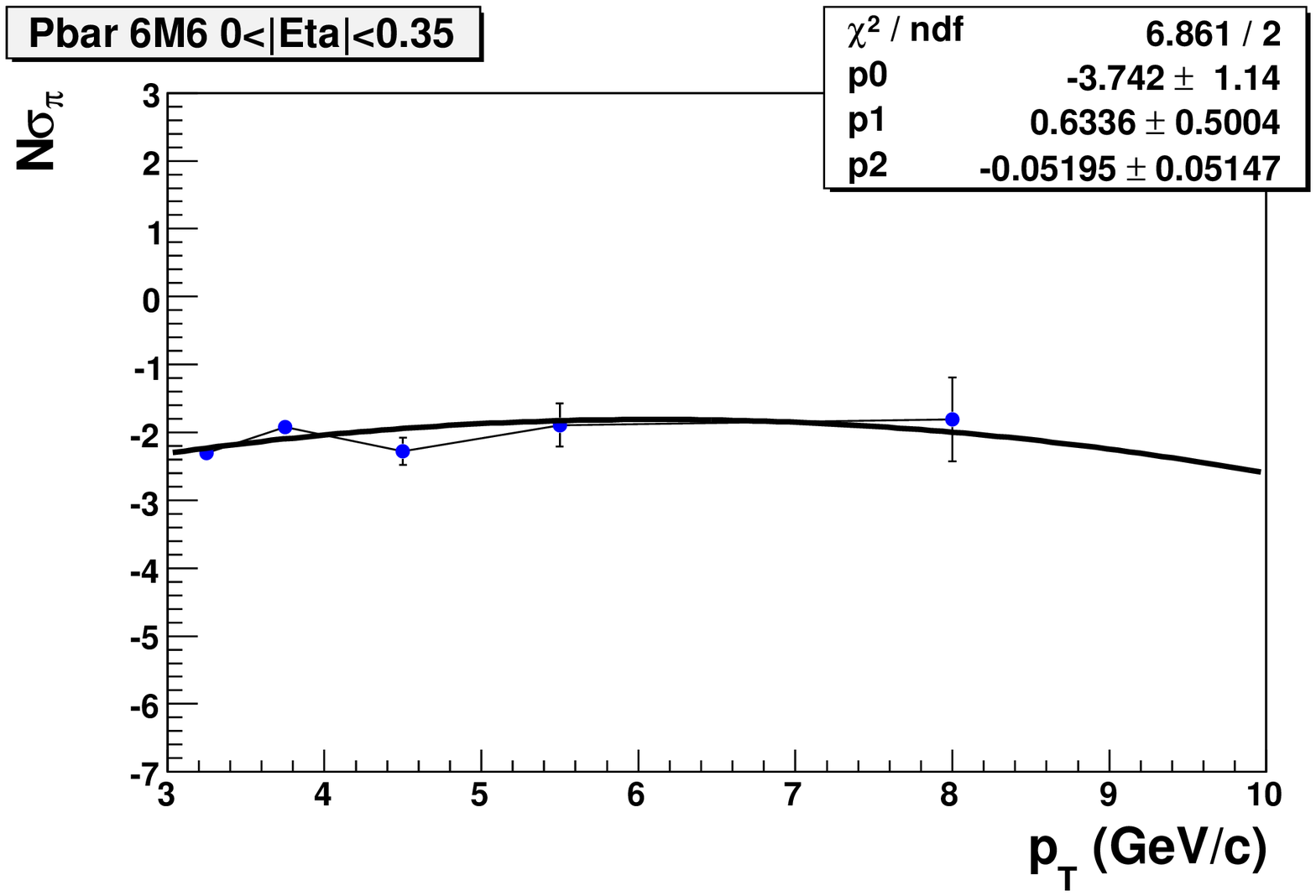}
		\includegraphics[width=1\textwidth]{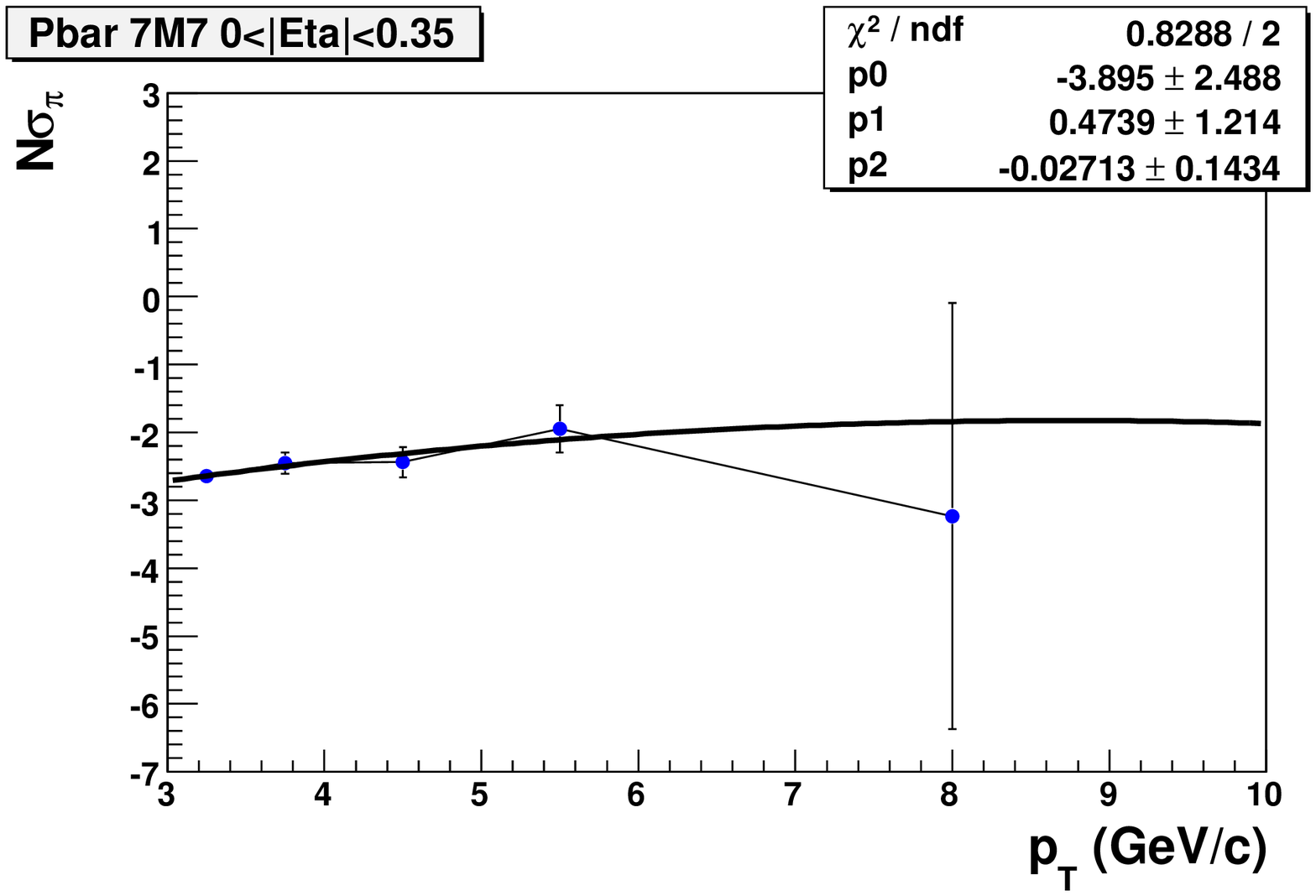}
		\includegraphics[width=1\textwidth]{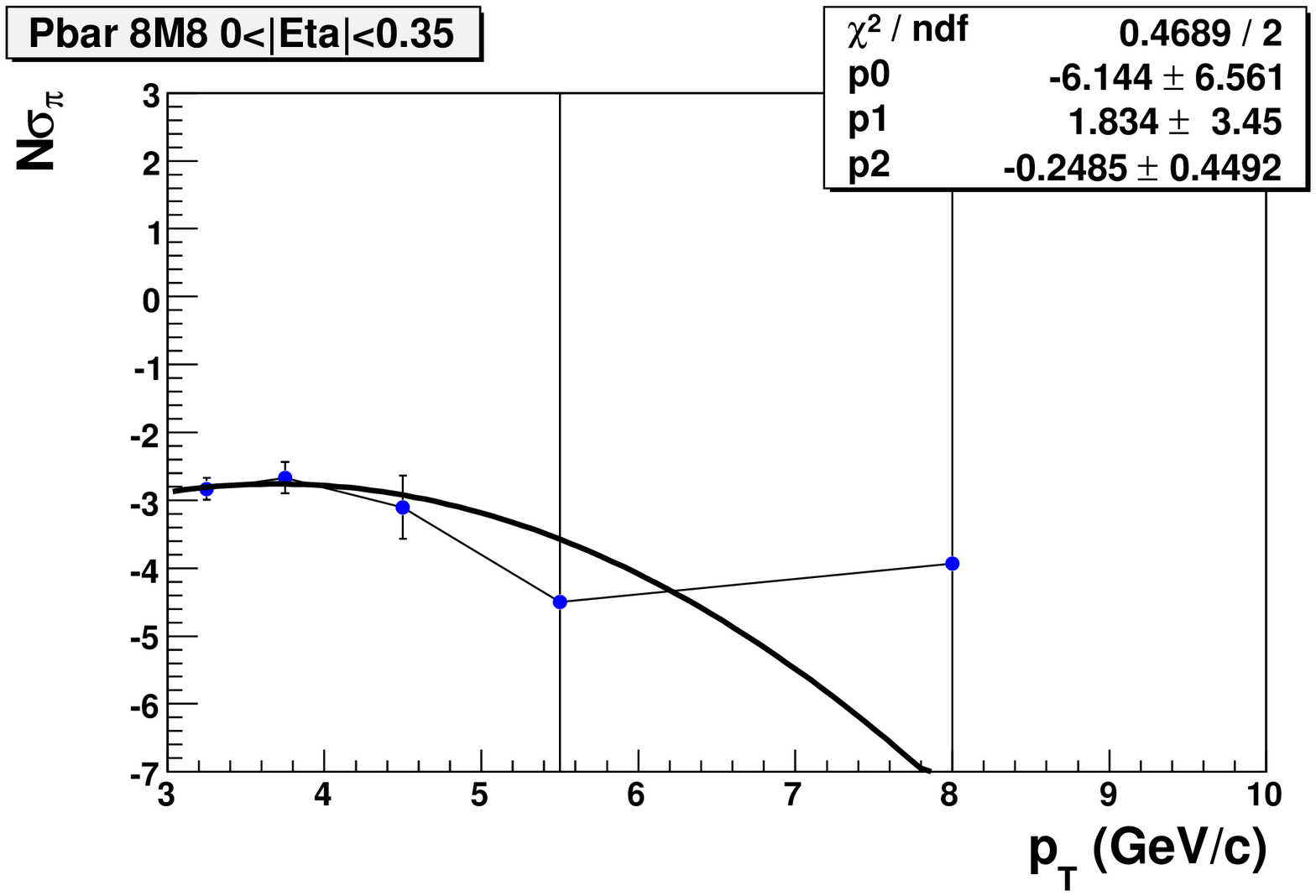}
		\includegraphics[width=1\textwidth]{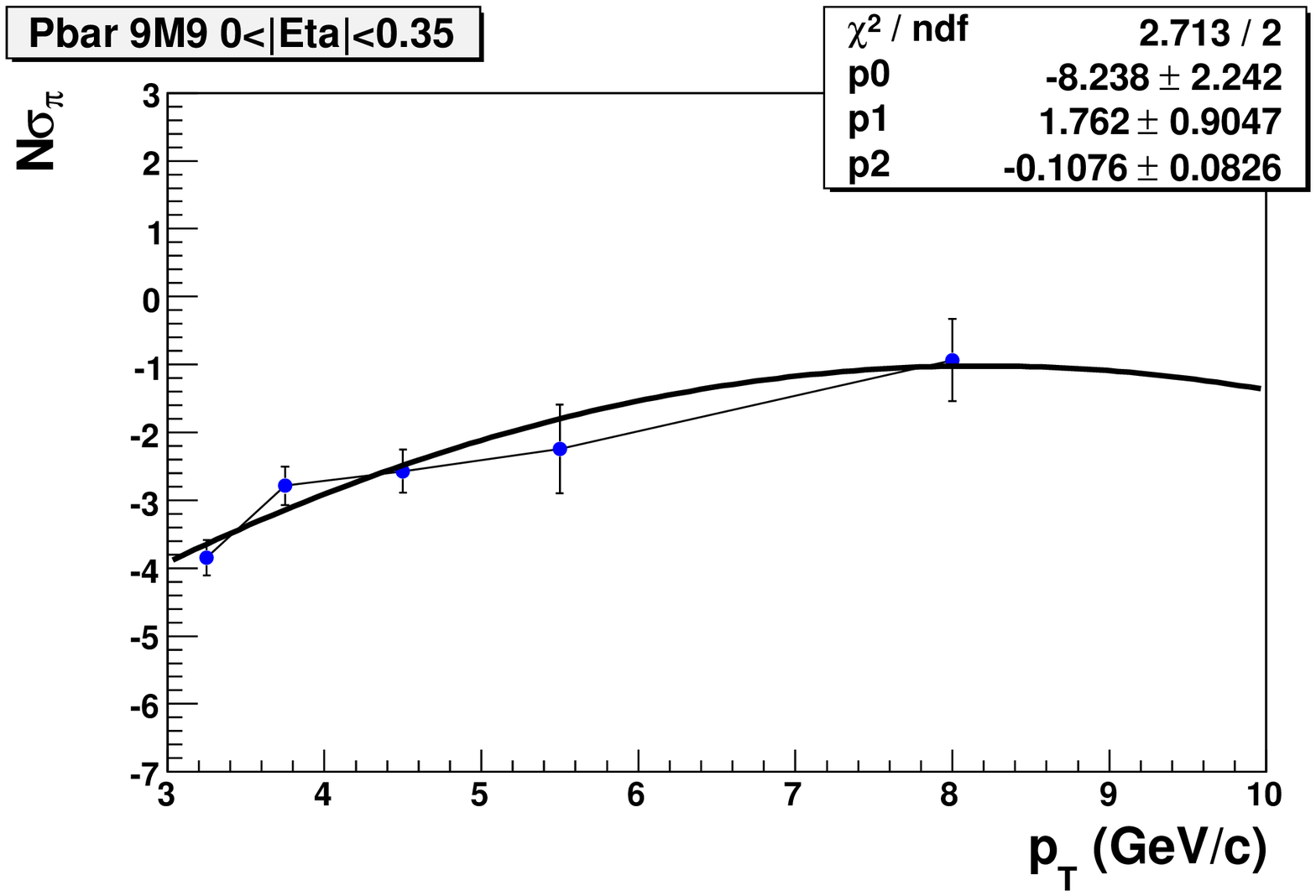}
										
			\end{minipage}
\hfill
\begin{minipage}[t]{.2\textwidth}
	\centering
		\includegraphics[width=1\textwidth]{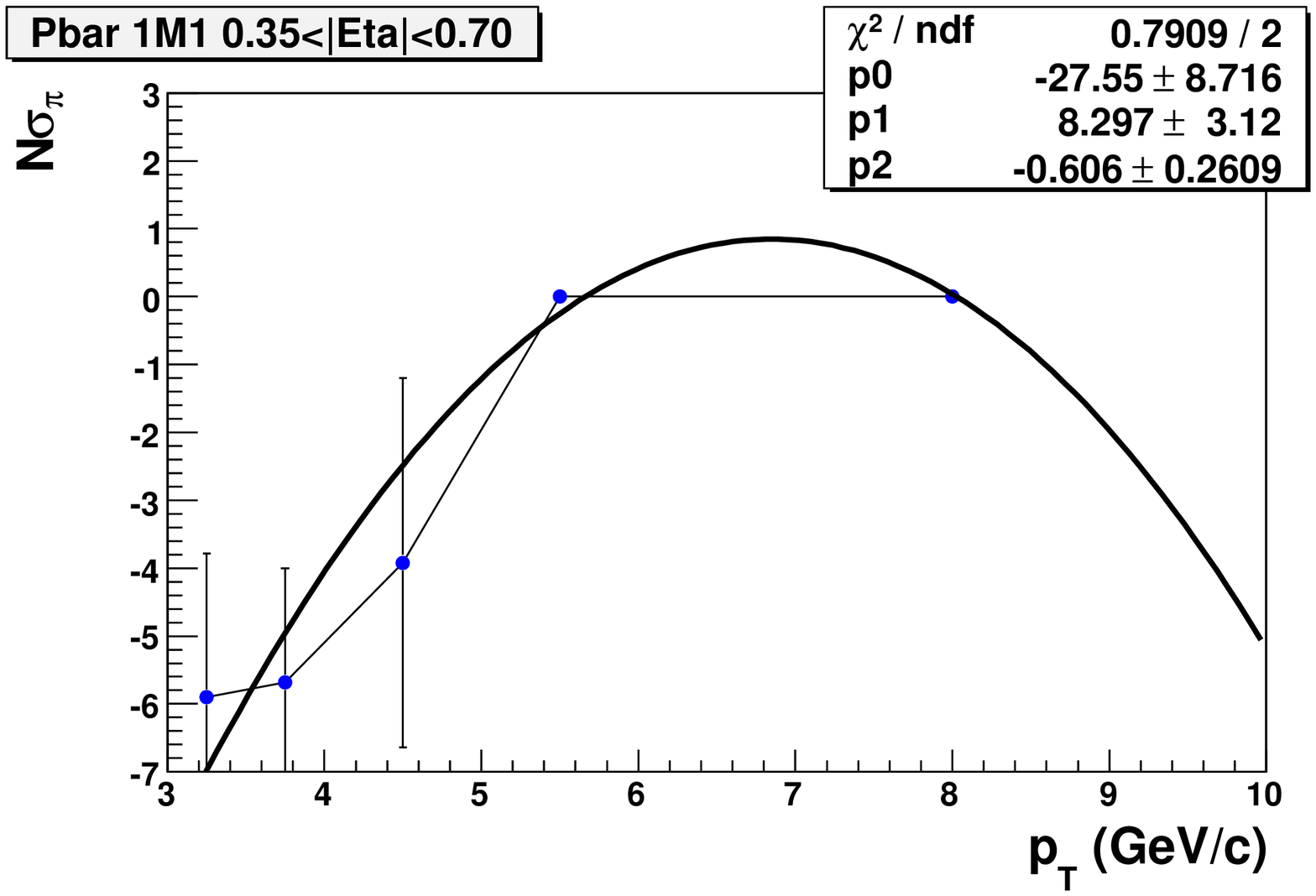}
		\includegraphics[width=1\textwidth]{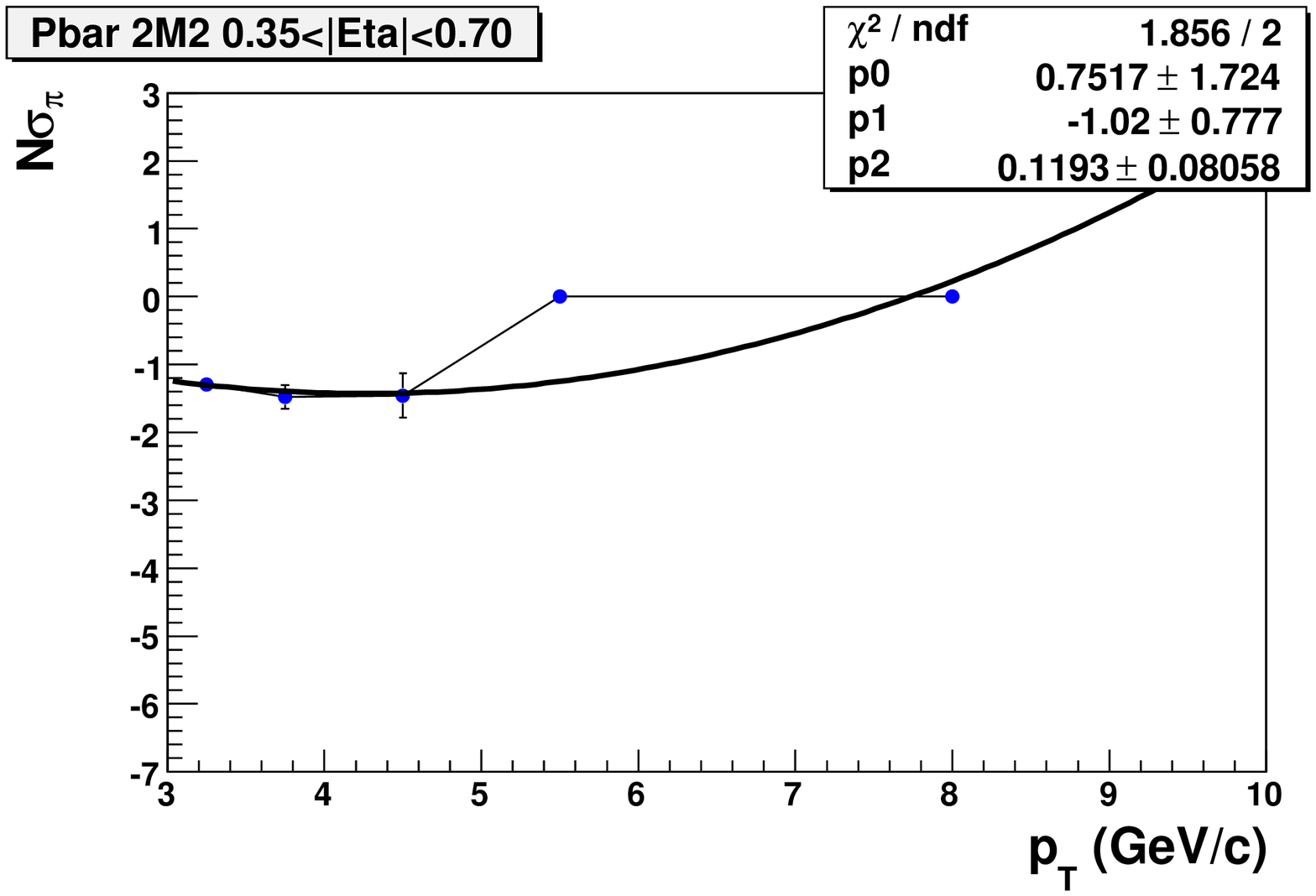}
		\includegraphics[width=1\textwidth]{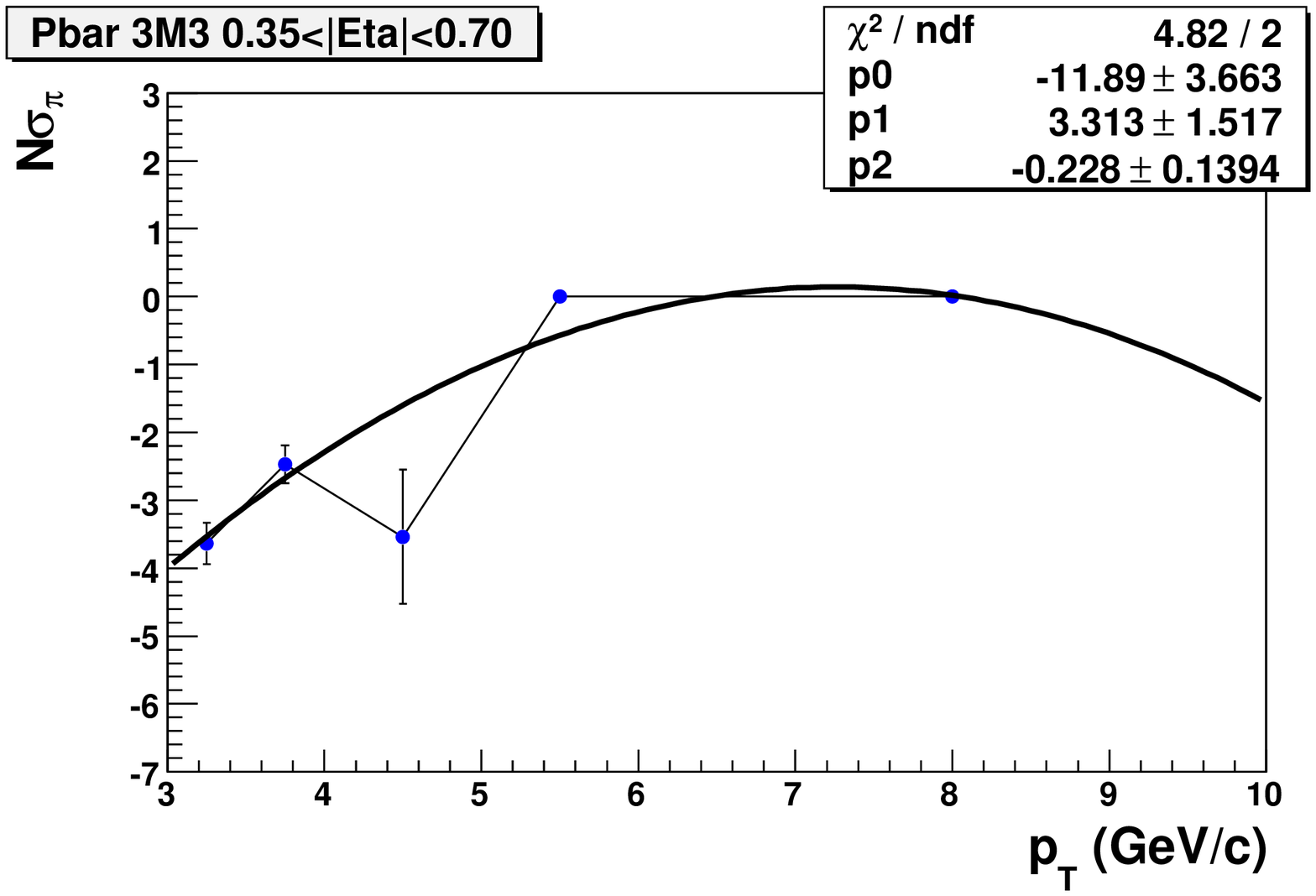}
		\includegraphics[width=1\textwidth]{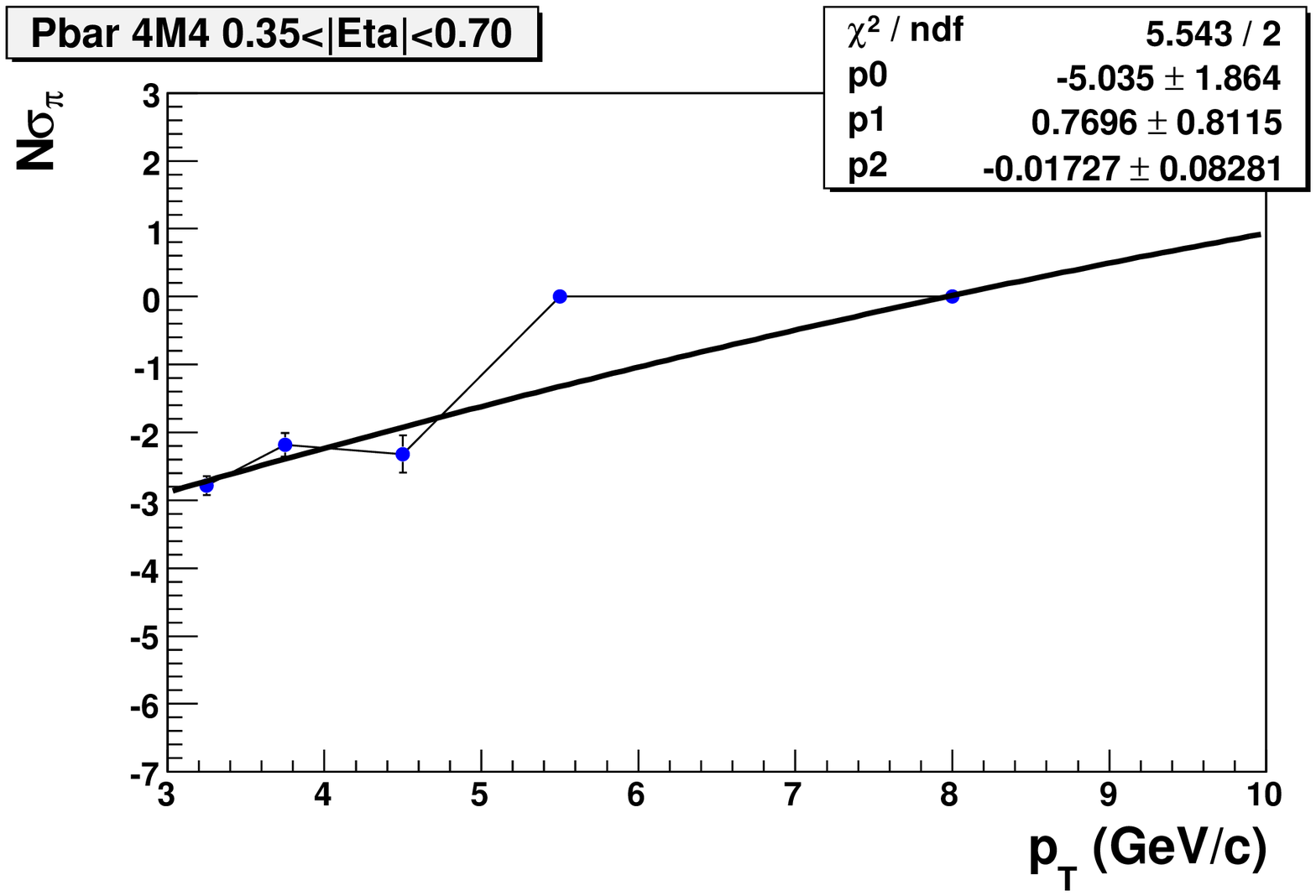}
		\includegraphics[width=1\textwidth]{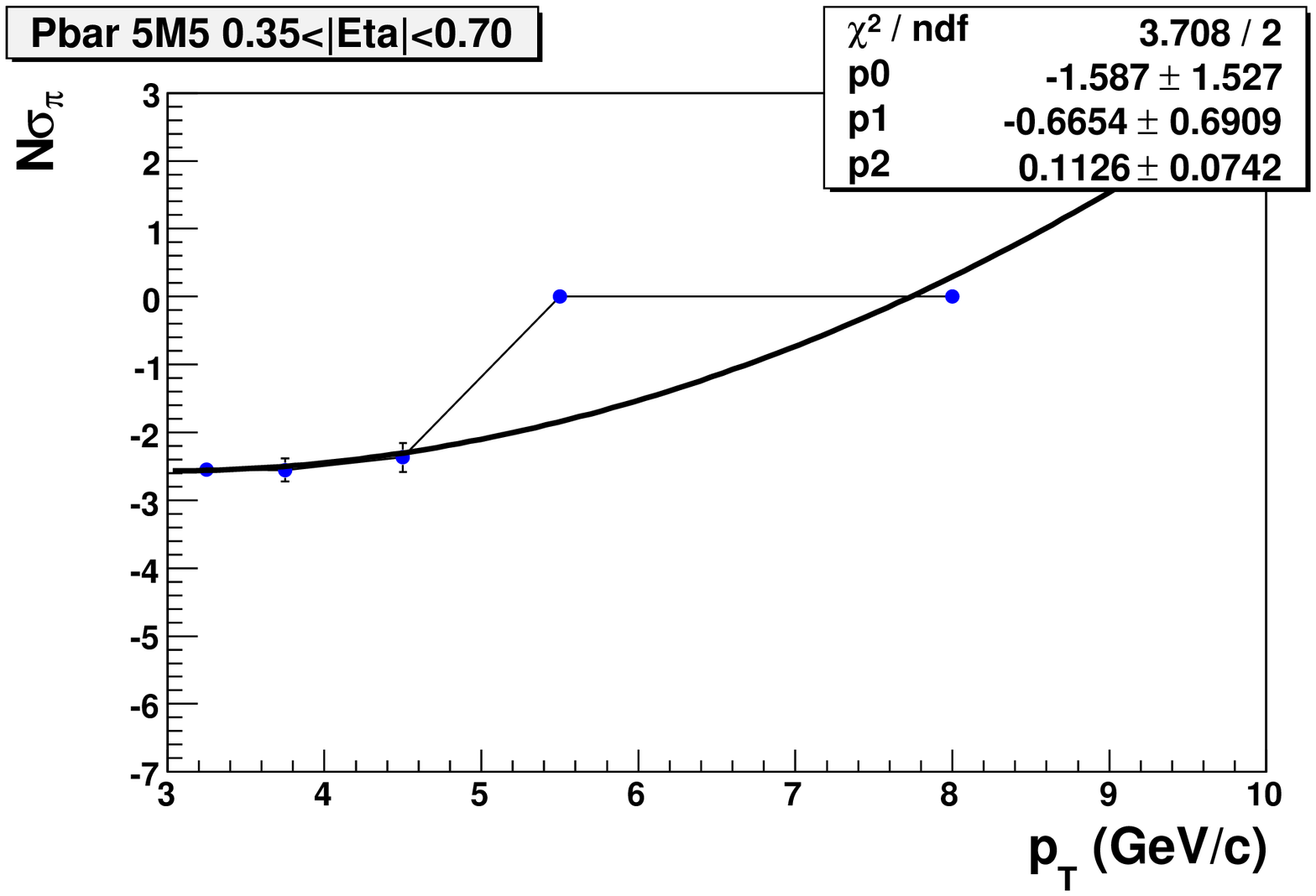}
		\includegraphics[width=1\textwidth]{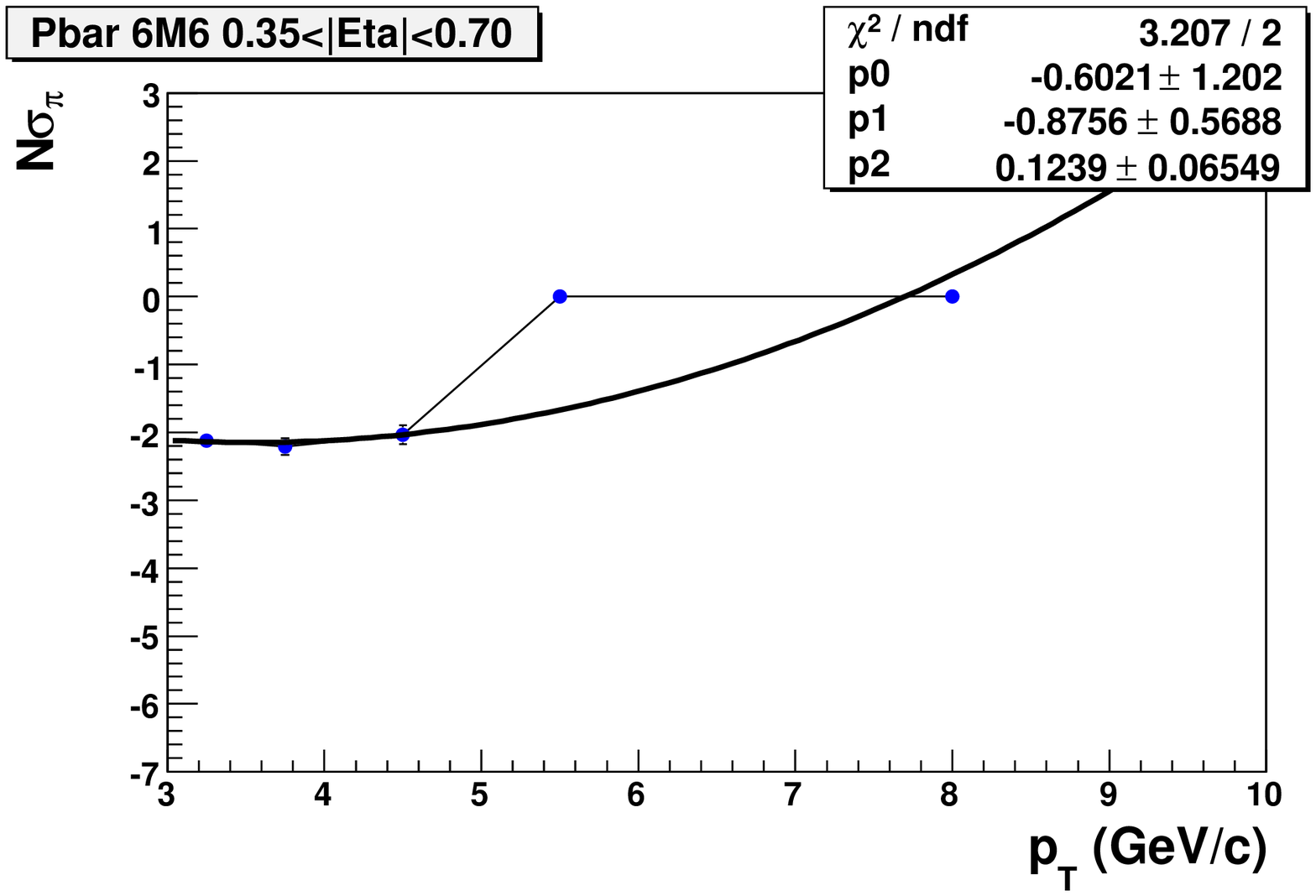}
		\includegraphics[width=1\textwidth]{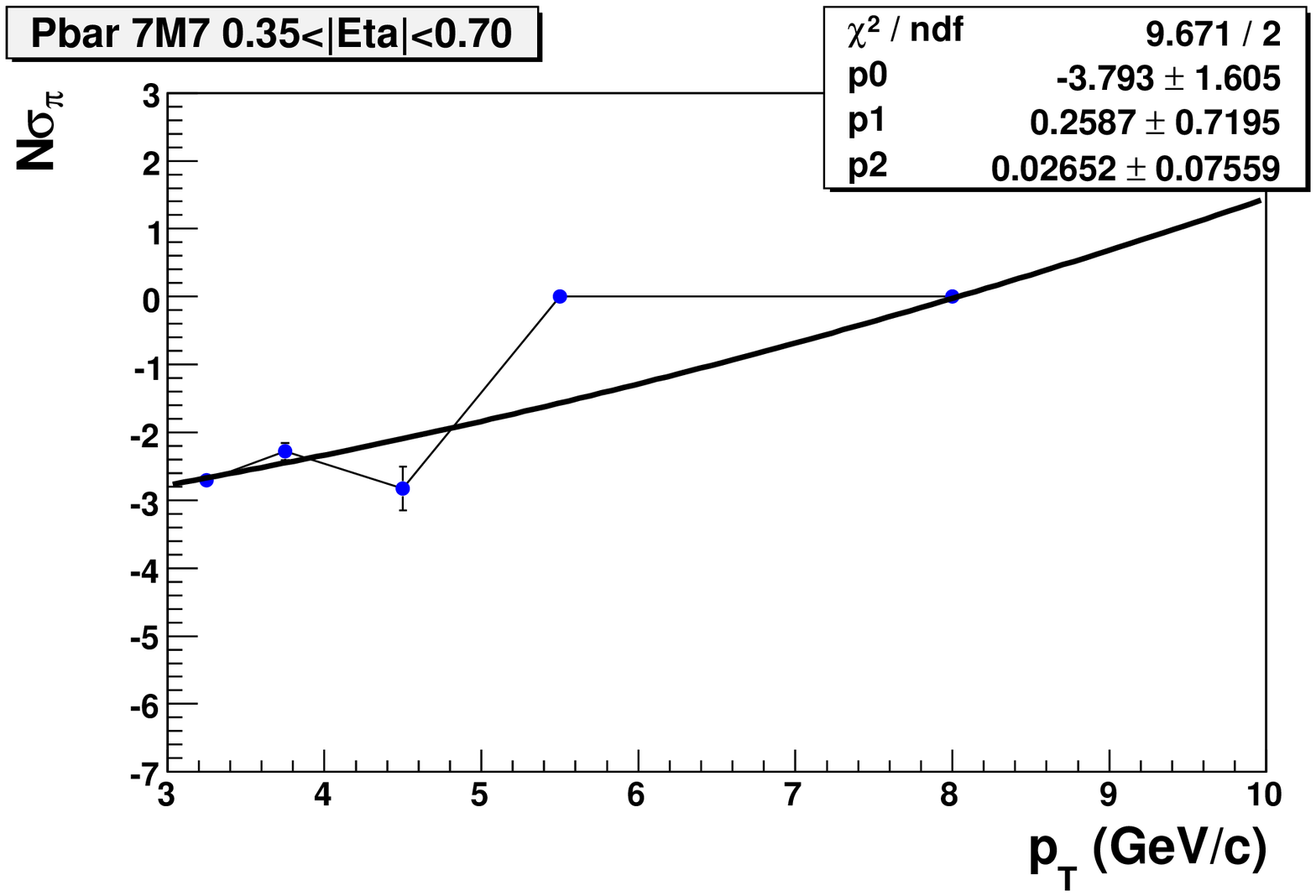}
		\includegraphics[width=1\textwidth]{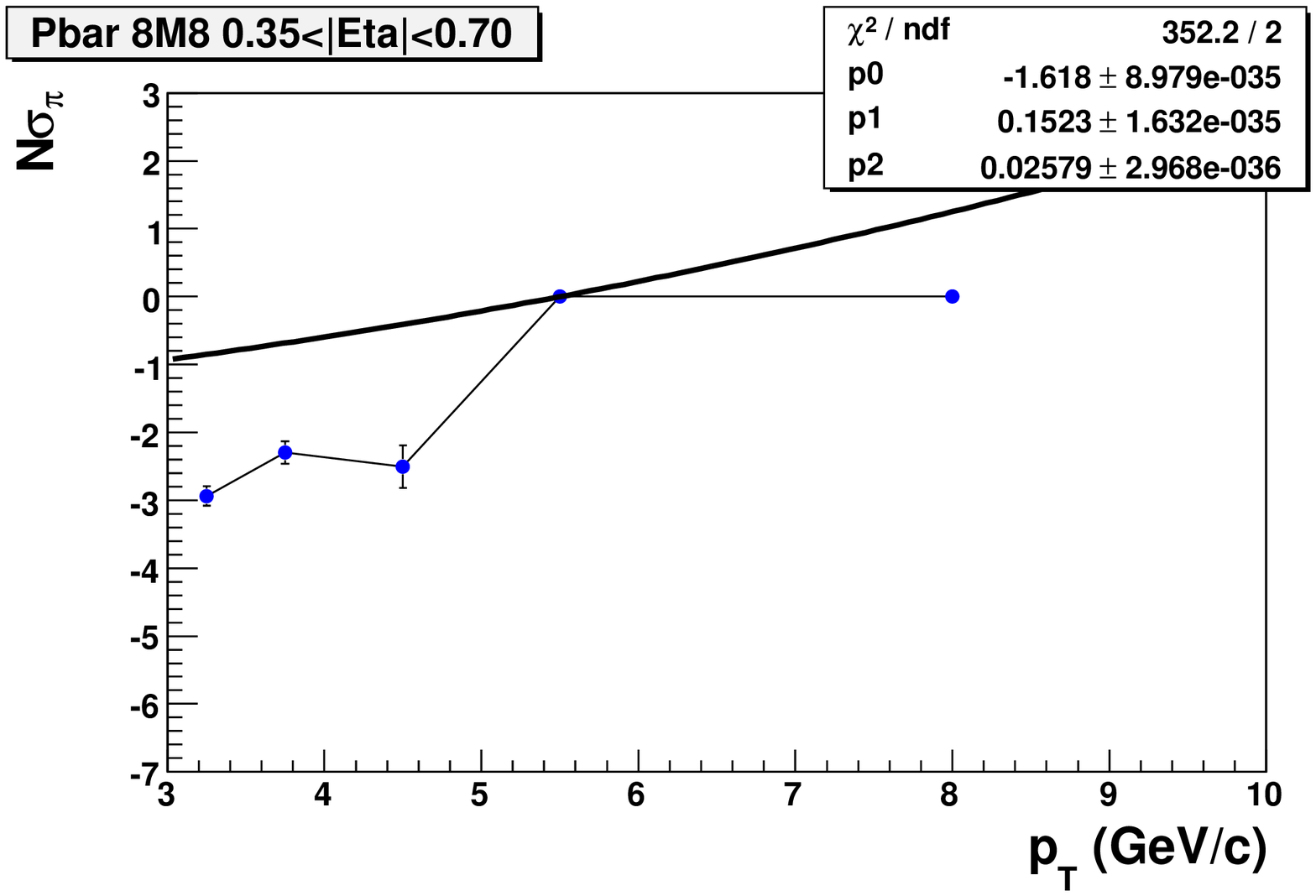}
		\includegraphics[width=1\textwidth]{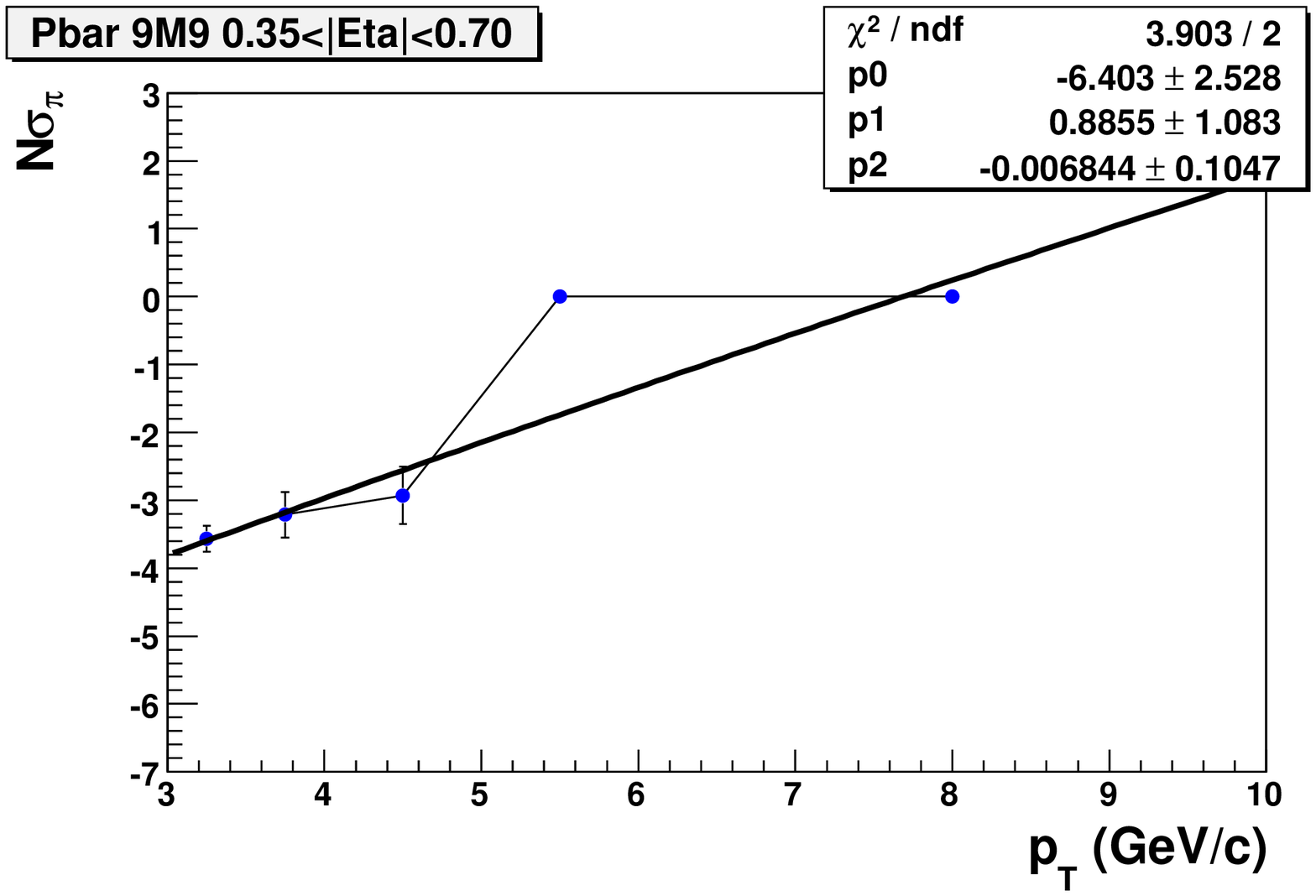}
		
			\end{minipage}	
					
	\caption{Same as Fig. 6.8 but for $p$($\bar{p}$).   Left: $p$ with $|\eta|<0.35$.  Left center:  $p$ with $0.35<|\eta|<0.7$.  Right center: $\bar{p}$ with $|\eta|<0.35$.  Right:  $\bar{p}$ with $0.35<|\eta|<0.7$.}
	\label{fig:fitcutsp}	
\end{figure}

\begin{figure}[H]
\hfill
\begin{minipage}[t]{.24\textwidth}
	\centering
		\includegraphics[width=1\textwidth]{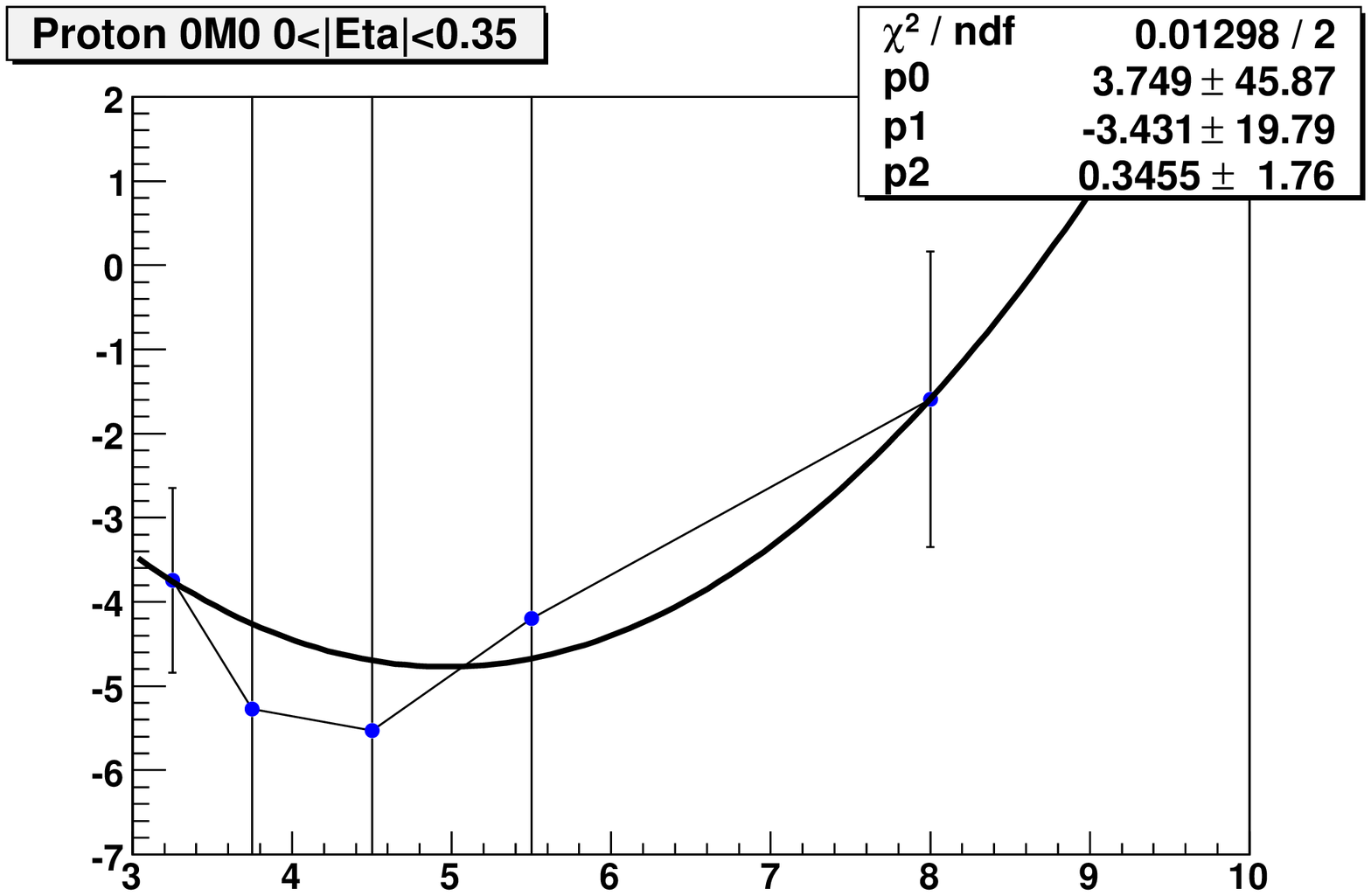}
		\includegraphics[width=1\textwidth]{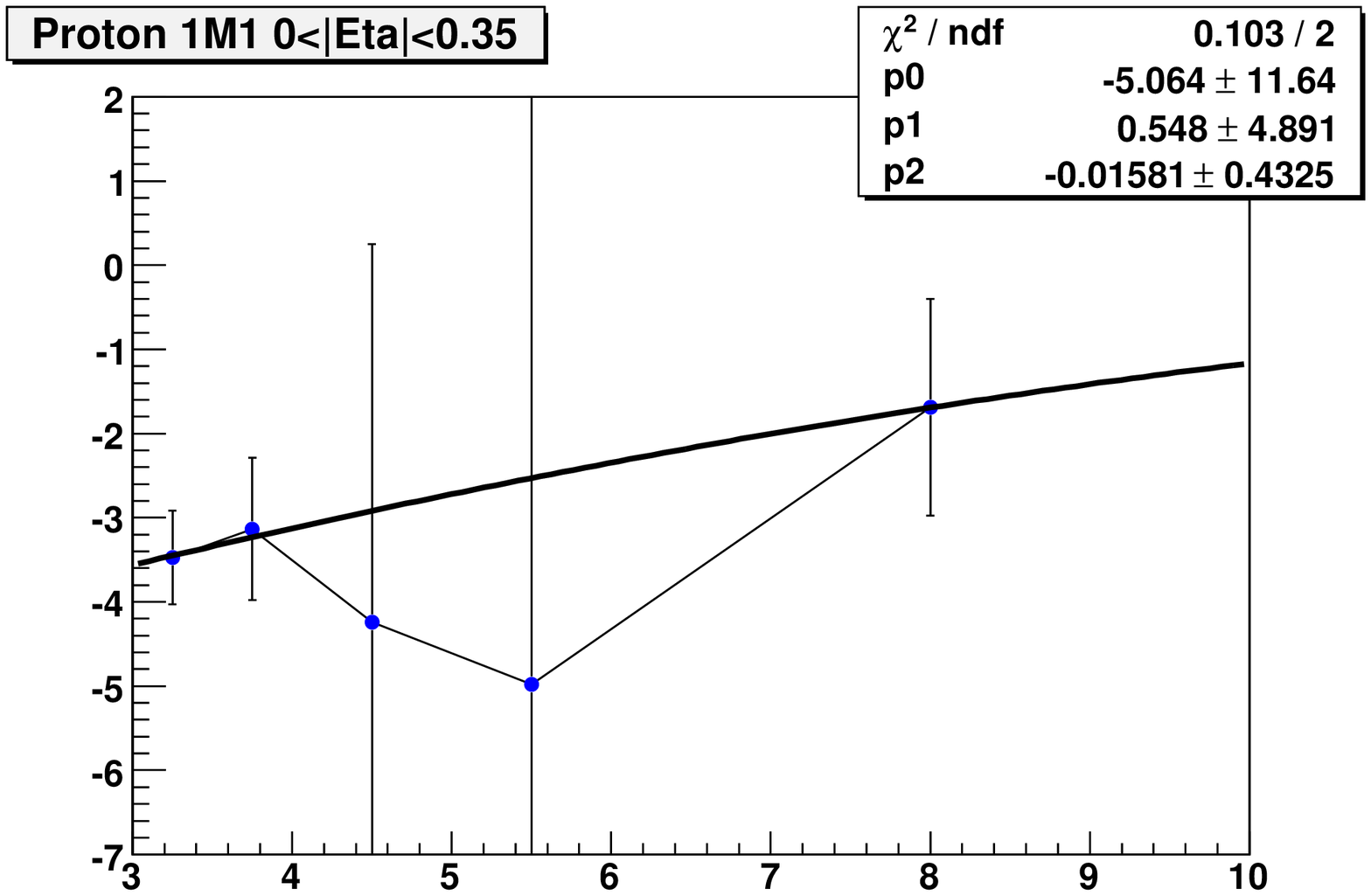}
		\includegraphics[width=1\textwidth]{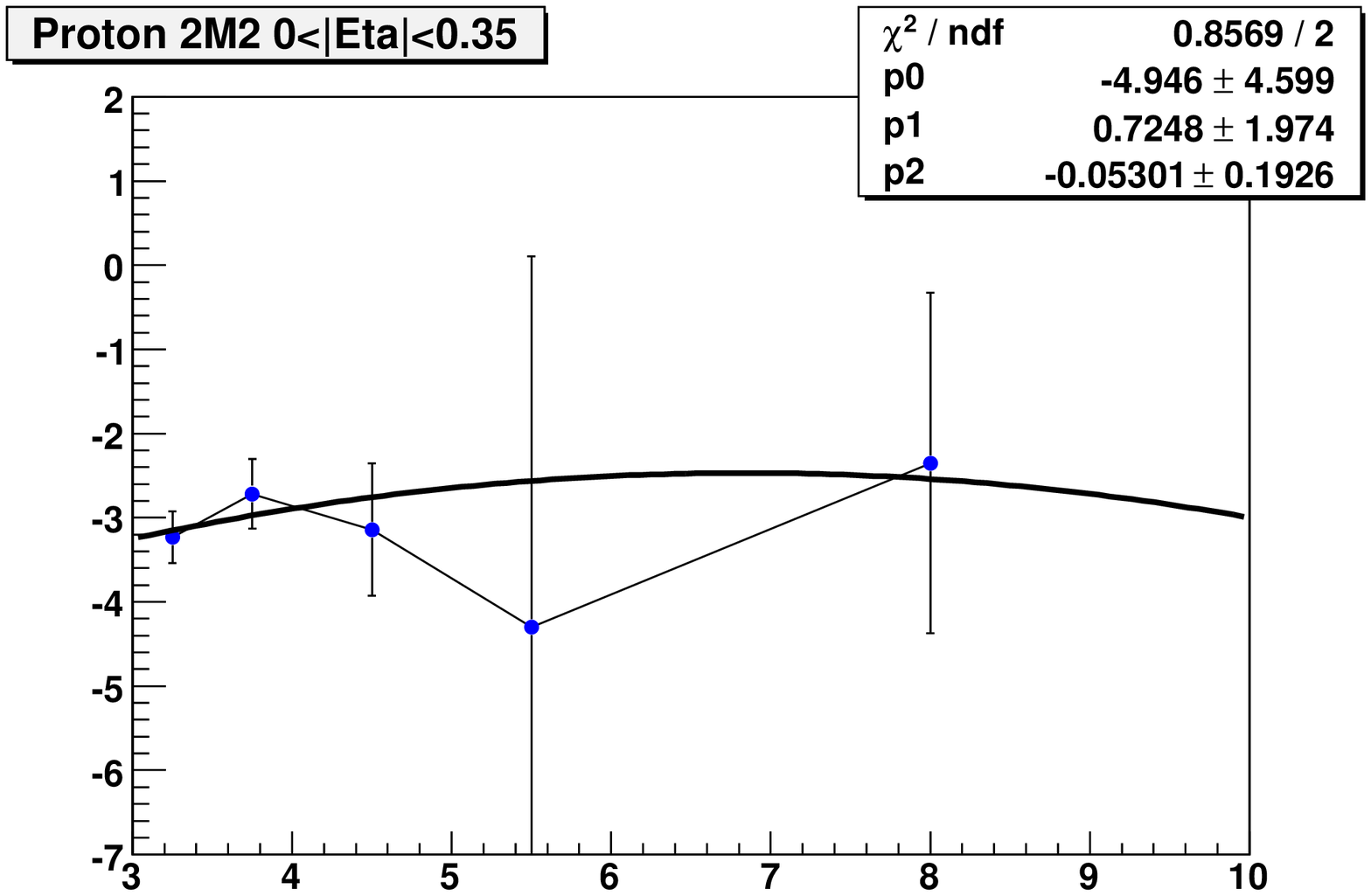}						
			\end{minipage}
\hfill
\begin{minipage}[t]{.24\textwidth}
	\centering
		\includegraphics[width=1\textwidth]{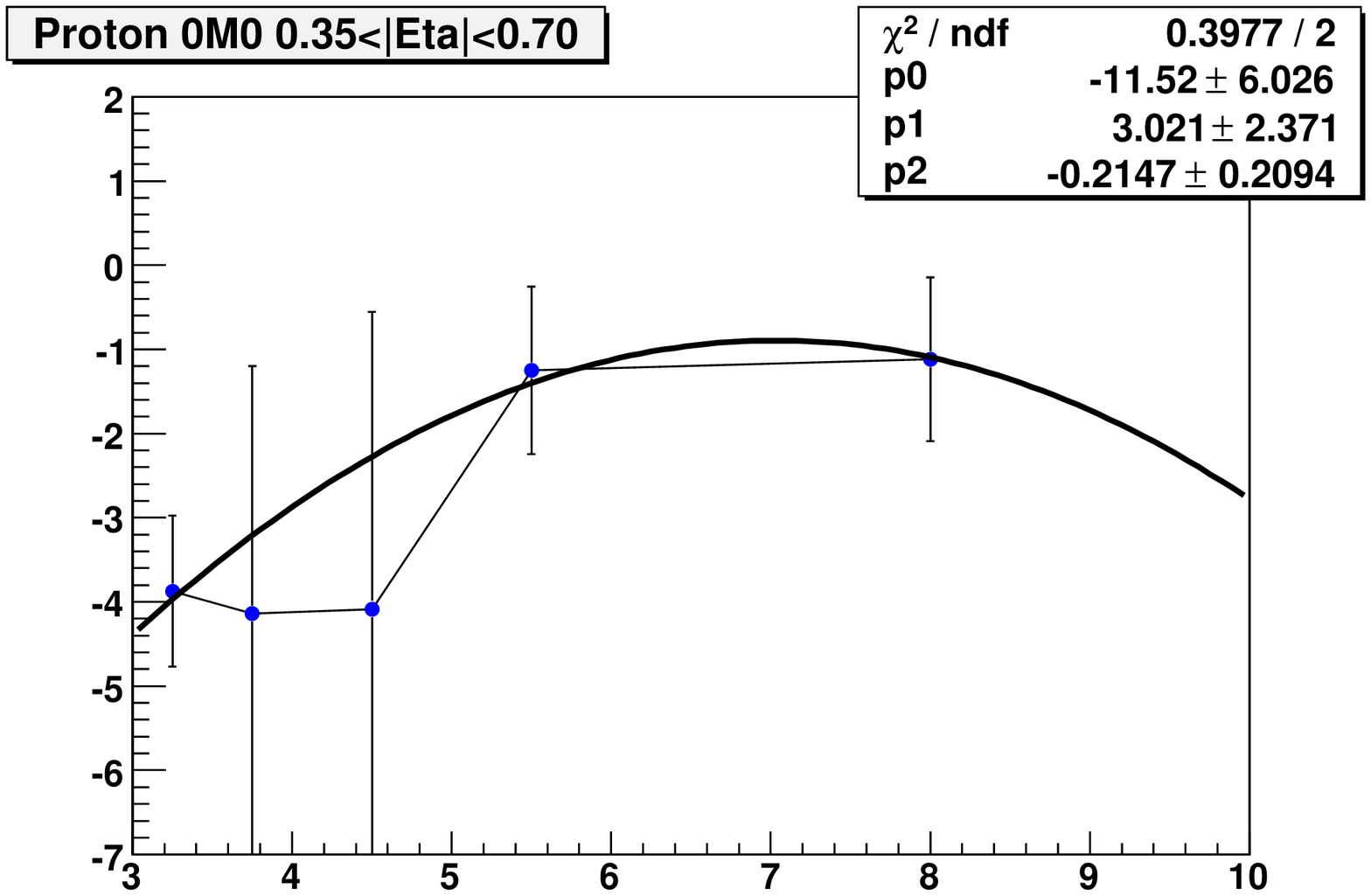}
		\includegraphics[width=1\textwidth]{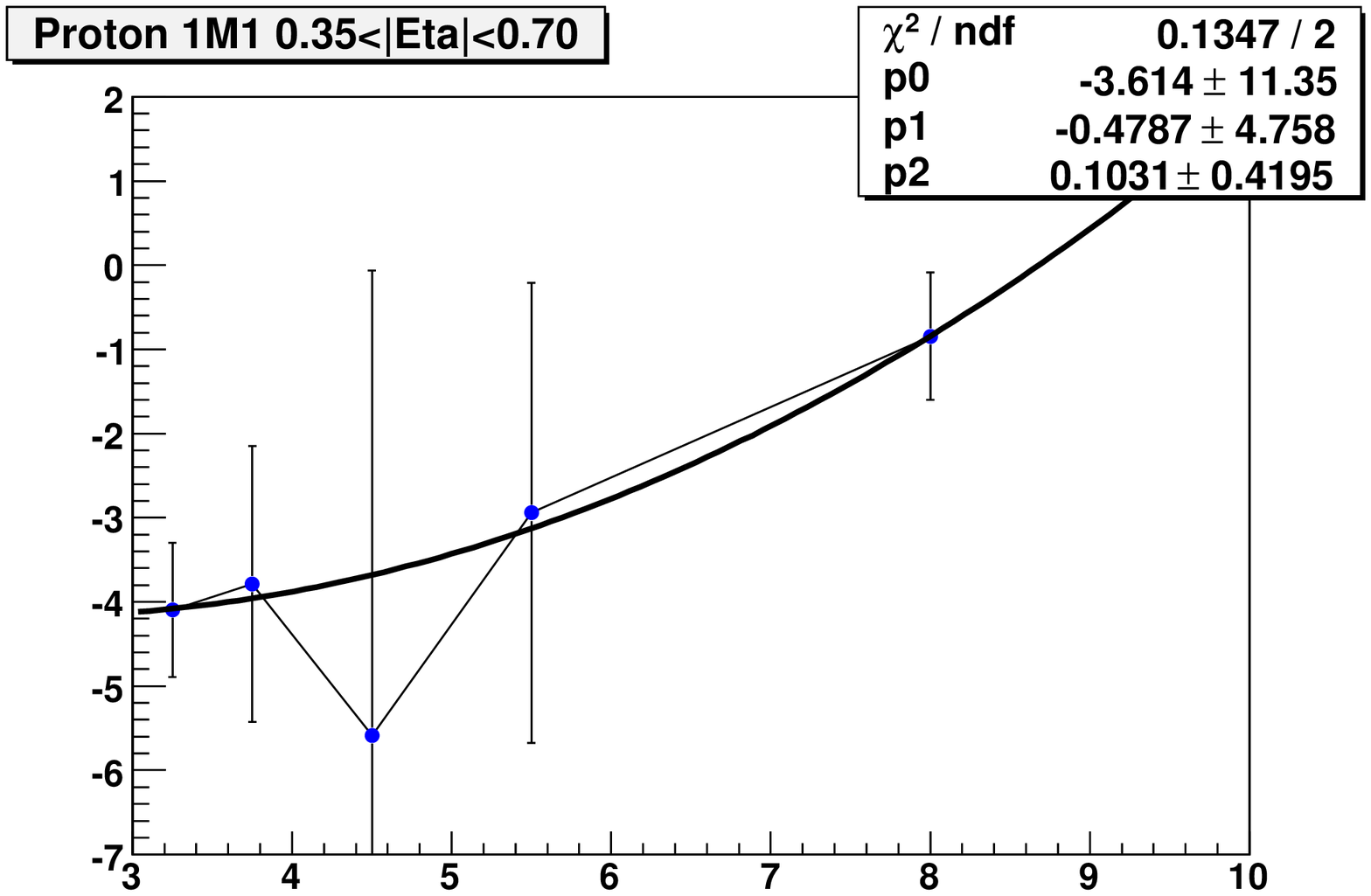}
		\includegraphics[width=1\textwidth]{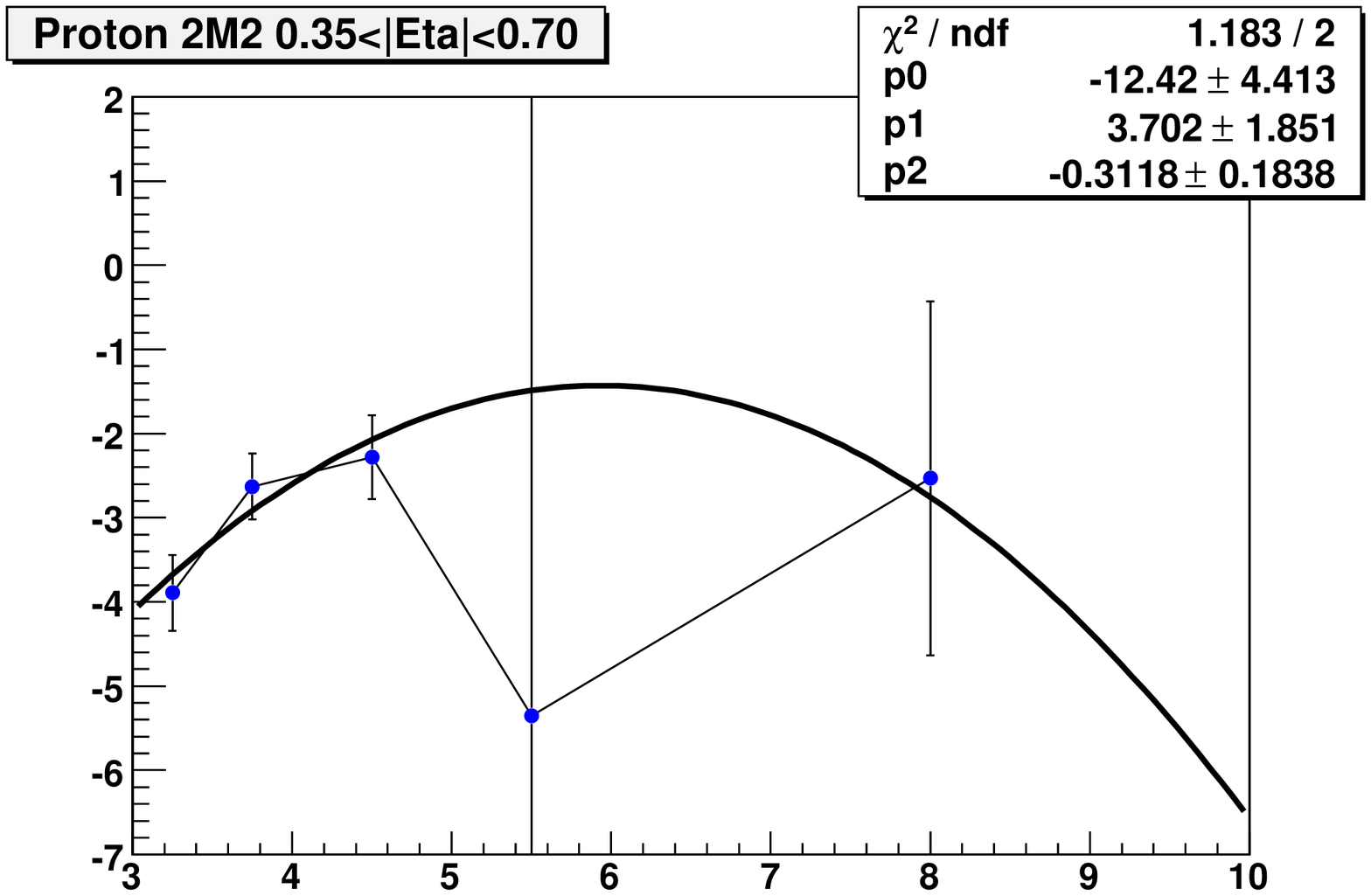}							
			\end{minipage}
\hfill
\begin{minipage}[t]{.24\textwidth}
	\centering
		\includegraphics[width=1\textwidth]{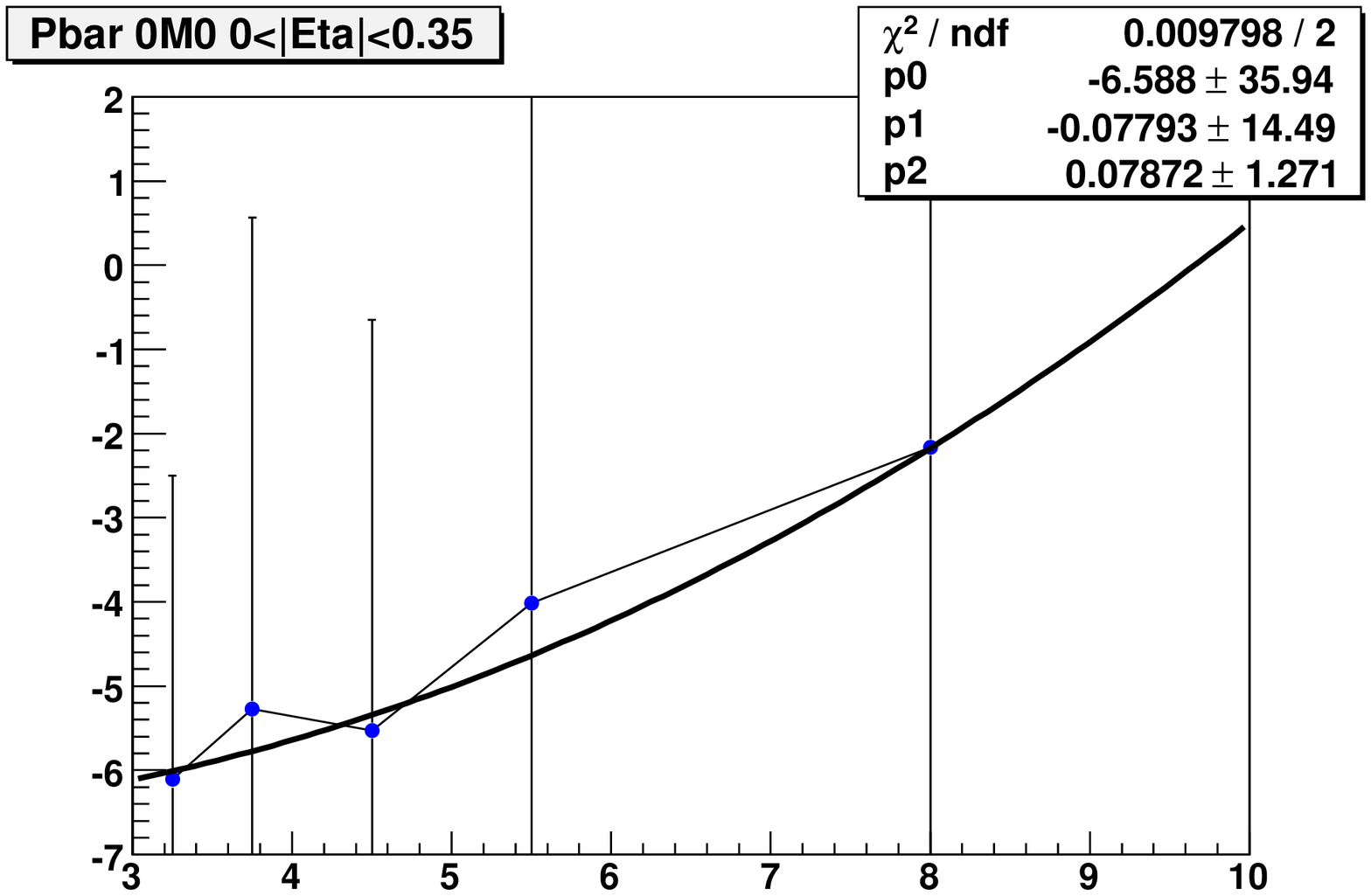}
		\includegraphics[width=1\textwidth]{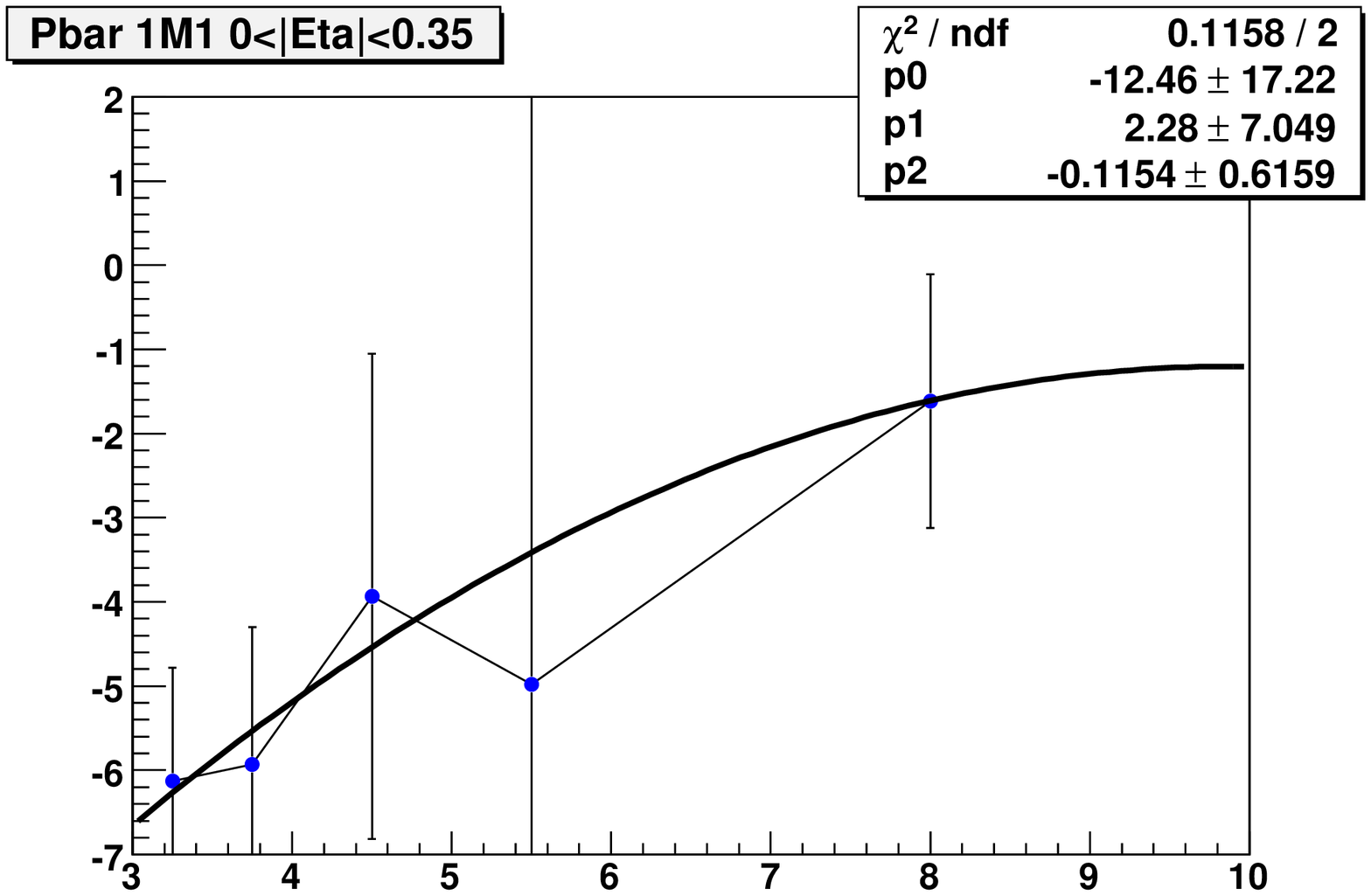}
		\includegraphics[width=1\textwidth]{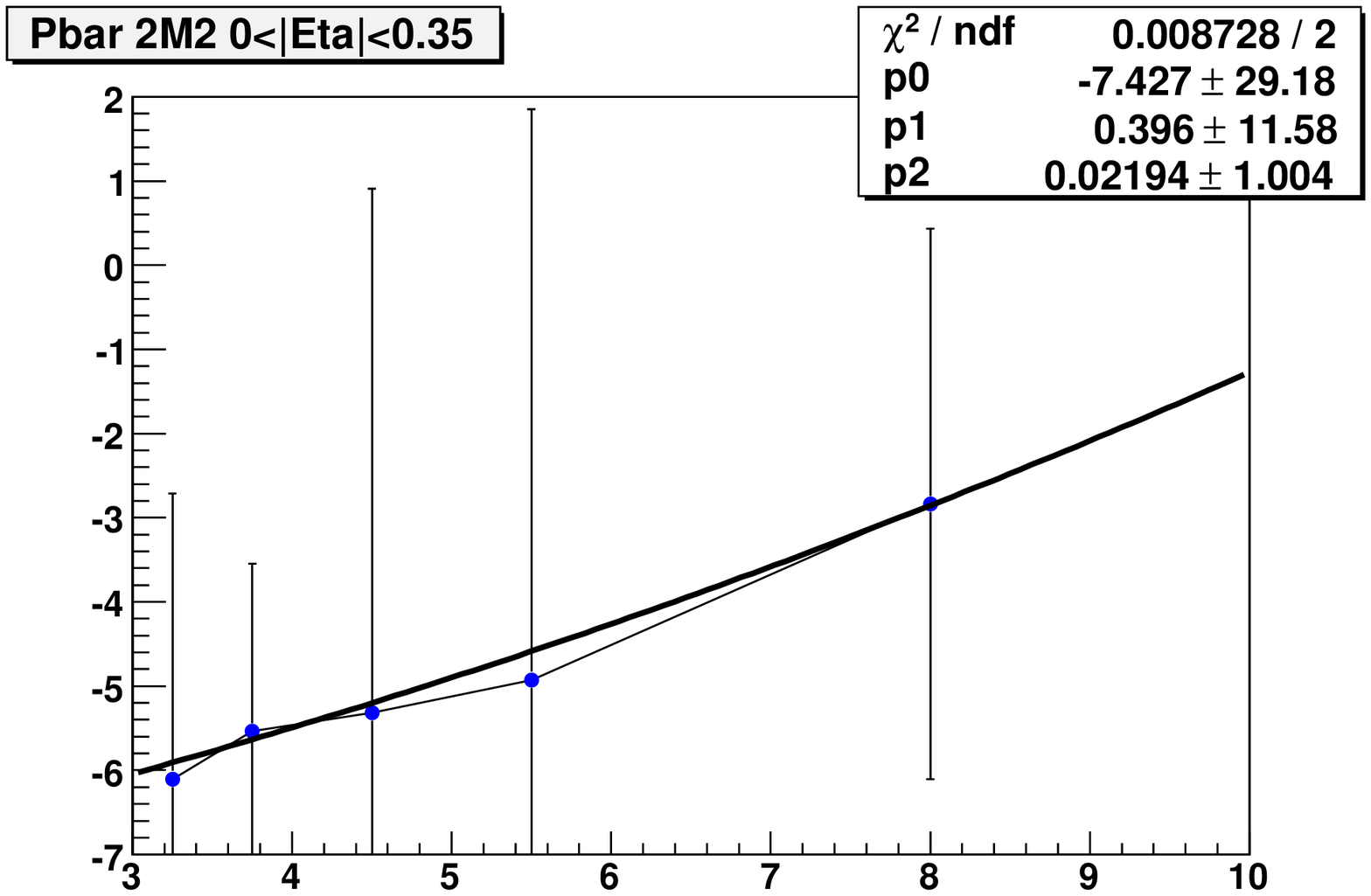}										
			\end{minipage}
\hfill
\begin{minipage}[t]{.24\textwidth}
	\centering
		\includegraphics[width=1\textwidth]{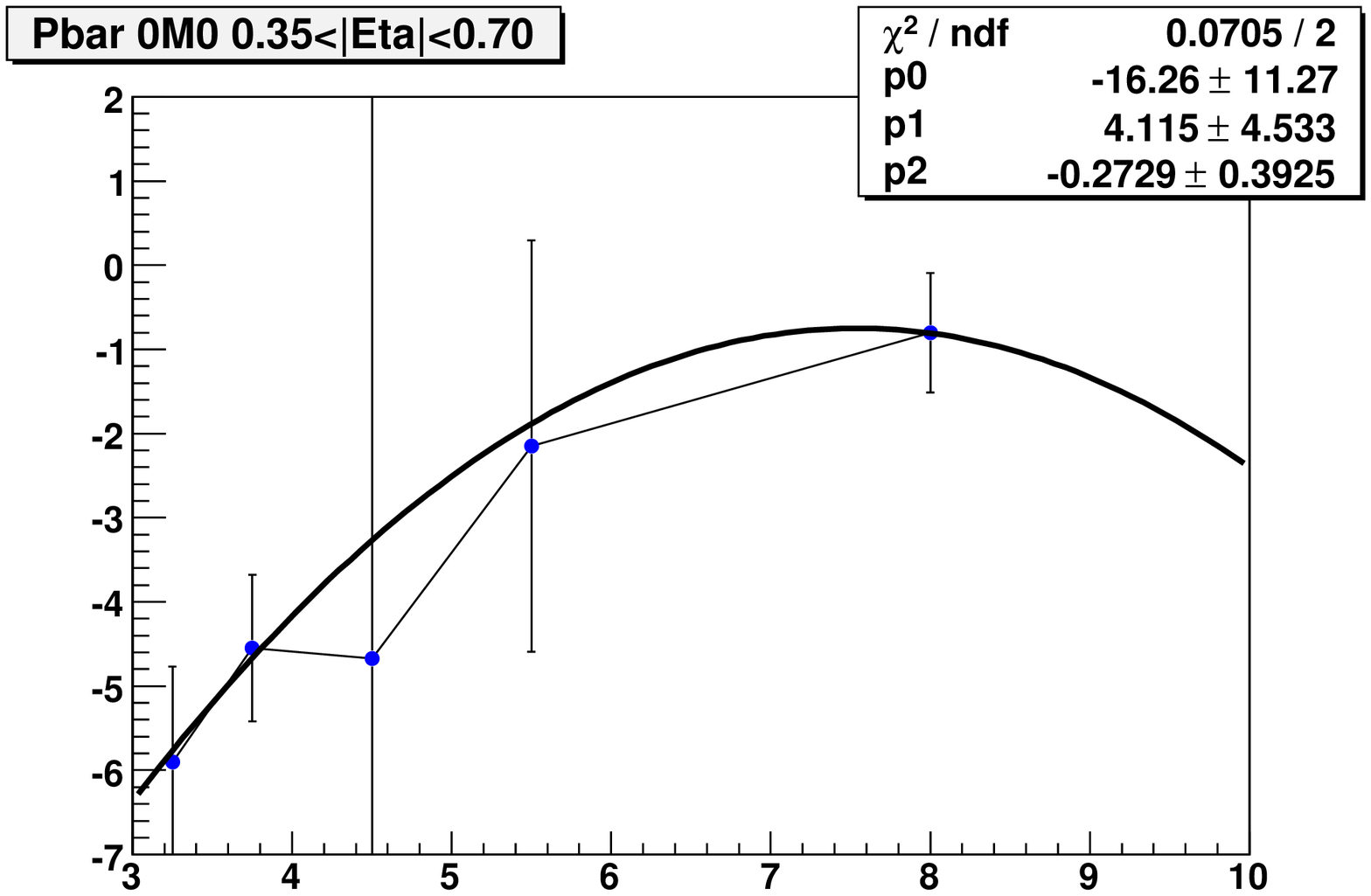}
		\includegraphics[width=1\textwidth]{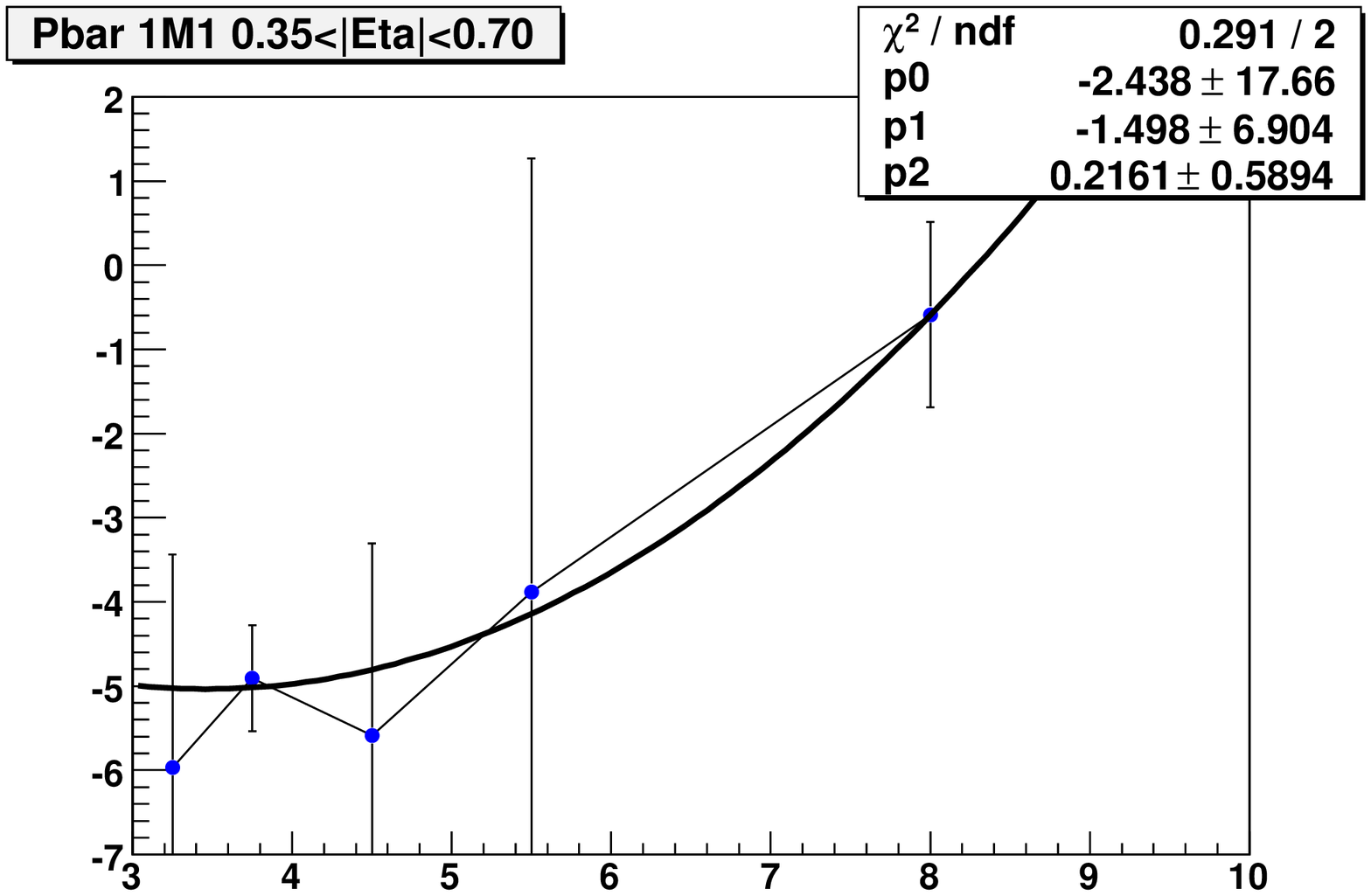}
		\includegraphics[width=1\textwidth]{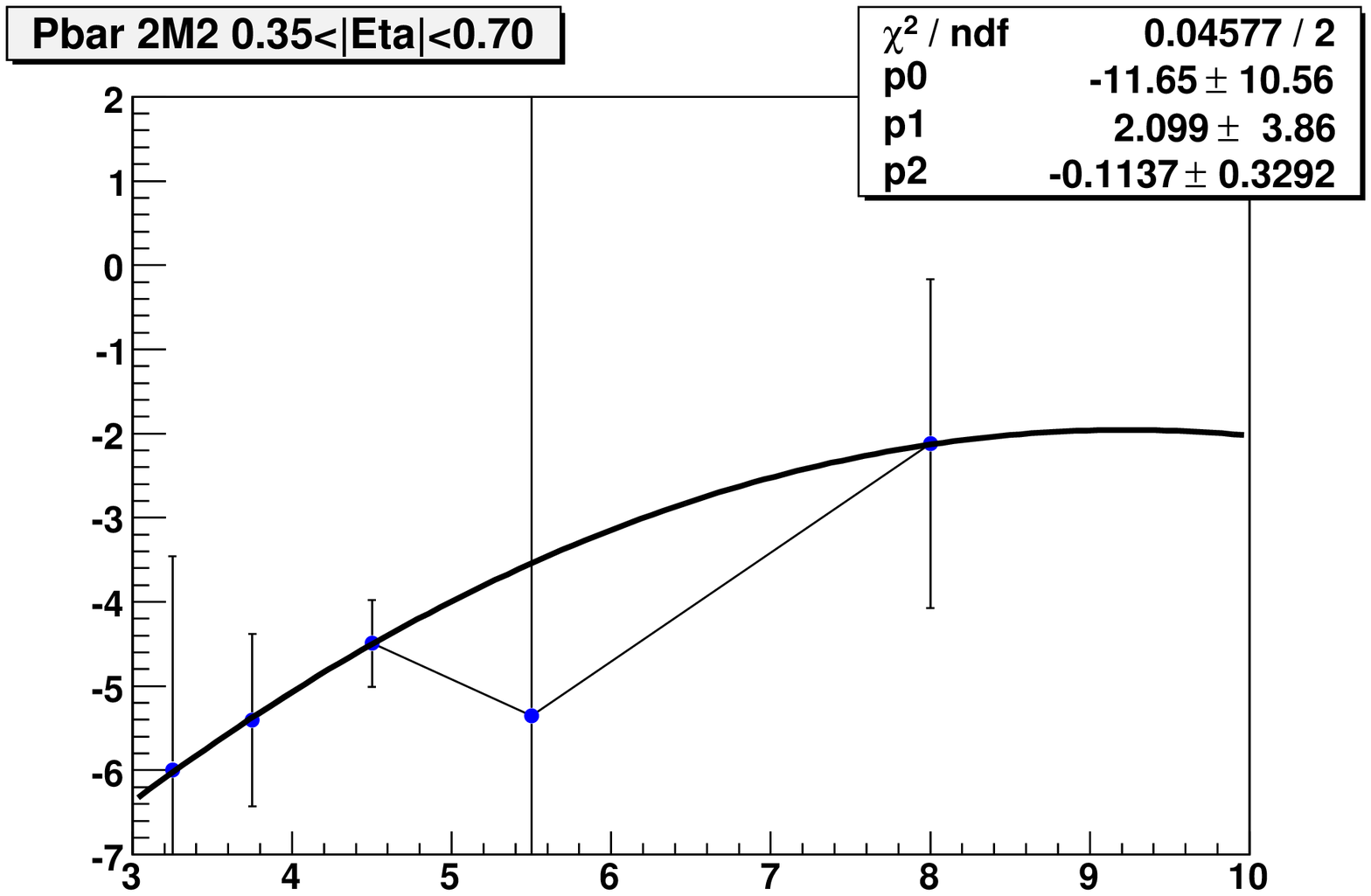}
			\end{minipage}	
					
	\caption{Same as Fig. 6.8 but for d+Au.  Rows correspond to the centrality bins 40-100\%, 20-40\%, and 0-20\% from top to bottom.}
	\label{fig:dAufitcutsp}	
\end{figure}

The $N\sigma_{\pi}$ cuts on the trigger particles are fit as a function of trigger particle $p_T$ to reduce the error on the fit.  Figures~\ref{fig:fitcutspi}-\ref{fig:dAufitcutsp} show these fits.  Although we fit the $N\sigma_{\pi}$ out to higher $p_T$ only trigger particles of $3<p_T<4$ GeV/c were used, due to limited statistics at high $p_T$.  The cuts are fit to a second order polynomial.

\begin{figure}[H]
\hfill
\begin{minipage}[t]{.24\textwidth}
	\centering
		\includegraphics[width=1\textwidth]{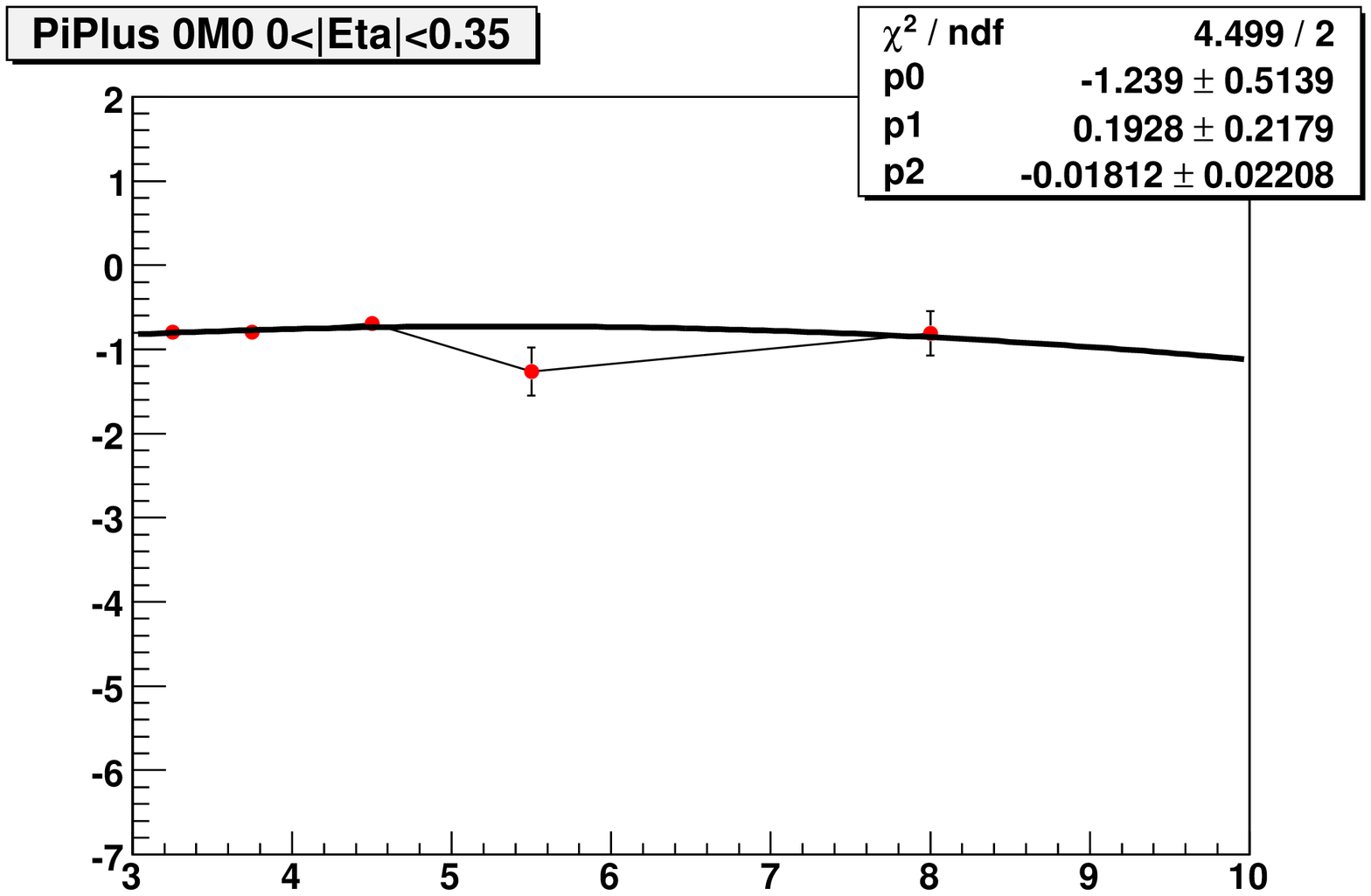}
		\includegraphics[width=1\textwidth]{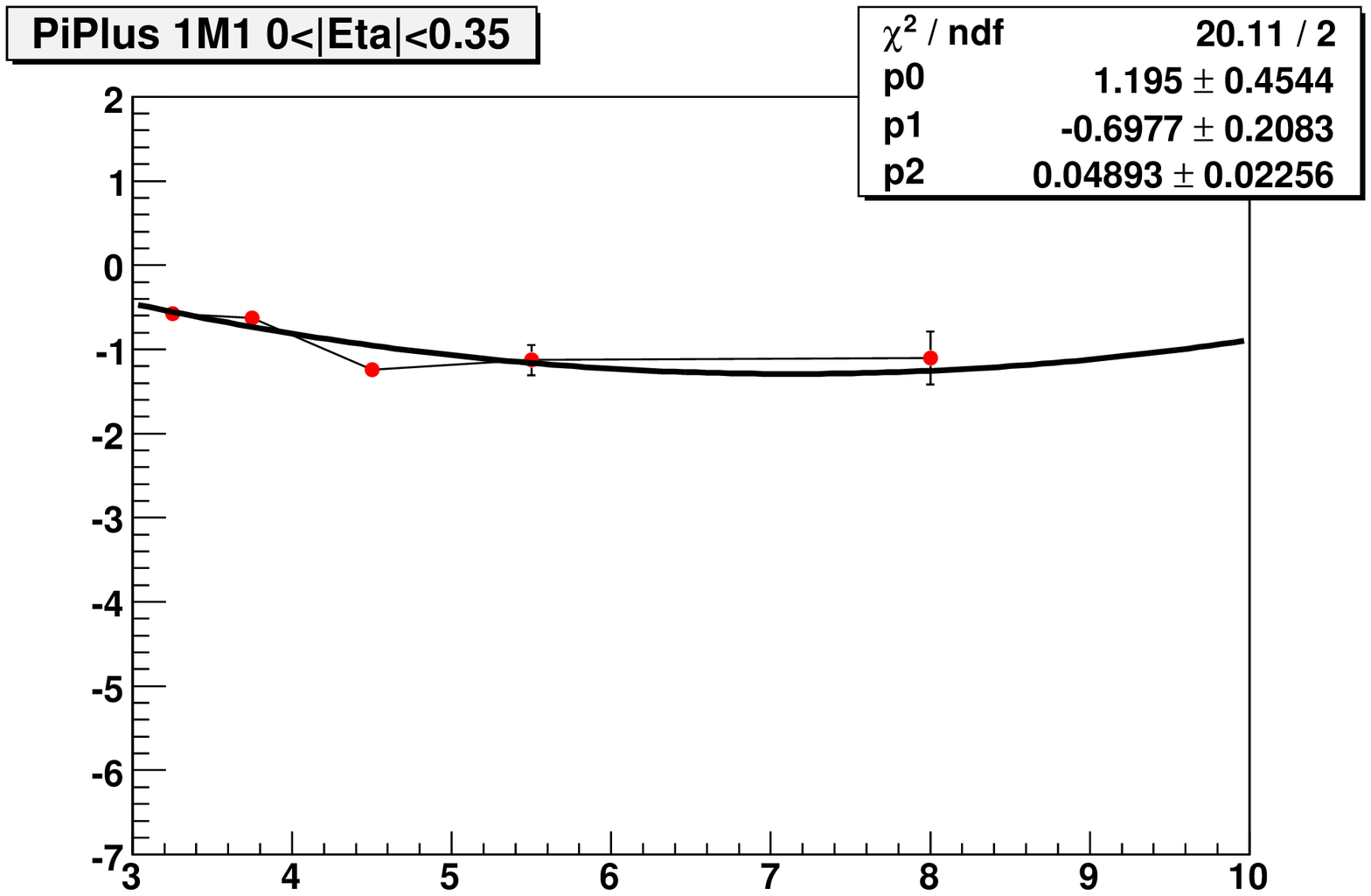}
		\includegraphics[width=1\textwidth]{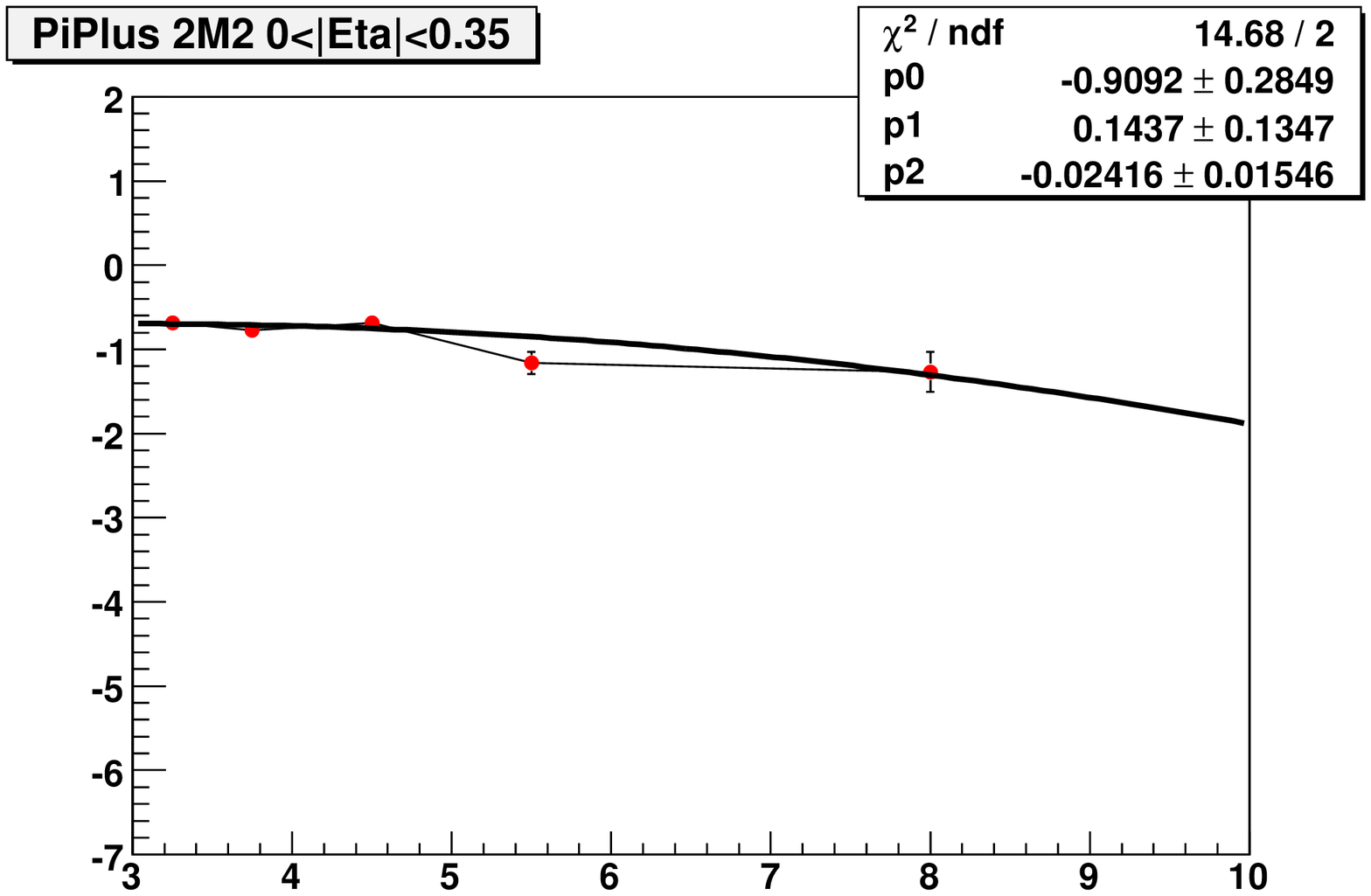}						
			\end{minipage}
\hfill
\begin{minipage}[t]{.24\textwidth}
	\centering
		\includegraphics[width=1\textwidth]{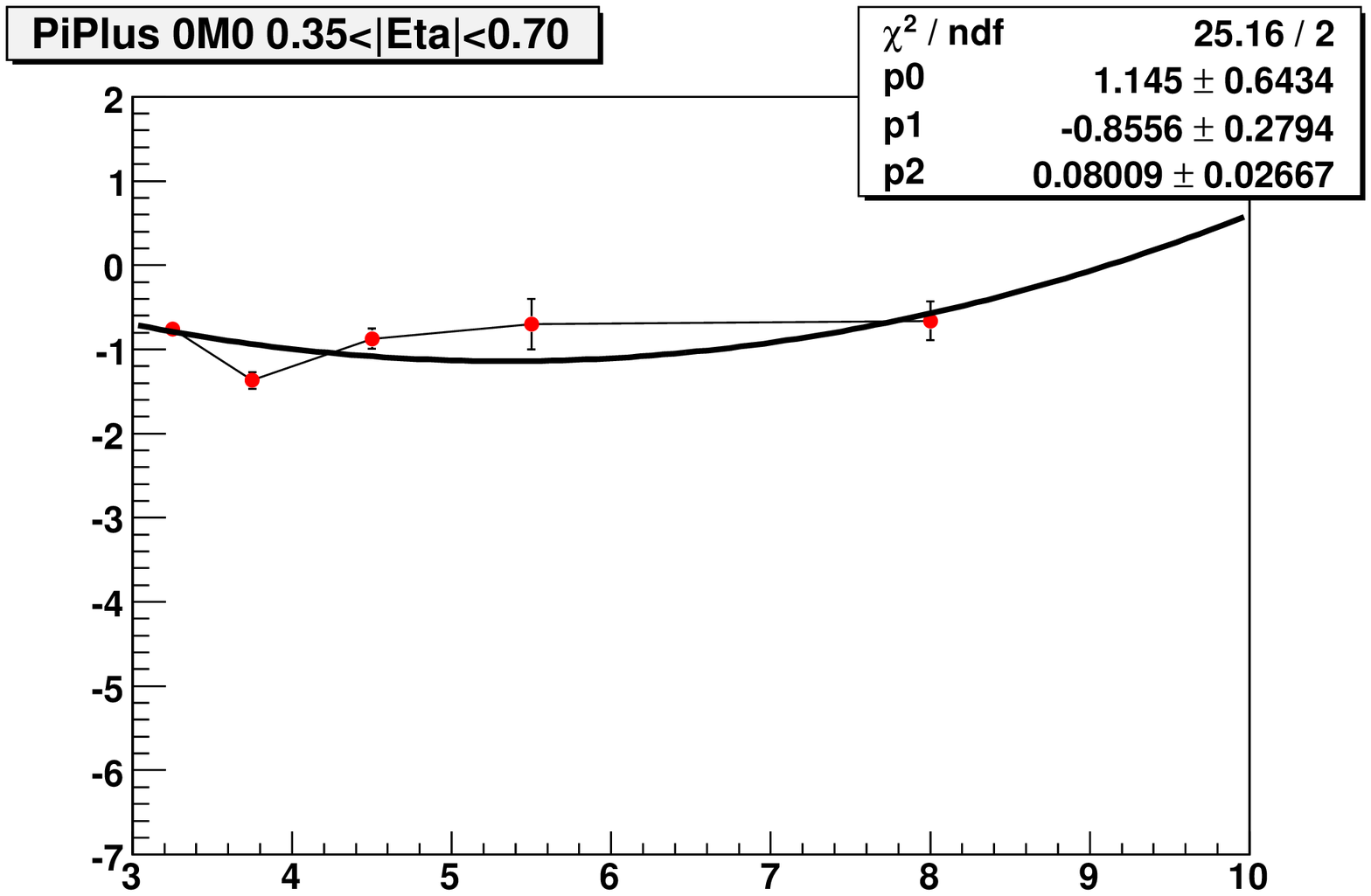}
		\includegraphics[width=1\textwidth]{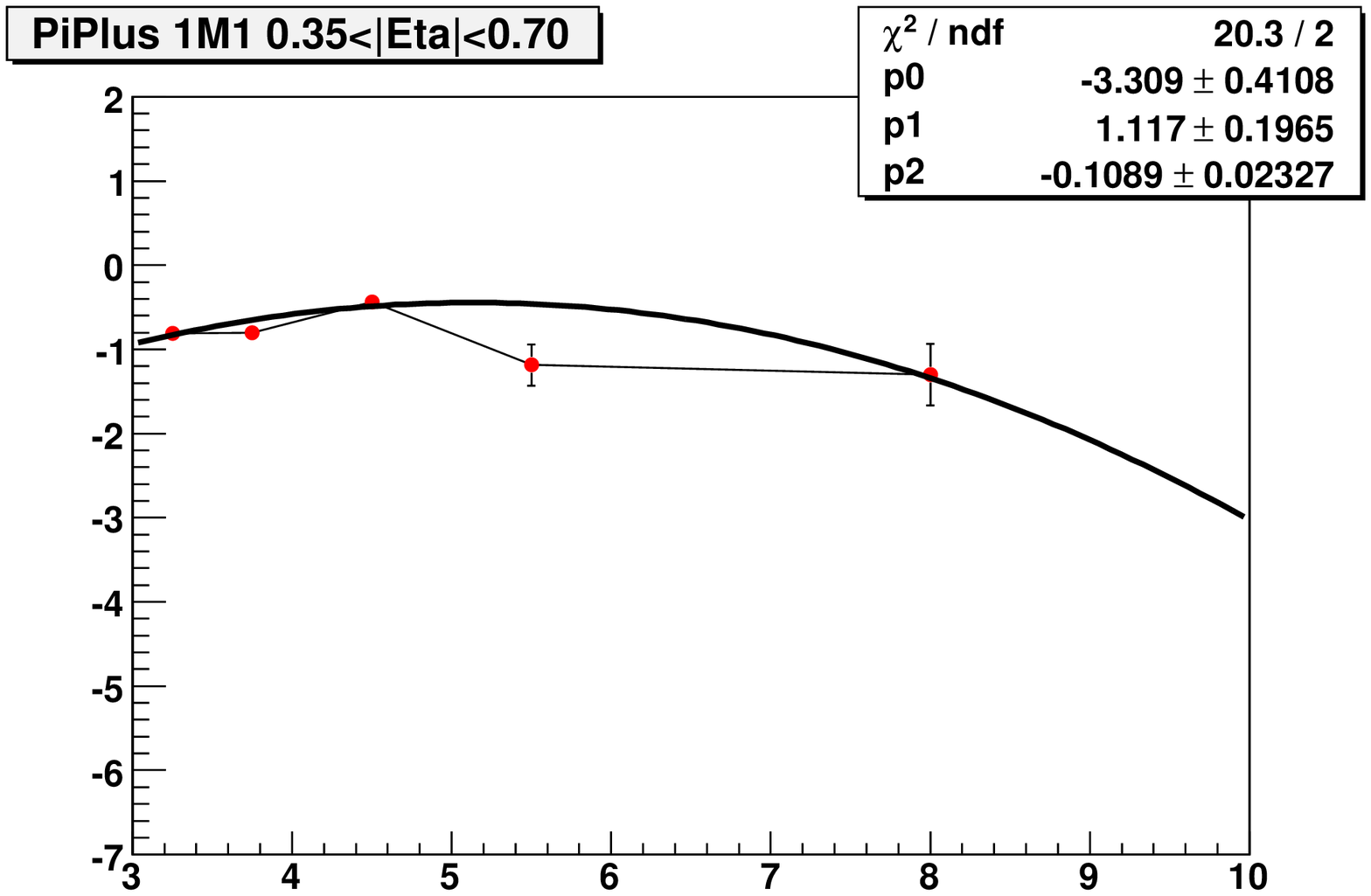}
		\includegraphics[width=1\textwidth]{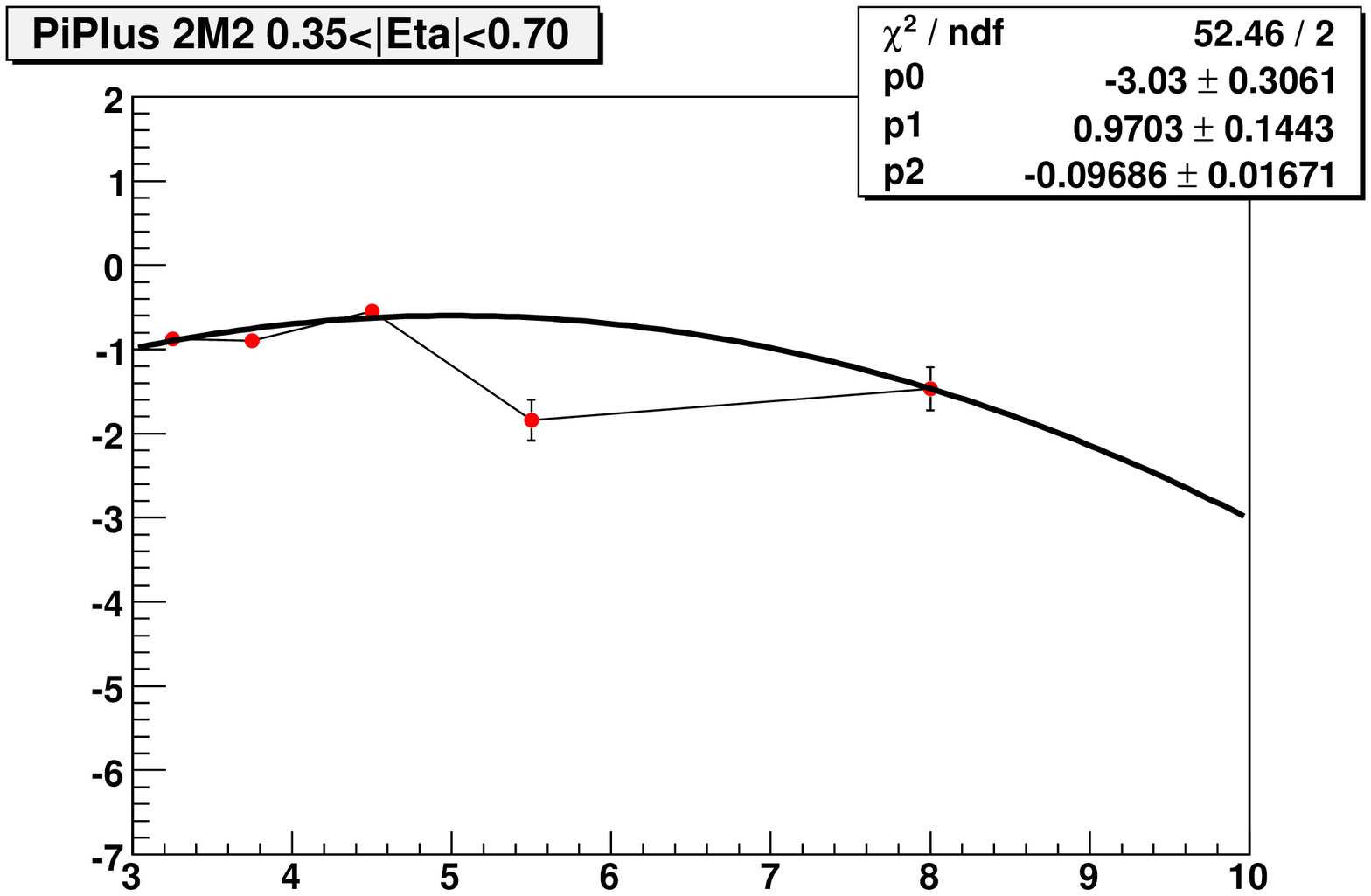}							
			\end{minipage}
\hfill
\begin{minipage}[t]{.24\textwidth}
	\centering
		\includegraphics[width=1\textwidth]{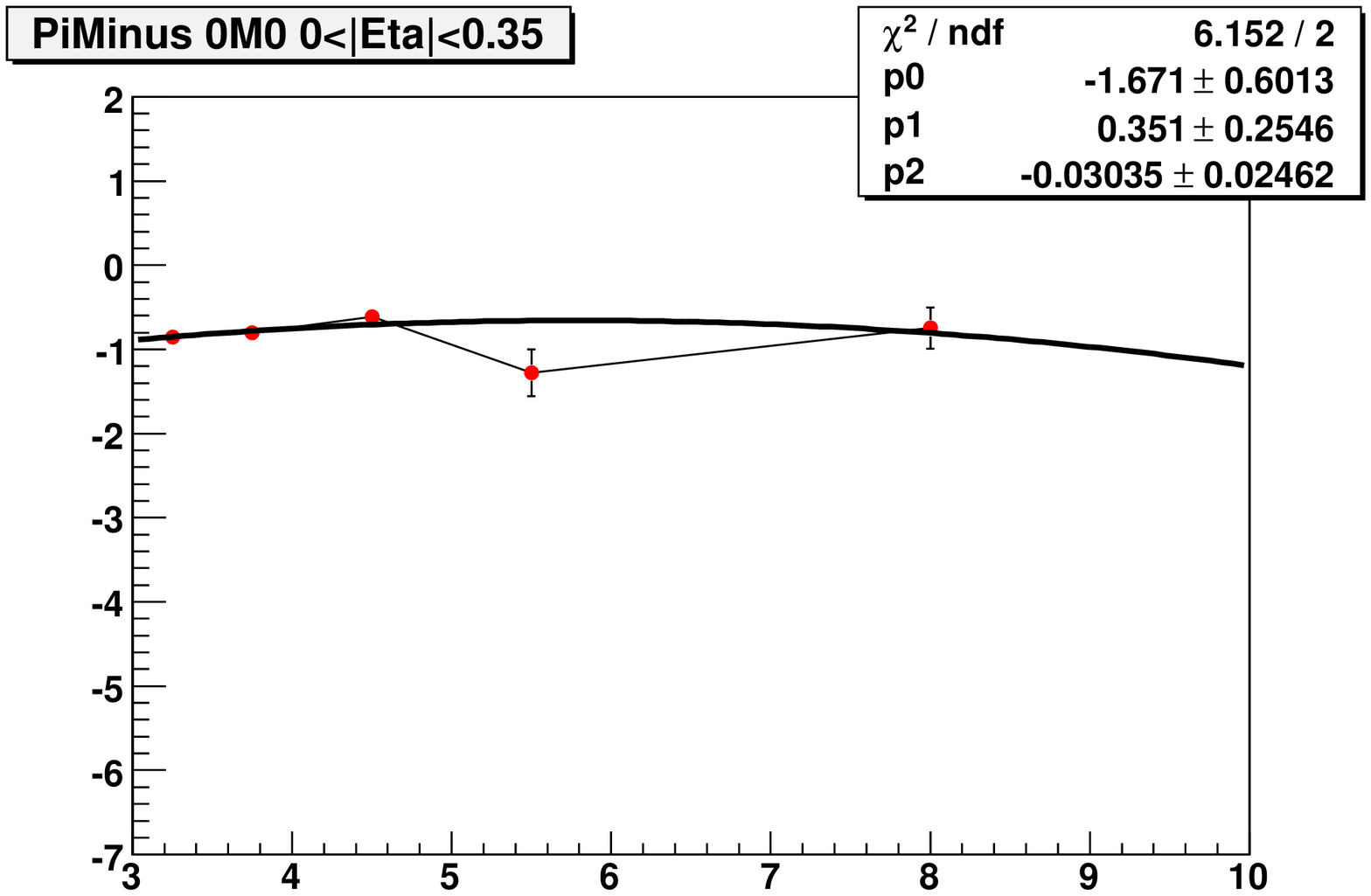}
		\includegraphics[width=1\textwidth]{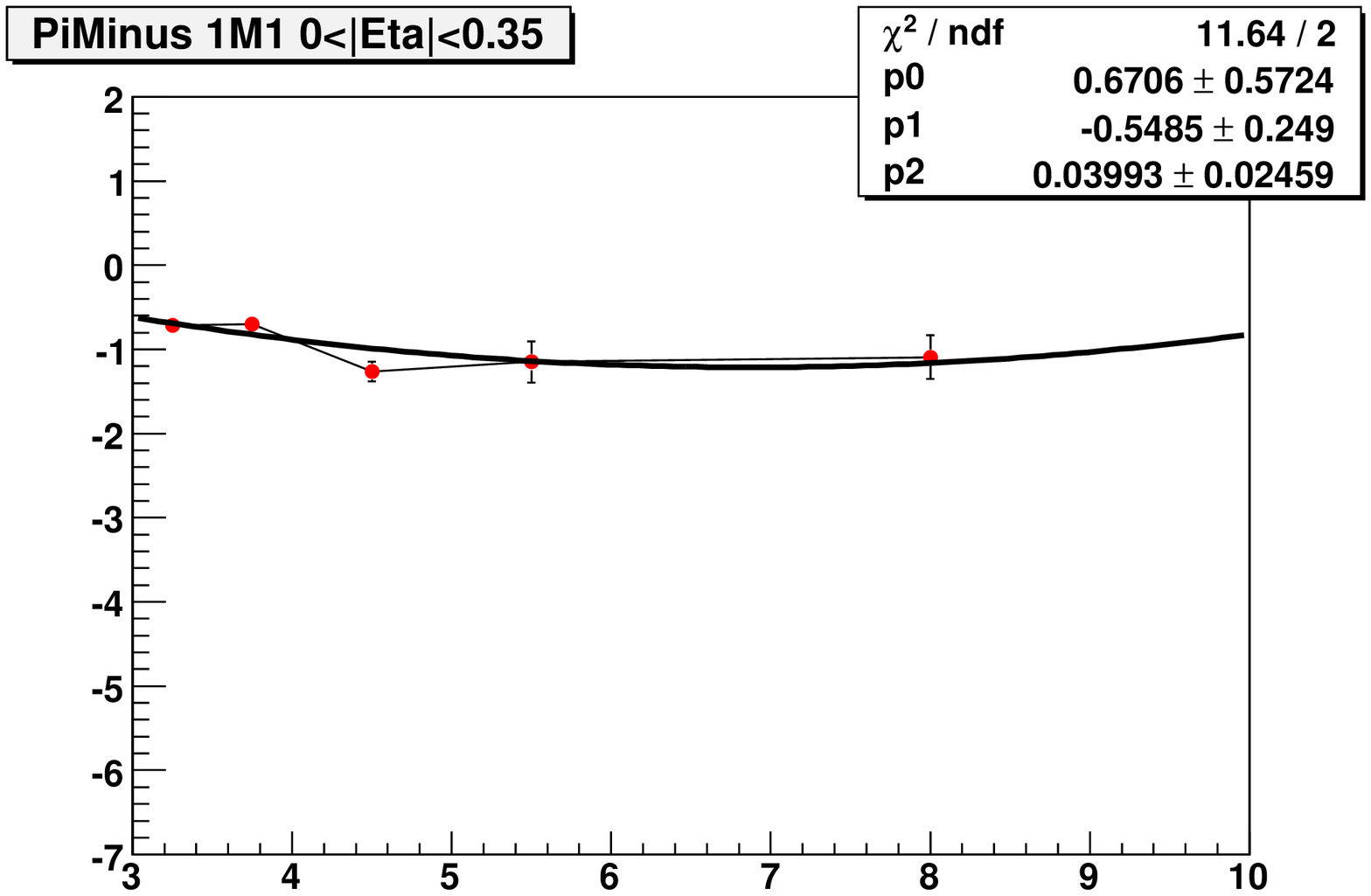}
		\includegraphics[width=1\textwidth]{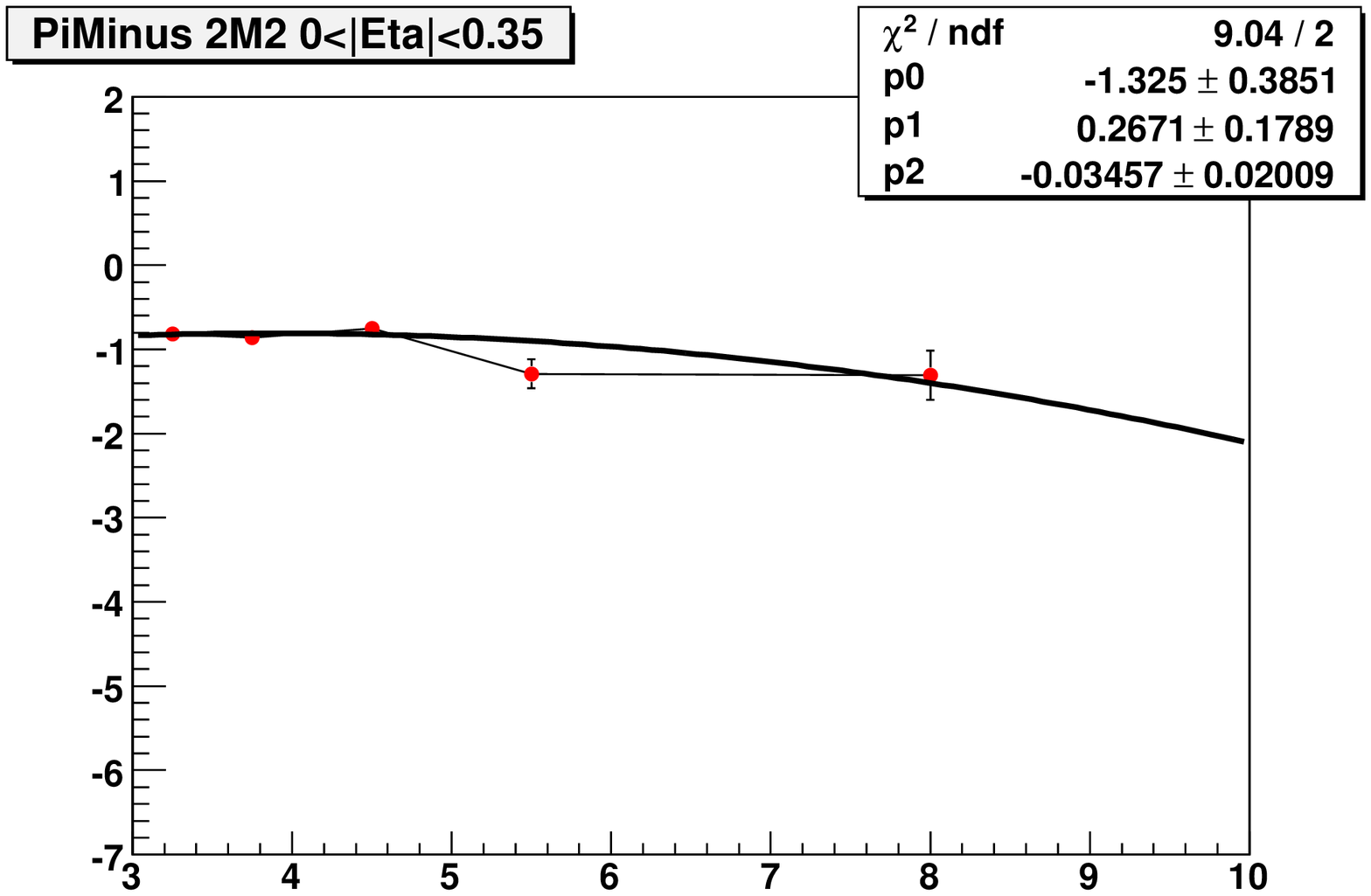}										
			\end{minipage}
\hfill
\begin{minipage}[t]{.24\textwidth}
	\centering
		\includegraphics[width=1\textwidth]{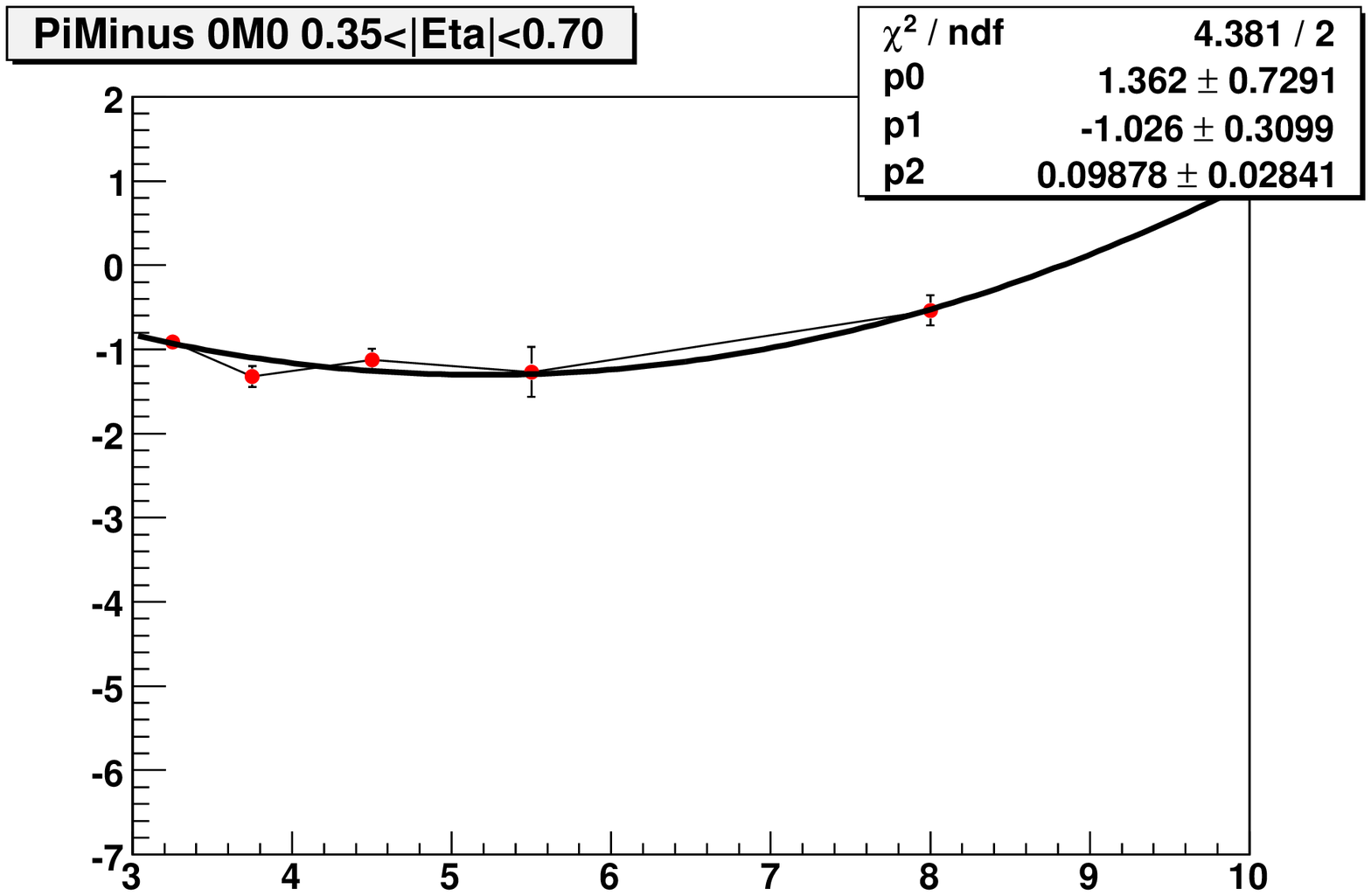}
		\includegraphics[width=1\textwidth]{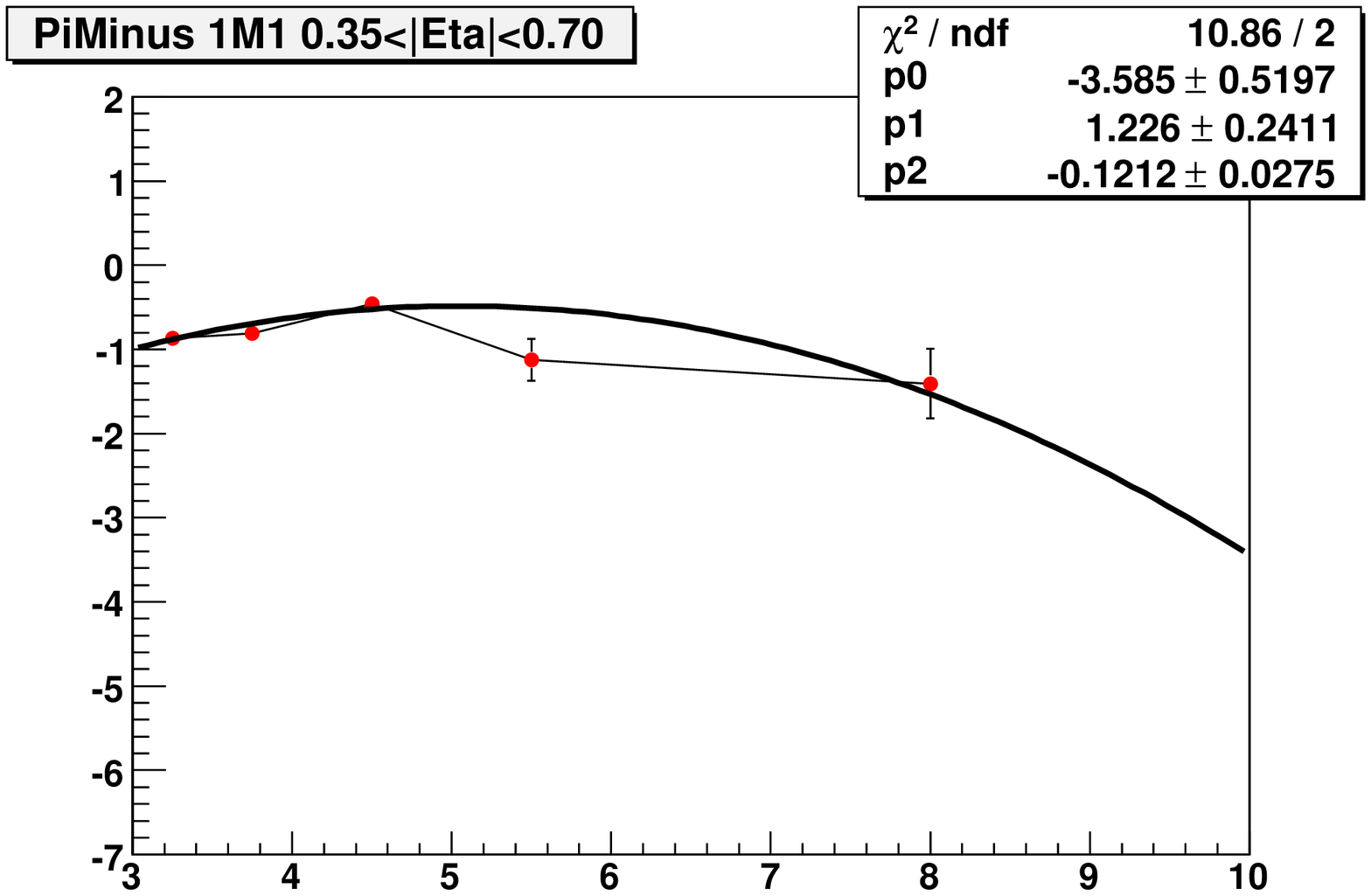}
		\includegraphics[width=1\textwidth]{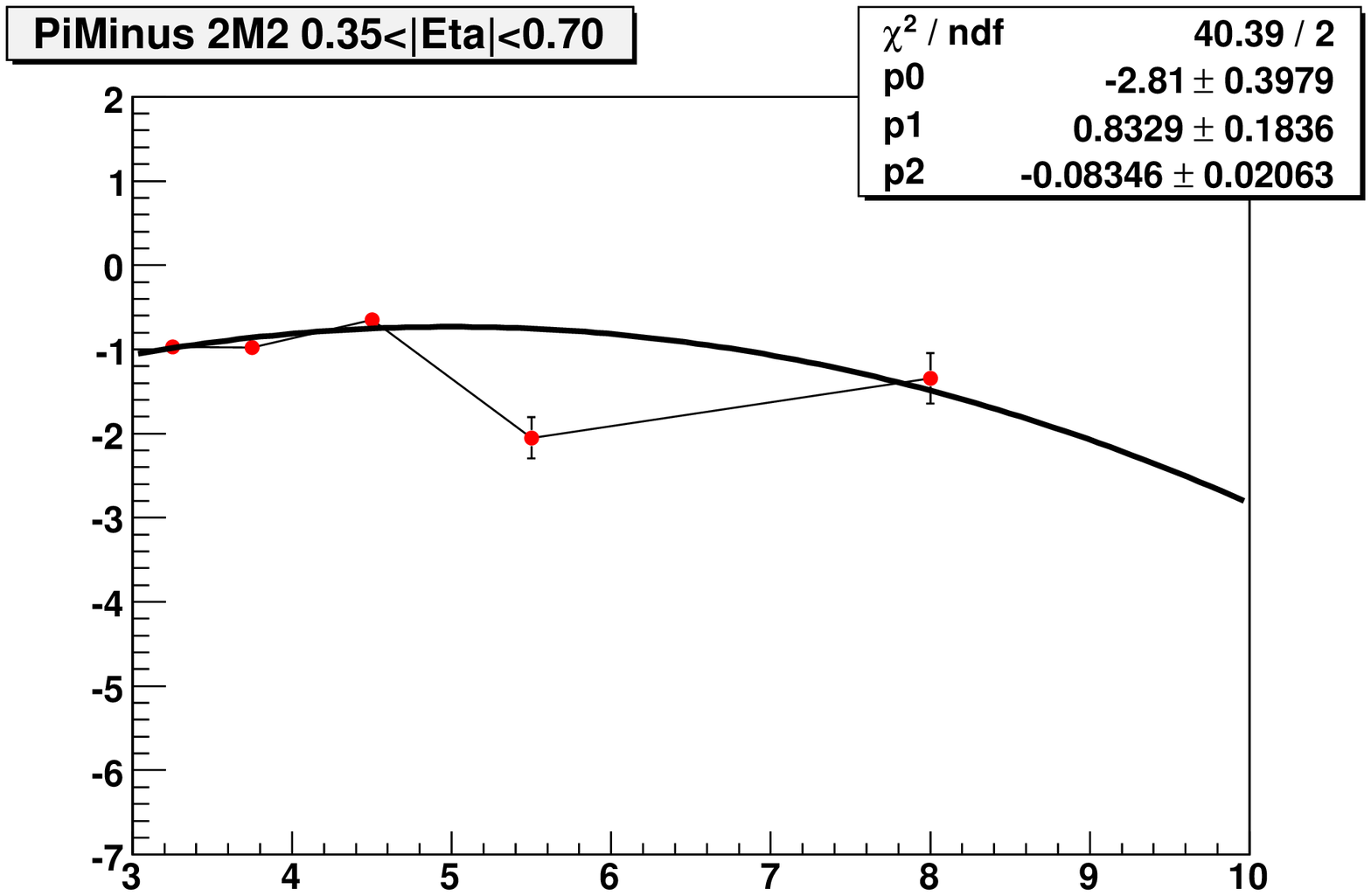}
			\end{minipage}	
					
	\caption{Same as Fig. 6.9 but for d+Au.  Rows correspond to the centrality bins 40-100\%, 20-40\% and 0-20\% from top to bottom.}
	\label{fig:dAufitcutspi}	
\end{figure}

\begin{figure}[H]
\hfill
\begin{minipage}[t]{0.32\textwidth}
\centering
\includegraphics[width=1.0\textwidth]{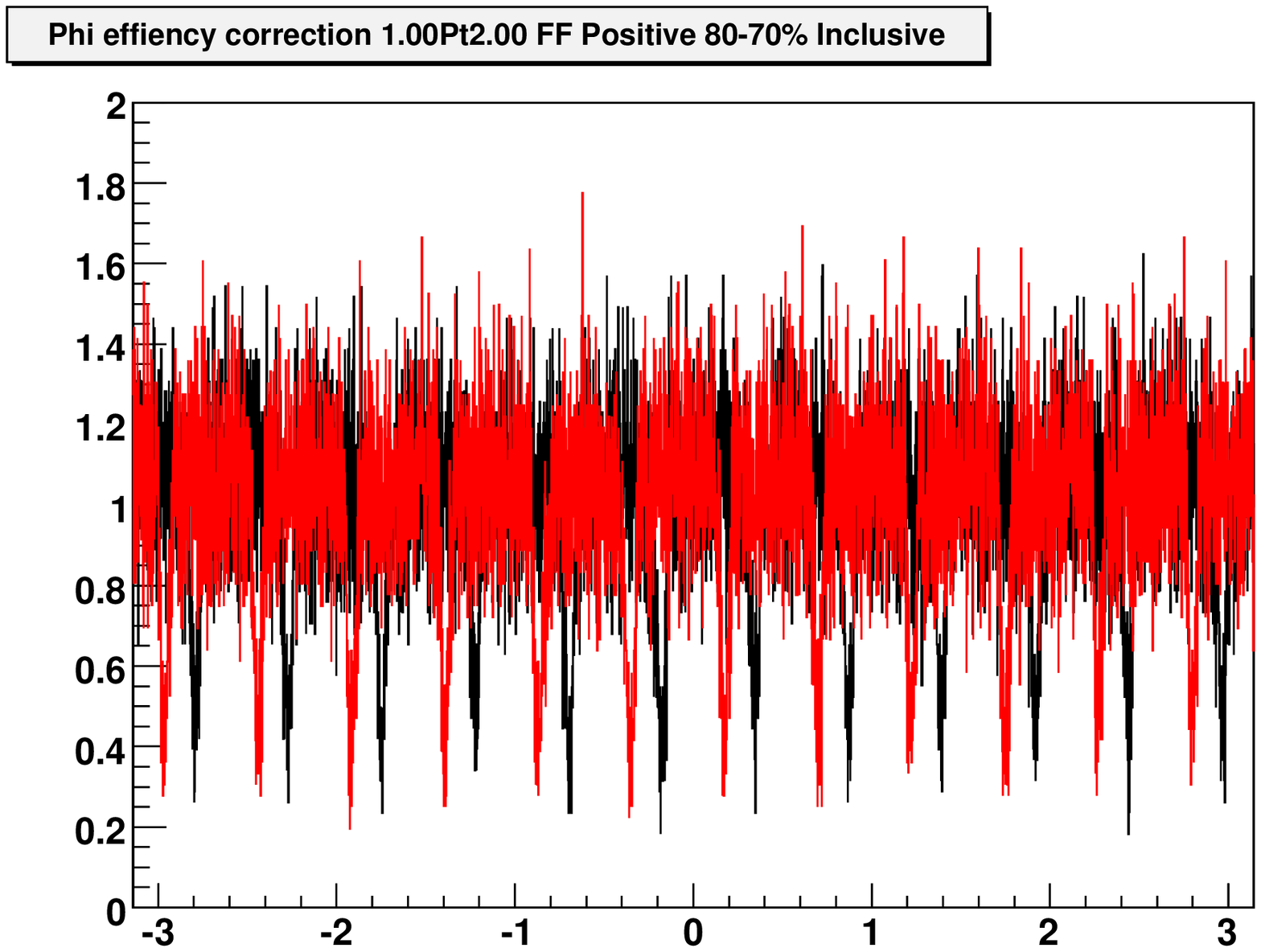}
\includegraphics[width=1.0\textwidth]{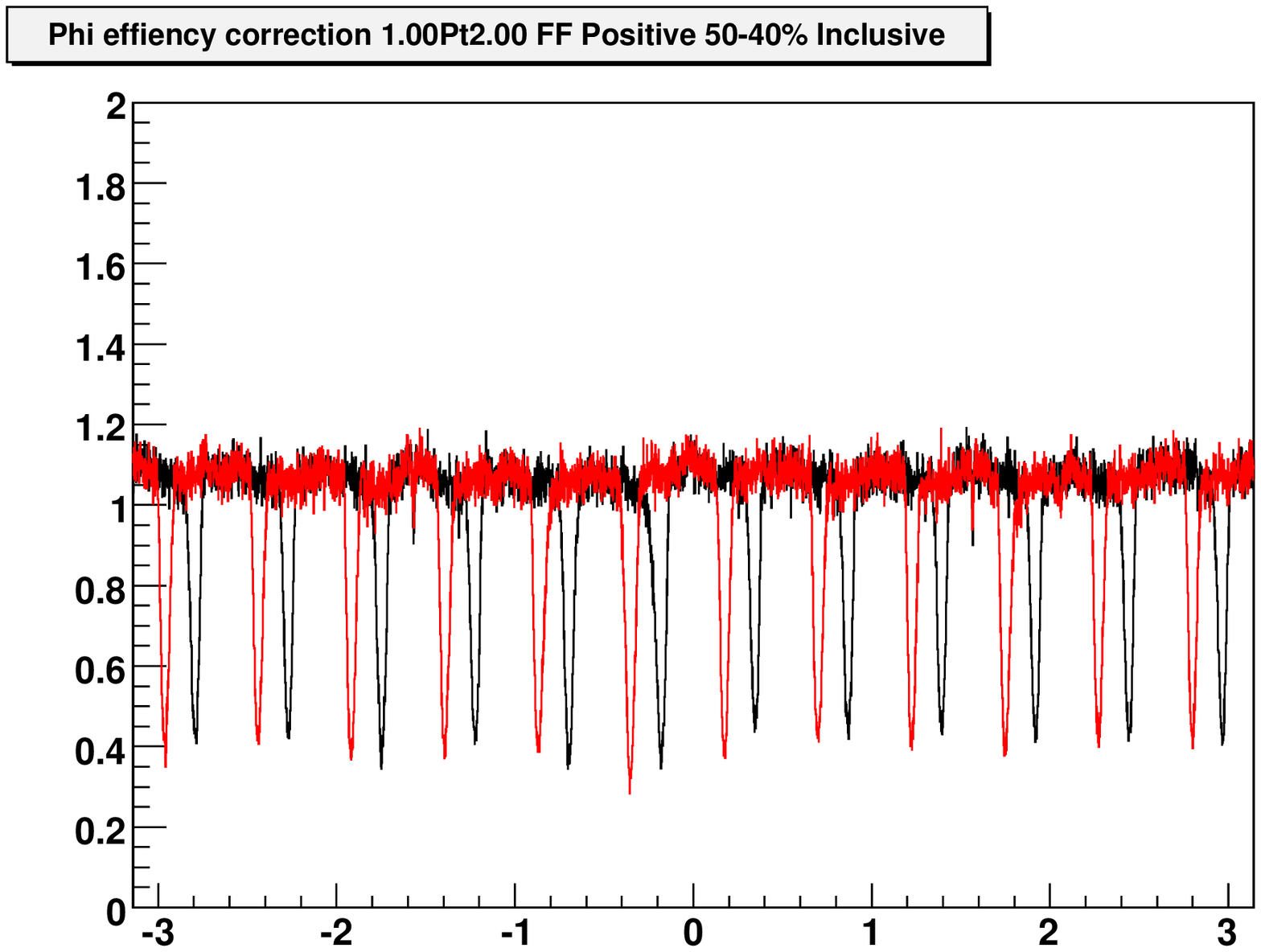}
\includegraphics[width=1.0\textwidth]{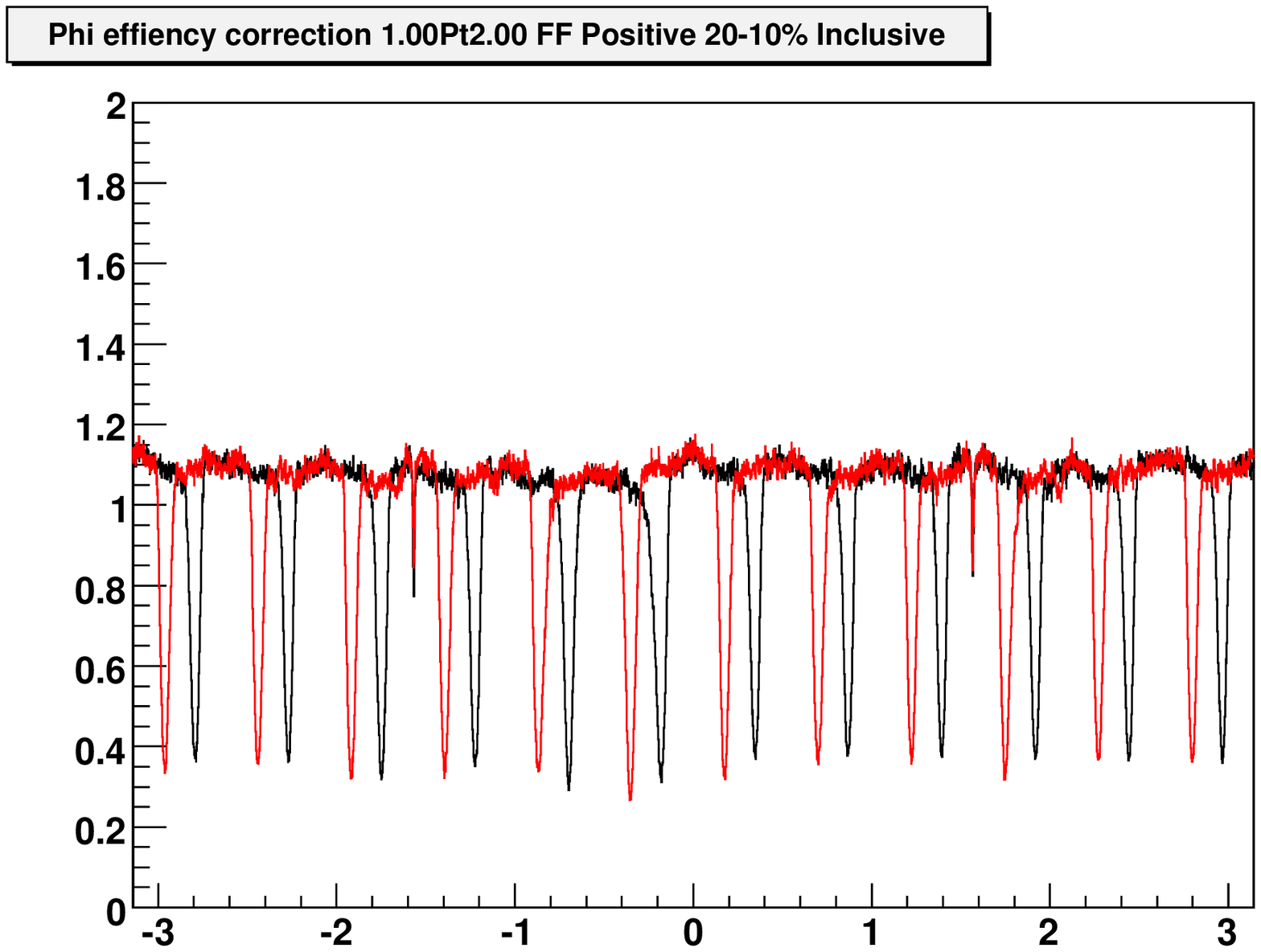}
\end{minipage}
\hfill
\begin{minipage}[t]{0.32\textwidth}
\centering
\includegraphics[width=1.0\textwidth]{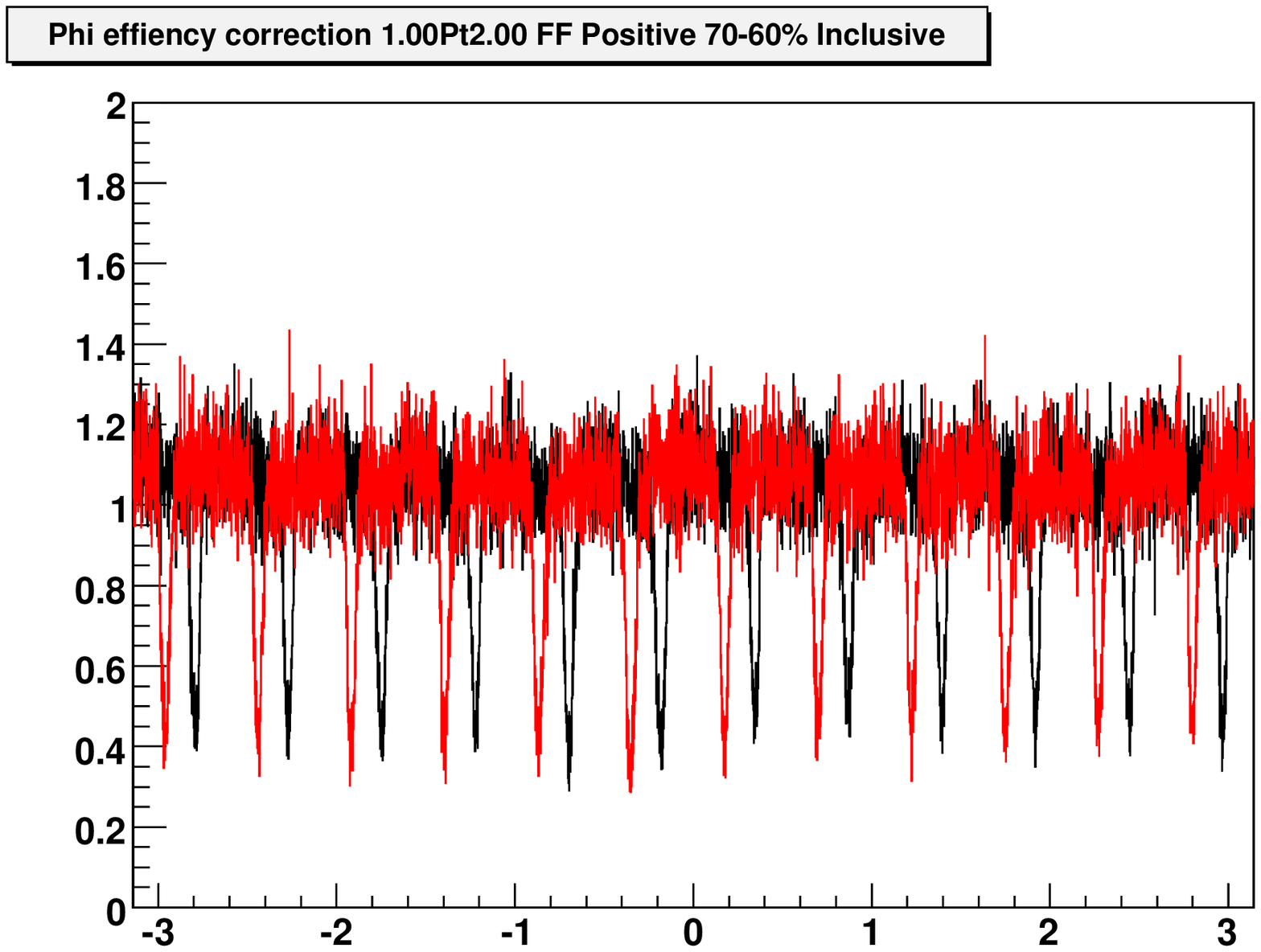}
\includegraphics[width=1.0\textwidth]{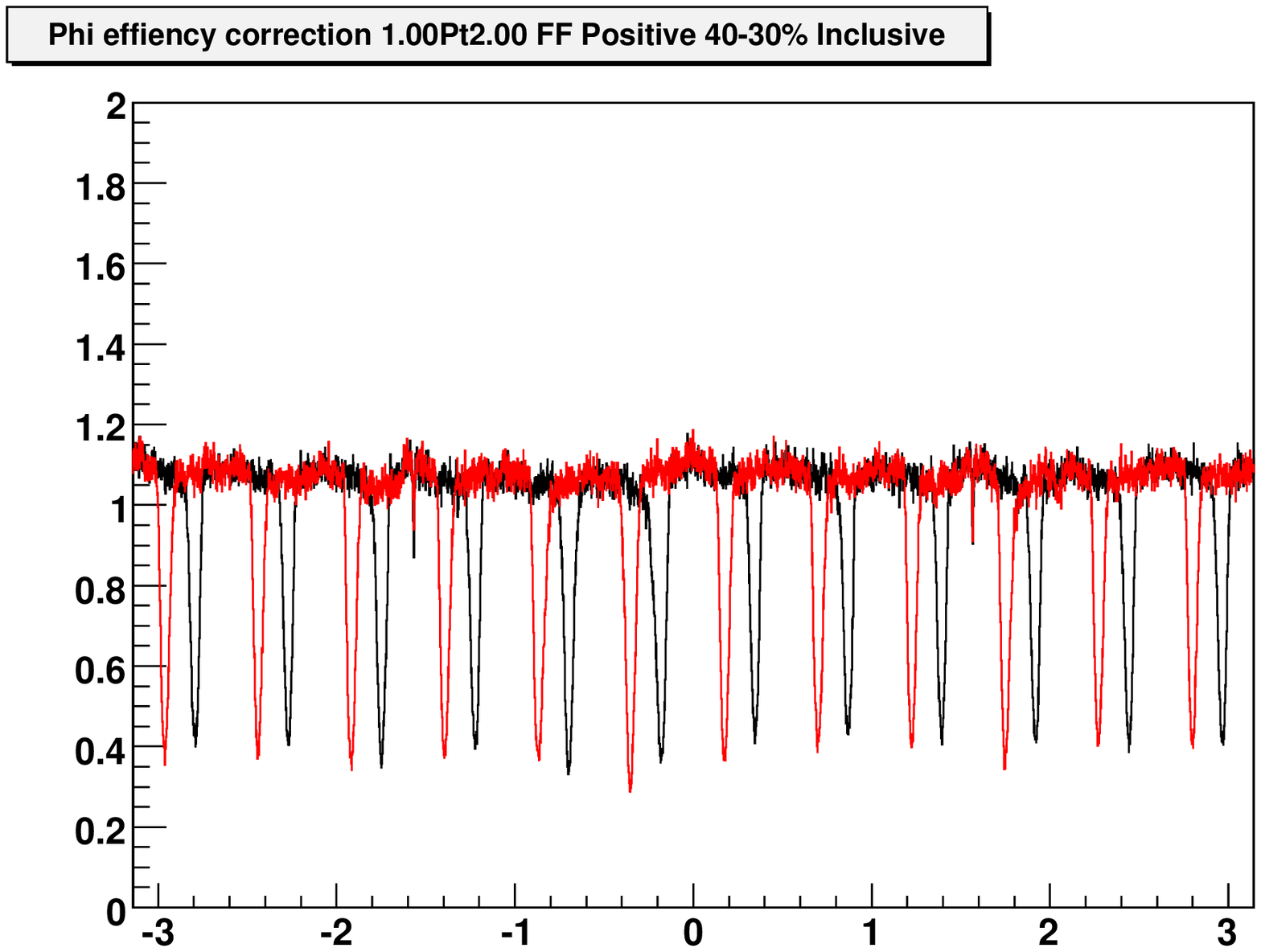}
\includegraphics[width=1.0\textwidth]{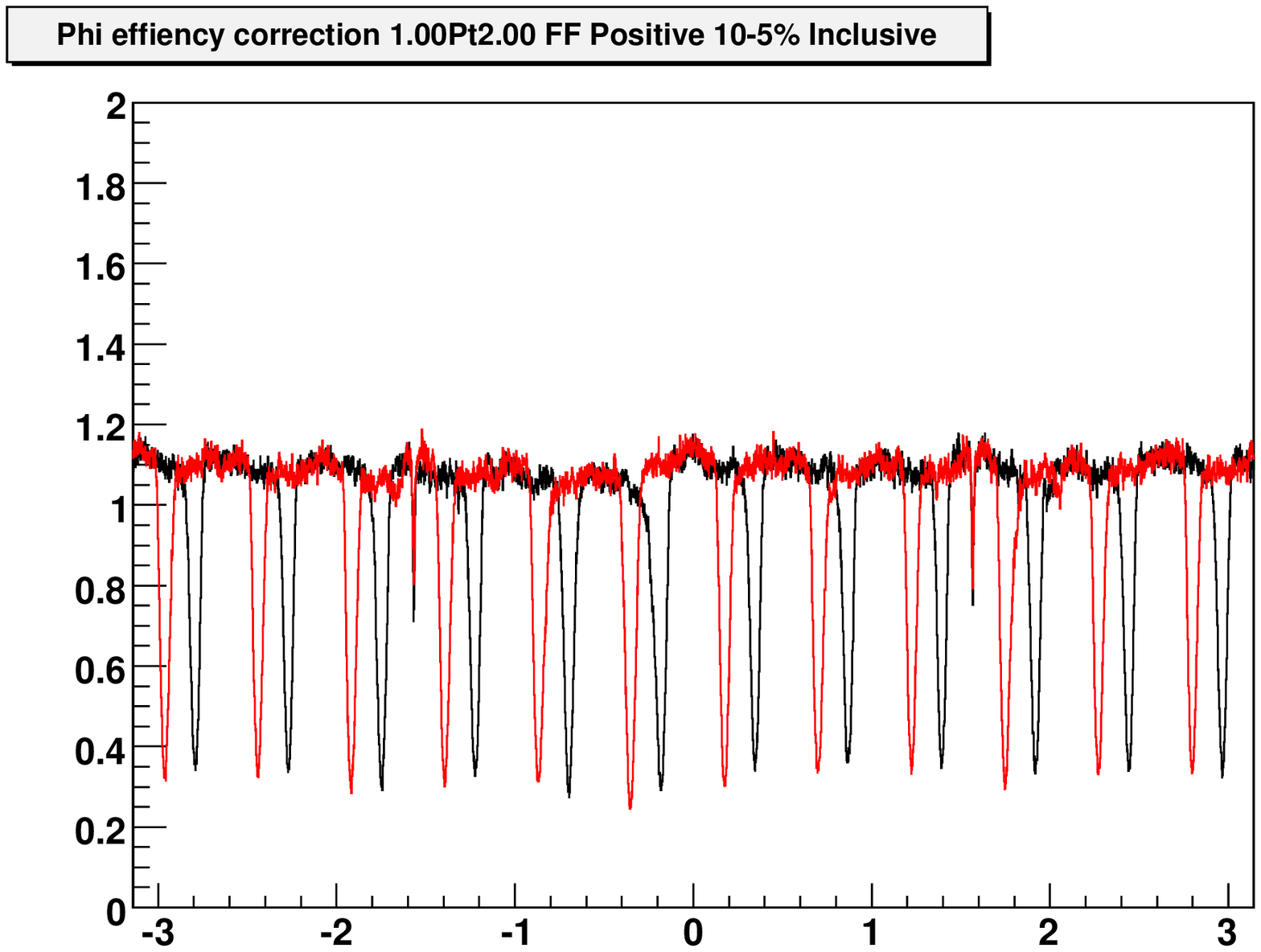}
\end{minipage}
\hfill
\begin{minipage}[t]{0.32\textwidth}
\centering
\includegraphics[width=1.0\textwidth]{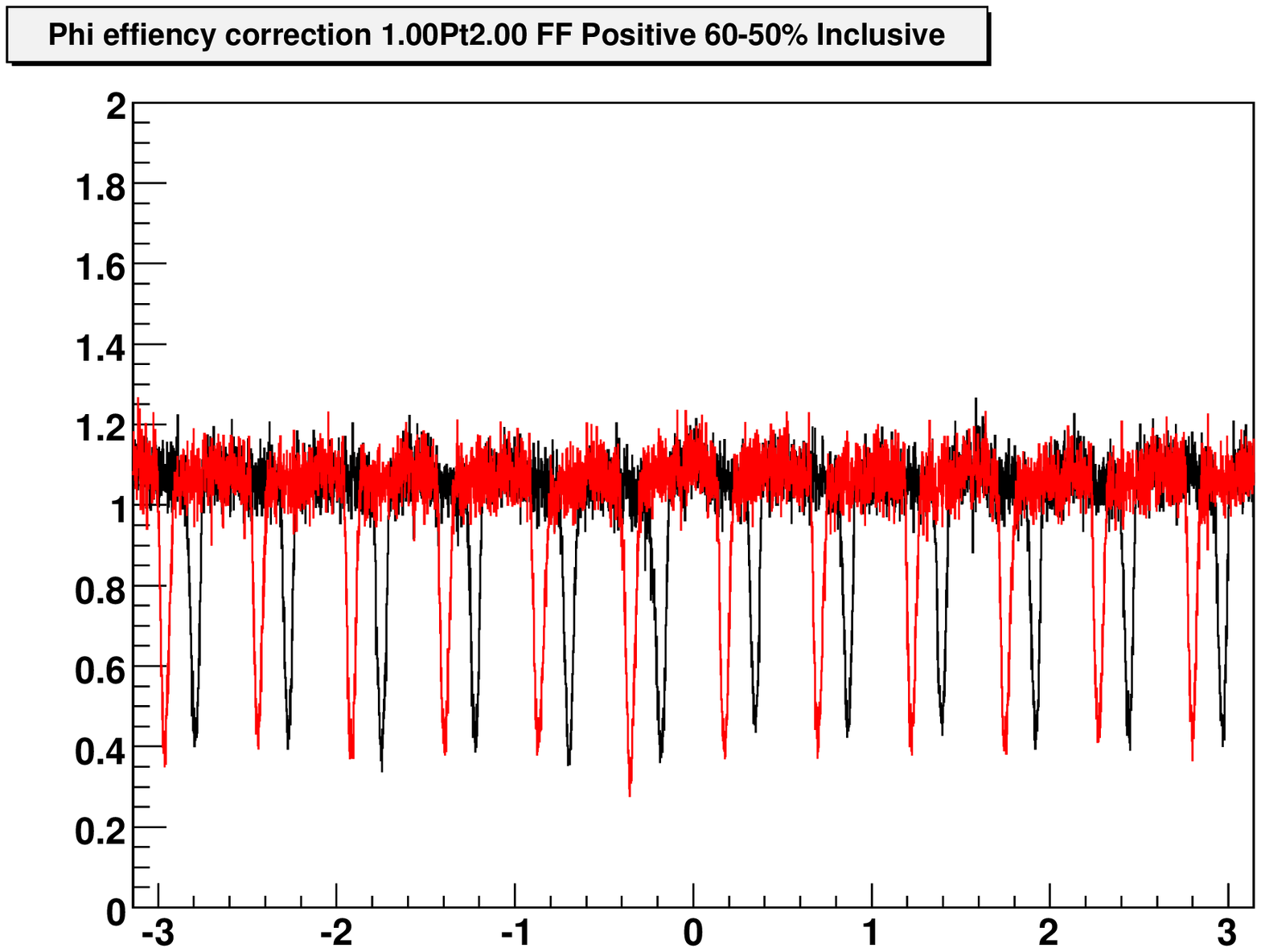}
\includegraphics[width=1.0\textwidth]{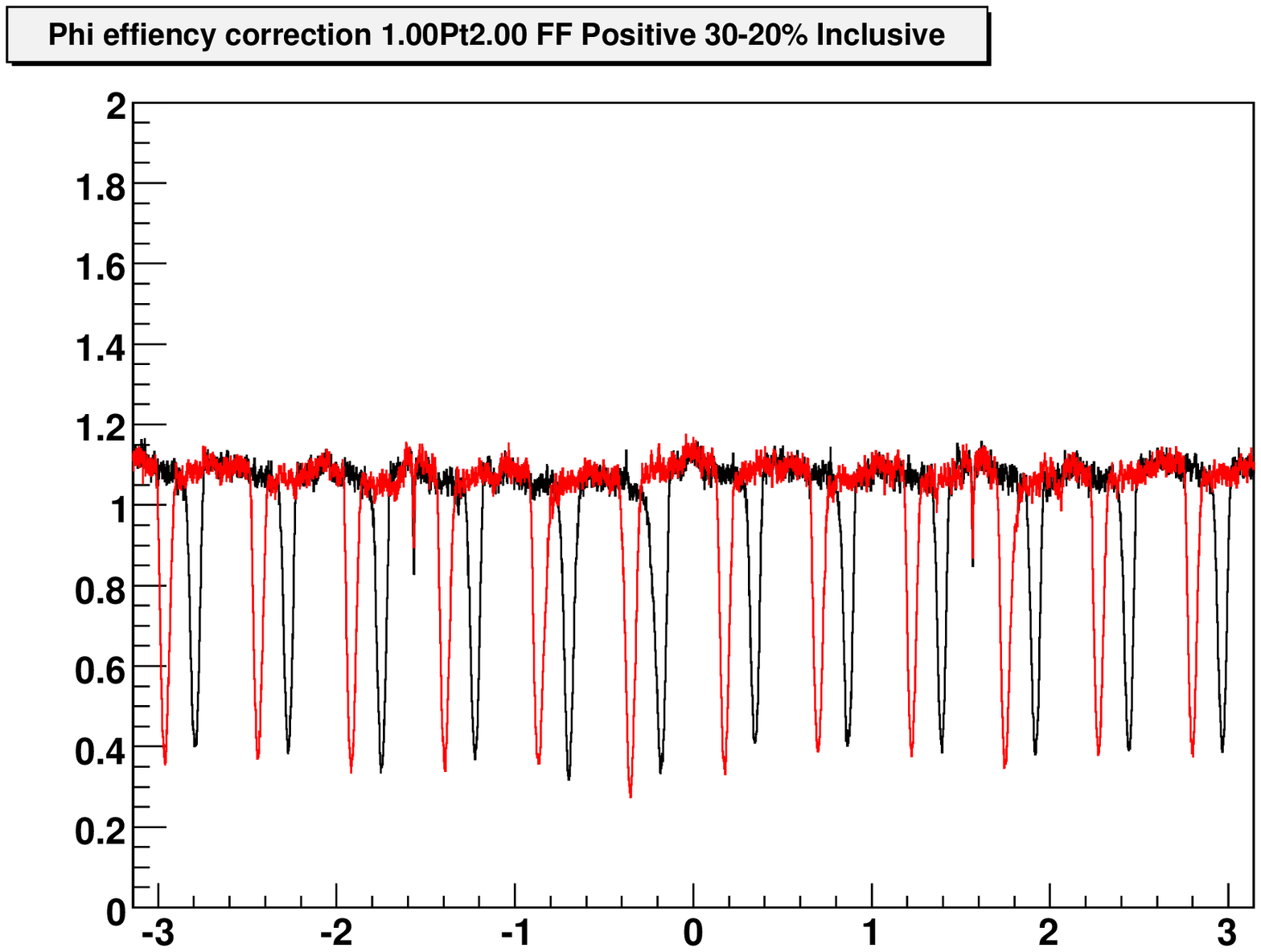}
\includegraphics[width=1.0\textwidth]{Plots/hacc_AuAuMB_Pt0B0M9.eps}
\end{minipage}

\caption{TPC acceptance in $\phi$ for particles of $1<p_T<2$ GeV/c with +0.5 Tesla magnetic field.  Positive particles are shown in black and negative particles are shown in red.  Plots (left to right and top to bottom) are for Au+Au collisions at $\sqrt{s_{NN}}=200$ GeV/c in centralities 70-80\%, 60-70\%, 50-60\%, 40-50\%, 30-40\%, 20-30\%, 10-20\%, 5-10\%, and 0-5\%.}  
\label{fig:accPt0B0}
\end{figure}

\begin{figure}[H]
\hfill
\begin{minipage}[t]{0.32\textwidth}
\centering
\includegraphics[width=1.0\textwidth]{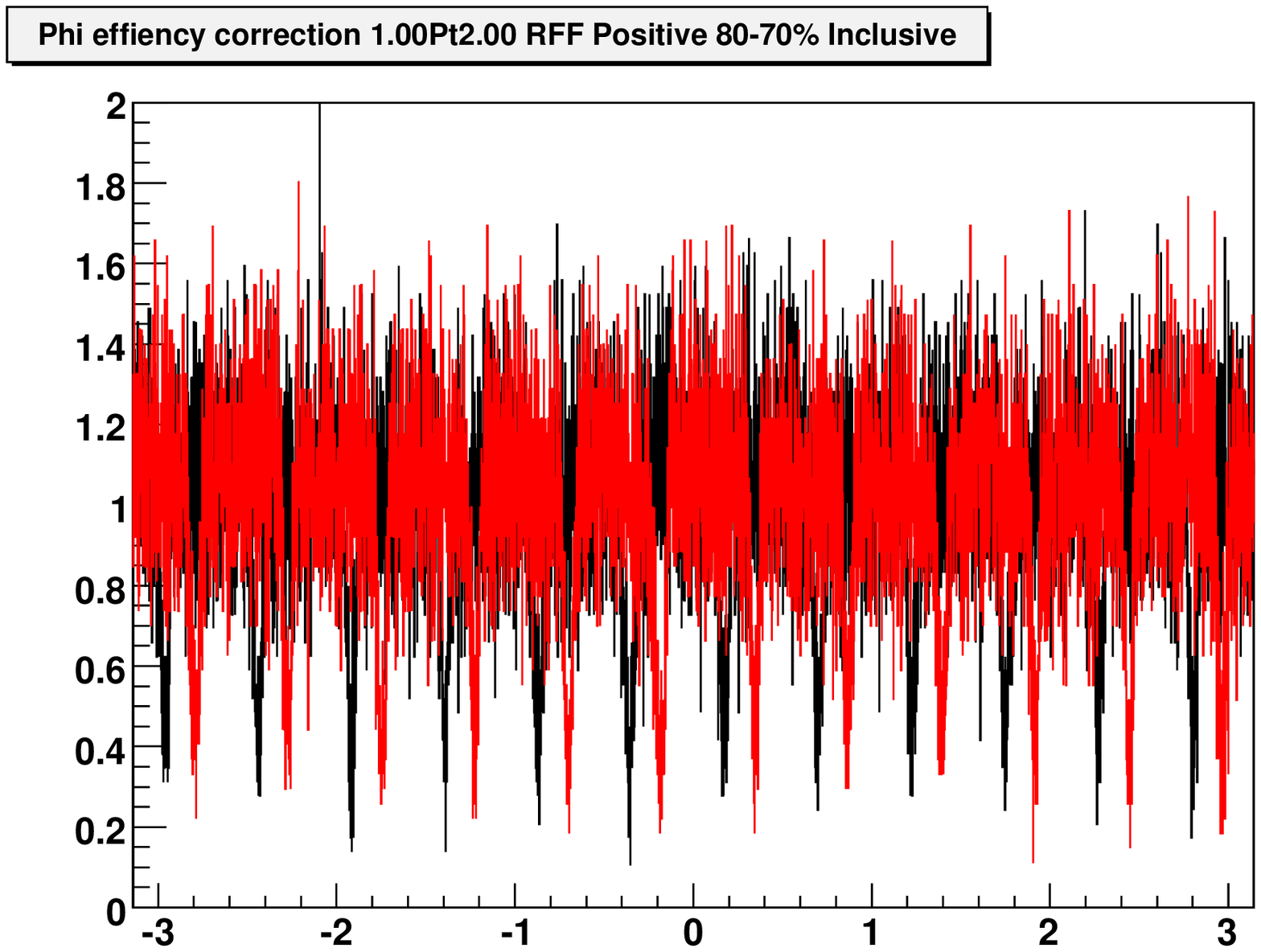}
\includegraphics[width=1.0\textwidth]{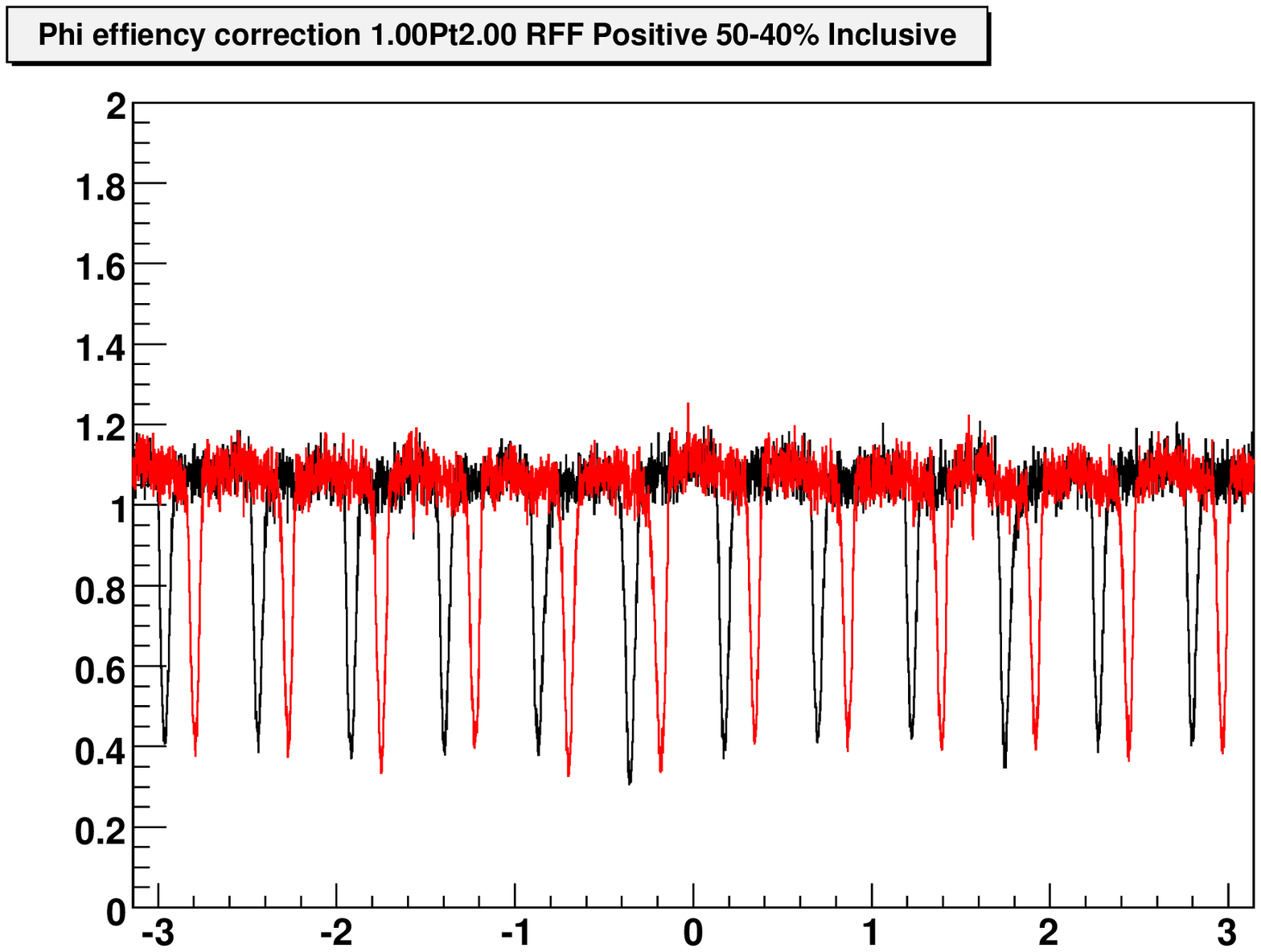}
\includegraphics[width=1.0\textwidth]{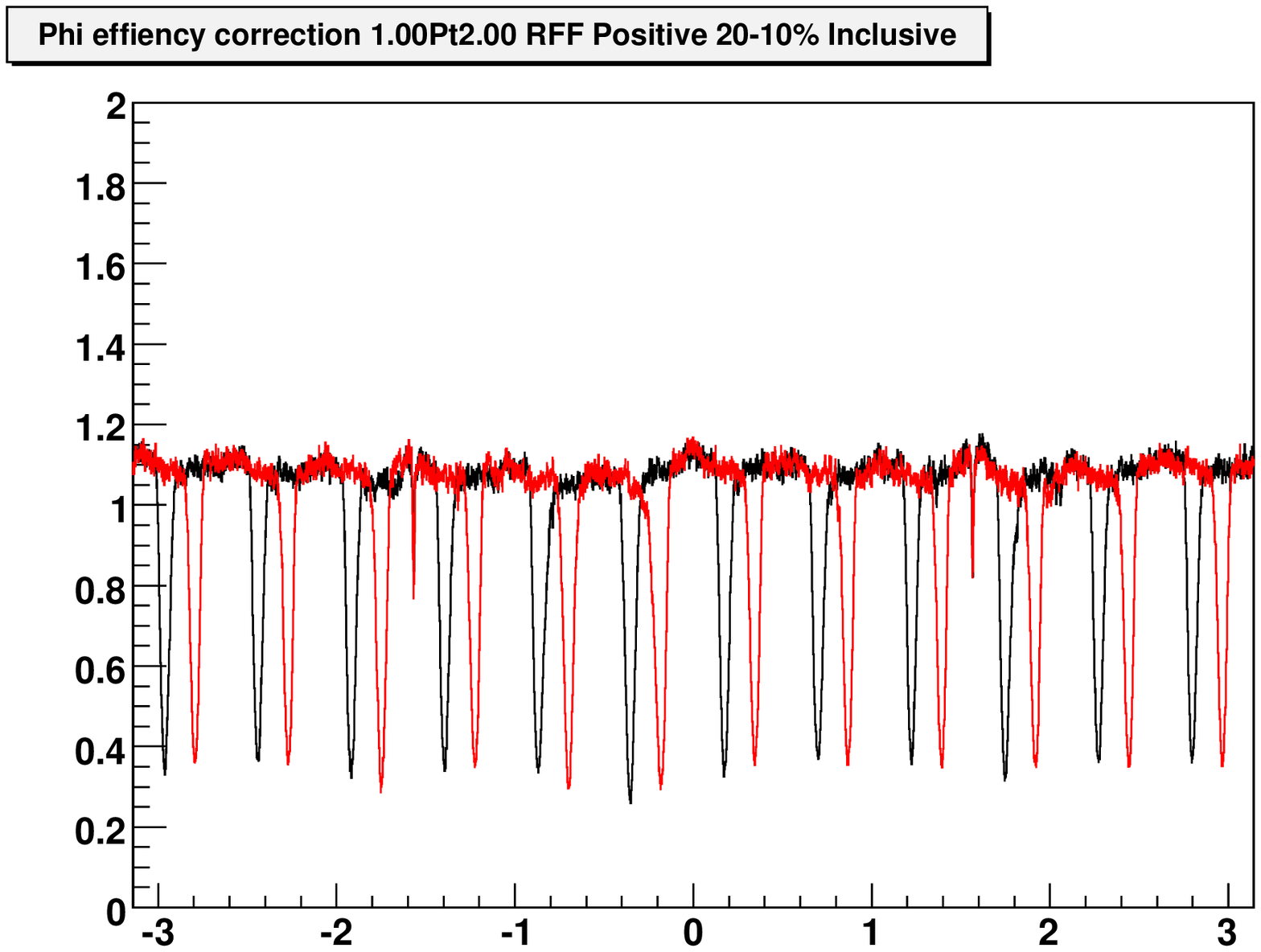}
\end{minipage}
\hfill
\begin{minipage}[t]{0.32\textwidth}
\centering
\includegraphics[width=1.0\textwidth]{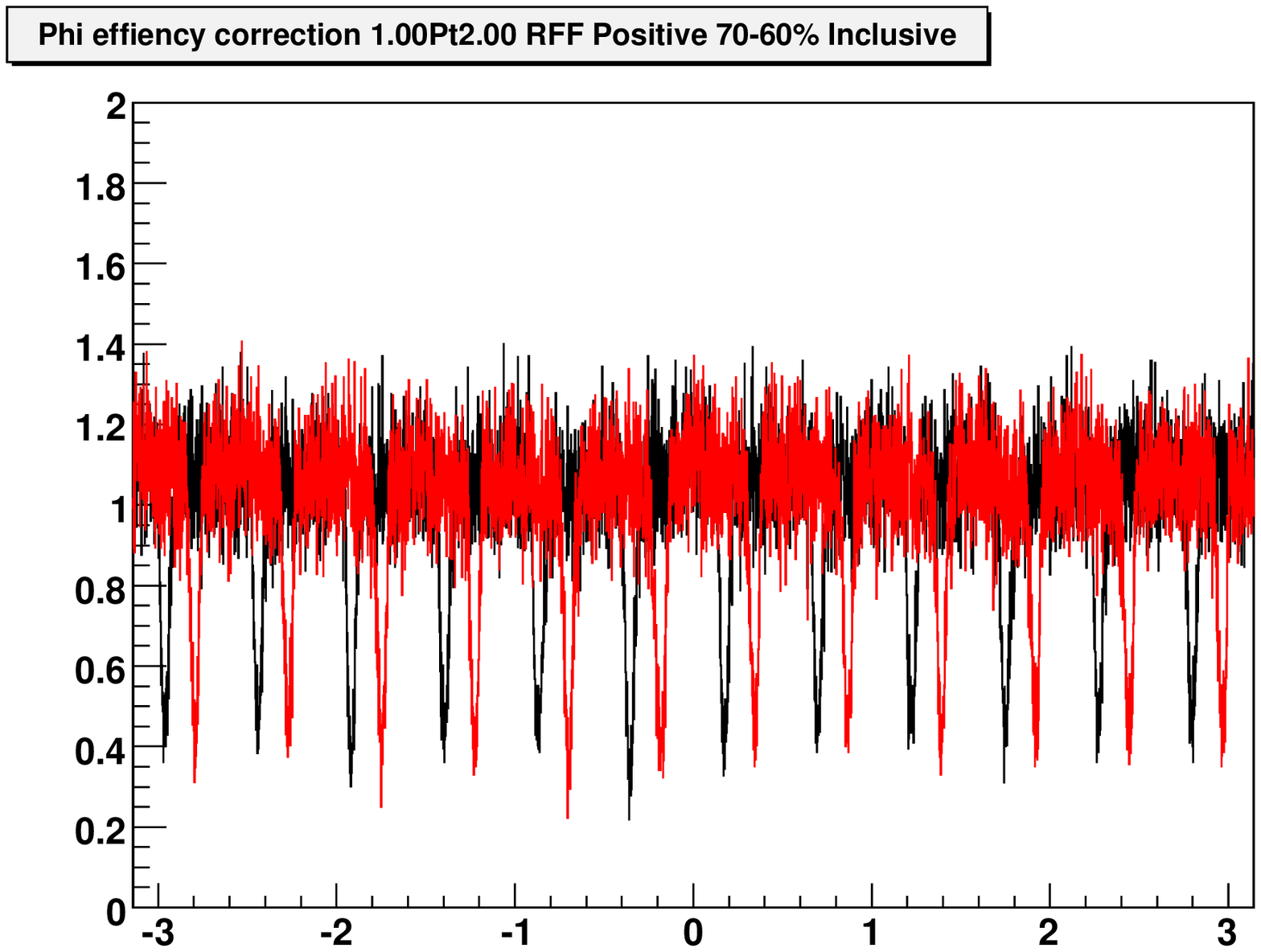}
\includegraphics[width=1.0\textwidth]{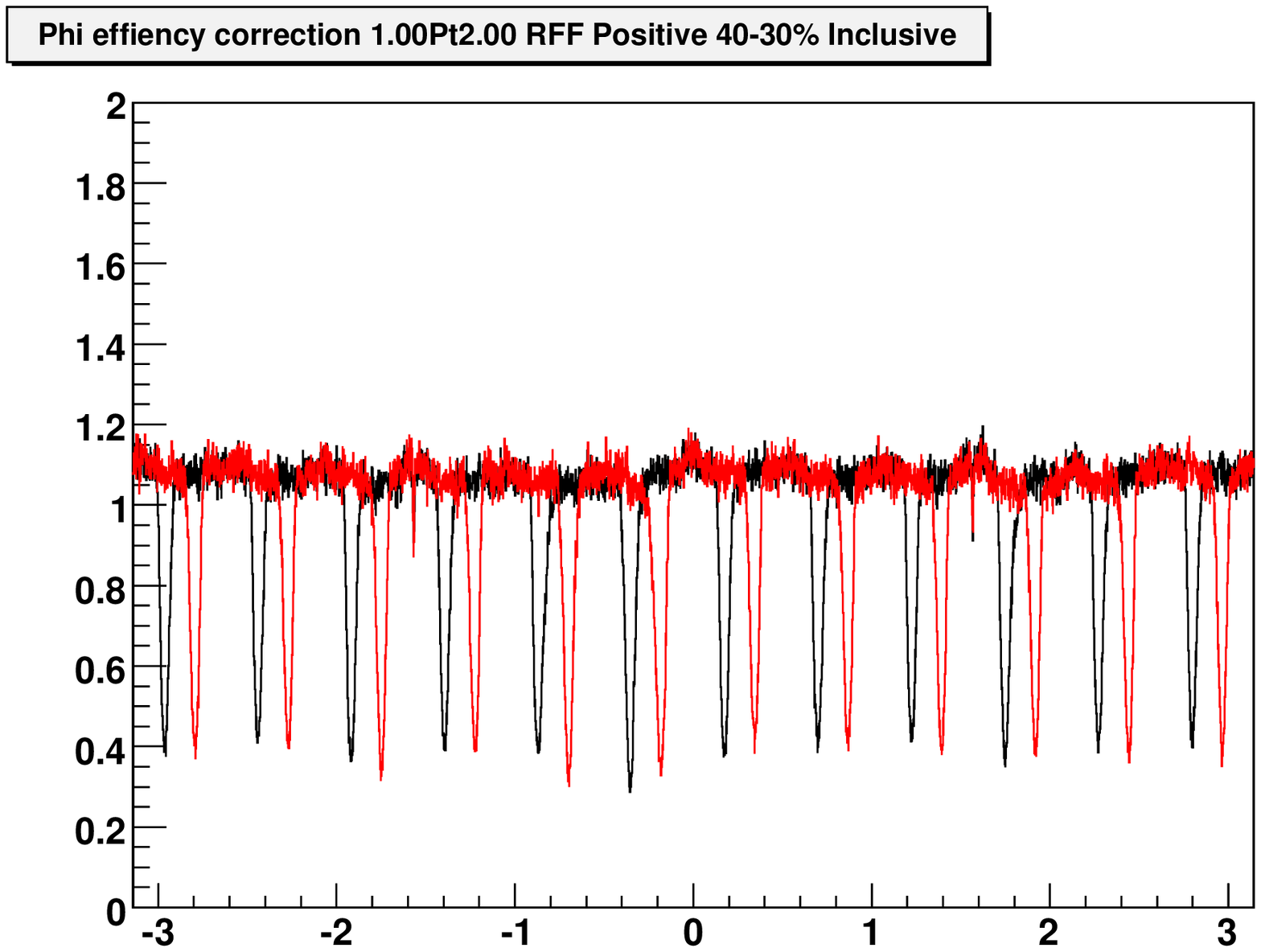}
\includegraphics[width=1.0\textwidth]{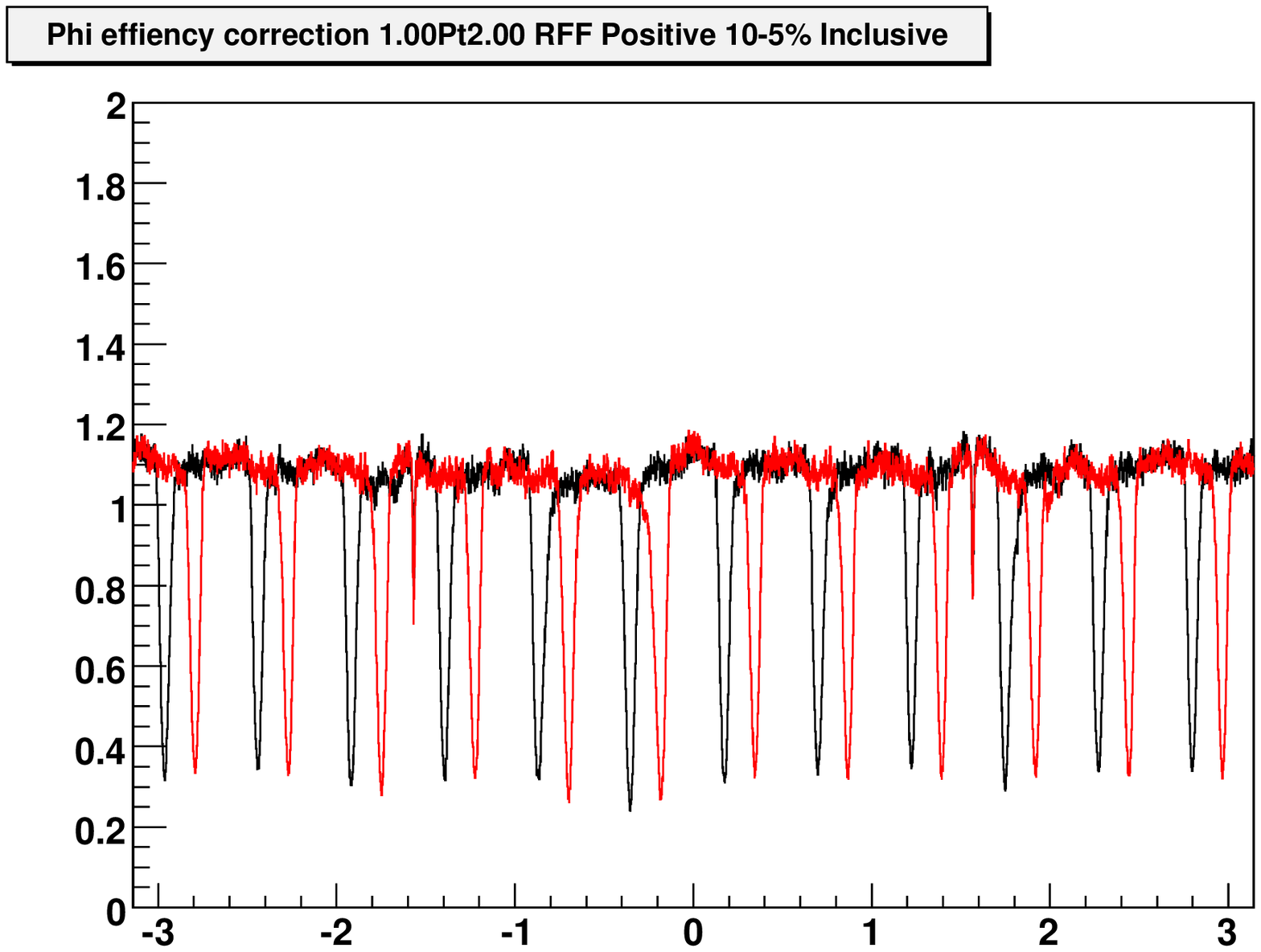}
\end{minipage}
\hfill
\begin{minipage}[t]{0.32\textwidth}
\centering
\includegraphics[width=1.0\textwidth]{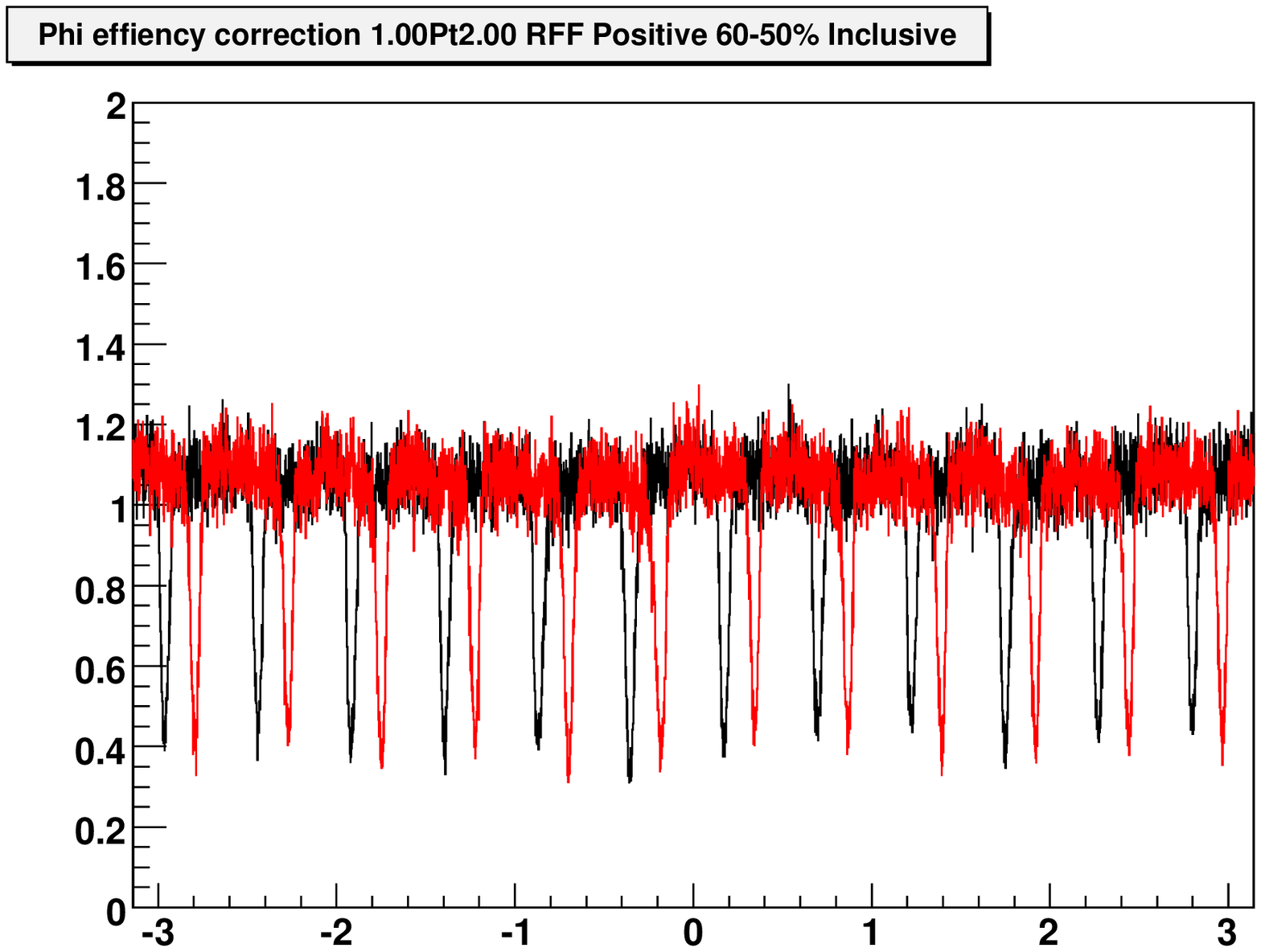}
\includegraphics[width=1.0\textwidth]{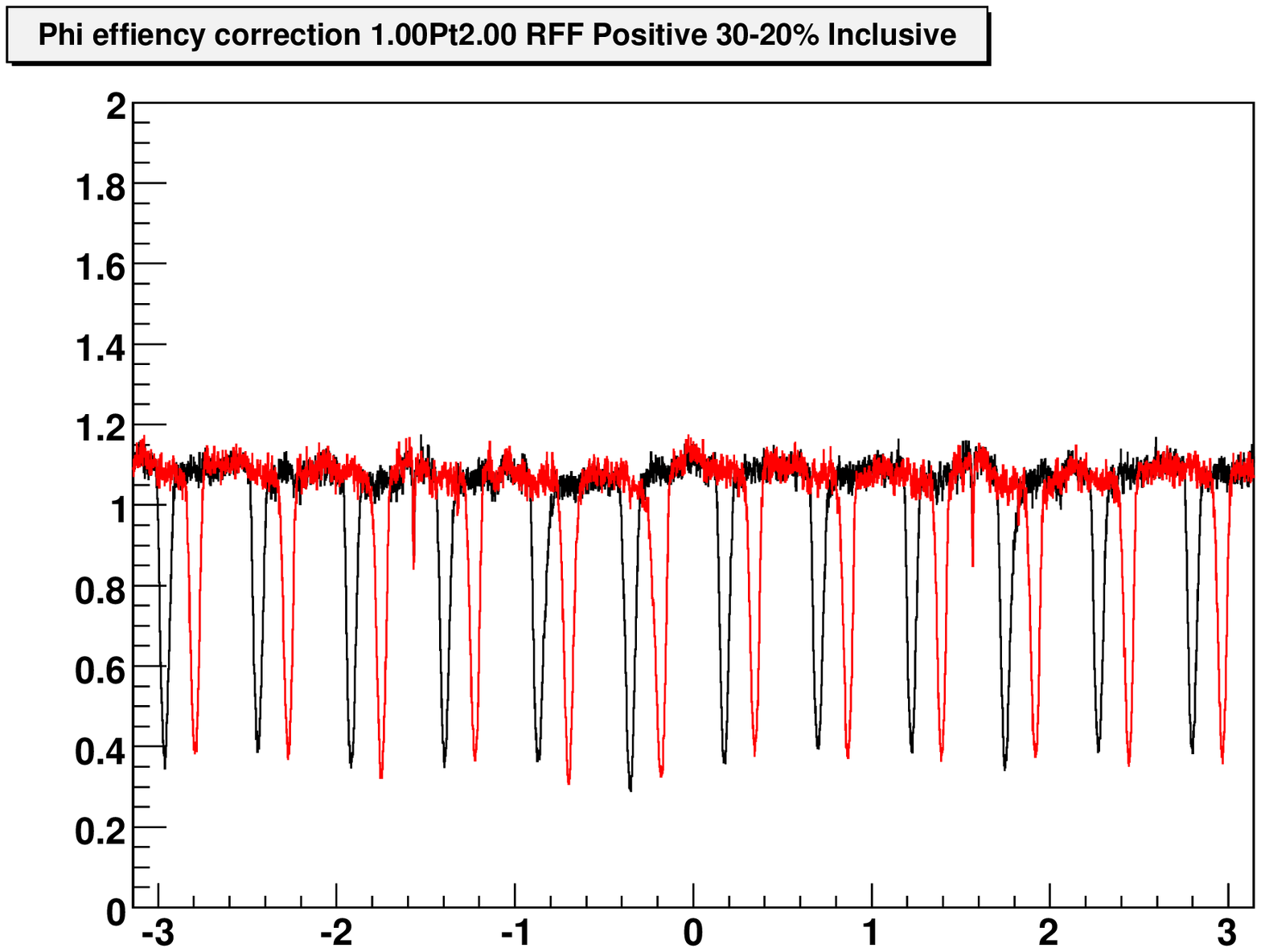}
\includegraphics[width=1.0\textwidth]{Plots/hacc_AuAuMB_Pt0B1M9.eps}
\end{minipage}

\caption{Same as Figure 6.12 but for $-0.5$ T magnetic field.}
\label{fig:accPt0B1}
\end{figure}

\begin{figure}[H]
\hfill
\begin{minipage}[t]{0.32\textwidth}
\centering
\includegraphics[width=1.0\textwidth]{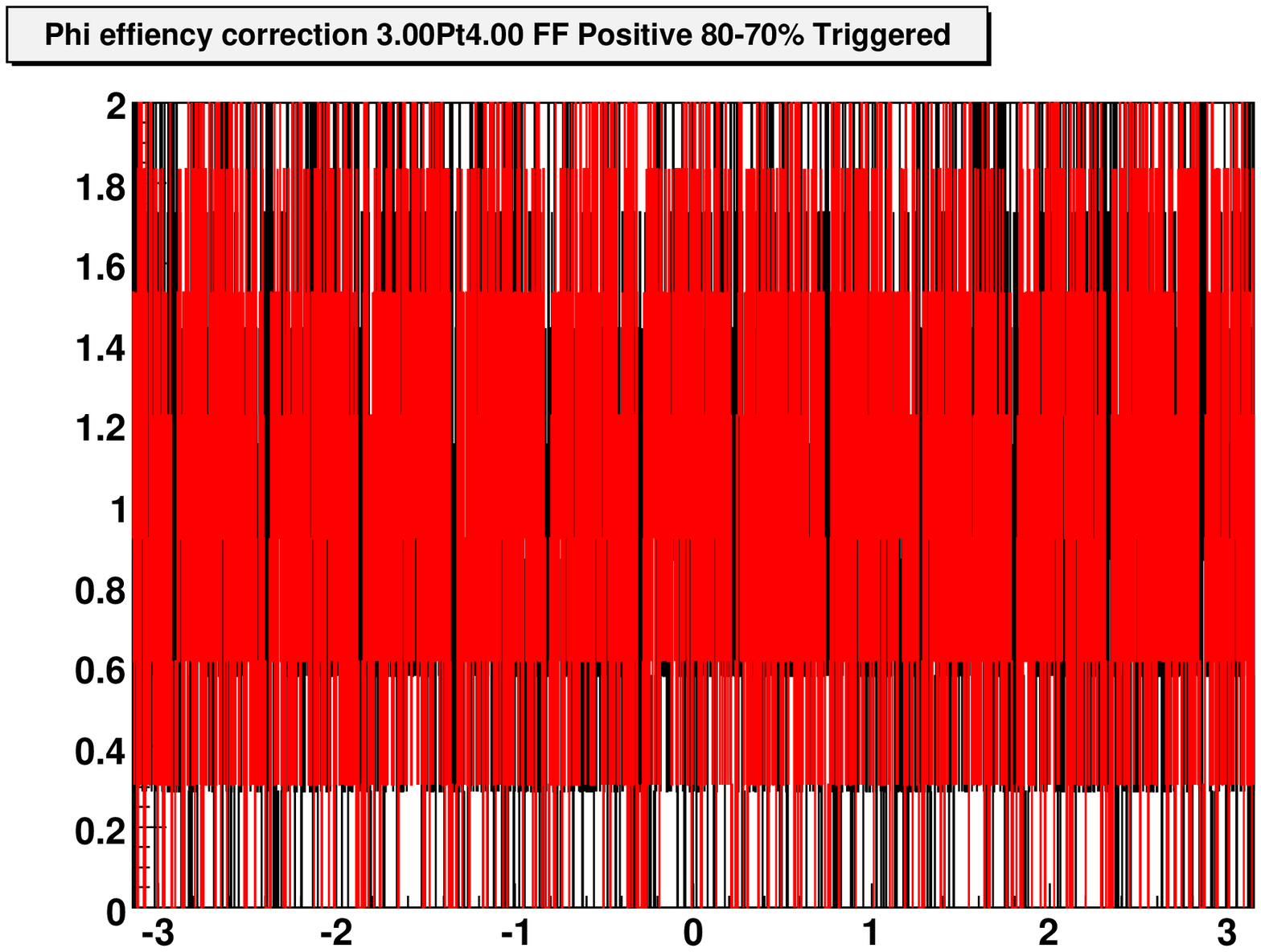}
\includegraphics[width=1.0\textwidth]{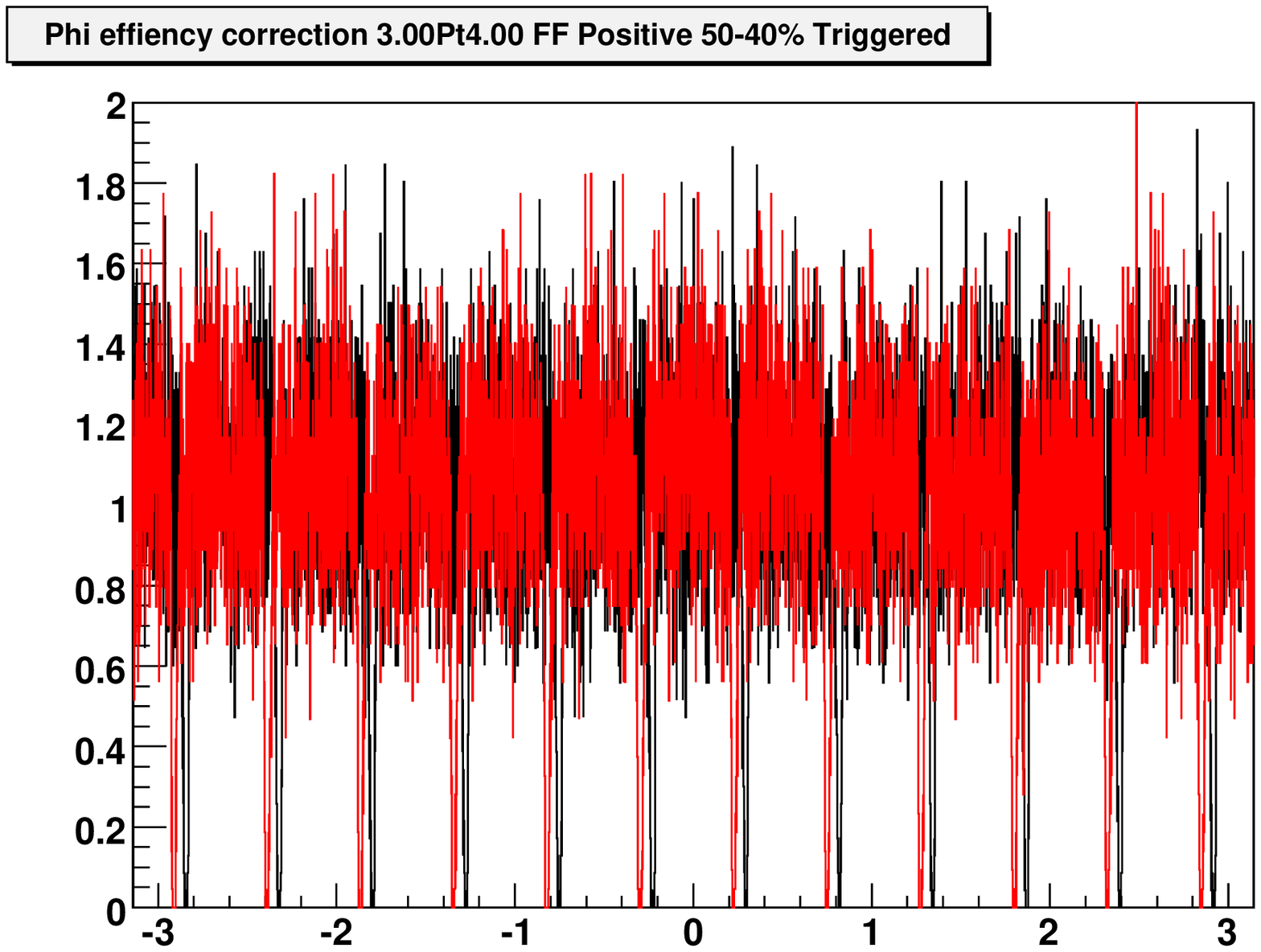}
\includegraphics[width=1.0\textwidth]{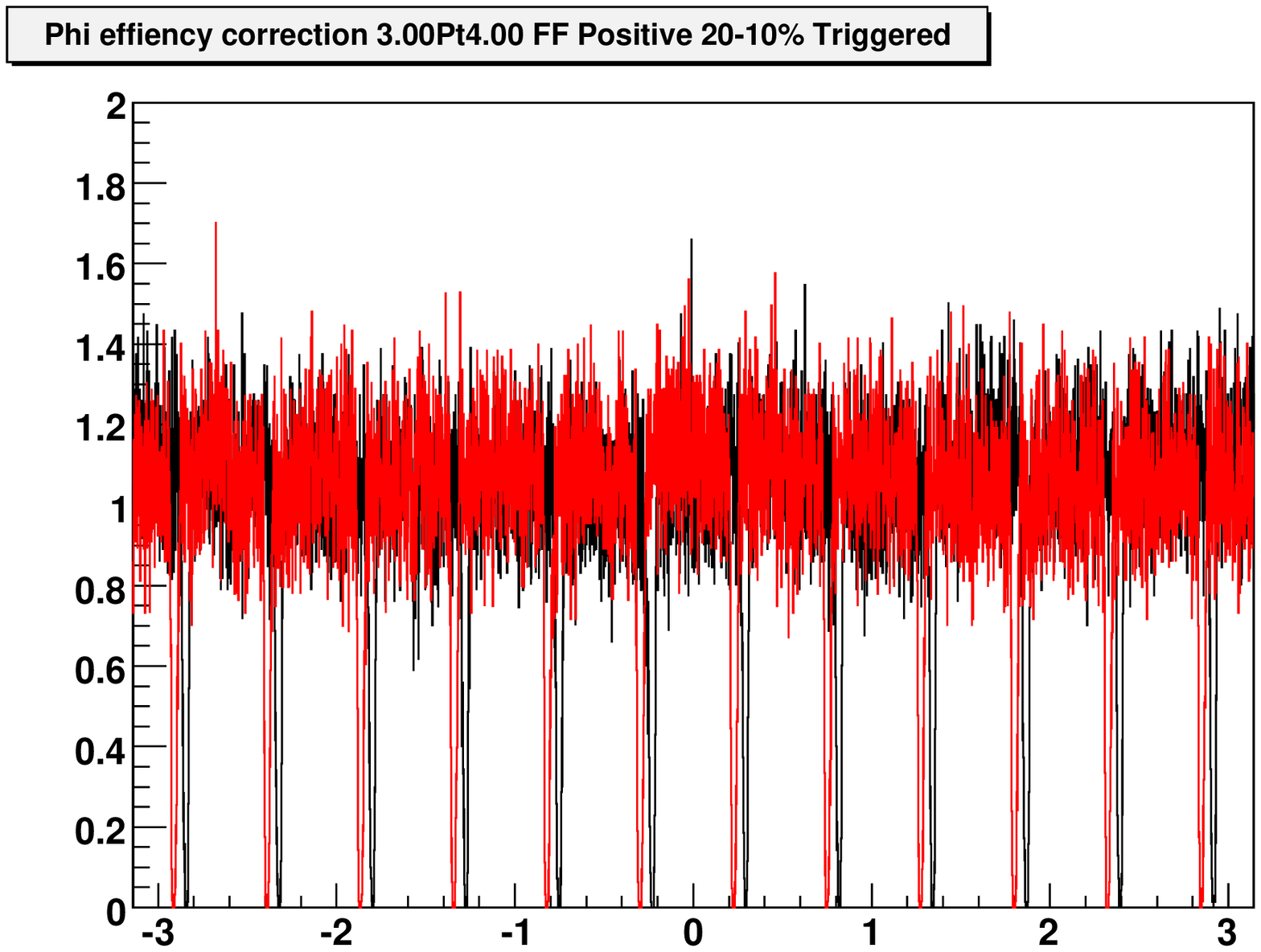}
\end{minipage}
\hfill
\begin{minipage}[t]{0.32\textwidth}
\centering
\includegraphics[width=1.0\textwidth]{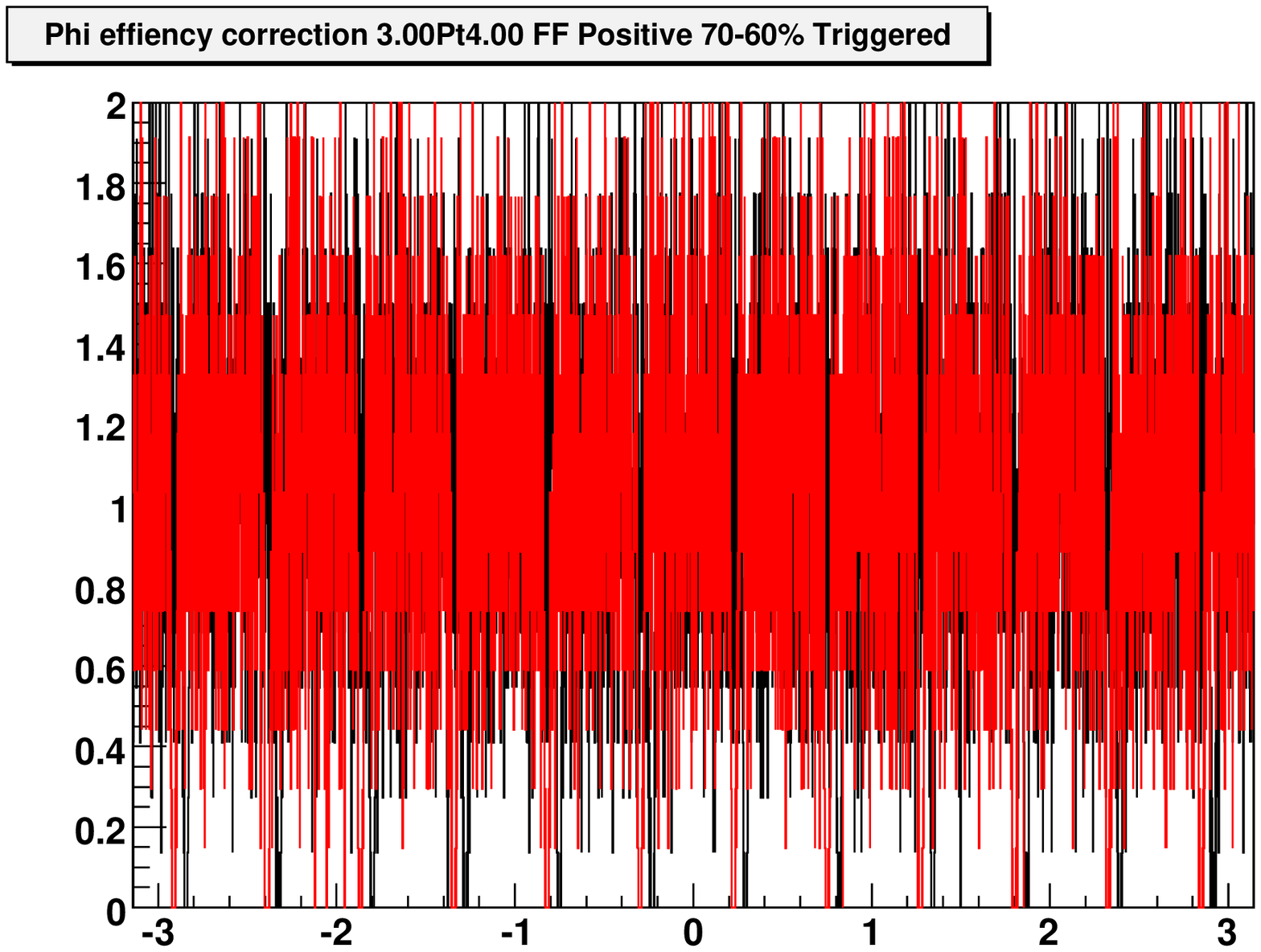}
\includegraphics[width=1.0\textwidth]{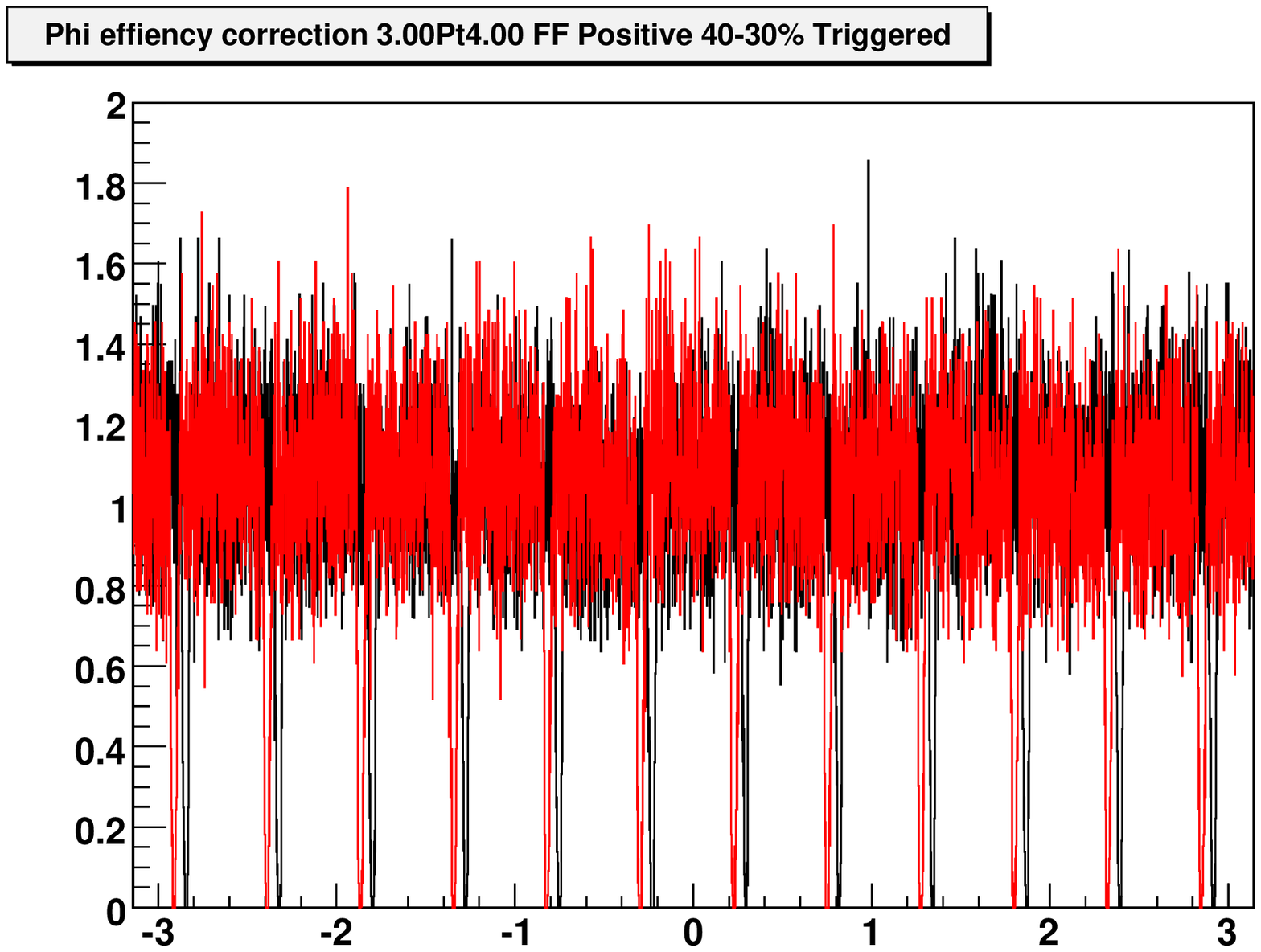}
\includegraphics[width=1.0\textwidth]{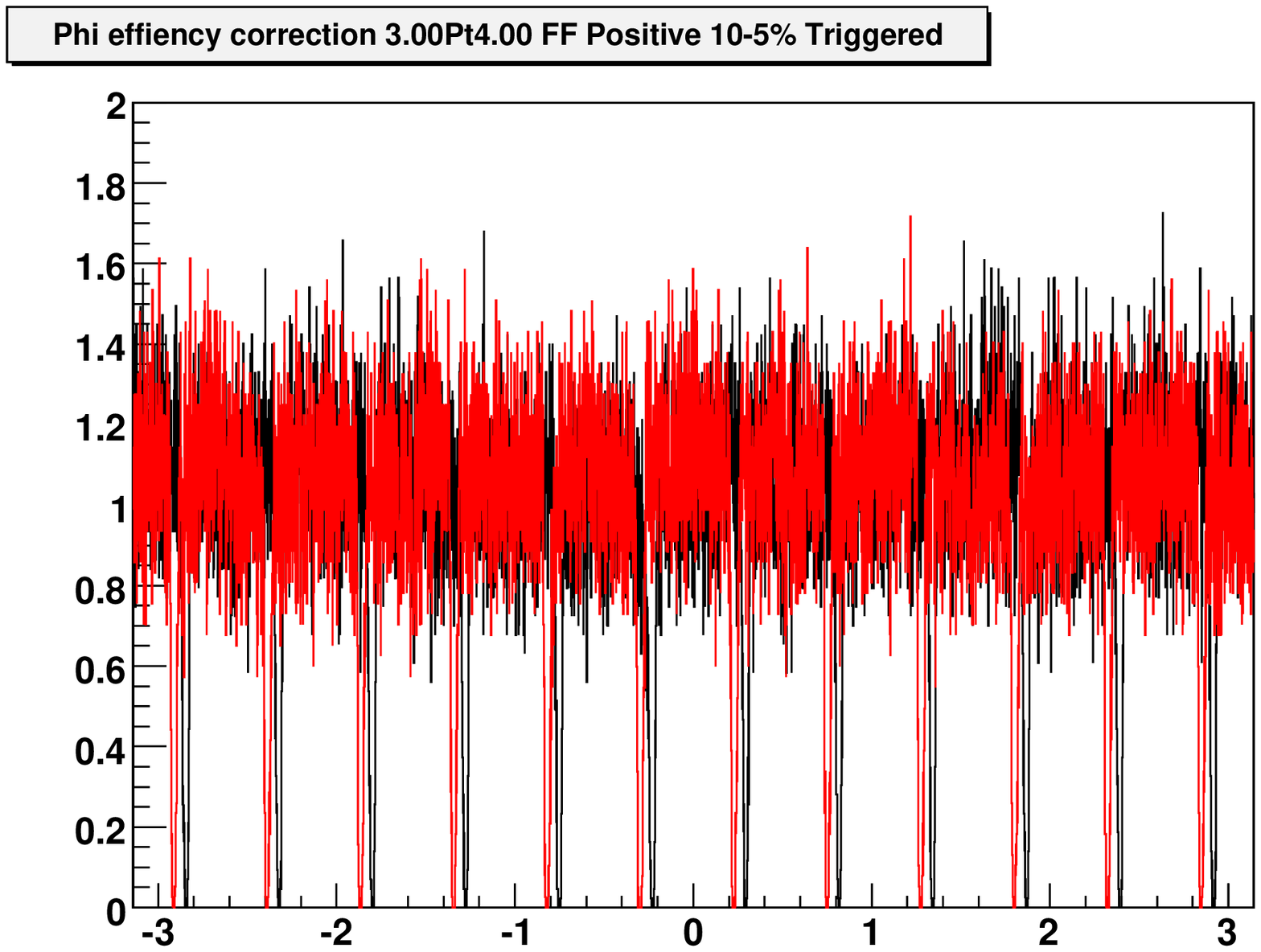}
\end{minipage}
\hfill
\begin{minipage}[t]{0.32\textwidth}
\centering
\includegraphics[width=1.0\textwidth]{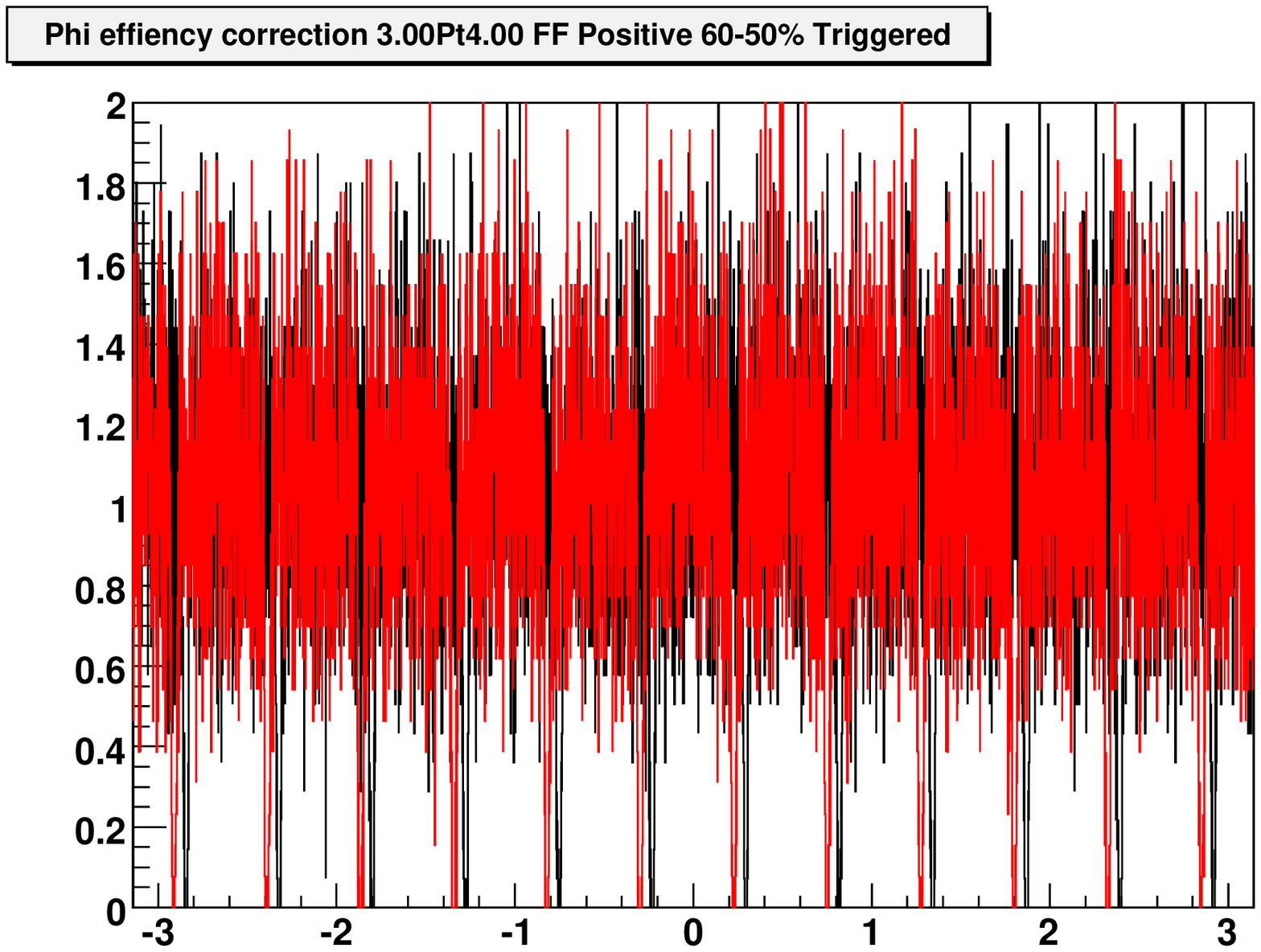}
\includegraphics[width=1.0\textwidth]{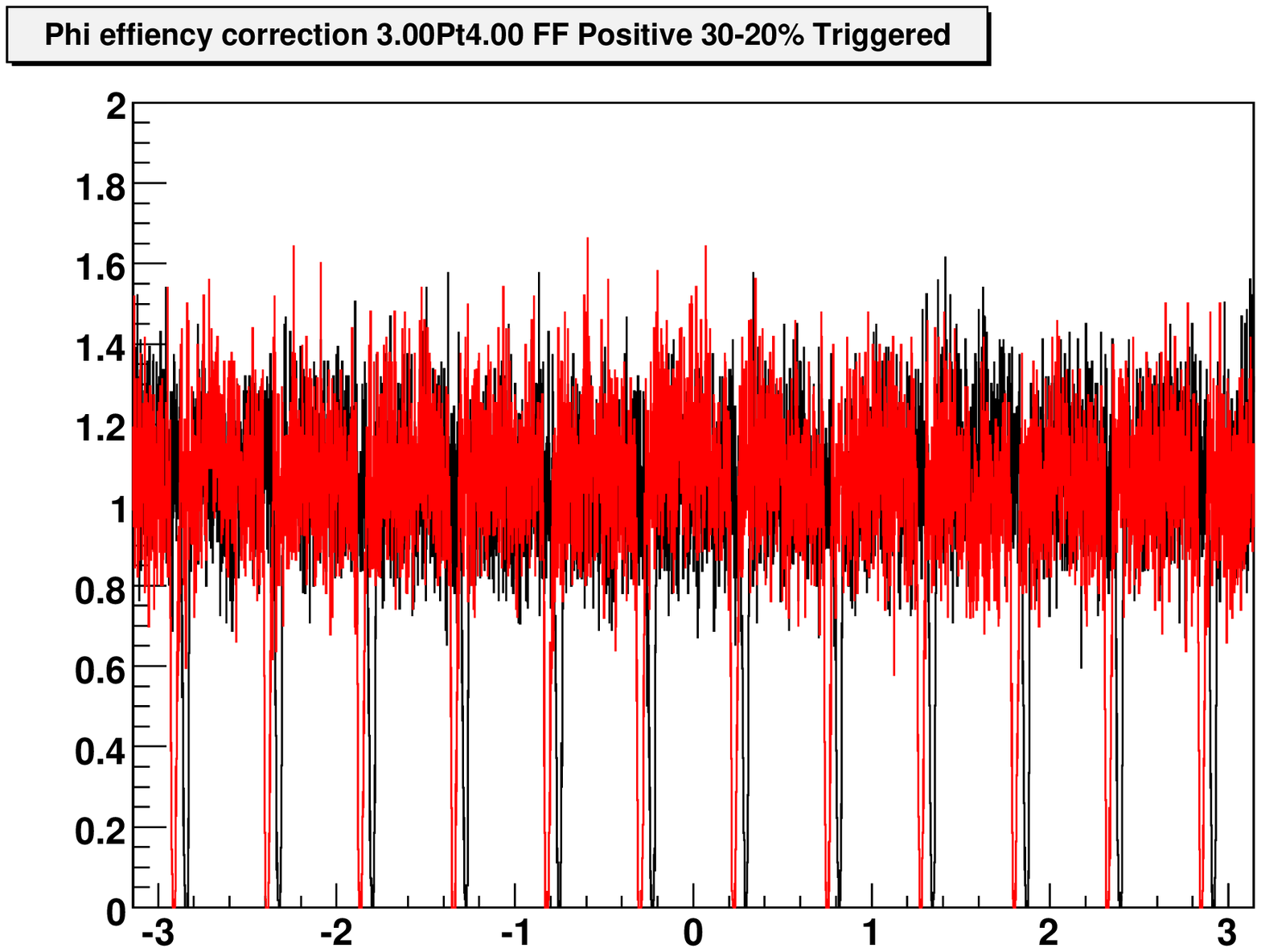}
\includegraphics[width=1.0\textwidth]{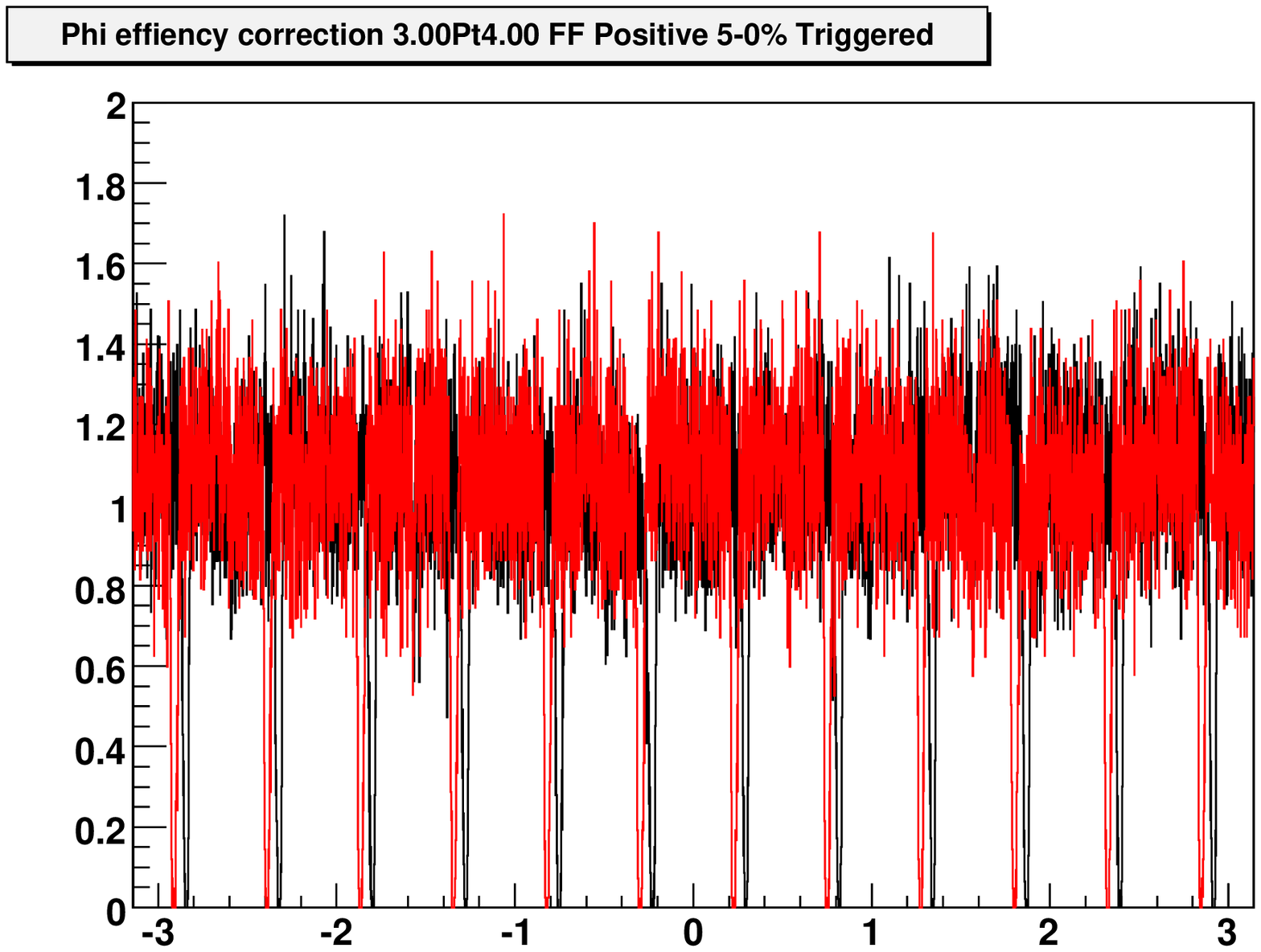}
\end{minipage}

\caption{Same as Fig. 6.12 but for $3<p_T<4$ GeV/c.}
\label{fig:accPt1B0}
\end{figure}

\begin{figure}[H]
\hfill
\begin{minipage}[t]{0.32\textwidth}
\centering
\includegraphics[width=1.0\textwidth]{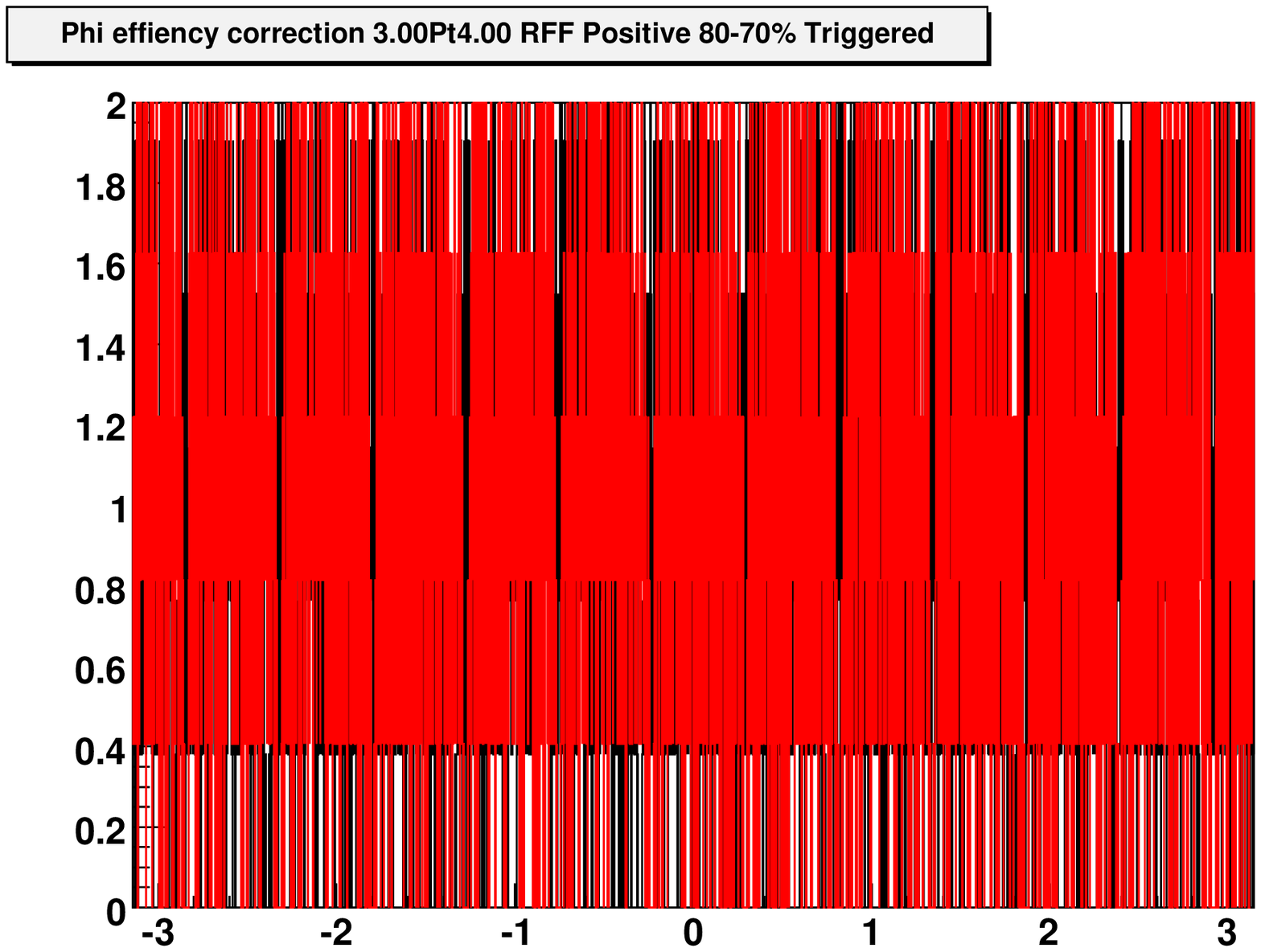}
\includegraphics[width=1.0\textwidth]{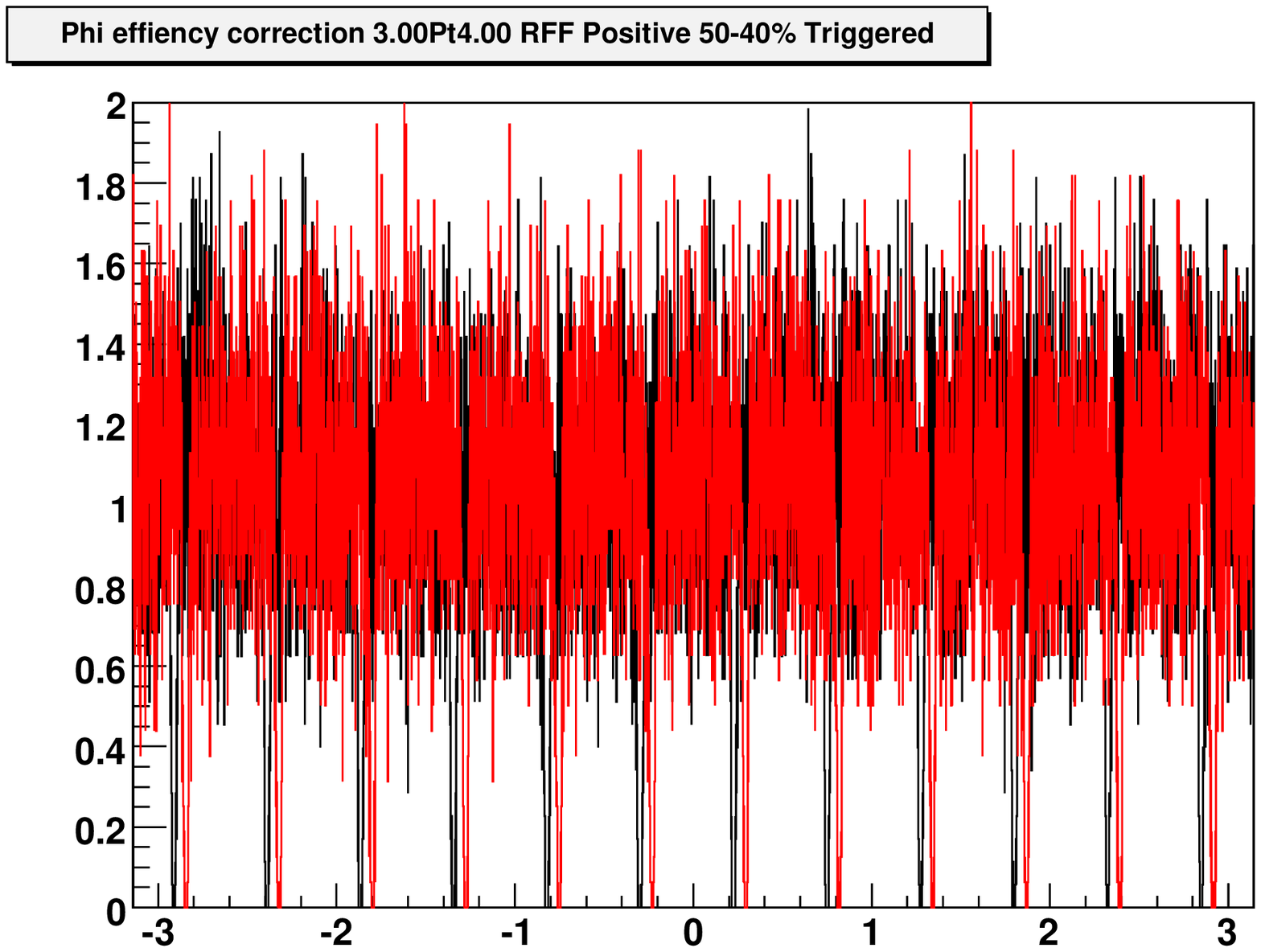}
\includegraphics[width=1.0\textwidth]{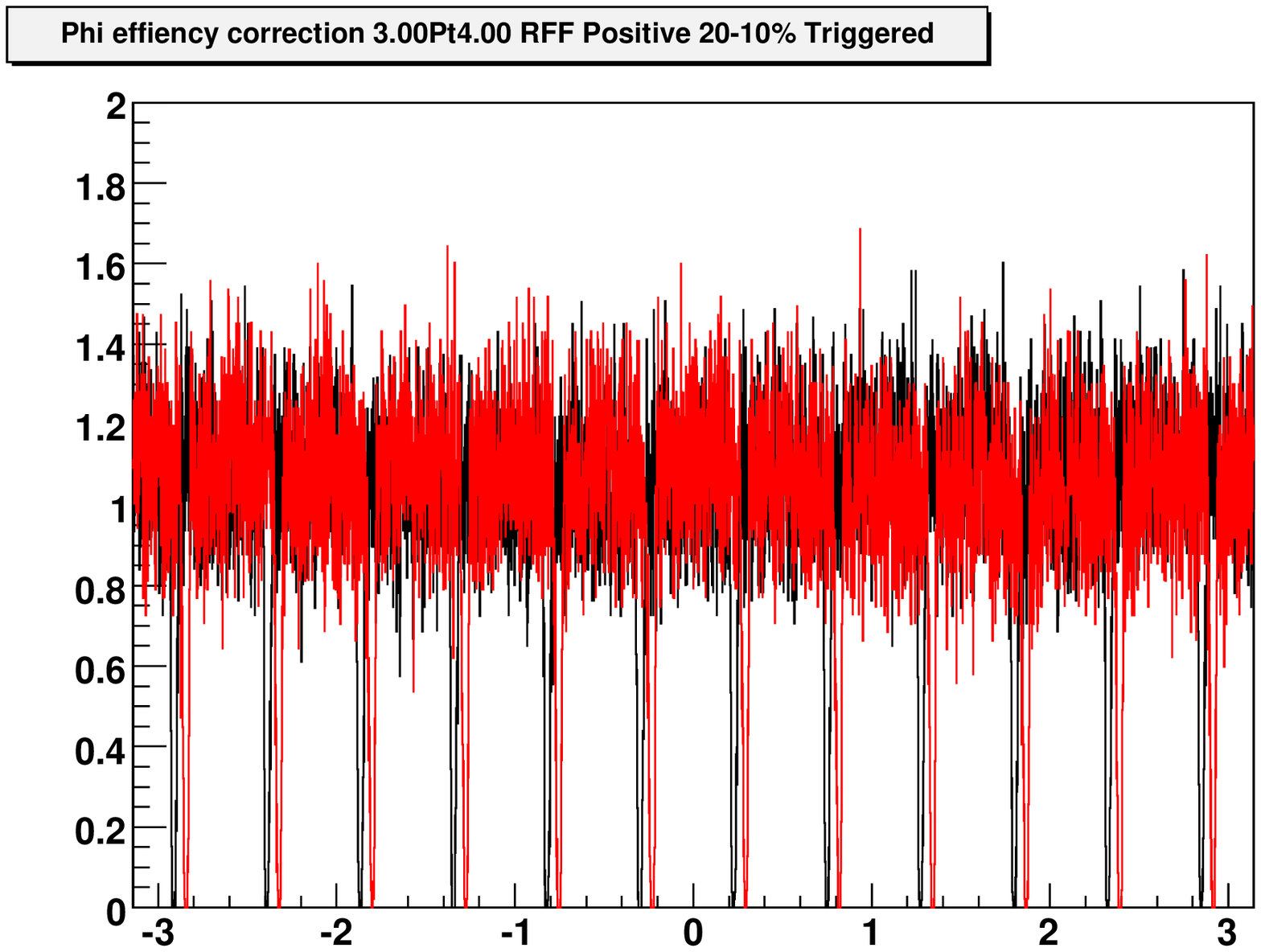}
\end{minipage}
\hfill
\begin{minipage}[t]{0.32\textwidth}
\centering
\includegraphics[width=1.0\textwidth]{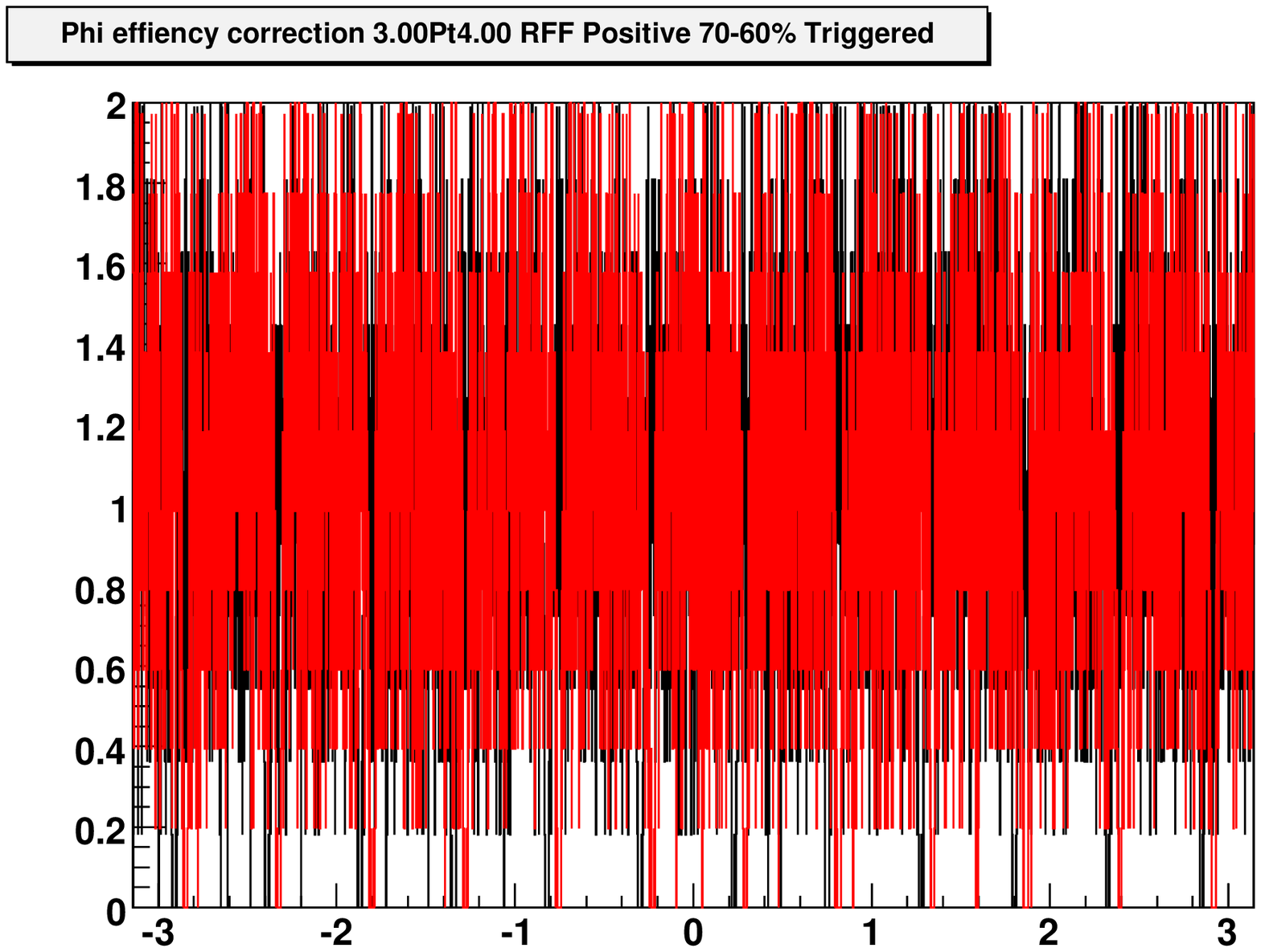}
\includegraphics[width=1.0\textwidth]{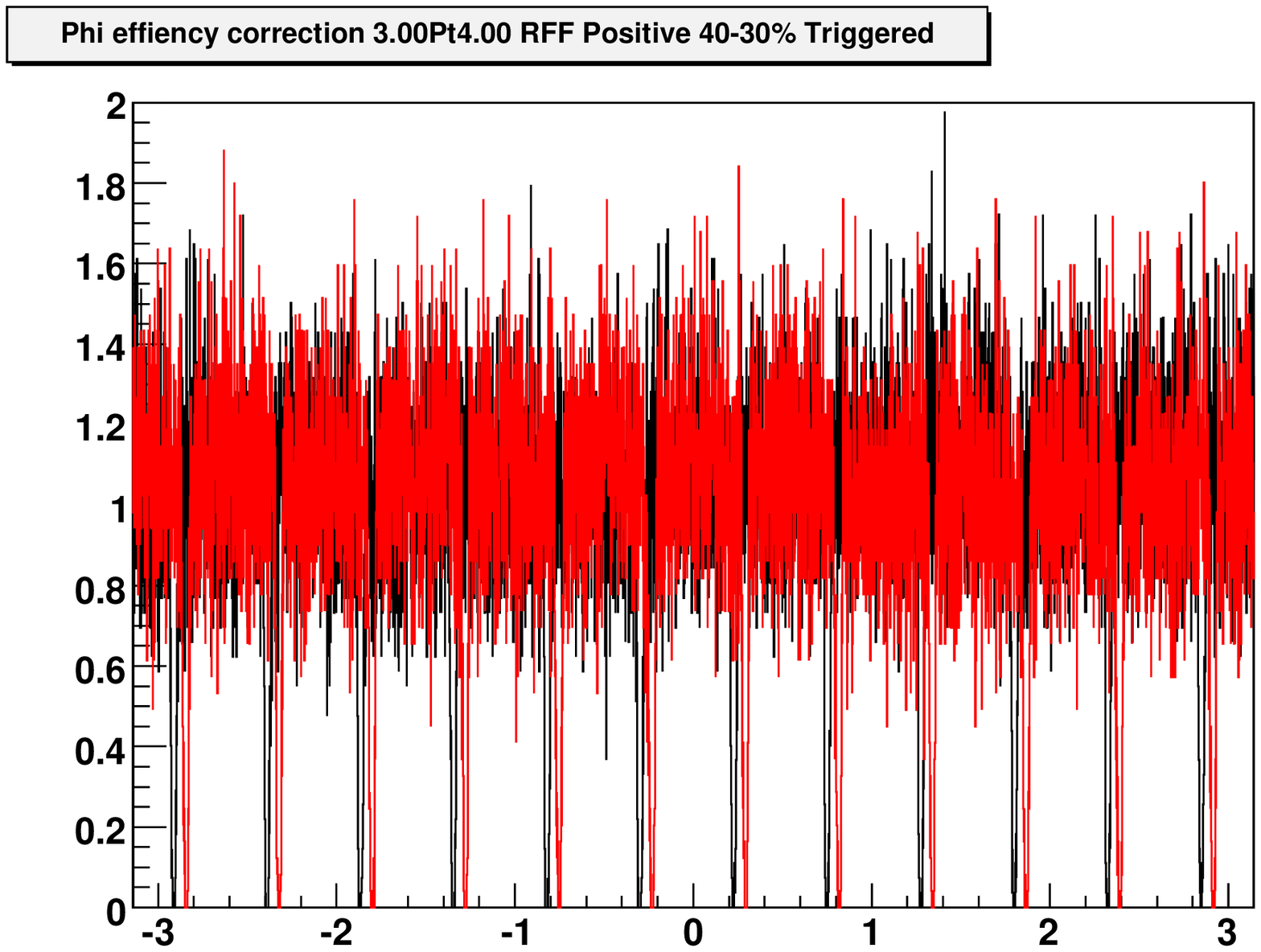}
\includegraphics[width=1.0\textwidth]{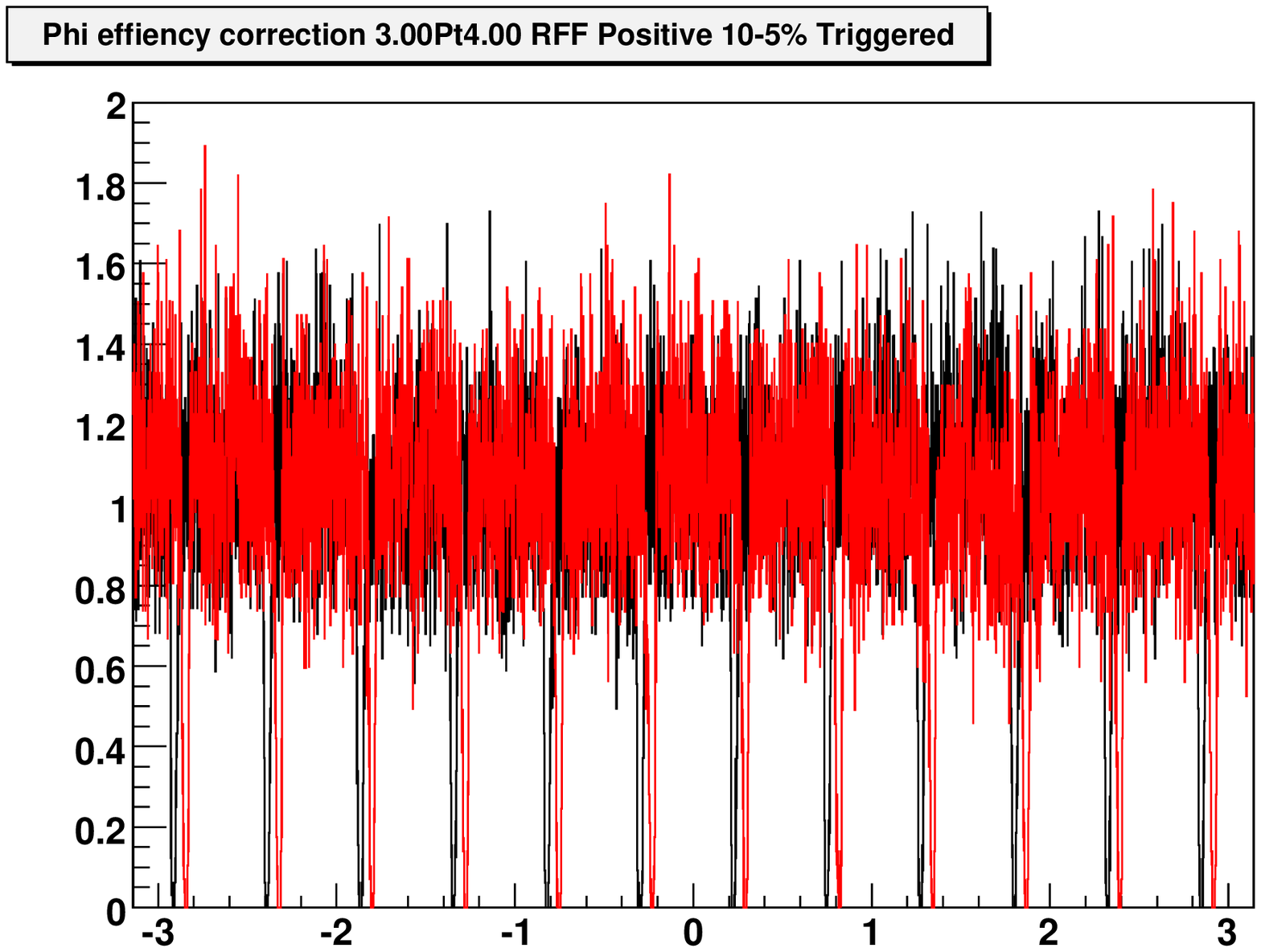}
\end{minipage}
\hfill
\begin{minipage}[t]{0.32\textwidth}
\centering
\includegraphics[width=1.0\textwidth]{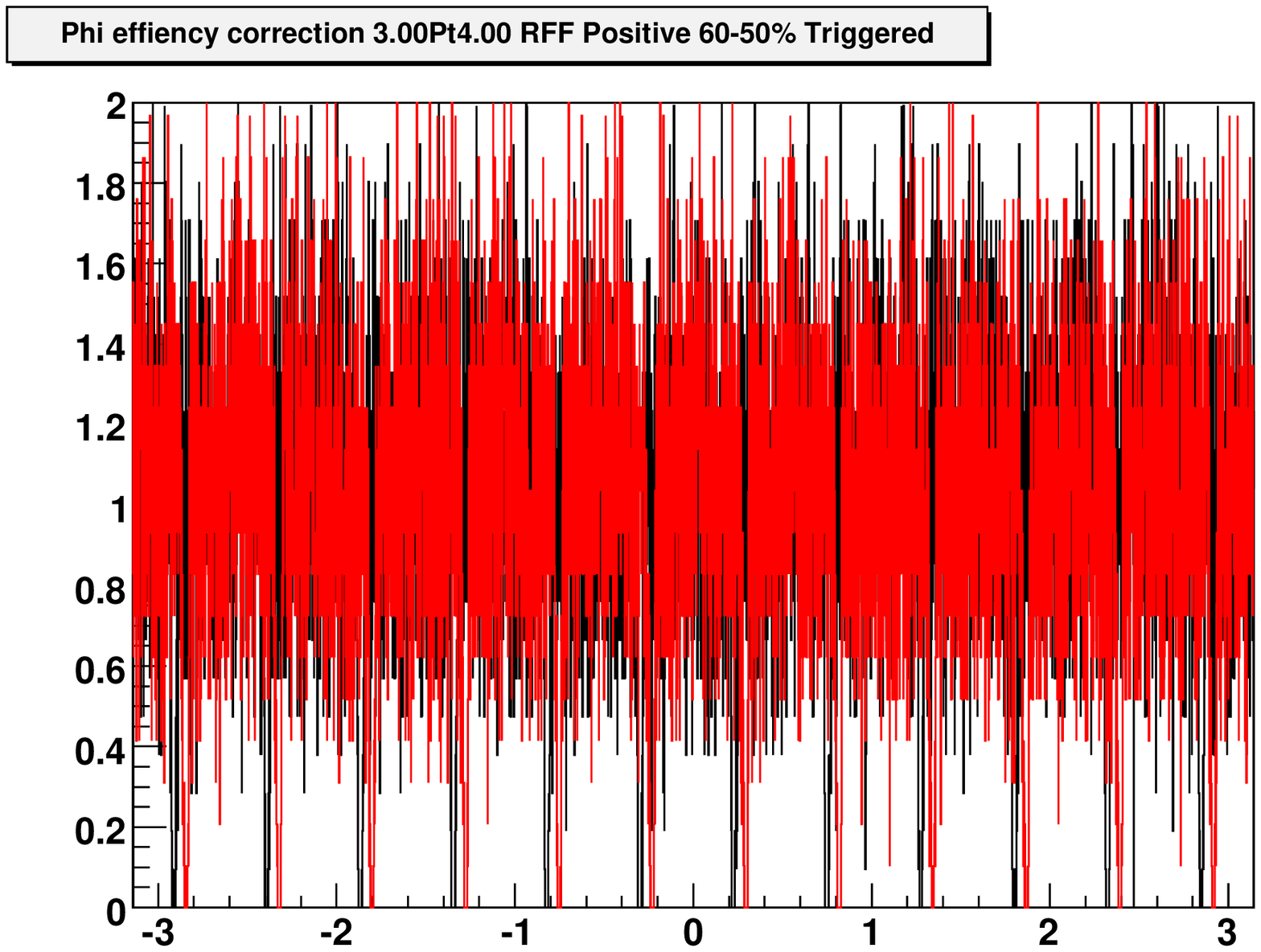}
\includegraphics[width=1.0\textwidth]{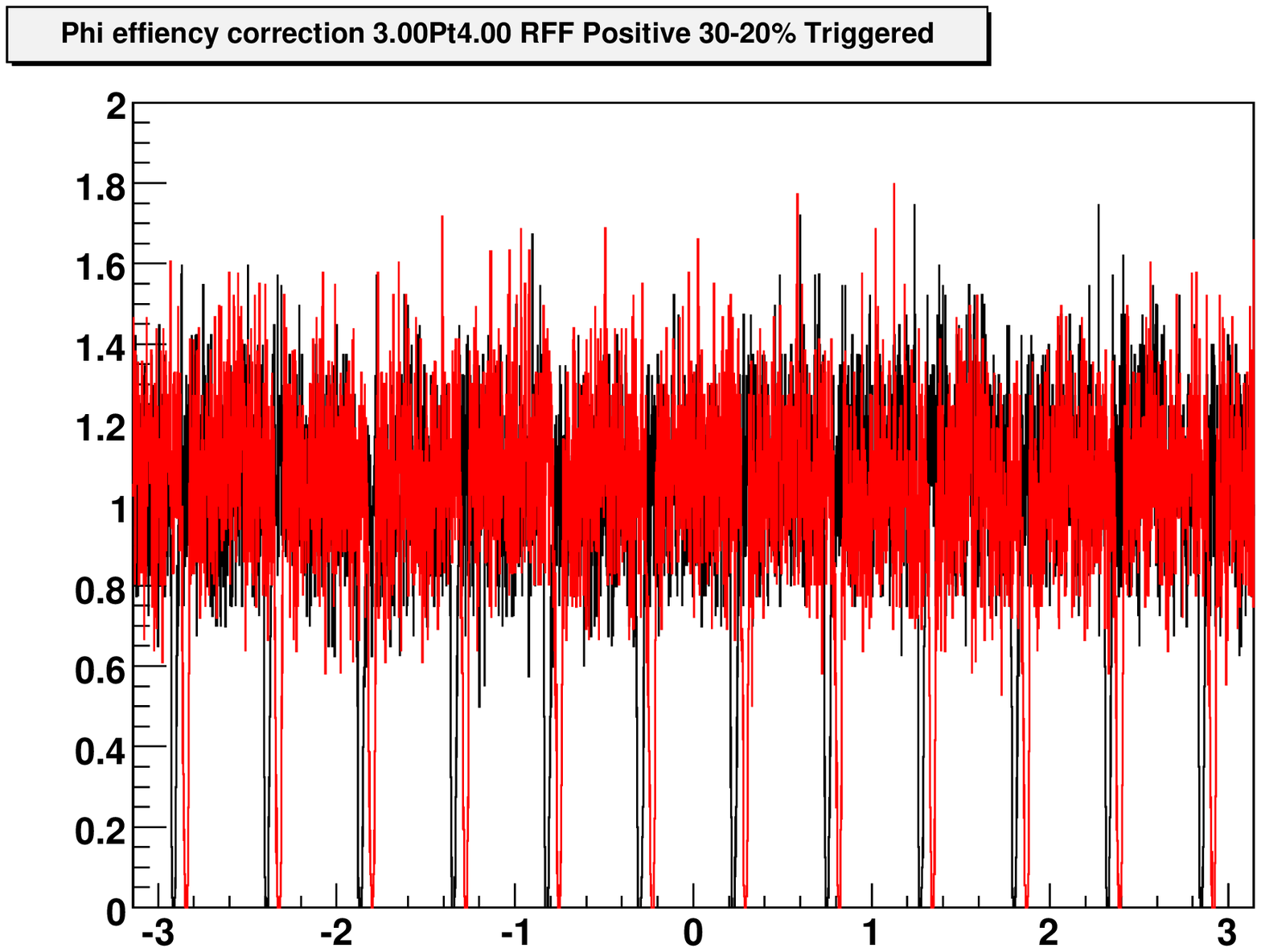}
\includegraphics[width=1.0\textwidth]{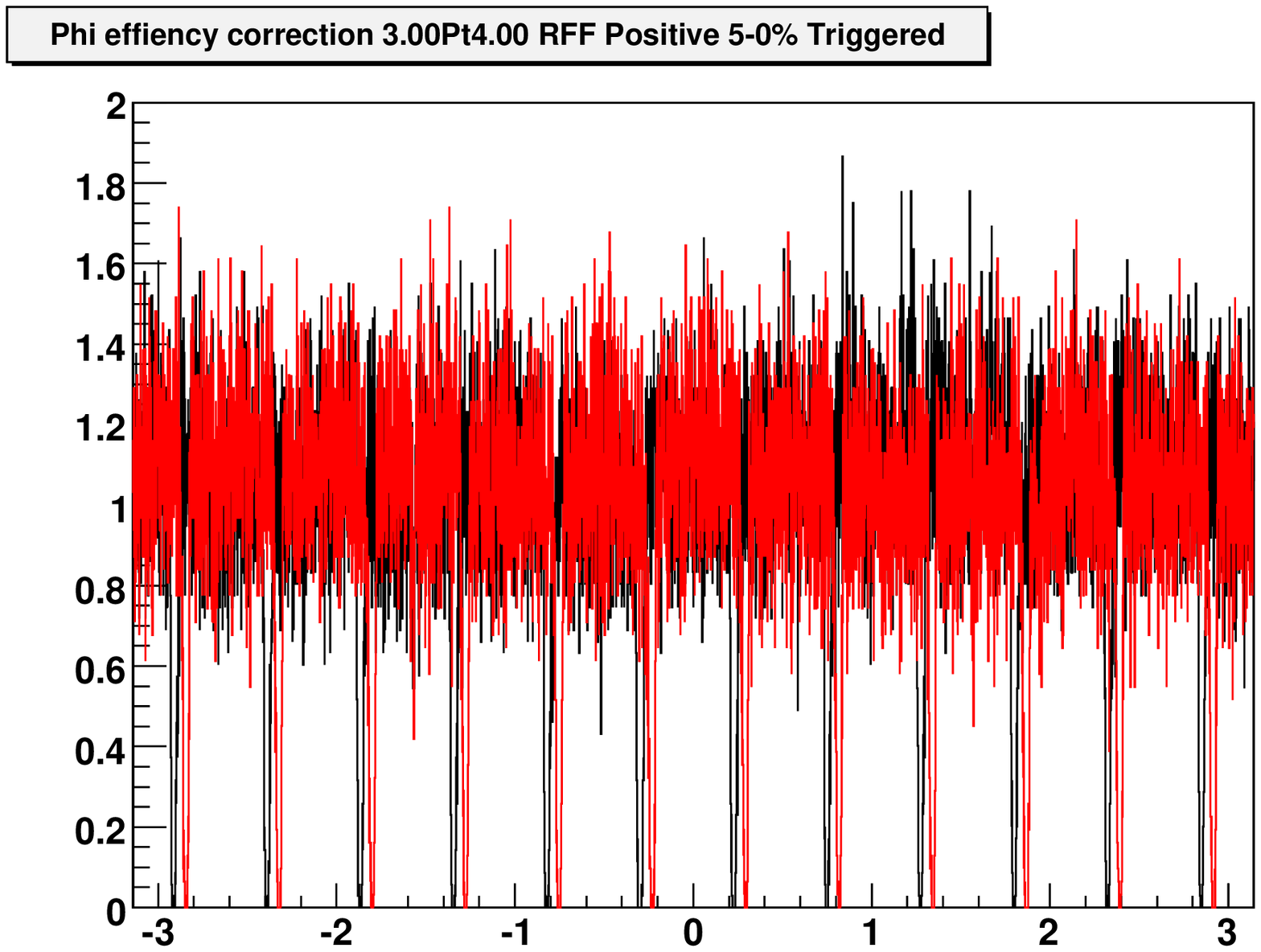}
\end{minipage}

\caption{Same as Fig. 6.12 but for $3<p_T<4$ GeV/c and $-0.5$ T magnetic field.}
\label{fig:accPt1B1}
\end{figure}

\begin{figure}[H]
\hfill
\begin{minipage}[t]{0.32\textwidth}
\centering
\includegraphics[width=1.0\textwidth]{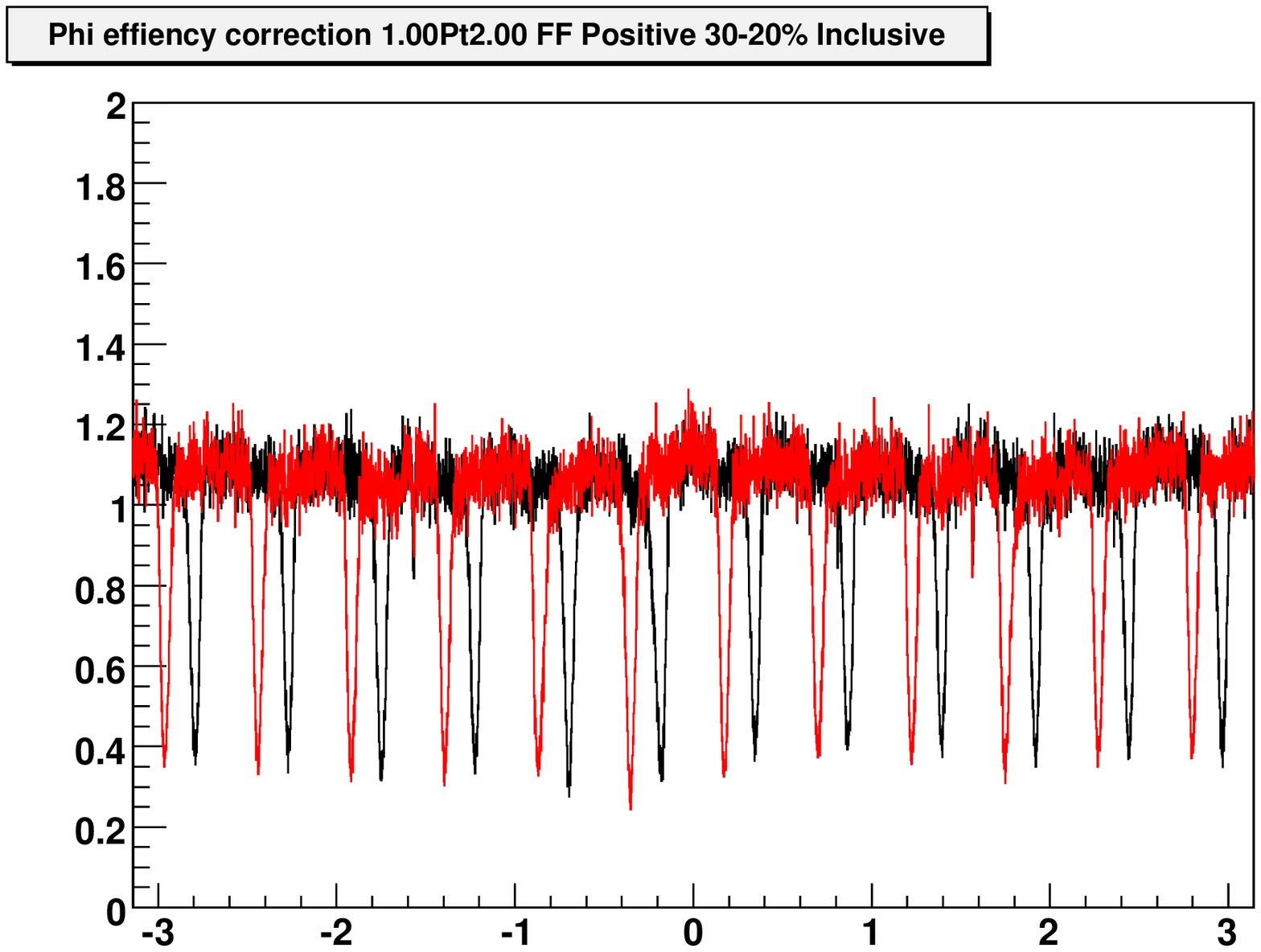}
\includegraphics[width=1.0\textwidth]{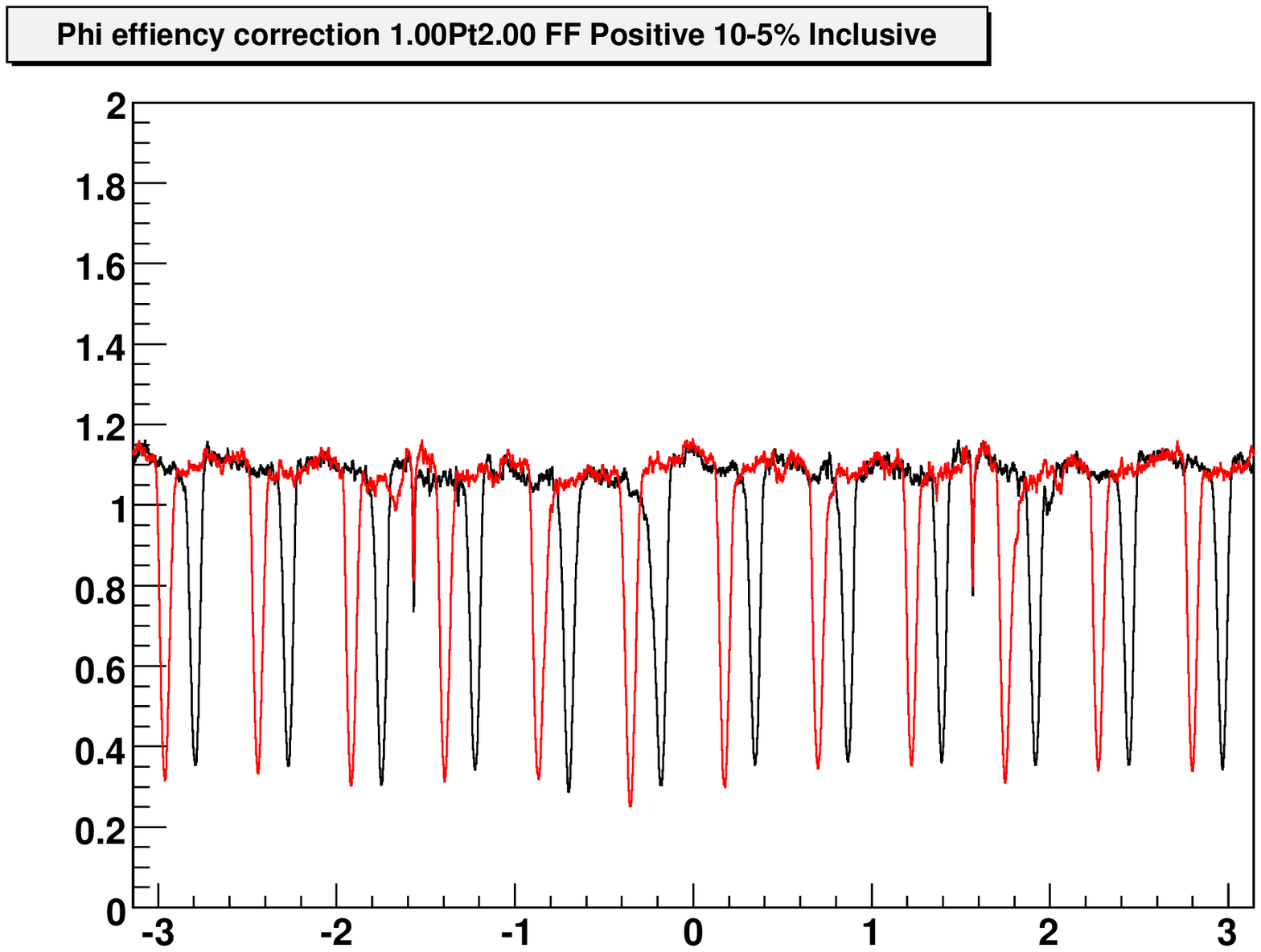}
\end{minipage}
\hfill
\begin{minipage}[t]{0.32\textwidth}
\centering
\includegraphics[width=1.0\textwidth]{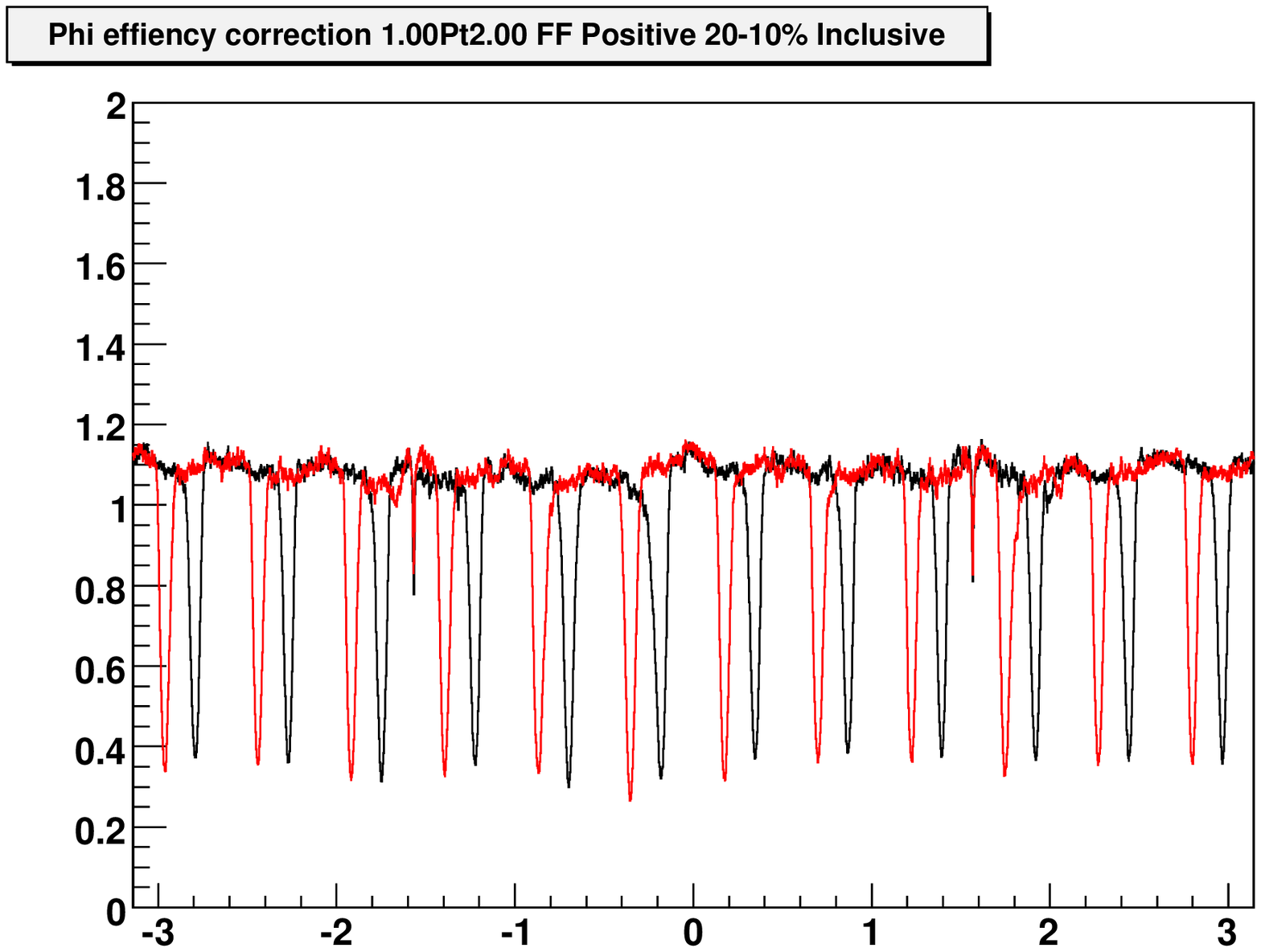}
\includegraphics[width=1.0\textwidth]{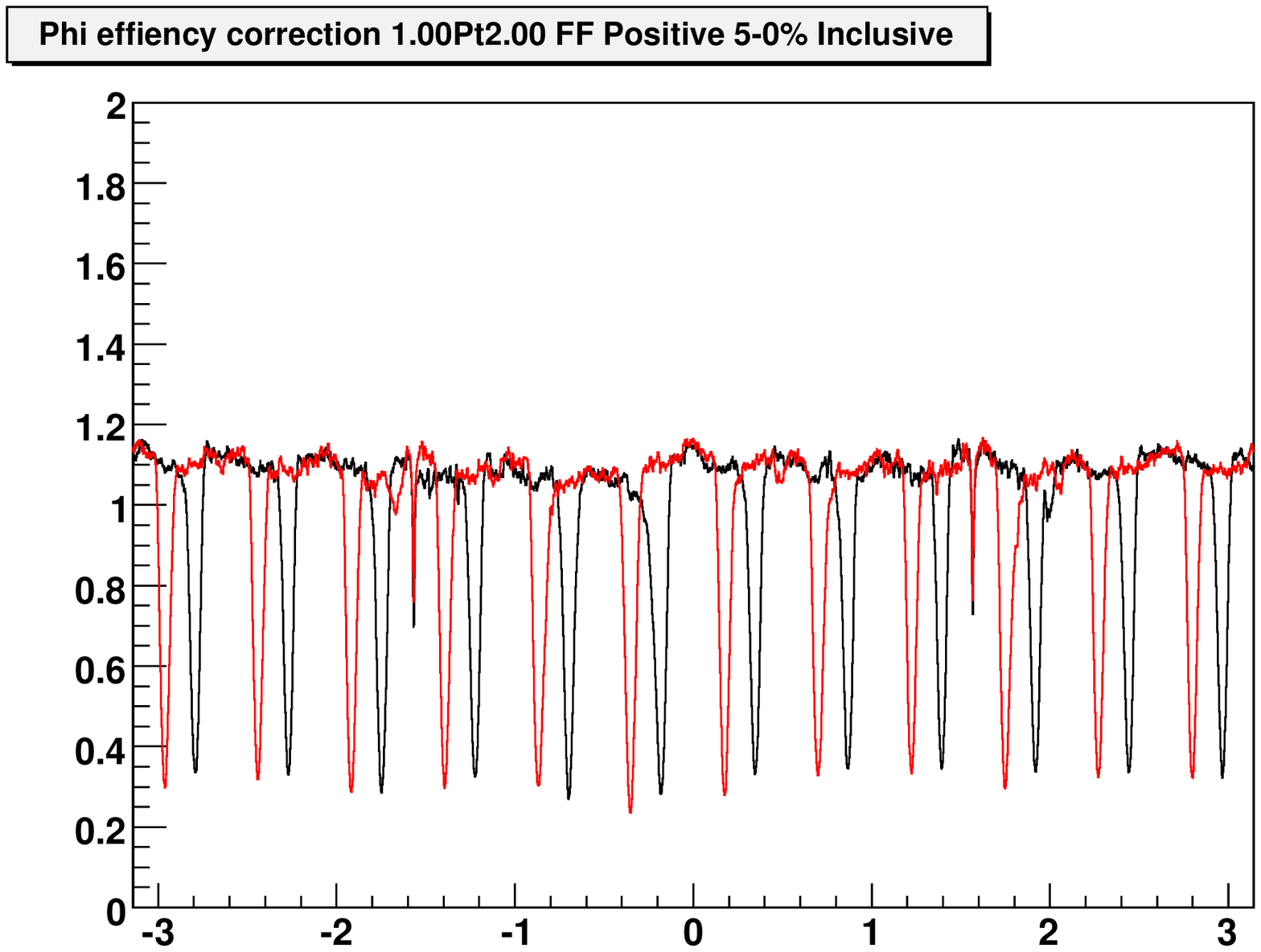}
\end{minipage}

\caption{Same as Fig. 6.12 but for ZDC trigger central Au+Au collisions in centralities (from left to right and top to bottom) ``20-30\%'', ``10-20\%'', ``5-10\%'', and 0-5\% most central where the first three centralites do not correspond to the actual percantage of cross section due to the ZDC trigger but have the same multiplicty cuts as the minimum bias events that do correspond to the given cross section.}
\label{fig:accCPt0B0}
\end{figure}

\begin{figure}[H]
\hfill
\begin{minipage}[t]{0.32\textwidth}
\centering
\includegraphics[width=1.0\textwidth]{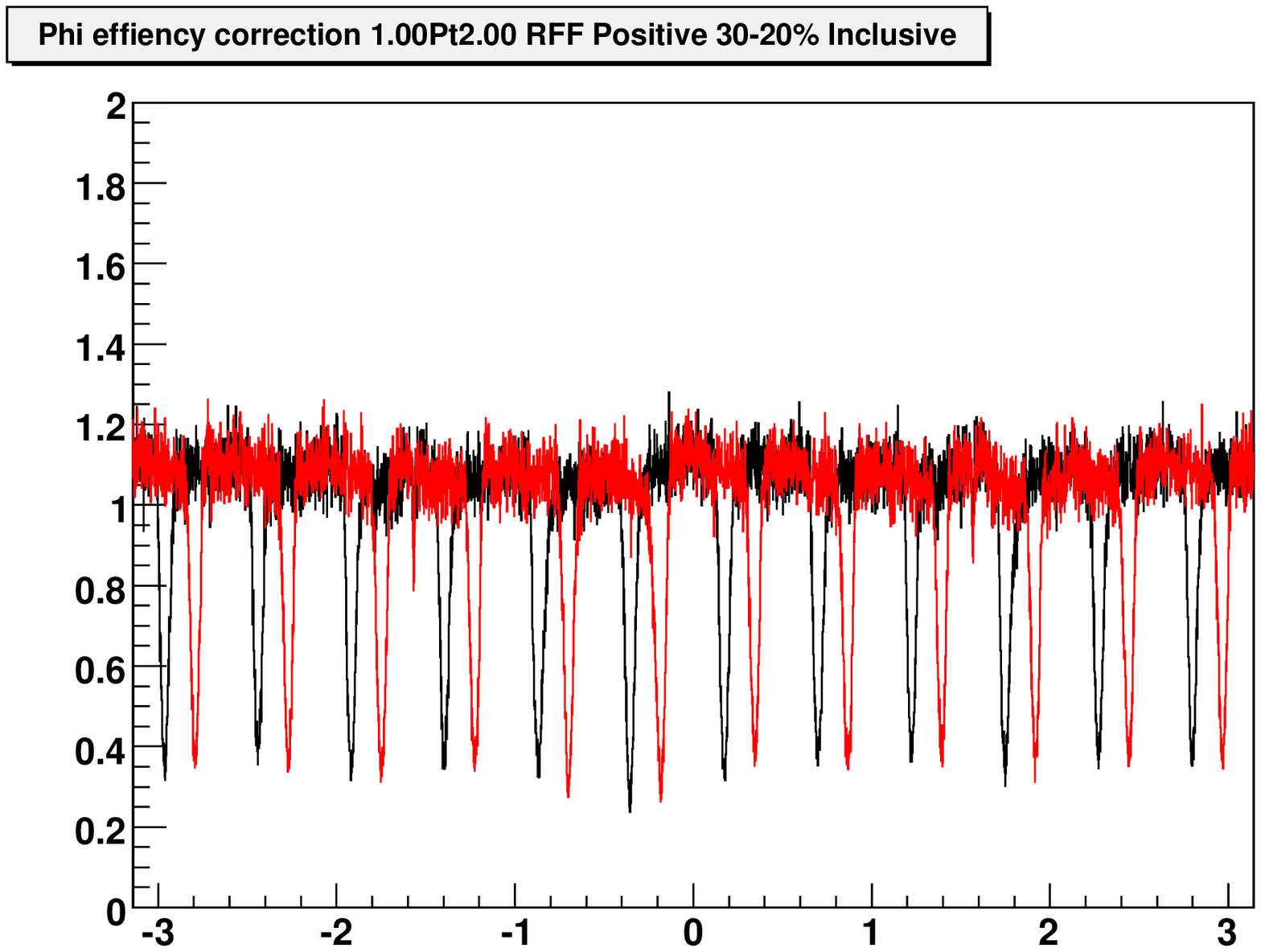}
\includegraphics[width=1.0\textwidth]{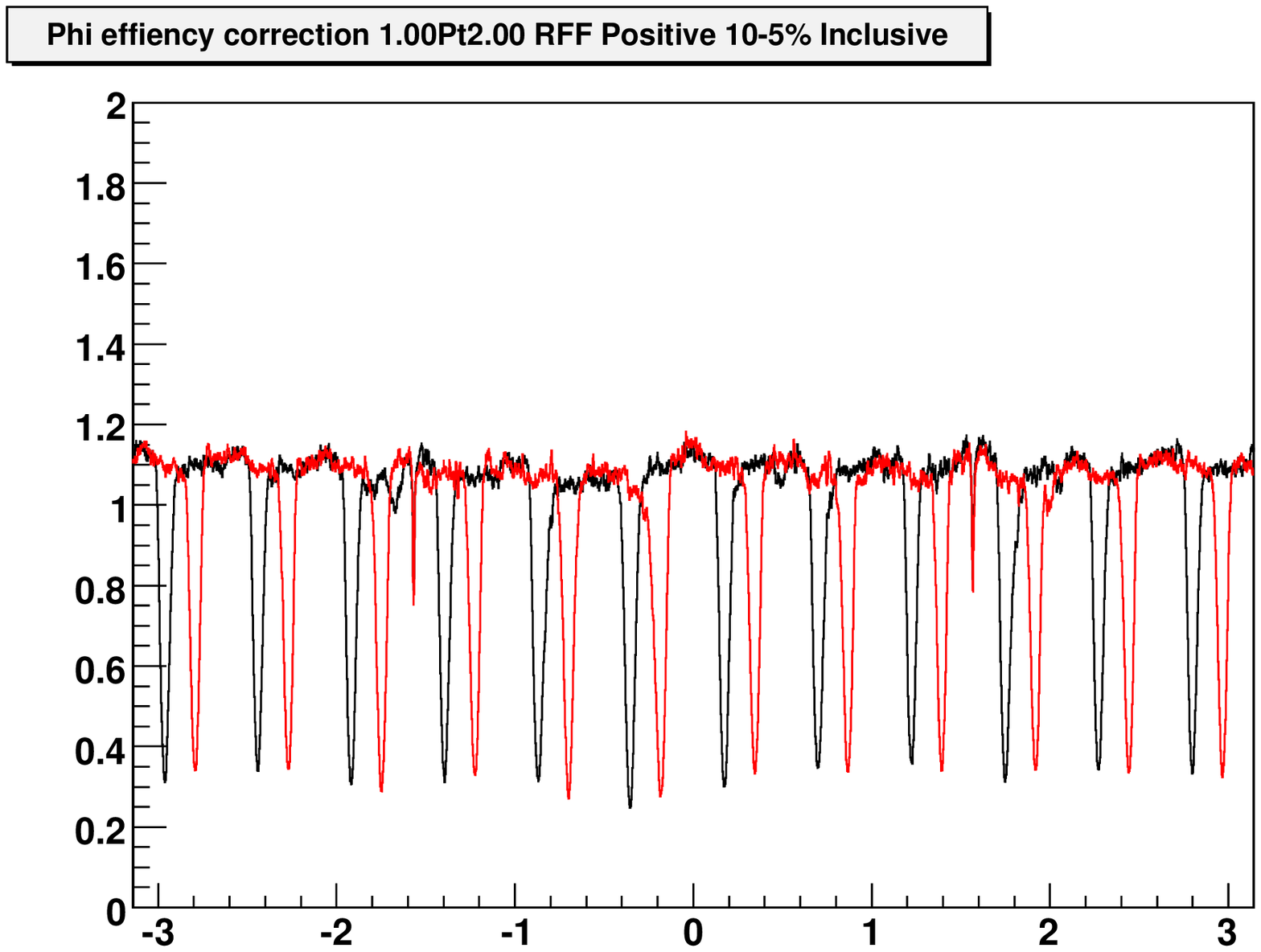}
\end{minipage}
\hfill
\begin{minipage}[t]{0.32\textwidth}
\centering
\includegraphics[width=1.0\textwidth]{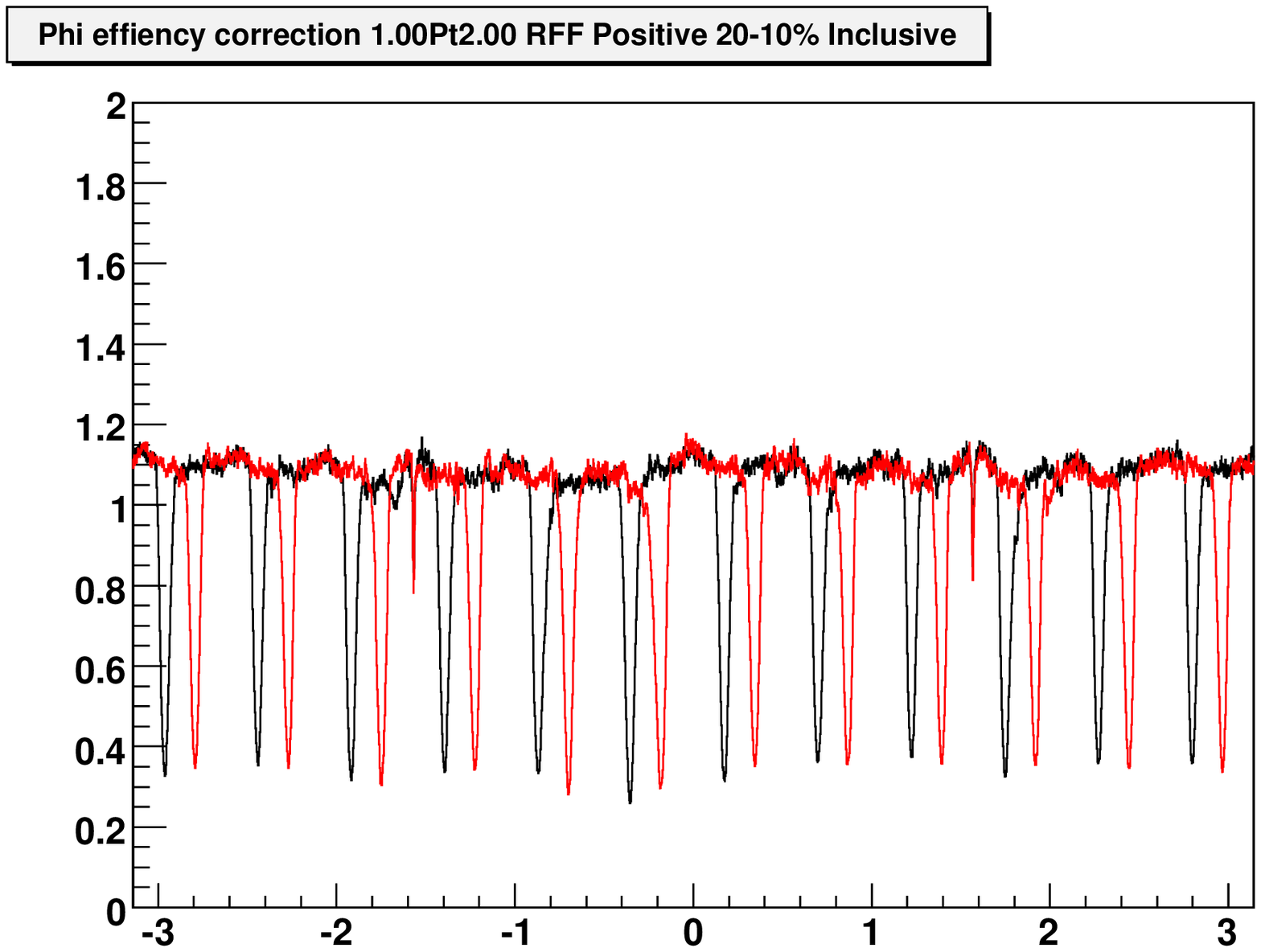}
\includegraphics[width=1.0\textwidth]{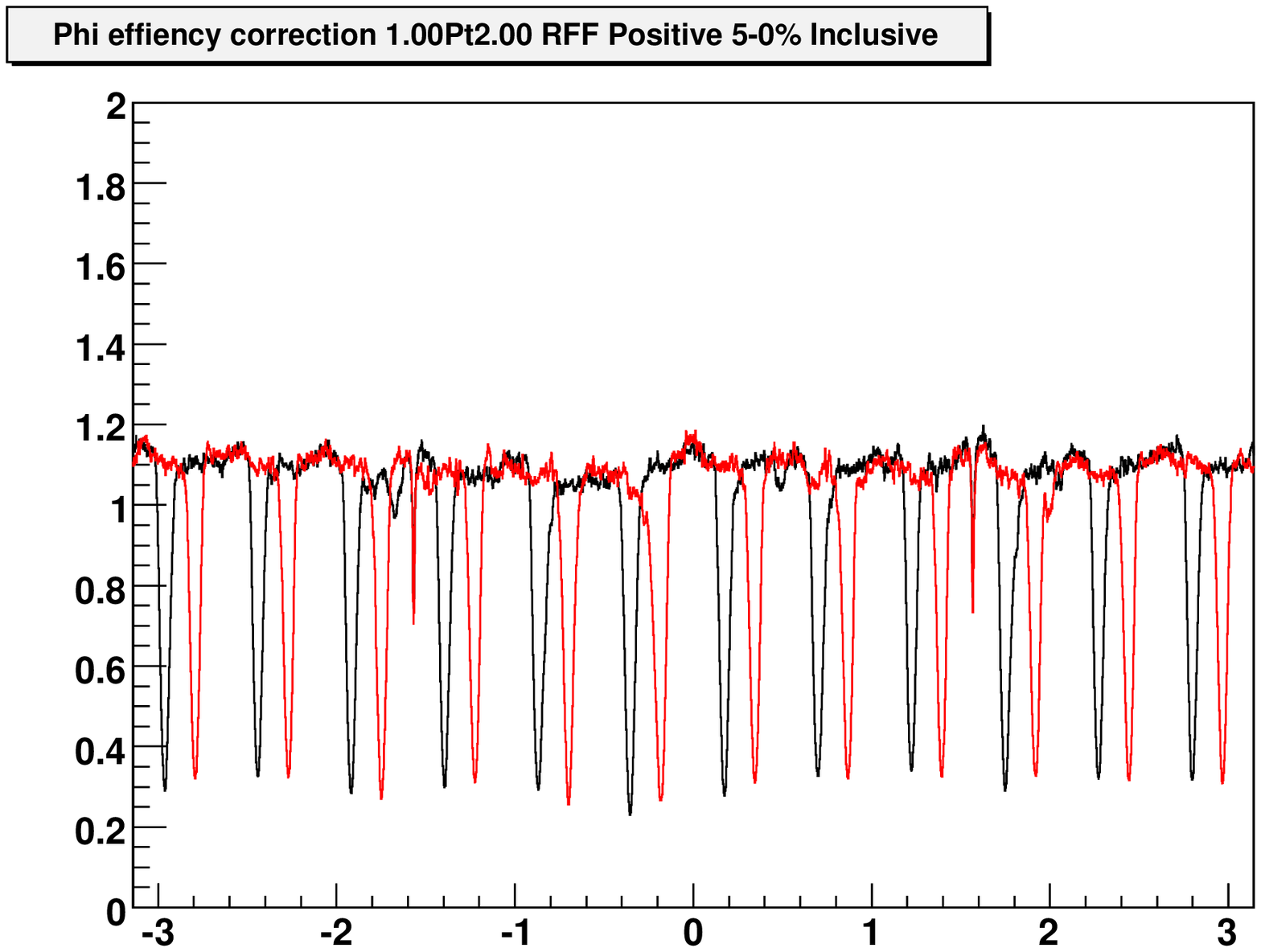}
\end{minipage}

\caption{Same as Fig. 6.16 but for $-0.5$ T magnetic field.}
\label{fig:accCPt0B1}
\end{figure}

\begin{figure}[H]
\hfill
\begin{minipage}[t]{0.32\textwidth}
\centering
\includegraphics[width=1.0\textwidth]{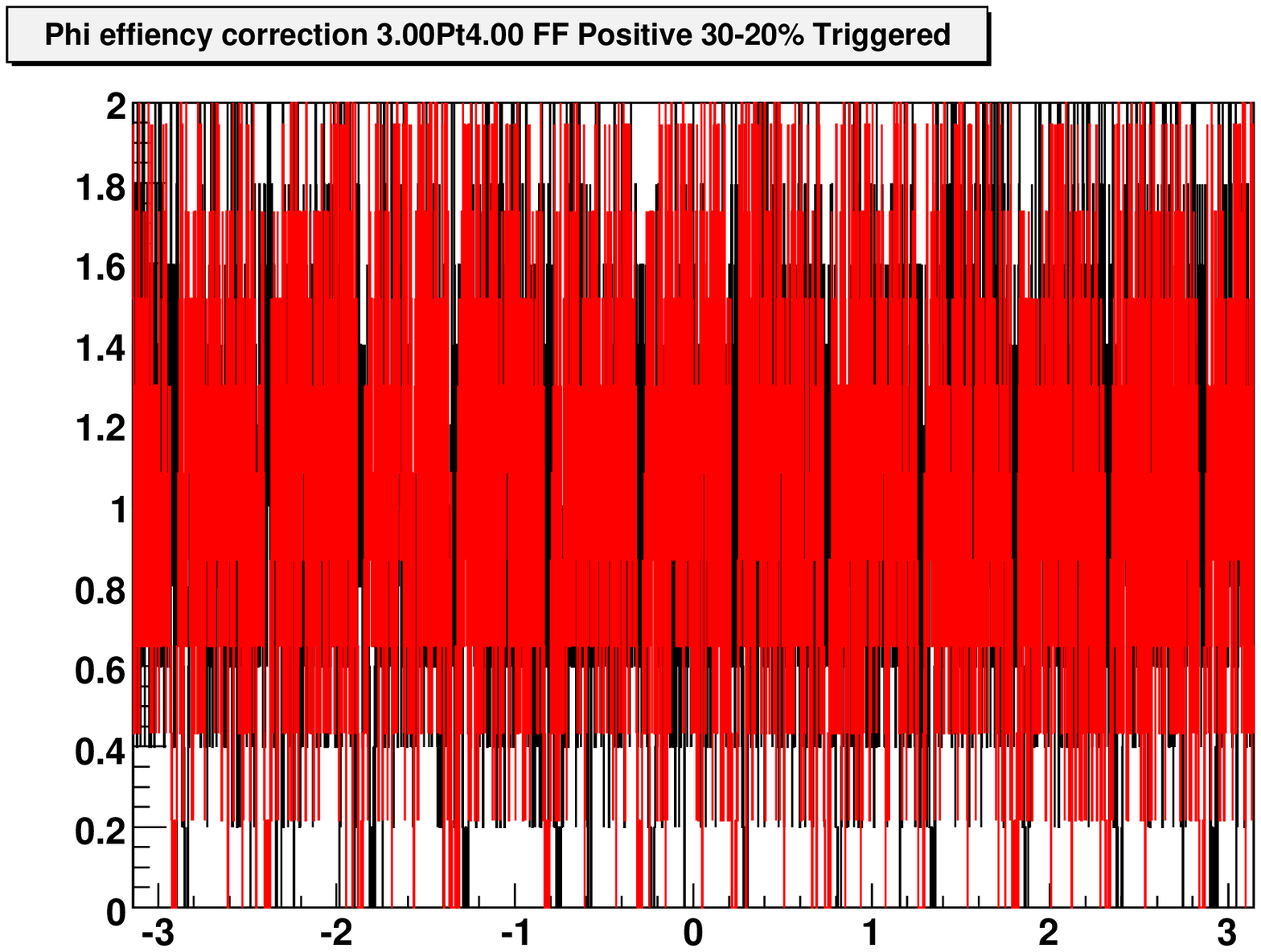}
\includegraphics[width=1.0\textwidth]{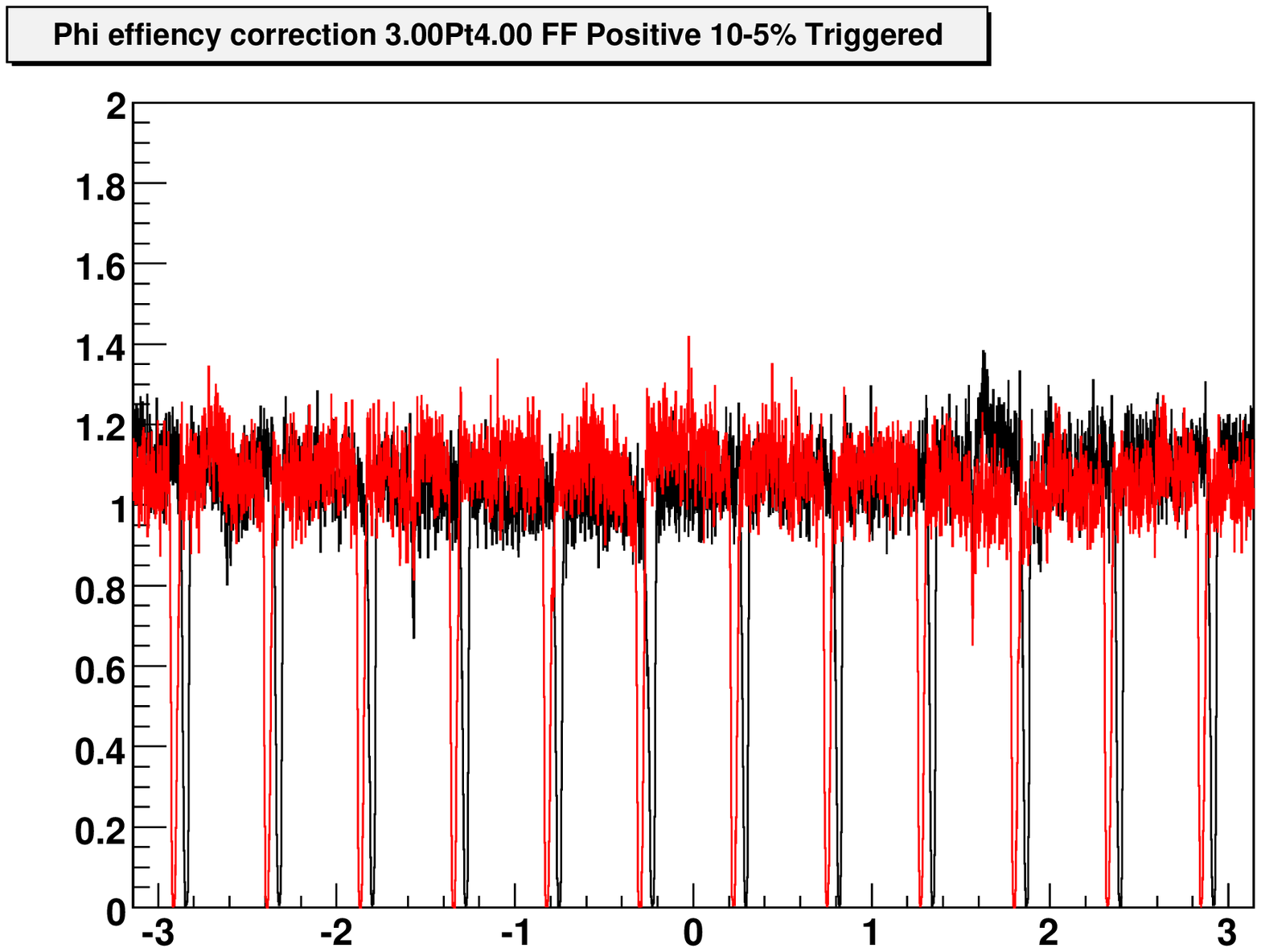}
\end{minipage}
\hfill
\begin{minipage}[t]{0.32\textwidth}
\centering
\includegraphics[width=1.0\textwidth]{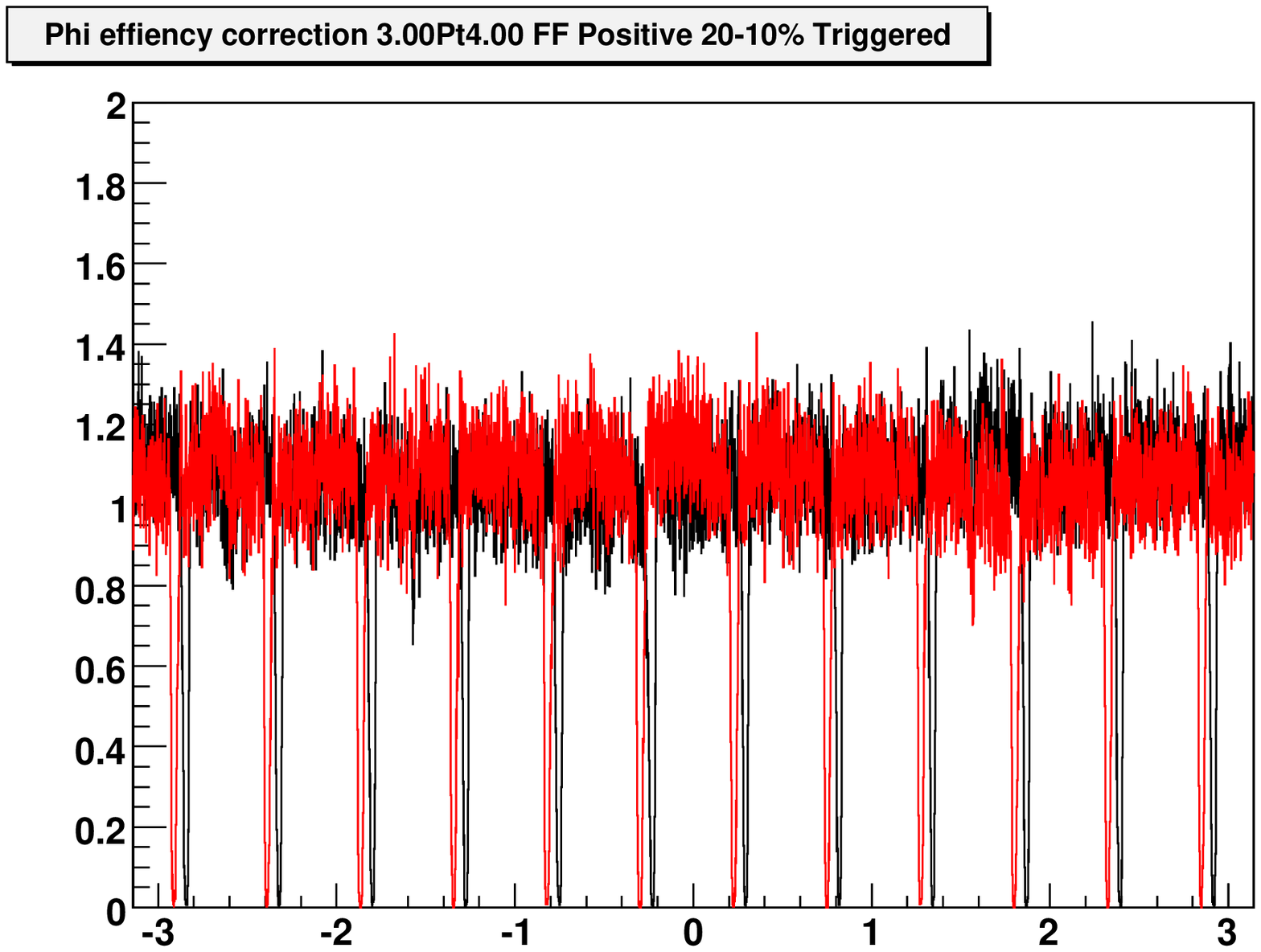}
\includegraphics[width=1.0\textwidth]{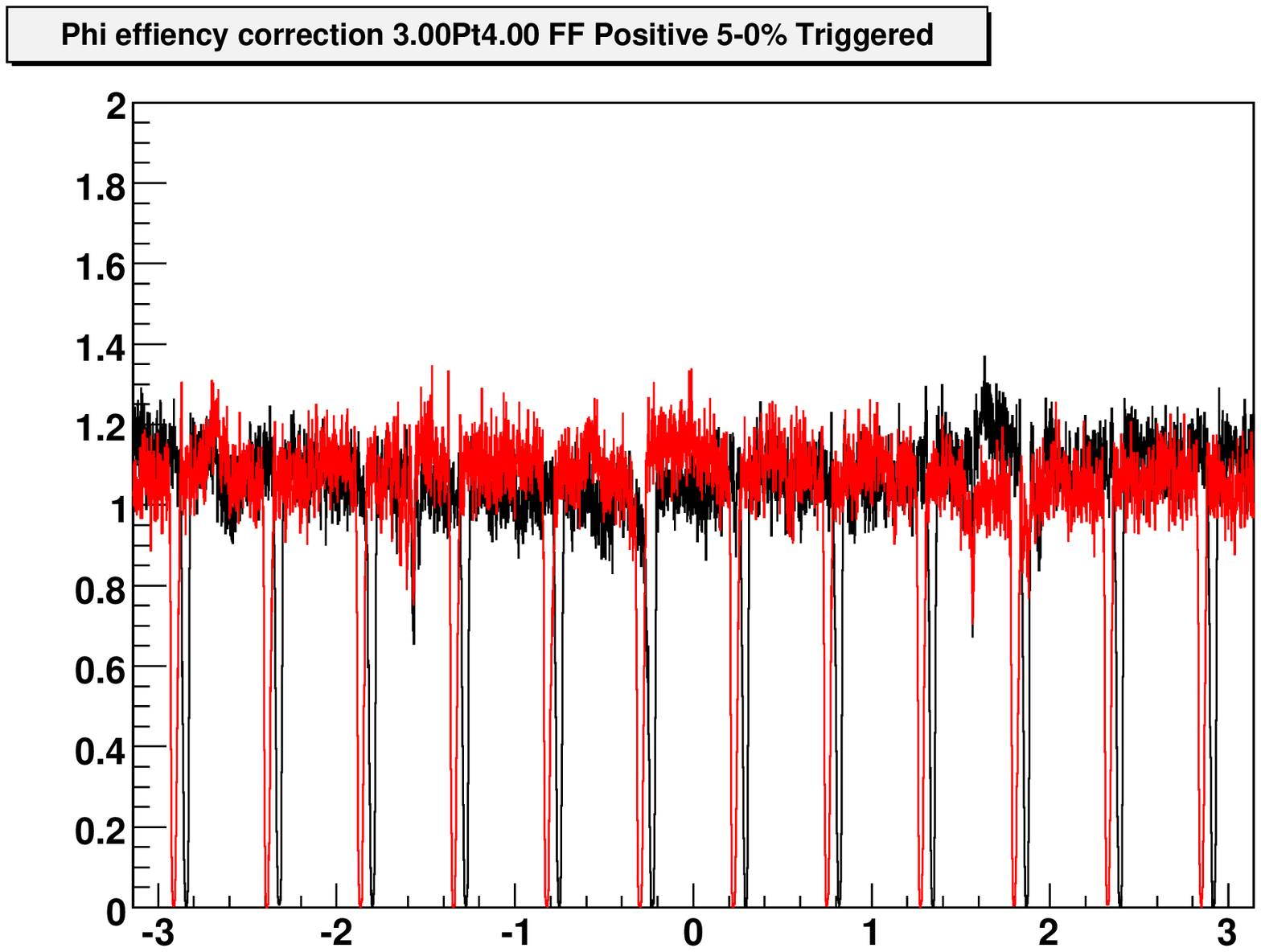}
\end{minipage}
\caption{Same as Fig. 6.16 but for $3<p_T<4$ GeV/c.}
\label{fig:accCPt1B0}
\end{figure}

\begin{figure}[H]
\hfill
\begin{minipage}[t]{0.32\textwidth}
\centering
\includegraphics[width=1.0\textwidth]{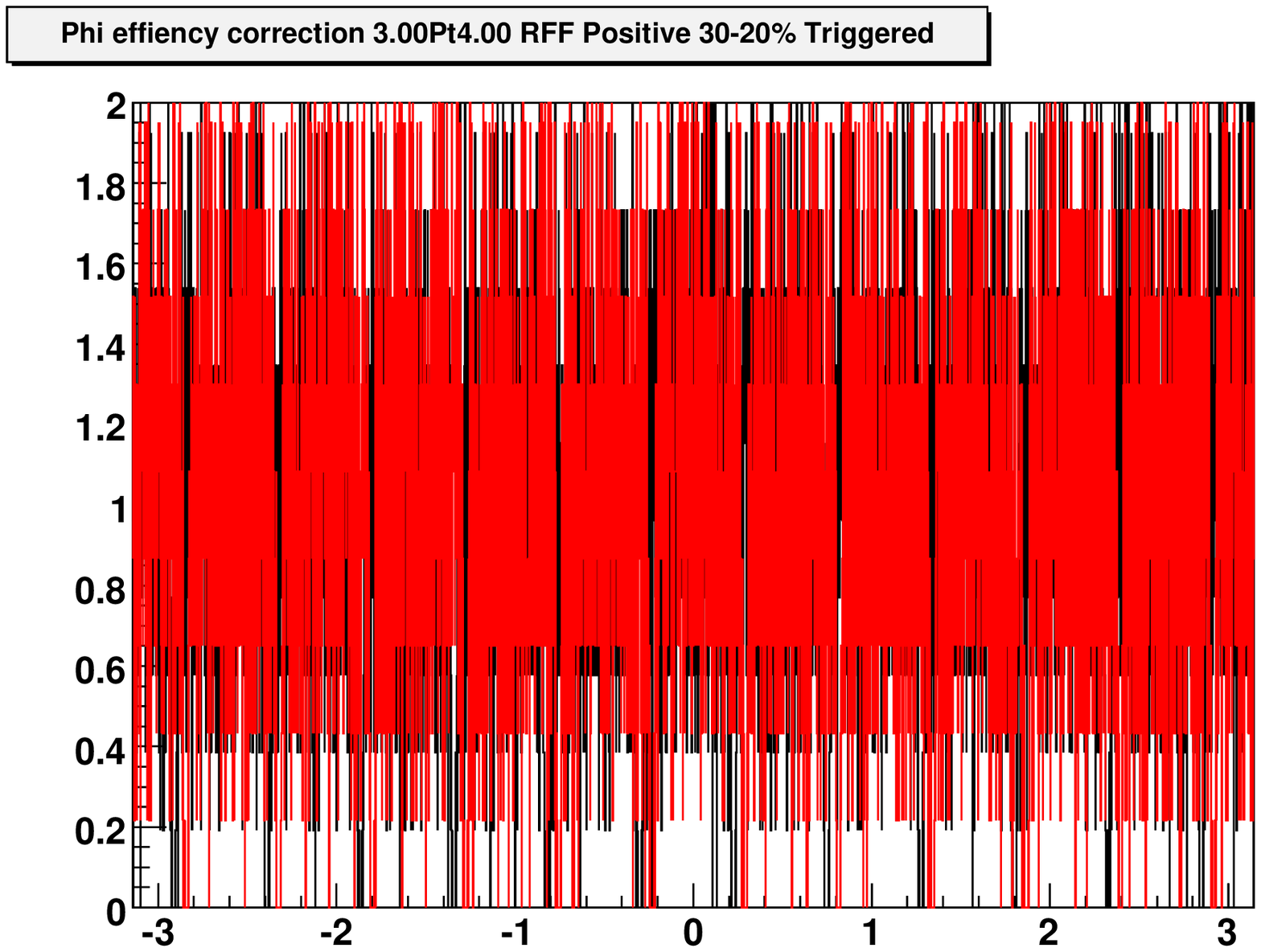}
\includegraphics[width=1.0\textwidth]{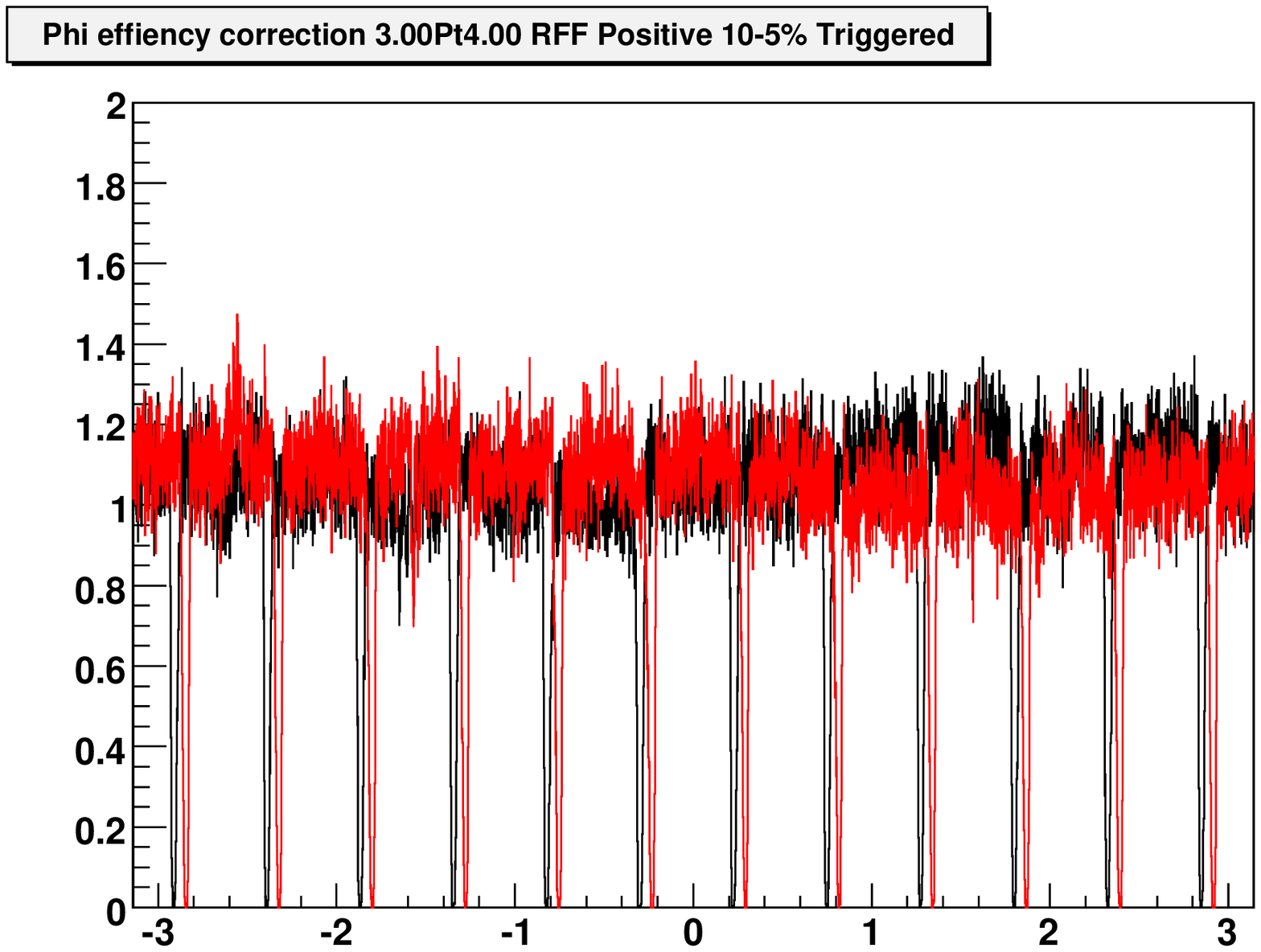}
\end{minipage}
\hfill
\begin{minipage}[t]{0.32\textwidth}
\centering
\includegraphics[width=1.0\textwidth]{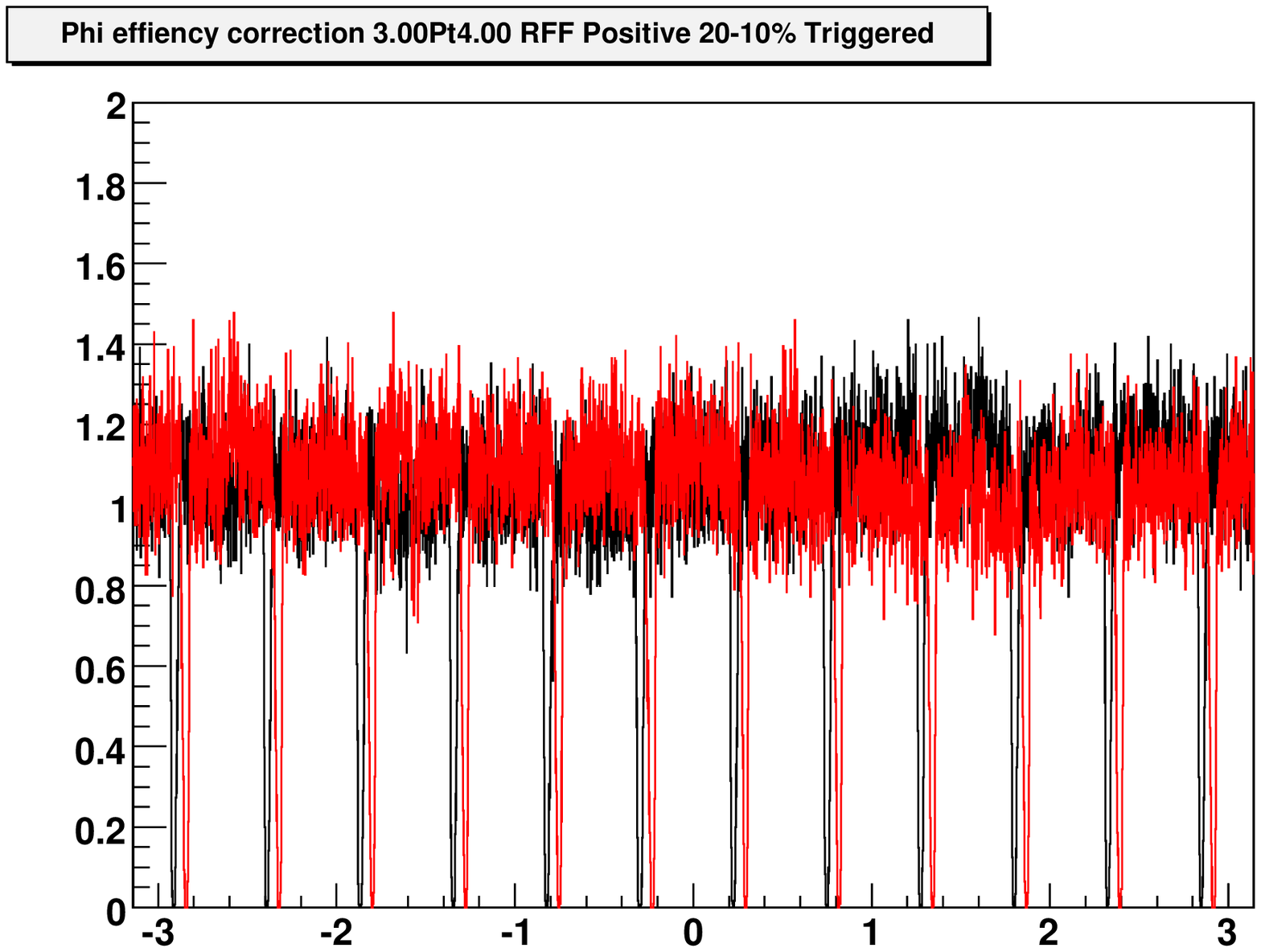}
\includegraphics[width=1.0\textwidth]{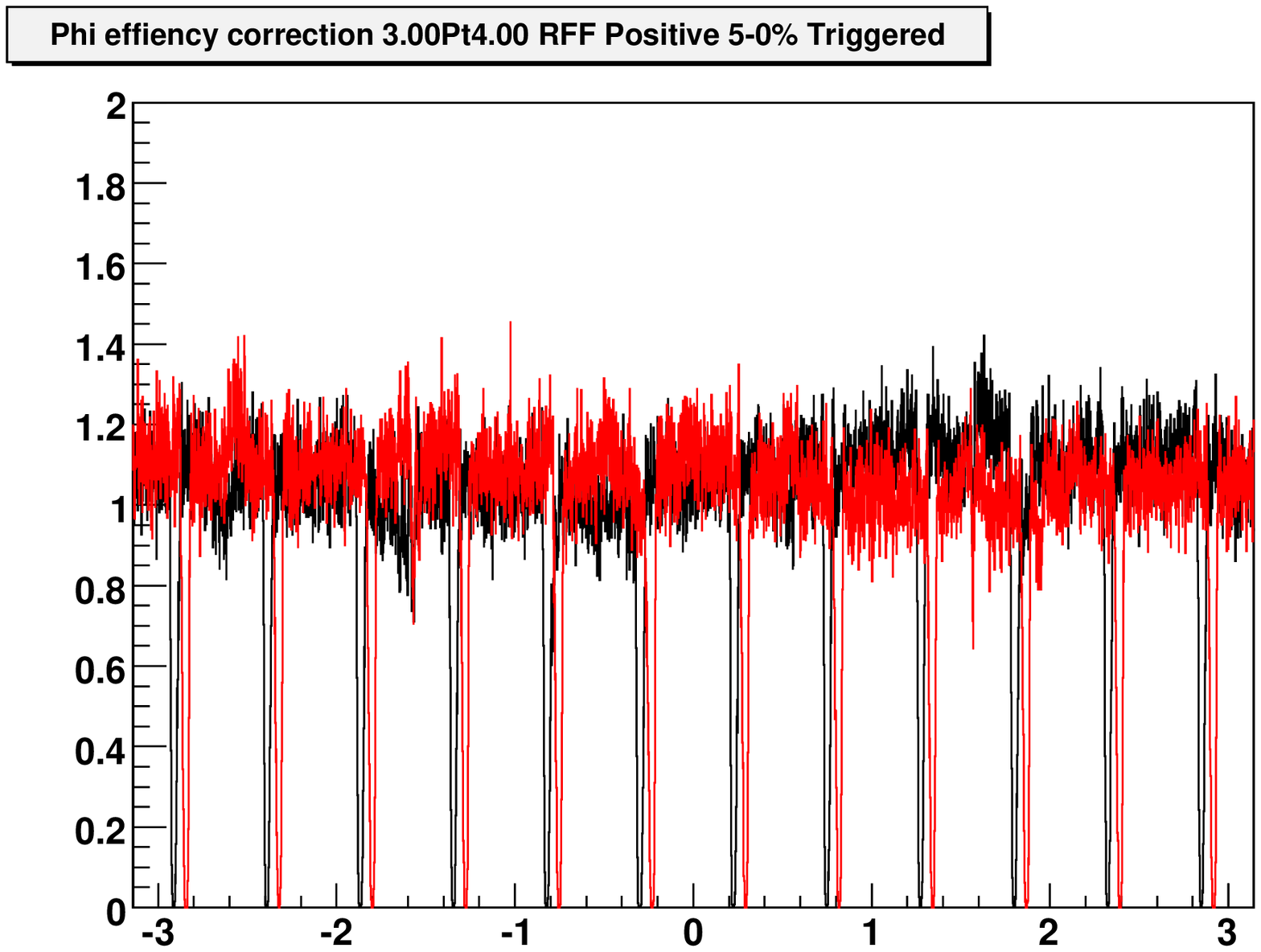}
\end{minipage}
\caption{Same as Fig. 6.16 but for $3<p_T<4$ GeV/c and $-0.5$ T magnetic field.}
\label{fig:accCPt1B1}
\end{figure}

Figures~\ref{fig:accPt0B0}-\ref{fig:accCPt1B1} show the $\phi$-dependence of the acceptance.  This $\phi$-dependence is due to the TPC sector boundries.  The plots are normalized such that the average is one since the $\phi$-average efficiency is obtianed seperately (Fig.~\ref{fig:Eff2}).  Trigger and associated particles were corrected for this $\phi$-dependence on the single particle level.  

\begin{figure}[H]
\centering
\includegraphics[width=1.0\textwidth]{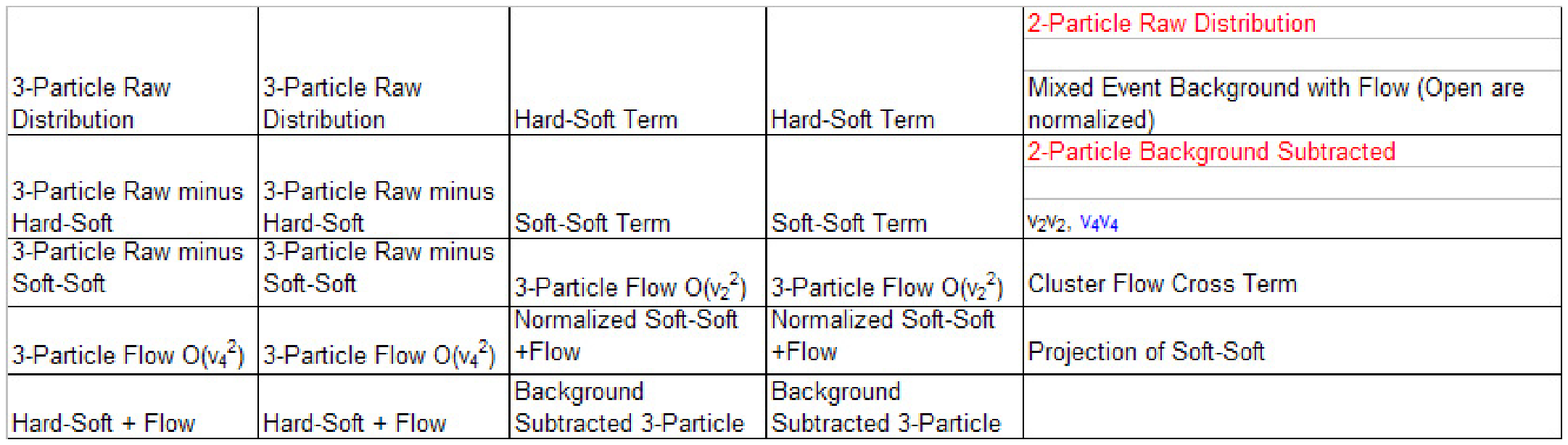}
\caption{Explanation of panels for 3-particle analysis plots.}
\label{fig:3parttable}
\end{figure}

\begin{figure}[H]
\centering
\includegraphics[width=1.0\textwidth]{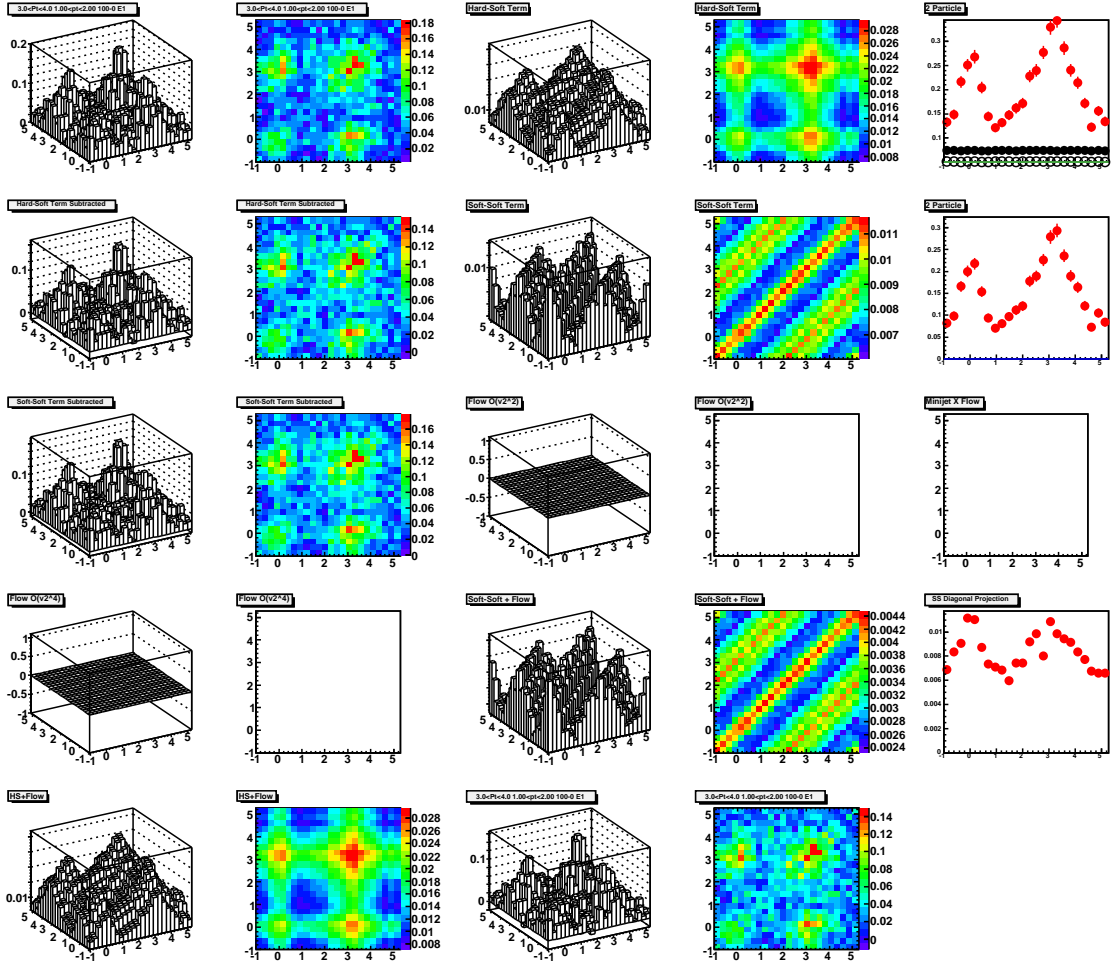}
\caption{Three-particle correlation analysis plots for $3<p_{T}^{trig}<4$ GeV/c and $1<p_{T}<2$ GeV/c in {\it pp} collisions at $\sqrt{s_{NN}}=200$ GeV/c.  Explanation of panels is in preceeding table and text.}
\label{fig:3partpp}
\end{figure}

\begin{figure}[H]
\centering
\includegraphics[width=1.0\textwidth]{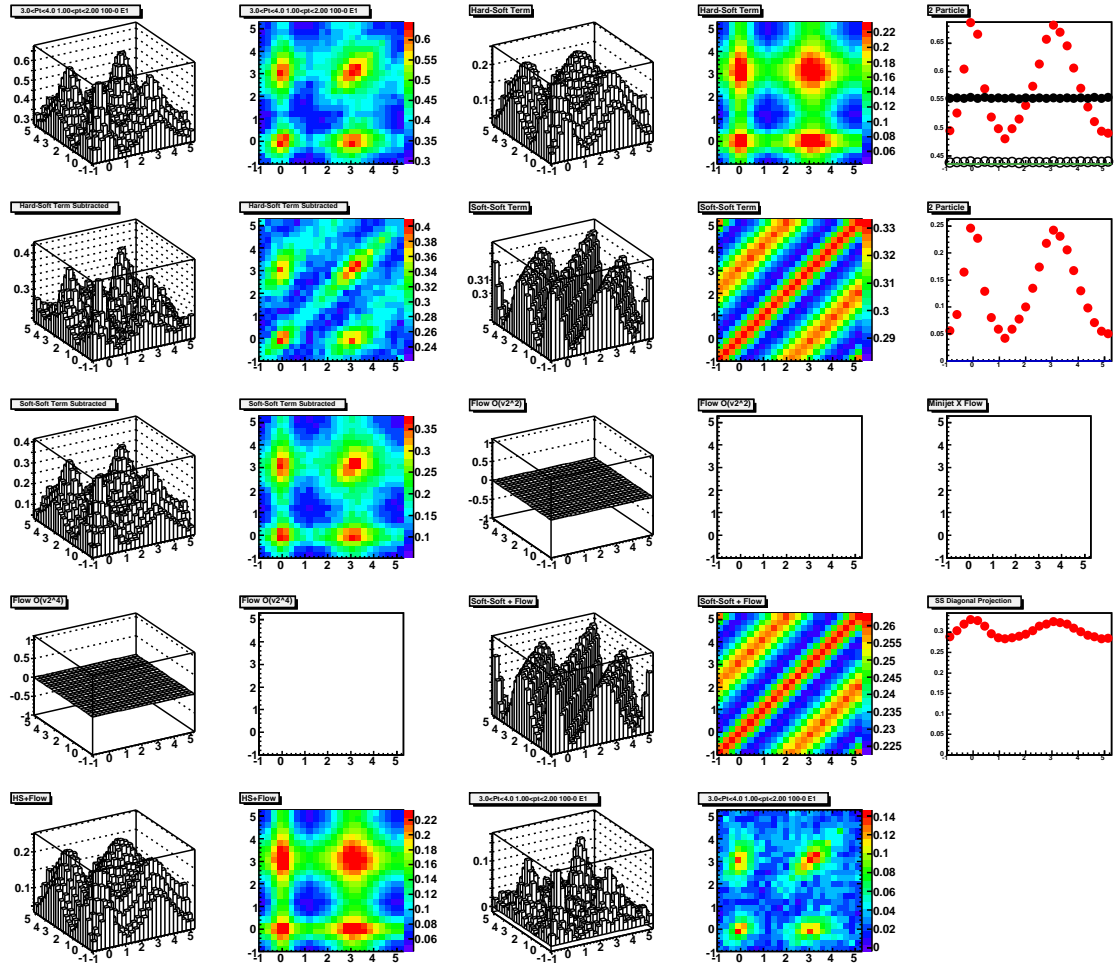}
\caption{Same as Fig. 6.21 but for d+Au collisions.}
\label{fig:3partdAu}
\end{figure}

\begin{figure}[H]
\centering
\includegraphics[width=1.0\textwidth]{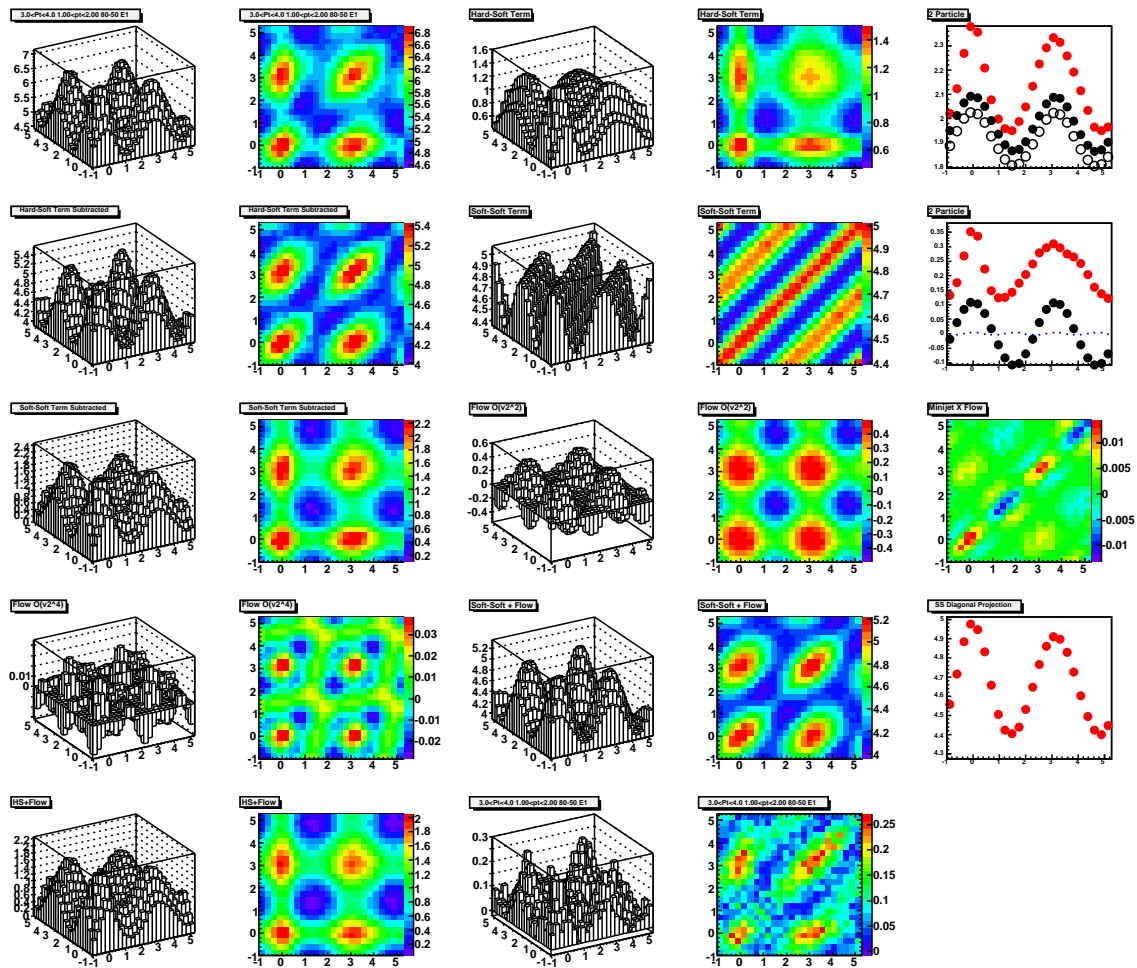}
\caption{Same as Fig. 6.21 but for 50-80\% Au+Au collisions.}
\label{fig:3partAuAu1}
\end{figure}

\begin{figure}[H]
\centering
\includegraphics[width=1.0\textwidth]{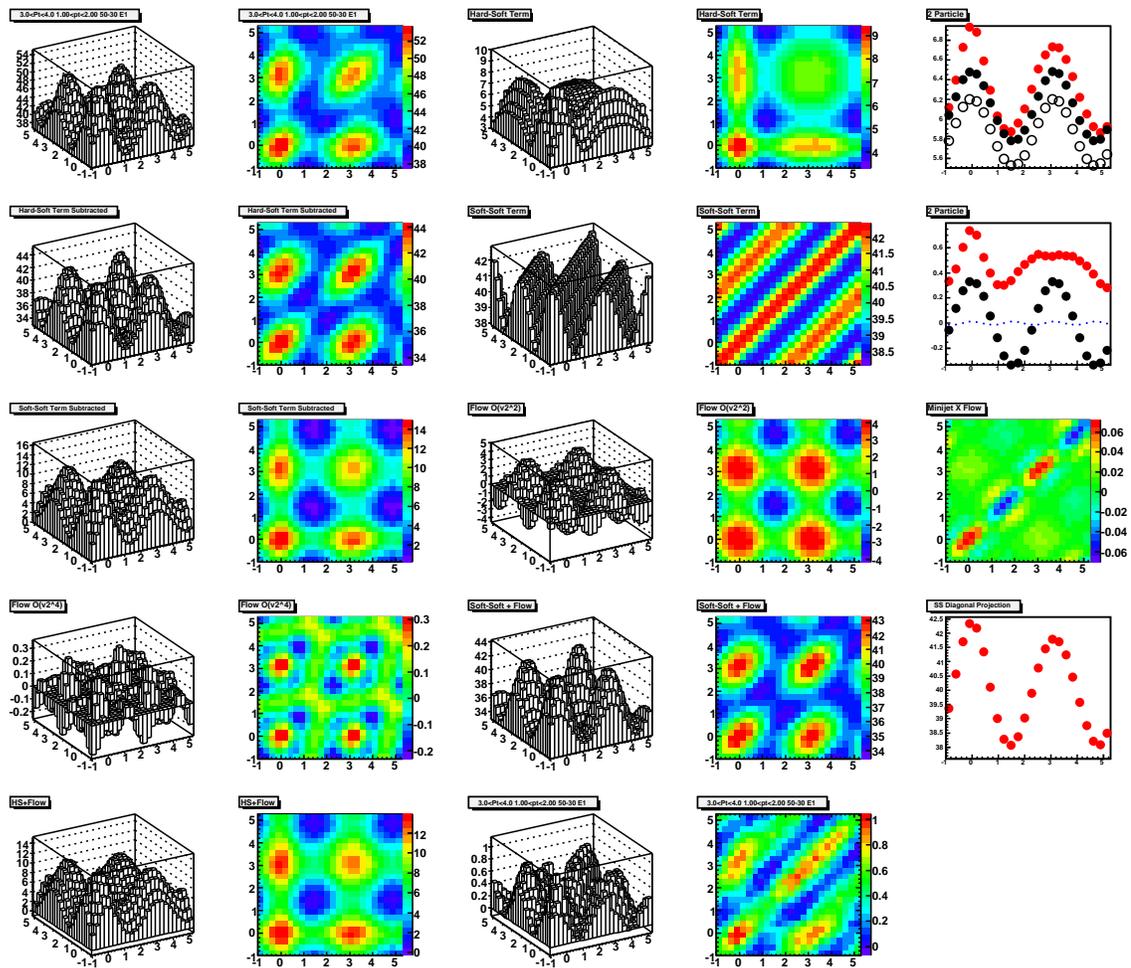}
\caption{Same as Fig. 6.21 but for 30-50\% Au+Au collisions.}
\label{fig:3partAuAu2}
\end{figure}

\begin{figure}[H]
\centering
\includegraphics[width=1.0\textwidth]{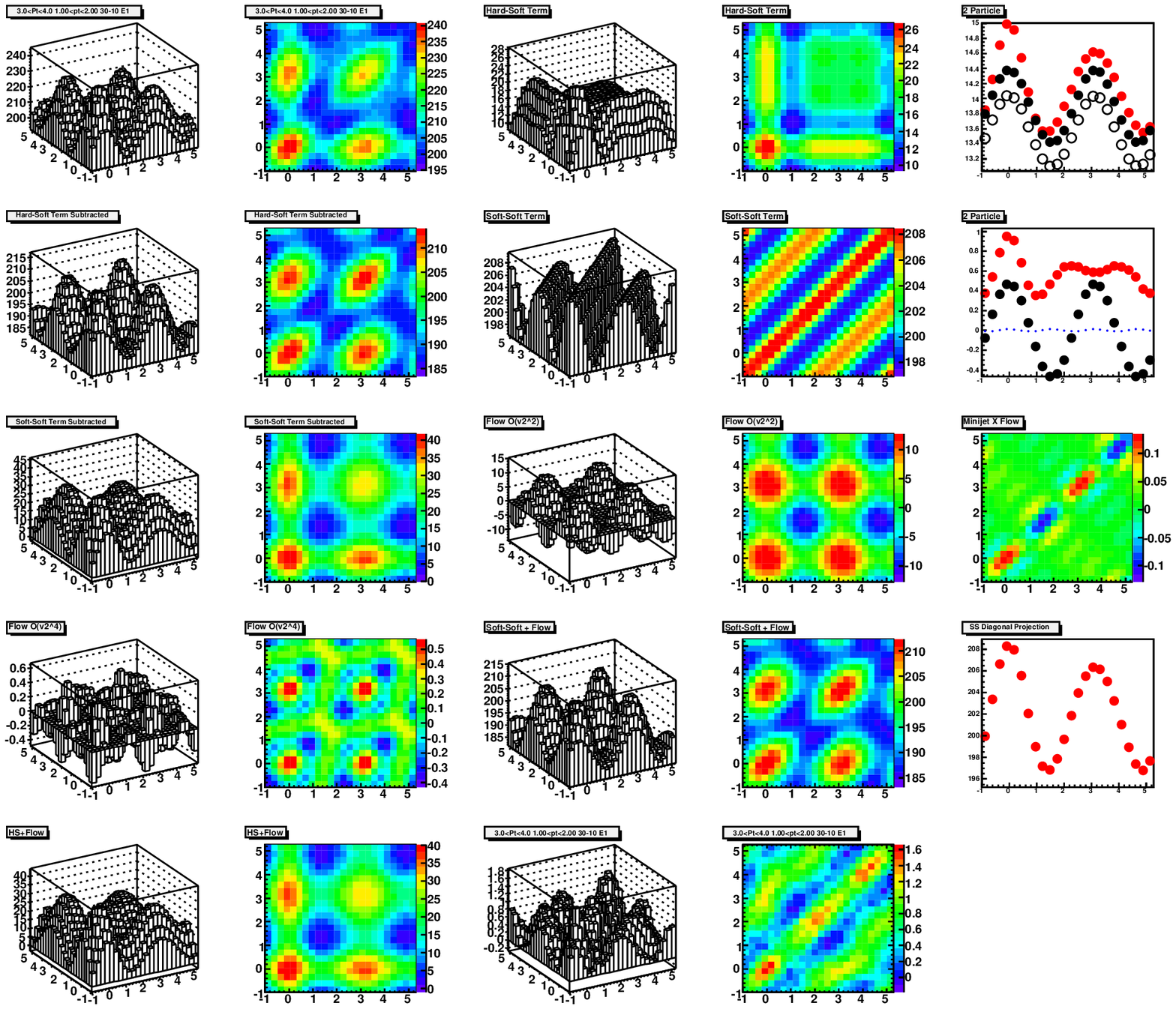}
\caption{Same as Fig. 6.21 but for 10-30\% Au+Au collisions.}
\label{fig:3partAuAu3}
\end{figure}

\begin{figure}[H]
\centering
\includegraphics[width=1.0\textwidth]{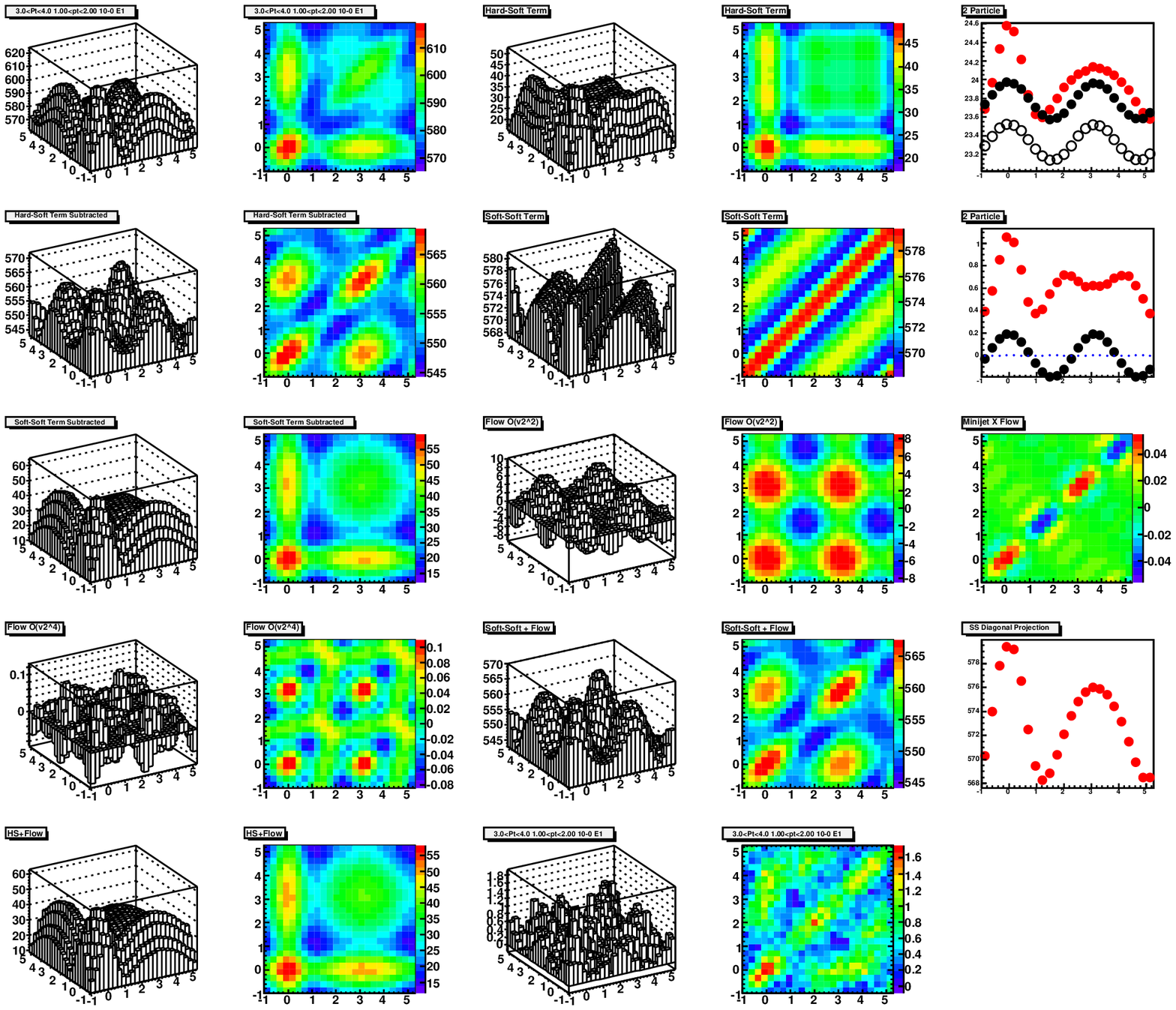}
\caption{Same as Fig. 6.21 but for 0-10\% Au+Au collisions.}
\label{fig:3partAuAu4}
\end{figure}

\begin{figure}[H]
\centering
\includegraphics[width=1.0\textwidth]{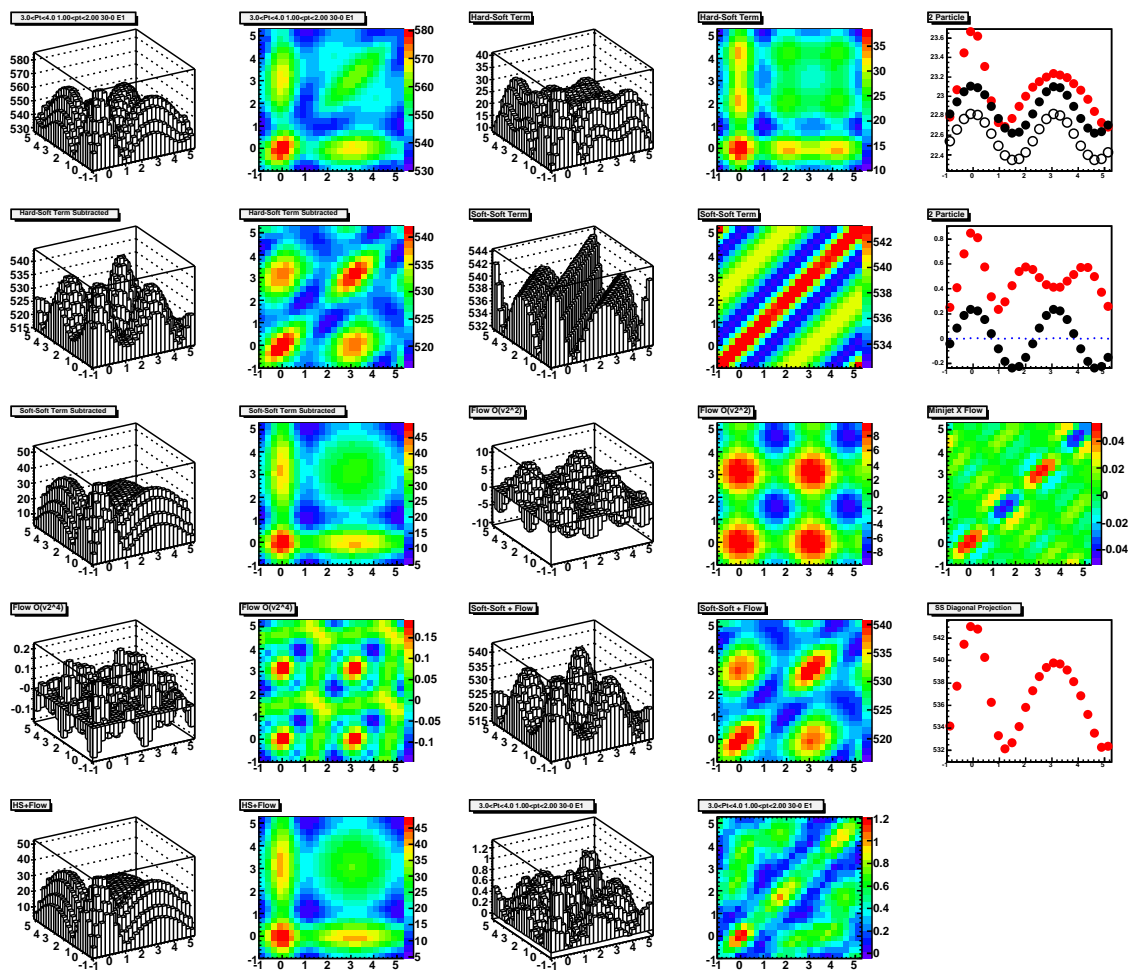}         
\caption{Same as Fig. 6.21 but for 0-12\% ZDC triggered Au+Au collisions.}
\label{fig:3partAuAu5}   
\end{figure}

\begin{figure}[H]
\centering
\includegraphics[width=1.0\textwidth]{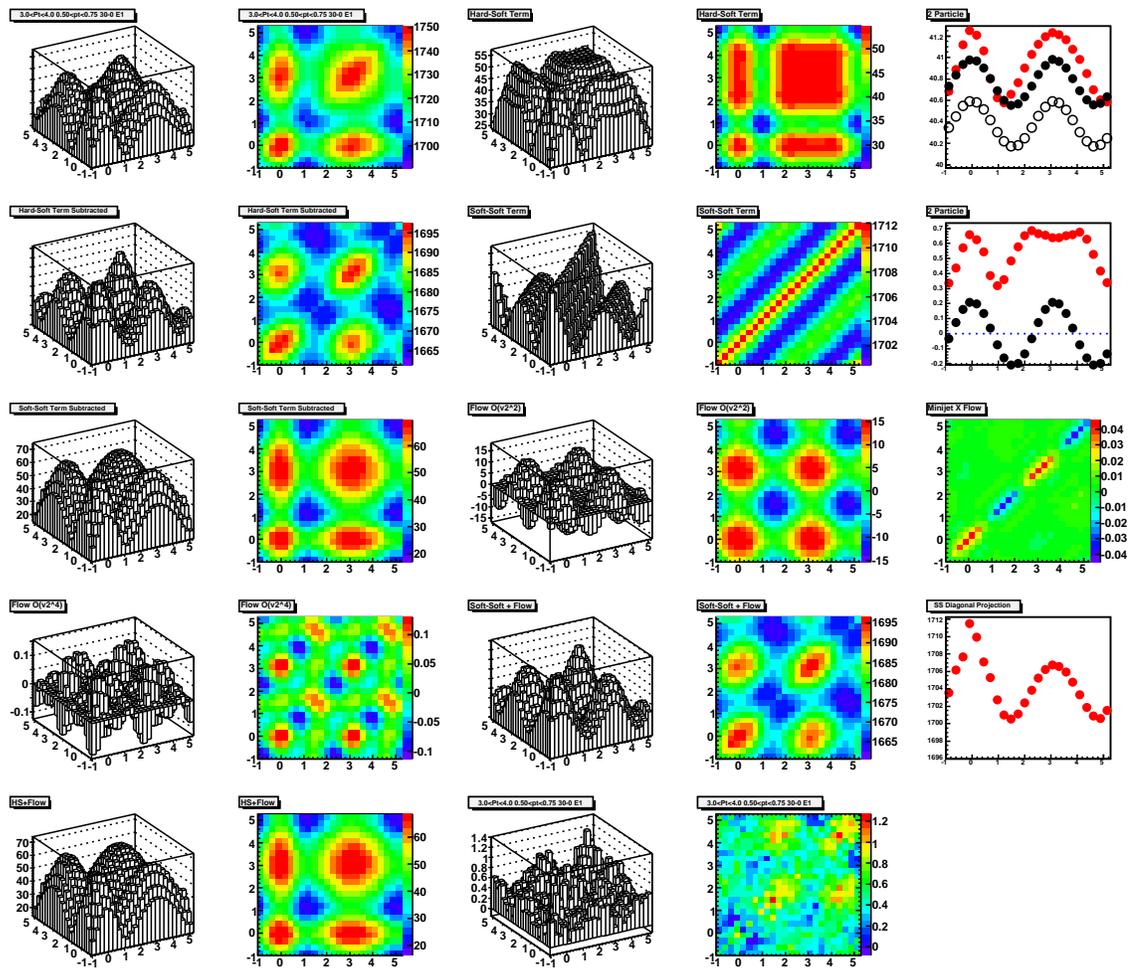}         
\caption{Same as Fig. 6.27 but for $0.5<p_{T}^{Assoc}<0.75$.}
\label{fig:3partAuAu6}   
\end{figure}

\begin{figure}[H]
\centering
\includegraphics[width=1.0\textwidth]{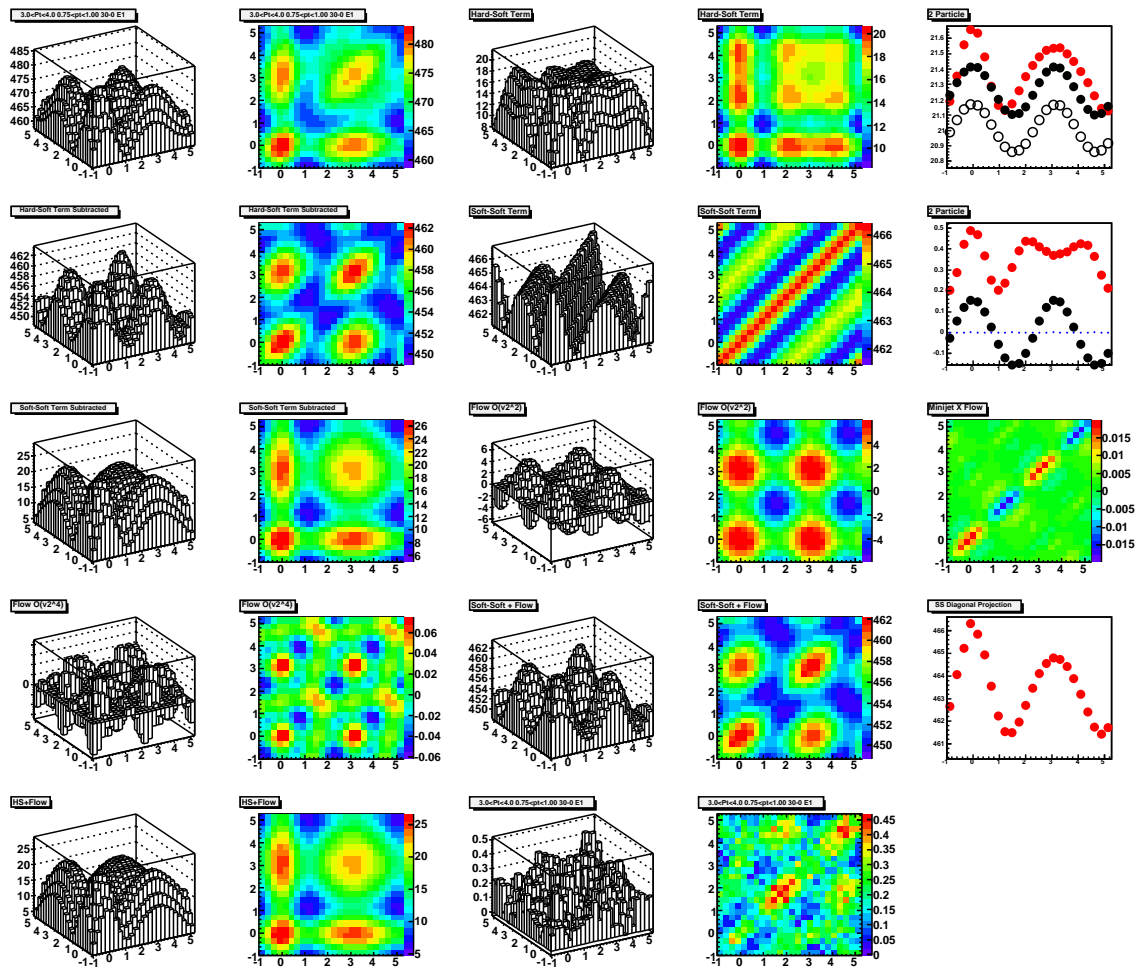}         
\caption{Same as Fig. 6.27 but for $0.75<p_{T}^{Assoc}<1.0$.}
\label{fig:3partAuAu7}   
\end{figure}

\begin{figure}[H]
\centering
\includegraphics[width=1.0\textwidth]{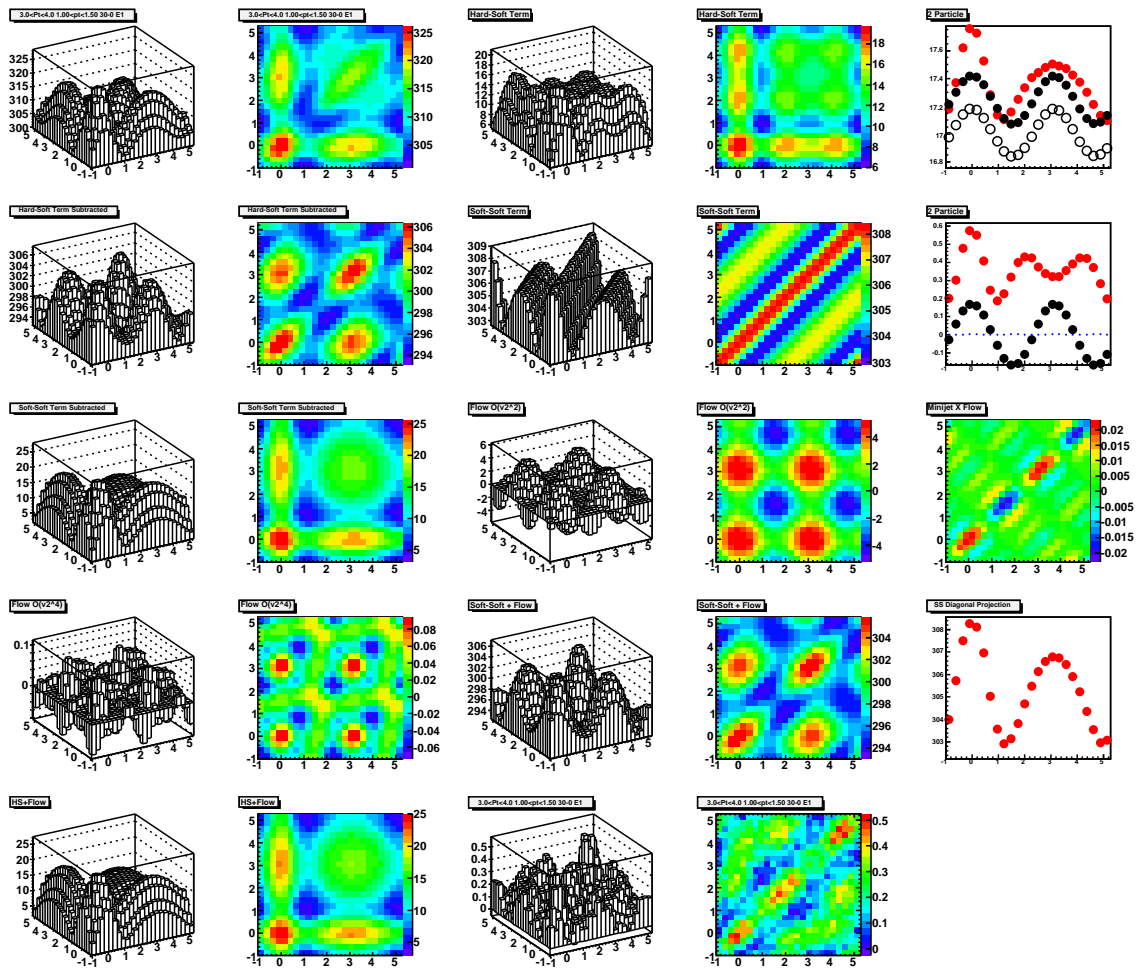}         
\caption{Same as Fig. 6.27 but for $1.0<p_{T}^{Assoc}<1.5$.}
\label{fig:3partAuAu8}   
\end{figure}

\begin{figure}[H]
\centering
\includegraphics[width=1.0\textwidth]{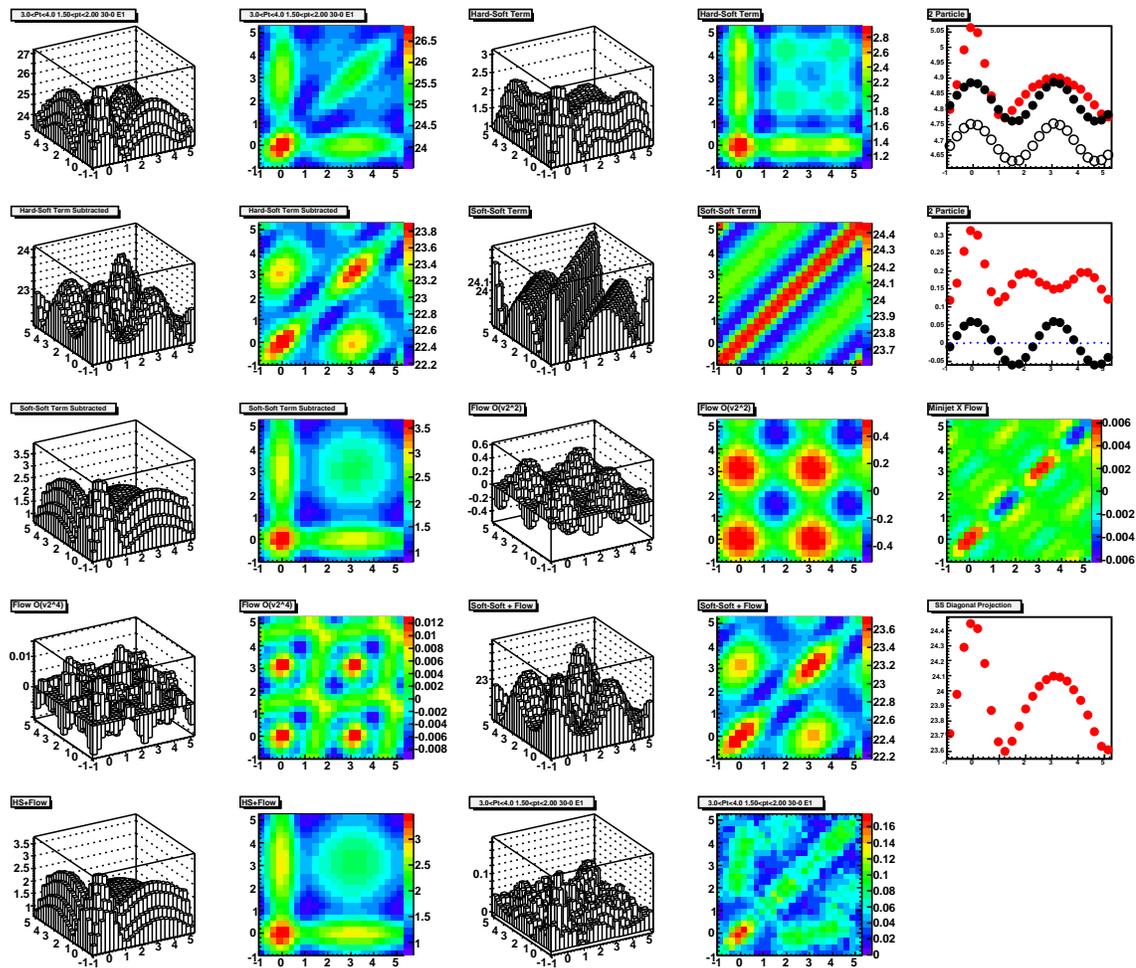}         
\caption{Same as Fig. 6.27 but for $1.5<p_{T}^{Assoc}<2.0$.}
\label{fig:3partAuAu9}   
\end{figure}

\begin{figure}[H]
\centering
\includegraphics[width=1.0\textwidth]{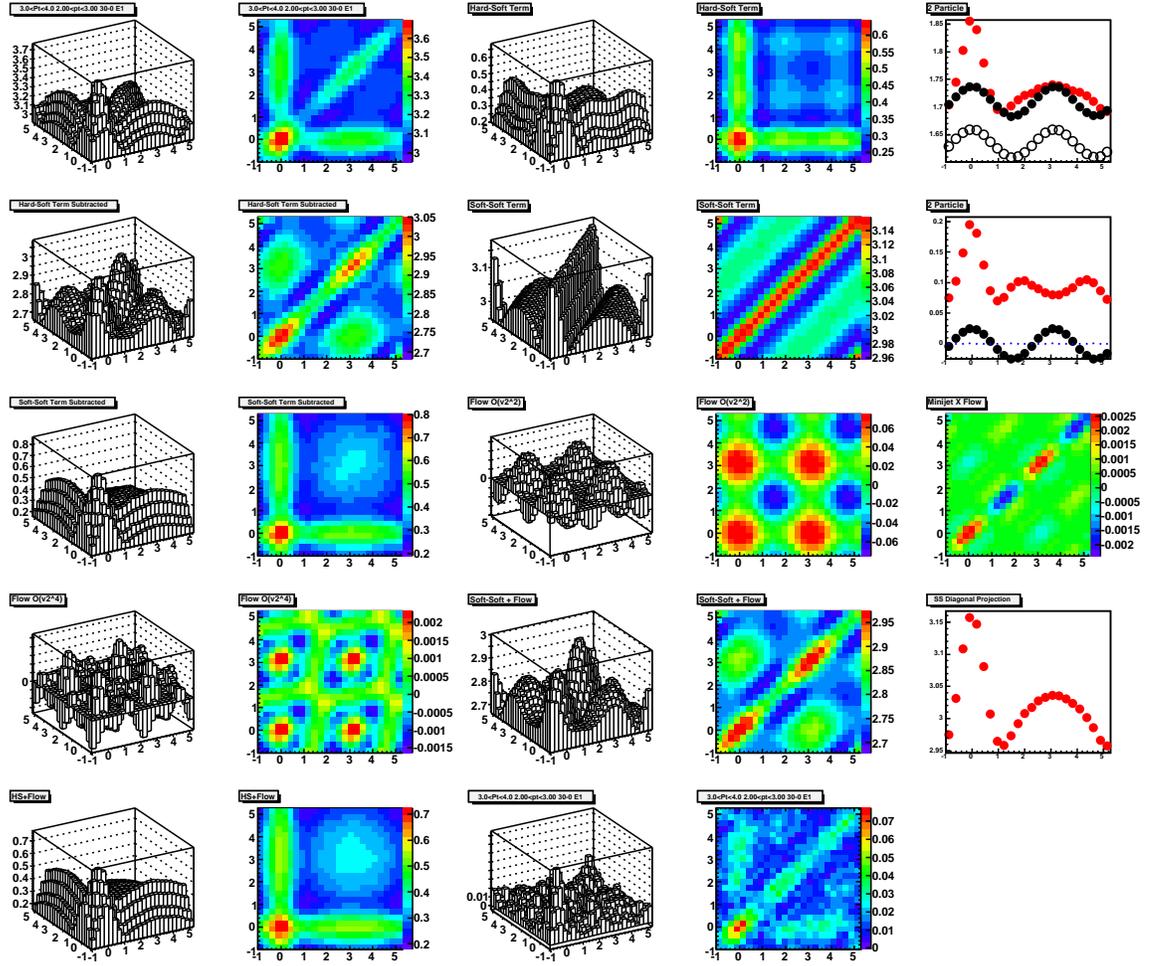}         
\caption{Same as Fig. 6.27 but for $2<p_{T}^{Assoc}<3$.}
\label{fig:3partAuAu10}   
\end{figure}

\begin{figure}[H]
\centering
\includegraphics[width=1.0\textwidth]{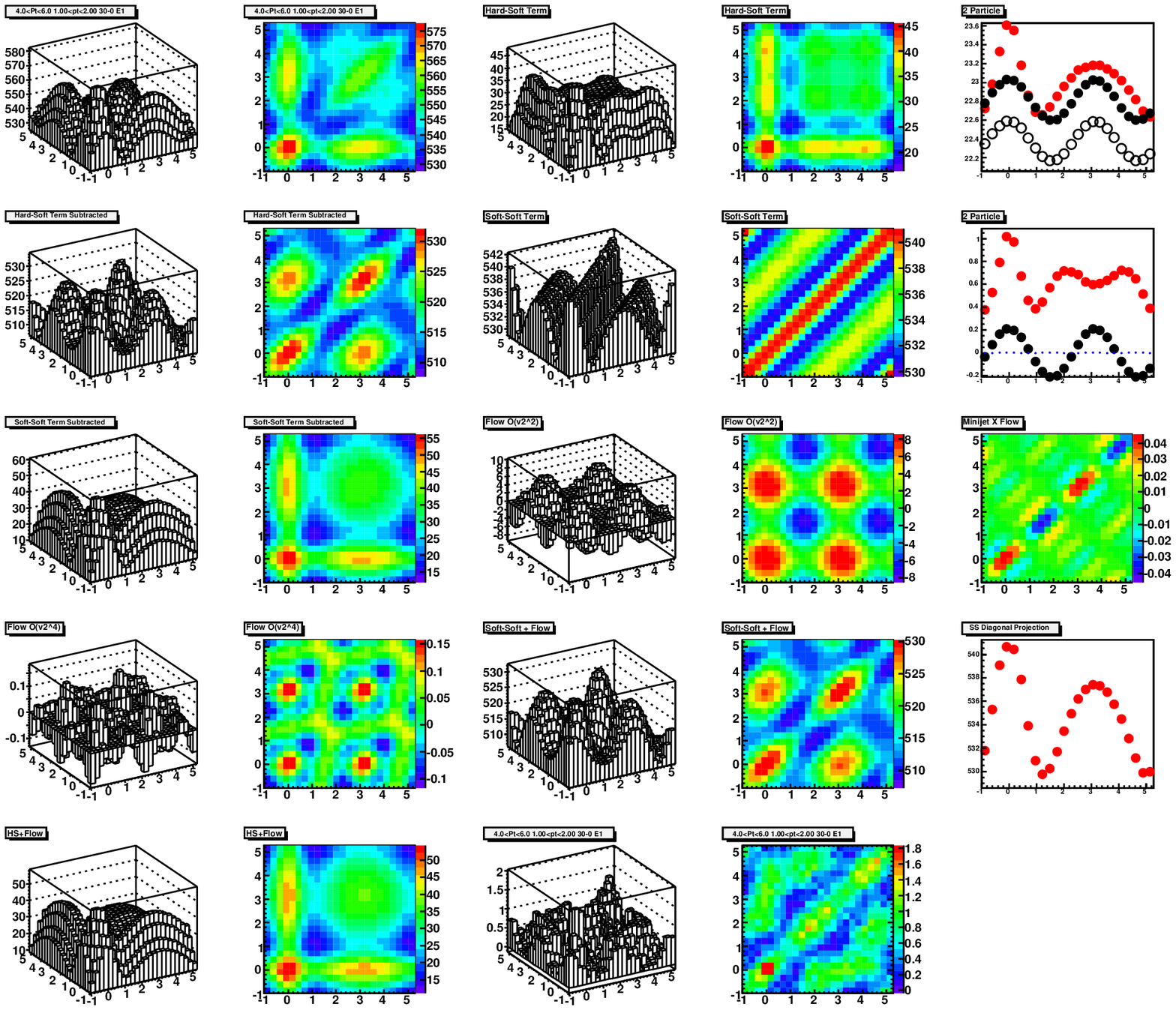}         
\caption{Same as Fig. 6.27 but for $4<p_{T}^{Trig}<6$.}
\label{fig:3partAuAu11}   
\end{figure}

\begin{figure}[H]
\centering
\includegraphics[width=1.0\textwidth]{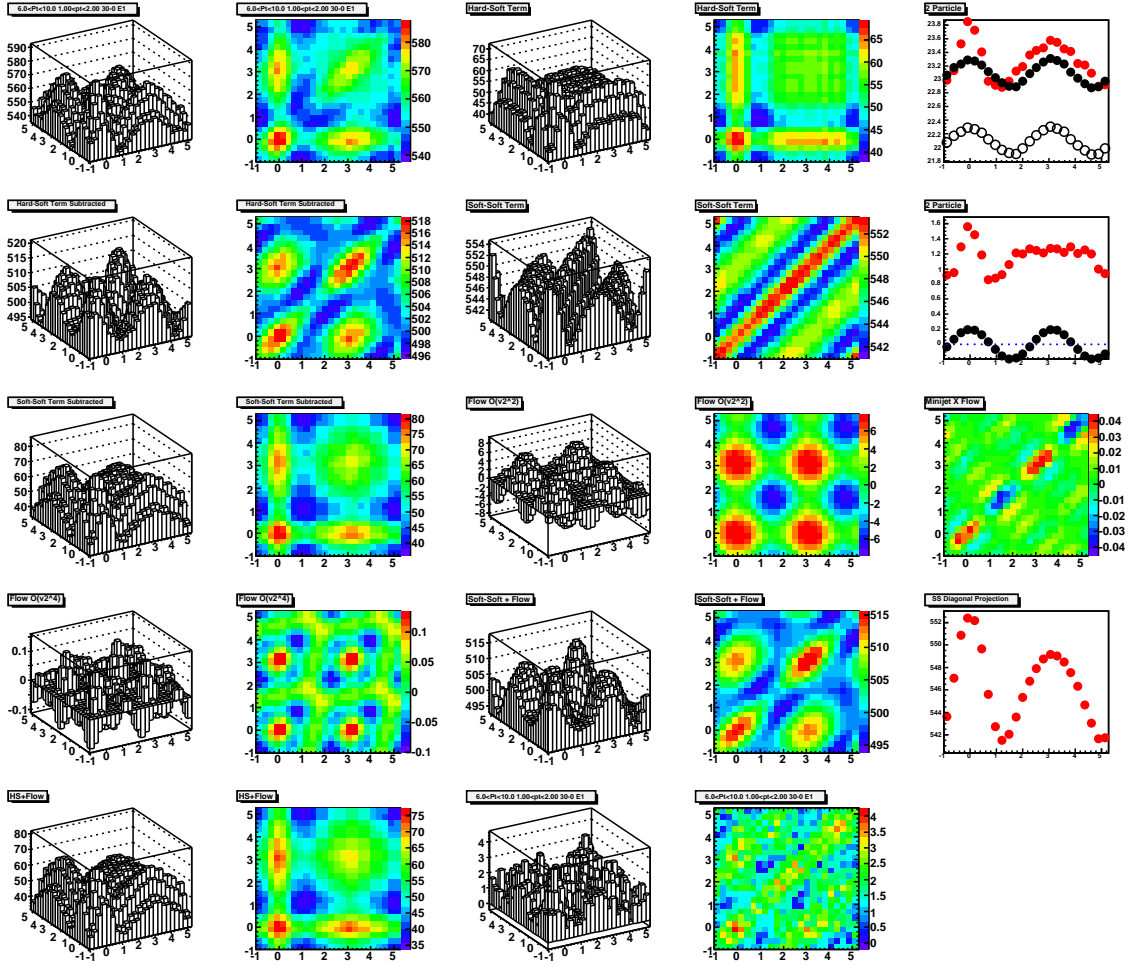}         
\caption{Same as Fig. 6.27 but for $4<p_{T}^{Trig}<6$.}
\label{fig:3partAuAu12}   
\end{figure}

Figure~\ref{fig:3parttable} is a graphical explanation of the various panels in Figures~\ref{fig:3partpp}-\ref{fig:3partAuAu12}.  These figures are the working plots for the 3-particle analysis in Chapt. 4.  In these figures, the top left and top left center panels show the raw 3-particle correlation function in two different representations.  The top center and top right center show the hard-soft background term.  The top right panel shows the raw 2-particle correlation function in red, the mixed event background with flow modulation from $v_2$ and $v_4$ in solid black, and the normalized to 3-particle ZYAM mixed event background with flow modulation in open black.  In the second row, the left and left center plots shown the raw 3-particle correlation function with the hard-soft background subtracted.  The center and right center plots show the soft-soft background term.  The right plots show the background subtracted 2-particle correlation function in red, the flow from $v_2$ in black, and the flow from $v_4$ in blue.  In the third row, the left and left center plots show the raw 3-particle correlation function with the soft-soft background subtracted.  The center and right center panel shows the flow contribution from elliptic flow, $v_2$, between the trigger and associated particles.  The right panel shows the flow background from the non-flow structure in the soft-soft term flowing with the trigger particle.  In the fourth row, the left and left center panels show the additional 3-particle flow between the trigger and associated particle when $v_4$ contributions are considered.  The center and right center panels show the normalized sum of the soft-soft and flow terms.  The right panel is an off-diagonal projection of the soft-soft term.  In the bottom row, the left and left center panels show the sum of the hard-soft terms and the flow terms.  The center and right center panels show the background subtracted 3-particle correlation.  The bottom right plot is not used.

\begin{figure}[H]
\hfill
\begin{minipage}[t]{0.19\textwidth}
\includegraphics[width=1\textwidth]{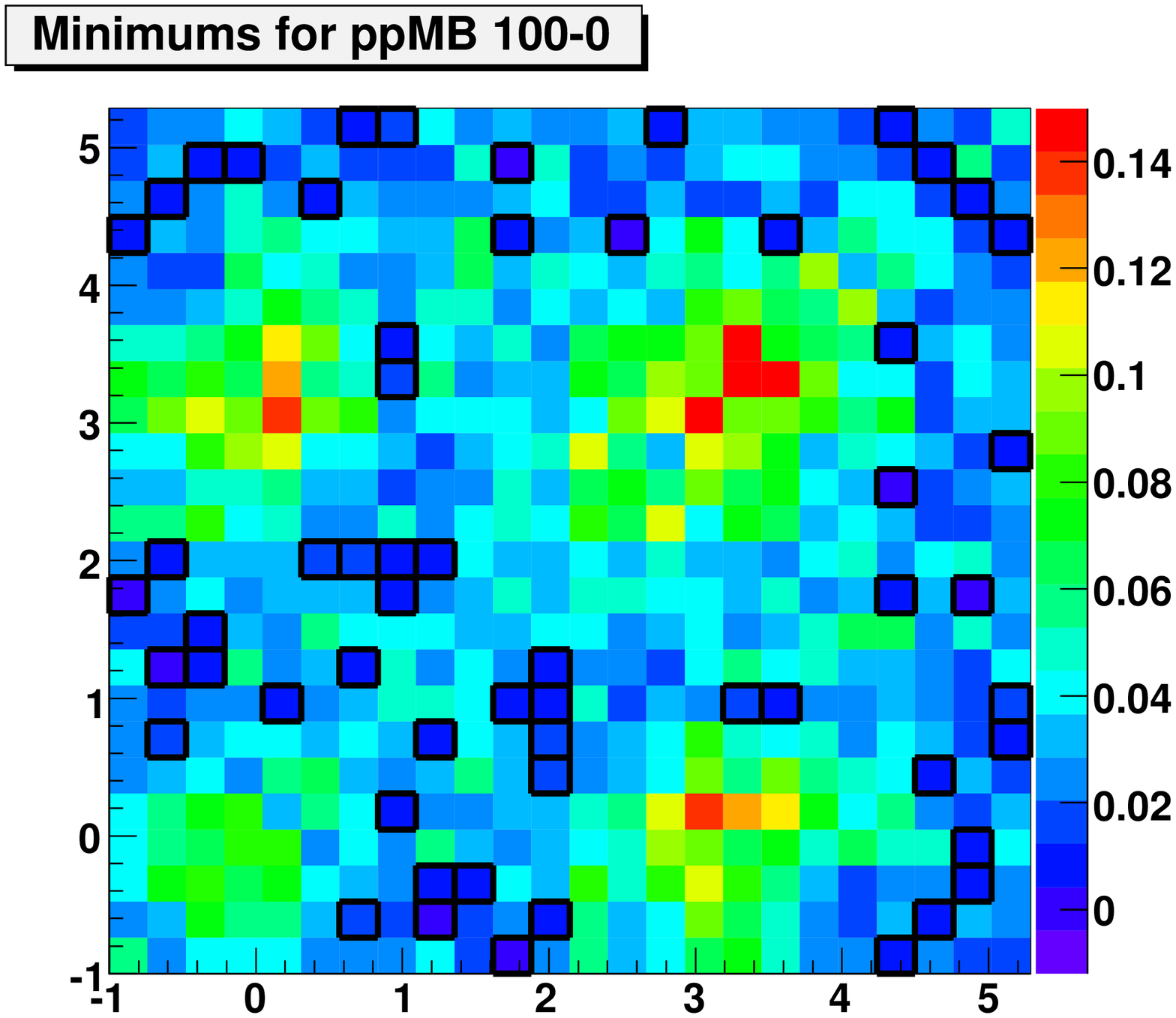}
\includegraphics[width=1\textwidth]{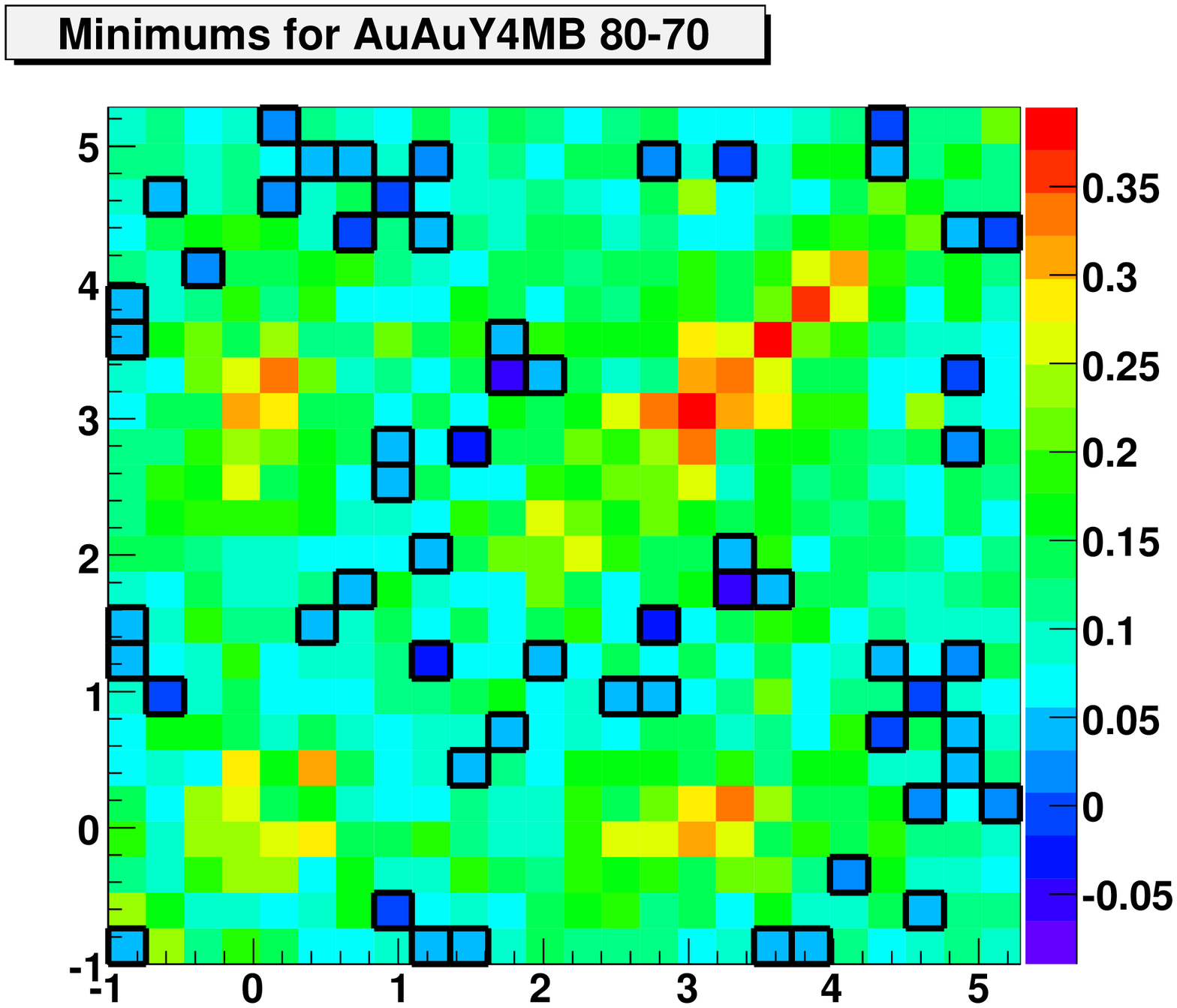}
\includegraphics[width=1\textwidth]{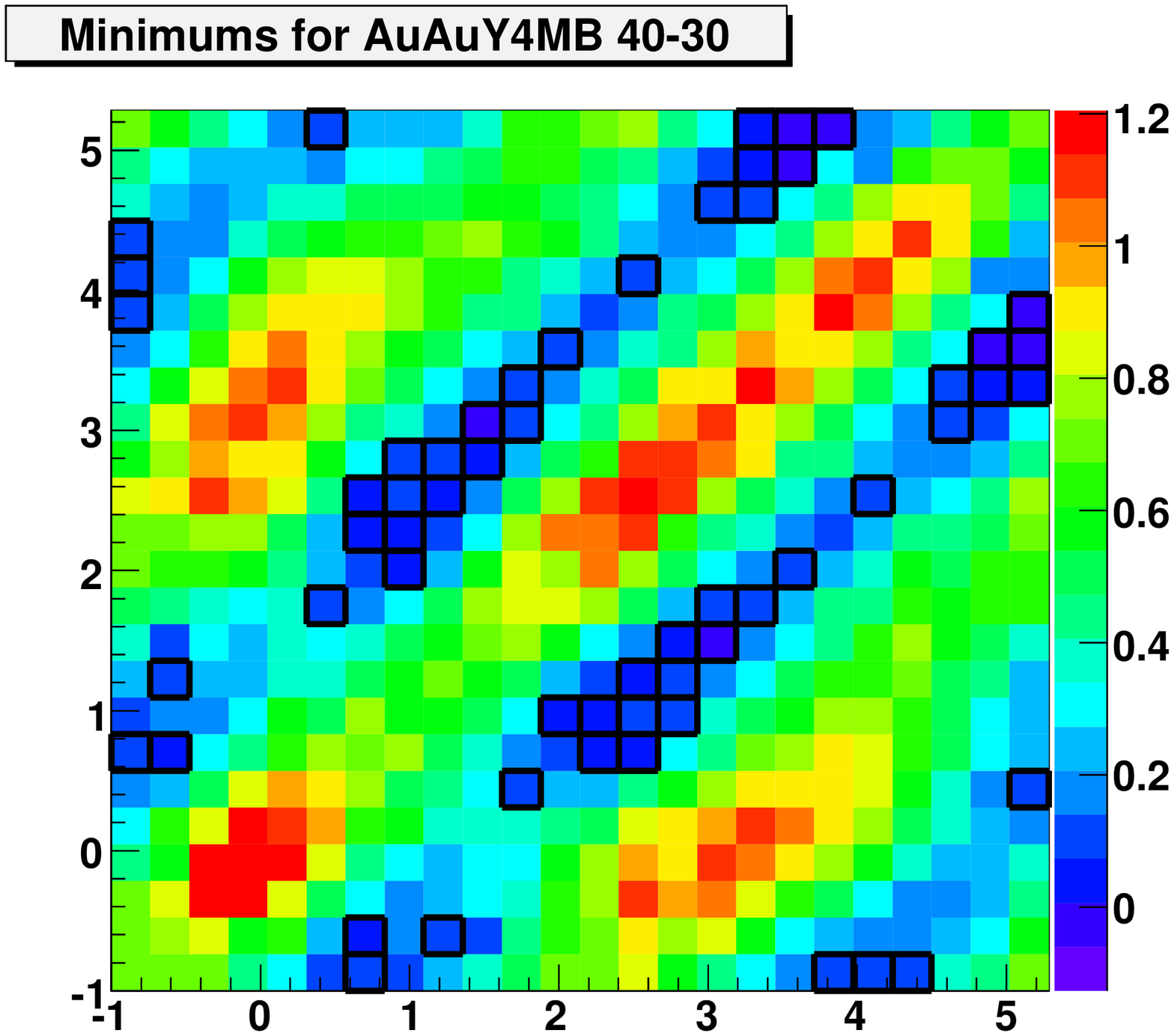}
\includegraphics[width=1\textwidth]{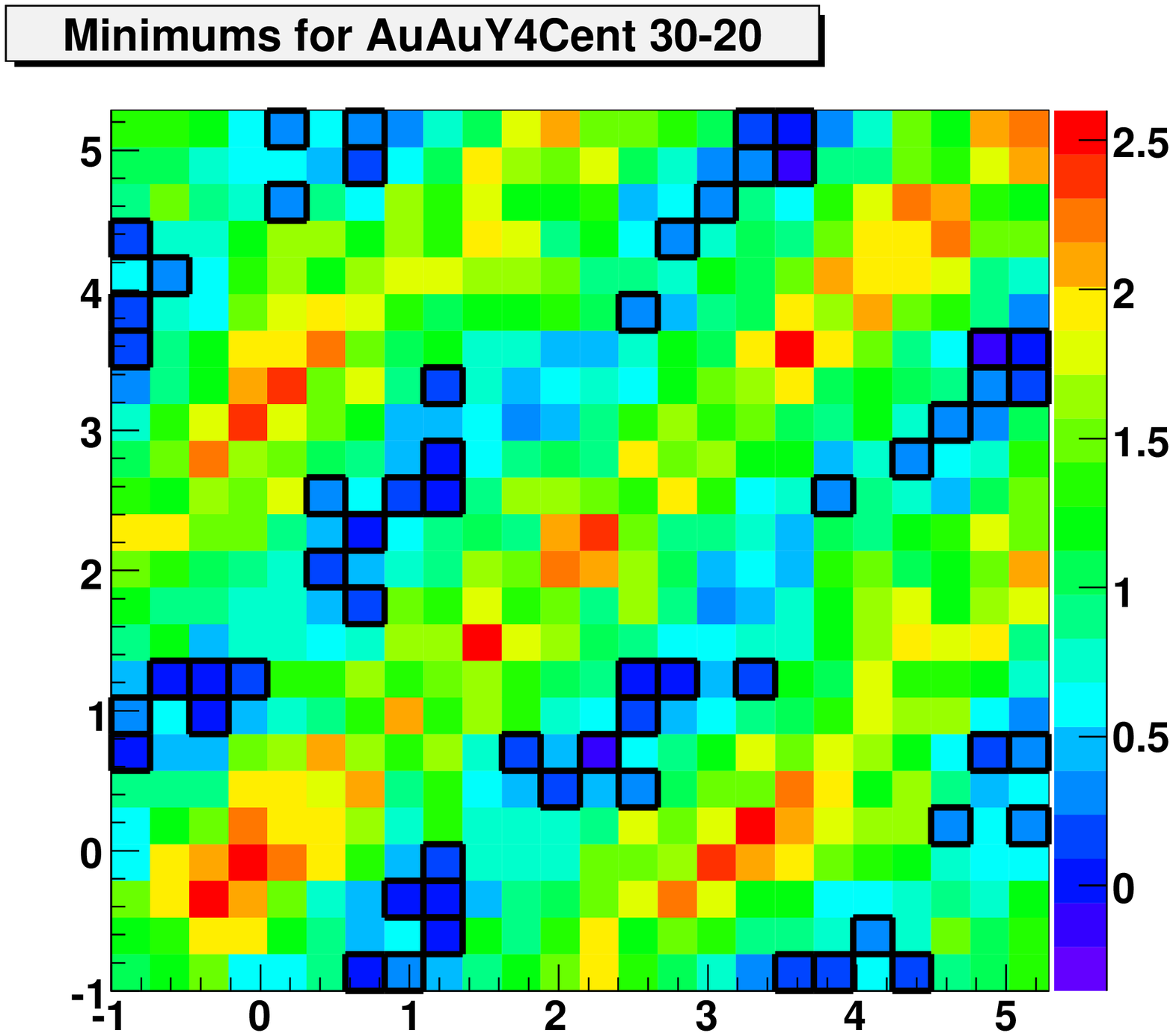}
\end{minipage}
\hfill
\begin{minipage}[t]{0.19\textwidth}
\includegraphics[width=1\textwidth]{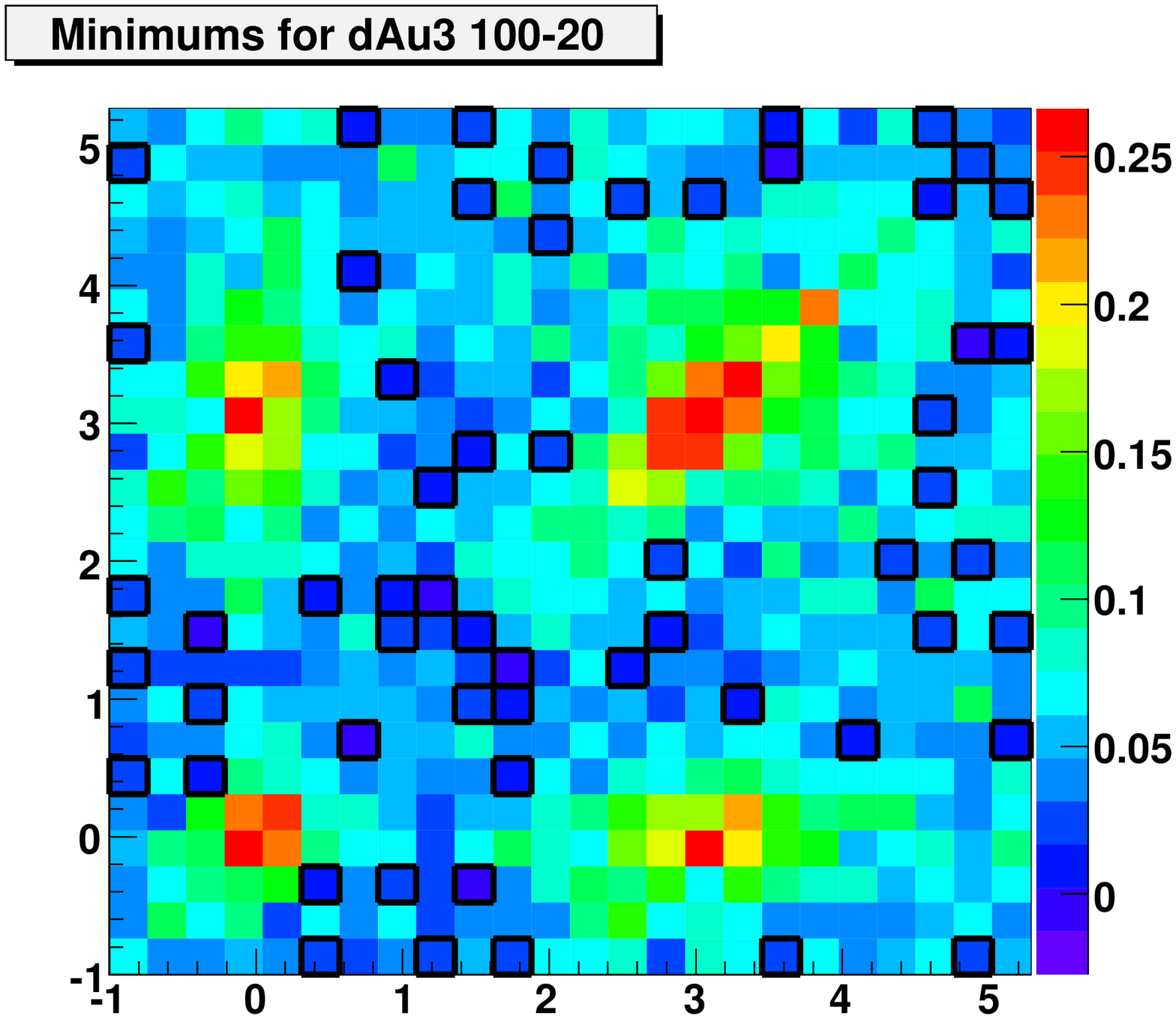}
\includegraphics[width=1\textwidth]{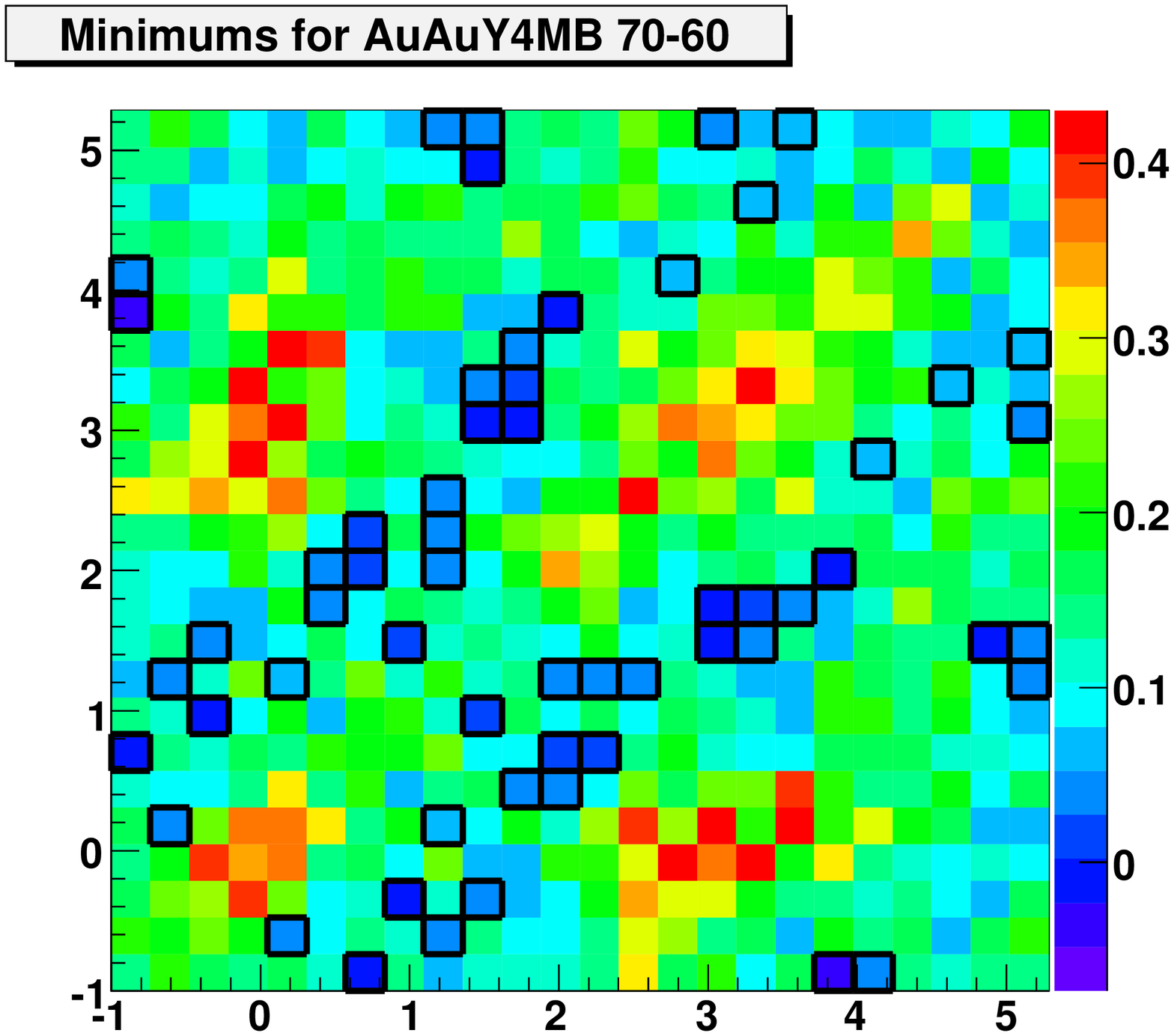}
\includegraphics[width=1\textwidth]{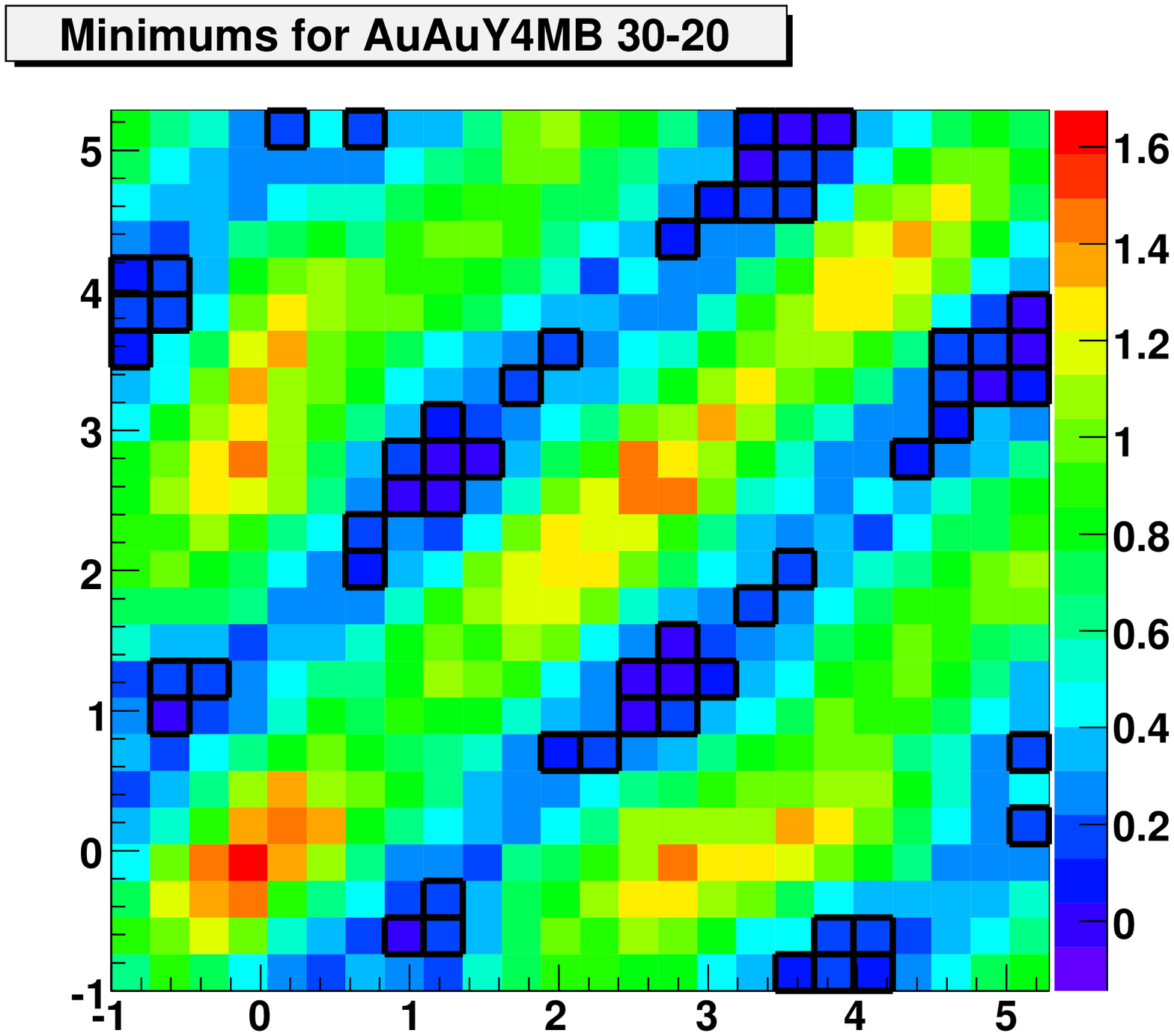}
\includegraphics[width=1\textwidth]{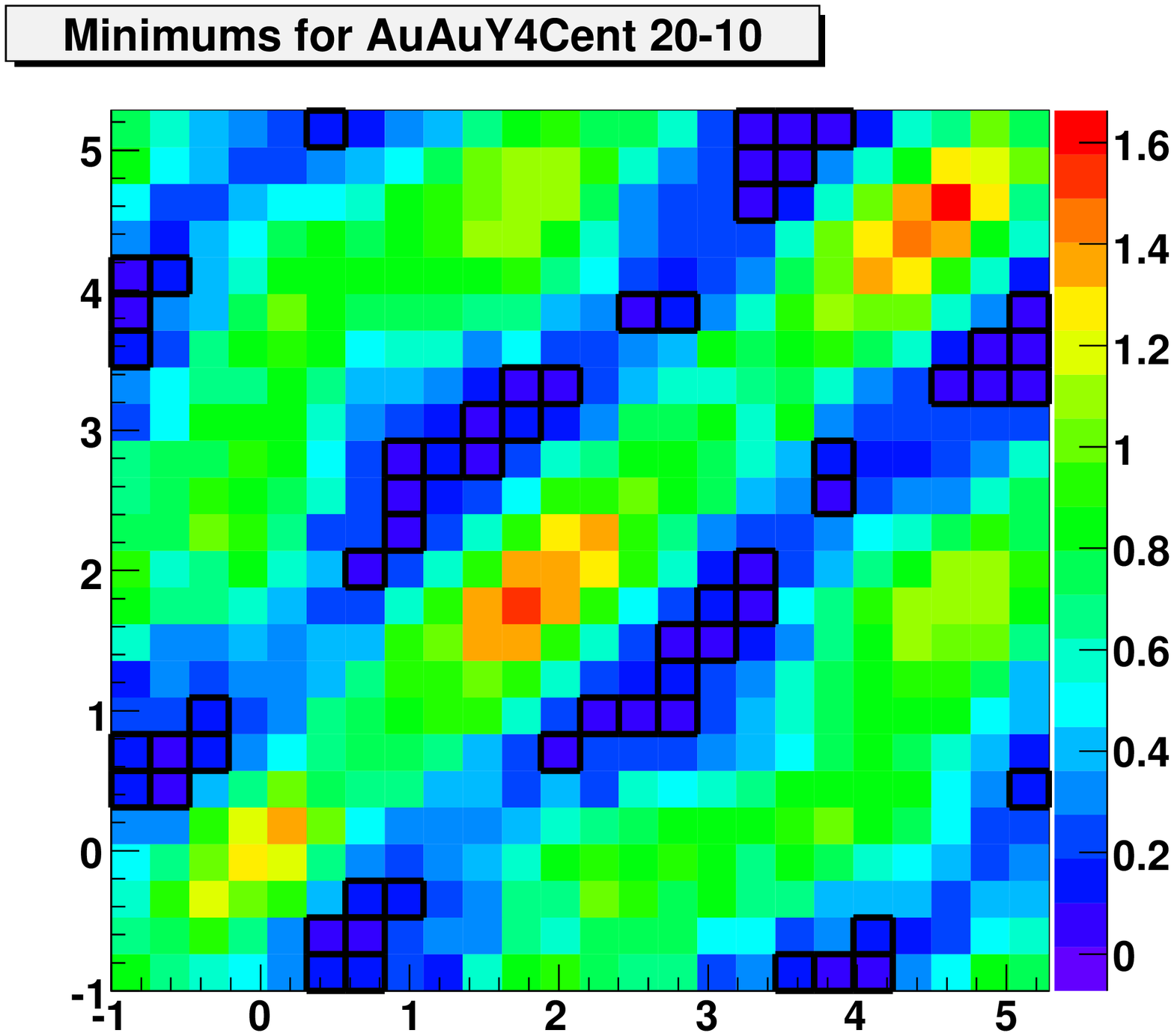}
\end{minipage}
\hfill
\begin{minipage}[t]{0.19\textwidth}
\includegraphics[width=1\textwidth]{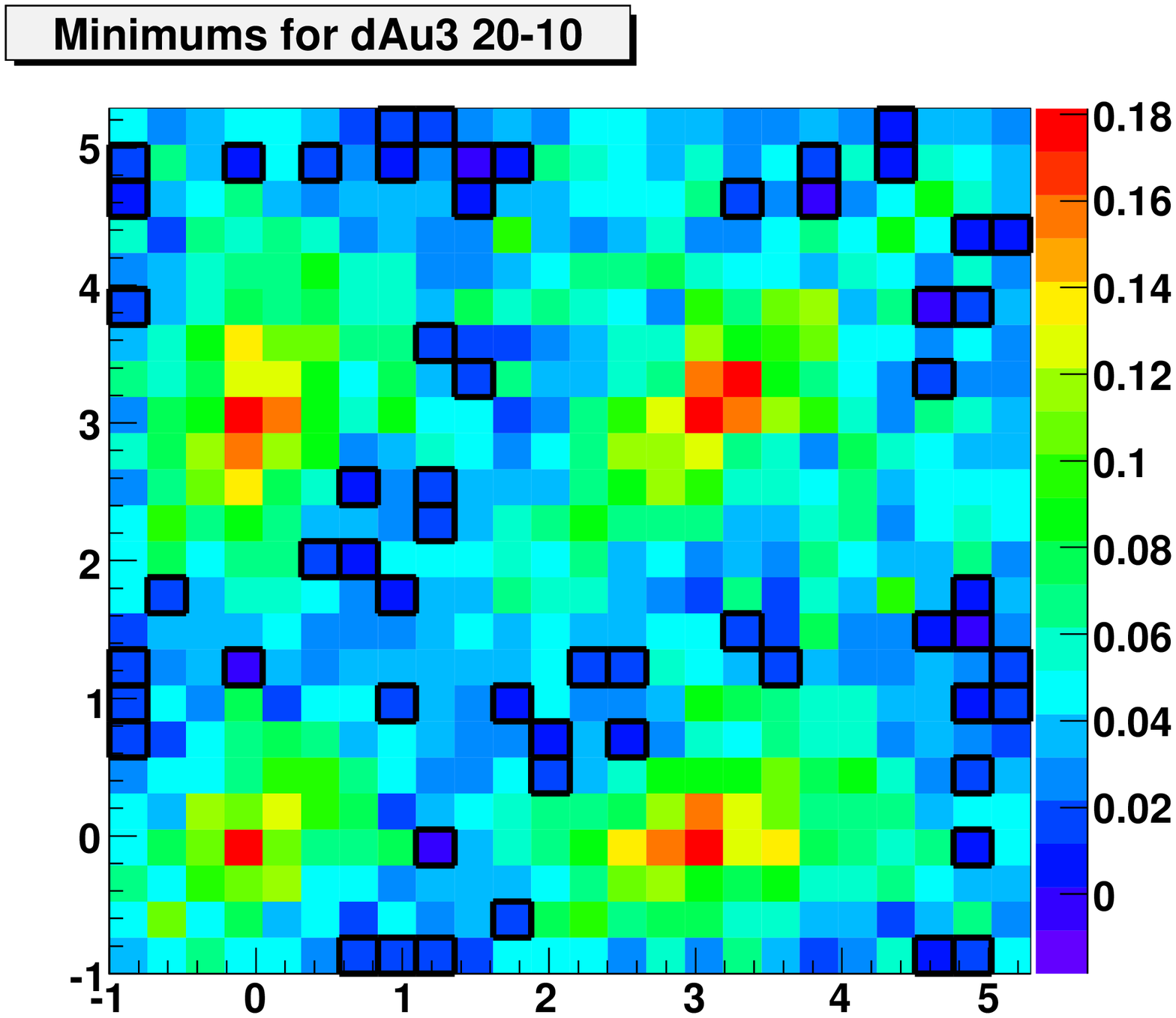}
\includegraphics[width=1\textwidth]{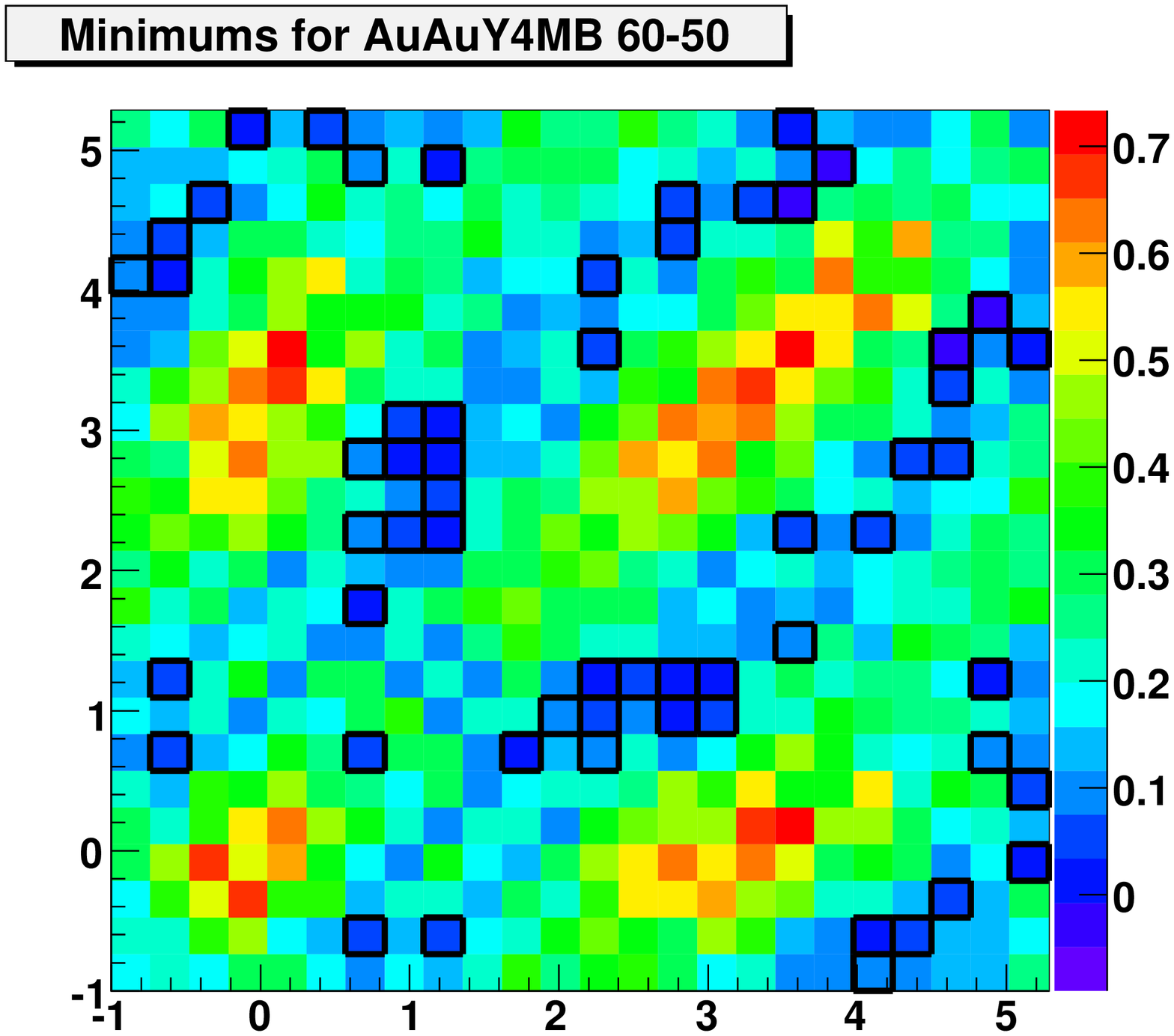}
\includegraphics[width=1\textwidth]{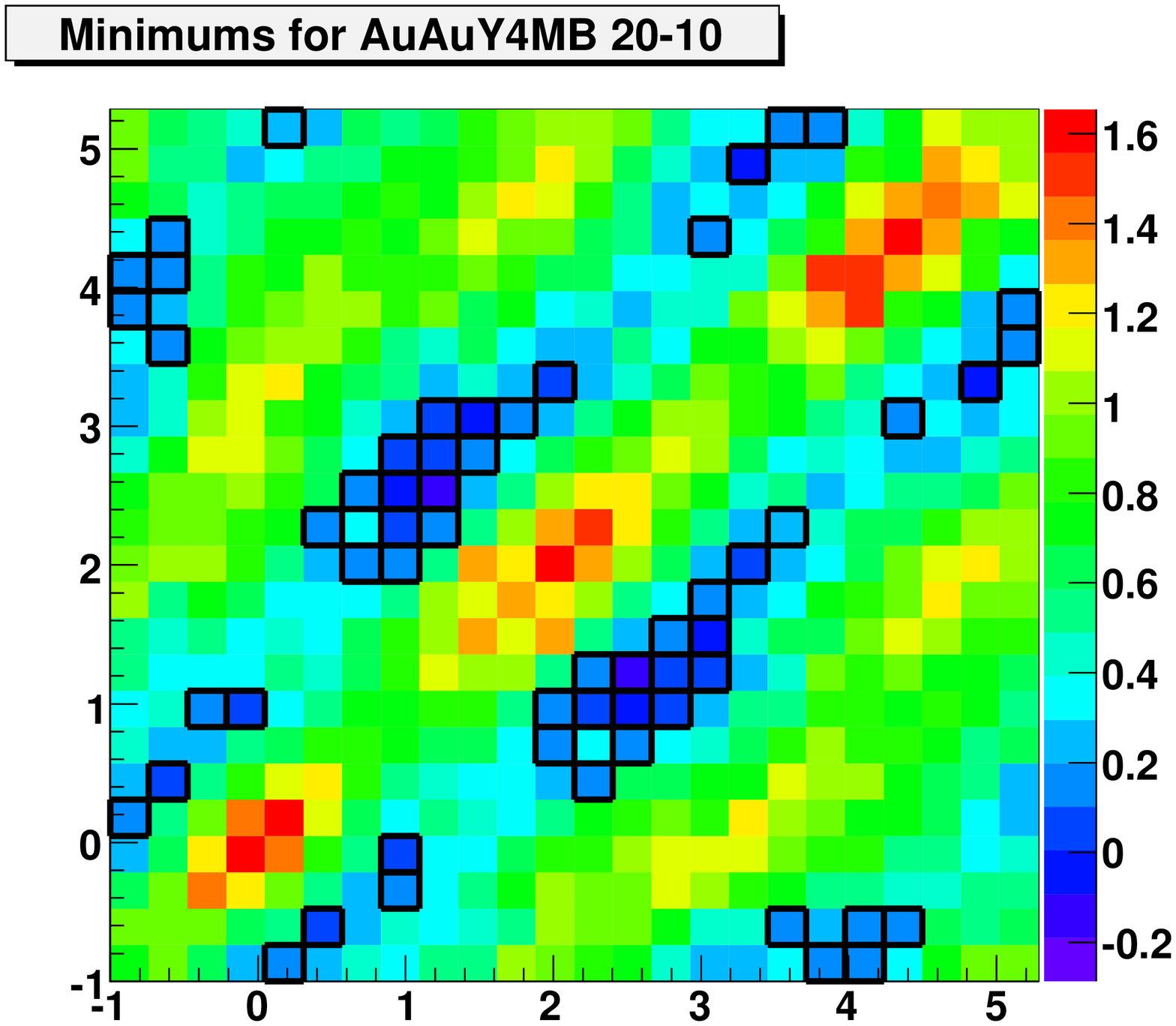}
\includegraphics[width=1\textwidth]{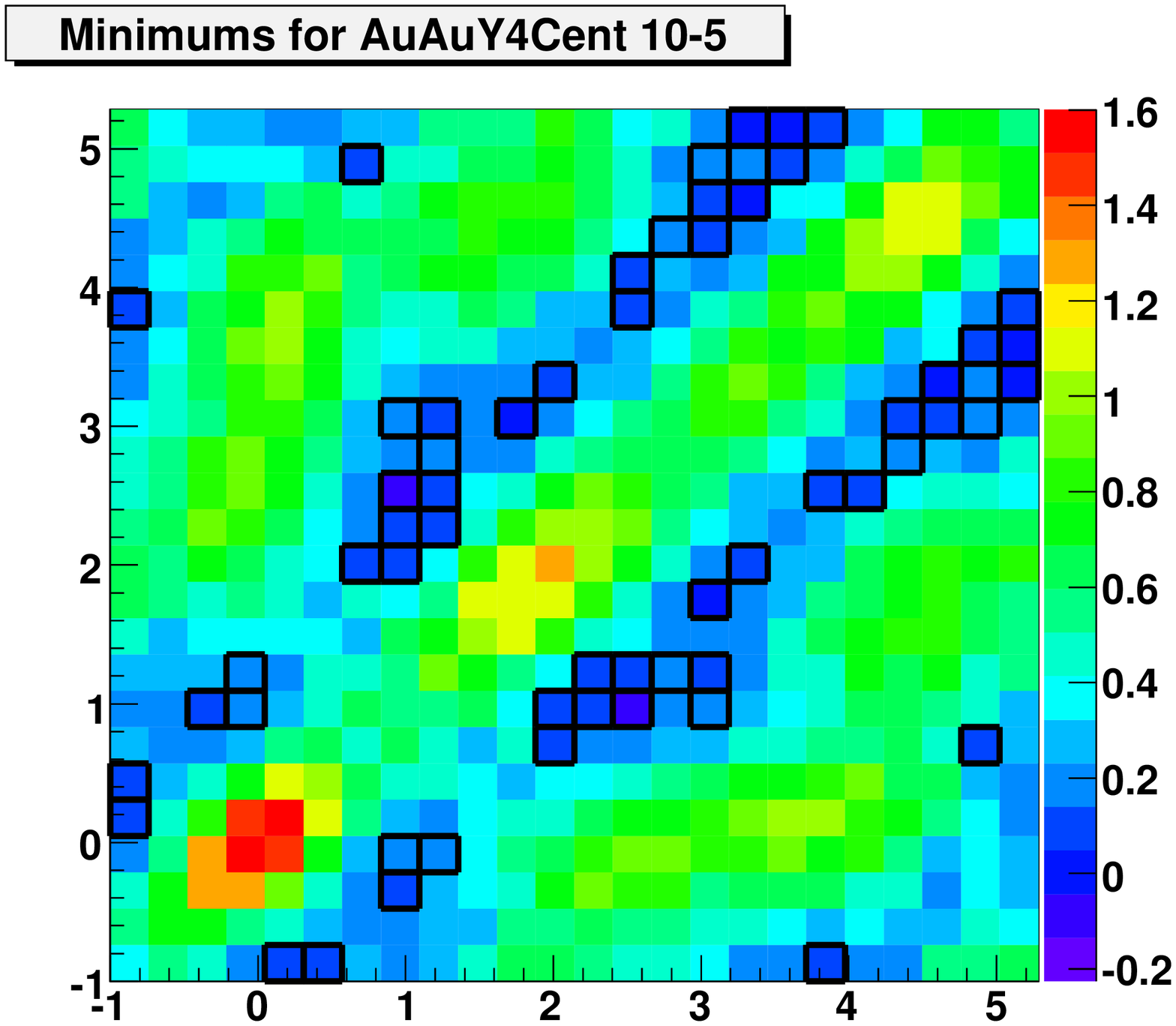}
\end{minipage}
\hfill
\begin{minipage}[t]{0.19\textwidth}
\includegraphics[width=1\textwidth]{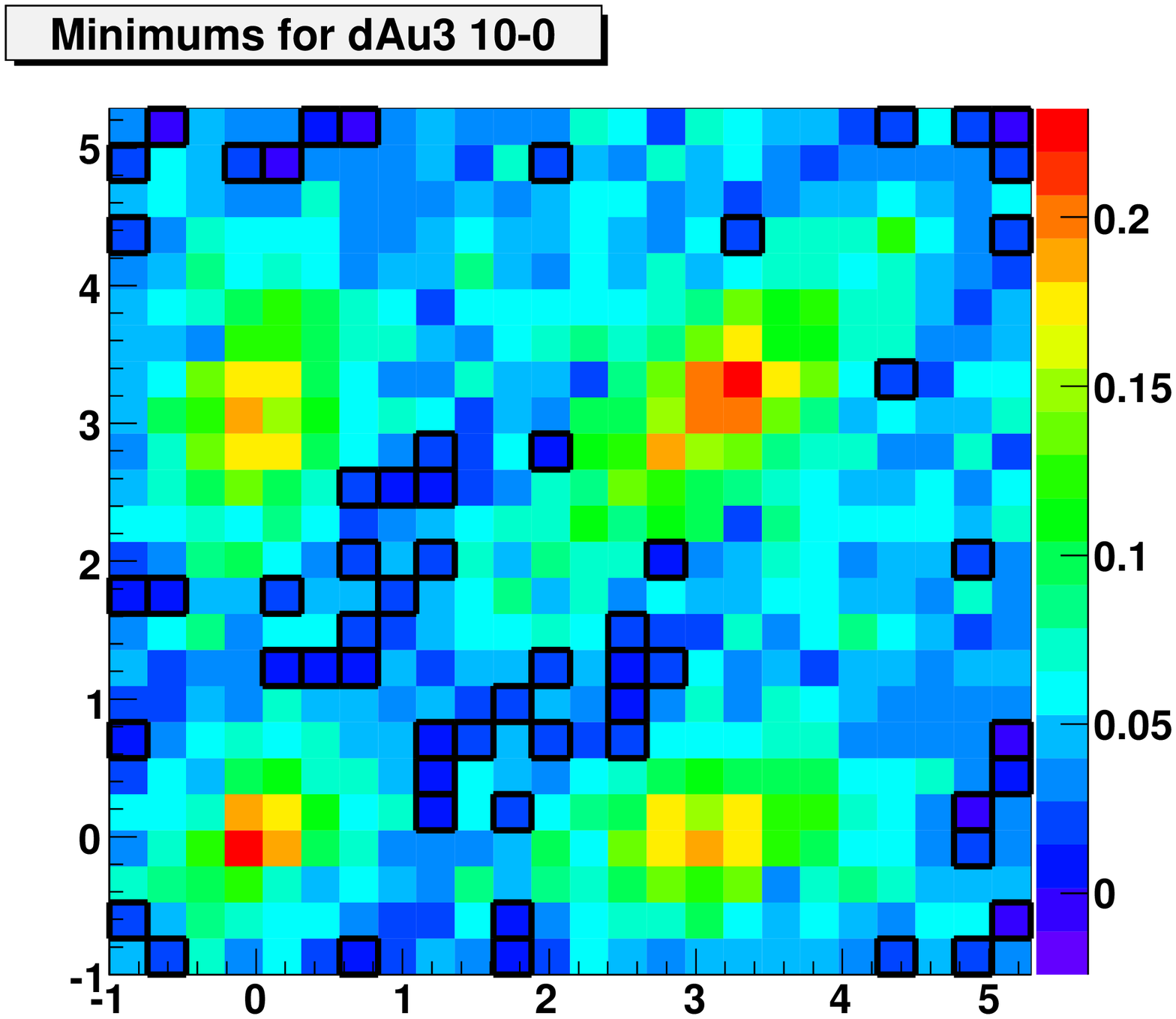}
\includegraphics[width=1\textwidth]{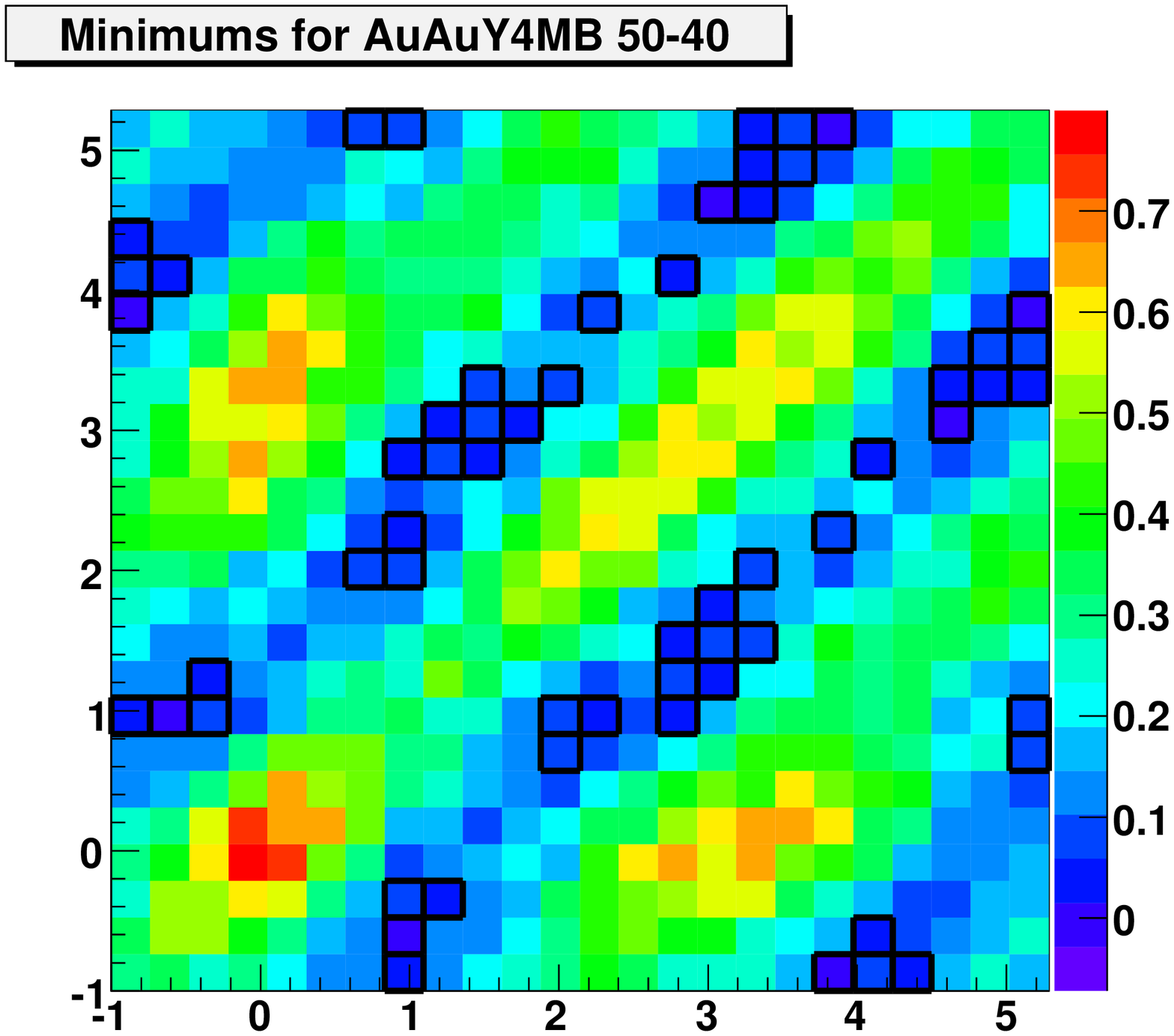}
\includegraphics[width=1\textwidth]{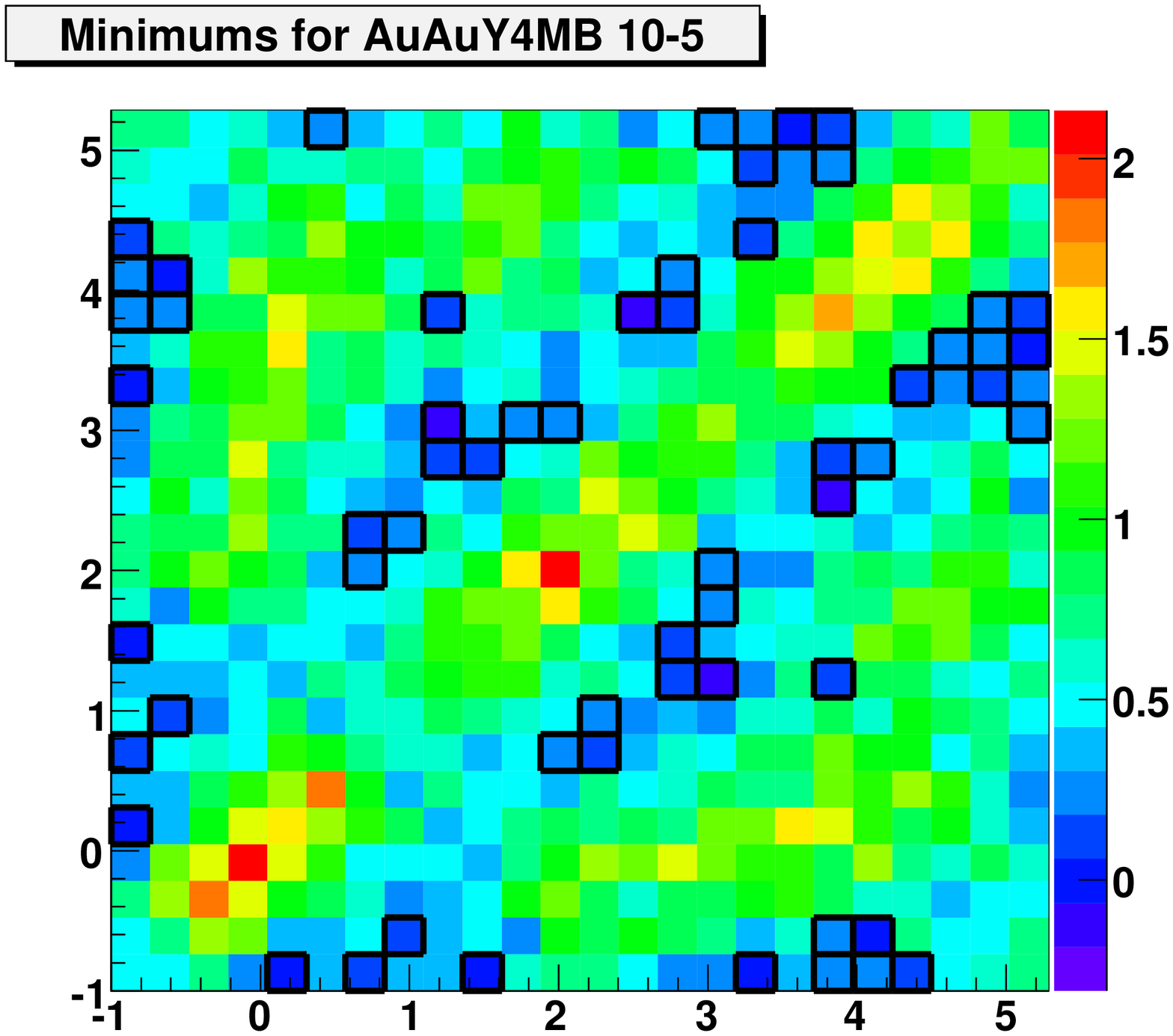}
\includegraphics[width=1\textwidth]{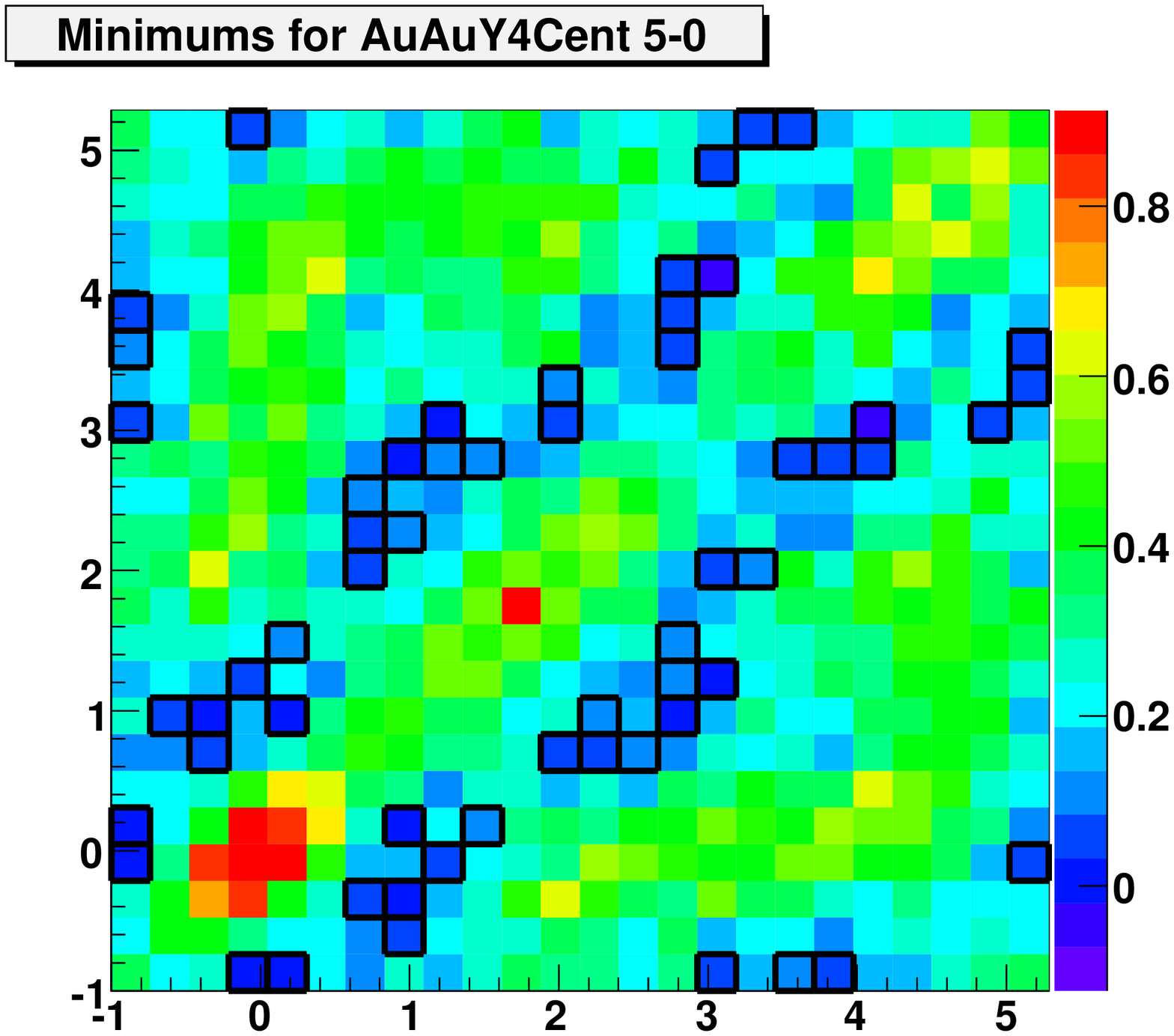}
\end{minipage}
\hfill
\begin{minipage}[t]{0.19\textwidth}
\includegraphics[width=1\textwidth]{Plots/minblank.eps}
\includegraphics[width=1\textwidth]{Plots/minblank.eps}
\includegraphics[width=1\textwidth]{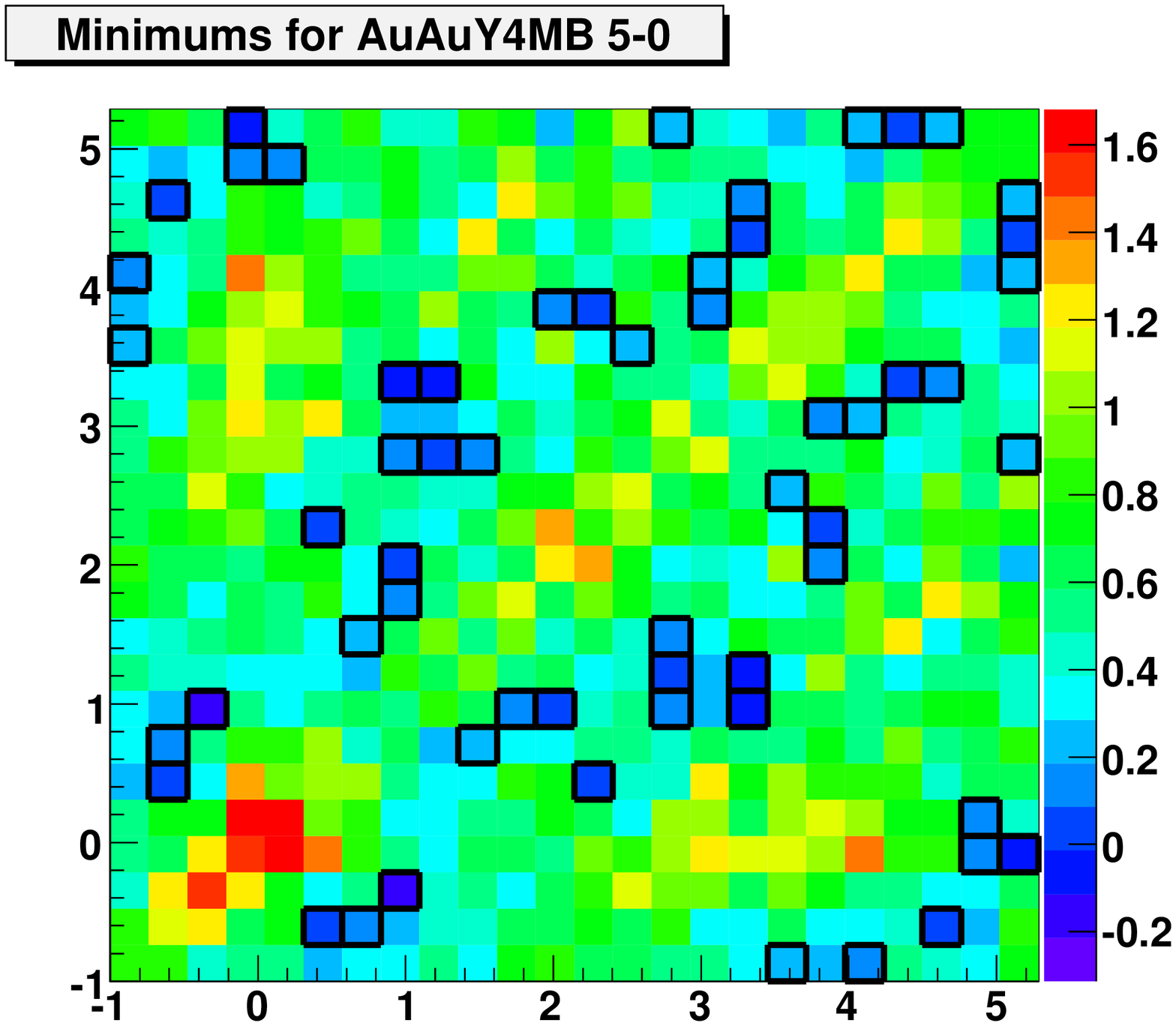}
\includegraphics[width=1\textwidth]{Plots/minblank.eps}
\end{minipage}
\caption{Background subtracted 3-particle correlations with the bins used for the 3-particle ZYAM highlighted in black.  Panels are, from left to right and top to bottom, $pp$ 0-100\%, d+Au 20-100\%, 10-20\%, and 0-10\%, Au+Au 70-80\%, 60-70\%, 50-60\%, 40-50\%, 30-40\%, 20-30\%, 10-20\%, 5-10\%, and 0-5\% and central ZDC triggered Au+Au ``20-30\%'', ``10-20\%'', ``5-10\%'', and 0-5\% collisions at $\sqrt{s_{NN}}=200$ GeV/c.  Where the cross sections in quotes are not the actual cross sections but the minimum bias cross section cuts on the centeral triggered data.}
\label{fig:mini}
\end{figure}

Figure~\ref{fig:mini} shows the background subracted 3-particle correlations in the centrality bins in which they are analyzed.  The bins used for the minimazition are outlined in black.  These bins are summed together with a number of trigger particle weighting to consturct the background subtracted 3-particle correlation plots in the wider centrality bins.

\begin{figure}[H]
\centering
\includegraphics[width=0.55\textwidth]{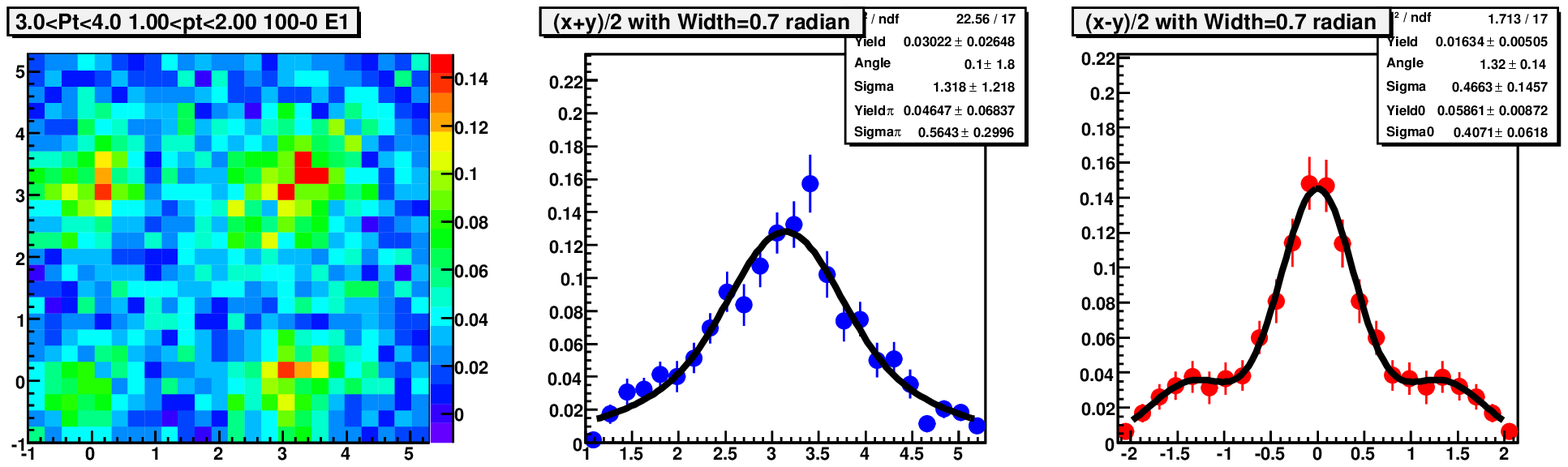}
\includegraphics[width=0.55\textwidth]{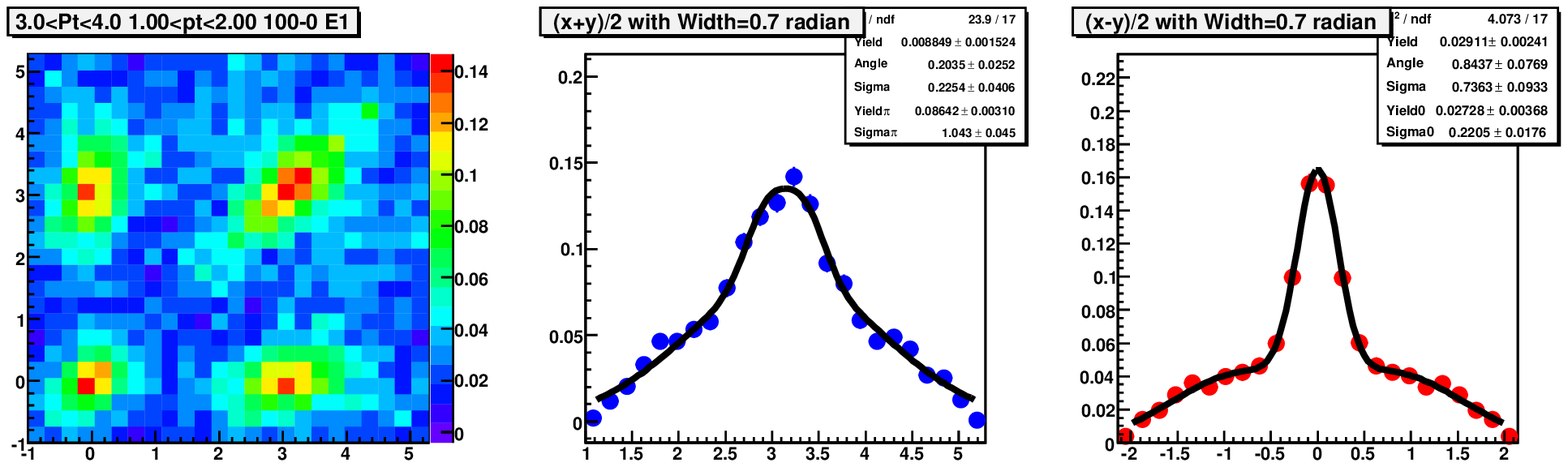}
\includegraphics[width=0.55\textwidth]{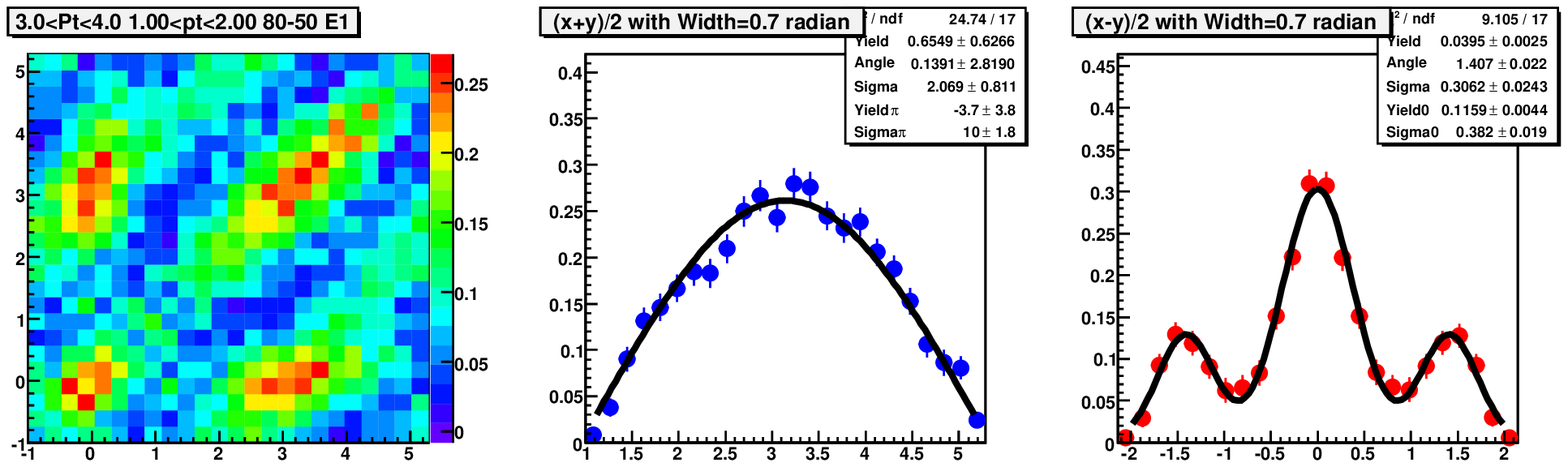}
\includegraphics[width=0.55\textwidth]{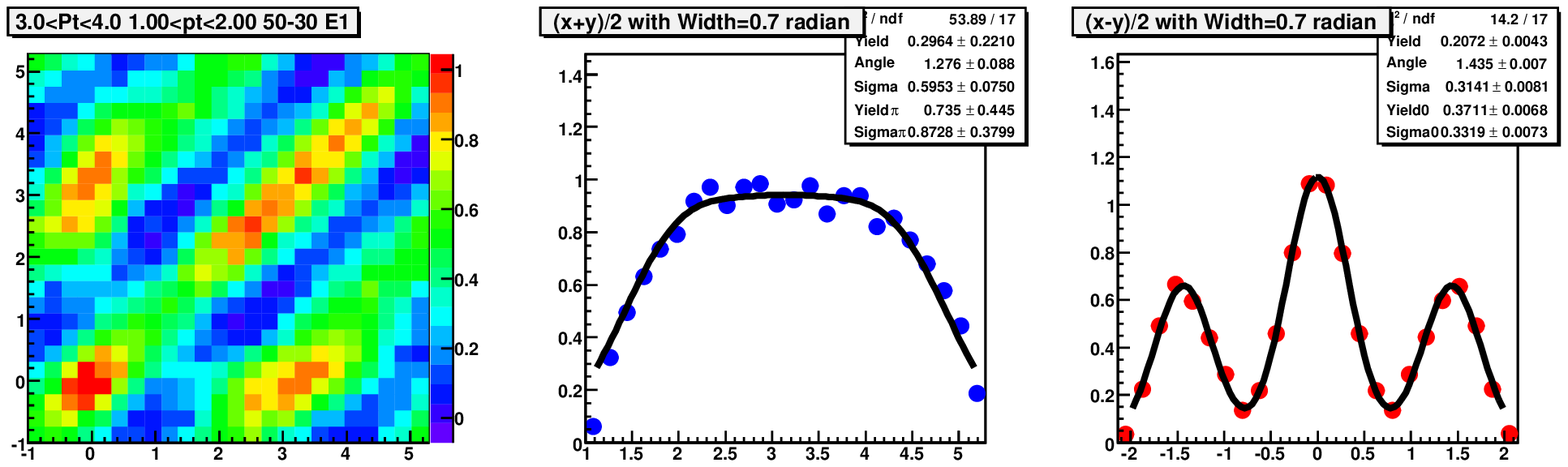}
\includegraphics[width=0.55\textwidth]{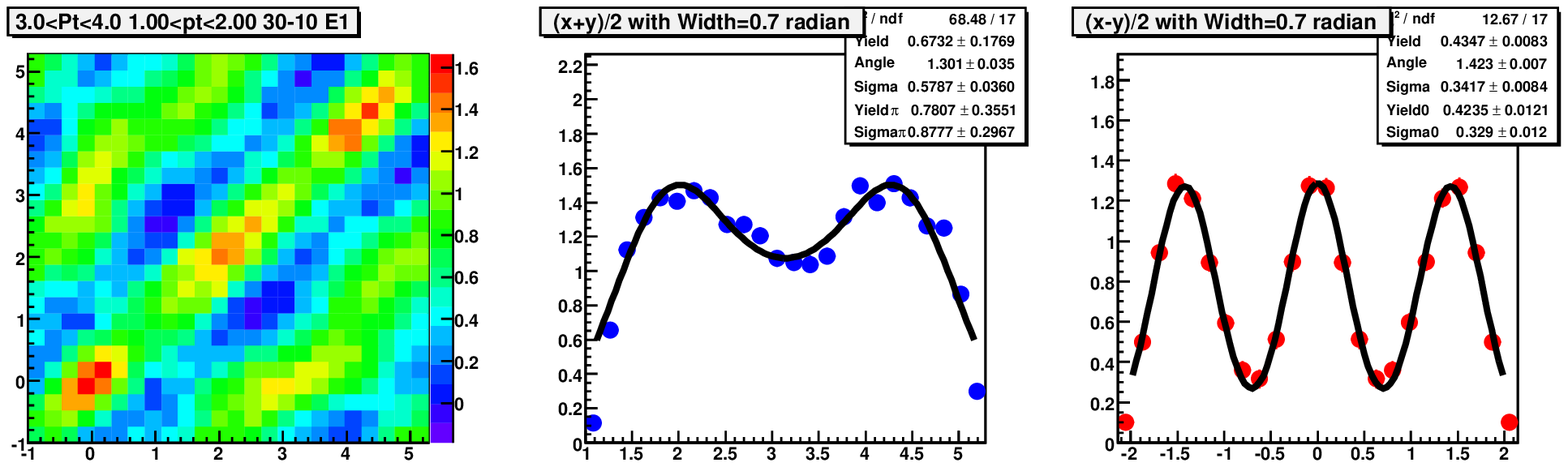}
\includegraphics[width=0.55\textwidth]{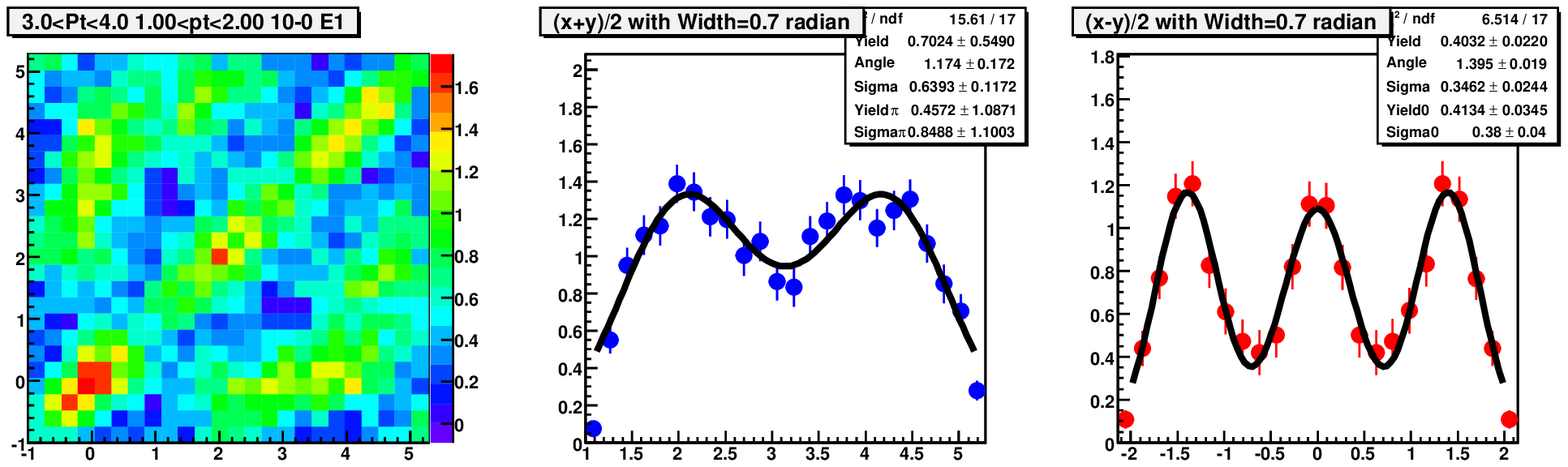}
\includegraphics[width=0.55\textwidth]{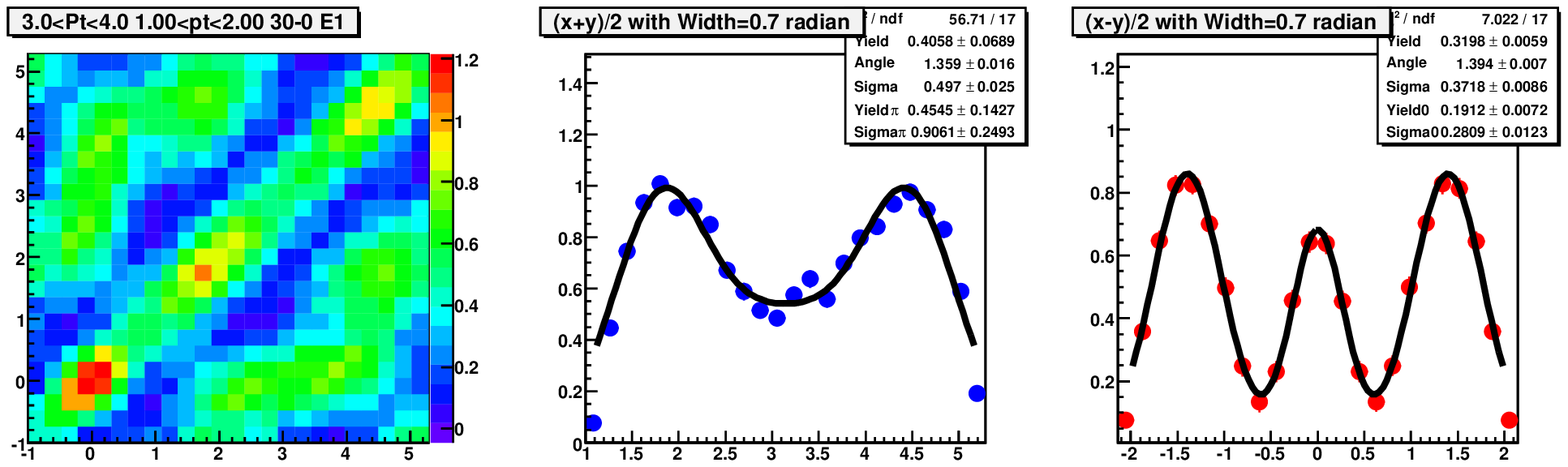}
\caption{Backgroud subtracted 3-particle correlations shown with on-diagonal (center) and off-diagonal (right) projections of the away-side distribution.  The curve represent fits to a central Gaussian and symmetric side Gaussians.   The rows are (from top to bottom) {\it pp}, d+Au, Au+Au 50-80\%, Au+Au 30-50\%, Au+Au 10-30\%, Au+Au 0-10\% and ZDC triggered Au+Au 0-12\% collisions at $\sqrt{s_{NN}}=200$ GeV/c.  The projections are of a strips of full width of 0.7 radians.  Errors are statistical.}
\label{fig:projfits}
\end{figure}

Figrue~\ref{fig:projfits} shows the on-diagonal and off-diagonal projections with the Gaussian fits.  Each projection is fit to a central Gaussian and symmetric side Gaussians.  The parameters listed for the fits are (from top to bottom) $\chi^2$/ndf, side Gaussian yield, side Gaussian distance from the center (in radians), side Gaussian width (in radians), central Gaussian yield, and central Gaussian width.  The central Gaussian is centered at $\pi$ (on-diagonal) or zero (off-diagonal).

\begin{figure}[H]
\hfill
\begin{minipage}[t]{0.32\textwidth}
\centering
\includegraphics[width=1.0\textwidth]{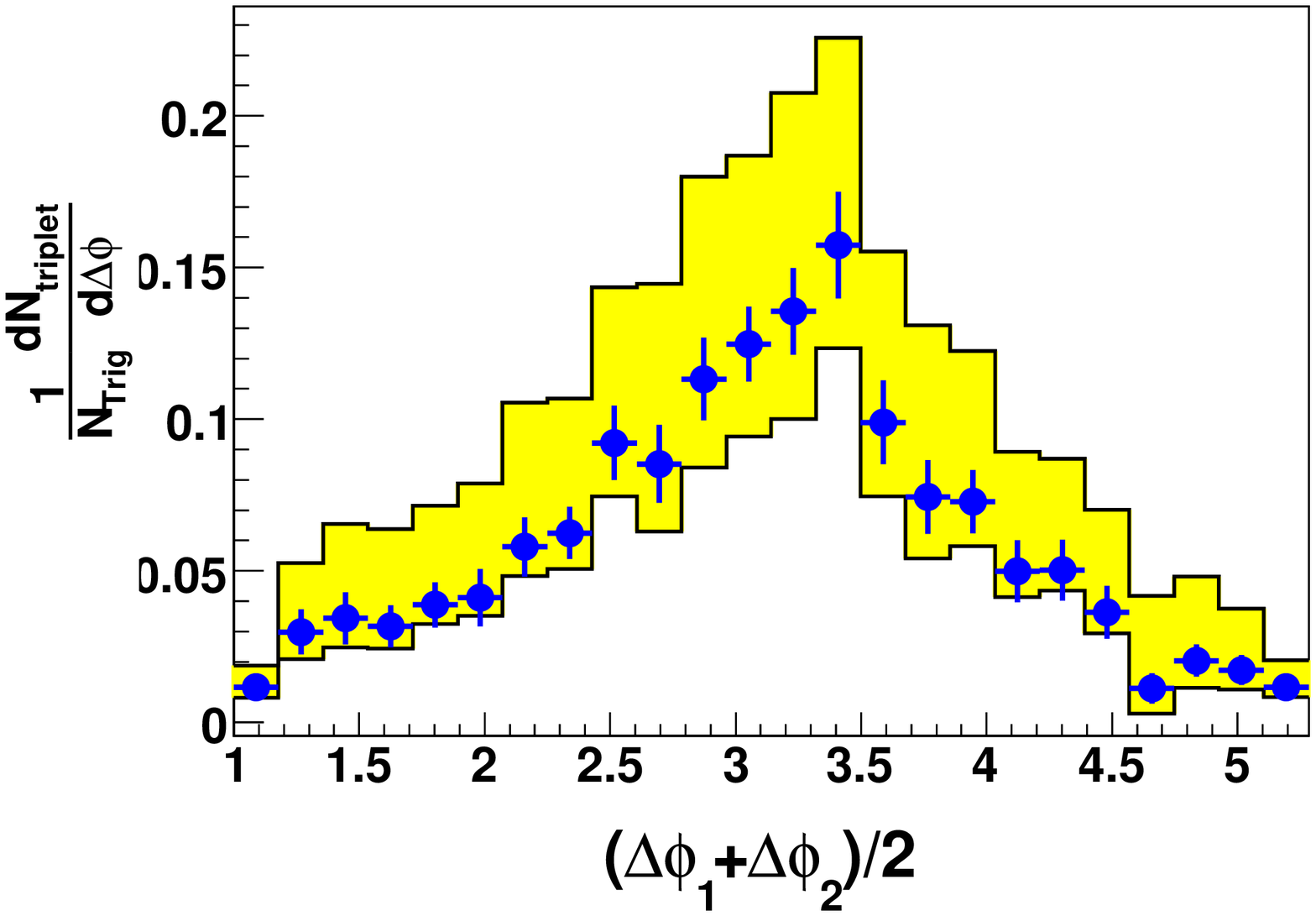}
\includegraphics[width=1.0\textwidth]{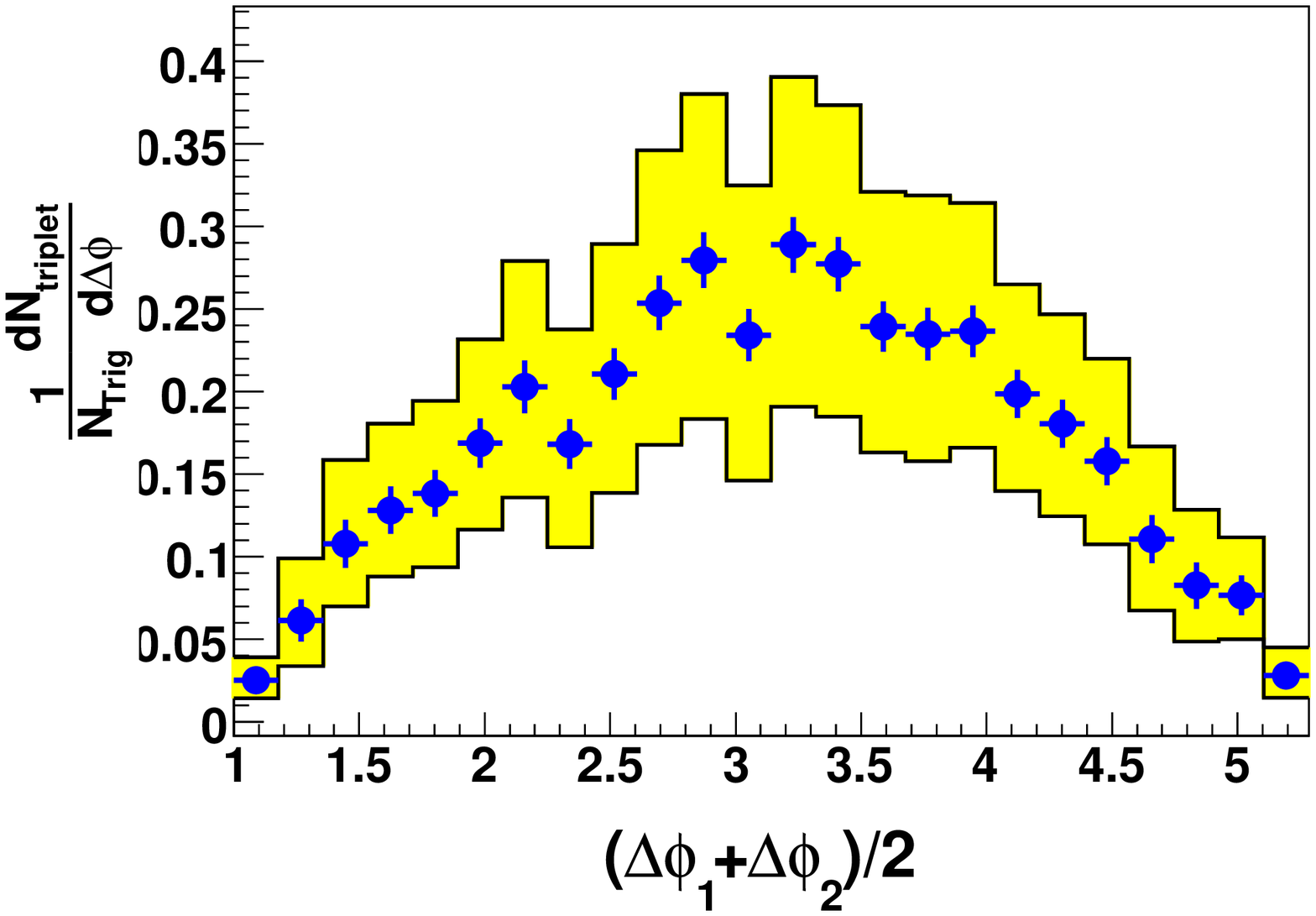}
\includegraphics[width=1.0\textwidth]{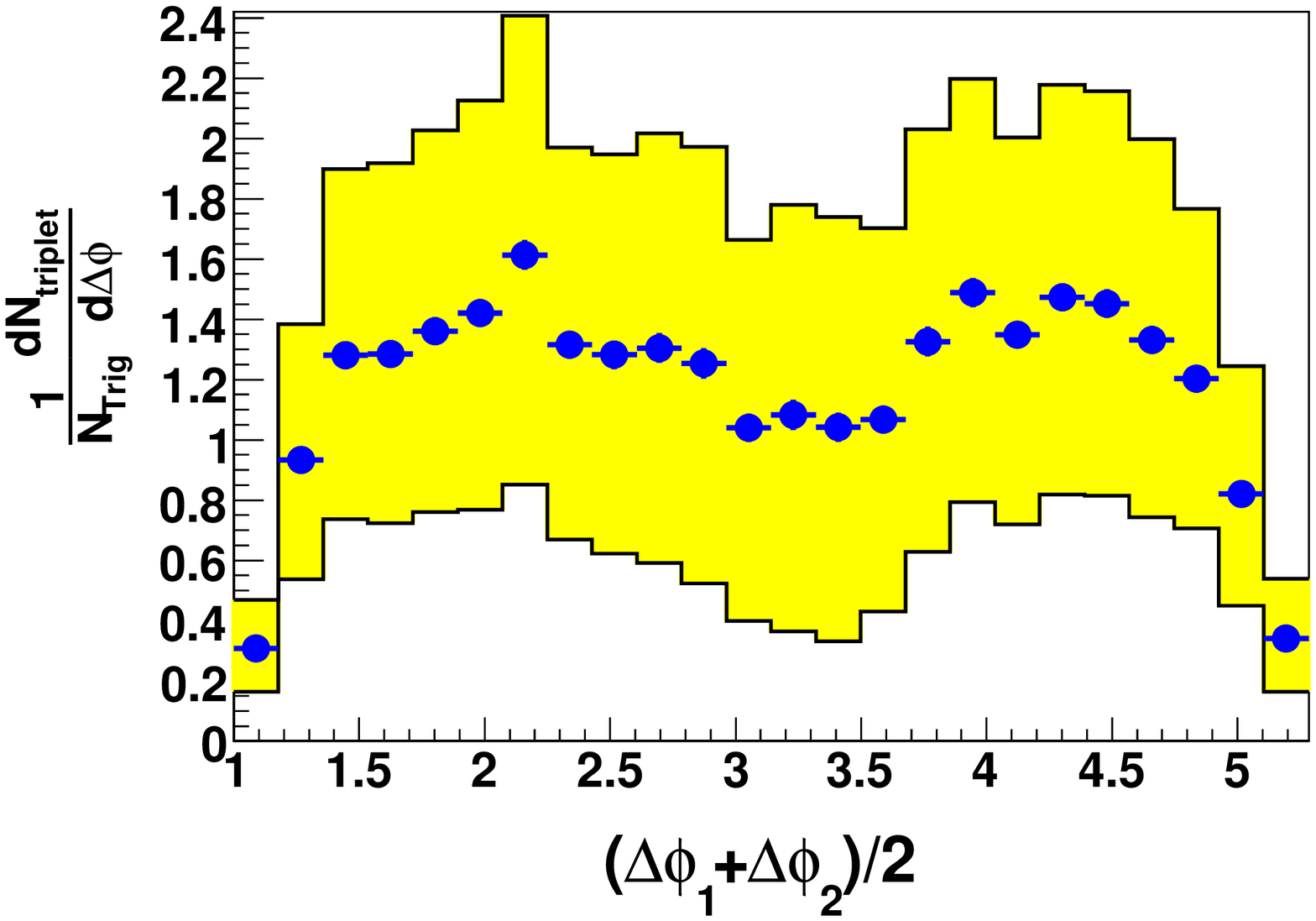}
\end{minipage}
\hfill
\begin{minipage}[t]{0.32\textwidth}
\includegraphics[width=1.0\textwidth]{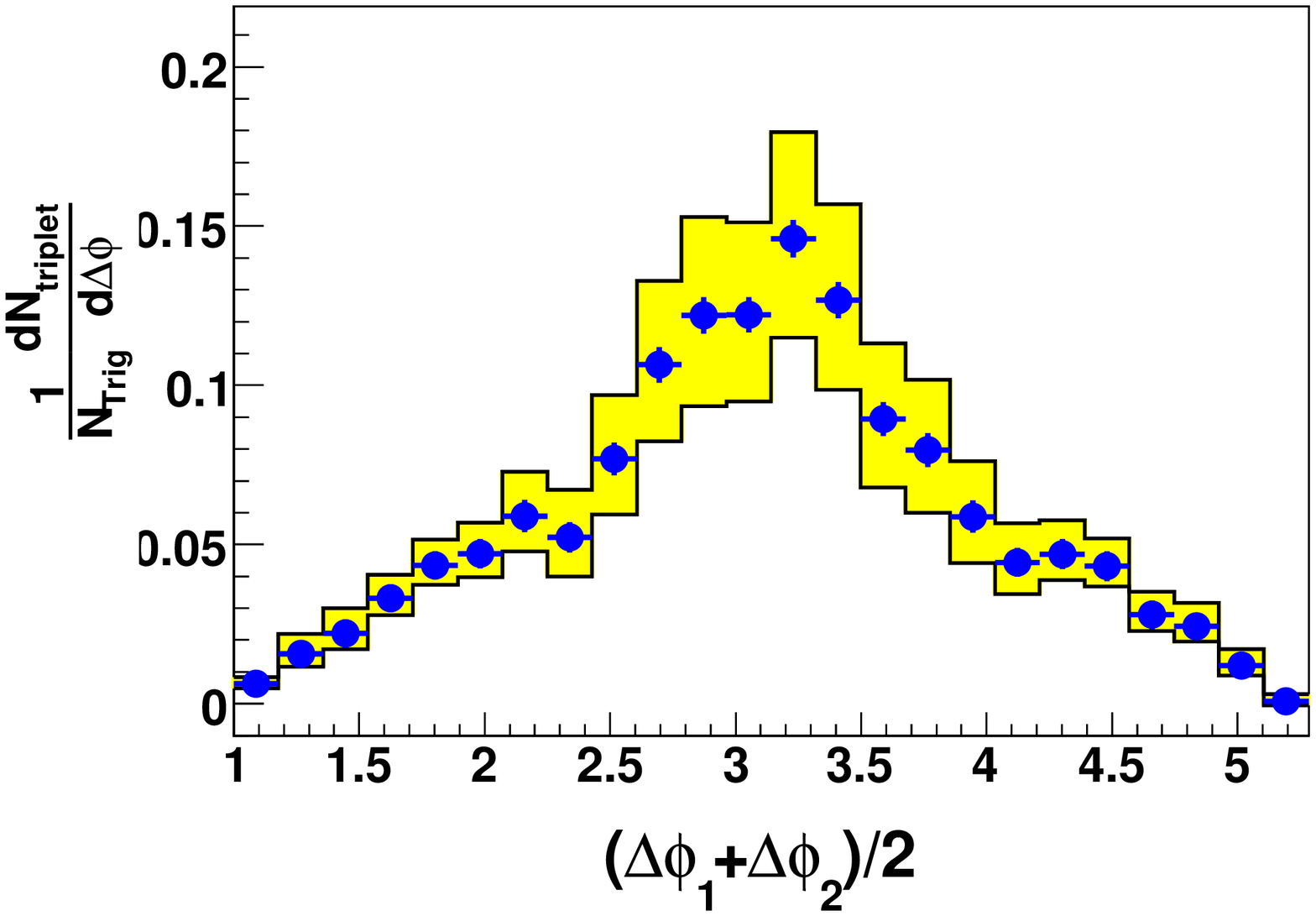}
\includegraphics[width=1.0\textwidth]{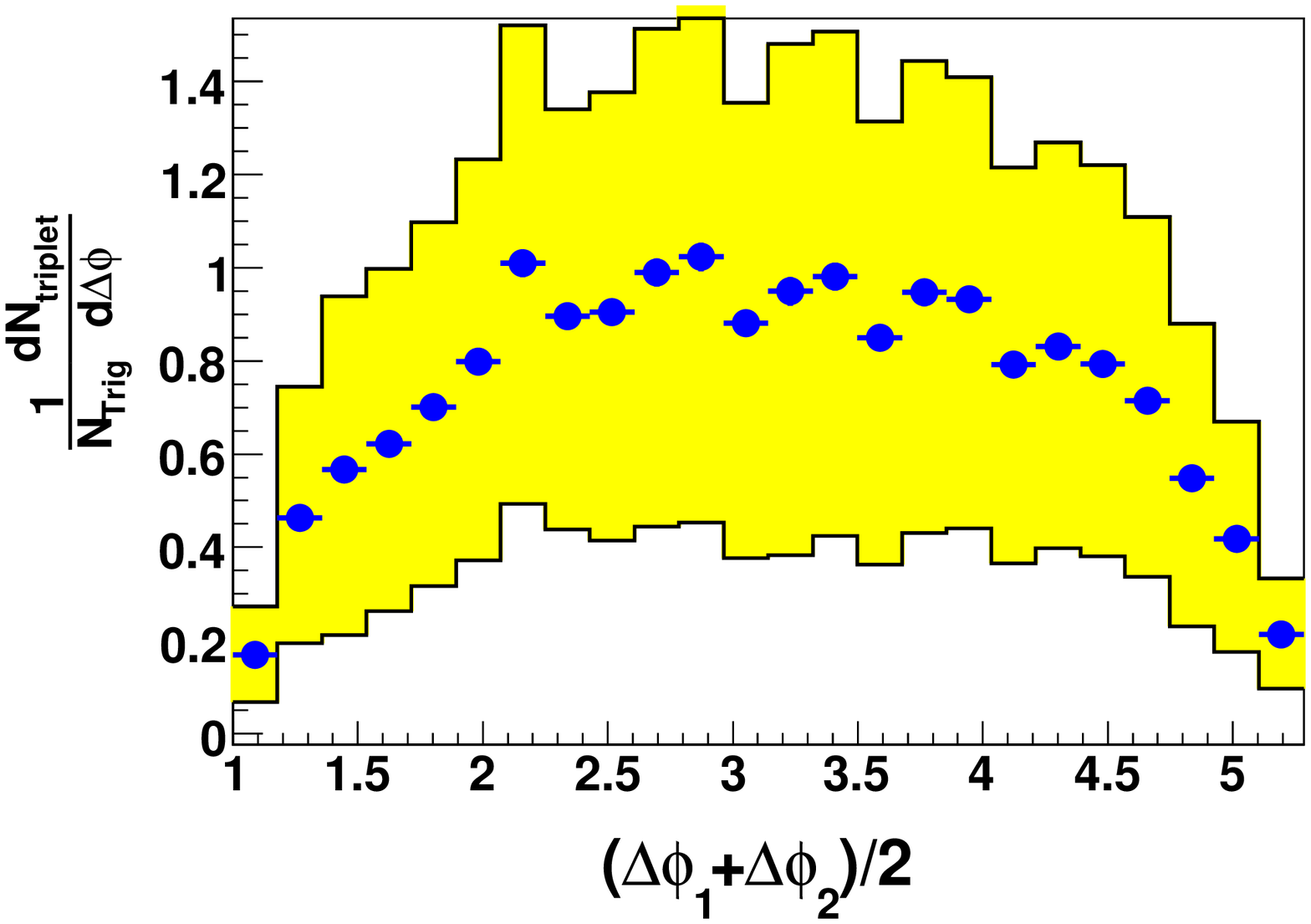}
\includegraphics[width=1.0\textwidth]{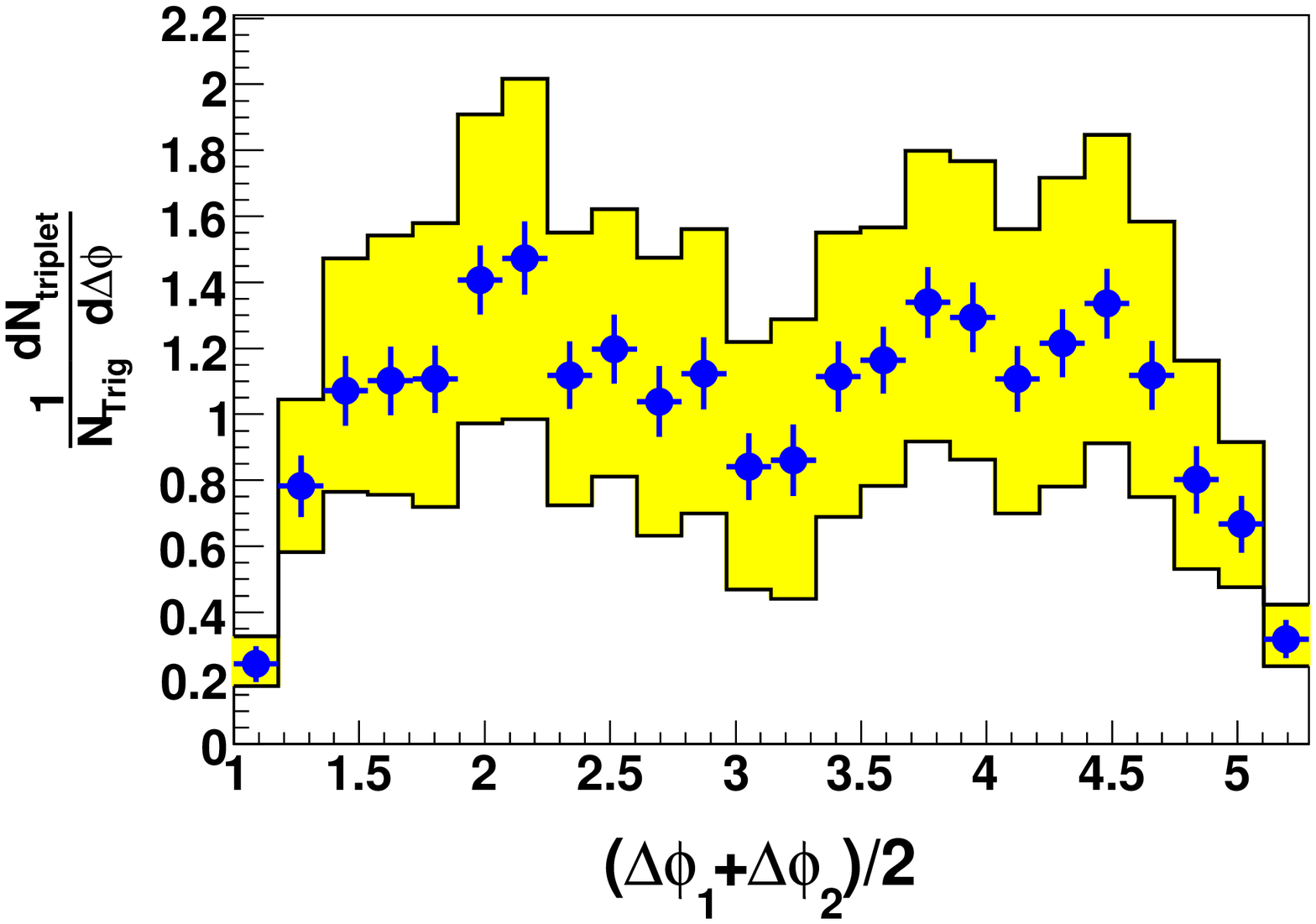}
\end{minipage}
\hfill
\begin{minipage}[t]{0.32\textwidth}
\includegraphics[width=1.0\textwidth]{Plots/ProjStripBlank.eps}
\includegraphics[width=1.0\textwidth]{Plots/ProjStripBlank.eps}
\includegraphics[width=1.0\textwidth]{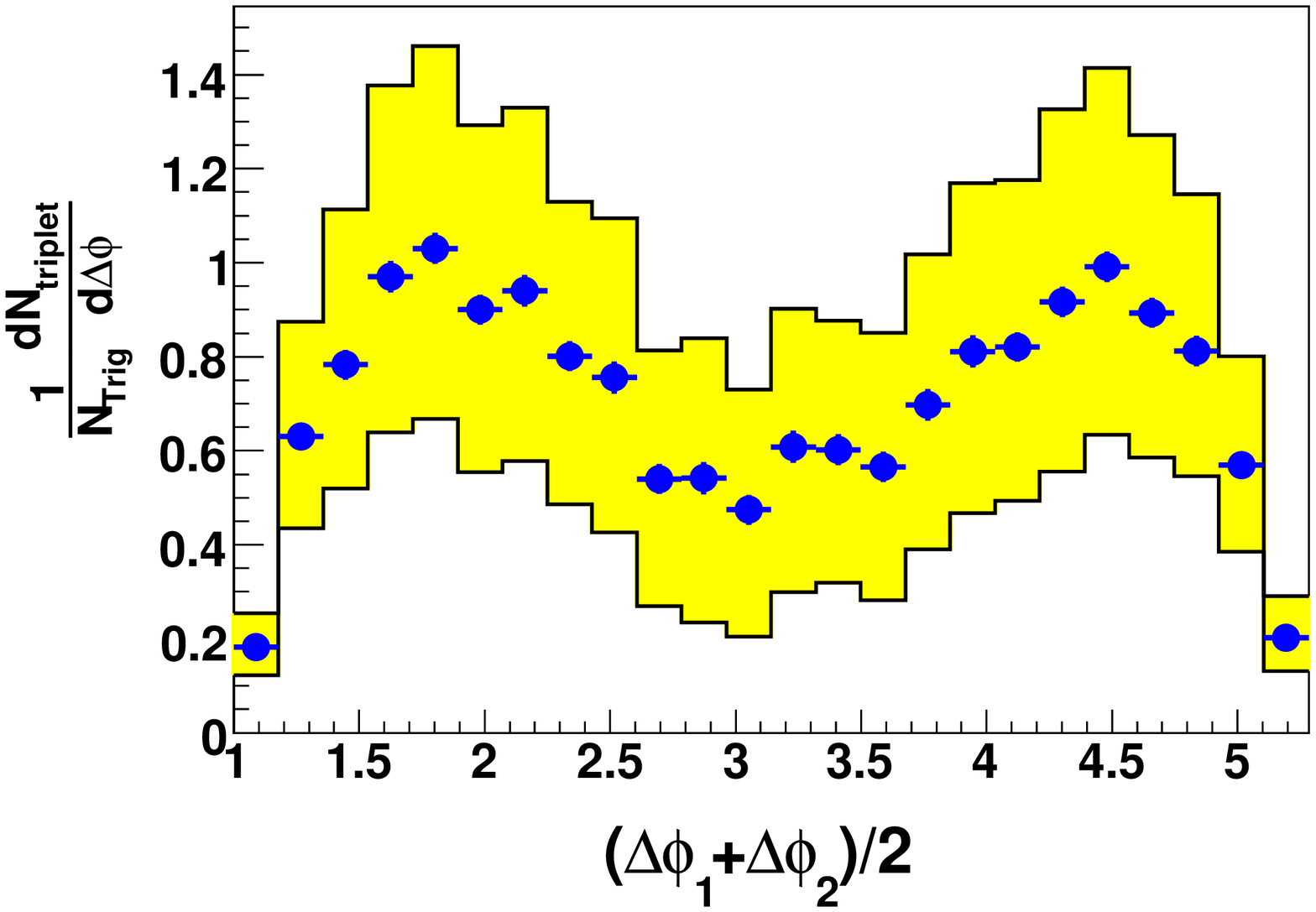}
\end{minipage}
\caption{On-diagonal projections of the background subtracted 3-particle correlations in strips of full width 0.7 radians on the away-side.  From left to right, top to bottom are {\it pp}, d+Au, Au+Au 50-80\%, Au+Au 30-50\%, Au+Au 10-30\%, Au+Au 0-10\% and ZDC triggered Au+Au 0-12\% collisions at $\sqrt{s_{NN}}=200$ GeV/c.  Systematic errors are shown in the yellow boxes.  Dashed black lines are at zero.}
\label{fig:projon}
\end{figure}

\begin{figure}[htbp]    
\hfill
\begin{minipage}{0.32\textwidth}
\centering
\includegraphics[width=1.0\textwidth]{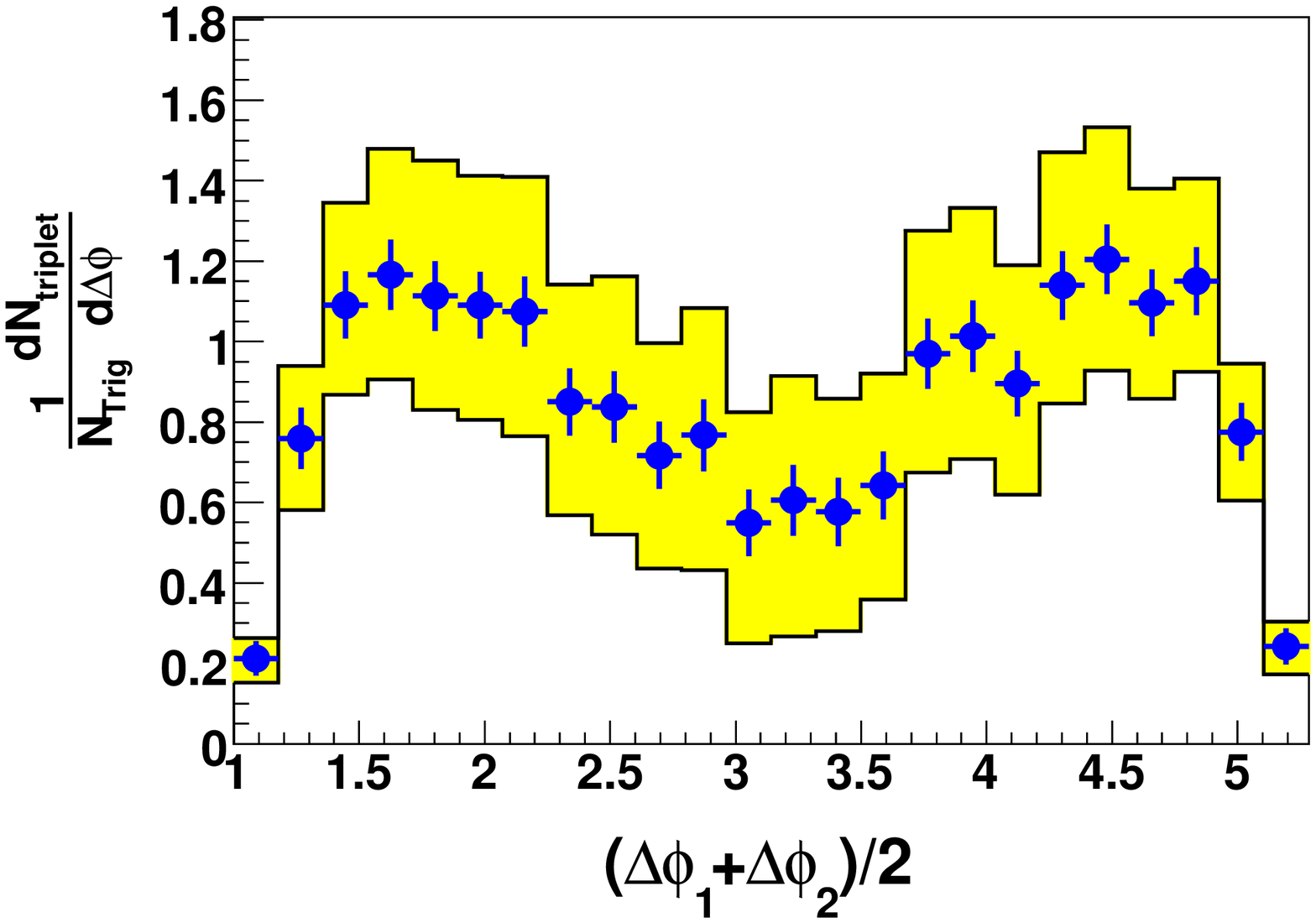}
\includegraphics[width=1.0\textwidth]{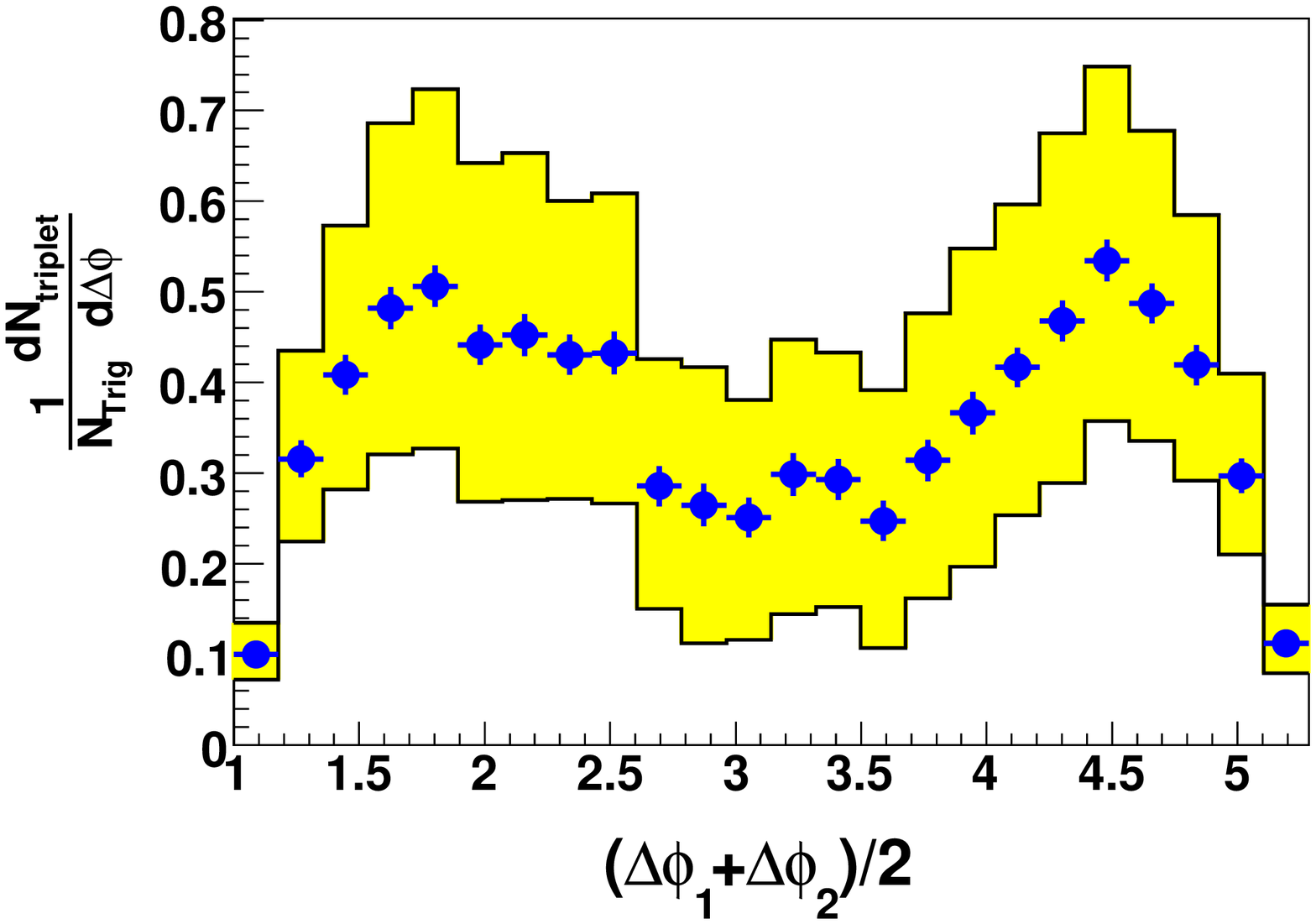}
\end{minipage}
\hfill
\begin{minipage}{0.32\textwidth}
\centering
\includegraphics[width=1.0\textwidth]{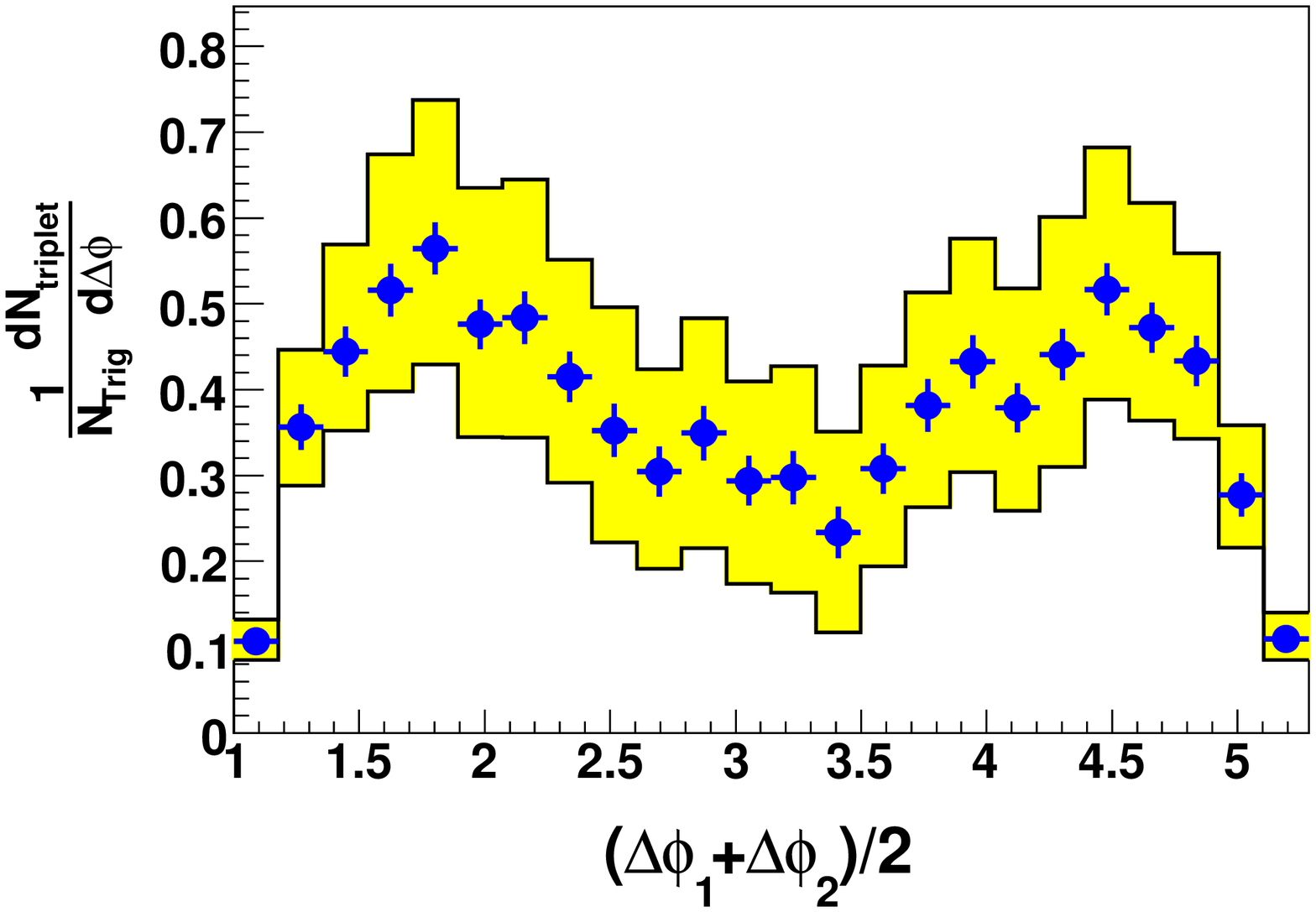}
\includegraphics[width=1.0\textwidth]{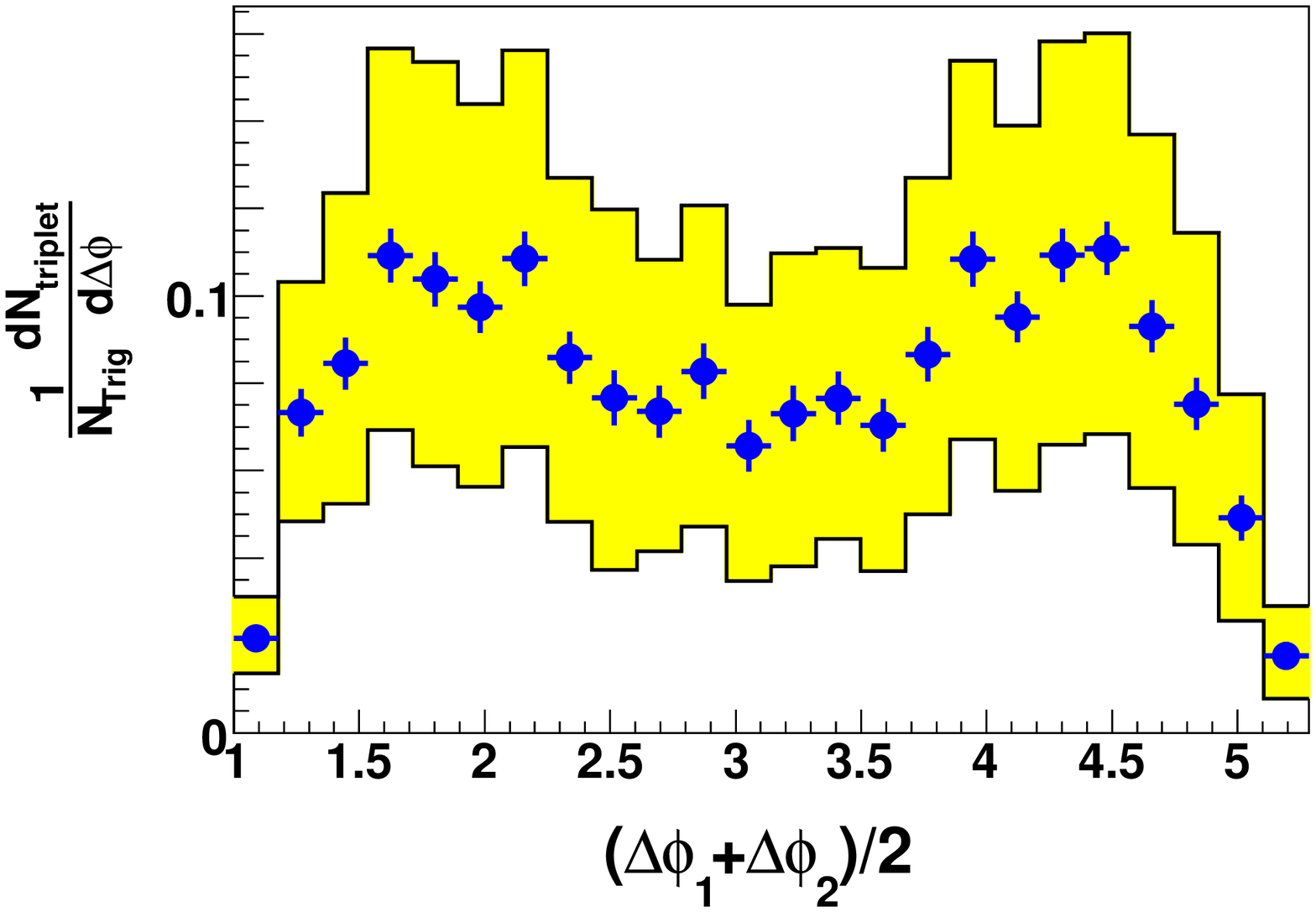}
\end{minipage}
\hfill
\begin{minipage}{0.32\textwidth}
\centering
\includegraphics[width=1.0\textwidth]{Plots/ProjStripBlank.eps}
\includegraphics[width=1.0\textwidth]{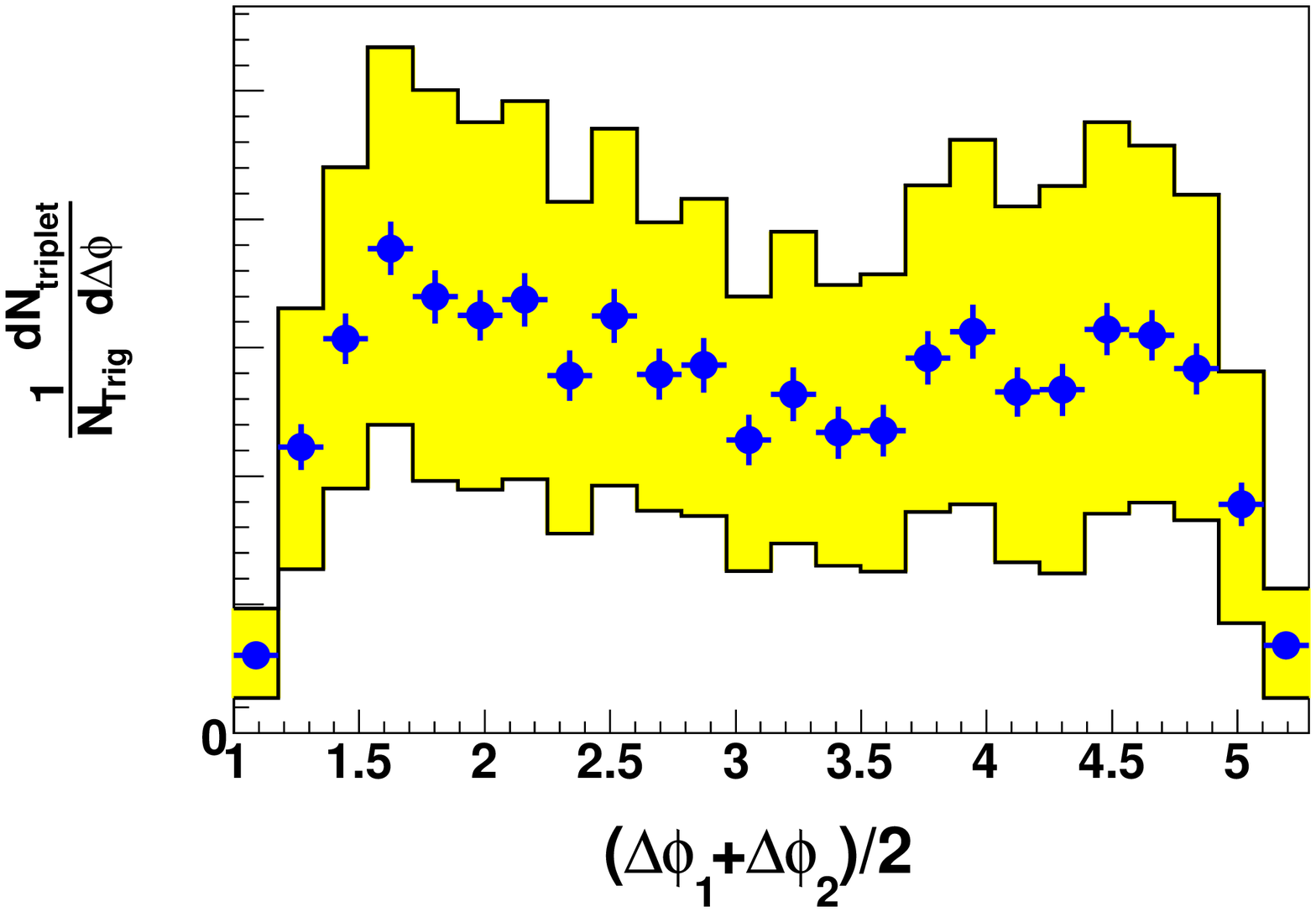}
\end{minipage}
\caption{Same as previous exept panels are from left to right and top to bottom, $0.5<p_{T}^{Assoc}<0.75$ GeV/c, $0.75<p_{T}^{Assoc}<1.0$ GeV/c, $1.0<p_{T}^{Assoc}<1.5$ GeV/c, $1.5<p_{T}^{Assoc}<2.0$ GeV/c, and $2.0<p_{T}^{Assoc}<3.0$ GeV/c for 0-12\% ZDC triggered Au+Au.}
\label{fig:projon2}
\end{figure}

Figures~\ref{fig:projon} and ~\ref{fig:projon2} show the on-diagonal projections of the away-side of the background subtracted 3-particle correlations.  The on-diagonal projections are shown here with systematic errors.  

\begin{figure}[H]
\hfill
\begin{minipage}[t]{0.32\textwidth}
\centering
\includegraphics[width=1.0\textwidth]{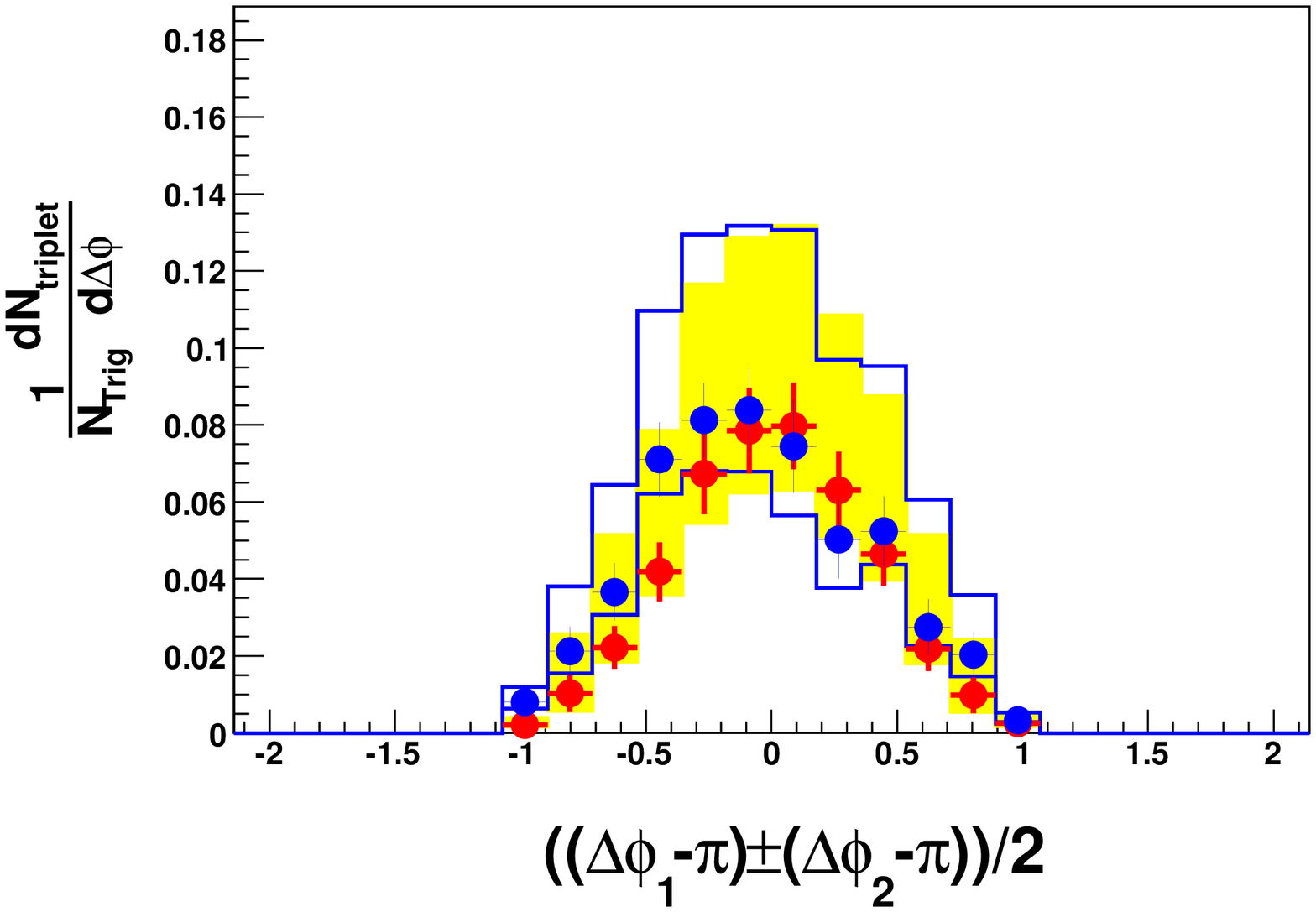}
\includegraphics[width=1.0\textwidth]{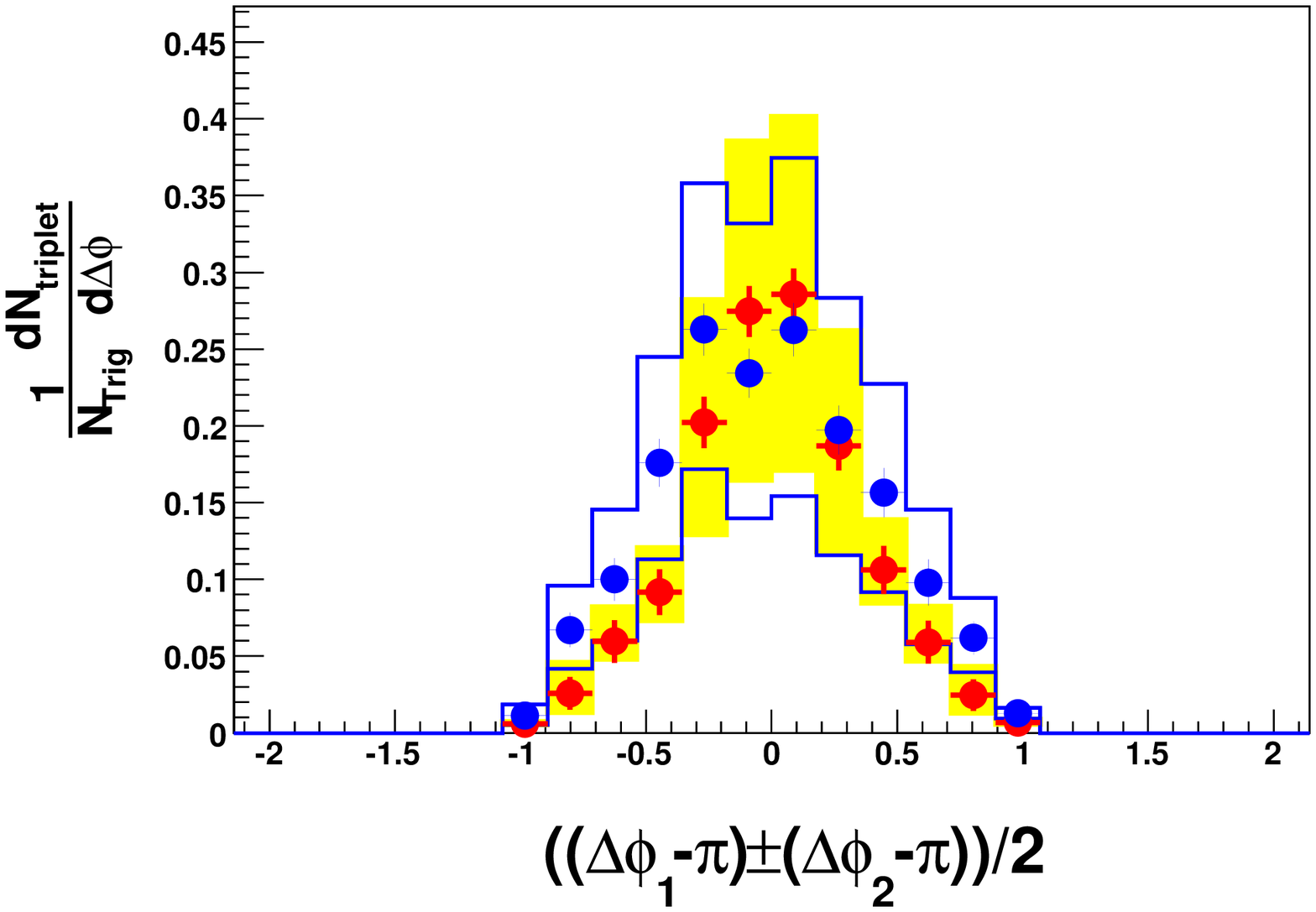}
\includegraphics[width=1.0\textwidth]{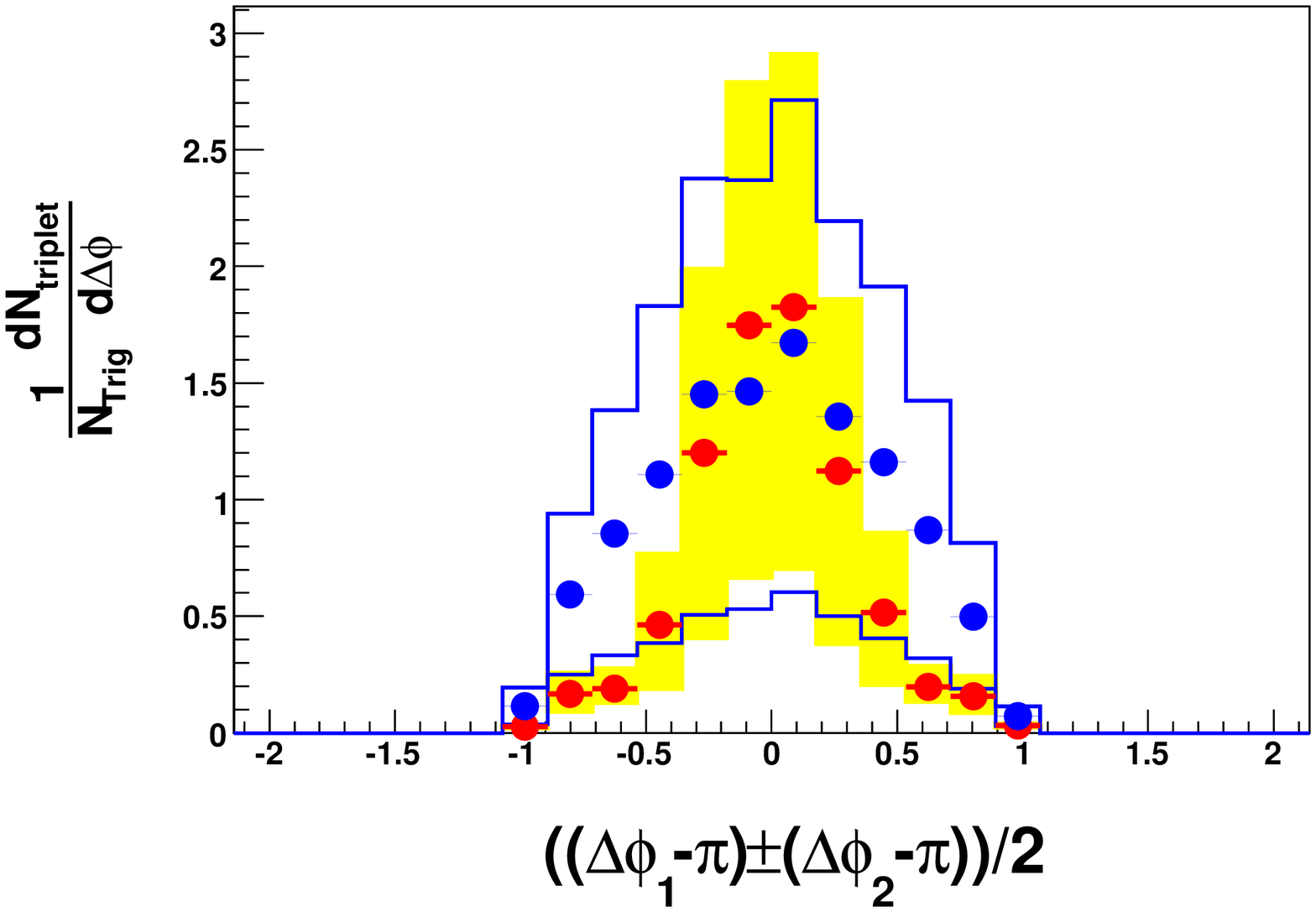}
\end{minipage}
\hfill
\begin{minipage}[t]{0.32\textwidth}
\includegraphics[width=1.0\textwidth]{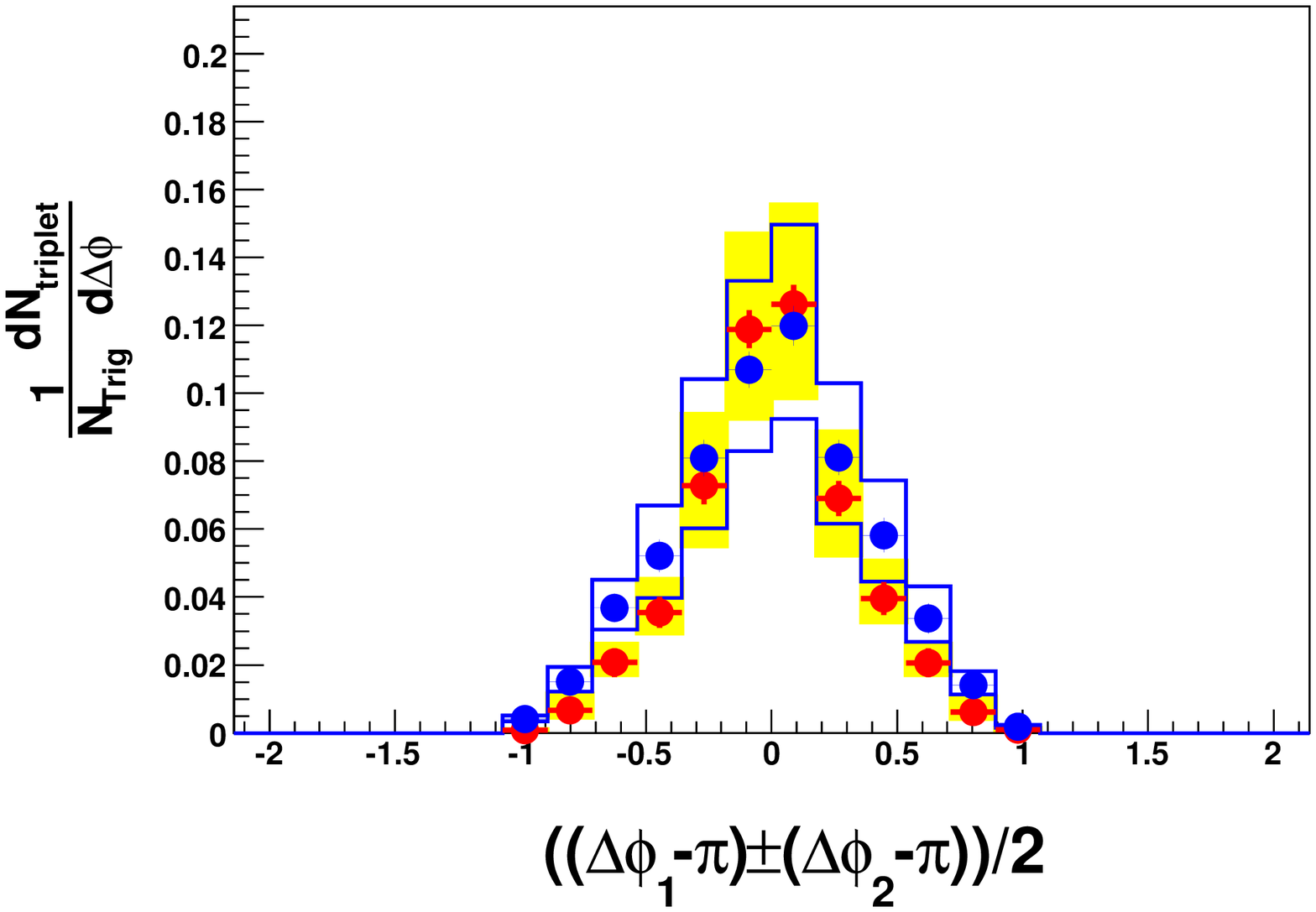}
\includegraphics[width=1.0\textwidth]{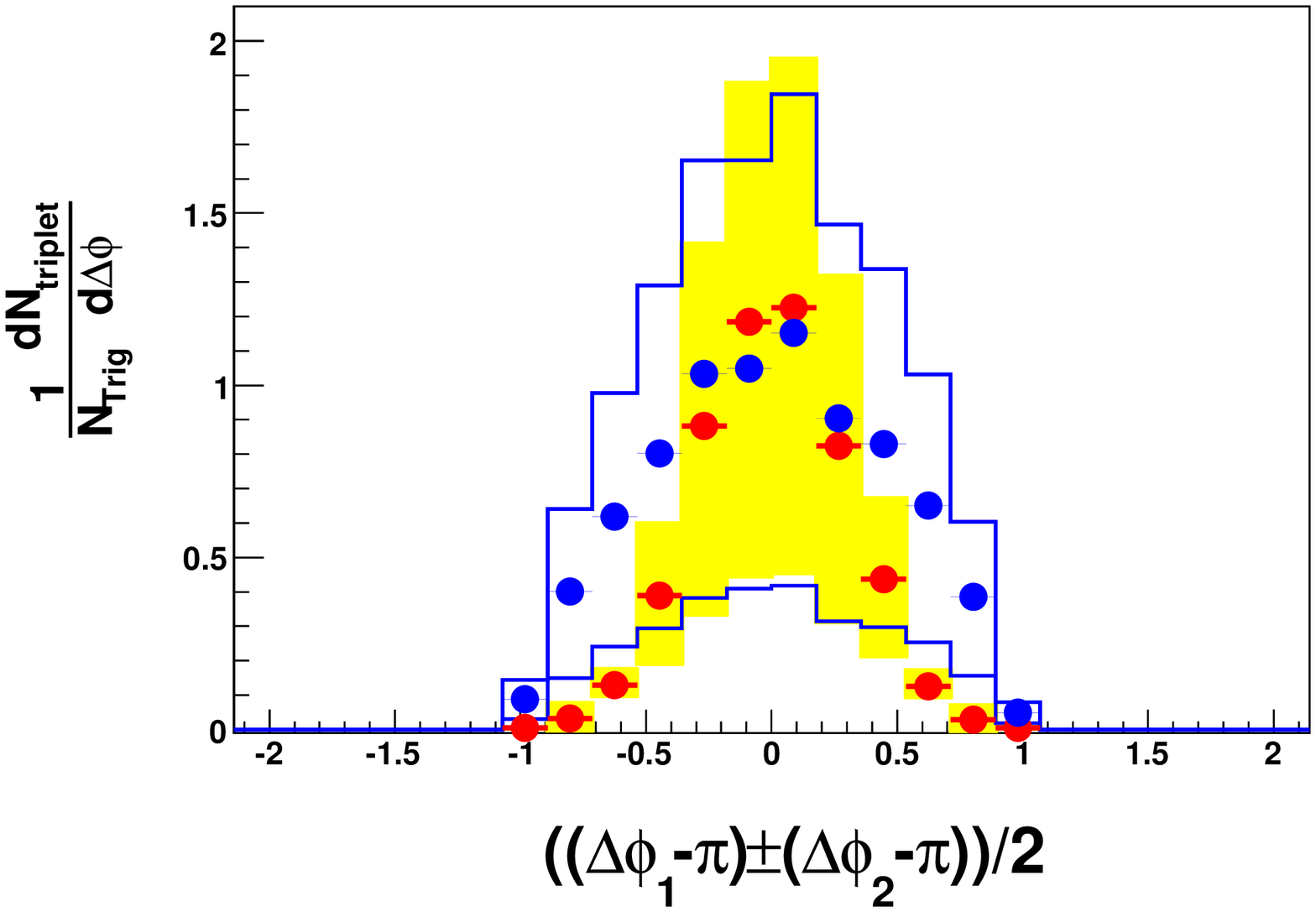}
\includegraphics[width=1.0\textwidth]{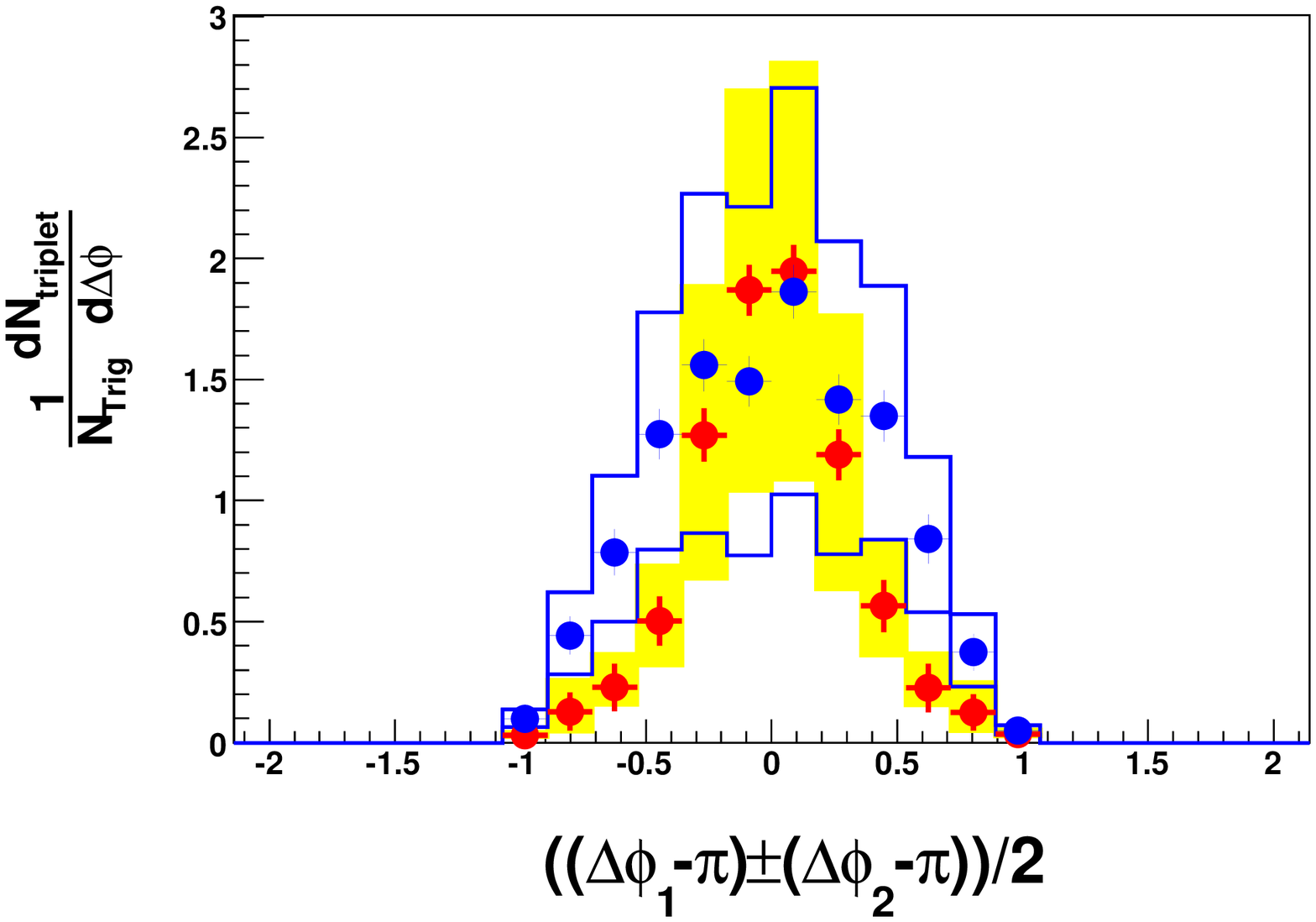}
\end{minipage}
\hfill
\begin{minipage}[t]{0.32\textwidth}
\includegraphics[width=1.0\textwidth]{Plots/ProjStripBlank.eps}
\includegraphics[width=1.0\textwidth]{Plots/ProjStripBlank.eps}
\includegraphics[width=1.0\textwidth]{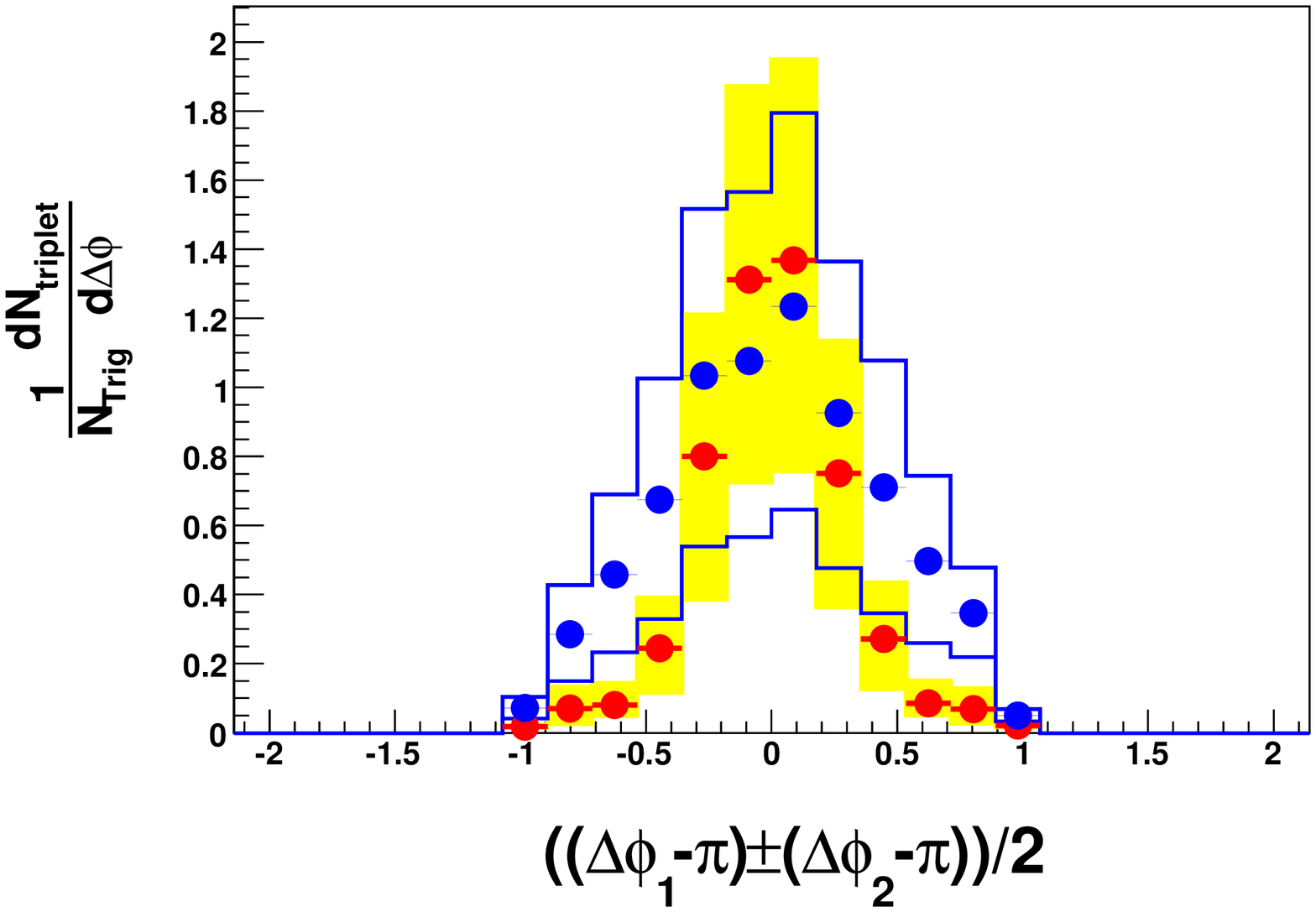}
\end{minipage}
\caption{On-diagonal projections (blue) and off-diagonal projection (red) of the background subtracted 3-particle correlations in strips of full width 0.7 radians on the away-side.  From left to right, top to bottom are {\it pp}, d+Au, Au+Au 50-80\%, Au+Au 30-50\%, Au+Au 10-30\%, Au+Au 0-10\% and ZDC triggered Au+Au 0-12\% collisions at $\sqrt{s_{NN}}=200$ GeV/c.  Systematic errors are shown with the blue histograms and yellow boxes for the on-diagaonl and off-diagonal projections, respectively.  Dashed black lines are at zero.}
\label{fig:projon3}
\end{figure}

\begin{figure}[htbp]    
\hfill
\begin{minipage}{0.32\textwidth}
\centering
\includegraphics[width=1.0\textwidth]{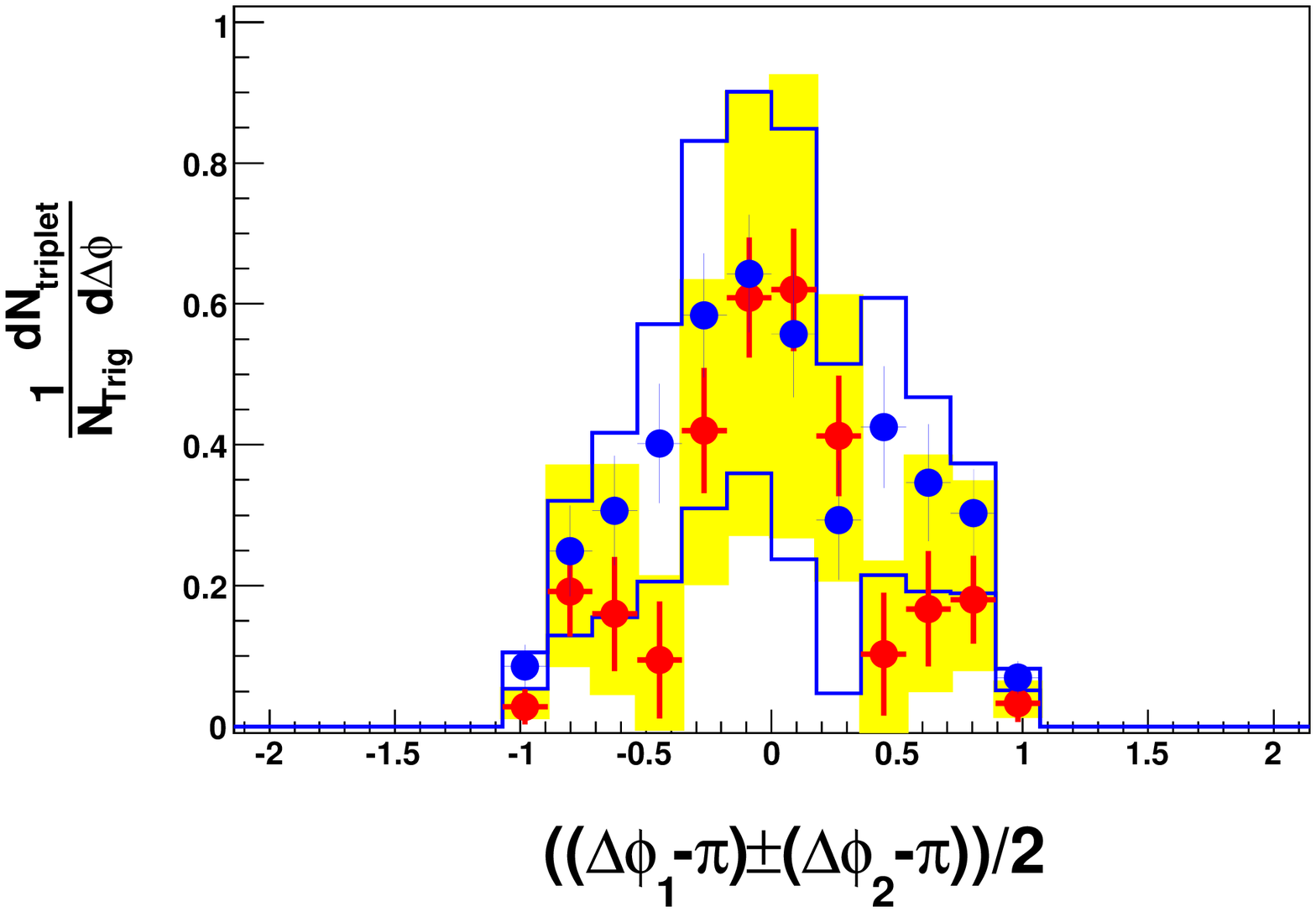}
\includegraphics[width=1.0\textwidth]{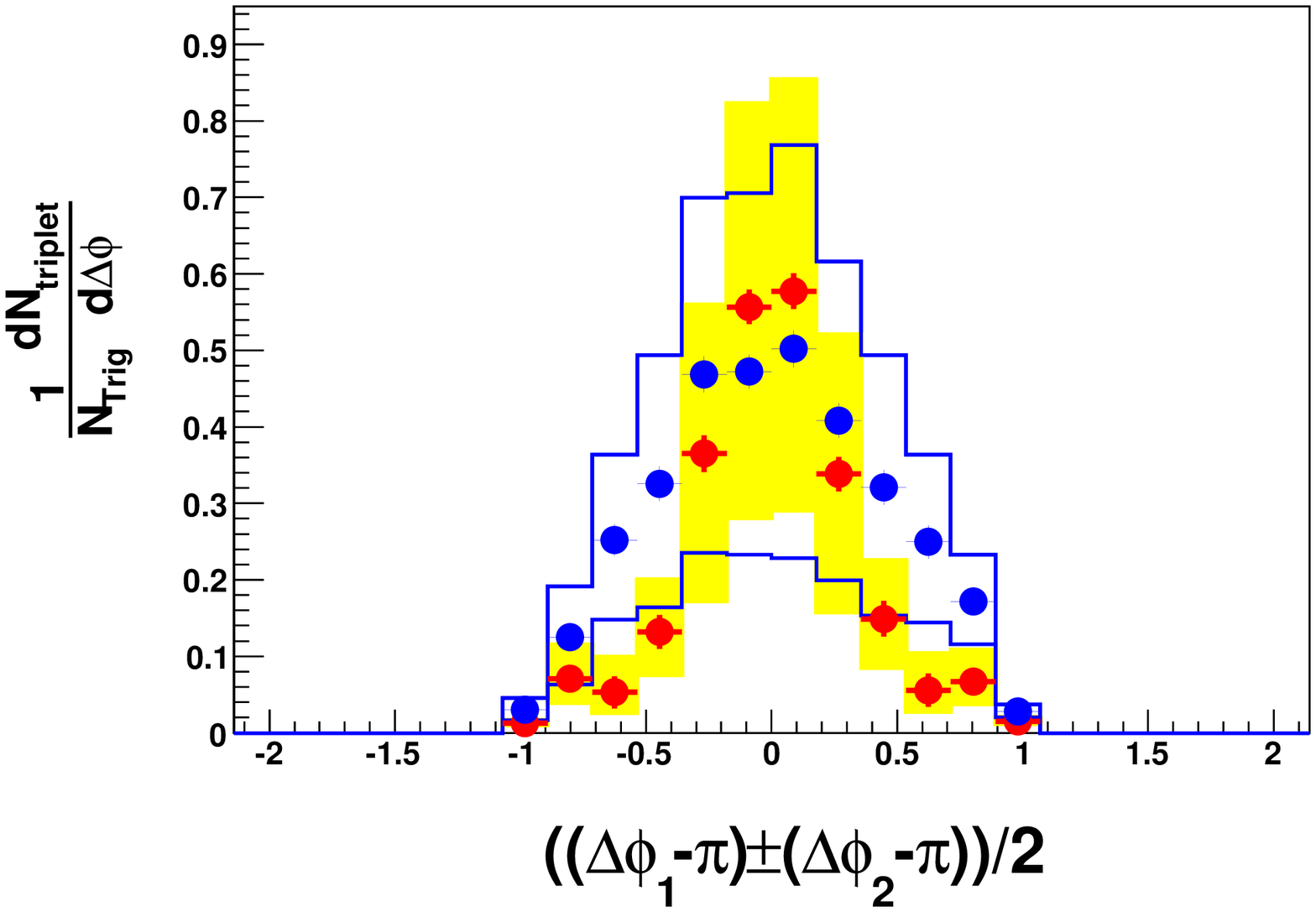}
\end{minipage}
\hfill
\begin{minipage}{0.32\textwidth}
\centering
\includegraphics[width=1.0\textwidth]{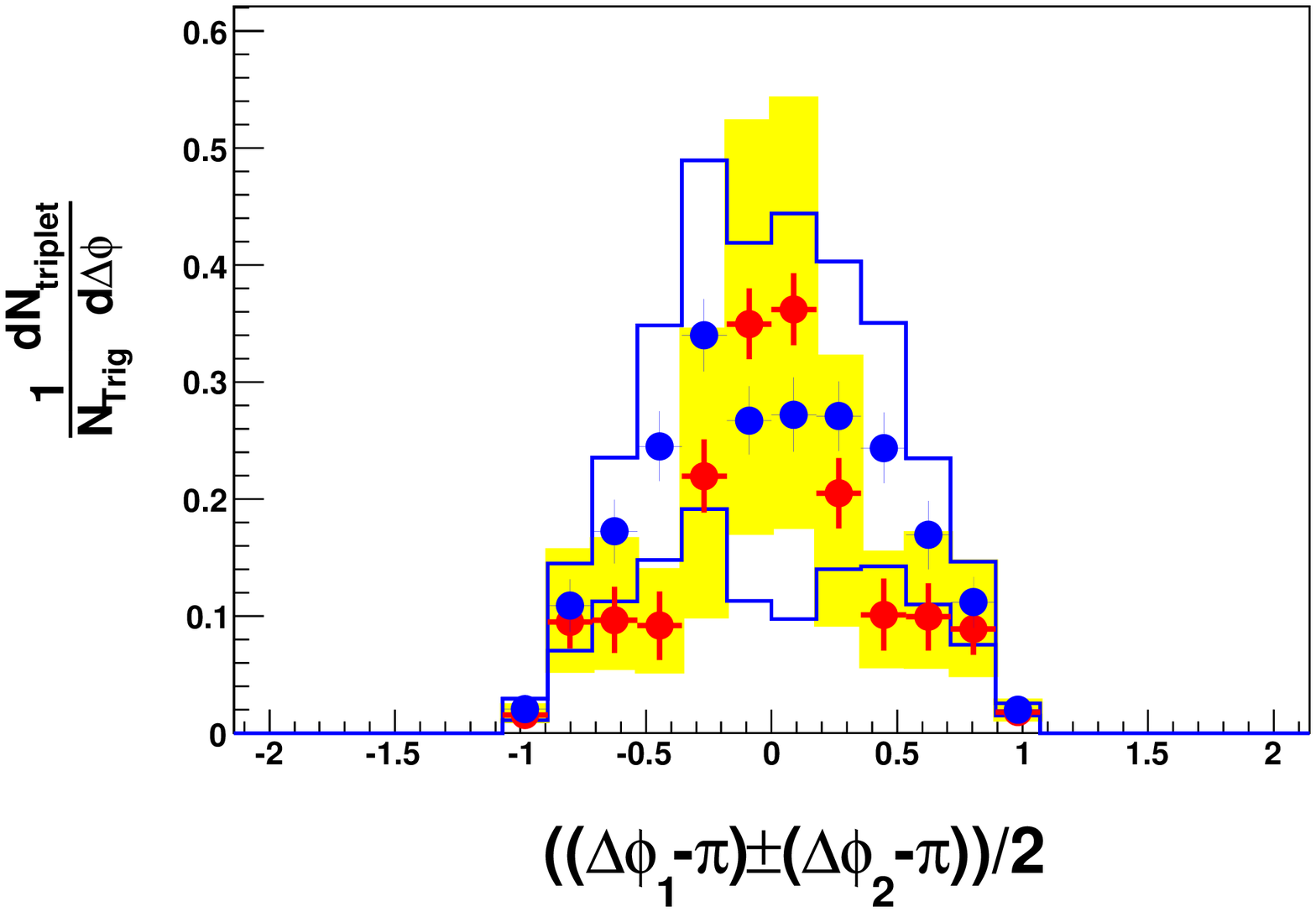}
\includegraphics[width=1.0\textwidth]{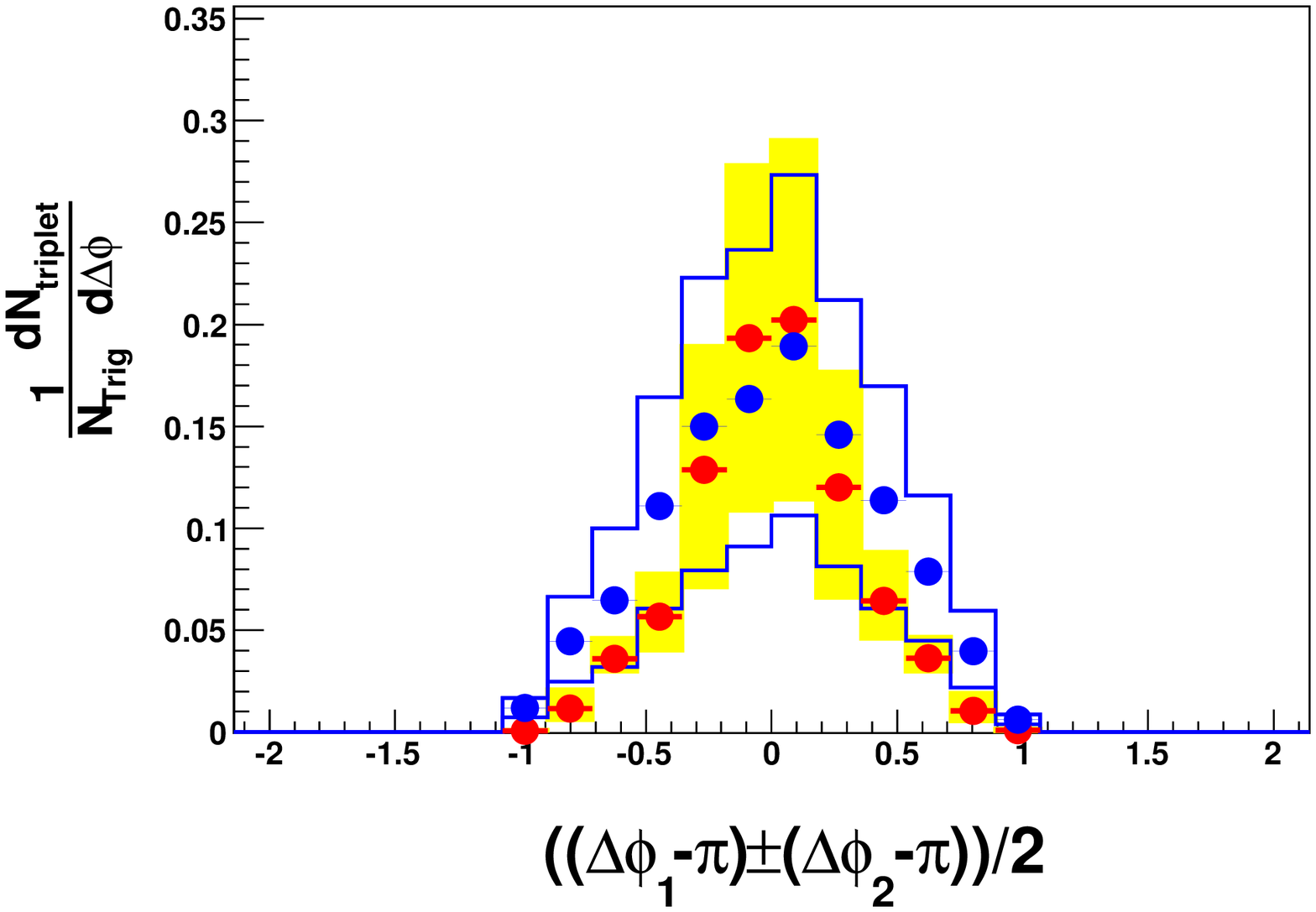}
\end{minipage}
\hfill
\begin{minipage}{0.32\textwidth}
\centering
\includegraphics[width=1.0\textwidth]{Plots/ProjStripBlank.eps}
\includegraphics[width=1.0\textwidth]{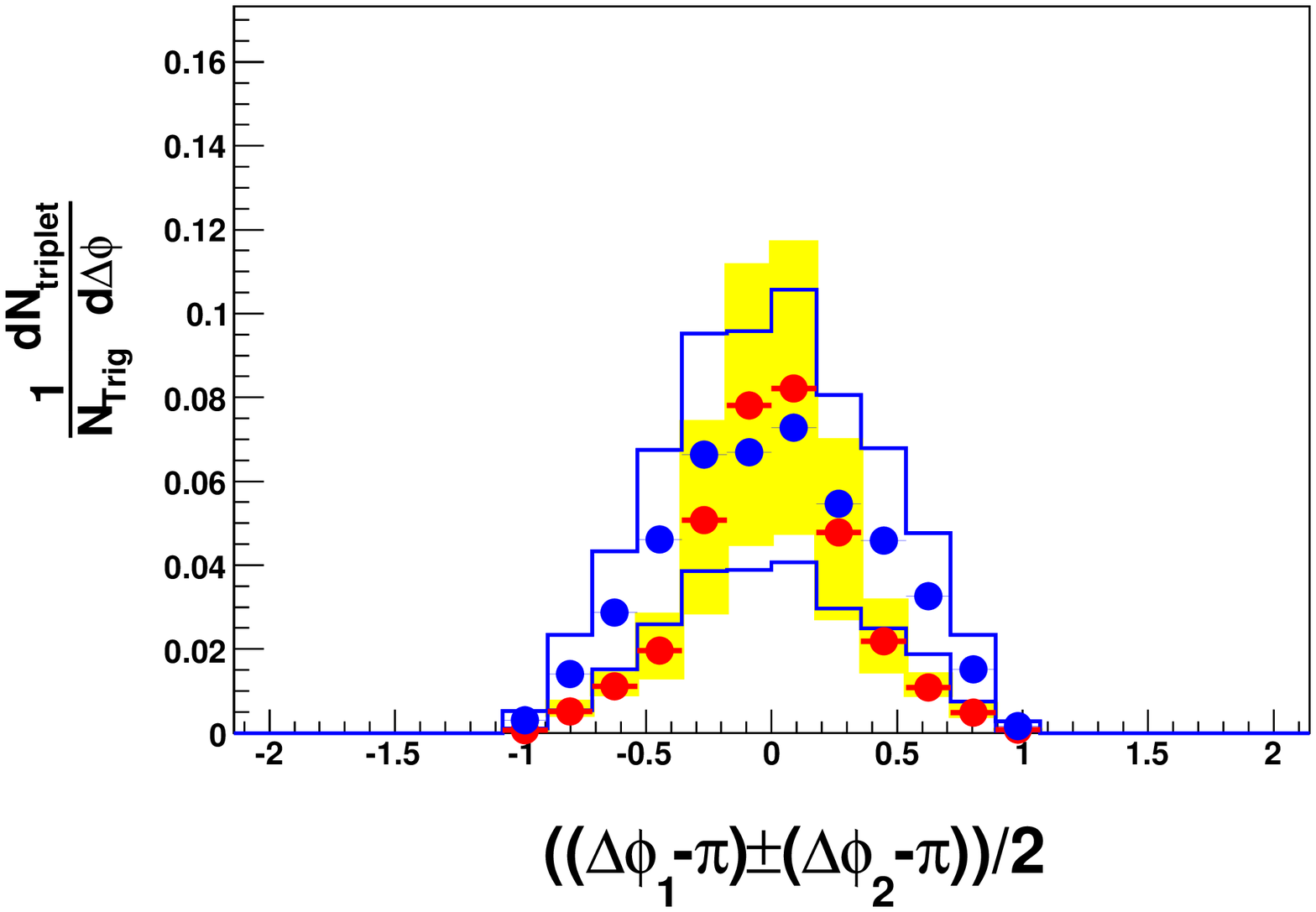}
\end{minipage}
\caption{Same as previous exept panels are from left to right and top to bottom, $0.5<p_{T}^{Assoc}<0.75$ GeV/c, $0.75<p_{T}^{Assoc}<1.0$ GeV/c, $1.0<p_{T}^{Assoc}<1.5$ GeV/c, $1.5<p_{T}^{Assoc}<2.0$ GeV/c, and $2.0<p_{T}^{Assoc}<3.0$ GeV/c for 0-12\% ZDC triggered Au+Au.}
\label{fig:projon4}
\end{figure}

Figures~\ref{fig:projon3} and ~\ref{fig:projon4} show the on-diagonal and off-diagonal projections of the near-side of the background subtracted 3-particle correlations.  The projections are shown here with systematic errors.  

\begin{figure}[H]
\centering
\includegraphics[width=0.55\textwidth]{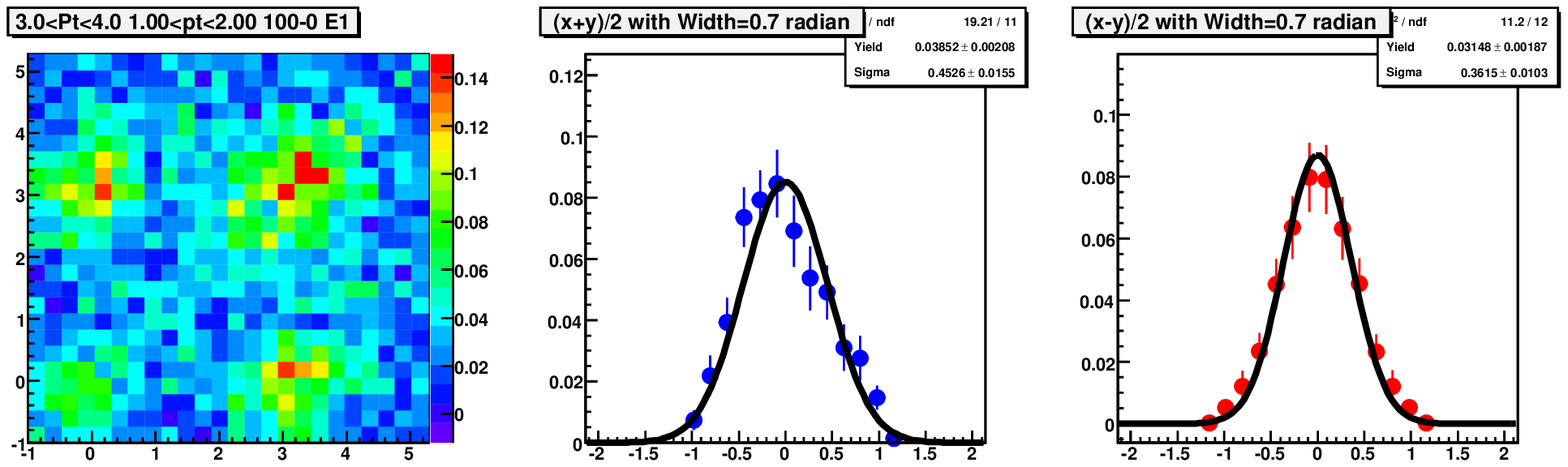}
\includegraphics[width=0.55\textwidth]{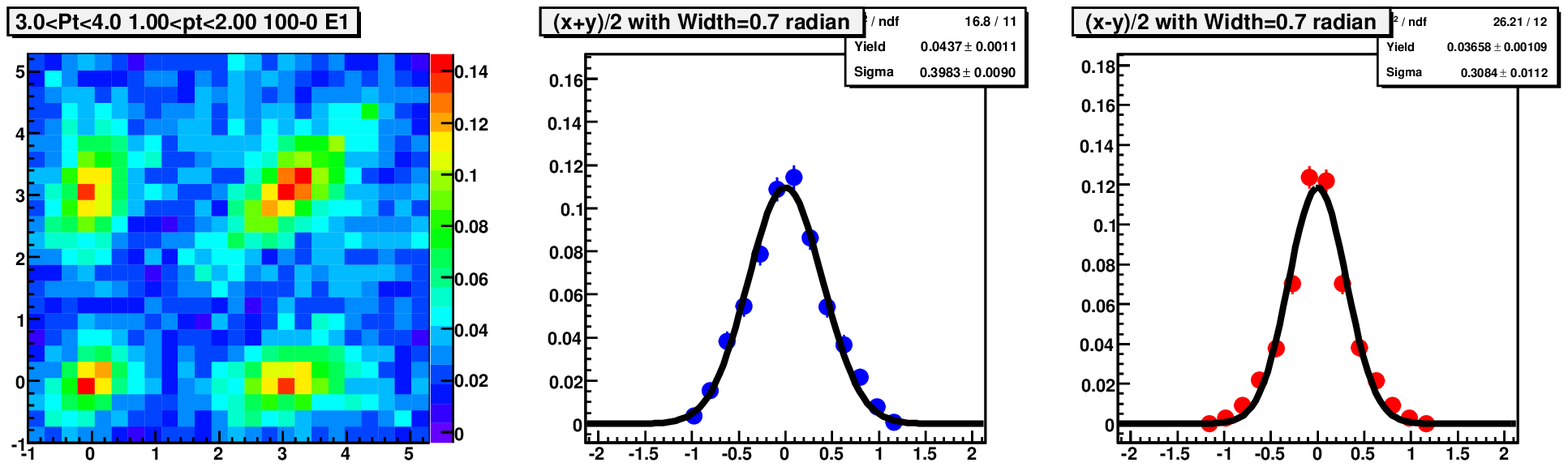}
\includegraphics[width=0.55\textwidth]{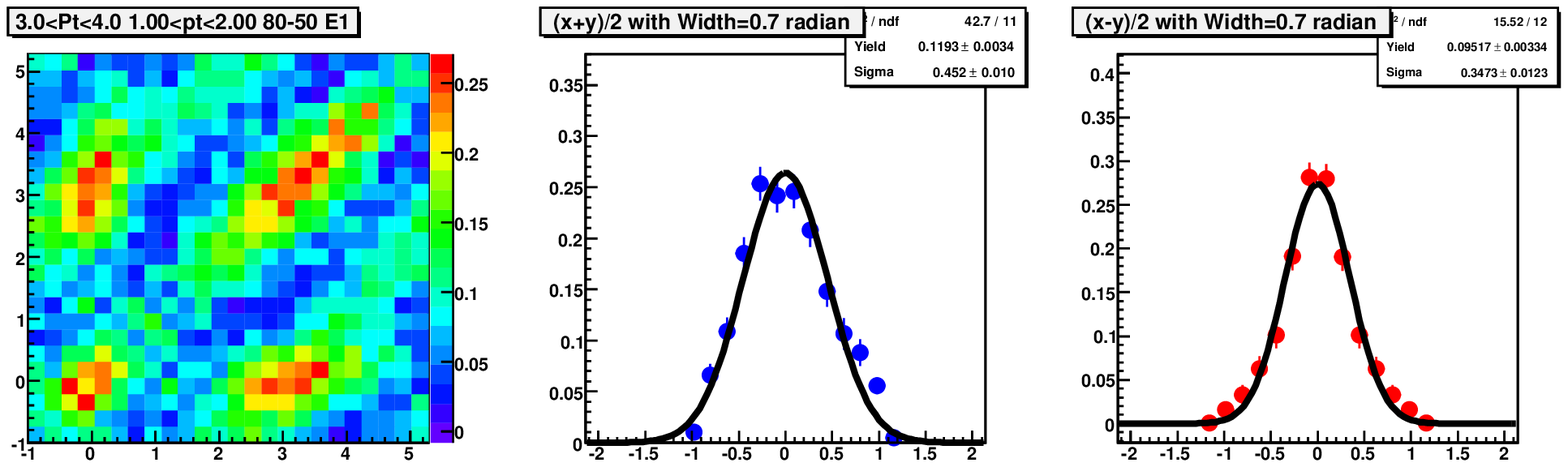}
\includegraphics[width=0.55\textwidth]{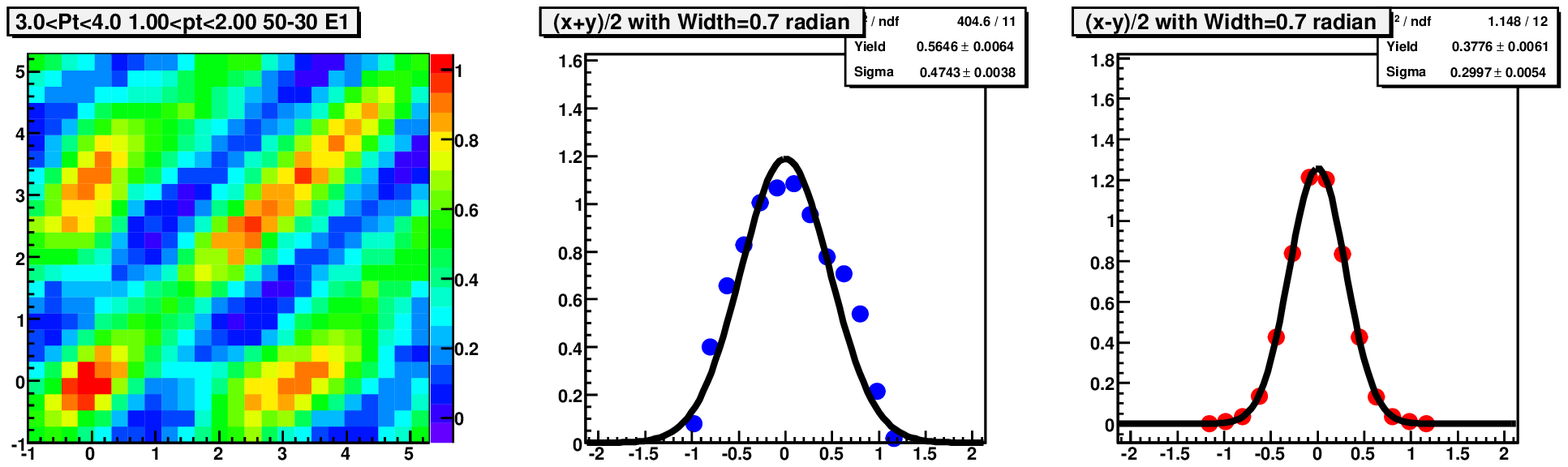}
\includegraphics[width=0.55\textwidth]{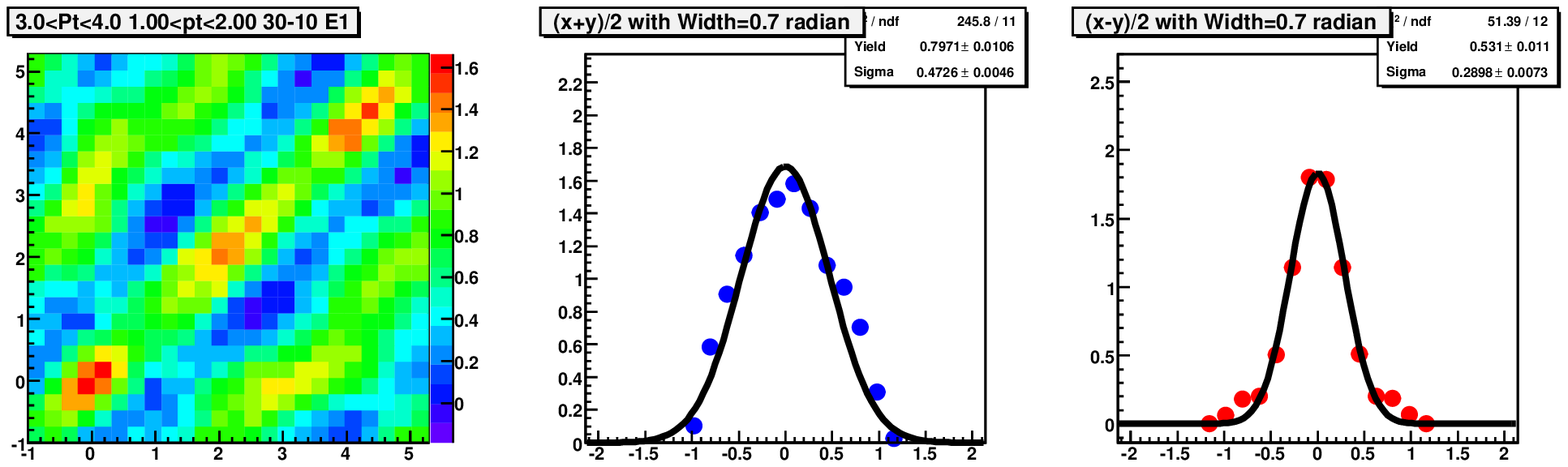}
\includegraphics[width=0.55\textwidth]{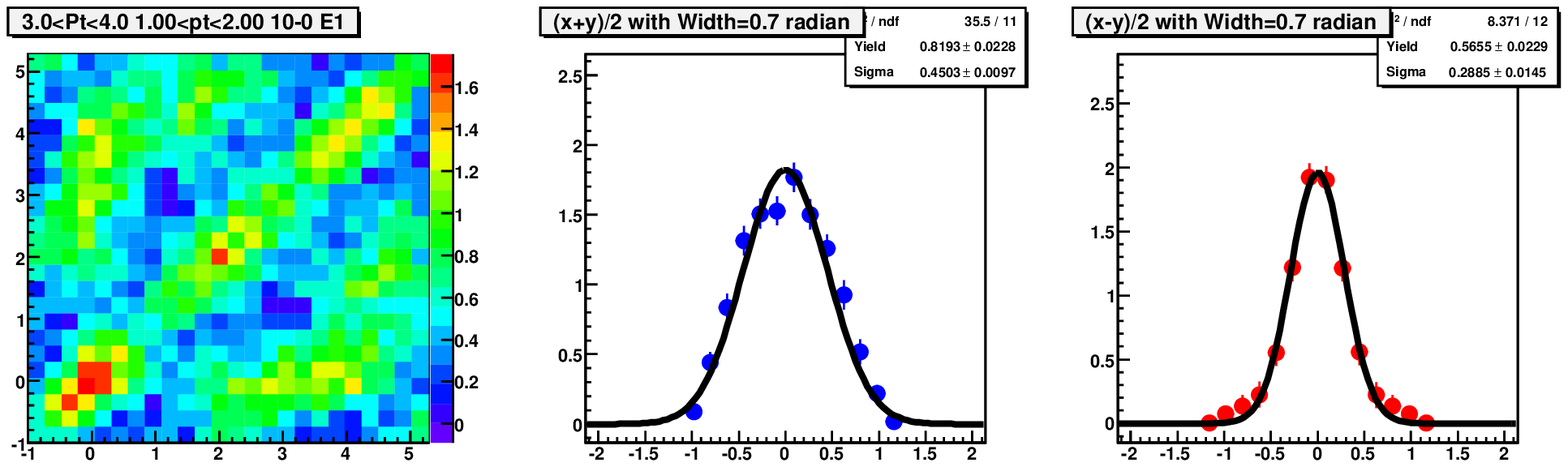}
\includegraphics[width=0.55\textwidth]{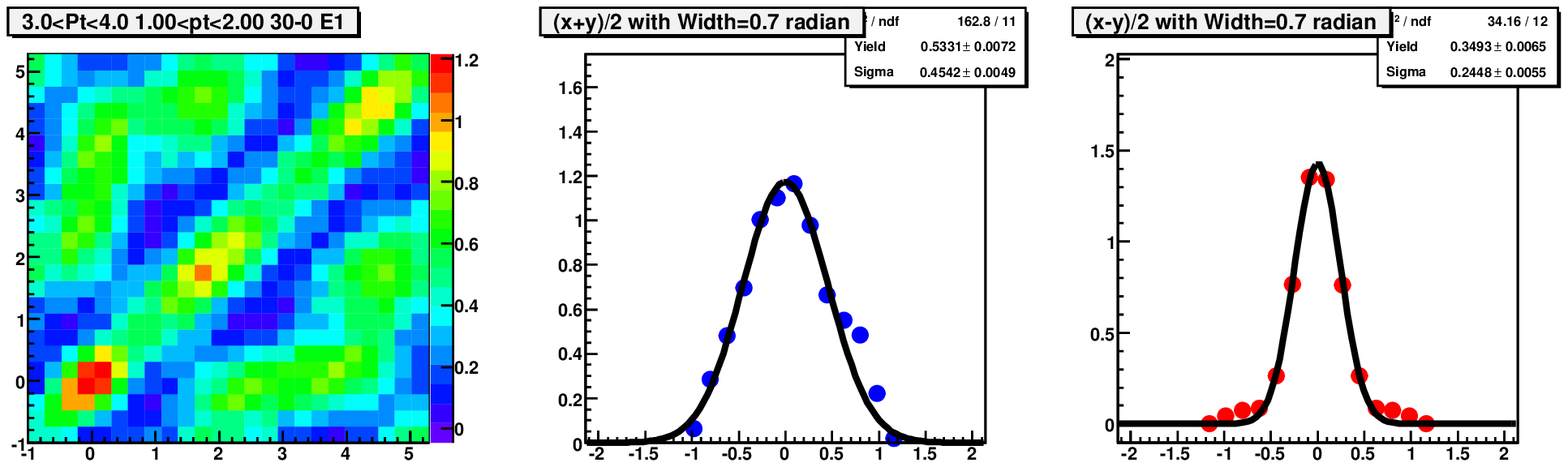}
\caption{Backgroud subtracted 3-particle correlations shown with on-diagonal (center) and off-diagonal (right) projections of the near-side peak.  The curve represent fits to a Gaussian centered at zero.   The rows are (from top to bottom) {\it pp}, d+Au, Au+Au 50-80\%, Au+Au 30-50\%, Au+Au 10-30\%, Au+Au 0-10\% and ZDC triggered Au+Au 0-12\% collisions at $\sqrt{s_{NN}}=200$ GeV/c.  The projections are of a strips of full width of 0.7 radians. Errors are statistical.}
\label{fig:projfitsnear}
\end{figure}

Figure~\ref{fig:projfitsnear} shows the on-diagonal and off-diagonal projections of the near-side peak.  The lines show fits to a single Gaussian centered at zero.  The parameters listed for the fits are (from top to bottom) $\chi^2$/ndf, Gaussian yield, and Gaussian width (in radians).

%
%
%
%

\begin{vita}

Jason Glyndwr Ulery was born in Elkhart, Indiana on April 15, 1978.  He received his Bachlor of Science Degree in Physics and Mathematics in May of 2001 from Manchester College.  He worked as an undergraduate physics labratory assistent duing his time at Manchester College from 1999-2001.  He also spent the summer of 2000 working on a low energy nuclear experiment as an REU summer researcher at Ball State University.  He served as a graduate student teaching assistent at Purdue University from 2001-2003.  He was employed as a gradute student research assistent from 2003-2007, working with the STAR experiment.  He received his Ph.D. in experimental high energy nuclear physics from Purdue University in December of 2007.  His thesis gives further understanding of jet-like correlations and contains the first 3-particle correlation results used to study conical emission.  It is titled ``Two- and Three-Particle Jet-Like Correlations''.

\end{vita}

\end{document}